\RequirePackage{lhchiggs}
\documentclass[letterpaper]{cernyrep}
\usepackage{rotating}
\usepackage{color}
\usepackage{hhline}
\usepackage{subfigure}
\usepackage{lscape}
\usepackage{floatpag}
\floatpagestyle{plain}
\usepackage{axodraw}
\usepackage{lineno}
\usepackage{lhchiggs,heppennames2,cernunits,hepparticles}
\providecommand{\HDECAY}{{\sc HDECAY}}
\providecommand{\HIGLU}{{\sc HIGLU}}
\providecommand{\Prophecy}{{\sc Prophecy4f}}
\providecommand{\CPsuperH}{{\sc CPsuperH}}
\providecommand{\FeynHiggs}{{\sc FeynHiggs}}

\providecommand{\lsim}
{\;\raisebox{-.3em}{$\stackrel{\displaystyle <}{\sim}$}\;}
\providecommand{\gsim}
{\;\raisebox{-.3em}{$\stackrel{\displaystyle >}{\sim}$}\;}
\providecommand{\orderx}[1]{\ensuremath{{\cal O}(#1)}}

\DeclareRobustCommand{\PA}{\HepParticle{A}{}{}\Xspace}
\DeclareRobustCommand{\PV}{\HepParticle{V}{}{}\Xspace}
\DeclareRobustCommand{\PX}{\HepParticle{X}{}{}\Xspace}
\DeclareRobustCommand{\Pf}{\HepParticle{f}{}{}\Xspace}
\DeclareRobustCommand{\PAf}{\HepAntiParticle{\Pf}{}{}\Xspace}
\DeclareRobustCommand{\PF}{\HepParticle{F}{}{}\Xspace}

\newcommand{\OS}{\mathrm{OS}}

\newcommand{\ssF}{{\mathrm{F}}}
\newcommand{\ssR}{{\mathrm{R}}}
\newcommand{\ssD}{{\mathrm{D}}}

\newcommand{\VV}{\mathrm{VV}}

\newcommand{\bqas}{\begin{eqnarray*}}
\newcommand{\eqas}{\end{eqnarray*}}

\newcommand{\bq}{\begin{equation}}                    % equationing
\newcommand{\eq}{\end{equation}}
\newcommand{\bqa}{\arraycolsep 0.14em\begin{eqnarray}}
\newcommand{\eqa}{\end{eqnarray}}
\newcommand{\ba}[1]{\begin{array}{#1}}
\newcommand{\ea}{\end{array}}
\newcommand{\ben}{\begin{enumerate}}
\newcommand{\een}{\end{enumerate}}
\newcommand{\bei}{\begin{itemize}}
\newcommand{\eei}{\end{itemize}}
\newcommand{\eqn}[1]{Eq.(\ref{#1})}
\newcommand{\eqns}[2]{Eqs.(\ref{#1})--(\ref{#2})}
\newcommand{\eqnss}[1]{Eqs.(\ref{#1})}
\newcommand{\eqnsc}[2]{Eqs.(\ref{#1}) and (\ref{#2})}

\newcommand{\bmid}{\Bigr|}

\newcommand{\mh}{\mathswitch {M_{\PH}}}

\newcommand{\mt}{\mathswitch {M_{\PQt}}}

\newcommand{\muh}{\mathswitch {\mu_{\PH}}}

\newcommand{\muhs}{\mathswitch {\mu^2_{\PH}}}
\newcommand{\gh}{\mathswitch {\gamma_{\PH}}}

\newcommand{\muR}{\mathswitch {\mu_{\ssR}}}
\newcommand{\muF}{\mathswitch {\mu_{\ssF}}}

\newcommand{\tot}{{\mbox{\scriptsize tot}}}

\newcommand{\ep}{\mathswitch \varepsilon}

\providecommand\POWHEG{{\sc POWHEG}}

\providecommand{\NLOacc}{N\scalebox{0.8}{LO}\xspace}

\providecommand{\MEPSatNLO}{M\scalebox{0.8}{E}P\scalebox{0.8}{S}@N\scalebox{0.8}{LO}\xspace}
\providecommand{\MCatNLO}{M\scalebox{0.8}{C}@N\scalebox{0.8}{LO}\xspace}
\providecommand{\Sherpa}{S\scalebox{0.8}{HERPA}\xspace}
\providecommand{\Gosam}{G\scalebox{0.8}{OSAM}\xspace}
\providecommand{\samurai}{S\scalebox{0.8}{AMURAI}\xspace}
\providecommand{\OpenLoops}{O\scalebox{0.8}{PEN}L\scalebox{0.8}{OOPS}\xspace}
\providecommand{\Rivet}{R\scalebox{0.8}{IVET}\xspace}
\providecommand{\ATLASexp}{{\sc ATLAS}\xspace}
\providecommand{\CMSexp}{{\sc CMS}\xspace}
\newcommand{\Collier}{C\scalebox{0.8}{OLLIER}\xspace}

\newcommand{\flfs}{\mu^+\nu_\mu \Pe^-\bar\nu_{\Pe}}
\newcommand{\rF}{\mathrm{F}}
\newcommand{\delres}{\Delta_\mathrm{res}}
\newcommand{\delqcd}{\Delta_\mathrm{QCD}}
\newcommand{\sigsr}{\sigma_\mathrm{S2s}}
\newcommand{\sigcr}{\sigma_\mathrm{S2c}}
\newcommand{\HEJ}{HEJ\Xspace}
\newcommand{\aMCatNLO}{aM\protect{C}@N\protect{LO}\Xspace}
\newcommand{\PowhegBox}{P\protect{OWHEG}B\protect{OX}\Xspace}
\newcommand{\MCFM}{M\protect{CFM}\Xspace}
\newcommand{\tmop}[1]{\ensuremath{\operatorname{#1}}}
\newcommand{\tmtexttt}[1]{{\ttfamily{#1}}}
\newcommand{\PythiaSix}{P\protect{YTHIA}6\xspace}
\newcommand{\NLOPS}{N{LO}P{S}\xspace}
\newcommand{\GoSam}{G\protect{O}S\protect{AM}\xspace}
\newcommand{\MEPS}{M{E}P{S}\xspace}
\newcommand{\ssHG}{{\mathrm{HG}}}
\newcommand{\diste}[1]{\frac{d\sigma^{#1}_{\mathrm{eff}}}{d x}} 
\newcommand{\dist}[1]{\frac{d\sigma^{#1}}{d x}} 
\DeclareRobustCommand{\PVB}{\HepParticle{V}{}{}\Xspace} 
\DeclareRobustCommand{\PVBL}{\HepParticle{V}{L}{}\Xspace} 
\DeclareRobustCommand{\PCF}{\HepParticle{F}{}{}\Xspace} 
\DeclareRobustCommand{\Pho}{\HepParticle{\Ph}{1}{}\Xspace}
\DeclareRobustCommand{\Pht}{\HepParticle{\Ph}{2}{}\Xspace}

\newcommand{\mT}{\ensuremath{m_{\rm T}}}

\def\ca{{\cal A}}
\def\sh{\hat{s}}
\def\Re{{\rm Re}}
\def\Im{{\rm Im}}

\providecommand{\sla}[1]{\ifmmode%
  \setbox0=\hbox{$#1$}%
  \setbox1=\hbox to\wd0{\hss$/$\hss}\else%
  \setbox0=\hbox{#1}%
  \setbox1=\hbox to\wd0{\hss/\hss}\fi%
  #1\hskip-\wd0\box1 }

\begin{document}

% --- line numbers and draft heading
%\linenumbers
%\draftheading
  
%%% from here take the YR front page:

\thispagestyle{empty}
{
\setlength{\unitlength}{1mm}
\begin{picture}(0.001,0.001)
%%\graphpaper(0,-280)(210,290)
%%\put(80,-15){\makebox(0,0){\LARGE\itshape Final draft version \today}}
\put(135,8){CERN--2013--004}
\put(135,3){29 July 2013}
%----
%%\put(-5,-40){\includegraphics[width=15cm]{CERNtitl.eps}}
\put(0,-40){\large\bf ORGANISATION EUROP\'EENNE POUR LA RECHERCHE NUCL\'EAIRE}
\put(0,-50){\huge\bf CERN}
\put(23.35,-50){\large\bf EUROPEAN ORGANIZATION FOR NUCLEAR RESEARCH}

%---
\put(18,-100){\LARGE\bfseries
            Handbook of LHC Higgs cross sections:}
\put(48,-110){\LARGE\bfseries
             3. Higgs Properties}
\put(9,-145){\Large\bfseries
             Report of the LHC Higgs Cross Section Working Group}
\put(117,-185){\Large Editors:}
\put(135,-185){\Large S.~Heinemeyer}
\put(135,-191){\Large C.~Mariotti}
\put(135,-197){\Large G.~Passarino}
\put(135,-203){\Large R.~Tanaka}

\put(70,-250){\makebox(0,0){GENEVA}}
\put(70,-255){\makebox(0,0){2013}}
\end{picture}
}

% --- print draft version
%\begin{flushright}
%   Draft version 2.0 \\
%\end{flushright}

\pagenumbering{roman}

%\maketitle % this produces the title block

%\vfill

% ------------------------------------

\newpage

\leftline{\bf Conveners}

%\begin{enumerate}

%\item 
\noindent \emph{Gluon-fusion process:}
         Y.~Gao, \,
         D.~de Florian, \,
         B.~di Micco, \,
         K.~Melnikov,\,  	 
         F.~Petriello
\vspace{0.1cm}

%\item 
\noindent \emph{Vector-boson-fusion process:} 	
         A.~Denner,\, 	
         P.~Govoni,\,
         C.~Oleari,\,
         D.~Rebuzzi 
\vspace{0.1cm}

%\item 
\noindent{$\PW\PH/\PZ\PH$ \emph{production mode}:} 	
         S.~Dittmaier,\, 	
         G.~Ferrera,\,
%         C.~Matteuzzi,\, 	
         J.~Olsen,\, 	
         G.~Piacquadio,\,
         A.~Rizzi
\vspace{0.1cm}

%\item 
\noindent{$\PQt\PQt\PH$ \emph{process}:} 	
         C.~Neu,\, 	  	
         C.~Potter,\, 	
         L.~Reina,\,
         A.~Rizzi, \, 	
         M.~Spira
\vspace{0.1cm}

%\item 
\noindent\emph{Light Mass Higgs}
         A.~David, \,
         A.~Denner, \, 
         M.~D\"uhrssen,\, 
         C.~Grojean, \, 
         M.~Grazzini,\,
         K.~Prokofiev, \,
         M.~Schumacher, \,
         G.~Weiglein, \,
         M.~Zanetti 
\vspace{0.1cm}
         
%\item 
\noindent\emph{MSSM Higgs:} 	
         M.~Flechl,\, 	
         M.~Kr\"amer,\, 	
         S.~Lehti,\,
         R.~Harlander,\,
         P.~Slavich,\,
         M.~Spira,\, 	
         M.~Vazquez Acosta,\,	  	
         T.~Vickey
\vspace{0.1cm}
         
% item
\noindent\emph{Heavy Higgs and BSM }
         S.~Bolognesi,\,
         S.~Diglio,\,
         C.~Grojean,\,
         M.~Kadastik, \,
         H.~E.~Logan, \,
         M.~M\"uhlleitner,\,
         K.~Peters
\vspace{0.1cm}

%\item 
\noindent\emph{Branching ratios:} 	
         S.~Heinemeyer,\,
         A.~M\"uck,\,
         I.~Puljak,\, 	  	
         D.~Rebuzzi
\vspace{0.1cm}

%\item 
\noindent\emph{Jets:} 	
         B.~Mellado,\,
         D.~Del Re,\,
         G.~P.~Salam,\,
         F.~Tackmann
         \vspace{0.1cm}
         
%\item 
\noindent\emph{NLO MC:}
%         S.~Frixione,\,
         F.~Krauss,\,
         F.~Maltoni,\, 	
         P.~Nason
\vspace{0.1cm}

%\item 
\noindent\emph{PDF:} 	
         S.~Forte,\, 	
         J.~Huston,\, 	
         R.~Thorne
\vspace{0.1cm}

%\end{enumerate}
\vfill
\begin{flushleft}%\large
\begin{tabular}{@{}l@{~}l}
ISBN & 978--92--9083--389--5 \\%<--- Change
ISSN & 0007--8328 \\ % Non-CERN Physics School Publication
DOI  & 10.5170/CERN--2013--004 \\
\end{tabular}\\[1mm]
Copyright \copyright{} CERN, 2013\\[1mm]
%\raisebox{-1mm}{\includegraphics[height=12pt]{cc.pdf}}
Creative Commons Attribution 3.0\\[1mm]
Knowledge transfer is an integral part of CERN's mission.\\[1mm]
CERN publishes this report Open Access under the Creative Commons
Attribution 3.0 license (\texttt{http://creativecommons.org/licenses/by/3.0/})
in order to permit its wide dissemination and use.\\[3mm]

This Report should be cited as:\\[1mm]
LHC Higgs Cross Section Working Group,
S.~Heinemeier, C.~Mariotti, G.~Passarino, R.~Tanaka (Eds.), \\
\emph{Handbook of LHC Higgs Cross Sections: 3. Higgs Properties}, \\
CERN--2013--004 (CERN, Geneva, 2013), DOI: 10.5170/CERN--2013--004\\[3mm]
\end{flushleft}

\newpage
\vspace*{10cm}
\begin{center} 
 {\bf Abstract}
\end{center}
\vspace{0.5cm}
This Report summarizes the results of the activities in 2012 and the first
half of 2013 of the LHC Higgs Cross Section Working Group.
The main goal of the working group was to present the state of the art of
Higgs Physics at the LHC, integrating all new results that have appeared in the
last few years.
This report follows the first working group report
{\it Handbook of LHC Higgs Cross Sections: 1.~Inclusive Observables} 
(CERN-2011-002) and the second working group report 
{\it Handbook of LHC Higgs Cross Sections: 2.~Differential Distributions} 
(CERN-2012-002). 
After the discovery of a Higgs boson at the LHC in mid-2012 this report
focuses on refined prediction of Standard Model (SM) Higgs phenomenology
around the experimentally observed value of $125-126 \UGeV$, refined
predictions for heavy SM-like Higgs bosons as well as predictions in the
Minimal Supersymmetric Standard Model and first steps to go beyond these
models.  
The other main focus is on the extraction of the characteristics and
properties of the newly 
discovered particle such as couplings to SM particles, spin and CP-quantum
numbers etc.

\newpage
\vspace*{10cm}
\begin{center}
We, the authors, would like to dedicate this Report to the memory of
\\
Ken Wilson.
\end{center}

%\newpage
%\mbox{}

\newpage
%\begin{center}
%\textsc{Men at work, please be patient!}
%\end{center}

\begin{flushleft}

 S.~Heinemeyer$^{1}$,\,   
 C.~Mariotti$^{2}$,\, 
 G.~Passarino$^{2,3}$\,and\,
 R.~Tanaka$^{4}$\,(eds.);\\
%- 
 J.~R.~Andersen$^{5}$,\,
 P.~Artoisenet$^{6}$,\,
%--
 E.~A.~Bagnaschi$^{7}$,\,
 A.~Banfi$^{8}$,\,
 T.~Becher$^{9}$,\,
 F.~U.~Bernlochner$^{10}$
% ...

\end{flushleft}
 
\begin{itemize}

\item[$^{1}$]  Instituto de F\'isica de Cantabria (IFCA), CSIC-Universidad de Cantabria, Santander, Spain

\item[$^{2}$]  INFN, Sezione di Torino, Via P. Giuria 1, 10125 Torino, Italy

\item[$^{3}$]  Dipartimento di Fisica Teorica, Universit\`a di Torino, Via P. Giuria 1, 10125 Torino, Italy

\item[$^{4}$]  Laboratoire de l'Acc\'el\'erateur Lin\'eaire, CNRS/IN2P3, F-91898 Orsay CEDEX, France 

  %---

\item[$^{5}$]  Institute for Particle Physics Phenomenology, Department of Physics, University of Durham, \\ 
               Durham DH1 3LE, UK

\item[$^{6}$]  Nikhef Theory Group, Science Park 105, 1098 XG Amsterdam, The Netherlands

\item[$^{7}$]  Laboratoire de Physique Th\'eorique et Hautes Energies
  (LPTHE), CNRS et Universit\'e Paris 6,  4 Place Jussieu, F-75252 Paris CEDEX 05, France

\item[$^{8}$]  Physikalisches Institut, Albert-Ludwigs-Universit\"at Freiburg, D-79104 Freiburg, Germany

\item[$^{9}$]  Universit\"at Bern - Albert Einstein Center for Fundamental Physics, \\
               Institute for Theoretical Physics Sidlerstrasse 5, CH-3012 Bern, Switzerland

\item[$^{10}$] University of Victoria, Victoria, British Columbia, V8W 3P, Canada 

\item[$^{11}$] Johns Hopkins University, Baltimore, Maryland. 410-516-8000, USA

\item[$^{12}$] II. Institut f{\"u}r Theoretische Physik, Universit{\"a}t Hamburg, \\ Luruper Chaussee 149, D-22761 Hamburg, Germany

\item[$^{13}$] High Energy Physics Division, Argonne National Laboratory, Argonne, IL 60439, USA

\item[$^{14}$] Centre for Cosmology, Particle Physics and Phenomenology (CP3),\\  
               Universit\'e Catholique de Louvain, B-1348 Louvain-la-Neuve, Belgium

\item[$^{15}$] Theoretical Physics Department, Fermi National Accelerator Laboratory,\\ MS 106, Batavia, IL 60510-0500, USA

\item[$^{16}$] University of Chicago - Enrico Fermi Institute, \\ 5640 S. Ellis Avenue, RI-183, Chicago, IL 60637, USA

\item[$^{17}$] University of Chicago - Kavli Institute for Cosmological Physics, \\
               Lab. for Astrophys.  Space Res. (LASR), 933 East 56th St., Chicago, IL 60637, USA

\item[$^{18}$] Institute for Theoretical Physics, University of Zurich, \\ Winterthurerstrasse 190, CH-8057 Zurich, Switzerland

\item[$^{19}$] Institute for Particle Physics, ETH Zurich, \\ Schafmattstrasse 16, CH-8093, Zurich, Switzerland

\item[$^{20}$] University of Florida, 215 Williamson Hall, P.O. Box 118440 Gainesville, FL 32611, USA

\item[$^{21}$] Chunbuk National University, Physics Department, Jeonju 561-756, South Korea

\item[$^{22}$] CERN, CH-1211 Geneva 23, Switzerland

\item[$^{23}$] Laborat\'{o}rio de Instrumenta\c{c}\~{a}o e F\'{i}sica Experimental de Part\'{i}culas, LIP-Lisboa, Portugal 

\item[$^{24}$] Theoretische Natuurkunde and IIHE/ELEM, Vrije Universiteit Brussel, \\
               and International Solvay Institutes, Pleinlaan 2, B-1050 Brussels, Belgium

\item[$^{25}$] Universit\`a degli Studi di "Roma Tre" Dipartimento di Fisica,\\ Via Vasca Navale 84, Rome, Italy

\item[$^{26}$] Universit\`a di Roma "Sapienza" and INFN Sezione di Roma, Italy

\item[$^{27}$] Institut f\"ur Theoretische Physik und Astrophysik, Universit\"at W\"urzburg,\\ Emil-Hilb-Weg 22, D-97074 W\"urzburg, Germany

\item[$^{28}$] Max-Planck-Institut f\"ur Physik, Werner-Heisenberg-Institut,\\ F\"ohringer Ring 6, D-80805 M\"unchen, Germany

\item[$^{29}$] School of Physics, University of Melbourne, Victoria, Australia

\item[$^{30}$] INFN Laboratori Nazionali di Frascati, Frascati, Italy

\item[$^{31}$] Dipartimento di Fisica, Universit\`a degli Studi di Milano and INFN,\\ 
               Sezione di Milano, Via Celoria 16, I-20133 Milan, Italy

\item[$^{32}$] Departamento de F\'isica, Facultad de Ciencias Exactas y Naturales, \\ 
               Universidad de Buenos Aires, Pabellon I, Ciudad Universitaria (1428),\\ Capital Federal, Argentina 

\item[$^{33}$] Institute for Theoretical Physics, University of Amsterdam, \\ Science Park 904, 1090 GL Amsterdam, The Netherlands

\item[$^{34}$] PH Department, TH Unit, CERN, CH-1211 Geneva 23, Switzerland

\item[$^{35}$] DESY, Notkestrasse 85, D-22607 Hamburg, Germany

\item[$^{36}$] Fermi National Accelerator Laboratory, MS 106, Batavia, IL 60510-0500, USA

\item[$^{37}$] University of Nova Gorica, Laboratory for Astroparticle Physics,\\ SI-5000 Nova Gorica, Slovenia 

\item[$^{38}$] Universit\`a di Milano-Bicocca and INFN, Sezione di Milano-Bicocca,\\ Piazza della Scienza 3, 20126 Milan, Italy

\item[$^{39}$] Department of Physics, Univ. of Texas at Arlington, SH108, University of Texas, \\ Arlington, TX 76019, USA

\item[$^{40}$] ICREA at IFAE, Universitat Aut\`onoma de Barcelona, Bellaterra, Spain

\item[$^{41}$] Department of Physics, University of Oxford, Denys Wilkinson Building,\\ Keble Road, Oxford OX1 3RH, UK
  
\item[$^{42}$] Bergische Universit\"at Wuppertal, D-42097 Wuppertal, Germany

\item[$^{43}$] SLAC National Accelerator Laboratory, Menlo Park, CA 94025, USA

\item[$^{44}$] Department of Physics and Astronomy, Michigan State University,\\ East Lansing, MI 48824, USA

\item[$^{45}$] National Institute of Chemical Physics and Biophysics, Tallinn, Estonia

\item[$^{46}$] Univ. of Wisconsin, Dept. of Physics, High Energy Physics,\\ 2506 Sterling Hall 1150 University Ave, Madison, WI 53706, USA

\item[$^{47}$] Department of Physics, Royal Holloway, University of London, Egham TW20 0EX, UK

\item[$^{48}$] Institut f\"ur Theoretische Teilchenphysik und Kosmologie, RWTH Aachen University,\\ D-52056 Aachen, Germany

\item[$^{49}$] Helsinki Institute of Physics, P.O. Box 64, FIN-00014 University of Helsinki, Finland

\item[$^{50}$] School of Physics, and State Key Laboratory of Nuclear Physics and Technology,\\ Peking University, China

\item[$^{51}$] Department of Physics \& Astronomy, Northwestern University, Evanston, IL 60208, USA

\item[$^{52}$] Ottawa-Carleton Institute for Physics (OCIP), Ottawa, Ontario K1S 5B6, Canada

\item[$^{53}$] Northern Illinois University (NIU) - Department of Physics DeKalb, IL 60115, USA

\item[$^{54}$] Dipartimento di Fisica e Astronomia, Universit\`a di Padova, and INFN Sezione di Padova, \\ via Marzolo 8, 35131 Padova, Italy

\item[$^{55}$] Department of Physics, University of Illinois at Urbana-Champaign, Urbana, IL 61801, USA

\item[$^{56}$] School of Physics, University of the Witwatersrand, Private Bag 3,\\ Wits 2050, Johannesburg, South Africa

\item[$^{57}$] SUPA, School of Physics and Astronomy, University of Glasgow, Glasgow, G12 8QQ, UK

\item[$^{58}$] Institute for Theoretical Physics, Karlsruhe Institute of Technology, 76128 Karlsruhe, Germany

\item[$^{59}$] University of Virginia, Charlottesville, VA 22906, USA

\item[$^{60}$] Johannes Gutenberg-Universit\"at Mainz (JGU) - Institut f\"ur Physik (IPH), \\ Staudingerweg 7, 55128 Mainz, Germany

\item[$^{61}$] Department of Physics, Princeton University, Princeton, NJ 08542, USA

\item[$^{62}$] New York City College of Technology, City University of New York, \\
               300 Jay Street, Brooklyn NY 11201, USA  \\
               and The Graduate School and University Center, City University of New York, \\
               365 Fifth Avenue, New York, NY 10016, USA

\item[$^{63}$] Department of Physics, University of Oregon, Eugene, OR 97403-1274, USA

\item[$^{64}$] Department of Physics, New York University, New York, NY, USA

\item[$^{65}$] University of Split, FESB, R. Boskovica bb, 21 000 Split, Croatia

\item[$^{66}$] Universit\`a di Pavia and INFN, Sezione di Pavia, Via A. Bassi, 6, 27100 Pavia, Italy

\item[$^{67}$] Physics Department, Florida State University, Tallahassee, FL 32306-4350, USA

\item[$^{68}$] Universit\`a and INFN of Pisa, via F. Buonarotti, PIsa, Italy

\item[$^{69}$] School of Physics and Astronomy, University of Edinburgh, \\ Mayfield Road, Edinburgh EH9 3JZ, UK

\item[$^{70}$] The Oskar Klein Centre, Deptartment of Physics Stockholm University, \\ SE-106 91 Stockholm, Sweden

\item[$^{71}$] Paul Scherrer Institut, CH--5232 Villigen PSI, Switzerland

\item[$^{72}$] Massachusetts Institute of Technology, \\ 77 Massachusetts Avenue, Cambridge, MA 02139-4307, USA

\item[$^{73}$] Institute of Physics, University of Debrecen, H-4010 Debrecen P.O.Box 105, Hungary

\item[$^{74}$] Department of Physics and Astronomy, University College London, \\ Gower Street, London WC1E 6BT, UK

\item[$^{75}$] Universit\`a di Napoli Federico II Dipartimento di Scienze Fisiche, via Cintia I-80126 Napoli, Italy

\item[$^{76}$] Institute of Nuclear Research of the Hungarian Academy of Sciences, Hungary

\item[$^{77}$] Univ. Libre de Bruxelles, B-1050 Bruxelles, Belgium

\item[$^{78}$] Physics Dept., Blackett Laboratory, Imperial College London, \\ Prince Consort Rd, London SW7 2BW, UK

\item[$^{79}$] Department of Physics, University of California at San Diego, La Jolla, CA 92093, USA

\item[$^{80}$] Department of Physics, SUNY at Buffalo, Buffalo, NY 14260-1500, USA 

\item[$^{81}$] Lawrence Berkeley National Laboratory (LBNL) - Physics Division, \\
               1 Cyclotron Road, 50R4049, Berkeley, CA 94720-8153, USA

\item[$\dagger$] On leave of absence from INFN, Sezione di Firenze, Italy

\item[$\ddag$] On leave from Department of Physics, Princeton University, Princeton, NJ 08544, USA

 \end{itemize}

% 
%\newpage
%\input{YRHXS3_Prologue}
%

%
\newpage
\tableofcontents
\newpage
\mbox{}

% --- Introduction
\newpage
\pagenumbering{arabic}
\setcounter{footnote}{0}
%

% Introduction
% -------------

\section{Introduction%
\footnote{S.~Heinemeyer, C.~Mariotti, G.~Passarino and R.~Tanaka.}} 

\newcommand{\ggF}{\ensuremath{\Pg\Pg \to \PH}}
\newcommand{\VBF}{\ensuremath{\PQq\PQq' \to \PQq\PQq'\PH}}
\newcommand{\VH}{\ensuremath{\PQq\PAQq \to \PW\PH/\PZ\PH}}
\newcommand{\ttH}{\ensuremath{\PQq\PAQq/\Pg\Pg \to \PQt\PAQt\PH}}

% Introduction and LHC

% Higgs search results 2012
% -------------

The 4th of July 2012 ATLAS and CMS announced that they had discovered
a new particle with a mass around $125 \UGeV$~\cite{Aad:2012tfa,Chatrchyan:2012ufa}.
This came after only a bit more than one year of data taken at center of
mass energies of $7$ and $8\UTeV$. 
The discovery has been made while searching for the Higgs boson, the particle
linked to the Brout-Englert-Higgs
mechanism~\cite{Englert:1964et,Higgs:1964ia,Higgs:1964pj,Guralnik:1964eu,Higgs:1966ev,Kibble:1967sv}.   
The outstanding performance of the LHC in 2012
delivered in total four times as much 8 TeV data 
as was used in the discovery analyses.  The  experiments were able
thus to present new preliminary results confirming the existence of
the particles and measuring the properties. 
The new particle is a boson, since it decays into two photons, $2\,\PZ$ and
$2\,\PW$ bosons, and it could possibly be the Standard Model (SM) Higgs boson.
More data and new theoretical approaches will help us in the future
to establish the nature of this particle and, more important, if there is new physics 
around the corner. 

\begin{figure}
\begin{center}
\includegraphics[width=0.45\textwidth]{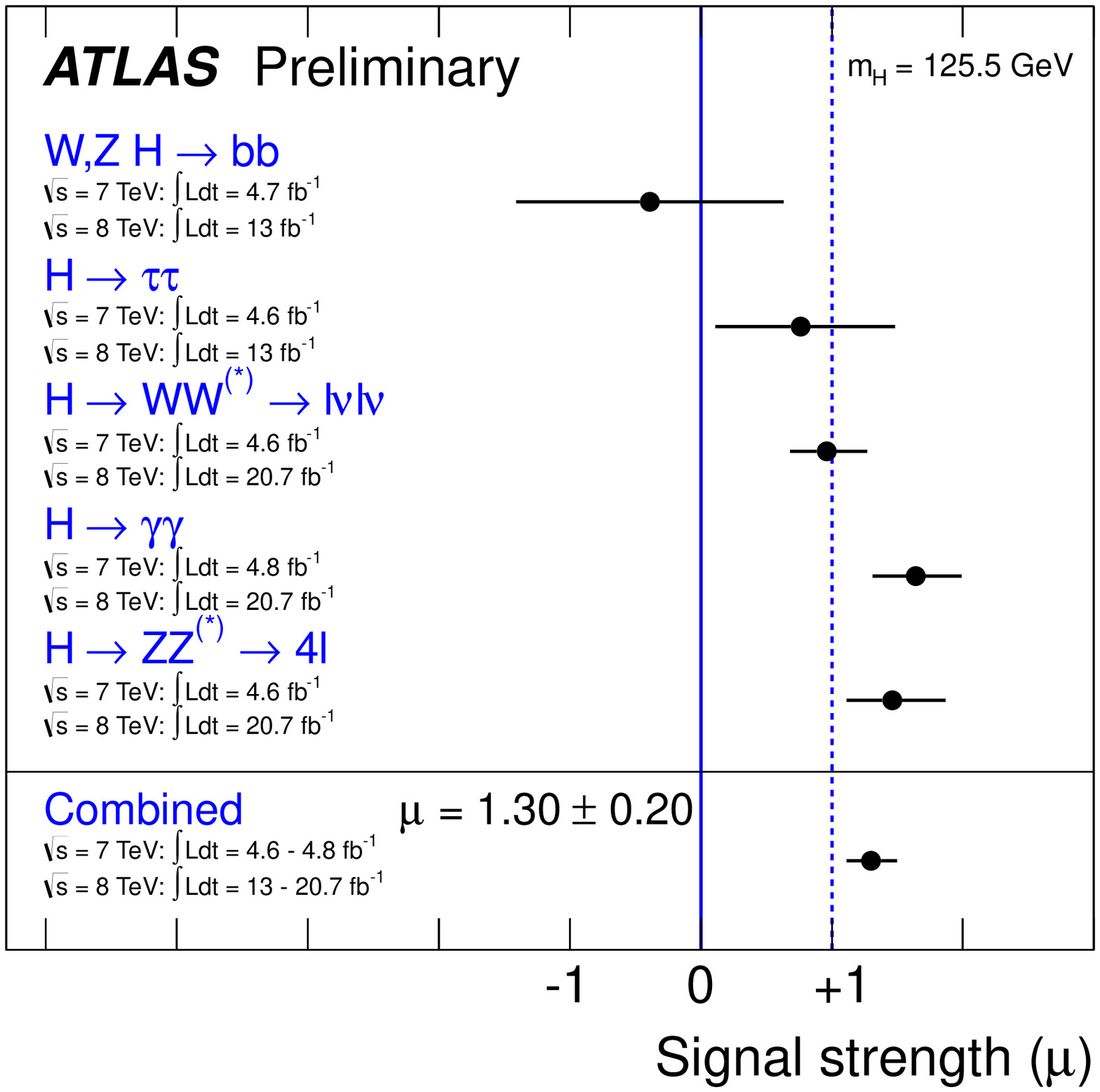}
\includegraphics[width=0.45\textwidth]{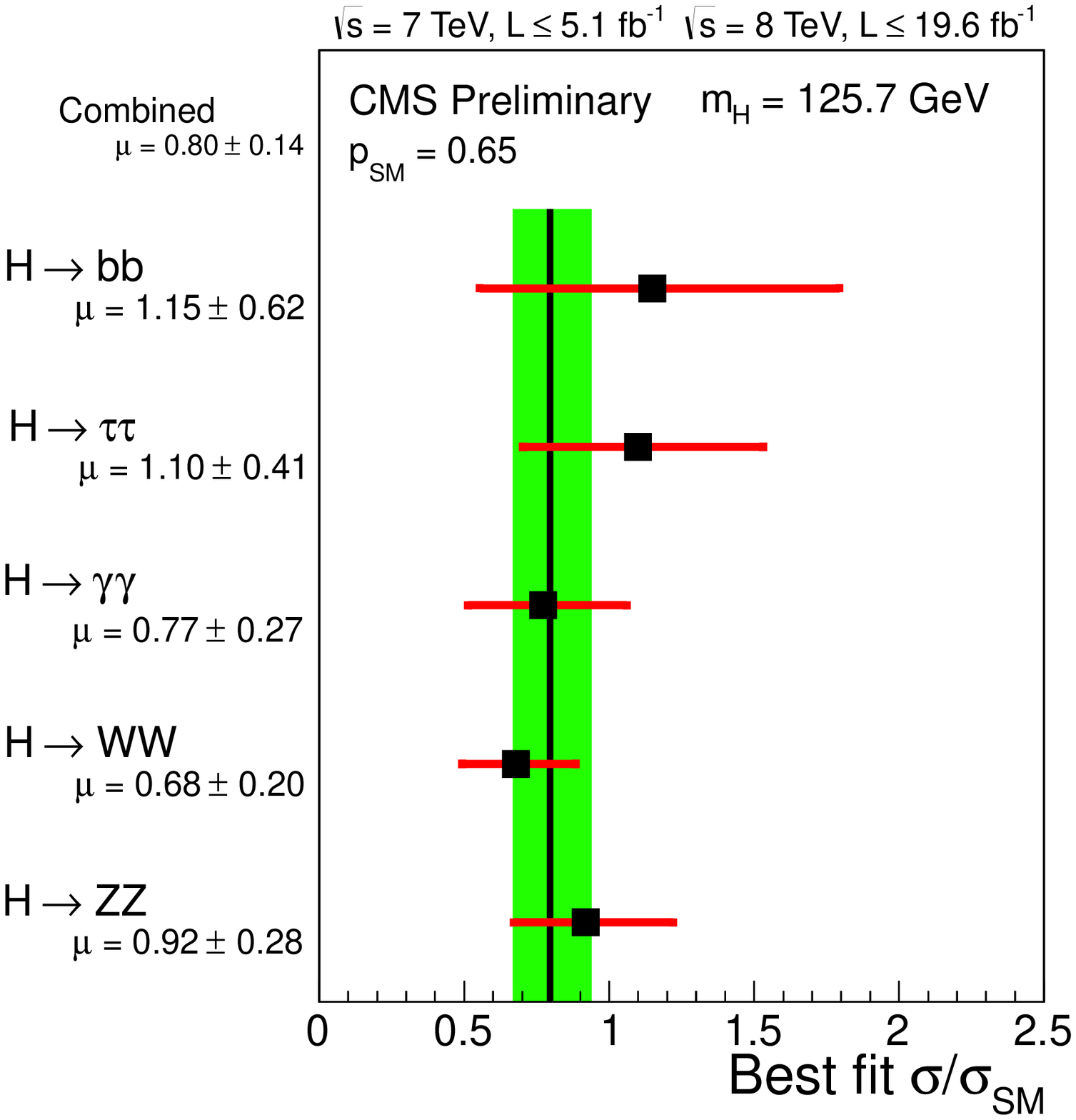}
\caption{The signal strength for the individual channel and their
  combination. The values of $\mu$ are given for $\MH=125.5 \UGeV$
  for ATLAS and for $\MH=125.7 \UGeV$ for CMS. }
\label{fig:mu_intro}
\end{center}
\end{figure}

At the Moriond conference this year, 2013, ATLAS and CMS presented 
the results on five main decay modes:
$\PH \to 2 \PGg$, $\PH \to \PZ\PZ \to 4\ell$,  $\PH \to \PW\PW \to
\ell\nu \ell\nu$, $\PH \to \PGt\PGt$, and $\PH \to \PAQb\PQb$ channels, with 
an integrated luminosities of up to 5 fb$^{-1}$ at $\sqrt{s}= 7\UTeV$ and up to
21 fb$^{-1}$ at $\sqrt{s} = 8\UTeV$.  
The $\PH \to 2\PGg$ and $\PH \to \PZ\PZ \to 4\ell$ channels allow to
measure the mass of the boson with very high precision. 
ATLAS measures $125.5 \pm 0.2 ({\rm stat.}) ^{+0.5}_{-0.6} ({\rm syst.}) \UGeV$~\cite{ATLAS-CONF-2013-014}, 
CMS measures a mass of $125.7 \pm 0.3 ({\rm stat.}) \pm 0.3 ({\rm
  syst.}) \UGeV$~\cite{CMS-PAS-HIG-13-005}.  
Figure~\ref{fig:mu_intro} shows the signal strength for the various
decay channels for ATLAS and CMS.
For ATLAS the combined signal strength is determined to be 
$\mu=1.30 \pm 0.13 ({\rm stat.}) \pm 0.14 ({\rm syst.})$ 
at the measured mass value~\cite{ATLAS-CONF-2013-034}.
For CMS the combined signal strength is determined to be 
$\mu = 0.80 \pm 0.14$ at the measured mass value~\cite{CMS-PAS-HIG-13-005}.  

Whether or not the new particle is a Higgs boson is demonstrated by how it 
interacts with other particles and its own quantum properties. For example, a 
Higgs boson is postulated to have no spin and in the SM its parity,
a measure of how its mirror image behaves, should be positive. ATLAS and CMS
have compared a number of alternative spin-parity ($\mathrm{J}^\mathrm{P}$) assignments for this
particle and, in pairwise hypothesis tests, the hypothesis of no spin and
positive parity  ($0^+$) is consistently favored against the
alternative hypotheses \cite{Chatrchyan:2012jja,CMS-PAS-HIG-13-005}. 

\medskip
This report presents improved inclusive cross section calculation at
$7$ and $8\UTeV$ in
the SM and its minimal supersymmetric extension, together with the calculation
of the relevant decay channels, in particular around the measured mass value
of $125 - 126 \UGeV$. 
Results and extensive discussions are also presented for
a heavy (SM-like) Higgs boson, including the correct treatment via the Complex
Pole Scheme (CPS) as well as interference effects in particular in the 
$\PW\PW$ and $\PZ\PZ$ channels.

In view of the newly discovered particle the property determination becomes
paramount. This report presents the interim recommendation to extract
couplings as well as the spin and the parity of the new particle. 
For the determination of the coupling strength factors the correlated
uncertainties for the decay calculations have to be taken into account and
corresponding descriptions and results for the uncertainties at the level of
partial widths are included in this report.
Given that now the theoretical uncertainties from QCD scale and
PDF+$\alphas$ dominate the experimental systematic 
uncertainties, a new proposal is presented on how to treat and evaluate the
theory uncertainties due to the QCD scale.

The report furthermore includes the state-of-the-art description of the relevant
Monte Carlo generators, in particular for the difficult task of
simulating the Higgs production together with (two) jets in gluon gluon fusion
as compared to Vector boson fusion. The treatment of jet bin uncertainties is
also discussed.

%The best way to consider the uncertainties on the partial widths.
%It presents the first ad interim recommendation to extract couplings and the spin and parity.
%It present also the status of the art of the Monte Carlo generator for
%the difficult task of simulating the H produced in gluongluon 
%fusion plus 2 jets, as  background to the measurements of the Higgs when produced in  Vector boson Fusion.
%An extensive discussion on the high mass Higgs is presented. CPS,
%interference etc that open the challenging chapter on BSM Higgs.
%Finally, given that the statistical uncertainties is not dominating
%anymore, a new proposal on how to treat more 
%correctly the theoretical uncertainty due to the QCD scale.
%The future:  HARD? IS THERE ANY?

\medskip
This report tries to lay the path towards a further exploration of the Higgs
sector of the (yet to be determined) underlying model. First, this includes the
(re)quest for even more refined calculations of inclusive and differential
cross sections in the SM, for heavy (SM-like) Higgs bosons, as well as in as
many BSM models as possible. Dedicated efforts are needed to match the
required accuracy in the SM, the MSSM and other BSM models. Methods for a
reliable estimate of remaining uncertainties are needed. 
Second, this includes prescriptions for the correct extraction of the
properties of the newly discovered particle. While we report on substantial
progress on these topics, it also becomes clear that a lot of dedicated effort
will be needed to match the challenges electroweak symmetry breaking holds for
us. 

% --- Higgs subsections

%\newpage
%\input{YRHXS3_Section}
%\clearpage

\newpage
\providecommand{\br}{{\mathrm{BR}}}
\providecommand{\Gatot}{\Gamma_{\mathrm H}}
\providecommand{\lsim}
{\;\raisebox{-.3em}{$\stackrel{\displaystyle <}{\sim}$}\;}
\providecommand{\gsim}
{\;\raisebox{-.3em}{$\stackrel{\displaystyle >}{\sim}$}\;}
\providecommand{\Pqb}{\bar{\Pq}}
\providecommand{\Pfb}{\bar{\Pf}}
\providecommand{\PV}{\mathrm{V}}
\providecommand{\tb}{\tan\beta}
\providecommand{\Mh}{M_\mathrm{h}}
\providecommand{\MH}{M_\mathrm{H}}
\providecommand{\MA}{M_\mathrm{A}}
\providecommand{\MHp}{M_{\mathrm{H}^\pm}}
\providecommand{\mhmaxx}{\ensuremath{m_{\Ph}^{\rm max}}}
\providecommand{\mhmodp}{\ensuremath{m_{\Ph}^{\rm mod+}}}
\providecommand{\mhmodm}{\ensuremath{m_{\Ph}^{\rm mod-}}}
\providecommand{\tauphobic}{$\PGt$-phobic}
\providecommand{\hhbb}{\Ph \to \PQb \PQb}
\providecommand{\htautau}{\Ph \to \PGtp\PGtm}
\providecommand{\hmumu}{\Ph \to \PGmp\PGmm}
\providecommand{\hss}{\Ph \to \PQs \PQs}
\providecommand{\hcc}{\Ph \to \PQc \PAQc}
\providecommand{\hgaga}{\Ph \to \PGg\PGg}
\providecommand{\hZga}{\Ph \to \PZ\PGg}
\providecommand{\hWW}{\Ph \to \PW\PW}
\providecommand{\hZZ}{\Ph \to \PZ\PZ}
\providecommand{\Hbb}{\PH \to \PQb \PQb}
\providecommand{\Htautau}{\PH \to \PGtp\PGtm}
\providecommand{\Hmumu}{\PH \to \PGmp\PGmm}
\providecommand{\Hss}{\PH \to \PQs \PQs}
\providecommand{\Hcc}{\PH \to \PQc \PAQc}
\providecommand{\Htt}{\PH \to \PQt \PAQt}
\providecommand{\Hgg}{\PH \to \Pg\Pg}
\providecommand{\Hgaga}{\PH \to \PGg\PGg}
\providecommand{\HZga}{\PH \to \PZ\PGg}
\providecommand{\HWW}{\PH \to \PW\PW}
\providecommand{\HZZ}{\PH \to \PZ\PZ}
\providecommand{\Abb}{\PA \to \PQb \PQb}
\providecommand{\Atautau}{\PA \to \PGtp\PGtm}
\providecommand{\Amumu}{\PA \to \PGmp\PGmm}
\providecommand{\Ass}{\PA \to \PQs \PQs}
\providecommand{\Acc}{\PA \to \PQc \PAQc}
\providecommand{\Att}{\PA \to \PQt \PAQt}
\providecommand{\Agg}{\PA \to \Pg\Pg}
\providecommand{\Agaga}{\PA \to \PGg\PGg}
\providecommand{\AZga}{\PA \to \PZ\PGg}
\providecommand{\AWW}{\PA \to \PW\PW}
\providecommand{\AZZ}{\PA \to \PZ\PZ}
\providecommand{\phibb}{\PH \to \PQb \PAQb}
\providecommand{\phitautau}{\PH \to \PGtp \PGtm}
\providecommand{\phimumu}{\PH \to \PGmp \PGmm}
\providecommand{\phiss}{\PH \to \PQs \PAQs}
\providecommand{\phicc}{\PH \to \PQc \PAQc}
\providecommand{\phitt}{\PH \to \PQt \PAQt}
\providecommand{\phigg}{\PH \to \Pg\Pg}
\providecommand{\phigaga}{\PH \to \PGg\PGg}
\providecommand{\phiZga}{\PH \to \PZ\PGg}
\providecommand{\phiVV}{\PH \to \PV^{(*)}\PV^{(*)}}
\providecommand{\phiWW}{\PH \to \PW^{(*)}\PW^{(*)}}
\providecommand{\phiZZ}{\PH \to \PZ^{(*)}\PZ^{(*)}}
\providecommand{\Hptb}{\PH^\pm \to \PQt \PQb}
\providecommand{\Hpts}{\PH^\pm \to \PQt \PQs}
\providecommand{\Hptd}{\PH^\pm \to \PQt \PQd}
\providecommand{\Hpcb}{\PH^\pm \to \PQc \PQb}
\providecommand{\Hpcs}{\PH^\pm \to \PQc \PQs}
\providecommand{\Hpcd}{\PH^\pm \to \PQc \PQd}
\providecommand{\Hpub}{\PH^\pm \to \PQu \PQb}
\providecommand{\Hpus}{\PH^\pm \to \PQu \PQs}
\providecommand{\Hpud}{\PH^\pm \to \PQu \PQd}
\providecommand{\Hptaunu}{\PH^\pm \to \PGt\PGnGt}
\providecommand{\Hpmunu}{\PH^\pm \to \PGm\PGnGm}
\providecommand{\Hpenu}{\PH^\pm \to \Pe\PGnGe}
\providecommand{\HphW}{\PH^\pm \to \Ph\PW}
\providecommand{\HpHW}{\PH^\pm \to \PH\PW}
\providecommand{\HpAW}{\PH^\pm \to \PA\PW}

\providecommand{\zehomi}[1]{$\cdot 10^{-#1}$}
\providecommand{\zehoze}{}
\providecommand{\zehopl}[1]{$\cdot 10^{#1}$}

\section{Branching Ratios \footnote{%
    S.~Heinemeyer, A.~M\"uck, I.~Puljak, D.~Rebuzzi (eds.);
    A.~Denner, S.~Dittmaier, M.~Spira, M. M\"uhlleitner. } }
\label{sec:br}

For a correct interpretation of experimental data, precise calculations
not only of the various production cross sections, but also for the
relevant decay widths are essential, including their respective
uncertainties. Concerning the SM Higgs boson, in
\Bref{Dittmaier:2011ti} a first precise prediction
of the branching ratios (BR) was presented. 
In \Bref{Dittmaier:2012vm,Denner:2011mq} the BR predictions were supplemented with an
uncertainty estimate including parametric uncertainties as well
as the effects of unknown higher-order corrections.
In \refS{sec:br-sm}, we
update these predictions with a fine step size around the mass of the
newly discovered Higgs-like particle at ATLAS~\cite{Aad:2012tfa}
and CMS~\cite{Chatrchyan:2012ufa}. 
We also present the error estimates in a form which is suitable for taking
error correlations into account. In \refS{sec:br-differential}, we discuss 
differential distributions for four-fermion final states. We show that
interference effects (already at LO) and higher-order corrections distort
the shapes of distributions at the level of 10\%.
For the lightest Higgs boson in the MSSM in
\Bref{Dittmaier:2012vm} first results for
$\br(\phitautau)$ ($\phi = \Ph,\PH,\PA$) were given in the
\mhmaxx\ scenario~\cite{Carena:2002qg}. 
In \refS{sec:br-mssm} we present a first prediction for all
relevant decay channels of the charged Higgs bosons in the
\mhmaxx\ scenario. We also provide first results for the BRs of all MSSM Higgs
bosons in the newly presented benchmark scenarios.

%%%%%%%%%%%%%%%%%%%%%%%%%%%%%%%%%%%%%%%%%%%%%%%%%%%%%%%%%%%%%%%%%%%%%%%%%%%%%%%

\subsection{SM Branching Ratios}
\label{sec:br-sm}

In this section we update the SM BR calculations presented in
\Brefs{Dittmaier:2011ti,Dittmaier:2012vm,Denner:2011mq}. The strategy and the
calculational tools are unchanged with respect to \Bref{Dittmaier:2012vm}.
Here, we focus on more detailed results for the BRs and the corresponding
uncertainties for a SM Higgs boson around the mass of the newly discovered 
Higgs-like particle, i.e.\ around $\MH = 126 \UGeV$, and correct small
inconsistencies in the tabulated error estimates in \Bref{Dittmaier:2012vm}. 
The BRs as well as the corresponding 
uncertainties are a crucial ingredient entering the phase of precision 
measurements in order to compare the properties of the new resonance to a 
SM Higgs boson in a reliable way. Moreover, for a Higgs mass around $126 \UGeV$, 
we give detailed results on the different parametric uncertainties (PU) and
theoretical uncertainties (THU), 
as introduced in \Brefs{Dittmaier:2012vm,Denner:2011mq}, 
of the relevant partial Higgs decay widths. The given results
facilitate the combination of Higgs measurements including error correlations
in the BRs. First, in \refS{sec:br-strategy}, 
we briefly review the evaluation of the decay
widths, the BRs, and the relevant uncertainties. More details can be found in
\Brefs{Dittmaier:2011ti,Dittmaier:2012vm,Denner:2011mq}. Results are presented
for the total width, $\Gatot$, and the BRs for the decay modes 
$\Hbb$, $\Hcc$, $\Htautau$, $\Hmumu$, $\Hgg$, $\Hgaga$, $\HZga$,
$\HWW$, and $\HZZ$ (including detailed results also for the various
four-fermion final states) in \refS{sec:br-126results}. 
For large Higgs masses also the decay mode $\Htt$ has been analyzed.
The various PUs and THUs on the level of the partial widths are 
presented in \refS{sec:br-correlations} for selected Higgs masses.

\subsubsection{Strategy and input for Branching Ratio Calculations}
\label{sec:br-strategy}

In this section we briefly summarize the strategy for the BR
calculations for the updates in this report. The calculations are performed
in exactly the same setup as the BR predictions
in \Bref{Dittmaier:2012vm}. A detailed description can be found there.

We use {\HDECAY} \cite{Djouadi:1997yw,Spira:1997dg,hdecay2} and {\Prophecy}
\cite{Bredenstein:2006rh,Bredenstein:2006ha,Prophecy4f} to calculate all 
the partial widths with the highest accuracy available. The included 
higher-order corrections and the remaining THUs have been discussed in detail 
in Section 2.1.3.2 of \Bref{Dittmaier:2012vm}. 
For the detailed results in the low-mass region, the total uncertainties in 
Table 2 of \Bref{Dittmaier:2012vm} are used as THUs for the different 
Higgs-boson decay modes. The uncertainty for the total width is derived by 
adding the uncertainties for the partial widths linearly. Concerning the BRs,
the variations of all branching ratios are calculated for each 
individual partial width being varied within the corresponding relative
error keeping all other partial widths fixed at their central value
(since each branching ratio depends on all partial widths,
scaling a single partial width modifies all branching ratios).
Hence, there is an individual THU of each branching 
ratio due to the THU of each partial width. We assume only 
all $\PH\to \PW\PW/\PZ\PZ\to4\Pf$ decays to be correlated and, hence, 
only consider the simultaneous scaling of all 4-fermion partial widths. 
The derived individual THUs for each branching ratios are added
linearly to obtain the corresponding total THU.

For our calculations, the input parameter set as defined in Appendix A of
\Bref{Dittmaier:2011ti} has been used (for quark masses see 
comments below). From the given PDG values of 
the gauge-boson masses, we derive the pole masses
$\MZ=91.15349\UGeV$ and $\MW= 80.36951\UGeV$ which are used as input. 
The gauge-boson widths have been calculated at NLO from
the other input parameters resulting in $\Gamma_{\PZ}=2.49581\UGeV$ and
$\Gamma_{\PW}=2.08856\UGeV$.
It should be noted again that for our numerical analysis we have used the
one-loop pole masses for the charm and bottom quarks and their
uncertainties, since these values do not exhibit a significant
dependence on the value of the strong coupling constant $\alphas$ in
contrast to the $\MSbar$ masses \cite{Narison:1994ag}. To be precise, 
we use $M_{\PQb} = 4.49 \UGeV$, $M_{\PQc} = 1.42 \UGeV$, and 
$M_{\PQs} = 0.10 \UGeV$. The small shifts with respect to the charm- and 
strange-quark masses used in \Bref{Dittmaier:2012vm} do not affect 
the BRs significantly at all.

Concerning the parametric uncertainties, we take only into account the 
uncertainties of the input parameters $\alphas$, $\Mc$, $\Mb$, and $\Mt$
as given in Table 1 of \Bref{Dittmaier:2012vm}, where also
the detailed reasoning leading to this specific choice is given.
Using these uncertainties, for each parameter
$p=\alphas,\Mc,\Mb,\Mt$ we have calculated the Higgs branching ratios
for $p$, $p+\Delta p$ and $p-\Delta p$ keeping all the other parameters 
fixed at their central values. The resulting error on each BR is then given 
by
\begin{eqnarray}
\Delta^p_+ \br &=&  \max \{\br(p+\Delta p),\br(p),\br(p-\Delta p)\} -
\br(p),\nonumber\\
\Delta^p_- \br &=&  \br(p) -\min \{\br(p+\Delta p),\br(p),\br(p-\Delta p)\},
\end{eqnarray}
which may lead to asymmetric errors.
The total PUs have been obtained by adding the calculated
errors due to the four parameters in quadrature. 
In analogy, the uncertainties of the partial and total decay widths 
are given by
\begin{eqnarray}
\label{eqn:deltagamma}
\Delta^p_+ \Gamma &=&  \max \{\Gamma(p+\Delta
p),\Gamma(p),\Gamma(p-\Delta p)\} - \Gamma(p),\nonumber\\
\Delta^p_- \Gamma &=&  \Gamma(p) -\min \{\Gamma(p+\Delta
p),\Gamma(p),\Gamma(p-\Delta p)\},
\end{eqnarray}
where $\Gamma$ denotes the partial decay width for each considered decay
channel or the total width, respectively. The total PUs have been 
calculated again by adding the individual PUs in
quadrature. 

The total uncertainties on the BRs, i.e.\ combining PUs and THUs, 
are derived by adding linearly the total parametric uncertainties and 
the total theoretical uncertainties. To allow for taking into account
correlations in the errors of the different BRs, we provide also 
the uncertainties for the different partial widths 
in \refS{sec:br-correlations} for selected Higgs masses.

For completeness, we repeat that the Higgs total width resulting from {\HDECAY} 
has been modified according to the prescription
\begin{equation}
\Gamma_{\PH} = \Gamma^{\mathrm{HD}} - \Gamma^{\mathrm{HD}}_{\PZ\PZ} 
            - \Gamma^{\mathrm{HD}}_{\PW\PW} + \Gamma^{\mathrm{Proph.}}_{4\Pf}~,
\end{equation}
where $\Gamma_{\PH}$ is the total Higgs width, $\Gamma^{\mathrm{HD}}$
the Higgs width obtained from {\HDECAY},
$\Gamma^{\mathrm{HD}}_{\PZ\PZ}$ and $\Gamma^{\mathrm{HD}}_{\PW\PW}$
stand for the partial widths to $\PZ\PZ$ and $\PW\PW$ calculated with
{\HDECAY}, while $\Gamma^{\mathrm{Proph.}}_{4\Pf}$ represents the
partial width of $\PH\to 4\Pf$ calculated with {\Prophecy}.  The
latter can be split into the decays into $\PZ\PZ$, $\PW\PW$, and the
interference,
\begin{equation}
\Gamma^{\mathrm{Proph.}}_{4\Pf}=\Gamma_{{\PH}\to \PW^*\PW^*\to 4\Pf}
+ \Gamma_{{\PH}\to \PZ^*\PZ^*\to 4\Pf}
+ \Gamma_{\mathrm{\PW\PW/\PZ\PZ-int.}}\,,
\end{equation} 
where the individual contributions are defined in terms of partial widths
with specific final states according to
\begin{eqnarray*}
\Ga_{{\mathrm{H}}\to \PW^*\PW^*\to 4f} &=&
  9 \cdot \Ga_{{\PH}\to\PGne\Pep\PGm\bar\PGnGm}
+ 12 \cdot \Ga_{{\PH}\to\PGne\Pep\PQd\PAQu}
+ 4 \cdot \Ga_{{\PH}\to\PQu\PAQd\PQs\PAQc}\, ,\\[1ex]
\Ga_{{\PH}\to \PZ^*\PZ^*\to 4f} &=&
  3 \cdot \Ga_{{\PH}\to\PGne\PAGne\PGnGm\PAGnGm}
+ 3 \cdot \Ga_{{\PH}\to\Pe\Pep\PGm\PGmp}
+ 9 \cdot \Ga_{{\PH}\to\PGne\PAGne\PGm\PGmp}\\
&&{}
+ 3 \cdot \Ga_{{\PH}\to\PGne\PAGne\PGne\PAGne}
+ 3 \cdot \Ga_{{\PH}\to\Pe\Pep\Pe\Pep}\\
&&{}+ 6 \cdot \Ga_{{\PH}\to\PGne\PAGne\PQu\PAQu}
+ 9 \cdot \Ga_{{\PH}\to\PGne\PAGne\PQd\PAQd}
+ 6 \cdot \Ga_{{\PH}\to\PQu\PAQu\Pe\Pep}
+ 9 \cdot \Ga_{{\PH}\to\PQd\PAQd\Pe\Pep}\\
&&{}+ 1 \cdot \Ga_{{\PH}\to\PQu\PAQu\PQc\PAQc}
+ 3 \cdot \Ga_{{\PH}\to\PQd\PAQd\PQs\PAQs}
+ 6 \cdot \Ga_{{\PH}\to\PQu\PAQu\PQs\PAQs}
+ 2 \cdot \Ga_{{\PH}\to\PQu\PAQu\PQu\PAQu}\\
&&{}+ 3 \cdot \Ga_{{\PH}\to\PQd\PAQd\PQd\PAQd}\, ,\\[1ex]
\Ga_{\mathrm{WW/ZZ-int.}} &=&
  3 \cdot \Ga_{{\PH}\to\PGne\Pep\Pe\PAGne}
- 3 \cdot \Ga_{{\PH}\to\PGne\PAGne\PGm\PGmp}
- 3 \cdot \Ga_{{\PH}\to\PGne\Pep\PGm\bar\PGnGm}\\
&&
  {}+2 \cdot \Ga_{{\PH}\to\PQu\PAQd\PQd\PAQu}
- 2 \cdot \Ga_{{\PH}\to\PQu\PAQu\PQs\PAQs}
- 2 \cdot \Ga_{{\PH}\to\PQu\PAQd\PQs\PAQc}\, .
\end{eqnarray*}

\subsubsection{BR Results for Higgs masses}
\label{sec:br-126results}

In this section we provide results for the BRs of the 
SM Higgs boson, using a particularly fine grid of mass points 
close to $\MH = 126 \UGeV$. The results 
are generated and presented in complete analogy to the predictions
in \Brefs{Dittmaier:2012vm}, including the error estimates
for each BR. In the error estimates, we have identified and removed 
inconsistencies in the calculation of the numbers presented 
in \Brefs{Dittmaier:2012vm}.
The corresponding changes in the error estimate are at the level of one percent
for $m_H>135 \UGeV$. For $m_H>500 \UGeV$ the changes increase
for some decay modes, in particular for $\Htt$. The central values 
of the BRs are not affected.

\begin{figure}[htb!]
\vspace{0.5cm}
\includegraphics[width=0.48\textwidth]{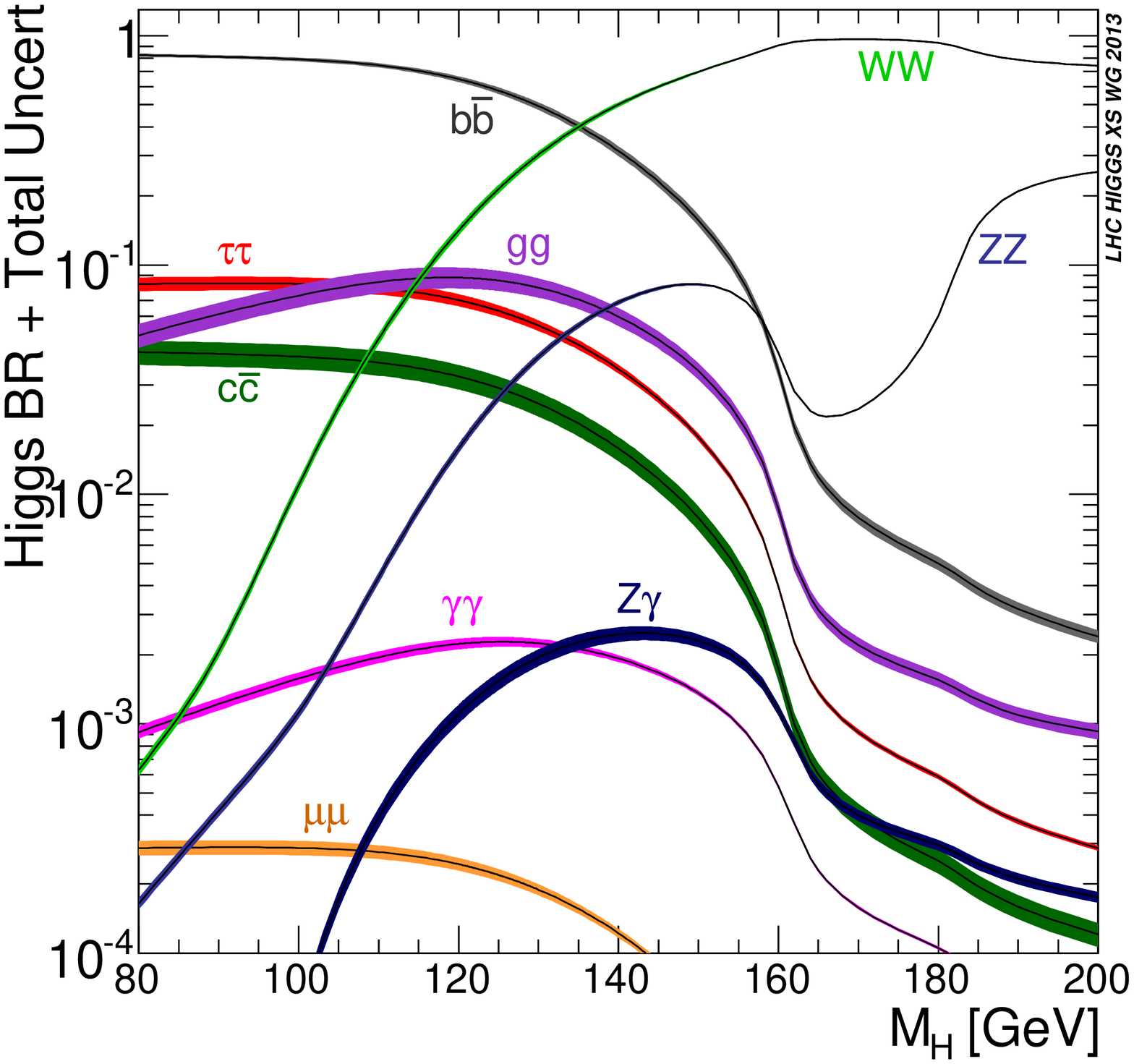}
\includegraphics[width=0.48\textwidth]{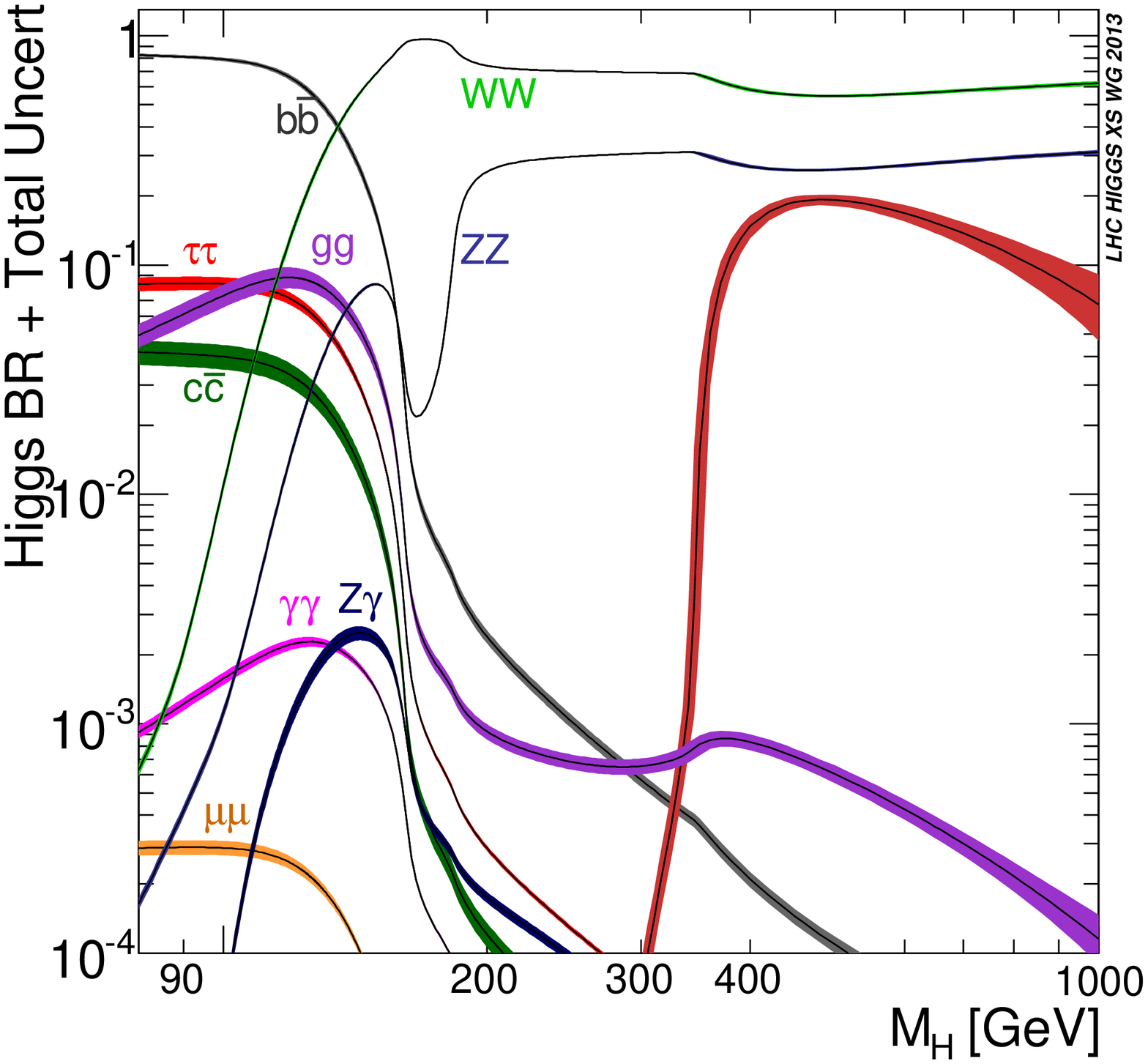}\\[0.5em]
\vspace{-0.7cm}
\caption{Higgs branching ratios and their uncertainties for the low
  mass range (left) and for the full mass range (right).}
\label{fig:YRHXS3_BR_plots}
\end{figure}

The fermionic decay modes are shown in 
\refT{tab:YRHXS3_2fermions.1} to \refT{tab:YRHXS3_2fermions.7}. The 
bosonic decay modes together with the total width are given in 
\refT{tab:YRHXS3_bosons-width.1} to \refT{tab:YRHXS3_bosons-width.7}. 
%BRs and error estimates for 
%$\PH\to \PW\PW/\PZ\PZ\to4\Pf$ are shown for various specific 
%four-fermion final states and combinations of those in 
%\refT{}--\refT{}. 
The same information (including the full uncertainty) is also presented 
graphically in \refF{fig:YRHXS3_BR_plots} for the low-mass region (left) and 
for the full mass range (right).
%{\bf References to the Tables and plots to be added!}

%%%%%%%%%%%%%%%%%%% F I G U R E %%%%%%%%%%%%%%%%%%%%%%%%%%%%%%%%%%%%%%%%%%%%%%%
%%%%%%%%%%%%%%%%%%% F I G U R E %%%%%%%%%%%%%%%%%%%%%%%%%%%%%%%%%%%%%%%%%%%%%%%

\subsubsection{BR Correlations for Higgs masses close to 126 GeV}
\label{sec:br-correlations}

In this section, we focus on the error correlations for the different
BRs. The reason for the correlations is two-fold: Varying the
input parameters within their error bands will induce shifts
of the different partial widths and the resulting BRs in a correlated way.
Moreover, there is trivial correlation between the BRs because
all the BRs have to add up to one. The shift in a single partial 
width will shift all BRs in a correlated way. 

In Table 3 of \Bref{Dittmaier:2012vm}, 
focusing on the errors of the BRs, we showed how
the PUs for each input parameter and the THUs affect the final BR 
predictions. To take correlations into account, it is beneficial to
present the same information directly on the level of the partial widths.
For the partial widths, the THUs can be assumed to be uncorrelated. Moreover, 
the correlated effect on each partial width from varying a parameter within its 
errors is disentangled from the additional trivial correlation when calculating 
the BRs. We show the results for the partial widths
in \refT{tab:br-correlation} for $\MH = 122 \UGeV$, $126 \UGeV$, 
and $130 \UGeV$. For each relevant partial width, we 
show the THU and the different PUs evaluated as before according to 
\refE{eqn:deltagamma}.

To be even more precise, as in Table 3 of \Brefs{Dittmaier:2012vm}, 
for each input parameter we show the induced shift on each partial width for 
the maximal and minimal choice of the input parameter as upper and lower
entry in the table, respectively. Hence, the table allows to read off the correlation
in the variation of the different partial widths. The associated error bands are slightly
asymmetric. However, it is a good approximation to symmetrize the error band and assume
a Gaussian probability distribution for the corresponding prediction.

The THUs on the partial widths of all the four-fermion
final-states can be considered to be fully correlated. All other THUs are considered to be
uncorrelated. Hence, for the BRs only the trivial correlation is present. However, it should 
be stressed again that in contrast to the PUs theory errors cannot be assumed to be Gaussian
errors. Assuming a Gaussian distribution and, hence, effectively adding THUs to the
PUs in quadrature will in general lead to underestimated errors. According to the recommendations 
in Section 12 of \Bref{Dittmaier:2011ti}, the THUs should be considered to have a 
flat probability distribution within the given range. Alternatively,
the envelope of extreme choices for the theory prediction on 
the partial widths should be used as an error estimate. (For all the presented errors 
on the BRs, we have added PUs and THUs of the resulting BRs linearly, as discussed 
in \refS{sec:br-strategy} before. Thereby we provide the envelope for each resulting BR, 
however, correlation is lost on the level of BRs.) 

In total, there are four input parameters to be varied corresponding to the PUs and 
one has to include eight uncorrelated THUs for the various partial widths.
Analyzing in detail the most interesting region around $\MH = 126 \UGeV$, the 
different uncertainties are of different importance. Aiming for a given accuracy, 
some uncertainties may be safely neglected, as can be inferred from \refT{tab:br-correlation}. 
Even sizeable uncertainties for a given partial width can be unimportant if
the decay mode has a small BR and does not contribute significantly to combined measurements.

Concerning the PUs, the variation of $\alphas$ and the bottom quark mass impact the 
BR predictions at the few percent level each. 
The charm quark mass is only relevant for $\Hcc$ and affects 
other BRs only at the few per mille level.
The dominant THU for most relevant channels is the one for $\Hbb$. 
The THU for $\Hgaga$ amounts to 1\% and is needed at this level of precision. 
The THU for  $\PH\to \PW\PW/\PZ\PZ\to4\Pf$ is estimated at 
0.5\% and thus also quite small. 
The THU for $\Hcc$, $\Hgg$, $\HZga$, $\Hmumu$, and $\Htautau$
only has sizeable effects if a measurement of the corresponding channel is included or 
errors of a few per mille are important. 

%%%%%%%%%%%%%%%%%%%%%%%%%% T A B L E %%%%%%%%%%%%%%%%%%%%%%%%%%%%%%%%%%%%%%%%
\begin{table}\footnotesize
\renewcommand{\arraystretch}{1.3}
\setlength{\tabcolsep}{1.0ex}
\caption{SM Higgs partial widths and their relative parametric (PU) and
  theoretical (THU) uncertainties for a selection of Higgs
  masses. For PU, all the single contributions are shown. For these
  four columns, the upper percentage value (with its sign) refers 
  to the positive variation of the parameter, while the lower one
  refers to the negative variation of the parameter. 
}
\label{tab:br-correlation}
\begin{center}
\begin{tabular}{lccrrrrr}
\hline
Channel & $\MH$ [GeV]  &  $\Gamma$ [MeV] &  $\Delta \alphas$ & $\Delta \Mb$ & $\Delta \Mc$ & $\Delta \Mt$ & THU \\
\hline
& 122  &2.30\zehoze           &$^{-2.3\%}_{+2.3\%}$ & $ ^{+3.2\%}_{-3.2\%}$ & $ ^{+0.0\%}_{-0.0\%}$ & $ ^{+0.0\%}_{-0.0\%}$& $ ^{+2.0\%}_{-2.0\%}$\\ $\Hbb$
& 126  &2.36\zehoze           &$^{-2.3\%}_{+2.3\%}$ & $ ^{+3.3\%}_{-3.2\%}$ & $ ^{+0.0\%}_{-0.0\%}$ & $ ^{+0.0\%}_{-0.0\%}$& $ ^{+2.0\%}_{-2.0\%}$\\
& 130  &2.42\zehoze           &$^{-2.4\%}_{+2.3\%}$ & $ ^{+3.2\%}_{-3.2\%}$ & $ ^{+0.0\%}_{-0.0\%}$ & $ ^{+0.0\%}_{-0.0\%}$& $ ^{+2.0\%}_{-2.0\%}$\\ 
\hline
& 122  &2.51\zehomi{1}        &$^{+0.0\%}_{+0.0\%}$ & $ ^{+0.0\%}_{-0.0\%}$ & $ ^{+0.0\%}_{-0.0\%}$ & $ ^{+0.0\%}_{-0.1\%}$& $ ^{+2.0\%}_{-2.0\%}$\\ $\Htautau$
& 126  &2.59\zehomi{1}        &$^{+0.0\%}_{+0.0\%}$ & $ ^{+0.0\%}_{-0.0\%}$ & $ ^{+0.0\%}_{-0.0\%}$ & $ ^{+0.1\%}_{-0.1\%}$& $ ^{+2.0\%}_{-2.0\%}$\\
& 130  &2.67\zehomi{1}        &$^{+0.0\%}_{+0.0\%}$ & $ ^{+0.0\%}_{-0.0\%}$ & $ ^{+0.0\%}_{-0.0\%}$ & $ ^{+0.1\%}_{-0.1\%}$& $ ^{+2.0\%}_{-2.0\%}$\\ 
\hline
& 122  &8.71\zehomi{4}        &$^{+0.0\%}_{+0.0\%}$ & $ ^{+0.0\%}_{-0.0\%}$ & $ ^{+0.0\%}_{-0.0\%}$ & $ ^{+0.1\%}_{-0.1\%}$& $ ^{+2.0\%}_{-2.0\%}$\\ $\Hmumu$
& 126  &8.99\zehomi{4}        &$^{+0.0\%}_{+0.0\%}$ & $ ^{+0.0\%}_{-0.0\%}$ & $ ^{-0.1\%}_{-0.0\%}$ & $ ^{+0.0\%}_{-0.1\%}$& $ ^{+2.0\%}_{-2.0\%}$\\
& 130  &9.27\zehomi{4}        &$^{+0.1\%}_{+0.0\%}$ & $ ^{+0.0\%}_{-0.0\%}$ & $ ^{+0.0\%}_{-0.0\%}$ & $ ^{+0.1\%}_{-0.0\%}$& $ ^{+2.0\%}_{-2.0\%}$\\ 
\hline
& 122  &1.16\zehomi{1}        &$^{-7.1\%}_{+7.0\%}$ & $ ^{-0.1\%}_{+0.1\%}$ & $ ^{+6.2\%}_{-6.0\%}$ & $ ^{+0.0\%}_{-0.1\%}$& $ ^{+2.0\%}_{-2.0\%}$\\ $\Hcc$
& 126  &1.19\zehomi{1}        &$^{-7.1\%}_{+7.0\%}$ & $ ^{-0.1\%}_{+0.1\%}$ & $ ^{+6.2\%}_{-6.1\%}$ & $ ^{+0.0\%}_{-0.1\%}$& $ ^{+2.0\%}_{-2.0\%}$\\
& 130  &1.22\zehomi{1}        &$^{-7.1\%}_{+7.0\%}$ & $ ^{-0.1\%}_{+0.1\%}$ & $ ^{+6.3\%}_{-6.0\%}$ & $ ^{+0.1\%}_{-0.1\%}$& $ ^{+2.0\%}_{-2.0\%}$\\ 
\hline
& 122  &3.25\zehomi{1}        &$^{+4.2\%}_{-4.1\%}$ & $ ^{-0.1\%}_{+0.1\%}$ & $ ^{+0.0\%}_{-0.0\%}$ & $ ^{-0.2\%}_{+0.2\%}$& $ ^{+3.0\%}_{-3.0\%}$\\ $\Hgg$
& 126  &3.57\zehomi{1}        &$^{+4.2\%}_{-4.1\%}$ & $ ^{-0.1\%}_{+0.1\%}$ & $ ^{+0.0\%}_{-0.0\%}$ & $ ^{-0.2\%}_{+0.2\%}$& $ ^{+3.0\%}_{-3.0\%}$\\
& 130  &3.91\zehomi{1}        &$^{+4.2\%}_{-4.1\%}$ & $ ^{-0.1\%}_{+0.2\%}$ & $ ^{+0.0\%}_{-0.0\%}$ & $ ^{-0.2\%}_{+0.2\%}$& $ ^{+3.0\%}_{-3.0\%}$\\ 
\hline
& 122  &8.37\zehomi{3}        &$^{+0.0\%}_{-0.0\%}$ & $ ^{+0.0\%}_{-0.0\%}$ & $ ^{+0.0\%}_{-0.0\%}$ & $ ^{+0.0\%}_{-0.0\%}$& $ ^{+1.0\%}_{-1.0\%}$\\ $\Hgaga$
& 126  &9.59\zehomi{3}        &$^{+0.0\%}_{-0.0\%}$ & $ ^{+0.0\%}_{-0.0\%}$ & $ ^{+0.0\%}_{-0.0\%}$ & $ ^{+0.0\%}_{-0.0\%}$& $ ^{+1.0\%}_{-1.0\%}$\\
& 130  &1.10\zehomi{2}        &$^{+0.1\%}_{-0.0\%}$ & $ ^{+0.0\%}_{-0.0\%}$ & $ ^{+0.0\%}_{-0.0\%}$ & $ ^{+0.0\%}_{-0.0\%}$& $ ^{+1.0\%}_{-1.0\%}$\\ 
\hline
& 122  &4.74\zehomi{3}        &$^{+0.0\%}_{-0.1\%}$ & $ ^{+0.0\%}_{-0.0\%}$ & $ ^{+0.0\%}_{-0.0\%}$ & $ ^{+0.0\%}_{-0.1\%}$& $ ^{+5.0\%}_{-5.0\%}$\\ $\HZga$
& 126  &6.84\zehomi{3}        &$^{+0.0\%}_{-0.0\%}$ & $ ^{+0.0\%}_{-0.0\%}$ & $ ^{+0.0\%}_{-0.1\%}$ & $ ^{+0.0\%}_{-0.1\%}$& $ ^{+5.0\%}_{-5.0\%}$\\
& 130  &9.55\zehomi{3}        &$^{+0.0\%}_{-0.0\%}$ & $ ^{+0.0\%}_{-0.0\%}$ & $ ^{+0.0\%}_{-0.0\%}$ & $ ^{+0.0\%}_{-0.0\%}$& $ ^{+5.0\%}_{-5.0\%}$\\ 
\hline
& 122  &6.25\zehomi{1}        &$^{+0.0\%}_{-0.0\%}$ & $ ^{+0.0\%}_{-0.0\%}$ & $ ^{+0.0\%}_{-0.0\%}$ & $ ^{+0.0\%}_{-0.0\%}$& $ ^{+0.5\%}_{-0.5\%}$\\ $\HWW$
& 126  &9.73\zehomi{1}        &$^{+0.0\%}_{-0.0\%}$ & $ ^{+0.0\%}_{-0.0\%}$ & $ ^{+0.0\%}_{-0.0\%}$ & $ ^{+0.0\%}_{-0.0\%}$& $ ^{+0.5\%}_{-0.5\%}$\\
& 130  &1.49\zehoze           &$^{+0.0\%}_{-0.0\%}$ & $ ^{+0.0\%}_{-0.0\%}$ & $ ^{+0.0\%}_{-0.0\%}$ & $ ^{+0.0\%}_{-0.0\%}$& $ ^{+0.5\%}_{-0.5\%}$\\ 
\hline
& 122  &7.30\zehomi{2}        &$^{+0.0\%}_{-0.0\%}$ & $ ^{+0.0\%}_{-0.0\%}$ & $ ^{+0.0\%}_{-0.0\%}$ & $ ^{+0.0\%}_{-0.0\%}$& $ ^{+0.5\%}_{-0.5\%}$\\ $\HZZ$
& 126  &1.22\zehomi{1}        &$^{+0.0\%}_{-0.0\%}$ & $ ^{+0.0\%}_{-0.0\%}$ & $ ^{+0.0\%}_{-0.0\%}$ & $ ^{+0.0\%}_{-0.0\%}$& $ ^{+0.5\%}_{-0.5\%}$\\
& 130  &1.95\zehomi{1}        &$^{+0.0\%}_{-0.0\%}$ & $ ^{+0.0\%}_{-0.0\%}$ & $ ^{+0.0\%}_{-0.0\%}$ & $ ^{+0.0\%}_{-0.0\%}$& $ ^{+0.5\%}_{-0.5\%}$\\ 
\hline

\end{tabular}\end{center}
\end{table}
%%%%%%%%%%%%%%%%%%%%%%%%%% T A B L E %%%%%%%%%%%%%%%%%%%%%%%%%%%%%%%%%%%%%%%%

%%%%%%%%%%%%%%%%%%%%%%%%%%%%%%%%%%%%%%%%%%%%%%%%%%%%%%%%%%%%%%%%%%%%%%%%%%%%%%%

\subsection{Differential prediction for the final state 
$\PH\to \PW\PW/\PZ\PZ\to4\Pf$}
\label{sec:br-differential}

In this section, we discuss differential distributions for $\PH\to
\PW\PW/\PZ\PZ\to4\Pf$ as calculated with \linebreak {\Prophecy} for a SM
Higgs boson with mass $\MH=126 \UGeV$. It is not our goal to provide
an analysis of the role of differential distributions
in the measurement of Higgs-boson properties as done in other 
chapters of this report. Here, we merely want to emphasize the impact 
of NLO corrections and in particular the impact of interference effects 
on distributions. These interference effects have
already been discussed for the branching rations in \Bref{Dittmaier:2012vm}.
They arise when the final-state fermions can pair up in 
more than one way to form intermediate vector bosons. Therefore, they
are not included in any approximation which relies on factorizing the 
Higgs decays into a decay to vector bosons $\PH\to \PW\PW/\PZ\PZ$, where the
vector bosons have definite momenta, and successive 
vector-boson decays $\PW/\PZ\to 2\Pf$. In contrast, they are included in {\Prophecy} which 
is based on the full $\PH\to4\Pf$ matrix elements including all interferences 
between different Feynman diagrams. To anticipate the results of this 
section, NLO corrections become important at the level of 5\% accuracy, while the
(LO) interference effects can distort distributions by more than 10\%. 

To be specific, we
exemplarily analyze the following differential distributions for a Higgs
decay with four charged leptons in the final state, for which 
the Higgs-boson rest frame is assumed to be reconstructed:
\begin{itemize}
\item
In the Higgs-boson rest frame, we investigate $\cos \theta_{\Pf^-\Pf^-}$,
where $\theta_{\Pf^-\Pf^-}$ is the angle between the two negatively charged 
leptons. This angle is unambiguously defined in any of the final states $\PH
\to 4\Pe$, $\PH \to 4\PGm$, and $\PH \to 2\PGm2\Pe$ so that interference
effects can be easily studied.
\item
In the Higgs-boson rest frame, we investigate the angle $\phi\prime$ 
between the two decay planes of the vector bosons, where
\[
\cos \phi\prime = 
\frac{({\bf k}_{12} \times {\bf k}_{1})\cdot({\bf k}_{12} \times {\bf k}_{3})}
     {|{\bf k}_{12} \times {\bf k}_{1}||{\bf k}_{12} \times {\bf k}_{3}|}
\quad
,
\quad
\sgn(\sin \phi\prime) = 
\sgn({\bf k}_{12} \cdot [({\bf k}_{12} \times {\bf k}_{1})\times({\bf k}_{12} \times {\bf k}_{3})]) \, ,
\quad
\]
and ${\bf k}_{12} = {\bf k}_{1} + {\bf k}_{2}$. In turn, ${\bf k}_{1,2}$ are
the three-momenta of the fermions forming the fermion pair which is closest in invariant mass to
an  on-shell \PZ-boson. Moreover, ${\bf k}_{1}$ and ${\bf k}_{3}$ correspond
to the momenta of the negatively charged fermions. For $\PH \to 2\PGm2\Pe$,
the fermion momenta could, of course,  be associated to intermediate \PZ\
bosons unambiguously without kinematic information simply according to their
flavour. However, only the
kinematic selection is possible for  $\PH \to 4\Pe$ and $\PH \to 4\PGm$.
Since we want to compare the different channels to investigate interference
effects, we use the kinematic identification for all channels. 
\item
For the fermion pair that is closest in invariant mass to
an on-shell \PZ\ boson, we investigate $\cos \Theta_{\PV \Pf^-}$, where
$\Theta_{\PV \Pf^-}$ is the angle between the vector boson momentum
(the sum of the momenta of the fermion pair) in the Higgs rest frame 
and the momentum of the negatively charged fermion associated to the 
vector boson in the vector-boson rest frame.
\end{itemize}
For a Higgs decay with 2 charged leptons and two neutrinos in the final state
the Higgs rest frame is not reconstructable. However, we assume that the
transverse momentum of the  Higgs boson is known so that one can boost into
the frame with vanishing transverse momentum. In this frame, we 
investigate
\begin{itemize}
\item 
the angle between the two charged leptons 
$\phi_{\Pf^-\Pf^+,\mathrm{T}}$ in the transverse plane, i.e.\
\[
\cos \phi_{\Pf^-\Pf^+,\mathrm{T}} = 
\frac{{\bf k}_{f^+,\mathrm{T}}\cdot{\bf k}_{f^-,\mathrm{T}}}
     {|{\bf k}_{f^+,\mathrm{T}}||{\bf k}_{f^-,\mathrm{T}}|}
\quad
\mathrm{and}
\quad
\sgn(\sin \phi_{\Pf^-\Pf^+,\mathrm{T}}) = 
\sgn({\bf e}_z \cdot ({\bf k}_{f^+,\mathrm{T}}\times{\bf k}_{f^-,\mathrm{T}})) \, ,
\quad
\]
where ${\bf k}_{f^\pm,\mathrm{T}}$ denote the transverse momentum vectors of
the charged fermions and ${\bf e}_z$ is the unit vector along one of the
beams in the lab frame. 
\end{itemize} 

\begin{figure}
\includegraphics[width=0.3\textwidth]{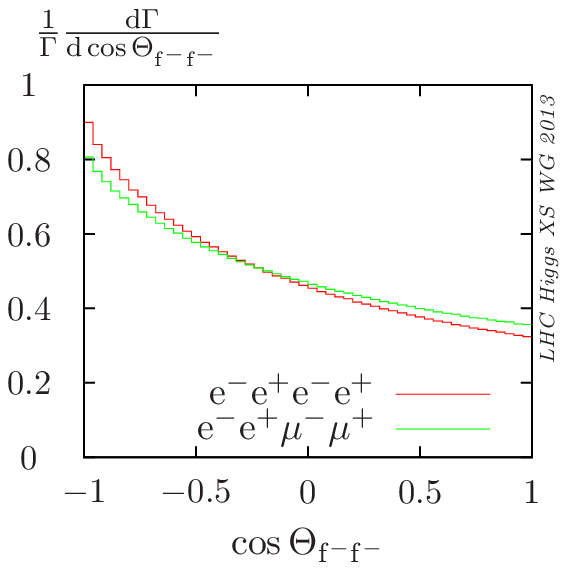}
\hfill
\includegraphics[width=0.3\textwidth]{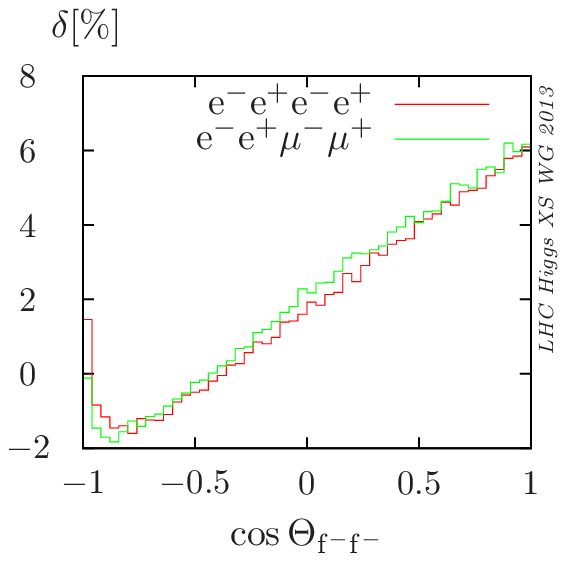}
\hfill
\includegraphics[width=0.3\textwidth]{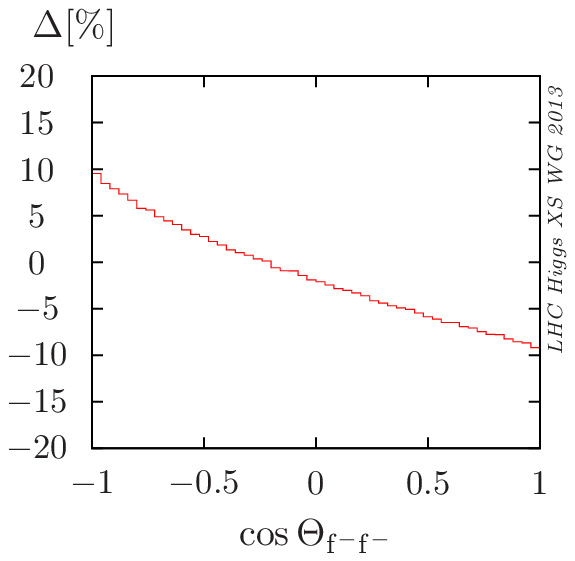}

\includegraphics[width=0.3\textwidth]{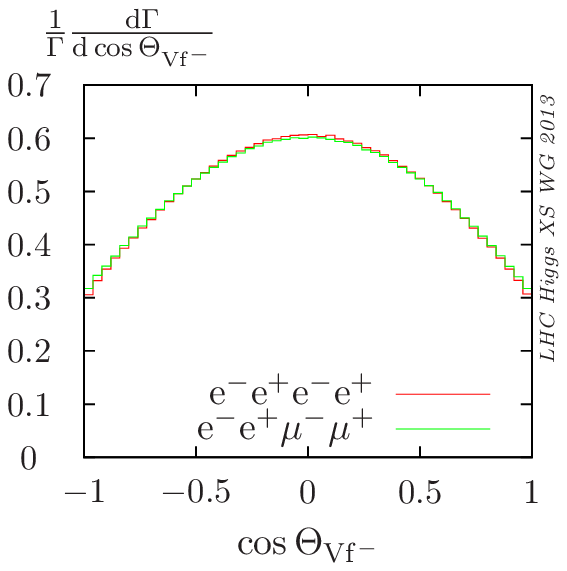}
\hfill
\includegraphics[width=0.3\textwidth]{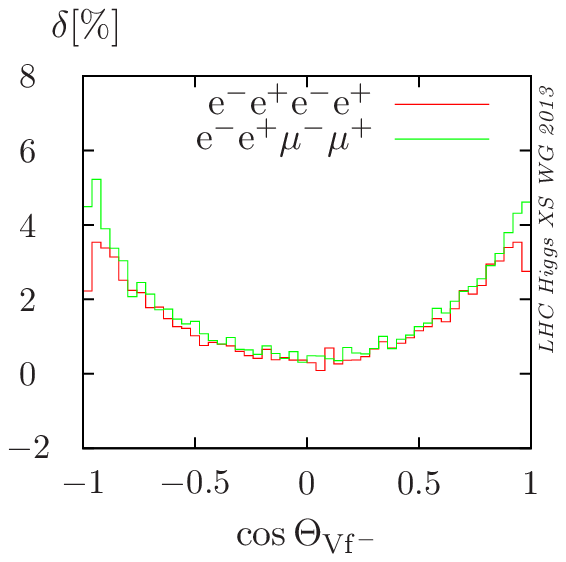}
\hfill
\includegraphics[width=0.3\textwidth]{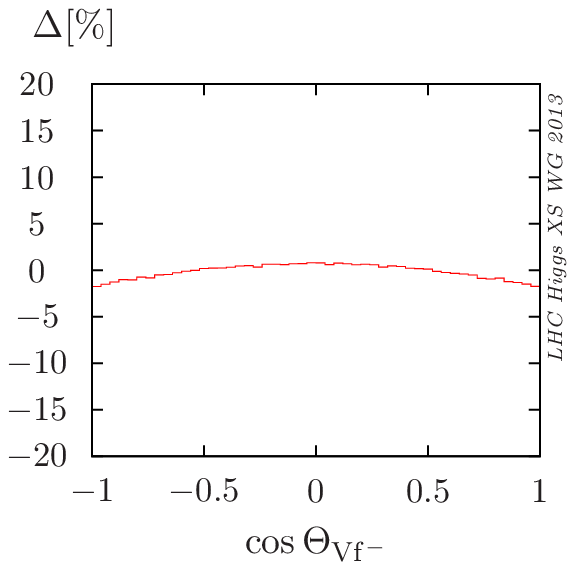}

\includegraphics[width=0.3\textwidth]{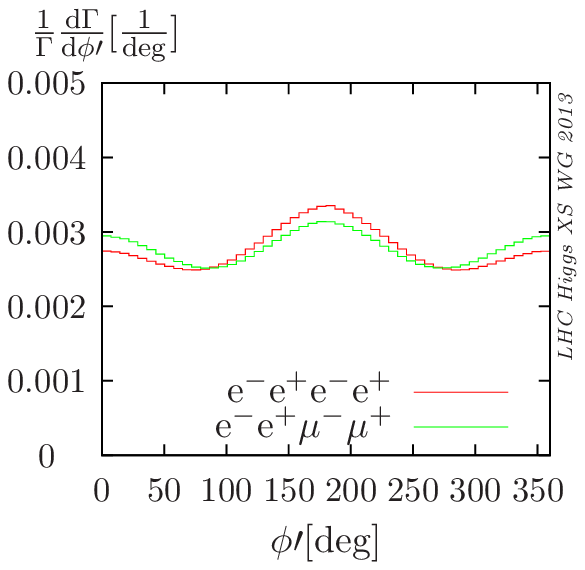}
\hfill
\includegraphics[width=0.3\textwidth]{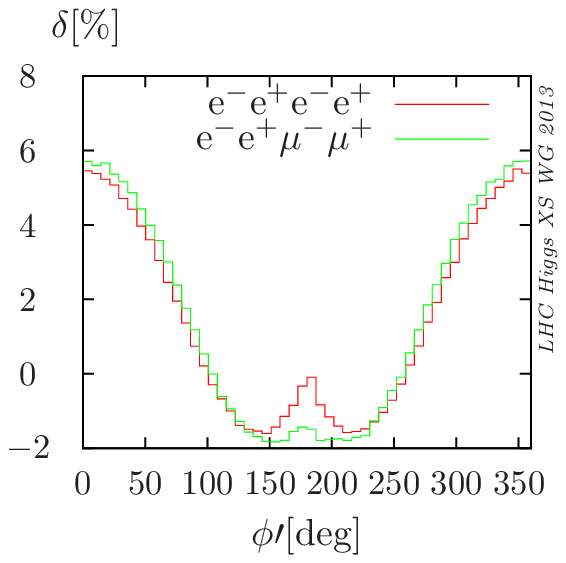}
\hfill
\includegraphics[width=0.3\textwidth]{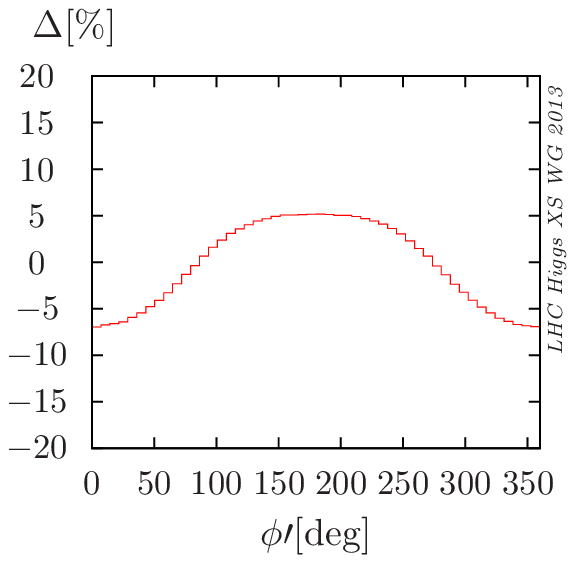}

\includegraphics[width=0.3\textwidth]{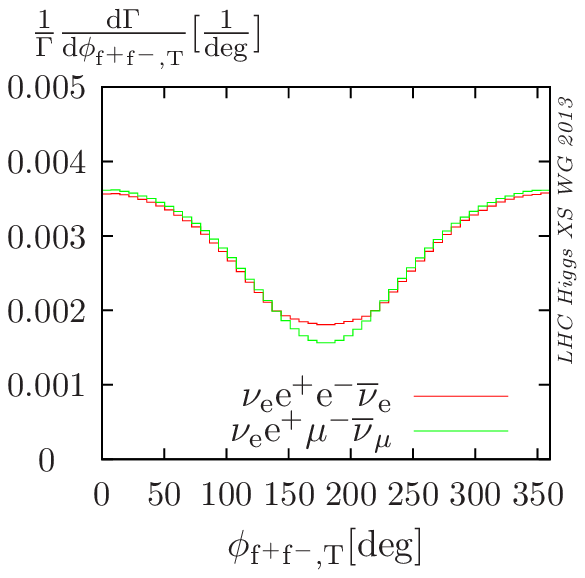}
\hfill
\includegraphics[width=0.3\textwidth]{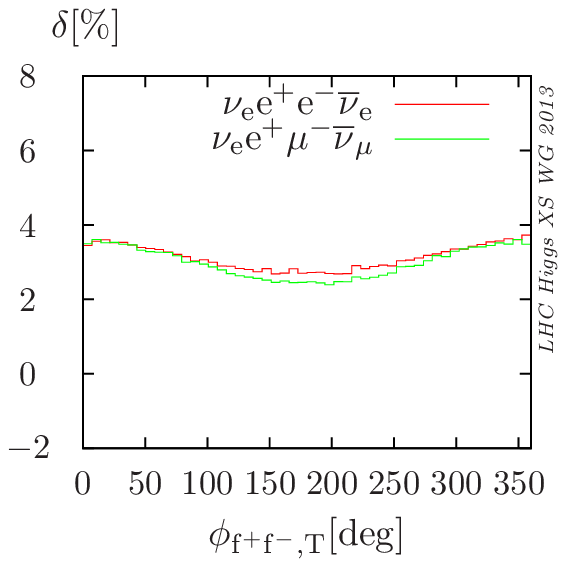}
\hfill
\includegraphics[width=0.3\textwidth]{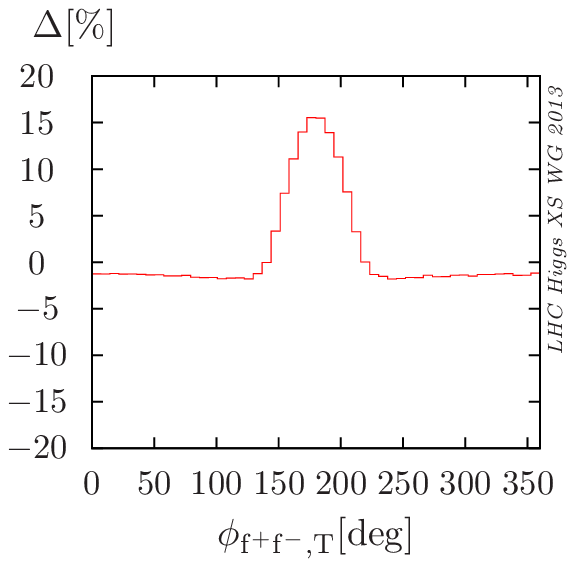}
\vspace*{-0.3cm}
\caption{\label{fig:BR_distributions} The normalized distributions for different observables 
(left column) is shown together with the relative NLO corrections to the 
unnormalized distributions (middle column) and 
the relative difference of the two investigated channels $\Delta$ due to 
interference effects (right column) for $\MH=126 \UGeV$ (calculated with {\Prophecy}).}
\end{figure}

Similar differential distributions have been analyzed with {\Prophecy}
already in \Bref{Bredenstein:2006rh} where additional discussions can be
found. In \refF{fig:BR_distributions} (left column), we show the 
distributions, normalized to the corresponding partial width, with four charged 
leptons in the final state for a Higgs-boson
mass $\MH=126 \UGeV$ at NLO accuracy for $\PH \to 4\Pe$ and $\PH \to
2\PGm2\Pe$. The electrons and positrons in $\PH \to 4\Pe$ can form 
intermediate \PZ\ bosons in two different ways and the corresponding 
amplitudes interfere. The interference contributions are absent for
fermion pairs of different flavour.
Concerning final states with neutrinos, we show the results for
the decay channels $\PH \to \PGne \Pe^+ \PAGne \Pe^-$ 
(with interference contributions) and 
$\PH \to \PGne \Pe^+ \PAGnGm \PGm$ (without interference contributions). 

%%%%%%%%%%%%%%%%%%% F I G U R E %%%%%%%%%%%%%%%%%%%%%%%%%%%%%%%%%%%%%%%%%%%%%%%
%%%%%%%%%%%%%%%%%%% F I G U R E %%%%%%%%%%%%%%%%%%%%%%%%%%%%%%%%%%%%%%%%%%%%%%%

Also in \refF{fig:BR_distributions} (middle column), we show the relative NLO
corrections $\delta$ to the different distributions (not to the normalized distributions
to also show the overall effects on the partial width). To
be precise, in the presence of bremsstrahlung photons, the fermion momenta
are defined after recombination with the photon if the invariant mass of the
lepton--photon pair is smaller than $5 \UGeV$. We use the invariant mass as a
criterion for recombination because it is independent of the lab frame 
which depends on the Higgs-production process. 
The results are similar to the NLO corrections
shown in \Bref{Bredenstein:2006rh} for different Higgs-boson masses. It is
also evident that the NLO corrections for channels with and without
interference are similar. 

Finally, in the right column of \refF{fig:BR_distributions}, the relative
difference 
\[
\Delta=\left(\frac{1}{\Gamma^{\mathrm{LO}}_\mathrm{w \, int.}}
        \frac{\mathrm{d} \Gamma^{\mathrm{LO}}_\mathrm{w. \, int.}}{\mathrm{d} \mathcal{O}} -
	\frac{1}{\Gamma^{\mathrm{LO}}_\mathrm{wo \, int.}}
        \frac{\mathrm{d} \Gamma^{\mathrm{LO}}_\mathrm{wo \, int.}}{\mathrm{d} \mathcal{O}}
	\right)/
	\left(\frac{1}{\Gamma^{\mathrm{LO}}_\mathrm{wo \, int.}}
        \frac{\mathrm{d} \Gamma^{\mathrm{LO}}_\mathrm{wo \, int.}}{\mathrm{d} \mathcal{O}}
	\right)
\]
between the channels with interference 
$\Gamma^{\mathrm{LO}}_\mathrm{w \, int.}$ and without interference 
$\Gamma^{\mathrm{LO}}_\mathrm{wo \, int.}$ is shown for the
various observables $\mathcal{O}$ introduced above. 
To stress, that the interference is already present at 
LO, we show the relative difference between the channels at LO.
However, the difference at NLO would hardly differ as can be seen from
the similarity of the NLO corrections in the middle column. 

When aiming at an accuracy at the 10\% level,
the interference effects cannot be neglected any more. In particular, any
approximation to the full process based on intermediate vector-bosons 
which neglects interference effects, must fail 
at this level of accuracy for final states like $\PH \to 4\Pe$ and
$\PH \to \PGne \Pep \PAGne \Pe$. Note that the interference effects as well 
as the NLO corrections distort the distributions and could be mistaken for anomalous
couplings in precision measurements if they are not taken into account in the predictions.
	
%%%%%%%%%%%%%%%%%%%%%%%%%%%%%%%%%%%%%%%%%%%%%%%%%%%%%%%%%%%%%%%%%%%%%%%%%%%%%%%

\subsection{MSSM Branching Ratios}
\label{sec:br-mssm}

In the MSSM the evaluation of cross sections and of branching ratios
have several common issues as outlined in ~\refS{sec:Introduction-sub} (see
also Sect.\,12.1 in \Bref{Dittmaier:2012vm}).
It was discussed that {\em before} any branching ratio calculation can be
performed in a first step the Higgs-boson masses, couplings and mixings
have to be evaluated from the underlying set of (soft SUSY-breaking)
parameters. For the case of real parameters in the MSSM the code 
(\FeynHiggs~\cite{Heinemeyer:1998yj,Heinemeyer:1998np,Degrassi:2002fi,Frank:2006yh,Hahn:2009zz}
was selected for the evaluations in this report.
(The case with complex parameters has not been investigated so far.)
The results for Higgs-boson masses and
couplings can be provided to other codes
(especially \HDECAY~\cite{Djouadi:1997yw,Spira:1997dg,hdecay2}) via the SUSY
Les Houches Accord~\cite{Skands:2003cj,Allanach:2008qq}. 

In the following subsections we describe how the relevant codes for the
calculation of partial decay widths, \FeynHiggs\ and \HDECAY, are
combined to give the most precise result for the Higgs-boson branching
ratios in the MSSM. Numerical results are shown for all MSSM Higgs
bosons (including the charged Higgs) within the updated benchmark
scenarios~\cite{Carena:2013qia}, see \refS{sec:sven-sub}.
It should be stressed that it would be desirable to 
interpret the model-independent results of various Higgs-boson searches
at the LHC also in other benchmark models, see for instance
\Bref{AbdusSalam:2011fc}. While we show exemplary plots for the
branching ratios of $\PA$ and $\PH^\pm$ for $\tb = 10, 50$,
detailed results for all Higgs BRs for $\tb = 0.5 \ldots 60$ and 
$\MA = 90 \UGeV \ldots 1000 \UGeV$ can be found in the working group web
page~\cite{MSSM-BR}.

%%%%%%%%%%%%%%%%%%%%%%%%%%%%%%%%%%%%%%%%%%%%%%%%%%%%%%%%%%%%%%%%%%%%%%%%%%%%%%%

\subsubsection{Combination of calculations}

After the calculation of Higgs-boson masses and mixings from the
original SUSY input the branching ratio calculation has to be
performed. 
%This can be done with the codes, \CPsuperH\ and
%\FeynHiggs\ for real or complex parameters, or
%\HDECAY\ for real parameters. The higher-order corrections included
%in the calculation of the various decay channels differ in the three
%codes. 
%
Here we concentrate on the MSSM with real parameters. We combine the
results from \HDECAY\ and \FeynHiggs\ on various decay channels to
obtain the most accurate result for the branching ratios currently
available. In a first
step, all partial widths have been calculated as accurately as
possible. Then the branching ratios have been derived from this full
set of partial widths. 
Concretely, we used \FeynHiggs\ for the evaluation of the
Higgs-boson masses and couplings from the original input
parameters, including corrections up to the two-loop level. 
The status of the various evaluations in \FeynHiggs\ and \HDECAY\ are
detailed in \Bref{Dittmaier:2012vm}.
The total decay width of the neutral Higgs bosons is calculated as,
\begin{align}
\Gamma_\phi &= 
  \Gamma^{\mathrm{FH}}_{\phitautau} 
+ \Gamma^{\mathrm{FH}}_{\phimumu} 
+ \Gamma^{\mathrm{FH/P4f}}_{\phiWW} 
+ \Gamma^{\mathrm{FH/P4f}}_{\phiZZ} \nonumber\\
&\quad 
+ \Gamma^{\mathrm{HD}}_{\phibb}
+ \Gamma^{\mathrm{HD}}_{\phitt}
+ \Gamma^{\mathrm{HD}}_{\phicc}
+ \Gamma^{\mathrm{HD}}_{\phigg}
+ \Gamma^{\mathrm{HD}}_{\phigaga}
+ \Gamma^{\mathrm{HD}}_{\phiZga}~,
\end{align}
followed by a corresponding evaluation of the respective branching
ratio. Decays to strange quarks or other lighter fermions have been neglected. 
Due to the somewhat different calculation compared to the SM case in
\refS{sec:br-strategy} no full decoupling of the decay widths and
branching ratios of the light MSSM Higgs to the respective SM values can
be expected.

The total decay width of the charged Higgs boson is calculated as,
\begin{align}
\Gamma_{\PH^\pm} &= 
  \Gamma^{\mathrm{FH}}_{\Hptaunu} 
+ \Gamma^{\mathrm{FH}}_{\Hpmunu} 
+ \Gamma^{\mathrm{FH}}_{\HphW} 
+ \Gamma^{\mathrm{FH}}_{\HpHW} 
+ \Gamma^{\mathrm{FH}}_{\HpAW} \nonumber\\
&\quad 
+ \Gamma^{\mathrm{HD}}_{\Hptb}
+ \Gamma^{\mathrm{HD}}_{\Hpts}
+ \Gamma^{\mathrm{HD}}_{\Hptd}
+ \Gamma^{\mathrm{HD}}_{\Hpcb}
+ \Gamma^{\mathrm{HD}}_{\Hpcs}
+ \Gamma^{\mathrm{HD}}_{\Hpcd} \nonumber \\
&\quad
+ \Gamma^{\mathrm{HD}}_{\Hpub}
+ \Gamma^{\mathrm{HD}}_{\Hpus}
+ \Gamma^{\mathrm{HD}}_{\Hpud}~,
\end{align}
followed by a corresponding evaluation of the respective branching
ratio. 

%%%%%%%%%%%%%%%%%%%%%%%%%%%%%%%%%%%%%%%%%%%%%%%%%%%%%%%%%%%%%%%%%%%%%%%%%%%%%%%

\subsubsection{Results in the new benchmark scenarios}

\begin{figure}[htb!]
%\vspace{3.2cm}
%\vspace*{-2.0cm}
\includegraphics[width=0.48\textwidth]{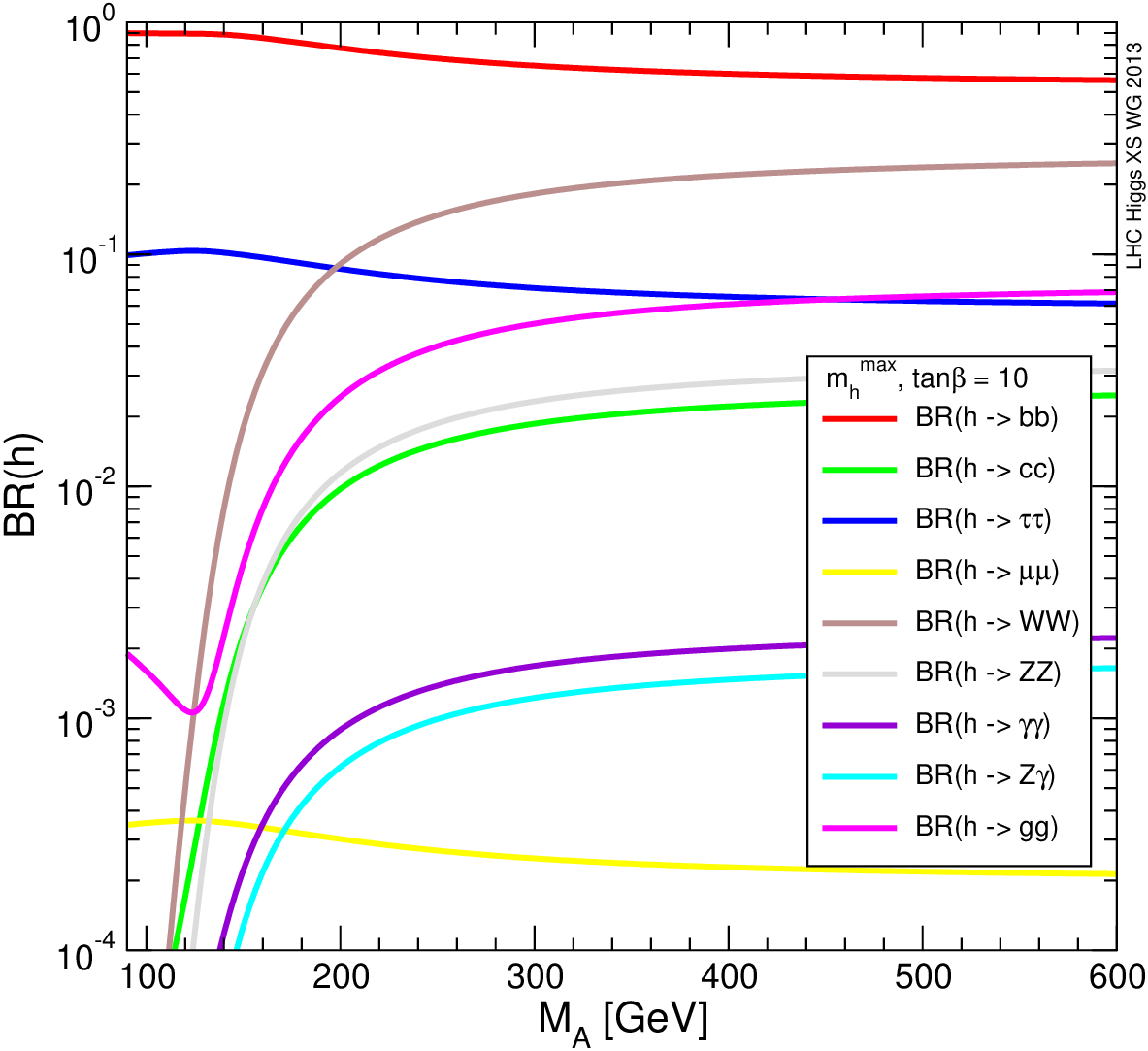}
\includegraphics[width=0.48\textwidth]{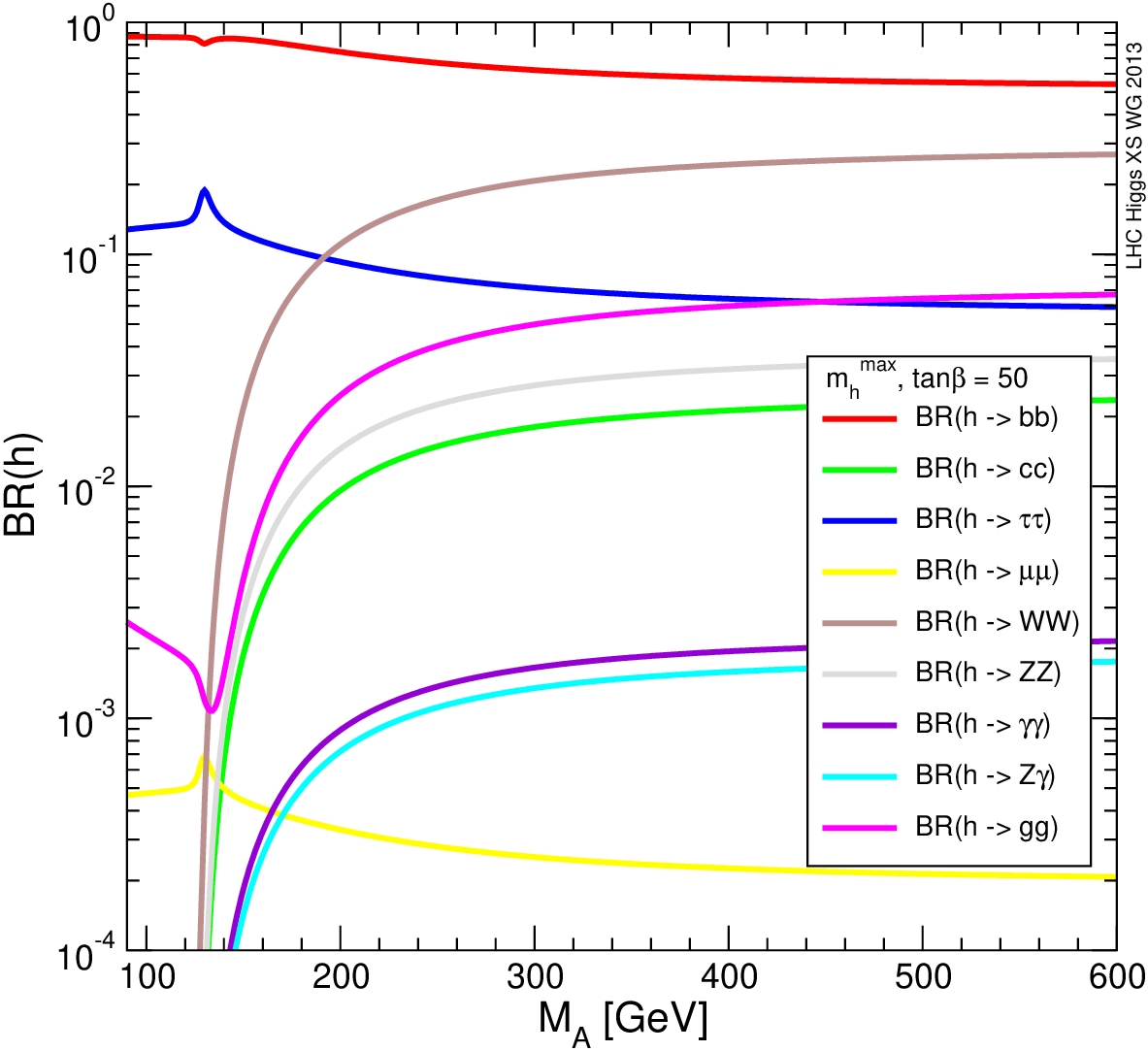}\\[.5em]
\includegraphics[width=0.48\textwidth]{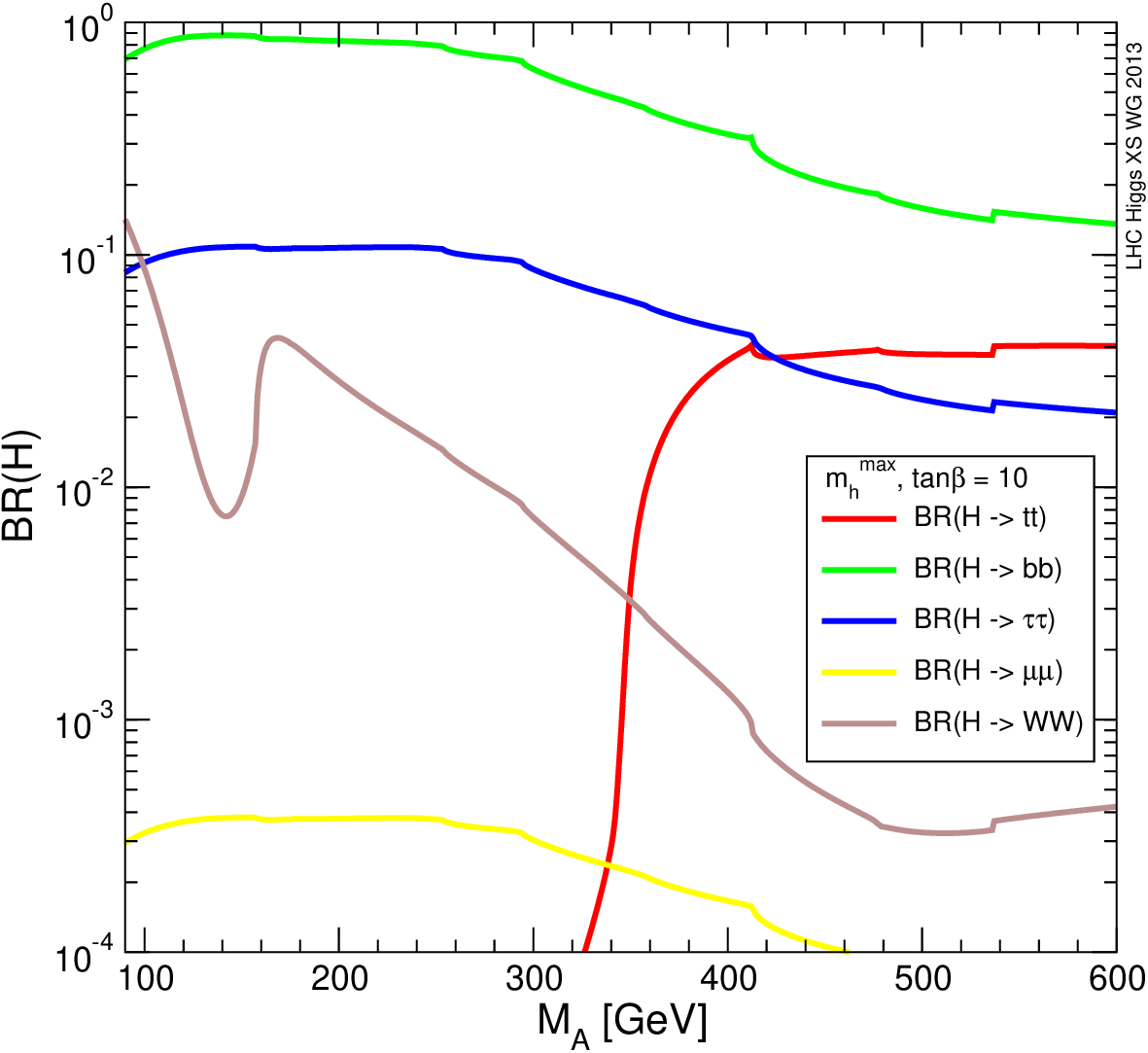}
\includegraphics[width=0.48\textwidth]{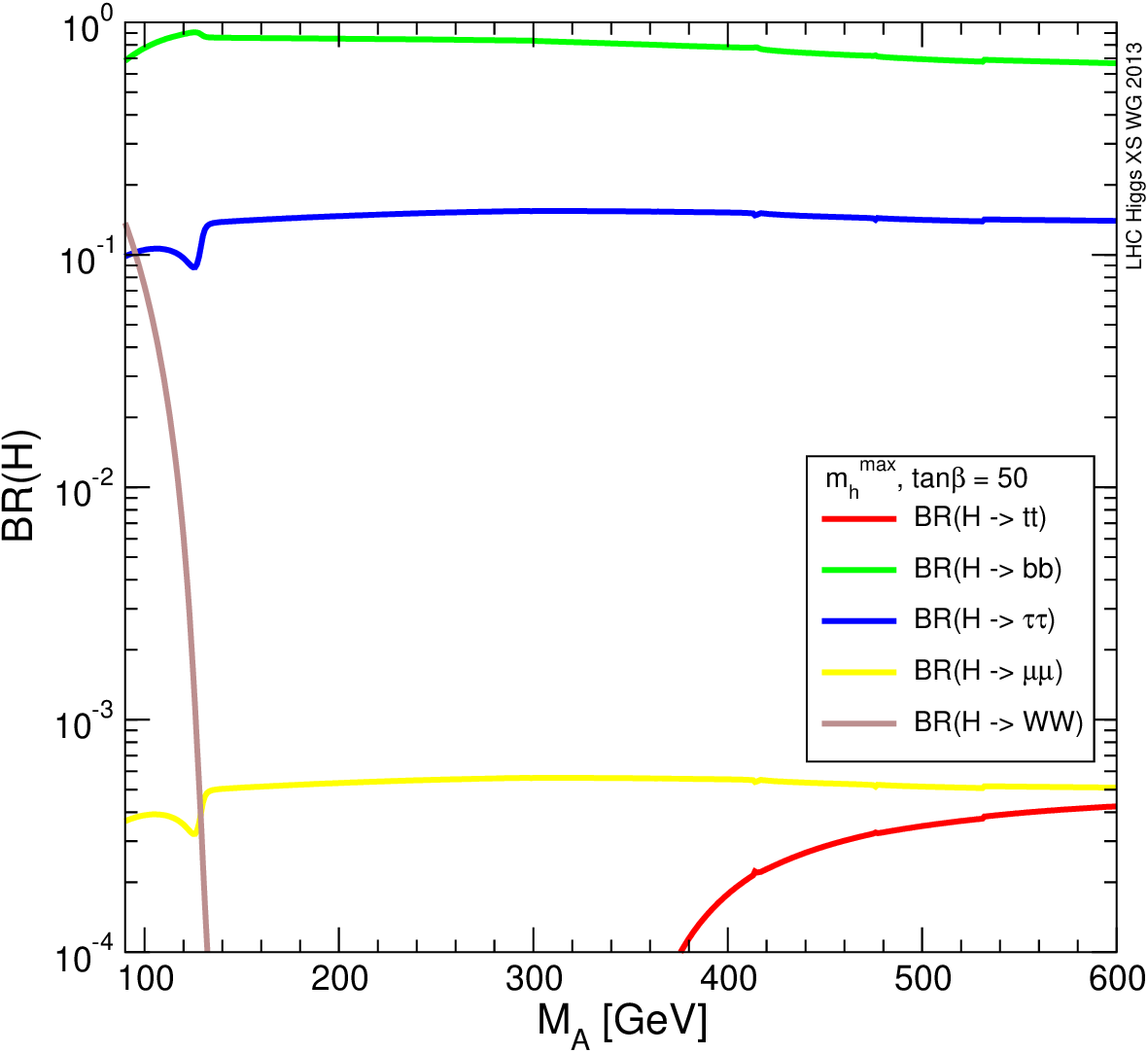}\\[0.5em]
\includegraphics[width=0.48\textwidth]{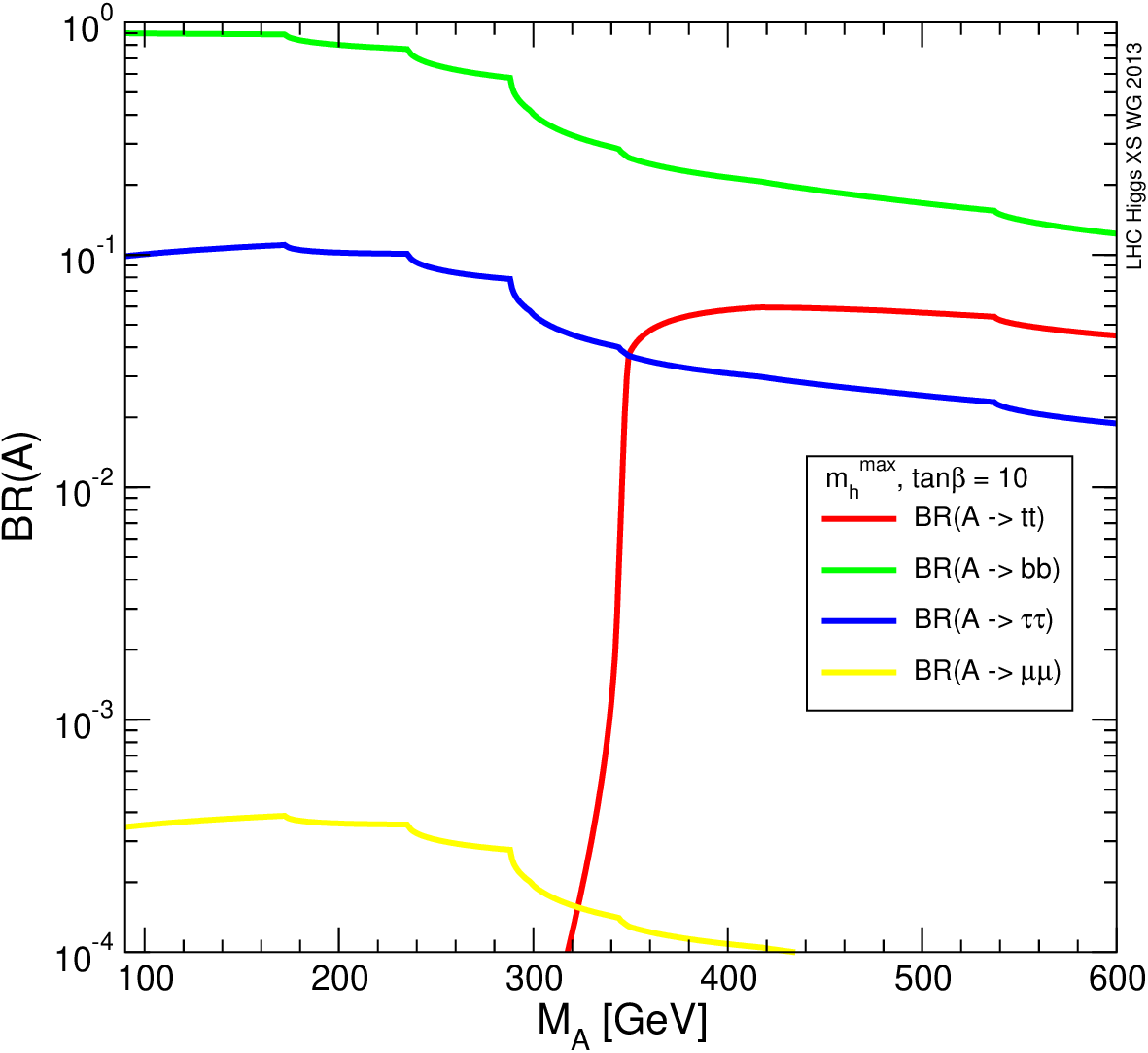}
\includegraphics[width=0.48\textwidth]{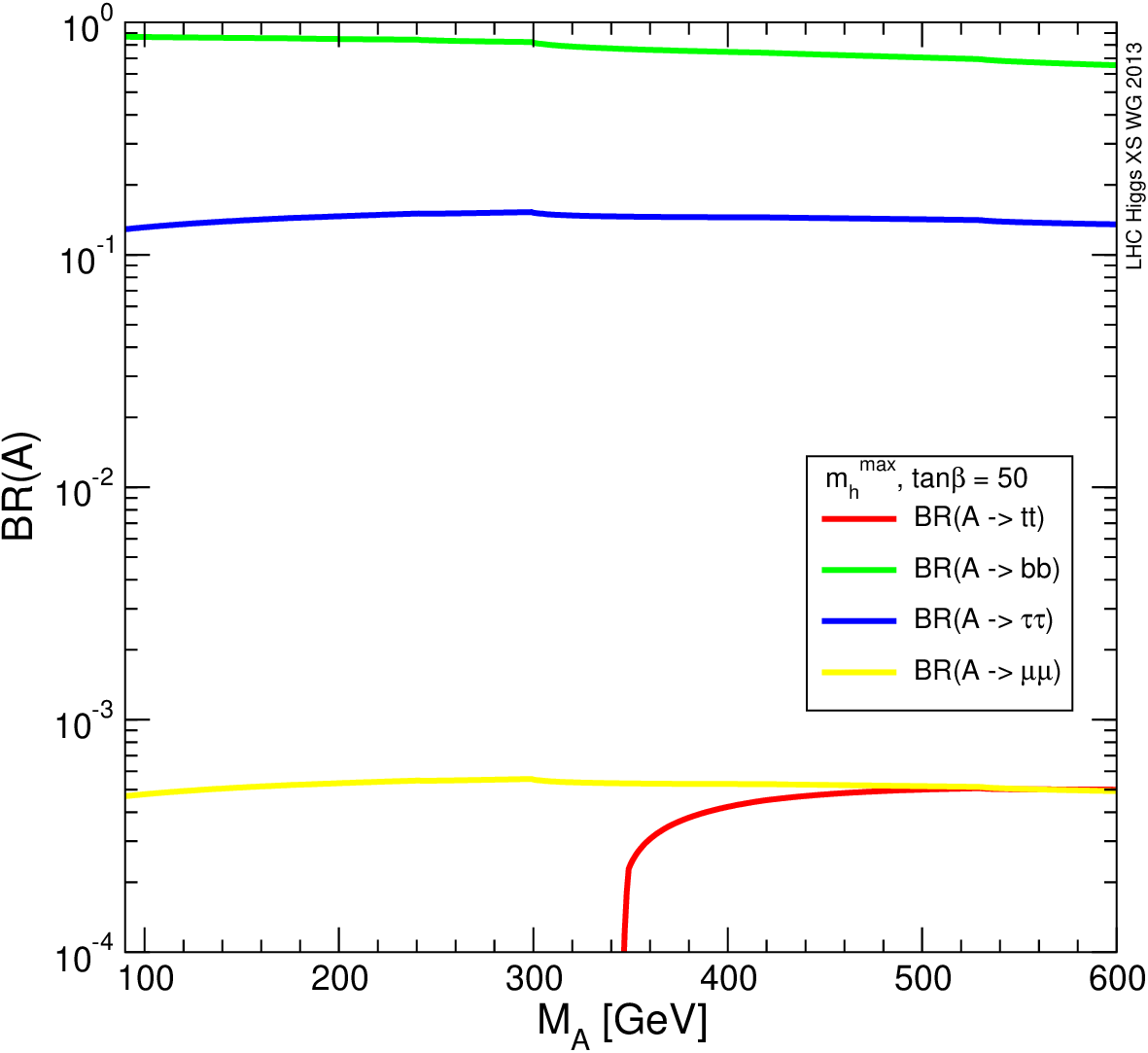}\\[0.5em]
%\vspace*{-0.3cm}
\caption{Branching ratios of the MSSM Higgs bosons $\Ph$ (upper row),
$\PH$ (middle row) and $\PA$ (lower row) in the \mhmaxx\
scenario as a function of $\MA$.
The left (right) column shows the results for $\tb = 10 (50)$.}
\label{fig:YRHXS3_BR_mhmax}
%\vspace*{-0.3cm}
%\vspace{3.2cm}
\end{figure}

\begin{figure}[htb!]
%\vspace{3.2cm}
\includegraphics[width=0.48\textwidth]{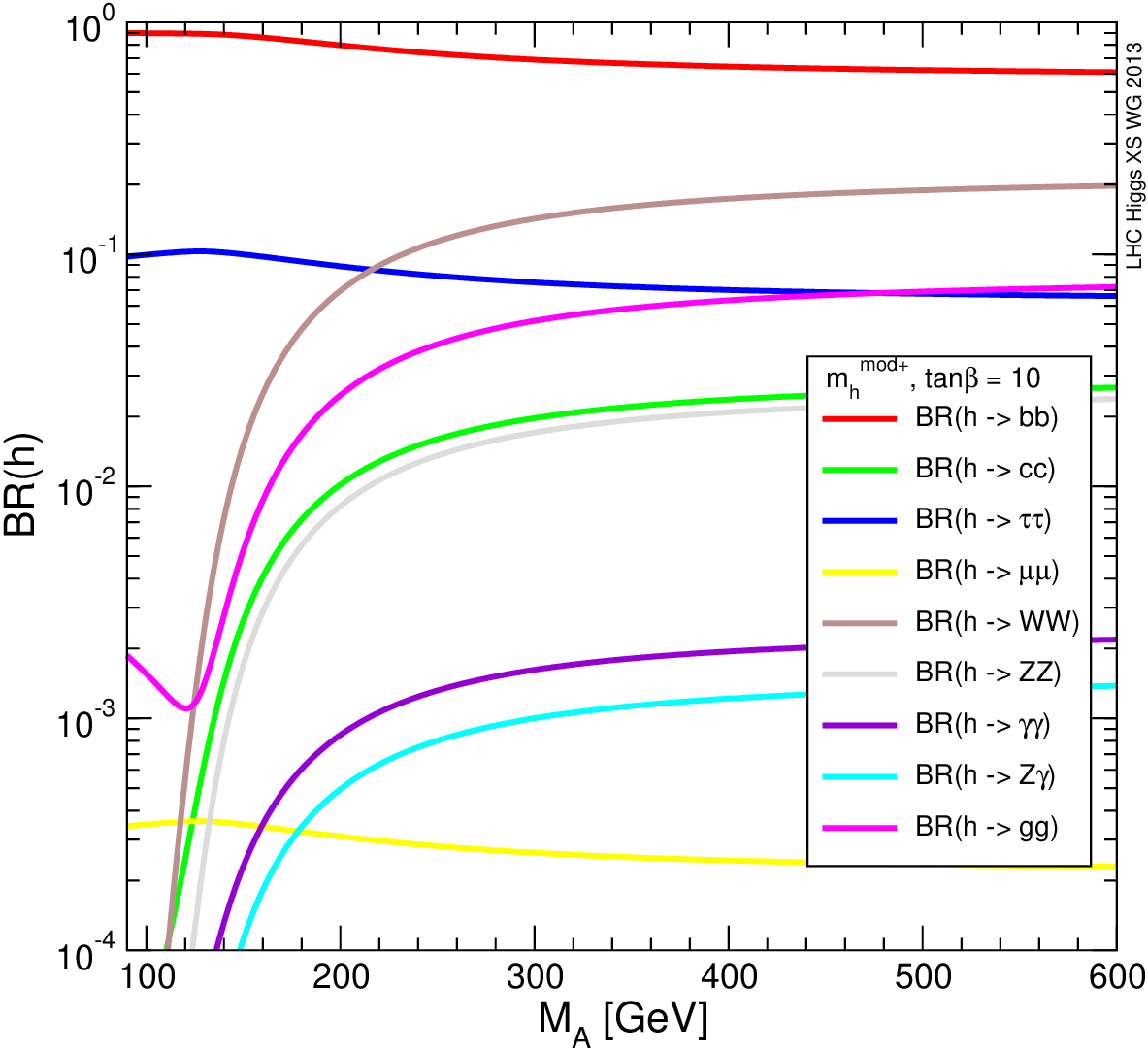}
\includegraphics[width=0.48\textwidth]{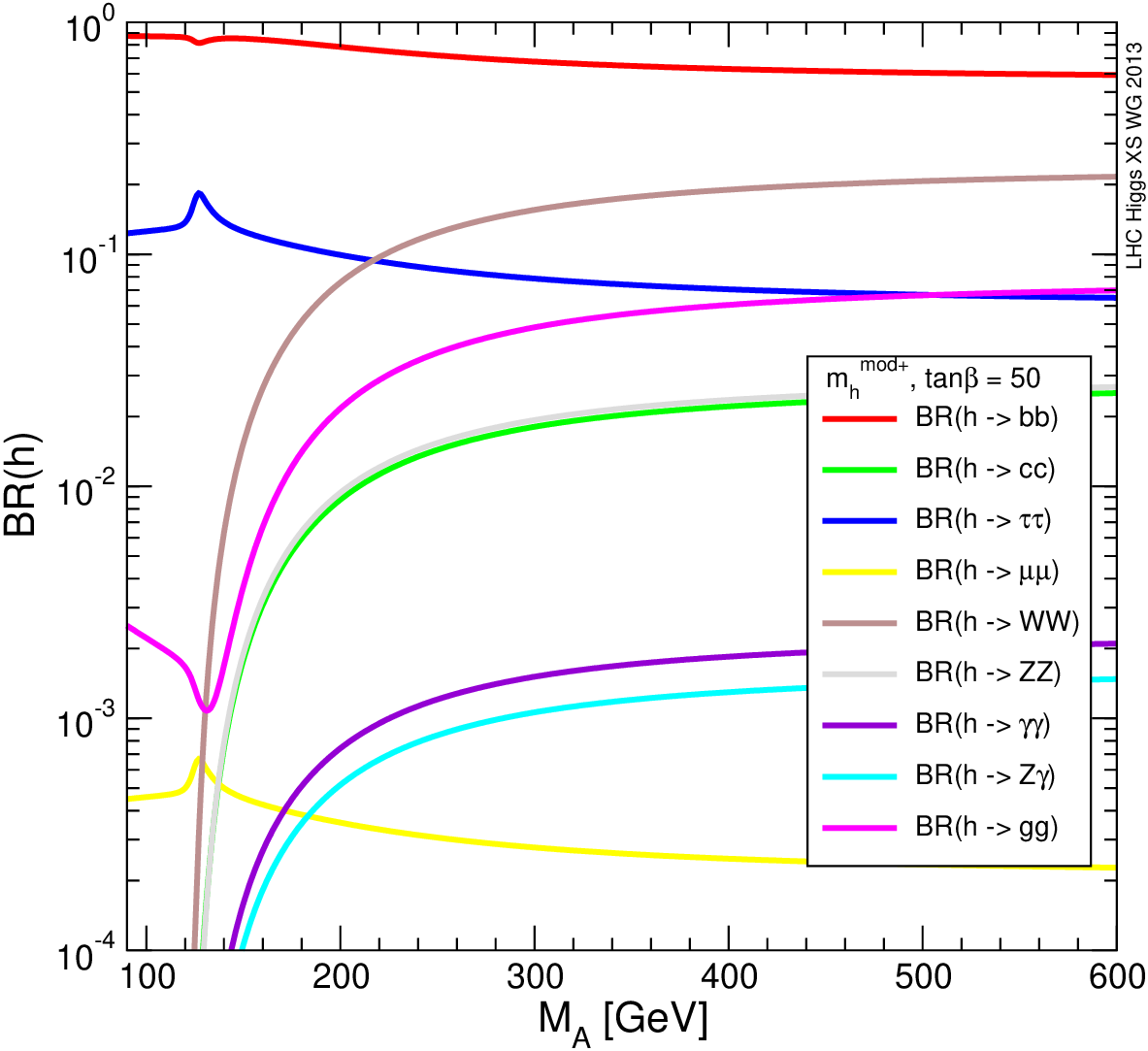}\\[.5em]
\includegraphics[width=0.48\textwidth]{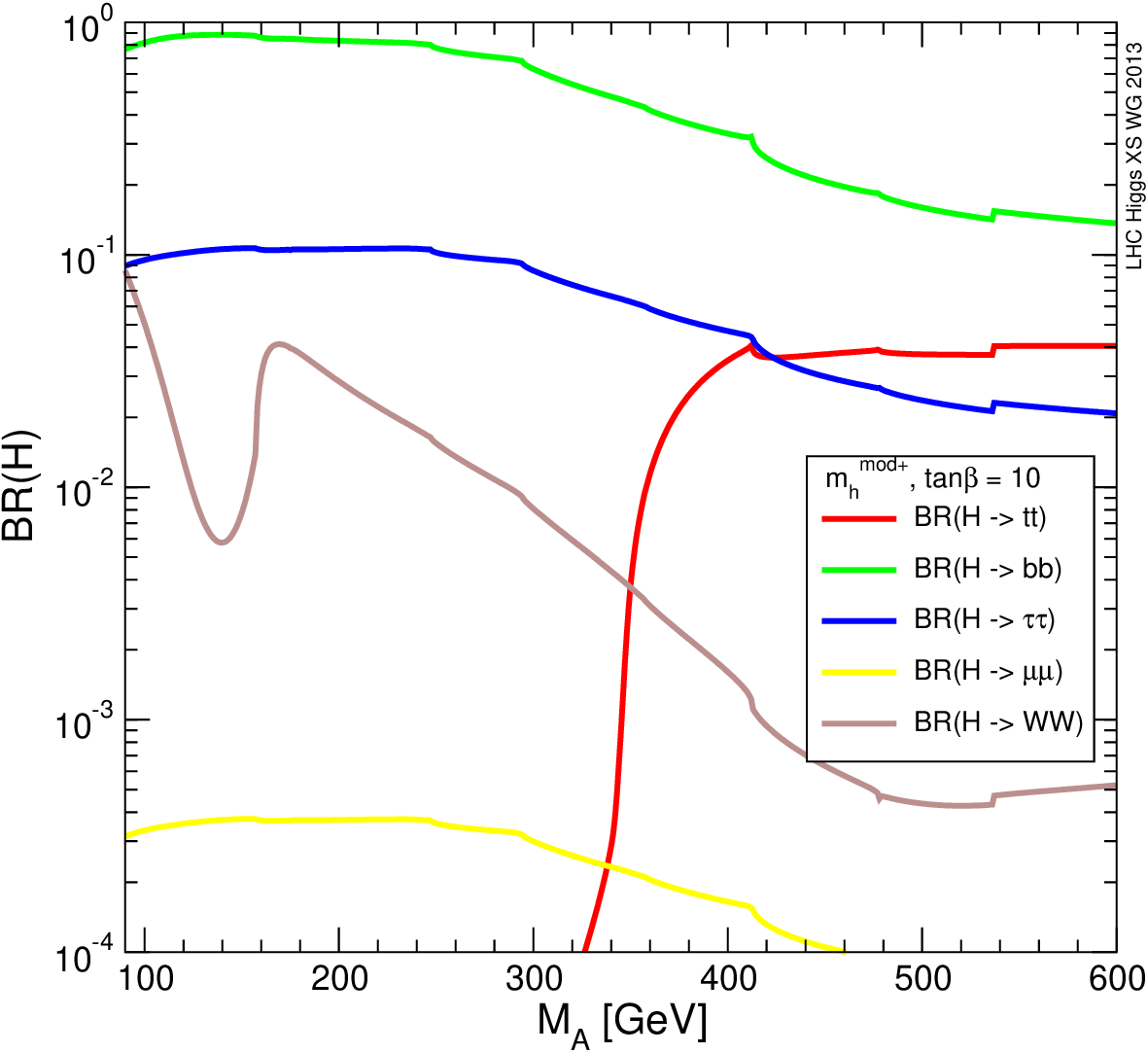}
\includegraphics[width=0.48\textwidth]{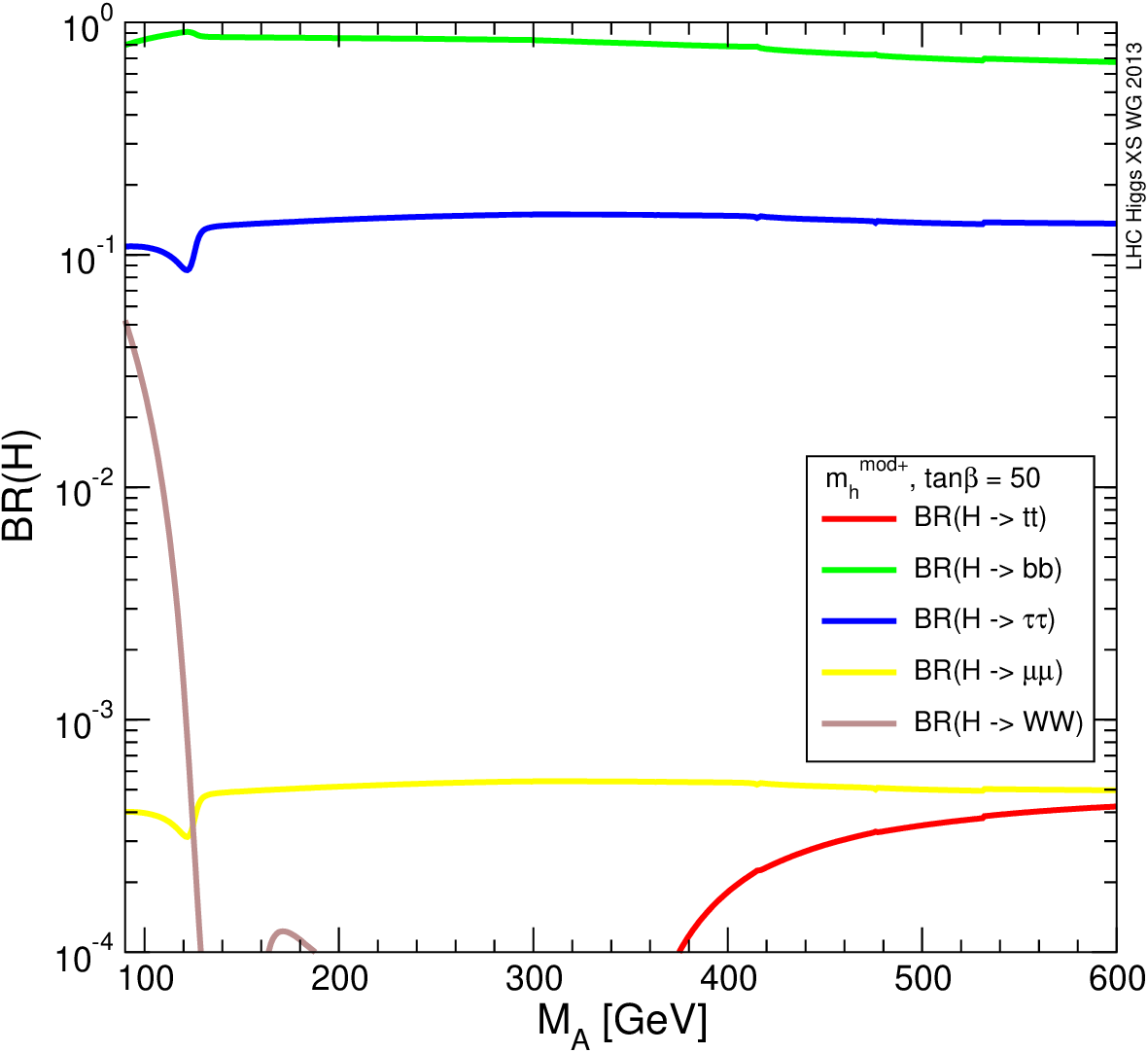}\\[0.5em]
\includegraphics[width=0.48\textwidth]{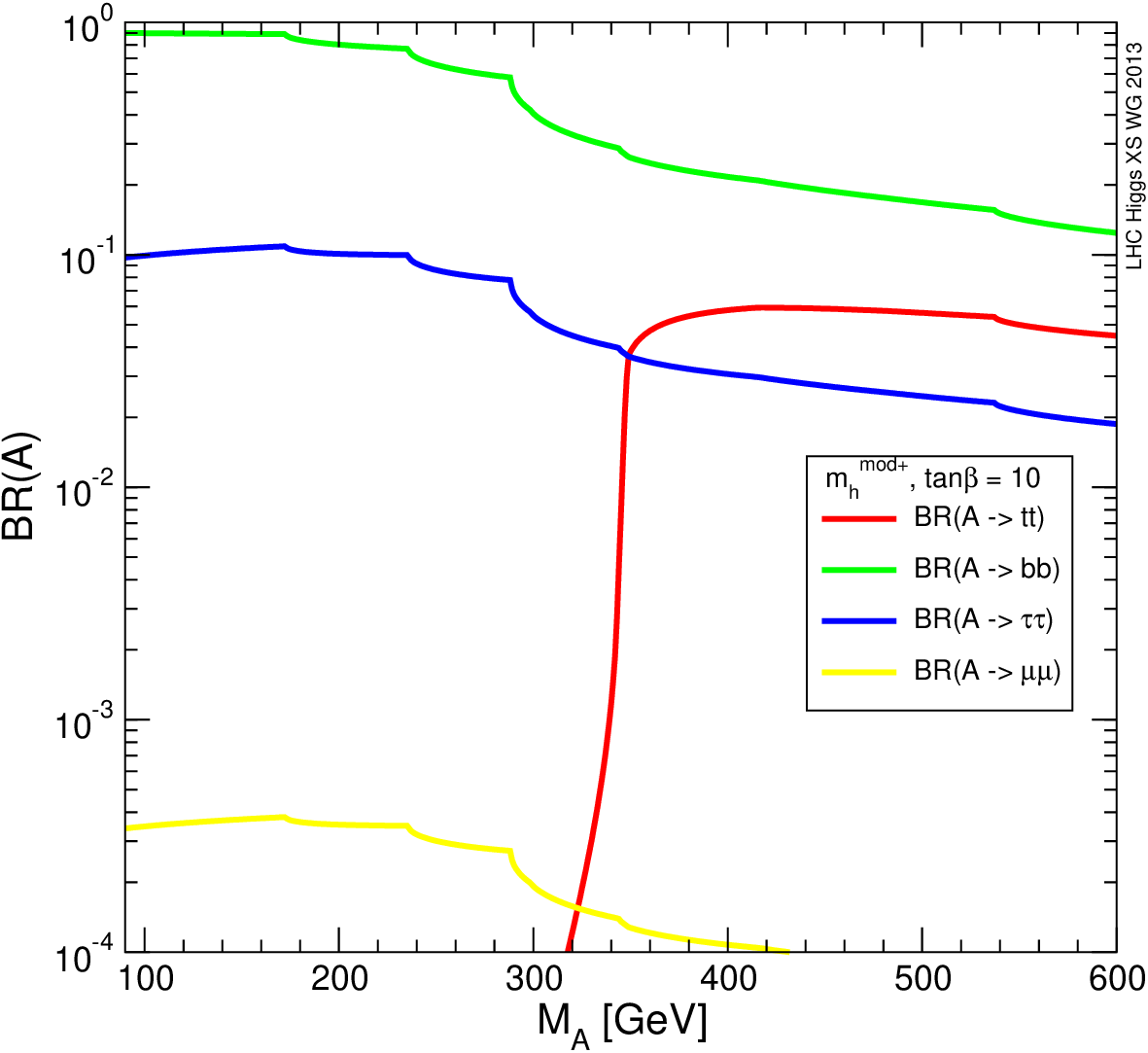}
\includegraphics[width=0.48\textwidth]{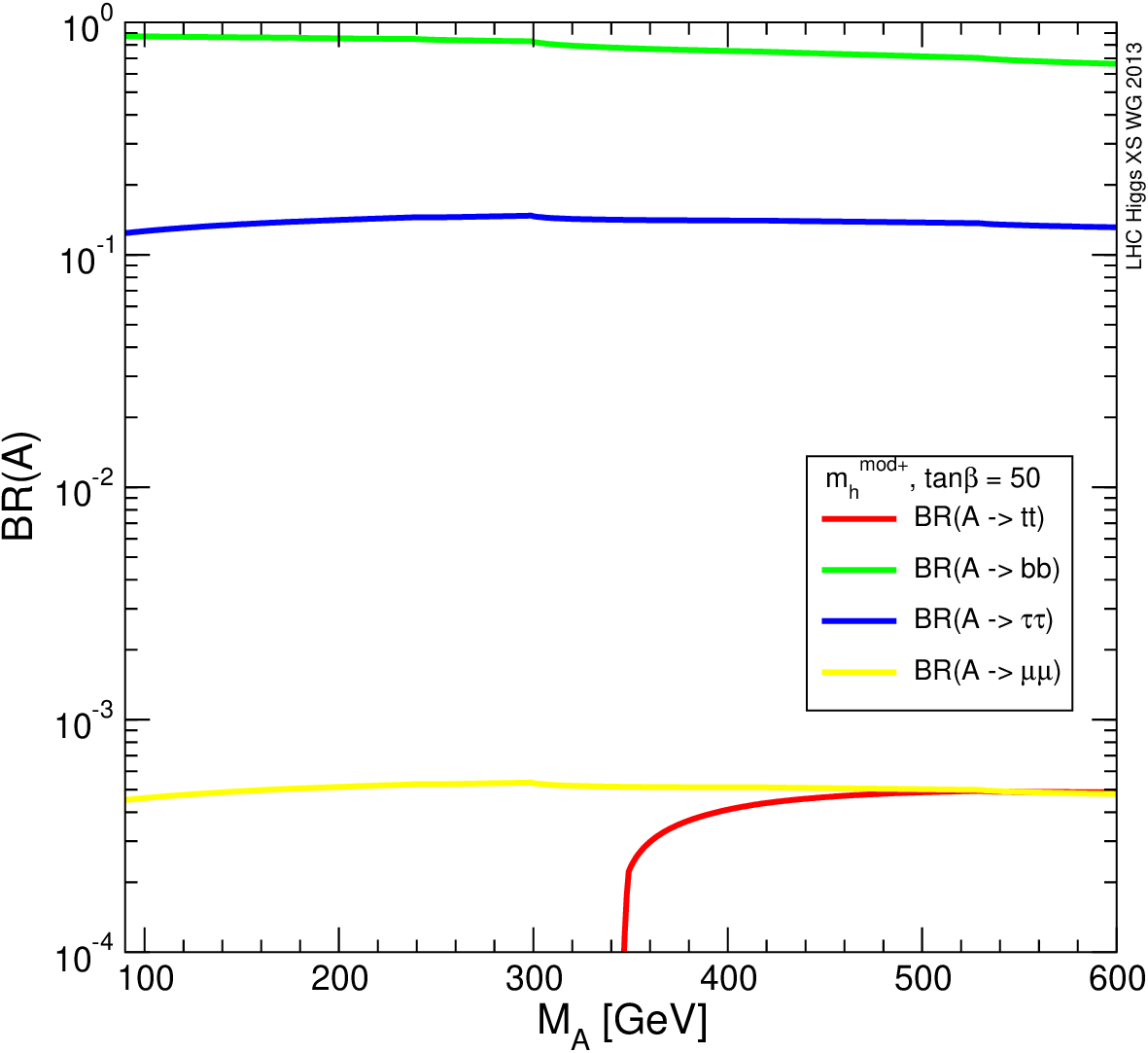}\\[0.5em]
%\vspace*{-0.3cm}
\caption{Branching ratios of the MSSM Higgs bosons $\Ph$ (upper row),
$\PH$ (middle row) and $\PA$ (lower row) in the \mhmodp\
scenario as a function of $\MA$.
The left (right) column shows the results for $\tb = 10 (50)$.}
\label{fig:YRHXS3_BR_mhmodp}
%\vspace{3.2cm}
\end{figure}

\begin{figure}[htb!]
%\vspace{3.2cm}
\includegraphics[width=0.48\textwidth]{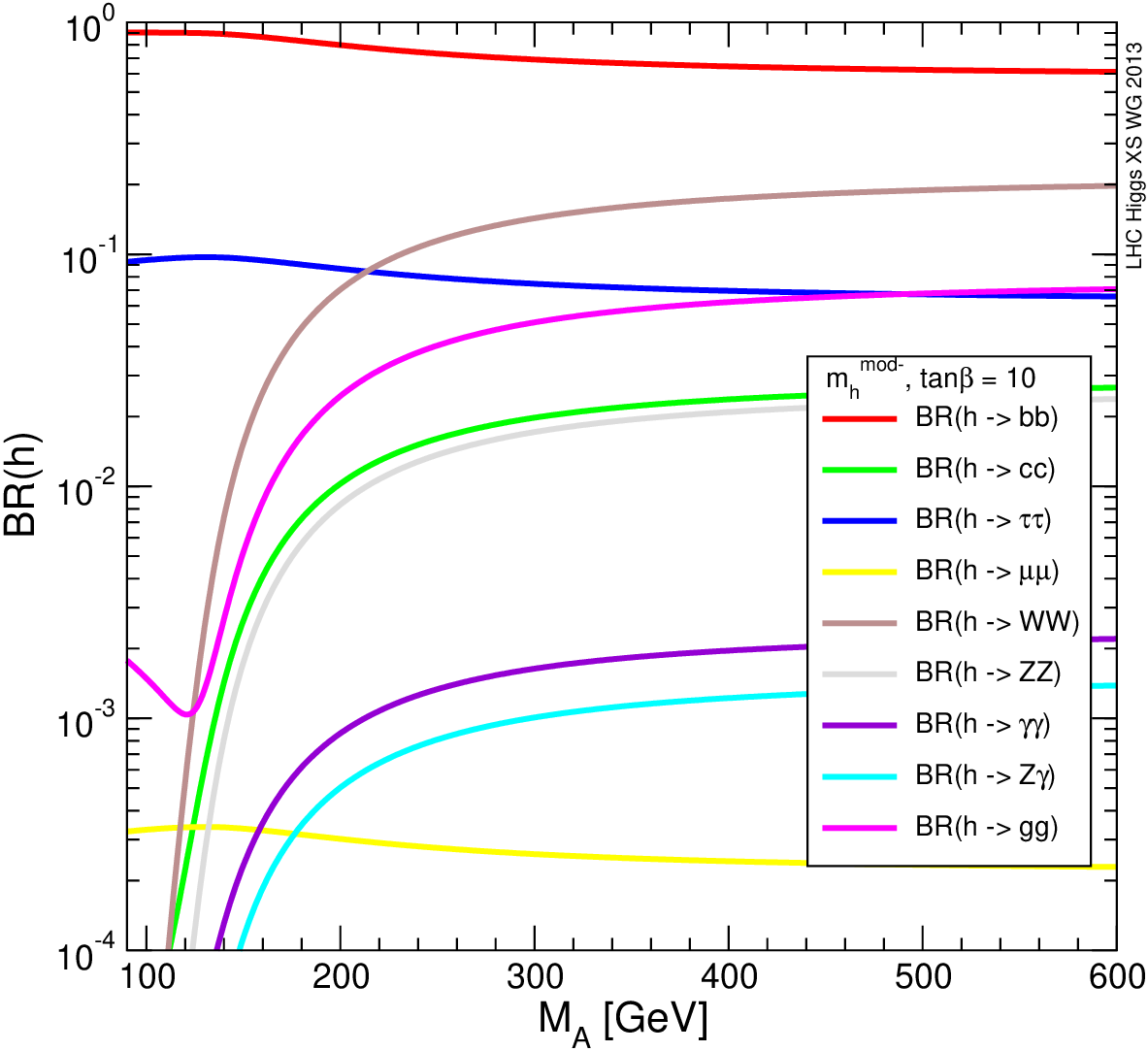}
\includegraphics[width=0.48\textwidth]{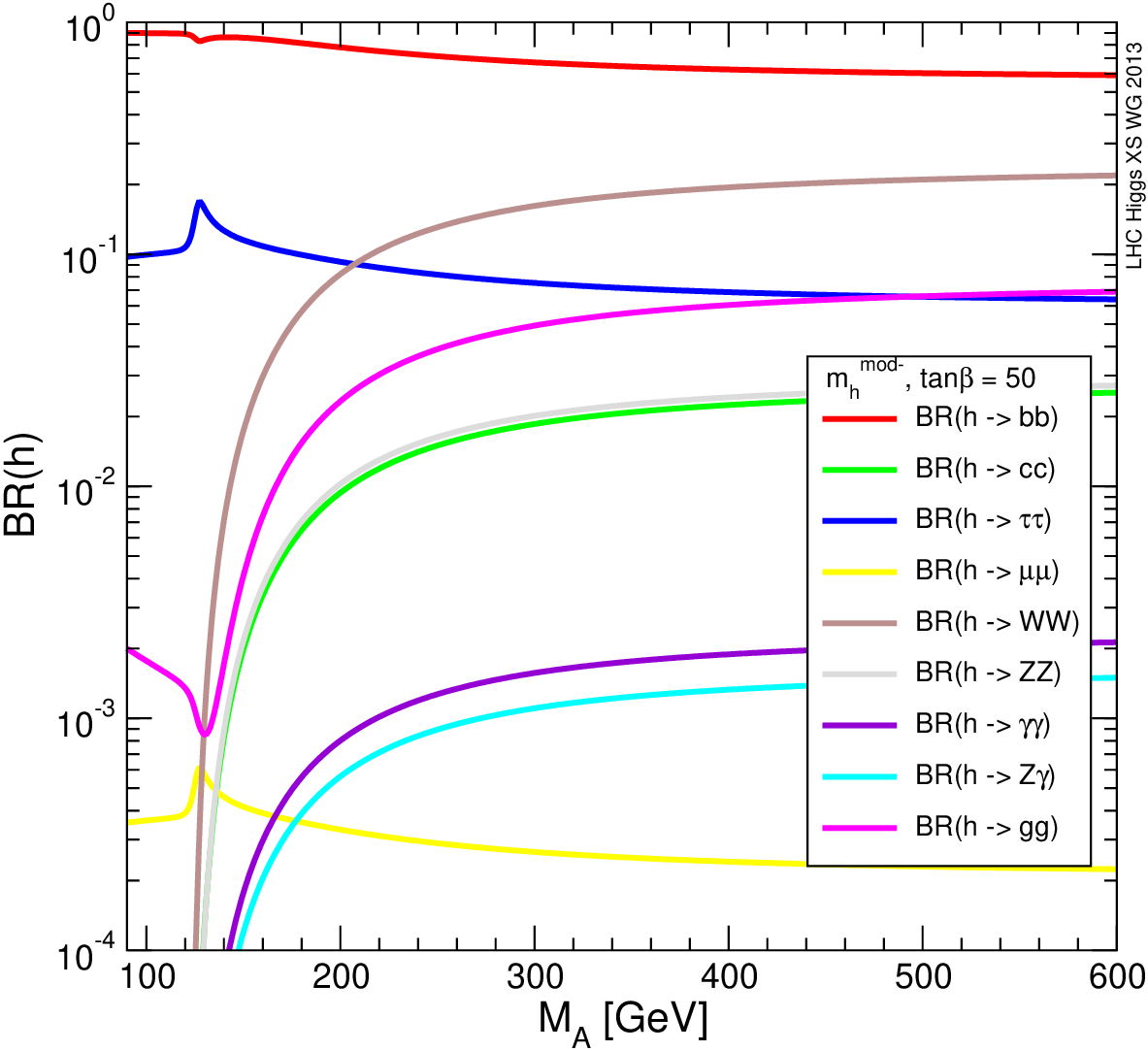}\\[.5em]
\includegraphics[width=0.48\textwidth]{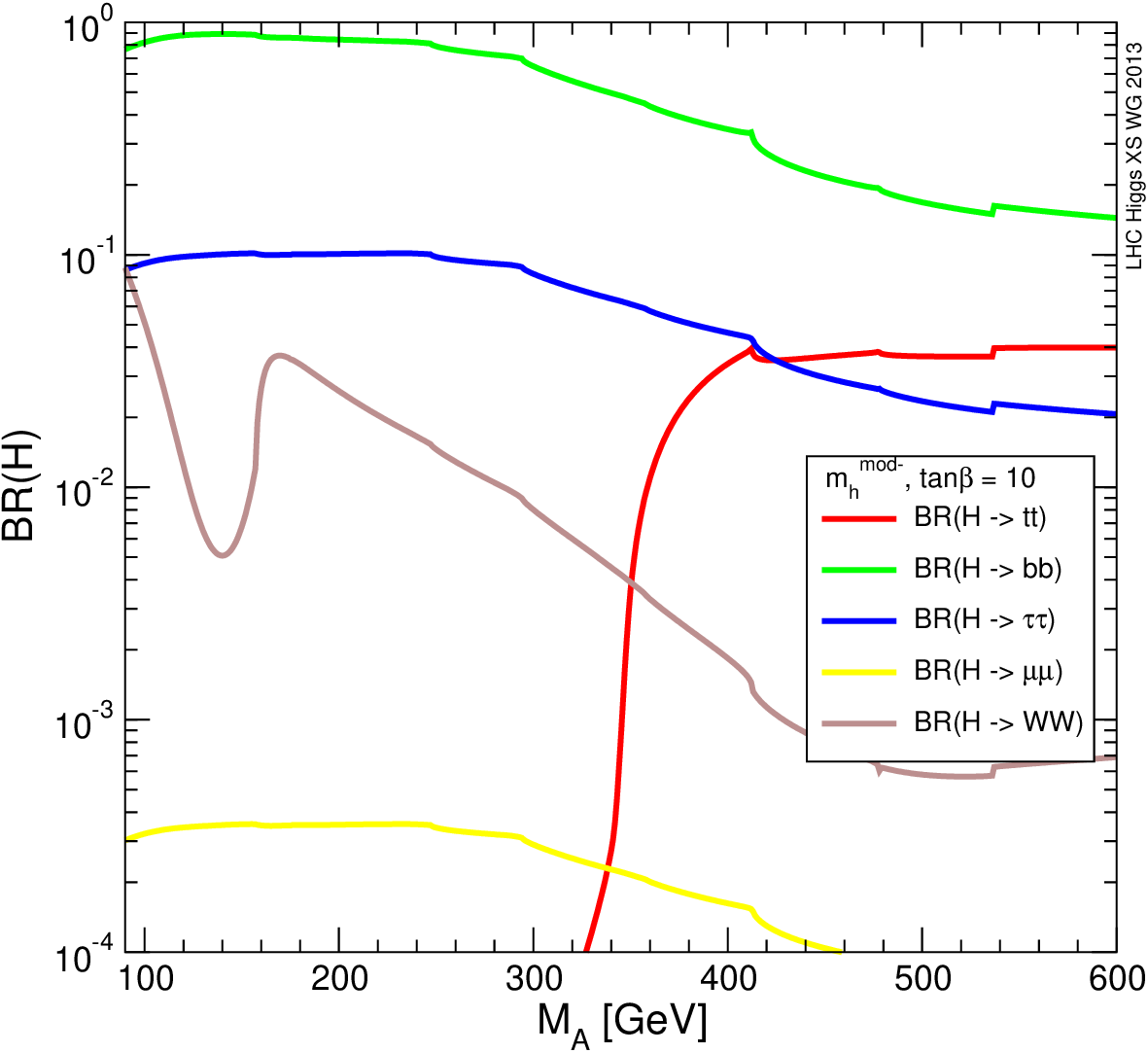}
\includegraphics[width=0.48\textwidth]{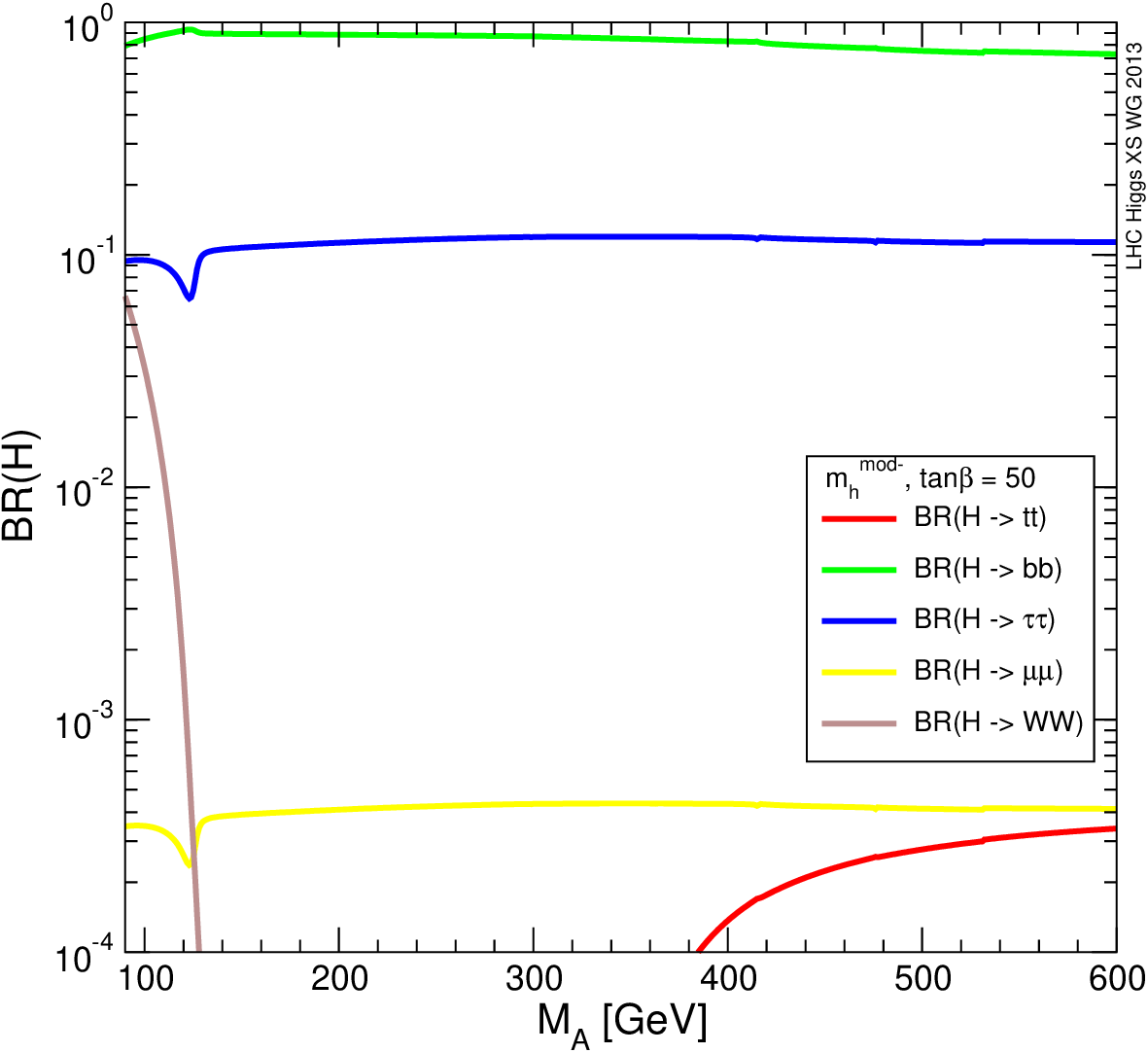}\\[0.5em]
\includegraphics[width=0.48\textwidth]{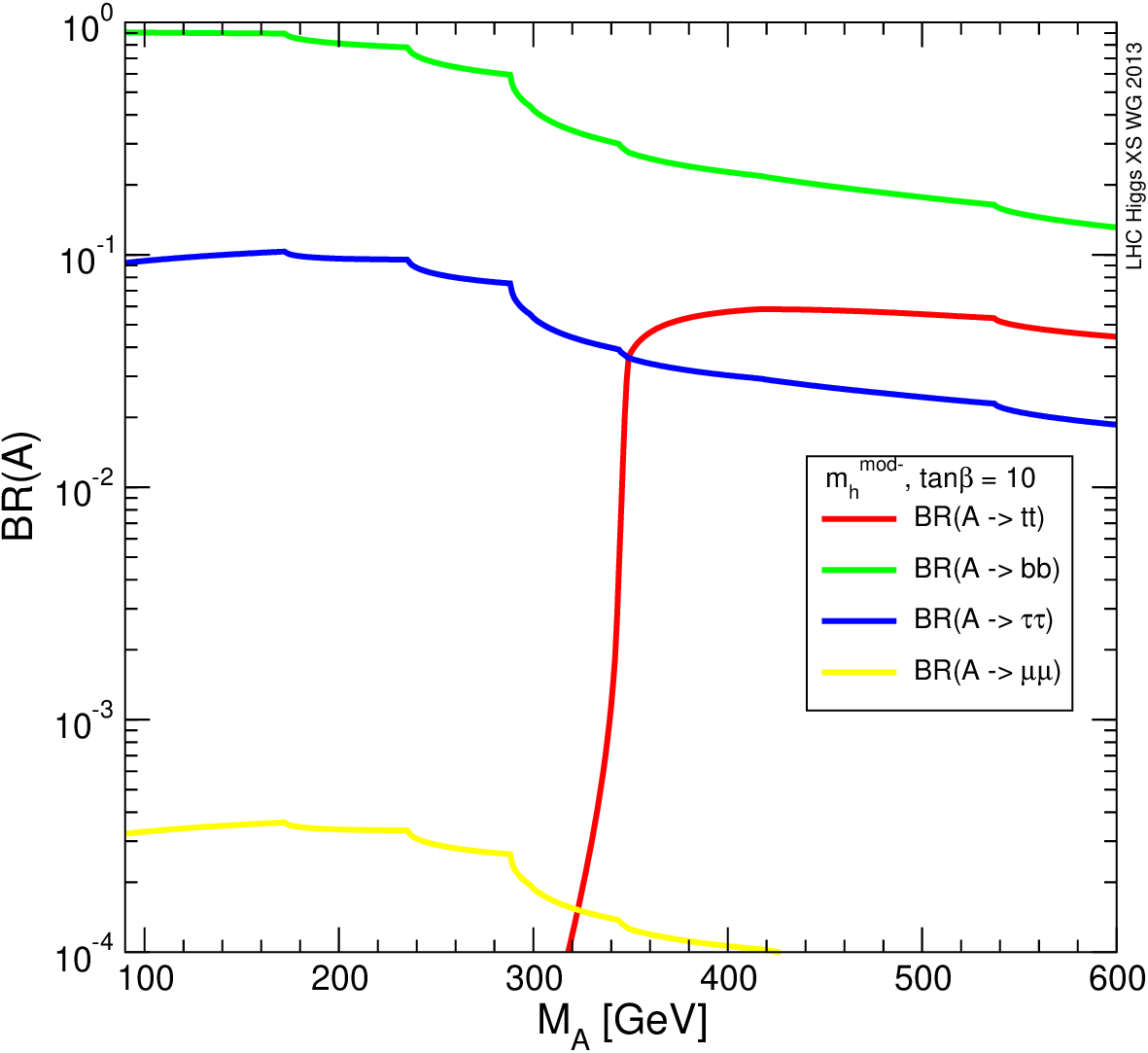}
\includegraphics[width=0.48\textwidth]{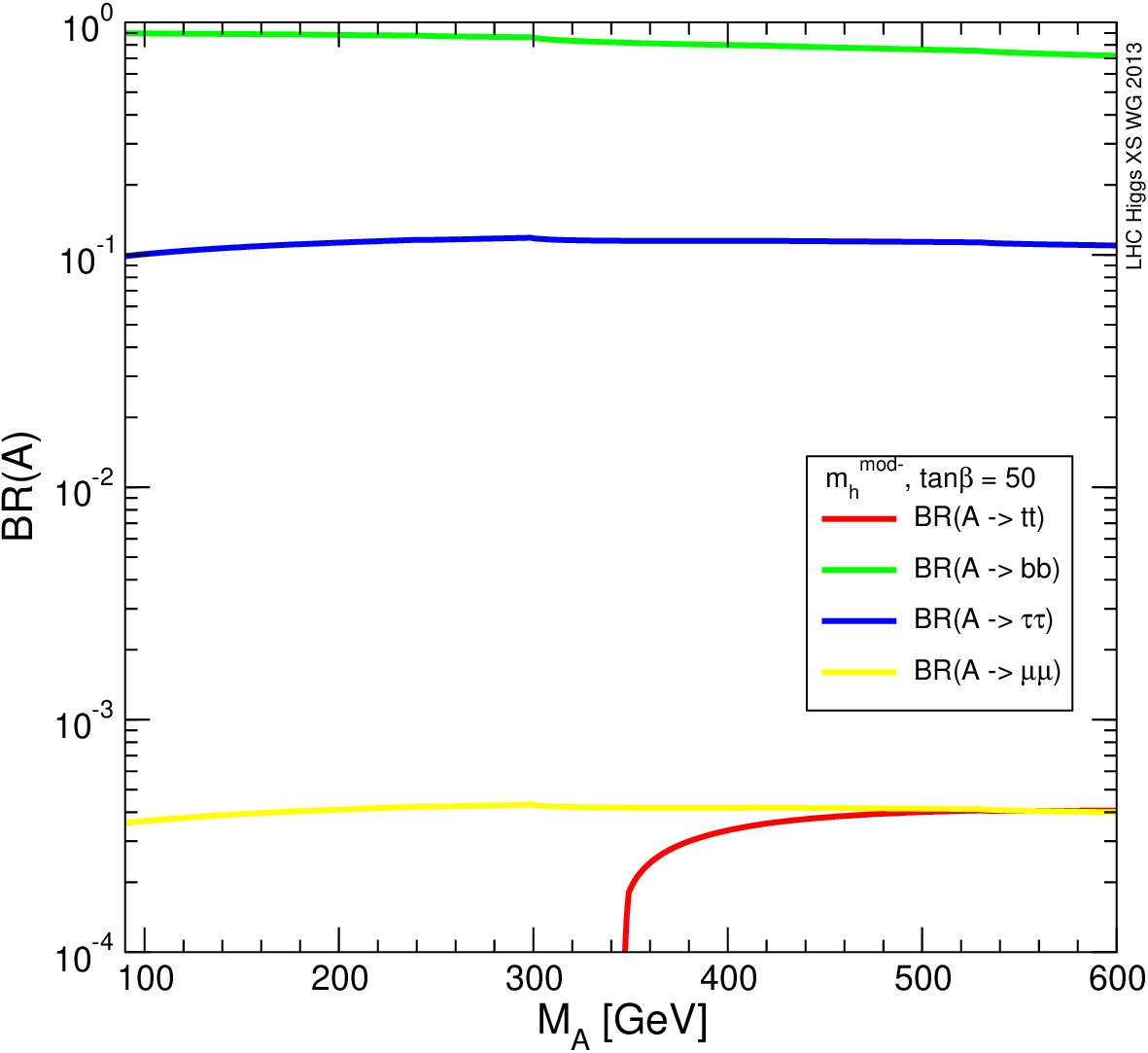}\\[0.5em]
%\vspace*{-0.3cm}
\caption{Branching ratios of the MSSM Higgs bosons $\Ph$ (upper row),
$\PH$ (middle row) and $\PA$ (lower row) in the \mhmodm\
scenario as a function of $\MA$.
The left (right) column shows the results for $\tb = 10 (50)$.}
\label{fig:YRHXS3_BR_mhmodm}
%\vspace{3.2cm}
\end{figure}

\begin{figure}[htb!]
%\vspace{3.2cm}
\includegraphics[width=0.48\textwidth]{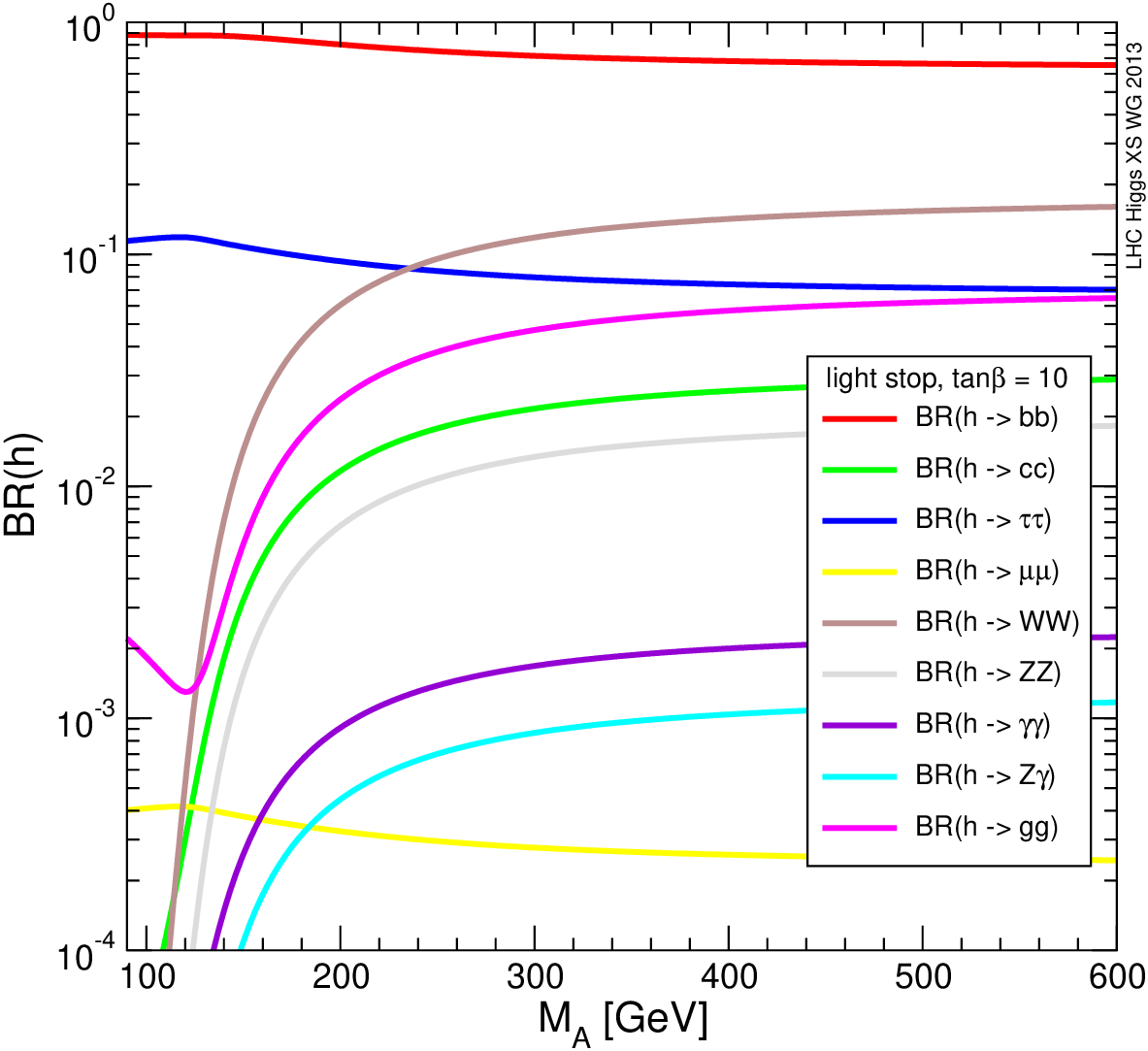}
\includegraphics[width=0.48\textwidth]{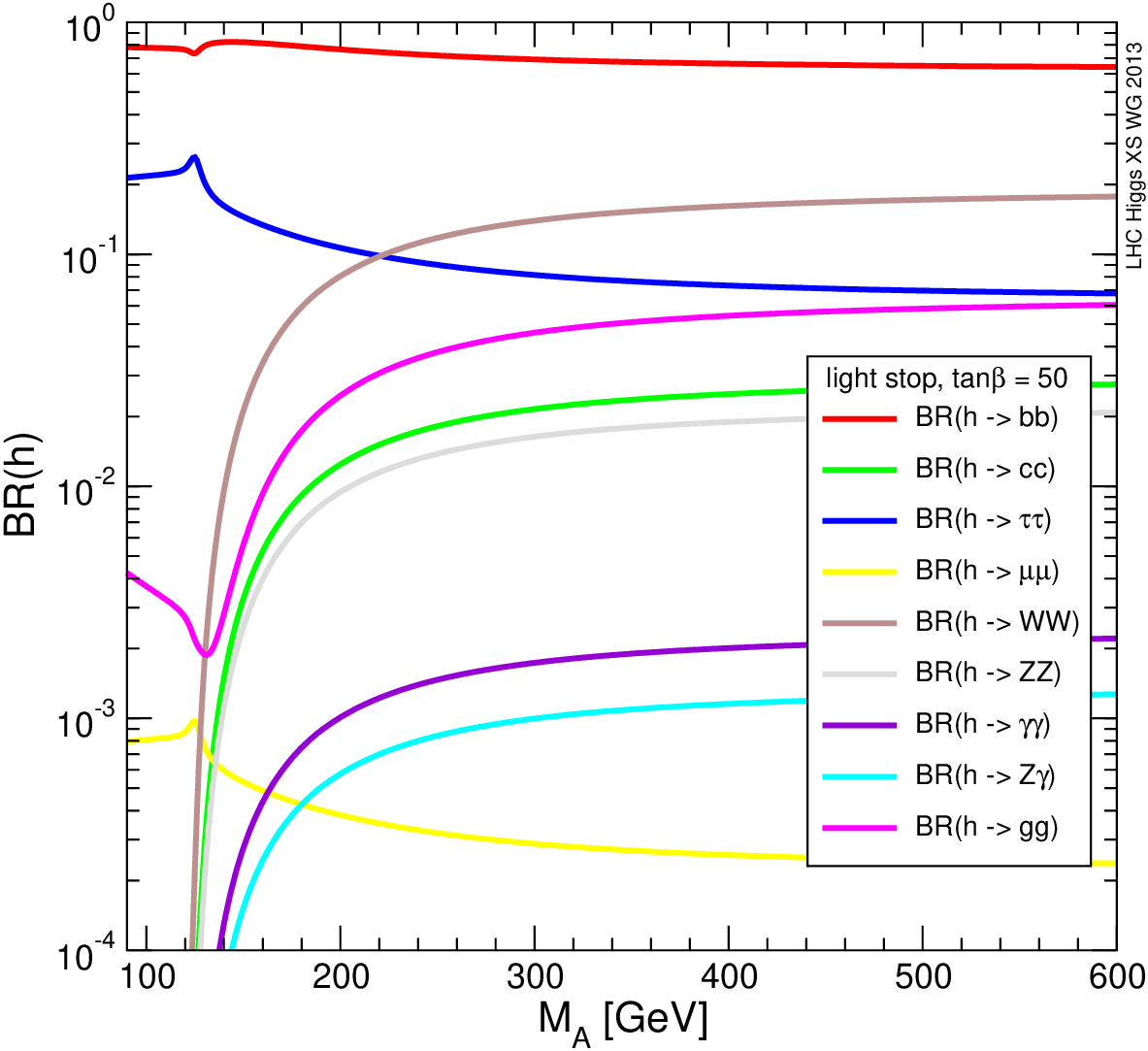}\\[.5em]
\includegraphics[width=0.48\textwidth]{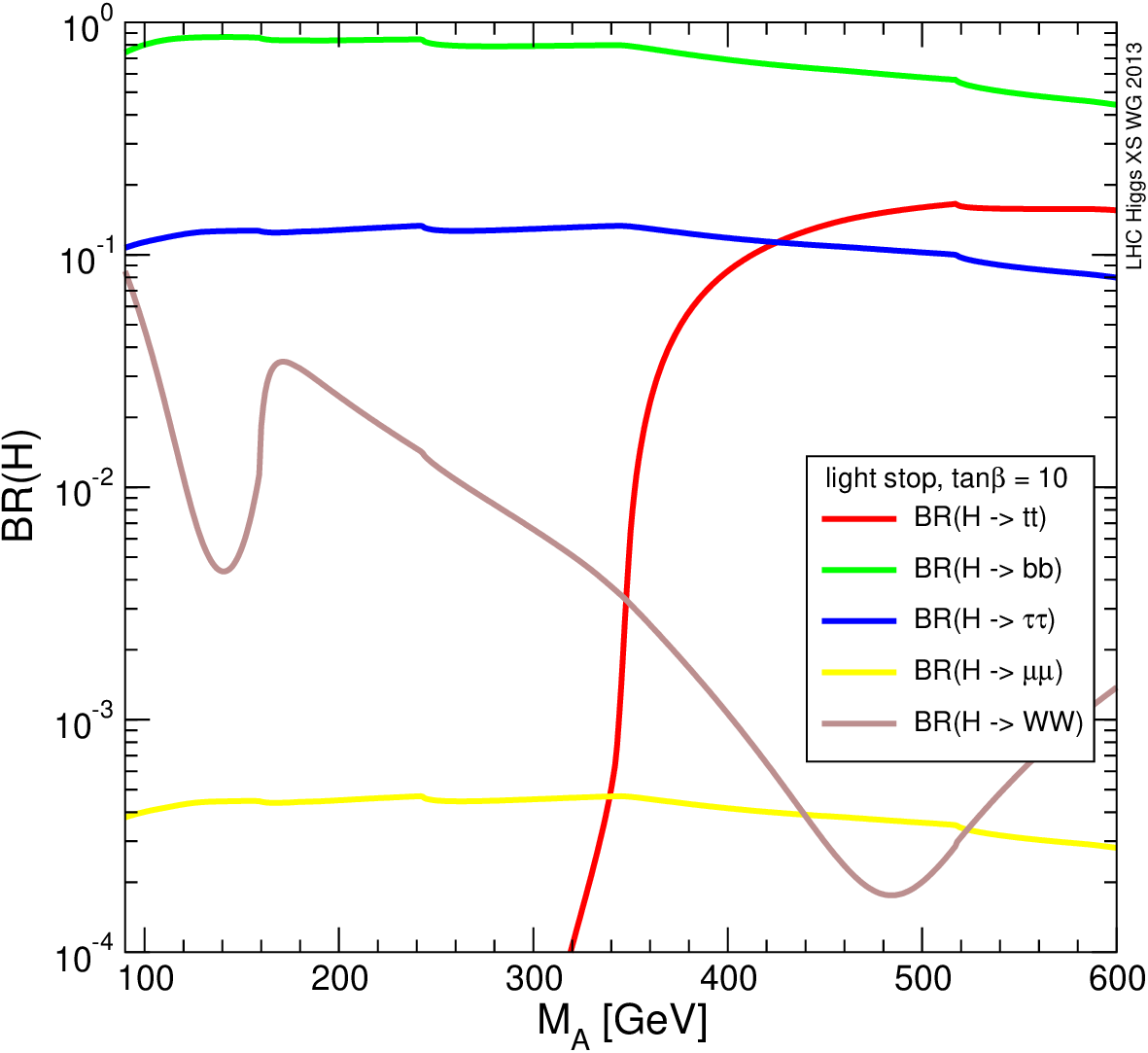}
\includegraphics[width=0.48\textwidth]{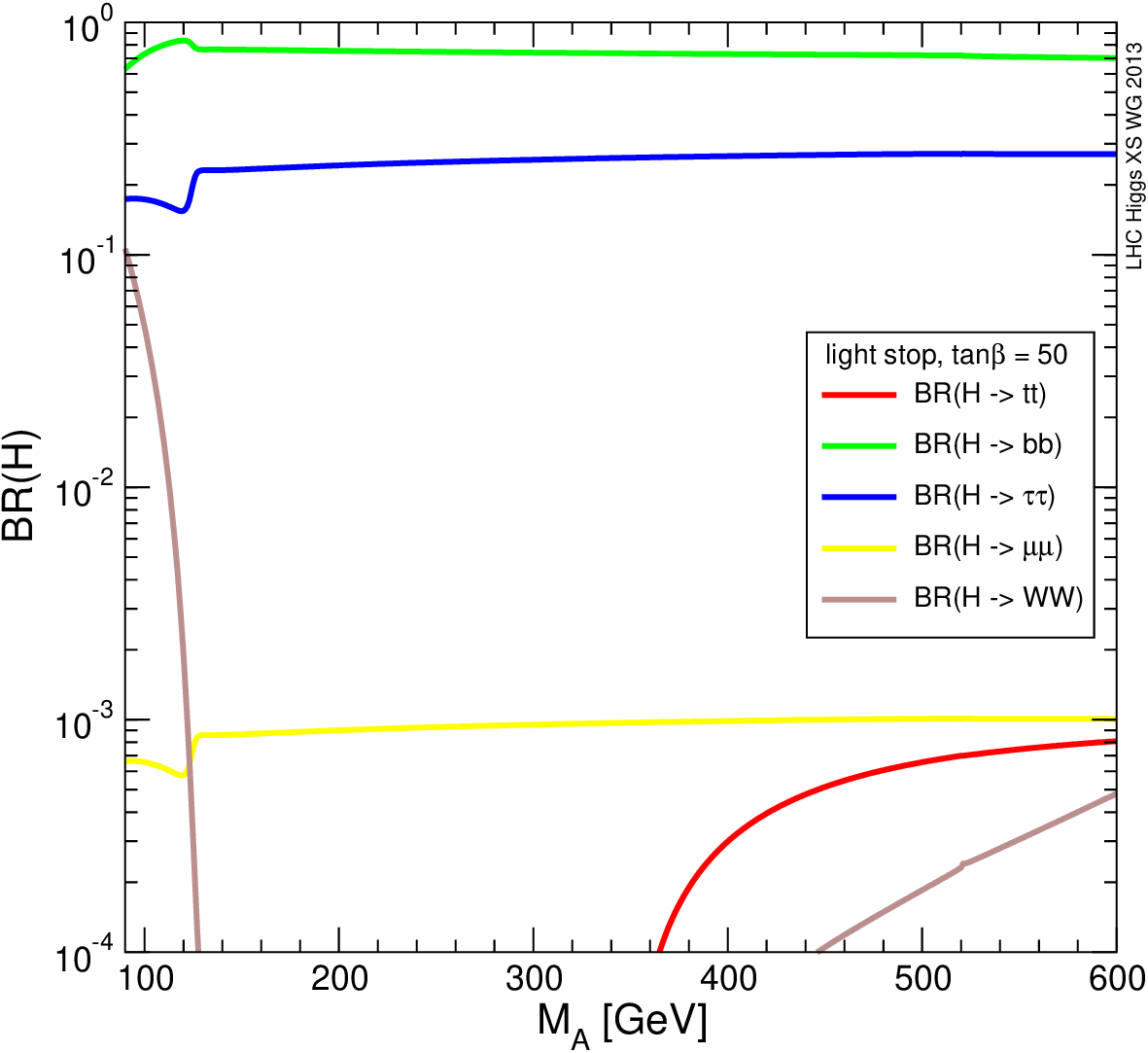}\\[0.5em]
\includegraphics[width=0.48\textwidth]{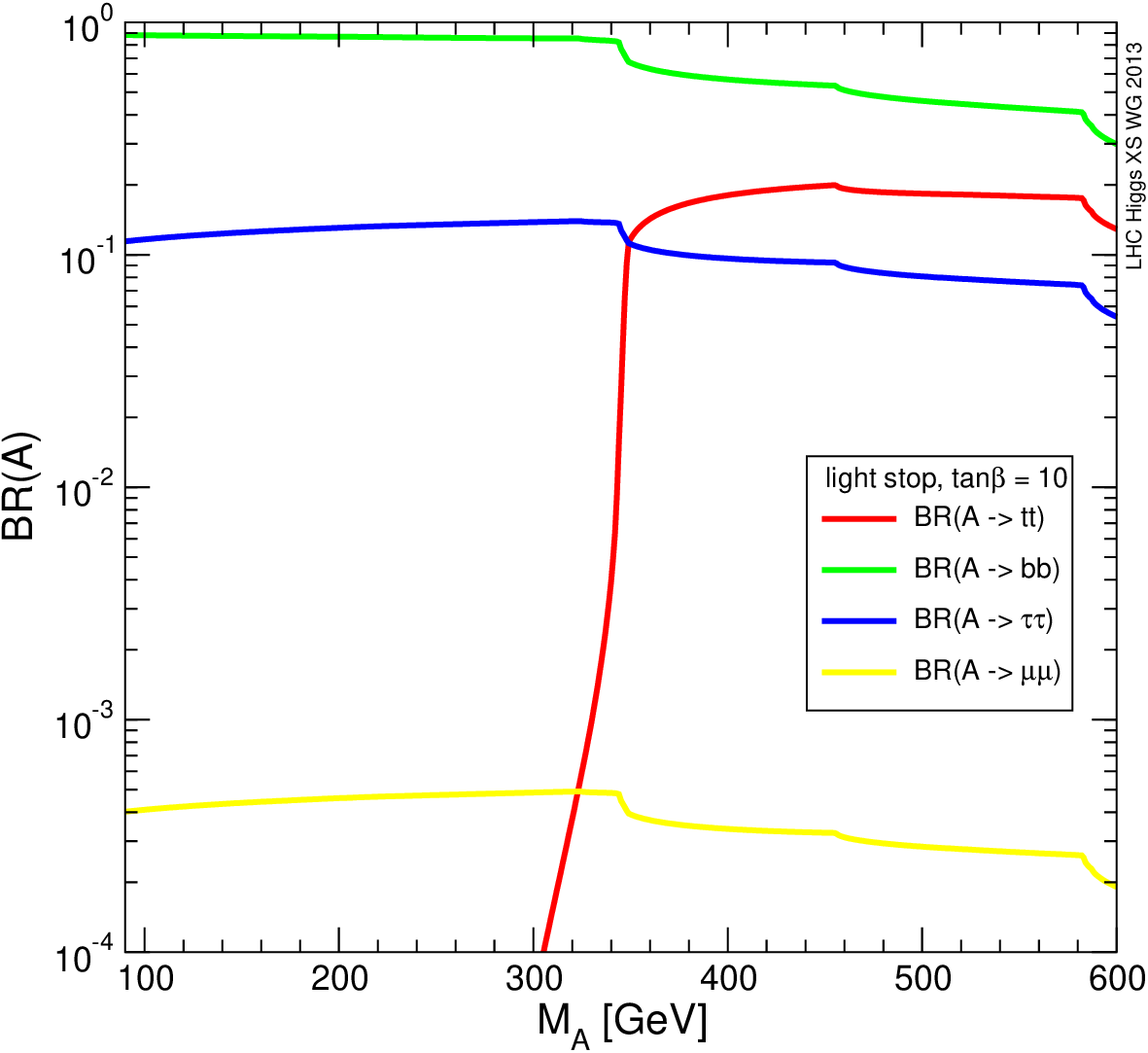}
\includegraphics[width=0.48\textwidth]{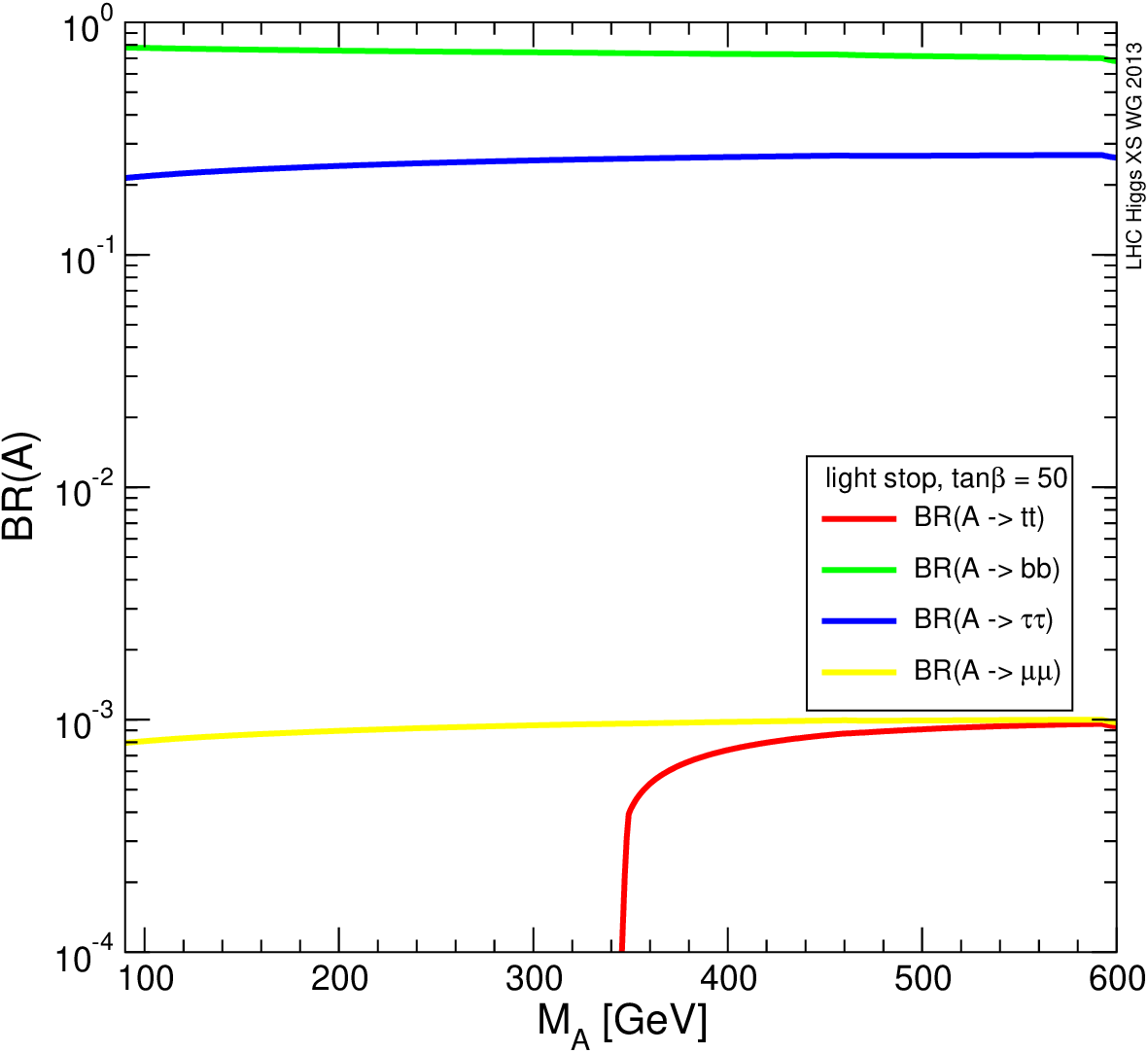}\\[0.5em]
%\vspace*{-0.3cm}
\caption{Branching ratios of the MSSM Higgs bosons $\Ph$ (upper row),
$\PH$ (middle row) and $\PA$ (lower row) in the light stop
scenario as a function of $\MA$.
The left (right) column shows the results for $\tb = 10 (50)$.}
\label{fig:YRHXS3_BR_lightstop}
%\vspace{3.2cm}
\end{figure}

\begin{figure}[htb!]
%\vspace{3.2cm}
\includegraphics[width=0.48\textwidth]{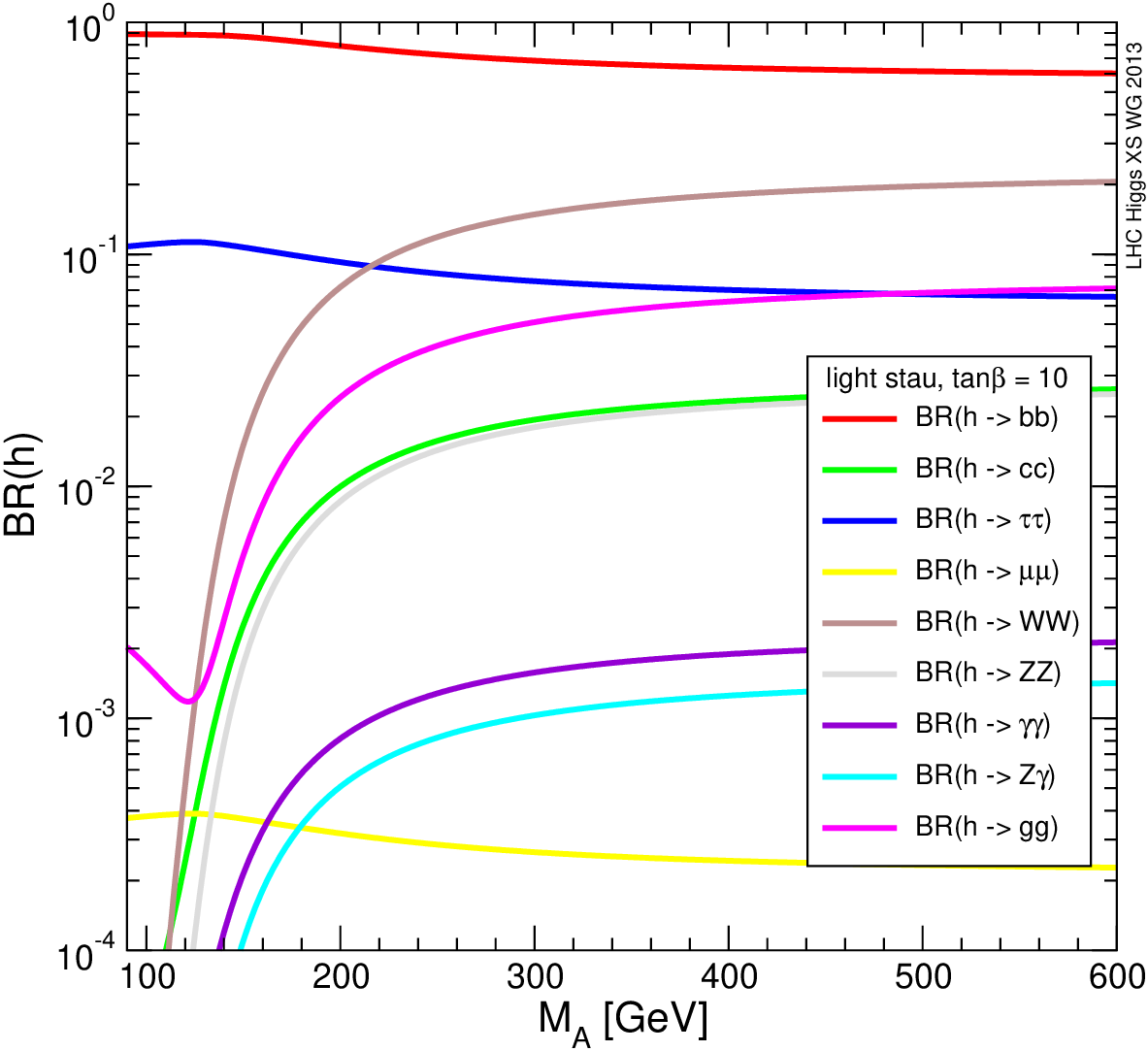}
\includegraphics[width=0.48\textwidth]{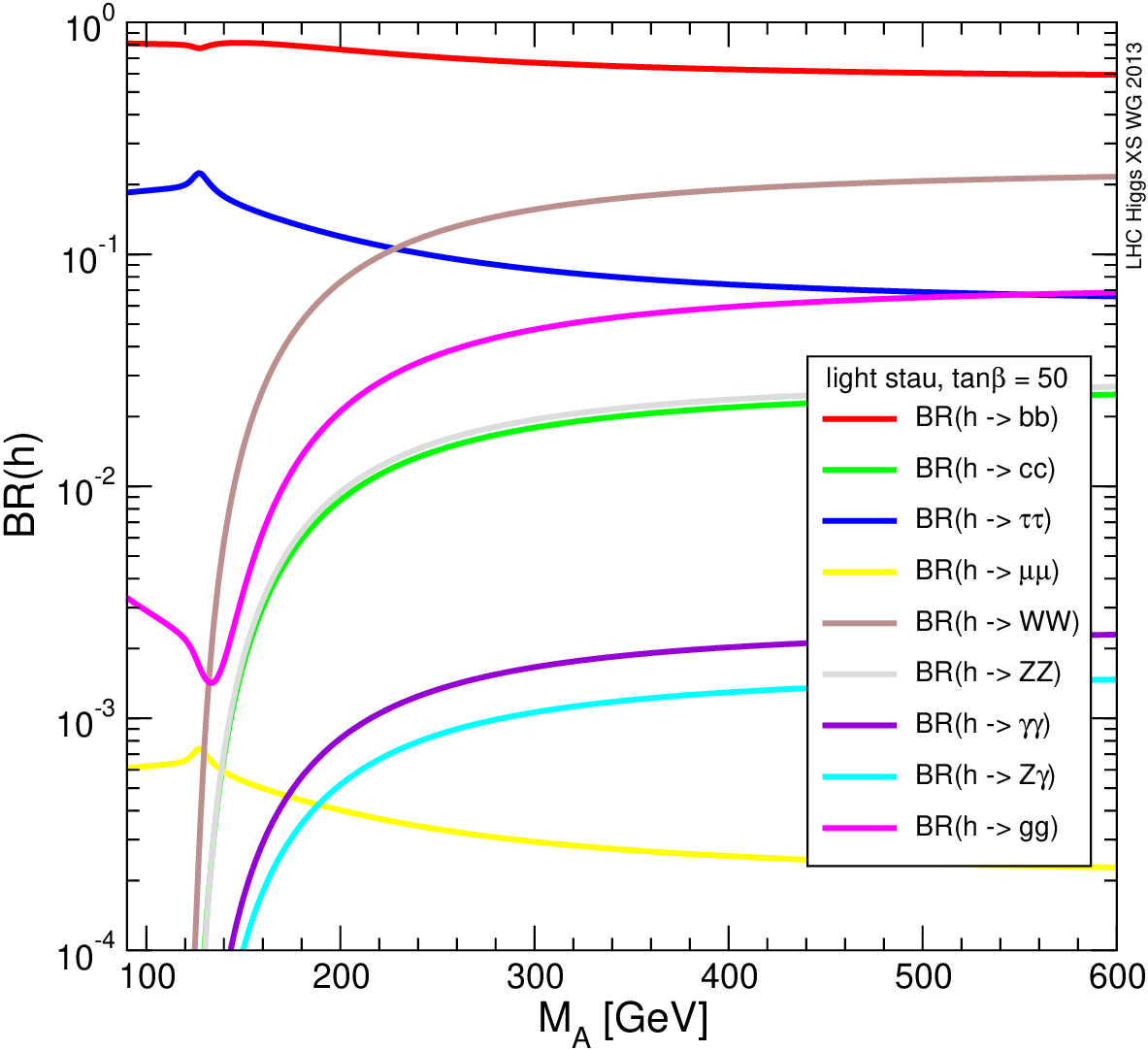}\\[.5em]
\includegraphics[width=0.48\textwidth]{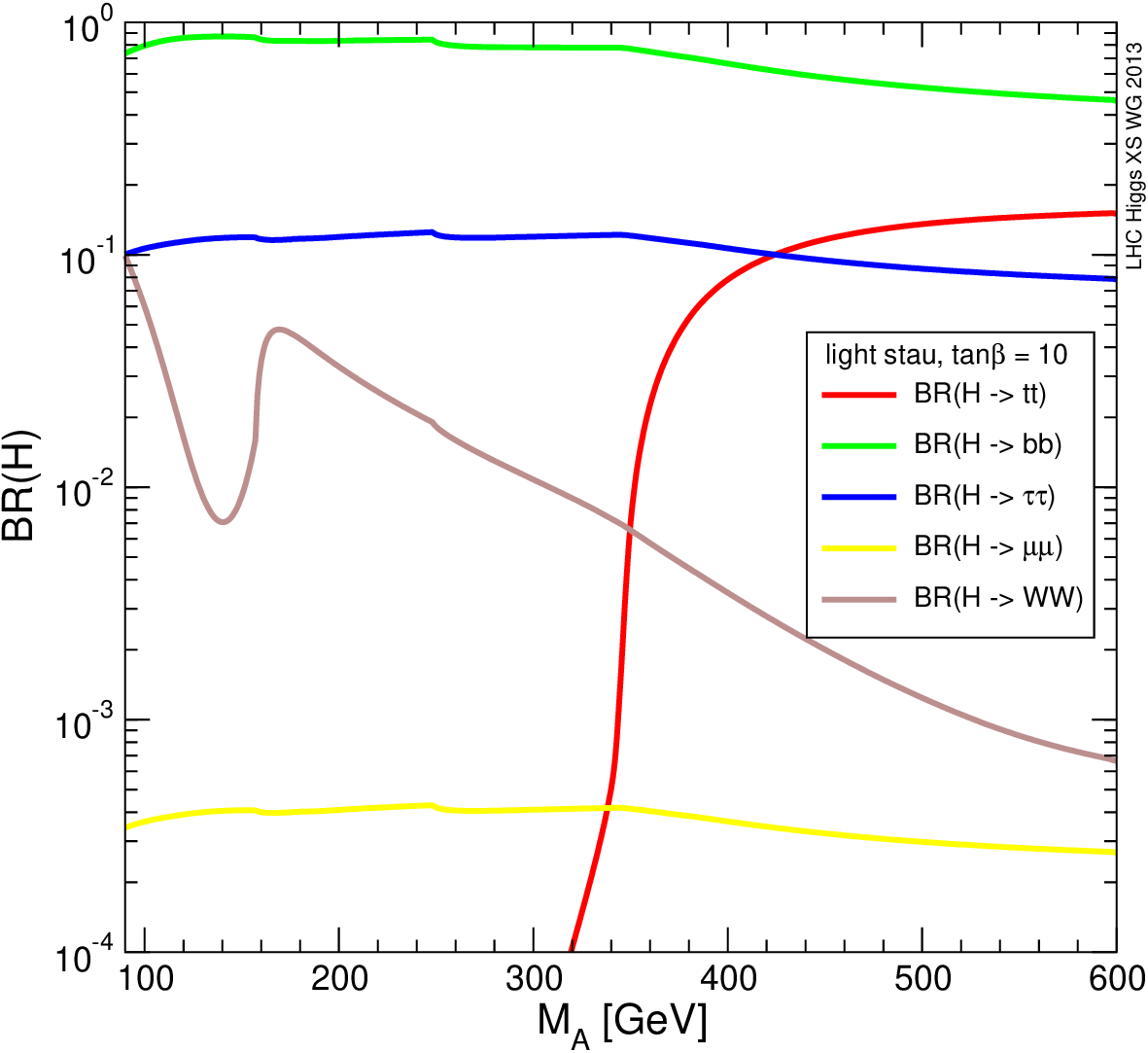}
\includegraphics[width=0.48\textwidth]{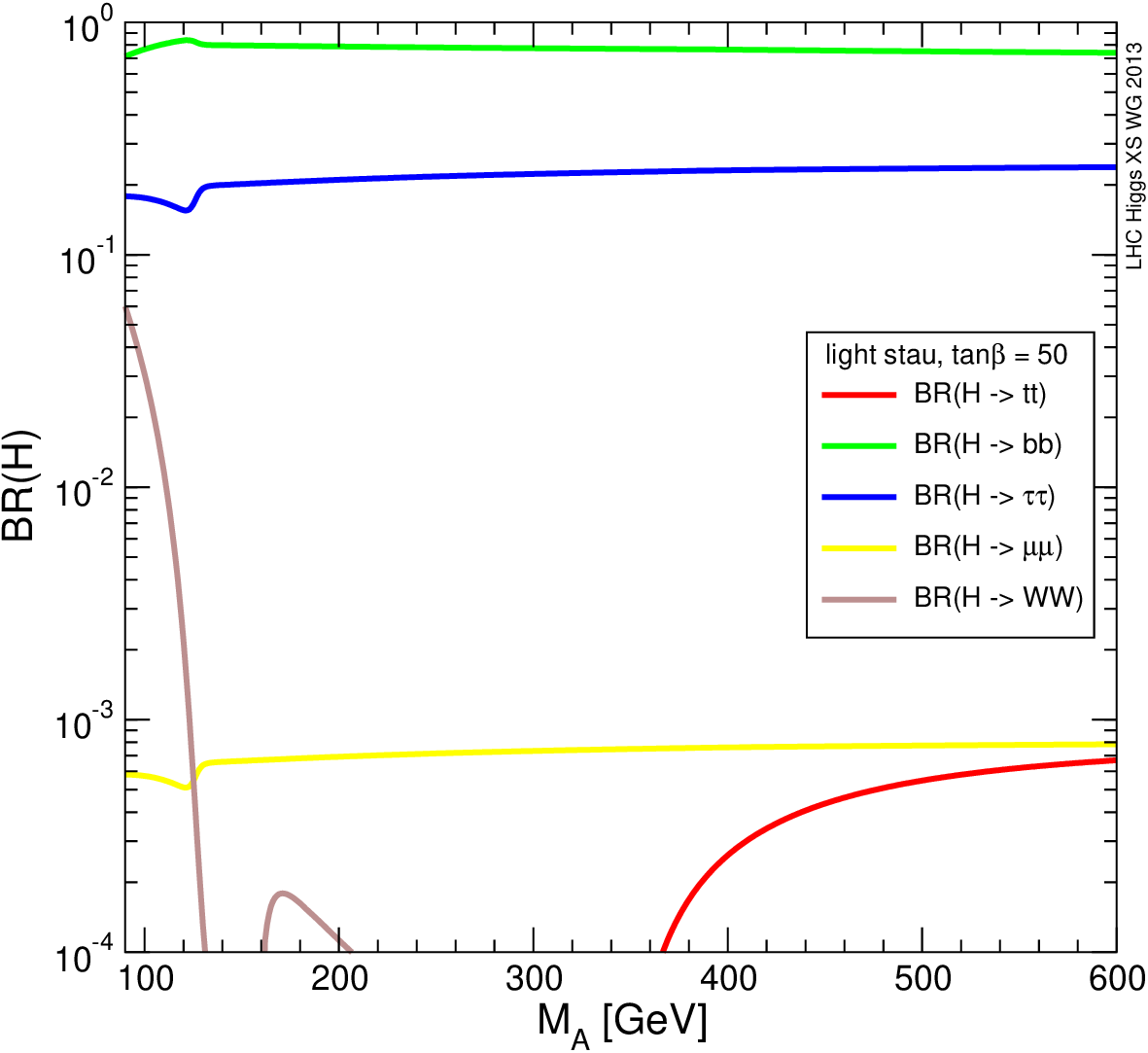}\\[0.5em]
\includegraphics[width=0.48\textwidth]{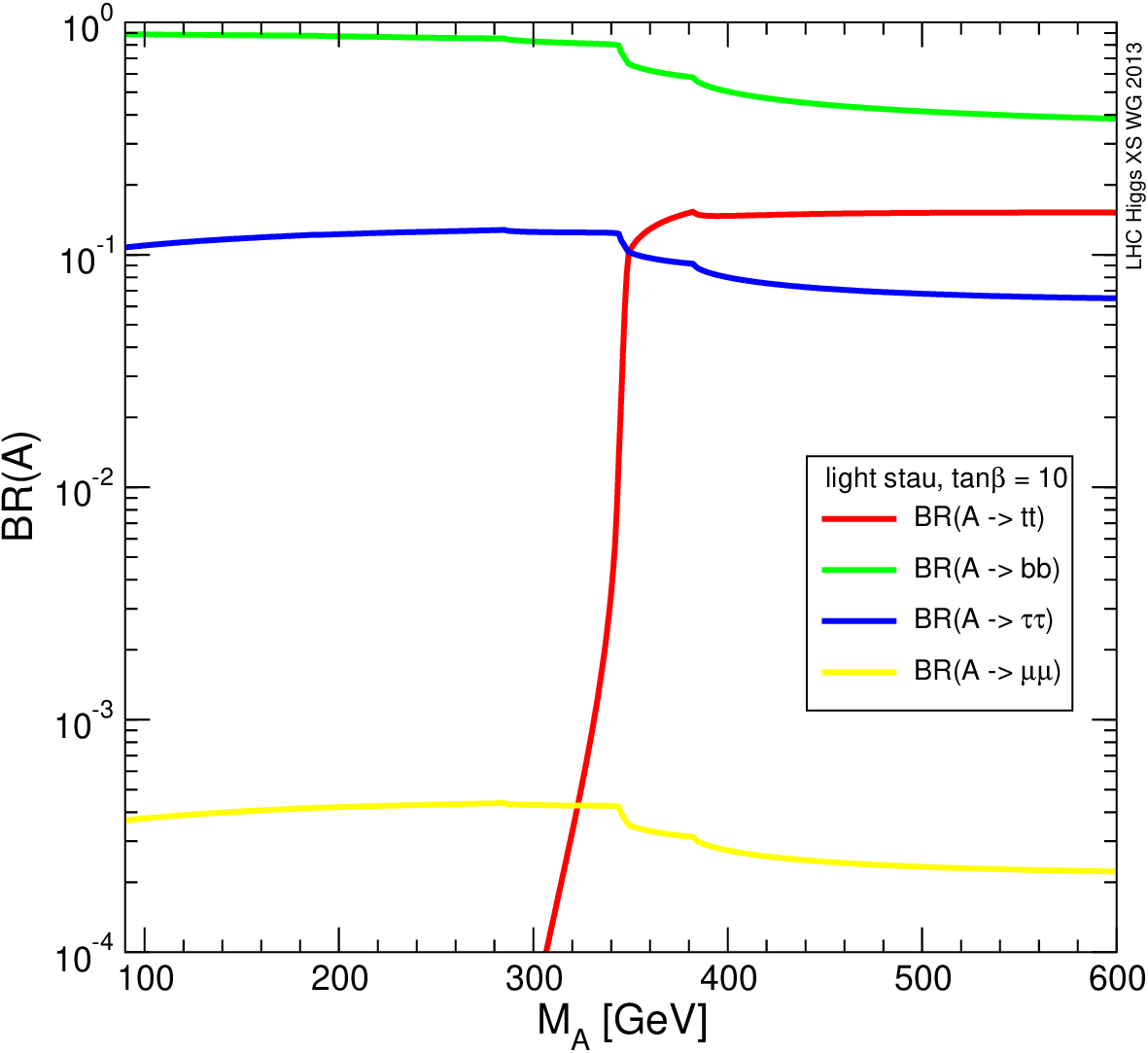}
\includegraphics[width=0.48\textwidth]{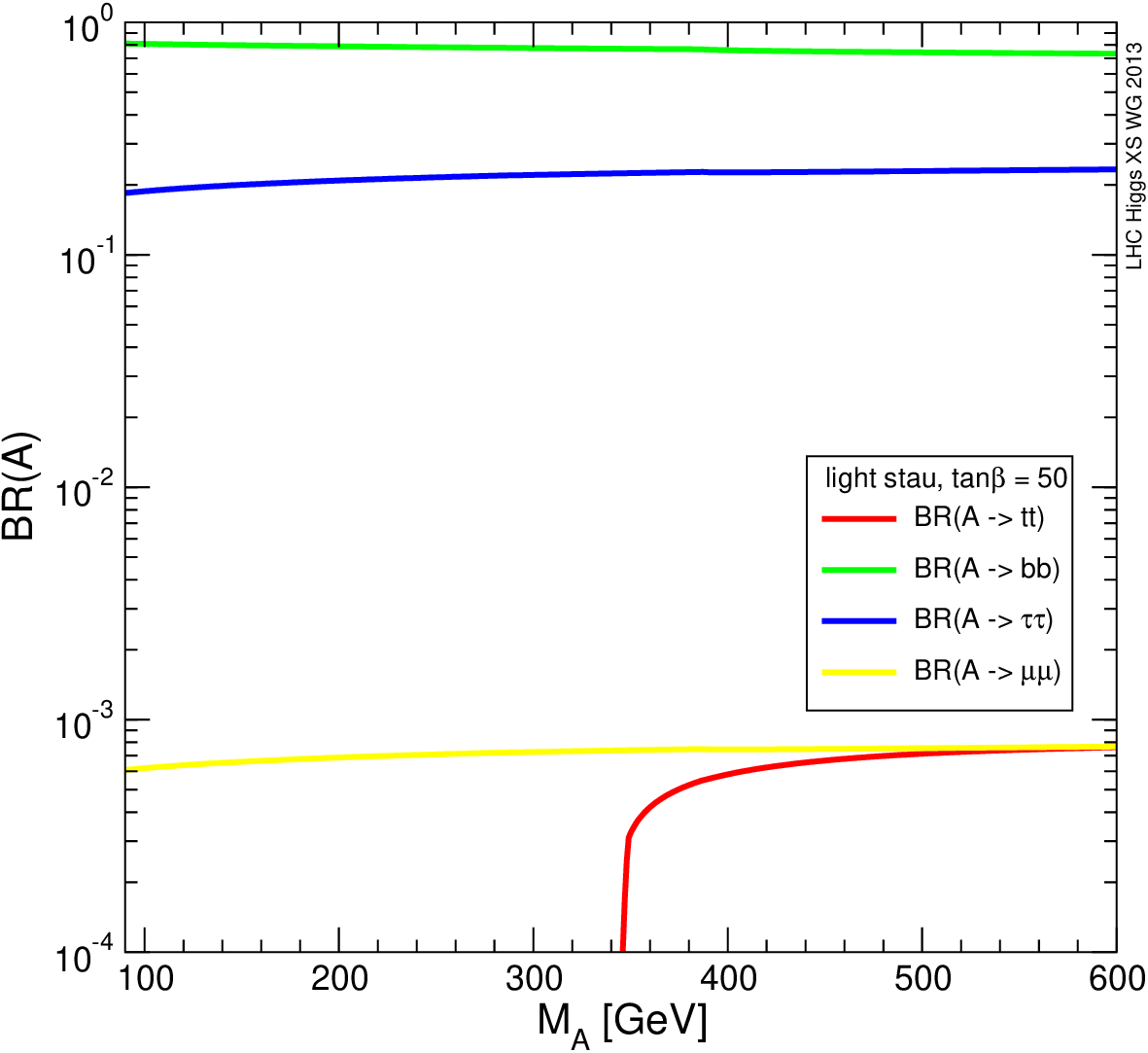}\\[0.5em]
%\vspace*{-0.3cm}
\caption{Branching ratios of the MSSM Higgs bosons $\Ph$ (upper row),
$\PH$ (middle row) and $\PA$ (lower row) in the light stau
scenario as a function of $\MA$.
The left (right) column shows the results for $\tb = 10 (50)$.}
\label{fig:YRHXS3_BR_lightstau}
%\vspace{3.2cm}
\end{figure}

\begin{figure}[htb!]
%\vspace{3.2cm}
\includegraphics[width=0.48\textwidth]{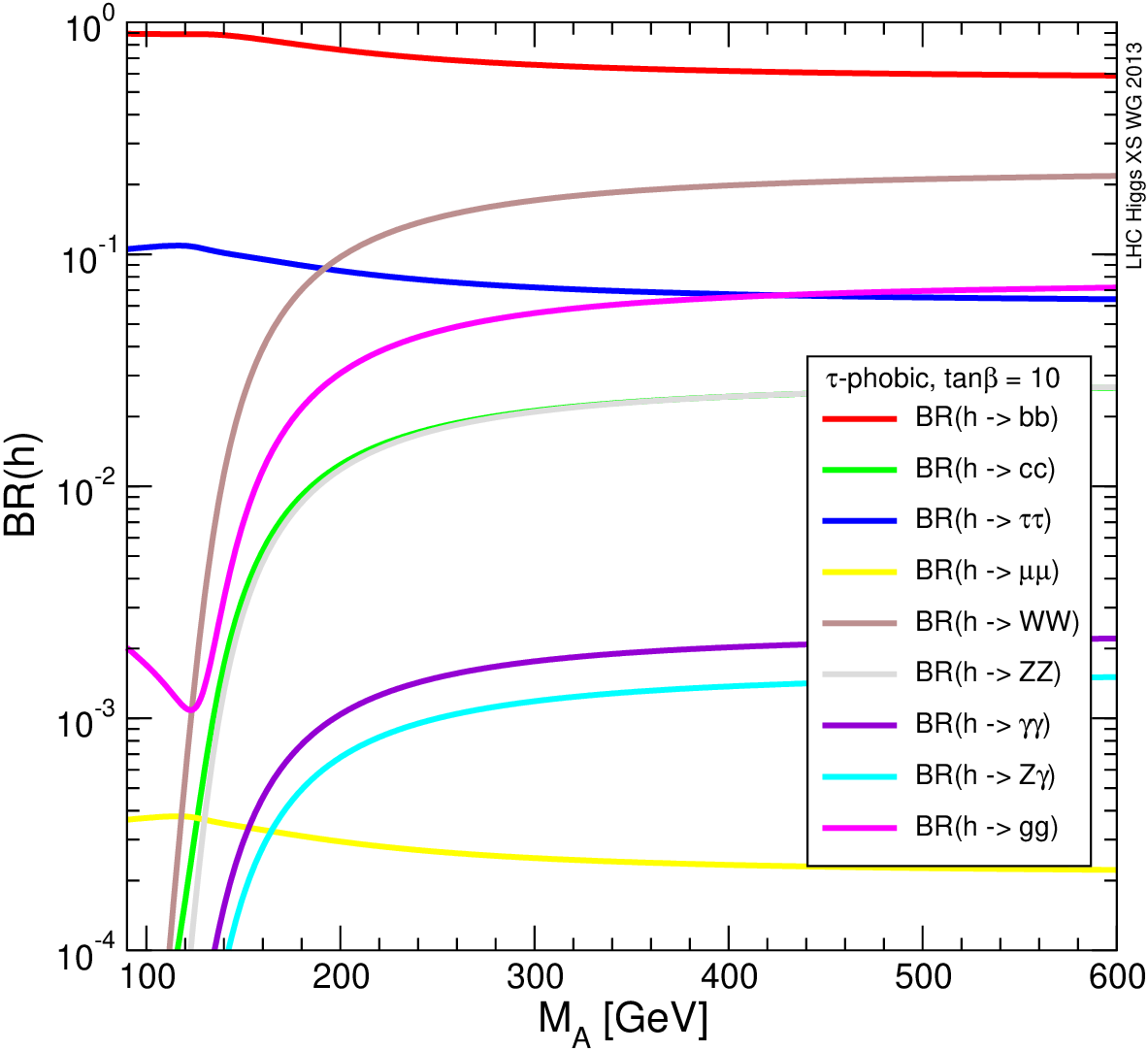}
\includegraphics[width=0.48\textwidth]{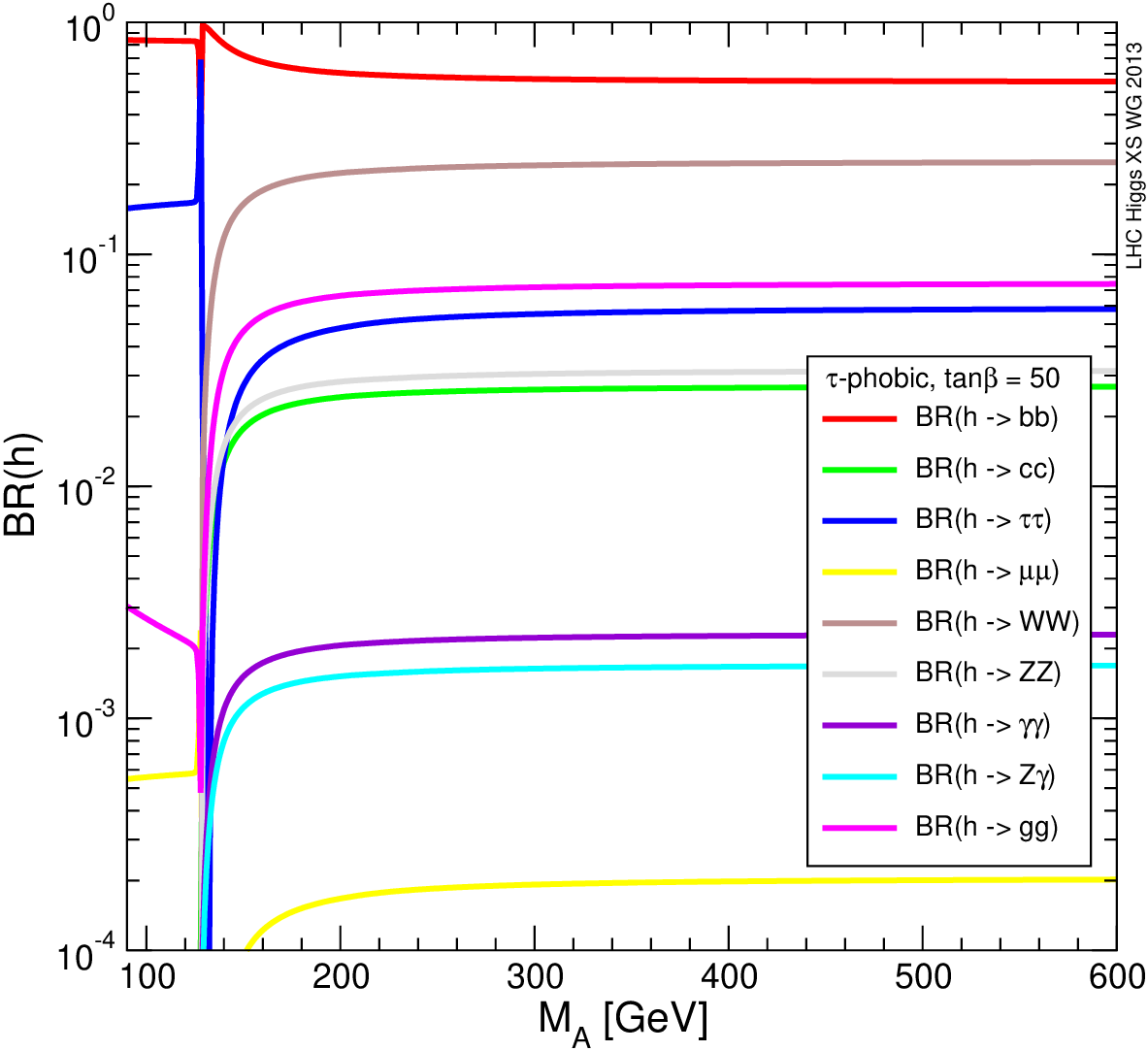}\\[.5em]
\includegraphics[width=0.48\textwidth]{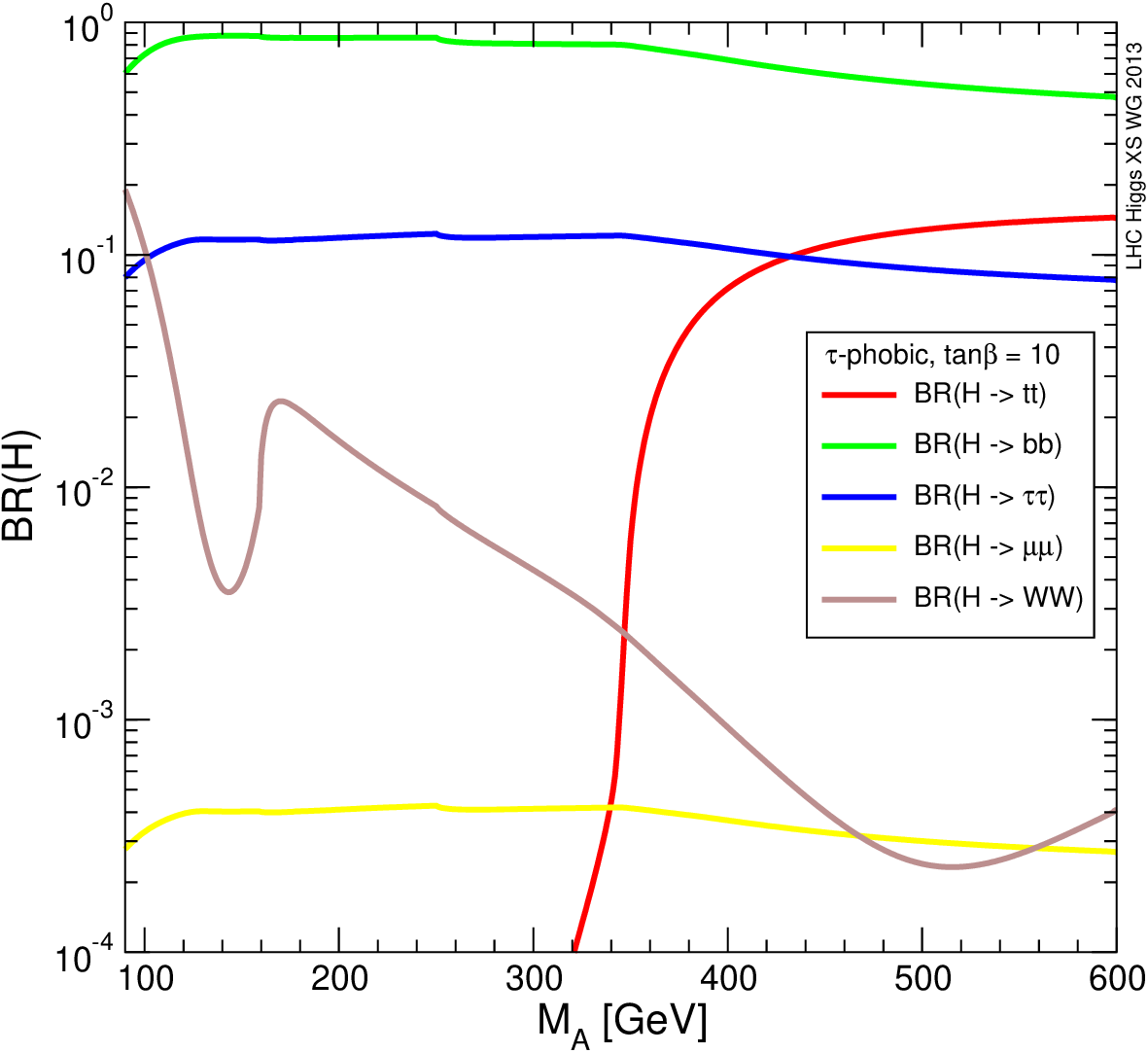}
\includegraphics[width=0.48\textwidth]{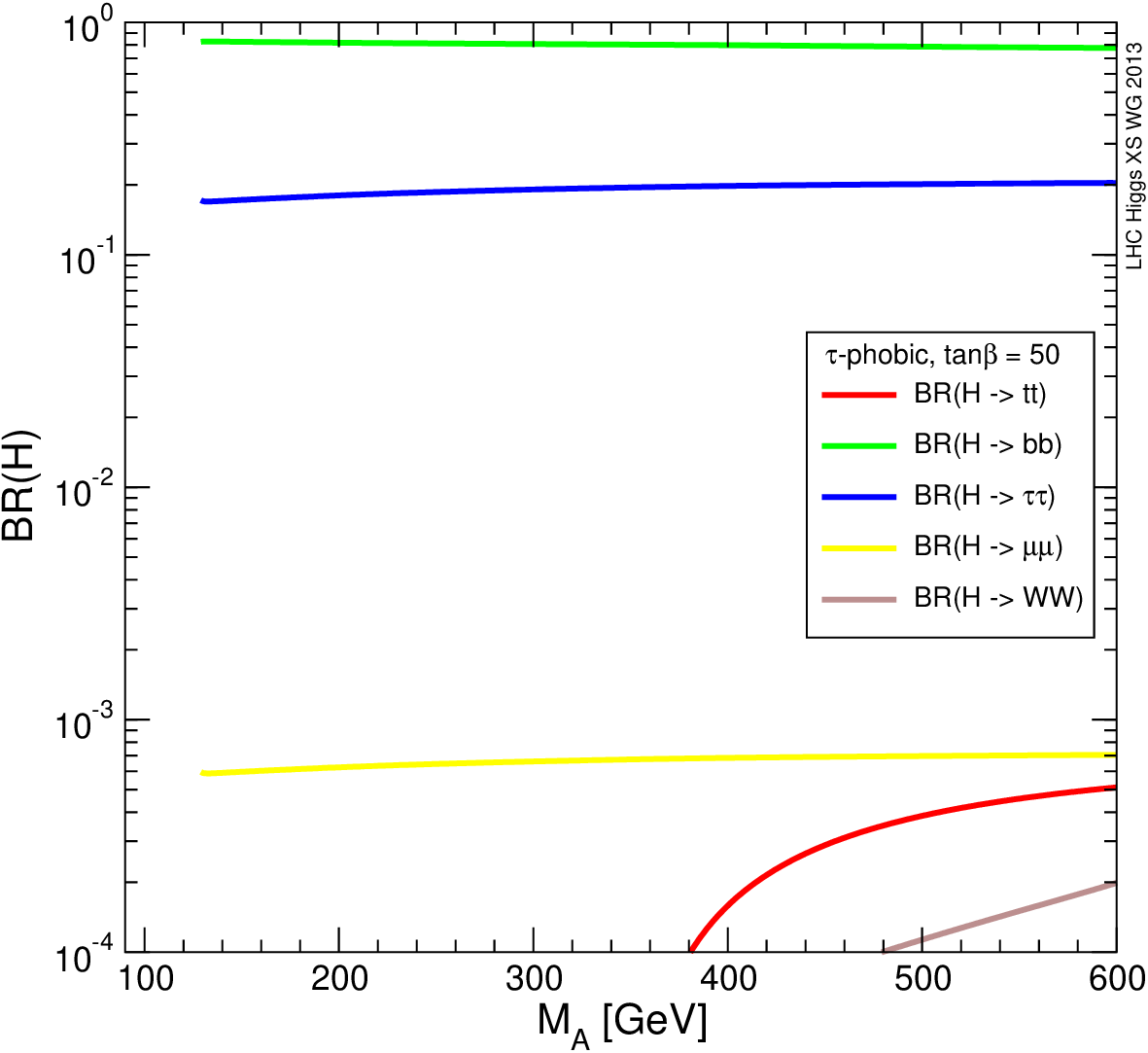}\\[0.5em]
\includegraphics[width=0.48\textwidth]{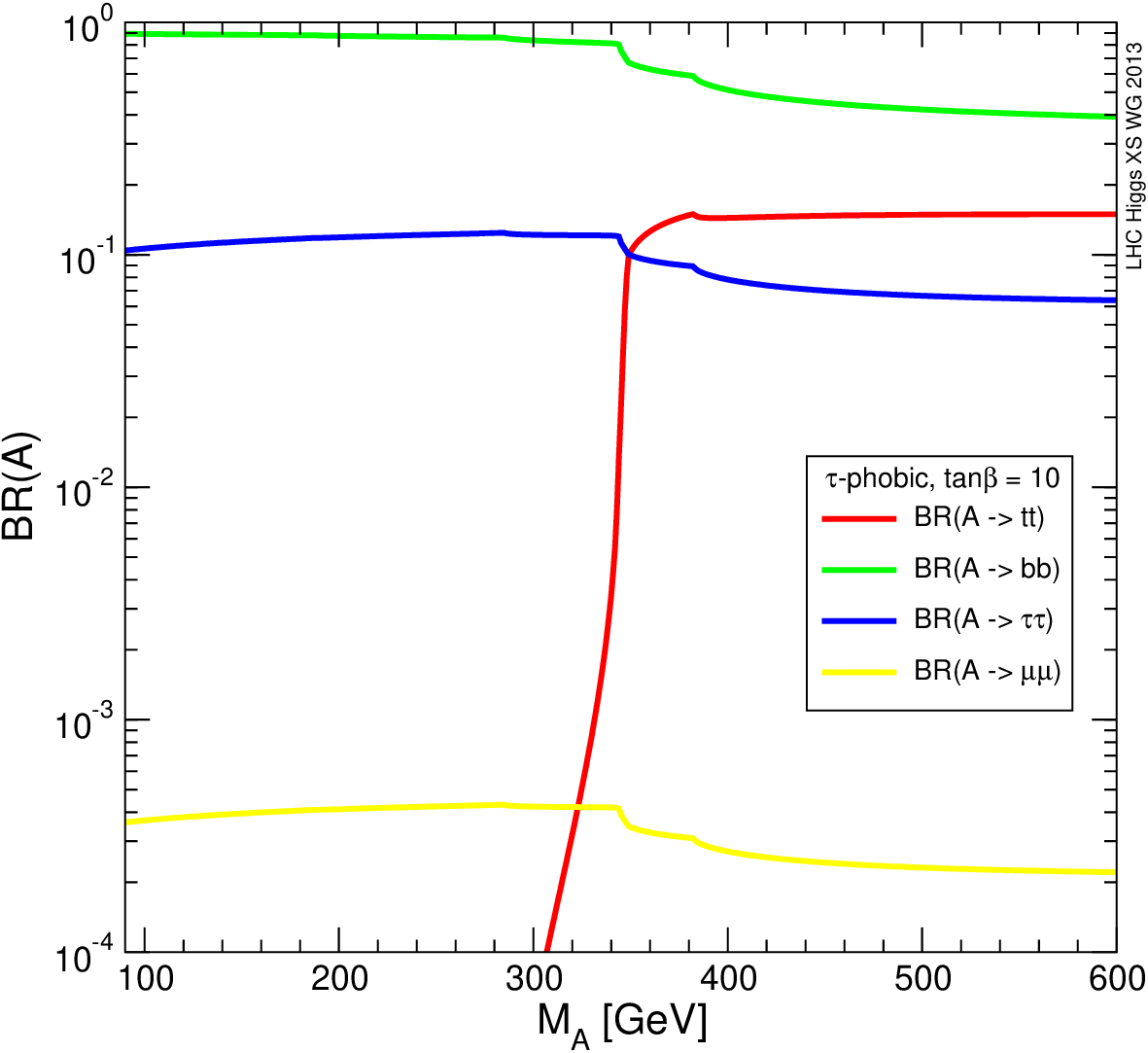}
\includegraphics[width=0.48\textwidth]{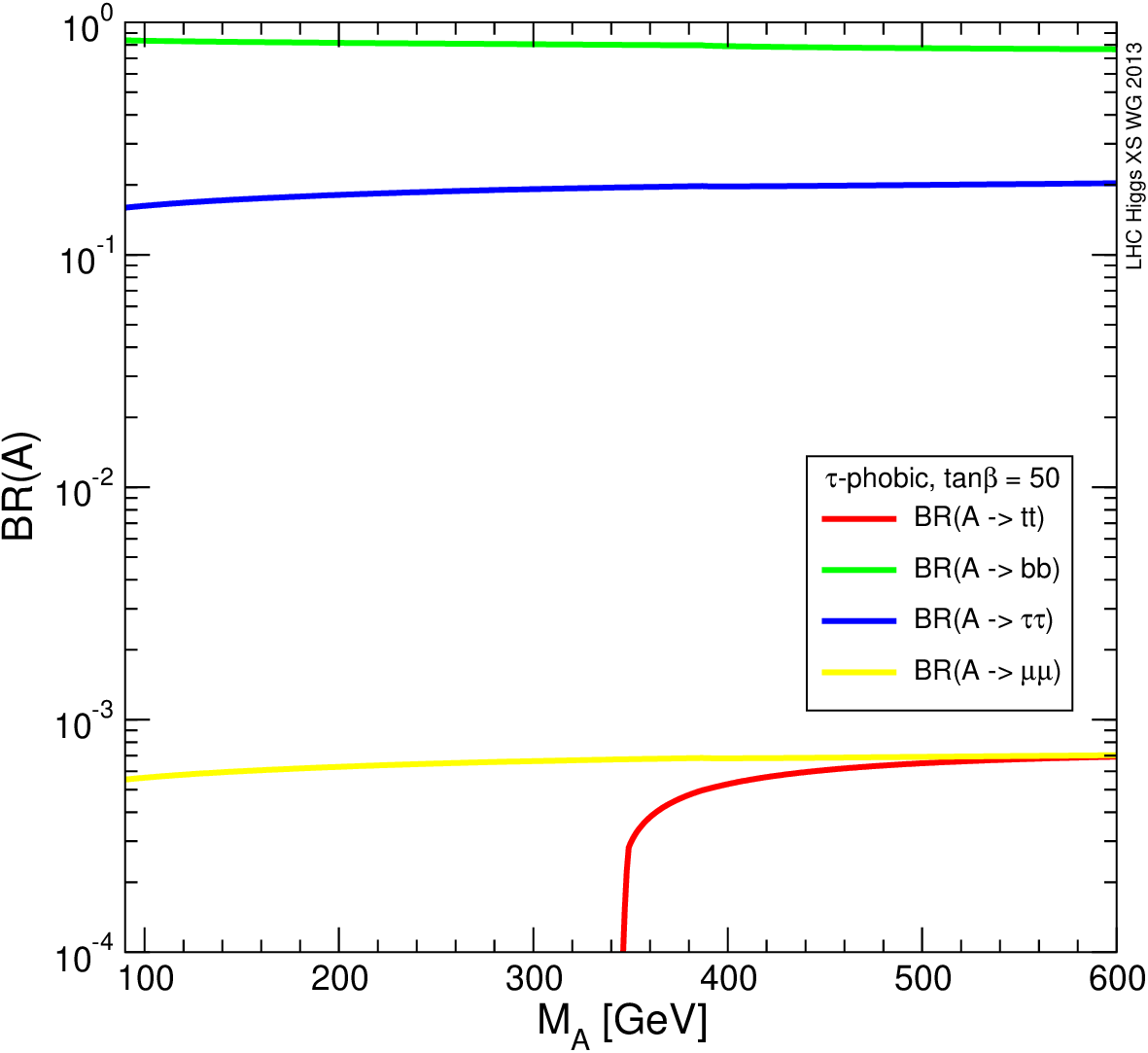}\\[0.5em]
%\vspace*{-0.3cm}
\caption{Branching ratios of the MSSM Higgs bosons $\Ph$ (upper row),
$\PH$ (middle row) and $\PA^\pm$ (lower row)in the \tauphobic\
scenario as a function of $\MA$.
The left (right) column shows the results for $\tb = 10 (50)$.}
\label{fig:YRHXS3_BR_tauphobic}
%\vspace{3.2cm}
\end{figure}

\begin{figure}[htb!]
%\vspace{3.2cm}
\includegraphics[width=0.48\textwidth]{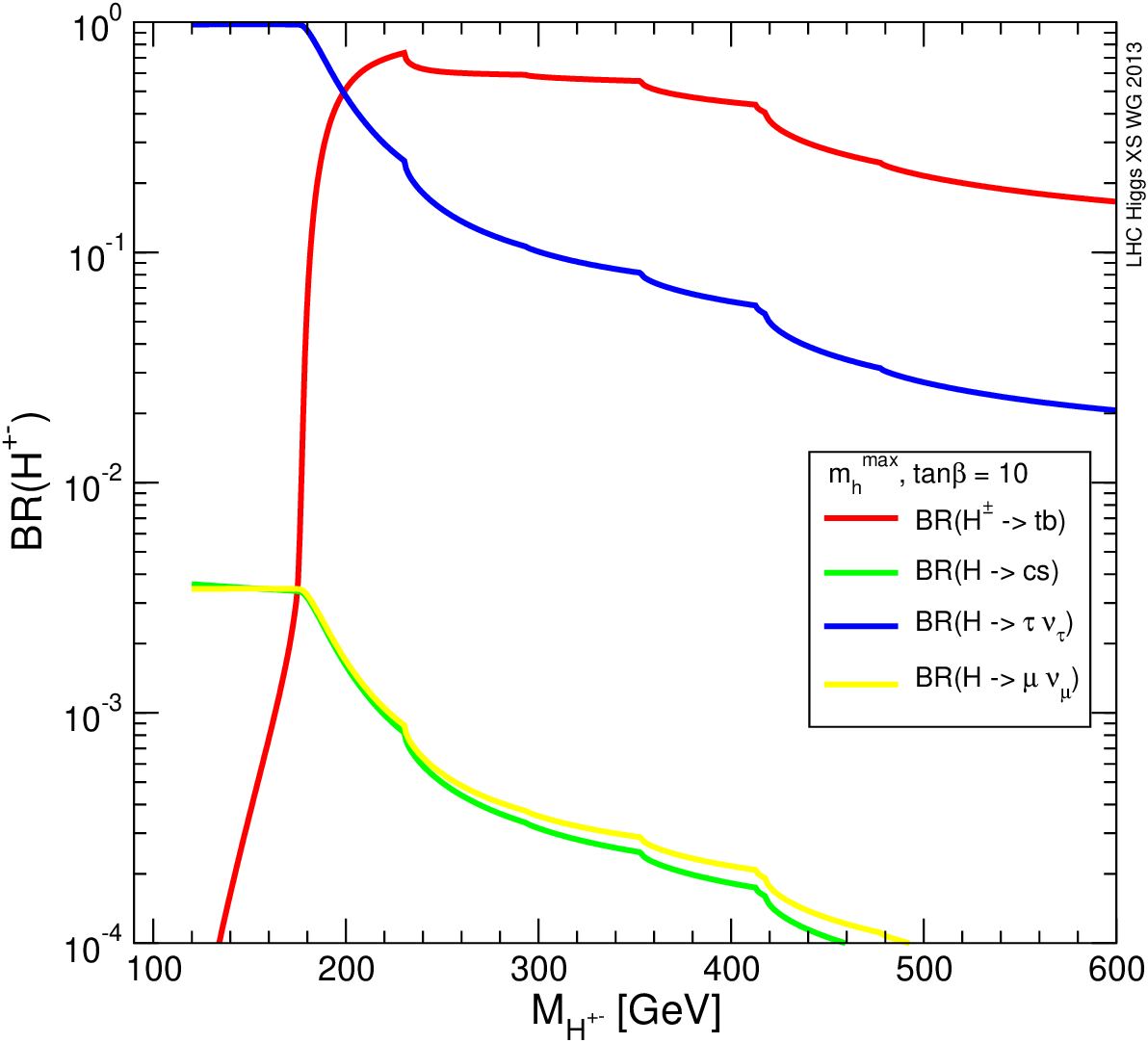}
\includegraphics[width=0.48\textwidth]{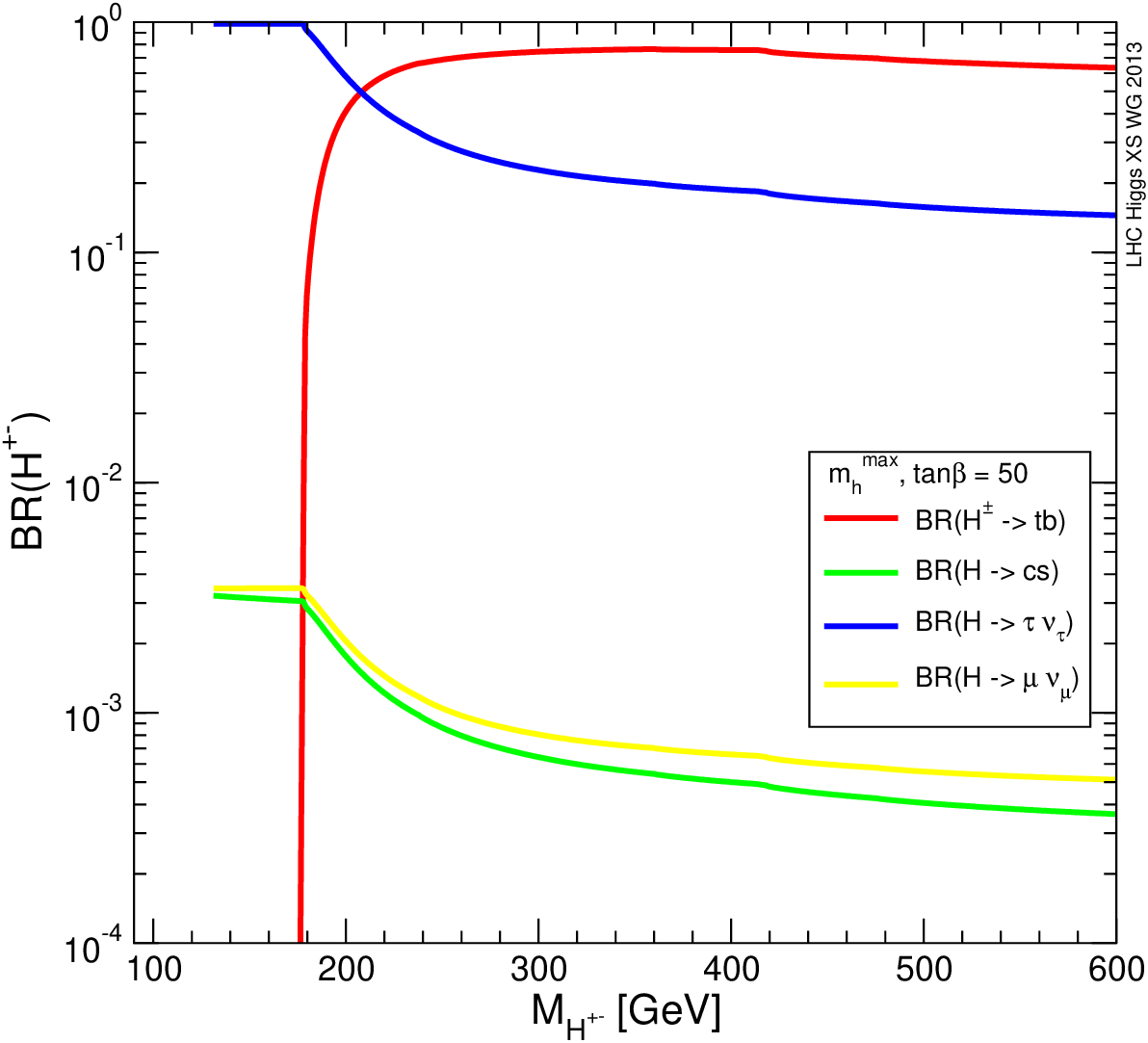}\\[0.5em]
\includegraphics[width=0.48\textwidth]{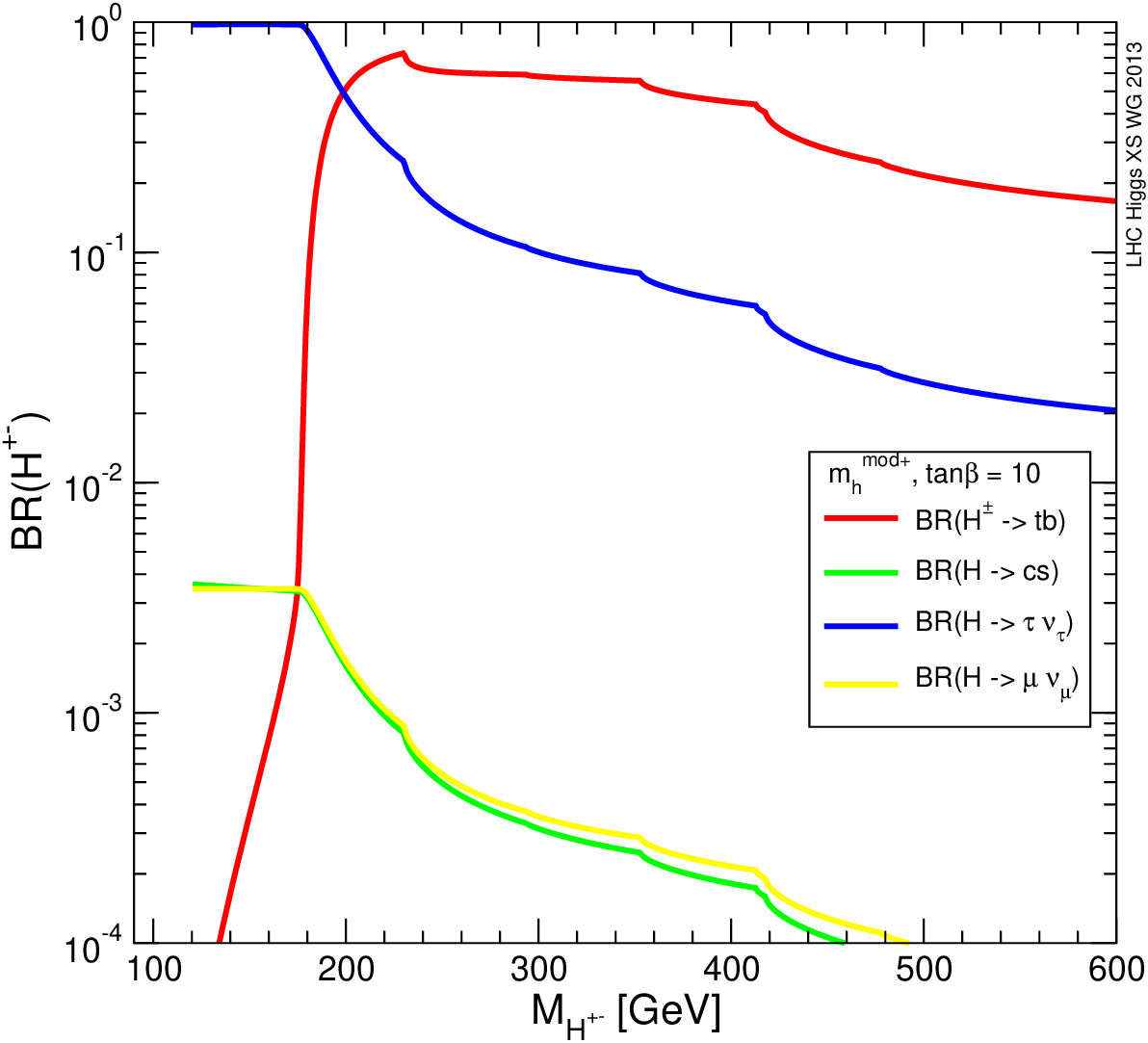}
\includegraphics[width=0.48\textwidth]{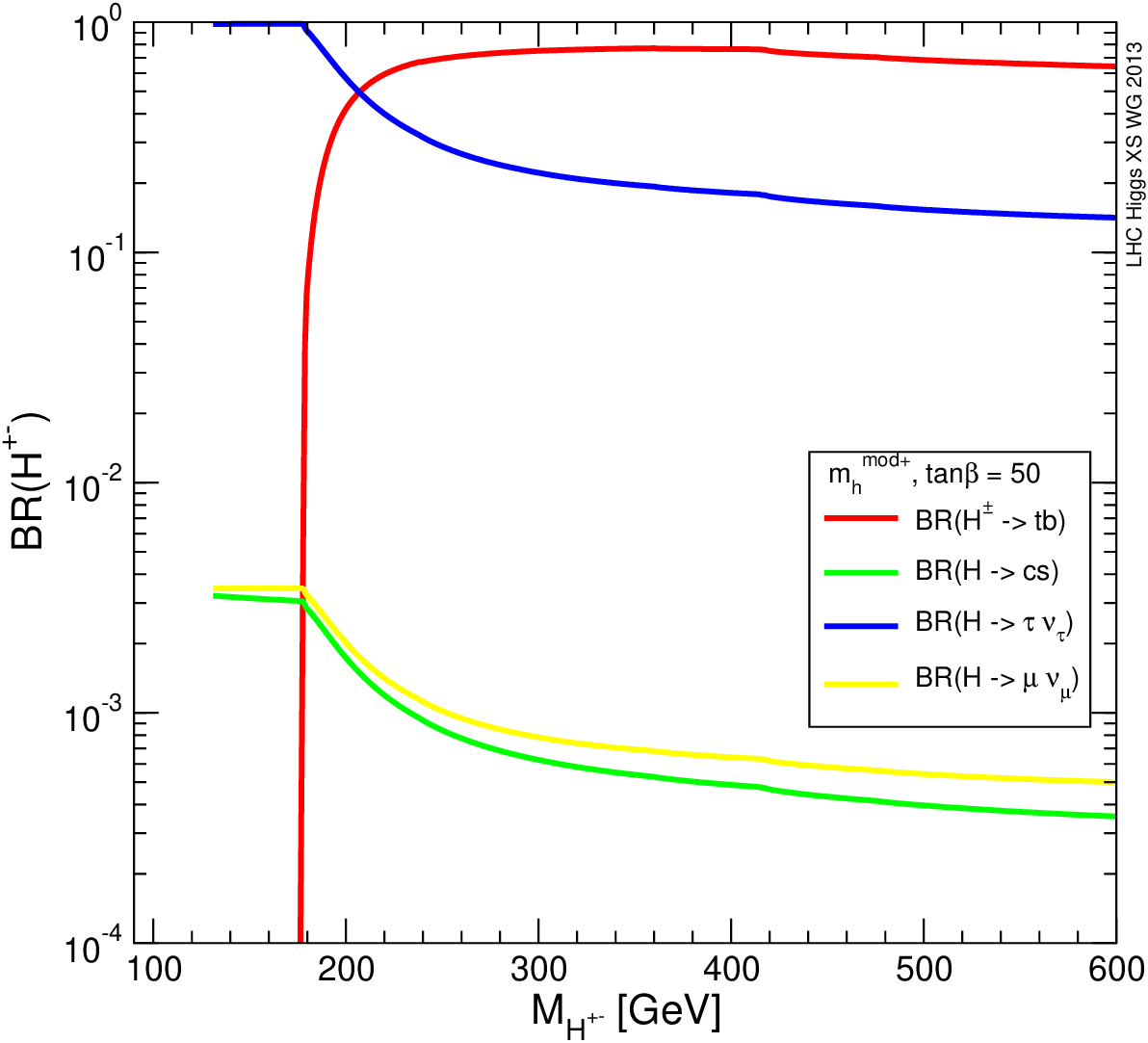}\\[0.5em]
\includegraphics[width=0.48\textwidth]{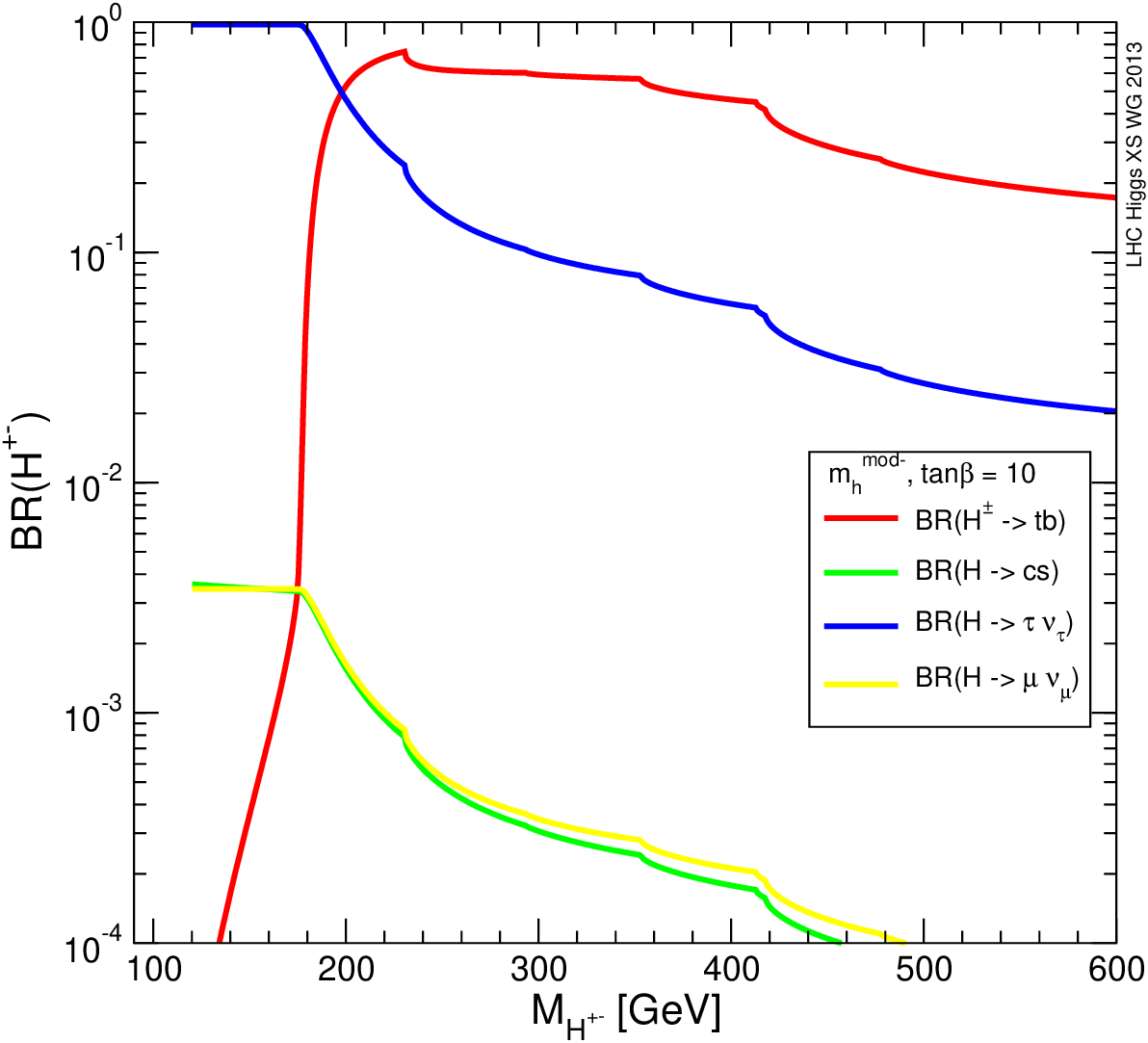}
\includegraphics[width=0.48\textwidth]{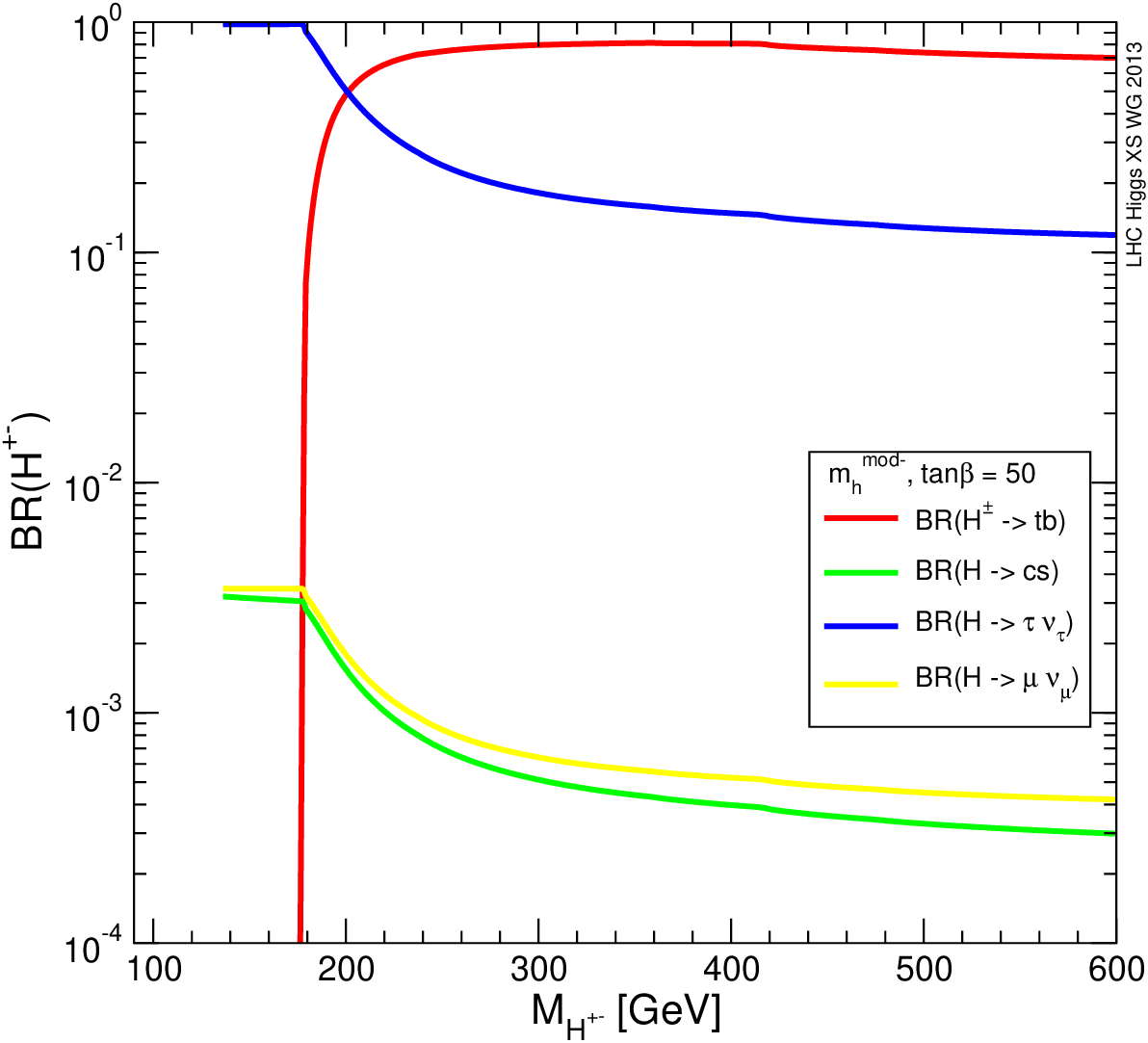}\\[0.5em]
%\vspace*{-0.3cm}
\caption{Branching ratios of the charged MSSM Higgs boson
in the \mhmaxx\ (upper row), \mhmodp\ (middle row) and the \mhmodm\ (lower
 row) scenario as a function of $\MHp$. 
The left (right) column shows the results for $\tb = 10 (50)$.}
\label{fig:YRHXS3_BR_MHp1}
%\vspace{3.2cm}
\end{figure}

\begin{figure}[htb!]
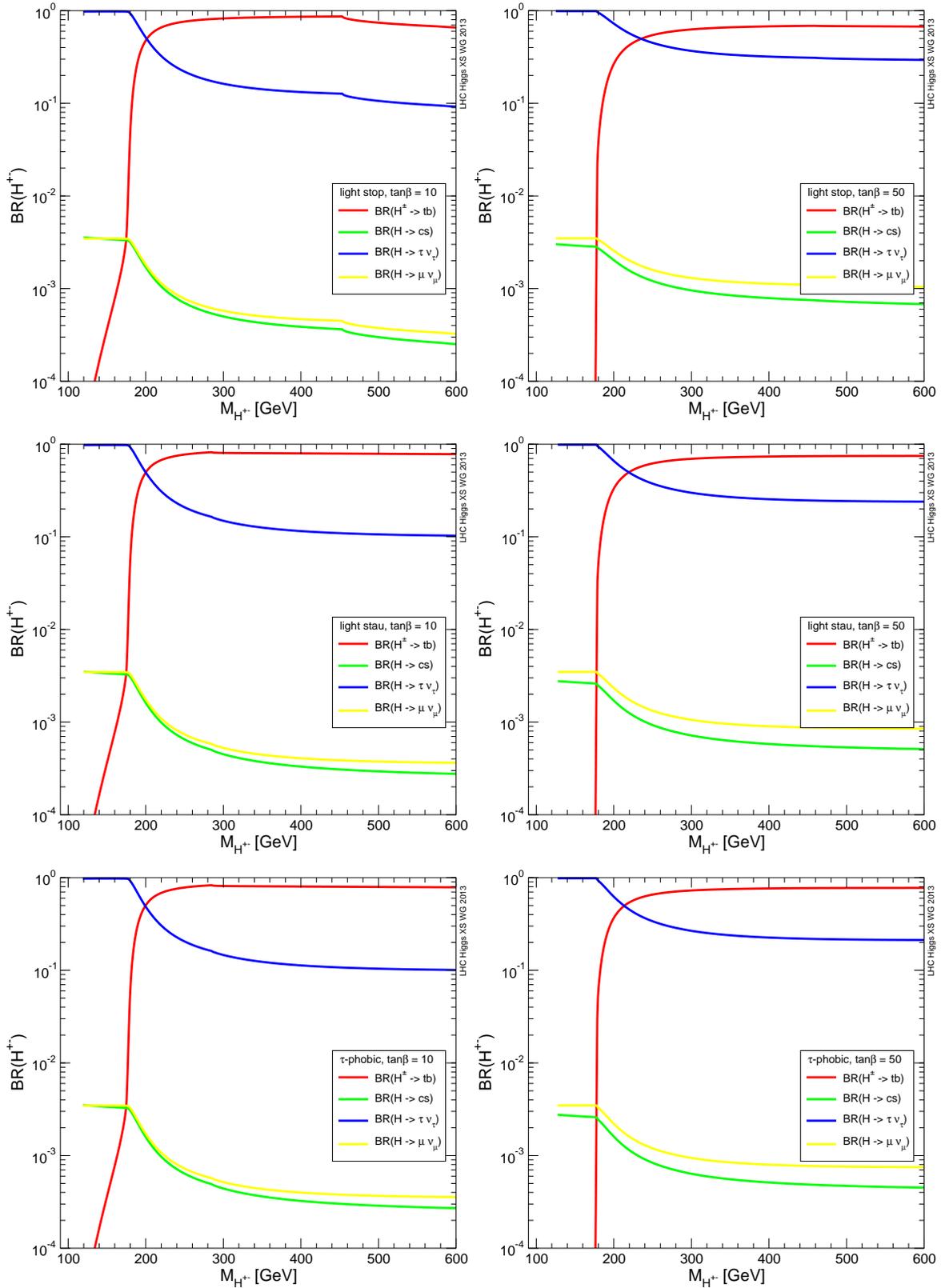

%\vspace{3.2cm}
\includegraphics[width=0.48\textwidth]{YRHXS3_BR/YRHXS3_BR_fig53}
\includegraphics[width=0.48\textwidth]{YRHXS3_BR/YRHXS3_BR_fig54}\\[0.5em]
\includegraphics[width=0.48\textwidth]{YRHXS3_BR/YRHXS3_BR_fig63}
\includegraphics[width=0.48\textwidth]{YRHXS3_BR/YRHXS3_BR_fig64}\\[0.5em]
\includegraphics[width=0.48\textwidth]{YRHXS3_BR/YRHXS3_BR_fig73}
\includegraphics[width=0.48\textwidth]{YRHXS3_BR/YRHXS3_BR_fig74}\\[0.5em]
%\vspace*{-0.3cm}
\caption{Branching ratios of the charged MSSM Higgs boson
in the light stop (upper row), light stau (middle row) and the \tauphobic\
 (lower row) scenario as a function of $\MHp$. 
The left (right) column shows the results for $\tb = 10 (50)$.}
\label{fig:YRHXS3_BR_MHp2}
%\vspace{3.2cm}
\end{figure}

The procedure outlined in the previous subsection can be applied to
arbitrary points in the MSSM parameter space. Here we show
representative results for the decay of the neutral MSSM Higgs bosons
in the (updated) \mhmaxx, \mhmodp, \mhmodm, light stop, light stau,
and \tauphobic\ scenario
in \refF{fig:YRHXS3_BR_mhmax} - \ref{fig:YRHXS3_BR_tauphobic}, respectively.
Shown are the branching ratios for the $\Ph$ in the upper row, for the 
$\PH$ in the middle row and for the
$\PA$ in the lower row with $\tb = 10 (50)$ in the left (right) column.
The results for the charged Higgs boson are given
in \refF{fig:YRHXS3_BR_MHp1}, (\ref{fig:YRHXS3_BR_MHp2}). The first plot shows
the \mhmaxx\ (upper row), \mhmodp\ (middle row) and \mhmodm\ (lower row)
scenario, while the second plot contains the light stop (upper row), light
stau (middle row) and \tauphobic\ Higgs (lower row) scenario.

%%%%%%%%%%%%%%%%%%% F I G U R E %%%%%%%%%%%%%%%%%%%%%%%%%%%%%%%%%%%%%%%%%%%%%%%
%%%%%%%%%%%%%%%%%%% F I G U R E %%%%%%%%%%%%%%%%%%%%%%%%%%%%%%%%%%%%%%%%%%%%%%%

%%%%%%%%%%%%%%%%%%% F I G U R E %%%%%%%%%%%%%%%%%%%%%%%%%%%%%%%%%%%%%%%%%%%%%%%
%%%%%%%%%%%%%%%%%%% F I G U R E %%%%%%%%%%%%%%%%%%%%%%%%%%%%%%%%%%%%%%%%%%%%%%%

%%%%%%%%%%%%%%%%%%% F I G U R E %%%%%%%%%%%%%%%%%%%%%%%%%%%%%%%%%%%%%%%%%%%%%%%
%%%%%%%%%%%%%%%%%%% F I G U R E %%%%%%%%%%%%%%%%%%%%%%%%%%%%%%%%%%%%%%%%%%%%%%%

%%%%%%%%%%%%%%%%%%% F I G U R E %%%%%%%%%%%%%%%%%%%%%%%%%%%%%%%%%%%%%%%%%%%%%%%
%%%%%%%%%%%%%%%%%%% F I G U R E %%%%%%%%%%%%%%%%%%%%%%%%%%%%%%%%%%%%%%%%%%%%%%%

%%%%%%%%%%%%%%%%%%% F I G U R E %%%%%%%%%%%%%%%%%%%%%%%%%%%%%%%%%%%%%%%%%%%%%%%
%%%%%%%%%%%%%%%%%%% F I G U R E %%%%%%%%%%%%%%%%%%%%%%%%%%%%%%%%%%%%%%%%%%%%%%%

%%%%%%%%%%%%%%%%%%% F I G U R E %%%%%%%%%%%%%%%%%%%%%%%%%%%%%%%%%%%%%%%%%%%%%%%
%%%%%%%%%%%%%%%%%%% F I G U R E %%%%%%%%%%%%%%%%%%%%%%%%%%%%%%%%%%%%%%%%%%%%%%%

%%%%%%%%%%%%%%%%%%% F I G U R E %%%%%%%%%%%%%%%%%%%%%%%%%%%%%%%%%%%%%%%%%%%%%%%
%%%%%%%%%%%%%%%%%%% F I G U R E %%%%%%%%%%%%%%%%%%%%%%%%%%%%%%%%%%%%%%%%%%%%%%%

%%%%%%%%%%%%%%%%%%% F I G U R E %%%%%%%%%%%%%%%%%%%%%%%%%%%%%%%%%%%%%%%%%%%%%%%
%%%%%%%%%%%%%%%%%%% F I G U R E %%%%%%%%%%%%%%%%%%%%%%%%%%%%%%%%%%%%%%%%%%%%%%%

The branching ratios of the light Higgs boson, $\Ph$, exhibit a strong
variation at low $\MA$, while for large $\MA$ the SM limit is reached.
The corresponding values of the light Higgs boson mass, $\Mh$, are indicated
in the corresponding plots in \refS{sec:sven-sub}.  %\refS{sec:MSSM-benchmark}. 
In particular in the $\PGt$-phobic scenario a reduction in the $\hhbb$ and
$\htautau$ channel can be observed, as analyzed in \Bref{Carena:2013qia}.

%For the two heavy neutral Higgs bosons only the results for $b \bar b$,
%$\tau^+\tau^-$, $\PGmp\PGm$ and $t \bar t$ are shown. While for 
%$\MA \lsim 350 \UGeV$ the $b \bar b$ channel is dominating at the 
%$\sim 90\%$ level, the $\tau^+\tau^-$ channel is subdominant at $\sim 10\%$.
%The $\PGmp\PGm$ channel remains well below the per-mille level. At the
%$t \bar t$ threshold the corresponding branching ratio rises up to the few
%per-cent level at $\tb = 10$, but remaining at or below the level of the
%$\PGmp\PGm$ channel for large $\tb$ due to the suppression of the coupling
%with $1/\tb$. 

The branching ratios of the $\PH$ and $\PA$ boson follow the same pattern in
all scenarios. For very low values of $\MA$ the $\HWW$ channel contributes,
but overall the $\Hbb$ channel is dominant, about 10 times larger than
the $\Htautau$ channel, which in turn is about 200 times larger than the
$\Hmumu$ channel. The size of the $\Htt$ channel, above the $\PQt \PAQt$
threshold is somewhat above the $\Htautau$ channel for $\tb = 10$, but
stays below even the $\Amumu$ channel for $\tb = 50$ due to the
suppression of the Higgs top coupling.
The same pattern, except the decay to $\PW^+\PW^-$ can be observed for the
$\PA$ boson.
It should be noted that the decays of the heavy Higgs bosons to charginos and
neutralinos, while taken into account in the BR evaluation, are not
shown. However, their effects are visible at the kinks in the lines of the
other channels, in particular for $\tb = 10$ (see
also \Bref{Carena:2013qia}).

The branching ratios of the $\PH^\pm$, shown as a function of $\MHp$, are also
very similar in the various scenarios. At low values only the channels
$\Hptaunu$, $\Hpcs$ and $\Hpmunu$ are open. At $\sim 180 \UGeV$ the channel
$\Hptb$ opens up and becomes dominant, while $\br(\Hpcs)$ and $\br(\Hpmunu)$
are very similar in size and more than two orders of magnitude smaller than
$\br(\Hptaunu)$. 
The decays of the charged Higgs boson to charginos and
neutralinos, again while taken into account in the BR evaluation, are not
shown. However, their effects are visible as for the neutral Higgs bosons 
at the kinks in the lines of the
other channels, in particular for $\tb = 10$.

%%%%%%%%%%%%%%%%%%%%%%%%%%%%%%%%%%%%%%%%%%%%%%%%%%%%%%%%%%%%%%%%%%%%%%%%%%%%%%%
%%%%%%%%%%%%%%%%%%%%%%%%%%%%%%%%%%%%%%%%%%%%%%%%%%%%%%%%%%%%%%%%%%%%%%%%%%%%%%%

%\clearpage

\clearpage

\newpage
\section{Gluon-gluon fusion production mode\footnote{
   D.~de Florian, B.~Di Micco  (eds.);  R.~Boughezal, F.~Caola, N.~Chanon, R.~Di~Nardo, G.~Ferrera, N.~Fidanza, M.~Grazzini, D.C.~Hall, C.~Hays, J.~Griffiths, R.~Hernandez-Pinto, N.~Kauer, H.~Kim, S.~Martin, J.~Mazzitelli, K.~Melnikov,  F.~Petriello, Y.~Rotstein-Habarnau, G.~Sborlini,  M.~Schulze, D.~Tommasini and J.~Yu
    }}

The first two volumes of this Handbook \cite{Dittmaier:2011ti,Dittmaier:2012vm} summarized the status of the inclusive and differential cross section for Higgs production in gluon fusion, respectively. Our goal in this volume is two-folded: we present an update on some of the relevant cross sections described in the previous volumes and continue the research on a number of issues relevant for Higgs boson production at the LHC.

The section is organised as follows. In \refS{sec:inclusive} we present the state of the art in the inclusive cross section for Higgs production. 
\refS{subsec:ggF_HjNNLO} presents the first calculation for $H+$jet production at second order in perturbation theory with phenomenological results arising from the gluon fusion channel. 
In \refSs{subsec:Hresggf}-\ref{subsec:Hqtggf} we compare the Higgs transverse momentum distribution outcome from different generators.
\refS{sec:interference} summarizes recent findings on the interference effects in light Higgs $\PV\PV$ modes. Finally, \refS{sec:WWbackground}
discusses theoretical  uncertainties on the $\Pp\Pp \rightarrow \PW\PW$  estimation in the Higgs search.

\subsection{Update on the inclusive Higgs boson production by gluon-gluon fusion\footnote{D.~de Florian, M.~Grazzini}}

\label{sec:inclusive}

The dominant mechanism for SM Higgs boson production at hadron colliders is gluon-gluon fusion \cite{Georgi:1977gs}, through a heavy-quark loop. The QCD radiative corrections to the total cross section have been computed at the next-to-leading order (NLO) in \Brefs{Dawson:1991zj,Djouadi:1991tka,Spira:1995rr} and at the next-to-next-to-leading order (NNLO accuracy) in \cite{Harlander:2002wh,Anastasiou:2002yz,Ravindran:2003um}. NNLO results at the exclusive level can be found in \Brefs{Anastasiou:2005qj,Anastasiou:2007mz,Catani:2007vq,Grazzini:2008tf}.

The main features of the different calculations for the inclusive cross section were discussed in \Bref{Dittmaier:2011ti}. For the 7 TeV run, both LHC experiments based their analysis on the combination \cite{Dittmaier:2011ti}  of the predictions of Anastasiou, Boughezal, Petriello and Stoeckli (ABPS) \cite{Anastasiou:2008tj} and  de Florian and Grazzini (dFG) \cite{deFlorian:2009hc}. 
Since the presentation of the First Yellow Report \cite{Dittmaier:2011ti} a number of advancements, including a better treatment of the effect of heavy quark masses and the Higgs boson line-shape were discussed. 
Considering that the ABPS and dFG agree within $1-2\%$, and that the improvements reported below were applied only to the calculation of dFG, 
here we focus on the improvements presented in \cite{deFlorian:2012yg}, which represented the theoretical prediction being used nowadays by the LHC collaborations.

The calculation of dFG is based on the resummation of soft-gluon contributions to next-to-next-to-leading logarithmic accuracy (NNLL) \cite{Catani:2003zt}, as a way to improve state of the art fixed-order predictions with the dominant effect from higher-order corrections. The implementation requieres the knowledge of  the Sudakov radiative factor, which depends only on the dynamics of soft gluon emission from the initial state partons and  the hard coefficient ($Cgg$) which includes  terms arising from both soft and hard gluon emission. The hard coefficient depends on the details of the coupling to the Higgs boson and, therefore, on the masses of the heavy quarks in the loop. 
\Bref{deFlorian:2012yg} presented the {\it exact} expression for this  coefficient up to next-to-leading logarithmic accuracy (NLL) with the explicit dependence on the heavy-quark masses in the loop, matching the precision reached in the fixed order calculation. Furthermore, the usually neglected contribution form the charm quark is also taken into account up to that order.
Corrections beyond NL accuracy are treated in the infinite quark mass limit, accounting only for the top quark contribution.
The inclusion of the exact dependence on the top- and bottom-quark masses up to NLL accuracy results in a decrease of the cross section
ranging from about $1.5\%$ at $\MH=125\UGeV$, to about $6\%$ at $\MH=800\UGeV$.
The usually neglected charm-quark contribution further decreases the cross section by about $1\%$ for a light Higgs, being  very small in the high-mass region.

 The second improvement with respect to the work of \Bref{deFlorian:2009hc} regards the treatment
of the Higgs boson width.
While the Zero Width Approximation (ZWA) can be considered sufficiently accurate for the evaluation of the {\it inclusive} cross section for a light Higgs boson, the increase of the Higgs boson width at large masses requires a proper implementation of the corresponding line-shape. We rely on the OFFP scheme described in \Bref{Actis:2008ug} as an effective implementation of the complex-pole scheme.
The calculation in \Bref{Actis:2008ug} provides a realistic estimate of the complex-pole width $\gamma_{\PH}$ above the $\PZ\PZ$ threshold but might introduce an artificial effect at low masses. In order to recover the ZWA for light Higgs we use an extrapolation of the value of $\gamma_{\PH}$ towards the on-shell decay width $\Gamma(\MH)$ below $\MH=200\UGeV$
%\footnote{Notice that effectively the OFFP scheme matches the naive Breit Wigner implementation below 200 GeV.}
.
The inclusion of finite-width effects results in an increase of the cross section with respect to the ZWA of about ${\cal O}(10\%)$ at $\MH=800\UGeV$.
The use of a naive Breit Wigner would give a smaller cross section with respect to the result in the complex-pole scheme, the difference ranging from $-3.5\%$ for $\MH=300\UGeV$ to $-18\%$ at $\MH=600\UGeV$, to $-27\%$ at $\MH=800\UGeV$.

 As discussed in \Bref{Dittmaier:2011ti}, the ensuing result is finally corrected for two-loop electro-weak (EW) contributions \cite{Aglietti:2004nj,Degrassi:2004mx,Aglietti:2006yd} as evaluated in \cite{Actis:2008ug} in the {\em complete factorization} scheme.
 %, in which the EW corrections are applied to the full QCD corrected cross section.

Uncertainties are estimated as in \Bref{Dittmaier:2011ti}. The updated numbers for the cross section at $7 \UTeV$ and $8 \UTeV$ are presented in 
\refTs{tab:YRHXS3_ggF_71}-\ref{tab:YRHXS3_ggF_77} and \refTs{tab:YRHXS3_ggF_81}-\ref{tab:YRHXS3_ggF_87}, respectively.

These results can be compared to those presented in 
\Bref{Anastasiou:2012hx}, where the impact of finite-width effects is also included. For a light Higgs boson, the main difference with the computation of dFG arises from the evaluation of higher-order QCD corrections, which are computed up to NNLO but choosing the factorization and renormalization scales $\muF=\muR=\MH/2$,
as an attempt to reproduce effects beyond NNLO, that, in the dFG calculation, are instead estimated through soft-gluon resummation.
For example, at $\MH=125\UGeV$ the result of \Bref{Anastasiou:2012hx} is about $7\%$ higher that the one of dFG, but still well within the corresponding  uncertainty bands.
Larger differences are observed in the high mass region, due to the different implementation of finite-width effects. In \Bref{Anastasiou:2012hx} a Breit-Wigner with running width is used
as the default implementation of the line-shape. At $\MH=400\UGeV$, the result 
of \Bref{Anastasiou:2012hx} turns out to be about $16\%$ smaller than the one of dFG.

\subsection{Higgs+jet at NNLO
\footnote{R.~Boughezal, F.~Caola, K.~Melnikov, F.~Petriello, M.~Schulze}}

\label{subsec:ggF_HjNNLO}
We describe in this section a first calculation of Higgs-boson production in association with a jet at next-to-next-to-leading order in perturbative QCD~\cite{Boughezal:2013uia}.  This result is urgently needed in order to reduce the theoretical uncertainties hindering a precise extraction of the Higgs properties at the LHC.  Currently, the theoretical errors in the one-jet bin comprise one of the largest systematic errors in Higgs analyses, particularly in the WW final state.  There are two theoretical methods one can pursue to try to reduce these uncertainties.  The first is to resum sources of large logarithmic corrections to all orders in QCD perturbation theory.  An especially pernicious source of large logarithmic corrections comes from dividing the final state into bins of exclusive jet multiplicities.  An improved theoretical treatment of these terms has been pursued in both the zero-jet~\cite{Berger:2010xi, Banfi:2012yh, Becher:2012qa, Tackmann:2012bt, Banfi:2012jm} and one-jet~\cite{Liu:2012sz, Liu:2013hba} bins (see \refS{sec:jets}).  The second, which we discuss here, is to compute the higher-order corrections to the next-to-next-to-leading order (NNLO) in perturbative QCD.  Both are essential to produce the reliable results necessary in experimental analyses.   In this contribution we give a brief overview of the calculational framework that was used to obtain the NNLO calculation for Higgs plus jet production, and present initial numerical results arising from gluon-fusion.

\subsubsection{Notation and setup}
\label{subsubsec:ggF_HjNNLO-setup}

We begin by presenting the basic notation needed to describe our calculation.  We use the QCD 
Lagrangian, supplemented with a dimension-five non-renormalizable 
operator that describes the interaction of the Higgs boson with gluons 
in the limit of very large top quark mass:
\begin{equation}
{\mathcal L} = -\frac{1}{4} G_{\mu \nu}^{(a)} G^{(a)\mu \nu}
- \lambda_{\PH\Pg\Pg} \PH G_{\mu \nu}^{(a)} G^{(a)\mu \nu}.
\label{ggF_HjNNLO:lag}
\end{equation}
Here, $G_{\mu \nu}^{(a)}$ is the field-strength tensor 
of the gluon field and $\PH$ is the Higgs-boson field. 
Matrix elements computed  with the Lagrangian of Eq.~(\ref{ggF_HjNNLO:lag}) 
need to be renormalized.  Two renormalization constants 
are required to do so: one which relates the bare and renormalized strong coupling constants, and another which ensures that matrix elements of the $\PH\Pg\Pg$ dimension-five operator are finite.  The expressions for these quantities are given in \Bref{Boughezal:2013uia}.  We note that the Lagrangian of Eq.~(\ref{ggF_HjNNLO:lag}) neglects light fermions, as will 
the initial numerical results presented.  We comment on the phenomenological impact of this approximation later in this section.

Renormalization of the strong coupling constant and of the effective 
Higgs-gluon coupling removes ultraviolet divergences from the matrix 
elements. The remaining divergences are of infrared origin. To 
remove them, we must both define and compute infrared-safe observables, 
and absorb the remaining collinear singularities by renormalizing 
parton distribution functions. Generic infrared safe observables are defined using jet algorithms. 
For the calculation described here we employ the $k_{\mathrm T}$-algorithm. 

Collinear singularities associated with 
gluon radiation by incoming partons must be removed by additional renormalization of parton 
distribution functions. We describe how to perform this renormalization in what follows. 
We denote the ultraviolet-renormalized 
partonic cross section by ${\bar \sigma}(x_1,x_2)$, and the collinear-renormalized 
partonic cross section by  $\sigma(x_1,x_2)$. Once we know $\sigma(x_1,x_2)$, we 
can compute the hadronic cross sections by integrating 
the product of $\sigma$  and 
the gluon  distribution  functions  over $x_1$ and $x_2$:
\begin{equation}
\sigma(\Pp + \Pp \to \PH+j) = \int {\mathrm d} x_1 {\mathrm d}x_2 \;g(x_1) g(x_2) \; \sigma(x_1,x_2).
\end{equation}
The relation between $\sigma$ and ${\bar \sigma}$ is given by the following formula:
\begin{equation}
\sigma = \Gamma^{-1} \otimes {\bar \sigma} \otimes \Gamma^{-1},
\label{ggF_HjNNLO:eq_basic}
\end{equation}
where the convolution sign stands for 
\begin{equation}
\left [ f \otimes g \right](x) = 
\int \limits_{0}^{1} {\mathrm d} z \, {\mathrm d} y \delta( x - yz) f(y) g(z).
\end{equation}
The collinear counterterm can be expanded in the strong coupling constant as
\begin{equation}
\Gamma = \delta(1-x) - \left ( \frac{\alphas}{2\pi} \right ) \Gamma_1 
+ \left ( \frac{\alphas}{2\pi} \right )^2  \Gamma_2.
\end{equation}
We write the ultraviolet-renormalized partonic cross section through NNLO as  
\begin{equation}
{\bar \sigma} = {\bar \sigma}^{(0)} 
+
\left ( \frac{\Gamma(1+\epsilon) \alphas}{2\pi} \right ) 
 {\bar \sigma}^{(1)} 
+ \left (\frac{ \Gamma(1+\epsilon)  \alphas}{2\pi} \right )^2 {\bar \sigma}^{(2)}, 
\end{equation}
and the collinear-renormalized partonic cross section as 
\begin{equation}
 \sigma = \sigma^{(0)} + 
\left ( \frac{ \alphas}{2\pi} \right )   {\sigma}^{(1)} 
+ \left ( \frac{ \alphas}{2\pi} \right )^2 {\sigma}^{(2)}. 
\end{equation}
Using these results, we can solve to find the following results for the finite cross section 
expanded in $\alphas$:
\begin{equation}
\begin{split} 
& 
\sigma^{(0)} = {\bar \sigma}^{(0)},\;\;\;\;\;\;\;\;\;\;\;
\sigma^{(1)} = {\bar \sigma}^{(1)}
 + \frac{\Gamma_1 \otimes {\sigma}^{(0)}}{\Gamma(1+\epsilon)} 
 + \frac{{\sigma}^{(0)} \otimes \Gamma_1 }{\Gamma(1+\epsilon)},
\\
& \sigma^{(2)} = {\bar \sigma}^{(2)} 
- \frac{\Gamma_2 \otimes {\sigma}^{(0)} }{\Gamma(1+\epsilon)^2} 
- \frac{{\sigma}^{(0)} \otimes \Gamma_2  }{\Gamma(1+\epsilon)^2} 
- \frac{  \Gamma_1 \otimes {\sigma}^{(0)} \otimes \Gamma_1  }{\Gamma(1+\epsilon)^2} 
%\\
%&~~~~~~~~~~~ 
+ \frac{ \Gamma_1 \otimes \sigma^{(1)} }{\Gamma(1+\epsilon)}   
+ \frac{\sigma^{(1)} \otimes \Gamma_1}{\Gamma(1+\epsilon)}.    
\end{split}
\label{ggF_HjNNLO:eq2}
\end{equation}
We note that although finite, the $\sigma^{(i)}$ still depend on unphysical renormalization and
factorization scales because of the truncation of the perturbative expansion. 
In the following, we will consider for simplicity 
the case of equal renormalization and factorization scales ,
$\muR=\muF=\mu$. 
The residual $\mu$ dependence is easily determined by solving the 
renormalization-group equation order-by-order in $\alphas$.

\subsubsection{Calculational framework}
\label{subsubsec:ggF_HjNNLO-framework}

It follows from the previous section that in order to 
obtain $\sigma^{(2)}$ at a generic scale, apart from lower-order results
we need to know the NNLO renormalized cross section ${\bar \sigma}^{(2)}$ 
and convolutions of NLO and 
LO cross sections with the various splitting functions which appear in the collinear counterterms.
Up to terms induced by the renormalization, 
there are three contributions to  ${\bar \sigma}^{(2)}$ that are  required:
\begin{itemize} 
\item the two-loop virtual corrections to $\Pg\Pg \to \PH\Pg$;
\item   the one-loop virtual corrections to 
$\Pg\Pg \to \PH\Pg\Pg$; 
\item the double-real contribution  $\Pg\Pg \to \PH\Pg\Pg\Pg$.  
\end{itemize} 
We note that helicity amplitudes for 
all of these processes  are available in the literature. The 
two-loop amplitudes for $\Pg\Pg \to \PH\Pg$ were recently 
computed  in 
\Bref{Gehrmann:2011aa}. 
The one-loop corrections to $\Pg\Pg \to \PH\Pg\Pg$ and the tree amplitudes for $ \Pg\Pg \to \PH\Pg\Pg\Pg$
are also known, and are available in the form 
of a Fortran code in the program 
MCFM~\cite{Campbell:2010ff}.  

Since all ingredients for the 
NNLO computation of $\Pg\Pg \to \PH+{\mathrm{jet}}$ have 
been available for some time, it is interesting  
to understand what has prevented this calculation from being performed.
The main
difficulties  with NNLO calculations appear when one attempts
to combine the different contributions, since integration 
over phase space introduces 
additional singularities if the required number of jets 
is lower than the parton multiplicity.  
To perform the phase-space  integration, 
we  must  first isolate singularities in tree- and loop amplitudes.  
%It required a long time to establish a convenient way to do this.

The computational method that we will explain shortly  is based 
on the idea that relevant singularities can be isolated 
using appropriate parameterizations of  phase space 
and expansions in plus-distributions.  To use this approach for computing   NNLO QCD corrections, we need 
to map the relevant phase space  to a unit hypercube 
in such a way that extraction of singularities  is 
straightforward.  %It is clear that t
The correct variables 
to use  are the re-scaled energies of unresolved 
partons and the relative angles between 
two unresolved collinear partons. However, 
the problem is that  different partons become unresolved in different parts of the phase space.  It is not immediately clear how to switch between different 
sets of coordinates and cover the full phase space.  

We note that for  NLO QCD computations, this  problem was 
solved in  \Bref{Frixione:1995ms}, where it was 
explained that the  full phase space can be partitioned into 
sectors in such a way 
that in each sector only one parton ($i$) can  produce 
a soft singularity and only one pair 
of partons ($ij$) can produce a collinear singularity. 
In each sector,  
the proper variables are  the energy of the parton $i$ and the relative 
angle between partons $i$ and $j$. 
Once the partitioning of the phase space is established and proper variables are chosen for each sector, 
we can use an expansion in plus-distributions to construct  
relevant subtraction terms for each 
sector. With the  subtraction terms in place,  
the Laurent expansion of  cross sections in  $\epsilon$ can be constructed, 
and each 
term in such an expansion can be integrated 
over the phase space independently. 
Therefore, partitioning of the 
phase space into suitable sectors and proper parameterization of the 
phase space in each of these sectors are the two crucial elements 
needed to extend this method to NNLO.  It was first suggested in \Bref{Czakon:2010td} how to construct this 
extension for double real-radiation processes at NNLO.  We give a brief overview of this technique in the next section.

\subsubsection{An example: double-real emission corrections}
\label{subsubsec:ggF_HjNNLO-doublereal}

We briefly present the flavor of our calculational methods by outlining how the double-real emission corrections are handled.  To start, we follow the logic used at NLO in \Bref{Frixione:1995ms} and partition the phase space for the $\Pg(p_1)\Pg(p_2) \to \PH \Pg(p_3) \Pg(p_4)\Pg(p_5)$ process into separate structures that we call `pre-sectors' where only a given set of singularities can occur:
\begin{equation}
\frac{1}{3!} {\mathrm d}{\mathrm{Lips}}_{12 \to \PH 345 }
= \sum \limits_{\alpha}^{} {\mathrm d}{\mathrm{Lips}}_{12 \to \PH 345 }^{(\alpha)},
\end{equation}
Here, $\alpha$ is a label which denotes which singularities can occur, and dLIPS denotes the standard Lorentz-invariant phase space.  At NNLO we can have at most two soft singularities and two collinear singularities in each pre-sector, so as an example there will be an $\alpha$ labeling where $p_4$ and $p_5$ can be soft, and where both can be collinear to $p_1$.  Within this particular pre-sector, which we label as a `triple-collinear' pre-sector, it is clear that the appropriate variables to describe the phase space are the energies of gluons $p_4$ and $p_5$, and the angles between these gluons and the direction of $p_1$. 

Our goal in introducing a parameterization is to have all singularities appear in the following form:
\begin{equation}
I(\epsilon) = \int \limits_{0}^{1} {\mathrm d} x x^{-1-a\epsilon} F(x), 
\label{ggF_HjNNLO:fact}
\end{equation}
where the function $F(x)$ has a well-defined limit $\lim \limits_{x \to 0}^{} F(x) = F(0)$.  Here, the $F(x)/x$ structure comes from the matrix elements, while the $x^{-a\epsilon}$ comes from the phase space.  When such a structure is obtained, we can extract singularities using the plus-distribution expansion
\begin{equation}
\frac{1}{x^{1+a\epsilon}} = -\frac{1}{a\epsilon} \delta(x) + \sum_{n=0}^{\infty}
\frac{(-\epsilon a)^n}{n!}\left[\frac{\ln^n(x)}{x}\right]_+ \, ,
\end{equation}
so that 
\begin{equation}
I(\epsilon) = \int \limits_{0}^{1} {\mathrm d} x \left ( -\frac{F(0)}{a\epsilon}  + \frac{F(x) - F(0)}{x}
 -a\epsilon  \frac{F(x) - F(0)}{x} \ln(x) +...\right ).
 \label{ggF_HjNNLO:ints}
\end{equation}
The above equation provides the required Laurent  expansion of the integral $I(\epsilon)$.  We note that each term in such an expansion can be calculated numerically, and independently from the other terms.   

Unfortunately, no phase-space parameterization for double-real emission processes can immediately achieve the structure of Eq.~(\ref{ggF_HjNNLO:fact}).  Each pre-sector must be further divided into a number of sectors using variable changes designed to produce the structure 
of Eq.~(\ref{ggF_HjNNLO:fact}) in each sector.  Following \Bref{Czakon:2010td}, we further split the triple-collinear pre-sector mentioned above into five sectors so that all singularities in the matrix elements appear in the form of Eq.~(\ref{ggF_HjNNLO:fact}).  Explicit details for these variables changes, and those for all other pre-sectors needed for the NNLO calculation of Higgs plus jet, are presented in \Bref{Boughezal:2013uia}.

Once we have performed the relevant variables changes, we are left with a set of integrals of the form shown in Eq.~(\ref{ggF_HjNNLO:ints}).  We now discuss how we evaluate the analogs of the $F(x)$ and $F(0)$ terms that appear in the full calculation.  When all $x_i$ variables that describe the final-state phase space are non-zero, we are then evaluating the $\Pg\Pg \to \PH\Pg\Pg\Pg$ matrix elements with all gluons resolved.  The helicity amplitudes for this process are readily available, as discussed above, and can be efficiently evaluated numerically.  When one or more of the $x_i$ vanish, we are then in a singular limit of QCD.  The factorization of the matrix elements in possible singular limits appearing in double-real emission corrections in QCD has been studied in detail~\cite{Catani:1999ss}, and we can appeal to this factorization to evaluate the analogs of the $F(0)$.  For example, one singular limit that appears in all pre-sectors is the so-called `double-soft' limit, in which both gluons $p_4$ and $p_5$ have vanishingly small energies.  The matrix elements squared factorize in this limit in the following way in terms of single $S_{ij}(p)$ and double $S_{ij}(p,q)$ universal eikonal factors~\cite{Catani:1999ss}:
\begin{equation}
\begin{split} 
& |{\mathcal M}_{\Pg_1 \Pg_2 \to \PH \Pg_3 \Pg_4 \Pg_5}|^2 
\approx  C_A^2 g_s^4 
 \Bigg [
\left ( \sum \limits_{ij \in S_p}^{} S_{ij}(p_4) \right ) 
\left ( \sum \limits_{kn \in S_p}^{} S_{kn}(p_5) \right ) 
\\
&
+ 
\sum \limits_{ij \in S_p}^{} S_{ij}(p_4,p_5) 
 - \sum \limits_{i=1}^{3}  S_{ii}(p_4,p_5) 
\Bigg ] |{\mathcal M}_{g_1 g_2 \to H g_3}|^2.
\end{split}
\label{ggF_HjNNLO:doublesoft}
\end{equation}
The advantage of using this factorization is that all structures on the right-hand side of Eq.~(\ref{ggF_HjNNLO:doublesoft}) are readily available in the literature and can be efficiently evaluated numerically; as discussed the helicity amplitudes for $\Pg\Pg \to \PH\Pg$ are known, and the $S_{ij}$ eikonal factors which appear are also well-known functions.  Using this and other such relations, the analogs of the integrals in Eq.~(\ref{ggF_HjNNLO:ints}) appearing in the full theory can be calculated using known results.  Similar techniques can be used to obtain the other structures needed for the NNLO computation of Higgs plus jet production.  For a discussion of all relevant details, we refer the reader to \Bref{Boughezal:2013uia}.

\subsubsection{Numerical results}
\label{subsubsec:ggF_HjNNLO-numerics}

We present here initial numerical results for Higgs production in association with one or more jets at NNLO, arising from the dominant gluon-fusion subprocess.  A detailed series of checks on the presented calculation were performed in \Bref{Boughezal:2013uia}, and we do not repeat this discussion here. 
We compute 
the hadronic cross section for the production of the Higgs boson 
in association with one or more jets  at the 8 TeV LHC 
through NNLO in perturbative QCD. 
We reconstruct jets using the $k_{\mathrm T}$-algorithm
with $\Delta R = 0.5$ and $\pT^j=30~\UGeV$. 
The Higgs mass is taken to be $\MH=125\UGeV$
and the top-quark mass $m_{\PQt}=172~\UGeV$. We use the 
latest NNPDF parton distributions~\cite{Ball:2011uy,Ball:2012cx} with the number of active fermion flavors 
set to five, and 
numerical values of the strong coupling constant $\alphas$ 
at various orders in QCD perturbation theory as provided 
by the NNPDF fit. We note that in this case  $\alphas(\MZ)=[0.130,0.118,0.118]$ 
at leading,  next-to-leading  and next-to-next-to-leading 
order, respectively. 
We choose the central renormalization
and factorization scales to be $\mu=\muR=\muF=\MH$. 

\begin{figure}[!t]
\begin{center}
\includegraphics[width=0.42\textwidth,angle=270]{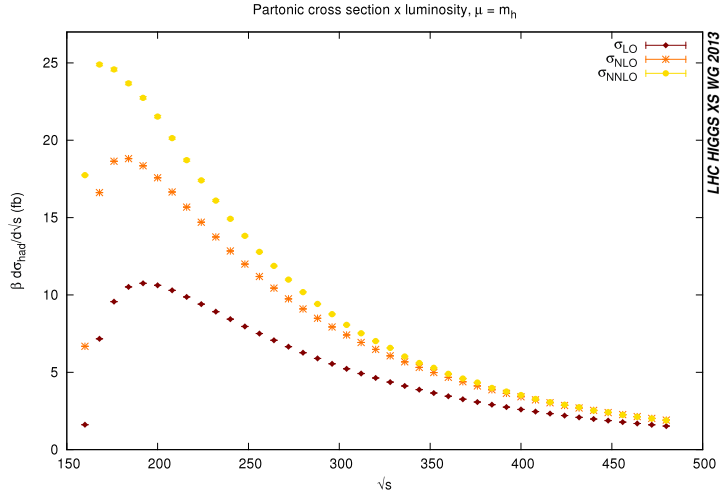}
\end{center}
\caption{Results for the product of partonic cross sections 
$\Pg\Pg \to \PH + {\mathrm{jet}}$ and  parton luminosity in consecutive  orders in perturbative QCD
at $\mu = \muR = \muF = \MH = 125\UGeV$. 
See the text for explanation.  }
\label{ggF_HjNNLO:fig:xsect}
\end{figure}

In \Fref{ggF_HjNNLO:fig:xsect} we show the partonic cross
section for $\Pg\Pg\to \PH + j$ multiplied by the gluon luminosity through NNLO in perturbative QCD:
\begin{equation}
\beta \frac{{\mathrm d}\sigma_{\mathrm{had}}}{{\mathrm d}\sqrt{s}} = \beta \frac{{\mathrm d}\sigma (s,\alphas,\muR,\muF)}{{\mathrm d}\sqrt{s}} \times
\mathcal L(\frac{s}{s_{\mathrm{had}}},\muF),
\end{equation}
where $\beta$ measures the distance from the partonic threshold,
\begin{equation}
\beta = \sqrt{1-\frac{E_{\mathrm{th}}^2}{s}},~~~~~~~ E_{\mathrm{th}} = \sqrt{\MH^2 + p_{\perp,j}^2} + p_{\perp,j}\approx 158.55~\UGeV.
\end{equation}
The partonic luminosity $\mathcal L$ is given by the integral 
of the product of two gluon distribution functions
\begin{equation}
\mathcal L(z,\muF) = \int_z^1 \frac{{\mathrm d}x}{x} 
g(x,\muF) g\left(\frac{z}{x},\muF\right).
\end{equation}
It follows from \Fref{ggF_HjNNLO:fig:xsect} that NNLO QCD corrections are significant in the 
region $\sqrt{s} < 500\UGeV$. In particular, 
close to  partonic threshold $ \sqrt{s} \sim E_{\mathrm{th}}$, radiative corrections are enhanced 
by threshold logarithms $\ln \beta$ that originate from the incomplete 
cancellation of virtual and real corrections.  There seems to be no significant enhancement 
of these corrections at higher  energies, where  the  NNLO QCD prediction for 
the partonic cross section becomes almost indistinguishable from 
the NLO QCD one.  

\begin{figure}[!t]
\begin{center}
\includegraphics[width=0.65\textwidth]{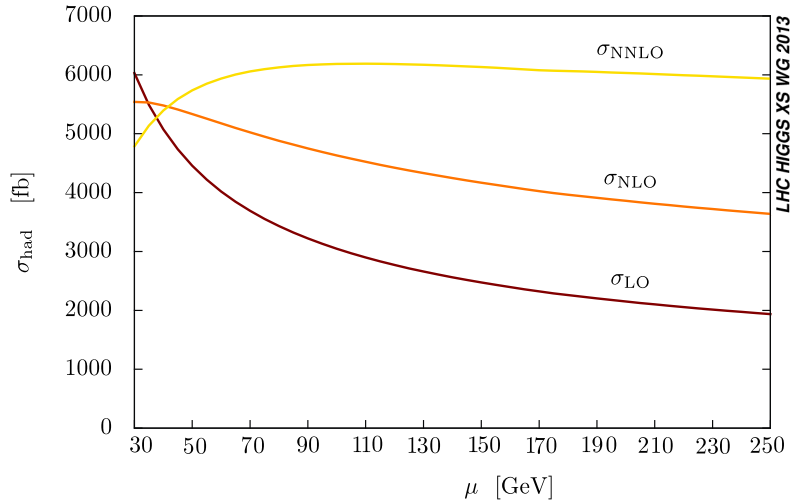}
\end{center}
\caption{Scale dependence of the hadronic cross section in consecutive  orders in perturbative QCD.
See the text for details. 
 }\label{ggF_HjNNLO:fig:scale}
\end{figure}

We now show the integrated hadronic cross sections in the all-gluon channel. 
We choose to vary the renormalization and factorization scale 
in the range $\mu = \muR=\muF = \MH/2,~\MH,~2\MH$. 
After convolution with the parton luminosities, we 
obtain
\begin{equation}
\begin{split} 
&\sigma_{\LO}(\Pp\Pp\to \PH j) = 2713^{+1216}_{-776}~ {\mathrm{fb}}, \\
&\sigma_{\NLO}(\Pp\Pp\to \PH j) = 4377^{+760}_{-738}~ {\mathrm{fb}}, \\
&\sigma_{\NNLO}(\Pp\Pp\to \PH j) = 6177^{+204}_{-242}~ {\mathrm{fb}}.
\end{split} 
\end{equation}
We note that NNLO corrections are sizable, as expected
from the large NLO $K-$factor, but the perturbative expansion shows  
marginal  convergence. We also evaluated PDF errors using the full set of NNPDF replicas,
and found it to be of order $5\%$ at LO, and of order $1-2\%$ at both NLO and NNLO, similarly
to the inclusive Higgs case~\cite{Ball:2012cx}. 
The cross section increases by about sixty percent when we move from LO to NLO 
and by thirty percent when we move from NLO to NNLO.  It is also clear that 
by accounting for the NNLO QCD corrections we reduce the dependence on the renormalization 
and factorization scales in a significant way.  The scale variation of the 
result decreases from almost $50\%$ at LO, to $20\%$ at NLO, to less than $5\%$ at NNLO. 
We also note that a perturbatively-stable result is obtained for the scale 
choice $\mu\approx \MH/2$. In this case the 
ratio of the NNLO over the LO cross section  is just 
$1.5$, to be compared 
with  $2.3$ for  $\mu=\MH$ and $3.06$ 
for $\mu=2\MH$, and the 
ratio of NNLO to  NLO is $1.2$.  A similar  trend was observed 
in the calculation of higher-order QCD corrections 
to the Higgs boson production cross section in gluon fusion. 
The reduced scale dependence is also apparent from \Fref{ggF_HjNNLO:fig:scale},
where we plot total cross section as a function of the renormalization and
factorization scale $\mu$ in the region $\pT^j<\mu<2 \MH$.

Finally, we comment on the phenomenological 
relevance of the results for the cross sections 
and $K$-factors reported here that refer only to the Higgs production through gluon-gluon collisions. We note that 
at leading and next-to-leading order, quark-gluon 
collisions increase the $\PH+j$ production cross section 
by about $30$ percent, for the input parameters that we use 
in this paper. At the same time, the NLO $K$-factors for the 
full $\PH + j$ cross section are smaller by about $10{-}15$\%
than the `gluons-only' $K$-factors,  presumably because quark color charges 
are smaller than the gluon ones.  Therefore, 
we conclude that the gluon-only results 
can be used   for reliable phenomenological estimates 
of perturbative $K$-factors but adding quark channels will be essential 
for achieving precise results for the $\PH + j$ cross section.

%\clearpage

\subsection{Higgs boson production by the gluon-gluon fusion and decay in the electroweak channels in the \textsc{HRes} code
\footnote{D.~de Florian, G.~Ferrera, M.~Grazzini, D.~Tommasini}}
\label{subsec:Hresggf}

\subsubsection{Introduction}
\label{subsec:introduction}

In the section we introduce the \textsc{HRes}\cite{deFlorian:2012mx} Monte Carlo program implementing the Higgs boson production by gluon-gluon fusion channel (at the NNLL+NNLO accuracy) and the electroweak decay
modes $\PH\rightarrow\PGg\PGg$, $\PH\rightarrow \PW\PW\rightarrow \Pl\PGnl \Pl\PGnl$ and $\PH\rightarrow \PZ\PZ\rightarrow 4l$. In the latter case
the user can choose between $\PH\rightarrow \PZ\PZ\rightarrow \PGmp\PGm\Pep\Pe$ and $\PH\rightarrow \PZ\PZ\rightarrow \Pep\Pe\Pep\Pe$, which includes
the appropriate interference contribution.
This \textsc{HRes} numerical
program embodies the features of both the \textsc{HNNLO}\cite{Catani:2007vq,Grazzini:2008tf} and
 \textsc{HqT}\cite{deFlorian:2011xf} numerical codes.
Here we compare some selected results with the fixed order ones, up to the NNLO accuracy, obtained with the \textsc{HNNLO}. The program can be downloaded from \cite{hqt}, together
with some accompanying notes.

%%%%%%%%%%%%%%%%%

\subsubsection{\textsc{HRes} example with the $\PGg\PGg$ decay channel}
\label{subsec:HRes_examples}

We present selected numerical results for the signal cross section at the LHC ($\sqrt{s} = 8\UTeV$), by using explicative cuts that can be applied in current Higgs boson searches by the ATLAS and CMS collaborations.
We consider the production of a SM Higgs boson
with mass $\MH = 125\UGeV$ and the decay channel into $\PGg\PGg$, but analogous studies can be performed by the other decay channels (see \Bref{deFlorian:2012mx}).

For
each event, we classify the photon transverse momenta according to their
minimum and maximum value, $p_{\mathrm{T min}}$ and $p_{mathrm{T max}}$ . 
We apply the following cuts on the photons: they are required
to be in the central rapidity region, $|\eta| < 2.5$, with $p_{{\mathrm{T min}},\PGg} > 25\UGeV$ and
$p_{{\mathrm{T max}},\PGg} > 40\UGeV$.
The corresponding inclusive cross sections are reported in \Tref{tab:gamgacentralscale}, where we show the resummed results obtained through the \textsc{HRes} code, and we compare them with the fixed order predictions obtained with the \textsc{HNNLO} code. We see that the NLL+NLO (NNLL+NNLO) inclusive cross section agrees with the NLO (NNLO) result to better than $1\%$.
We recall that the resummation does not affect the total cross section for the Higgs boson production, but when geometrical cuts are applied, their effect can act in a different way on fixed order and resummed calculations. In \Tref{tab:gamgacentralscale} we compare the accepted cross sections, obtained by the fixed order and resummed calculations,
and the corresponding efficiencies. 
\begin{table}[!ht]
\caption{ Fixed order and resummed cross sections for $\Pp\Pp\to \PH+\PX\to \PGg\PGg+\PX$ at the LHC, before and after geometrical acceptance cuts.}
\label{tab:gamgacentralscale}
\begin{center}
\begin{tabular}{lcccc}
\hline
Cross section & NLO & NLL+NLO & NNLO & NNLL+NNLO\\ \hline 
Total [fb]& 30.65 $\pm$ 0.01 & 30.79 $\pm$ 0.03 & 38.47 $\pm$ 0.15 & 38.41 $\pm$ 0.06\\ 
With cuts [fb]& 21.53 $\pm$ 0.02 & 21.55 $\pm$ 0.01 & 27.08 $\pm$ 0.08 & 26.96 $\pm$ 0.04\\ 
Efficiency [\%] & 70.2 & 70.0 & 70.4 & 70.2\\
\hline
\end{tabular}
\end{center}
\end{table}
The numerical errors estimate the statistical uncertainty of the Monte Carlo integration.
Comparing resummed and fixed order predictions, we see that there are no substantial differences on the accepted cross section, due to the fact that the integration is performed over a wide kinematical range. It is also possible to study the accepted cross section for different choices of the scales. After selection cuts, by varying the scales $2Q=\muF=\muR$ in the range $\MH/2\leq2Q\leq2\MH$, the scale uncertainty is about $\pm 15\%$ ($\pm 18 \%$) at NLL+NLO (NLO) and $\pm 9\%$ ($\pm 10\%$) at NNLL+NNLO (NNLO).

\begin{figure}[!ht]
\centering
\includegraphics[width=0.485\textwidth]{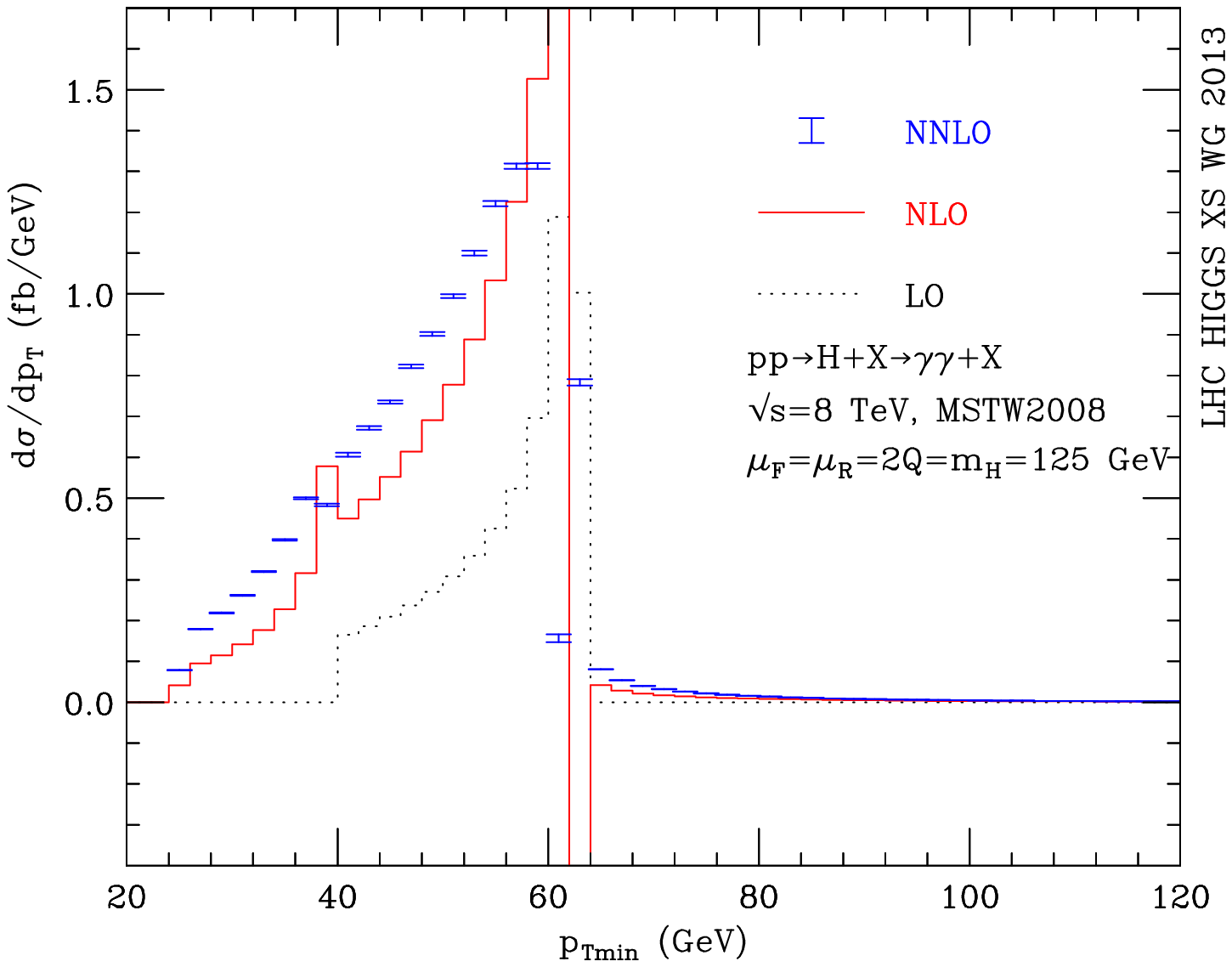}
\includegraphics[width=0.485\textwidth]{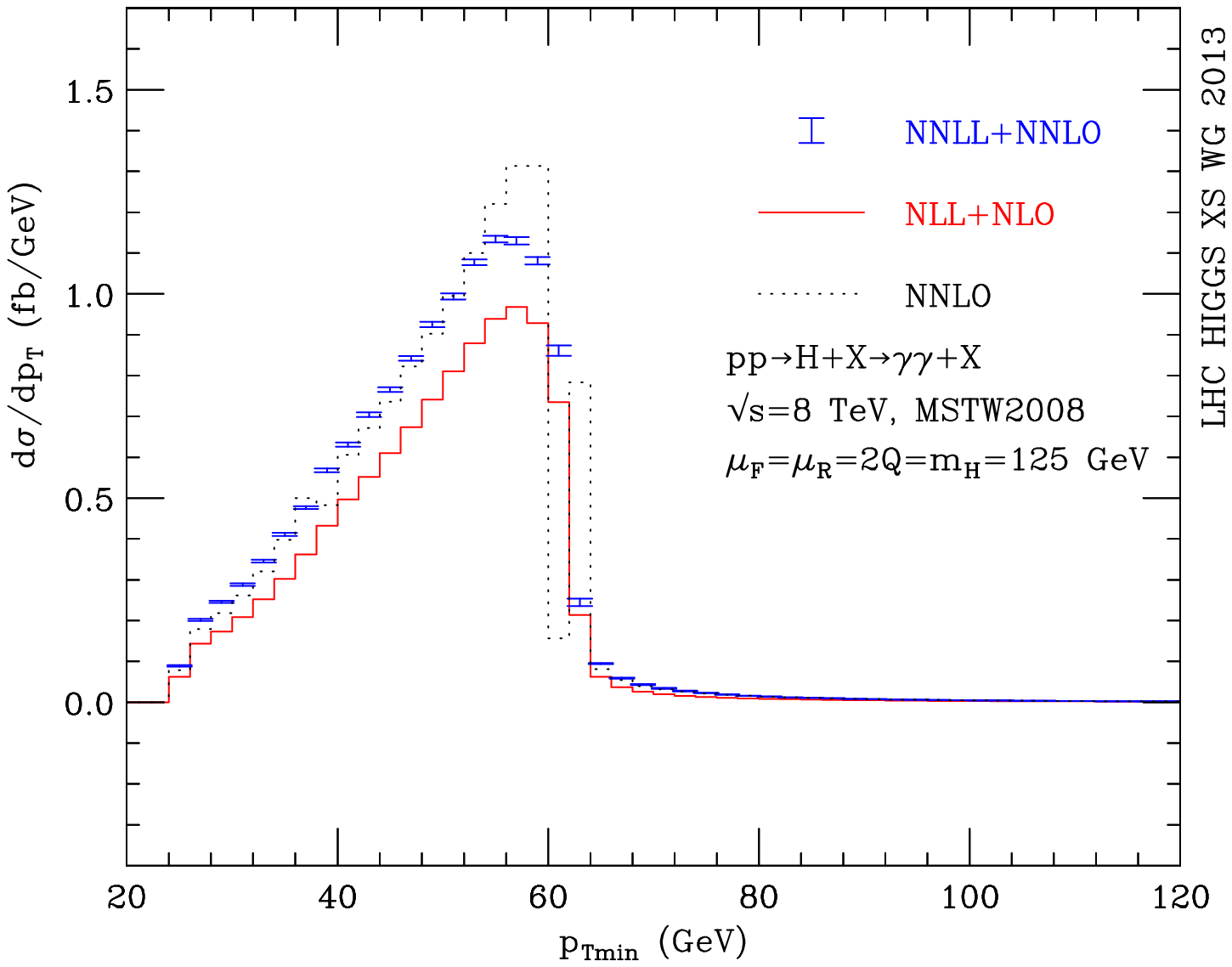}
\caption{\label{fig:pTphoton}{Distribution  $\pT^{{\mathrm{max}},\PGg}$ for the
$\PH\to \PGg\PGg$ signal at the LHC, obtained by resummed (right plot) calculations and fixed order (left plot) for comparison.}}
\end{figure}

In \Fref{fig:pTphoton}-left we plot the photon $\pT^{{\mathrm{min}},\PGg}$ distribution (
the $\pT^{{\mathrm{max}},\PGg}$ distribution has similar features). These distribution are enhanced when going from LO to NLO to NNLO according to the increase of the total cross section. We note that, as pointed out in \Bref{Catani:2007vq}, the shape of these distributions sizeable differs when going from LO to NLO and to
NNLO. In particular, at the LO the two photons are emitted with the same $\pT^{\PGg}$ because the Higgs boson is produced with zero transverse momentum, hence the LO $\pT^{{\mathrm{min}},\PGg}$ and $\pT^{{\mathrm{max}},\PGg}$ are exactly identical. Furthermore the LO distribution has a kinematical boundary at $\pT^{\PGg} = \MH /2$ (Jacobian peak), which is due to the use of the narrow width approximation. Such condition is released once extra radiation is accounted for. Thus higher order predictions suffer of perturbative instabilities, i.e. each higher-order perturbative contribution produces (integrable) logarithmic singularities in the vicinity of that boundary, as explained in \Bref{Catani:1997xc}.
The same  $\pT^{{\mathrm{min}},\PGg}$ predictions are shown in \Fref{fig:pTphoton}(right); in this case the NNLO result is compared with the resummed result at the NLL+NLO and NNLL+NNLO accuracy.
As expected \cite{Catani:1997xc}, resummed results do not suffer of such instabilities in the vicinity of the LO kinematical boundary; the resummed distributions are smooth and the shape is rather stable when going from NLL+NLO to NNLL+NNLO.

The \textsc{HRes} code provides all the momenta of final state particles and studies on other variables can be performed.
For example a variable that is often studied is $\cos(\theta^*)$, where $\theta^*$ is the
polar angle of one of the photons with respect to the beam axis in the Higgs
boson rest frame. A cut on the photon transverse momentum $\pT^{\PGg}$ implies
a maximum value for $\cos(\theta^*)$ at LO. For example for $\MH = 125\UGeV$ and
$\pT^{\PGg} \geq 40\UGeV$ we obtain
 $|\cos(\theta^*)| \leq |\cos(\theta^*_{\mathrm{cut}})|\simeq0.768$. At the NLO and
NNLO the Higgs transverse momentum is non vanishing and events with $|\cos(\theta^*)| > |\cos(\theta^*_{\mathrm{cut}})|$
are kinematically allowed. In the region of the kinematical boundary
higher-order perturbative distributions suffer of logarithmic
singularities. As expected \cite{Catani:1997xc}, resummed results do not suffer of such
instabilities in the vicinity of the LO kinematical boundary; the resummed
distributions are smooth and the shape is rather stable when going from
NLL+NLO to NNLL+NNLO. In Fig. 2 we report both the distributions
(normalized to unity) obtained by fixed order and the resummed calculations.
We see that the resummed results are smooth in the region around the
kinematical boundary. Away from such region, fixed order and resummed
results show perfect agreement.

\begin{figure}[!ht]
\centering
\includegraphics[width=0.485\textwidth]{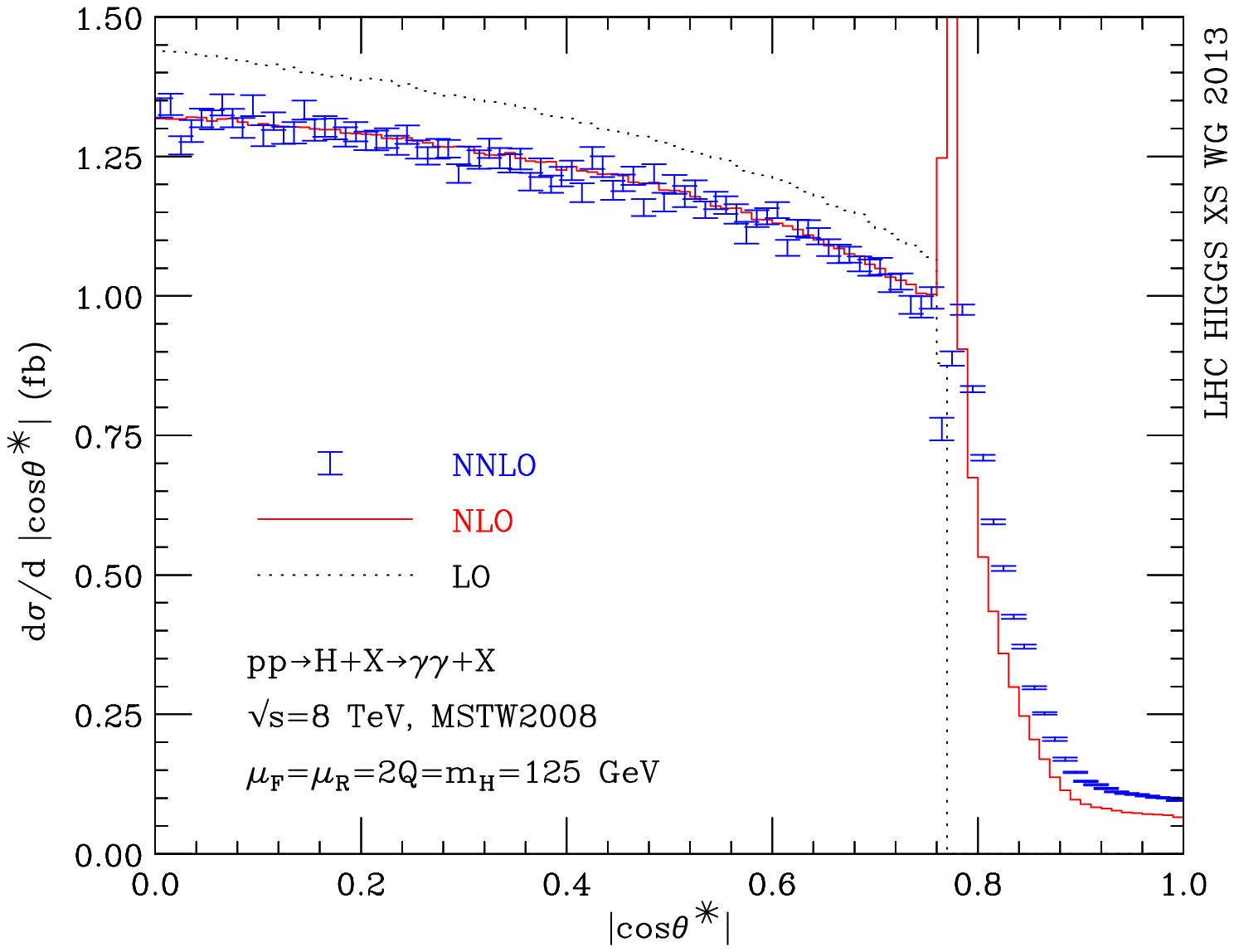}
\includegraphics[width=0.485\textwidth]{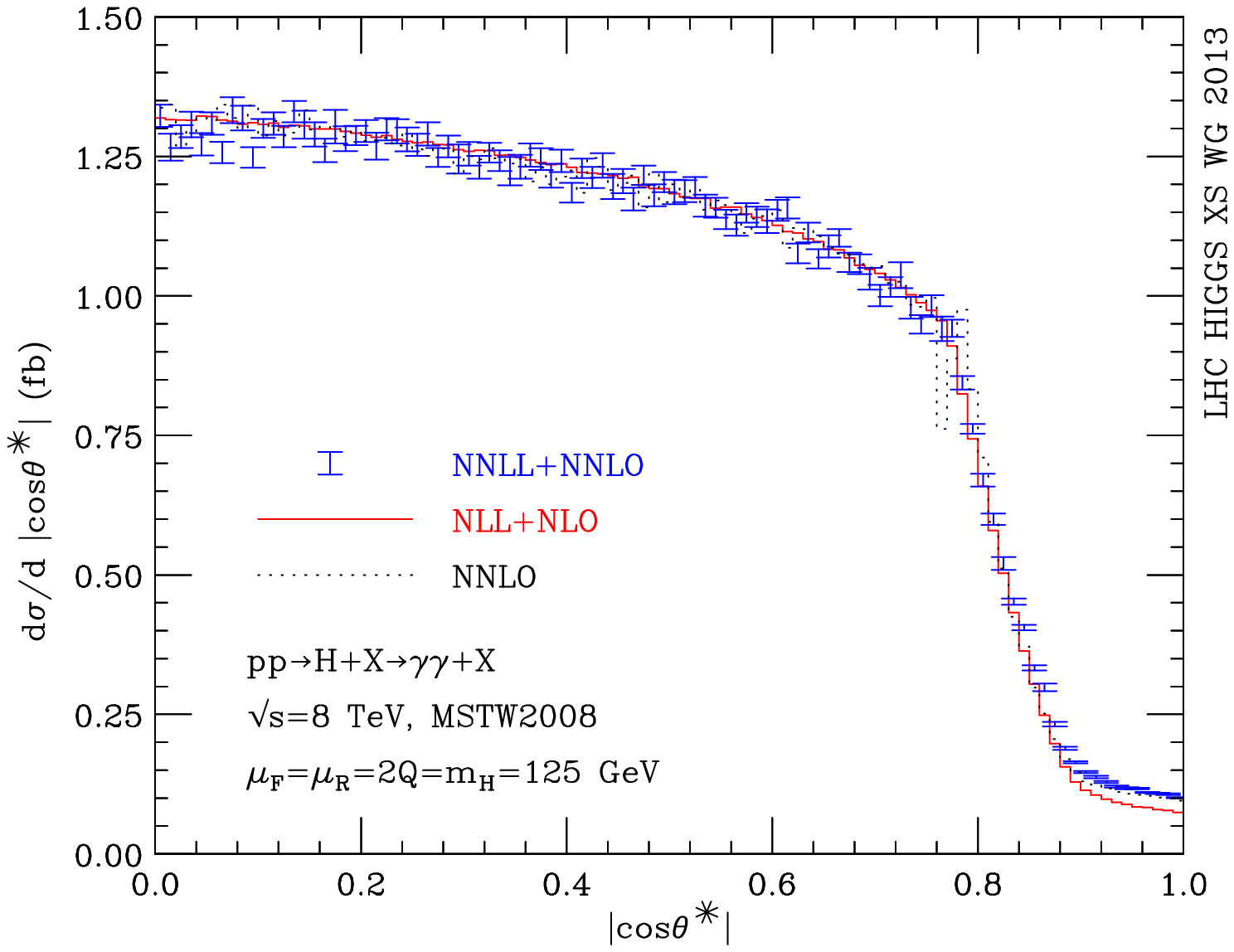}
\caption{Normalized $\cos\theta^*$ distribution at the LHC. On the left: LO, NLO and NNLO results. On the right: resummed predictions at NLL+NLO and NNLL+NNLO accuracy are compared with the NNLO result.}
\label{fig:norm_costheta}
\end{figure}

\subsection{Higgs $\pT$ distribution using  different generators} 
\label{subsec:Hqtggf}

\subsubsection{Comparison among \textsc{HRes}, \textsc{POWHEG}, \textsc{Madgraph}, \textsc{aMC@NLO}, \textsc{Sherpa} without Heavy Quark mass effect
\footnote{N. Chanon}}

Results of Higgs boson searches performed at 7 TeV were using \textsc{POWHEG} as main generator for the gluon fusion process. In both CMS and ATLAS collaborations the Higgs boson $\pT$ was reweighted event by event to match the NNLO+NNLL $\pT$ spectrum predicted by HqT. Higgs boson searches performed at $8\UTeV$ are using \textsc{POWHEG} where the dampening factor \emph{hfact}   was tuned to reproduce HqT spectrum, therefore no reweighting was needed. In this section we perform a comparison of different generators able to generate the gluon fusion process, with the following setup.

We consider a Higgs boson mass of $125\UGeV$. \textsc{HRes} predictions are computed with a factorization and renormalization scales set to $125\UGeV$, and resummation scale set to 62.5 GeV, using the PDF set MSTW2008. The theory uncertainty bands are computed by multiplying/dividing by 2 the renormalization and factorization scales independently. \textsc{HRes} spectrum was cross-checked against HqT and found to be identical. \textsc{POWHEG} predictions are computed at NLO with \emph{hfact}$=\MH/1.2$ \cite{Campbell:2012am,Dittmaier:2012vm} and the showering is performed with \textsc{Pythia 6}. \textsc{Madgraph 5} predictions are computed with up to 3 extra jets using \textsc{Pythia 6} shower, the PDF set CTEQ6L1 is used and the matching scale, obtained by looking at the differential jet rate which needs to be smoothed, was found to be $Q_{\mathrm{cut}}=26\UGeV$. We also compare two generators with matching/merging of extra-jets at NLO. aMC@NLO predictions are performed on sample of matched and merged 0/1/2 jets at NLO and the third jet at LO with herwig6 shower, the PDF used is MSTW2008 and the matching scale is set to 50 GeV. Sherpa predictions are performed with a matched/merged sample of 0/1 jets at NLO, the second jet a LO, with hadronization and MPI, the PDF set used is CT10 and the matching scale is set to 30 GeV. Generated samples are compared inclusively (no further selection is applied). All are normalized to the cross-section computed with \textsc{HRes} with central renormalization and factorization scales. 

\begin{figure}[!ht]
\centering
\includegraphics[width=0.485\textwidth]{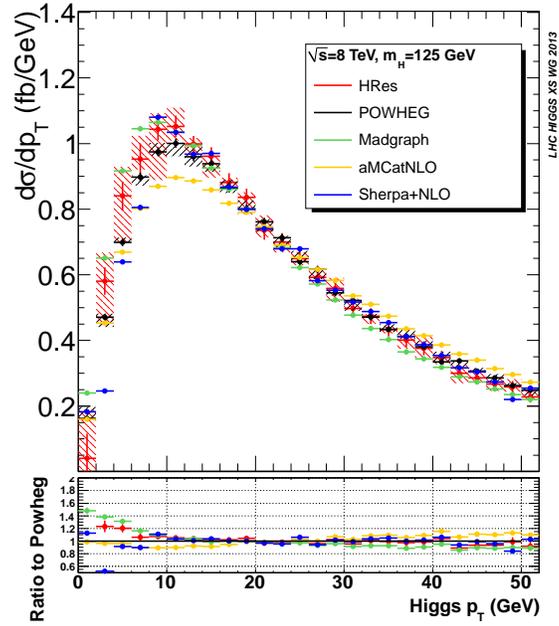}
\caption{\label{fig:HResPowhegOthers_fig1}{ Distribution  of Higgs $\pT$ for different generators for Higgs boson of $\MH=125\UGeV$ at $8\UTeV$.}}
\end{figure}

\begin{figure}[!ht]
\centering
\includegraphics[width=0.485\textwidth]{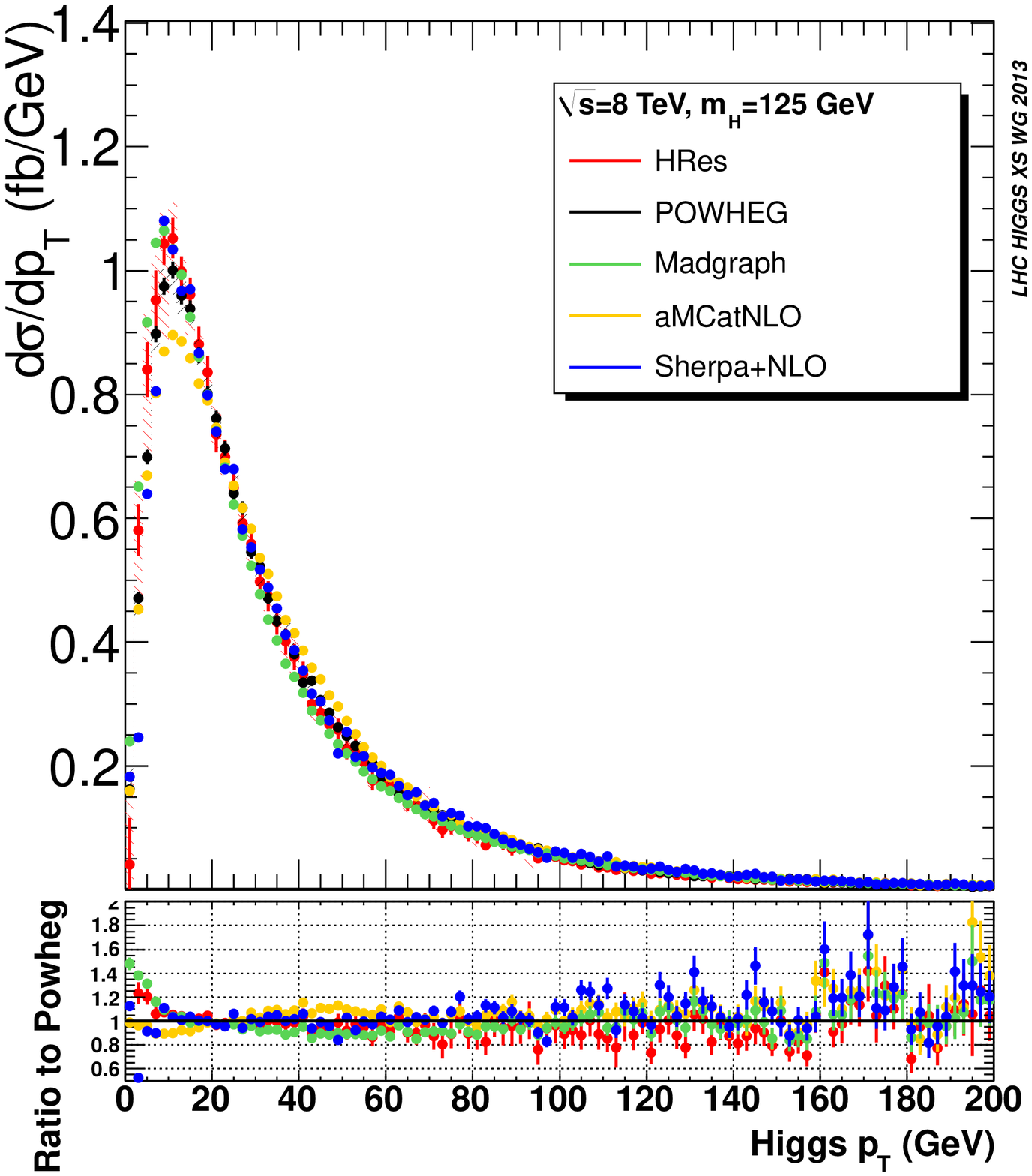}
\includegraphics[width=0.485\textwidth]{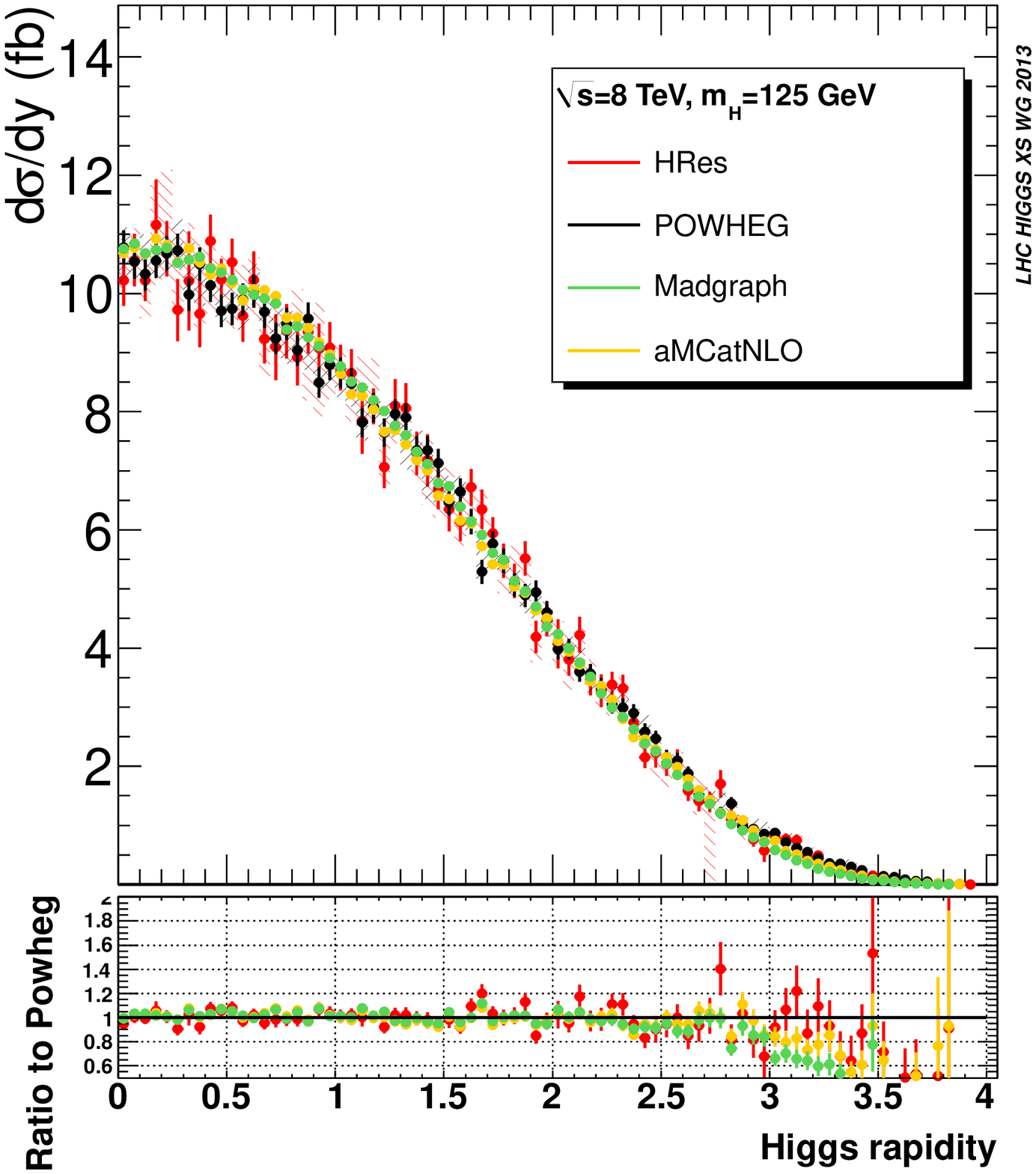}
\caption{\label{fig:HResPowhegOthers_fig2}{ Distribution  of Higgs $\pT$ (left) and rapidity (right) for different generators for Higgs boson of $\MH=125\UGeV$ at $8\UTeV$.}}
\end{figure}

\begin{figure}[!ht]
\centering
\includegraphics[width=0.485\textwidth]{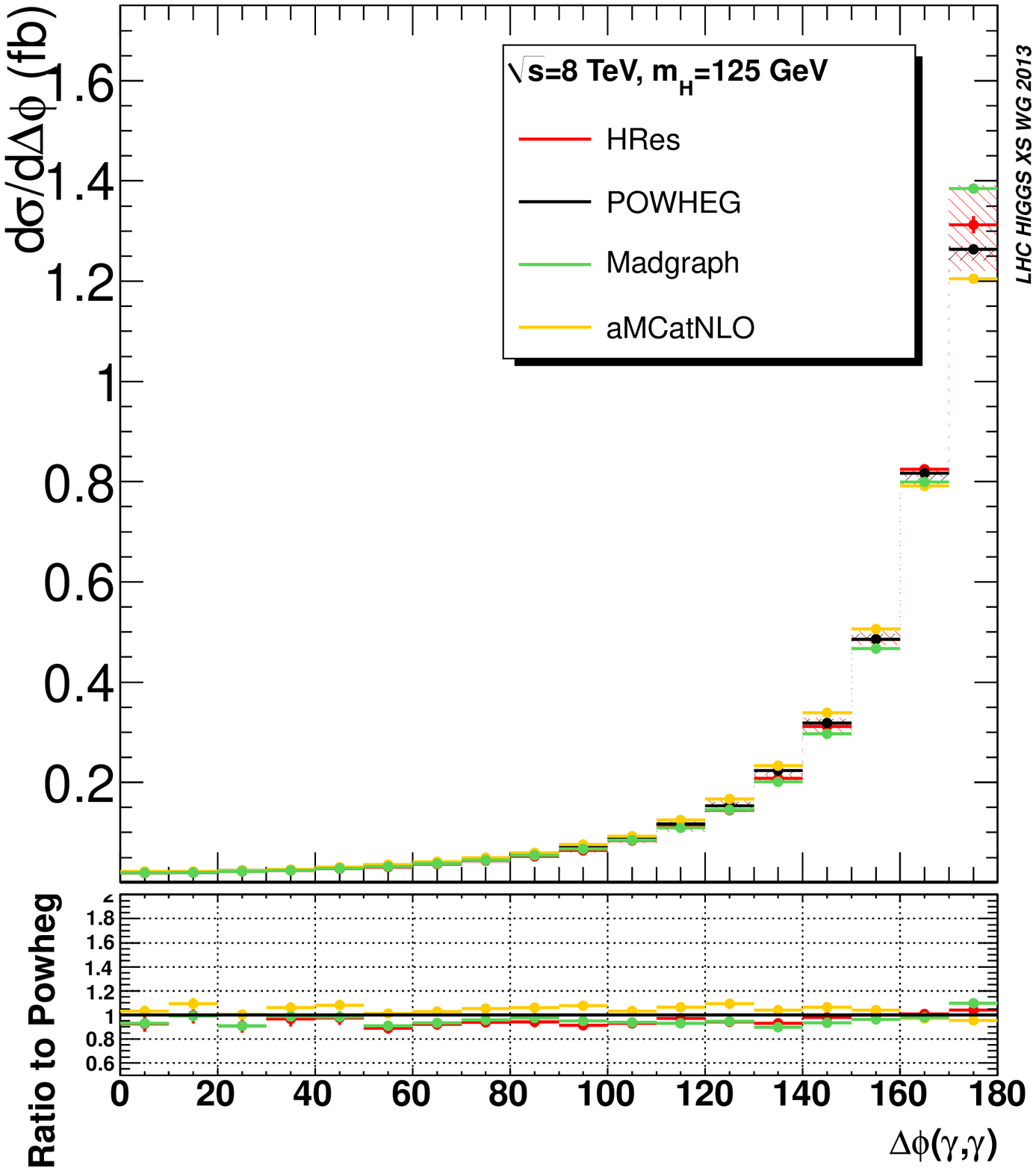}
\includegraphics[width=0.485\textwidth]{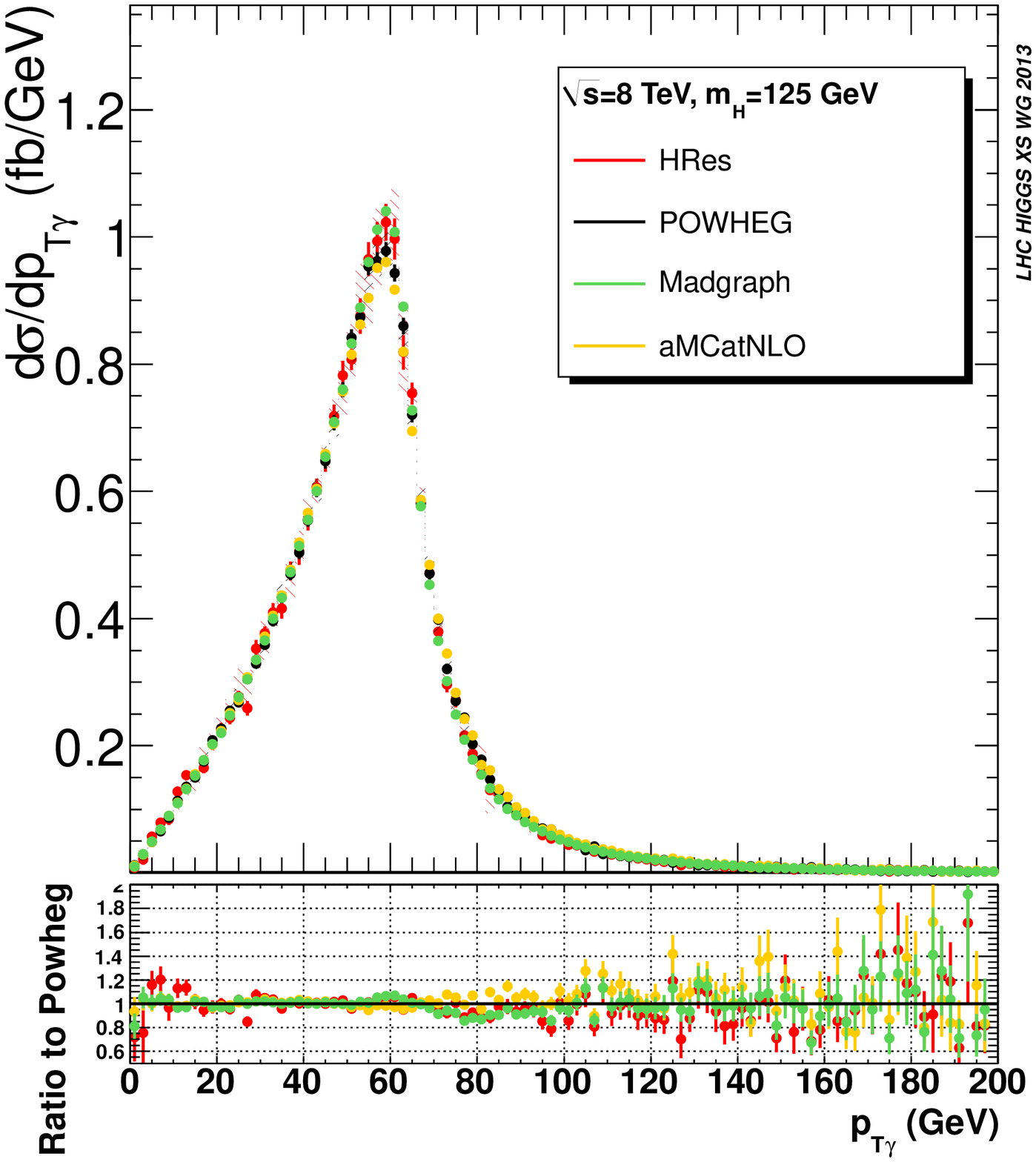}
\caption{\label{fig:HResPowhegOthers_HiggsPTfig3}{ Distribution  of diphoton $\Delta\phi(\PGg,\PGg)$ and $p_{T,\PGg}$ for different generators for Higgs boson of $\MH=125\UGeV$ at $8\UTeV$.}}
\end{figure}

The Higgs boson $\pT$ distribution zoomed in the range [0,50] GeV is shown \Fref{fig:HResPowhegOthers_fig1}. As previously noted and shown on \Fref{fig:HResPowhegOthers_fig1}, the $\pT$ spectrum of \textsc{POWHEG} with \emph{hfact} tuned is slightly harder than the one predicted with \textsc{HRes}. It can also be observed that Madgraph has a somehow softer $\pT$ than \textsc{POWHEG}. On the other hand, aMC@NLO seems to predict a mildly harder $\pT$. A larger range of $\pT$ is shown on \Fref{fig:HResPowhegOthers_fig2} along with the Higgs boson rapidity. Decay to photons was included in all generators but Sherpa. We show on \Fref{fig:HResPowhegOthers_HiggsPTfig3} the distributions of $\Delta\phi(\PGg,\PGg)$ and $p_{T\PGg}$.

%\textbf{FIXME}: generate more Sherpa events for Higgs rapidity plot.

\clearpage
\subsubsection{Higgs $\pT$ distribution, MC@NLO-Powheg comparison using finite HQ mass effect
\footnote{B. Di Micco, R. Di Nardo}}
In the search  of the Higgs boson  and the measurement of the Higgs boson
production yield in the $\PH \to \PGg\PGg$ and $\PH \to \PZ\PZ^* \to 4\Pl$
channel, 
the Higgs transverse momentum is of particular interest because it affects the
signal acceptance due to the cuts applied on the photon and lepton
momenta. Moreover the signal purity can be increased by cutting on the
transverse momentum of the di-boson system  that is on average larger in the
Higgs decay than the non resonant background due to the higher  jet activity
in the Higgs gluon fusion production process than the di-photon and $\PZ\PZ$ non
resonant backgrounds. Such techniques are applied in the $\PH \to \PGg \PGg$
search \cite{ATLAS-CONF-2012-091} and under study in the $\PH \to \PZ\PZ \to 4\Pl$ and
$\PH \to \PGt \PGt$ searches. It is therefore important to understand the impact
of several contributions to the Higgs $\pT$ and how they affect the
$\pT$ spectrum predicted  by the MC generators. 

In the present section we show a comparison of the Higgs $\pT$
distribution between \textsc{ MC@NLO 4.09} \cite{Frixione:2003ei},
using \textsc{ HERWIG} 6.5 \cite{Corcella:2000bw} for the showering,
and  \textsc{POWHEG} \cite{Bagnaschi:2011tu,Nason:2004rx,Frixione:2007vw,Alioli:2010xd} that has been
showered with both \textsc{PYTHIA8} \cite{Sjostrand:2006za,Sjostrand:2007gs}
and \textsc{HERWIG 6.5}. The signal process is $\Pp\Pp \to \PH \to \PZ\PZ^{*}$ at the mass $\MH = 125.5\UGeV$. This value has been chosen being the 
last ATLAS best fit value \cite{ATLAS-CONF-2013-014}. \textsc{ HERWIG} 6.5 has been interfaced to  \textsc{ Jimmy} \cite{Butterworth:1996zw} for the underlying event simulation.
The ATLAS AUET2\cite{ATLAS-PHYS-PUB-2011-008} tune using the CTEQ 10 NLO pdfs in the showering algorithm has been used. Both \textsc{ MC@NLO 4.09} and \textsc{ POWHEG} include the heavy quark mass effect
in the gluon gluon loop for the Higgs $\pT$ determination. The contribution is available for t,b and  c quarks in  \textsc{ POWHEG} and for the $t$ and  $b$ quarks in \textsc{ MC@NLO}.
In the present section the contribution from the $c$ quark has been switched off in  \textsc{ POWHEG} so that the quark mass effects refer to the contribution 
from the top and bottom quarks.
The $\pT$ dampening factor \emph{hfact} has been set to $m_H/1.2$ in the \textsc{ POWHEG} case. A configuration  without the \emph{hfact} has also been studied.
The top mass has been set at $m_{\PQt} = 172.5\UGeV$   and the bottom quark mass at $m_{\PQb} = 4.75\UGeV$. The generation has been performed using the CT10 pdf set.

\begin{figure}
\begin{center}
\includegraphics[width=0.45\textwidth]{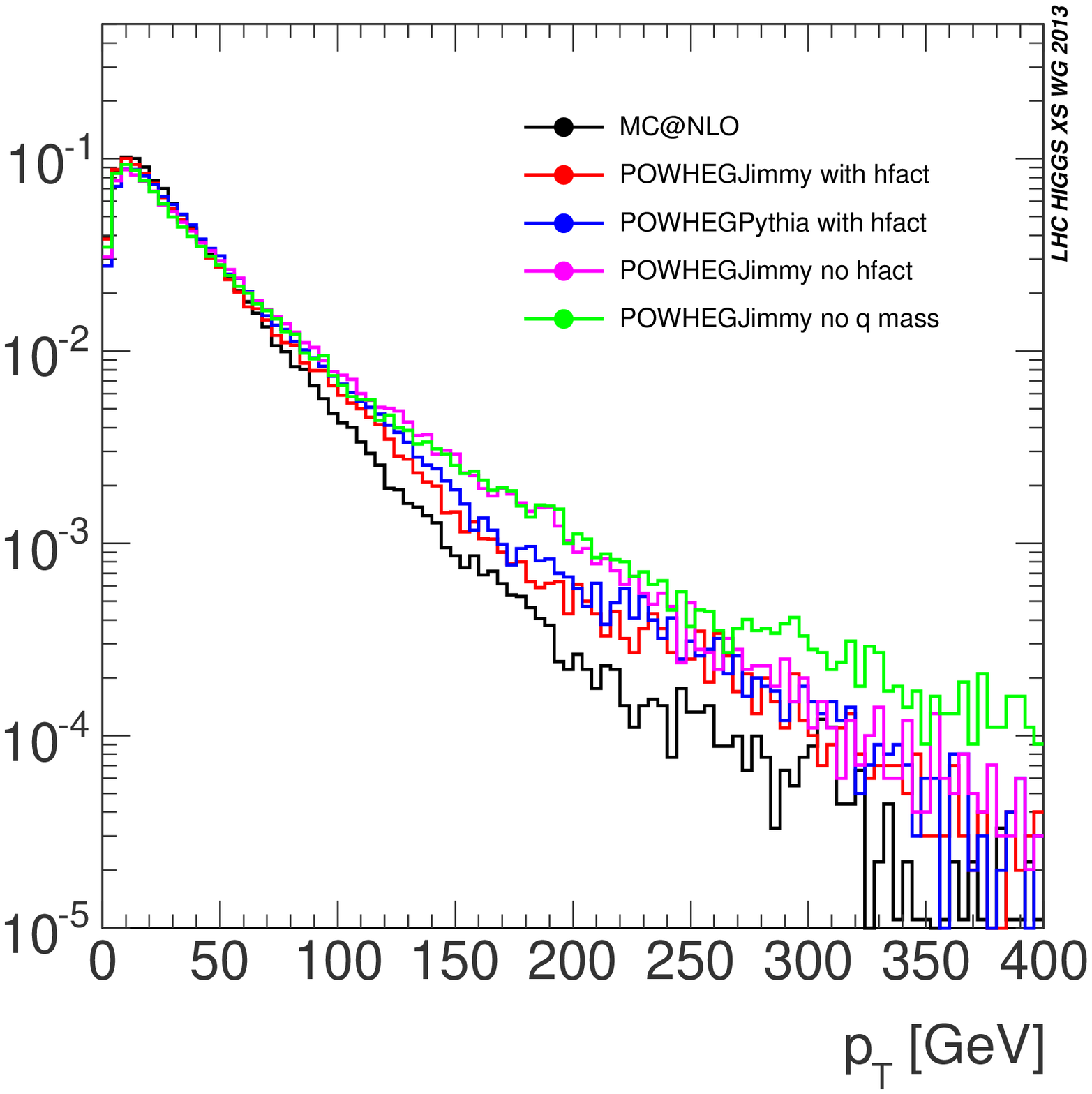}
\includegraphics[width=0.45\textwidth]{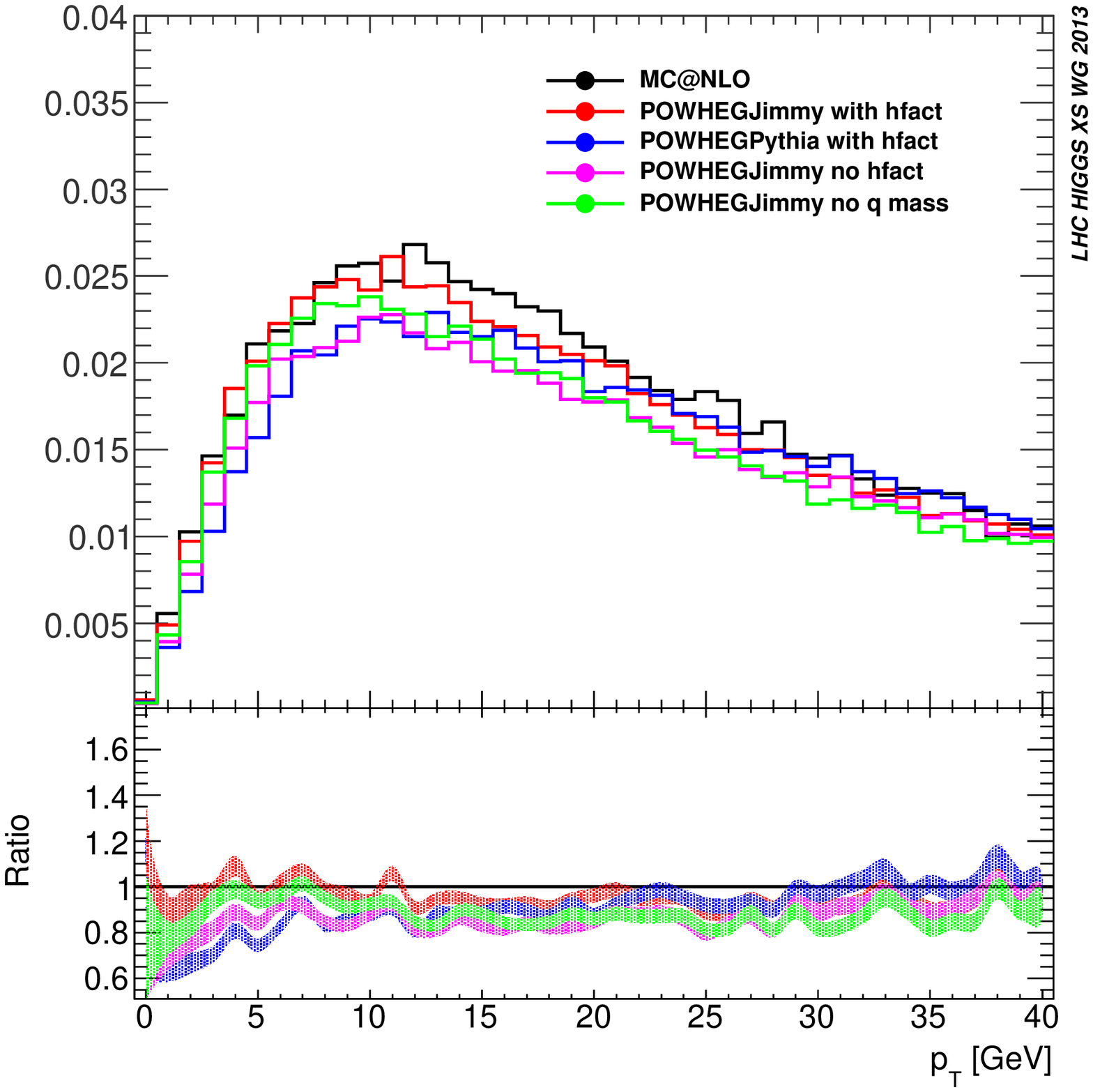}
\end{center}
\caption{Higgs $\pT$ distribution using different MC generators and
parton shower configurations. POWHEGPythia indicates \textsc{POWHEG + PYTHIA8}, POWHEGJimmy indicates \textsc{POWHEG + HERWIG6} and \textsc{Jimmy} for the underlying event simulation. The \emph{no q mass} distribution corresponds to a configuration where both \emph{hfact} and the Heavy Quark mass effect are switched off.} \label{fig:pT_long_range}
\end{figure}

In \Fref{fig:pT_long_range} the Higgs $\pT$ distribution is shown with different configurations.
In the  range, $0{-}400\UGeV$, we observe that the \textsc{ MC@NLO} spectrum is softer than the \textsc{ POWHEG + PYTHIA8} one. Comparing \textsc{ POWHEG+HERWIG}  with \textsc{ POWHEG + PYTHIA8} is possible to see that the parton shower doesn't affect the high Higgs $\pT$ tail
as expected.  The use of the \emph{hfact} dampening factor 
makes the \textsc{ POWHEG} spectrum closer to the \textsc{ MC@NLO} one at high $\pT$ but still significantly harder. The heavy quark mass effect has been switched off
in the \textsc{ POWHEG + HERWIG} sample in order to estimate the size of the effect. The contribution  is visible in the very high $\pT$ tail ($\pT > 250\UGeV$) but doesn't seem responsible for the main
high $\pT$ behaviour. In order to compare the generators at low $\pT$ the same figure has been zoomed in the range 0-40 GeV. For $\pT < 15\UGeV$ differences between the \textsc{ HERWIG} and \textsc{ PYTHIA8} showering are visible and in the very low $\pT$ region ($\pT < 10\UGeV$), \textsc{ MC@NLO + HERWIG} and \textsc{ POWHEG + HERWIG} with the \emph{hfact} are compatible while both the \textsc{ PYTHIA8} showering and the \emph{no hfact} configuration are different from the \textsc{ MC@NLO} prediction. This shows that at low $\pT$ the parton shower has a relevant role but the dampening factor correction is still important and brings the \textsc{ POWHEG + HERWIG} spectrum in  agreement with \textsc{ MC@NLO 4.09}.

%%%%%%%%%%%%%%%%%%%%%%%%%%%%%%%%%%%%%%%%%%%%%%%%%%%%%%%%%%%%%%%%%%%%%%%%%%%%%%%

\subsection{{\boldmath Interference in light Higgs $\PV\PV$ modes}}
\label{sec:interference}

Interferences between signal and background amplitudes are known to be relevant for heavy Higgs production. However, they can play a significant role also for a light Higgs, by considerably modifying the cross section and eventually contributing to a shift in the signal peak. In this section we analyse the interference effects for the $\PW\PW$, $\PZ\PZ$ and $\gamma\gamma$ channels.

In \refS{sec:ggF_ggVV_ZWA_interference}, the findings of \Bref{Kauer:2012hd} 
about the inadequacy of zero-with approximation and importance 
of signal-background interference are summarised.  
In \refSs{sec:ggF_ggVV_WWZZ} and \ref{sec:ggF_ggVV_peakshift},
results for $\Pg\Pg\ (\to \PH)\to 4$ leptons including 
Higgs-continuum interference effects calculated with {\sc gg2VV} \cite{gg2VV}
are presented.  The complex-pole scheme \cite{Goria:2011wa} with $\MH=125\UGeV$ and 
$\GH=4.434\UMeV$ is used.  
%Further details can be found in \refS{sec:HH_ggVV}.

\refSs{sec:interference_2gamma} and \ref{sec:interference_2gammajet} summarize the recent findings of \cite{deFlorian:2013psa,Martin:2013ula}
about the effect on the  diphoton invariant mass shift due to the inclusion of the contribution from the $\Pg\PQq$ and $\PAQq\PQq$ channels in the signal-background interference. \refS{sec:interference_2gamma} considers the case of inclusive 
 $\PGg\PGg$ production while \refS{sec:interference_2gammajet} discusses the effect on  $\PGg\PGg$ + jet production.

%%%%%%%%%%%%%%%%%%%%%%%%%%%%%%%%%%%%%%%%%%%%%%%%%%%%%%%%%%%%%%%%%%%%%%%%%%%%%%%

\subsubsection{Inadequacy of zero-with approximation and importance 
of signal-background interference
\footnote{N.~Kauer}}

\label{sec:ggF_ggVV_ZWA_interference}

\begin{figure}
\centering\includegraphics[width=0.48\textwidth]{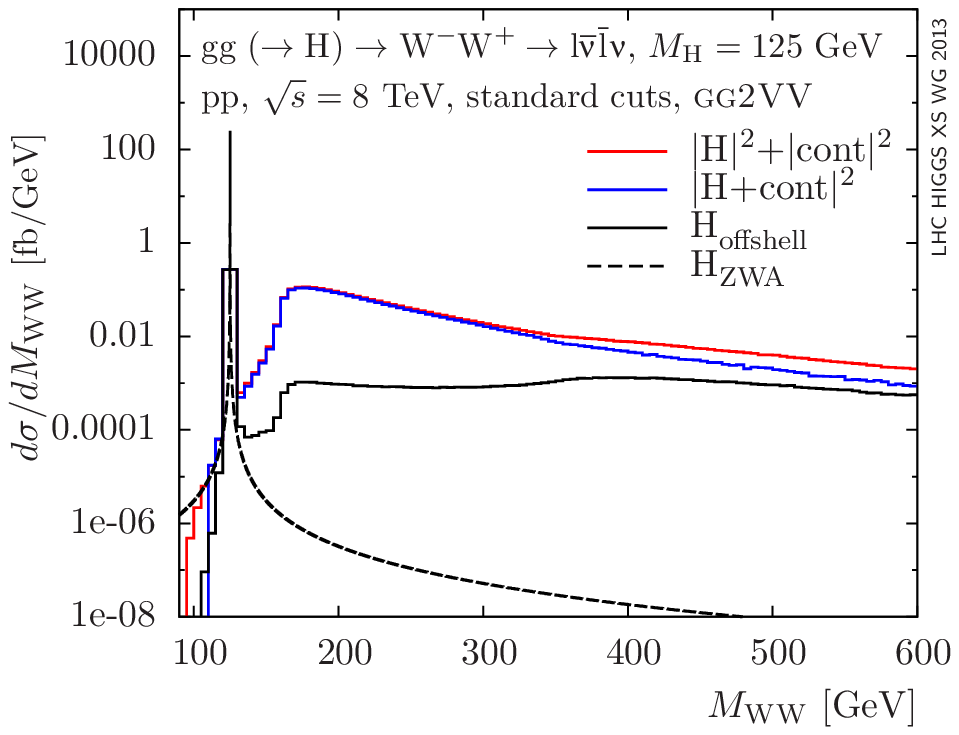}
\vspace*{0.cm}
\caption{$M_{\PW\PW}$ distributions for 
$\Pg\Pg\ (\to \PH)\to \PWm\PWp\to \Pl\PAGn\PAl\PGn$ 
in $\Pp\Pp$ collisions at $\sqrt{s} = 8\UTeV$
for $\MH=125$\,GeV.
The ZWA distribution (black, dashed) as defined in Eq.~(3.2)
in \Bref{Kauer:2012hd}, the off-shell 
Higgs distribution (black, solid), the 
$d\sigma(|{\cal M}_{\PH} + {\cal M}_\mathrm{cont}|^2)/dM_{\PW\PW}$ 
distribution (blue) and the 
$d\sigma(|{\cal M}_{\PH}|^2 + |{\cal M}_\mathrm{cont}|^2)/dM_{\PW\PW}$ 
distribution (red) are shown.
Standard cuts are applied: 
$p_{T\Pl} > 20\UGeV$, $|\eta_{\Pl}| < 2.5$, 
$\sla{p}_{\mathrm T} > 30\UGeV$, $M_{\Pl\Pl} > 12\UGeV$.
No flavour summation is carried out for charged leptons or neutrinos.
}
\label{fig:ggF_ggVV_ZWA}
\end{figure}

For the SM Higgs boson with $\MH\approx 125\UGeV$, one has
$\GH/\MH < 10^{-4}$, which suggests an excellent accuracy 
of the zero-width approximation (ZWA).  However, as shown in 
\Bref{Kauer:2012hd} for inclusive cross sections and 
cross sections with experimental selection cuts, the ZWA is 
in general not adequate and the error estimate 
$\Ord(\GH/\MH)$ is not reliable for a light Higgs boson.
The inclusion of off-shell contributions is essential to obtain 
an accurate Higgs signal normalisation at the $1\%$ precision level
as well as correct kinematic distributions.  ZWA deviations are 
particularly large for $\PH\to \PV\PV^\ast$ processes ($\PV=\PW,\PZ$).
To be more specific, without optimised selection cuts they are of 
\mbox{$\Ord(5\,$--$\,10\%)$}.  
The ZWA caveat also applies 
to Monte Carlos that approximate off-shell effects with an 
ad hoc Breit-Wigner reweighting of the on-shell propagator, as can 
be seen by comparing the $\PH_\mathrm{ZWA}$ and $\PH_\mathrm{offshell}$ 
distributions in \refF{fig:ggF_ggVV_ZWA}.
The ZWA limitations are also relevant for the extraction of Higgs 
couplings, which is initially being performed using the ZWA.  
The findings of \Bref{Kauer:2012hd} make clear that off-shell 
effects have to be included in future Higgs couplings analyses.

The unexpected off-shell effect can be traced back to the dependence of the 
decay matrix 
element on the Higgs invariant mass $\sqrt{q^2}$.  For $\PH\to \PV\PV^\ast$ 
decay modes one finds that  the $q^4$ dependence of the decay matrix 
element for $q^2 > (2M_{\PV})^2$ compensates the 
$q^2$-dependence of the Higgs propagator, which 
causes a strongly enhanced off-shell cross section in 
comparison to the ZWA up to invariant masses of about $600\UGeV$ 
(see \refF{fig:ggF_ggVV_ZWA}).
The total $\Pg\Pg\to \PH \to \PV\PV^\ast$ cross section thus receives an 
${\cal O}(10\%)$ off-shell correction.
Furthermore, in the region above $2M_{\PV}$ the Higgs signal is affected 
by ${\cal O}(10\%)$ 
signal-background interference effects, which, 
due to the enhanced off-shell tail, can have a significant impact also for 
$\MH \mathchar"321C 2M_{\PV}$.  On the other hand, in the vicinity of the 
Higgs resonance 
finite-width and Higgs-continuum interference effects are negligible if 
$\MH \mathchar"321C 2M_{\PV}$.
For weak boson decays that permit the reconstruction 
of the Higgs invariant mass, the experimental procedure focuses on the 
Higgs resonance region and for $\MH \mathchar"321C 2M_{\PV}$ the 
enhanced off-shell region is thus typically excluded.
For $\PH\to \PV\PV^\ast$ channels that do not allow to reconstruct the Higgs 
invariant mass,
the tail can nevertheless be effectively excluded by applying a $M_{\mathrm T}<M_{\PH}$ cut
on a suitable transverse mass observable 
$M_{\mathrm T}$ which approximates the Higgs invariant mass.
Finally, in addition to gluon fusion the $\PH\to \PV\PV^\ast$ modes in other Higgs 
production 
channels also exhibit an enhanced off-shell tail, since the effect is caused by 
the decay matrix element.

% ----------------------------------------------------------------------------

%%%%%%%%%%%%%%%%%%%%%%%%%%%%%%%%%%%%%%%%%%%%%%%%%%%%%%%%%%%%%%%%%%%%%%%%%%%%%%%

\subsubsection{\boldmath $\PW\PW^\ast/\PZ\PZ^\ast$ interference in $\Pg\Pg\ (\to \PH)\to 
\Pl\PAGnl\PAl\PGnl$
%\footnote{N.~Kauer}
}
\label{sec:ggF_ggVV_WWZZ}

In this section, $\PZ\PZ^\ast$ corrections (including $\PGg^\ast$ contributions) to $\Pg\Pg\ (\to \PH)\to \PW\PW^\ast\to \Pl\PAGnl\PAl\PGnl$
with same-flavour final state in $\Pp\Pp$ collisions at $8\UTeV$ are studied for 
$\MH=125\UGeV$ including $\PW\PW^\ast/\PZ\PZ^\ast$ interference.
%\footnote{ See also \refS{sec:HH_ggVV_WWZZ}.}
%
To quantify the signal-background interference effect the
$S$+$B$-inspired measure $R_1$ and $S/\sqrt{B}$-inspired measure $R_2$
are used:
\beq
\label{eqn:ggF_ggVV_WWZZ_1}
R_1 = \frac{\sigma(|{\cal{M}}_{\PH}+{\cal{M}}_\text{cont}|^2)}{
\sigma(|{\cal{M}}_{\PH}|^2)+\sigma(|{\cal{M}}_\text{cont}|^2)}\;,\quad
R_2 = \frac{\sigma(|{\cal{M}}_{\PH}|^2 + 2\,\mbox{Re}({\cal{M}}_{\PH}{\cal{M}}^\ast_\text{cont}))}{\sigma(|{\cal{M}}_{\PH}|^2)}\;,
\eeq
where ${\cal{M}}_{\PH}$ and ${\cal{M}}_\text{cont}$ are the $\Pg\Pg\ \to \PH\to 
\Pl\PAGnl\PAl\PGnl$ and $\Pg\Pg\ \to 
\Pl\PAGnl\PAl\PGnl$ amplitudes, respectively.
Off-shell Higgs contributions are included.
The selection cuts choice (see \refT{tab:ggF_ggVV_WWZZ_1}) 
follows \Bref{Kauer:2012hd}.
A cut on the transverse mass 
\beq
\label{eqn:ggF_ggVV_WWZZ_2}
M_{\mathrm T}=\sqrt{(M_{T,\Pl\Pl}+\sla{p}_{\mathrm T})^2-({\bf{p}}_{T,\Pl\Pl}+{\sla{\bf{p}}}_{\mathrm T})^2}\quad\mathrm{with}\quad M_{T,\Pl\Pl}=\sqrt{p_{T,\Pl\Pl}^2+M_{\Pl\Pl}^2}
\eeq
is applied.

Results with $\PZ\PZ^\ast$ contribution including 
$\PW\PW^\ast/\PZ\PZ^\ast$ interference are given in 
\refT{tab:ggF_ggVV_WWZZ_1}.  To assess the importance of including
the $\PZ\PZ^\ast$ contribution for the same-flavour final state, 
for the observables in \refT{tab:ggF_ggVV_WWZZ_1} the ratio with/without 
$\PZ\PZ^\ast$ correction (including interference) is displayed in 
\refT{tab:ggF_ggVV_WWZZ_2}.  It shows that the cross sections are affected
by the $\PZ\PZ^\ast$ correction at the few percent level, but the 
interference measures $R_{1,2}$ are essentially unchanged.

\begin{table}
\caption{
Cross sections in $\UfbZ$ for $\Pg\Pg\ (\to \PH)\to \PW\PW^\ast/\PZ\PZ^\ast\to 
\Pl\PAGnl\PAl\PGnl$ (same flavour) in $\Pp\Pp$ collisions at $8\UTeV$ for $\MH=125\UGeV$.  
Results are given for signal ($|\PH|^2$), $\Pg\Pg$ continuum background 
($|\mathrm{cont}|^2$) and signal+background+interference 
($|\mathrm{H}$+$\mathrm{cont}|^2$).  
Off-shell Higgs contributions are included.
$R_{1,2}$ as defined in 
\eqn{eqn:ggF_ggVV_WWZZ_1} are also displayed.  
Standard cuts as in \refF{fig:ggF_ggVV_ZWA}.  Higgs search cuts: standard 
cuts and $M_{\Pl\Pl} < 50\UGeV$ , $\Delta\phi_{\Pl\Pl} < 1.8$.
$M_{\mathrm T}$ is defined in \eqn{eqn:ggF_ggVV_WWZZ_2}.
No flavour summation is carried out for charged leptons or neutrinos.
The integration error is given in brackets.
}
\label{tab:ggF_ggVV_WWZZ_1}%
\renewcommand{\arraystretch}{1.2}%
\setlength{\tabcolsep}{1.5ex}%
\begin{center}
\begin{tabular}{lccccc}
\hline
\multicolumn{1}{c}{selection cuts} & $|\PH|^2$ & $|\mathrm{cont}|^2$ & $|\mathrm{H}$+$\mathrm{cont}|^2$ & $R_1$ & $R_2$ \\
\hline
standard cuts & $3.225(4)$ & $11.42(5)$ & $12.95(8)$ & $0.884(6)$ & $0.47(3)$ \\
Higgs search cuts & $1.919(3)$ & $2.711(7)$ & $4.438(8)$ & $0.958(3)$ & $0.900(6)$ \\
$+(0.75\MH < M_{\mathrm T} < \MH)$ & $1.736(2)$ & $0.645(2)$ & $2.335(4)$ & $0.981(2)$ & $0.974(3)$ \\
\hline
\end{tabular}
\end{center}
\end{table}

\begin{table}
\caption{
As \refT{tab:ggF_ggVV_WWZZ_1}, but the ratio with/without 
$\PZ\PZ^\ast$ correction (including interference) is shown
for cross sections and $R_{1,2}$.  The results without 
$\PZ\PZ^\ast$ correction are taken from Table 4 in \Bref{Kauer:2012hd}.
}
\label{tab:ggF_ggVV_WWZZ_2}%
\renewcommand{\arraystretch}{1.2}%
\setlength{\tabcolsep}{1.5ex}%
\begin{center}
\begin{tabular}{lccccc}
\hline
\multicolumn{1}{c}{selection cuts} & $|\PH|^2$ & $|\mathrm{cont}|^2$ & $|\mathrm{H}$+$\mathrm{cont}|^2$ & $R_1$ & $R_2$ \\
\hline
standard cuts & $1.000(2)$ & $1.088(5)$ & $1.058(7)$ & $0.991(7)$ & $0.88(6)$ \\
Higgs search cuts & $0.969(2)$ & $1.002(3)$ & $0.987(2)$ & $0.998(3)$ & $0.994(6)$ \\
$+(0.75\MH < M_{\mathrm T} < \MH)$ & $0.976(2)$ & $1.001(3)$ & $0.980(2)$ & $0.997(3)$ & $0.996(3)$ \\
\hline
\end{tabular}
\end{center}
\end{table}

%%%%%%%%%%%%%%%%%%%%%%%%%%%%%%%%%%%%%%%%%%%%%%%%%%%%%%%%%%%%%%%%%%%%%%%%%%%%%%%

\subsubsection{$\PH\to \PZ\PZ^\ast$ invariant mass peak shift due to signal-background interference
%\footnote{N.~Kauer}
}
\label{sec:ggF_ggVV_peakshift}

\begin{figure}
\centering\includegraphics[width=0.48\textwidth]{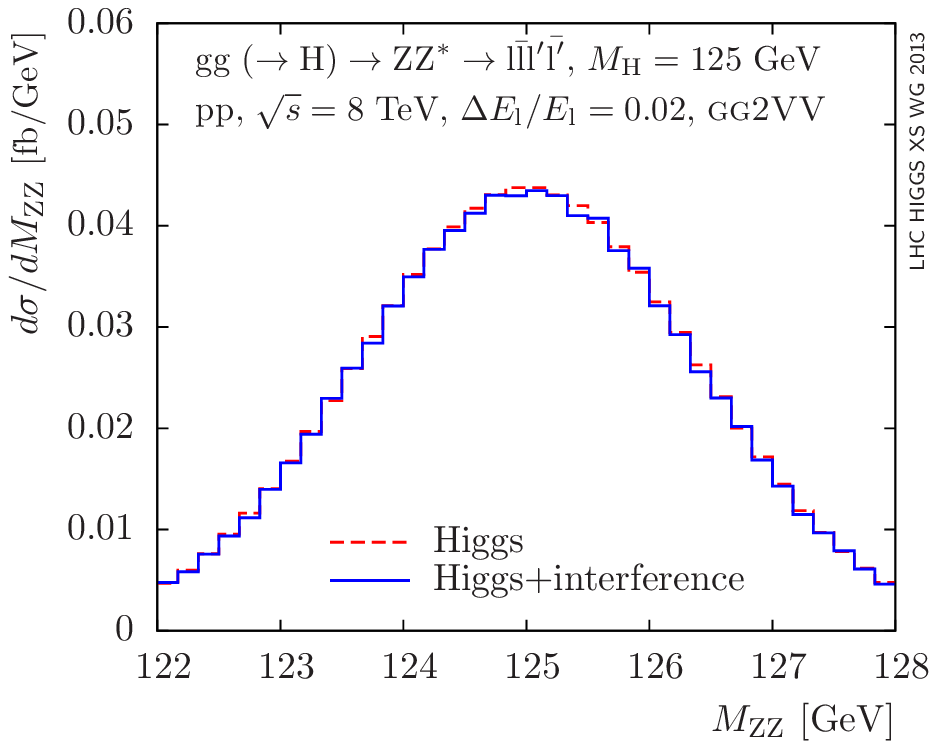}
\vspace*{0.cm}
\caption{$\PH\to \PZ\PZ^\ast$ invariant mass peak shift due to signal-background interference.}
\label{fig:ggF_ggVV_peakshift}
\end{figure}

The prediction of a $\Ord(-100\UMeV)$ Higgs invariant mass peak shift 
in $\Pg\Pg\to \PH\to \PGg\PGg$ for $\MH=125\UGeV$ in \Bref{Martin:2012xc} 
raises the question if a similar effect occurs in the $\Pg\Pg\to \PH\to \PZ\PZ^\ast$ 
mode.  That the deformation of the $\PH\to \PZ\PZ^\ast$ Breit-Wigner peak at 
$\MH=125\UGeV$ due to Higgs-continuum interference is negligible at parton level 
can, for example, be seen in Figure 17 in \Bref{Kauer:2012hd}, which displays 
the $M_{\PZ\PZ}$ distribution in the range $\MH\pm 3\GH$ for the process 
$\Pg\Pg\to \PH\to \PZ\PZ^\ast\to \Pl\PAl\PGn\PAGn$ in $\Pp\Pp$ collisions at $8\UTeV$.
No difference is perceptible between the signal and signal+interference+background 
distributions.  At $\MH\approx 125\UGeV$, the SM Higgs width is several orders of 
magnitude smaller than the $M_{\PZ\PZ}$ resolution.  Hence, detector 
resolution effects need to be taken into account to obtain a realistic prediction.  
For $\Pg\Pg\to \PH\to \PZ\PZ^\ast\to \Pl\PAl\Pl'\PAl'$ in $\Pp\Pp$ collisions at $8\UTeV$
with $\MH= 125\UGeV$, one obtains the $M_{\PZ\PZ}$ distributions shown 
in \refF{fig:ggF_ggVV_peakshift} when a Gaussian smearing of 
$\Delta E_{\Pl}/E_{\Pl} = 0.02$ is applied to the charged lepton momenta.  
\refF{fig:ggF_ggVV_peakshift} demonstrates that any shift between the
signal and signal+interference distribution is tiny compared to the histogram 
bin width of $167\UMeV$.  That the peak shift effect is much smaller 
for $\PH\to \PZ\PZ^\ast$ than for $\PH\to \PGg\PGg$ can be traced back to the 
tree-level-enhanced Higgs decay process and suppressed background 
process for a $\PZ\PZ^\ast$ versus $\PGg\PGg$ final state.

\subsubsection{Interference effects in $\PGg\PGg$
\footnote{ D.~de Florian, N.~Fidanza, R.~Hernandez-Pinto, J.~Mazzitelli, Y.~Rotstein-Habarnau, G.~Sborlini }}

\label{sec:interference_2gamma}

The resonance observed in the reconstruction of the diphoton invariant mass in proton proton collisions at the LHC turns out to be one of the main discovery channels and, therefore, requests precise theoretical calculations for the corresponding cross section. 

The  high precision achieved for the signal in gluon-gluon fusion $\Pg\Pg\rightarrow \PH$, the main production mechanism, is discussed in detail in \refS{sec:inclusive}. 
The rare decay $\PH\rightarrow \PGg\PGg$ is also mediated by loops. Corrections for the corresponding branching ratio are known to NLO accuracy for both QCD \cite{Spira:1995rr,Zheng:1990qa,Djouadi:1990aj,Dawson:1992cy,Djouadi:1993ji,Melnikov:1993tj,Inoue:1994jq} and electroweak \cite{Actis:2008ug} cases. Missing higher orders are estimated to be below $1\%$.

The corresponding background for diphoton production has been recently computed also up to NNLO \cite{Catani:2011qz}, but the interference between {\it signal} and {\it background} has not been evaluated to such level of accuracy yet.

The interference of the resonant process $ij \to X+\PH  \to \PGg \PGg $ with the continuum QCD background $ij \to X+\PGg\PGg $ induced by quark loops can be expressed at the level of the partonic cross section as:
\begin{eqnarray}
\delta\hat{\sigma}_{ij\to X+ \PH\to \PGg\PGg} &=& 
-2 (\sh-\MH^2) { \Re \left( \ca_{ij\to X+\PH} \ca_{\PH\to\PGg\PGg} 
                          \ca_{\mathrm{cont}}^* \right) 
        \over (\sh - \MH^2)^2 + \MH^2 \Gamma_{\PH}^2 }
\nonumber\\
&& %\hskip-0.3cm
-2 \MH \Gamma_{\PH} { \Im \left( \ca_{ij\to X+\PH} \ca_{\PH\to\PGg\PGg} 
                          \ca_{\mathrm{cont}}^* \right)
        \over (\sh - \MH^2)^2 + \MH^2 \Gamma_{\PH}^2 } \,,
\label{intpartonic}
\end{eqnarray}
where $\sh$ is the partonic invariant mass, $\MH$ and $\Gamma_{\PH}$ are the Higgs mass and decay width respectively \footnote{The details on the implementation of the lineshape \cite{Goria:2011wa} have a very small effect on the light Higgs discussed in this section. We rely here on a naive Breit-Wigner prescription.}.

As pointed out in \cite{Dixon:2003yb,Dicus:1987fk}, given that the contribution arising from the real part of the amplitudes is odd in $\sh$ around $\MH$, its effect on the total $\PGg\PGg$ rate is subdominant. For the gluon-gluon partonic subprocess, Dicus and Willenbrock \cite{Dicus:1987fk} found that the imaginary part of the corresponding one-loop amplitude has a quark mass suppression for the relevant helicity combinations. 
Dixon and Siu \cite{Dixon:2003yb} computed the main contribution of the interference to the cross-section, which originates on the two-loop imaginary part of the continuum amplitud $\Pg\Pg \to \PGg\PGg$. 
Recently, Martin \cite{Martin:2012xc} showed that even though the real part hardly contributes at the cross-section level, it has a quantifiable effect on the position of the diphoton invariant mass peak, producing a shift of about $ {\mathcal O}(100\,\UMeV)$ towards lower mass region once the smearing effect of the detector is taken into account.

The $\Pg\Pg$ interference channel considered in \cite{Martin:2012xc} is not the only ${\mathcal O}(\alphas^2)$ contribution that has to be considered for a full understanding of the interference term, since other partonic subprocesses initiated by $\Pg\PQq$ and $\PQq\PAQq$ can contribute at the same order.
At variance with the $\Pg\Pg$ subprocess that necessarily requests at least a one-loop amplitude for the background, the contribution from the remaining channels arises from tree-level amplitudes and can therefore only contribute to the real part of the interference in Eq.\ref{intpartonic} \footnote{Apart from a small imaginary part originated on the heavy-quark loops.}. 

In this section we present the results obtained for the remaining $\Pg\PQq \rightarrow \PQq\PGg\PGg$ and $\PQq\PQq \rightarrow \Pg\PGg\PGg$ channels, finalizing a full (lowest order) ${\mathcal O}(\alphas^2)$ calculation of the interference between Higgs diphoton decay amplitude and the corresponding continuum background \cite{deFlorian:2013psa}. We concentrate on the effect of the new interference channels on the position of the diphoton invariant mass peak.

It is worth noticing that, compared to the $\Pg\Pg \rightarrow \PGg\PGg$ subprocess, there is one more parton in the final state in the new channels. This parton  has to be integrated out to evaluate the impact on the cross section and its appearance might provide the wrong impression that the contribution is next-to-leading order-like. However, since signal and background amplitudes develop infrared singularities in different kinematical configurations, the interference is finite after phase space integration and behave as a true tree-level contribution, with exactly the same power of the coupling constant as the one arising from gluon-gluon interference channel.

For a phenomenological analysis of the results, we need to perform a convolution of the partonic cross-section with the parton density functions. We use the MSTW2008 LO set \cite{Martin:2009iq}  (five massless flavours are considered), and the one-loop expression of the strong coupling constant, setting the factorization and renormalization scales to the diphoton invariant mass $\muF=\muR= M_{\PGg\PGg}$. For the sake of simplicity, the production amplitudes are computed within the effective Lagrangian approach for the $\Pg\Pg\PH$ coupling (relying in the infinite top mass limit), approximation known to work at the few percent level for the process of interest. The decay into two photons is treated exactly and we set $\alpha=1/137$.
For the Higgs boson we use $\MH=125\,\UGeV$ and $\Gamma_{\PH}=4.2\,\UMeV$. For all the histograms we present in this section, an asymmetric cut is applied to the transverse momentum of the photons: $p_{T, \PGg}^{hard (soft)} \geq 40(30) \, \UGeV$. Their pseudorapidity is constrained to $|\eta_\PGg | \leq 2.5$.
We also implement the standard isolation prescription for the photons,  
requesting that the transverse hadronic energy deposited within a cone of size
$R=\sqrt{\Delta \phi^2+\Delta \eta^2}<0.4$ around the photon should satisfy $
p_{T,had} \leq 3\,\UGeV$. Furthermore, we reject all the events with $R_{\PGg\PGg} < 0.4$. 

\begin{figure} 
\begin {center}
\includegraphics[scale=1.4]{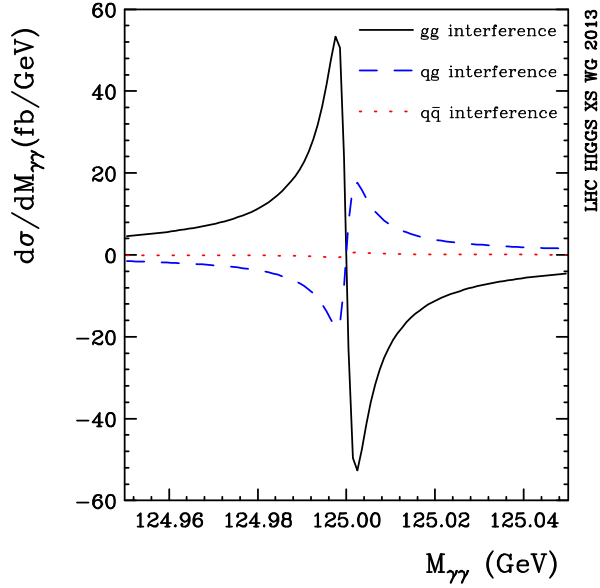}
\hspace{0.5cm} 
\caption{\label{fig:int_singauss} \small{Diphoton invariant mass distribution for the interference terms. The solid line is the $\Pg\Pg$ channel contribution, the dotted one the $\Pg\PQq$ channel, and dashed the $\PAQq\PQq$.}}
\end {center}
\end{figure}

In Figure \ref{fig:int_singauss} we show the three contributions to the full signal-background interference as a function of the diphoton invariant mass $M_{\PGg\PGg}$ after having implemented all the cuts mentioned above. The $\Pg\Pg$ term (solid line) represents the dominant $\Pg\Pg$ channel, while the $\Pg\PQq$ contribution (dashed) is about 3 times smaller in absolute magnitude, but as we can observe, has the same shape but opposite sign to the $\Pg\Pg$ channel. The $\PAQq\PQq$ contribution (dotted) is a couple of orders of magnitude smaller than the $\Pg\Pg$ one. The position of the maximum and minimum of the distribution are located near $M_{\PGg\PGg}=\PH \pm \Gamma_{\PH}/2$, with a shift at this level that remains at ${\mathcal O}(1\, \UMeV)$.

To simulate the smearing effects introduced by the detector, we convolute the cross-section with a Gaussian function of mass resolution width $\sigma_{{\mathrm{MR}}}=1.7\,\UGeV$ following the procedure \Bref{Martin:2012xc}.
\begin{figure} 
\begin {center}
\includegraphics[scale=1.4]{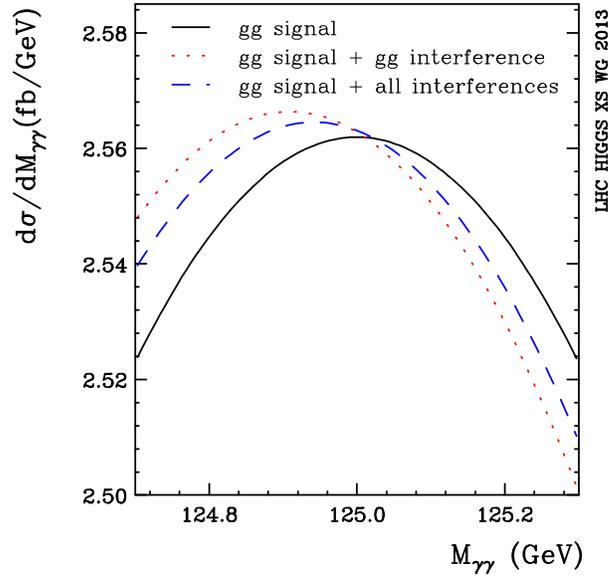}
\hspace{0.5cm} 
\caption{\label{fig:shift} \small{Diphoton invariant mass distribution including the smearing effects of the detector (Gaussian function of width $1.7 \, \UGeV$). The solid line corresponds to the signal-only contribution. The dotted line corresponds to the distribution  after adding the $\Pg\Pg$ interference term, and the dashed line represents the complete Higgs signal plus all three interference contributions ($\Pg\Pg$, $\Pg\PQq$ and $\PAQq\PQq$).}}
\end {center}
\end{figure}

In order to quantify the physical effect of the interferences in the diphoton invariant mass spectrum, we present in Figure \ref{fig:shift} the corresponding results after adding the Higgs signal. The solid curve corresponds to the signal cross-section, without the interference terms, but including the detector smearing effects. As expected, the (signal) Higgs peak remains at $M_{\PGg\PGg}=125 \, \UGeV$. 
When adding the $\Pg\Pg$ interference term, we observe a shift on the position of the peak of about $90 \, \UMeV$ towards a lower mass (dotted), as found in \Bref{Martin:2012xc}. If we also add the $\Pg\PQq$ and $\PAQq\PQq$ contributions (dashed), the peak is shifted around $30 \, \UMeV$ back towards a higher mass region because of the opposite sign of the amplitudes.

Given the fact that $\PAQq\PQq$ and $\Pg\PQq$ channels involve one extra particle in the final state, one might expect their contribution to be even more relevant for the corresponding interference in the process $\Pp\Pp \rightarrow \PH(\rightarrow \PGg\PGg) + \text{jet}$, since the {\it usually dominant} $\Pg\Pg$ channel \cite{deFlorian:1999tp} starts to contribute at the next order in the strong coupling constant for this observable.

It is worth noticing that the results presented here are plain LO in QCD. Given the fact that very large K-factors are observed in both the signal and the background, one might expect a considerable increase in the interference as well. While reaching NNLO accuracy for the interference looks impossible at the present time, a prescription to estimate the uncertainty on the evaluation of the interference and a way to combine it with more precise higher order computations for signal and background for $gg\rightarrow \PZ\PZ$ was recently presented in \cite{Passarino:2012ri}. The procedure can be easily extended to the case presented here by including all possible initial state channels.

Finally, we would like to emphasize that a more realistic simulation of the detector effects should be performed in order to obtain reliable predictions and allow for a direct comparison with the experimental data.

\subsubsection{Interference effects in $\PGg\PGg$ + jet 
\footnote{ S.~Martin}}

\label{sec:interference_2gammajet}

The Higgs diphoton signal at the LHC is in principle affected by interference 
between the Higgs resonant amplitudes and the continuum background amplitudes with the same
initial and final states.
Because the continuum amplitude 
$\Pg\Pg \rightarrow \PGg\PGg$ is of one-loop order
while the resonant amplitude $\Pg\Pg\rightarrow \PH\rightarrow \PGg\PGg$
is effectively of two-loop order, the interference need not be negligible.
It was shown in \cite{Dicus:1987fk,Dixon:2003yb} that the effect on the cross-section is 
very small at leading order, but the imaginary part of the two-loop amplitude leads to a suppression of 
the diphoton rate of order a few per cent for $\PH$ near $126\UGeV$ \cite{Dixon:2003yb}. 
The interference also produces a 
shift in the position of the diphoton mass peak. The diphoton lineshape can be written in the form
\begin{eqnarray}
\frac{d\sigma_{\Pp\Pp\rightarrow \PGg\PGg + X}}{d(\sqrt{h})} &=& C(h) +
\frac{1}{D(h)}\left [P(h) + (h - \PH^2) I(h) \right]
.   
\label{eq:genform}
\end{eqnarray}
Here, $h = M_{\PGg\PGg}^2$, and $C(h)$, $P(h)$, and $I(h)$ are smooth functions of $h$
near the resonance, and the Breit-Wigner function
$
D(h) \equiv (h - \PH^2)^2 + \PH^2 \Gamma^2_{\PH}
$
defines the Higgs mass $\PH$.
The function $C(h)$ comes from the pure continuum amplitudes not containing the Higgs, and 
$P(h)$ comes mainly from the pure Higgs-mediated contribution, while $I(h)$ comes from the interference.
The integral of the $I(h)$ term over the whole lineshape nearly vanishes.
However, at leading order 
it produces an excess of events below the Higgs mass $\PH$, and a corresponding 
deficit above, that is potentially observable \Bref{Martin:2012xc}.

%%%%%%%%%%

\begin{figure}[!htb]
\begin{minipage}[]{0.49\linewidth}
\includegraphics[width=7.2cm,angle=0]{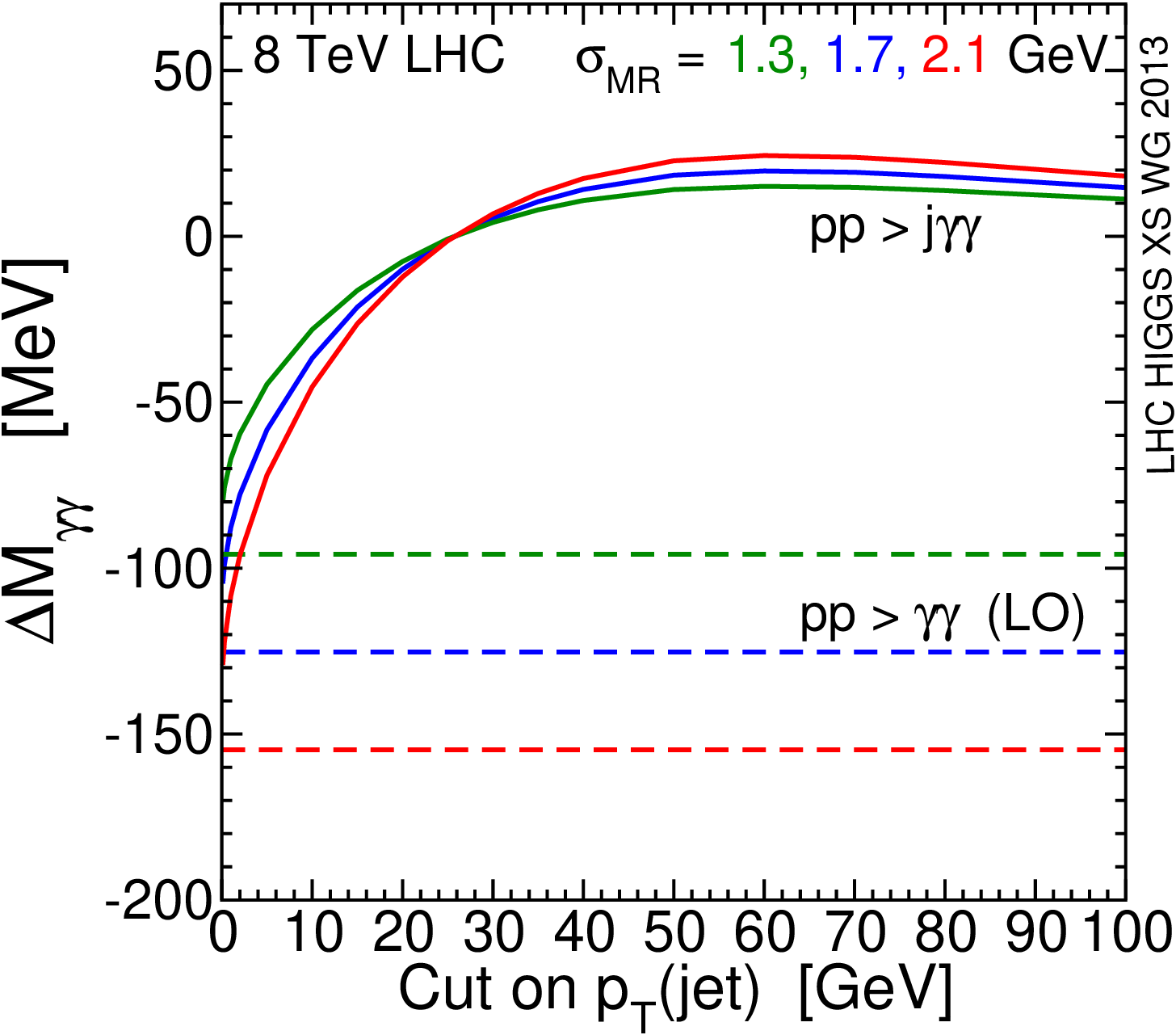}
\end{minipage}
\begin{minipage}[]{0.49\linewidth}
\begin{flushright}
\includegraphics[width=7.2cm,angle=0]{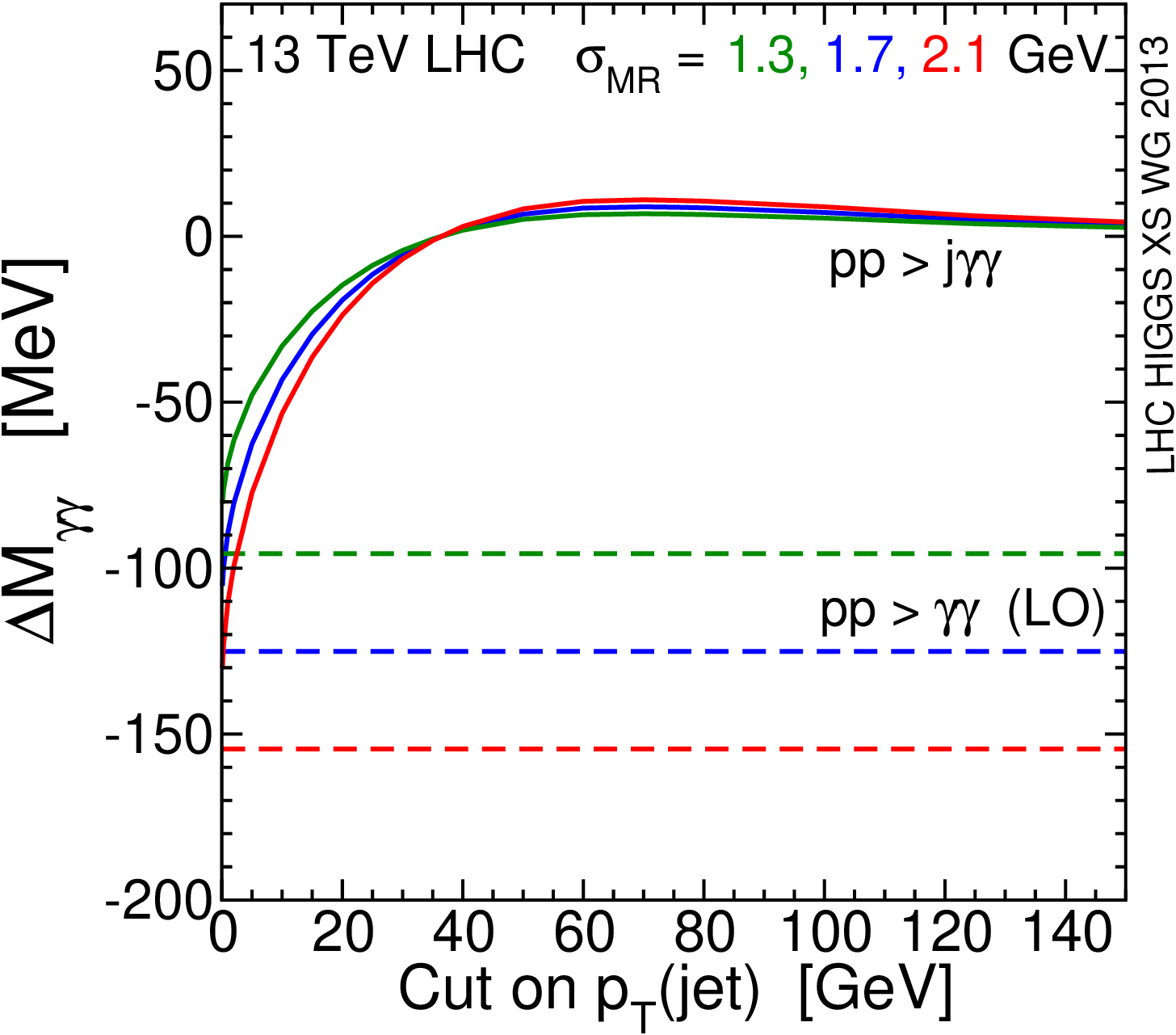}
\end{flushright}
\end{minipage}
\caption{\label{fig:shift2} The solid lines show the shifts in the diphoton mass peak, 
$\Delta M_{\PGg\PGg} \equiv M_{\PGg\PGg}^{\mathrm{peak}} - \PH$,
for $\Pp\Pp \rightarrow j\PGg\PGg$, as a function of the cut on the transverse 
momentum of the jet, $p^j_{{\mathrm T},{\mathrm{cut}}}$, for $\sigma_{\mathrm{MR}} = 1.3, 1.7\UGeV$, and 
$2.1\UGeV$ (from top to bottom on the left). The dashed lines shows the results for $\Pp\Pp \rightarrow \PGg\PGg$ at leading order without a jet requirement,
again for $\sigma_{\mathrm{MR}} = 1.3, 1.7\UGeV$, and $2.1\UGeV$ (from top to bottom).
The left (right) panel is for 
$\sqrt{s} = 8(13)\UTeV$. From \Bref{Martin:2013ula}.
} 
\end{figure}

The magnitude of the mass shift is potentially larger than the eventual experimental uncertainty in 
the mass. It is important to note that this shift will be different for different final states. 
For example, the interference should be much smaller for the $\PZ\PZ \rightarrow 4\ell$ final state; 
this makes the shift observable. For the diphoton case, it is reasonable to expect that the shift will 
be greatly affected by higher order effects, and a full NLO calculation (at least) should be done. As a 
precursor to this, in \cite{deFlorian:2013psa} and \cite{Martin:2013ula} 
the interference has been evaluated for the case of a 
final state including an extra jet, $\Pp\Pp \rightarrow j\PGg\PGg$. This follows from the 
parton-level subprocesses $\Pg\Pg \rightarrow g\PGg\PGg$ and $Qg \rightarrow Q\PGg\PGg$ (together 
with $\overline Qg \rightarrow \overline Q\PGg\PGg$) and a numerically very small contribution
from $Q\overline Q \rightarrow g\PGg\PGg$. As was pointed out in \cite{deFlorian:2013psa}, the shift
for the processes involving quarks has the opposite sign from the leading order, while in 
\cite{Martin:2013ula} it was shown that the $\Pg\Pg$-initiated process shift has the same sign, but smaller 
relative magnitude, than the leading order. The combined effects of this are shown in Figure
\ref{fig:shift2}, for cuts $\pT^{\PGg} \mbox{{(leading, sub-leading)}} > (40, 30)\>{\rm GeV}$,
$|\eta_{\PGg}| < 2.5$, $|\eta_j| < 3.0$, and $\Delta R_{\PGg\PGg}, \Delta R_{j\PGg} > 0.4$.
The cut on the transverse momentum of the jet is varied, and used as the horizontal axis of the plot.
%%%%%%%%%%%%%%%%%%%%%%%%%%%%%%%%%%%%%%%%%%%%%%%
%%%%%%%%%%%%%%%%%%%%%%%%%%%%%%%%%%%%%%%%%%%%%%%
For an experimentally reasonable cut $p^j_{T,{\mathrm{cut}}} > 25\UGeV$, 
the magnitude of the mass shift is much less than at leading order and is positive. This is in contrast 
to the negative shift
of about $(-95,-125,-155)$ MeV for $\sigma_{\mathrm{MR}} = (1.3, 1.7, 2.4)\UGeV$ 
from the leading order $\Pp\Pp \rightarrow \PGg\PGg$ case with no jet. The plot also shows the 
calculated shift for very low values of $p^j_{T,{\mathrm{cut}}}$, where the process $\Pg\Pg \rightarrow \Pg\PGg
\PGg$ dominates due to logarithmically enhanced soft gluons attached to the leading order diagrams. 
The shift in the limit of extremely small $p^j_{{\mathrm T},{\mathrm{cut}}}$ therefore approaches 
the leading order case.

The fact that $\Delta M_{\PGg\PGg}$ 
depends on the transverse momentum of the diphoton system is potentially useful, because it allows a measurement of the effect entirely within the sample of diphoton events, which have different systematics from the $\PZ\PZ \rightarrow 4 \ell$. So far, the interference effects of genuine virtual corrections has not been done, and a full NLO analysis would be interesting.

\subsection{Theoretical  uncertainties on the $\Pp\Pp \rightarrow \PW\PW$  estimation in the Higgs search
\footnote{B.~Di~Micco, R.Di~Nardo, H.~Kim, J.~Griffiths, D. C. Hall, C.~Hays  and J.~Yu}}
\label{sec:WWbackground}

The most relevant background to the $H \to \PWp\PWm \to \Pl\PGnl \Pl\PGnl$ channel is the non resonant $\Pp\Pp \to \PWp\PWm \to \Pl\PGnl \Pl\PGnl$ process.
 
In the decay of the Higgs boson to W pairs, $\PW$ bosons 
have opposite spin orientation, since the Higgs has spin zero.  
In the weak decay of the $\PW$ boson, due to the V-A nature of the interaction, the 
positive lepton is preferably emitted in the direction of the $\PWp$ spin and the 
negative lepton in the opposite direction of the $\PWm$ spin.  Therefore the two 
leptons are emitted close to each other and their invariant mass $m_{\Pl\Pl}$ is small.  
This feature is used  in the ATLAS\cite{ATLAS-CONF-2013-030} analysis to define a low-signal control region 
(CR) through a cut on $m_{\Pl\Pl}$ of $50{-}100\UGeV$.  The event yield of the $\PW\PW$ background 
is computed in the control region  and extrapolated to 
the signal regions.  The signal region is divided in two $m_{\Pl\Pl}$ bins when the outcoming leptons belong to different families (different flavour in the following),  each one defining one signal region (SR1,2 in the following), while if the outcoming leptons belong to the same family only one signal region (SR) is defined.
 
The $\PW\PW$ yield in the signal regions  is therefore:

\begin{eqnarray}
  \label{eqn:alpha}
  N^{\mathrm{WW 0j}}_{\mathrm{SR} (1,2)} & = & \alpha^{1,2}_{\mathrm{0j}} N^{\mathrm{WW 0j}}_{\mathrm{CR}}, \nonumber\\
  N^{\mathrm{WW 1j}}_{\mathrm{SR (1,2)}} & = & \alpha_{\mathrm{1j}}
  N^{\mathrm{WW 1j}}_{\mathrm{CR}} \nonumber
\end{eqnarray}

where alpha is the ratio (evaluated with the MC simulation) of expected events in the signal region and the $\PW\PW$ control region. 
The uncertainty on $\alpha$ is dominated by theoretical uncertainties, since it is 
defined using only well-measured charged-lepton quantities. There are two separate 
$\PW\PW$ production processes to consider: $\PAQq\PQq \rightarrow \PW\PW$ and $gg \rightarrow 
\PW\PW$.  Since the $\PAQq\PQq \rightarrow \PW\PW$ process contributes $95\% (93\%)$ of the 
total $\PW\PW$ background in the 0-jet (1-jet) channel, uncertainties on this process 
are the most important and are evaluated in Sec.~\ref{sec:alpha}.

\subsubsection{Uncertainties for the 0-jet and 1-jet analyses}
\label{sec:alpha}

The $\PW\PW$ background is estimated  using event counts in a 
control region (CR) defined using cuts on the $m_{\Pl\Pl}$ variable.  In \Tref{tab:WW_preselection} we describe the preselection cuts, and in \Tref{tab:WW_cuts} we show the cuts used to define the signal regions (SR) and the 
$\PW\PW$ CR. Note that for both the different flavour (DF) and same flavour (SF) 
analyses, a DF CR is used to normalise to data.

\begin{table}
  \begin{center}
  \caption{Definition of the preselection cuts for the $\PW\PW$ studies.} 
  \label{tab:WW_preselection}
    \begin{tabular}{lcc}
      \hline
      & Different flavour & Same flavour \\
      \hline
      Exactly 2 leptons      & \multicolumn{2}{c}{\textbf{lepton:} $\pT > 15\UGeV$, $|\eta| < 2.47$} \\
      Leading lepton $\pT$   & \multicolumn{2}{c}{$> 25\UGeV$} \\
      $m_{\Pl\Pl}$               & $> 10\UGeV$     & $> 12\UGeV$ \\
      $\ET^{\mathrm miss}$       & $> 25\UGeV$     & $> 45\UGeV$ \\
      Jet binning            & \multicolumn{2}{c}{\textbf{jet:} $\pT > 25\UGeV$, $|\eta| < 4.5$} \\
      $\pT^{\Pl\Pl}$ (0~jet only) & \multicolumn{2}{c}{$> 30\UGeV$} \\
      \hline
    \end{tabular}
  \end{center}
\end{table}

\begin{table}
  \begin{center}
   \caption{Definitions of the signal regions (SR), control regions (CR), and validation 
    region (VR).  The cuts are in addition to the preselection cuts described in 
    Table~\ref{tab:WW_preselection}. Note that, for both the different-flavour and same-flavour 
    analyses, the CR is always defined in the different flavour channel.} 
  \label{tab:WW_cuts}
   \begin{tabular}{lcc}
      \hline
      & $m_{\Pl\Pl}$ & $\Delta \phi_{\Pl\Pl}$ \\
      \hline
      SR1 (DF)    & $12 - 30\UGeV$     & $< 1.8$ \\
      SR2 (DF)    & $30 - 50\UGeV$  & $< 1.8$ \\
      SR  (SF)    & $12 - 50\UGeV$     & $< 1.8$ \\      
      CR (DF 0j)  & $50 - 100\UGeV$ & -- \\
      CR (DF 1j)  & $> 80\UGeV$     & -- \\
      VR (DF 0j)  & $>100\UGeV$     & -- \\
      \hline
    \end{tabular}
  \end{center}
\end{table}

The parameter $\alpha$ defined by:
\begin{equation}
\alpha_{\PW\PW} = \frac{N_{\PW\PW}^{\mathrm SR/VR}}{N_{\PW\PW}^{\mathrm{CR}}}
\label{eq:alpha}
\end{equation}

is used to predict the amount of $\PW\PW$ background in each signal or validation region (VR) from the 
data counts in the control region. The Validation Region is a signal free region non overlapping with the $\PW\PW$ CR, this region is used to test the validity of the extrapolation procedure on data. The $\alpha$ parameters are evaluated independently for the 0-jet and 1-jet bins, the same-flavor and 
different-flavor analyses, and in the two signal regions in the different-flavor analysis.  \\

\noindent
The non resonant $\Pp\Pp \to \PW\PW^{(*)}$ process is simulated with the \textsc{Powheg} Monte Carlo program interfaced 
to \textsc{pythia}{}8 for parton showering. \textsc{Powheg} computes the process $\Pp\Pp \to \PW\PW^{(*)} \to \Pl\PGnl \Pl\PGnl$ 
at NLO including off-shell  contributions.
The calculation includes the ``single-resonant'' process where the process $\Pl \to \PW \PGnl, 
\PW\to \Pl \PGnl$  
happens from a lepton of a ``single resonant'' $\PZ$ boson decay.  Uncertainties on 
the $\alpha$ parameters arise from PDF modelling, missing orders in the perturbative 
calculation, parton shower modelling, and the merging of the fixed-order calculation with 
the parton-shower model.  

\subsubsection{PDF uncertainties}
In order to evaluate the PDF uncertainties, we used $90\%$ C.L. CT10 PDF eigenvectors
and the  PDF parametrizations from MSTW2008 and NNPDF2.3. The last two are significantly smaller than the CT10 uncertainty. 
We take the quadrature sum of the 
CT10 uncertainties and the differences with respect to other parametrizations as the PDF uncertainty.  
A summary of $\alpha$ values and uncertainties are shown in \Tref{tab:pdfs}. This methods gives uncertainty bands close enough to the envelope method but allows
to compute the spread respect to the central PDF set that is used in the full MC simulation. The envelope method cannot be applied because it is not possible to generate a
sample with a PDF set that exactly matches the central value.

\begin{table}
  \begin{center}
  \caption{The $\alpha$ parameters for the standard analysis computed using 
    different PDF sets and the spread obtained using the CT10 error set.  The signs 
    indicate the difference with respect to the CT10 central value and show the 
    correlated differences in the different regions.  }
  \label{tab:pdfs}
    \begin{tabular}{lccc}
      \hline
      & CT10 error set & MSTW2008 & NNPDF2.3 \\
      \hline
      $\alpha^{\mathrm{DF}}_{\mathrm{0j}}$ (SR1) & 1.4\% & 0.01\%  & -0.5\% \\
      $\alpha^{\mathrm{DF}}_{\mathrm{0j}}$ (SR2) & 1.0\% & -0.02\% & -0.4\% \\
      $\alpha^{\mathrm{SF}}_{\mathrm{0j}}$       & 1.1\% & -0.01\% & -0.4\% \\
      $\alpha^{\mathrm{DF}}_{\mathrm{1j}}$ (SR1) & 1.8\% &  0.6\%  & -0.5\% \\
      $\alpha^{\mathrm{DF}}_{\mathrm{1j}}$ (SR2) & 1.6\% &  0.5\%  & -0.4\% \\
      $\alpha^{\mathrm{SF}}_{\mathrm{1j}}$       & 1.6 \% & 0.5\%  & -0.4\% \\
      \hline
    \end{tabular}
  \end{center}
\end{table}

\begin{table}
  \begin{center}
  \caption{The $\alpha$ parameters for the low $\pT$~analysis computed using different PDF 
    sets and the spread obtained using the CT10 error set.}
  \label{tab:pdfslowpt}
    \begin{tabular}{lccc}
      \hline
      & CT10 error set & MSTW2008 & NNPDF2.3 \\
      \hline
      $\alpha^{\mathrm{DF}}_{\mathrm{0j}}$ (SR1) & 1.7\% & 0.00\% & -0.6\% \\
      $\alpha^{\mathrm{DF}}_{\mathrm{0j}}$ (SR2) & 1.2\% & 0.04\% & -0.4\% \\
      $\alpha^{\mathrm{SF}}_{\mathrm{0j}}$       & 1.4\% & 0.01\% & -0.6\% \\
      $\alpha^{\mathrm{DF}}_{\mathrm{1j}}$ (SR1) & 1.9\% & 0.4\%  & -0.5\% \\
      $\alpha^{\mathrm{DF}}_{\mathrm{1j}}$ (SR2) & 1.7\% & 0.6\%  & -0.4\% \\
      $\alpha^{\mathrm{SF}}_{\mathrm{1j}}$       & 1.7\% & 0.4\%  & -0.6\% \\
      \hline
    \end{tabular}
  \end{center}
\end{table}

\subsubsection{Renormalization and factorization scales}
The uncertainty from the missing higher-order terms in the calculation of the production process 
can affect the $\pT$ distribution of the $\PW\PW^{*}$ pair and the $\alpha$ values.
The MCFM generator was used to estimate their effect  producing samples with renormalisation 
($\muR$) and factorisation ($\muF$) scale variations. The renormalisation scales are defined
as $\muR = \xi_{\mathrm R} \mu_{0}$ and $\muF = \xi_{\mathrm F} \mu_0$, where $\mu_0$ is a dynamic scale 
defined as $\mu_0 = m_{\PW\PW}$.

The scale variations were calculated using 20  million `virtual' integrations 
and $\approx 100$ million `real' integrations. In order to obtain an estimate of the 
statistical uncertainty on the $\alpha$ value produced by each scale variation, each sample 
was used individually to give an estimate of $\alpha$ and the central limit theorem was 
used to evaluate the uncertainty on the mean value. This gives a statistical uncertainty 
of around $0.4\%$ for the 0-jet case (the 1 jet case has negligible statistical uncertainty).

The nominal scale is obtained with $\xi_{\mathrm R} = \xi_{\mathrm F} = 1$ and the scale uncertainties
are obtained by varying $\xi_{\mathrm R}$ and $\xi_{\mathrm F}$ in the range $1/2 - 2$ while keeping
$\xi_{\mathrm R}/\xi_{\mathrm F}$ between $1/2$ and $2$; the maximum deviation from the nominal value 
is then taken as the scale uncertainty.  The scale uncertainties on $\alpha$ are shown 
in \Tref{tab:scale_and_pdfs}, where we summarize also the PDF, parton-shower, and 
modelling uncertainties.  The correlation between the $\alpha$ parameters of the 0-jet 
and 1-jet analyses is also evaluated in the calculation and their values are found to be 
fully correlated.

In order to ensure that we are not missing a localised large deviation in $\alpha$ on 
the $\muR - \muF$ plane and hence underestimating the scale uncertainty, we used the MCFM 
to simulate cases where $\xi_R$ and $\xi_F$ were equal to $3/4$ and $3/2$, but still 
fulfilling the same requirements as mentioned previously.  This corresponded to 12 more 
points in the $\muR - \muF$ plane, and for each point we generated 5 files with the 
same number of events as before.  Although this resulted in maximum deviations that were 
slightly larger than in the nominal variations, these deviations are within the statistical 
uncertainty evaluated using the central limit theorem. \Tref{tab:WW-extrascaleunc-alphaDF1}, 
\ref{tab:WW-extrascaleunc-alphaDF2} and \ref{tab:WW-extrascaleunc-alphaSF} summarise the results.

\begin{table}
  \begin{center}
  \caption{Values of $\alpha^{\mathrm{DF}}_{\mathrm{0j}}$ (SR1) as the renormalisation scale 
    (columns) and factorisation scale (rows) were varied. Statistical uncertainties
    in these vales are also shown.}
  \label{tab:WW-extrascaleunc-alphaDF1}
    \makebox[\linewidth]{
      \begin{tabular}{lccccc}
        & $\mu_0/2$         & $3\mu_0/2$        & $\mu_0$           & $3\mu_0/2$        & $2\mu_0$          \\
        \hline
        $\mu_0/2$  & $0.2261\pm0.21\%$ & $0.2250\pm0.26\%$ & $0.2248\pm0.17\%$ & --                & --                \\
        $3\mu_0/2$ & $0.2274\pm0.67\%$ & $0.2250\pm0.42\%$ & $0.2244\pm0.22\%$ & $0.2242\pm0.36\%$ & --                \\
        $\mu_0$    & $0.2264\pm0.32\%$ & $0.2254\pm0.26\%$ & $0.2247\pm0.22\%$ & $0.2243\pm0.26\%$ & $0.2232\pm0.23\%$ \\
        $3\mu_0/2$ & --                & $0.2248\pm0.24\%$ & $0.2255\pm0.48\%$ & $0.2237\pm0.19\%$ & $0.2233\pm0.19\%$ \\
        $2\mu_0$   & --                & --                &
  $0.2249\pm0.21\%$ & $0.2240\pm0.43\%$ & $0.2239\pm0.24\%$ \\
\hline
      \end{tabular}
    }
  \end{center}
\end{table}

\begin{table}
  \begin{center}
  \caption{Values of $\alpha^{\mathrm{DF}}_{\mathrm{0j}}$ (SR2) as the renormalisation scale 
    (columns) and factorisation scale (rows) were varied. Statistical uncertainties
    in these vales are also shown.}
  \label{tab:WW-extrascaleunc-alphaDF2}
    \makebox[\linewidth]{
      \begin{tabular}{lccccc}
        & $\mu_0/2$ & $3\mu_0/4$ & $\mu_0$ & $3\mu_0/2$ & $2\mu_0$\\
        \hline
        $\mu_0/2$ & $0.3846\pm0.22\%$ & $0.3816\pm0.19\%$ & $0.3828\pm0.23\%$ & -- & --\\
        $3\mu_0/4$ & $0.3856\pm0.64\%$ & $0.3839\pm0.40\%$ & $0.3821\pm0.29\%$ & $0.3801\pm0.37\%$ & --\\
        $\mu_0$ & $0.3849\pm0.24\%$ & $0.3819\pm0.42\%$ & $0.3833\pm0.24\%$ & $0.3805\pm0.30\%$ & $0.3810\pm0.24\%$\\
        $3\mu_0/2$ & -- & $0.3810\pm0.28\%$ & $0.3823\pm0.41\%$ & $0.3805\pm0.32\%$ & $0.3794\pm0.24\%$\\
        $2\mu_0$ & -- & -- & $0.3848\pm0.43\%$ & $0.3808\pm0.45\%$ & $0.3826\pm0.24\%$\\
\hline
      \end{tabular}
    }
  \end{center}
\end{table}

\begin{table}
  \begin{center}
  \caption{Values of $\alpha^{\mathrm{SF}}_{\mathrm{0j}}$ as the renormalisation scale 
    (columns) and factorisation scale (rows) were varied. Statistical uncertainties
    in these vales are also shown.}
  \label{tab:WW-extrascaleunc-alphaSF}
    \makebox[\linewidth]{
      \begin{tabular}{lccccc}
        & $\mu_0/2$ & $3\mu_0/4$ & $\mu_0$ & $3\mu_0/2$ & $2\mu_0$\\
        \hline
        $\mu_0/2$ & $0.4694\pm0.45\%$ & $0.4665\pm0.43\%$ & $0.4675\pm0.39\%$ & -- & --\\
        $3\mu_0/4$ & $0.4713\pm0.63\%$ & $0.4667\pm0.43\%$ & $0.4650\pm0.42\%$ & $0.4640\pm0.42\%$ & --\\
        $\mu_0$ & $0.4698\pm0.40\%$ & $0.4675\pm0.47\%$ & $0.4689\pm0.52\%$ & $0.4645\pm0.39\%$ & $0.4657\pm0.48\%$\\
        $3\mu_0/2$ & -- & $0.4663\pm0.39\%$ & $0.4659\pm0.44\%$ & $0.4647\pm0.37\%$ & $0.4639\pm0.40\%$\\
        $2\mu_0$ & -- & -- & $0.4674\pm0.41\%$ & $0.4655\pm0.59\%$ & $0.4659\pm0.40\%$\\
\hline
      \end{tabular}
    }
  \end{center}
\end{table}

Scale uncertainties were alternatively evaluated with the \textsc{aMC@NLO} generator 
 varying $\xi_{\mathrm R}$ and $\xi_{\mathrm F}$ in the range $1/2 - 2$ while keeping
$\xi_{\mathrm R}/\xi_{\mathrm F}$ between $1/2$.  These uncertainties are summarized in 
\Tref{tab:WW_amcatnlo_scale_unc} and are statistically consistent with those of 
MCFM.  

\begin{table}
  \begin{center}
  \caption{The uncertainty on the $\PW\PW$ extrapolation parameters, $\alpha$, calculated by varying the renormalisation
    and factorisation scales with the a\textsc{MC@NLO} generator.  The statistical uncertainty is included as an uncertainty
    on the uncertainty.}
  \label{tab:WW_amcatnlo_scale_unc}
    \begin{tabular}{lc}
      \hline
      & Maximum Deviation \\
      \hline
      $\alpha^{\mathrm{DF}}_{\mathrm{0j}}$ (SR1) & $1.7 \pm 0.7\%$ \\
      $\alpha^{\mathrm{DF}}_{\mathrm{0j}}$ (SR2) & $0.6 \pm 0.6\%$ \\
      $\alpha^{\mathrm{SF}}_{\mathrm{0j}}$       & $1.0 \pm 0.5\%$ \\
      $\alpha^{\mathrm{DF}}_{\mathrm{1j}}$ (SR1) & $3.4 \pm 1.1\%$ \\
      $\alpha^{\mathrm{DF}}_{\mathrm{1j}}$ (SR2) & $1.4 \pm 0.9\%$ \\
      $\alpha^{\mathrm{SF}}_{\mathrm{1j}}$       & $2.3 \pm 0.8\%$ \\
      \hline
    \end{tabular}
  \end{center}
\end{table}

\subsubsection{Generator modelling}
The $\alpha$ predictions using various generators was studied to get a range
encompassing different orders of perturbative calculation and different models of
parton showering and the associated merging with the fixed-order calculation. 
We conservatively assign a modelling uncertainty on the difference in $\alpha$ predicted 
between the best available pair of generators in terms of the fixed-order calculation.
\textsc{Powheg + Pythia8} and \textsc{MCFM} have been compared for this study. 
\textsc{MCMF} is a pure parton level MC that is not matched to a parton showering algorithm.  However, other effects included in this uncertainty 
might involve e.g. different renormalisation and factorisation scales, or different 
electroweak schemes.  The differences in $\alpha$ between the generators and the 
assigned uncertainties are shown in \Tref{tab:WW_powheg_mcfm_comparison}.

\begin{table}
  \begin{center}
  \caption{The $\PW\PW$ extrapolation parameters, $\alpha$, calculated using \textsc{powheg + pythia 8} 
    and MCFM.  The 0-jet and 1-jet, different-flavour (DF) and same-flavour (SF) values are each 
    calculated.}
  \label{tab:WW_powheg_mcfm_comparison}
    \begin{tabular}{lccc}
      \hline
      & POWHEG + Pythia 8 & MCFM & $\delta\alpha$ \\
      \hline
      $\alpha^{\mathrm{DF}}_{\mathrm{0j}}$ (SR1) & 0.2277  & 0.225 &  -1.2\% \\
      $\alpha^{\mathrm{DF}}_{\mathrm{0j}}$ (SR2) & 0.3883  & 0.383 &  -1.4\% \\
      $\alpha^{\mathrm{SF}}_{\mathrm{0j}}$       & 0.4609  & 0.469 &  +1.7\% \\
      $\alpha^{\mathrm{DF}}_{\mathrm{1j}}$ (SR1) & 0.1107  & 0.105 &  -5.1\% \\
      $\alpha^{\mathrm{DF}}_{\mathrm{1j}}$ (SR2) & 0.1895  & 0.180 &  -5.0\% \\
      $\alpha^{\mathrm{SF}}_{\mathrm{1j}}$       & 0.2235  & 0.217 &  -3.1\% \\
      \hline
    \end{tabular}
  \end{center}
\end{table}

\begin{table}
  \begin{center}
  \caption{The $\PW\PW$ extrapolation parameters, $\alpha$, calculated using \textsc{powheg + herwig} 
    and \textsc{aMC@NLO}.  The 0-jet and 1-jet, different-flavour (DF) and same-flavour (SF) values 
    are each calculated.}
  \label{tab:WW_powheg_amcatnlo_comparison}
    \begin{tabular}{lccc}
      \hline
      & POWHEG + Herwig & aMC@NLO & $\delta\alpha$ \\
      \hline
      $\alpha^{\mathrm{DF}}_{\mathrm{0j}}$ (SR1) & 0.2277 & 0.2274 & -0.4\% \\
      $\alpha^{\mathrm{DF}}_{\mathrm{0j}}$ (SR2) & 0.3914 & 0.3845 & -1.7\% \\
      $\alpha^{\mathrm{SF}}_{\mathrm{0j}}$       & 0.4623 & 0.4581 & -0.9\% \\
      $\alpha^{\mathrm{DF}}_{\mathrm{1j}}$ (SR1) & 0.1113 & 0.1064 & -4.3\% \\
      $\alpha^{\mathrm{DF}}_{\mathrm{1j}}$ (SR2) & 0.1904 & 0.1840 & -3.4\% \\
      $\alpha^{\mathrm{SF}}_{\mathrm{1j}}$       & 0.2247 & 0.2122 & -5.6\% \\
      \hline
    \end{tabular}
  \end{center}
\end{table}

The modelling uncertainties were computed also for the extrapolation from the signal region to an higher $m_{\Pl\Pl}$ region (validation region)
used to test the uncertainty prescription. The validation region is  defined by $m_{\Pl\Pl} > 100\UGeV$ after preselection. The results are 
shown in \Tref{tab:WW_powheg_mcfm_comparison_validation}.

\begin{table}
  \begin{center}
  \caption{The $\PW\PW$ extrapolation parameters, $\alpha$, calculated for the validation 
    region using \textsc{Powheg + Pythia8} and \textsc{MCFM}.  The 0 jet and 1 jet, different 
    flavour (DF) values are each calculated. }
  \label{tab:WW_powheg_mcfm_comparison_validation}
    \begin{tabular}{lccc}
      \hline
      & \textsc{POWHEG + Pythia} 8 & \textsc{MCFM} & $\delta\alpha$ \\
      \hline
      $\alpha^{\mathrm{DF}}_{\mathrm{0j}}$ (VR) & 0.9420 & 0.961 & 2.0\% \\
      \hline
    \end{tabular}
  \end{center}
\end{table}

\begin{figure}[h!]
  \centering
  \includegraphics[width=0.475\textwidth]{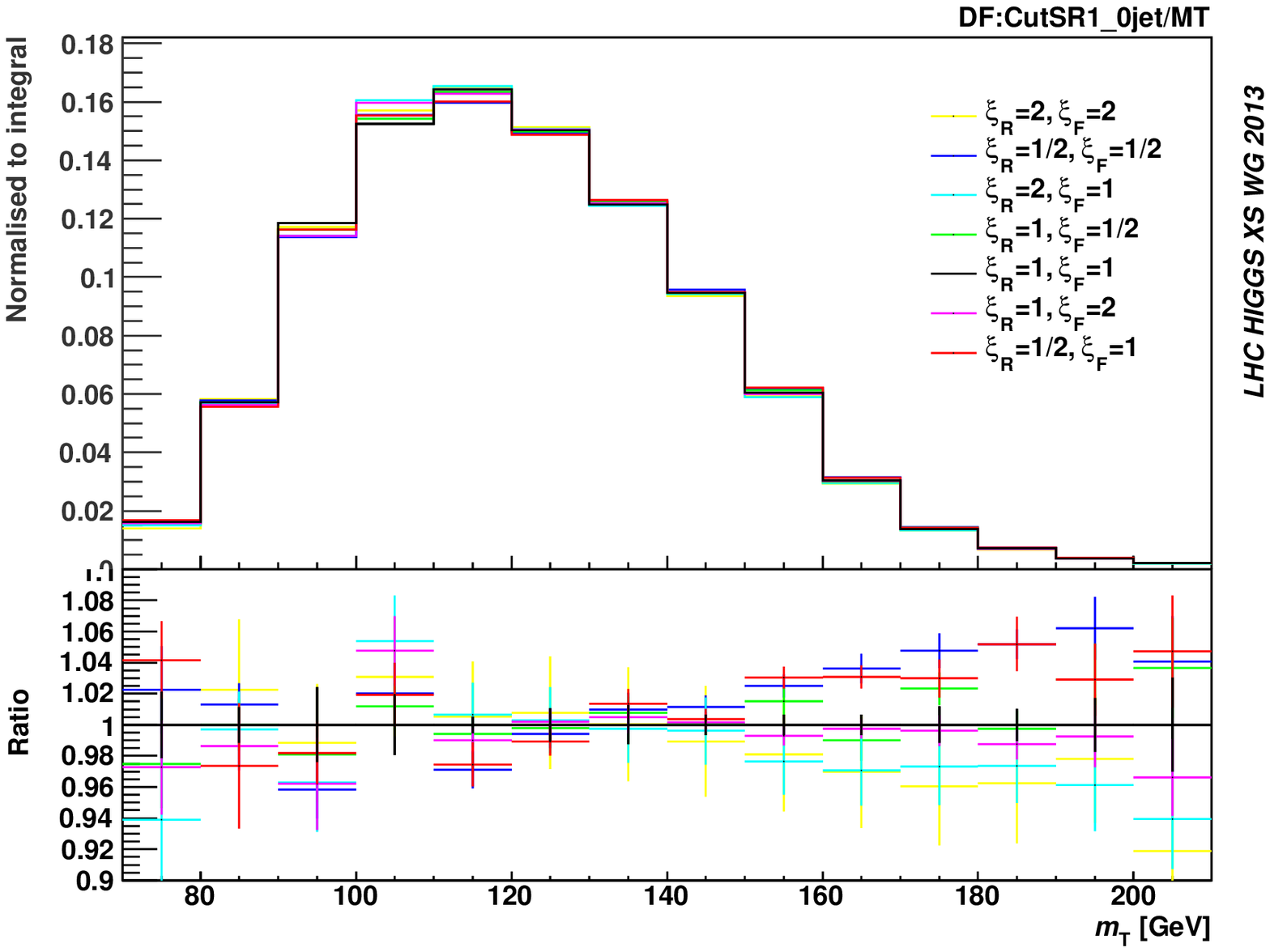}
  \includegraphics[width=0.475\textwidth]{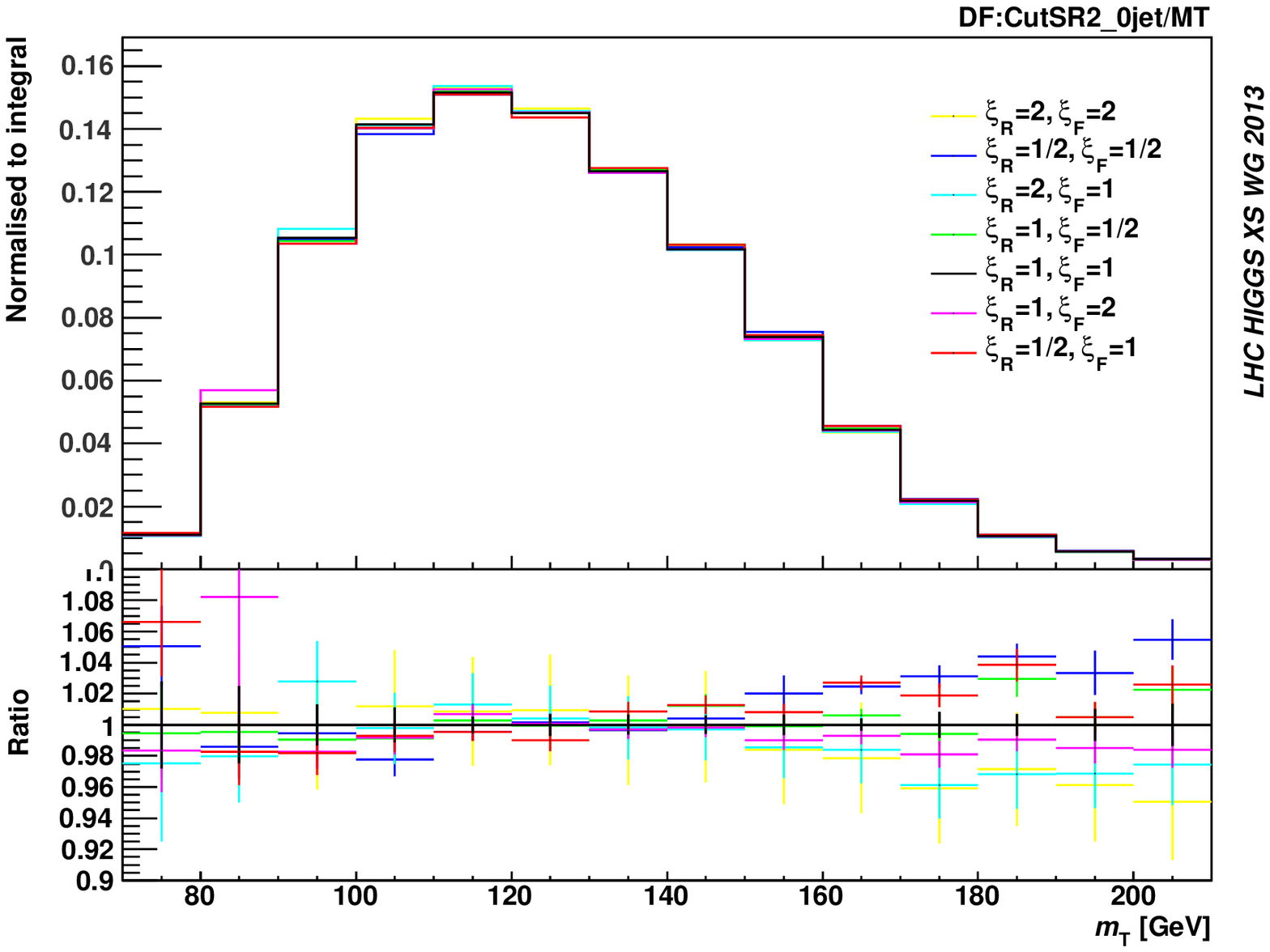}
  \includegraphics[width=0.475\textwidth]{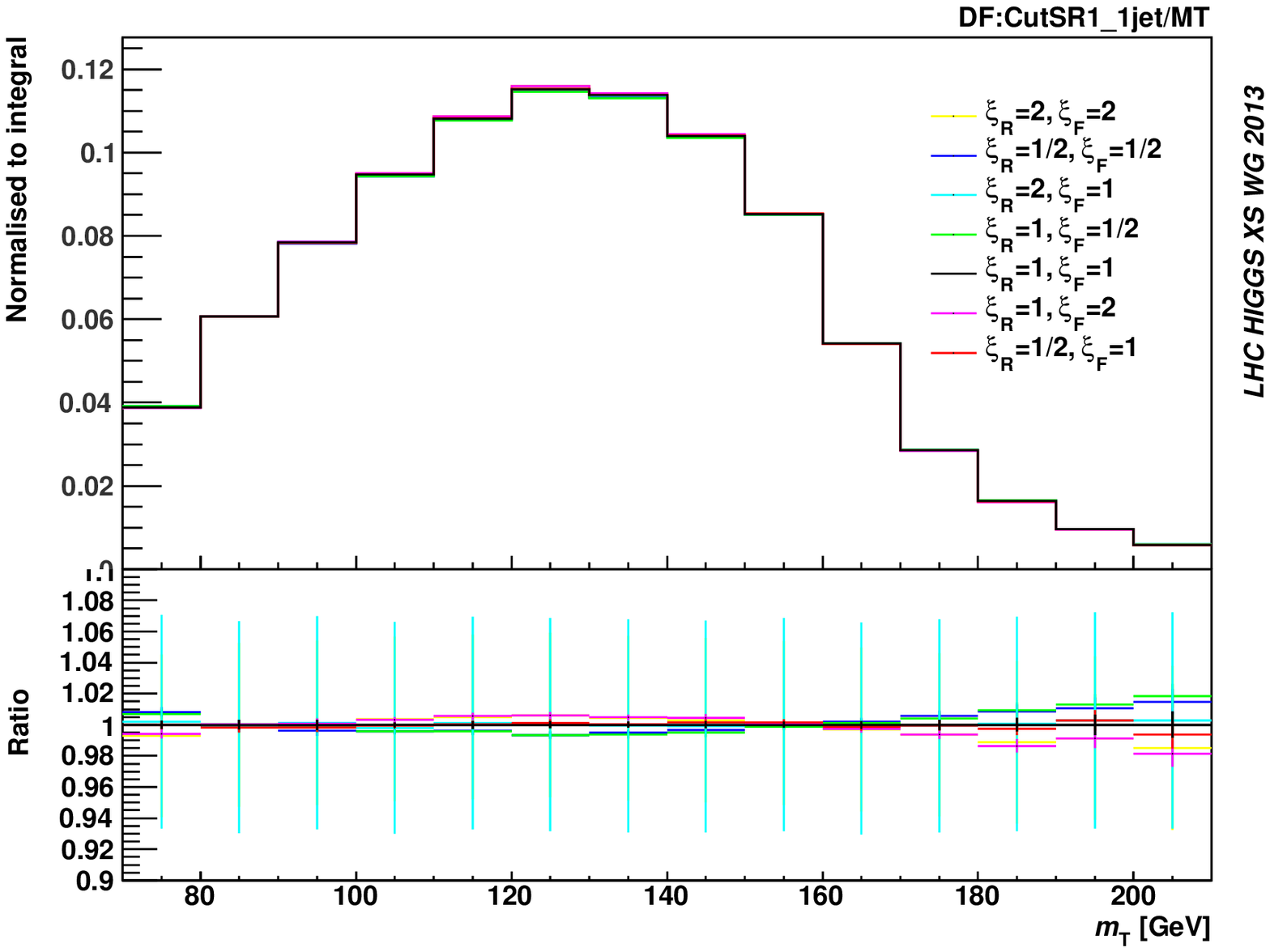}
  \includegraphics[width=0.475\textwidth]{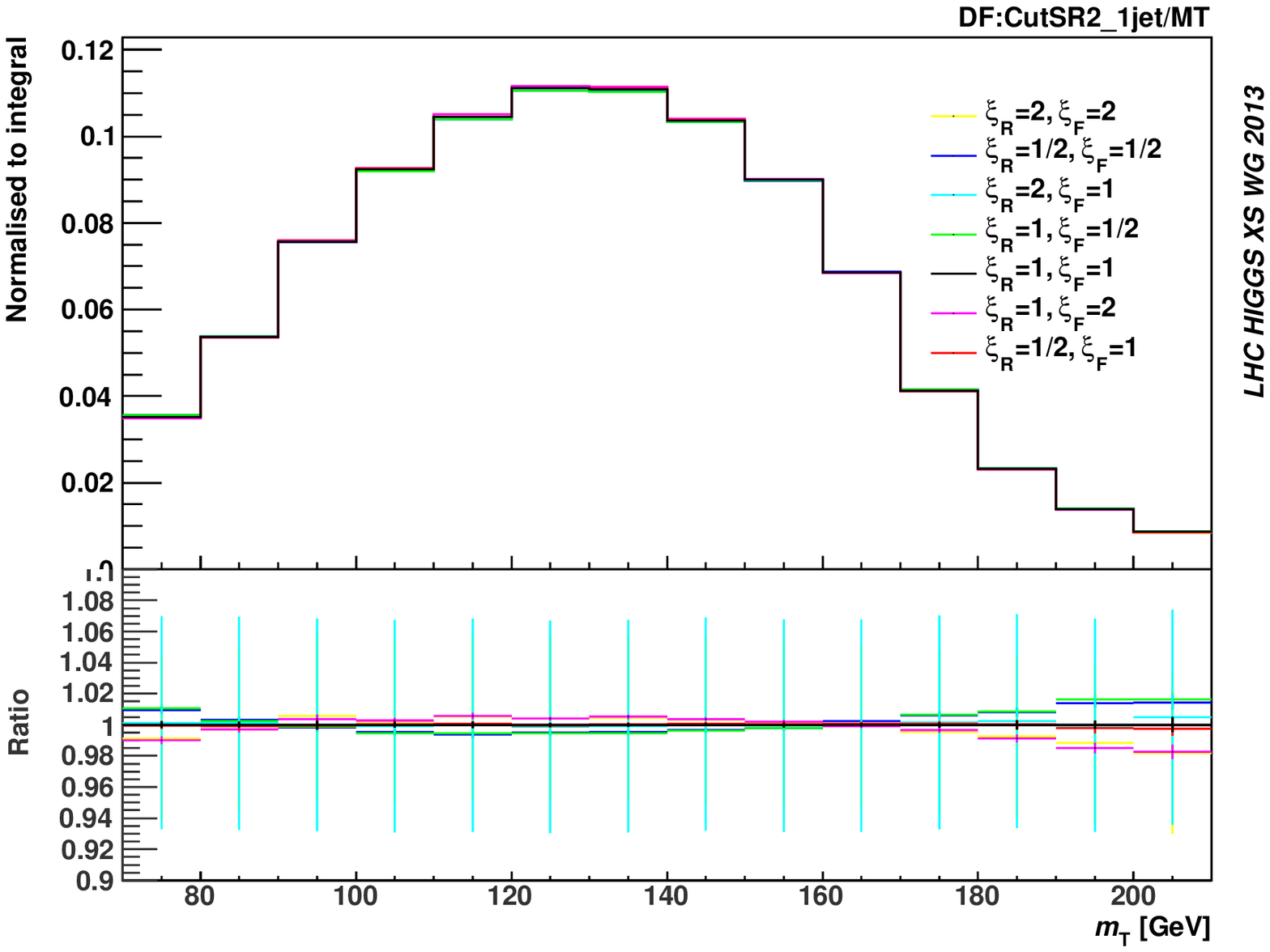}
  \includegraphics[width=0.475\textwidth]{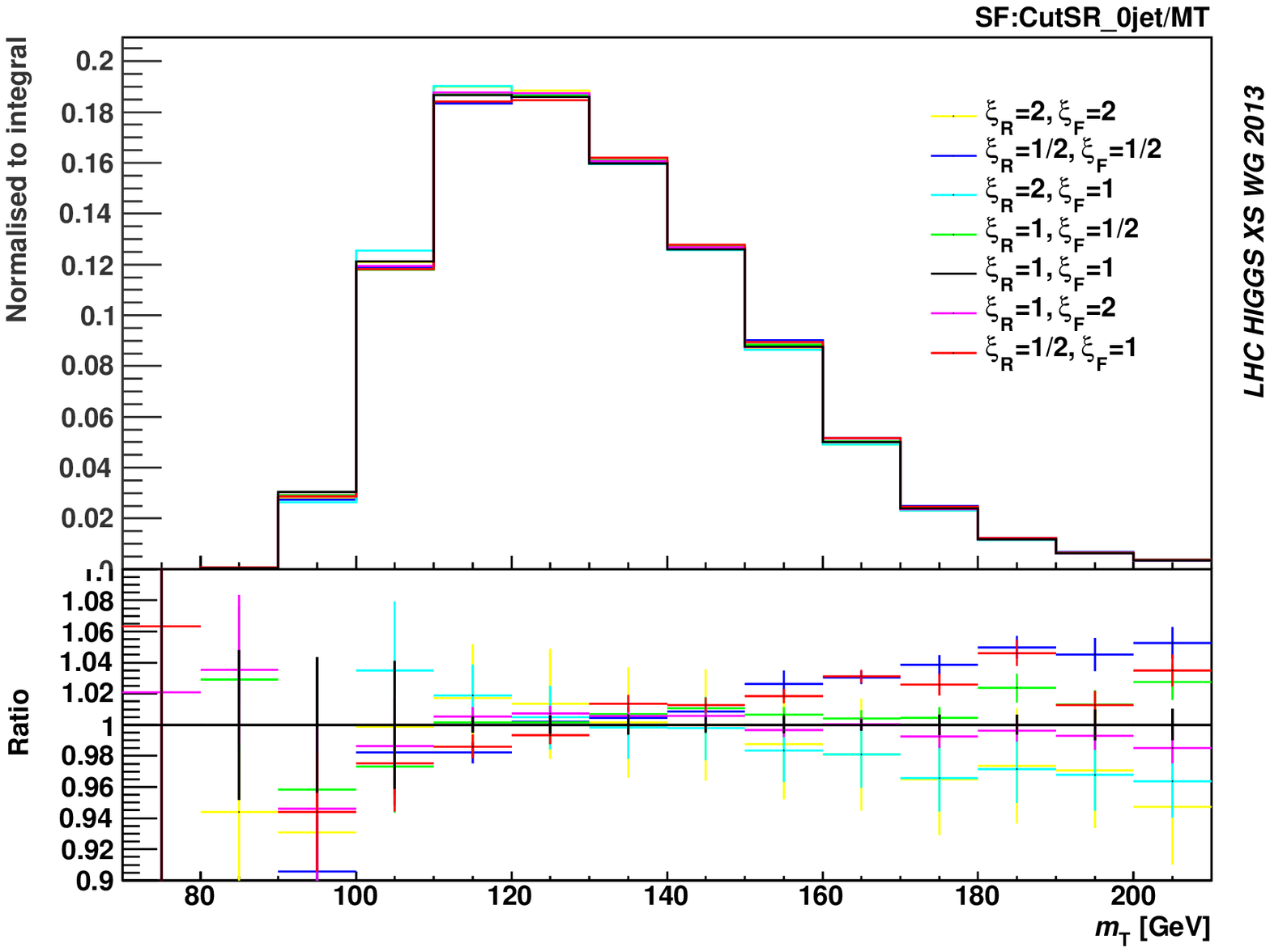}
  \includegraphics[width=0.475\textwidth]{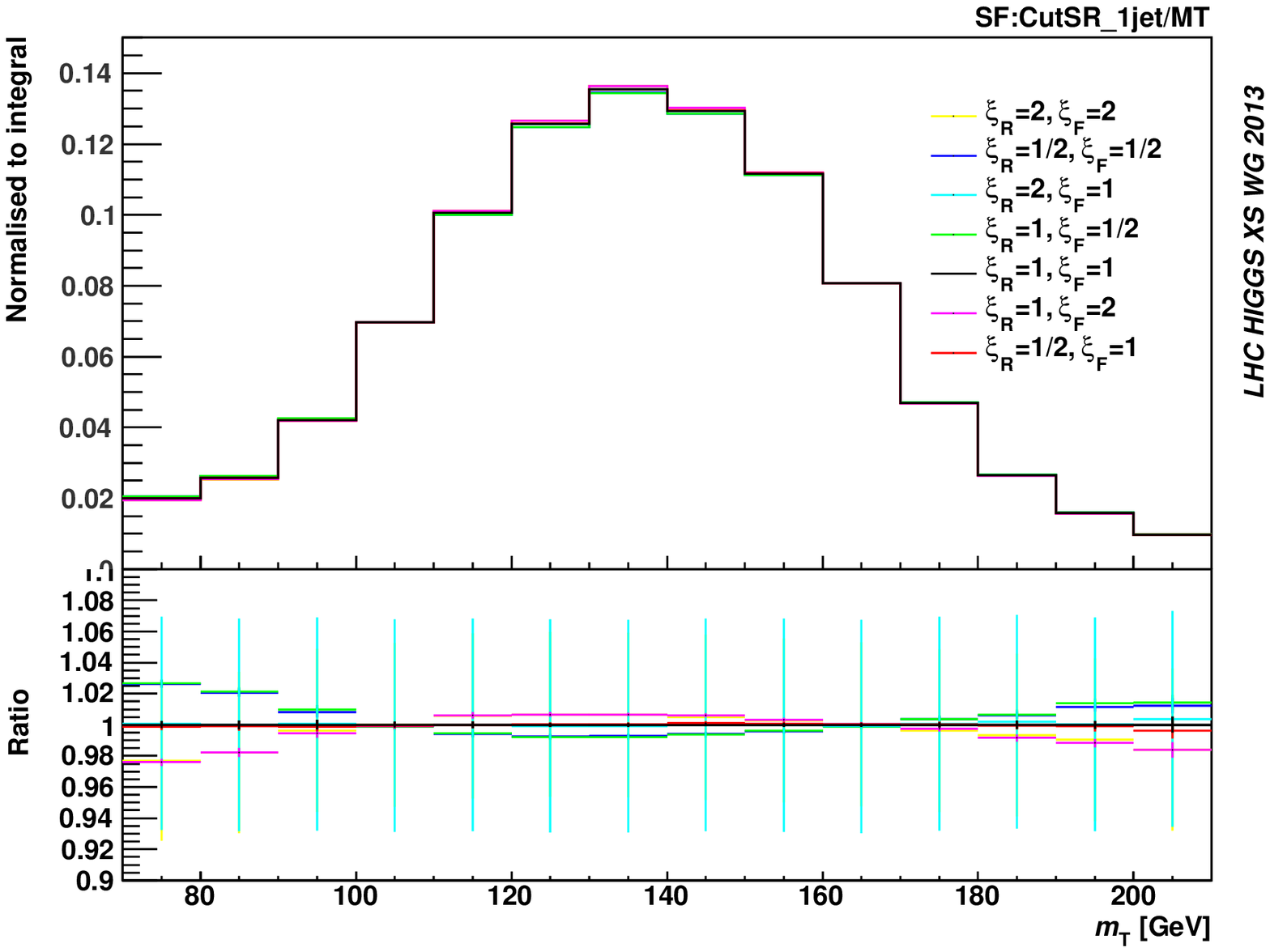}
  \caption{Scale uncertainties in the $m_{\mathrm T}$ distribution in (top left) SR1 in DF 0j 
    analysis, (top right) SR2 in DF 0j analysis, (middle left) SR1 in DF 1j analysis, 
    (middle right) SR2 in DF 1j analysis, (bottom left) SR in SF 0j analysis, and (bottom 
    right) SR in SF 1j analysis.}
  \label{fig:ww_mT_scaleunc}
\end{figure}

\begin{figure}[h!]
  \centering
  \includegraphics[width=0.475\textwidth]{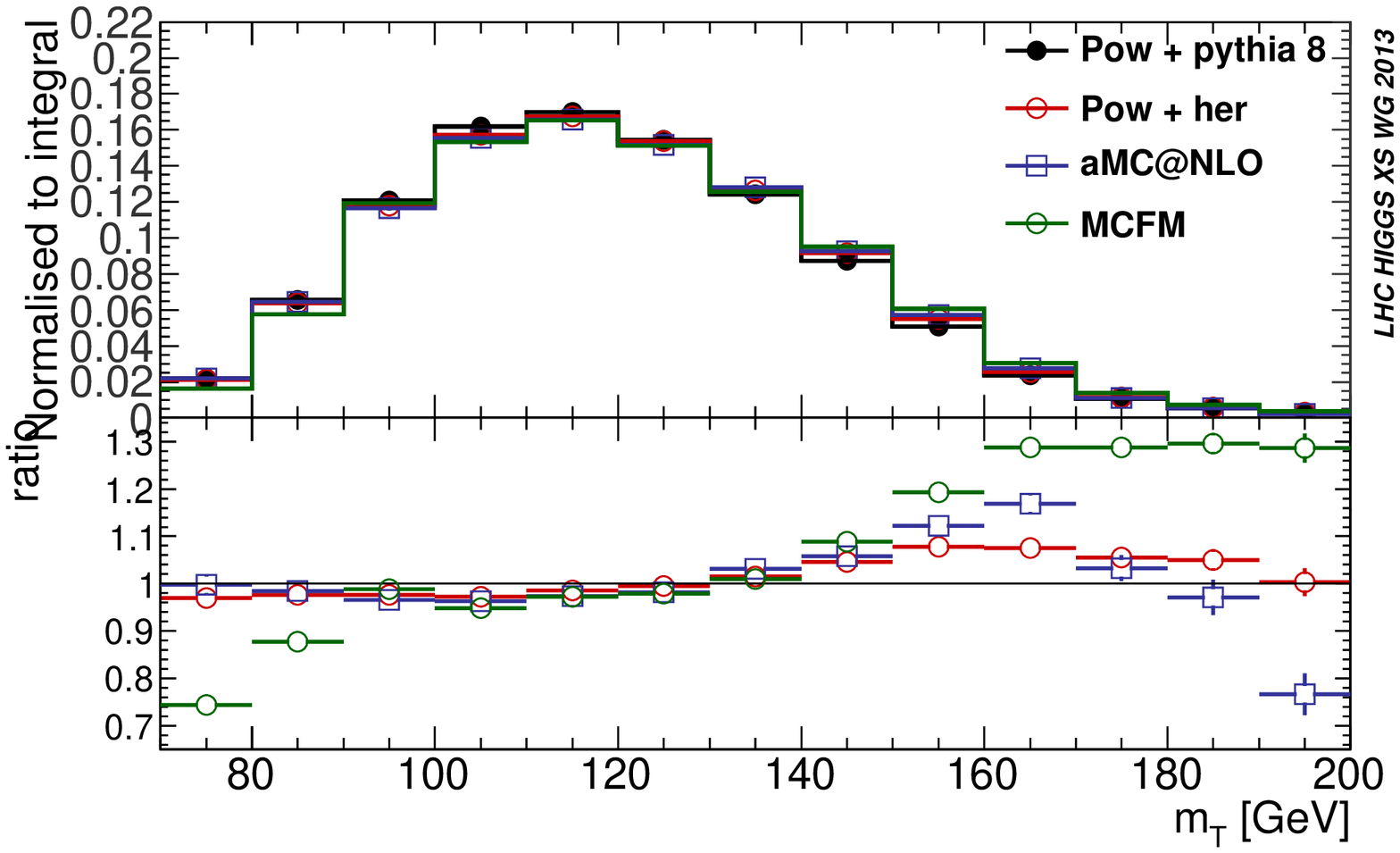}
  \includegraphics[width=0.475\textwidth]{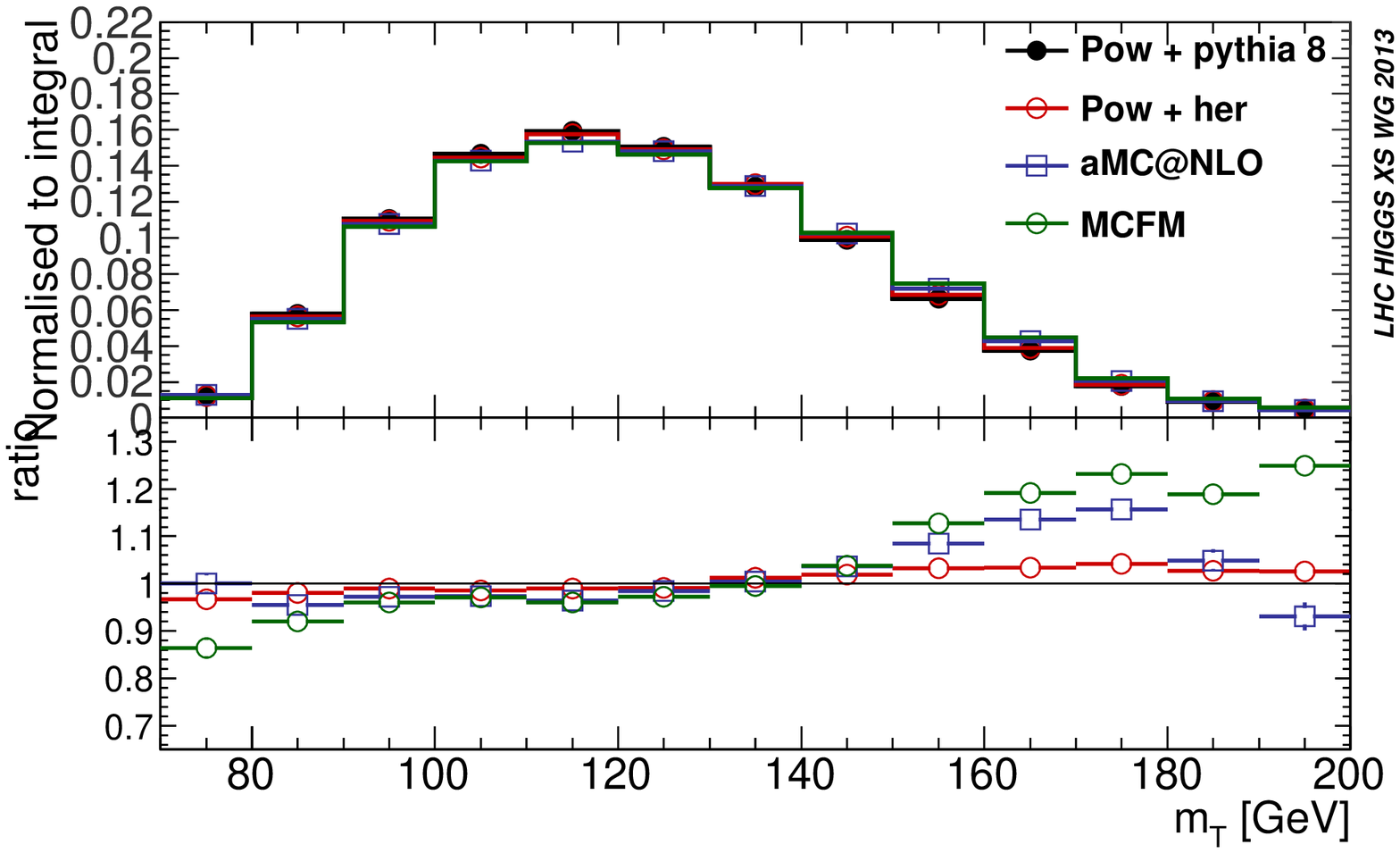}
  \includegraphics[width=0.475\textwidth]{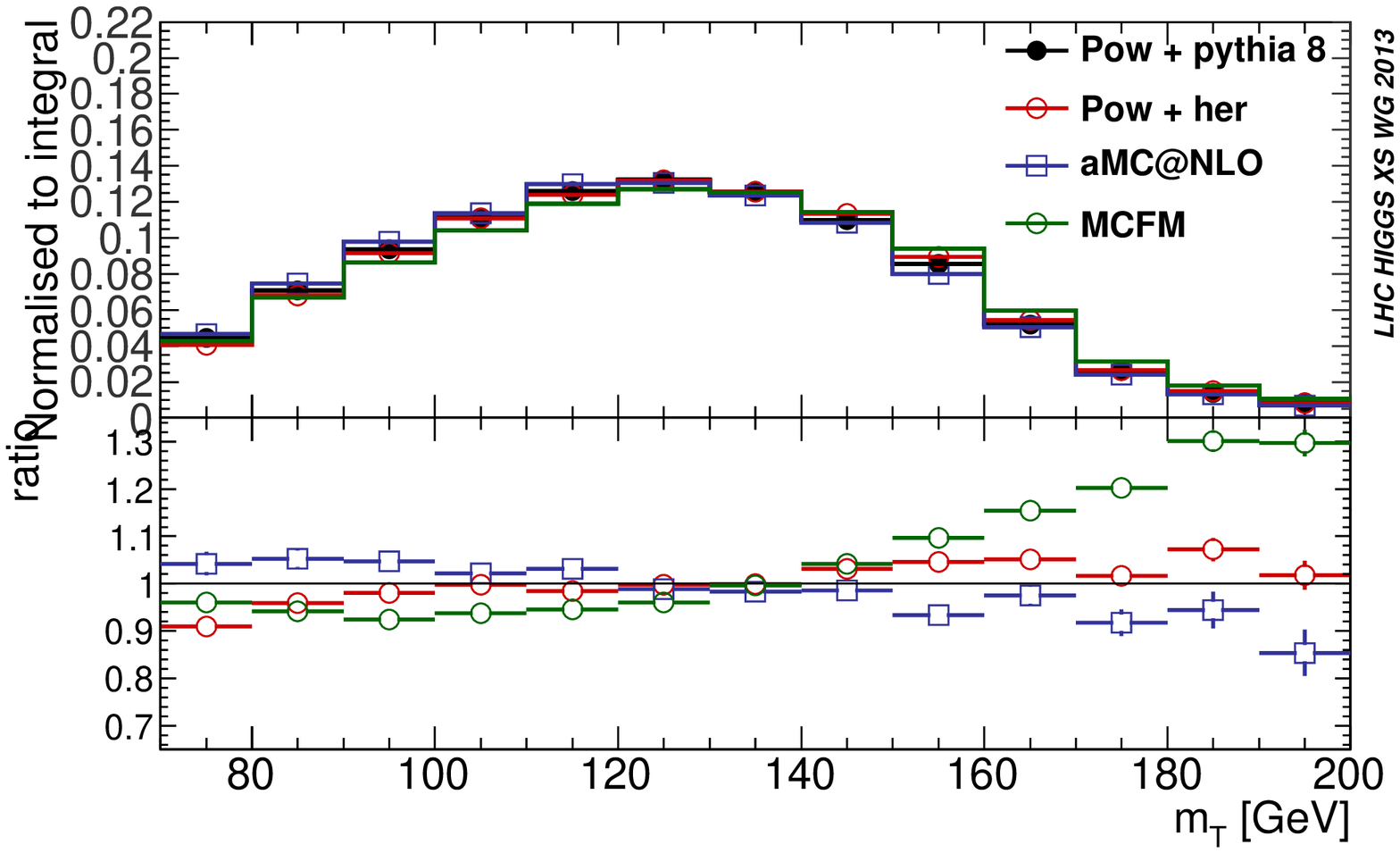}
  \includegraphics[width=0.475\textwidth]{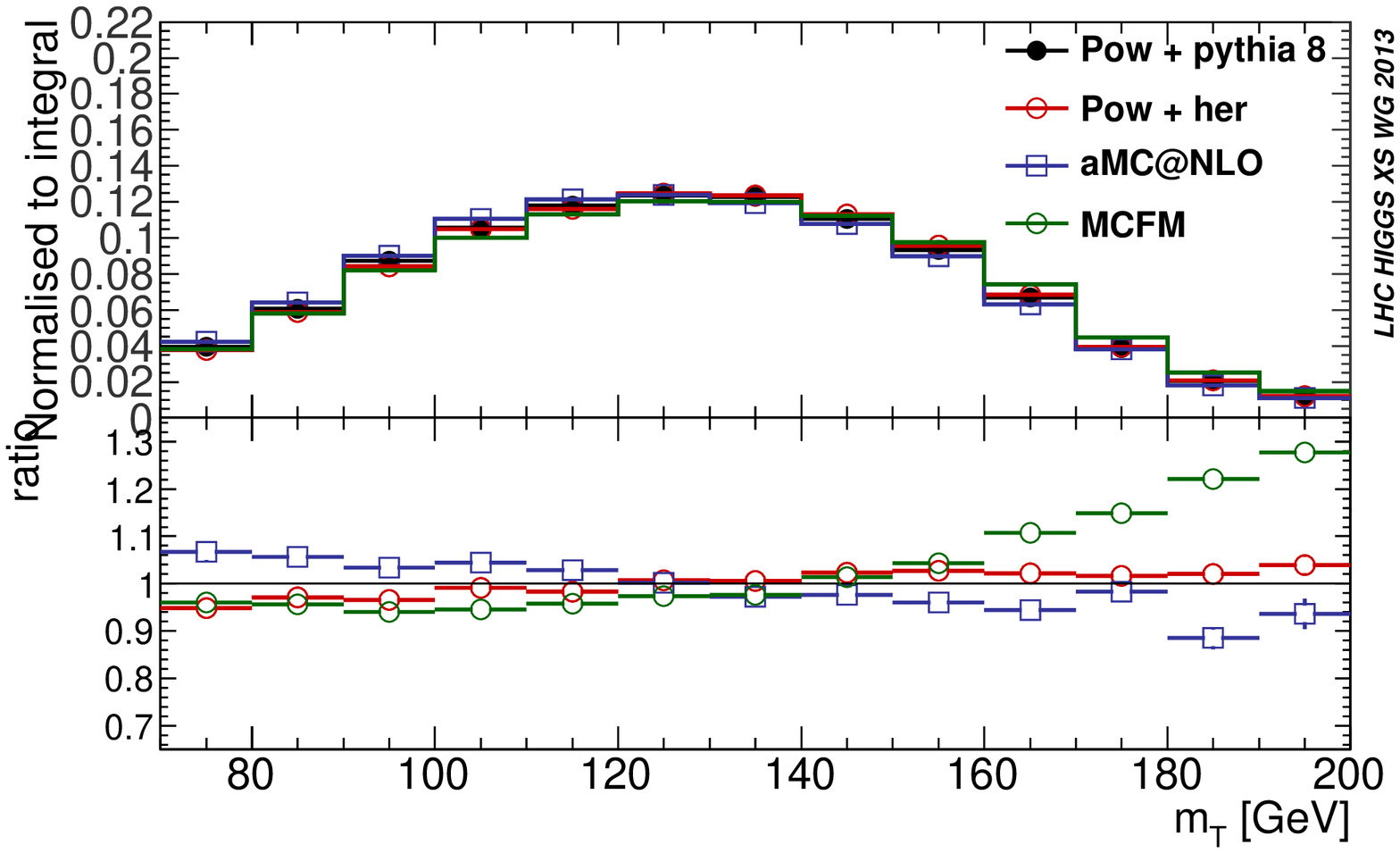}
  \includegraphics[width=0.475\textwidth]{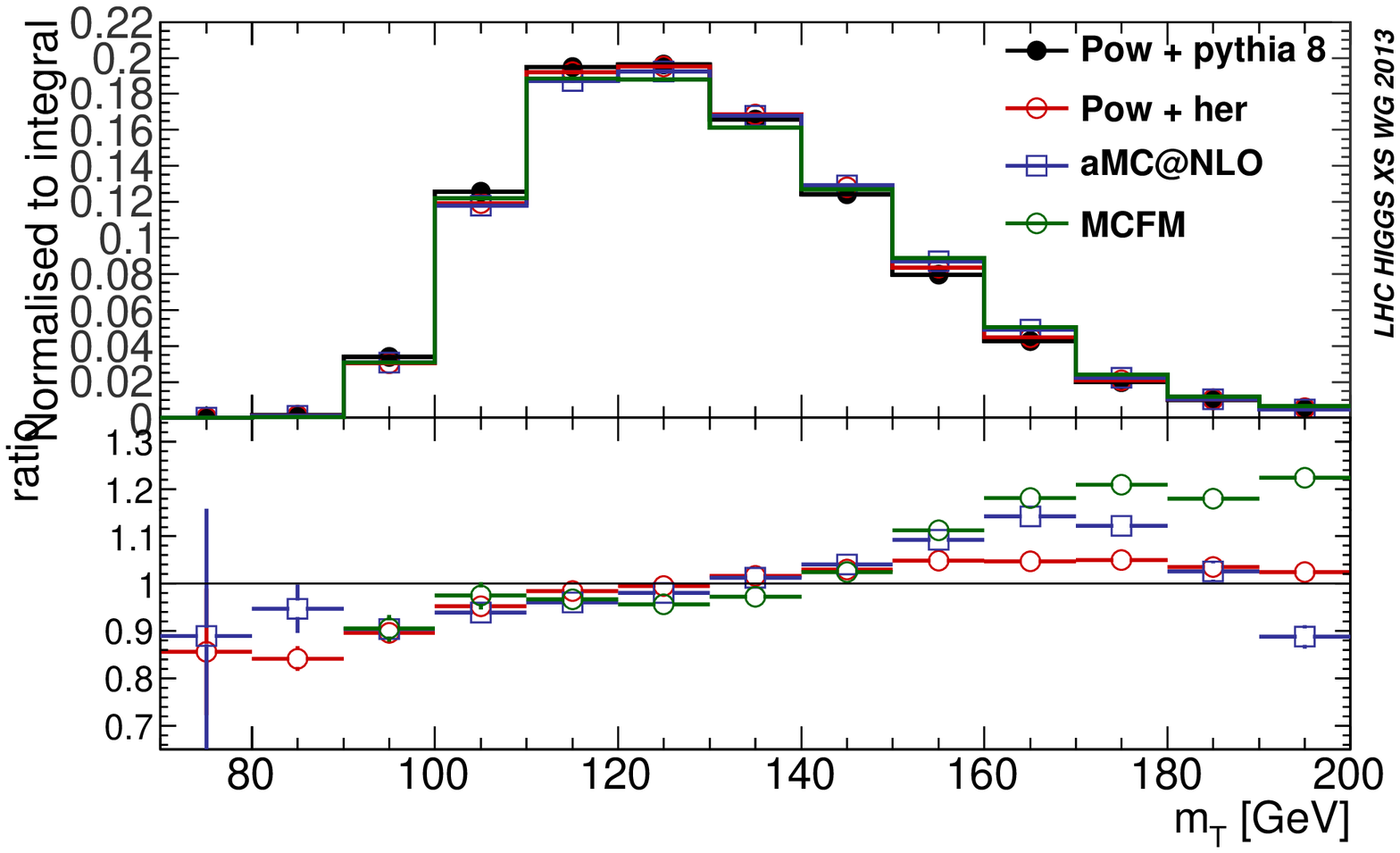}
  \includegraphics[width=0.475\textwidth]{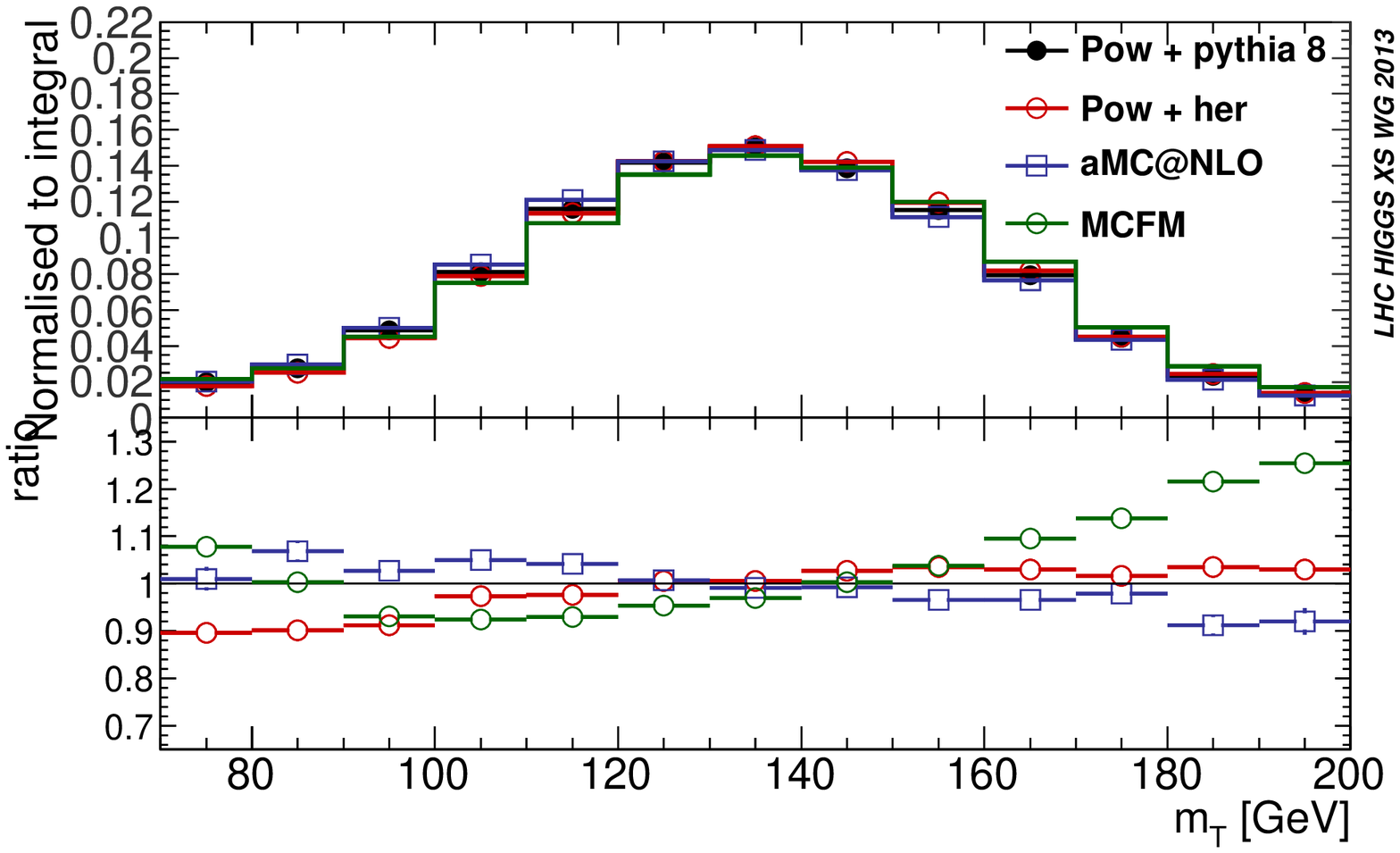}
  \caption{Parton showering/modelling uncertainties in the $m_{\mathrm T}$ distribution in (top left) SR1 
    in DF 0j analysis, (top right) SR2 in DF 0j analysis, (middle left) SR1 in DF 1j analysis, 
    (middle right) SR2 in DF 1j analysis, (bottom left) SR in SF 0j analysis, and (bottom 
    right) SR in SF 1j analysis.  \textsc{powheg} + \textsc{pythia}{}8/\textsc{Herwig} is shown along with \textsc{mcfm}. 
    All ratios are with respect to \textsc{powheg} + \textsc{pythia}{}8.}
  \label{fig:ww_mT_psueunc}
\end{figure}

\begin{figure}[h!]
  \centering
  \includegraphics[width=0.475\textwidth]{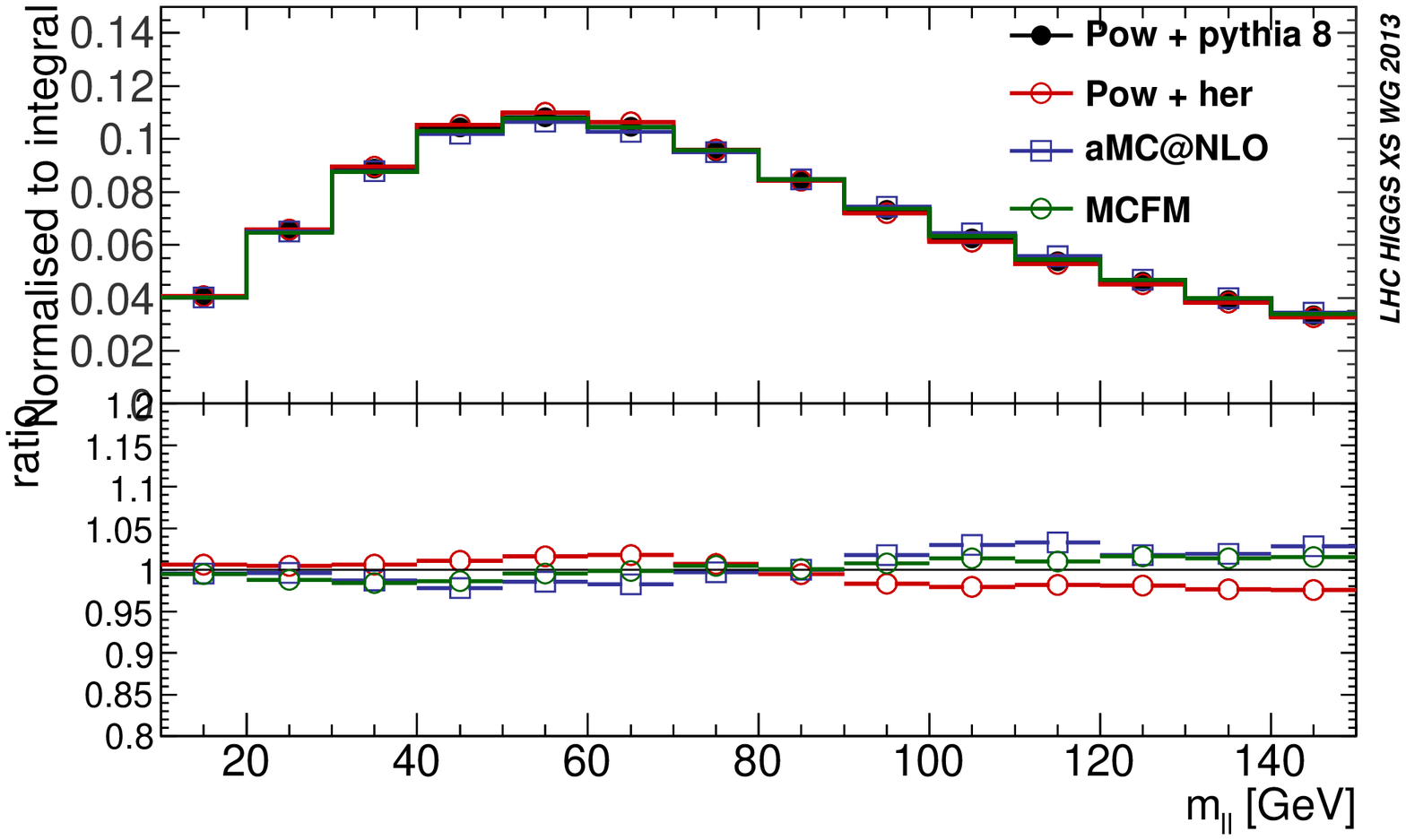}
  \includegraphics[width=0.475\textwidth]{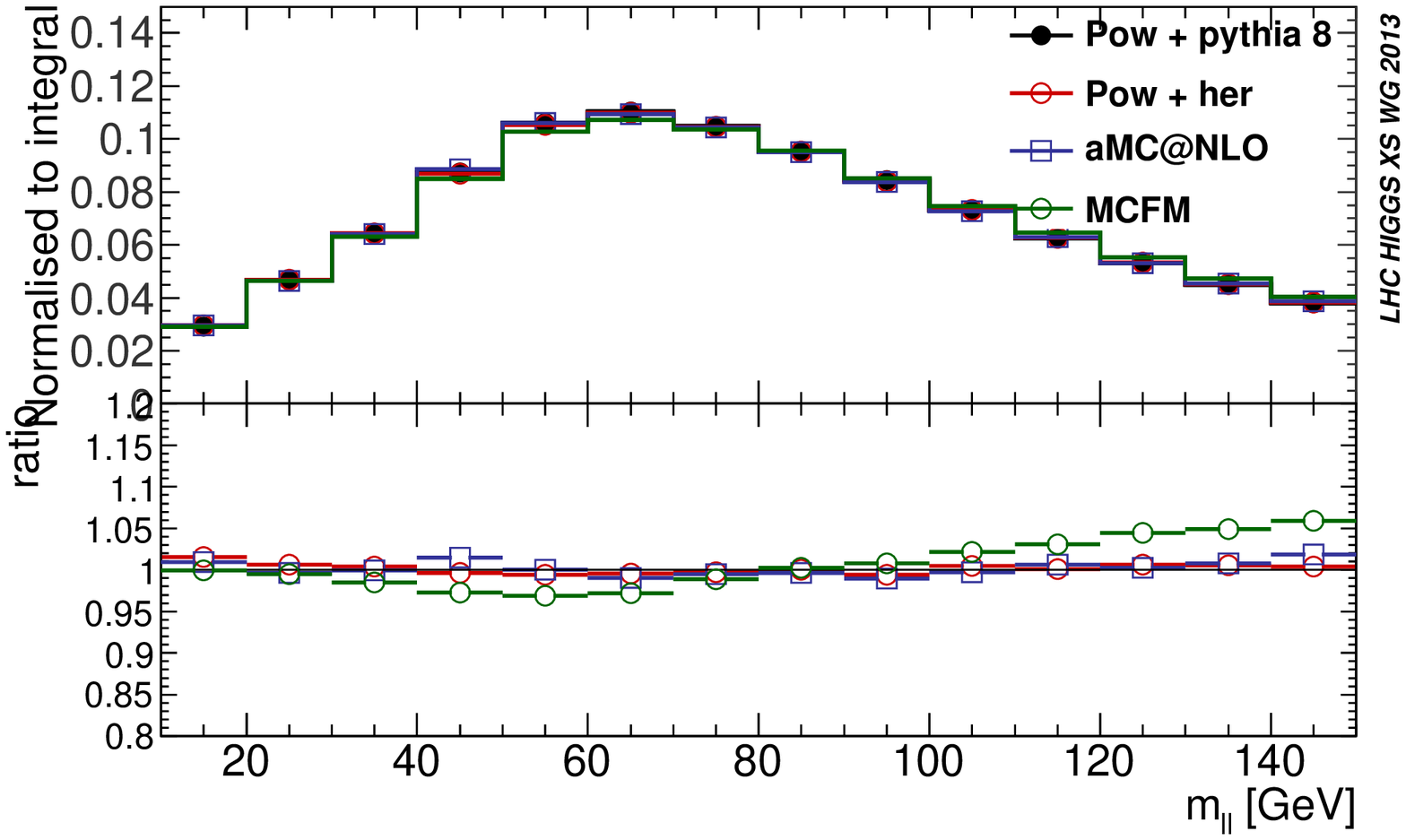}
  \includegraphics[width=0.475\textwidth]{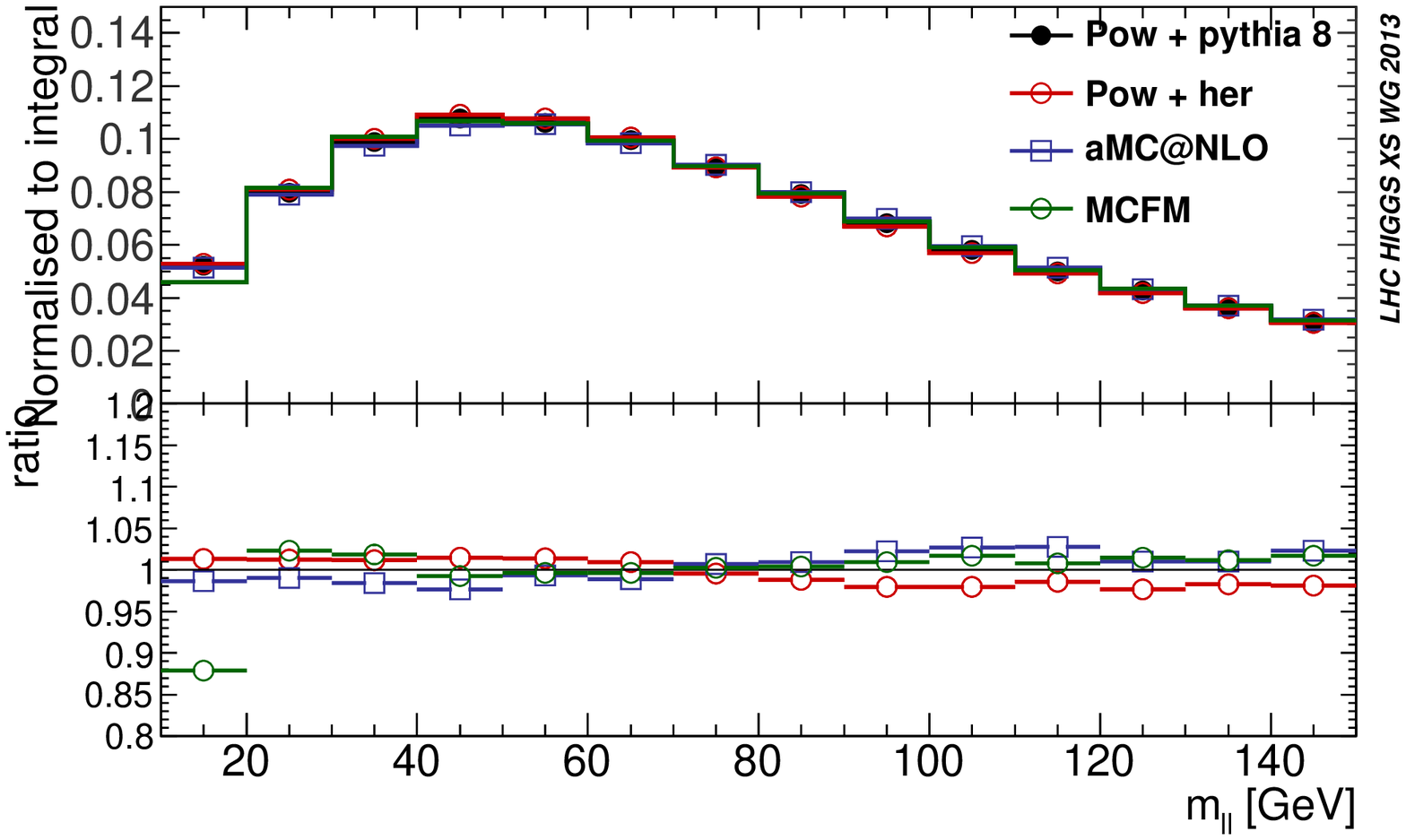}
  \includegraphics[width=0.475\textwidth]{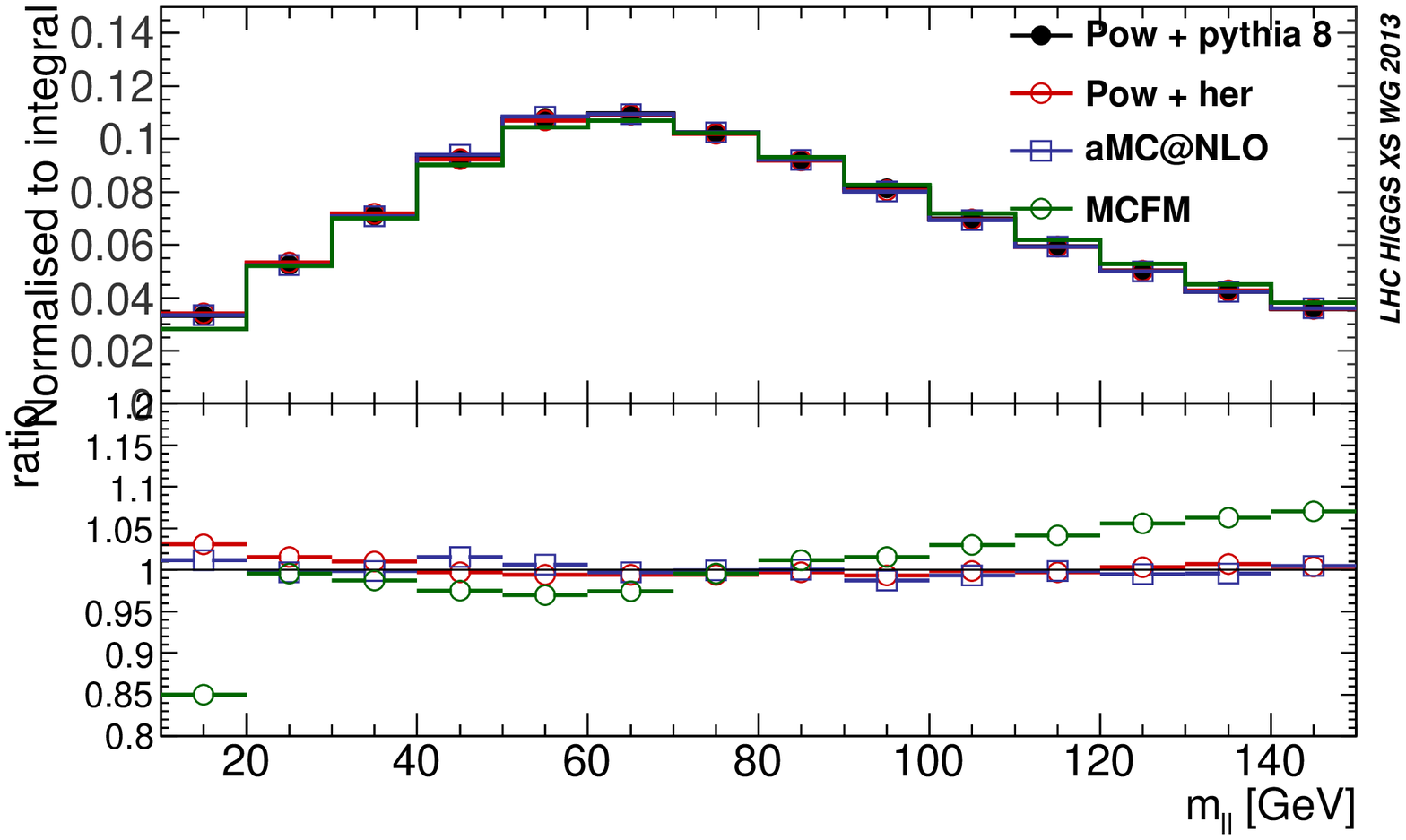}
  \caption{Parton showering/modelling uncertainties in the $m_{\Pl\Pl}$ distribution in (top left) SR1/2 
    in DF 0j analysis, (top right) SR1/2 in DF 1j analysis, (bottom left) SR in SF 0j analysis, and (bottom 
    right) SR in SF 1j analysis.  \textsc{powheg} + \textsc{pythia}{}8/\textsc{Herwig} is shown along with \textsc{mcfm}.
    All ratios are with respect to \textsc{powheg} + \textsc{pythia}{}8.}
  \label{fig:ww_mll_psueunc}
\end{figure}

\begin{table}
  \begin{center}
  \caption{The $\PW\PW$ extrapolation parameters, $\alpha$, calculated with respect to the 
    validation region using \textsc{powheg + Herwig} and \textsc{aMC@NLO}.  The 0 jet and 1 jet, 
    different flavour (DF) values are each calculated. }
  \label{tab:WW_powheg_amcatnlo_comparison_validation}
    \begin{tabular}{lccc}
      \hline
      & POWHEG + Herwig & aMC@NLO & $\delta\alpha$ \\
      \hline
      $\alpha^{\mathrm{DF}}_{\mathrm{0j}}$ (VR) &  0.9177 & 0.9738 & 6.1\% \\
      \hline
    \end{tabular}
  \end{center}
\end{table}
The uncertainty in the shape of the $m_{\mathrm T}$ distribution in the signal region arising from 
higher order corrections was estimated by varying the renormalisation and factorisation 
scales in MCFM, in a similar fashion to that used when calculating the uncertainty in 
$\alpha$. In addition, the uncertainty in the shape of the \mT~distribution due to the 
underlying event and parton showering is evaluated by comparing events generated with 
\textsc{Powheg} and showered with \textsc{pythia}{}8 to those same events showered with \textsc{herwig}.
These shape systematics are shown in Figures~\ref{fig:ww_mT_scaleunc},  
~\ref{fig:ww_mT_psueunc}, and ~\ref{fig:ww_mll_psueunc}. The largest
systematics are at low $m_{\mathrm T}$ and at high $m_{\mathrm T}$, on the tails of the distribution, while the core
part and in particular the region around the Higgs mass $m = 125 \UGeV$ is not affected by large systematics.

Various additional generators have been used to investigate the modelling of the $\PW\PW$ background,
Events generated with \textsc{POWHEG} are compared to those generated with Sherpa, \textsc{MC@NLO} + \textsc{Herwig}, and \textsc{MCFM}.
Between those generators, some differences are existing.
The generators have the following features:
\begin{itemize}
\item \textsc{Powheg}: NLO calculation matched to Sudakov factor for first emission
\item \textsc{mcfm}: NLO calculation with no parton shower
\item \textsc{mc@nlo} + \textsc{Herwig}: NLO calculation with parton shower but no
  ``singly resonant'' diagrams
\item Sherpa: LO calculation with parton shower
\end{itemize}
These studies offer insight into the effect of including NLO contributions, 
single-resonant diagrams, or the parton shower in the model.

\begin{figure}[h!]
  \begin{center}
    \includegraphics[width=0.475\textwidth]{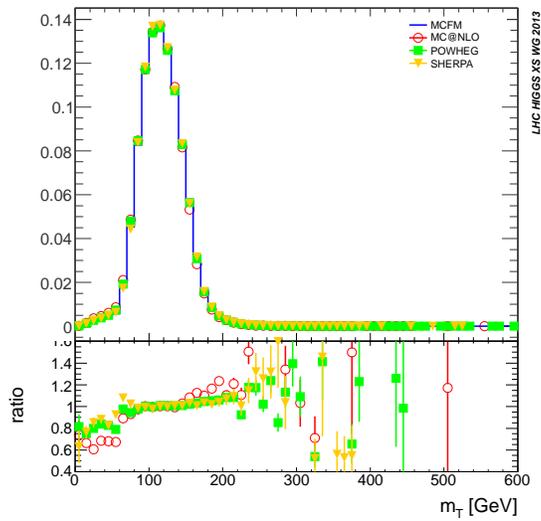}
  \end{center}
%%   \caption{Distributions of $m_{\mathrm T}$ in the signal region (left) and $\Delta \phi(ll)$ in the
%%     control region (right), for the four generators considered.}
  \caption{Distributions of $m_{\mathrm T}$ in the signal region for the four generators considered.}
  \label{fig:mtsr_dphicr}
\end{figure}

\noindent
The events are generated at a center of mass energy of 7 TeV using CTEQ 6.6 parton
distribution functions (CTEQ 6.1 for Sherpa).  
Figure~\ref{fig:mtsr_dphicr} shows the $m_{\mathrm T}$ distribution in the signal region.
%%  and
%% the $\Delta\phi (ll)$ distribution in the control region.

The ratio of the \textsc{mc@nlo} to \textsc{mcfm} $m_{\mathrm T}$ distributions is taken as an uncertainty
in the final $m_{\mathrm T}$ shape fit (the uncertainty is symmetrized).  
\clearpage
\subsubsection{Parton shower modelling}

\begin{figure}[h!]
  \includegraphics[width=0.9\textwidth]{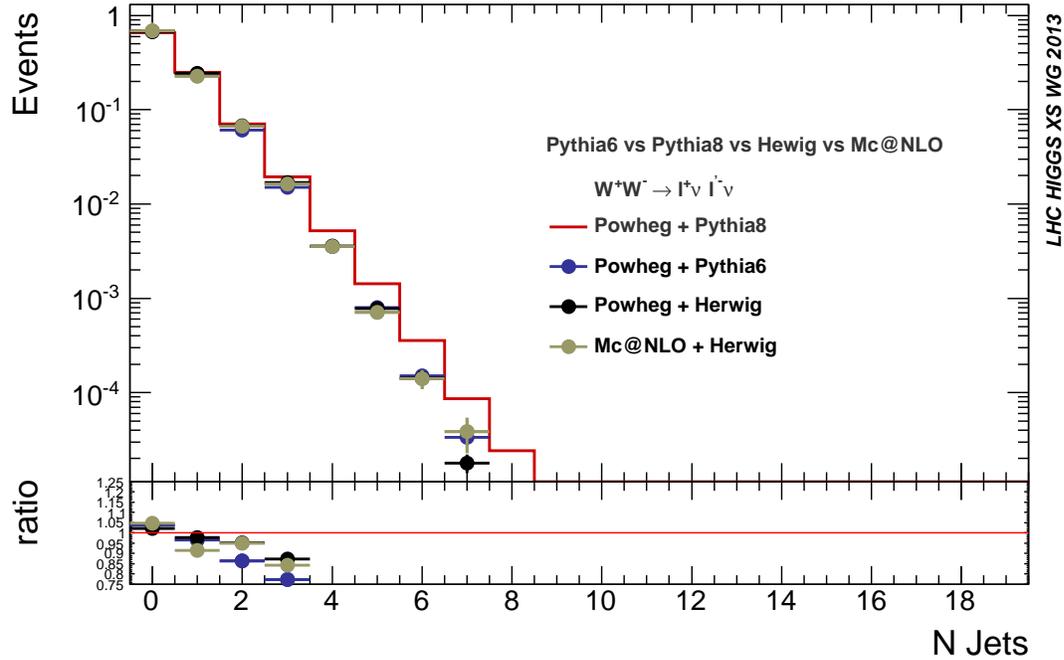}
  \caption{The jet multiplicity distributions for the four generators considered. }
  \label{fig:jetmult}
\end{figure}

Parton showering effect is studied by comparing the measured $\alpha$ values 
(eqn.~\ref{eqn:alpha}) when using the nominal matrix element event
generator, \textsc{PowHeg}, interfaced with either \textsc{pythia}{}8, \textsc{pythia}{}6, or
\textsc{Herwig}.
One million events were generated with \textsc{PowHeg} in each of the
$\PW\PW\rightarrow \Pe\Pe/\PGm\PGm/\Pe\PGm/\PGm \Pe$ final states, yielding a total of 4 million events.
The $\alpha$ values are computed in both the same flavor and different flavor analyses
separately for 0, 1, and $\ge 2$ jet events.  The jet multiplicity distributions after
preselection for the different generators are shown in Fig.~\ref{fig:jetmult}.

\Tref{tab:alpha_shower} shows the measured alpha values for each showering
program split into the two jet bins for both the same flavor and opposite flavor
analyses.  \Tref{tab:alpha_shower_diff} shows the relative difference
between each showering program and the nominal, \textsc{pythia}{}8. In general,
\textsc{pythia}{}{}8 and \textsc{pythia}{}6 are consistent with each other.  
The measured systematic uncertainties for choice of parton showering is determined from
the difference between \textsc{pythia}{}8 and \textsc{Herwig}.  For both analyses, the
observed shift is about 4.5\% for the 0-jet bin, while for the other jet bins, the
observed shift is consistent with zero (except for the $\ge 2$-jet bin in the
opposite flavor analysis).  We conservatively take 4.5\% as a correlated uncertainty
for both 0-jet and 1-jet bins, since further study is required to appropriately assess
the uncertainty in the 1-jet bin.

\begin{table}[htbp]
  \begin{center}
  \caption{The $\alpha_{\PW\PW}$ values for \textsc{PowHeg} generated events showered with
    \textsc{pythia}{}6, \textsc{pythia}{}8, or \textsc{Herwig} are shown.  The values are shown both
    for same flavor and opposite flavor analyses in either the 0, 1, or $\ge$ 2 jet-bins.
    For reference, the $\alpha$ value is also shown in events generated with \textsc{mc@nlo}
    and showered with \textsc{Herwig}.}
  \label{tab:alpha_shower}
    \makebox[\linewidth]{
      \begin{tabular}{llccc}
        \hline
            {SR vs CR} & UEPS & \multicolumn{2}{c}{DF} &   SF \\
            &      &                                  SR1   & SR2   &    \\
            \hline
            \multirow{3}{*}{0-jet} & \textsc{PowHeg}+\textsc{pythia}{}6 & $0.2275 \pm 0.0006$ &$0.3877 \pm 0.0008$ & $0.4602 \pm 0.0009$\\
            & \textsc{PowHeg}+\textsc{pythia}{}8 & $0.2277 \pm 0.0003$ &$0.3883 \pm 0.0004$ & $0.4609 \pm 0.0004$ \\
            & \textsc{PowHeg}+\textsc{Herwig}  & $0.2277 \pm 0.0003$ &$0.3914 \pm 0.0004$ & $0.4623 \pm 0.0004$ \\
            & \textsc{mc@nlo}+\textsc{Herwig}  & $0.229 \pm 0.003$   &$0.385 \pm 0.004$   & $0.4597 \pm 0.0043$ \\
            \hline                                                                                       
            \multirow{3}{*}{1-jet} & \textsc{PowHeg}+\textsc{pythia}{}6 & $0.1098 \pm 0.0004$ &$0.1895 \pm 0.0006$ & $0.2232 \pm 0.0007$ \\
            & \textsc{PowHeg}+\textsc{pythia}{}8 & $0.1107 \pm 0.0002$ &$0.1895 \pm 0.0003$ & $0.2235 \pm 0.0003$ \\
            & \textsc{PowHeg}+\textsc{Herwig}  & $0.1113 \pm 0.0002$ &$0.1904 \pm 0.0003$ & $0.2247 \pm 0.0003$ \\
            & \textsc{mc@nlo}+\textsc{Herwig}  & $0.1071 \pm 0.0021$ &$0.179 \pm 0.003$   & $0.211 \pm 0.003$ \\
            \hline
      \end{tabular}
    }
  \end{center}
\end{table}

\begin{table}
  \begin{center}
  \caption{Systematic uncertainty between nominal (\textsc{PowHeg} + \textsc{pythia}{}8) vs. the
    other PDF showering tools and \textsc{mc@nlo}+\textsc{Herwig}.  The systematic uncertainty is
    computed by subtracting \textsc{pythia}{}6 or \textsc{Herwig} from \textsc{pythia}{}8 and then dividing
    the difference by \textsc{pythia}{}8.}
  \label{tab:alpha_shower_diff}
    \makebox[\linewidth]{
      \begin{tabular}{llcccc}
        \hline
            {SR vs CR} & UEPS & \multicolumn{2}{c}{DF} &   SF \\
            &      &                                  SR1   & SR2   &          \\
            \hline
            \multirow{3}{*}{0-jet} & \textsc{PowHeg}+\textsc{pythia}{}6 &  $0.1   \pm 0.3 \%$  & $0.2  \pm 0.2 \%$   & $0.14 \pm 0.2 \%$\\
            & \textsc{PowHeg}+\textsc{Herwig}  &  $-0.02 \pm 0.2 \%$  & $-0.8 \pm 0.1 \%$   & $-0.3 \pm 0.1 \%$\\
            & \textsc{mc@nlo}+\textsc{Herwig}  &  $-0.3  \pm 1.2 \%$  & $0.8 \pm 1.0 \%$    &  $0.3 \pm 0.9 \%$\\
            \hline
            \multirow{2}{*}{1-jet} & \textsc{PowHeg}+\textsc{pythia}{}6 & $0.8 \pm 0.4 \%$  & $-0.1 \pm 0.4 \%$ &  $0.1 \pm 0.3 \%$\\
            & \textsc{PowHeg}+\textsc{Herwig} &  $-0.5 \pm 0.3 \%$ & $-0.5 \pm 0.2 \%$ &  $-0.6 \pm 0.2 \%$\\
            & \textsc{mc@nlo}+\textsc{Herwig} &  $3.2 \pm 2.0 \%$  & $5.8 \pm 1.6 \%$  &  $5.6 \pm 1.5 \%$\\
            \hline
      \end{tabular}
    }
  \end{center}
\end{table}

\subsubsection{Summary}

The theoretical uncertainties on the normalization in the signal and validation regions are 
summarized in Tables~\ref{tab:scale_and_pdfs} and~\ref{tab:alpha_validation_uncertainties}.  
The modelling uncertainty has been checked with aMC@NLO; the differences between \textsc{Powheg} 
and aMC@NLO are generally smaller than, or the same within uncertainties, those between 
\textsc{Powheg} and \textsc{mcfm}.

\begin{table}
  \begin{center}
  \caption{Scale, PDF, parton-shower/underlying event, and modelling uncertainties on the $\PW\PW$
    extrapolation parameters $\alpha$ for the NLO $\PAQq\PQq,\PQq\Pg \to \PW\PW$ processes; the errors are taken
    to be fully correlated between the 0-jet and 1-jet bins.  The correlations in the parton-shower
    and modelling variations are shown explicitly by including the signed difference in the comparison.}
  \label{tab:scale_and_pdfs}
    \begin{tabular}{lcccc}
      \hline
      & Scale & Parton-shower & PDFs & Modelling  \\
      \hline
      $\alpha^{\mathrm{DF}}_{\mathrm{0j}}$ (SR1) & 0.9\% & +0.2 \% & 1.5\% & -1.2\% \\
      $\alpha^{\mathrm{DF}}_{\mathrm{0j}}$ (SR2) & 0.9\% & +0.8\% & 1.1\% & -1.4\% \\
      $\alpha^{\mathrm{SF}}_{\mathrm{0j}}$       & 1.0\% & +0.3\% & 1.1\% & 1.7\% \\
      $\alpha^{\mathrm{DF}}_{\mathrm{1j}}$ (SR1) & 1.6\% & +0.5\% & 2.0\% & -5.1\% \\
      $\alpha^{\mathrm{DF}}_{\mathrm{1j}}$ (SR2) & 1.5\% & +0.5\% & 1.8\% & -5.0\% \\
      $\alpha^{\mathrm{SF}}_{\mathrm{1j}}$       & 1.4\% & +0.6\% & 1.7\% & -3.1\% \\
      \hline
      $\alpha^{\mathrm{0j}}_{\mathrm{WW}}$, $\alpha^{\mathrm{1j}}_{\mathrm{WW}}$ correlation & \multicolumn{4}{c}{1} \\
      \hline
    \end{tabular}
  \end{center}
\end{table}

\begin{table}
  \begin{center}
  \caption{Scale, PDF, parton-shower/underlying event, and modelling uncertainties on the $\PW\PW$
    extrapolation parameters $\alpha$ calculated relative to the validation region
    for the NLO $\PQq\PQq,\PQq\Pg \to \PW\PW$ processes.}
  \label{tab:alpha_validation_uncertainties}
    \begin{tabular}{lcccc}
      \hline
      & scale & parton-shower & PDFs & modelling  \\
      \hline
      $\alpha^{\mathrm{DF}}_{\mathrm{0j}}$ (VR) & 1.0\% & 2.6\%  & 2.2\% & 2.0\% \\
      \hline
    \end{tabular}
  \end{center}
\end{table}

\clearpage

\newpage
\providecommand{\muR}{\mathswitch {\mu_{\mathrm{R}}}}
\providecommand{\muF}{\mathswitch {\mu_{\mathrm{F}}}}
\providecommand\HAWK{{\sc HAWK}}
\providecommand\VBFNLO{{\sc VBFNLO}}
\providecommand\POWHEG{{\sc POWHEG}}
\providecommand\POWHEGBOX{{\sc POWHEG BOX}}
\providecommand\HERWIG{{\sc HERWIG}}
\providecommand\PYTHIA{{\sc PYTHIA}}
\providecommand{\kT}{\ensuremath{k\sb{\scriptstyle\mathrm{T}}}}

\section{VBF production mode\footnote{%
    A.~Denner, P.~Govoni, C.~Oleari, D.~Rebuzzi (eds.);
    P.~Bolzoni, S.~Dittmaier, S.~Frixione,
         %B.~Quayle, 
    F.~Maltoni, C.~Mariotti, \mbox{S.-O.~Moch},
    A.~M\"uck, P.~Nason, 
         % G.~Passarino, 
    M.~Rauch, 
         % R.~Tanaka, 
    P.~Torrielli, 
         %G.~Steele,
    M.~Zaro.}} 
\label{sec:VBF}

\subsection{Programs and Tools} 
\label{sec:programs-sub}

\subsubsection{HAWK}
\label{sec:HAWK-sub-sub}

\HAWK{} is a parton-level event generator for Higgs production in
vector-boson fusion~\cite{Ciccolini:2007jr, Ciccolini:2007ec,HAWK},
$\Pp\Pp\to\PH jj$, and Higgs-strahlung \cite{Denner:2011id},
$\Pp\Pp\to\PW\PH\to\PGn_{\Pl}\,\Pl\,\PH$ and
$\Pp\Pp\to\PZ\PH\to\Pl^-\Pl^+\PH/\PGn_{\Pl}\PAGn_{\Pl}\PH$. Here we
summarize its most important features for the VBF channel. \HAWK{}
includes the complete NLO QCD and EW corrections and all weak-boson
fusion and quark--antiquark annihilation diagrams, i.e.~$t$-channel
and $u$-channel diagrams with VBF-like vector-boson exchange and
$s$-channel Higgs-strahlung diagrams with hadronic weak-boson decay,
as well as all interferences. External fermion masses are neglected
and the renormalization and factorization scales are set to $\MW$ by
default. Recently also anomalous Higgs-boson--vector-boson couplings
have been implemented.

For the results presented below, $s$-channel contributions have been
switched off.  These contributions are at the level of a few per mille
once VBF cuts are applied \cite{Ciccolini:2007ec}. Interferences
between $t$- and $u$-channel diagrams are included at LO and NLO,
while contributions of $\PQb$-quark parton distribution functions
(PDFs) and final-state $\PQb$ quarks, which can amount to about $2\%$,
are taken into account at LO.  Contributions from photon-induced
processes, which are part of the real EW corrections and at the level
of $1{-}2\%$, can be calculated by HAWK as well, but have not been
included, since photon PDFs are not supported by the PDF sets used.
\HAWK{} allows for an on-shell Higgs boson or for an off-shell Higgs
boson (with optional decay into a pair of gauge singlets). For an
off-shell Higgs boson, besides the fixed-width scheme [see
\Eref{eq:VBF_BW_PWG_fix}] the complex-pole scheme (CPS)
\cite{Passarino:2010qk,Goria:2011wa,Anastasiou:2011pi} is supported
with a Higgs propagator of the form
\begin{equation} 
\label{eq:VBF_BW}
\frac{1}{\pi} \frac{M \, \GH(M)} {\left(M^2 -
\MH^2\right)^2 + ( \MH\GH )^2 },
\end{equation} 
where $\MH$ and $\GH$ are the Higgs mass and width in the complex-mass
scheme and $\GH(M)$ is the Higgs width for the off-shell mass $M$ as taken
from the tables in \Bref{Dittmaier:2011ti}.

\subsubsection{VBFNLO}
\label{sec:VBFNLO-sub-sub}

\VBFNLO{} \cite{Arnold:2012xn,Arnold:2011wj,Arnold:2008rz} is a
parton-level Monte Carlo event generator which can simulate vector-boson
fusion, double and triple vector-boson production in hadronic collisions
at next-to-leading order in the strong coupling constant, as well as
Higgs-boson plus two jets and double vector-boson production via gluon
fusion at the one-loop level. 
For Higgs-boson production via VBF, both the NLO QCD and EW
corrections to the $t$-channel can be included in the SM (and also in
the (complex) MSSM), and the NLO QCD corrections are included for
anomalous couplings between the Higgs and a pair of vector bosons.

\VBFNLO{} can also simulate the Higgs decays $\PH \rightarrow
\gamma\gamma, \PGmp \PGmm, \PGtp \PGtm, \PQb \overline{\PQb}$ in the
narrow-width approximation, either calculating the appropriate branching
ratios internally at LO (or, in the case of the MSSM not considered in
this section, taking them from an input SLHA file).
The Higgs-boson decays $\PH \rightarrow \PW^{+} \PW^{-} \rightarrow
\Pl^{+} \PGnl \Pl^{-} \PAGn_{\Pl}$ and $\PH \rightarrow \PZ\PZ
\rightarrow \Pl^{+} \Pl^{-} \Pl^{+} \Pl^{-}, \Pl^{+} \Pl^{-} \PGnl
\overline{\PGnl}$ are calculated using a Breit--Wigner distribution for
the Higgs boson and the full LO matrix element for the $\PH \rightarrow
4f$ decay. Initial- and final-state $\PQb$ quarks can be included at NLO
QCD for the neutral-current diagrams, where no external top quarks
appear. For PDF sets which support photon PDFs, their effects are
automatically included as well. As the used PDF sets do not include
photon PDFs, this does however not play any role here.
Interference effects between VBF-Higgs and continuum production in
leptonic or semi-leptonic $\PW^{+} \PW^{-}$ or $\PZ\PZ$ VBF processes
are contained in VBFNLO as part of the di-boson VBF processes.

For the results presented here, a modified version of \VBFNLO{} was used,
that simulates off-shell Higgs boson using the complex-pole
scheme (see Eq.~\ref{eq:VBF_BW}).

\subsubsection{POWHEG}
\label{sec:POWHEG-sub-sub}
The \POWHEG{} method is a prescription for interfacing NLO calculations with
parton-shower generators, like \HERWIG{} and \PYTHIA{}. It was first
presented in~\Bref{Nason:2004rx} and was described in great detail
in~\Bref{Frixione:2007vw}.  In~\Bref{Nason:2009ai}, Higgs-boson production in
VBF has been implemented in the framework of
the \POWHEGBOX{}~\cite{Alioli:2010xd}, a computer framework that implements
in practice the theoretical construction of~\Bref{Frixione:2007vw}.

All the details of the implementation can be found
in~\Bref{Nason:2009ai}. Here we briefly recall that, in the
calculation of the partonic matrix elements, all partons have been
treated as massless.  This gives rise to a different treatment of
quark flavors for diagrams where a $\PZ$ boson or a $\PW$ boson is
exchanged in the $t$-channel.  In fact, for all $\PZ$-exchange
contributions, the $\PQb$ quark is included as an initial and/or
final-state massless parton.  For the (dominant) $\PW$-exchange
contributions, no initial $\PQb$ quark has been considered, since it
would have produced mostly a $\PQt$ quark in the final state, which would
have been misleadingly treated as massless.  The
 Cabibbo--Kobayashi--Maskawa~(CKM) matrix $V_{\rm \scriptscriptstyle
  CKM}$ elements can be assigned by the user. The default setting used to
  compute the results in this section is
\begin{equation}
\begin{array}{c}
\\
V_{\rm  \scriptscriptstyle CKM}=
\end{array}
\begin{array}{c c}
& \PQd\quad\quad\quad \ \PQs\ \quad \quad\quad \PQb\ \\
\begin{array}{c}
\PQu\\
\PQc\\
\PQt
\end{array} 
&
\left(
\begin{array}{c c c}
$0.9748$ & $0.2225$ & $0.0036$\\
$0.2225$ & $0.9740$ & $0.041$ \\
$0.009$ & $0.0405$ & $0.9992$
\end{array}
\right).
\end{array}
\end{equation}
We point out that, as long as no final-state hadronic flavor is
tagged, this is practically equivalent to the result obtained using
the identity matrix, due to unitarity.

The Higgs-boson virtuality $M^2$ can be generated according to three
different schemes:
\begin{enumerate}
\item the fixed-width scheme
\begin{equation}
\label{eq:VBF_BW_PWG_fix}
\frac{1}{\pi} \frac{\MH \GH}{\left(M^2 - \MH^2\right)^2 + \MH^2 \GH^2  }\,,
\end{equation}
 with fixed decay  width $\GH$,
\item the running-width scheme
\begin{equation}
\label{eq:VBF_BW_PWG_run}
\frac{1}{\pi} \frac{M^2 \,\GH
  / \MH}{\left(M^2 - \MH^2\right)^2 + ( M^2\, \GH / \MH)^2 }\,,
\end{equation}

\item the complex-pole scheme, as given in \Eref{eq:VBF_BW}.

\end{enumerate}
As a final remark, we recall that the renormalization $\muR$ and
factorization $\muF$ scales used in the calculation of the \POWHEG{}
$\bar{B}$ function (i.e.~the function that is used to generate the underlying
Born variables to which the \POWHEGBOX{} attaches the radiation ones) are
arbitrary and can be set by the user.  For the results in this section, they
have been taken equal to $\MW$.  The renormalization scale for the evaluation
of the strong coupling associated to the radiated parton is set to its
transverse momentum, as the \POWHEG{} method requires. The transverse
momentum of the radiated parton is taken, in the case of initial-state
radiation, as exactly equal to the transverse momentum of the parton with
respect to the beam axis. For final-state radiation one takes instead
\begin{equation}
\pT^2=2E^2(1-\cos\theta),
\end{equation}
where $E$ is the energy of the radiated parton and $\theta$ the angle it
forms with respect to the final-state parton that has emitted it, both taken
in the partonic centre-of-mass frame.

\subsubsection{VBF@NNLO}
\label{sec:VBFNNLO-sub-sub}

{\sc VBF@NNLO}~\cite{Bolzoni:2010xr,Bolzoni:2011cu} computes VBF total Higgs cross sections 
at LO, NLO, and NNLO in QCD via the structure-function approach.  This
approach~\cite{Han:1992hr} consists  in considering VBF process as a
double deep-inelastic scattering (DIS) attached to the colorless pure
electroweak vector-boson fusion into a Higgs boson.  According to this
approach one can include NLO QCD corrections to the VBF process employing the
standard DIS structure functions $F_i(x,Q^2);\,i=1,2,3$ at
NLO~\cite{Bardeen:1978yd} or similarly the corresponding structure functions 
at NNLO~\cite{Kazakov:1990fu,Zijlstra:1992kj,Zijlstra:1992qd,Moch:1999eb}.

The effective factorization underlying the structure-function approach does not include all types of contributions.
At LO an additional contribution arises from the
interferences between identical final-state quarks (e.g.,\
$\PQu\PQu\rightarrow \PH\PQu\PQu$) or between processes where either a $\PW$
or a $\PZ$ can be exchanged (e.g.,\ $\PQu\PQd\rightarrow \PH\PQu\PQd$).  These
LO contributions have been added to the NNLO results presented here, even if they are very small. 
Apart from such contributions, the structure-function approach is exact also at NLO.  
At NNLO, however, several types of diagrams violate the underlying factorization.
Their impact on the total rate has been computed or estimated in~\Bref{Bolzoni:2011cu} and found to be negligible.
Some of them are color suppressed and kinematically
suppressed~\cite{vanNeerven:1984ak,Blumlein:1992eh,Figy:2007kv}, others have
been shown in \Bref{Harlander:2008xn} to be small enough not to produce
a significant deterioration of the VBF signal.

At NNLO QCD, the theoretical QCD uncertainty is reduced to less than
$2\%$. Electroweak corrections, which are at the level of $5\%$, are
not included in {\sc VBF@NNLO}.
The Higgs boson can either be produced on its mass-shell, or off-shell effects can be included in the 
complex-pole scheme.

\subsubsection{\textsc{aMC@NLO}}
\label{sec:aMCNLO-sub-sub}

The \textsc{aMC@NLO} generator \cite{Alwall:2013xx,website} is a program that
implements the matching of a generic NLO QCD computation to parton-shower
simulations according to the MC@NLO formalism \cite{Frixione:2002ik}; its
defining feature is that all ingredients of such matching and computation are
fully automated.  The program is developed within the \textsc{MadGraph5}
\cite{Alwall:2011uj} framework and, as such, it does not necessitate of any
coding by the user, the specification of the process and of its basic physics
features (such as particle masses or phase-space cuts) being the only external
informations required: the relevant computer codes are then generated
on-the-fly, and the only practical limitation is represented by CPU
availability.

\textsc{aMC@NLO} is based on different building blocks, each devoted to the
generation and evaluation of a specific contribution to an NLO-matched
computation. \textsc{MadFKS} \cite{Frederix:2009yq} deals with the Born and
real-emission terms, and in particular it performs in a general way the
analytical subtraction of the infrared singularities that appear in the
latter matrix elements according to the FKS prescription
\cite{Frixione:1995ms,Frixione:1997np}; moreover, it is also responsible for
the process-independent generation of the so-called Monte Carlo subtraction
terms, namely the contributions ensuring the avoidance of double-counting in
the MC@NLO cross sections; \textsc{MadLoop} \cite{Hirschi:2011pa} computes
the finite part of the virtual contributions, based on the OPP
\cite{Ossola:2006us} one-loop integrand-reduction method and on its
implementation in \textsc{CutTools} \cite{Ossola:2007ax}.

The first applications of \textsc{aMC@NLO}\footnote{These results were still
based on version 4 of \textsc{MadGraph} \cite{Alwall:2007st}.} have been
presented in \Brefs{Frederix:2011zi,Frederix:2011qg,Frederix:2011ss,
Frederix:2011ig,Frederix:2012dh,Frederix:2012dp},
and a novel technique for merging different multiplicities of NLO-matched
samples has been developed in the \textsc{aMC@NLO} environment and documented
in \Bref{Frederix:2012ps}.

As the MC@NLO method is Monte-Carlo dependent (through the Monte Carlo
subtraction terms), a different subtraction has to be performed for each
parton shower one wants to interface a computation to. So far this has been
achieved in \textsc{aMC@NLO} for the \textsc{HERWIG6} and \textsc{HERWIG++}
event generators (which amounts to the automation of the implementations
detailed in \Brefs{Frixione:2002ik,Frixione:2010ra}, respectively), 
and for the virtuality-ordered version of \textsc{PYTHIA6} (the proof of
concept of which was given in \Bref{Torrielli:2010aw}). The present
publication is the first in which \textsc{aMC@NLO} results are presented for
\textsc{PYTHIA6} in a process involving final-state radiation, and for
\textsc{HERWIG++}.

In the \textsc{aMC@NLO} framework, $t$-channel VBF can be generated with NLO 
accuracy in QCD.
%with the following commands in the \textsc{MadGraph5} command 
%interface:
%\begin{verbatim}
% import model loop_sm-no_b_mass
% define p = u u~ d d~ c c~ s s~ b b~ g
% define j = p
% generate p p > h j j $$ w+ w- z [QCD]
%\end{verbatim}
%where the first three lines are to import the NLO model with massless
%$\PQb$ quark and to include the latter in the definition of proton and jet,
%respectively. The '\texttt{\$\$}' sign forbids diagrams with either a
%W or a Z boson in the $s$-channel.  
A $V_{\tiny{\mbox{CKM}}}=1$ is
employed, as there is no flavor tagging in our analysis.  This
computation includes all interferences between $t$- and $u$-channel
diagrams, such as those occurring for same-flavor quark scattering
and for partonic channels that can be obtained by the exchange of
either a Z or a W boson (e.g. \texttt{u d > h u d}).  Only vertex
loop-corrections are included (the omitted loops are however totally
negligible \cite{Ciccolini:2007ec}).  The Les-Houches parton-level
event file thus generated also contains the necessary information for
the computation of scale and PDF uncertainties without the need of
extra runs \cite{Frederix:2011ss}.  The Higgs boson is considered as
stable.

\subsection{VBF Parameters and Cuts}
\label{sec:cuts-sub}

The numerical results presented in \refS{sec:results-sub} have been
computed using the values of the EW parameters given in Appendix A of
\Bref{Dittmaier:2011ti}. The electromagnetic coupling is fixed in the
$\GF$ scheme, to be
\beq
 \alpha_{\GF} = \sqrt{2}\GF\MW^2(1-\MW^2/\MZ^2)/\pi = 1/132.4528\ldots\,.
\eeq
The weak mixing angle is defined in the on-shell scheme,
\begin{equation}
\sin^2\theta_{\mathrm{w}}=1-\MW^2/\MZ^2=0.222645\ldots \,.
\end{equation}
The renormalization and factorization scales are set equal to the
$\PW$-boson mass,
\begin{equation}
\label{eq:VBF_ren_fac_scales}
\muR = \muF= \MW,
\end{equation}
and both scales are varied independently in the range $\MW/2 < \mu < 2\MW$.

In the calculation of the inclusive cross sections, we have used the
MSTW2008~\cite{Martin:2009iq}, CT10 \cite{Lai:2010vv}, and NNPDF2.1
\cite{Cerutti:2011au} PDFs, for the calculation of the differential
distributions we employed MSTW2008nlo PDFs~\cite{Martin:2009iq}.

Contributions from $s$-channel diagrams and corresponding
interferences have been neglected throughout.

For the differential distributions presented in
\refS{sec:differential-sub-sub} the following reconstruction scheme
and cuts have been applied.
Jets are constructed according to the anti-$\kT$ algorithm, with
the rapidity--azimuthal angle separation $\Delta R=0.5$, using the default
recombination scheme ($E$ scheme).  Jets are subject to the
cuts
\begin{equation}
\label{eq:VBF_cuts1}
{\pT}_j > 20\UGeV, \qquad  |y_j| < 4.5\,,
\end{equation}
where $y_j$ denotes the rapidity of the (massive) jet.
% of pseudorapidity $|\eta|<5$. 
Jets are ordered according to their $\pT$ in decreasing
progression.  The jet with highest $\pT$ is called leading jet, the
one with next highest $\pT$ subleading jet, and both are the tagging
jets.  Only events with at least two jets are kept.  They must satisfy
the additional constraints
\begin{equation}
\label{eq:VBF_cuts2}
|y_{j_1} - y_{j_2}| > 4\,, \qquad m_{jj} > 600\UGeV.
\end{equation}
%The Higgs boson is generated off shell
%according to \Erefs{eq:VBF_BW} or~\refE{eq:VBF_BW_PWG}, 
%and there are no cuts applied to its decay products.

%For the calculation of EW corrections, real photons are recombined
%with jets according to the same  anti-$\kT$ algorithm as used
%for jet recombination.

\subsection{Results}
\label{sec:results-sub}

\subsubsection{Inclusive Cross Sections with CPS}
\label{sec:inclusive-sub-sub}

In the following we present results for VBF inclusive cross sections
at $7\UTeV$ (for Higgs-boson masses below $300\UGeV$) and $8\UTeV$ (over the full Higgs-boson mass range)
calculated in the CPS, as described above. 

Tables \ref{tab:YRHXS3_7VBF_NNLO1}--\ref{tab:YRHXS3_7VBF_NNLO7} list the
VBF inclusive cross sections at $7\UTeV$ as a function of the
Higgs-boson mass, while Tables \ref{tab:YRHXS3_VBF_NNLO1}--\ref{tab:YRHXS3_VBF_NNLO7}
display results at $8\UTeV$.
The pure NNLO QCD results (second column), obtained
with {\sc VBF@NNLO}, the relative NLO EW corrections (fourth column),
obtained with {\HAWK}, and the combination of NNLO QCD and NLO EW
corrections (third column) are given, together with uncertainties from
PDF + $\Oas$ and from QCD scale variations.  The combination has been
performed under the assumption that QCD and EW corrections factorize
completely, i.e.\
\begin{equation}
\sigma^{\mathrm{NNLO}+\mathrm{EW}} =
\sigma^{\mathrm{NNLO}}_{\mathrm{VBF@NNLO}}
\times (1 + \delta^{\mathrm{EW}}_{\mathrm{HAWK}})\,,
\end{equation} 
where $\sigma^{\mathrm{NNLO}}_{\mathrm{VBF@NNLO}}$ is the NNLO QCD result and
$\delta^{\mathrm{EW}}_{\mathrm{HAWK}}$ the relative EW correction determined in the limit
$\alphas=0$.

\begin{figure}
\includegraphics[width=0.5\textwidth]{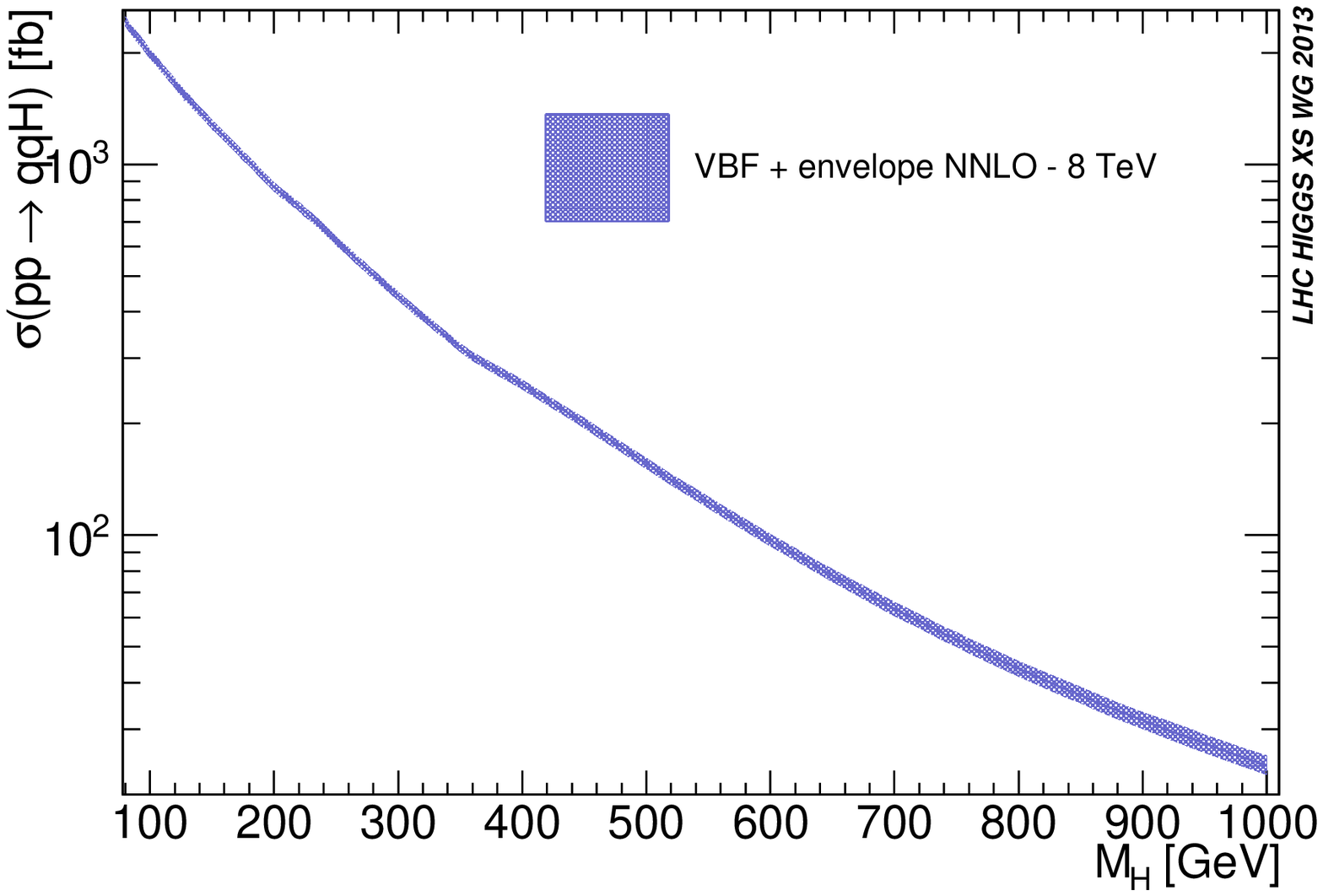}
\includegraphics[width=0.5\textwidth]{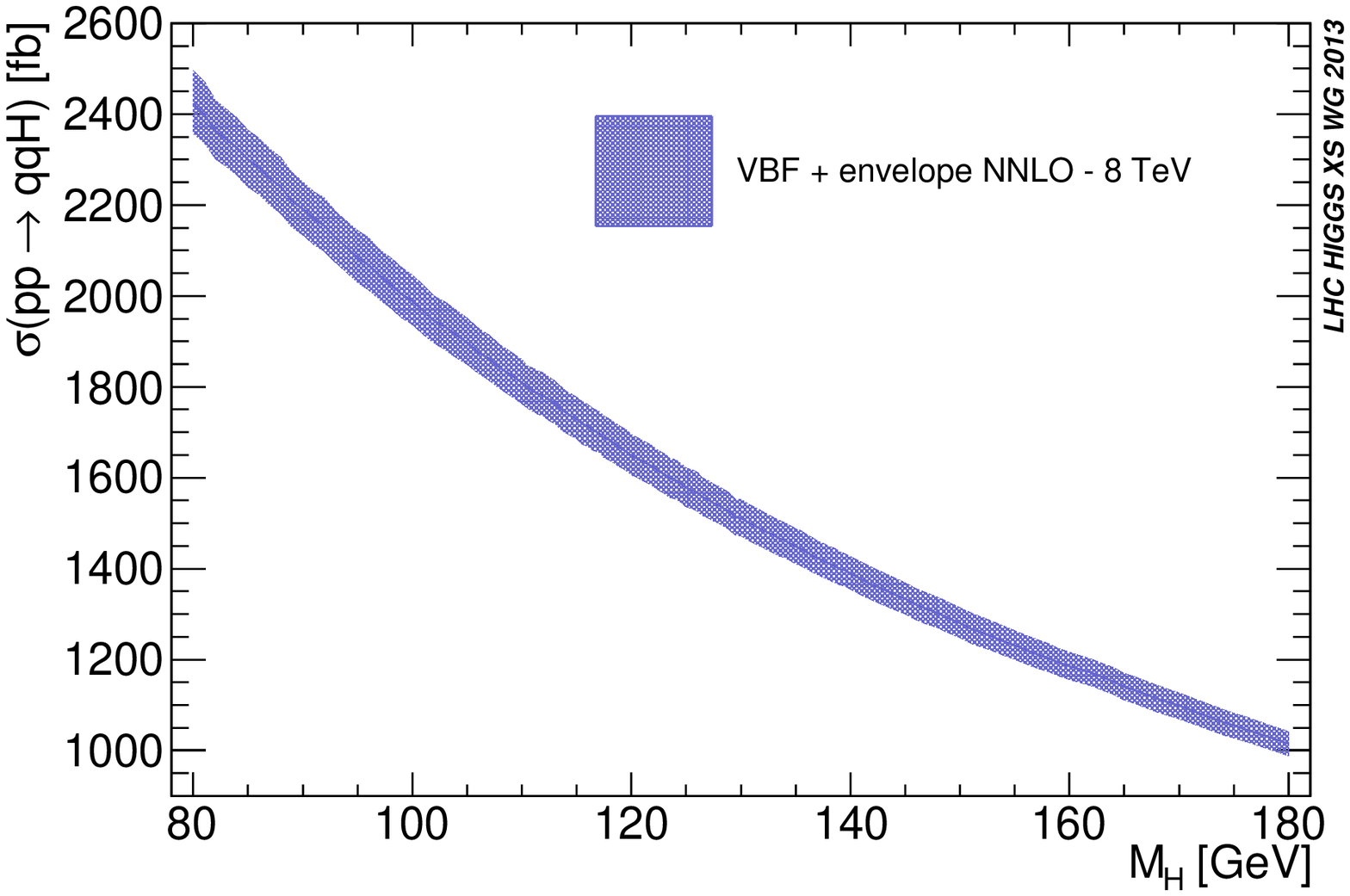}
%\vspace*{-0.3cm}
\caption{VBF inclusive cross sections at the LHC at $8\UTeV$ for the
  full (left) and low (right) Higgs-boson mass 
  range. The uncertainty band represents the PDF + $\Oas$ envelope
  estimated according to the PDF4LHC prescription \cite{Botje:2011sn}.}
\label{fig:YRHXS3_VBF_xsec}
\end{figure}

Figure \ref{fig:YRHXS3_VBF_xsec} summarizes the VBF results at
$8\UTeV$ for the full and for the low mass range. The inclusive cross
section including the combined NNLO QCD and NLO EW corrections are
shown as a function of the Higgs-boson mass. The band represents the
PDF + $\Oas$ uncertainties, calculated from the envelope of the three
PDF sets, CT10, MSTW2008, and NNPDF2.1, according to the PDF4LHC
prescription.

In \refT{tab:YRHXS3_VBF_NNLO8_13} we compare the predictions for the
Higgs cross section in VBF from different calculations at NLO QCD and
including EW corrections at $8$ and $13\UTeV$.  Using MSTW2008 PDFs,
results are presented at NLO from VBF@NNLO, HAWK, and VBFNLO.  The NLO
QCD corrections agree within 0.6\% between HAWK and VBF@NLO.  The NLO
EW corrections amount to $-5\%$ while the NNLO QCD corrections are
below 0.3\%.

All results are obtained in the CPS. Note that for Higgs-boson masses
in the range $120{-}130\UGeV$ the cross section calculated in the
complex-pole scheme is larger than the one for an on-shell Higgs boson
by $\sim1.4\%$ at $8\UTeV$ and $\sim2.2\%$ at $13\UTeV$. The
presented results in the CPS assume that the Higgs distribution is
integrated over all kinematically allowed invariant Higgs masses. If
cuts on the invariant mass are applied, the cross section decreases
down to the on-shell value.

%In Tables \ref{tab:YRHXS3_VBF_NNLO7}--\ref{tab:YRHXS3_VBF_NNLO12}
%comparisons among different NLO computations are given for $8\UTeV$. HAWK
%results are compared to VBFNLO and VBF@NNLO ones for the pure QCD part
%and for NLO QCD + EW corrections.

\subsubsection{Differential Distributions}
\label{sec:differential-sub-sub}

In this section we present results relevant to the production
of a $125\UGeV$ Standard Model Higgs boson through a VBF mechanism
at the $8\UTeV$ LHC. These have been obtained with
\textsc{aMC@NLO}~\cite{Frixione:2013mta} and 
\textsc{POWHEG}~\cite{Nason:2009ai,Alioli:2010xd}. As \textsc{aMC@NLO} and
\textsc{POWHEG} perform the NLO matching with parton showers by means of two
different prescriptions (see~\Brefs{Frixione:2002ik}
and~\cite{Nason:2004rx,Frixione:2007vw}, respectively), the comparison
of the two allows one to assess the matching systematics that affects
VBF Higgs production; it also constitutes a non-trivial validation 
of the public, fully-automated code \textsc{aMC@NLO}.
Our predictions are
obtained with the MSTW2008nlo68cl PDF set~\cite{Martin:2009iq}, and by 
setting the
renormalization and factorization scales equal to the $\PW$-boson mass $\MW$. Matching
with \textsc{HERWIG6}~\cite{Corcella:2000bw}, virtuality-ordered
\textsc{PYTHIA6}~\cite{Sjostrand:2006za}, and
\textsc{HERWIG++}~\cite{Bahr:2008pv} has been considered in both schemes with
default settings for such event generators; the only
exception is for \textsc{aMC@NLO+PYTHIA6}, where \texttt{PARP(67)} and
\texttt{PARP(71)} have been set to \texttt{1D0}.
No simulation of the underlying event has been performed. 

%In order to improve numerical stability, \textsc{aMC@NLO} events have been generated 
%requiring at least two jets with transverse momentum larger than $2\UGeV$ 

%\footnote{These technical generation cuts have been checked not to bias in any 
%way the results shown in the following.}. Moreover, the \textsc{aMC@NLO} default 
%behavior has been slightly modified in order to integrate the finite part of the 
%virtual corrections together with the rest of the cross section: since VBF virtuals 
%are relatively fast to evaluate, this has the advantage of reducing the amount of 
%negative weights without sizably affecting the generation speed.\\

Plots for the most relevant distributions are shown using the following
pattern. The curves in the main frame represent \textsc{aMC@NLO} matched with
\textsc{HERWIG6} (solid, black), \textsc{PYTHIA6} (dashed, red), and
\textsc{HERWIG++} (dot-dashed, blue). The upper inset carries the same
information as the main frame, displayed as ratio of the three
\textsc{aMC@NLO} curves (with the same colors and patterns as described
above) over \textsc{aMC@NLO+HERWIG6}. The middle inset contains the ratio of
the \textsc{POWHEG} events showered with the three Monte Carlos (again, using
the same colors and patterns) over \textsc{aMC@NLO} +\textsc{HERWIG6}. Finally, the
lower inset displays the PDF (dot-dashed, black) and scale (dashed, red) 
uncertainties obtained automatically \cite{Frederix:2011ss} after the
\textsc{aMC@NLO+HERWIG6} run. The scale band is the envelope of the nine
curves corresponding to $(\muR,\muF)=(\kappa_R,\kappa_F) \MW$, with
$\kappa_R$ and $\kappa_F$ varied independently in the range $\{1/2,1,2\}$.
The parton-distribution band is obtained by following the asymmetric Hessian
prescription given by the PDF set in~\Bref{Martin:2009iq}. We remind the reader
that all results presented in this section are obtained by imposing the cuts
introduced in \refS{sec:cuts-sub}.

\begin{figure}
\begin{minipage}{0.49\textwidth}
\centering
\includegraphics[width=1.03\textwidth]{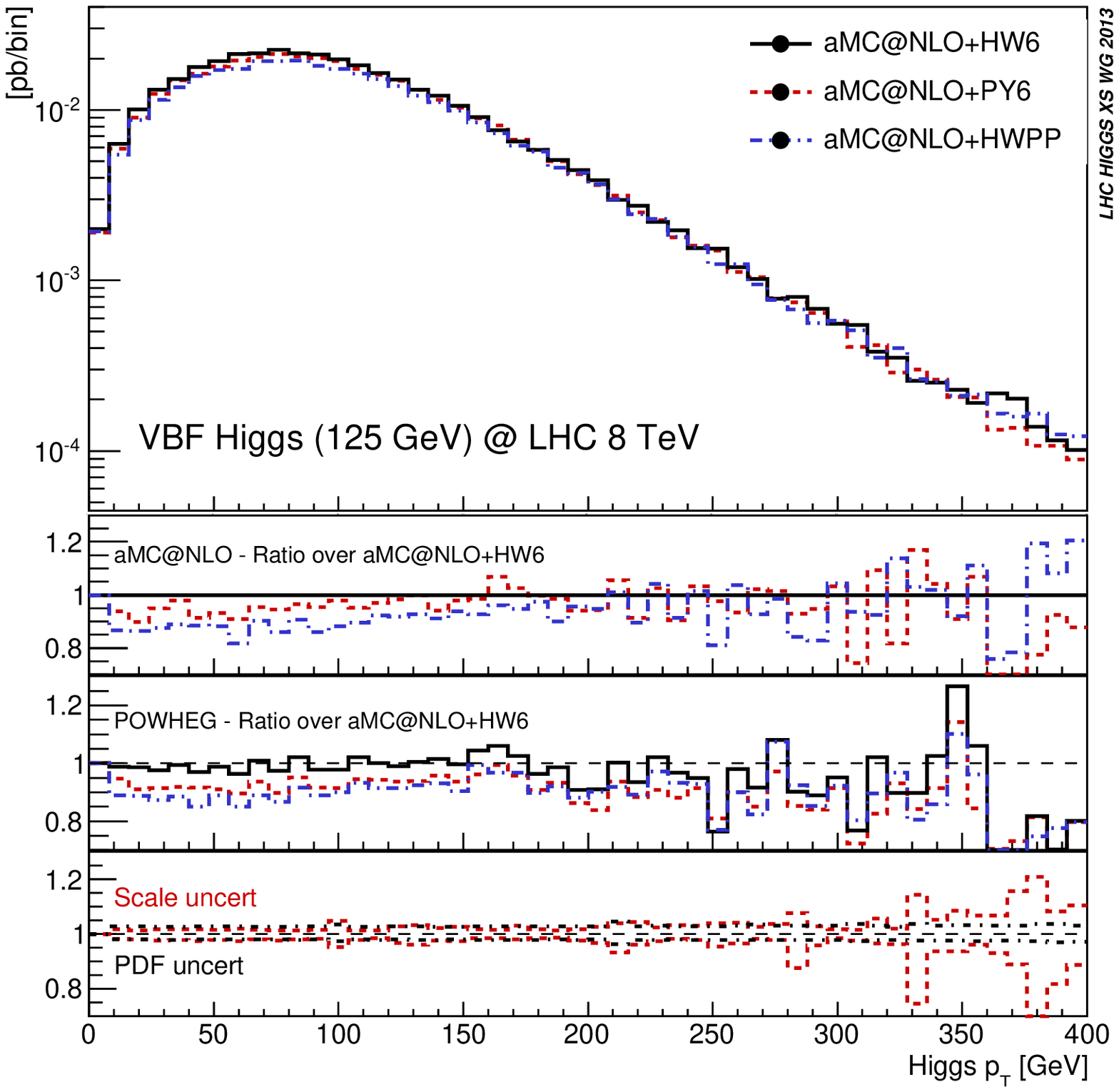}
\end{minipage}
\hspace{1mm}
\begin{minipage}{0.49\textwidth}
\centering
\includegraphics[width=1.08\textwidth]{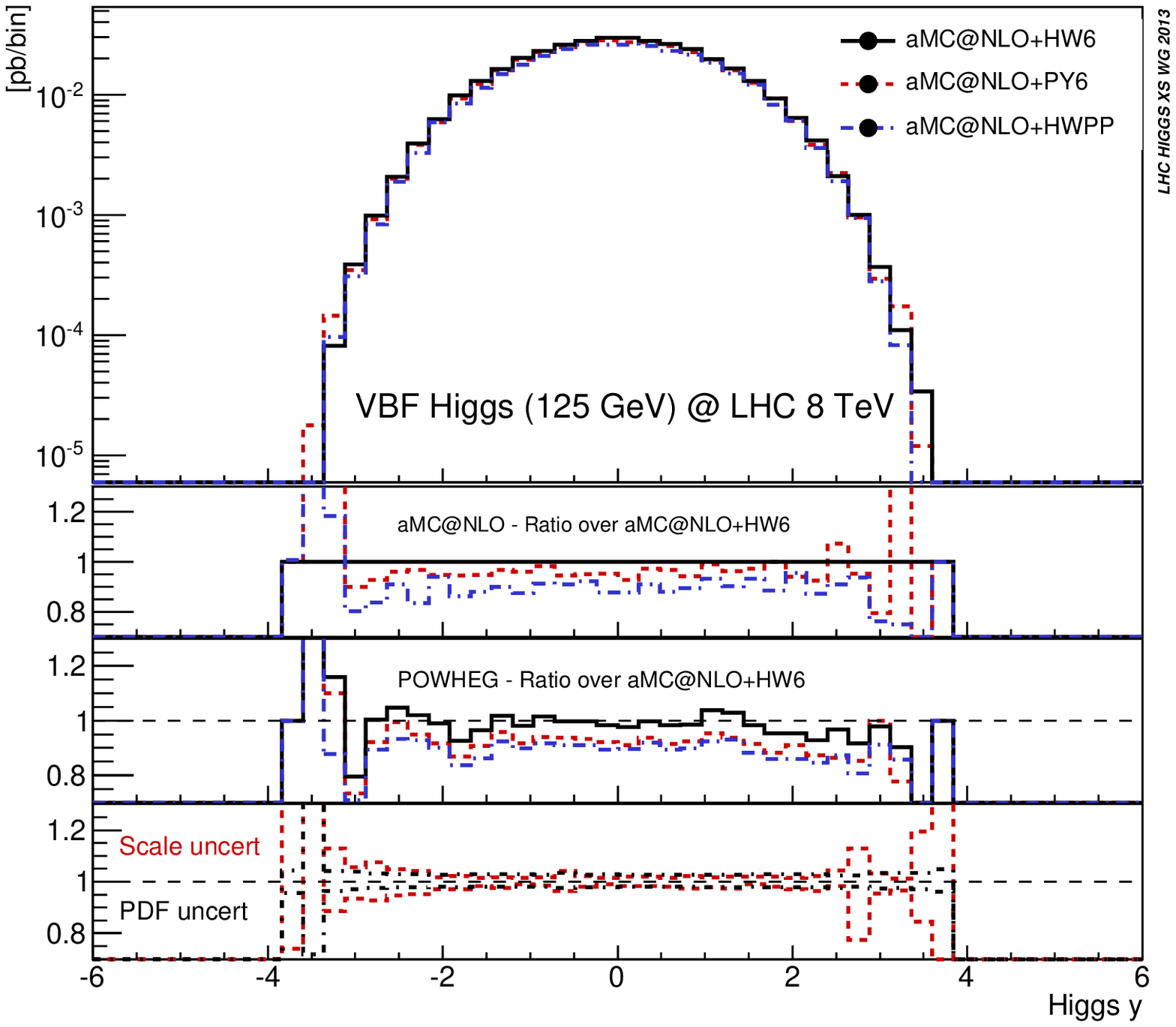}
\end{minipage}
\vspace{-1.8cm}
\caption{Higgs-boson transverse-momentum (left) and rapidity (right) distributions. Main frame and upper inset: \textsc{aMC@NLO} results; middle inset: ratio of \textsc{POWHEG} results over \textsc{aMC@NLO+HERWIG6}; lower inset: \textsc{aMC@NLO+HERWIG6} scale and PDF variations.}
\label{figvbf:ptyh}
\end{figure}

In \refF{figvbf:ptyh} the transverse-momentum and rapidity
distributions of the Higgs boson are shown. These observables are
mildly affected by extra QCD radiation, essentially because they are
genuinely NLO (in the sense that they are non-trivial in their full
kinematical ranges already at the Born level ${\cal
  O}(\alpha_\mathrm{s}^0)$). This explains why there is a general good
agreement (in terms of shape) among the results obtained with
different parton showers (main frame and upper inset), and between
those obtained with \textsc{aMC@NLO} and \textsc{POWHEG} (middle
inset). The discrepancy in the normalization of the various curves is
due to the impact of the VBF cuts on the radiation generated by the
different showers, as can be deduced by looking at the numbers
reported in Table \ref{tablevbf:ratios}. The scale and PDF bands are
fairly constant and both of the order of $\pm 3\%$, with a marginal
increase of the former at large rapidities and transverse momentum.
%%%%%%%%%%%%%%%%%%%%%%%%%%%%%%%%%%%%
%%%%%%%%%%%%%%%%%%%%%%%%%%%%%%%%%%%%

%%%%%%%%%%%%%%%%%%%%%%%%%%%%%%%%%%%%
\begin{table}
\centering
\caption{Cross section after VBF cuts normalized to \textsc{aMC@NLO+HERWIG6}.}
\label{tablevbf:ratios}
\begin{tabular}{cccc}
\hline
 & \textsc{HERWIG6} & \textsc{PYTHIA6} & \textsc{HERWIG++} \\ \hline
\textsc{aMC@NLO} & 1.00 & 0.96 & 0.90 \\
\textsc{POWHEG} & 0.99 & 0.93 & 0.90 \\
\hline
\end{tabular}
\end{table}
%%%%%%%%%%%%%%%%%%%%%%%%%%%%%%%%%%%%

\begin{figure}
%    \vspace{5cm}
    \begin{minipage}{0.5\textwidth}
    \centering
   \includegraphics[width=1\textwidth]{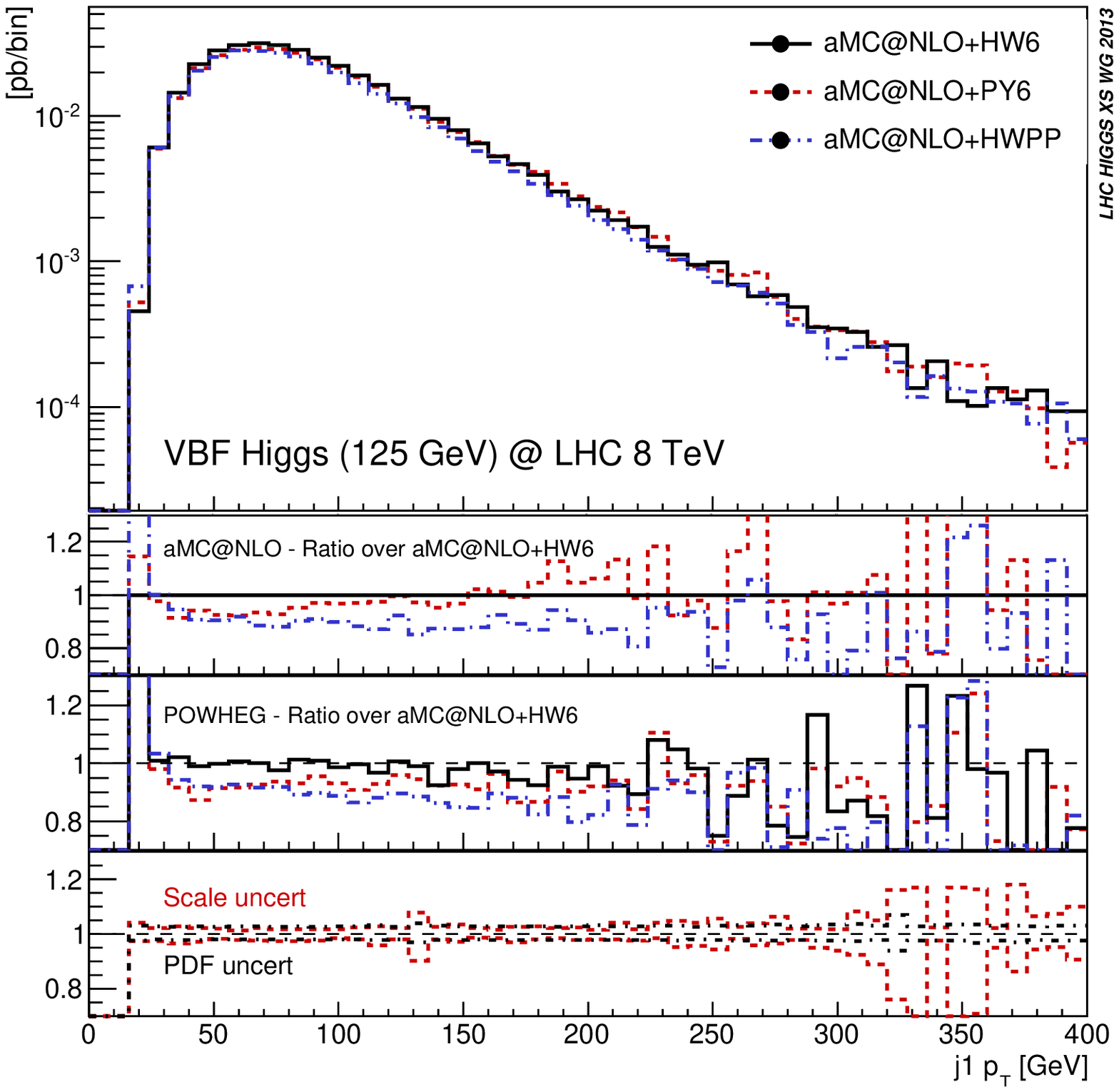}
    \end{minipage}
    \hspace{1mm}
    \begin{minipage}{0.5\textwidth}
    \centering
   \includegraphics[width=1\textwidth]{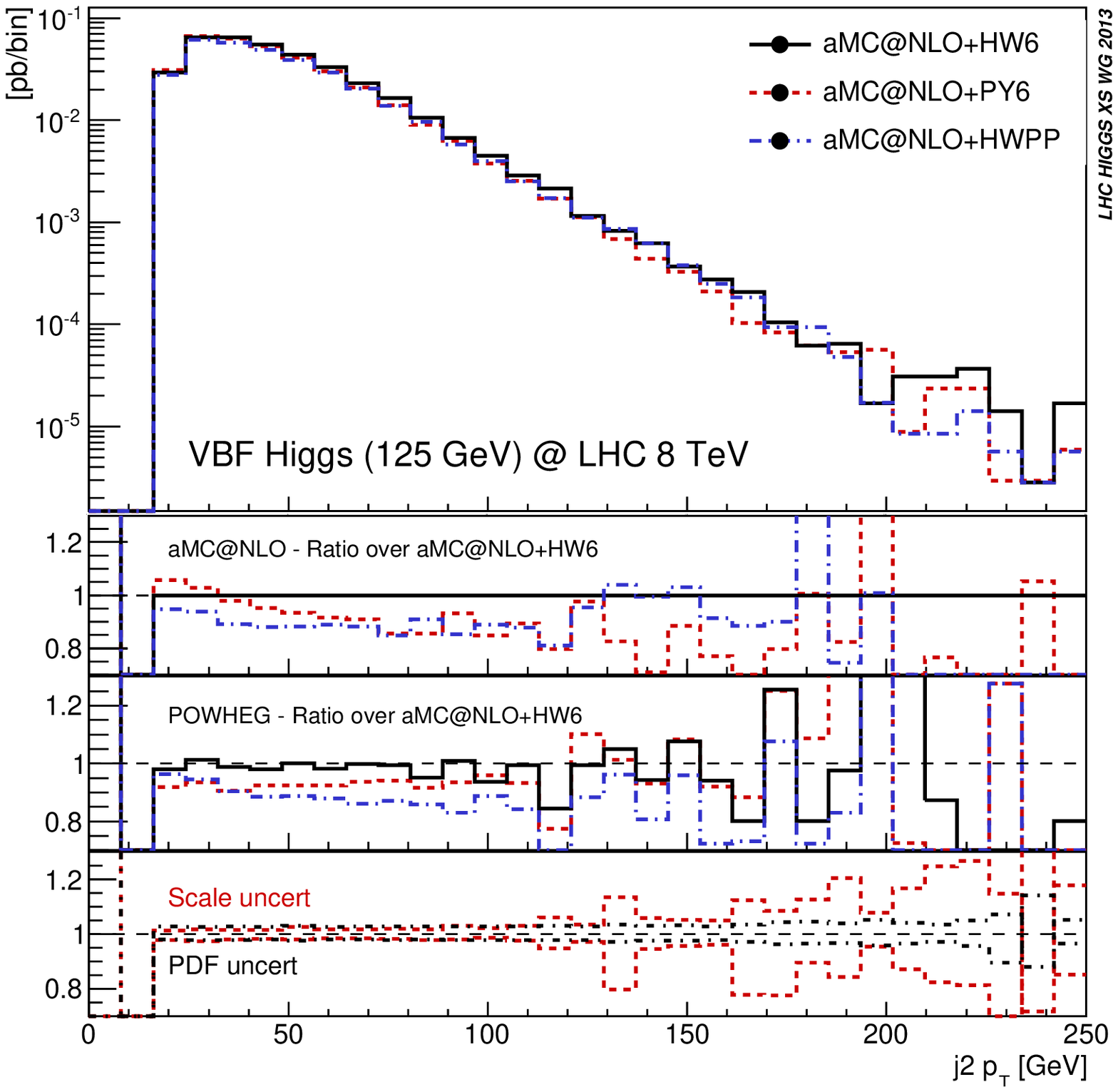}
    \end{minipage}
\vspace{-1.8cm}
    \caption{Transverse-momentum distributions for the hardest (left) and next-to-hardest  (right) jets. Main frame and upper inset: \textsc{aMC@NLO} results; middle inset: ratio of \textsc{POWHEG} results over \textsc{aMC@NLO+HERWIG6}; lower inset: \textsc{aMC@NLO+HERWIG6} scale and PDF variations.}
    \label{figvbf:pt12}
\end{figure}

\begin{figure}
%    \vspace{5cm}
    \begin{minipage}{0.5\textwidth}
    \centering
   \includegraphics[width=1\textwidth]{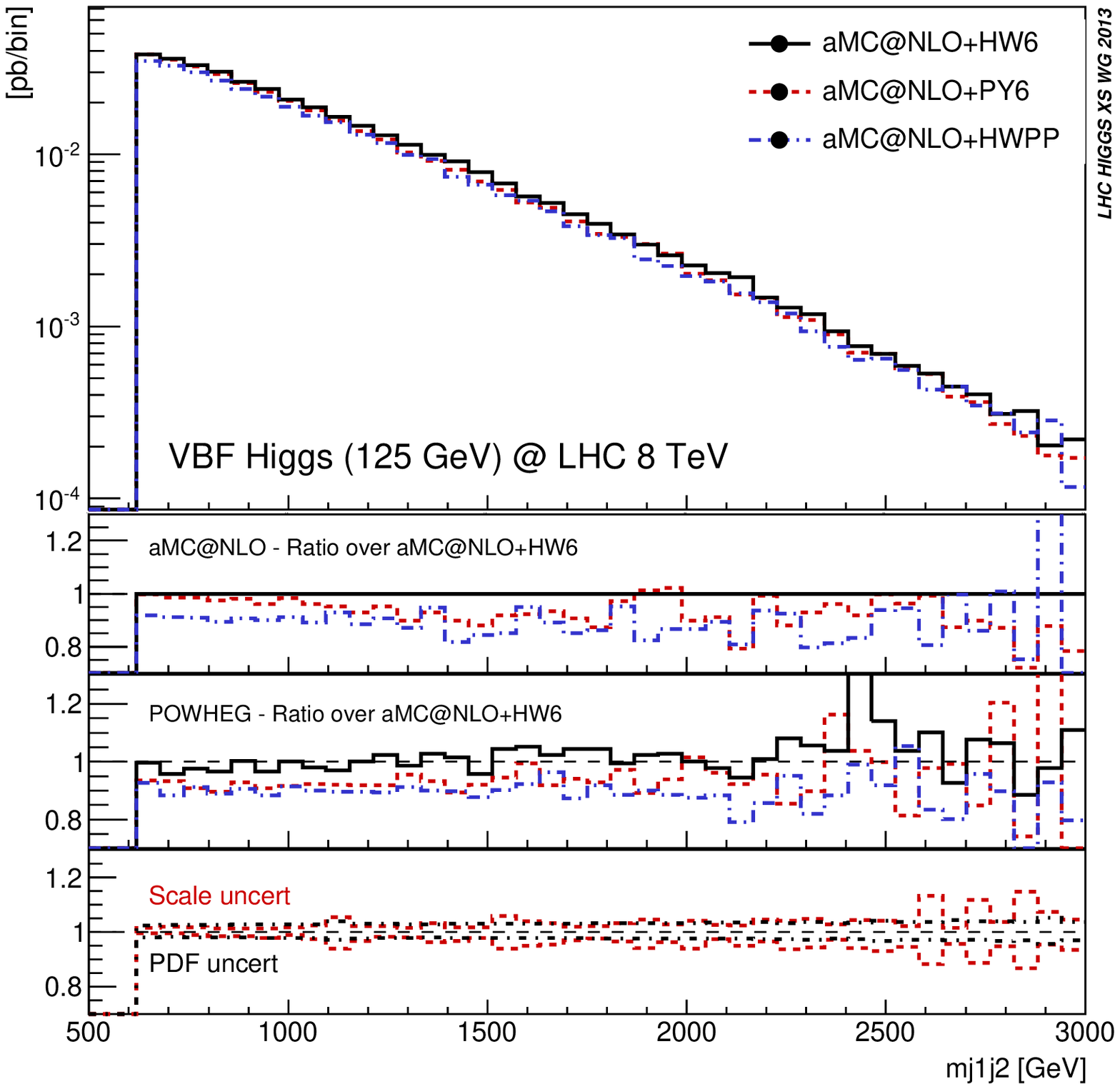}
    \end{minipage}
    \hspace{1mm}
    \begin{minipage}{0.5\textwidth}
    \centering
   \includegraphics[width=1\textwidth]{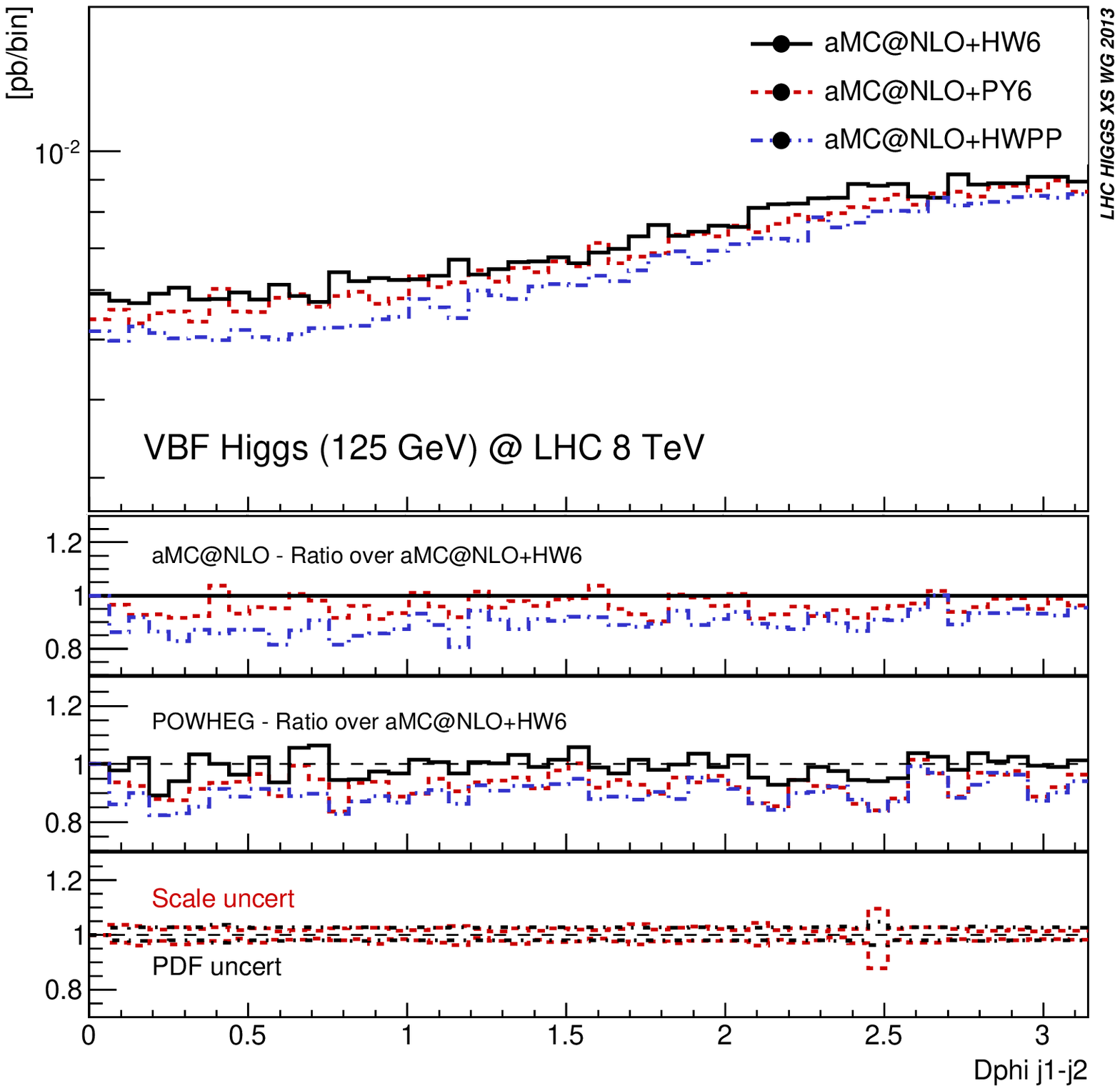}
    \end{minipage}
\vspace{-1.8cm}
    \caption{Invariant mass (left) and azimuthal separation (right) of the two tagging jets. Main frame and upper inset: \textsc{aMC@NLO} results; middle inset: ratio of \textsc{POWHEG} results over \textsc{aMC@NLO+HERWIG6}; lower inset: \textsc{aMC@NLO+HERWIG6} scale and PDF variations.}
    \label{figvbf:mphi12}
    \centering
\end{figure}

\begin{figure}
%    \vspace{5cm}
    \begin{minipage}{0.5\textwidth}
    \centering
   \includegraphics[width=1\textwidth]{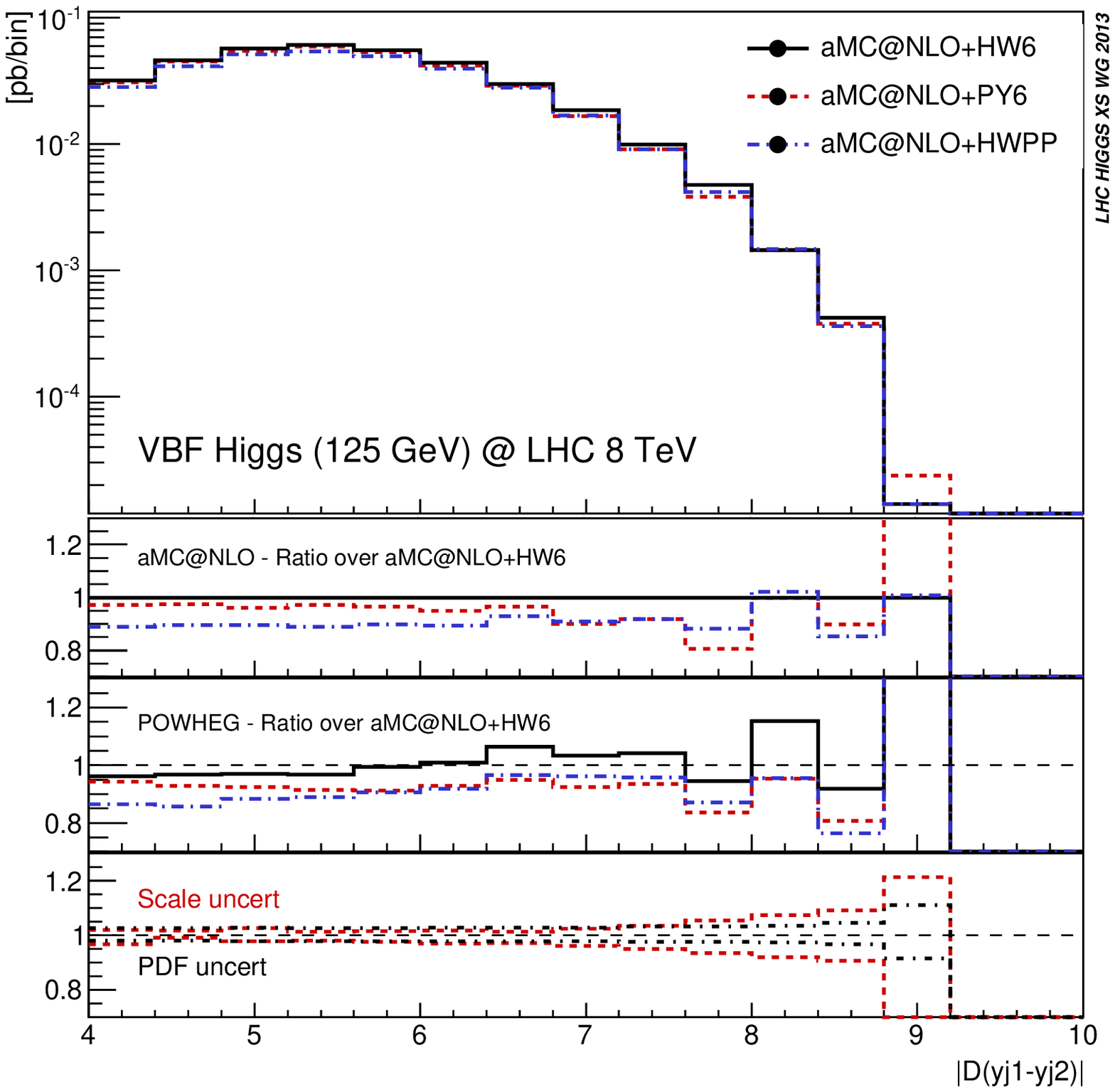}
    \end{minipage}
    \hspace{1mm}
    \begin{minipage}{0.5\textwidth}
    \centering
   \includegraphics[width=1\textwidth]{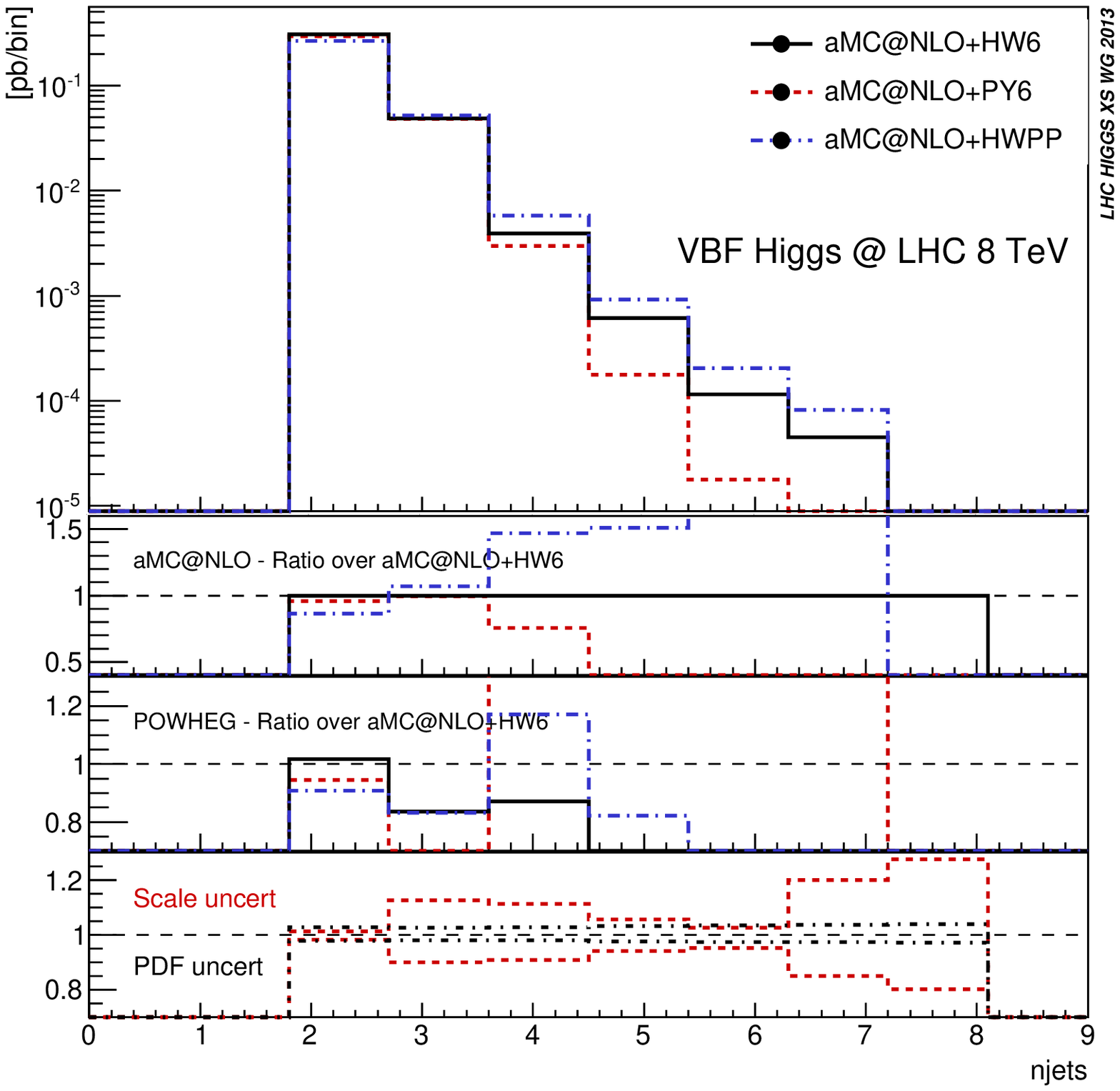}
    \end{minipage}
\vspace{-1.8cm}
        \caption{Rapidity separation of the two tagging jets (left) and exclusive jet-multiplicity (right). Main frame and upper inset: \textsc{aMC@NLO} results; middle inset: ratio of \textsc{POWHEG} results over \textsc{aMC@NLO+HERWIG6}; lower inset: \textsc{aMC@NLO+HERWIG6} scale and PDF variations.}
    \label{figvbf:dy12n}
    \centering
\end{figure}

Figures~\ref{figvbf:pt12}, \ref{figvbf:mphi12}, and the left panel of
\Fref{figvbf:dy12n} display quantities related to the two
hardest (tagging) jets, namely their transverse momenta, the pair
invariant mass and azimuthal distance, and the absolute value of their
rapidity difference, respectively. In spite of their being directly
related to QCD radiation, these observables are described at the NLO,
which translates in the overall agreement, up to the normalization
effect discussed above, of the results. Predictions obtained with
\textsc{PYTHIA6} tend to be marginally harder with respect to the two
\textsc{HERWIG} curves for the first-jet transverse momentum both in
\textsc{POWHEG} and in \textsc{aMC@NLO}, while the second-jet
transverse momentum is slightly softer in \textsc{aMC@NLO}. The
theoretical-uncertainty bands are constant with a magnitude of about
$\pm 3\%$ for all observables, with the exceptions of a slight
increase at large pair invariant mass ($\pm 5\%$), and of a more
visible growth, up to $\pm 10{-}15\%$, towards the upper edge of the
rapidity-difference range. For the parton distributions, this is due
to the larger uncertainties in the high-$x$ region, while the growth
in the scale error may be related to $\MW$ not being a representative
scale choice for such extreme kinematical configurations.
%%%%%%%%%%%%%%%%%%%%%%%%%%%%%%%%%%%%
%%%%%%%%%%%%%%%%%%%%%%%%%%%%%%%%%%%%
%%%%%%%%%%%%%%%%%%%%%%%%%%%%%%%%%%%%
%%%%%%%%%%%%%%%%%%%%%%%%%%%%%%%%%%%%
%%%%%%%%%%%%%%%%%%%%%%%%%%%%%%%%%%%%
%%%%%%%%%%%%%%%%%%%%%%%%%%%%%%%%%%%%

The right panel \Fref{figvbf:dy12n} shows the exclusive
jet-multiplicity. This observable is described with NLO accuracy in the 2-jet
bin, and at the LO in the 3-jet bin. For the 2-jet bin, the pattern displayed
in the upper and middle inset closely follows that of Table
\ref{tablevbf:ratios}, while in the 3-jet bin discrepancies are of the order
of $10{-}20\%$, with \textsc{POWHEG} predicting less events than
\textsc{aMC@NLO}. Consistently with this picture, and with the formal accuracy
of the two bins, scale uncertainties are $\pm3\%$ and $\pm10\%$,
respectively. The differences between \textsc{POWHEG} and \textsc{aMC@NLO}, of
the same order of (or slightly larger than) the scale variations, can be
considered as an independent way of estimating the theoretical uncertainty
associated with higher-order corrections.

From the 4-jet bin onwards, the description is completely driven by the
leading-logarithmic (LL) accuracy of the showers, and by the tunes employed,
which is reflected in the large differences that can be observed in the
predictions given by the different event generators (for both
\textsc{aMC@NLO} and \textsc{POWHEG}). For such jet multiplicities,
theoretical-uncertainty bands are completely unrepresentative.

%%%%%%%%%%%%%%%%%%%%%%%%%%%%%%%%%%%%
\begin{figure}
%    \vspace{5cm}
    \begin{minipage}{0.5\textwidth}
    \centering
   \includegraphics[width=1\textwidth]{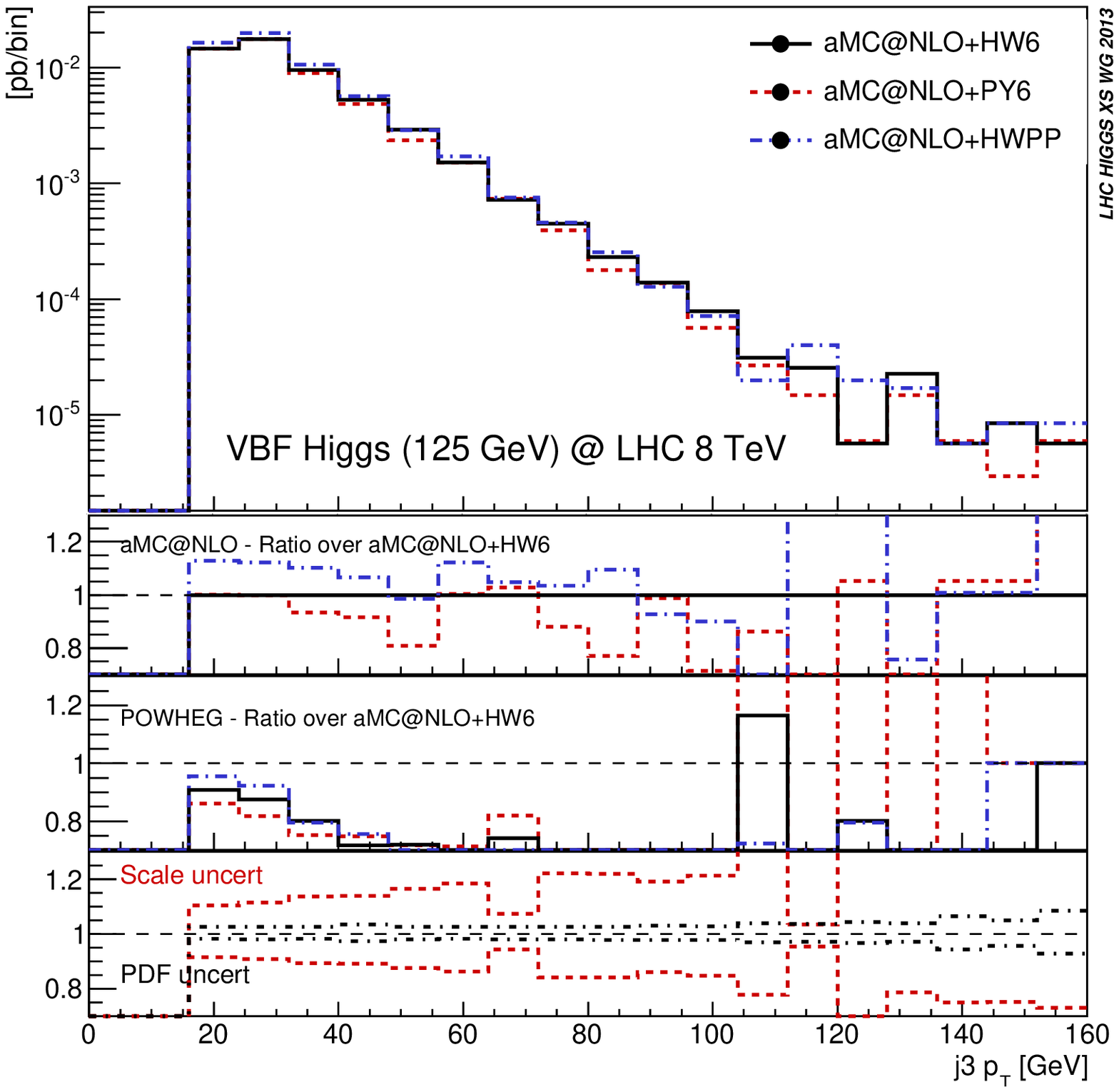}
    \end{minipage}
    \hspace{1mm}
    \begin{minipage}{0.5\textwidth}
    \centering
   \includegraphics[width=1.1\textwidth]{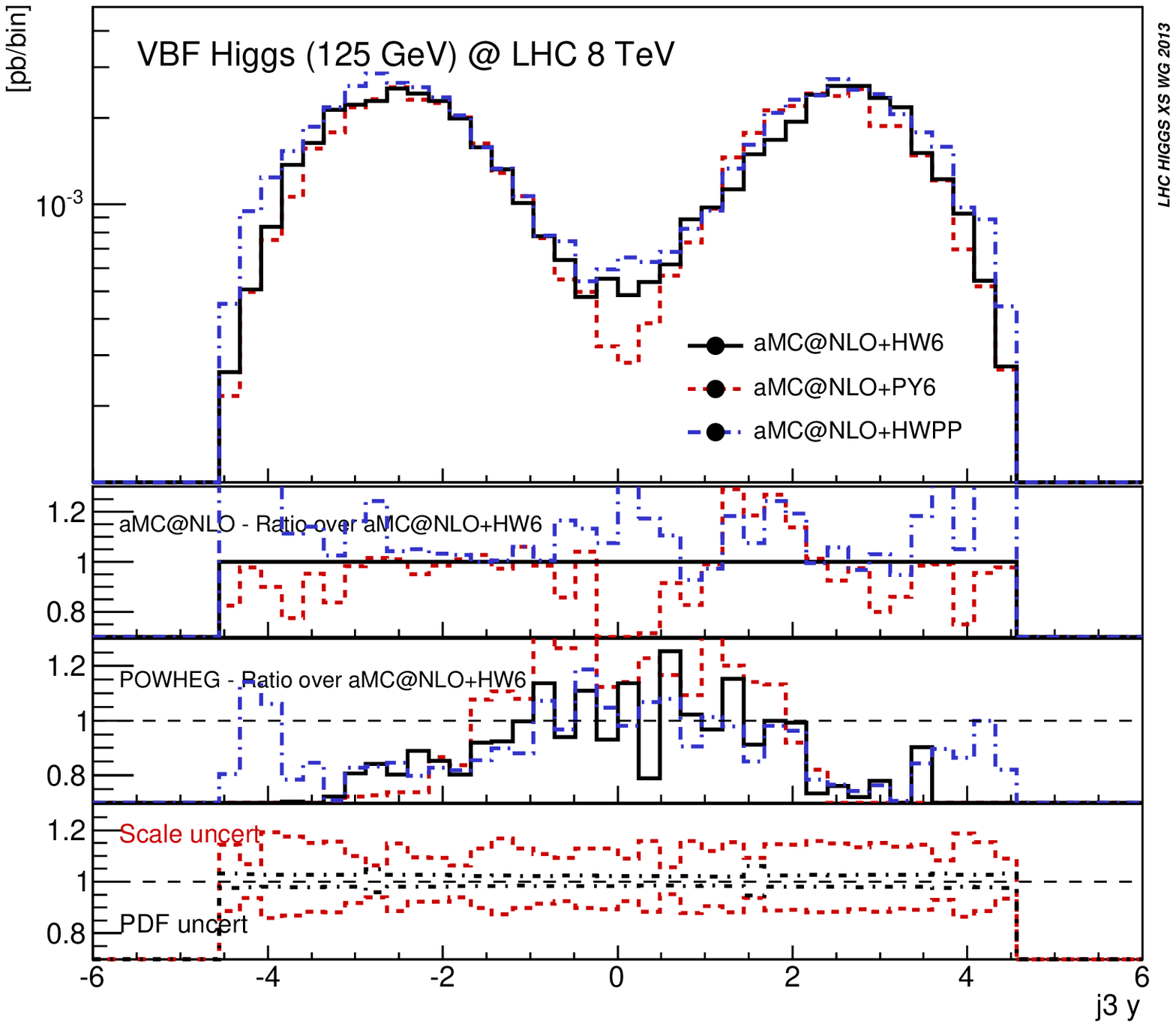}
    \end{minipage}
\vspace{-1.8cm}
        \caption{Third-hardest jet transverse momentum (left) and rapidity (right) distributions. Main frame and upper inset: \textsc{aMC@NLO} results; middle inset: ratio of \textsc{POWHEG} results over \textsc{aMC@NLO+HERWIG6}; lower inset: \textsc{aMC@NLO+HERWIG6} scale and PDF variations.}
    \label{figvbf:pty3}
    \centering
\end{figure}
%%%%%%%%%%%%%%%%%%%%%%%%%%%%%%%%%%%%

The left panel of \Fref{figvbf:pty3} shows the
transverse-momentum distribution of the third-hardest jet, which is a
LO variable that can be affected by the different radiation patterns
produced by the Monte Carlos. This is indeed what can be observed in
the plot. In particular the \textsc{aMC@NLO} results display a $10{-}15\%$
dependence on the parton shower adopted. The \textsc{POWHEG} curves
are slightly closer to each other, since in this case the hardest
extra emission (the most relevant for quantities related to the
overall third-hardest jet) is performed independently of the actual
Monte Carlo employed. Furthermore, the two matching schemes differ
quite sizably for this observable, up to $\pm20{-}25\%$ at large
transverse momentum, with \textsc{POWHEG} predicting a softer spectrum
with respect to \textsc{aMC@NLO}. This discrepancy is consistent with
the normalizations of the 3-jet bin in the right panel of
\Fref{figvbf:dy12n}. Different settings in \textsc{PYTHIA6} have
been checked to induce a variation in the curves within the previously
mentioned discrepancy range, which is thus to be considered as a
measure of the matching systematics affecting the prediction of this
quantity. Scale uncertainties are quite large, compatibly with the LO
nature of this observable, and growing from $\pm10\%$ to $\pm25\%$
with increasing transverse momentum.  The rapidity distribution of the
third-hardest jet, shown in the right panel of
\Fref{figvbf:pty3} has similar features, with the
\textsc{POWHEG} samples more central than the \textsc{aMC@NLO} ones.

\begin{figure}
%    \vspace{5cm}
    \begin{minipage}{0.5\textwidth}
    \centering
   \includegraphics[width=1\textwidth]{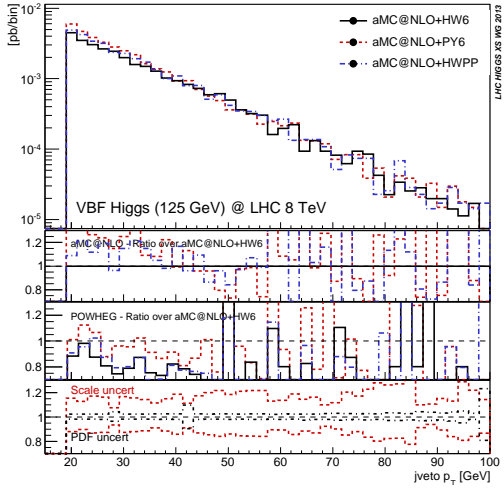}
    \end{minipage}
    \hspace{1mm}
    \begin{minipage}{0.5\textwidth}
    \centering
   \includegraphics[width=1.1\textwidth]{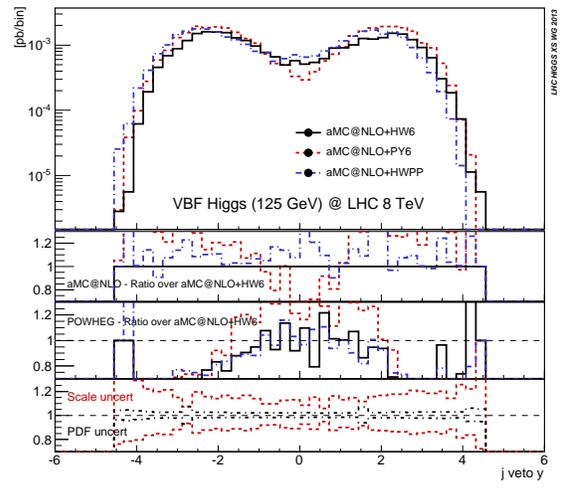}
    \end{minipage}
    \vspace{-1.8cm}
        \caption{Veto-jet transverse momentum (left) and rapidity (right) distributions. Main frame and upper inset: \textsc{aMC@NLO} results; middle inset: ratio of \textsc{POWHEG} results over \textsc{aMC@NLO+HERWIG6}; lower inset: \textsc{aMC@NLO+HERWIG6} scale and PDF variations.}
    \label{figvbf:ptyv}
    \centering
\end{figure}

\begin{figure}
%    \vspace{5cm}
    \begin{minipage}{0.98\textwidth}
    \centering
   \includegraphics[width=1\textwidth]{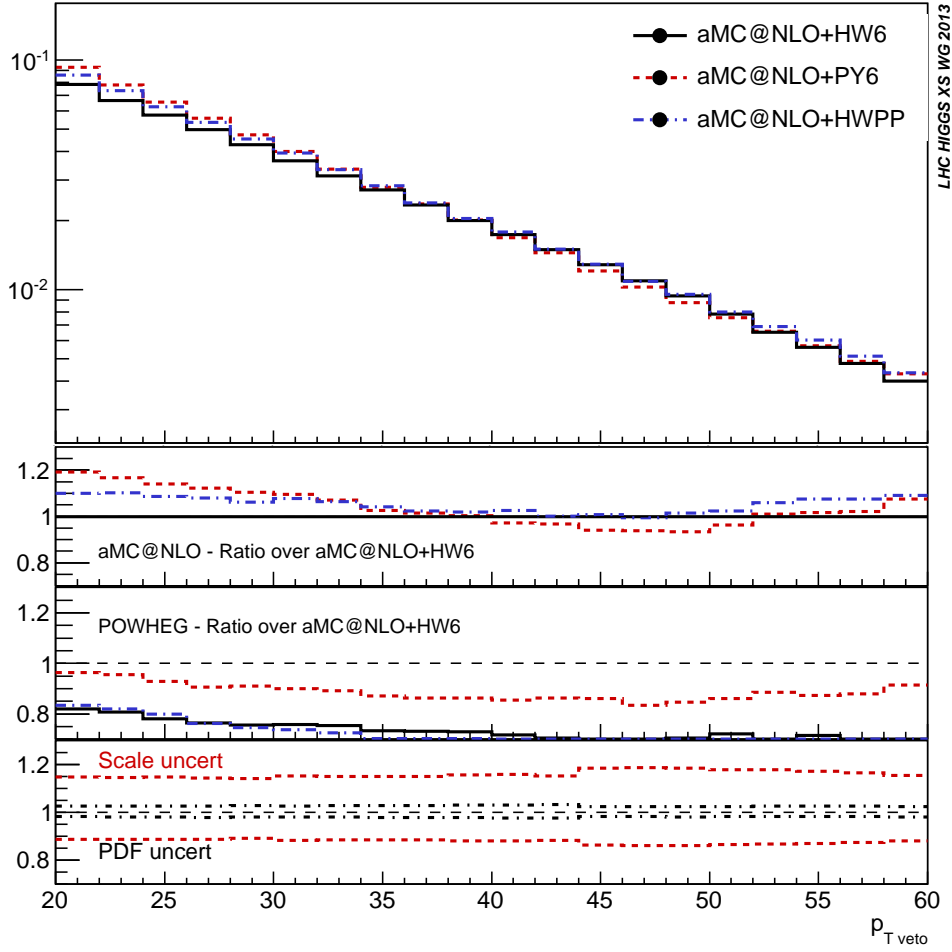}
    \end{minipage}
\vspace{-3cm}
        \caption{Veto probability. Main frame and upper inset: \textsc{aMC@NLO} results; middle inset: ratio of \textsc{POWHEG} results over \textsc{aMC@NLO+HERWIG6}; lower inset: \textsc{aMC@NLO+HERWIG6} scale and PDF variations.}
    \label{figvbf:pve}
    \centering
\end{figure}

\Fref{figvbf:ptyv} displays features of the veto jet, defined as the hardest jet 
with rapidity ($y_{j_{\tiny\mbox{veto}}}$) lying between those ($y_{j
_1}$ and $y_{j_2}$) 
of the two tagging jets:
%%%%%%%%%%%%%%%%%%%%%%%%%%%%%%%%%%%%
\begin{equation}
    \min\left(y_{j_1},y_{j_2}\right) < y_{j_{\tiny\mbox{veto}}} < \max\left(y_{j_1},y_{j_2}\right).
\end{equation}%
%%%%%%%%%%%%%%%%%%%%%%%%%%%%%%%%%%%%
This definition implies that the more central the third jet, the larger the probability 
that it be the veto jet. Since \textsc{POWHEG} predicts a more central third jet with 
respect to \textsc{aMC@NLO}, the veto condition has
the effect that the two predictions for the veto jet
are closer to each other than for the third jet; this is interesting
in view of the fact that the differences between the two approaches
can to a large extent be interpreted as matching systematics.\\
To further investigate this point, \Fref{figvbf:pve} shows the veto
probability     
$P_{\tiny\mbox{veto}}(p_{\tiny\mbox{T,veto}})$, defined as
\begin{equation}
P_{\tiny\mbox{veto}}(p_{\tiny\mbox{T,veto}})=\frac{1}{\sigma_{\tiny\mbox{NLO}}}\int_{p_{\tiny\mbox{T,veto}}}^\infty
\rd p_{\tiny\mbox T} \,\frac{\rd\sigma}{\rd p_{\tiny\mbox T}},
\end{equation}
where $\sigma_{\tiny\mbox{NLO}}=(0.388\pm0.002)\Upb$ is the (fixed-order) 
NLO cross section within VBF cuts \footnote{The \textsc{aMC@NLO+HERWIG6} cross 
section within VBF cuts is $0.361\Upb$. The cross sections relevant to the other parton showers and to \textsc{POWHEG} 
can be deduced using the ratios in Table \ref{tablevbf:ratios}.}.\\
The excess of events between $p_{\tiny\mbox T}\sim20\UGeV$ and
$p_{\tiny\mbox T}\sim40\UGeV$ in \textsc{aMC@NLO} matched to
\textsc{PYTHIA6} and \textsc{HERWIG++} translates in a slightly larger
cumulative probability of passing the veto cut. The \textsc{POWHEG}
curves are lower, with up to $20{-}25\%$ discrepancy with respect to
\textsc{aMC@NLO} at medium-high transverse momentum. Scale variation
are compatible with a LO prediction, with a fairly constant magnitude
of $\pm15\%$.
%%%%%%%%%%%%%%%%%%%%%%%%%%%%%%%%%%%%
%%%%%%%%%%%%%%%%%%%%%%%%%%%%%%%%%%%%

 \begin{table}
\caption{VBF inclusive cross sections at $8$ and $13\UTeV$ calculated
  using MSTW2008NLO PDF set.  Comparison among different calculations 
at NLO QCD for a selected number of Higgs-boson masses. The
$t$--$u$-channel interferences and the $\PQb$-quark
contributions are included in the computation.}
\label{tab:YRHXS3_VBF_NNLO8_13}%
\setlength{\tabcolsep}{1.5ex}%
\begin{center}
\begin{tabular}{lcccc}
\hline \rule{0ex}{2.7ex}
 $ \MH $  &   $ \sigma^{\mathrm{NLO QCD}}_{\mathrm{VBF@NNLO}} $ &  $
 \sigma^{\mathrm{NLO QCD}}_{\mathrm{HAWK}}  $  & $ \sigma^{\mathrm{NLO
     QCD} + \mathrm{NLO EW}}_{\mathrm{HAWK}} $  &  $
 \sigma^{\mathrm{NLO QCD}}_{\mathrm{VBFNLO}}  $   \\ 
  $\mbox{[GeV]}$   &   [fb]  &  [fb]   & [fb] &  [fb]   \\ 
 \hline 
\multicolumn{4}{l}{$8\UTeV$} \\
\hline
120 &  1737 $\pm$       0.6 & 1732 $\pm$ 0.8 & 1651 $\pm$ 0.9  & 1743 $\pm$ 0.5  \\
122 &  1706 $\pm$       0.6 & 1701 $\pm$ 0.7 & 1623 $\pm$ 0.8  & 1711 $\pm$ 0.5  \\
124 &  1675 $\pm$       0.5 & 1671 $\pm$ 0.7 & 1594 $\pm$ 0.8  & 1681 $\pm$ 0.5  \\
126 &  1646 $\pm$       0.5 & 1641 $\pm$ 0.7 & 1566 $\pm$ 0.8  & 1651 $\pm$ 0.5  \\
128 &  1617 $\pm$       0.5 & 1613 $\pm$ 0.7 & 1539 $\pm$ 0.8  & 1622 $\pm$ 0.5  \\
130 &  1588 $\pm$       0.5 & 1585 $\pm$ 0.7 & 1513 $\pm$ 0.8  & 1593 $\pm$ 0.5  \\
\hline
\multicolumn{4}{l}{$13\UTeV$} \\
\hline
120 & 4164 $\pm$    2.5 & 4162 $\pm$ 2.2 & 3938 $\pm$ 2.5 & 4168 $\pm$ 2.1  \\
122 & 4096 $\pm$    2.1 & 4098 $\pm$ 2.2 & 3878 $\pm$ 2.5 & 4103 $\pm$ 2.0  \\
124 & 4034 $\pm$    1.8 & 4035 $\pm$ 2.2 & 3820 $\pm$ 2.4 & 4040 $\pm$ 1.8  \\
126 & 3975 $\pm$    2.0 & 3974 $\pm$ 2.2 & 3762 $\pm$ 2.4 & 3979 $\pm$ 1.8  \\
128 & 3914 $\pm$    2.3 & 3913 $\pm$ 2.3 & 3705 $\pm$ 2.4 & 3919 $\pm$ 1.7  \\
130 & 3854 $\pm$    1.6 & 3857 $\pm$ 2.0 & 3651 $\pm$ 2.2 & 3860 $\pm$ 1.6  \\
\hline

\end{tabular}

\end{center}
\end{table}

%\subsubsubsection{aMC@NLO Comparisons}
%\label{sec:aMCNLOcomp-sub-sub-sub}

%\newpage

\clearpage

\newpage
\section{$\PW\PH$/$\PZ\PH$ production mode\footnote{%
    S.~Dittmaier, G.~Ferrera, A.~Rizzi, G.~Piacquadio (eds.);
    A.~Denner, M.~Grazzini, R.V.~Harlander, S.~Kallweit, A.~M\"uck, 
    F.~Tramontano and T.J.E.~Zirke.}}

\subsection{Theoretical developments}
\label{sec:YRHX3_WHZH_th}

In the previous two working group reports \cite{Dittmaier:2011ti} and \cite{Dittmaier:2012vm}
the state-of-the-art predictions for $\Pp\Pp\to\PW\PH/\PZ\PH$ were summarized and numerically
discussed for the total and differential cross sections, respectively, taking into account
all available higher-order corrections of the strong and electroweak interactions.
The remaining uncertainties in the cross-section predictions, originating from missing higher
orders and parametric errors in $\alphas$ and PDFs, were estimated to be smaller
than $5\%$ for integrated quantities, with somewhat larger uncertainties for differential
distributions. In the meantime the predictions have been refined and supplemented upon
including further higher-order corrections that are relevant at this level of accuracy as well.
In the following we, thus, update the cross-section predictions accordingly.

Current state-of-the-art predictions are based on the following ingredients:
\begin{itemize}
\item
QCD corrections are characterized by the similarity of $\PW\PH/\PZ\PH$ production to the
Drell--Yan process. While the NLO QCD corrections to these two process classes are
analytically identical, the NNLO QCD corrections to $\PW\PH/\PZ\PH$ production
also receive contributions that have no counterparts in Drell--Yan production.
The Drell--Yan-like contributions comprise the bulk of the QCD corrections of $\sim30\%$
and are completely known to NNLO both for 
integrated~\cite{Brein:2003wg} and 
fully differential~\cite{Ferrera:2011bk,Ferrera_ZH}
observables for the $\PW\PH$ and $\PZ\PH$ channels, where the NNLO corrections to
the differential $\PZ\PH$ cross section were not yet available in \Bref{Dittmaier:2012vm}.

QCD corrections beyond NLO that are not of Drell--Yan type are widely known only for 
total cross sections. The most prominent contribution of this kind comprises $\PZ\PH$ production
via gluon fusion which is mediated via quark loops. This part shows up first at the 
NNLO level~\cite{Brein:2003wg}
where it adds $\sim3\%(5\%)$ to the total cross section at $7\UTeV(14\UTeV)$ for $\MH=126\UGeV$
with a significant scale uncertainty of $\sim 30{-}60\%$.
This uncertainty has been reduced recently~\cite{Altenkamp:2012sx} upon adding the NLO corrections
to the $\Pg\Pg$ channel in the heavy-top-quark limit (which is the dominant contribution of NNNLO
to the $\Pp\Pp$ cross section). This contribution, which roughly doubles the impact of the $\Pg\Pg$ channel%
\footnote{The fact that this $100\%$ correction exceeds the above-mentioned scale uncertainty of
$\sim60\%$ is similar to the related well-known situation observed for $\Pg\Pg\to\PH$, where the LO scale
uncertainty underestimates the size of missing higher-order corrections as well.}
and mildly reduces its scale uncertainty to $\sim20{-}30\%$, was not yet taken into account
in the predictions documented in \Bref{Dittmaier:2011ti}, but is included in the
results on total cross sections below.

\begin{sloppypar}
Both $\PW\PH$ and $\PZ\PH$ production receive non-Drell--Yan-like corrections in the quark/antiquark-initiated
channels at the NNLO level where the Higgs boson is radiated off a top-quark loop. After the completion of 
report~\cite{Dittmaier:2011ti}, they were calculated
in \Bref{Brein:2011vx} and amount to $1{-}2\%$ for a Higgs-boson mass of $\MH\lsim150\UGeV$.
These effects are taken into account in the total-cross-section predictions below.
\end{sloppypar}

\item
Electroweak corrections, in contrast to QCD corrections, are quite different from the ones to 
Drell--Yan processes already at NLO and, in particular, distinguish between the various leptonic
decay modes of the $\PW^\pm$ and $\PZ$ bosons.
The NLO corrections to total cross sections already revealed EW effects of the size of
$-7\%(-5\%)$ for $\PW\PH$ ($\PZ\PH$) production~\cite{Ciccolini:2003jy}, 
almost independent from the collider energy for $\MH=126\UGeV$.
The NLO EW corrections to differential cross sections~\cite{Denner:2011id}, which were calculated 
with the {\HAWK} Monte Carlo program~\cite{HAWK,Ciccolini:2007jr,Ciccolini:2007ec} for the full processes
$\Pp\Pp\to\PW\PH\to\PGn_{\Pl}\Pl\PH$ and
$\Pp\Pp\to\PZ\PH\to\Pl^-\Pl^+\PH/\PGn_{\Pl}\PAGn_{\Pl}\PH$, i.e.\ including the W/Z decays,
get even more pronounced in comparison to the ones for the total cross sections.
Requiring a minimal transverse momentum of the Higgs boson of $200\UGeV$ in the so-called
``boosted-Higgs regime'' leads to EW corrections of about $-(10{-}15)\%$ with a trend of
further increasing in size at larger transverse momenta.

The EW corrections depend only very weakly on the hadronic environment, i.e.\ on the PDF choice and 
factorization scale, which suggests to include them in the form of
relative corrections factors to QCD-based predictions as detailed below.
\end{itemize}

The following numerical results are based on the same input parameters as used for the
total and differential cross sections in 
\Brefs{Dittmaier:2011ti} and \cite{Dittmaier:2012vm}, respectively, if not stated otherwise.
The same applies to the theoretical setup of the calculations, such as the EW
input parameter scheme, scale choices, etc.

\subsection{Predictions for total cross sections}
\label{sec:YRHX2_WHZH_txs}

\begin{sloppypar}
The following numerical results for the total cross sections
are obtained with the program {\sc VH@NNLO}~\cite{Brein:2012ne},
which includes the full QCD corrections up to NNLO, the NLO corrections to the $\Pg\Pg$ channel,
and the NLO EW corrections (with the latter taken from \Bref{Ciccolini:2003jy} in parametrized form).
In detail the QCD and EW corrections are combined as follows,
\begin{equation}
\sigma_{\PV\PH} = \sigma_{\PV\PH}^{\mbox{\footnotesize\sc NNLO QCD(DY)}}
\times (1 + \delta_{\PV\PH,\rm EW})
+\sigma_{\PV\PH}^{\mbox{\footnotesize\sc NNLO QCD(non-DY)}},
\label{eq:sigwhzh}
\end{equation}
i.e.\ the EW corrections are incorporated as relative correction factor to
the NNLO QCD cross section based on Drell--Yan-like corrections,
$\sigma_{\PV\PH}^{\mbox{\footnotesize\sc NNLO QCD(DY)}}$.
Electroweak corrections induced by initial-state photons, which are at the
level of $1\%$ (see below), are not included here.

\Trefs{tab:YRHXS3_WHZH_wh71}--\ref{tab:YRHXS3_WHZH_wh76} and
\ref{tab:YRHXS3_WHZH_wh81}--\ref{tab:YRHXS3_WHZH_wh86} display numerical
values for the $\PW\PH$ production cross section as evaluated according
to (\ref{eq:sigwhzh}).  Note that the cross sections for $\PW^+\PH$ and
$\PW^-\PH$ production are added here. The scale uncertainty is obtained
by varying the renormalization and the factorization scale independently
within the interval $[Q/3,3Q]$, where $Q\equiv\sqrt{Q^2}$ is the
invariant mass of the $\PV\PH$ system. The PDF uncertainties are
calculated by following the PDF4LHC recipe, using
MSTW2008~\cite{Martin:2009iq}, CT10~\cite{Lai:2010vv}, and
NNPDF2.3~\cite{Ball:2012cx}; the total uncertainties are just the
linear sum of the PDF and the scale uncertainties.

Similarly, \Trefs{tab:YRHXS3_WHZH_zh71}--\ref{tab:YRHXS3_WHZH_zh76}
and \ref{tab:YRHXS3_WHZH_zh81}--\ref{tab:YRHXS3_WHZH_zh86}
show up-to-date results for $\PZ\PH$ production. The gluon-fusion
channel $\sigma_{\Pg\Pg\to \PZ\PH}$ is listed separately in the last
column. It is obtained by calculating the radiative correction factor of
this channel through order $\alphas^3$ in the heavy-top limit, and
multiplying it with the exact LO result, as described in
\Bref{Altenkamp:2012sx}.  The scale uncertainty of $\sigma_{\Pg\Pg\to
  \PZ\PH}$ is obtained by varying the renormalization and the
factorization scales of the NLO term simultaneously by a factor of three
around $\sqrt{Q^2}$. The PDF uncertainty of $\sigma_{\Pg\Pg\to \PZ\PH}$
is evaluated only at LO, and its total uncertainty is simply the sum of
the scale and the PDF uncertainty.  The uncertainties arising from all
terms except for gluon fusion are obtained in analogy to the $\PW\PH$
process, see above. 
%they are listed in columns ``Scale'' and ``PDF''.
%Adding these linearly to the gluon fusion uncertainties results in column ``Total''.
The sum of all scale and PDF uncertainties are listed in columns
``Scale'' and ``PDF''. Adding them linearly results in column ``Total''. 
The second columns in
\Trefs{tab:YRHXS3_WHZH_zh71}--\ref{tab:YRHXS3_WHZH_zh76} and
\ref{tab:YRHXS3_WHZH_zh81}--\ref{tab:YRHXS3_WHZH_zh86}
contain the cross section including all
available radiative corrections.

Note that the uncertainties are symmetrized around the central values
which in turn are obtained with the MSTW2008 PDF set and by setting the
central renormalization and the factorization scales equal to
$Q$, the invariant mass of the $\PV\PH$ system.

\end{sloppypar}

\subsection{Predictions for differential cross sections}
\label{sec:YRHX2_WHZH_dxs}

We first briefly recall the salient features in the definition of the cross sections
with leptonic $\PW/\PZ$ decays. A detailed description can be found in Section~7.2
of \Bref{Dittmaier:2012vm}.
All results are given for a specific leptonic decay mode without
summation over lepton generations. For charged leptons $\Pl$ in the final state
we distinguish two different treatments of photons that are collinear to those leptons.
While the ``bare'' setup assumes perfect isolation of photons and leptons, which is reasonable
only for muons, in the ``rec'' setup we mimic electromagnetic
showers in the detector upon recombining photons and leptons to ``quasi-leptons'' for
$R_{\Pl\PGg}<0.1$, where $R_{\Pl\PGg}$ is the usual distance in the plane spanned by
rapidity and the azimuthal angle in the transverse plane. 
After the eventual recombination procedure the following cuts are applied if not stated otherwise,
\begin{eqnarray}
&& p_{\mathrm{T},\Pl} > 20\UGeV, \qquad
|y_{\Pl}|< 2.5, \qquad
p_{\mathrm{T,miss}} > 25\UGeV, \\
&& p_{\mathrm{T},\PH} > 200\UGeV, \qquad
p_{\mathrm{T},\PW/\PZ} > 190\UGeV,
\label{eq:VHpTcuts}
\end{eqnarray}
where $p_{\mathrm{T}}$ is the transverse momentum of the respective particle and
$p_{\mathrm{T,miss}}$ the total transverse momentum of the involved neutrinos.

Similar to the procedure for the total cross section, QCD-based predictions are dressed
with relative EW correction factors,
\begin{equation}
\sigma=\sigma^{\mbox{\footnotesize\sc NNLO QCD(DY)}}\times
\left( 1 + \delta_{\mathrm{EW}}^\mathrm{bare/rec} \right)
+ \sigma_\gamma\, ,
\end{equation}
where $\sigma^{\mbox{\footnotesize\sc NNLO QCD(DY)}}$ is the NNLO QCD cross-section prediction of 
\Brefs{Ferrera:2011bk,Ferrera_ZH} and 
$\delta_{\mathrm{EW}}^\mathrm{bare/rec}$ the EW correction factor obtained 
with {\HAWK}~\cite{HAWK,Ciccolini:2007jr,Ciccolini:2007ec,Denner:2011id}.
Note that the relative EW correction is not included on an event-by-event basis during the 
phase-space integration, but used as reweighting factor in the histograms bin by bin.
The contribution $\sigma_\gamma$, which is induced by processes with photons
in the initial state, also delivered by {\HAWK}, is found to be at the level
of $1\%$ (see \Brefs{Denner:2011id,Dittmaier:2012vm}).
All cross-section predictions of this section are based on the
MSTW2008 NNLO PDF set~\cite{Martin:2009iq}, but the EW correction factor
hardly depends on the PDF choice.
We recall that the non-Drell--Yan-like corrections, which are included in the predictions
for total cross sections (see previous section), are not (yet) available for differential quantities.

\begin{figure}
\includegraphics[width=7.5cm]{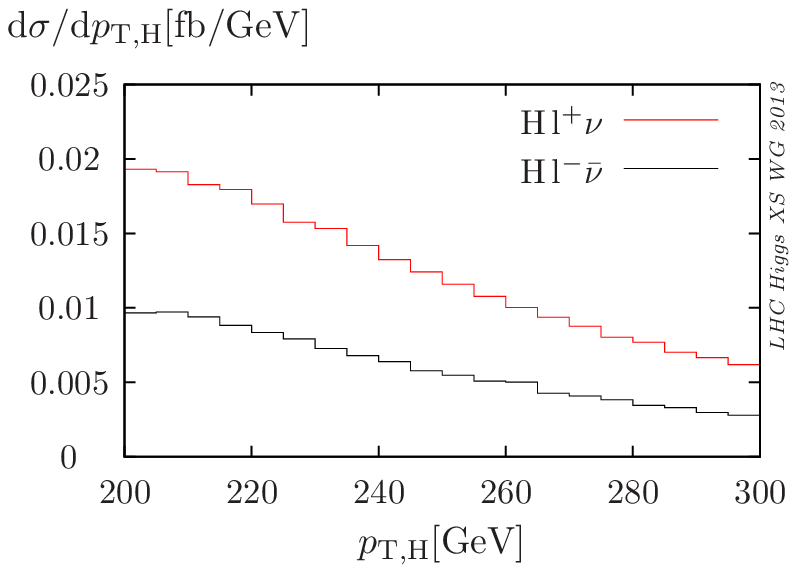}
\hfill
\includegraphics[width=7.5cm]{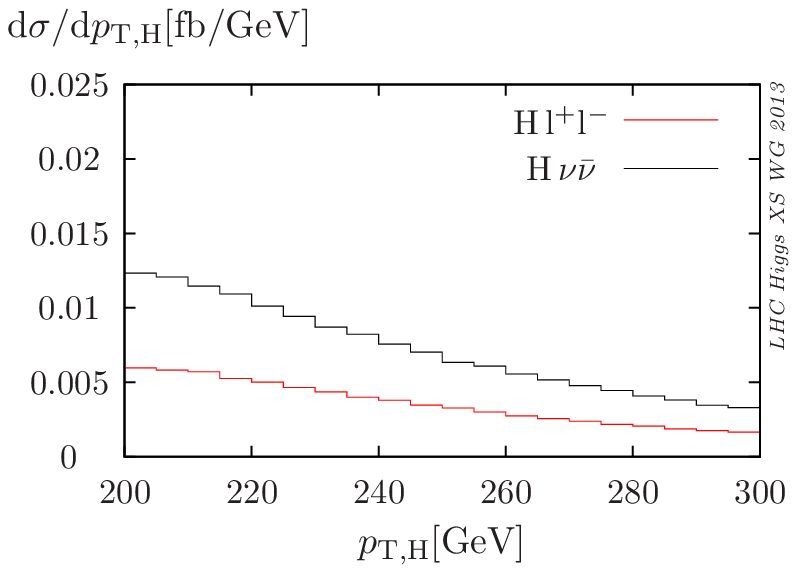}
\includegraphics[width=7.5cm]{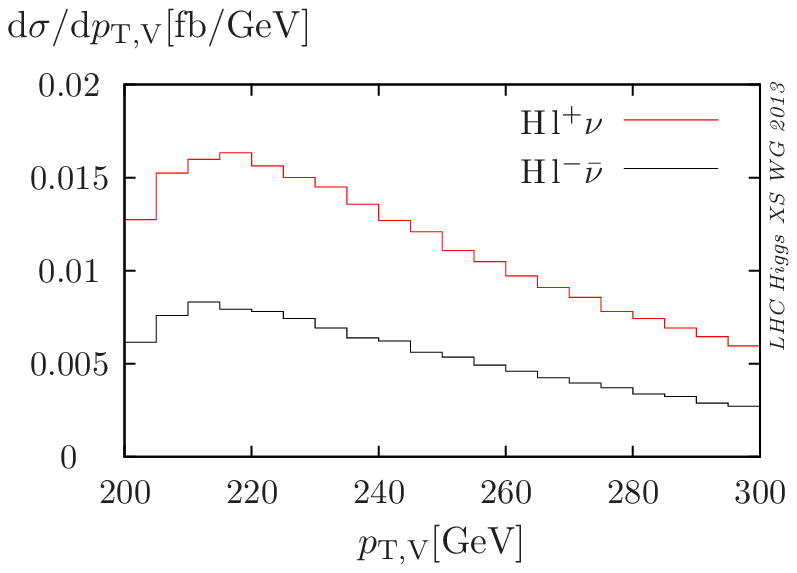}
\hfill
\includegraphics[width=7.5cm]{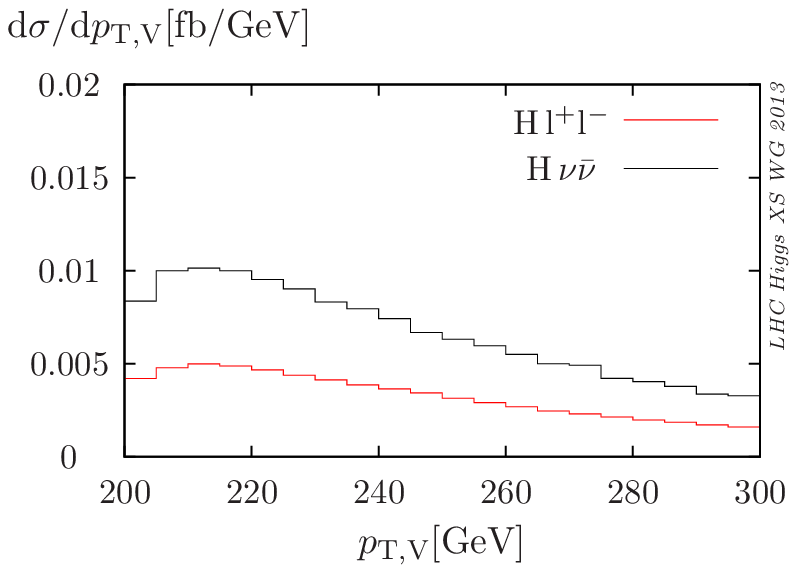}
\includegraphics[width=7.5cm]{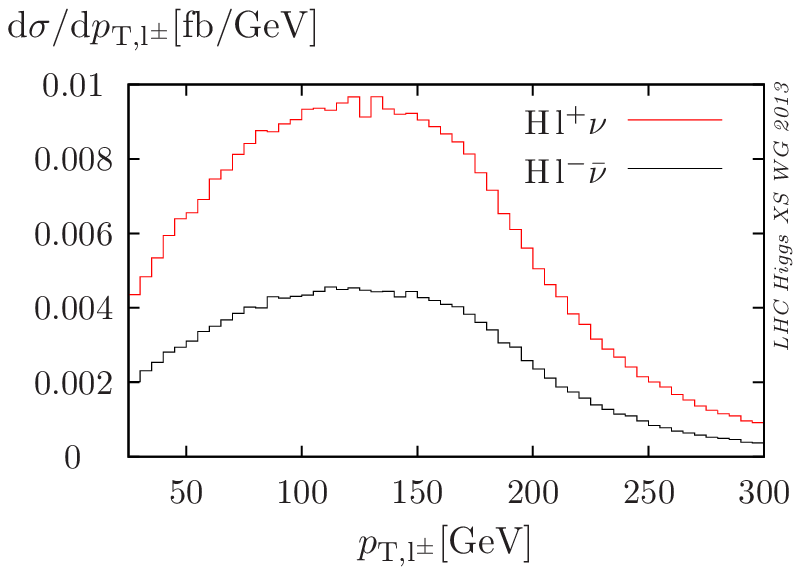}
\hfill
\includegraphics[width=7.5cm]{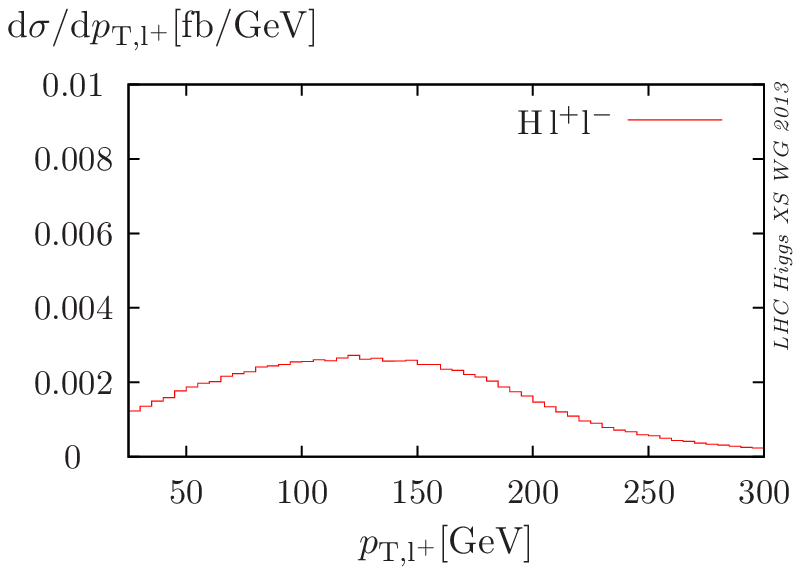}
\includegraphics[width=7.5cm]{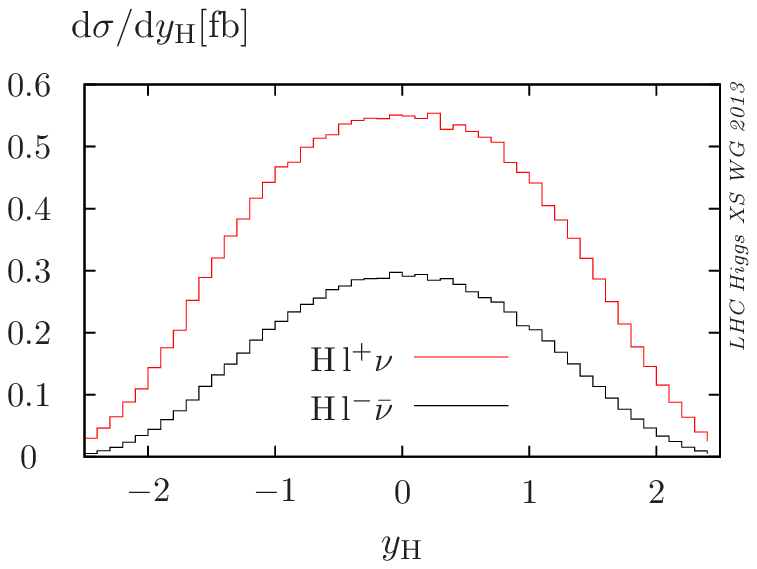}
\hfill
\includegraphics[width=7.5cm]{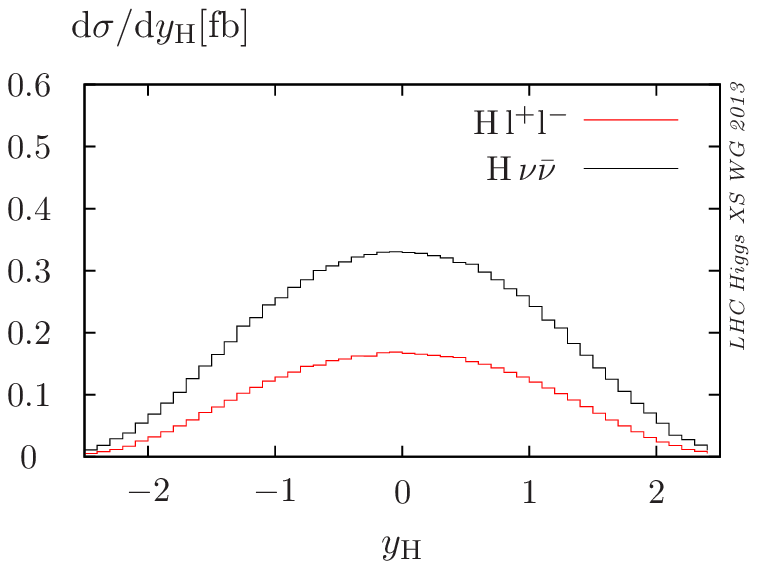}
\vspace{0.5cm}
\caption{
Predictions for the
$p_{\mathrm{T},\PH}$, $p_{\mathrm{T},\PV}$,
$p_{\mathrm{T},\Pl}$, and $y_{\PH}$ distributions (top to bottom)
for Higgs strahlung off \PW\ bosons (left) and \PZ\ bosons (right)
for boosted Higgs bosons at
the $8\UTeV$ LHC for $\MH=126\UGeV$. 
}
\label{fig:WHZH-dxs-abs}
\end{figure}

\begin{figure}
\includegraphics[width=7.5cm]{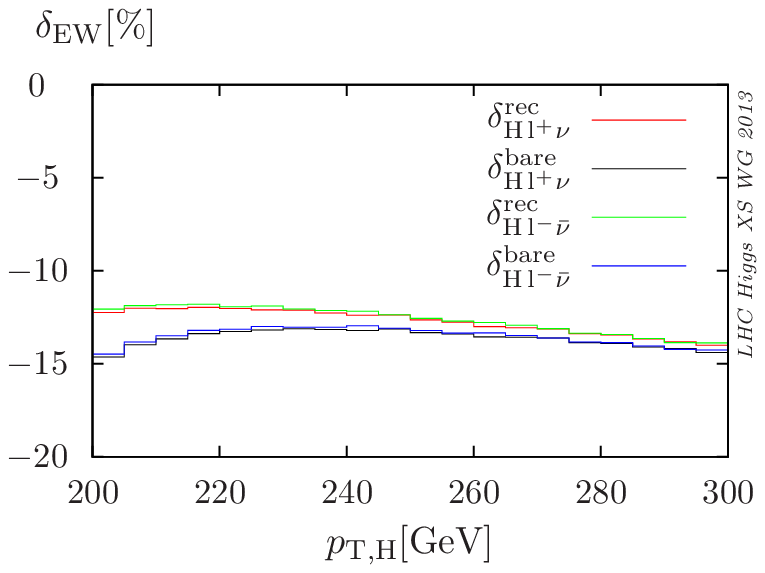}
\hfill
\includegraphics[width=7.5cm]{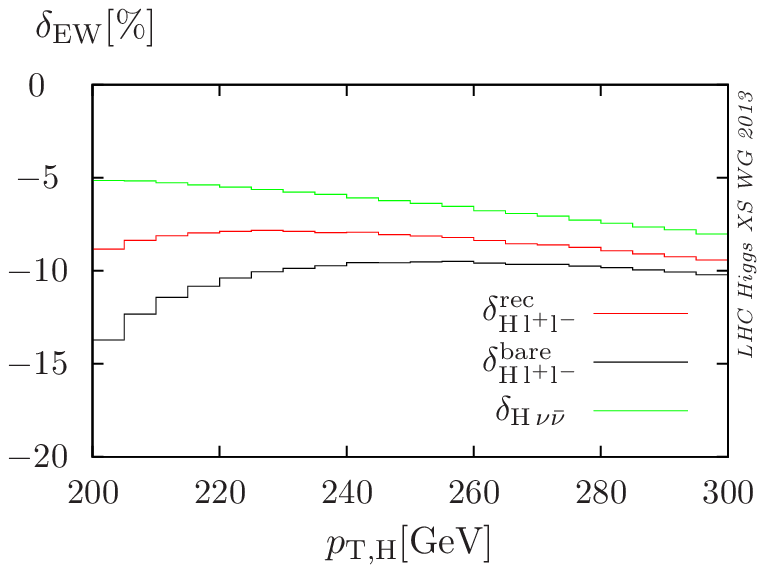}
\includegraphics[width=7.5cm]{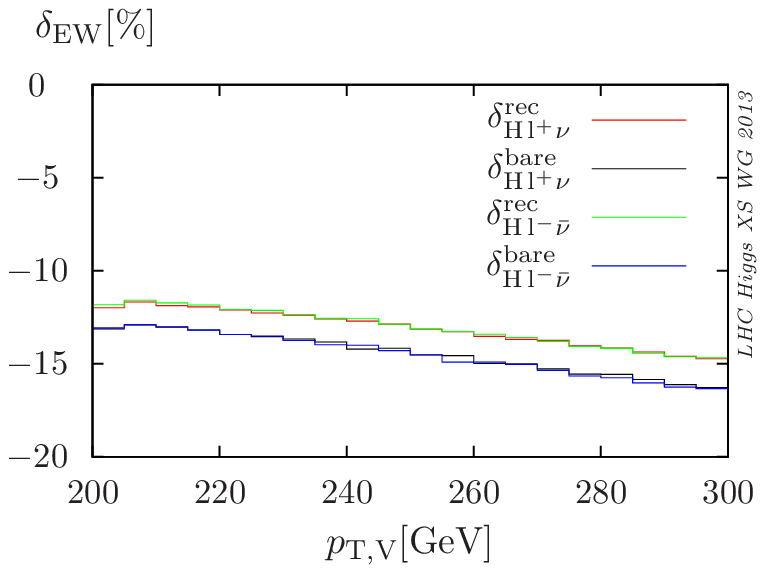}
\hfill
\includegraphics[width=7.5cm]{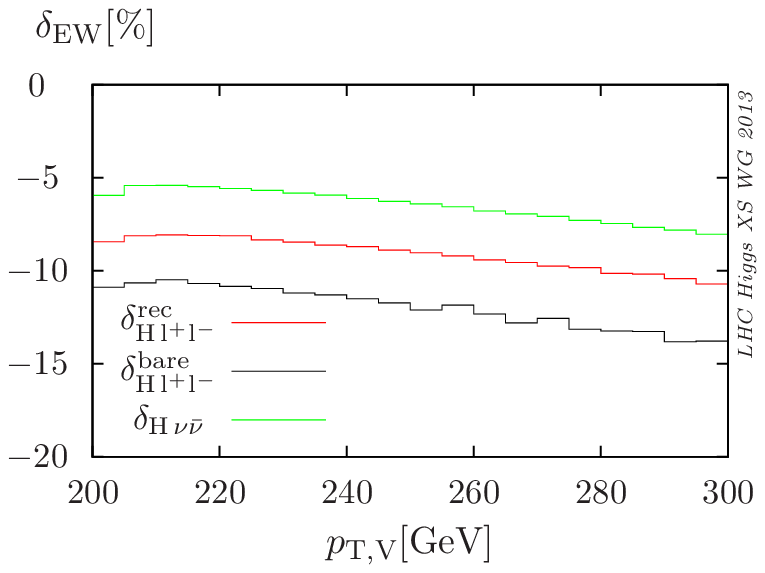}
\includegraphics[width=7.5cm]{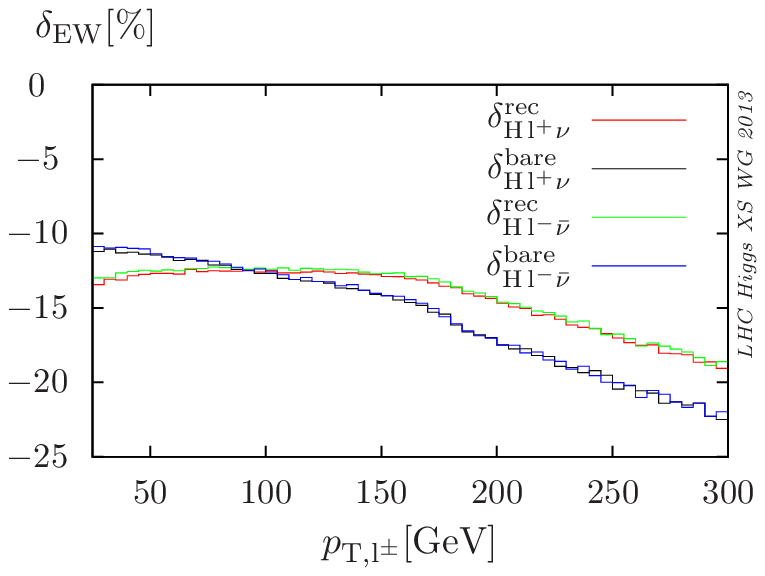}
\hfill
\includegraphics[width=7.5cm]{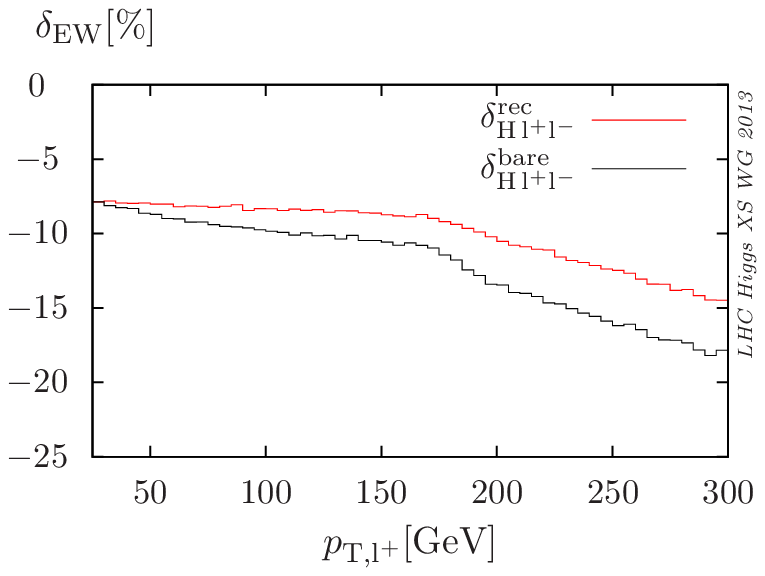}
\includegraphics[width=7.5cm]{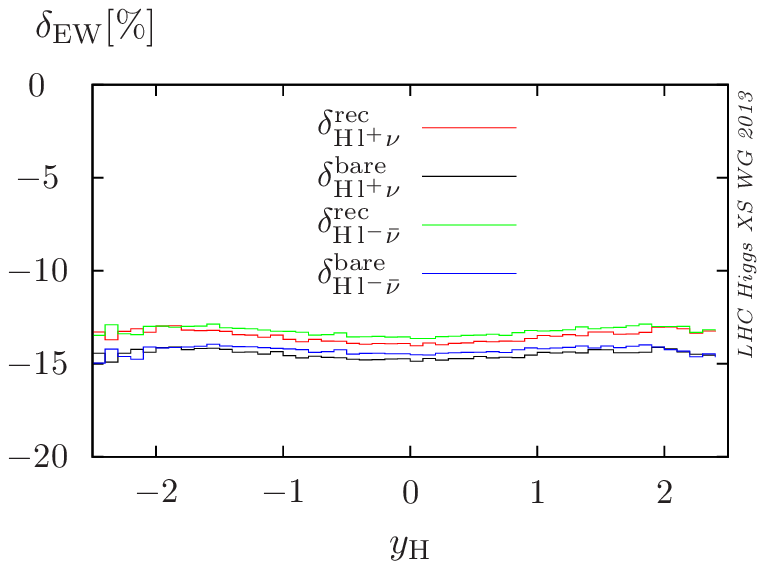}
\hfill
\includegraphics[width=7.5cm]{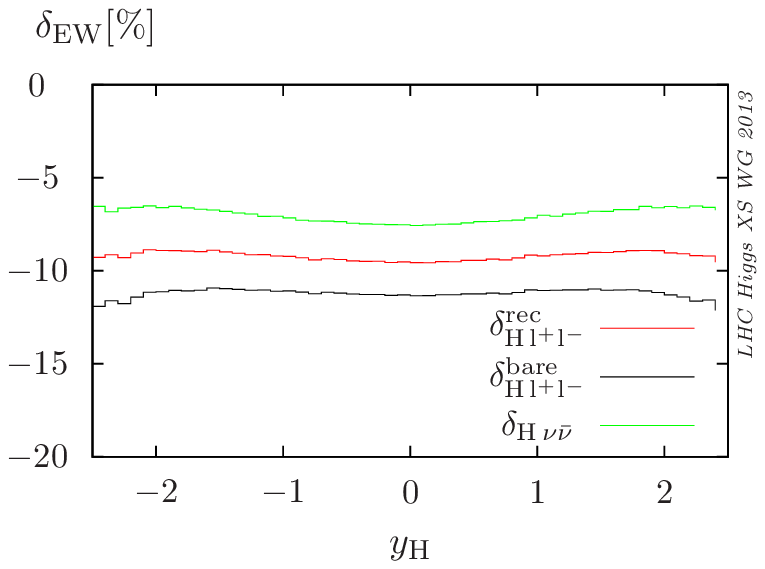}
\vspace*{1.5cm}
\caption{
Relative EW corrections for the
$p_{\mathrm{T},\PH}$, $p_{\mathrm{T},\PV}$,
$p_{\mathrm{T},\Pl}$, and $y_{\PH}$ distributions (top to bottom)
for Higgs strahlung off \PW\ bosons (left) and \PZ\ bosons (right)
for boosted Higgs bosons at
the $8\UTeV$ LHC for $\MH=126\UGeV$.
}
\label{fig:WHZH-dxs-ew}
\end{figure}

\begin{figure}
\includegraphics[width=7.5cm]{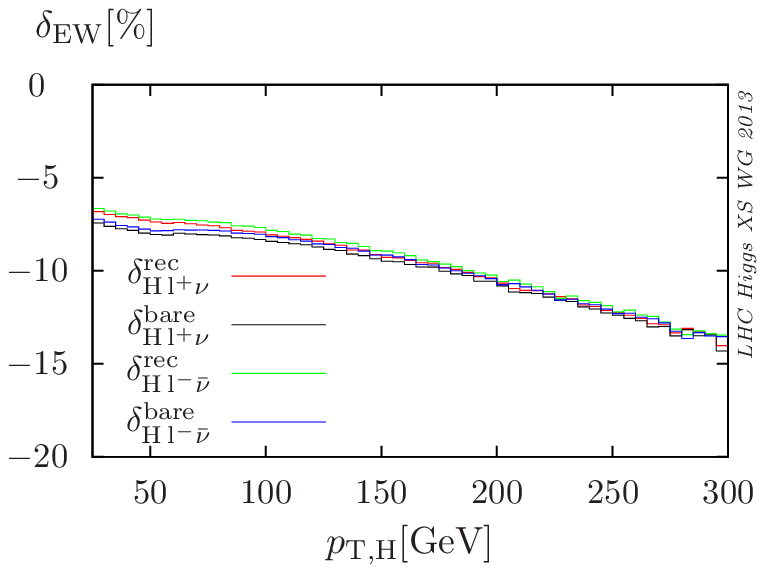}
\hfill
\includegraphics[width=7.5cm]{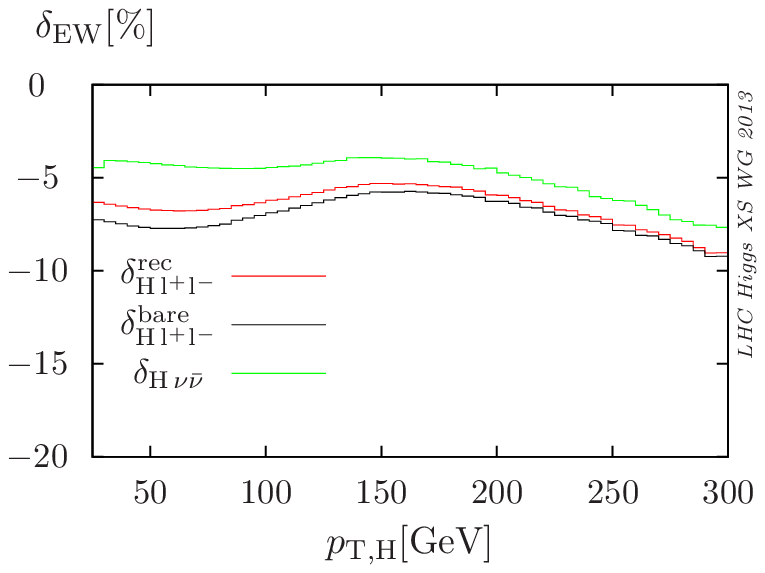}
\includegraphics[width=7.5cm]{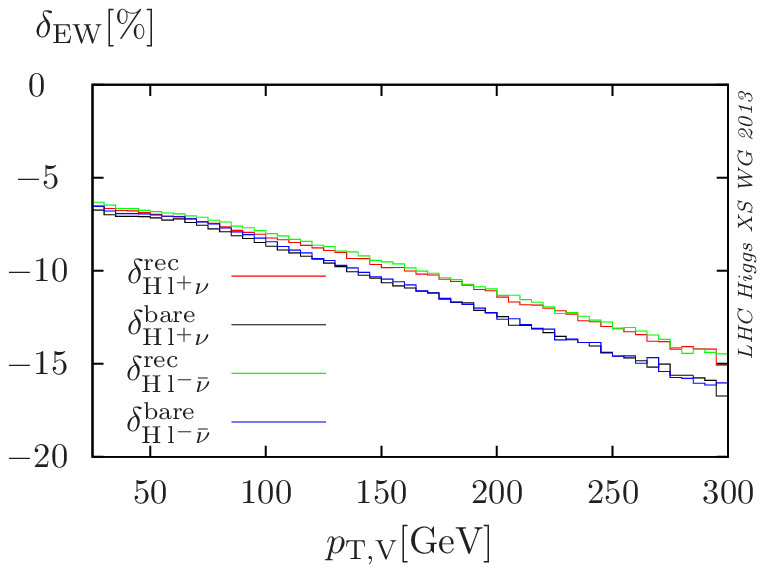}
\hfill
\includegraphics[width=7.5cm]{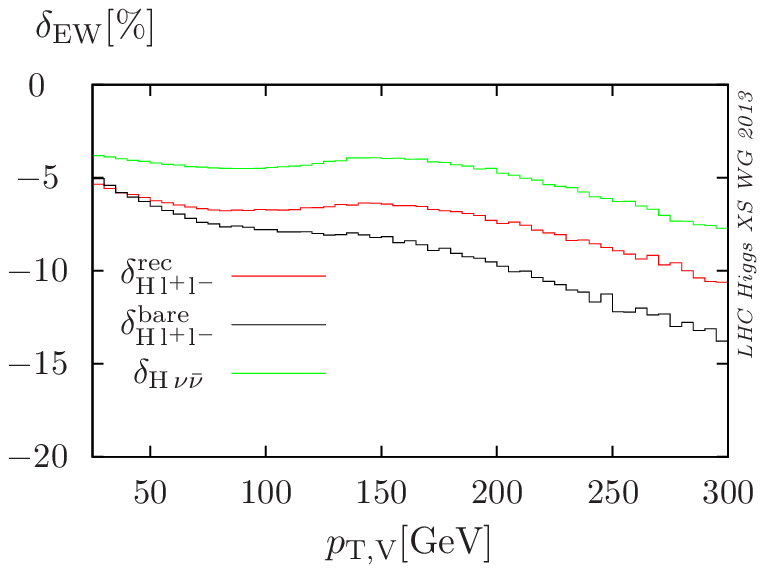}
\includegraphics[width=7.5cm]{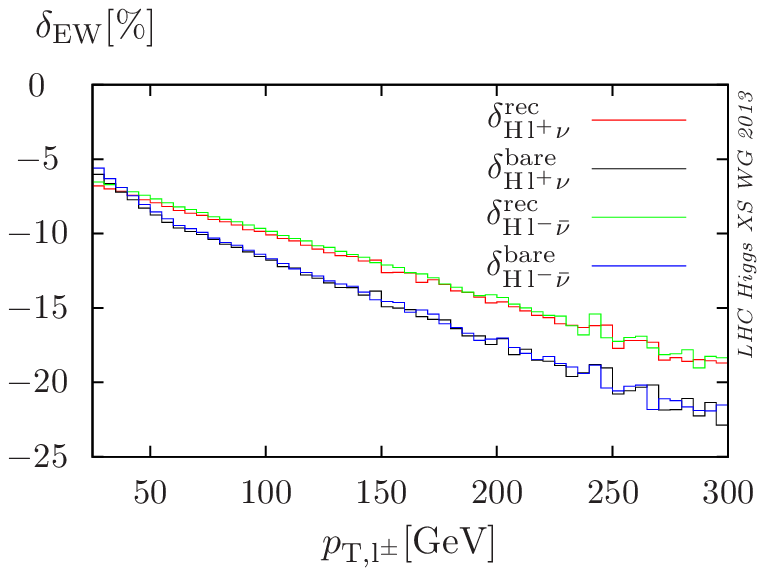}
\hfill
\includegraphics[width=7.5cm]{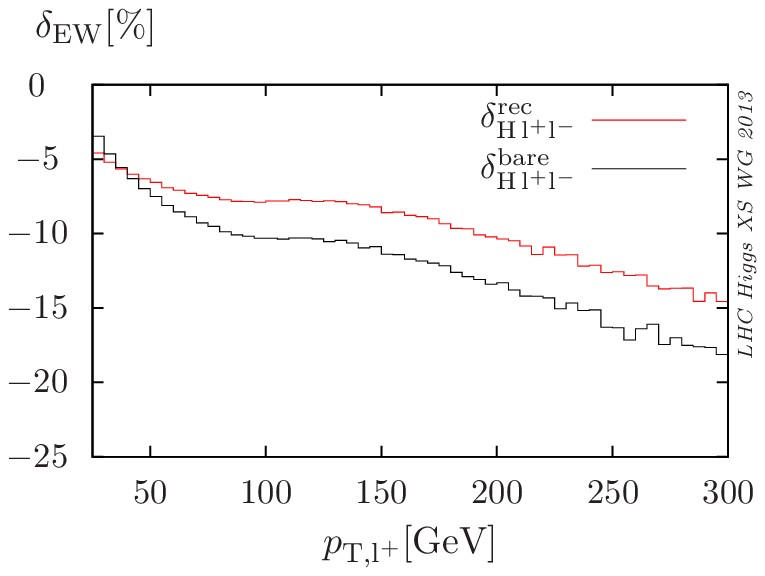}
\includegraphics[width=7.5cm]{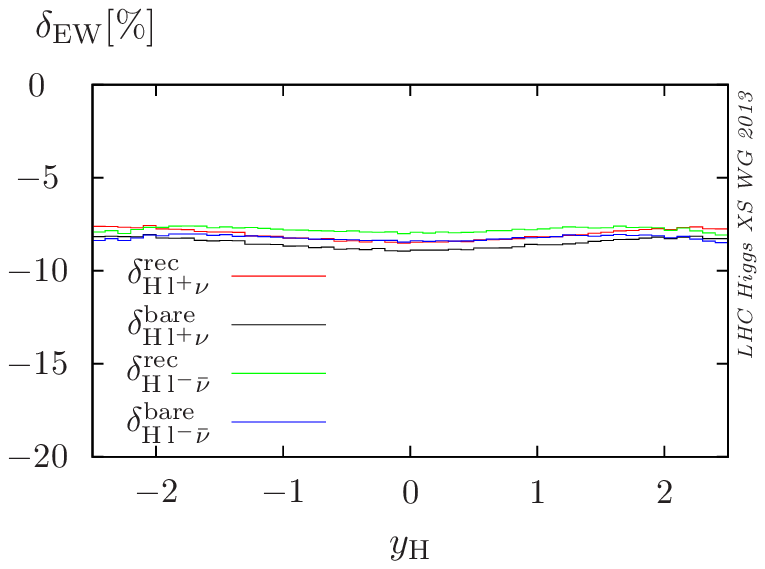}
\hfill
\includegraphics[width=7.5cm]{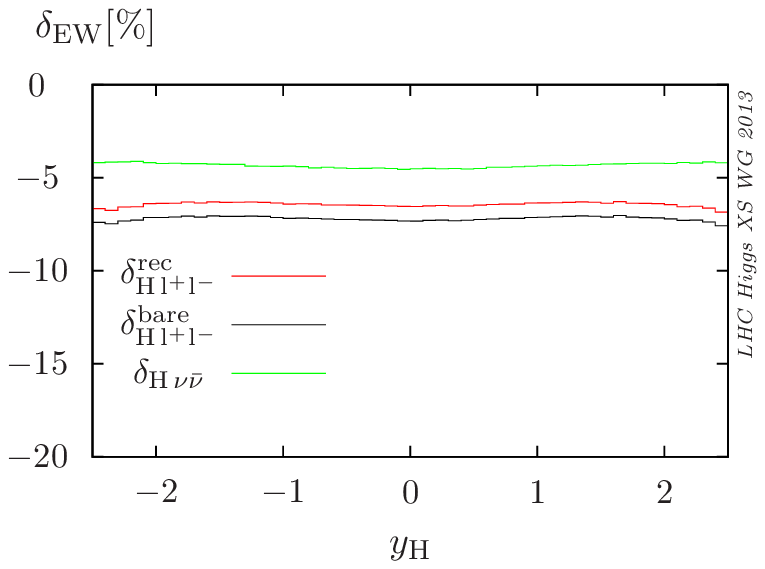}
\vspace*{1.5cm}
\caption{
Relative EW corrections for the
$p_{\mathrm{T},\PH}$, $p_{\mathrm{T},\PV}$,
$p_{\mathrm{T},\Pl}$, and $y_{\PH}$ distributions (top to bottom)
for Higgs strahlung off \PW\ bosons (left) and \PZ\ bosons (right)
for the basic cuts at the $8\UTeV$ LHC for $\MH=126\UGeV$.
}
\label{fig:WHZH-dxs-ew2}
\end{figure}

Figure~\ref{fig:WHZH-dxs-abs} shows the
distributions for the various $\PV\PH$ production channels at the LHC with a CM
energy of $8\UTeV$
for a Higgs-boson mass of $\MH=126\UGeV$ in the boosted-Higgs scenario, where the cuts
(\ref{eq:VHpTcuts}) on the Higgs and $\PW/\PZ$ transverse momenta
are applied. The only differences to the results
shown in Fig.~56 of \Bref{Dittmaier:2012vm} concerns the new value of $\MH$
and the transition from NLO QCD to NNLO QCD for $\PZ\PH$ production.
Qualitatively the results look very similar, so that the discussion 
presented in \Bref{Dittmaier:2012vm} still holds.
This applies, in particular, to the
respective EW corrections which are depicted in \refF{fig:WHZH-dxs-ew} for
the ``bare'' and ``rec'' treatments of radiated photons. 
The smallness of the difference between the two variants, which is about
$1{-}3\%$, shows that the bulk of the EW corrections, which are typically 
$-(10{-}15)\%$, is of pure weak origin.
While the EW corrections to rapidity distributions are flat and resemble the 
ones to the respective integrated cross sections, the corrections to $\pT$
distributions show the typical tendency to larger negative values with
increasing $\pT$ (weak Sudakov logarithms).
Finally, for comparison we show the EW corrections in the boosted-Higgs regime,
where the transverse momenta of the Higgs and $\PW/\PZ$ bosons are
$\gsim200\UGeV$, to the scenario of \refF{fig:WHZH-dxs-ew2}
where only basic isolation cuts are kept,
i.e.\ the cuts (\ref{eq:VHpTcuts}) on $p_{\mathrm{T},\PH}$ and $p_{\mathrm{T},\PW/\PZ}$ are dropped.
As already noted in \Bref{Dittmaier:2012vm},
switching from the basic cuts to the boosted-Higgs scenario
increases the size of the EW corrections by about $5\%$ in the negative 
direction. 

In spite of the theoretical improvement by the transition from NLO QCD to
NNLO QCD in the $\PZ\PH$ channel, the estimate of the relative uncertainties
shown in Table~19 of \Bref{Dittmaier:2012vm} remains valid, because in
the predictions for the differential cross sections the contribution of the
$\Pg\Pg\to\PZ\PH$ channel are not (yet) included. The change from $\MH=120\UGeV$
to $\MH=126\UGeV$ leaves the error estimate untouched as well.

\clearpage

\newpage
%\documentclass[12pt]{cernyrep}
%
%\usepackage{epsf,amsmath,amssymb,graphicx,longtable}
%\usepackage{lhchiggs,heppennames2,cernunits}
%\usepackage{rotating}
%\usepackage{color}
%\usepackage{subfigure}
%\usepackage{lscape}
%\usepackage{axodraw}
%\usepackage{lineno}
%% for GKT
%\usepackage{fullpage}
%\usepackage{multirow}
%\usepackage{slashed}
%\usepackage{cite}
%
%\begin{document}
%
\providecommand{\muR}{\mu_{\mathrm{R}}}
\providecommand{\muF}{\mu_{\mathrm{F}}}
\providecommand{\pt}{p_\mathrm{T}}
\providecommand{\ie}{{\it i.e.~}}
\providecommand{\eg}{{\it e.g.~}}
\providecommand{\etal}{{\it et al.~}}

%%%%%%%%%%%%%%%%% personal special commands (Pozzorini) %%%%%%%%%
\providecommand{\muBDDP}{\mu_{\mathrm{dyn}}}
\providecommand{\muMG}{\mu_{\mathrm{MG}}}
\providecommand{\bind}{\mathrm{b}}
\providecommand{\rF}{\mathrm{F}}
\providecommand{\Pbh}{{\mathrm{b}_\mathrm{h}}}
\providecommand{\Pbs}{{\mathrm{b}_\mathrm{s}}}
\providecommand{\Pbp}{{\mathrm{b}_+}}
\providecommand{\Pbm}{{\mathrm{b}_-}}
\providecommand{\sregi}{{S1}}
\providecommand{\sregii}{{S2}}
%%%%%%%%%%%%%%%%%%%%%%%%%%%%%%%%%%%%%%%%%%%%%%%%%%%%%%%%%%%%%%%%%
% More space on page for figures and tables
\renewcommand{\topfraction}{0.85}
\renewcommand{\textfraction}{0.1}
%%%%%%%%%%%%%%%%%%%%%%%%%%%%%%%%%%%%%%%%%%%%%%%%%%%%%%%%%%%%%%%%%
%Abbreviations (GKT)
%
\newcommand{\powhel}{\textsc{PowHel}}
\newcommand{\helac}{\textsc{HELAC}}
\newcommand{\helacdipoles}{\textsc{HELAC-Dipoles}}
\newcommand{\helaconeloop}{\textsc{HELAC-Oneloop}}
\newcommand{\helacphegas}{\textsc{HELAC-PHEGAS}}
\newcommand{\helacnlo}{\textsc{HELAC-NLO}}
\newcommand{\mcatnlo}{\textsc{MC@NLO}}
\newcommand{\amcatnlo}{\textsc{aMC@NLO}}
\newcommand{\pythia}{\textsc{PYTHIA}}
\newcommand{\herwig}{\textsc{HERWIG}}
\newcommand{\lhapdf}{\textsc{LHAPDF}}
\newcommand{\madgraph}{\textsc{MADGRAPH}}
\newcommand{\madloop}{\textsc{MadLoop}}
\newcommand{\madspin}{\textsc{MadSpin}}
\newcommand{\madfks}{\textsc{MadFKS}}
\newcommand{\openloops}{\textsc{OpenLoops}}
\newcommand{\collier}{\textsc{COLLIER}}
\newcommand{\powheghelac}{\textsc{PowHel}}
\newcommand{\powhegbox}{\textsc{POWHEG-Box}}
\newcommand{\powheg}{\textsc{POWHEG}}
\newcommand{\sherpa}{\textsc{Sherpa}}
\newcommand{\fastjet}{\textsc{FastJet}}
\newcommand{\mctruth}{\textsc{MCtruth}}
\newcommand{\pb}{\,\mathrm{pb}}
\newcommand{\fb}{\,\mathrm{fb}}
\newcommand{\mev}{\ensuremath{\,\mathrm{MeV}}}
\newcommand{\gev}{\ensuremath{\,\mathrm{GeV}}}
\newcommand{\tev}{\ensuremath{\,\mathrm{TeV}}}
\newcommand{\ud}{\mathrm{d}}
\newcommand{\pTl}{\ensuremath{p_{\perp}^{\ell}}}
\newcommand{\pTlp}{\ensuremath{p_{\perp}^{\ell^+}}}
\newcommand{\pTlm}{\ensuremath{p_{\perp}^{\ell^-}}}
\newcommand{\pTj}{\ensuremath{p_{\perp,j_1}}}
\newcommand{\pTt}{\ensuremath{p_{{\rm t}\perp}}}
\newcommand{\pTmiss}{\ensuremath{p_{\perp}^{\rm miss}}}
\newcommand{\ptc}{\ensuremath{p_{\perp}^{\rm t.c.}}}
%\newcommand{\LO}{{\rm LO}}
%\newcommand{\NLO}{{\rm NLO}}
%
% Referencing
%
%\newcommand\Ref[1] {Ref.\,\cite{#1}}
%\newcommand\Refs[1]{Refs.\,\cite{#1}}
%\newcommand\eqn[1] {Eq.\,(\ref{#1})}
%\newcommand\eqns[2]{Eqs.\,(\ref{#1}) and~(\ref{#2})}
%\newcommand\eqnss[2]   {Eqs.\,(\ref{#1})--(\ref{#2})}
%\newcommand\fig[1] {Fig.\,{\ref{#1}}}
%\newcommand\figs[2]{Figs.\,{\ref{#1}} and ~\ref{#2}}
%\newcommand\figss[2]   {Figs.\,{\ref{#1}}--\ref{#2}}
%\newcommand\sect[1]{Sect.\,{\ref{#1}}}
%\newcommand\sects[2]   {Sects.\,(\ref{#1}) and~(\ref{#2})}
%\newcommand\app[1] {Appendix~\ref{#1}}
%\newcommand\tab[1] {Table~\ref{#1}}
%\newcommand\tabs[2]{Tables~{\ref{#1}} and ~\ref{#2}}
%%%%%%%%%%%%%%%%%%%%%%%%%%%%%%%%%%%%%%%%%%%%%%%%%%%%%%%%%%%%%%%%%%%%

\section{$\PQt\PAQt\PH$ process \footnote{
    C.~Neu, C.~Potter, L.~Reina, A.~Rizzi, M.~Spira, (eds.);
    P.~Artoisenet, F.~Cascioli, R.~Frederix, M.~V.~Garzelli,
    S.~Hoeche, A.~Kardos, F.~Krauss, P.~Maierh\"ofer, O.~Mattelaer,
    N.~Moretti, S.~Pozzorini, R.~Rietkerk, J.~Thompson, Z.~Tr\'ocs\'anyi,
    D.~Wackeroth }}
\label{sec:ttH}

The experimental discovery of a Higgs boson with mass around
$125-126~\UGeV$ casts new light on the role played by the associated
production of a Higgs boson with a pair of top quarks, i.e.
$\PQq\PAQq/\Pg\Pg\to\PH\PQt\PAQt$. Detailed studies of the properties
of the discovered Higgs boson will be used to confirm or exclude the
Higgs mechanism of electroweak symmetry breaking as minimally
implemented in the SM. In this context, the measurement of
the $\PQt\PAQt\PH$ production rate can provide direct information on
the top-Higgs Yukawa coupling, probably the most crucial coupling
to fermions.

Using the NLO codes developed %by the Authors of
\Brefs{Beenakker:2001rj,Beenakker:2002nc,Reina:2001sf,Dawson:2002tg},
in a previous report~\cite{Dittmaier:2011ti} we studied the inclusive
$\PQt\PAQt\PH$ production at both $\sqrt{s}=7$ and $14\UTeV$ and we
provided a breakdown of the estimated theoretical error from
renormalization- and factorization-scale dependence, from $\alphas$,
and from the choice of Parton Distribution Functions (PDFs). The total
theoretical errors were also estimated combining the uncertainties
from scale dependence, $\alphas$ dependence, and PDF dependence
according to the recommendation of the LHC Higgs Cross Section Working
Group~\cite{Dittmaier:2011ti}. For low Higgs-boson masses, the
theoretical errors typically amount to $10-15\%$ of the corresponding
cross sections. In Sect.~\ref{subsec:ttH-sigma-8-TeV} we will repeat
the same exercise for $\sqrt{s}=8\UTeV$ using finer
Higgs-mass steps, in particular around $125-126\UGeV$ and adding some
steps above $300\UGeV$.

In the context of this workshop, we continued the study of the
$\PQt\PAQt\PH$ signal and provided in a second
report~\cite{Dittmaier:2012vm} a thorough study of the interface of
the NLO QCD calculation of $\PQt\PAQt\PH$ with parton shower Monte
Carlo programs (\pythia{} and \herwig{}). We compared results obtained
using the \mcatnlo{} method via \amcatnlo~\cite{Frederix:2011zi} and
results obtained using the \powheg{} method via
\powhel~\cite{Garzelli:2011vp}. The two implementations were
found to be in very good agreement and should be considered as the
state-of-the-art tools to obtain theoretical predictions and
theoretical uncertainties on total and differential cross sections for
$\PQt\PAQt\PH$ production that also include experimental cuts and
vetos on the final-state particles and their decay products. 

In this report we expand on the studies presented
in~\cite{Dittmaier:2012vm} by presenting in
\Sref{subsec:ttH-sherpa-ps-uncertainty} a new study of
parton-shower uncertainty obtained by interfacing the NLO calculation
of \Bref{Reina:2001sf,Dawson:2002tg} with the \sherpa{} Monte-Carlo
program~\cite{Gleisberg:2008ta}, and in
\Sref{subsec:ttH-spin_corr_ttH_madspin} a novel implementation of
the decay of heavy resonances in NLO Monte-Carlo events as implemented
in \amcatnlo{} via the \madspin~\cite{Artoisenet:2012st} framework.

Finally, we focus on one of the main backgrounds for $\PQt\PAQt\PH$
production, namely $\PQt\PAQt\PQb\PAQb$ production and we present two
dedicated studies. In \Sref{subsec:ttH-NLO_ttbb} we present new
parton-level NLO predictions for $\PQt\PAQt\PQb\PAQb$ production at
$8\UTeV$, obtained within \textsc{OpenLoops} \cite{Cascioli:2011va} and
\textsc{Sherpa}~\cite{Krauss:2001iv,Gleisberg:2008ta,Gleisberg:2007md}.
On the other hand, in \Sref{subsec:ttH-ttH_ttbb_powhel+powheg}
the NLO calculation of $\PQt\PAQt\PQb\PAQb$ is for the first time
consistently interfaced with the \pythia{} parton-shower Monte Carlo
using the \powhel{} framework~\cite{Kardos:2013vxa} and compared with
the $\PQt\PAQt\PH$ signal at $14\UTeV$.

\subsection{Theoretical uncertainty on the parton level $\PQt\PAQt\PH$ total cross
  section at $7$ and $8\UTeV$}
\label{subsec:ttH-sigma-8-TeV}

In this section we provide results for the inclusive NLO signal
cross section at $\sqrt{s}=7$ and $8\UTeV$ for different values of Higgs
masses. The mass steps have been chosen following the recommendation of
the Higgs Cross Section Working Group, up to $\MH=400\UGeV$.  For
consistency, we have kept the same setup used in
\Bref{Dittmaier:2011ti} where analogous results for $\sqrt{s}=7$
and $14\UTeV$ were presented.  In summary, we used the on-shell
top-quark mass fixed at $\Mt=172.5\UGeV$ and did not include the
parametric uncertainties due to the experimental error on the
top-quark mass. Loop diagrams with a bottom-quark loop were calculated
using the $\PQb$-quark pole mass. The top-quark Yukawa coupling was
set to $y_{\PQt}=\Mt(\sqrt{2}\GF)^{1/2}$.  The central scale has been
chosen as $\muR=\muF=\mu_0=\Mt+\MH/2$.  We have used the
MSTW2008 \cite{Martin:2009iq,Martin:2009bu}, CTEQ6.6
\cite{Pumplin:2002vw} and NNPDF2.0 \cite{Ball:2010de} sets of parton
density functions.  The central values of the strong coupling constant
have been implemented according to the corresponding PDFs for the sake
of consistency.

The uncertainties due to scale variations of a factor of two around
the central scale $\mu_0$ as well as the 68\% CL uncertainties due to
the PDFs and the strong coupling $\alphas$ have been calculated and
are given explicitly in 
\refT{tab:YRHXS_ttH_7TeV_1} to \refT{tab:YRHXS_ttH_7TeV_6} for
$\sqrt{s}=7\UTeV$ and in 
\refT{tab:YRHXS_ttH_8TeV_1} to \refT{tab:YRHXS_ttH_8TeV_6} for $\sqrt{s}=8\UTeV$.
We exhibit the central values and the PDF+$\alphas$ uncertainties according to
the envelope method of the PDF4LHC recommendation and the relative
scale variations using MSTW2008 PDFs.

\subsection{Theory uncertainties in the simulation of $\PQt\PAQt\PH$}
\label{subsec:ttH-sherpa-ps-uncertainty}

\begin{figure}
\begin{center}
\begin{tabular}{cc}
\includegraphics[width=0.48\textwidth]{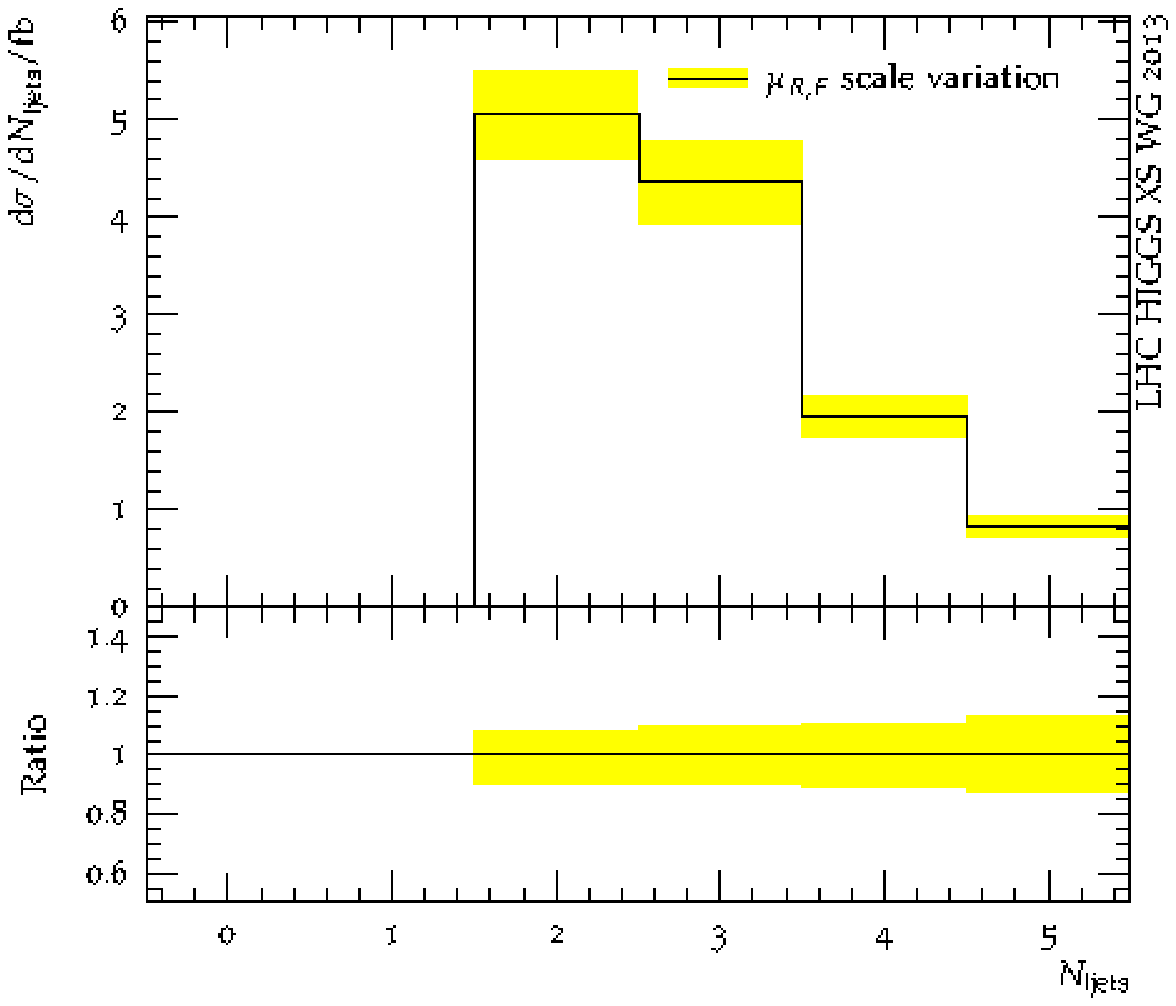} &
\includegraphics[width=0.48\textwidth]{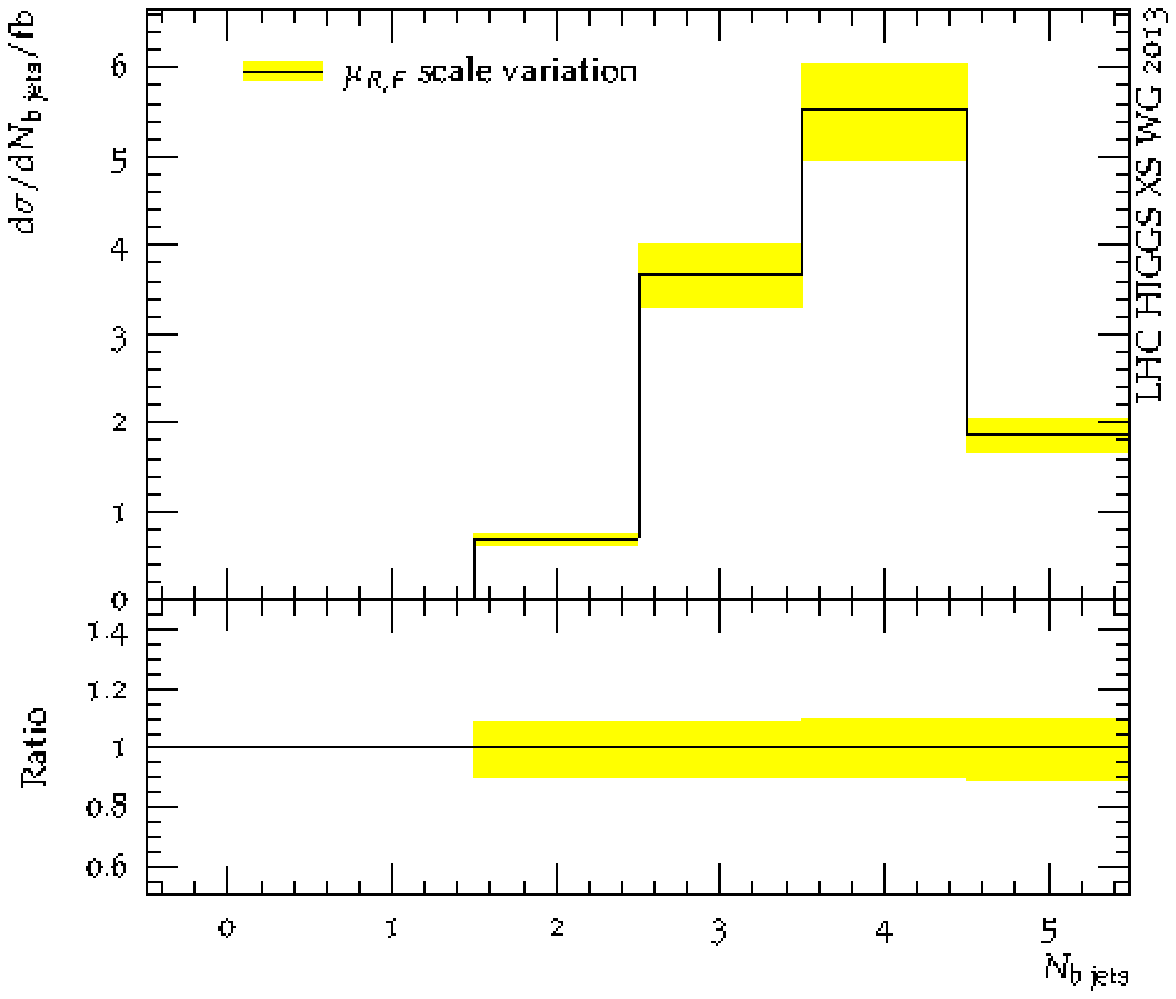}
\end{tabular}
\caption{The number of light (left panel) and $\PQb$ jets (right panel)
         in the acceptance region for $\PQt\PAQt\PH$ events at the 
         $8\UTeV$ LHC.  Uncertainties due to renormalization and factorization
         scale variations (yellow band) as given in an \protect\MCatNLO 
         simulation are also indicated.} 
         \label{fig:ttH_uncert_njets}
\end{center}
\end{figure}

\begin{figure}
\begin{center}
\begin{tabular}{cc}
\includegraphics[width=0.48\textwidth]{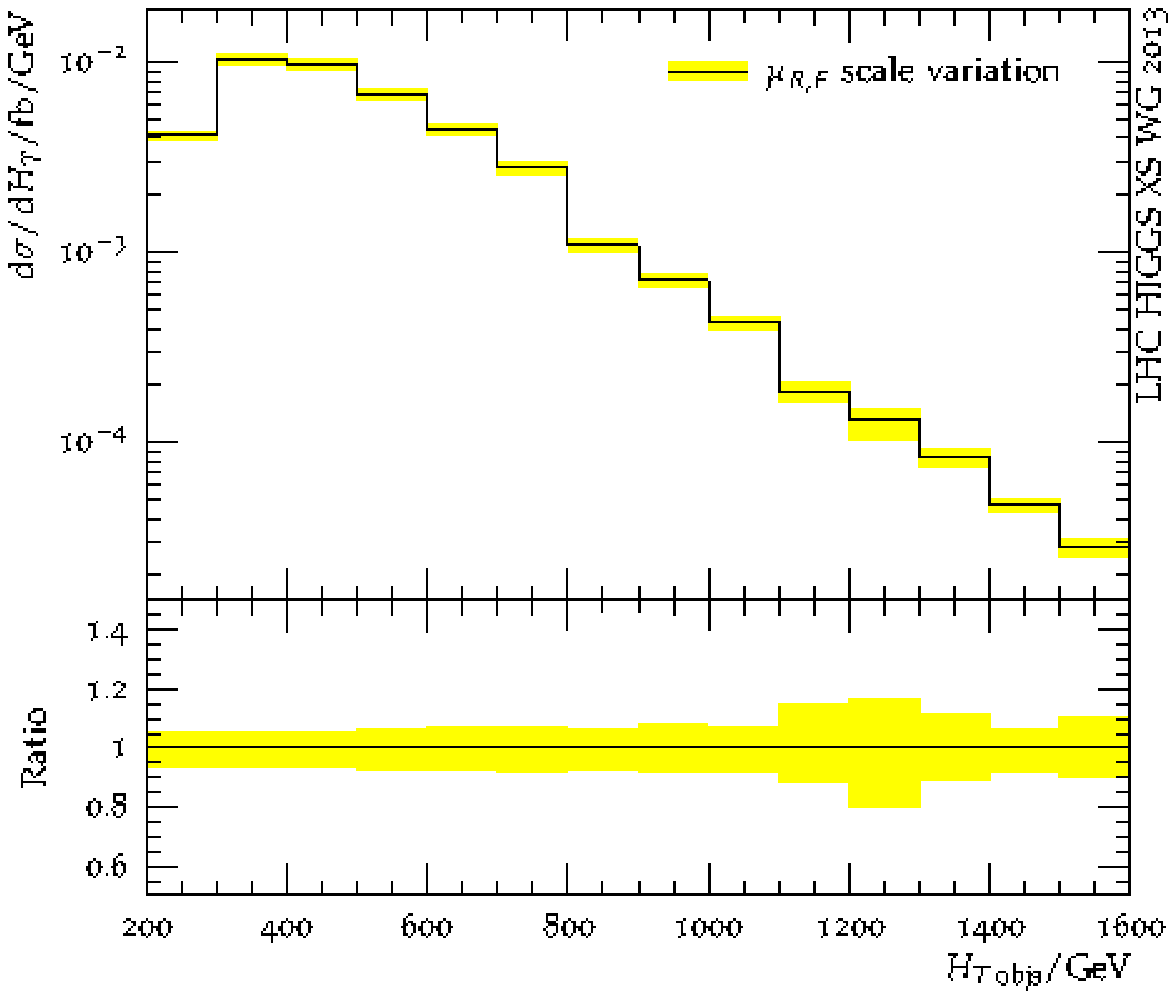} &
\includegraphics[width=0.48\textwidth]{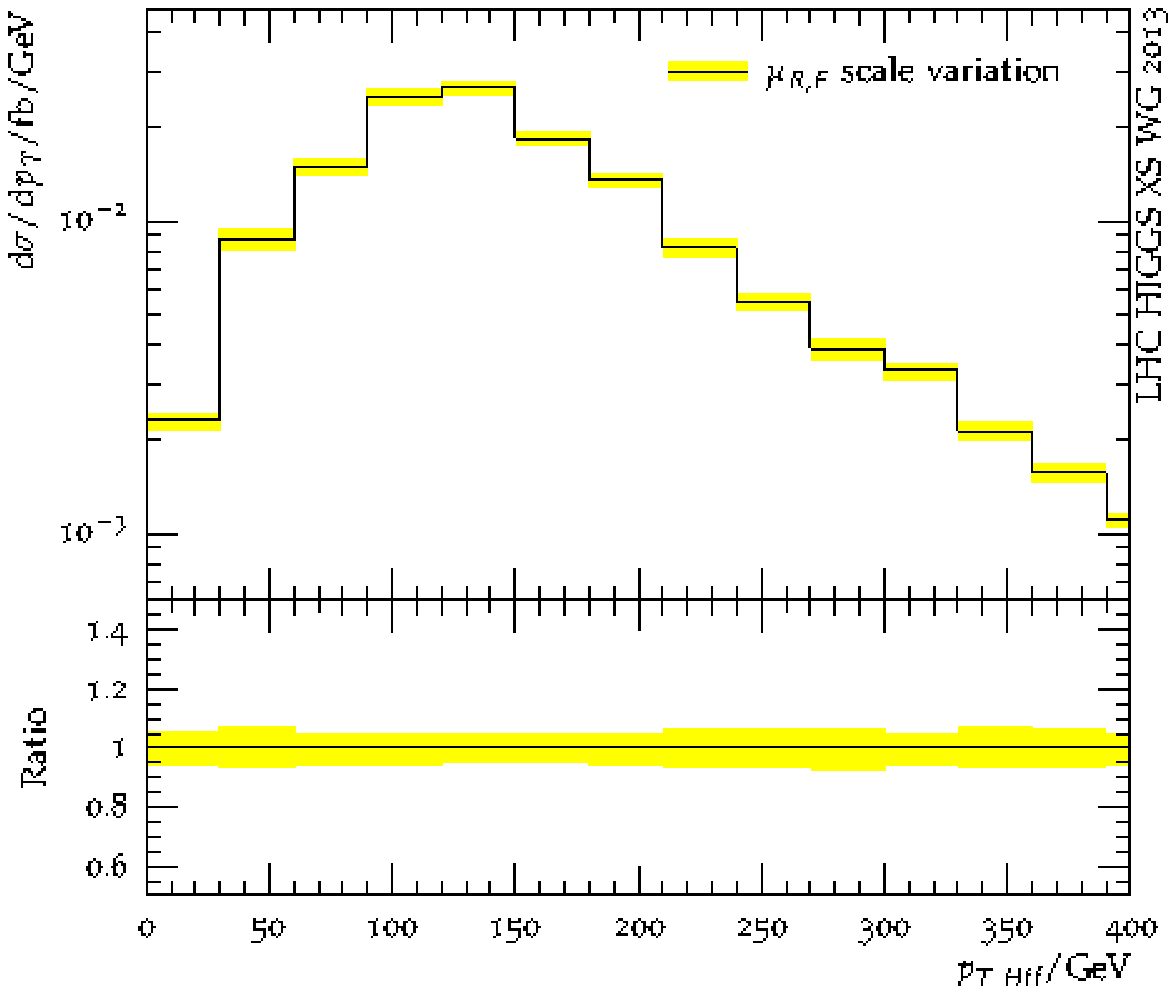}
\end{tabular}
\caption{$H_{\mathrm{T}}$(left panel) and $\pT^{\PQt\PAQt\PH}$
(right panel) distribution in the acceptance region for $\PQt\PAQt\PH$
         events at the $8\UTeV$ LHC.  Uncertainties due to
         renormalization and factorization scale variations (yellow
         band) as given in an \protect\MCatNLO simulation are also
         indicated.}  \label{fig:ttH_uncert_incl_2}
\end{center}
\end{figure}

\begin{figure}
\begin{center}
\begin{tabular}{ccc}
\includegraphics[width=0.32\textwidth]{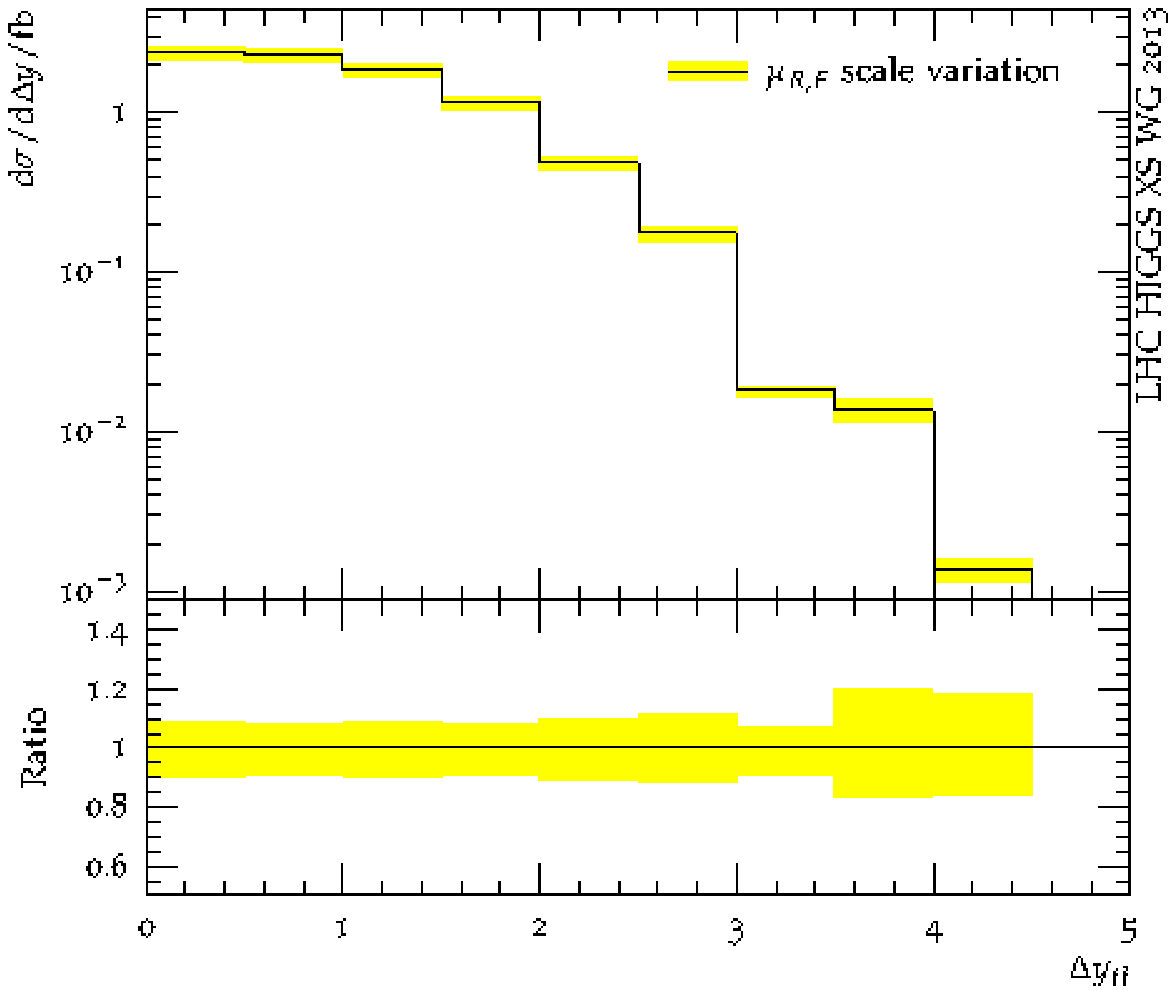}  &
\includegraphics[width=0.32\textwidth]{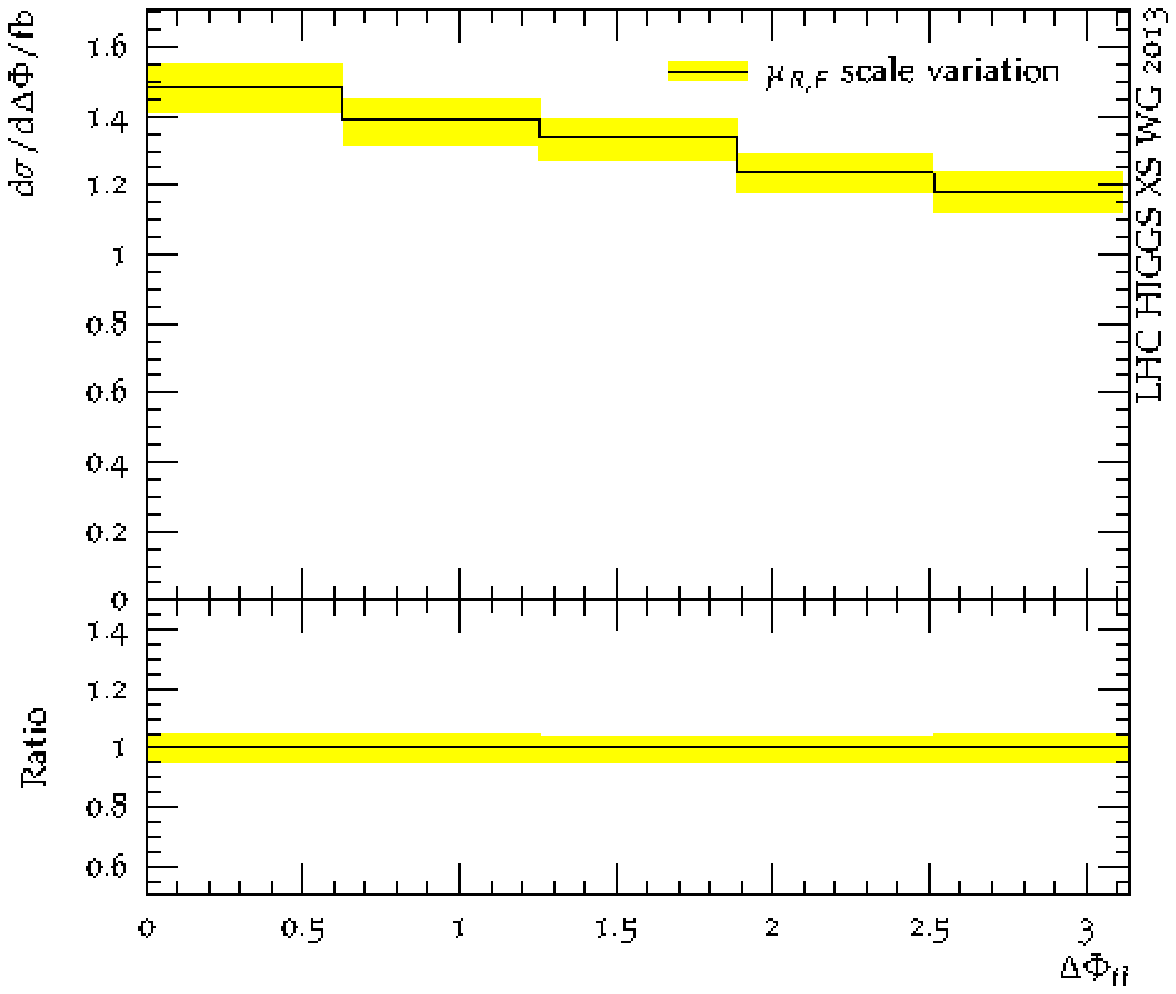} &
\includegraphics[width=0.32\textwidth]{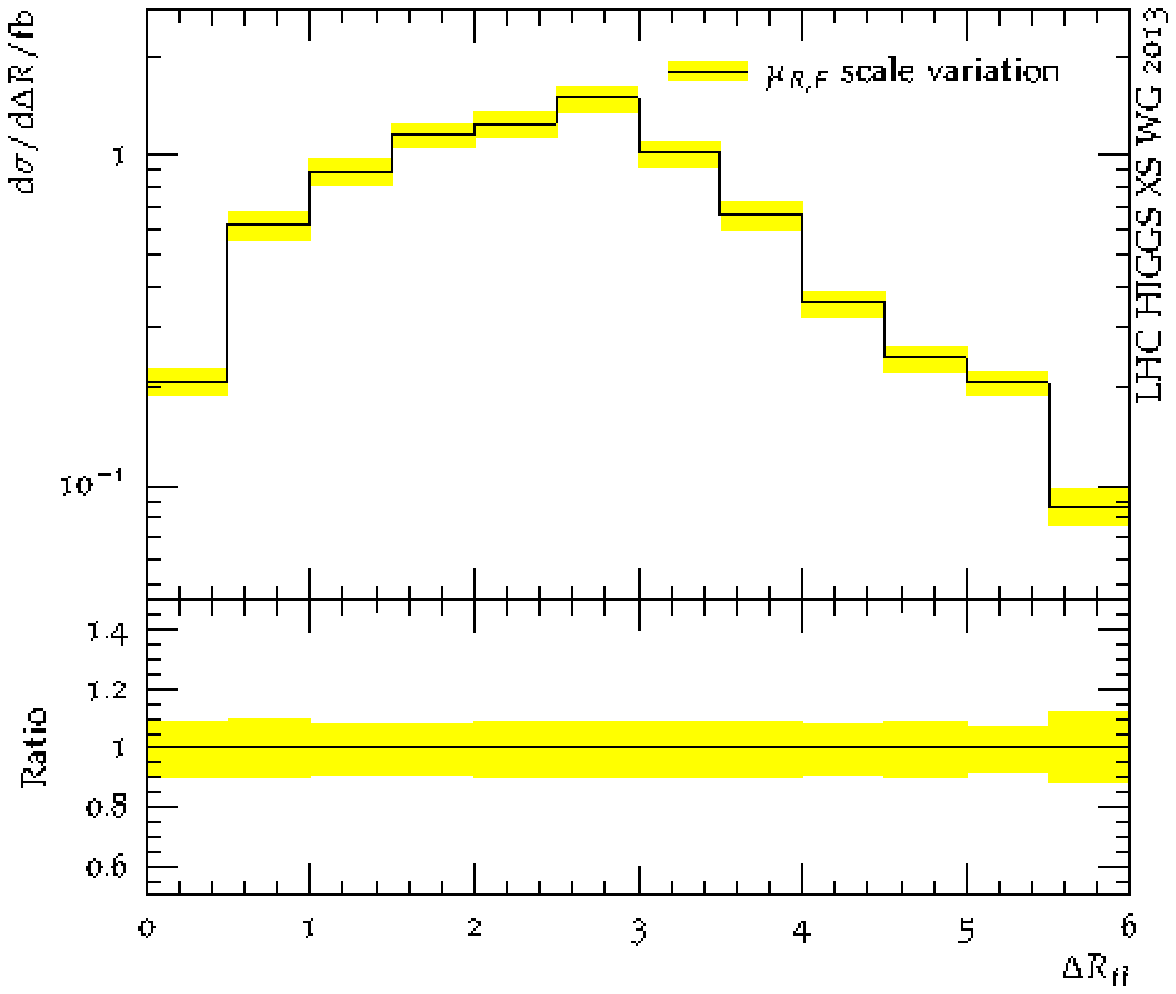}
\end{tabular}
\caption{Distance of the two reconstructed top quarks in rapidity
(left panel), azimuthal angle (central panel), and $R$ (right
panel) for $\PQt\PAQt\PH$ events at the $8\UTeV$ LHC.  Uncertainties
due to renormalization and factorization scale variations (yellow
band) as given in an \protect\MCatNLO simulation are also
indicated.}  \label{fig:ttH_uncert_tt}
\end{center}
\end{figure}

\begin{figure}
\begin{center}
\begin{tabular}{cc}
\includegraphics[width=0.48\textwidth]{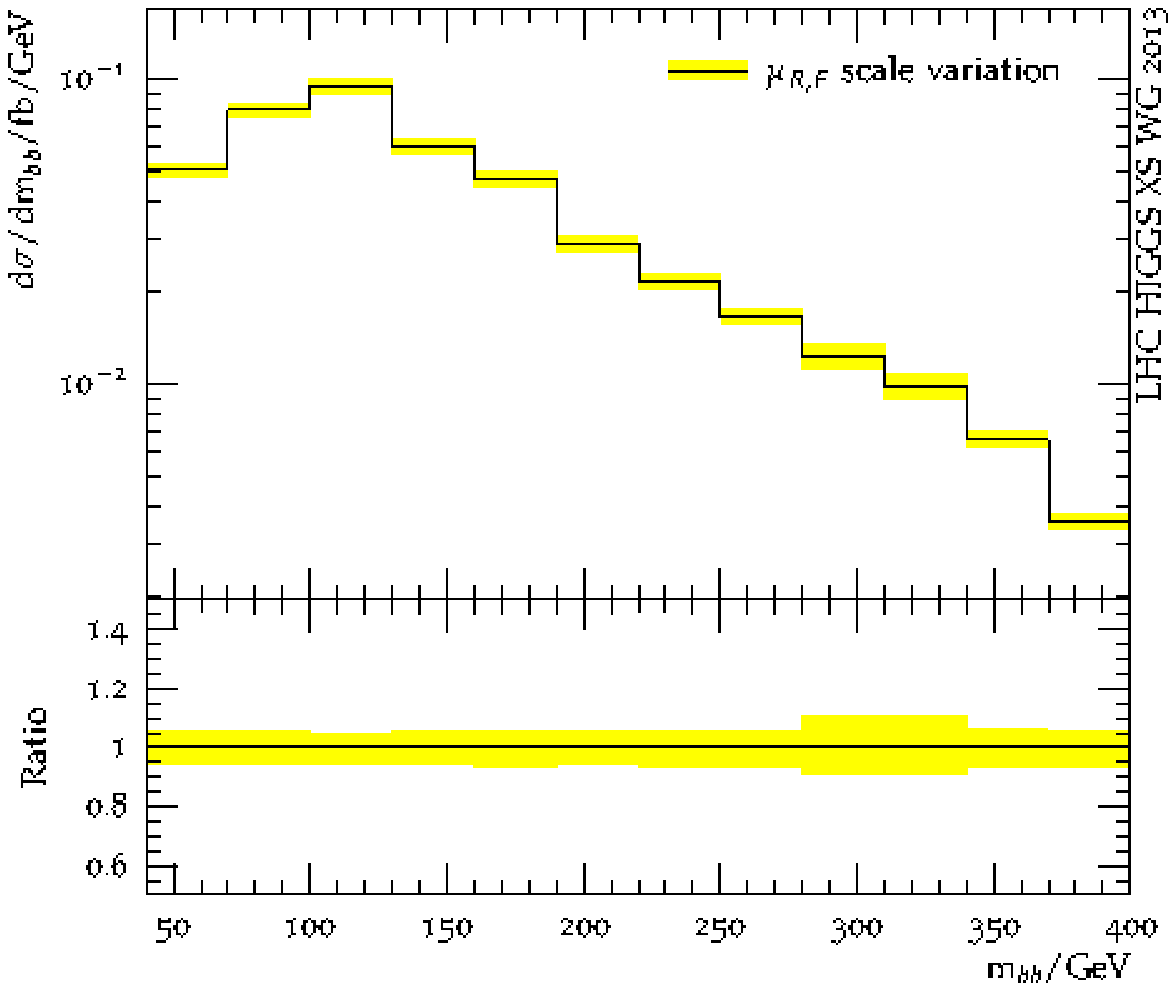} &
\includegraphics[width=0.48\textwidth]{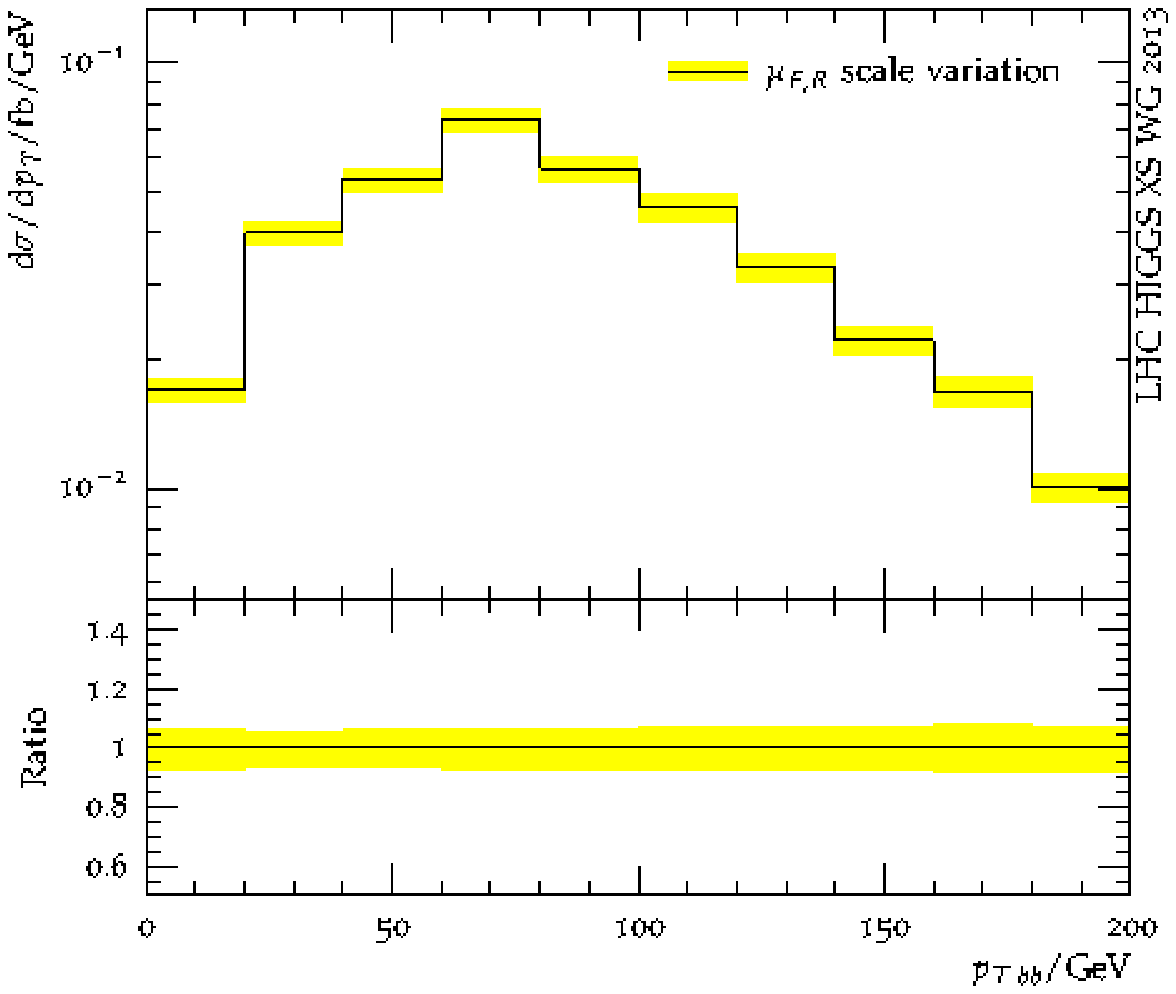}
\end{tabular}
\caption{Invariant mass (left panel) and transverse momentum (right 
         panel) distribution of the pair of $\PQb$ jets not associated
         with top decays in the acceptance region for $\PQt\PAQt\PH$
         events at the $8\UTeV$ LHC.  Uncertainties due to
         renormalization and factorization scale variations (yellow
         band) as given in an \protect\MCatNLO simulation are also
         indicated.}  \label{fig:ttH_uncert_bb}
\end{center}
\end{figure}

In this section we discuss the estimate of the theoretical
uncertainties on $\PQt\PAQt\PH$ production at the $8\UTeV$ LHC, using
the example of a few key distributions.  We consider the case in which
the top quarks decay semi-leptonically and the Higgs boson decays into
a $\PQb\PAQb$ pair.  For the simulation of the process at the hadron
level we use the \Sherpa event generator~\cite{Gleisberg:2008ta}, and
NLO matrix elements from~\cite{Reina:2001sf,Dawson:2002tg}, which we
also cross-checked with those from \OpenLoops~\cite{Cascioli:2011va}.
The latter uses loop integrals provided by the \Collier library which
implements the numerically stable reduction methods
of \Brefs{Denner:2002ii,Denner:2005nn} and the scalar integrals
of \Bref{Denner:2010tr}.  They are connected with the \Sherpa parton
shower~\cite{Schumann:2007mg}, based on Catani-Seymour splitting
kernels~\cite{Catani:1996vz,Nagy:2005aa}, through an \MCatNLO-type
matching~\cite{Frixione:2002ik} in the fully color-correct version
of~\cite{Hoeche:2011fd}.  Spin correlations have been included in the
full chain of production and decays, where the former is treated at
NLO accuracy and the latter is treated at LO accuracy.  In addition,
we employ the following conventions and settings for the simulation.
\begin{itemize}
\item We use  $m_t=172\UGeV$, $m_H=125\UGeV$, and $m_W=80.4\UGeV$. 
 The $\PQb$ quark is treated as massive with an on-shell mass of
      $m_b=4.79\UGeV$.  The $4$-flavor MSTW2008
      set of PDF~\cite{Martin:2009iq} is being used.
\item We use a dynamical renormalization and factorization scale of
      \begin{equation}
      \muR = \muF = \frac{m_{\mathrm{T},\PQt}+m_{\mathrm{T},\PAQt}+m_{\PH}}{2}\,,
      \end{equation}
      which also fixes the parton-shower starting scale $\mu_Q$.
\item In order to estimate the theoretical uncertainty we vary $\muR$ and $\muF$ in
      parallel by a factor of $2$ up and down. A full estimate of the
      theoretical uncertainty from parton showering should also
      consider the independent variation of $\mu_Q$, but we do not
      include it in the present study.
\item PDF uncertainties and those related to other perturbative parameters
      such as $\alphas(\MZ)$ or particle masses are ignored.
\item Similarly, uncertainties related to hadronization and the underlying
      events are not taken into account.
\end{itemize}
For the simulation, cuts similar to the ones typically used in ongoing 
analyses have been implemented in a \Rivet~\cite{Buckley:2010ar} analysis, namely:
\begin{itemize}
\item one isolated lepton with $\pT\ge 25\UGeV$ and $|\eta|\,\le\,2.5$.
      The isolation criterion requires the summed $\pT$ of tracks within 
      $\Delta R=0.3$ not to exceed $0.1$ of the lepton $\pT$.  In addition, 
      all further visible particles within $\Delta R=0.2$ of the lepton 
      deposit less transverse momentum than $0.1$ of the lepton $\pT$.
\item For events with electrons, $E\!\!\!/_{\mathrm{T}}\,\ge\,35\UGeV$, while for
      event with muons, $E\!\!\!/_{\mathrm{T}}\,\ge\,20\UGeV$.
\item Anti-$k_{\mathrm{T}}$ jets with $R=0.4$ are reconstructed using 
      FastJet~\cite{Cacciari:2011ma} with a minimum $\pT$ of $20\UGeV$ in
      $|\eta|\,\le\,2.5$.        
\item Jets containing at least one $\PQb$-hadron are considered $\PQb$-jets.
\end{itemize}
Typically at least two light and two $\PQb$ jets are required in
addition to an isolated lepton and missing transverse momentum in
order to reconstruct the $\PQt\PAQt$ system, plus further jets related
to the Higgs boson.  The tops are reconstructed by first finding the
hadronic $\PW$ from the light jets, and combining it with the
$\PQb$-tagged jet giving the best mass for the resulting top quark.
Then the leptonic $\PW$ is reconstructed from the lepton and the
missing transverse momentum, the degeneracy is resolved through
combination  with the remaining $\PQb$ jets and picking the best mass
of the resulting top quark.  It is only then that the Higgs-boson
candidate is reconstructed from the remaining jets.  The distribution
of light and $\PQb$ jets and their uncertainties are displayed in
\Fref{fig:ttH_uncert_njets}.
In all further plots shown here we will concentrate on events with 4
$\PQb$ jets and at least two light jets.  In
\Fref{fig:ttH_uncert_incl_2} we display inclusive observables
such as $H_{\mathrm{T}}$, the scalar sum of the transverse momenta of
all hard objects -- isolated lepton, $E\!\!\!/_{\mathrm{T}}$, and jets
-- and the transverse momentum of the overall $\PQt\PAQt\PH$ system as
reconstructed from its decay products.
In \Fref{fig:ttH_uncert_tt} we focus on the $\PQt\PAQt$ system
and depict the distance of the two reconstructed top quarks in rapidity,
azimuthal angle, and in $R$ (distance in the rapidity-azimuthal angle
plane).  Finally, in \Fref{fig:ttH_uncert_bb} we show
observables related to the two $\PQb$ jets not associated with top-quark
decays, namely the invariant mass of the pair of jets and their joint
transverse momentum.
The emerging picture of the uncertainties is consistent:
\begin{itemize}
\item the traditional scale variation of a factor of $2$ applied in
      parallel to both renormalization and factorization scale to the
      NLO matrix element impact  distributions at the level of
      $10\%-20\%$.  It is interesting to note that also observables like
      the transverse momentum of the produced (and reconstructed) $\PQt\PAQt\PH$
      system or the number of light jets, which are sensitive to further
      QCD radiation, are fairly stable under these variations, which may 
      indicate that a parallel shift in the scales could underestimate the
      actual scale uncertainty.  
\item The picture will certainly change somewhat when considering variations 
      in the parton shower starting scale, which also defines the hard regime 
      of radiation where no $K$ factor is applied to the 
      configurations.  In addition, uncertainties due to the underlying
      event will impact the tagging of $\PQb$ jets and, through varying
      QCD activity have an effect on, for instance, the overall acceptance.
\end{itemize}

\subsection{Spin correlation effects in $\PQt\PAQt\PH$ using \madspin}
\label{subsec:ttH-spin_corr_ttH_madspin}

Monte Carlo generators have now entered the era of fully automated NLO
event generators~\cite{Alioli:2010xd,Frederix:2011zi,Hoeche:2011fd},
which open several perspectives in hadron collider phenomenology by
allowing the simulation of a new class of processes at
next-to-leading-order accuracy.  Even though these automated NLO Monte
Carlo generators feature, in principle, no restrictions on complexity
of the process and particle multiplicity, in practice the CPU cost
becomes enormous for high-multiplicity final states.  Most of the
current tools cannot simulate the full production and decay at NLO
accuracy in a reasonable amount of time; only the generation of
undecayed events at next-to-leading order is feasible.

In this context a generic framework dubbed
\madspin~\cite{Artoisenet:2012st} has recently been proposed to decay
heavy resonances in next-to-leading-order Monte Carlo events.  The
method includes not only \textit{decay} spin correlation effects
(which induce angular correlations among the final-state particles
inside a given decay branch) but also \textit{production} spin
correlation effects (which induce angular correlations among the
particles from distinct decay branches) by unweighting decay
configurations with tree-level matrix elements associated with the
decayed process.  Generating the decay of a specific event typically
requires only a few evaluations of matrix elements, so that the
algorithm is in general very fast.  Although only tree-level matrix
elements are used to unweight the decay configurations, for specific
processes this procedure was shown to capture essentially all spin
correlation effects as predicted by a full next-to-leading-order
calculation.

The approach in \madspin{} is based on the narrow-width approximation,
as the production of events is factorized from the decay. However,
off-shell effects are partly recovered by smearing the mass of each
resonance in undecayed events according to a Breit-Wigner
distribution, and by applying the unweighting procedure also with
respect to the generated virtualities of the resonances.  The other
momenta in undecayed events are reshuffled in an optimal way with the
use of diagram-based information of the tree-level scattering
amplitude associated with the undecayed events.

In order to illustrate the capabilities of the tool, we apply it to
the case of top-quark pair production in association with a light
Higgs boson at the LHC (running at $8\UTeV$), considering both the
scalar and pseudo-scalar hypotheses for the Higgs boson.  Due to the
large irreducible QCD background, any search strategy for this Higgs
production process relies strongly on the accuracy of the Monte-Carlo
predictions.  QCD correction to these processes has been analysed by
two groups~\cite{Frederix:2011zi,Garzelli:2011vp} and a comparison
between these independent calculations has appeared in
\Bref{Dittmaier:2012vm}. In these works it was shown that the NLO
corrections are very mild, in particular on shapes of distributions.
 
To the best of our knowledge, the problem of retaining
spin-correlation effects in events generated at NLO accuracy has not
been addressed yet for these processes. This problem is trivially
solved using the scheme proposed in this section: NLO parton-level
events are generated with \amcatnlo{}, (LHC at $8\UTeV$, $\textrm{PDF
  set}=\textrm{MSTW2008(n)lo68cl}$~\cite{Martin:2009iq},
$\MH=\MA=125\UGeV$, $\muR=\muF=(m_\mathrm{T}(H/A)m_\mathrm{T}(t)m_\mathrm{T}(\bar{t}))^{(1/3)}$,
no cuts) and then decayed with \madspin{} before they are passed to
\herwig{} for showering and hadronization.  In this illustration, top
and anti-top quarks are decayed semi-leptonically, whereas the Higgs
is decayed into a pair of $\PQb$ quarks.

\begin{figure}[htb]
\includegraphics[scale=0.6]{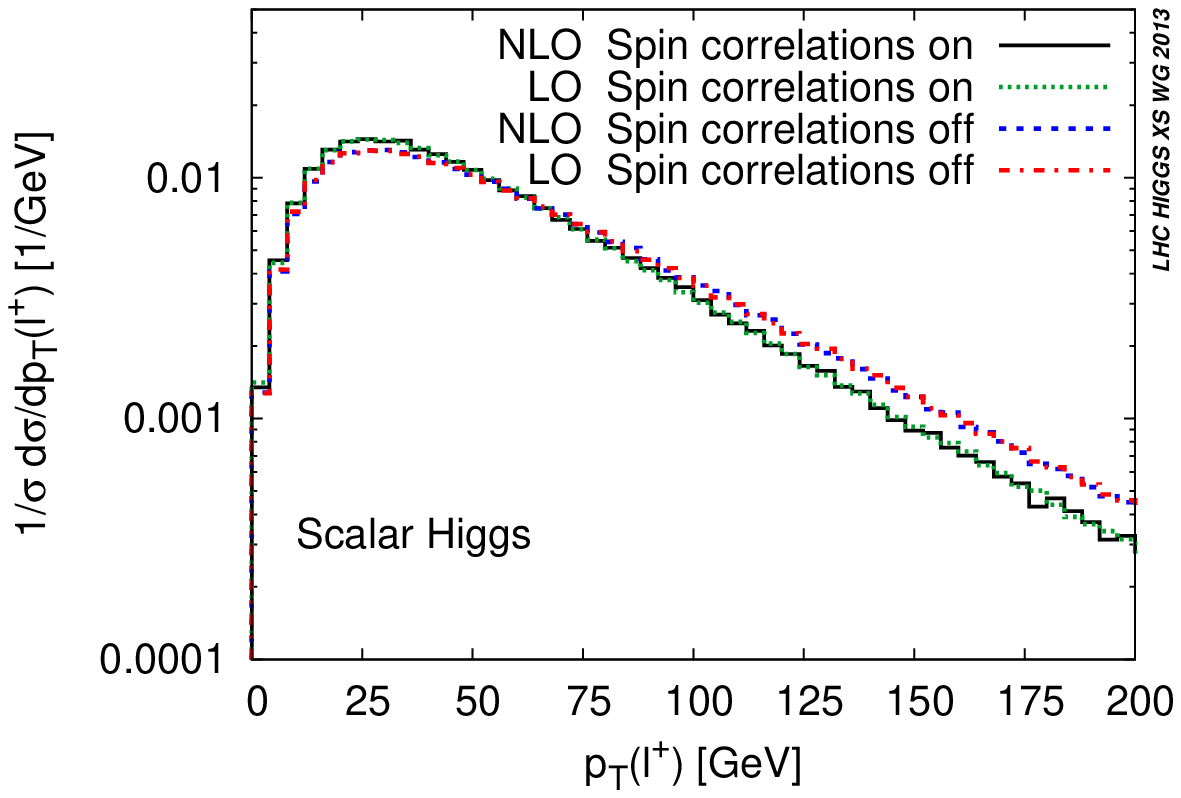}
\includegraphics[scale=0.6]{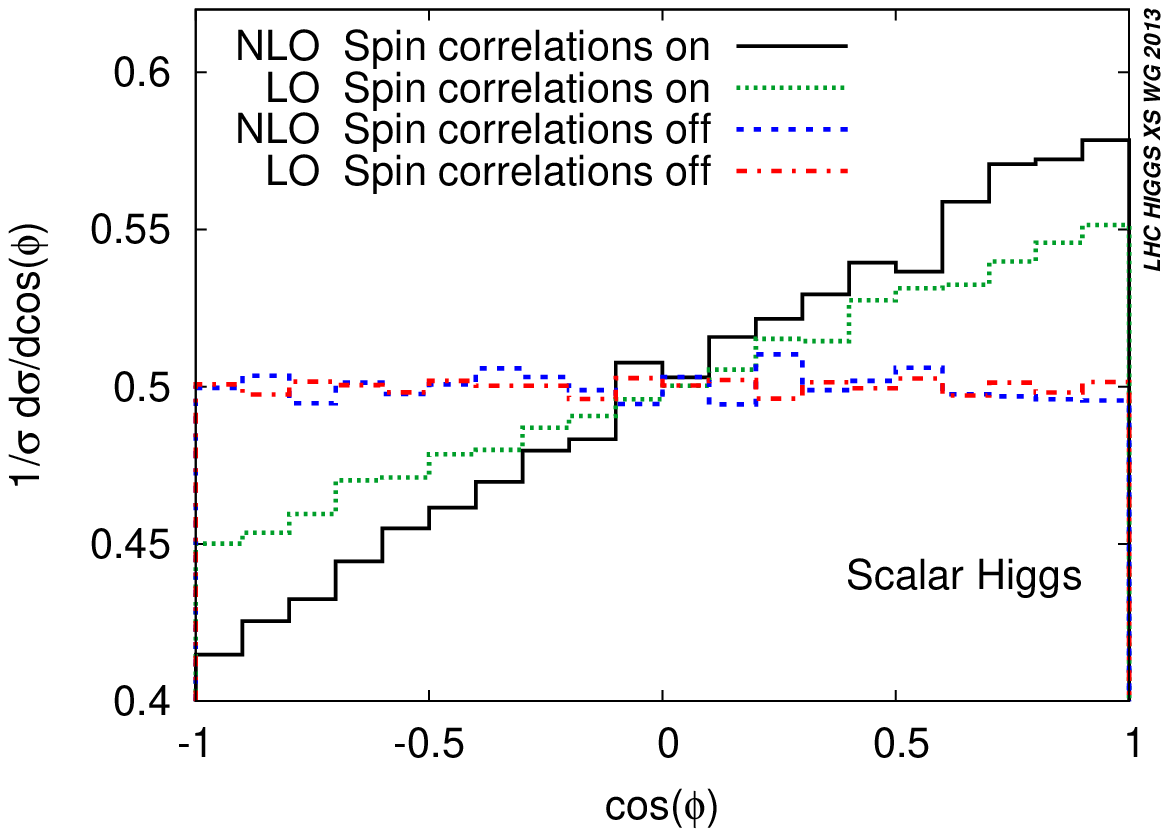}
\caption{Next-to-leading-order cross sections differential in
  $\pT(l^+)$ (left panel) and in and $\cos\phi$ (right panel) for
  $\PQt \PAQt\PH$ events with or without spin correlation effects. For
  comparison, also the leading-order results without spin-correlation
  effects are shown.  Events were generated with \amcatnlo{}, then
  decayed with \madspin{}, and finally passed to \herwig{} for
  parton showering and hadronization.}
\label{fig:ttH-htt_distr}
\end{figure}

\refF{fig:ttH-htt_distr} shows the normalized distribution of
events with respect to $\cos(\phi)$ (where $\phi$ is the angle between
the direction of flight of $l^+$ in the $\PQt$ rest frame and $l^-$ in
the $\PAQt$ rest frame), and with respect to the transverse
momentum of the hardest positively-charged lepton. Although
spin-correlation effects significantly distort the distribution of
events with respect to $\cos(\phi)$, their impact on the $\pT$
spectrum of the leptons is milder, except at large transverse
momentum. The relatively larger effect in the tail of this
distribution can easily be understood from the fact that the inclusion
of the spin correlations is a unitary procedure: a small change at low
$\pT$, where the cross section is large, needs to be compensated by a
larger (relative) effect at high $\pT$.

It is interesting that spin correlations have a much more dramatic
influence on the shape of these distributions than  NLO
corrections: the leading order results fall directly on top of the NLO
results for these normalized distributions (both without spin
correlations), as can be seen by comparing the dotted blue and
dash-dotted red curves. We can therefore conclude that preserving spin
correlations is more important than including NLO corrections for these
observables. Naturally, the inclusion of both, as is done here, is
preferred: it retains the good features of a NLO calculation,
\ie reduced uncertainties due to scale dependence (not shown), while
keeping the correlations between the top-quark decay products.

\begin{figure}[htb]
\includegraphics[scale=0.6]{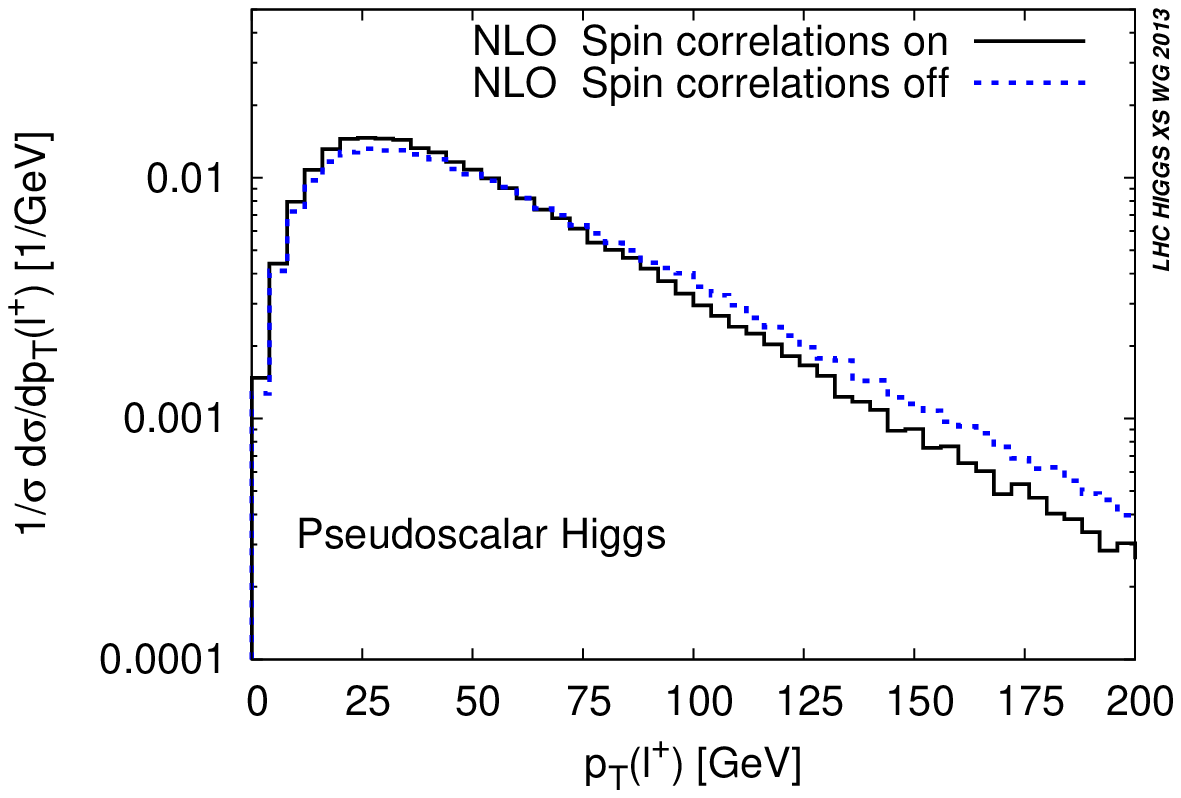}
\includegraphics[scale=0.6]{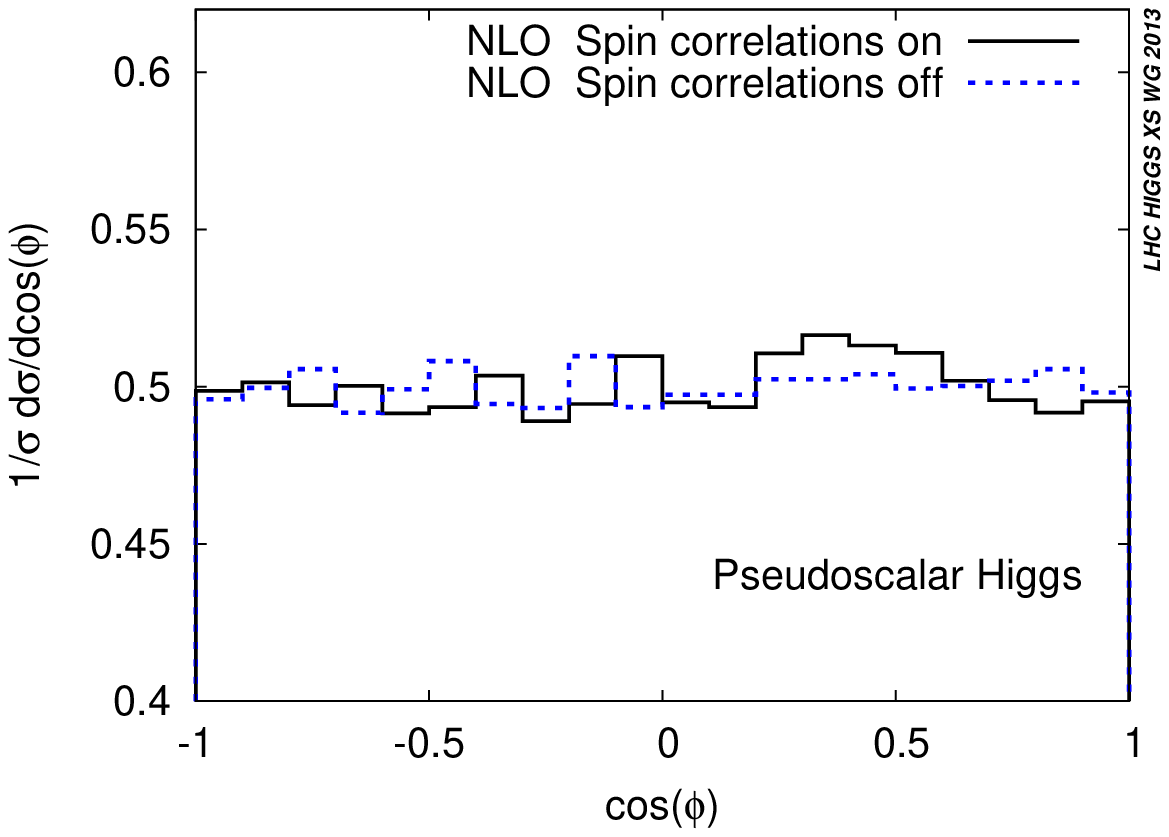}
\caption{Next-to-leading-order cross sections differential in
  $\pT(l^+)$ (left panel) and in and $\cos(\phi)$ (right panel) for
  $\PQt\PAQt\PA$ events with or without spin-correlation effects.
  Events were generated with \amcatnlo, then decayed with \madspin,
  and finally passed to \herwig{} for parton showering and
  hadronization.  }
\label{fig:ttH-Att_distr}
\end{figure}

The results for the pseudoscalar Higgs boson are shown in
\refF{fig:ttH-Att_distr}. The effects of the spin correlations on
the transverse momentum of the charged lepton are similar as in the
case of a scalar Higgs boson: about $10\%$ at small $\pT$, increasing to
about $40\%$ at $\pT=200\UGeV$. On the other hand, the $\cos(\phi)$ does
not show any significant effect from the spin-correlations. Therefore
this observable could possibly help in determining the CP nature of
the Higgs boson, underlining the importance of the inclusion of the
spin correlation effects.

\providecommand{\jj}{\mathrm{jj}}

\subsection{NLO parton-level predictions for $\PQt\PAQt\PQb\PAQb$ production at $8\UTeV$}
\label{subsec:ttH-NLO_ttbb}

The experimental observation of $\PQt\PAQt\PH(\PQb\PAQb)$ production
is notoriously very challenging due to large QCD backgrounds and the
non-trivial signature, involving at least four $\PQb$-jets.  The
selection of $\PH\to\PQb\PAQb$ candidates is contaminated by more than
$70\%$ combinatorial background, where one of the selected $\PQb$-jets
is a top-quark decay product or a misidentified light jet.  As a
consequence, the Higgs-boson mass peak is strongly diluted and
contaminated by QCD backgrounds.  The dominant background
contributions are given by $\PQt\PAQt\PQb\PAQb$ and $\PQt\PAQt \jj$
production.  The reducible $\PQt\PAQt \jj$ component constitutes more
than $95\%$ of the background cross section and can be estimated using
a data-driven approach.  Moreover, $\PQt\PAQt \jj$ can be strongly
suppressed by means of $\PQb$-tagging.  In contrast, the lack of
sufficiently distinctive kinematic features and the much smaller cross
section do not permit to determine the normalization of the
irreducible $\PQt\PAQt\PQb\PAQb$ background in a signal-free control
region. Thus theory predictions play a key role in the modeling of
$\PQt\PAQt\PQb\PAQb$ production.

The $\PQt\PAQt\PH(\PQb\PAQb)$ searches by ATLAS \cite{ATLAS-CONF-2012-135}
and CMS \cite{CMS-PAS-HIG-12-025,Chatrchyan:2013yea} are based on a
simultaneous fit of signal and backgrounds in $\PQt\PAQt$+jets
sub-samples with different light-jet and $\PQb$-jet multiplicities.  The
$\PQt\PAQt\PQb\PAQb$ background enters only the sub-samples with high
$\PQt\PAQt\PH$ sensitivity, where it can not be separated from the
signal, and experimental data are fitted using LO $\PQt\PAQt\PQb\PAQb$
predictions assuming an \textit{ad hoc} theory error of
$50\%$~\cite{ATLAS-CONF-2012-135,Chatrchyan:2013yea}. This uncertainty,
which lies between the typical scale dependence of LO and NLO
$\PQt\PAQt\PQb\PAQb$ predictions, constitutes the dominant systematic
error of $\PQt\PAQt\PH(\PQb\PAQb)$ searches.  The use of state-of-the
art theory predictions and error estimates for $\PQt\PAQt\PQb\PAQb$
(and $\PQt\PAQt \jj$) production is thus a key prerequisite to improve
the sensitivity to $\PQt\PAQt\PH(\PQb\PAQb)$.

Parton-level NLO studies of $\PQt\PAQt\PQb\PAQb$
\cite{Bredenstein:2008zb,Bredenstein:2009aj,Bredenstein:2010rs,
  Bevilacqua:2009zn} and $\PQt\PAQt \jj$
\cite{Bevilacqua:2010ve,Bevilacqua:2011aa} production at $14\UTeV$ 
indicate that NLO corrections reduce scale uncertainties in a drastic
way.  In the case of $\PQt\PAQt\PQb\PAQb$ production at $14\UTeV$, the
scale dependence goes down from roughly $80\%$ at LO to $20$-$30\%$ at
NLO, depending on the central-scale choice.  In order to further
increase the accuracy of theory predictions and render them applicable
to the experimental analyses, various major improvements are needed.
Apart from updating existing calculations to $7$ and $8\UTeV$, it will
be crucial to include top-quark decays, match NLO predictions to
parton showers, and merge $\PQt\PAQt +$jets final states with
different jet multiplicities at NLO.  Also finite $\PQb$-quark mass
effects and the relevance of $\PQb$-quark induced contributions at NLO
should be considered.

Thanks to recent developments in NLO automation, these goals are now
within reach. To illustrate progress in this direction, in the
following new NLO predictions for $\Pp\Pp\to\PQt\PAQt\PQb\PAQb$ at
$8\UTeV$, obtained with \openloops~\cite{Cascioli:2011va} and
\sherpa~\cite{Krauss:2001iv,Gleisberg:2008ta,Gleisberg:2007md},
are presented. The \openloops{} program is a fully automated
generator of one-loop QCD corrections to SM processes.
Loop amplitudes are evaluated with a numerical recursion that
guarantees high CPU efficiency for many-particle processes. Tensor
integrals are computed with the \collier{} library, which
implements the numerically stable reduction methods of
\Brefs{Denner:2002ii,Denner:2005nn} and the scalar integrals of
\Bref{Denner:2010tr}.  \openloops{} and \sherpa{} are interfaced
in a fully automated way, which allows to steer NLO simulations via
standard \sherpa{} run cards. Matching to parton shower within the
MC@NLO framework~\cite{ Frixione:2002ik,Hoeche:2011fd} and multi-jet
merging at NLO~\cite{Hoeche:2012yf} are also fully supported within
\textsc{Sherpa.2.0}.

\subsubsection{Input parameters and selection cuts}
\label{subsubsec:ttHib_setup}

Parton-level NLO results for $\Pp\Pp\to\PQt\PAQt\PQb\PAQb$ at
$\sqrt{s}=8\UTeV$ are presented for stable top quarks with mass
$\Mt=173.2\UGeV$.  Top-quark decays will be addressed in a forthcoming
study.  In NLO\,(LO) QCD the LHAPDF implementation of the
MSTW2008NLO\,(LO) parton distributions~\cite{Martin:2009iq} and the
corresponding running $\alphas$ are employed.  Contributions from
initial-state $\PQb$ quarks are discarded, otherwise the five-flavor
scheme with massless $\PQb$ quarks is consistently used.  Top-quark loops
are included in the virtual corrections but decoupled from the running
of $\alphas$ via zero-momentum subtraction.  All massless QCD partons
(including $\PQb$ quarks) are recombined into IR-safe jets using the
anti-$k_\rT$ algorithm~\cite{Cacciari:2008gp} with jet-resolution
parameter $R=0.5$.  The jet algorithm is not applied to top quarks.

Similarly as in the ATLAS and CMS $\PQt\PAQt\PH(\PQb\PAQb)$ analyses, 
$\PQt\PAQt\PQb\PAQb$ events are selected that contain
\begin{itemize}
\item[(\sregi)] $2$ $\PQb$ jets with $\pT,\PQb>20\UGeV$ and
  $|\eta_{\PQb}|<2.5$.
\end{itemize}
In addition to these rather inclusive cuts, to investigate NLO effects
in the Higgs-signal region an invariant-mass window around the Higgs
resonance is considered,
\begin{itemize}
\item[(\sregii)] $|m_{\PQb\PAQb}-\MH|< \Delta M$, with $\MH=126\UGeV$
  and $\Delta M=15\UGeV$.
\end{itemize}
The width $\Delta M$ is taken of the order of the the experimental
$m_{\PQb\PAQb}$ resolution, and $m_{\PQb\PAQb}$ is identified with the
invariant mass of the $\PQb\PAQb$ Higgs candidate, \ie of the two
$\PQb$ jets that do not arise from top decays.  From the experimental
viewpoint this selection is unrealistic, since the large combinatorial
background resulting from incorrect $\PQb$-jet assignments is not
taken into account.  Nevertheless, it is instructive to quantify the
background contamination of the signal region in the ideal limit of
exact $\PQb$-jet combinatorics.  On the one hand, this permits to
assess the potential sensitivity improvement that might be achieved
with a strong reduction of the combinatorial background.  On the other
hand, it is interesting to investigate if NLO corrections feature
significance differences in the \sregi~and \sregii~regions.

\subsubsection{Cross section results at $\sqrt{s}=8\UTeV$}
\label{subsubsec:ttHib_XS}

%%%%%%%%%%%%%%%%%%% TABLE 1: 8TeV XS  %%%%%%%%%%%%%%%%%%%%%%%%%%%%%%%%%%%%%%%%%%%%%%%%%
\begin{table}
\caption{Cross sections for $\Pp\Pp\to\PQt\PAQt\PQb\PAQb$ at $\sqrt{s}=8\UTeV$ with
  loose (\sregi) or signal (\sregii) selection cuts.  Predictions with
  fixed ($\mu_0=\Mt$) and dynamical ($\mu_0=\muBDDP$) scales are
  compared.  The upper and lower uncertainties correspond to scale
  variations $\mu/\mu_0=0.5$ and $2$, respectively. Statistical errors
  are given in parenthesis.  The $K$ factor,
  {$K=\sigma_\NLO/\sigma_\LO$}, is evaluated at $\mu=\mu_0$.  While
  $\sigma_\LO$ is computed with LO $\alphas$ and PDFs,
  $\tilde\sigma_{\LO}$ is obtained with NLO inputs, and the
  corresponding $K$ factor is denoted as $\tilde
  K=\sigma_\NLO/\tilde\sigma_\LO$.}
\label{tab:YRHXS_ttH_ttbb}%
\renewcommand{\arraystretch}{1.5}%
\setlength{\tabcolsep}{1.5ex}%
\begin{center}
\begin{tabular}{llllllll}
\hline
{cuts}	&  {$\mu_0$} 	& {$\sigma_\LO[\Ufb]$} 		& {$\tilde\sigma_\LO[\Ufb]$}
												& {$\sigma_\NLO[\Ufb]$}	& {$K$} & {$\tilde K$} \\
\hline
$S1$	&  $\Mt$	& $503(1)^{+84\%}_{-42\%}$	& $342(1)^{+74\%}_{-39\%}$	& $671(3)^{+33\%}_{-28\%}$  	& $1.34 $	& $1.96 $	\\
$S1$ 	&  $\muBDDP$	& $861(1)^{+95\%}_{-45\%}$	& $556(1)^{+83\%}_{-42\%}$ 	& $900(2)^{+23\%}_{-27\%}$	& $1.04 $	& $1.62 $	\\
\hline
$S2$	&  $\Mt$	& $37.35(2)^{+86\%}_{-43\%}$	& $25.33(2)^{+76\%}_{-40\%}$	& $45.5(1)^{+29\%}_{-26\%}$	& $1.22 $	& $1.79 $	\\
$S2$ 	&  $\muBDDP$	& $54.9(1)^{+94\%}_{-45\%}$	& $36.0(1)^{+82\%}_{-42\%}$	& $54.3(2)^{+17\%}_{-24\%}$	& $0.99 $	& $1.50 $	\\
\hline
\end{tabular}
\end{center}
\end{table}
%%%%%%%%%%%%%%%%%%% %%%%%%%%%%%%%%%%%%%%%%%%%%%%%%%%%%%%%%%%%%%%%%%%%

Cross section predictions for $\Pp\Pp\to\PQt\PAQt\Pb\PAQb$ at $8\UTeV$
with loose cuts (\sregi) and in the signal region (\sregii) are shown
in \refT{tab:YRHXS_ttH_ttbb}.
Perturbative uncertainties are estimated by uniform factor-two
variations of the renormalization and factorization scales,
$\mu_\rR=\mu_\rF=\xi\mu_0$ with $\xi=0.5,1,2$.\footnote{ Antipodal
  rescalings with $\mu_\rF=\mu_0/\xi$ are not considered, since at
  $14\UTeV$ it was shown that they induce smaller variations than
  uniform rescalings \cite{Bredenstein:2010rs}.}  For the central
scale $\mu_0$ two different options are compared:
\begin{itemize}
\item[(i)] a fixed scale choice $\mu_0^2=\Mt^2$;
\item[(ii)] the dynamical scale 
  $\mu_0^2=\muBDDP^2=\Mt\sqrt{p_{\mathrm{T},\PQb}p_{\mathrm{T},{\PAQb}}}$\;,
  introduced in \Bref{Bredenstein:2010rs}.
\end{itemize}
Here $p_{\mathrm{T},\PQb}$ and $p_{\mathrm{T},\PAQb}$ are the transverse momenta of
the two $\PQb$ jets that do not originate from top decays.  Absolute LO and
NLO cross sections are complemented by respective $K$ factors,
$K=\sigma_\NLO/\sigma_\LO$, which might be employed to correct the
normalization of LO event samples used in experimental studies.  In
this respect it is important to keep in mind that $K$ factors can be
applied to LO cross sections only if both are computed at the same
scale.  This is crucial since $K$ factors feature a similarly large
scale dependence as LO quantities.  Another issue, relevant for the
consistent rescaling of LO quantities by $K$ factors, is that LO cross
sections can be computed using PDFs and $\alphas$ either in LO or NLO
approximation, depending on the convention.  Obviously, combining LO
predictions and $K$ factors based on different conventions leads to
inconsistent results.  To point out the quantitative importance of a
consistent combination, in \refT{tab:YRHXS_ttH_ttbb} we show results
corresponding to both conventions: LO cross sections based on NLO
inputs and related $K$ factors are denoted as $\tilde\sigma_{\LO}$ and
$\tilde K=\sigma_{\NLO}/\tilde\sigma_{\LO}$, while the standard
notation is used when LO quantities are computed with LO inputs as
usual.

LO results in \refT{tab:YRHXS_ttH_ttbb} feature a huge scale
dependence, which results from the \mbox{$\alphas^4$-scaling} of the
$\PQt\PAQt\PQb\PAQb$ cross section and can reach $95\%$.  The LO cross
sections corresponding to different scale choices ($\Mt,\muBDDP$) and
conventions ($\sigma_\LO,\tilde\sigma_\LO$) can differ by even more
than $100\%$.  These sizable effects are clearly visible also in the
differences between the respective $K$ and $\tilde K$ factors.  An
inconsistent combination of $K$ factors and LO predictions, as
discussed above, can induce errors of tens of percent.

The impact of NLO corrections turns out to be rather mild as compared
to $14\UTeV$~\cite{Bredenstein:2009aj,Bredenstein:2010rs,
Bevilacqua:2009zn}.  Using $\mu_0=\muBDDP$ results in a $K$ factor
very close to one and a residual scale dependence of about $25\%$.
This applies to both kinematic regions and is consistent with the good
perturbative convergence observed at $14\UTeV$ with dynamical scale
choice.  As already pointed out in~\Bref{Bredenstein:2010rs}, a hard
fixed scale $\mu_0=\Mt$ is less adequate to the multi-scale nature of
$\PQt\PAQt\PQb\PAQb$ production and results in a slower
convergence. In fact corrections and scale uncertainties turn out to
be larger with $\mu_0=\Mt$.  Results based on the dynamical scale and
respective uncertainties can thus be regarded as better predictions.
The goodness of this scale choice is also supported by the fairly
little $K$-factor differences in the \sregi~and \sregii~regions.
\begin{figure}
\begin{center}
\includegraphics[width=0.48\textwidth]{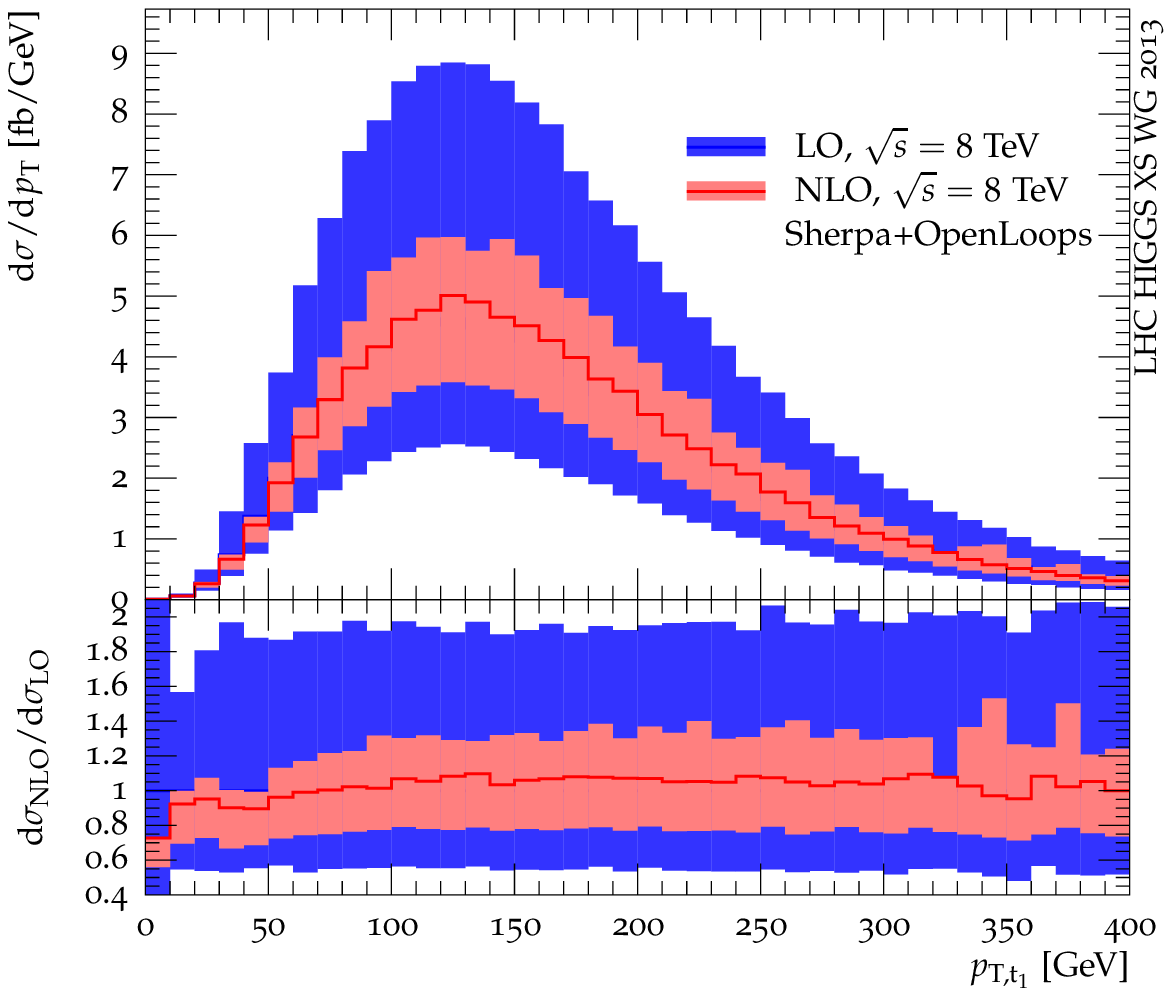}
\includegraphics[width=0.48\textwidth]{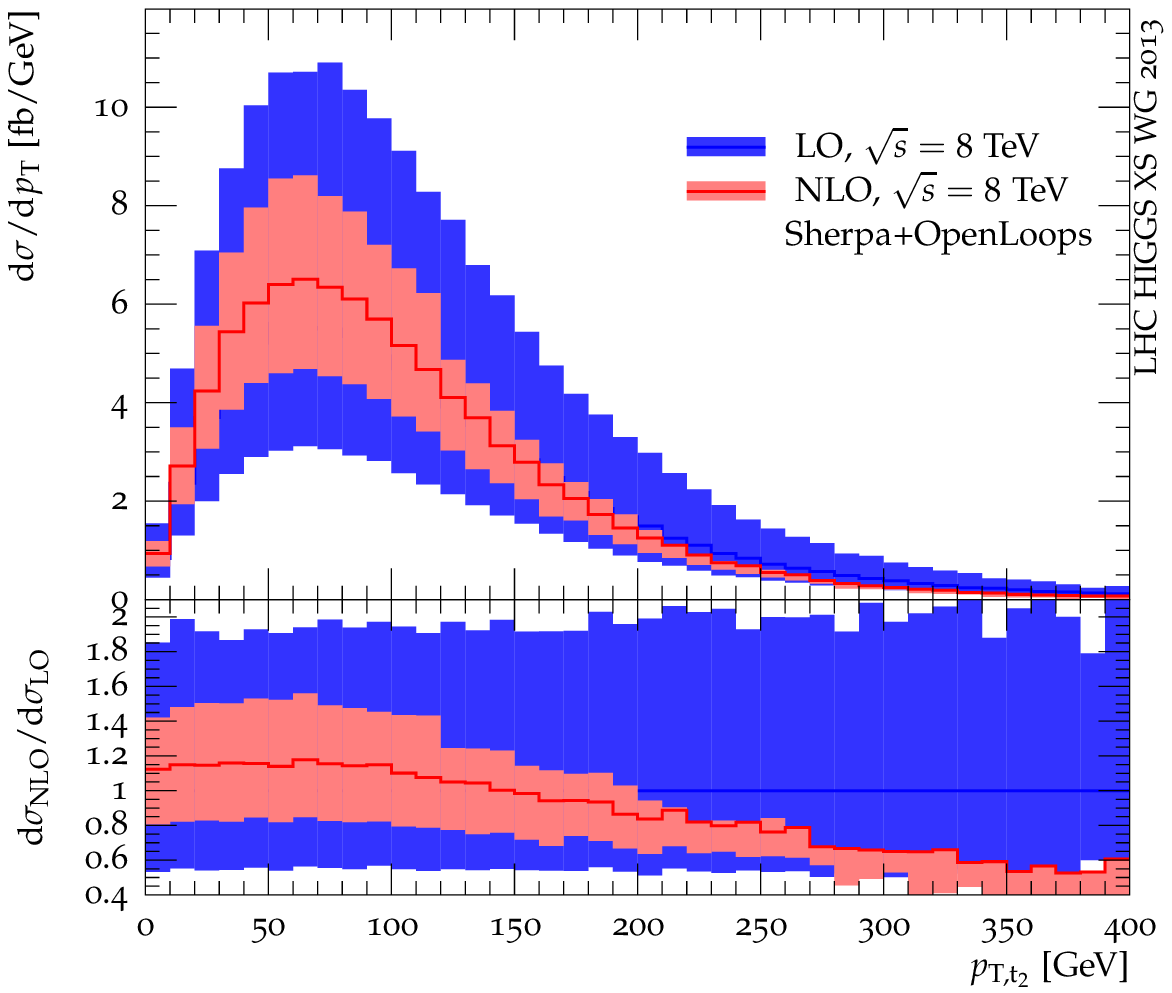}
\end{center}
\vspace*{-0.3cm}
\caption{Distributions in the transverse momenta of the harder (left)
  and softer (right) top quark in $\Pp\Pp\to\PQt\PAQt\PQb\PAQb$ at
  $\sqrt{s}=8\UTeV$.  Predictions at LO (blue) and NLO (red) are
  evaluated at the dynamical scale $\mu_0=\muBDDP$. The respective
  bands correspond to variations of the renormalization and
  factorization scales by a factor of $2$ around the central value
  $\mu_0$.  In the lower frame, LO and NLO bands are normalized to LO
  results at the central scale. The \sregi~selection is applied.  }
\label{fig:ttH-ttHib_ptt}%\refF{fig:ttH-ttHib_ptt}
\end{figure}
\begin{figure}
\begin{center}
\includegraphics[width=0.48\textwidth]{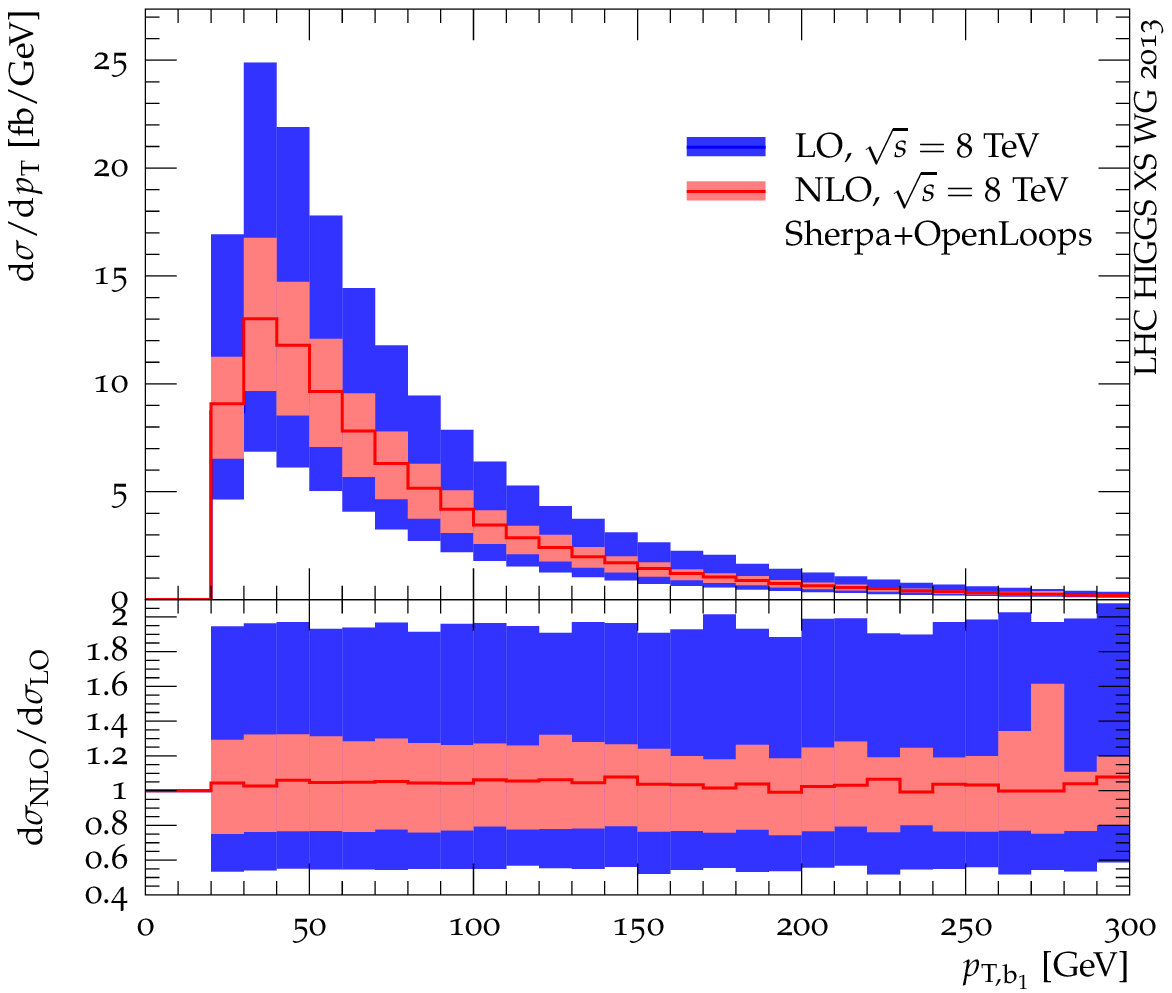}    
\includegraphics[width=0.48\textwidth]{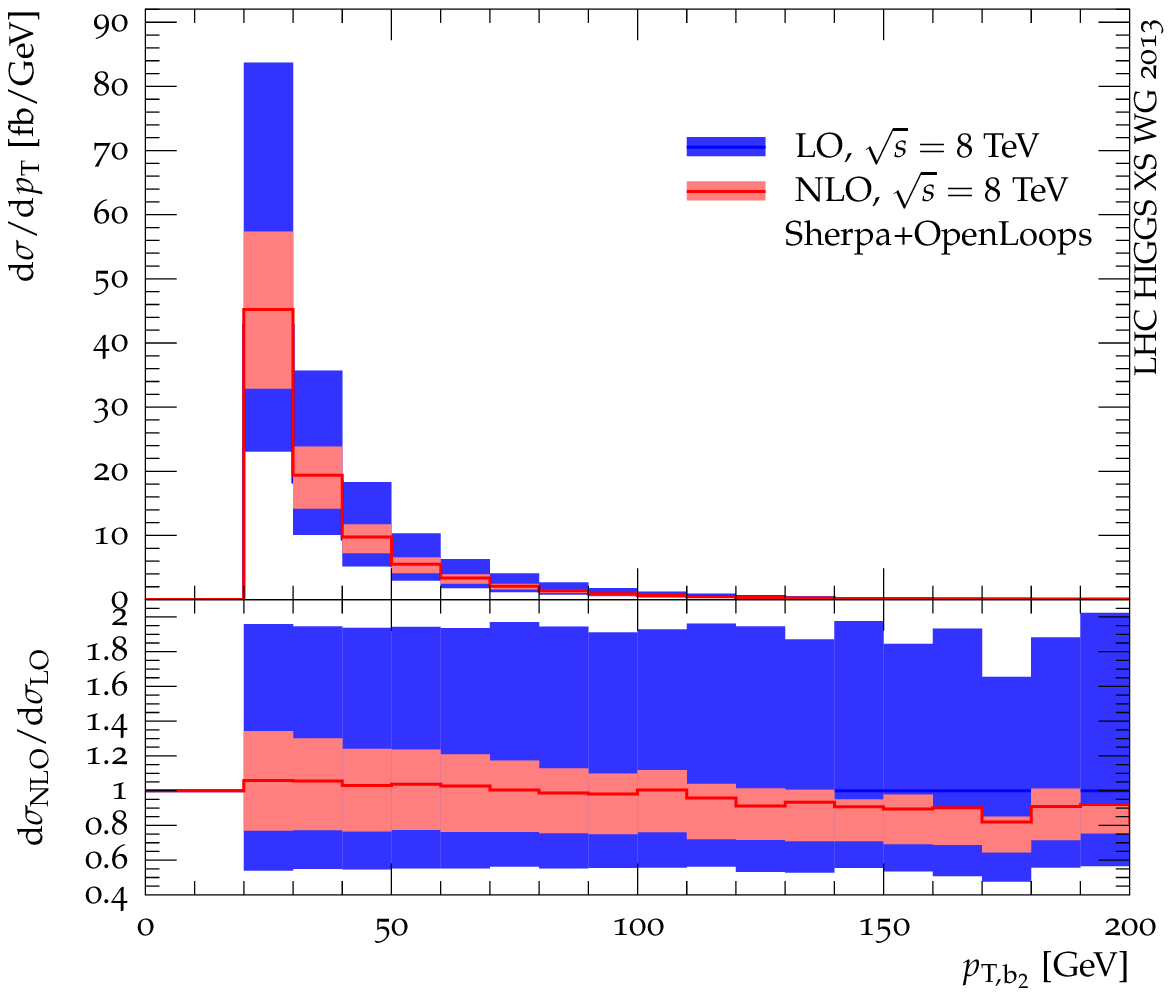}  
\end{center}
\vspace*{-0.3cm}
\caption{Distributions in the transverse momenta of the harder (left)
  and softer (right) $\PQb$ jet in $\Pp\Pp\to\PQt\PAQt\PQb\PAQb$ at
  $\sqrt{s}=8\UTeV$.  Same setup and conventions as in
  \refF{fig:ttH-ttHib_ptt}.  }
\label{fig:ttH-ttHib_ptb}%\refF{fig:ttH-ttHib_ptb}
\end{figure}

\subsubsection{Differential distributions}
\label{sec:ttHib_obs}

\begin{figure}
\begin{center}
\includegraphics[width=0.48\textwidth]{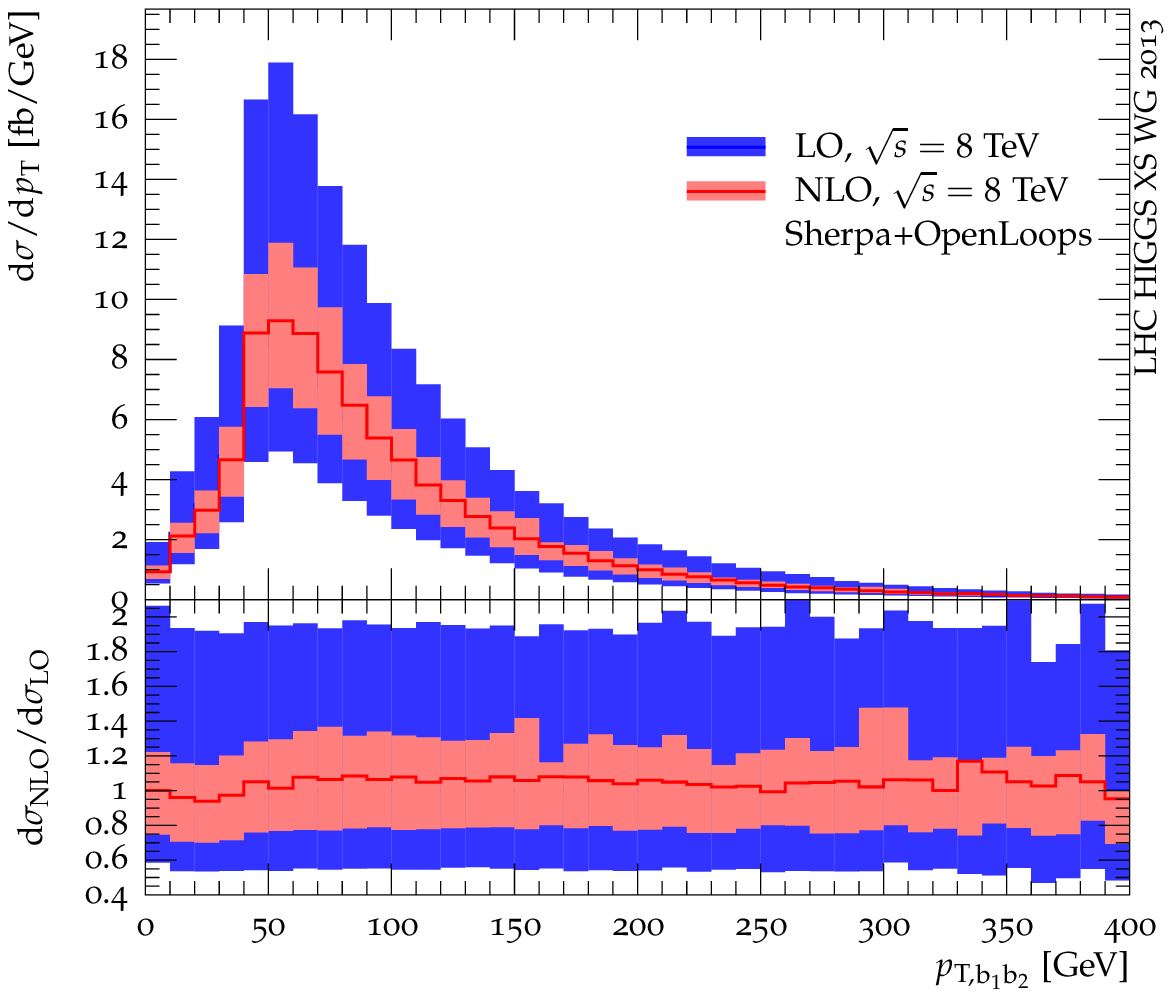}  
\includegraphics[width=0.48\textwidth]{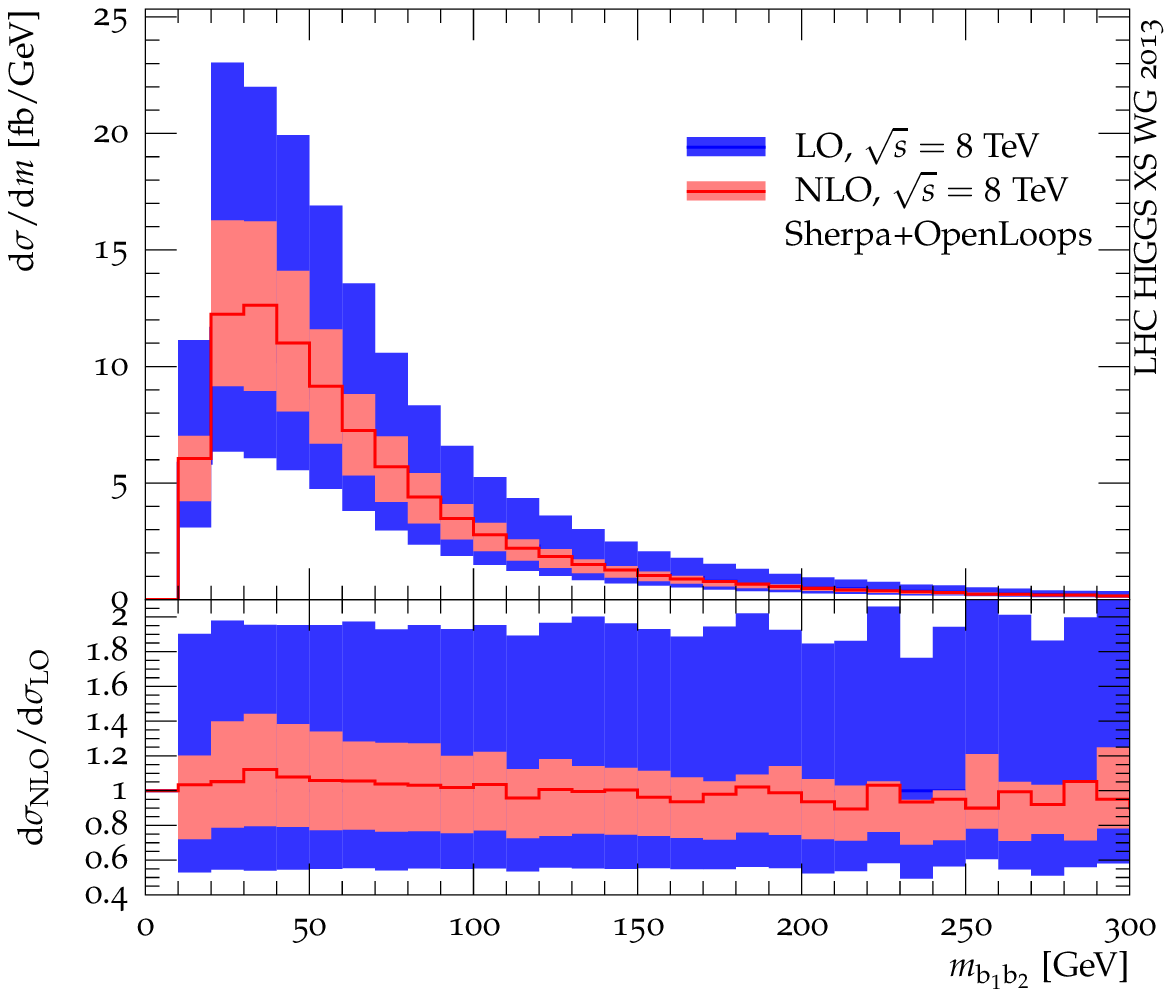}  
\end{center}
\vspace*{-0.3cm}
\caption{Distributions in the total transverse momentum (left) and the
  invariant mass (right) of the $\PQb$-jet pair in
  $\Pp\Pp\to\PQt\PAQt\PQb\PAQb$ at $\sqrt{s}=8\UTeV$.  Same setup and
  conventions as in \refF{fig:ttH-ttHib_ptt}.  }
\label{fig:ttH-ttHib_bbpair}%\refF{fig:ttH-ttHib_bbpair}
\end{figure}

\begin{figure}
\begin{center}
\includegraphics[width=0.48\textwidth]{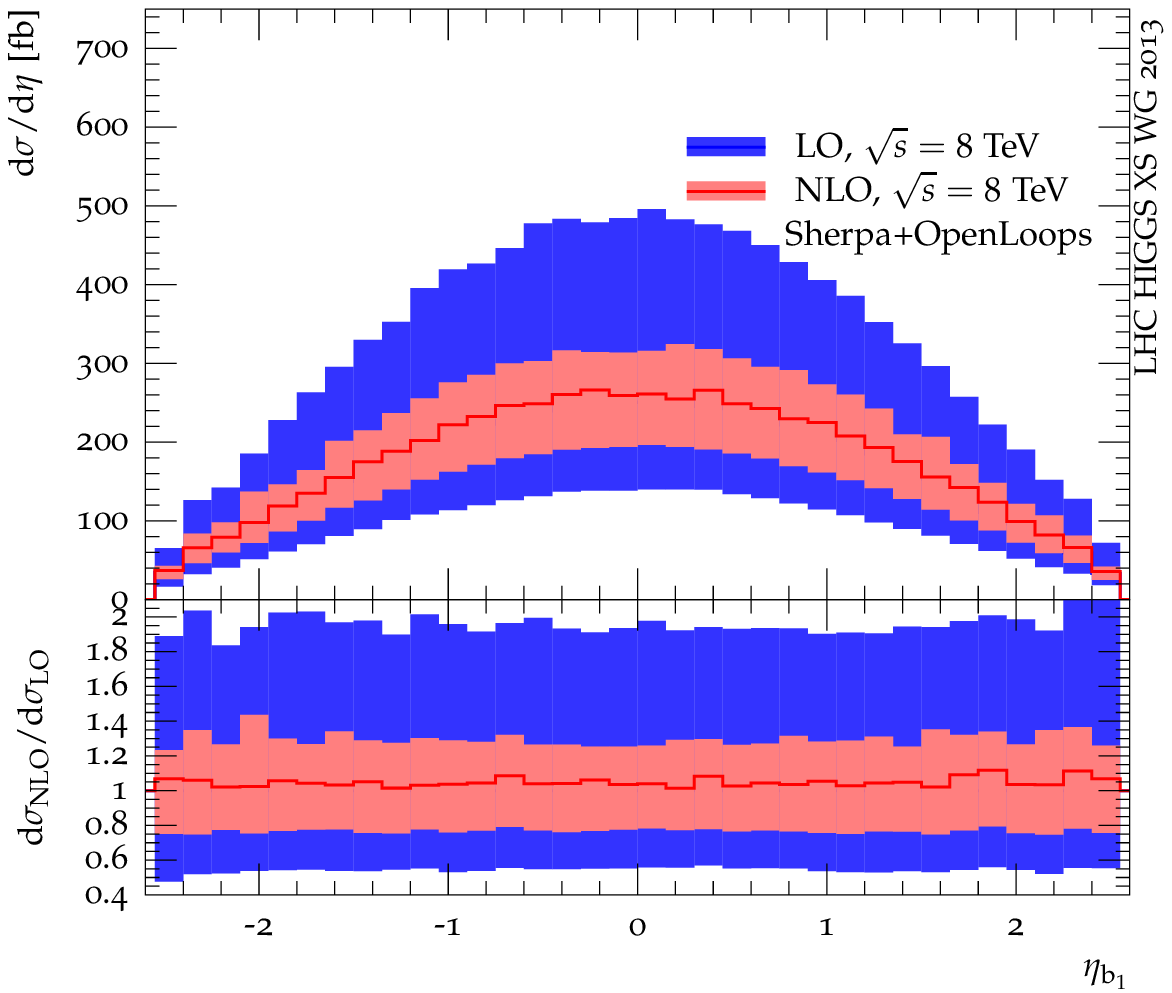}  
\includegraphics[width=0.48\textwidth]{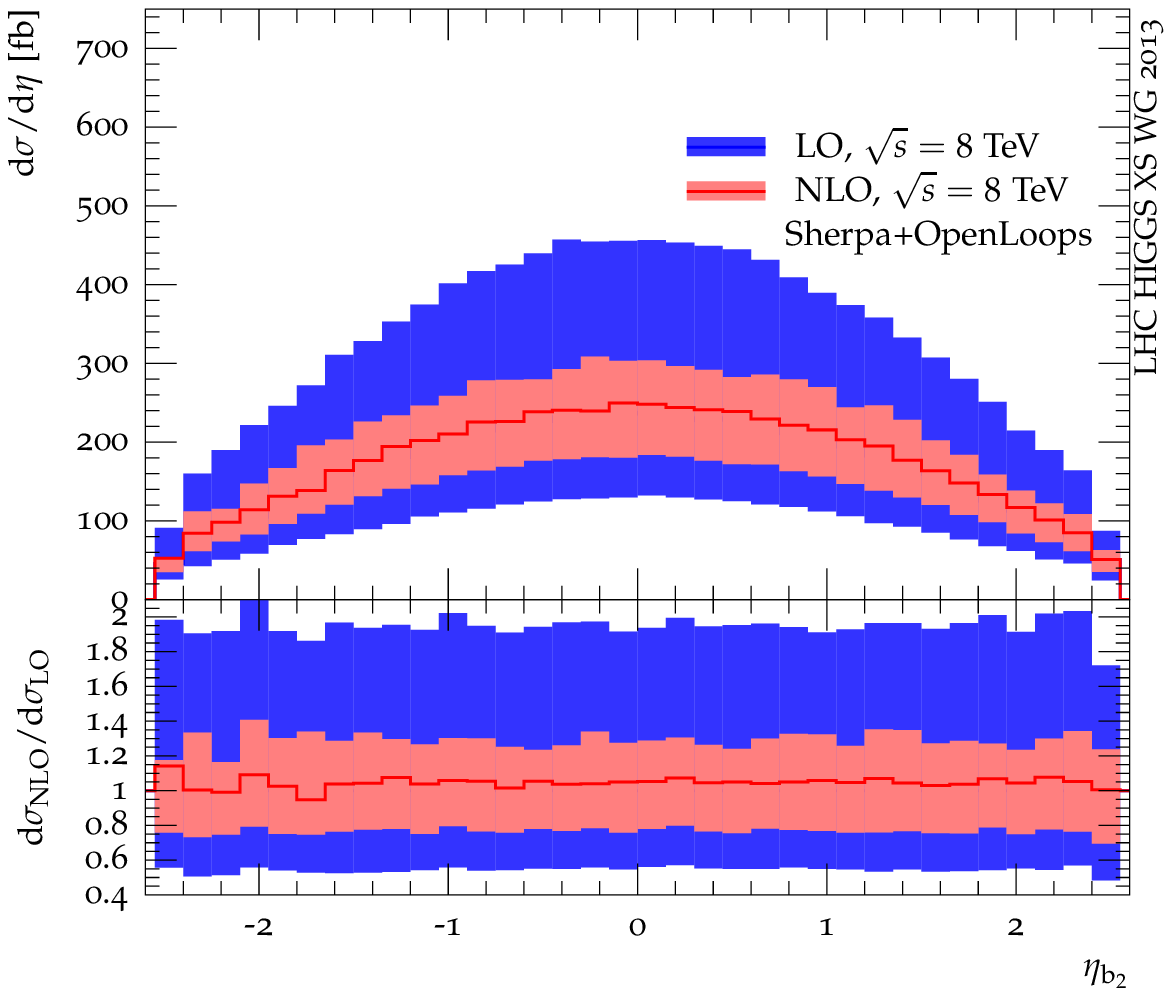}  
\end{center}
\vspace*{-0.3cm}
\caption{Distributions in pseudo-rapidity of the harder (left) and
  softer (right) $\PQb$ jet in $\Pp\Pp\to\PQt\PAQt\PQb\PAQb$ at
  $\sqrt{s}=8\UTeV$.  Same setup and conventions as in
  \refF{fig:ttH-ttHib_ptt}.  }
\label{fig:ttH-ttHib_etab}%\refF{fig:ttH-ttHib_etab}
\end{figure}

\begin{figure}
\begin{center}
\includegraphics[width=0.48\textwidth]{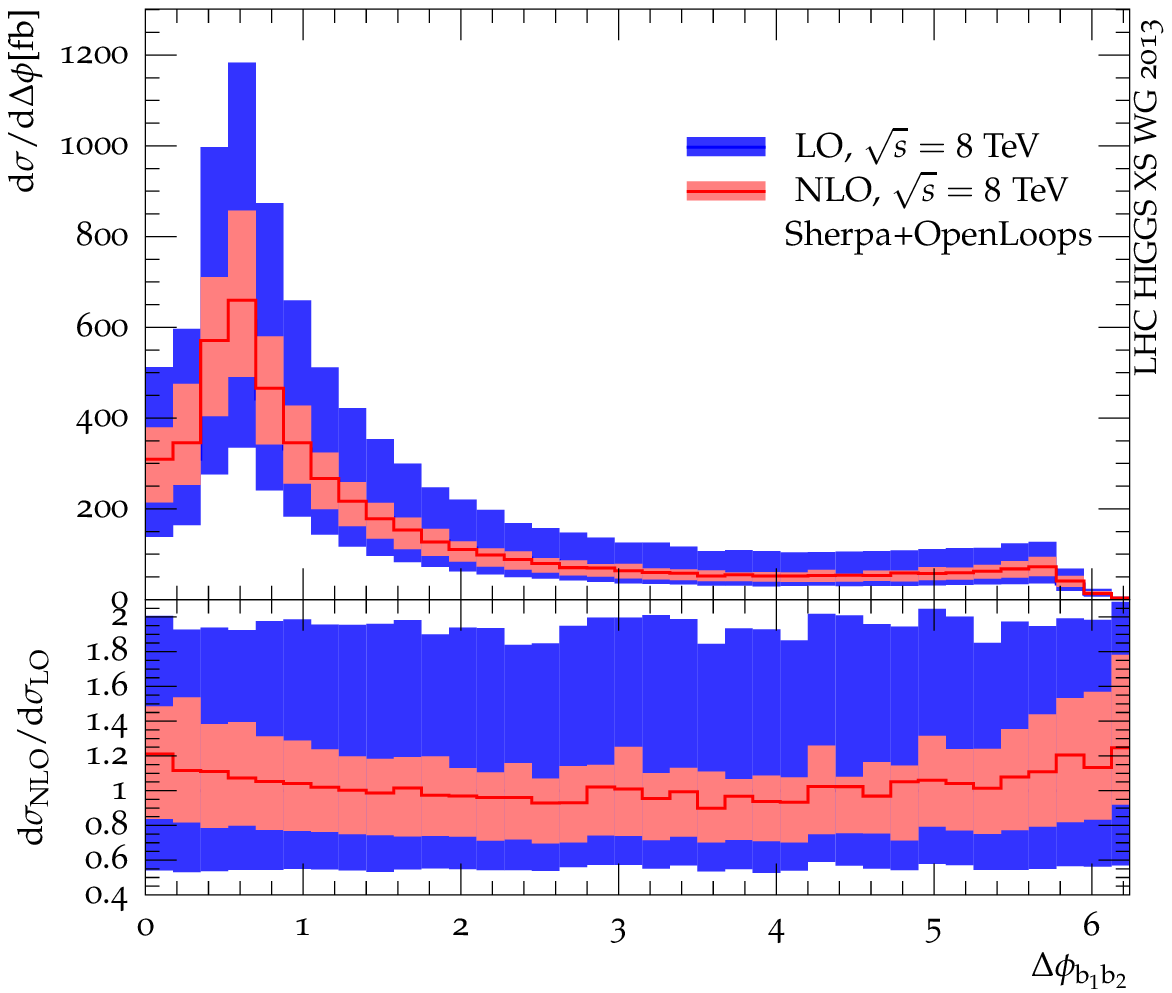}  
\includegraphics[width=0.48\textwidth]{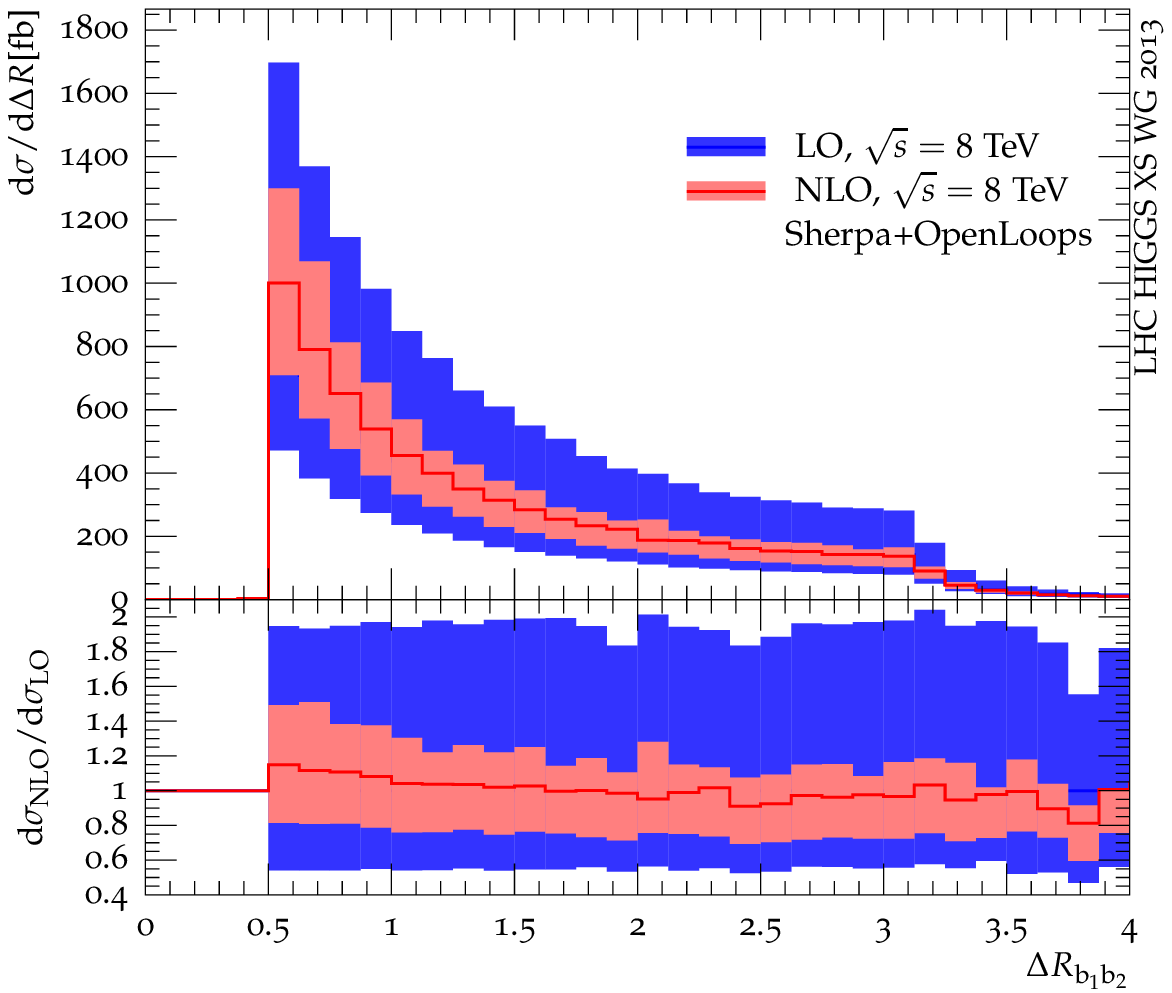}    
\end{center}
\vspace*{-0.3cm}
\caption{Distributions in azimuthal-angle (left) and $R$ separation
  (right) of the $\PQb$-jet pair in $\Pp\Pp\to\PQt\PAQt\PQb\PAQb$ at
  $\sqrt{s}=8\UTeV$.  Same setup and conventions as in
  \refF{fig:ttH-ttHib_ptt}.  }
\label{fig:ttH-ttHib_angsep}%\refF{fig:ttH-ttHib_angsep}
\end{figure}

Various top-quark and $\PQb$-jet distributions relevant for the
$\ttbar\PH$ analyses at $8\UTeV$ are shown in
\refFs{fig:ttH-ttHib_ptt}--\ref{fig:ttH-ttHib_angsep}.  Predictions at LO and
NLO are based on PDFs and $\alphas$ at the respective perturbative
order, and the dynamical scale $\mu_0=\muBDDP$ is used.  The $S1$
selection is applied. Variations of the renormalization and
factorization scales by a factor of $2$ around the central value
$\mu_0$ are displayed as uncertainty bands.  In the lower frames, LO
and NLO bands are normalized to LO results at the central scale.

In general the size of the corrections as well as the NLO scale
dependence turn out to be fairly stable with respect to all considered
kinematical variables. The most noticeable exception is given by the
top-quark transverse-momentum distributions. In particular, shape
distortions in the tail of the $\pT$ distribution of the softer top
quark (\refF{fig:ttH-ttHib_ptt}b) reach up to $40\%$. These
corrections might be reduced by using the top-quark transverse mass
instead of $\Mt$ in $\muBDDP$.  Non-negligible $10\%$-level
distortions are visible also in the $p_{\mathrm{T},\PQb\PAQb}$ and in
the $m_{\PQb\PAQb}$ distributions (\refF{fig:ttH-ttHib_bbpair}), while
the $\Delta\phi_{\PQb\PAQb}$ and $\Delta R_{\PQb\PAQb}$ distributions
receive somewhat larger shape corrections, up to $15-20\%$
(\refF{fig:ttH-ttHib_angsep}).

In summary, we have presented new NLO results for $\PQt\PAQt\PQb\PAQb$
production at $8\UTeV$, which confirm that the dynamical scale
introduced in~\Bref{Bredenstein:2010rs} guarantees a stable
perturbative description of this irreducible background to
$\PQt\PAQt\PH(\PQb\PAQb)$.  The $K$ factor turns out to be
surprisingly close to one, both for loose cuts and in the
$\PQt\PAQt\PH(\PQb\PAQb)$ signal region, while, similarly to what
originally seen at $\sqrt{s}=14\UTeV$, NLO corrections reduce the
factorization and renormalization scale uncertainty to about $25\%$.
Typical transverse-momentum, invariant-mass, and angular distributions
receive moderate but non-negligible shape corrections, which do not
exceed $15-20\%$ in general.  More pronounced kinematic distortions
are found only at large top-quark transverse momenta.

\subsection{$\PQt\PAQt\PH$ vs.\ $\PQt\PAQt\PQb\PAQb$: predictions
by \powhel{} plus Shower Monte Carlo}
\label{subsec:ttH-ttH_ttbb_powhel+powheg}

The $\PQt\PAQt\PQb\PAQb$ and $\PQt\PAQt \jj$ hadroproduction processes
represent important backgrounds to $\PQt\PAQt\PH$ production at the
LHC, when the Higgs particle decays hadronically. 

In order to allow for realistic phenomenological studies we have
started the effort of studying both signal ($\PQt\PAQt\PH$) and
background ($\PQt\PAQt\PQb\PAQb$ and $\PQt\PAQt \jj$) at the hadron
level, interfacing the corresponding NLO QCD calculations with Parton
Shower generators in the \powhel{} framework.  In a previous Working
Group report~\cite{Dittmaier:2012vm}, predictions at the hadron level
for $\PQt\PAQt\PH$ hadroproduction at the NLO QCD~+~Parton Shower
accuracy were presented by considering both a SM scalar Higgs boson
and a pseudoscalar one at $\sqrt{s}= 7\UTeV$. While predictions at the
parton level for $\PQt\PAQt\PQb\PAQb$ and $\PQt\PAQt \jj$ production at
the NLO QCD accuracy have been presented in the
literature~\cite{Bredenstein:2008zb, Bredenstein:2009aj,
  Bevilacqua:2009zn, Bevilacqua:2010ve, Bevilacqua:2011aa} and in
Section~\ref{subsec:ttH-NLO_ttbb} of this report , yet the extension
of these computations to the hadron level, by a proper NLO matching to
a Parton Shower approach, is a highly non-trivial task. This section
presents the status of our efforts in this direction, based on
developments in the \powhel{} event generator.

\powhel{} is an event generator resulting from the interface of the
\helacnlo{} set of codes~\cite{Bevilacqua:2011xh, Bevilacqua:2010mx},
that are publicly available to compute various SM processes at the NLO
QCD accuracy, and \powhegbox~\cite{Alioli:2010xd}, a
publicly available computer framework to match NLO QCD calculations to
shower Monte-Carlo (SMC) programs using the \powheg{}
method~\cite{Nason:2004rx, Frixione:2007vw}.  In \powhel, we use the
\helacnlo{} set of codes to produce all matrix elements required as
input in \powhegbox.  The SMC codes are used to compute all shower
emissions except the hardest one, already generated according to the
\powheg{} matching formalism, and to simulate hadronization and hadron-decay 
effects.  Thus \powhel{} can be used to produce both predictions at
the parton level, retaining NLO accuracy, and at the hadron level.  In
particular, one of the possible outputs of \powhel{} are Les Houches
event (LHE) files~\cite{Alwall:2006yp}, at the first radiation
emission level, ready to be further showered by SMC. So far,
\powhel{} (+ SMC) has been used to produce predictions for several
processes involving the production of a $\PQt\PAQt$ pair in association with
a third hard object ($\PQt\PAQt+$~jet~\cite{Kardos:2011qa}, $\PQt\PAQt\PH$
\cite{Garzelli:2011vp}, $\PQt\PAQt\PZ$ \cite{Garzelli:2011is}, $\PQt\PAQt\PW$
\cite{Garzelli:2012bn}), accompanied by the corresponding sets of LHE
files publicly available on our web-site~\cite{Garzelli:powhelweb}.

As for $\PQt\PAQt\PH$, a very recent update consisted in the
production of \powhel{} events at both $\sqrt{s}=8$ and $14~\UTeV$,
for different values of $\MH$. At present, our LHE files at both~$7$
and $8\UTeV$ are used by the ATLAS collaboration.  Here we present
predictions for $14\UTeV$ collider energy including a preliminary
comparison to the $\PQt\PAQt\PQb\PAQb$ background, simulated using
massless $\PQb$-quarks. More complete phenomenological analysis will
be published elsewhere. At $14\UTeV$, predictions for
$\PQt\PAQt\PQb\PAQb$ at the NLO QCD accuracy are known
\cite{Bredenstein:2009aj,Bevilacqua:2009zn}.  For the sake of
comparison, we consider the same configuration as in
\Bref{Bredenstein:2009aj} and we verify that \powhel{} predictions at
NLO accuracy agree with those published
in~\cite{Bredenstein:2009aj}. This is a non-trivial check, also taking
into account that \powhel, like \helacnlo, uses the OPP
method~\cite{Ossola:2006us,Ossola:2007ax} complemented by rational
terms of kind $\mathrm{R}_2$~\cite{Ossola:2008xq} to compute one-loop
amplitudes, whereas the authors of~\cite{Bredenstein:2009aj} rely on
tensor-reduction techniques.

As a further step, we used \powhel{} to produce $\PQt\PAQt\PQb\PAQb$
events and predictions at the first-radiation emission and at the SMC
level. The delicate issues in this respect are the following:
\begin{enumerate}
\item
the choice of a set of technical cuts, that we have to implement for
the generation of LHE events in \powhegbox, given that the
$\PQt\PAQt\PQb\PAQb$ process with massless $\PQb$-quarks is already
singular at LO; in particular, for $\PQb$ quarks, we impose a cut on
the minimum transverse momenta of the $\PQb$-quarks and on the minimum
invariant mass of the $\PQb$-quark pair, $p_{\mathrm{T},\PQb}$,
$p_{\mathrm{T},\PAQb}$, $m_{\PQb\PAQb} >2\UGeV$;
\item
the choice of suitable suppression factors to suppress the phase-space
regions less interesting from the physical point of view (i.e. the
regions that will be cut by the selection cuts);
\item
the correct estimate of the contribution of the remnants, taking into
account that in \powhegbox{} the phase space of the real emission is
split in two parts (singular and remnant regions);
\item
the stability of the computation, that requires the use of higher than
double precision in a non-negligible fraction of the phase space
points;
\item
the computing time: on the one hand, the high number of final-state
particles at the parton level, and, as a consequence, the complexity
of the virtual and real correction diagrams, and on the other, the use
of higher than double-precision arithmetic, requires intensive CPU
resources.
\end{enumerate}     
We discuss these points in detail in \Bref{Kardos:2013vxa}.

For the predictions shown in this section, we adopted the
parameters of \Bref{Bredenstein:2009aj}, in the generation of both the
signal and background events: the CTEQ6M PDF set~\cite{Pumplin:2002vw}
available in the LHAPDF interface~\cite{Whalley:2005nh}, with 5 active
flavors and a $2$-loop running $\alphas$, $\Mt=172.6\UGeV$, $\MW =
80.385\UGeV$, $\MZ = 91.1876\UGeV$, $\Mb=0$.  We use $\MH=125\UGeV$,
close to the measured mass of the newly discovered boson at the LHC
\cite{ATLAS:2012ae,Chatrchyan:2012tx}.  The factorization and
renormalization scales were set equal, $\muF = \muR \equiv \mu_0$ and
chosen to be $\mu_0 = \Mt + \MH/2$ for the signal and $\mu_0 = \Mt$
for the background.

As for SMC, we used \pythia-6.4.26.  The masses of heavy bosons in
\pythia{} were set to be the same as in \powhel. For simplicity the
contribution of photon emission from leptons was switched off and
B-hadron stability was enforced, whereas all other particles and
hadrons were assumed to be stable or to decay according to the
\pythia{} default implementation. This implies that both top quarks
and the Higgs boson were allowed to decay in all possible channels
available in the SMC. The effect of multiple interactions was
neglected.

We produced predictions for both $\PQt\PAQt\PH$ and
$\PQt\PAQt\PQb\PAQb$, by applying the following selection cuts at the
hadron level:
\begin{enumerate}
\item All hadronic tracks with $|\eta| <5$ were used to build jets, by
  means of the anti-$k_\perp$ algorithm~\cite{Cacciari:2008gp}, with a
  recombination parameter $R = 0.4$, as implemented in
  \fastjet-3.0.3~\cite{Cacciari:2011ma}.
\item We required at least six jets with $p_{\mathrm{T},j} > 20\UGeV$ and
$|\eta_j| < 5$. Jets not satisfying these constraints were not considered.
\item Among the jets we identify $\PQb$-jets by tagging them according to
the {\textsc{MCTRUTH}} information and we required at least four
$\PQb$-jets (two tagged with $\PQb$ and two with $\PAQb$) with
$|\eta_{\PQb}| < 2.7$.  Jets not satisfying the $|\eta_{\PQb}|$ constraint
were classified as non-$\PQb$ jets.
\item We required at least one isolated lepton with $p_{\mathrm{T},\ell} >
20\UGeV$ and $|\eta_\ell| < 2.5$. The lepton was isolated from all jets
by requiring a minimum separation in the pseudorapidity-azimuthal angle
plane $\Delta R=0.4$.
\item We required an event missing energy \pTmiss\ $>15\UGeV$.
\end{enumerate}
The aim of these cuts is selecting $\PQt\PAQt\PH$ and
$\PQt\PAQt\PQb\PAQb$ events with $\PH$ decaying in $\PQb\PAQb$ and
with a $\PQt\PAQt$ pair decaying semileptonically.
\begin{figure}[htb]
\includegraphics[width=0.49\linewidth] {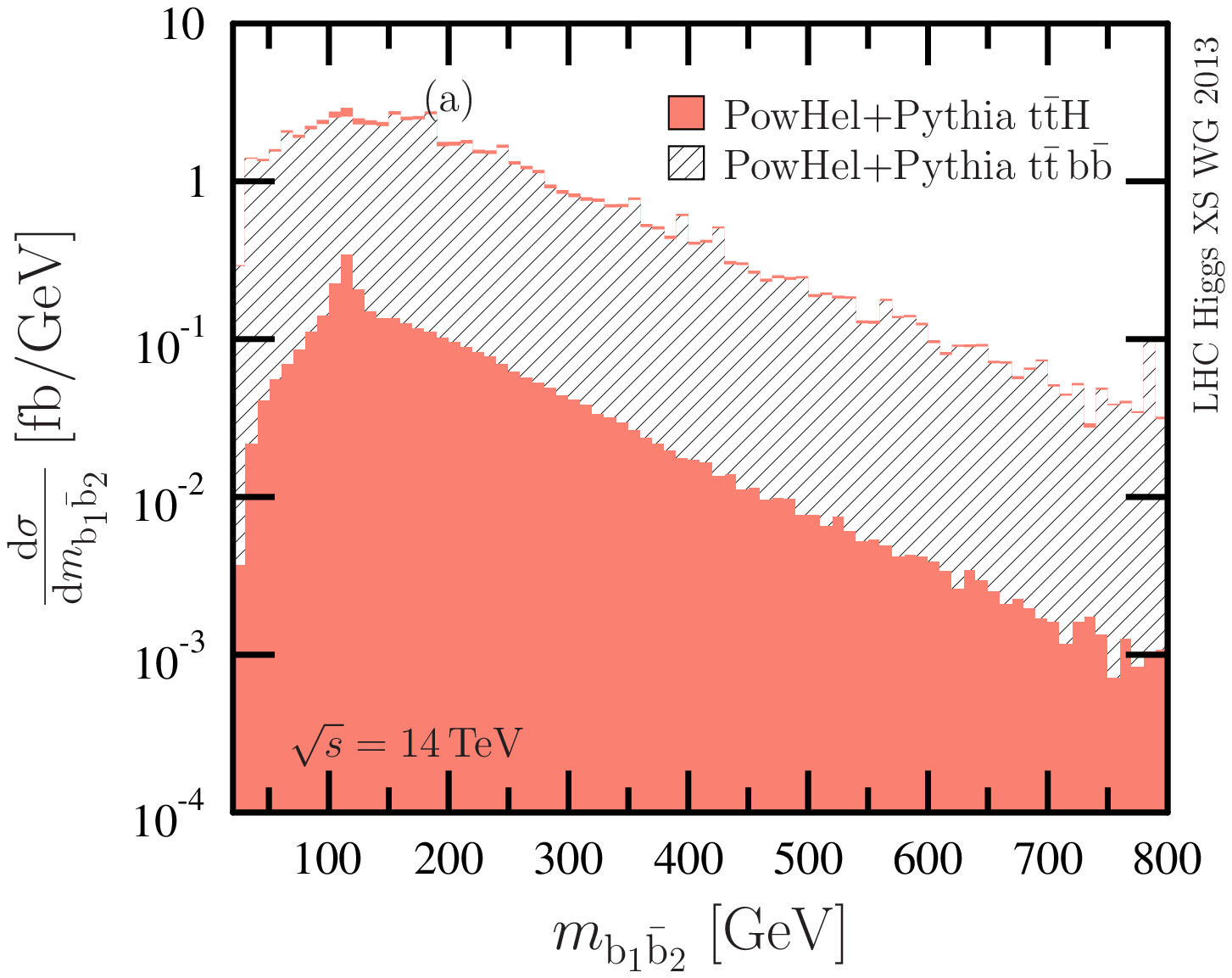}
\includegraphics[width=0.49\linewidth] {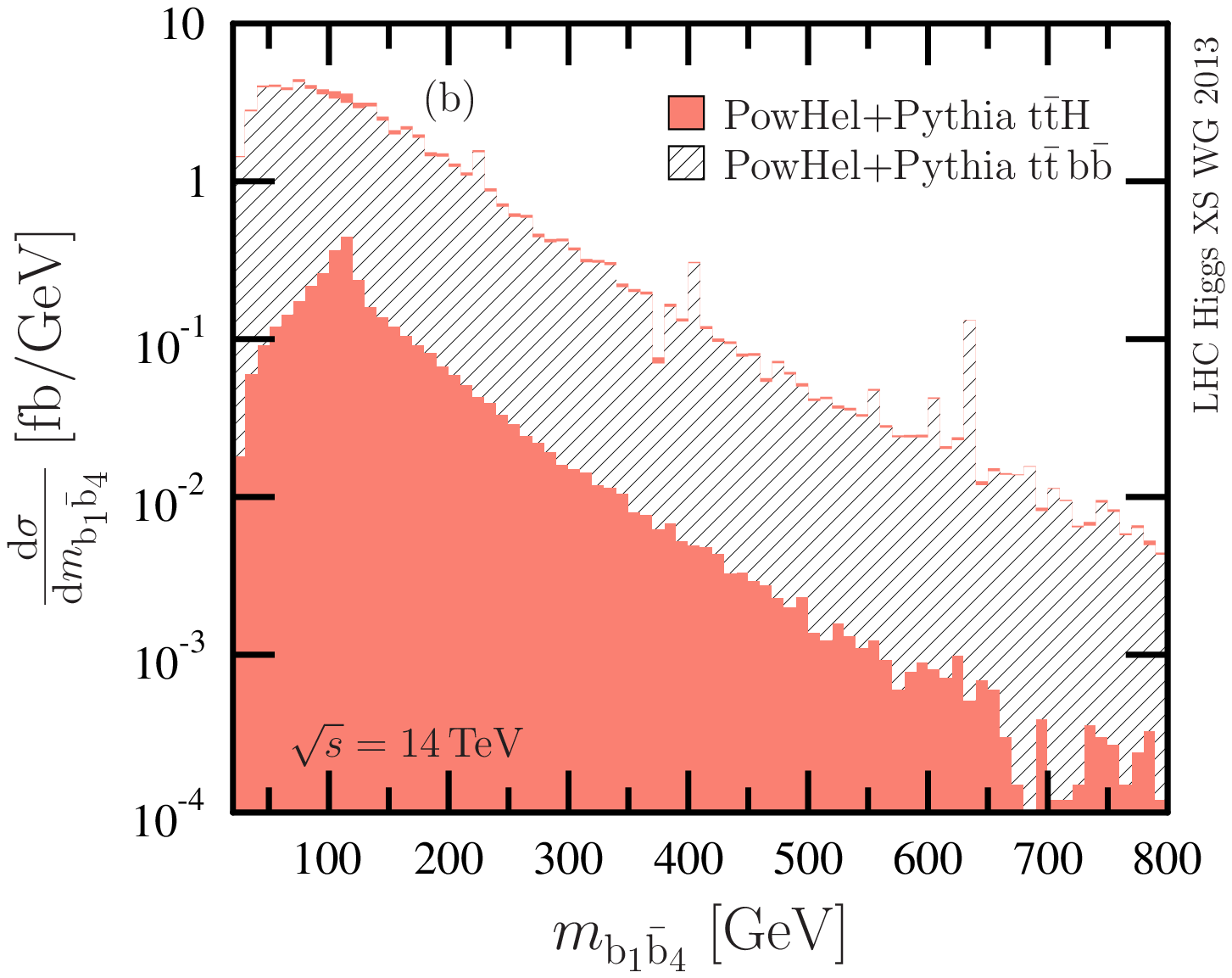}\\
\includegraphics[width=0.49\linewidth] {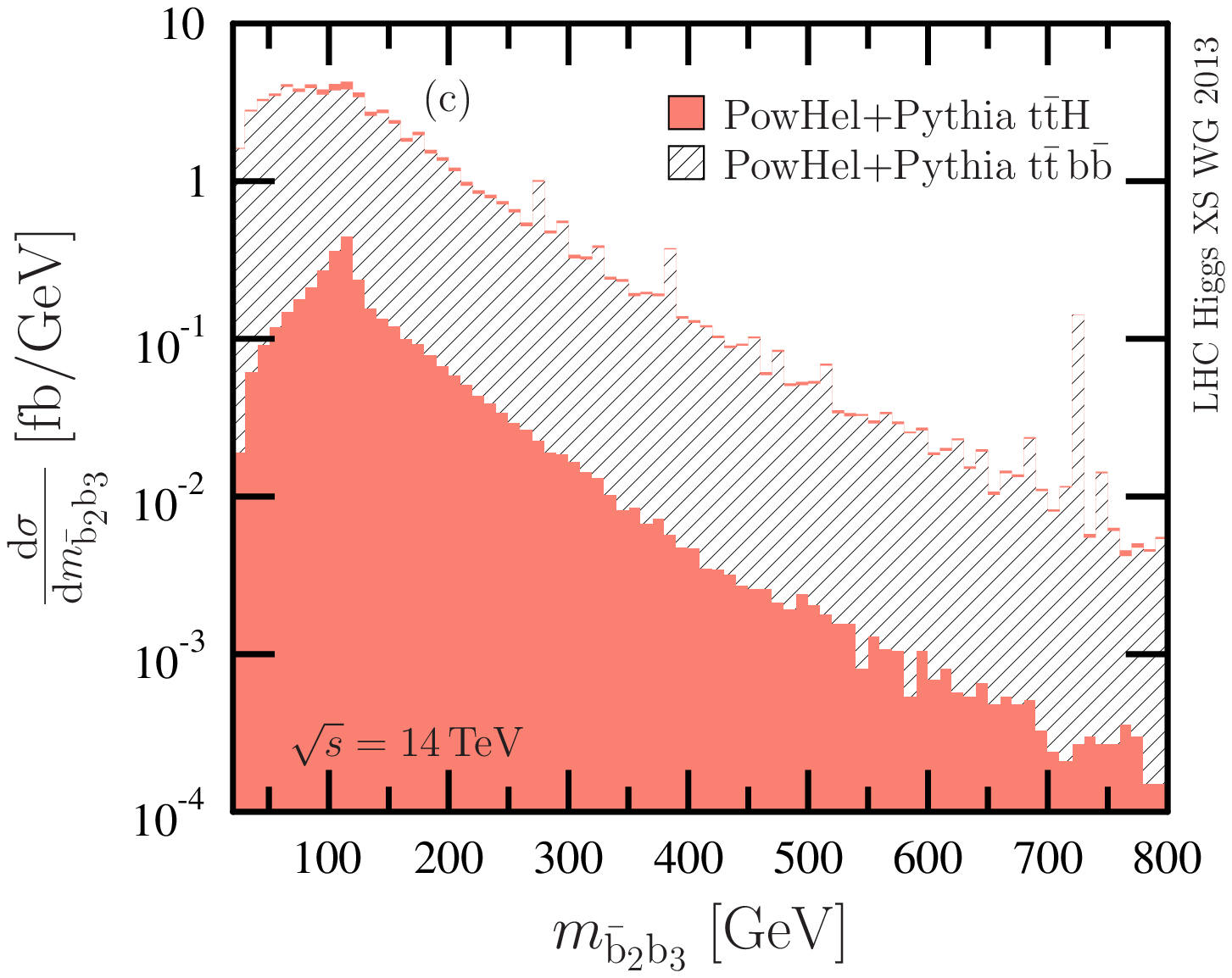}
\includegraphics[width=0.49\linewidth] {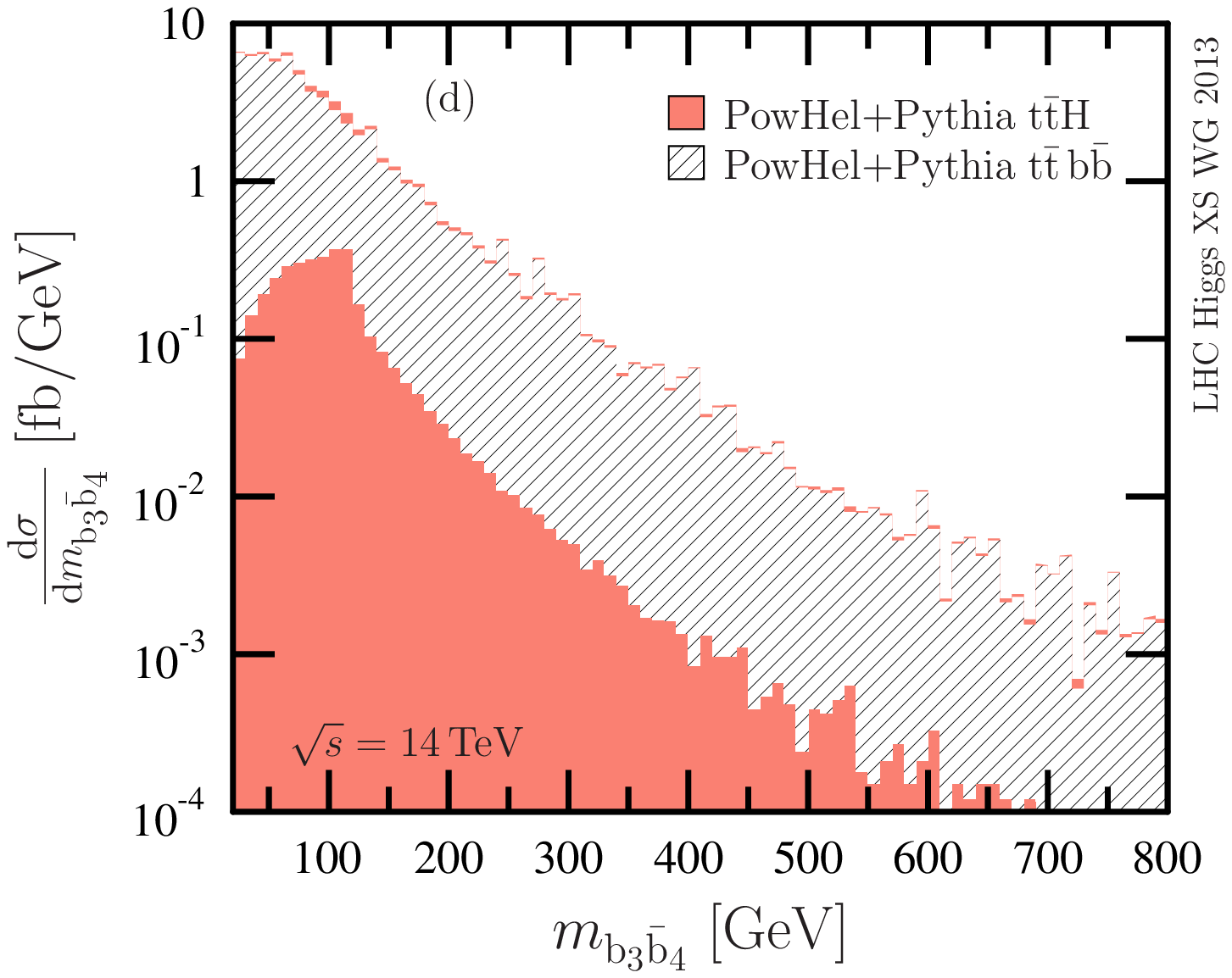}
\caption{\label{fig:ttH-fig1-powhel} Invariant mass of the four possible $\PQb$-jet
  pairs with one $\PQb$- and one $\PAQb$-tagged jet in the pairs. The
  $\PQt\PAQt\PH$ and the $\PQt\PAQt\PQb\PAQb$ distributions after
  \powhel+\pythia{} are shown by pink and shaded black histograms,
  respectively.  Peaks around the Higgs mass are visible in the
  $\PQt\PAQt\PH$ distributions.}
\end{figure}

The total cross-section of the signal after cuts is $34\Ufb$ and that
for the background is about $18$ times larger than that of the
signal. The differential distributions for $\PQt\PAQt\PH$ always lie
below those for $\PQt\PAQt\PQb\PAQb$. It is thus important to look for
differences in their shapes, to understand if it is possible to
disentangle effects of the signal in the cumulative contribution of
the signal and background. Here we concentrate on the distributions
where a difference of shape between signal and background turned out
to be more evident. In~\refF{fig:ttH-fig1-powhel} we show the
invariant masses of all possible pairings of the four $\PQb$-jets with
one $\PQb$- and one $\PAQb$-tagged jet in the pair: (a) the hardest
$\PQb$- and $\PAQb$-tagged jets, (b) the hardest $\PQb$- and second
hardest $\PAQb$-tagged jets, (c) the second hardest $\PQb$- and
hardest $\PAQb$-tagged jets, and (d) the second hardest $\PQb$- and
$\PAQb$-tagged jets. On each plot, the solid histogram is the signal
and the shaded one is the background (both based on about $200$~k
events). The signal is also
shown added to the background cumulatively.  In all cases, a peak in
vicinity of $m_H$ is clearly visible in the distributions from
$\PQt\PAQt\PH$, whereas it is absent in case of $\PQt\PAQt\PQb\PAQb$.
We emphasize that this is true not only for the pair of the hardest
$\PQb$-jets, but also for the other combinations.  It is interesting
to see how the effects of shower and hadronization do not eliminate
the presence of the $\PQt\PAQt\PH\to\PQt\PAQt\PQb\PAQb$ peak, at least
when our selection cuts are applied.  Seeing this peak in the
experimental analysis requires good reconstruction of the $\PQb$-jets,
with a low mistagging probability.
\begin{figure}[htb]
\includegraphics[width=0.49\linewidth] {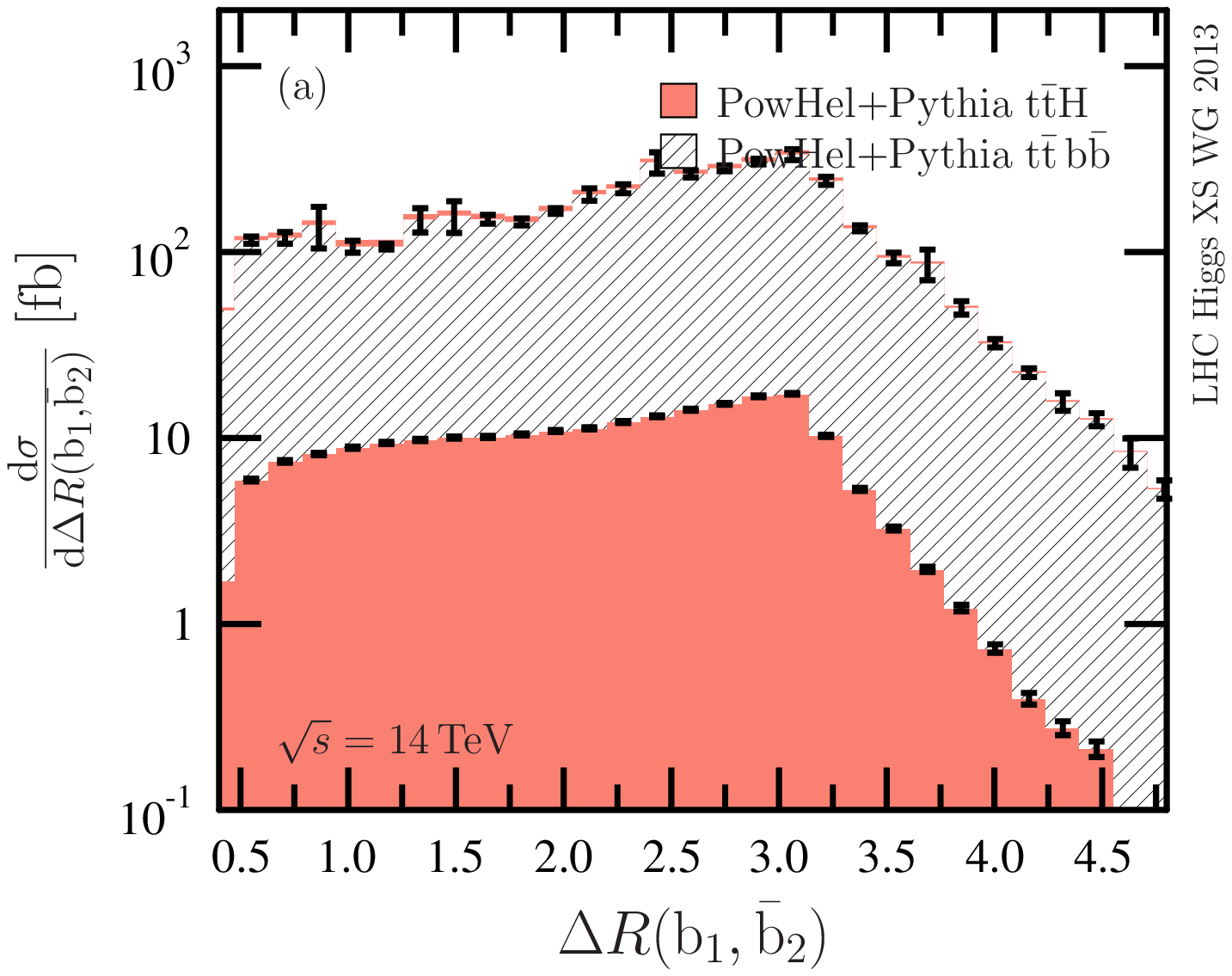}
\includegraphics[width=0.49\linewidth] {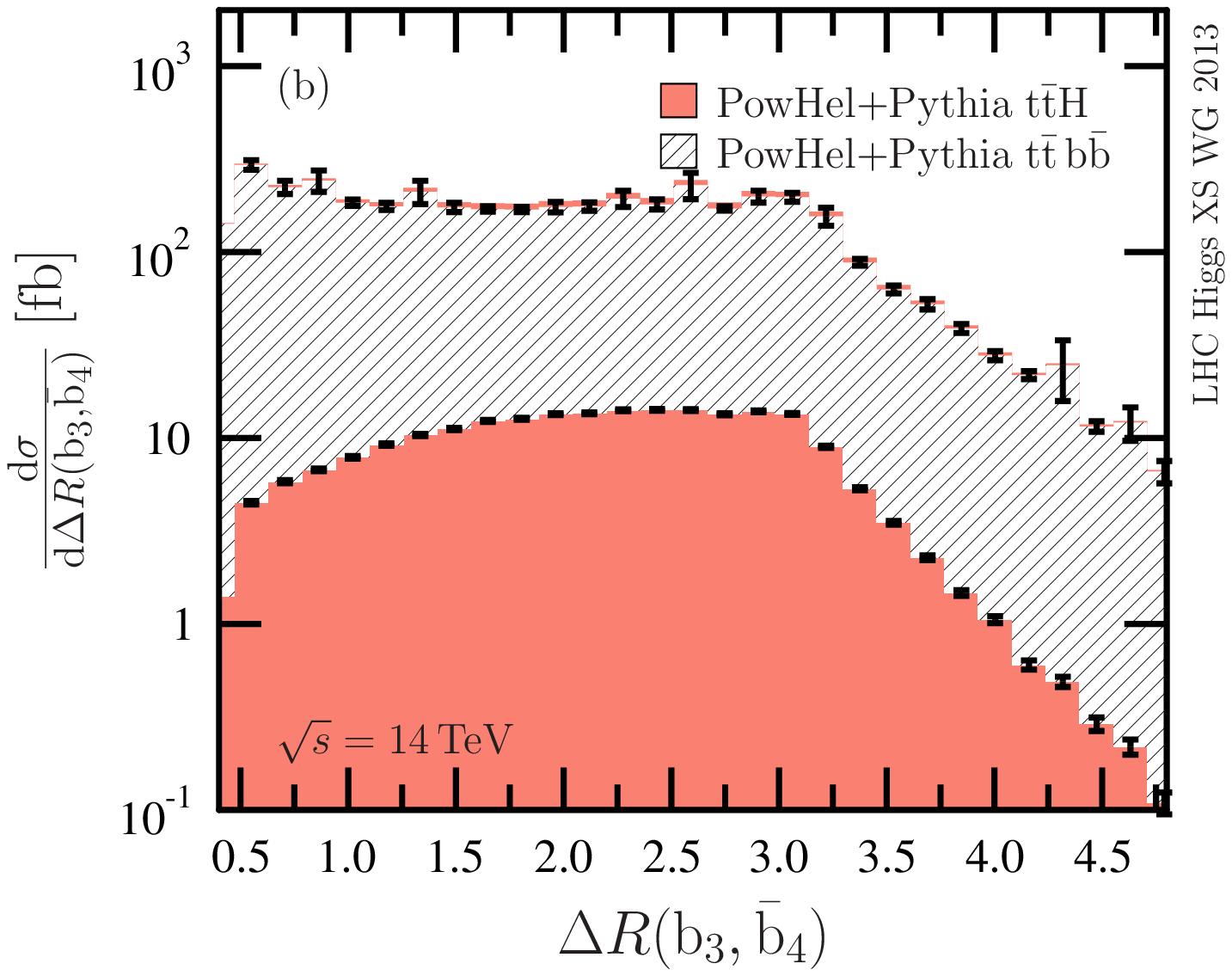}
\caption{\label{fig:ttH-fig2-powhel} $\Delta R$ distance in the
  pseudorapidity--azimuthal angle plane between the two hardest
  $\PQb$-jet pairs, and between the two second-hardest $\PQb$-jet pairs
  with one $\PQb$- and one $\PAQb$-tagged jet in the pairs. The
  $\PQt\PAQt\PH$ and the $\PQt\PAQt\PQb\PAQb$ distributions after
  \powhel+\pythia{} are shown by pink and shaded black histograms,
  respectively. When comparing the two panels a difference in the
  shape of the distributions is clearly visible. The error bars
  represent the statistical uncertainty of the events.}
\end{figure}

Other interesting distributions are the $\Delta R$-distributions of
the $\PQb$-jet pairs with one $\PQb$- and one $\PAQb$-tagged jet
in the pair. In par\-ti\-cular, when looking at the shape of the
$\Delta R$-distribution between the two second-hardest $\PQb$-jets,
shown in \refF{fig:ttH-fig2-powhel}.b, we see an almost flat
distribution in the region $\Delta R\in [0.4,3]$ in the case of the
background, as opposed to a growing behavior in the case of the
signal. However, when observing this same plot in non-log scale, it is
not easy to disentangle the $\PQt\PAQt\PH$ contribution from the
$\PQt\PAQt\PQb\PAQb$ one due to the large difference in the
cross-sections. The $\Delta R$-distribution between the two hardest
$\PQb$-jets, shown in~\refF{fig:ttH-fig2-powhel}.a, has a different
shape, and previous conclusions do not apply. The differences
between~\refF{fig:ttH-fig2-powhel}.a and~\refF{fig:ttH-fig2-powhel}.b
show that the $\Delta R$-distributions can be quite sensitive to the
experimental $\PQb$-jet reconstruction procedures and to the precision
of the b-tagging algorithms.

As for the $\pT$- and $\eta$-distributions of single $\PQb$-jets
and single leptons, our simulations do not bring evidence of big
changes of shape when comparing $\PQt\PAQt\PH$ and
$\PQt\PAQt\PQb\PAQb$. Analyses under different systems of cuts are
also under way.

Besides looking for systems of cuts that allow to increase the
signal/background ratio, further developments of this work will
consist in providing estimates of the theoretical uncertainties on our
predictions at the hadron level. In this respect, it is important to
assess the role of renormalization and factorization scale and pdf
variation, as well as to provide an estimate of the uncertainties
intrinsic to the NLO+PS matching. A first step in this direction, when
dealing with the POWHEG matching scheme, as implemented in the
\texttt{POWHEG-Box} (and also in \texttt{PowHel}), is quantifying the
impact of the variation of the starting scale for SMC emissions. In
the \texttt{POWHEG-Box}, this information is encoded in a prescription
for the choice of the \texttt{SCALUP} value of each event. The
original \texttt{POWHEG-Box}, prescription, also applied in this
section, fixes the \texttt{SCALUP} value to the relative
transverse momentum of the first radiation emission, with respect to
the emitting particle.  Recently, a modification of this prescription
has been proposed in Ref.~\cite{Nason:2013uba}.  This can be
optionally applied during the analysis of the LHE files created by
POWHEG-BOX and PowHel. The application of the new prescription has
no effect on predictions for t$\bar{\rm t}$H, whereas our preliminary
studies show an uncertainty of about 20\,\% of the
t$\bar{\rm t}$b$\bar{\rm b}$ differential cross-sections presented in
this contribution (decrease of the cross-sections after SMC for
hadroproduction of t$\bar{\rm t}$b$\bar{\rm b}$ events), whose exact
amount depends on the system of cuts. A complete quantitative study of
the uncertainties will be presented elsewhere.

%\bibliographystyle{atlasnote}
%\bibliography{YRHXS3_ttH}

%\end{document}

\clearpage

\newpage
\section{PDF \footnote{%
    S.~Forte, J.~Huston, R.~Thorne}}
\newcommand{\lp}{\left(}
\newcommand{\rp}{\right)}

Several of the PDF sets (at both NLO and NNLO) discussed in the previous Yellow
Reports~\cite{Dittmaier:2011ti,Dittmaier:2012vm}, and specifically
some of those recommended for inclusion in the PDF4LHC
prescription~\cite{Botje:2011sn,PDF4LHCwebpage} for the computation of
central value and uncertainty of a process have been
updated. Furthermore, a number of relevant studies and results from PDF
groups have become available in the last year. 

The NNLO CT10 PDFs have become available~\cite{Gao:2013xoa}. 
These PDFs use essentially the same data sets as in the
already existing CT10 NLO PDFs~\cite{Lai:2010vv}. In this new
analysis, the effects of finite quark masses have been implemented in
the S-ACOT-$\chi$ scheme (see e.g. \Bref{Binoth:2010ra},
Sect.~22 for a review)
at NNLO accuracy. A similar quality of
agreement with the fitted experimental data sets is obtained at NNLO
as at NLO. At low $x$ (<$10^{-2}$), the NNLO gluon distribution is
found to be suppressed, while the quark distributions increase,
compared to the same distributions at NLO. The GM-VFN scheme used in
the NNLO fit yields in changes to both the charm and bottom quark
distributions.

Two of the MSTW group~\cite{Watt:2012tq} have
investigated the generation of their PDFs as a Monte Carlo set
obtained from  fits to
data replicas (as done by NNPDF, see
e.g. \Bref{Dittmaier:2011ti}, Sect.~8.2.2),
 rather than the Hessian eigenvector approach
used by default. For given $\Delta \chi^2$ the results
are compatible with the eigenvector approach. Furthermore, it was shown that 
Monte Carlo PDF sets could also be generated starting from PDF determined
using the Hessian approach. This has the advantage that 
it is then possible to determine combined uncertainties from
different PDF sets by merging Monte Carlo sets. Also, it is then
possible to use for all sets the reweighting 
approach for estimating the effect of new data on PDFs introduced by NNPDF~\cite{Ball:2010gb,Ball:2011gg}.    
In \Bref{Martin:2012da}, the MSTW group also presented in  
new PDF sets based on a modification and 
extension of their standard parameterization to one using Chebyshev 
polynomials (MSTW2008CP) (and also including modified deuteron corrections - MSTW08CPdeut), 
and studied the effect of LHC data on the 
MSTW2008 PDFs~\cite{Martin:2009iq} and the new PDFs using the reweighting procedure. 
The modifications to the PDFs result in a change to the low-$x$ valence quark 
distributions, and improves the comparison to data for central 
rapidity lepton asymmetry from $\PW$ decays. Little else of significance is 
changed, including $\alphas(\MZ)$, and for the overwhelming majority of 
processes a new PDF release 
would be redundant.

NNPDF have presented a new set, NNPDF2.3~\cite{Ball:2012cx} at NLO and
NNLO, 
which, besides introducing some
small methodological improvements, is  the first set to fully include
available  LHC data. It turns out, however, that the impact of LHC
data is only moderate for the time being, and thus differences in
comparison to the previous  NNPDF2.1
set~\cite{Ball:2011mu,Ball:2011uy} 
are small (consequently, the LO NNPDF2.1 set has not been updated and
its usage together with NNPDF2.3 NLO and NNLO is recommended by
NNPDF). The only significant impact is that of the CMS $\PW$ asymmetry
data on the up-down separation,
which leads to a slightly more accurate prediction of the $\PWp$/$\PWm$
cross section ratio, besides of course more precise predictions for
the $\PW$ asymmetry itself. This is in agreement with the findings of
\Bref{Martin:2009iq}, and indeed the prediction for the up-down
valence ratio in the $x\sim0.001$ region obtained using reweighted 
MSTW08 PDFs is
in much better agreement with NNPDF2.3 (and also
NNPDF2.1)~\cite{Forte:2013wc}.  NNPDF has also presented a first PDF
determination using only collider data (hence avoiding both nuclear
target and lower-energy data): whereas these PDFs are less subject to
theoretical uncertainties related to nuclear an higher twist
corrections, their statistical uncertainties are still not competitive.
A similar conclusion, though based on studies without the recent LHC 
data appeared in \cite{Watt:2012tq}.

The updated PDF sets to be used with the NLO PDF4LHC prescriptions are
thus currently CT10, MSTW08, and NNPDF2.3.

A new NLO and NNLO set, ABM11, has
 been made available by the ABM group~\cite{Alekhin:2012ig}, and a 
benchmarking exercise performed. As well as updating for the combined HERA
data~\cite{Aaron:2009aa} this fit uses the $\overline{\mathrm{MS}}$ renormalization 
scheme for heavy quark masses. In all important respects these PDFs remain
similar to those of ABKM09 \cite{Alekhin:2009ni}, though the gluon 
distribution is a little larger at small $x$. As before Tevatron jet data is
not included directly in the fit, though a comparison to this data 
is presented by
the same authors in~\cite{Alekhin:2012ce}. The authors make the PDFs 
available for a wide variety of $\alphas(\MZ)$ values, though the extracted
value, which is recommended by the authors, is $\alphas(\MZ)=0.1134\pm 0.0011$ at NNLO.      

The HERAPDF collaboration have 
released the HERAPDF1.5 NLO and 
NNLO PDF set~\cite{Radescu:2010zz,CooperSarkar:2011aa}, 
which in addition to the combined HERA-I dataset uses 
the inclusive HERA-II data 
from H1~\cite{Aaron:2012qi} and 
ZEUS~\cite{Abramowicz:2012bx} (though the PDF set is partially based on
a yet-unpublished combined data set). 

Furthermore, within the HERAPDF-HERAFITTER
framework~\cite{Aaron:2009aa}, 
the ATLAS collaboration performed~\cite{Aad:2012sb}
a study of the impact  on the strange quark PDF of the
inclusion of their data on differential $\PW$ and $\PZ$
production~\cite{Aad:2011dm}. This implied a significant increase
of the strange quark contribution to the sea. However, NNPDF2.3
instead finds that whereas the ATLAS do tend to pull the strange
distribution upwards, the effect is negligible withing current
uncertainties; MSTW find similar results. 
There is also a study of the sensitivity $\PW$ and $\PZ$ production to
the strange quark distribution in~\cite{Kusina:2012vh}, but no
explicit examination of the effect of the published data. 

A variety of studies  of theoretical uncertainties on PDFs have recently
appeared. As mentioned above, the study of
extended parameterizations 
in \Bref{Martin:2012da} has been generalized to include 
variation and optimization of the nuclear corrections to deuteron
structure functions. 
Hence, a further modified version of the MSTW08 set, MSTW2008CPdeut
was obtained.   
A study of nuclear corrections using CTEQ PDFs has also 
been presented~\cite{Owens:2012bv}, with broadly similar conclusions,
i.e. a slight increase 
(and increased uncertainty) on the high-$x$ down quark. An increase of
the $\PQd/\PQu$ ratio at the one-sigma level 
for $0.1\lsim x\lsim0.5$ as a consequence of the
inclusion of deuterium corrections was similarly found in
\Bref{Ball:2013gsa}, where it was also shown that their impact is
however otherwise negligible (and in particular negligible for larger
$x$) in the scale of current PDF uncertainties.

In~\cite{JimenezDelgado:2012zx}
it was shown that there can be sensitivity to input scale for PDFs, though 
this will always depend on the flexibility of the input
parameterization. In~\cite{Thorne:2012az} the uncertainty associated
with choices of  
general mass variable flavour number scheme (GM-VFNS) was studied. 
At NLO this can be 
a couple of percent, but as with other scheme choices it diminishes at 
increasing order and becomes $<1\%$ at NNLO. These changes were less than
those obtained in using either the zero mass approximation or using 
fixed flavour number scheme (FFNS). There were implications that the
differences in  
PDFs and the value of $\alphas(\MZ)$ obtaining using either (GM-VFNS) and FFNS 
can be significant, and this requires further (ongoing) study.      
Similar conclusions were reached in~\Bref{Ball:2013gsa}, where it
was shown that use of a FFN scheme affects significantly the total
quark and gluon PDFs, and leads to significant loss in fit quality,
especially to difficulties in reproducing the high
$Q^2$, low $x$ HERA data. In the same Ref. it was also shown that
higher twist corrections to the DIS data included in the NNPDF PDF
determination  have a negligible impact on all PDFs, similar to 
previous conclusions by MRST~\cite{Martin:2003sk} 
and more recent studies involving MSTW PDFs. 

Largely as a part of the work on CT10 NNLO, a number of theoretical
uncertainties related to the treatment of jet cross sections
has been examined, notably those that may have impact on
processes involving gluon scattering. A benchmark comparison of NLO
computations for inclusive jet production constraining the gluon PDF
has been performed ~\cite{Gao:2012he,Ball:2012wy}. A new version of
the program EKS for NLO jet cross sections was
developed~\cite{Ellis:1992en,Gao:2012he} that is entirely independent
from \textsc{NLOJET++}~\cite{Nagy:2001fj} as well as
\textsc{APPLGRID}~\cite{Carli:2010rw} and \textsc{FastNLO}~\cite{Kluge:2006xs} programs
that interpolate the NLOJET++ input. Specific input settings that
produce agreement of all these codes were identified, and
uncertainties in the gluon PDFs associated with fitting the jet data
were examined. It was pointed out, for example,  that correlated
systematic errors published by the jet experiments 
are interpreted differently by the various PDF fitting groups, which
leads to non-negligible differences between the PDF sets. This issue
is not widely known and will be considered in the future to avoid
biases in PDF fits. 

Hence, overall, although there have been a significant number of important 
and interesting updates, there has been no dramatic change in 
PDFs in the past year, mainly because it is clear that LHC data so far
published do not add add a great deal of extra constraint.
A comparison of results using NLO PDFs would be little
different to a year ago. However, especially in the quark sector,
there is some gradual improvement in
agreement between the sets included in the PDF4LHC prescription, which
follows prior 
improvement at the time of the previous report~\cite{Dittmaier:2012vm}
since the original prescription~\cite{Botje:2011sn}. 

Also, a more complete comparison of results 
using NNLO PDFs is now possible. Therefore, 
in this section, we compare NNLO PDF
luminosities (and their uncertainties) from five PDF fitting groups, 
i.e. those that are made available for a variety of $\alphas(\MZ)$
values including those similar to the world average,  
and the resulting predictions for both standard model and Higgs boson
cross sections at $8\UTeV$ at the LHC. We follow the recent benchmarking
exercise in \cite{Ball:2012wy}. 

Following \Bref{Campbell:2006wx}, we define the parton luminosity 
for production of a final state with mass $M_X$ as

\begin{equation}
\Phi_{ij}\lp M_X^2\rp = \frac{1}{s}\int_{\tau}^1
\frac{dx_1}{x_1} f_i\lp x_1,M_X^2\rp f_j\lp \tau/x_1,M_X^2\rp \ ,
\label{eq:lumdef}
\end{equation}
where $f_i(x,M^2)$ is a PDF at a scale $M^2$, 
and $\tau \equiv M_X^2/s$. 

\begin{figure}[ht]
    \begin{center}
      \includegraphics[width=0.48\textwidth]{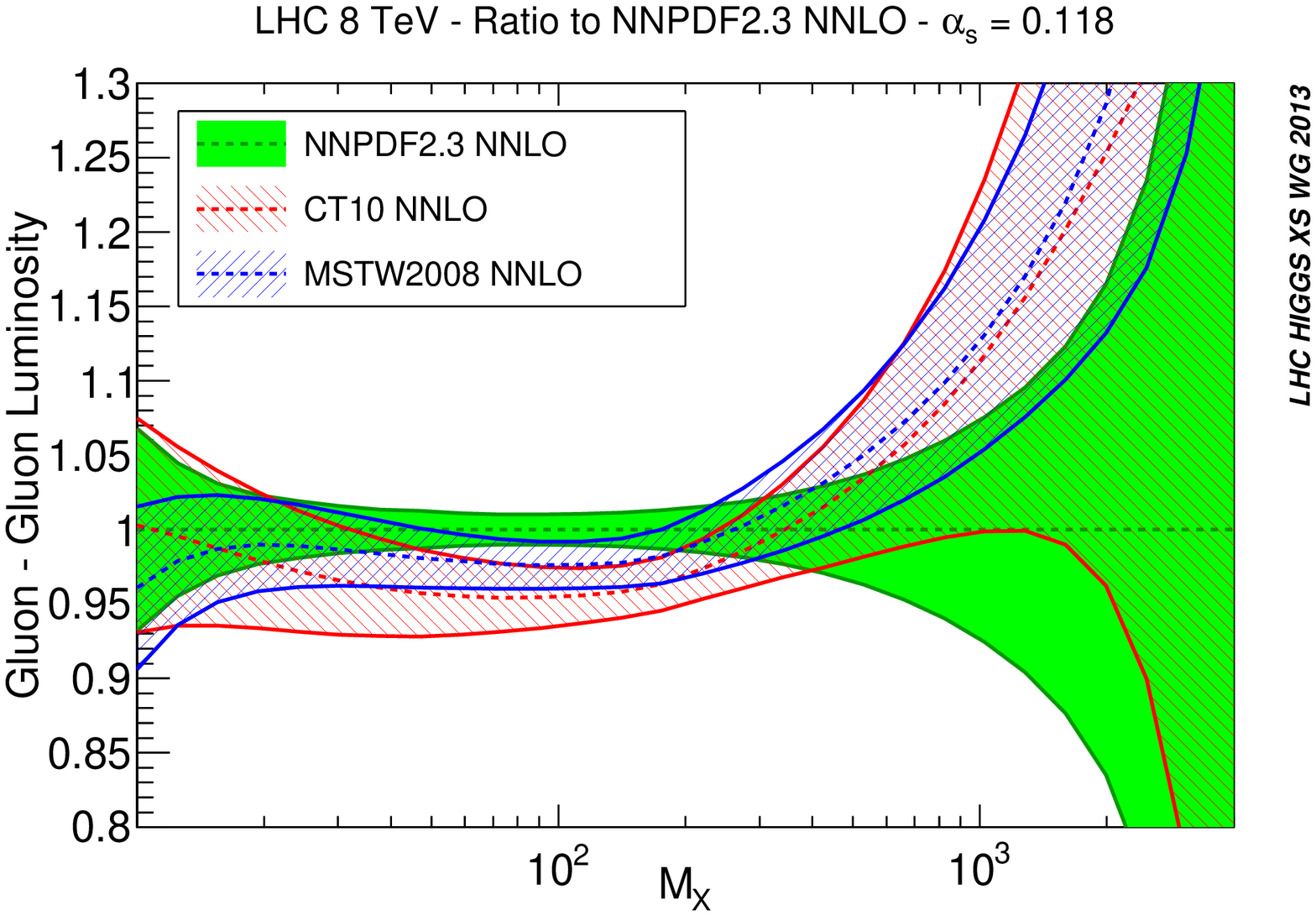}\quad
\includegraphics[width=0.48\textwidth]{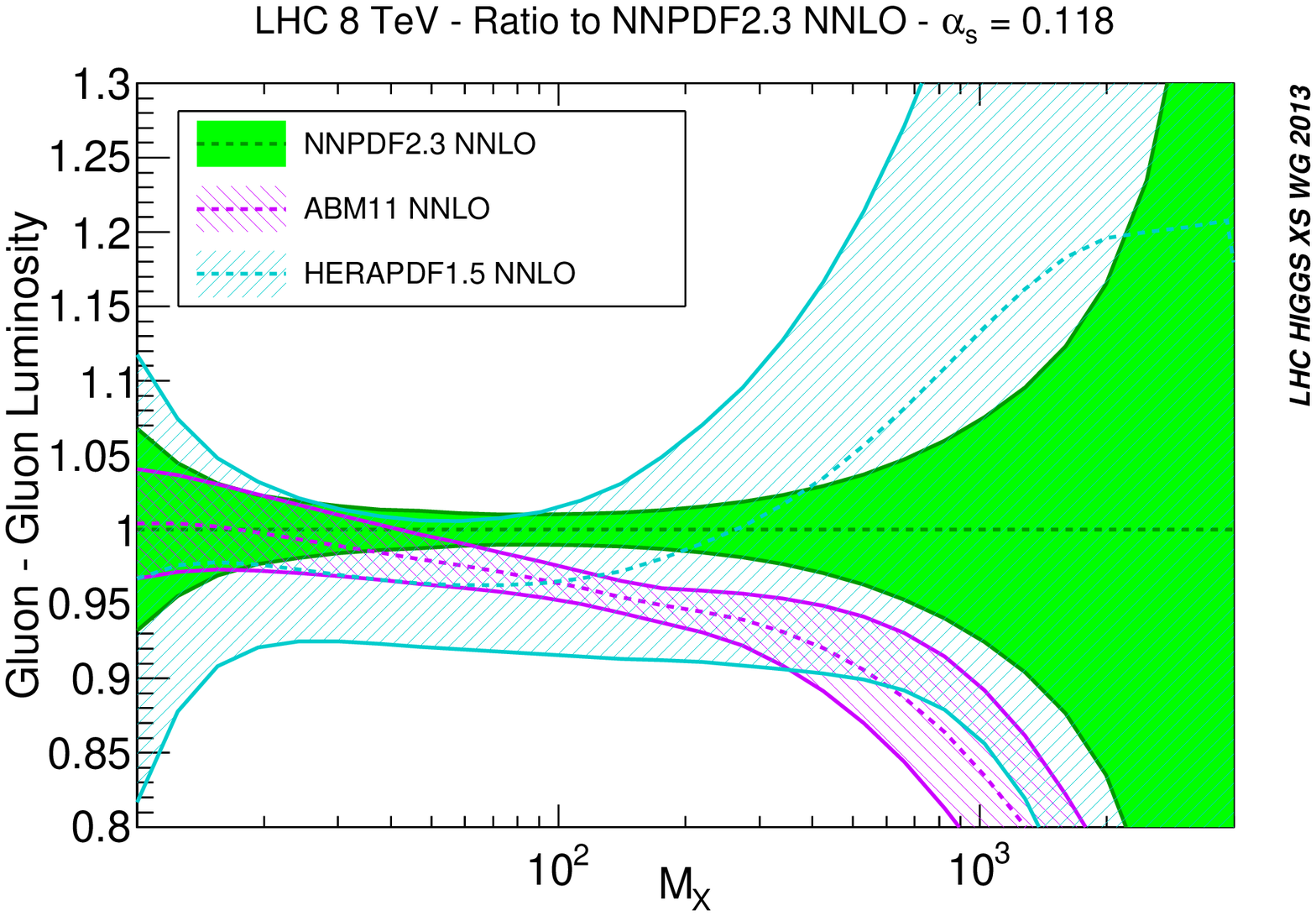}\\
  \includegraphics[width=0.48\textwidth]{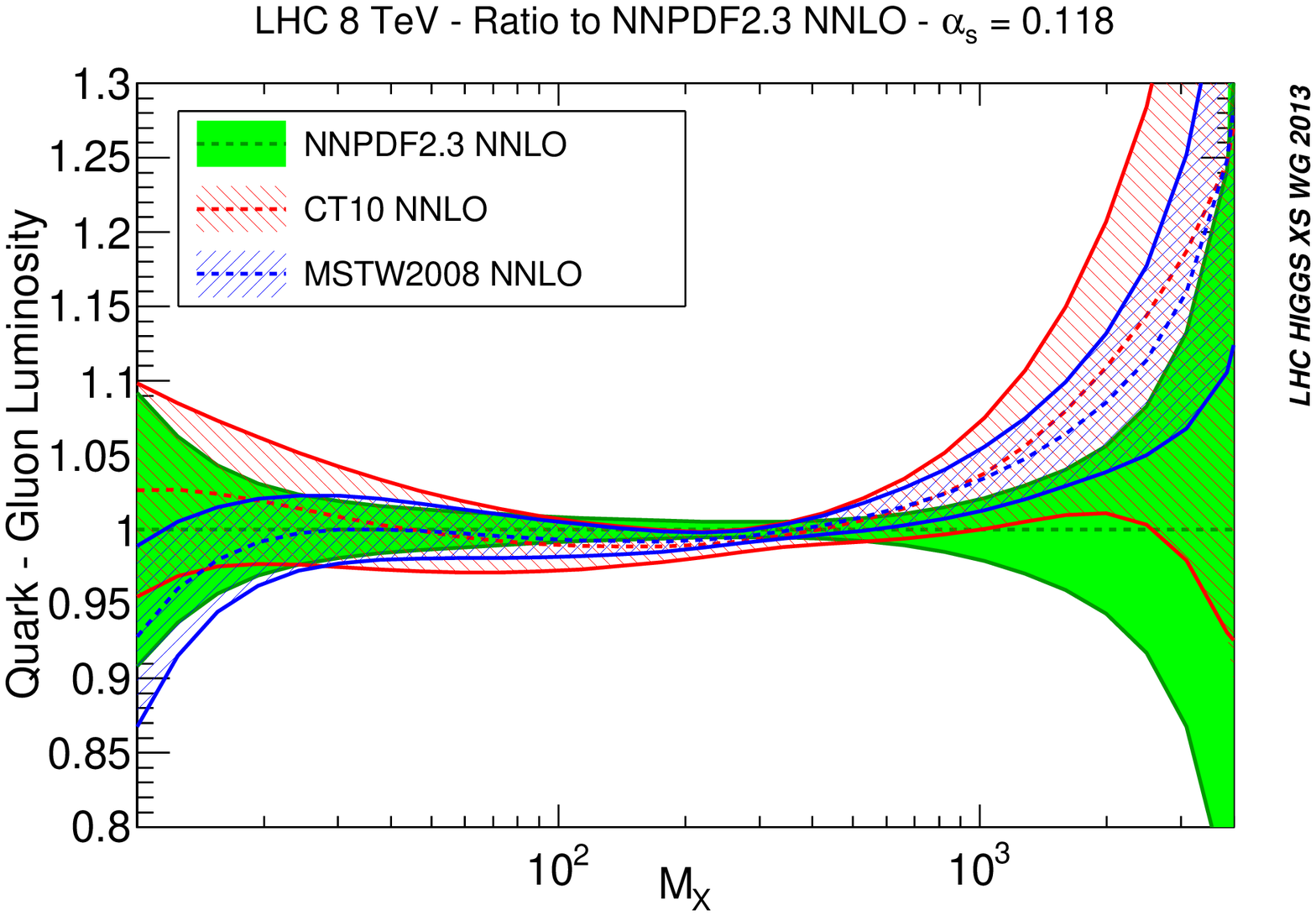}\quad
\includegraphics[width=0.48\textwidth]{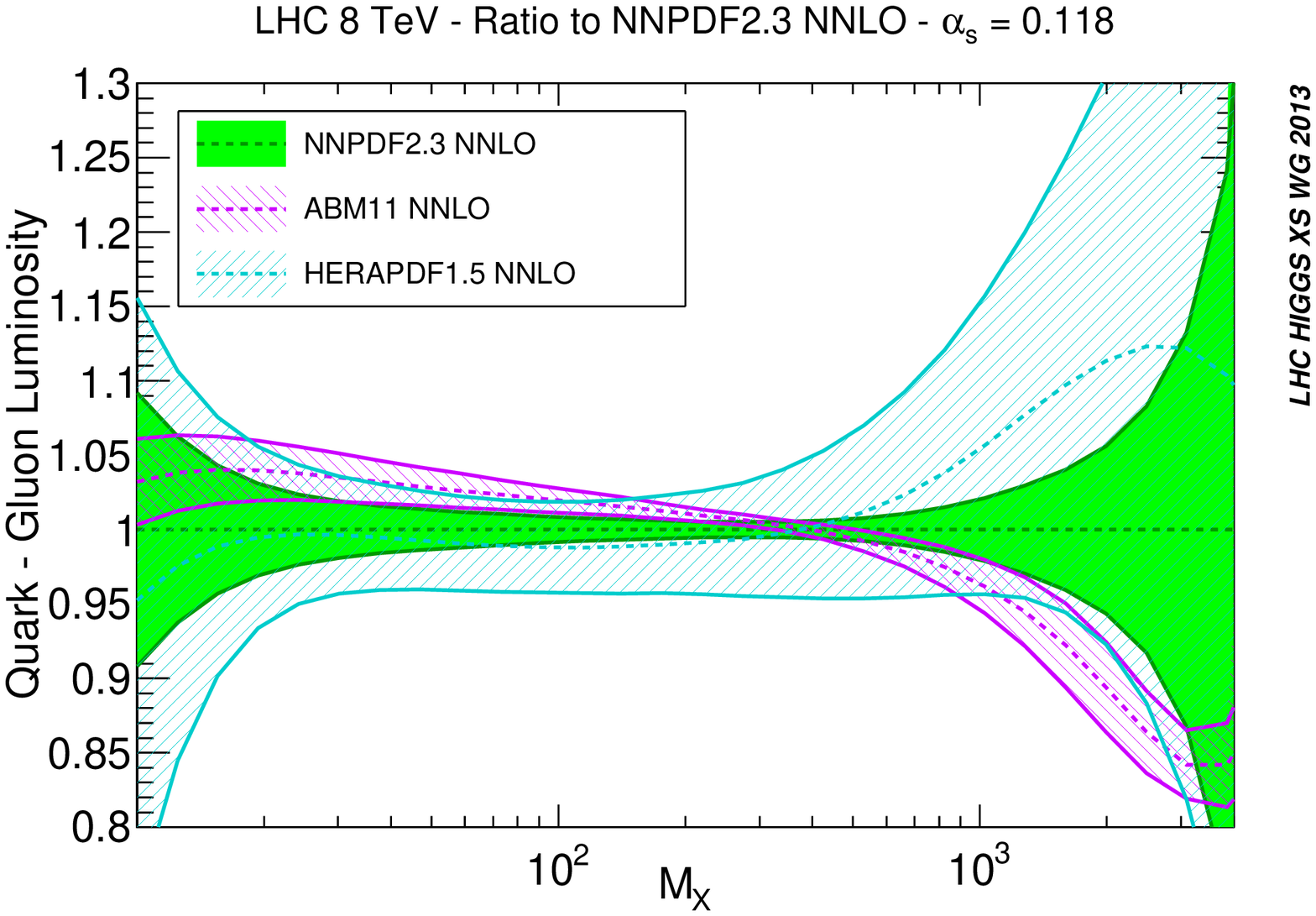}
      \end{center}
     \caption{\small
The gluon-gluon (upper plots)
and quark-gluon (lower plots) 
luminosities, Eq.~(\ref{eq:lumdef}), for the production
of a final state of invariant mass $M_X$ (in GeV) at LHC $8\UTeV$.  The left plots
show the comparison between NNPDF2.3, CT10 and MSTW08, while
in the right plots we compare NNPDF2.3, HERAPDF1.5 and MSTW08. 
All luminosities are computed at a common value of $\alphas(\MZ)=0.118$.
    \label{fig:PDFlumi-gg} }
\end{figure}

In \Fref{fig:PDFlumi-gg}, the gluon-gluon (top) and gluon-quark
(bottom) parton luminosities from five PDF groups are plotted for the
production of a state of  mass $M_X$ (GeV), as a ratio to the PDF
luminosity of NNPDF2.3. For these comparisons, a common value of
$\alphas(\MZ)$ of $0.118$ has been used. In the region of the Higgs
mass ($125\UGeV$), the $\Pg\Pg$ luminosities (and uncertainties) of NNPDF2.3,
CT10 and MSTW2008 are reasonably close, with the error bands
overlapping, but the total uncertainty range, defined from the bottom
of the CT10 error band to the top of the NNPDF2.3 error band, is of
the order of 8\%. This is approximately twice the size of the error
bands of either CT10 or MSTW2008 (and a bit more than twice that of
NNPDF2.3). The $\Pg\Pg$ luminosities of HERAPDF1.5 and ABM11 are similar
to the three PDFs sets discussed above in the Higgs mass range, although the
$\Pg\Pg$ PDF luminosity for ABM11 falls faster with mass than any of the
other PDFs. The HERAPDF PDF uncertainty is larger, due to the more
limited data sets included in the fit.  

%%%%%%%%%%%%%%%%%%%%%%%%%%%%%%%%%%%%%%%%%%%%%%%%%
%%%%%%%%%%%%%%%%%%%%%%%

\begin{figure}[ht]
    \begin{center}
      \includegraphics[width=0.48\textwidth]{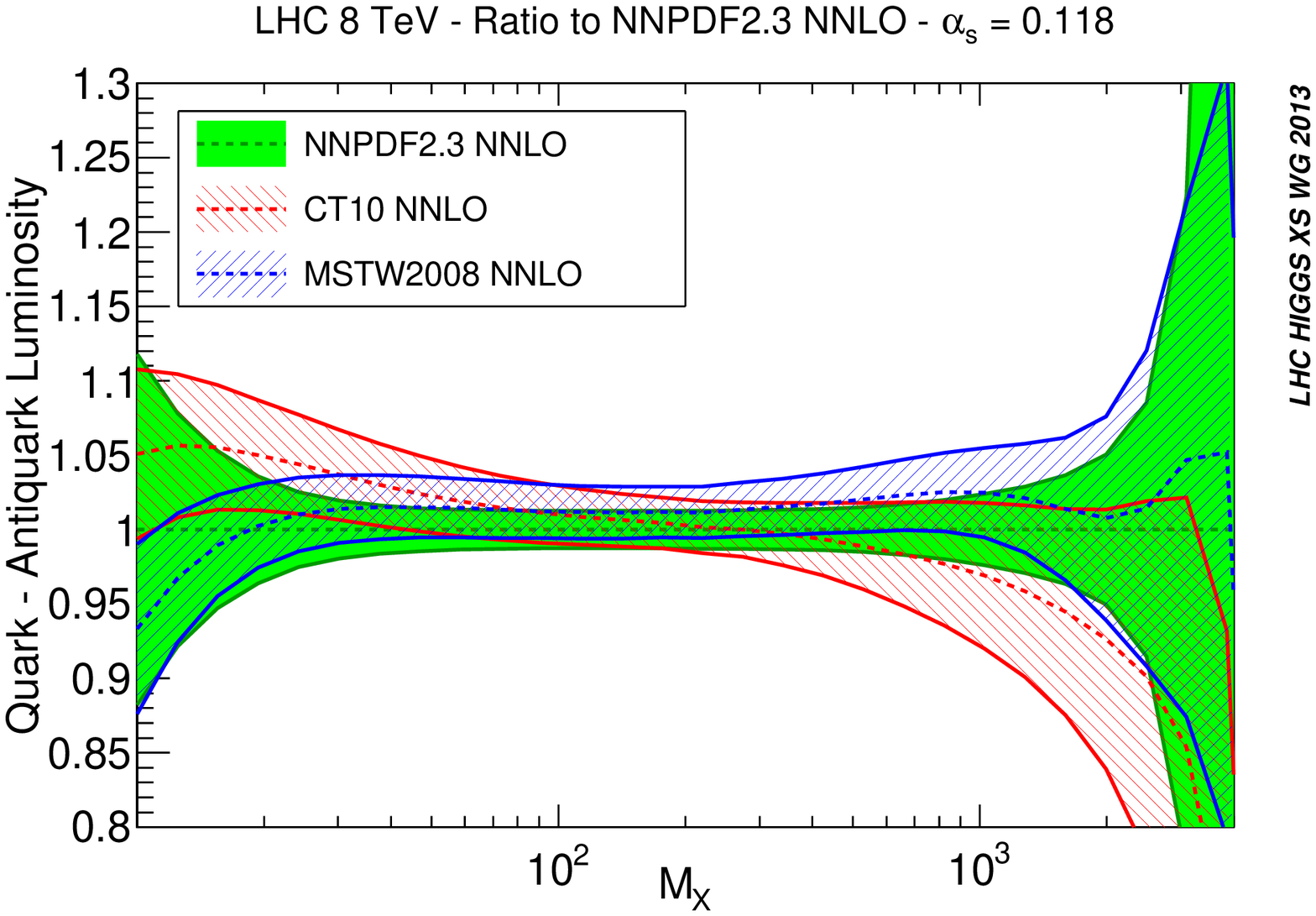}\quad
\includegraphics[width=0.48\textwidth]{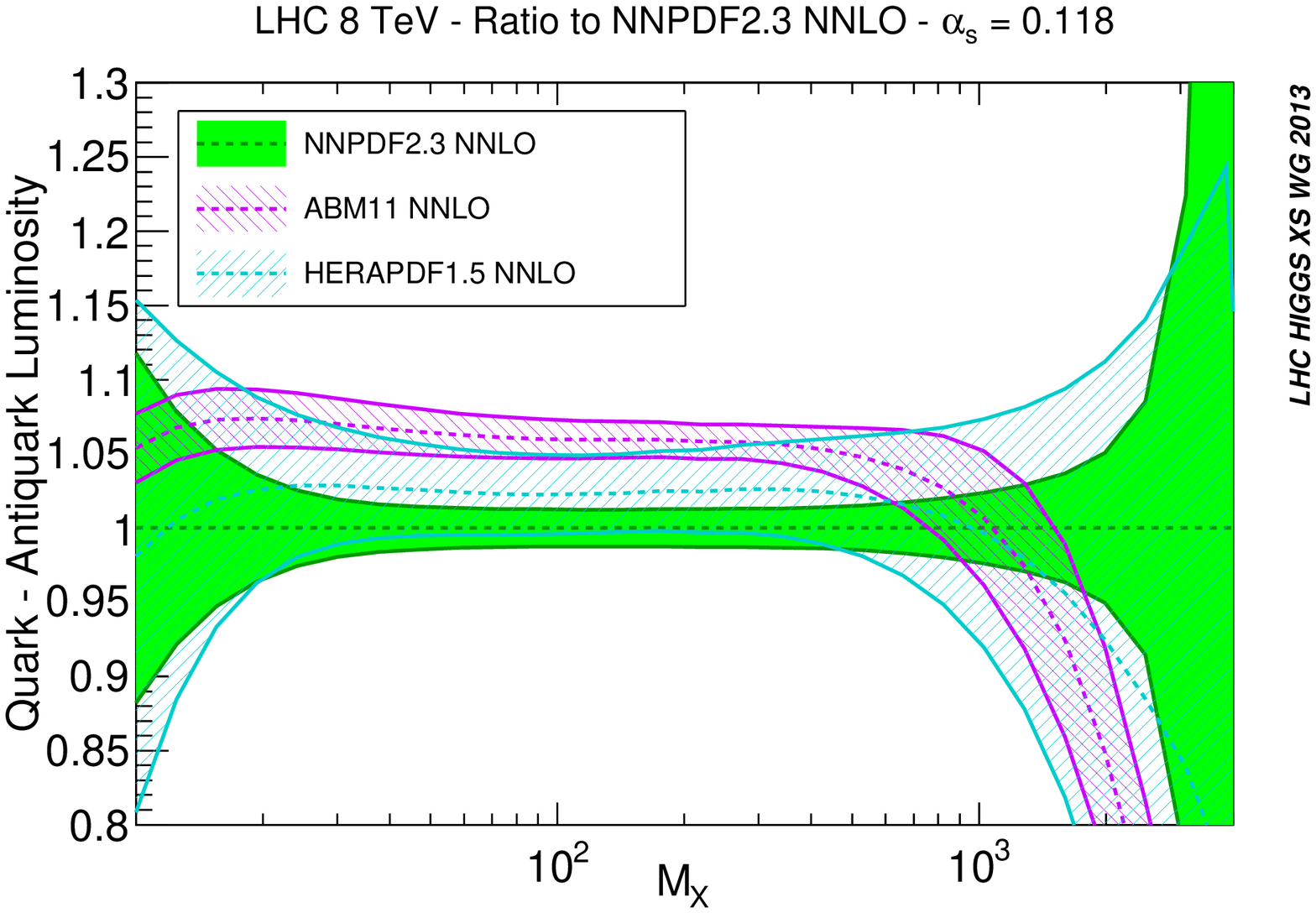}\\
  \includegraphics[width=0.48\textwidth]{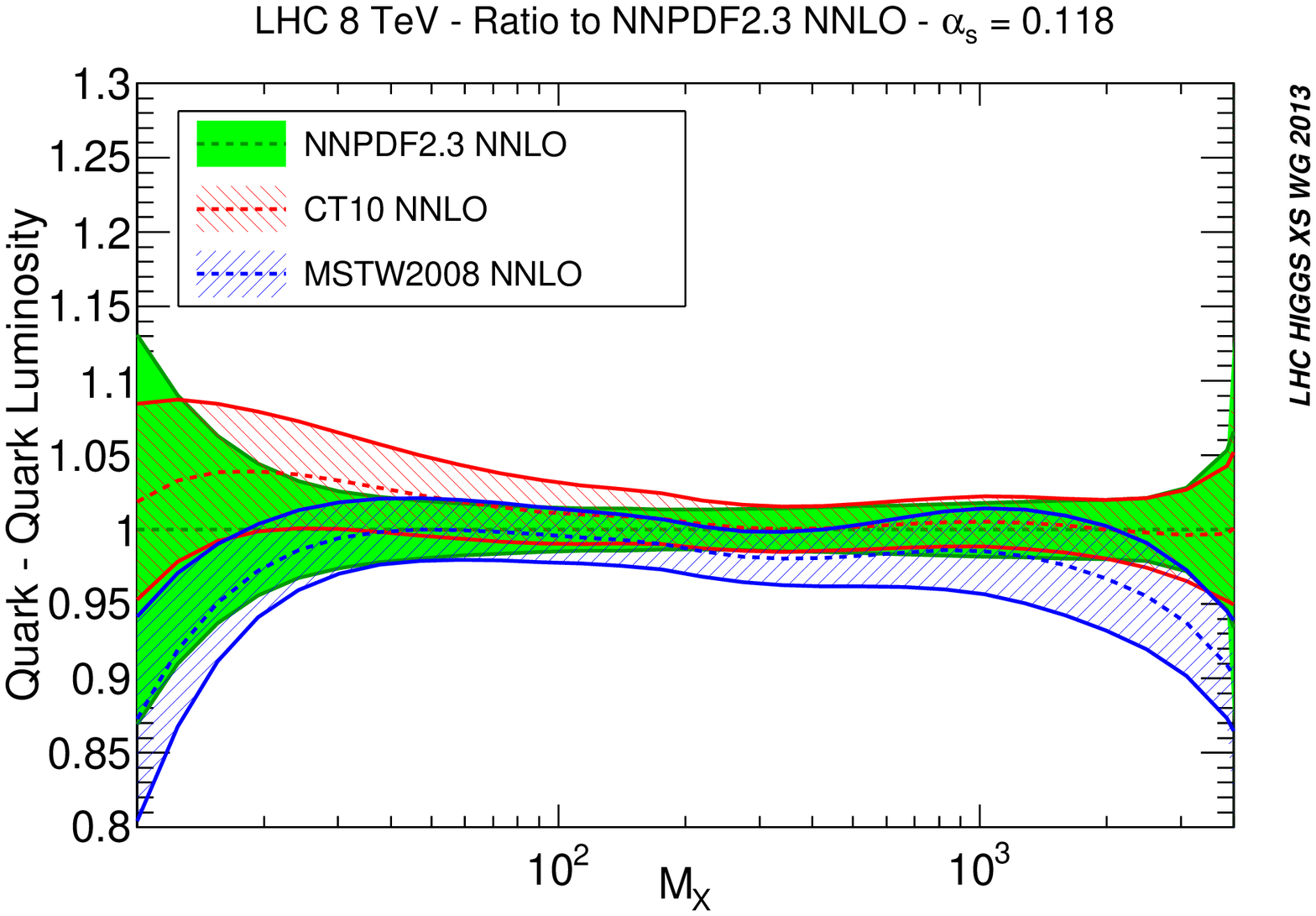}\quad
\includegraphics[width=0.48\textwidth]{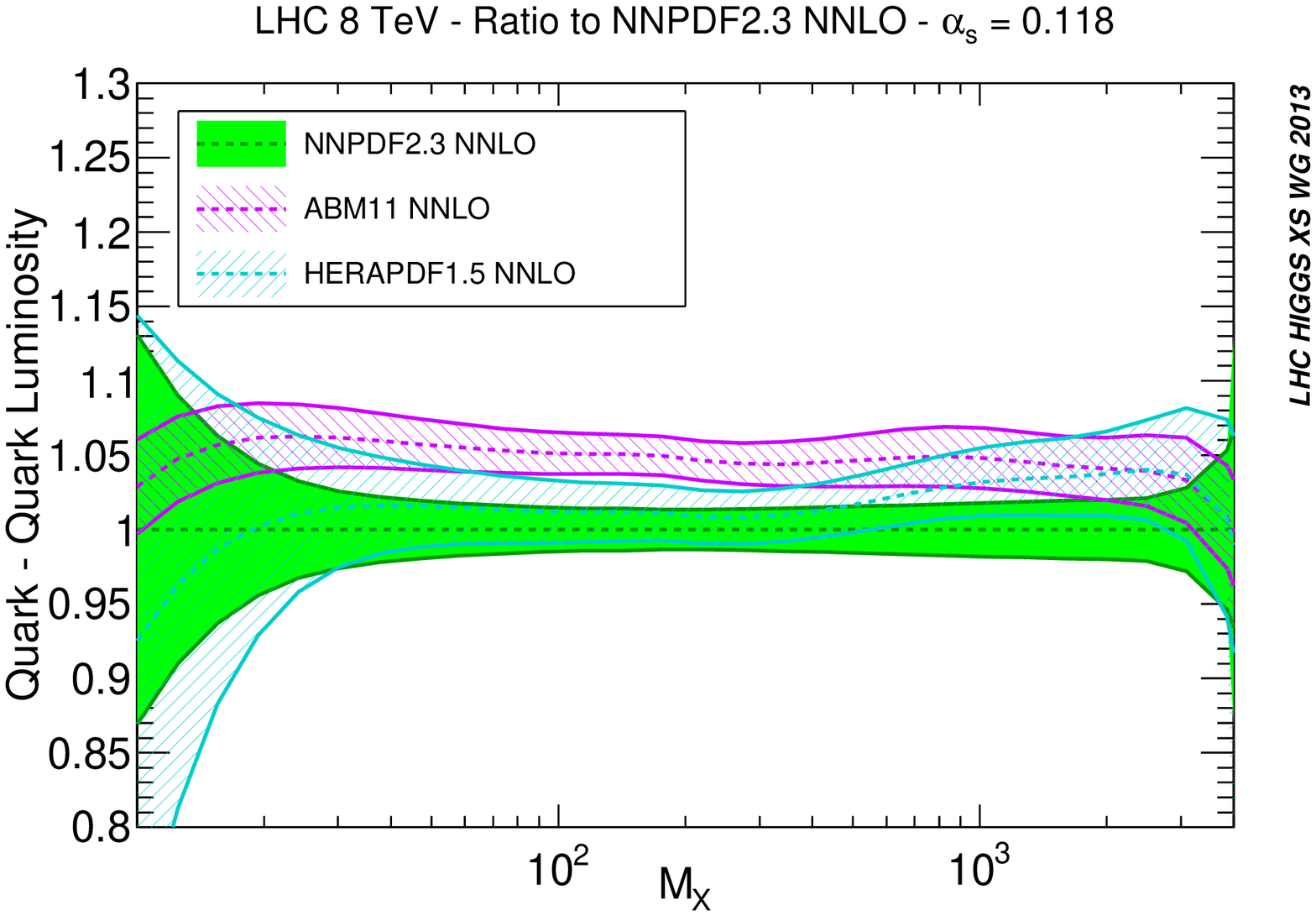}
      \end{center}
     \caption{\small
The same as  \Fref{fig:PDFlumi-gg} 
for the quark-antiquark (upper plots)
and quark-quark (lower plots) luminosities.
    \label{fig:PDFlumi-qq}} 
\end{figure}

The $\Pg\PQq$ PDF luminosity error bands overlap very well for CT10,
MSTW2008 and NNPDF2.3 in the Higgs mass range, and indeed at all masses except, 
to some extent, well above $1\UTeV$.  
Again the HERAPDF1.5 uncertainty band is larger. The $\PAQq\PQq$ (top) and
$\PQq\PQq$ (bottom) PDF luminosity comparisons are shown in
\Fref{fig:PDFlumi-qq}. For both $\PAQq\PQq$ and $\PQq\PQq$, there is a
very good overlap of the CT10, MSTW2008 and NNPDF2.3 error bands. The
central predictions of HERAPDF1.5 also agree well, with the
uncertainty band again being somewhat larger. The ABM11 luminosities
are somewhat higher in the low to medium mass range and fall more
quickly at high mass.  

%%%%%%%%%%%%%%%%%%%%%%%%%%%%%%%%%%%%%%%%%%%%%%%%%
%%%%%%%%%%%%%%%%%%%%%%%%%%%%%%%%%%%%%%%%%%%%%%%%%%

\begin{figure}[ht]
    \begin{center}
      \includegraphics[width=0.48\textwidth]{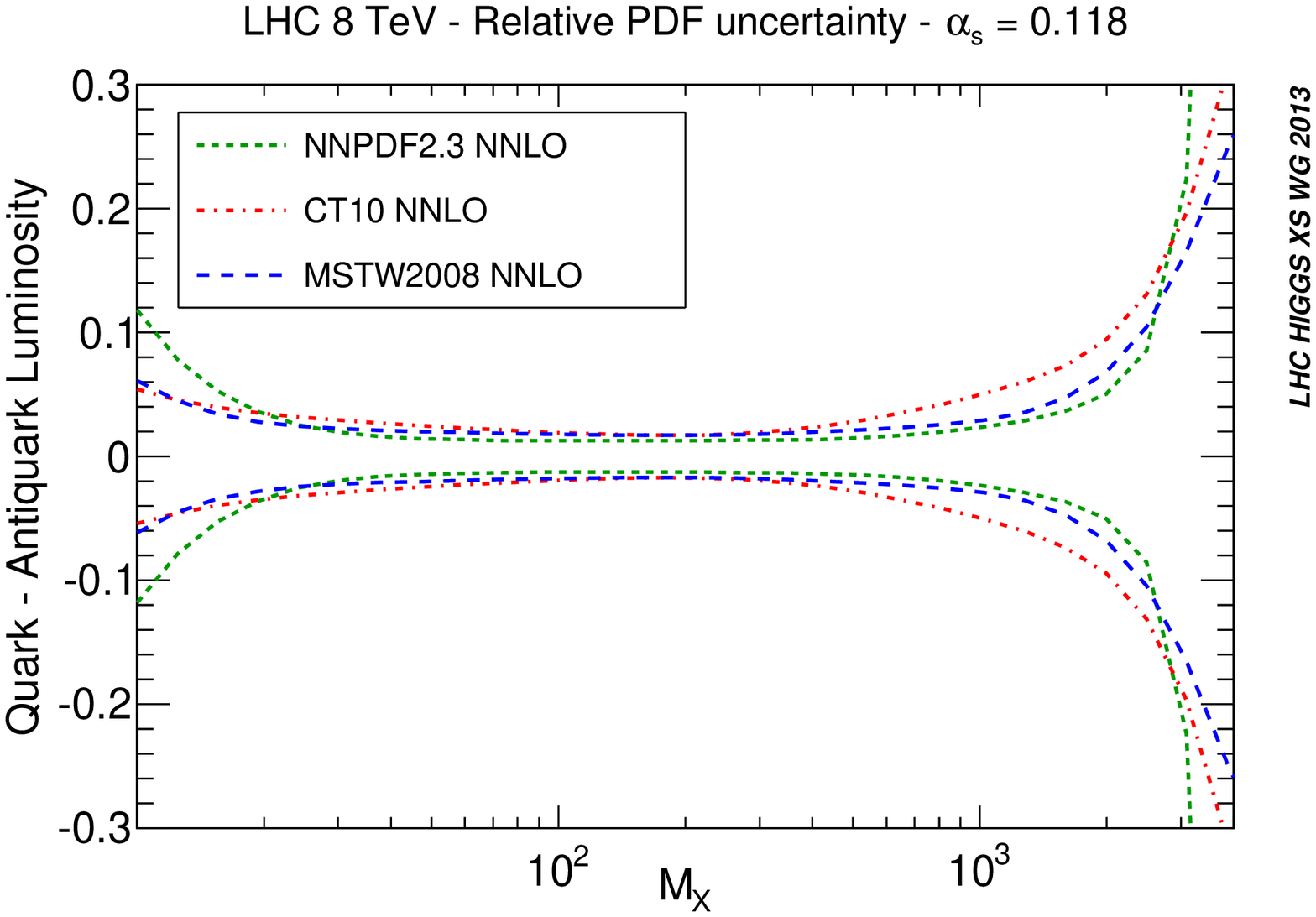}\quad
\includegraphics[width=0.48\textwidth]{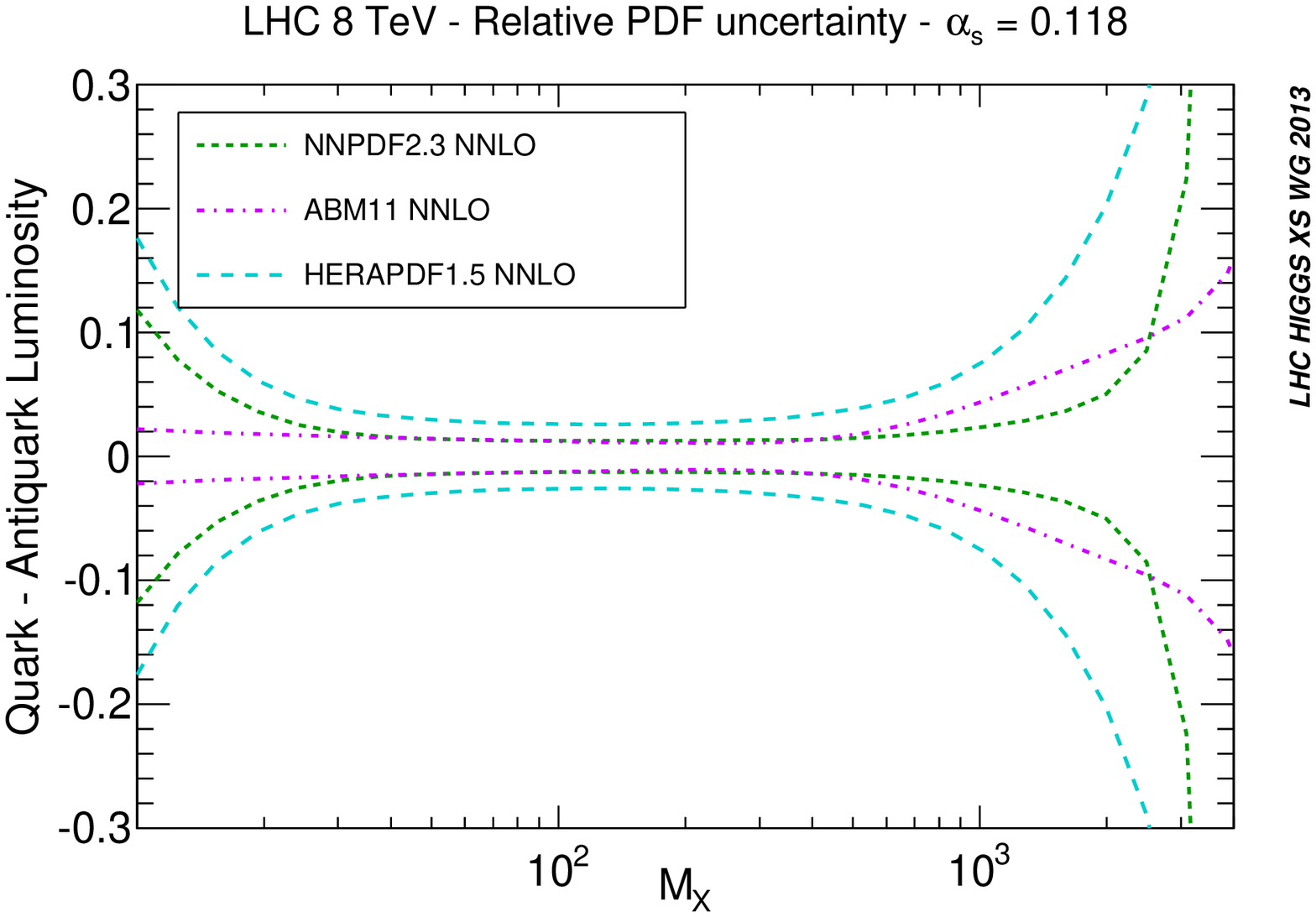}\\
  \includegraphics[width=0.48\textwidth]{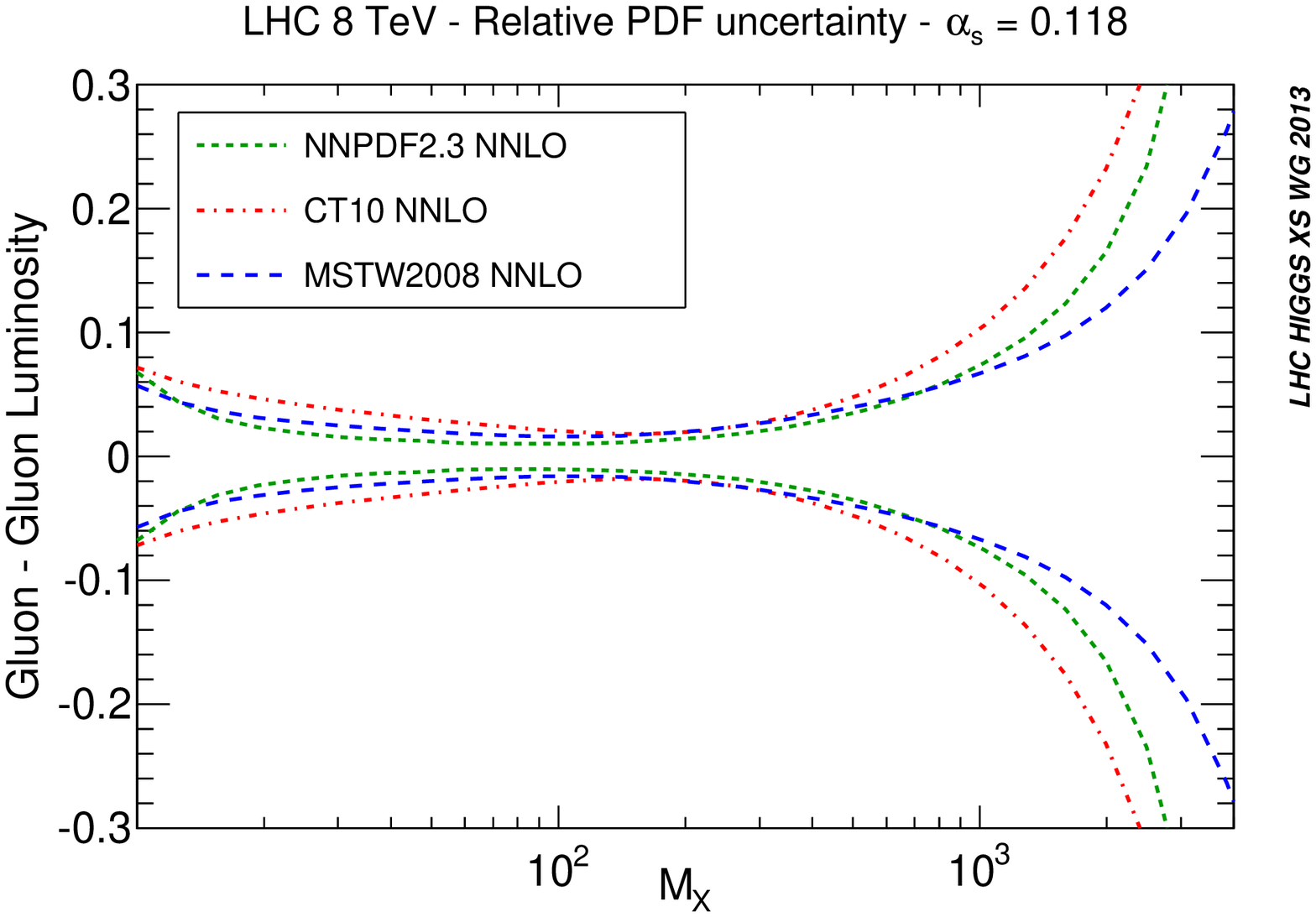}\quad
\includegraphics[width=0.48\textwidth]{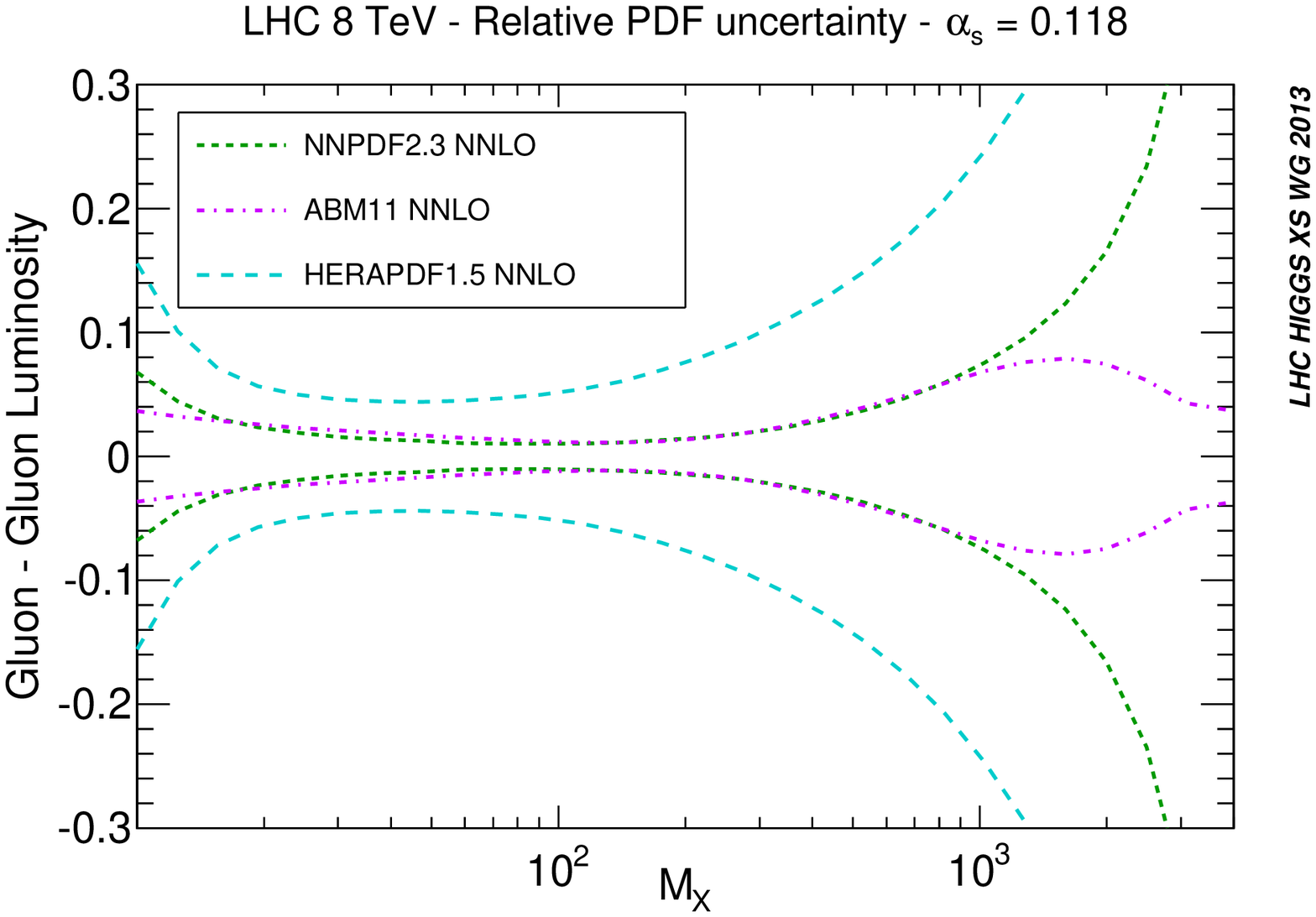}
      \end{center}
     \caption{\small
The relative PDF uncertainties in the quark-antiquark 
luminosity (upper plots)
and in the gluon-gluon luminosity (lower plots), 
for the production
of a final state of invariant mass $M_X$ (in GeV) at the LHC $8\UTeV$.  
All luminosities are computed at a common value of $\alphas=0.118$.
    \label{fig:PDFlumi-rel} }
\end{figure}

In the PDF4LHC report \cite{Botje:2011sn}, published at a time when NNLO PDFs were not
available from either CT or NNPDF, the recommendation at NNLO was to
use the MSTW2008 central prediction, and to multiply the MSTW2008 PDF
uncertainty by a factor of 2. This  factor of 2 appears to be an
overestimate  in this new comparison of the three global PDFs, for
$\PAQq\PQq$, $\PQq\PQq$ and $\Pg\PQq$ initial states, but is still needed for $\Pg\Pg$
initial states.  
A direct comparison of the parton luminosity uncertainties is shown in \Fref{fig:PDFlumi-rel}
 for $\PAQq\PQq$ (top) and $\Pg\Pg$ (bottom), where the points made
 previously about the relative size of the uncertainties can be more
 easily observed.  

%%%%%%%%%%%%%%%%%%%%%%%%%%%%%%%%%%%%%%%%%%%%%%%%%
%%%%%%%%%%%%%%%%%%%%%%%%%%%%%%%%%%%%%%%%%%%%%%%%%%

\begin{figure}[ht!]
\centering
\includegraphics[width=0.47\textwidth]{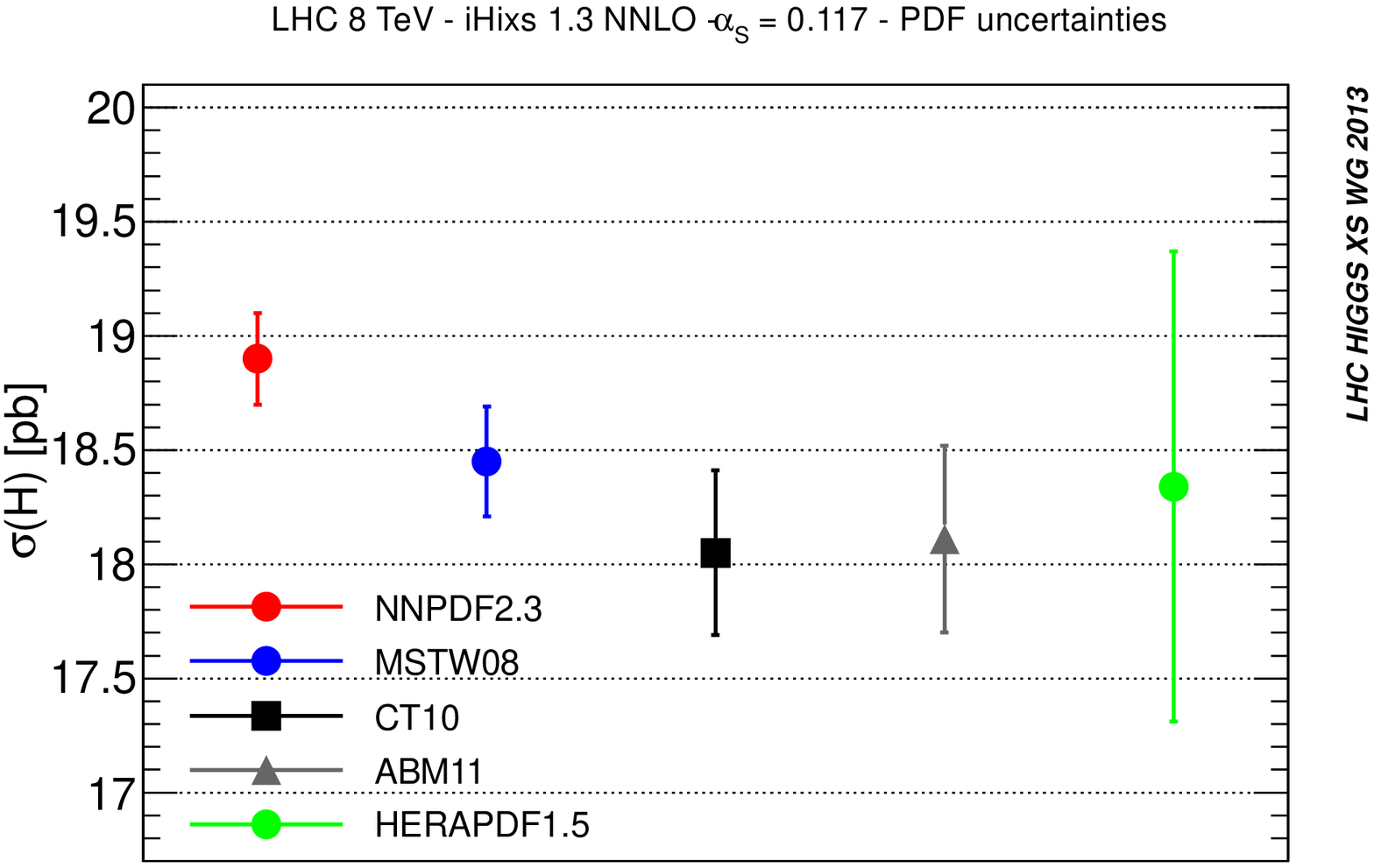}\quad
\includegraphics[width=0.47\textwidth]{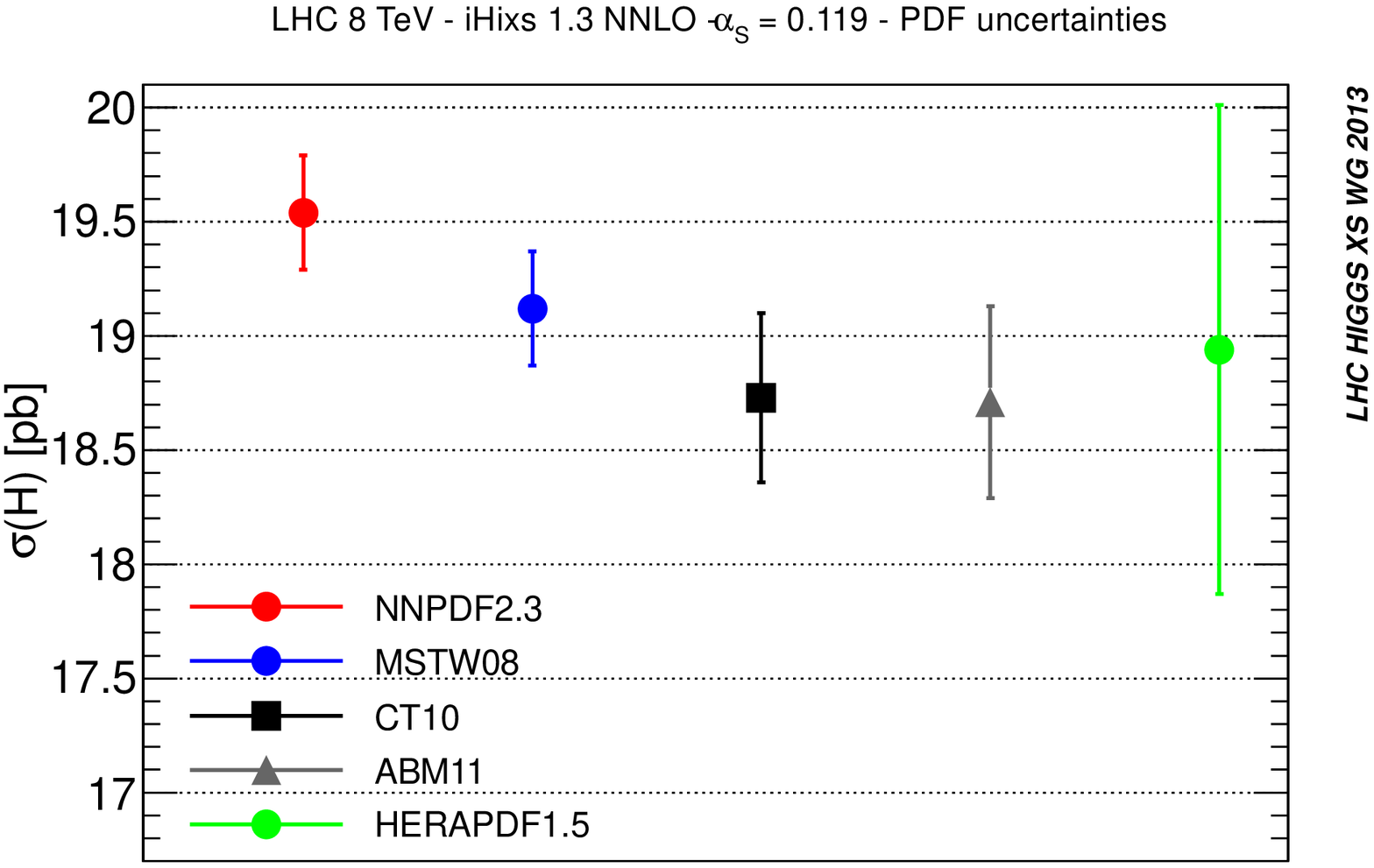}
\includegraphics[width=0.47\textwidth]{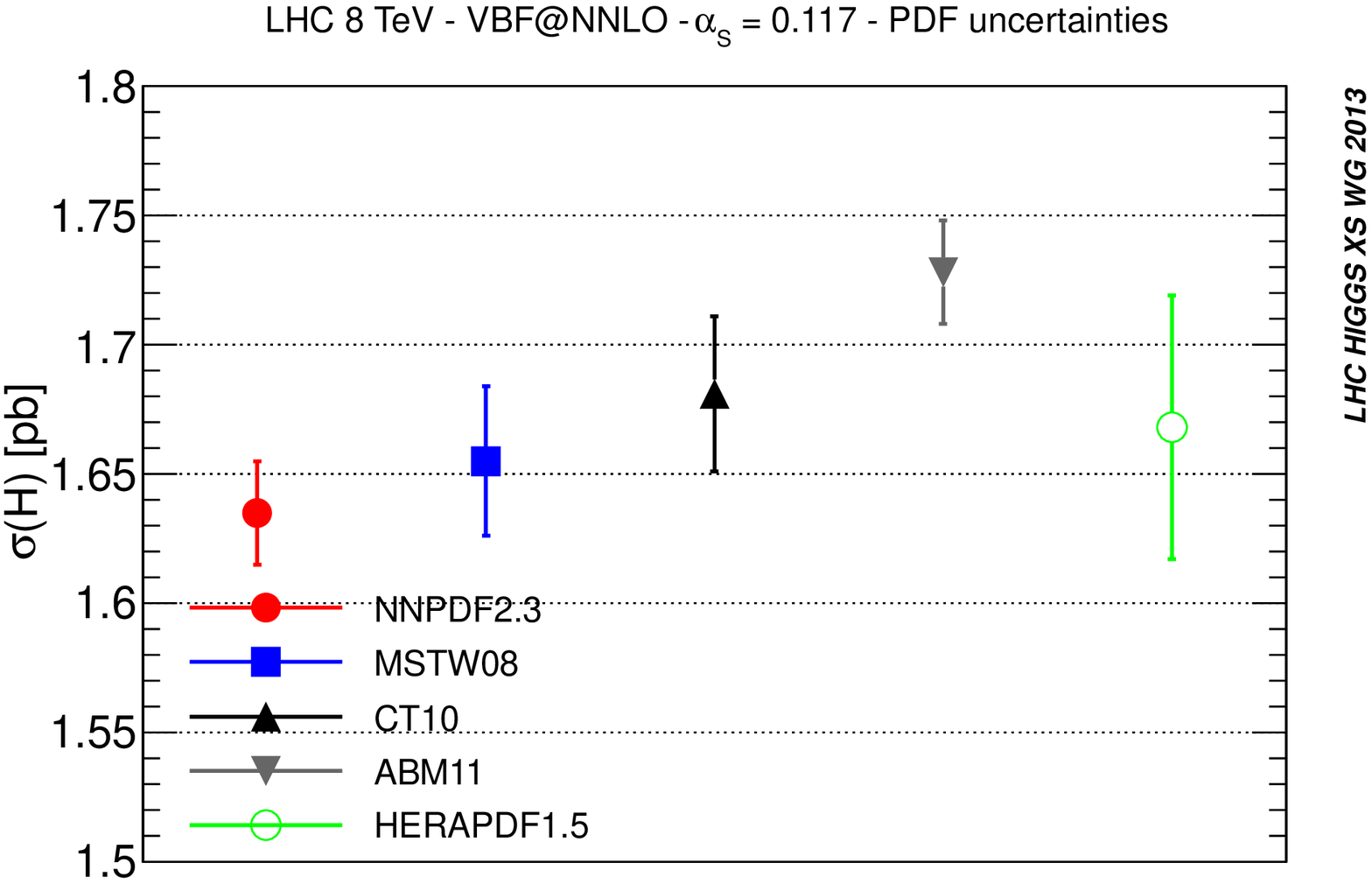}\quad
\includegraphics[width=0.47\textwidth]{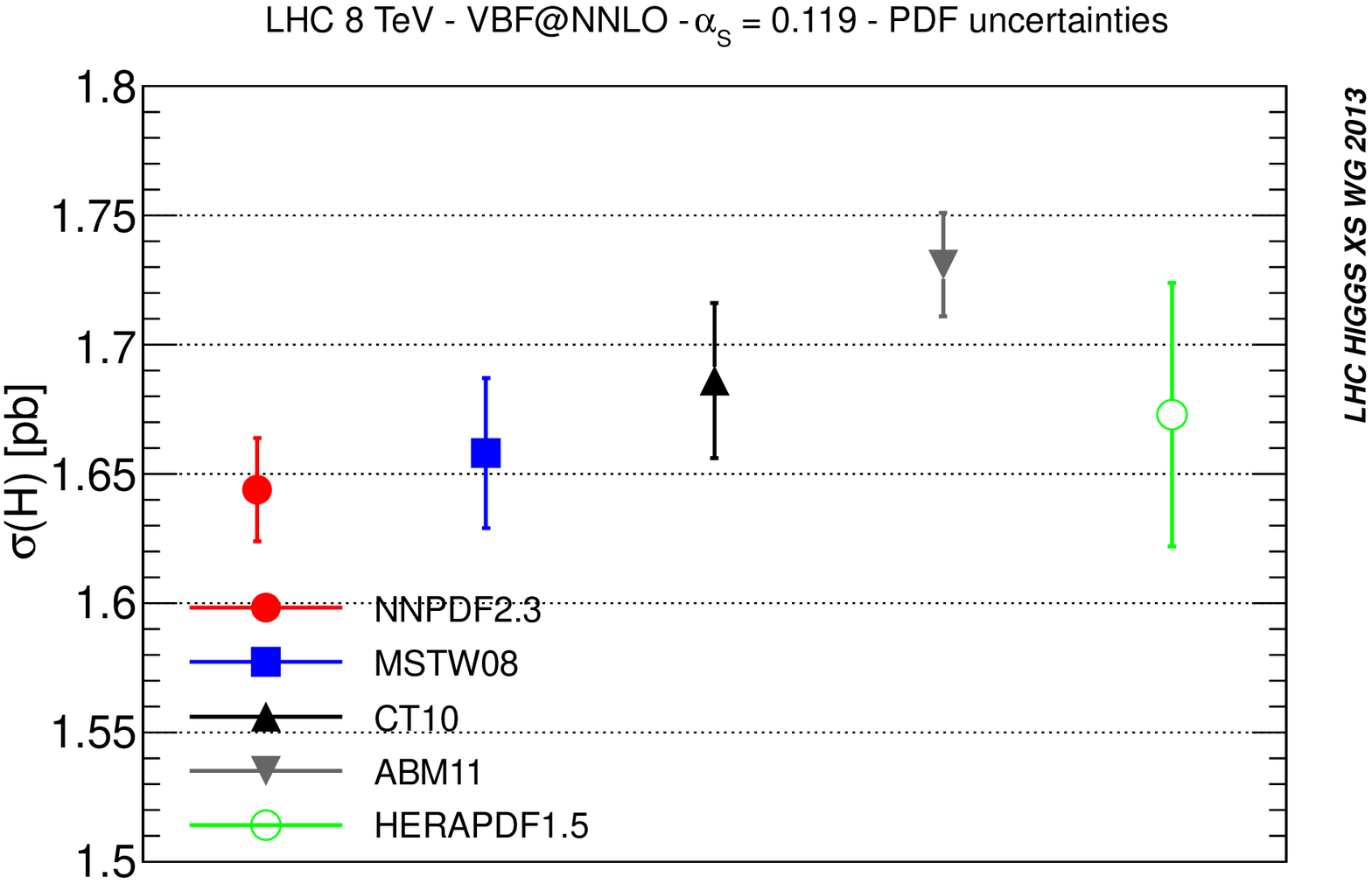}
\includegraphics[width=0.47\textwidth]{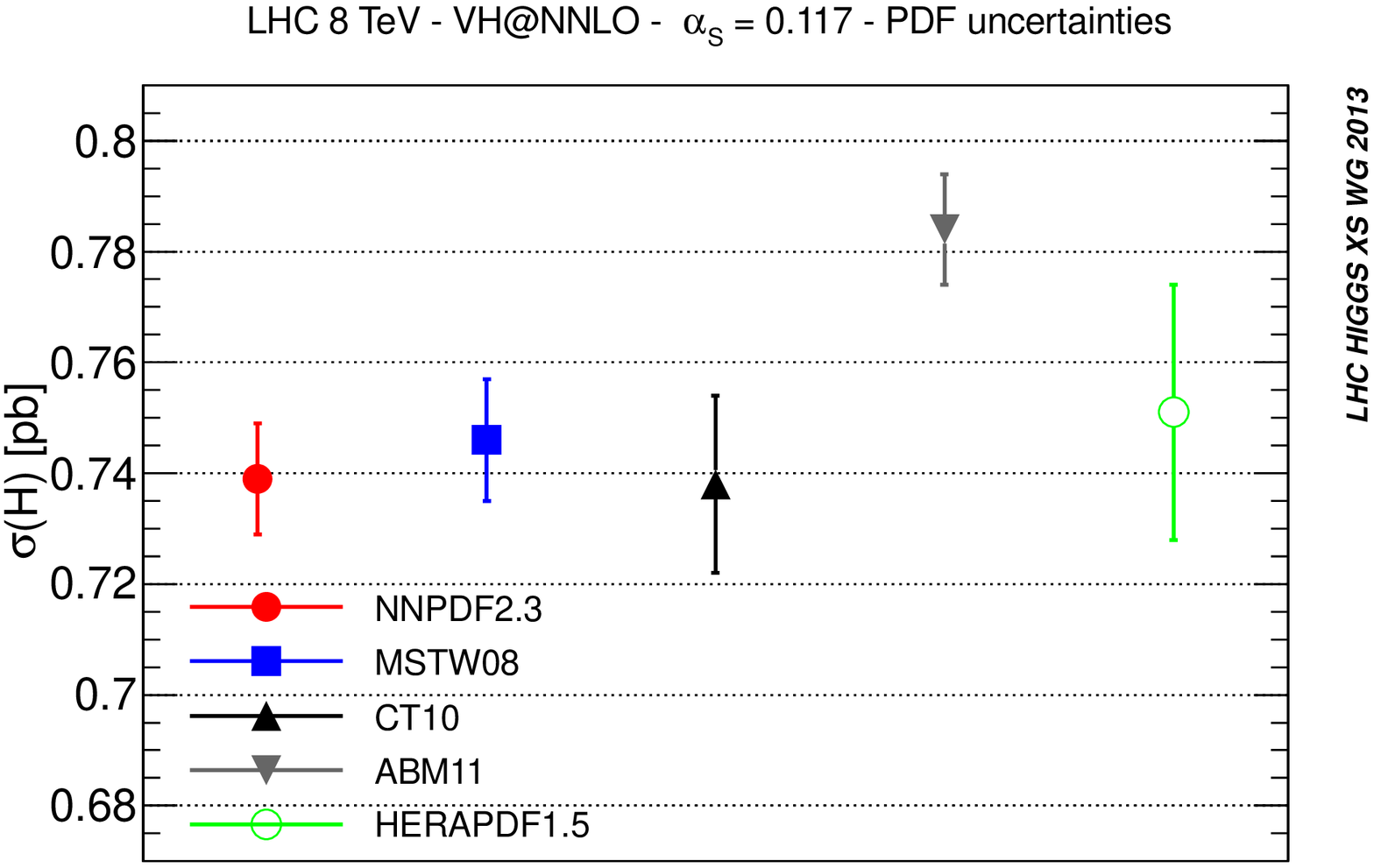}\quad
\includegraphics[width=0.47\textwidth]{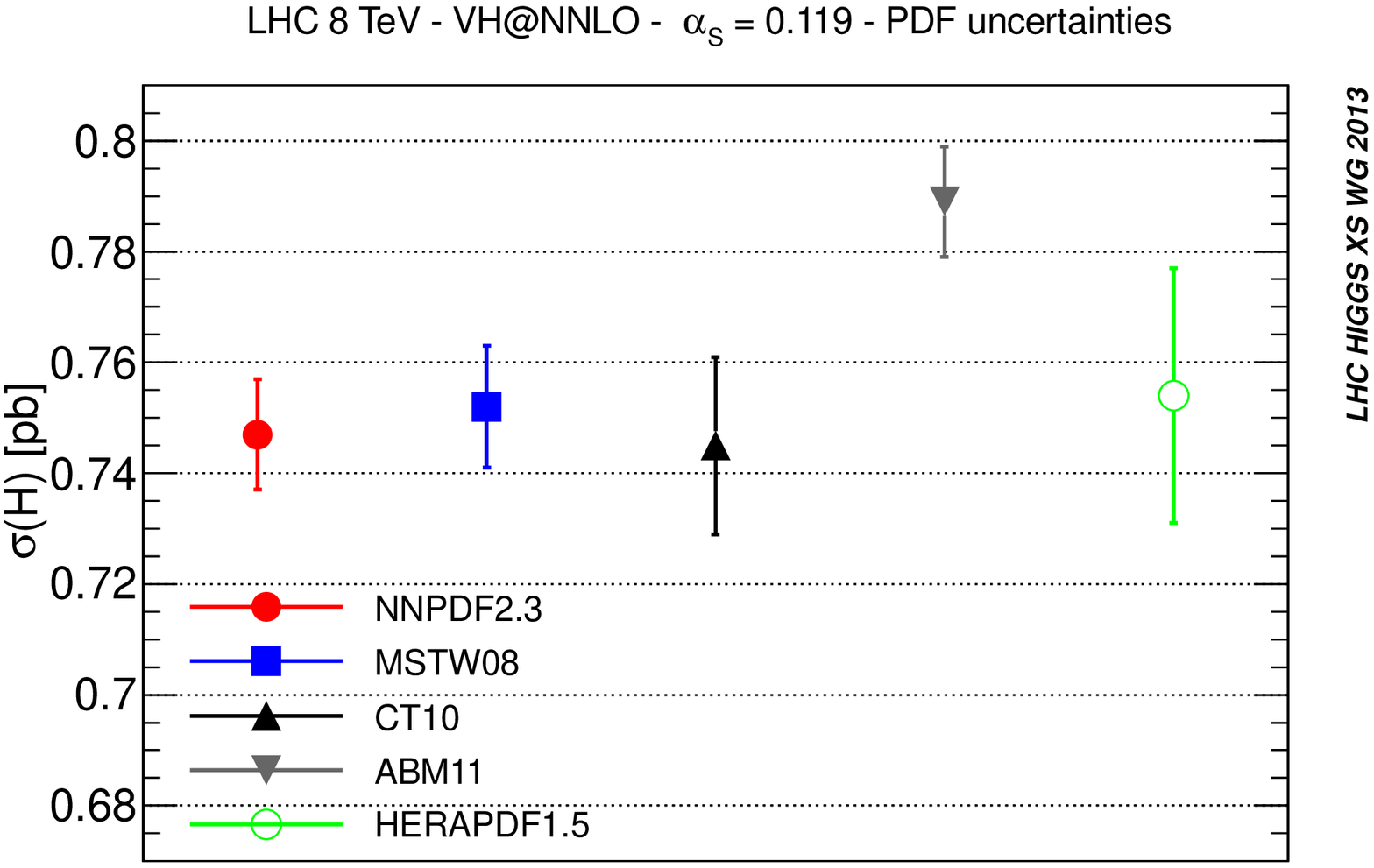}
\includegraphics[width=0.47\textwidth]{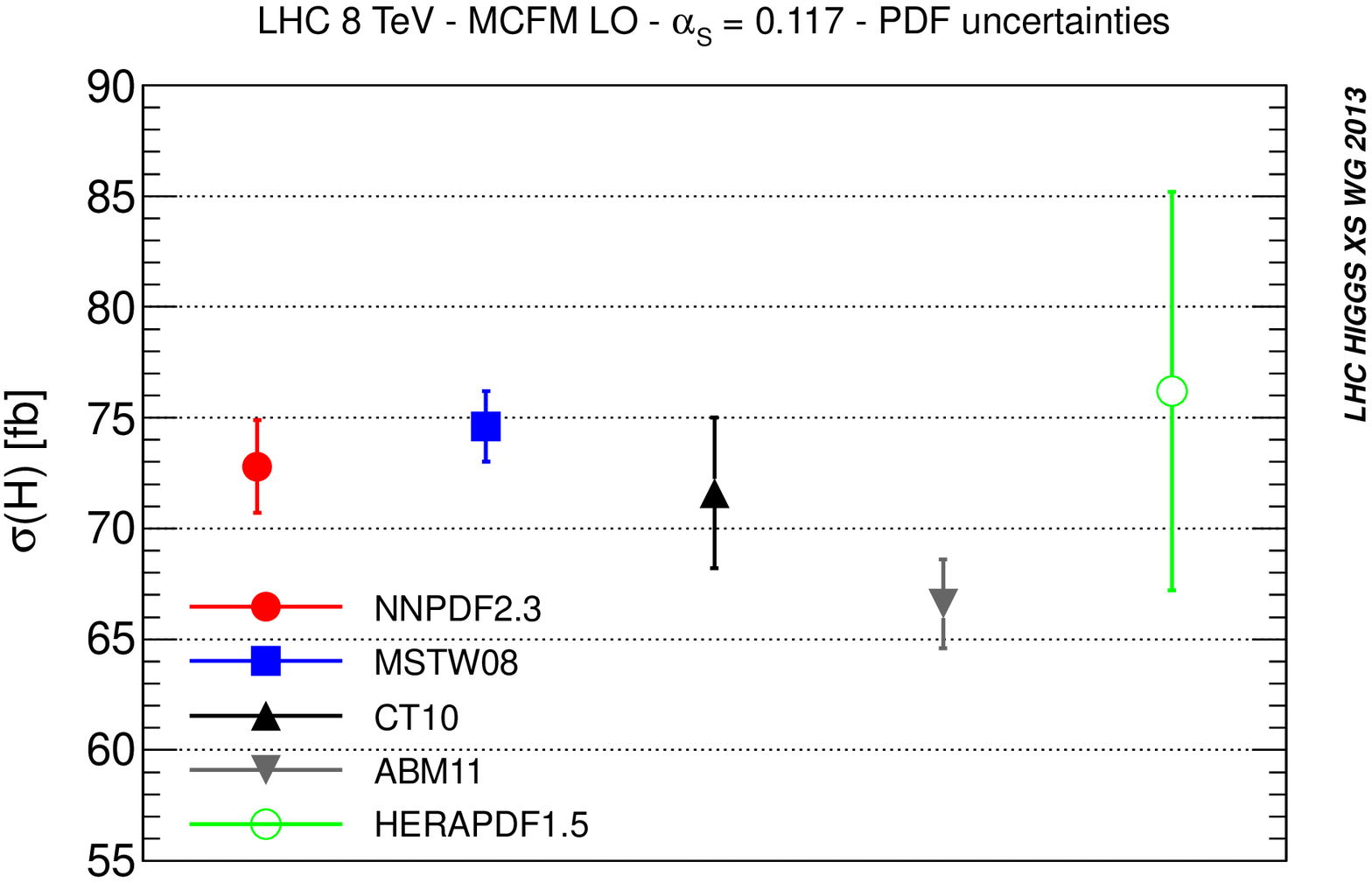}\quad
\includegraphics[width=0.47\textwidth]{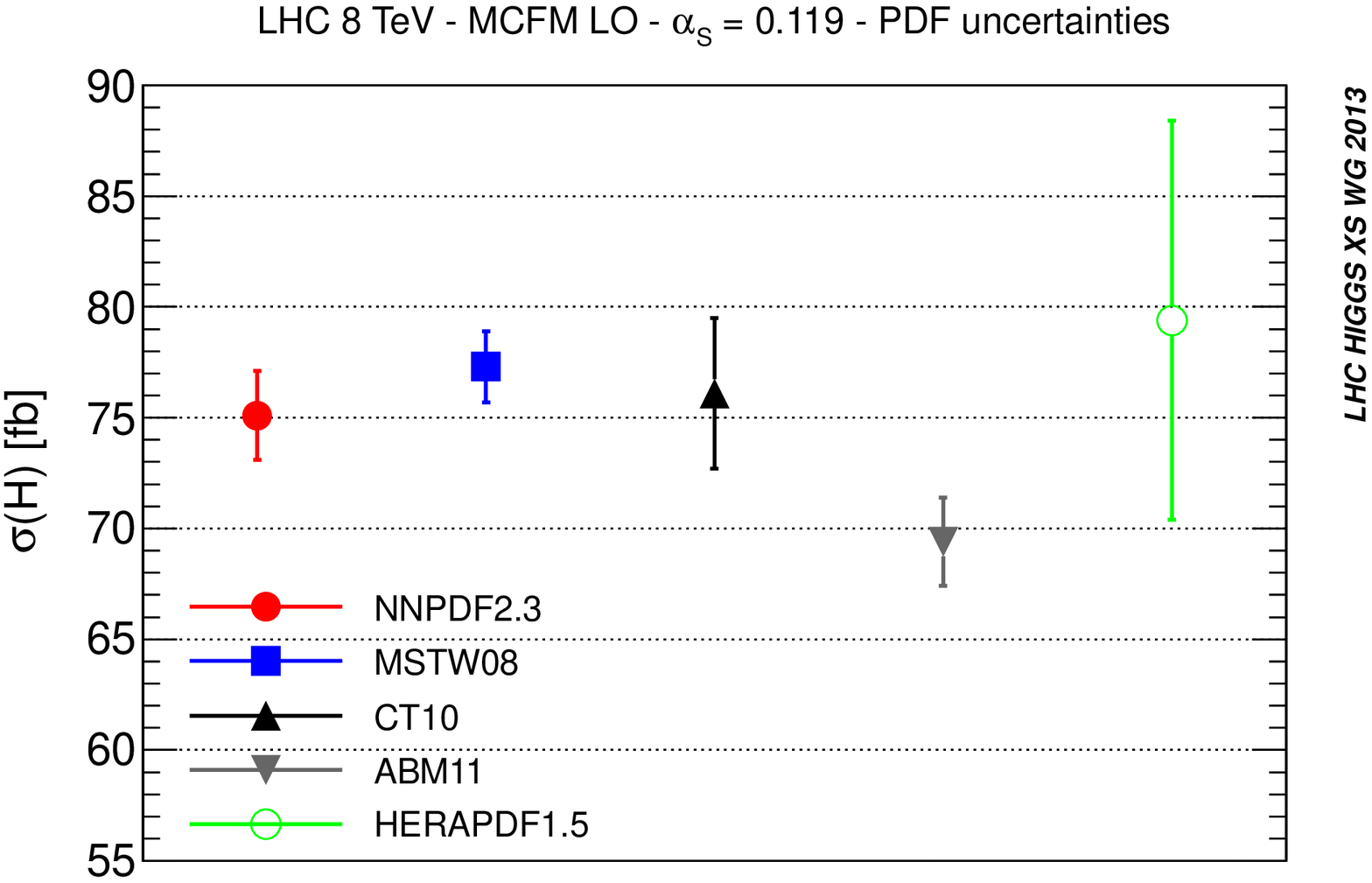}
\caption{\small Comparison of the 
predictions for the LHC Standard Model Higgs boson
cross sections at $8\UTeV$ obtained using
various NNLO PDF sets. From top to bottom we show gluon fusion, vector 
boson fusion, associated production (with $\PW$), and associated 
production with a $\PAQt\PQt$ pair.
The left hand plots show 
results for $\alphas(\MZ)=0.117$, while on the right we have 
$\alphas(\MZ)=0.119$.
\label{fig:8tev-higgs}}
\end{figure}

In \Fref{fig:8tev-higgs}, we show predictions for Higgs production
at $8\UTeV$ in various channels, for $\alphas(\MZ)$ values of $0.117$
(left) and  $0.119$ (right) for the 5 different PDFs being
considered. As expected, the cross section predictions follow the
trends discussed for the PDF luminosities. The strongest
disagreements, perhaps, are from  the ABM11 predictions for VBF and
associated  (VH) Higgs production, though if the ABM11 PDFs with $\alphas(\MZ)=0.1134$
are used the disagreement in these channels is reduced, but increases for the gluon fusion 
channel.   

%%%%%%%%%%%%%%%%%%%%%%%%%%%%%%%%%%%%%%%%%%
%%%%%%%%%%%%%%%%%%%%%%%%%%%%%%%%%%%%%%%%%

\begin{figure}[h]
    \begin{center}
\includegraphics[width=0.48\textwidth]{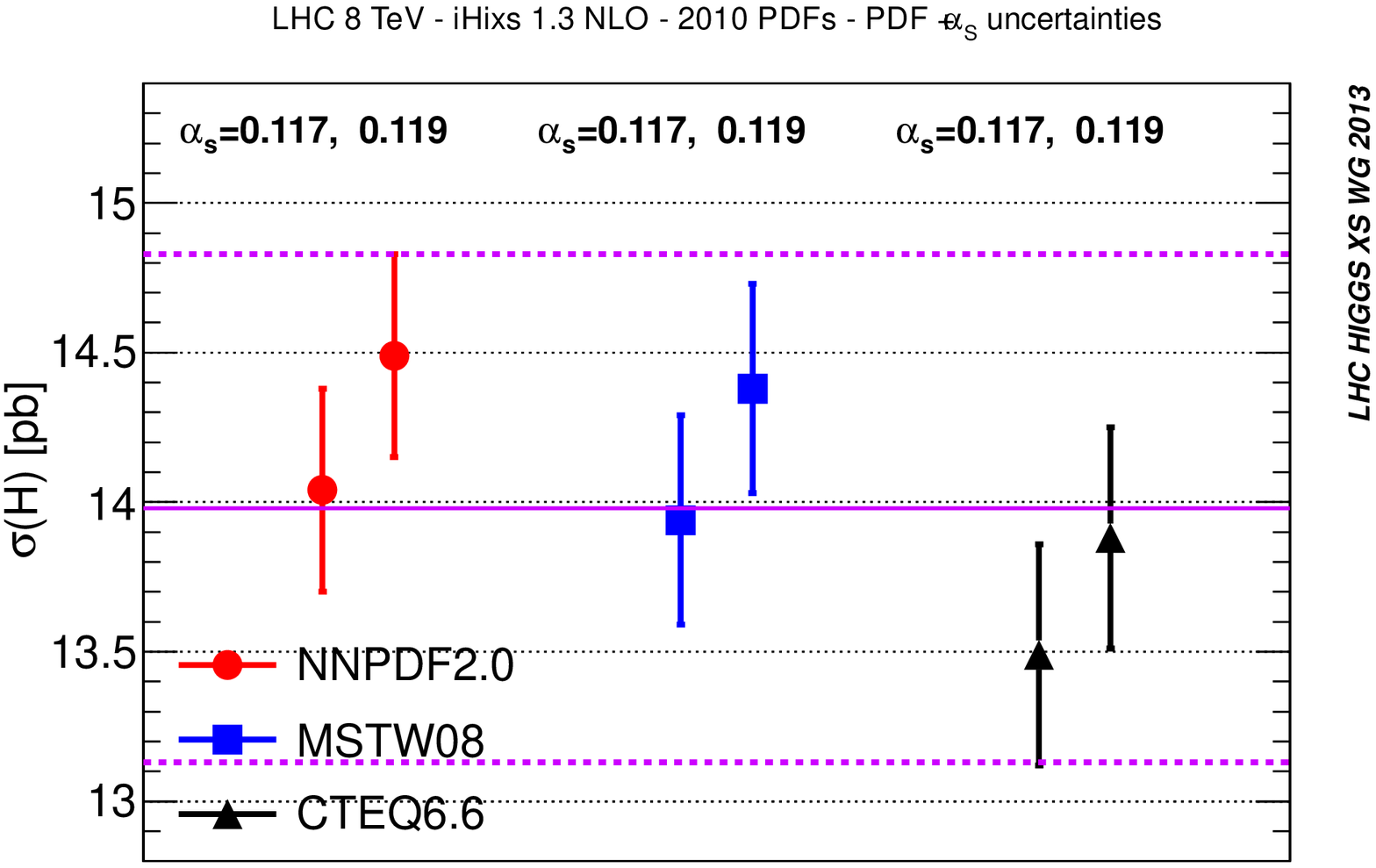}\quad
\includegraphics[width=0.48\textwidth]{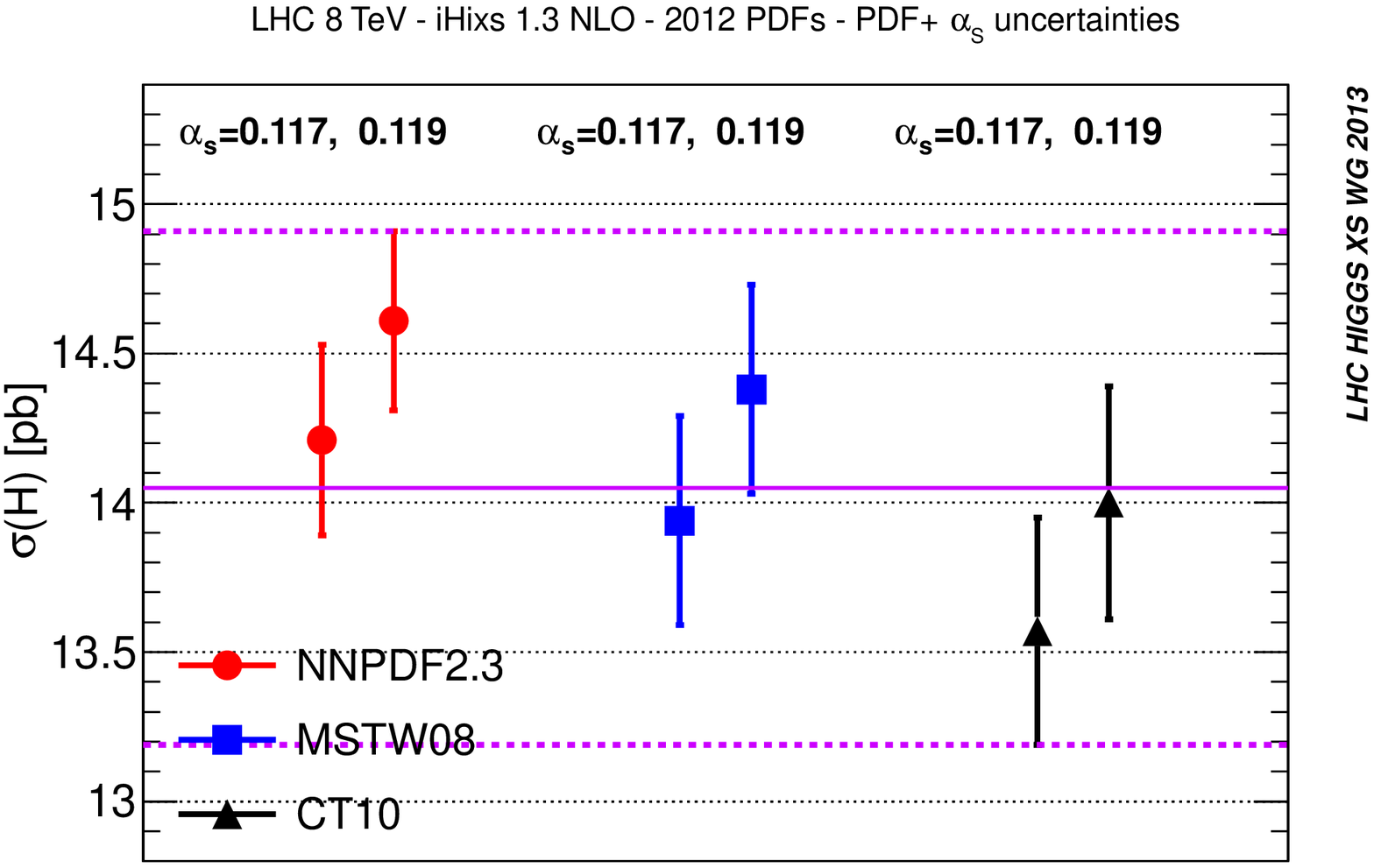}
      \end{center}
     \caption{\small The Higgs boson production
cross section in the gluon fusion channel using the 
NLO PDF sets included in the PDF4LHC prescription for $\alphas(\MZ)=0.117$ and
$0.119$. The left plot has been computed with 2010 PDFs and
the right plot with 2012 PDF sets. The envelope (dashed violet
horizontal lines) is defined by the upper and lower values
of the predictions from all the three PDF sets and the two values
of $\alphas$. The solid violet horizontal line is the midpoint
of the envelope.
    \label{fig:h8nlo} }
\end{figure}

\begin{figure}[h]
    \begin{center}
\includegraphics[width=0.49\textwidth]{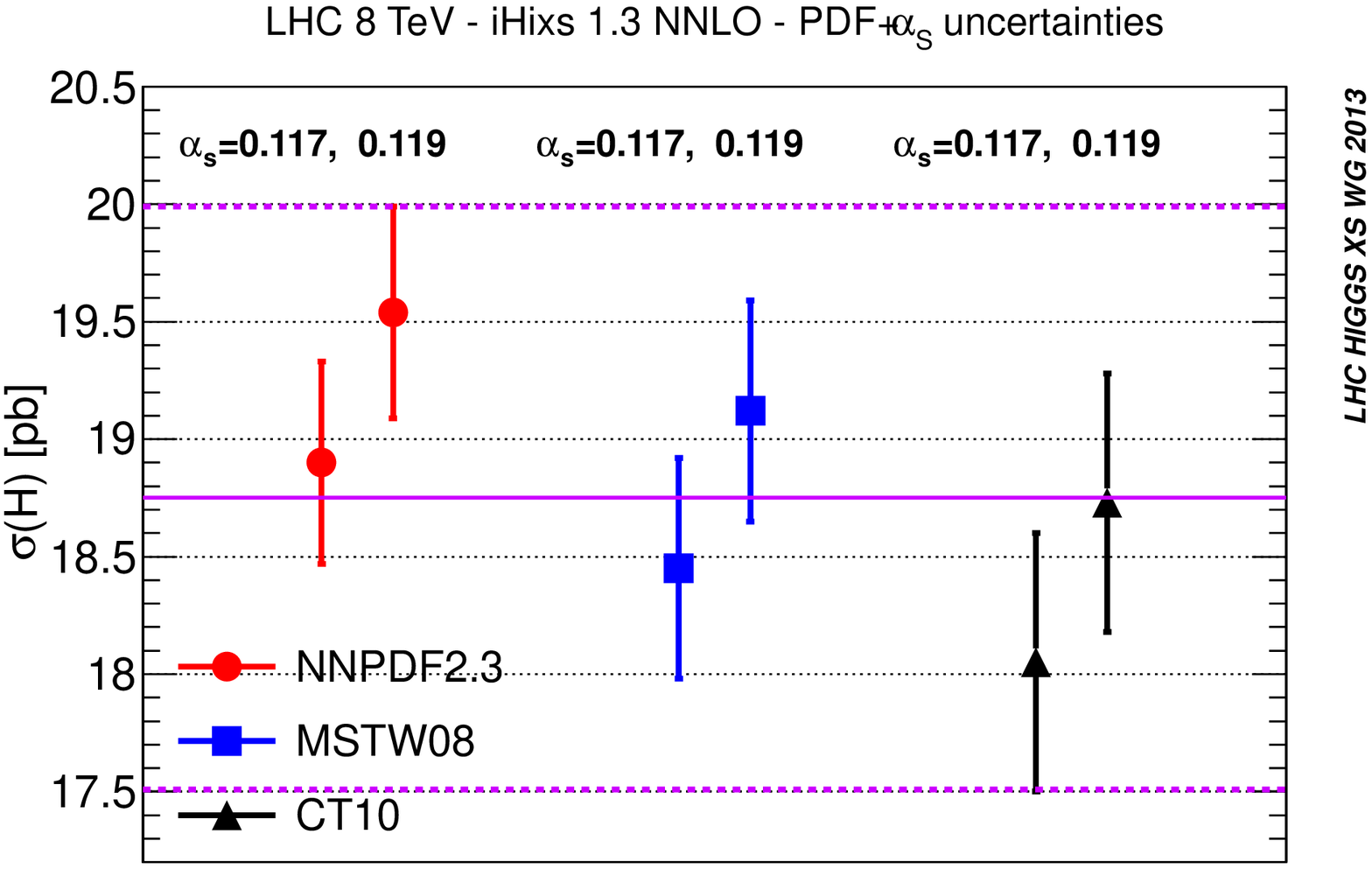}
      \end{center}
     \caption{\small The same as \Fref{fig:h8nlo}, but using 2012 NNLO PDFs.
    \label{fig:h8nnlo} }
\end{figure}

Cross section predictions for Higgs production (in the $\Pg\Pg$ channel at
$8\UTeV$, for values of $\alphas(\MZ)$ of $0.117$ and $0.119$) are shown in
\Fref{fig:h8nlo} and  
\Fref{fig:h8nnlo} for CTEQ, MSTW and NNPDF PDFs. In
\Fref{fig:h8nlo}, the predictions are at NLO, but using the PDFs
available in 2010 (left) and in 2012 (right). In
\Fref{fig:h8nnlo}, the cross sections are plotted using the 2012
NNLO PDFs. Here, we estimate the PDF+$\alphas(\MZ)$ uncertainty using
a small variation of the original PDF4LHC rubric; we take the envelope
of the predictions from CT/CTEQ, MSTW and NNPDF including their PDF
uncertainties, and using values of $\alphas(\MZ)$ of $0.117$ and
$0.119$. The uncertainty bands are given by the dashed lines. There is
little change at NLO with the evolution from CTEQ6.6 and NNPDF2.0 to
CT10 and NNPDF2.3, and the uncertainty at NNLO is very similar to the
uncertainty estimated for NLO. 

%%%%%%%%%%%%%%%%%%%%%%%%%%%%%%%%%%%%%%%%%%%%%%%%%
%%%%%%%%%%%%%%%%%%%%%%%%%%%%%%%%%%%%%%%%%%%%%%%%%%

%%%%%%%%%%%%%%%%%%%%%%%%%%%%%%%%%%%%%%%%%%%%%%%%%
%%%%%%%%%%%%%%%%%%%%%%%%%%%%%%%%%%%%%%%%%%%%%%%%%%

\begin{figure}[h]
    \begin{center}
\includegraphics[width=0.48\textwidth]{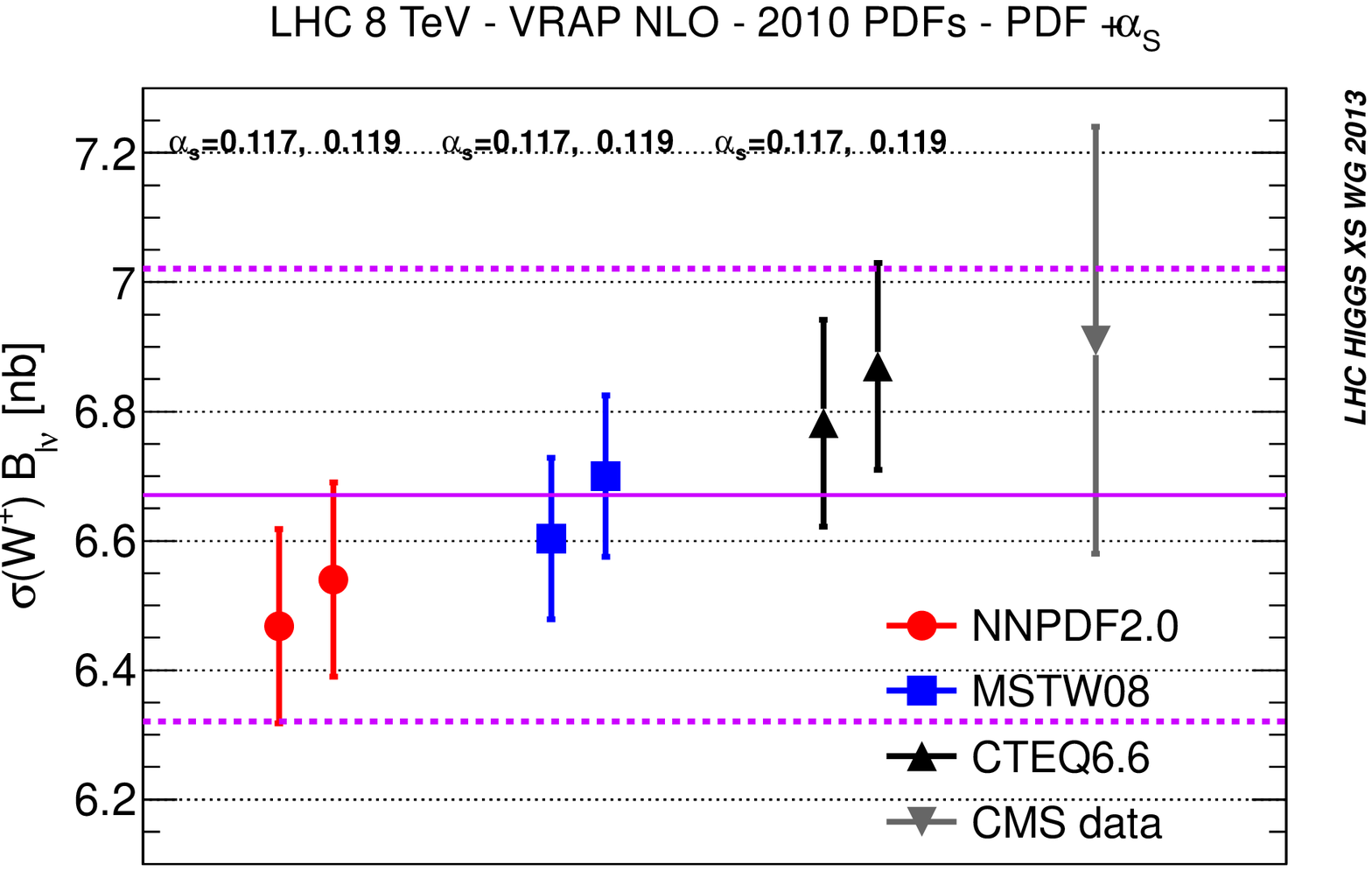}\quad
\includegraphics[width=0.48\textwidth]{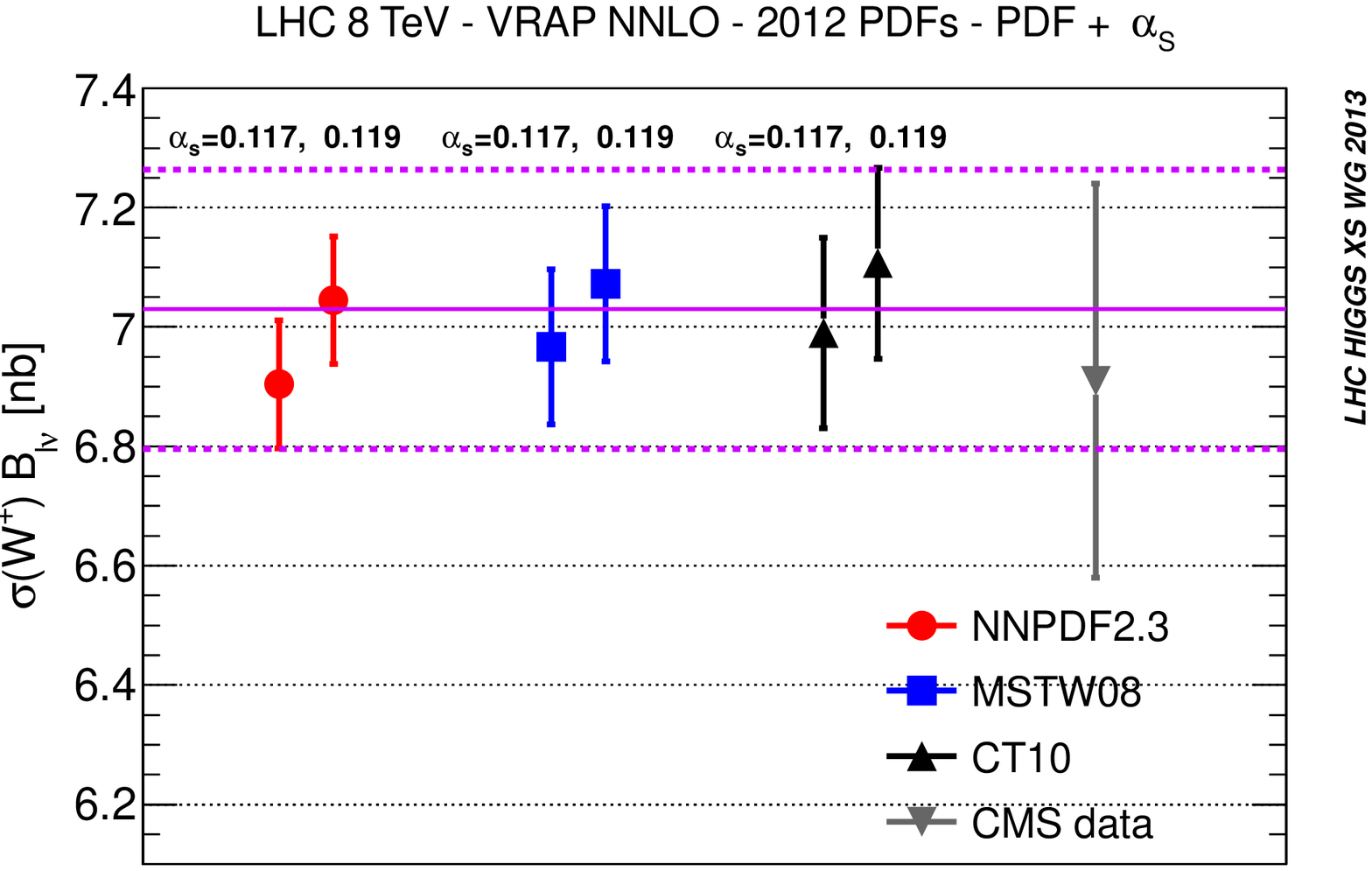}
      \end{center}
     \caption{\small The
$\PWp$ production
cross sections determined using the same PDFs and envelope as in
%Figs.~\ref{fig:h8nlo}-\ref{fig:h8nnlo}. The left plot shows 2010 NLO
\Frefs{fig:h8nlo}{fig:h8nnlo}. The left plot shows 2010 NLO
PDFs, the right plot 2012 NNLO PDFs. The recent $8\UTeV$ CMS measurement
is also shown.
    \label{fig:wpbench} }
\end{figure}

As a contrast, we show in  \Fref{fig:wpbench} predictions for
$\PWp$ production based on the 2010 NLO and 2012 NNLO PDFs from
CT/CTEQ, MSTW and NNPDF. The relative PDF+$\alphas(\MZ)$ uncertainty estimated
with the same prescription used for Higgs production has a sizable
decrease from the 2010 NLO predictions to the 2012 NNLO
predictions. Similar improvements should be expected for all
quark-initiated processes, including those involved in Higgs
production.  

%%%%%%%%%%%%%%%%%%%%%%%%%%%%%%%%%%%%%%%%%%%%%%%%%
%%%%%%%%%%%%%%%%%%%%%%%%%%%%%%%%%%%%%%%%%%%%%%%%%%

\begin{figure}[h]
    \begin{center}
\includegraphics[width=0.48\textwidth]{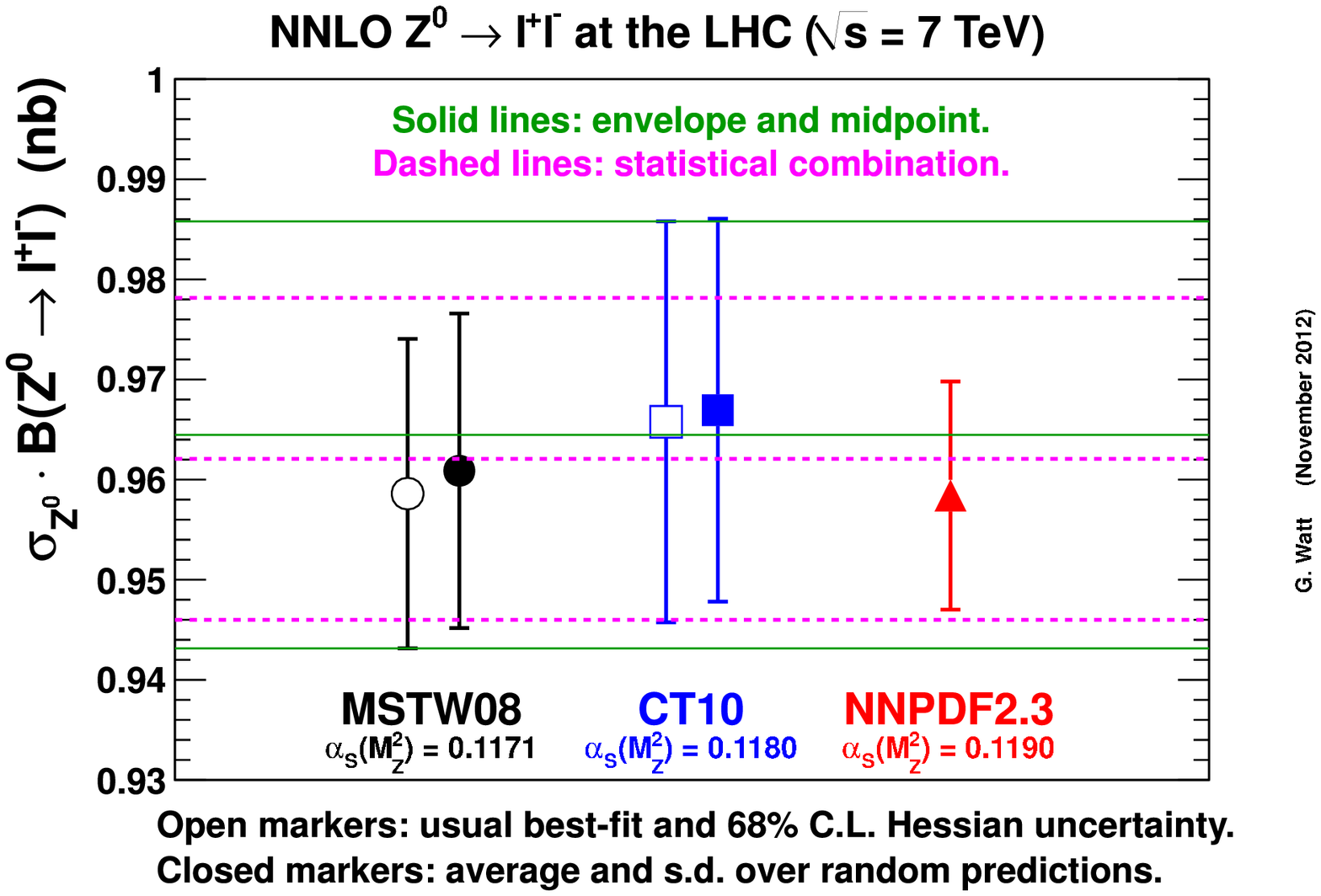}\quad
\includegraphics[width=0.48\textwidth]{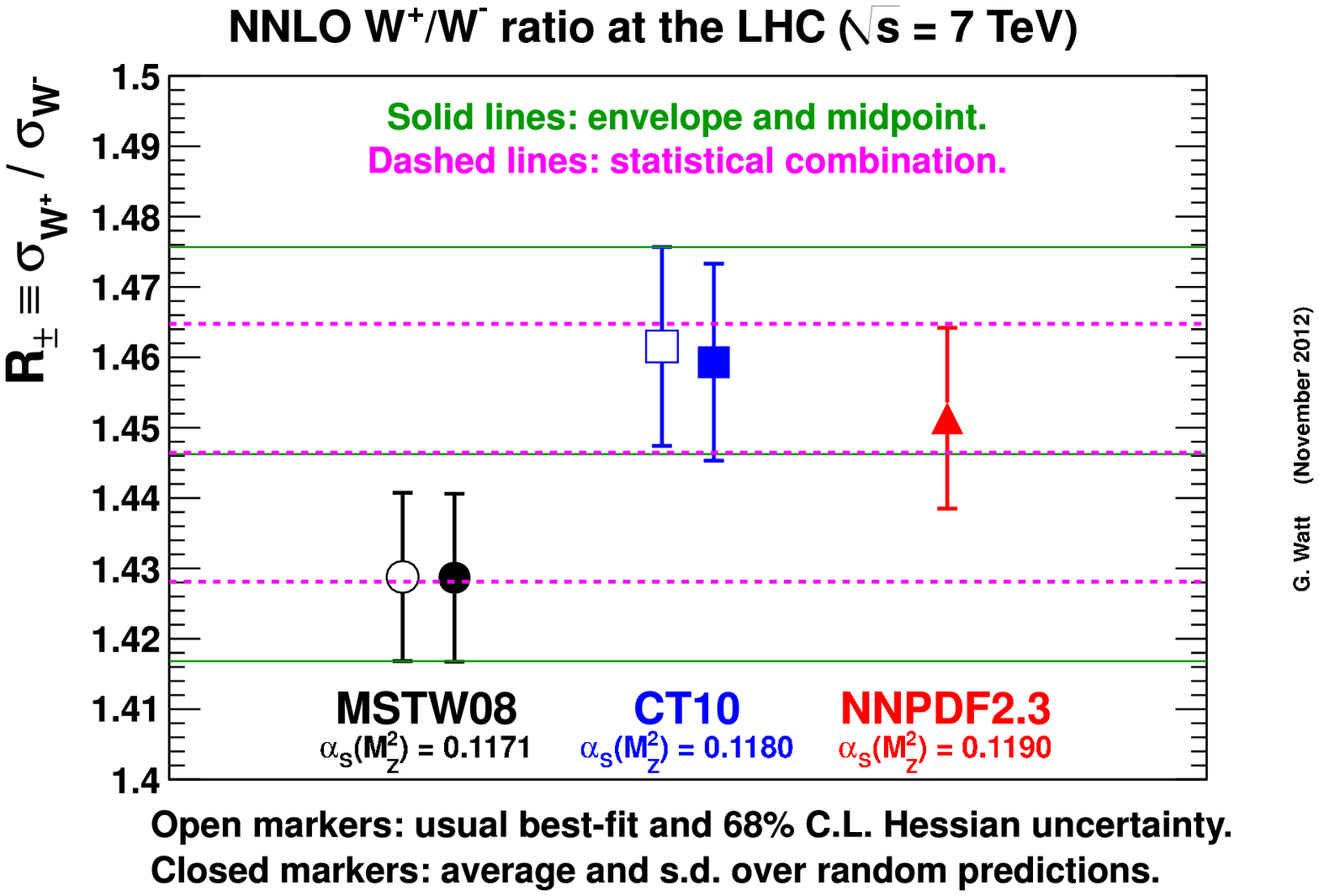}\\
\includegraphics[width=0.48\textwidth]{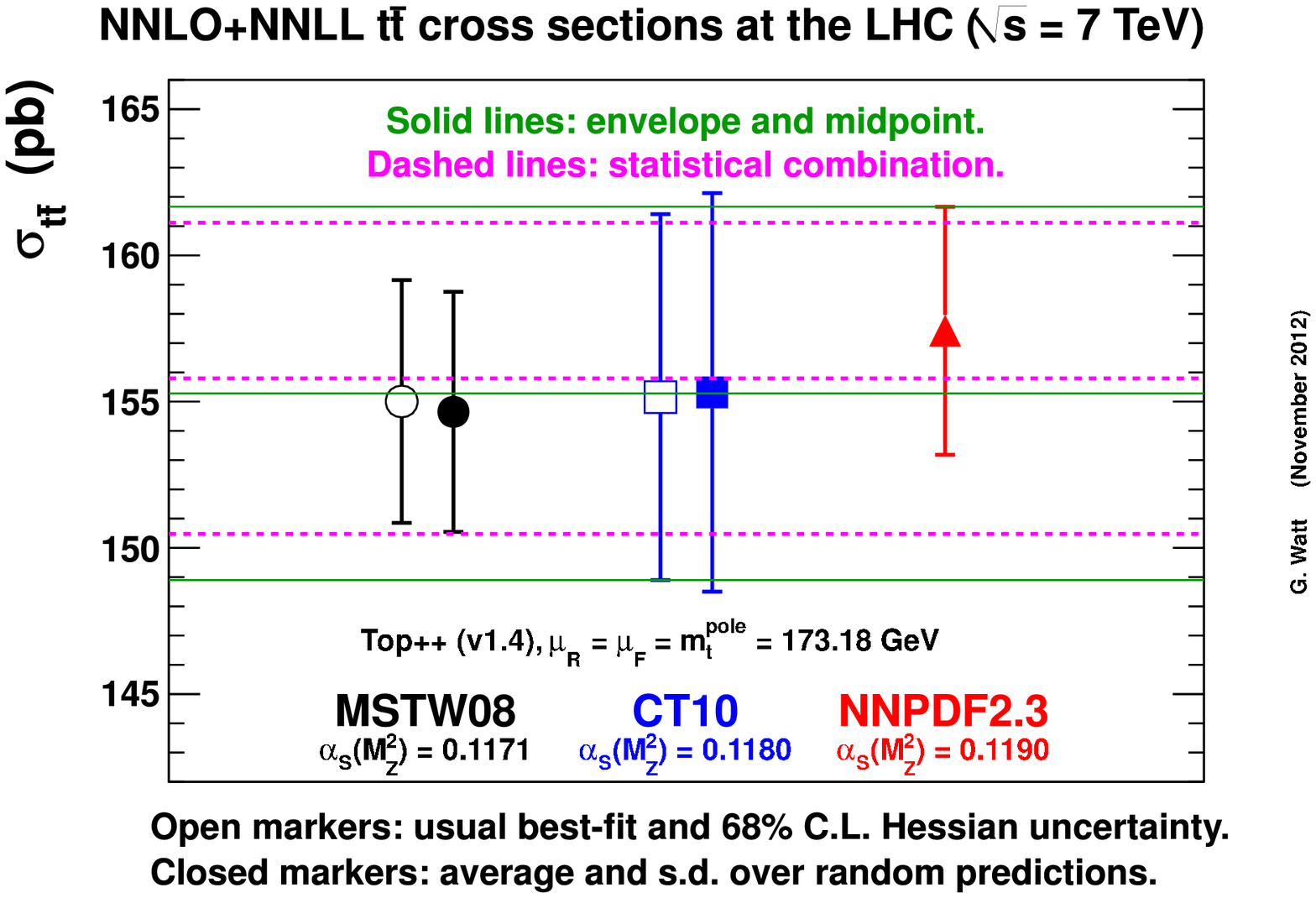}\quad
\includegraphics[width=0.48\textwidth]{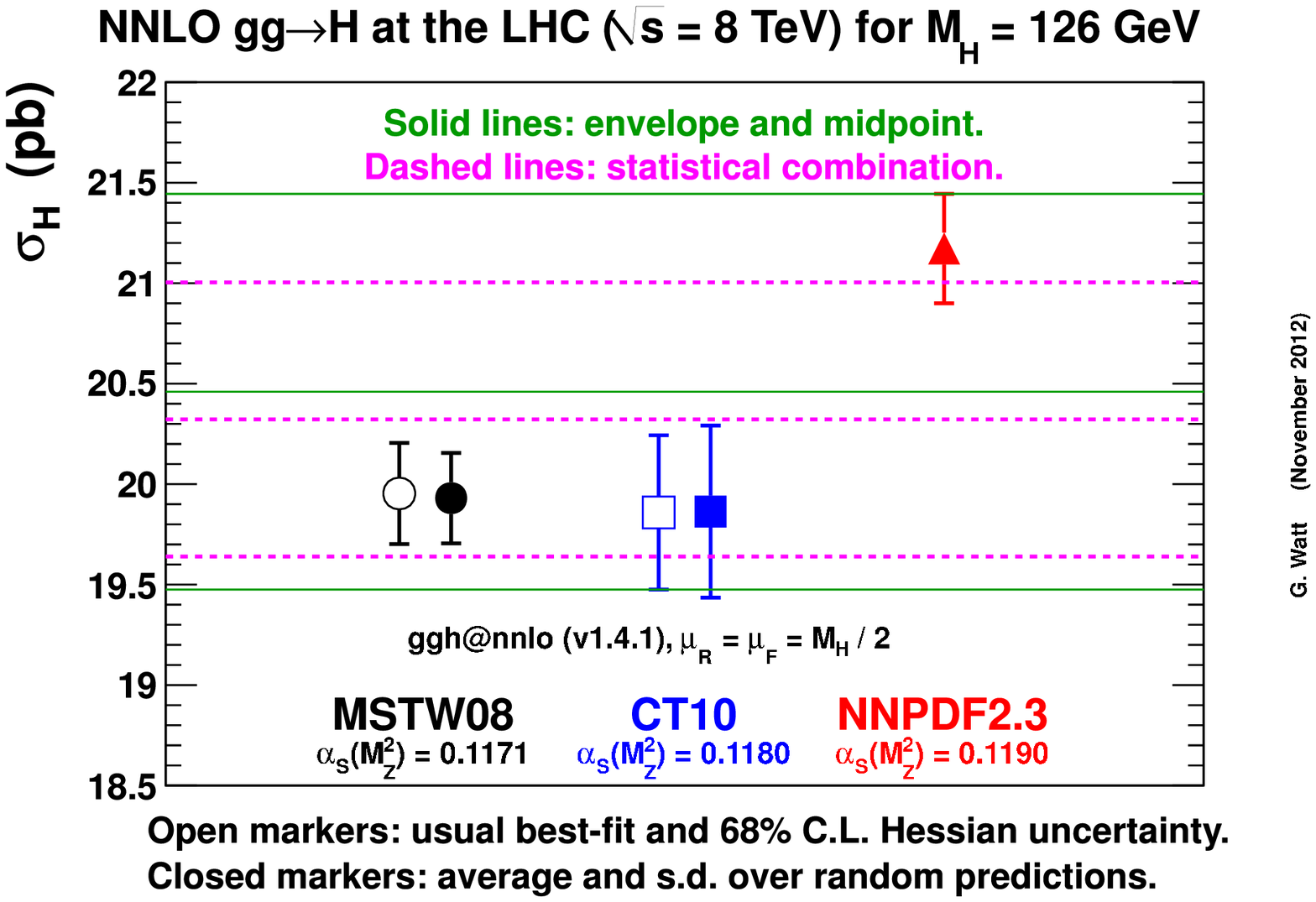}
      \end{center}
     \caption{\small NNLO (a) $\PZ$, (b) $\PWp/\PW$, (c) $\PAQt\PQt$ and (d) $\Pg\Pg \to \PH$ cross sections from MSTW08,
CT10 and NNPDF2.3, combined either by taking the envelope of the three predictions, or from the
statistical combination of 100 random predictions from each group
(from \Bref{Forte:2013wc}).
    \label{fig:random} }
\end{figure}

Finally, we demonstrate that although the previous proposal in the PDF4LHC
recommendation to use the envelope of the predictions using three PDF sets
does not strictly have a solid statistical basis, it certainly produces sensible results. Using
the techniques for generating random PDFs sets in \cite{Watt:2012tq} it was 
shown in section 4.1.3 of~\cite{Forte:2013wc} that similar results are 
obtained from combining the results from 100 random sets from MSTW2008, 
NNPDF2.3 and CT10 as from taking the envelopes. The results are shown in 
\Fref{fig:random}. The envelope procedure can be seen to be a little 
more conservative, and becomes more-so in comparison to the combination
of random sets as any discrepancy between sets becomes more
evident. However, for  
predictions using these three PDF sets there is generally not much 
differences between the two methods of calculation. In order to maintain
a conservative uncertainty the continuation of the envelope method is 
probably preferred.

%%%%%%%%%%%%%%%%%%%%%%%%%%%%%%%%%%%%%%%%%%%%%%%%%
%%%%%%%%%%%%%%%%%%%%%%%%%%%%%%%%%%%%%%%%%%%%%%%%%%

\clearpage

\newpage
% math abbreviations
\providecommand{\df}{\mathrm{d}}
\providecommand{\cut}{\mathrm{cut}}
\providecommand{\veto}{\mathrm{veto}}
\providecommand{\NLL}{\mathrm{NLL}}
\providecommand{\pTHjj}{p_{THjj}}
\providecommand{\dphi}{\Delta\phi_{\PH\!-\!jj}}

%%======================================================================
%% Draft structure of Jets section
%% --------------------------------
%%
%%
%% 1) Status of predictions for H+0 jets:
%% --------------------------------------
%% [to be signed by the three groups doing the resummations]
%% 
%% half a page with a brief discussion of the fact that
%% 
%% a) we now have NNLL+NNLO predictions for this and
%% b) that the three groups agree on the structure of the NNLL terms
%% c) that we intend to carry out comparisons in the near future
%% 
%% 2) Predictions for H+1 jet exclusive
%% --------------------------------------
%% [to be signed by Petriello and Liu]
%% 
%% Text as sent by Frank P. & Xiaohui
%% 
%% 3) Uncertainties of fixed-order predictions in H+2 jet studies
%% --------------------------------------------------------------
%% [to be signed by Frank, Shireen & some ATLAS + CMS authors]
%% 
%% A merging of Frank & Shireen's existing paper + extra contributions from Dag et al.
%% 
%% 4) (or 3b) Underlying event uncertainties in H+2-jet studies
%% ------------------------------------------------------------
%% [to be signed by whoever wishes to sign it!]
%% 
%% A recall that large
%% (30%) uncertainties had been found in the past & the interpretation
%% that this is due to changes not just in the UE but also the shower.
%% 
%% A summary of our findings that switching UE on/off has a 10% effect.
%% 
%% (Maybe) discussion of possible caveats.
%%
%%======================================================================
\section{Jets in Higgs physics\footnote{%
    D.~Del Re, B.~Mellado, G.~P.~Salam, F.~J.~Tackmann (eds.); 
A.~Banfi, T.~Becher, F.U.~Bernlochner, S.~Gangal, D.~Gillberg, X.~Liu,
M.~Malberti, P.~Meridiani, P.~Monni, P.~Musella, M.~Neubert, F.~Petriello,
I.~W.~Stewart, J.~R.~Walsh, G.~Zanderighi, S.~Zuberi
}}
\label{sec:jets}

Jets are relevant in multiple contexts in Higgs studies. 
The separation of the data into ``jet bins'', each with a specific number of
jets in the final state, can be useful to help discriminate Higgs production from
backgrounds.
Cuts on the kinematics of jets can also help to separate different Higgs
production mechanisms. This is of particular importance for
discriminating between vector-boson fusion, which tends to be
accompanied by two forward jets, from gluon fusion.
Finally, jets may be produced from the fragmentation of Higgs decay products, as
in the search for $\PH \to b\bar b$, or in analyses of $\PH \to VV$ when one
of the vector bosons decays hadronically.

The main type of question that we address in this section is how we can
reliably estimate cross sections for a given jet bin and the
kinematic distributions used to discriminate between different signal
production mechanisms.
We consider uncertainties in fixed-order predictions, their potential
reduction with the help of resummations, and non-perturbative
uncertainties from the underlying event.

%----------------------------------------------------------------------
\subsection{Resummation for Higgs production with a jet veto
\footnote{A.~Banfi, T.~Becher, P.~Monni, M.~Neubert, G.~P.~Salam,
I.~W.~Stewart, F.~J.~Tackmann, J.~R.~Walsh, G.~Zanderighi, S.~Zuberi}}
\label{sec:jets_H0jets}

In the study of $\PH \to \PW\PW$, with leptonic $\PW$ decays, it is customary
to distinguish events according to their number of jets.
In particular, selecting events with zero jets, i.e. imposing a jet
veto, significantly reduces the background from $\PAQt\PQt$ production.
To relate experimental observations with the jet veto to the total Higgs
production cross section, one must then be able to evaluate the
expected number of signal events that pass the jet veto requirement.

In practice the transverse-momentum threshold for identifying jets,
$p_{T,\text{min}} \simeq 25$--$30\UGeV$, is substantially smaller than
the Higgs mass, $\MH$.
As a result, perturbative calculations of the cross section with the
jet veto~\cite{Catani:2001cr,Anastasiou:2005qj,Grazzini:2008tf}
involve terms enhanced by up to two powers of $\ln
\MH/p_{T,\text{min}}$ for each power of $\alphas$ beyond the
leading-order cross section.
A large value for the logarithm degrades the convergence of the
perturbative series.
Currently the experiments account for the resulting additional
perturbative uncertainty in the fixed-order, NNLO, calculations via the
Stewart--Tackmann procedure~\cite{Stewart:2011cf}.
In the results with the full 2012 dataset~\cite{ATLAS-CONF-2013-030}, the
theoretical uncertainty in $\PH \to \PWp\PWm$ results is comparable to the
statistical error from a single experiment, and significantly larger
than the experimental systematic uncertainty.

The use of fixed-order predictions in the presence of large logarithms
is known not to be optimal, and one can usually obtain significantly
improved predictions by resumming the large logarithms to all
orders.
Previous work in this context~\cite{Anastasiou:2008ik, Anastasiou:2009bt, AlcarazMaestre:2012vp}
considered resummations for related cases, such as the Higgs boson $\pT$ spectrum
(the $\pT$ resummations themselves can be found in
\Bref{Bozzi:2003jy, Bozzi:2005wk, deFlorian:2011xf, Becher:2010tm, Chiu:2012ir, Wang:2012xs, Becher:2012yn}) or
a global jet veto using beam thrust~\cite{Stewart:2009yx, Berger:2010xi}, and using Monte Carlo
generators to establish the relation with the experimental jet veto.

In the past year substantial progress has been made in carrying out a
resummation directly for a veto based explicitly on a jet algorithm.
\Bref{Banfi:2012yh} presented a calculation at next-to-leading
logarithmic (NLL) accuracy,\footnote{Terminology for resummation
  accuracies differs according to the context. Here, N$^p$LL accuracy
  is defined to mean that in the \emph{logarithm} of the jet veto
  efficiency, one accounts for terms $\alphas^n \ln^{n+1-p}
  \MH/p_{\mathrm{T},\text{min}}$ for all $n$.}
where it was observed that the resummation structure at this order is simple
and reduces to a Sudakov form factor.
It also derived the subset of NNLL terms associated with the jet
radius ($R$) dependence, for the generalized-$k_{\mathrm T}$ family of jet
algorithms~\cite{Catani:1993hr,Ellis:1993tq,Dokshitzer:1997in,Wobisch:1998wt,Cacciari:2008gp},
which includes the anti-$k_{\mathrm T}$ algorithm used experimentally.
Subsequently, in \Bref{Becher:2012qa} a calculation at next-to-next-to-leading logarithmic (NNLL) order was presented, incorporating the $R$-dependent NNLL corrections of \Bref{Banfi:2012yh}, but relying on an extrapolation of these results to large $R$.
In addition, it proposed an all-order factorization theorem for the cross section in Soft Collinear Effective Theory (SCET)~\Brefs{Bauer:2000ew, Bauer:2000yr, Bauer:2001ct, Bauer:2001yt, Bauer:2002nz, Beneke:2002ph}.
\Bref{Tackmann:2012bt} discussed questions related to the all-order
factorization theorem. In addition, it pointed out that logarithms of the jet
radius $R$ can lead to sizable corrections for the moderately small values, $R=0.4$--$0.5$,
used experimentally. If these $\ln R$ terms are included
in the logarithmic counting, then they would constitute an NLL contribution, which is
not resummed at present.
\Bref{Banfi:2012jm} took a complementary approach to the
derivation of the full NNLL resummation, accompanied by cross checks
of the NNLL terms at orders $\alphas^2$ and $\alphas^3$ (relative to the total
cross section), obtained with
\textsc{MCFM}~\cite{Campbell:2002tg,Campbell:2006xx,Campbell:2010cz}.
While the results of \Bref{Becher:2012qa} and \Bref{Banfi:2012jm}
are identical in structure at NNLL accuracy, initially one of the
coefficients differed by a numerically relevant term.
The difference was traced in \Bref{Banfi:2012jm} to an extra term which
spoils the extrapolation used in \Bref{Becher:2012qa}.
Subsequent discussions between the groups have led to a consensus that
the NNLL terms are as given in \Bref{Banfi:2012jm}.

Numerically, \Bref{Banfi:2012jm} found that the resummation,
matched with the NNLO calculation, had only a small (percent-level)
impact on the predicted jet-veto efficiency relative to the pure NNLO
prediction, but with a reduction in the perturbative uncertainty from
about $15\%$ to $9\%$. 
The remaining uncertainty is partially associated with the choice of
prescription for matching the resummed and NNLO results.
The code to reproduce these results is available in the form of the
\textsc{jetvheto} program~\Bref{jetvhetoweb}.
A further reduction appears to be possible if one uses a jet radius
$R$ of order $1$, rather than the smaller values currently in use
experimentally, an observation that is consistent with the discussion
in \Bref{Tackmann:2012bt}.

Numerical results from the authors of
\Bref{Becher:2012qa} and \cite{Tackmann:2012bt} are currently available in 
preliminary form and are expected to be published soon.
We look forward to comparisons between all three sets of results in
the near future and also to the experiments taking advantage of the
corresponding reduced uncertainties.
%

%----------------------------------------------------------------------
\subsection{Resummation for exclusive Higgs plus 1-jet production\footnote{X.~Liu, F.~Petriello}}
\label{subsec:jets_Hj}

In this section we study the resummation of a class of large Sudakov logarithms affecting Higgs boson production in the exclusive one-jet bin at the LHC.  These large logarithms occur when the Higgs signal cross section is divided into
bins of exclusive jet multiplicity.  This division is an experimental necessity; for example, in the zero-jet bin of the $\PW\PW$ final state the background is dominated by continuum $\PW\PW$ production, while in the one-jet and two-jet bins, top-pair production becomes increasingly important.  The optimization of this search requires cuts dependent on the number of jets observed, and therefore also on theoretical predictions for exclusive jet multiplicities.

The theoretical community has invested significant recent effort in resumming jet-veto logarithms to all orders in perturbation theory in order to more accurately model the LHC Higgs signal.
As summarized in \refS{sec:jets_H0jets}, a significant reduction of the residual theoretical uncertainties
is obtained in the zero-jet bin by resumming the jet-veto logarithms.  Given that the theoretical uncertainties are currently
one of the largest systematic errors affecting the one-jet bin analyses of the Higgs-like particle properties, it is desirable to formulate the resummation when final-state jets are also present.
(Inclusion of the NNLO Higgs+1-jet prediction,
  calculated recently for the purely gluonic contributions to the
  process~\cite{Boughezal:2013uia} and described in \refS{subsec:ggF_HjNNLO}
  can also be expected to bring an improvement.)

The specific logarithms that we address in this section occur when the transverse momenta of the hard jet in the exclusive one-jet bin is larger than the veto scale.  We calculate contributions through next-to-leading order in the exponent of the Sudakov form factor and include the full one-loop functions describing hard, soft, and collinear emissions.  This implies that we correctly obtain the first three logarithmic corrections at each order in the QCD coupling constant:  $\alphas L^2$, $\alphas L$ and $\alphas$; $\alphas^2 L^4$, $\alphas^2 L^3$, and $\alphas^2 L^2$; $\alphas^3 L^6$, $\alphas^3 L^5$, $\alphas^3 L^4$; and so on.  We have set $L=\text{ln}(Q/\pT^\mathrm{veto})$, where $Q \sim \MH$ denotes any hard scale in the problem.  We match the results to fixed-order to obtain a $\mathrm{NLL}^{\prime}+{\NLO}$ prediction (using the order counting of \Bref{Berger:2010xi}), and present numerical results for use in LHC analyses.  We first demonstrate that the region of phase space where the leading-jet transverse momentum is of order the Higgs mass accounts for nearly half of the error in the fixed-order NLO prediction for Higgs plus one jet, and is therefore a prime candidate for an improved theoretical treatment.  We then perform a detailed study of the residual theoretical uncertainties using our prediction that accounts for the variation of all unphysical scales remaining in the prediction.  Even with a very conservative treatment of the errors, a significant reduction of the residual uncertainty as compared to the fixed-order estimate is found; the estimated uncertainties decrease by up to a quarter of their initial values.  The results, and the improvements in the zero-jet bin obtained previously, should form the basis for future theoretical error estimates in experimental analyses of Higgs properties.  In this section we briefly review the salient features of our formalism, and present numerical results for use in LHC searches.  Further details can be found in \Bref{Liu:2012sz, Liu:2013hba}.

\subsubsection{Review of the formalism}
\label{subsubsec:jets_Hj-formalism}

We use effective-field theory techniques to derive a factorization theorem for exclusive Higgs plus jet production.  The factorization of the cross section into separate hard, soft, and collinear sectors is complicated by the 
presence of the jet algorithm needed to obtain an infrared-safe observable.  Following the experimental analyses, 
we use the anti-$k_{\mathrm T}$ algorithm~\cite{Cacciari:2008gp} to define jets.  We demand that the final state contain only a single jet with $\pT^{\mathrm J} > \pT^\mathrm{veto} \sim 25-30\UGeV$.  Other jets with a transverse momentum above this threshold are vetoed. Since $\pT^\mathrm{veto}$ is usually substantially lower than the partonic center-of-mass energy $\hat{s}$, such that $\lambda \equiv \pT^\mathrm{veto}/\sqrt{\hat{s}} \muchless 1$, the vetoed observables are usually very sensitive to soft and collinear emissions.  We will make the following assumptions in order to proceed in our analysis:
\begin{equation}
\pT^{\mathrm J} \sim \MH \sim \sqrt{\hat{s}}
\,,\qquad
1\gg R^2\gg \lambda^2
\,,
\qquad
\frac{\alphas}{2\pi} \ln^2 R \muchless 1 \,.
\end{equation}
Given that $\pT^\mathrm{veto} \approx 25-30\UGeV$ and $R \approx 0.4-0.5$, when the leading jet $\pT^{\mathrm J} \sim \MH$, the second two assumptions are justified.  The first assumption restricts us to specific region of phase space, which we later show contributes roughly half of the uncertainty in the full one-jet bin.

We are able to utilize an effective-theory framework because of how the anti-$k_{\mathrm T}$ algorithm clusters soft and collinear emissions.  The initial clustering combines the final-state hard emissions
into a jet, so that the soft radiation sees only the jet direction and does not probe its internal structure.  The mixing between the soft and beam sectors is power-suppressed, as is the mixing between the beam and jet sectors.  Denoting the measurement function that imposes the jet clustering and vetoing as $\hat{{\mathcal M}}$, these
facts imply that we can factor the full $\hat{{\mathcal M}}$ into the product of individual measurement functions acting separately on the
soft ($\hat{\mathcal M}_{\mathrm S}$), jet ($\hat{\mathcal M}_{\mathrm J}$), and the two beam sectors ($\hat{\mathcal M}_a$ and $\hat{\mathcal M}_b$),
\begin{equation}
\hat{\mathcal M} = \hat{\mathcal M}_{\mathrm J}\hat{\mathcal M}_{\mathrm S}\hat{\mathcal M}_a\hat{\mathcal M}_b,
\end{equation}
up to power-suppressed corrections in $\pT^\mathrm{veto}$ and $R$~\cite{Becher:2012qa, Tackmann:2012bt}.
For more details we refer the reader to \Bref{Liu:2012sz,Liu: 2013hba}.

The remaining steps in the derivation of the factorization theorem are presented in detail in \Bref{Liu:2012sz}.  The final result for the cross section for exclusive Higgs plus one-jet production takes the following form:
\begin{align}\label{jets:Hj-factgen}
\mathrm{d}\sigma_{\mathrm{NLL}^{\prime}} &= \mathrm{d}\Phi_{H}\mathrm{d}\Phi_{J}\,
{\mathcal F}(\Phi_{H},\Phi_{J})
\,
\sum_{a,b}\int \mathrm{d}x_{a} \mathrm{d}x_b \frac{1}{2\hat{s}}\,
 (2\pi)^4 \delta^4\left(q_a + q_b - q_{J} -q_{H}\right)
\nn\\ & \quad \times
\bar{\sum_{\mathrm{spin}}}
\bar{\sum_{\mathrm{color}}}
{\rm Tr}(H\cdot S)\,
({\mathcal I}_{a,i_aj_a} \otimes f_{j_a})(x_a)\,
({\mathcal I}_{b,i_bj_b} \otimes f_{j_b})(x_b)
J_{J}(R)\,.
\end{align}
Here, $H$, $S$, and $J_J$ denote hard, soft, and final-state jet functions. The convolutions $({\mathcal I}_{ij} \otimes f_j)(x)$ give the initial-state beam functions for beams $a$ and $b$ in terms of the usual PDFs, $f_j$, with $i$, $j$ labeling the incoming parton types. We have denoted explicitly by the subscript that we will evaluate this cross section to the $\mathrm{NLL}^{\prime}$ level, in the language of \Bref{Berger:2010xi}.  We again remind the reader that this implies that we correctly obtain the leading three logarithmic corrections in the cross section at each order in the strong coupling constant.  $\mathrm{d}\Phi_{H}$ and $\mathrm{d}\Phi_{j_i}$ are the phase space measures for the Higgs and
the massless jet $J$, respectively. ${\mathcal F}(\Phi_{H_c},\Phi_{J})$ includes all additional
phase-space cuts other than the $\pT$ veto acting on the Higgs boson and the hard jet.  $H$ is the hard function that 
comes from matching full QCD onto the effective theory, and $S$ describes soft final-state emissions.  The trace is over the color indices.  The functions ${\mathcal I}$ and $J$ describe collinear emissions along the beam axes and along the final-state jet direction, respectively.  The measured jet $\pT^{\mathrm J}$ should be much larger than $\pT^\mathrm{veto}$.  For more details on this formula and the objects contained within, we refer the reader to \Bref{Liu:2012sz,Liu:2013hba}.

We briefly comment here on non-global logarithms~\cite{Dasgupta:2001sh} that first occur at the $\mathrm{NLL}^{\prime}$ level.  Although they are not included in our current factorization theorem, to estimate their numerical effect we use the large-$N_c$ resummation of these terms derived in \Bref{Dasgupta:2001sh}.  We include them as a multiplicative correction to our factorization formula.  Their numerical effect is small, at or below one percent of the total exclusive Higgs plus one-jet production rate for the relevant values of $\MH$ and $\pT^\mathrm{veto}$.  To check the robustness of this result we vary the hard scale appearing in these corrections by a factor of two around their nominal value of $\MH$, and find similarly small corrections.  We therefore believe that it is numerically safe to neglect these terms in our $\mathrm{NLL}^{\prime}$ result, although they should be further investigated in the future.

\subsubsection{Matching $\mathrm{NLL}^{\prime} $ with NLO}
\label{subsubsec:jets_Hj-matching}

We begin our presentation of the numerical results by matching our resummed expression with the fixed-order NLO result to obtain a $\mathrm{NLL}^{\prime}+{\NLO}$ prediction.  We use the NLO predictions for Higgs plus one-jet contained in \textsc{MCFM}~\cite{Campbell:2010ff}.  We obtain our prediction by setting
\begin{eqnarray}\label{jets:Hj-rgimproved}
\sigma_{\mathrm{NLL}^{\prime}+{\NLO}} = \sigma_{\mathrm{NLL}'} + \,
\sigma_{\NLO}-\sigma^\mathrm{exp}_{\mathrm{NLL}'}.
\end{eqnarray}
In this equation, $\sigma_{\NLO}$ is the fixed-order NLO 
cross section obtained from \textsc{MCFM}, and $\sigma_\mathrm{NLL}^{\prime}$ is the resummed
cross section up to $\mathrm{NLL}^{\prime}$ accuracy presented in Eq.~(\ref{jets:Hj-factgen}).
$\sigma^\mathrm{exp}_{\mathrm{NLL}'}$
captures the singular features of $\sigma_{\NLO}$, and is obtained by expanding 
$\sigma_{\mathrm{NLL}'}$ with all scales set to a common value $\mu$.  Schematically, we have
\begin{eqnarray}
&&\sigma_{\mathrm{NLL}'} = \,
\sigma_\LO \left(1 + \alphas  g_0\right)
 e^{-L g_{\mathrm{LL}}(\alphas L)-g_\mathrm{NLL}(\alphas L)}\, \nn \\
&&\sigma^\mathrm{exp}_{\mathrm{NLL}'} = \,
\sigma_\LO\left(1 + \alphas \left[ -g_2 L^2 - g_1 L \,
+ g_0\right] \right)\,,
\end{eqnarray}
where $L\,g_{\rm LL}$ and $g_\mathrm{NLL}$ resum the leading and next-to-leading logarithms,
respectively.   The difference between
$\sigma_{\NLO}$ and the
expanded $\mathrm{NLL}^{\prime}$ result $\sigma^\mathrm{exp}_{\mathrm{NLL}'}$ only
contains power-suppressed contributions for large values of $Q$:
\begin{eqnarray}
\sigma_{\rm non-singular}\equiv \sigma_{\NLO}-\sigma^\mathrm{exp}_{\mathrm{NLL}'}
\sim {\mathcal O}\left(R^2 L,\frac{\pT^\mathrm{veto}}{Q}L\,,
\frac{\pT^\mathrm{veto}}{Q} \ln R \,,
\cdots \right) \,,
\end{eqnarray}
with $L = \ln \left(Q/\pT^\mathrm{veto}\right)$, and $Q$ stands for any kinematic quantity of order $\MH$. Since
the scale $Q R$ is used to define the jet mode, the $R^2L$ terms are regarded as power suppressed.  We have demonstrated explicitly in \Bref{Liu:2013hba} that our formalism correctly captures the singular terms at NLO as $L \to 0$.

\subsubsection{Validity of the effective theory}
\label{subsubsec:jets_Hj-valid}

We comment here briefly on the expected range of validity of our effective theory approach.  In our derivation of the factorization theorem, we assumed that the signal jet $\pT^{\mathrm J}$ is of
order $\MH$.  This configuration contributes
a non-negligible fraction, roughly $30\%$, of the 
experimentally-interesting
total cross section for $\pT^\mathrm{veto} \sim 30\UGeV$ and $\pT^{\mathrm J} > \pT^\mathrm{veto}$.
Our factorization theorem holds for $\pT^\mathrm{veto}\muchless \pT^{\mathrm J} \sim Q $, but breaks down when
$\pT^{\mathrm J} \sim \pT^\mathrm{veto} \muchless \MH$.  Additional large logarithms of the form $\text{ln}^2 \MH/\pT^{\mathrm J}$ and $L \times \, \pT^\mathrm{veto}/\pT^{\mathrm J}$ are not resummed in our formalism.  We describe these terms only as well as a fixed NLO calculation.  A different effective theory is needed for this regime to correctly sum the large logarithms.  We do not consider
this theory in this contribution; our goal here is to consistently apply the currently available formalism at $\mathrm{NLL}^{\prime}+{\NLO}$ to see to what extent we can reduce the theoretical uncertainty.

Interestingly, the $\pT^{\mathrm J} \sim \MH$ region contributes roughly $50\%$ of the uncertainty in the one-jet bin, larger than might be expected.  We show this by computing the NLO cross section for an example parameter choice.  We set $\MH=126\UGeV$ and $\pT^\mathrm{veto}=25\UGeV$, and divide the Higgs plus one-jet cross section, whose inclusive value is $\sigma^{1j}_{{\NLO}} = 5.75^{+2.03}_{-2.66} \Upb$,
into two bins: the first with $\pT^{\mathrm J} < \MH/2$, and the second with 
$\pT^{\mathrm J} > \MH/2$.  As explained in detail later in \Sref{subsubsec:jets_Hj-uncert}, we use the fixed-order cross section in the first bin since our effective-theory analysis does not hold, and turn on resummation in the second bin.  Computing the cross section at NLO in each bin, and estimating the uncertainties as described in detail in \Sref{subsubsec:jets_Hj-uncert}, we find
\begin{align}
\sigma^{1j}_{{\NLO}} (\pT^{\mathrm J} < \MH/2) &= 4.74^{+1.31}_{-1.29} \Upb\,,
\nn \\
\sigma^{1j}_{{\NLO}} (\pT^{\mathrm J} > \MH/2) &= 1.01^{+0.85}_ {-1.51} \Upb\,.
\end{align}
The central values have been obtained using the scale choice for $\mu=\MH/2$.\footnote{We note that using a larger central scale choice leads to the same conclusions regarding the relative uncertainties of the two bins.}  Although it accounts for less than $25\%$ of the cross section, the region where our effective-theory analysis can improve the uncertainties contributes roughly half of the error in the full one-jet bin. 

\subsubsection{Scale choices and uncertainty estimation}
\label{subsubsec:jets_Hj-uncert}

Since the resummation holds only for $\pT \sim \MH$, we wish to turn it off and recover the fixed-order NLO result as 
$\pT^{\mathrm J}$ becomes small.   To do so, we note that the fixed order cross section $\sigma_{\NLO}$ 
and the expanded $\mathrm{NLL}^{\prime}$ cross section
$\sigma^\mathrm{exp}_{\mathrm{NLL}'}$ depend only on the scale $\muR = \muF = \mu$, while $\sigma_{\mathrm{NLL}'} $ also depends on the various scales $\mu_{\mathrm H}$, $\mu_{\mathrm J}$, $\mu_{\mathrm B}$, $\mu_{\mathrm S}$, $\nu_{\mathrm B}$ and $\nu_{\mathrm S}$ at which the hard, jet, beam, and soft functions are evaluated.
The optimal choice for each scale can be determined by minimizing the higher order corrections
to each separate component.  These functions are then RG-evolved from their respective starting scales $\mu_i$ and $\nu_i$ to the common scales $\mu$ and $\nu$.  Consequently, the resummation can be turned off by setting all scales
to $\mu$, so that the full $\mathrm{NLL}^{\prime}+{\NLO}$ result reduces to the NLO one.  We adopt a conservative scheme to turn off the resummation as soon as possible,
as suggested in \Bref{Berger:2010xi}.  In the region where
$\pT^{\mathrm J} \gg \pT^\mathrm{veto}$, we keep the resummation on.   When $\pT^{\mathrm J} \sim \pT^\mathrm{veto}$, we switch off the resummation by setting all scales to $\mu$, which leads to the fixed-order prediction. We interpolate between these two regions smoothly using
\begin{eqnarray}\label{jets:Hj-inter}
\mu_i^{\mathrm{int.}} = \mu + 
(\mu_i - \mu)\left[ \,
1+\tanh\left( \kappa\left(\pT^{\mathrm J}-p_{\mathrm{off}}\right) \right)\,
 \right]/2\,,
\end{eqnarray}
where the index $i = \{\mathrm{H}, \mathrm{J}, \mathrm{B}, \mathrm{S}\}$ runs over all appearing scales.  We use similar expressions for the $\nu$'s.  Our numerical predictions
are obtained using the $\mu_i^{int.}$ expressions in our code.   When $\pT^{\mathrm J} < p_{\mathrm{off}}$, the resummation starts to vanish. We set
$p_{\mathrm{off}} = \max(2\pT^\mathrm{veto},\frac{\MH}{2})$ to be the default value.\footnote{The reason for this choice is that our EFT is valid when
$\pT^{\mathrm J}$ is located in the hard domain whose lower boundary is 
estimated to be $\MH/2$, and it entirely breaks down
when $\pT^{\mathrm J}$ falls into the ``soft" regime whose upper boundary is 
roughly $2\pT^\mathrm{veto}$.}
When making uncertainty estimations, we vary each scale separately. In the resummation region, the cross section is relatively insensitive to the variation of $\mu$.  In the fixed-order range, it is insensitive to $\mu_i$ and $\nu_i$.
The slope $\kappa$ controls how smoothly we turn off the resummation. We find that the interpolated cross section is insensitive to the choice of $\kappa$. Varying $\kappa$ in a reasonable range from $0.04$ to $0.2$, the effect on the cross section is much smaller than our estimated uncertainties.

To derive the uncertainties in both the fixed-order and RG-improved results, we vary all  scales appearing in the cross section around their central values by factors of two in both directions in order to estimate the theoretical error. To avoid an underestimate of the uncertainty of the fixed-order calculation, we follow the procedure suggested by Stewart and Tackmann~\cite{Stewart:2011cf}.  We split the exclusive one-jet cross section into the difference of one-jet inclusive and two-jet inclusive results:
\begin{eqnarray}
\sigma_{1j} = \sigma_{\ge 1j} - \sigma_{\ge 2j} \,.
\end{eqnarray}
We estimate the scale uncertainty for each piece separately and 
add them in quadrature to obtain the scale uncertainty for the 
exclusive cross section:
\begin{eqnarray}\label{jets:Hj-UNLO}
\delta^2_{1j,{\NLO}} = \delta^2_{\ge 1j,{\NLO}} \,
 + \delta^2_{\ge 2j,{\NLO}} \,.
\end{eqnarray}
For the $\mathrm{NLL}^{\prime}+{\NLO}$ result, the uncertainty is derived by adding in quadrature the separate variations of all scales which enter~\cite{Stewart:2011cf}:
\begin{eqnarray}\label{jets:Hj-UNLL}
\delta^2_{1j} = \delta^2_{{\rm non-singular},\mu} \,
 + \delta^2_{{\mathrm{NLL}'},\mu} \,
 + \delta^2_{{\mathrm{NLL}'},\mu_{\PH}} \,
 + \delta^2_{{\mathrm{NLL}'},\mu_{\mathrm J}} \,
 + \delta^2_{{\mathrm{NLL}'},\mu_{\mathrm B},\nu_{\mathrm B}} \,
 + \delta^2_{{\mathrm{NLL}'},\mu_{\mathrm S},\nu_{\mathrm S}} \,.
\end{eqnarray}
Before continuing we comment briefly on the structure of Eq.~(\ref{jets:Hj-UNLL}).  In order to perform the matching to 
fixed order in Eq.~(\ref{jets:Hj-rgimproved}), we RG-evolve the $\mathrm{NLL}^{\prime}$ result so that all scales are set to the common scale $\mu$.  We then add on the non-singular NLO terms via the difference between the full NLO cross section and the expanded $\mathrm{NLL}^{\prime}$ results.  This explains the first two contributions to the above equations.  As the hard and jet functions live at the scales
$\sqrt{\MH \pT^{\mathrm J}}$ and $\pT^{\mathrm J} R$ respectively~\cite{Liu:2012sz}, these scale variations are treated as uncorrelated in 
Eq.~(\ref{jets:Hj-UNLL}).  Finally, the variations of beam and soft functions, which live at the scale $\pT^\mathrm{veto}$, are added to this.

When we apply this formalism and assume actual LHC kinematic cuts, a large fraction of the cross section comes from
the low-$\pT^{\mathrm J}$ regime where $\pT^{\mathrm J} < p_{\mathrm{off}}$, and the fixed-order
calculation dominates. In this situation, we split the 
cross section into two regions, one with
$\pT^\mathrm{veto}<\pT^{\mathrm J} < p_{\mathrm{off}}$ and
the other with $\pT^{\mathrm J} > p_{\mathrm{off}}$. For the former region, we use
Eq.~(\ref{jets:Hj-UNLO}) to estimate the uncertainty and for the latter one,
we utilize Eq.~(\ref{jets:Hj-UNLL}). We combine these two linearly to 
estimate the scale dependence for the RG-improved cross section:
\begin{eqnarray}\label{jets:Hj-UMIX}
\delta_{1j}(p^{\mathrm J}_{\mathrm T}>\pT^\mathrm{veto}) = \,
\delta_{1j,{\NLO}}(p^{\mathrm J}_{\mathrm T}<p_{\mathrm{off}}) \,
+ \delta_{1j}(p^{\mathrm J}_{\mathrm T}>p_{\mathrm{off}}) \,.
\end{eqnarray}
Since the resummation in the result used for $\pT^{\mathrm J} > p_{\mathrm{off}}$ is turned off quickly by using the interpolation in Eq.~(\ref{jets:Hj-inter}), and the uncertainty of the fixed-order cross section used for $\pT^{\mathrm J} < p_{\mathrm{off}}$ is obtained using the Stewart-Tackmann prescription, we believe that this leads to a very conservative estimate of the theoretical error after performing our RG-improvement.

\subsubsection{Numerics for the LHC}
\label{subsubsec:jets_Hj-numerics}

We now present predictions and uncertainty estimates for use in LHC analyses.  For the following numerical results, and those shown above, we use the MSTW 2008 parton distribution functions~\cite{Martin:2009bu} at NLO.  We assume an $8\UTeV$ LHC, and $\MH=126\UGeV$ unless stated otherwise.  We demand that the leading jet be produced with rapidity $|y_{\mathrm J}|< 4.5$, and veto all other jets with $\pT > \pT^\mathrm{veto}$ over the entire rapidity range.  The following central values are used for the scales which appear:
\begin{eqnarray}
\label{jets:Hj-cscale}
&&\mu = \sqrt{(m^{\mathrm T}_{\PH})_{\mathrm{min}} (p_{T}^{\mathrm J})_{\mathrm{min}}},
\hspace{5.ex} \mu_{\PH} = \sqrt{m^{\mathrm T}_{\PH} \pT^{\mathrm J}} \,,\nn \\
&&\mu_{\mathrm J} = \pT^{\mathrm J} R \,,
\hspace{5.ex} \mu_{\mathrm B} = \mu_{\mathrm S} = \pT^\mathrm{veto} \,, \nn \\
&&\nu_{B_{a,b}}  = x_{a,b}\sqrt{ s}\,,
\hspace{5.ex} \nu_{\mathrm S} = \pT^\mathrm{veto}\,.
\end{eqnarray}
where $m^{\mathrm T}_{\PH} = \sqrt{\MH^2+\pT^{{\mathrm J},2}}$.  We note that these central scale values, as well as the variations up and down by a factor of two, are used as the $\mu_i$ on the right-hand side of Eq.~(\ref{jets:Hj-inter}).  The actual numerical scale choices used in the code are the $\mu_i^{int.}$ appearing on the left-hand side of Eq.~(\ref{jets:Hj-inter}).  We use $\kappa=0.2$ to 
produce all numerical results, although we have checked that their dependence on $\kappa$ is negligible.

\begin{figure}[htbp]
\begin{center}
  \includegraphics[width=3.7in,angle=0]{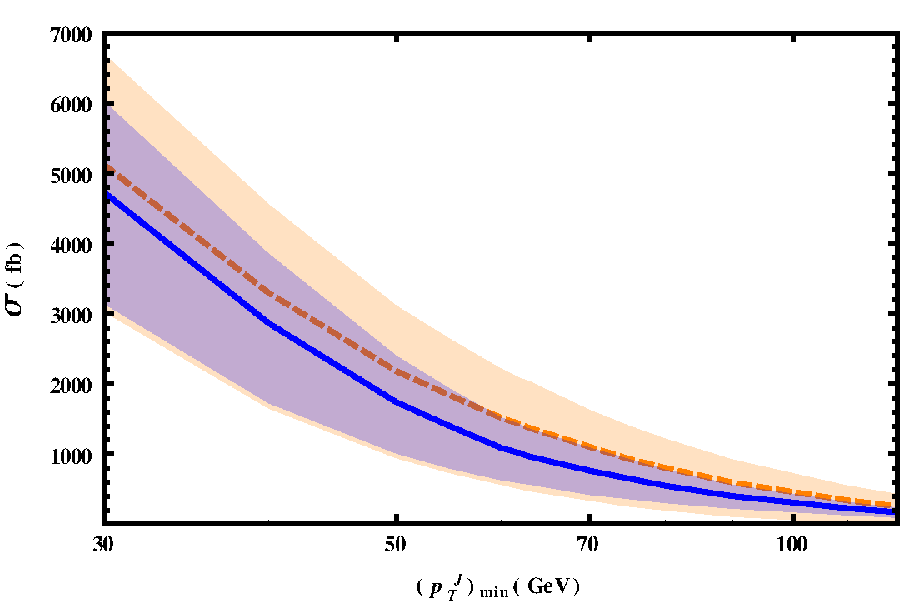}
\end{center}
\vspace{-0.5cm}
\caption{Shown are the $\mathrm{NLL}^{\prime}+ {\NLO}$ (blue band) and NLO (yellow band) cross sections for fixed $\pT^\mathrm{veto} = 30\UGeV$ as a function of the lower cut on $\pT^{\mathrm J}$.}
\label{jets:Hj-30ptveto}
\end{figure}

We show in Fig.~\ref{jets:Hj-30ptveto} the cross section as a function of the lower cut on $\pT^{\mathrm J}$ for a fixed $\pT^\mathrm{veto}=30\UGeV$.  The solid line and blue band show the $\mathrm{NLL}^{\prime}+ {\NLO}$ result together with its perturbative uncertainty, which can be compared with the dashed line and yellow band showing the fixed NLO result with its uncertainty. Even for values of the lower $\pT^{\mathrm J}$ cut near $\pT^\mathrm{veto}$, a sizeable reduction of the uncertainty occurs when the $\mathrm{NLL}^{\prime}+ {\NLO}$ result is used.  The reason for this is discussed in \Sref{subsubsec:jets_Hj-valid}; roughly half of the uncertainty comes from the high-$\pT^{\mathrm J}$ region, which is exactly the parameter space improved by our effective-theory description.

\begin{table}
\begin{center}
\caption{Shown are the central values and uncertainties for the NLO cross section, the resummed cross section, and the event fractions in the one-jet bin using both the fixed-order and the resummed results.  Numbers are given for several Higgs masses and for $\pT^\mathrm{veto}=25,30\UGeV$.}
\label{jets:Hj-nums}
\begin{tabular}{lccccc}
\hline
$\MH$ (GeV) & $\pT^\mathrm{veto}$ (GeV) & $\sigma_{{\NLO}}$ (pb) & $\sigma_{\mathrm{NLL}^{\prime}+{\NLO}}$ (pb) &
	$f^{1j}_{{\NLO}}$ & $f^{1j}_{\mathrm{NLL}^{\prime}+{\NLO}}$ \\
\hline 
124 & 25 & $5.92^{+35\%}_{-46\%}$ & $5.62^{+29\%}_{-30\%}$ & $0.299^{+38\%}_{-49\%}$ &  
	$0.283^{+33\%}_{-34\%}$ \\ 
125 & 25 & $5.85^{+34\%}_{-46\%}$ & $5.55^{+29\%}_{-30\%}$ & $0.300^{+37\%}_{-49\%}$ &  
	$0.284^{+33\%}_{-33\%}$ \\ 
126 & 25 & $5.75^{+35\%}_{-46\%}$ & $5.47^{+30\%}_{-30\%}$ & $0.300^{+38\%}_{-49\%}$ &  
	$0.284^{+34\%}_{-33\%}$ \\ 
\hline
124 & 30 & $5.25^{+31\%}_{-41\%}$ & $4.83^{+29\%}_{-29\%}$ & $0.265^{+35\%}_{-43\%}$ &  
	$0.244^{+33\%}_{-33\%}$ \\ 
125 & 30 & $5.19^{+32\%}_{-41\%}$ & $4.77^{+30\%}_{-29\%}$ & $0.266^{+35\%}_{-43\%}$ &  
	$0.244^{+33\%}_{-33\%}$ \\ 
126 & 30 & $5.12^{+32\%}_{-41\%}$ & $4.72^{+30\%}_{-29\%}$ & $0.266^{+35\%}_{-43\%}$ &  
	$0.246^{+33\%}_{-32\%}$ \\ 
\hline
\end{tabular}
\end{center}
\end{table}

We present in \Tref{jets:Hj-nums} numerical results for both the cross sections and the fraction of events in the one-jet bin, $f^{1j}$.  We define the event fraction as
\begin{equation}
f^{1j}_x = \frac{\sigma_x}{\sigma_{\mathrm{inc}}},
\end{equation}
where $x$ denotes either the NLO or the $\mathrm{NLL}^{\prime}+{\NLO}$ cross section in the one-jet bin.  We note that our values for $f^{1j}_{NLO}$ are consistent with those obtained by the ATLAS collaboration, which provides a cross-check of our results.The total cross section $\sigma_{\mathrm{inc}}$, as well as its estimated uncertainty, is taken from the LHC Higgs cross section working group.  The uncertainties shown are calculated as discussed in \Sref{subsubsec:jets_Hj-uncert}.  Results are given for $\MH=124-126\UGeV$, and for $\pT^\mathrm{veto}=25$ and $30\UGeV$.  The reductions of the uncertainties are significant for both values of $\pT^\mathrm{veto}$.  Symmetrizing the error for this discussion, the estimated uncertainty on the cross section improves from $\pm 40\%$ at NLO to $\pm 30\%$ at $\mathrm{NLL}^{\prime}+{\NLO}$, a reduction of one quarter of the initial value.  The one-jet fraction uncertainty decreases from $\pm 44\%$ to $\pm 34\%$.  For $\pT^\mathrm{veto}=30\UGeV$, the error on the cross section decreases from $\pm 36\%$ to $\pm 29\%$ when resummation is included, while the error on $f^{1j}$ decreases from $\pm 39\%$ to $\pm 33\%$. We note that these are extremely conservative error estimates, as discussed in \Sref{subsubsec:jets_Hj-uncert}.  We default to the Stewart-Tackmann prescription over a large region of the relevant parameter space, and turn off the resummation at a relatively high value of $\pT^{\mathrm J}$.  Enough of the error comes from the high $\pT^{\mathrm J}$ region that our RG-improvement is effective in taming the uncertainty.

%%%%%%%%%%%%%%%%%%%%%%%%%%%%%%%%%%%%%%%%%%%%%%%%%%%%%%%%%%%%%%%%%%%%%%%%%%%%%%%

\subsection{Perturbative uncertainties in the Higgs plus 2-jet VBF selection\footnote{F.~U.~Bernlochner,
S.~Gangal, D.~Gillberg, F.~J.~Tackmann}}
\label{sec:jets-H2jets}

With the typical kinematic cuts used by the ATLAS and CMS collaborations to select events from vector-boson fusion (VBF), the VBF sample is contaminated by a $\sim 25\%$ fraction from Higgs + 2 jet production via gluon fusion (ggF). In this section we discuss the perturbative uncertainties in this contribution, which tend to be sizable and therefore require a reliable estimate. Typical VBF selections include indirect restrictions or explicit vetoes on additional jet activity, primarily to reduce non-Higgs backgrounds but also to reduce the amount of contamination from ggF. Such exclusive restrictions constitute a nontrivial jet binning, where the inclusive Higgs plus 2-jet cross section is effectively divided into an exclusive 2-jet bin and a remaining inclusive 3-jet bin. With such a jet binning one has to account for two sources of perturbative uncertainties: In addition to the absolute yield uncertainty which is correlated between the jet bins, there is also a migration uncertainty which is anticorrelated and drops out in the sum of the bins. This migration uncertainty is associated with the additional perturbative uncertainty induced by the exclusive binning cut. In practice, the experimentally relevant region typically lies inside a transition region between the fully inclusive region (no binning) and the extreme exclusive region (very tight binning). In this region, fixed-order perturbation theory can still be applied however since the logarithms in the perturbative series induced by the binning are already sizeable their effect on the migration uncertainty must be taken into account. This can be achieved using the Stewart-Tackmann (ST) method~\cite{Stewart:2011cf}.

We discuss in detail the application of the ST method to estimate the perturbative uncertainties in the fixed NLO predictions for $\Pp\Pp \to \PH +2$ jets via ggF from \textsc{MCFM}~\cite{Campbell:2006xx, Campbell:2010cz}%
\footnote{For simplicity we denote the process as $\Pg\Pg\to \PH +2j$ below, where a sum over all possible partonic channels with a ggF vertex is implied.}.
To be specific we will concentrate on the VBF selection of the current $\PH\to\PGg\PGg$ analyses. Qualitatively, our results apply equally to other decay channels with similar VBF selection cuts.
After reviewing the general setup in \refS{sec:jets-H2jets-setup} and the jet-binning uncertainties in \refS{sec:jets-H2jets-binning}, we first discuss the fixed-order perturbative uncertainties in a cut-based setup closely following \Bref{Gangal:2013nxa}. In \refS{sec:jets-H2jets-MVA} we then discuss a simple method to propagate the theory uncertainties into a multivariate selection.

%~~~~~~~~~~~~~~~~~~~~~~~~~~~~~~~~~~~~~~~~~~~~~~~~~~~~~~~~~~~~~~~~~~~~~~~~~~~~~~~
\subsubsection{Setup and inclusive 2-jet cross section}
\label{sec:jets-H2jets-setup}
%~~~~~~~~~~~~~~~~~~~~~~~~~~~~~~~~~~~~~~~~~~~~~~~~~~~~~~~~~~~~~~~~~~~~~~~~~~~~~~~

We use \textsc{\textsc{MCFM}}~\cite{Campbell:2006xx, Campbell:2010cz} to compute the NLO cross section, with the $\Pg\Pg\PH$ effective vertex in the infinite top mass limit, and then rescale the cross section with the exact $m_{\PQt}$ dependence of the total LO cross section, $\sigma_{\mathrm{LO}}(m_{\PQt})/\sigma_{\mathrm{LO}}(\infty) = 1.0668$ for $\MH = 125\UGeV$. We take $\sqrt{s} = 8\UTeV$, $\MH = 125\UGeV$, and use the MSTW2008~\cite{Martin:2009bu} NLO PDFs with their corresponding value of $\alphas(m_Z) = 0.12018$. In our analysis we implement the 2-jet selection and VBF selection cuts from the current ATLAS and CMS $\PH\to\PGg\PGg$ analyses (summarized in \refT{tab:VBFcuts} below). However, note that we consider the cross section for the production of an on-shell Higgs boson, without including any branching ratios or cuts on the Higgs decay products.

\begin{figure*}
\includegraphics[width=0.5\columnwidth]{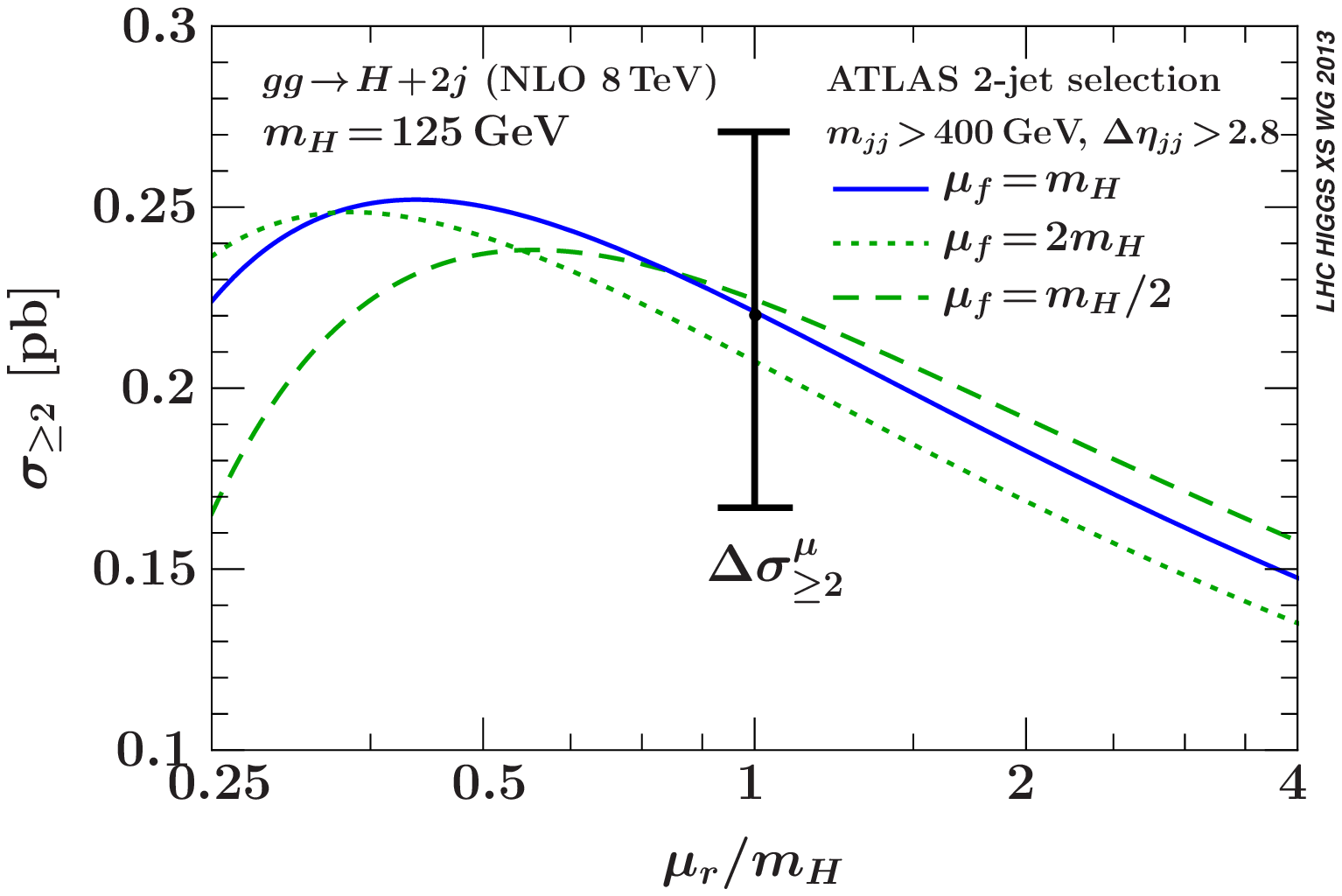}%
\hfill%
\includegraphics[width=0.5\columnwidth]{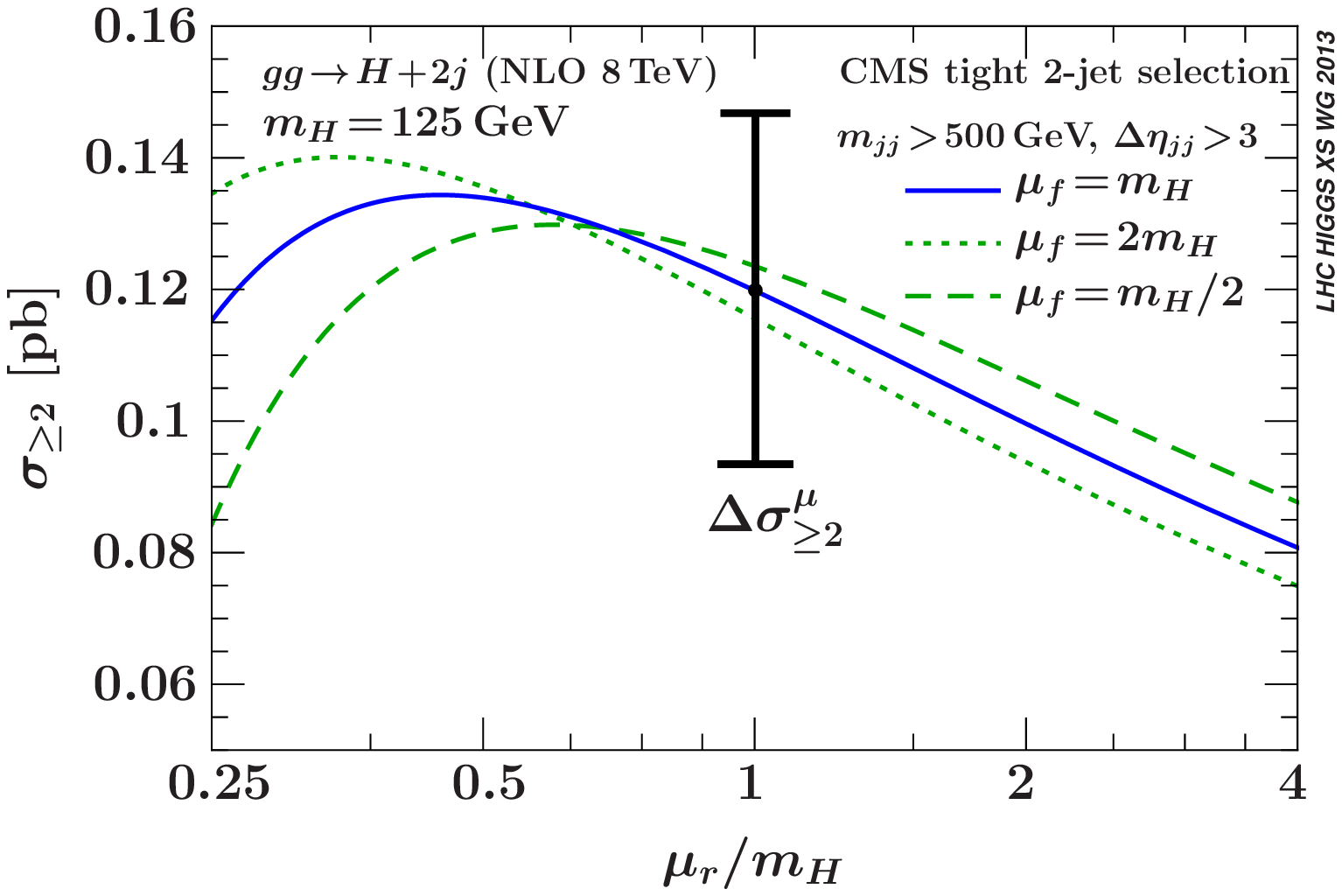}%
\caption{Inclusive 2-jet cross section over a range of $\muR/\MH$ for ATLAS VBF selection (left panel) and  CMS tight selection (right panel). The three curves show different values of $\muF$. The blue solid, green dotted, and green dashed curves correspond to $\muF=\MH$, $\muF=2\MH$, and $\muF=\MH/2$, respectively. The uncertainty bars show the inclusive 2-jet scale variation uncertainty.}
\label{fig:murmuf2}
\end{figure*}

For all our central value predictions we use $\muR = \muF = \MH$, which was also used in \Bref{Campbell:2006xx, Campbell:2010cz}.  In \refF{fig:murmuf2}, we show the scale dependence of the inclusive 2-jet cross section, $\sigma_{\geq 2}$, where we plot it over a range of $1/4<\muR/\MH<4$ for three different values of $\muF$.
To estimate the scale uncertainty we consider the range $0.5 \leq \muR/\MH \leq 2$. The maximum deviation from the central value is given by the green dotted curve for $\muF = \muR = 2\MH$. We use this maximum variation to construct a symmetric scale uncertainty for the inclusive 2-jet cross section, denoted as $\Delta\sigma_{\geq 2}^\mu$, and shown by the uncertainty bar in \refF{fig:murmuf2}. It corresponds to a relative uncertainty at NLO of $21\%$, which is similar to what was found in earlier studies~\cite{Campbell:2006xx, Campbell:2010cz} where a somewhat looser VBF selection was used.  We also require scale uncertainties for the inclusive 3-jet cross section, $\Delta\sigma_{\geq 3}^\mu$, for which we symmetrize the scale variation by taking half of the difference between the $\mu=\MH/2$ and $\mu=2\MH$ variations, as discussed in \Bref{Gangal:2013nxa}.

%~~~~~~~~~~~~~~~~~~~~~~~~~~~~~~~~~~~~~~~~~~~~~~~~~~~~~~~~~~~~~~~~~~~~~~~~~~~~~~~
\subsubsection{Review of jet binning uncertainties}
\label{sec:jets-H2jets-binning}
%~~~~~~~~~~~~~~~~~~~~~~~~~~~~~~~~~~~~~~~~~~~~~~~~~~~~~~~~~~~~~~~~~~~~~~~~~~~~~~~

We consider the \emph{inclusive} $N$-jet cross section, $\sigma_{\geq N}$, for some process containing at least $N$ jets, and assume that $\sigma_{\geq N}$ is a sufficiently inclusive quantity such that it can be computed in fixed-order perturbation theory. We are interested in the case where $\sigma_{\geq N}$ is divided up into a corresponding \emph{exclusive} $N$-jet cross section, $\sigma_N$, and a remainder $\sigma_{\geq N+1}$,
%%%
\begin{equation} \label{sigmaNgeneral}
\sigma_{\geq N}
= \sigma_{N}(\text{excl. cut}) + \sigma_{\geq N+1}(\text{inverse excl. cut})
\,.\end{equation}
%%%
All three cross sections here have the \emph{same} selection cuts applied that identify the leading $N$ signal jets. What defines $\sigma_N$ to be ``exclusive'' is that the additional exclusive cut applied to it restricts the phase space of additional emissions in such a way that $\sigma_N$ is dominated by configurations close to the $N$-parton Born kinematics. In particular, at leading order (LO) in perturbation theory $\sigma_{\geq N}^\LO = \sigma_N^\LO$, while relative to these $\sigma_{\geq N+1}$ is suppressed by $\orderx{\alphas}$. In other words, $\sigma_{\geq N+1}$ requires at least one additional emission to be non-vanishing. Hence, we can consider it an inclusive $(N+1)$-jet cross section with at least $N+1$ jets.%
\footnote{Note that that $\sigma_{\geq N+1}$ is defined by inverting the exclusive cut that defines $\sigma_N$ and so (as in the examples we consider) does not necessarily require the explicit identification of another well-separated jet via a jet algorithm.}

In the simplest case, $\sigma_{\geq N}$ is divided into the two jet bins $\sigma_N$ and $\sigma_{\geq N+1}$ by using a fixed cut on some kinematic variable, $p_{N+1}$, which characterizes additional emissions, with $p_{N+1} = 0$ for a tree-level $N$-parton state. (Two examples we will consider are $\pTHjj$ and $\dphi$, defined below.) The two jet bins then correspond to the integrals of the differential spectrum $\df\sigma/\df p_{N+1}$ above and below some cut,
%%%
\begin{align} \label{pcut}
\sigma_{\geq N} &= \int_0^{p^\cut}\!\df p_{N+1}\, \frac{\df\sigma_{\geq N}}{\df p_{N+1}} + \int_{p^\cut}\!\df p_{N+1}\,
\frac{\df\sigma_{\geq N}}{\df p_{N+1}}
% \nn\\
\equiv  \sigma_N(p^\cut) + \sigma_{\geq N+1}(p^\cut)
\,.\end{align}
%%%
In general, the jet bin boundary can be a much more complicated function of phase space, as in the multivariate analysis considered in \refS{sec:jets-H2jets-MVA}.

We are interested in the uncertainties involved in the binning. The covariance matrix for $\{\sigma_N, \sigma_{\geq N+1}\}$ is a symmetric $2\times 2$ matrix with three independent parameters. A convenient and general parameterization is to write it in terms of two components,
%%%
\begin{align} \label{Cgeneral}
C &=
\begin{pmatrix}
(\Delta^{\mathrm y}_{N})^2 &  \Delta^{\mathrm y}_{N}\,\Delta^{\mathrm y}_{\geq N+1}  \\
\Delta^{\mathrm y}_{N}\,\Delta^{\mathrm y}_{\geq N+1} & (\Delta^{\mathrm y}_{\geq N+1})^2
\end{pmatrix}
+
\begin{pmatrix}
 \Delta_\cut^2 &  - \Delta_\cut^2 \\
-\Delta_\cut^2 & \Delta_\cut^2
\end{pmatrix}
.\end{align}
%%%
Here, the first term is an absolute ``yield'' uncertainty, denoted with a superscript ``y'', which (by definition) is 100\% correlated between the two bins $\sigma_N$ and $\sigma_{\geq N+1}$. The second term is a ``migration'' uncertainty between the bins and corresponds to the uncertainty introduced by the binning cut. It has the same absolute size, $\Delta_\cut$, for both bins and is 100\% anticorrelated between them, such that it drops out when the two bins are added. Hence, the total uncertainty for each bin is given by
%%%
\begin{align} \label{DeltaN}
\Delta_{N}^2 &= (\Delta_N^{\mathrm y})^2 + \Delta_\cut^2
\,,\qquad
%\nn\\
\Delta_{\geq N+1}^2 = (\Delta_{\geq N+1}^{\mathrm y})^2 + \Delta_\cut^2
\,,\end{align}
%%%
while the total uncertainty on their sum, i.e. on $\sigma_{\geq N}$, is given by the total yield uncertainty,
%%%
\begin{equation} \label{DeltageqN}
\Delta_{\geq N} = \Delta^{\mathrm y}_{\geq N} = \Delta^{\mathrm y}_N + \Delta^{\mathrm y}_{\geq N+1}
\,.\end{equation}
%%%

Considering the perturbative uncertainties, the basic question is how each of the uncertainties in \eqn{Cgeneral} can be evaluated. The fixed-order prediction provides us with two independent pieces of information, namely the variations obtained by the standard scale variations, which we denote as $\Delta^\mu_{\geq N}$, $\Delta^\mu_{N}$, $\Delta^\mu_{\geq N+1}$, and which satisfy $\Delta^\mu_{\geq N} = \Delta^\mu_{N} + \Delta^\mu_{\geq N+1}$. The usual assumption that the standard fixed-order scale variations can be used to obtain a reliable estimate of the total uncertainties in the \emph{inclusive} cross sections imposes the two conditions
%%%
\begin{equation}
\Delta_{\geq N} = \Delta^\mu_{\geq N}
\,,\qquad
\Delta_{\geq N+1} = \Delta^\mu_{\geq N+1}
\,.\end{equation}
%%%
Together with \eqnsc{DeltaN}{DeltageqN} these lead to
%%%
\begin{align} \label{conditions}
\text{i)} && \Delta^\mu_{\geq N} &= \Delta^{\mathrm y}_N + \Delta^{\mathrm y}_{\geq N+1}
\,,\nn\\
\text{ii)} && (\Delta^\mu_{\geq N+1})^2 &= (\Delta_{\geq N+1}^{\mathrm y})^2 + \Delta_\cut^2
\,.\end{align}
%%%
Thus, the basic question is how to divide up $\Delta^\mu_{\geq N+1}$ between $\Delta_{\geq N+1}^{\mathrm y}$ and $\Delta_\cut$ in order to satisfy condition ii). Condition i) then determines $\Delta_N^{\mathrm y}$. The nontrivial effect $\Delta_\cut$ can have is on the size of $\Delta_N$ as well as on the off-diagonal entries in \eqn{Cgeneral}, which determine the correlation between $\Delta_N$ and $\Delta_{\geq N+1}$.

Clearly, the simplest is to neglect the effect of $\Delta_\cut$ altogether and to directly use the scale variations to estimate the uncertainties, i.e., to take
%%%
\begin{align} \label{direct}
\Delta_N^{\mathrm y} &= \Delta_{\geq N}^\mu - \Delta_{\geq N + 1}^\mu \equiv \Delta_N^\mu
\,,\quad \Delta_{\geq N + 1}^{\mathrm y} = \Delta_{\geq N + 1}^\mu
\,,\quad
\Delta_\cut = 0
\,.\end{align}
%%%
which leads to
%%%
\begin{equation} \label{Cdirect}
\mspace{-40mu}\text{direct:} \mspace{40mu}
C =
\begin{pmatrix}
(\Delta^\mu_{N})^2 &  \Delta^\mu_{N}\,\Delta^\mu_{\geq N+1}  \\
\Delta^\mu_{N}\,\Delta^\mu_{\geq N+1} & (\Delta^\mu_{\geq N+1})^2
\end{pmatrix}
.\end{equation}
%%%
Note that since $\sigma_{\geq N+1}$ starts at higher order in perturbation theory than $\sigma_{\geq N}$, its relative uncertainty $\Delta_{\geq N+1}^\mu/\sigma_{\geq N+1}$ will typically be (much) larger than $\sigma_{\geq N}$'s relative uncertainty $\Delta_{\geq N}^\mu/\sigma_{\geq N}$. This means one cannot simply apply the latter as the relative yield uncertainty in each bin, which means one \emph{cannot} take $\Delta_i^{\mathrm y} = (\Delta_{\geq N}^\mu/\sigma_{\geq N}) \sigma_i$, as this would violate the condition $\Delta_{\geq N+1} = \Delta_{\geq N+1}^\mu$. This point has already been emphasized in earlier studies~\cite{Anastasiou:2009bt}.

The direct scale variation choice is reasonable as long as the effect of $\Delta_\cut$ is indeed negligible. It is certainly justified if numerically $\Delta^\mu_{\geq N} \gg \Delta^\mu_{\geq N+1}$, since any uncertainty due to migration effects can be at most as large as $\Delta^\mu_{\geq N+1}$ (by virtue of condition ii) above). This can happen, for example, when $\Delta^\mu_{\geq N}$ is sizable due to large perturbative corrections in $\sigma_{\geq N}$ and/or the binning cut is very loose (i.e., is cutting out only a small fraction of phase space) such that $\sigma_{\geq N+1}$ is numerically small to begin with.

In perturbation theory, the effect of the binning cut is to introduce Sudakov double logarithms in the perturbative series of $\sigma_N$ and $\sigma_{\geq N+1}$, which have opposite sign and cancel in the sum of the two bins, schematically
%%%
\begin{align}
\sigma_{\geq N} &\simeq \sigma_{\mathrm B} [1 + \alphas + \alphas^2 + \orderx{\alphas^3} \bigr]
\,,\nn\\
\sigma_{\geq N+1} &\simeq \sigma_{\mathrm B} \bigl[\alphas (L^2 + L + 1)
+ \alphas^2 (L^4 + L^3 + L^2 + L + 1) + \orderx{\alphas^3 L^6} \bigr]
\,,\nn\\
\sigma_{N} &= \sigma_{\geq N} - \sigma_{\geq N+1}
\,,\end{align}
%%%
where $\sigma_{\mathrm B}$ denotes the Born cross section and $L$ is a Sudakov logarithm, e.g. for \eqn{pcut} $L = \ln(p^\cut/Q)$ where $Q\sim \MH$ is a typical hard scale.
The perturbative migration uncertainty $\Delta_\cut$ can be directly associated with the perturbative uncertainty in the logarithmic series induced by the binning, and so should not be neglected once the logarithms have a noticeable effect. In particular, as demonstrated in \Bref{Stewart:2011cf}, the simple choice in \eqnsc{direct}{Cdirect} can easily lead to an underestimate of $\Delta_N$ in the region where there are sizable numerical cancellations between the two series in $\sigma_{\geq N}$ and $\sigma_{\geq N+1}$. Since in this region the dominant contribution to $\sigma_{\geq N+1}$ comes from the logarithmic series, varying the scales in $\sigma_{\geq N+1}$ directly tracks the size of the logarithms, which means we can use $\Delta_\cut = \Delta^\mu_{\geq N+1}$ as an estimate for the binning uncertainty, as proposed in \Bref{Stewart:2011cf}. From \eqn{conditions}, we then find
%%%
\begin{align} \label{ST}
\Delta_N^{\mathrm y} &= \Delta_{\geq N}^\mu
\,,\qquad \Delta_{\geq N + 1}^{\mathrm y} = 0
\,,\qquad
\Delta_\cut = \Delta_{\geq N+1}^\mu
\,,\end{align}
%%%
such that
%%%
\begin{equation} \label{CST}
\text{ST:} \quad
C =
\begin{pmatrix}
(\Delta^\mu_{\geq N})^2 + (\Delta^\mu_{\geq N+1})^2 &  - (\Delta^\mu_{\geq N+1})^2 \\
- (\Delta^\mu_{\geq N+1})^2 & (\Delta^\mu_{\geq N+1})^2
\end{pmatrix}
.\end{equation}
%%%
Since $\Delta_{\geq N+1}^\mu$ is now used as $\Delta_\cut$, the effective outcome is that one treats $\Delta^\mu_{\geq N}$ and $\Delta^\mu_{\geq N+1}$ as uncorrelated.
More generally, we can introduce a parameter $0 \leq \rho \leq 1$, which controls the fraction of $\Delta_{\geq N+1}^\mu$ assigned to $\Delta^{\mathrm y}_{\geq N+1}$, such that
%%%
\begin{align}
\Delta^{\mathrm y}_{N} &= \Delta_{\geq N}^\mu - \rho\, \Delta_{\geq N+1}^\mu
\,,\qquad
\Delta^{\mathrm y}_{\geq N+1} = \rho\, \Delta_{\geq N+1}^\mu
\,,\qquad
\Delta_\cut = \sqrt{1 - \rho^2}\, \Delta^\mu_{\geq N+1}
\,,\end{align}
%%%
which leads to
%%%
\begin{equation} \label{CSTrho}
\text{ST ($\rho$):} \qquad
C =
\begin{pmatrix}
(\Delta_{\geq N}^\mu)^2 + (\Delta_{\geq N+1}^\mu)^2 - 2\rho\, \Delta_{\geq N}^\mu \Delta_{\geq N+1}^\mu
&  (\rho\, \Delta_{\geq N}^\mu - \Delta^\mu_{\geq N+1}) \Delta_{\geq N+1}^\mu
 \\
(\rho\, \Delta_{\geq N}^\mu - \Delta^\mu_{\geq N+1}) \Delta_{\geq N+1}^\mu
& (\Delta^\mu_{\geq N+1})^2
\end{pmatrix}
.\end{equation}
%%%
From this one can easily see that $\rho$ corresponds the correlation between $\Delta^\mu_{\geq N}$ and $\Delta^\mu_{\geq N+1}$. The choice $\rho = 1$ would be equivalent to the case in \eqn{Cdirect}, while $\rho = 0$ reproduces \eqnsc{ST}{CST}. Hence, from the above arguments one should take $\rho$ to be small. The dependence on $\rho$ was explored in \Bref{Gangal:2013nxa} where it was found that for $\rho \lesssim 0.4$ the results are not very sensitive to the precise value of $\rho$, so we take $\rho = 0$ as our default choice.

\begin{figure*}[t!]
\includegraphics[width=0.515\columnwidth]{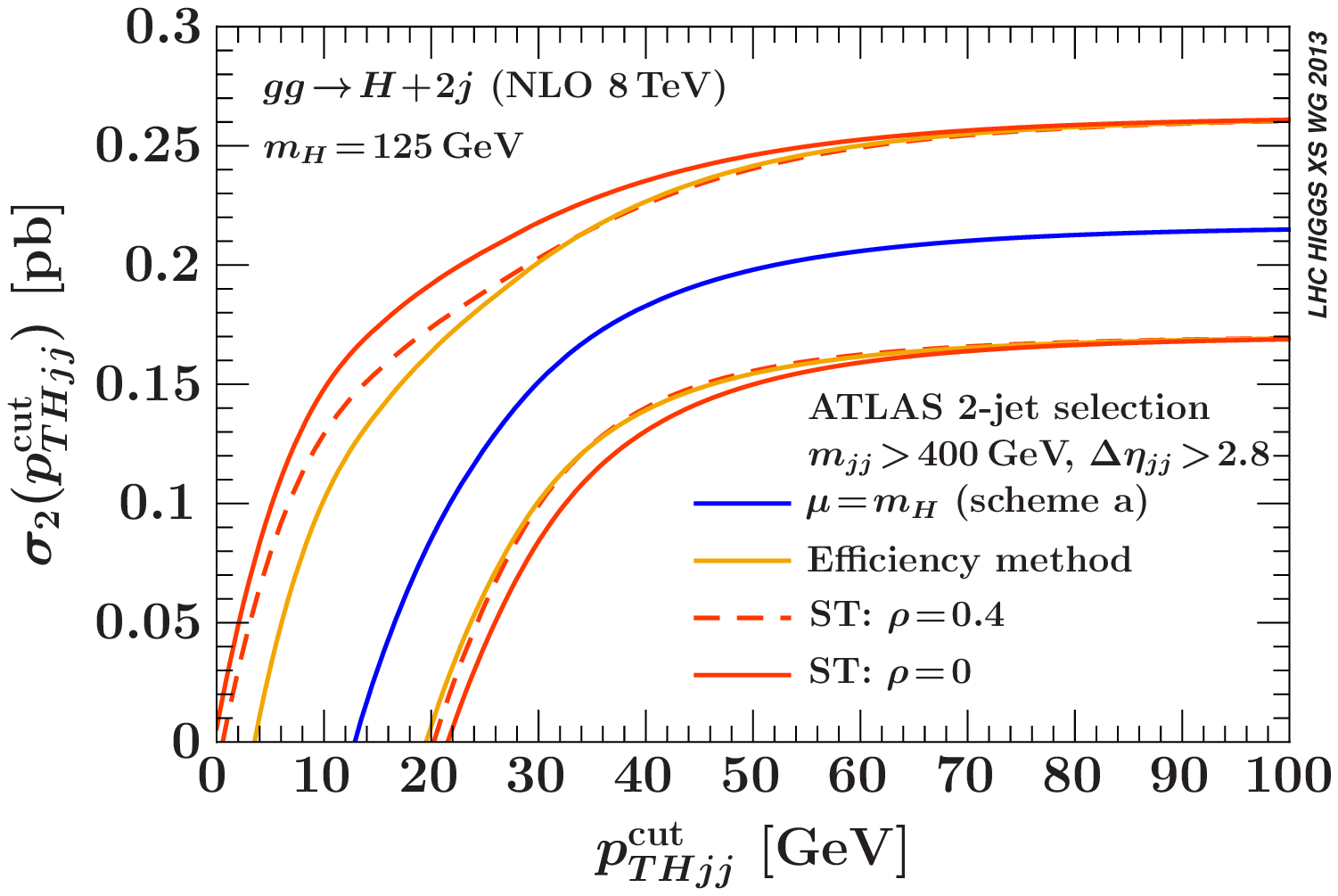}%
\hfill%
\includegraphics[width=0.5\columnwidth]{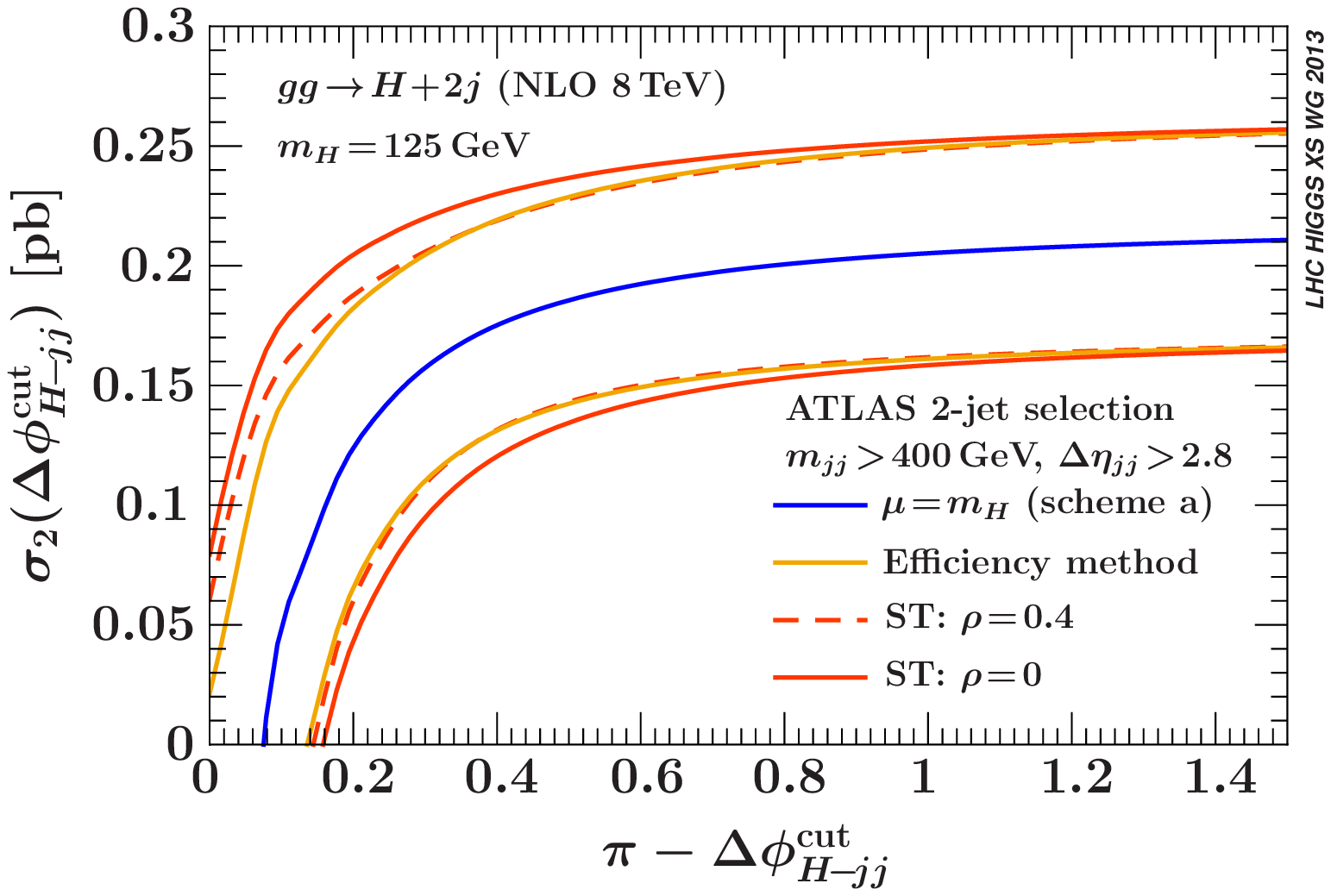}%
\caption{Comparison of the ST method with the efficiency method for $\pTHjj^\cut$ (left panel) and $\pi-\dphi^\cut$ (right panel) using the ATLAS VBF selection. The exclusive scale uncertainties from both methods are consistent with each other. The uncertainties from the efficiency method are very close to those from the ST method with $\rho = 0.4$.}
\label{fig:effST}
\end{figure*}

Another prescription to obtain fixed-order uncertainty estimates for exclusive jet cross section, which is based on using veto efficiencies, was applied in \Bref{Banfi:2012yh} to the $0$-jet case at NNLO. We will refer to it as ``efficiency method''. In \Bref{AlcarazMaestre:2012vp} it was shown that for the case of $\PH +0$ jets at NNLO the ST and efficiency methods yield very similar uncertainties, providing a good cross check on both methods. The starting point in the efficiency method is to write the exclusive jet cross section in terms of the corresponding inclusive jet cross section times the corresponding exclusive efficiency, i.e., applied to the $N=2$ case, one has
%%%
\begin{align}
\sigma_2 &= \sigma_{\geq 2} \Bigl(1 - \frac{\sigma_{\geq 3}}{\sigma_{\geq 2}} \Bigr)
\equiv \sigma_{\geq 2} \times \epsilon_2
\,,\qquad
\sigma_{\geq 3} = \sigma_{\geq 2}\, (1 - \epsilon_2)
\,.\end{align}
%%%
The basic assumption made in \Bref{Banfi:2012yh} is to treat the perturbative uncertainties in $\sigma_{\geq 2}$ and $\epsilon_2$ as uncorrelated. Since the 2-jet efficiency $\epsilon_2 = 1 - \sigma_{\geq 3}/\sigma_{\geq 2}$ is still an exclusive quantity, similar cancellations between the two perturbative series for $\sigma_{\geq 2}$ and $\sigma_{\geq 3}$ can happen in their ratio rather than in their difference, so the direct scale variation for $\epsilon_2$ might not provide a reliable uncertainty estimate. To circumvent this, the perturbative uncertainty in $\epsilon_2$ is instead estimated by using three different schemes for how to write the perturbative result for $\epsilon_2$, which differ by uncontrolled higher-order terms in $\alphas$, as follows:%
\footnote{As originally motivated in \Bref{Banfi:2012yh} in scheme (c) one strictly re-expands the ratio of cross section to a given order in $\alphas$, which at NLO yields the same result as scheme (b). To produce another expression with differing higher order terms, the analog to scheme (c) we use here is to keep the $\orderx{\alphas^2}$ cross term that comes from expanding the denominator.}
%%%
\begin{align}
&\text{Scheme (a):} &
\epsilon_2^{(a)}
&= 1 - \frac{\sigma_{\ge3}}{\sigma_{\ge 2}}
= 1 - \frac{\alphas \sigma^{(0)}_{\ge 3}}{\sigma^{(0)}_{\ge 2} + \alphas\sigma^{(1)}_{\ge2}} + \orderx{\alphas^2}
\,,\nn\\
&\text{Scheme (b):} &
\epsilon_2^{(b)} &= 1- \alphas\,\frac{\sigma^{(0)}_{\ge 3}}{\sigma^{(0)}_{\ge 2}} + \orderx{\alphas^2}
\,,\nn\\
&\text{Scheme (c):} &
\epsilon_2^{(c)}
&= 1 - \alphas\, \frac{\sigma^{(0)}_{\ge 3}}{\sigma^{(0)}_{\ge 2}}
\biggl(1 - \alphas\, \frac{\sigma_{\geq 2}^{(1)}}{\sigma^{(0)}_{\geq 2}} \biggr) + \orderx{\alphas^2}
\,,\end{align}
%%%
where we expanded the inclusive 2-jet and 3-jet cross sections as
%%%
\begin{align}
\sigma_{\ge2} &= \alphas^2 \bigl[\sigma_{\ge2}^{(0)} + \alphas\, \sigma_{\ge2}^{(1)} + \alphas^2\, \sigma_{\ge2}^{(2)} + \orderx{\alphas^3} \bigr]
\,,\nn\\
\sigma_{\ge3} &= \alphas^2 \bigl[\alphas \sigma_{\ge3}^{(0)} + \alphas^2\, \sigma_{\ge3}^{(1)}  + \orderx{\alphas^3} \bigr]
\,.\end{align}
%%%

In \refF{fig:effST} we compare the results of the ST and efficiency methods for the exclusive 2-jet cross section $\sigma_2$ using cuts on $\pTHjj$ and $\dphi$ for the ATLAS VBF selection (see \refS{sec:jets-H2jets-cutbased} below). The blue solid curve shows the default NLO central value (equivalent to the scheme (a) central value). The light orange solid curves are the uncertainties obtained in the efficiency method. They are obtained by combining in quadrature the inclusive scale uncertainties $\Delta_{\geq2}^\mu$ with the direct scale variations in $\epsilon_2^{(a)}$ treating both as uncorrelated. (Here, the values for $\epsilon_2$ in schemes (b) and (c) lie within the scheme (a) scale variations, so we use the latter as uncertainty estimate.) The dark orange solid curves show the ST uncertainties for $\rho = 0$, which are slightly larger. The dashed lines show the ST uncertainties for $\rho = 0.4$, which are in close agreement with the efficiency method. This can be understood by noting that by varying the scales in $\epsilon_2$ one effectively varies the scales correlated in $\sigma_{\geq2}$ and $\sigma_{\geq 3}$, which has the effect of reintroducing a certain amount of correlation between $\Delta_{\geq 2}^\mu$ and $\Delta_{\geq 3}^\mu$ when computing $\sigma_2$, which is also what a nonzero value of $\rho$ does. Overall, the good consistency between the various methods gives us confidence in the reliability of our uncertainty estimates.

%~~~~~~~~~~~~~~~~~~~~~~~~~~~~~~~~~~~~~~~~~~~~~~~~~~~~~~~~~~~~~~~~~~~~~~~~~~~~~~~
\subsubsection{Cut-based analyses}
\label{sec:jets-H2jets-cutbased}
%~~~~~~~~~~~~~~~~~~~~~~~~~~~~~~~~~~~~~~~~~~~~~~~~~~~~~~~~~~~~~~~~~~~~~~~~~~~~~~~

We now study the uncertainties in the exclusive $H + 2$ jet cross section as a function of two kinematic variables, $\pTHjj$ and $\dphi$. Here, $\pTHjj$ is the magnitude of the total transverse momentum of the Higgs-dijet system,
%%%
\begin{equation}
\pTHjj = \lvert \vec p_{Tj1} + \vec p_{Tj2} + \vec p_{TH} \rvert
\,.\end{equation}
%%%
At Born level, $\pTHjj = 0$ and so applying a cut $\pTHjj < \pTHjj^\cut$ restricts the phase space to the exclusive 2-jet region,
and induces Sudakov logarithms of the form $L =\ln(\pTHjj^\cut/\MH)$ in the perturbative series of $\sigma_2$ and $\sigma_{\geq 3}$.
At NLO $\pTHjj$ is equivalent to the $\pT$ of the third jet, so it is a useful reference variable for a $\pT$-veto on additional emissions, such as the central jet vetoes applied in the $\PH\to \PW\PW$ and $\PH\to\tau\tau$ VBF analyses.
The VBF category in the $\PH \rightarrow \PGg\PGg$ analyses by ATLAS and CMS includes a cut $\dphi > \dphi^\cut$. Taking the beam direction along the $z$-axes, $\dphi$ is defined as
%%%
\begin{equation}
\cos \dphi = \frac{(\vec p_{Tj1} + \vec p_{Tj2}) \cdot \vec p_{TH}}{\lvert\vec p_{Tj1}+ \vec p_{Tj2}\rvert \lvert\vec p_{TH}\rvert}
\,,\end{equation}
%%%
with the Higgs momentum given by the total momentum of the diphoton system. Events with only two jets always have $\dphi \approx \pi$, so the constraint $\dphi > \dphi^\cut$ forces the kinematics into the exclusive 2-jet region and restricts additional emissions. Hence, it behaves similar to $\pTHjj^\cut$ and for $\pi - \dphi^\cut\!\to 0$ induces large logarithms in the perturbative series.
The exclusive 2-jet bins defined in terms of these variables are written as
%%%
\begin{align}
\sigma_2(\pTHjj < \pTHjj^\cut) &= \sigma_{\geq 2} - \sigma_{\geq 3}(\pTHjj > \pTHjj^\cut)
\nn\\
\sigma_2(\dphi > \dphi^\cut) &= \sigma_{\geq 2} - \sigma_{\geq 3}(\dphi < \dphi^\cut)
\,,\end{align}
%%%
where in all cross sections the remaining VBF selection cuts in \refT{tab:VBFcuts} are applied (excluding the cut on $\dphi$ in case of $\pTHjj$).

\begin{table}[t!]
\caption{VBF selection cuts we use, corresponding to the $\PH\to\PGg\PGg$ analyses by
ATLAS~\cite{ATLAS-CONF-2012-091, ATLAS-CONF-2012-168} and CMS~\cite{CMS-PAS-HIG-12-015} (as of last year).
The cut on $\dphi$ in the last row is treated special as an exclusive binning cut.}
\label{tab:VBFcuts}
\centering
\begin{tabular}{lcc}
\hline
& ATLAS  & CMS tight  \\
\hline
& anti-$k_{\mathrm T}$ $R = 0.4$  & anti-$k_{\mathrm T}$ $R = 0.5$
\\
2-jet selection & $p_{Tj}\!>\! 25\UGeV$ for $\lvert\eta_{j}\rvert\! <\! 2.5$
& $p_{Tj}\!>\! 30\UGeV$, $\lvert\eta_{j}\rvert \!<\! 4.7 $
\\
& $p_{Tj}\!>\! 30\UGeV$ for $2.5\! <\! \lvert\eta_{j}\rvert \!<\! 4.5 $ &
\\\hline
$\Delta\eta_{jj} = \lvert\eta_{j1} - \eta_{j2}\rvert$ & $ > 2.8$   & $ > 3.0 $
\\
$m_{jj}$ &  $> 400\UGeV$ & $ >500\UGeV$
\\
$|\eta_{H} - (\eta_{j1} + \eta_{j2})/2|$ & -  & $<2.5 $ \\

$\Delta\phi_{H\!-\!jj}$ & $>  2.6 $  & $>  2.6 $\\
\hline
\end{tabular}
\end{table}

\begin{figure*}[t!]
\subfigure[ATLAS VBF selection]
{\includegraphics[width=0.515\columnwidth]{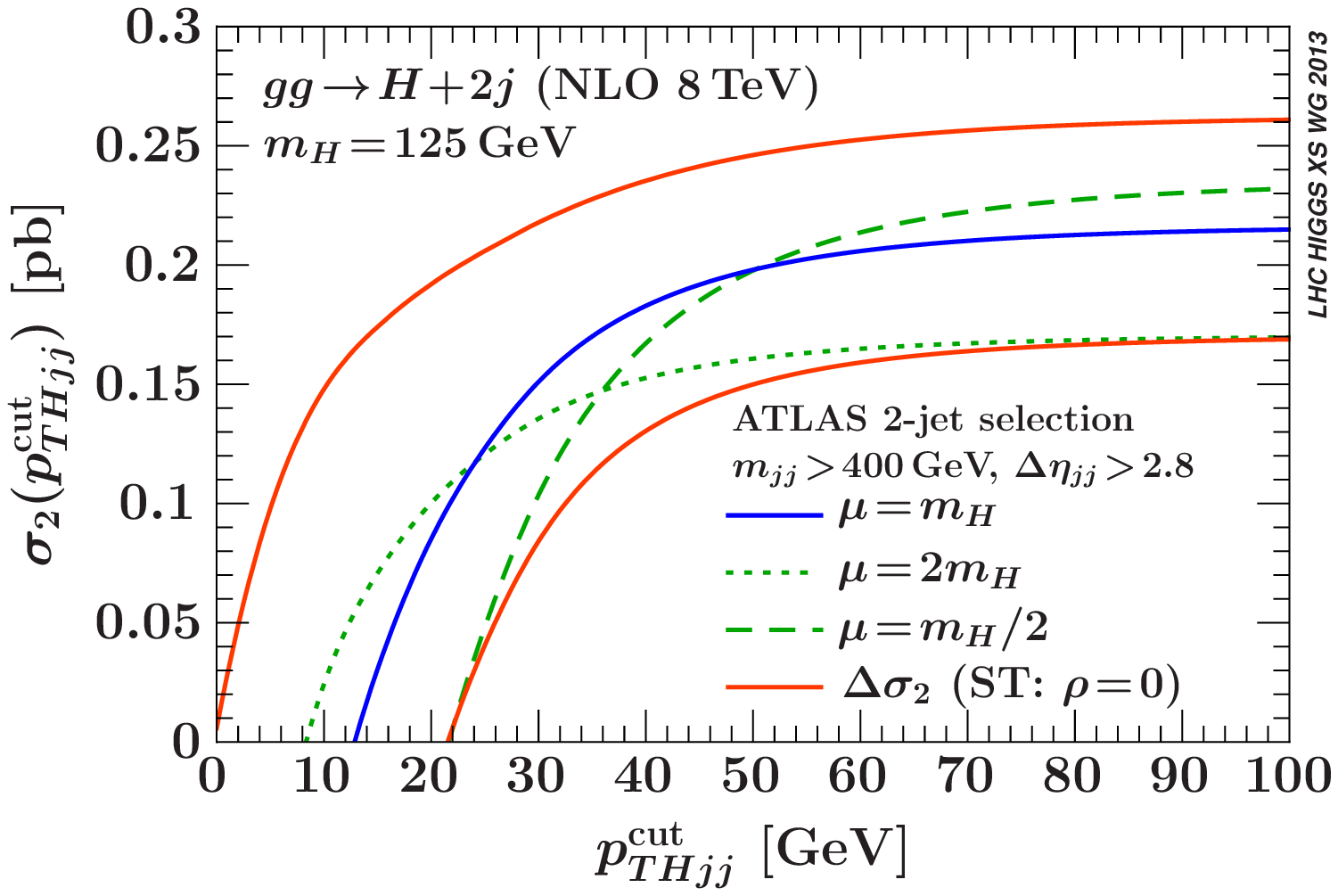}\label{fig:atlaspt}}%
\hfill%
\subfigure[ATLAS VBF selection]
{\includegraphics[width=0.5\columnwidth]{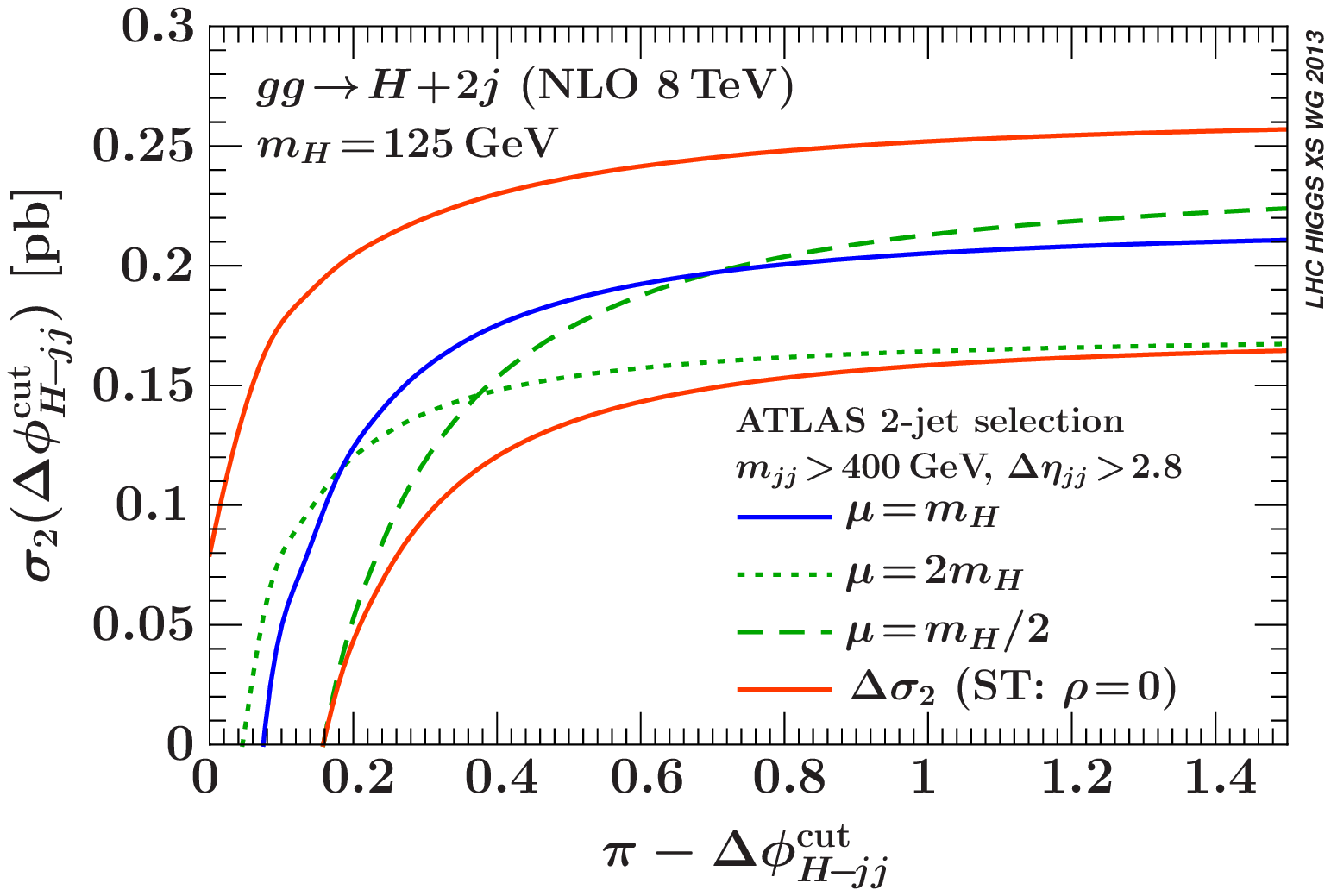}\label{fig:atlasphi}}%
\\
\subfigure[CMS tight VBF selection]
{\includegraphics[width=0.515\columnwidth]{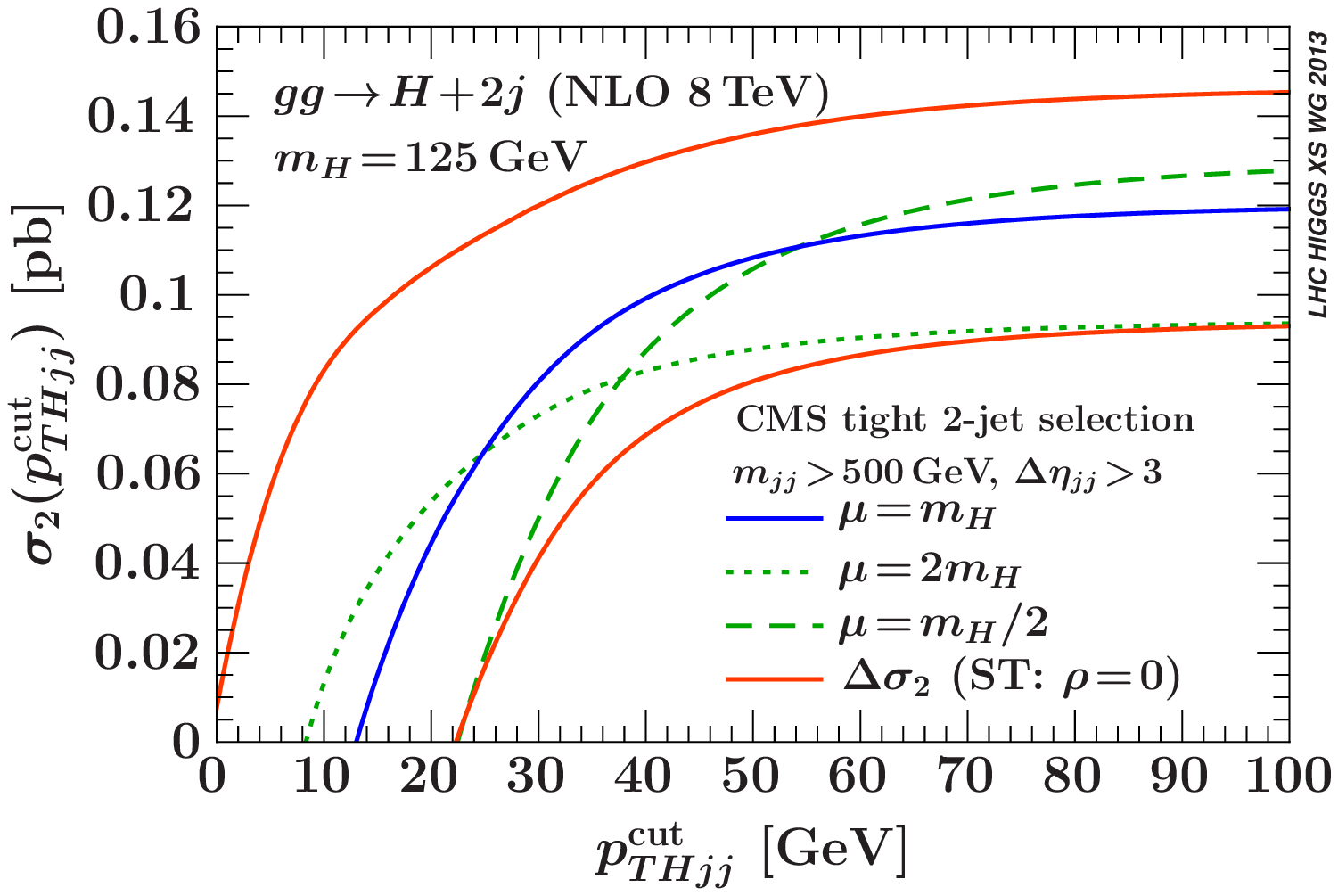}\label{fig:cmstightpt}}%
\hfill%
\subfigure[CMS tight VBF selection]
{\includegraphics[width=0.5\columnwidth]{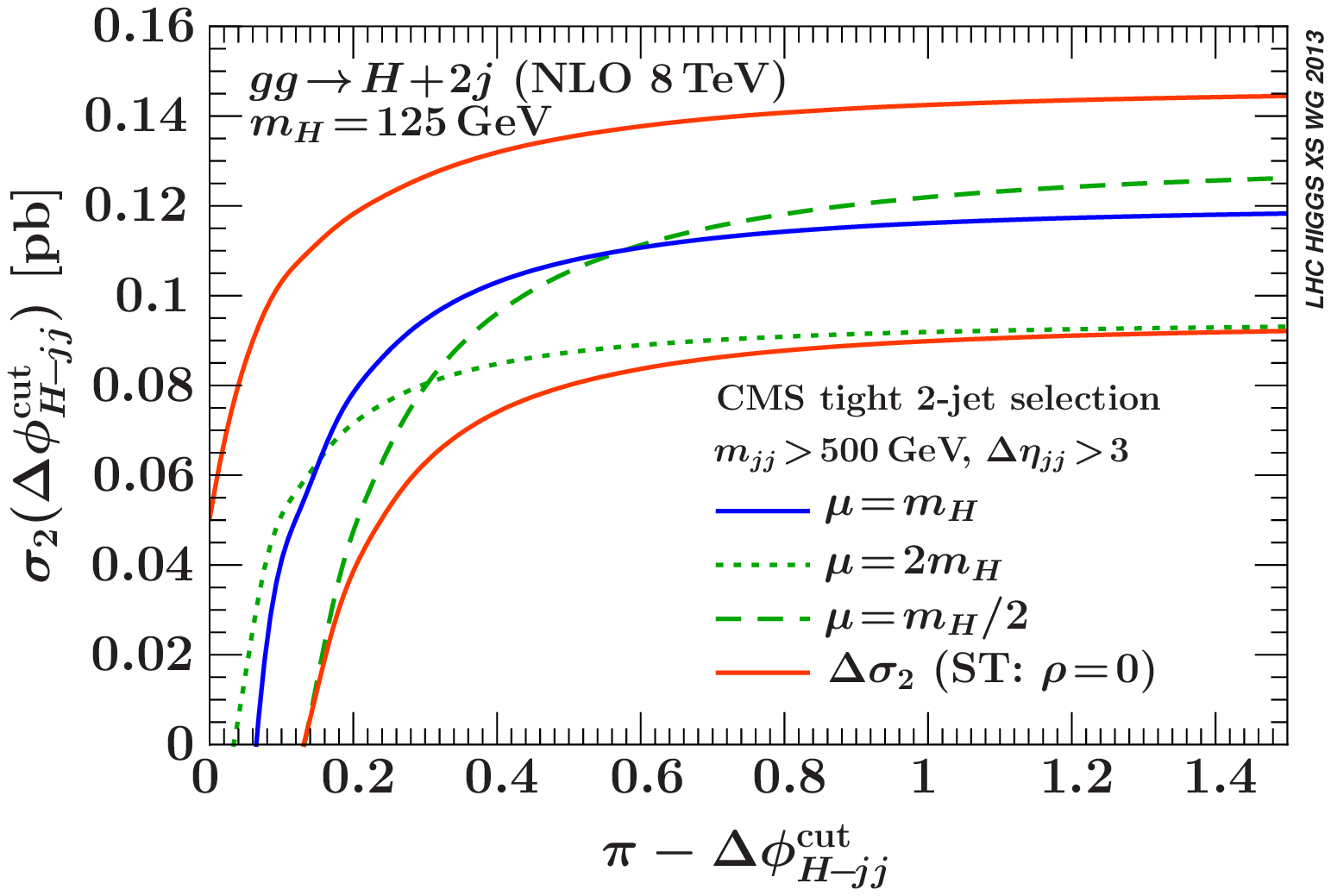}\label{fig:cmstightphi}}%
\caption{Exclusive $\Pp\Pp \to \PH + 2$ jet cross section via ggF at NLO for as function of $\pTHjj^\cut$ (left panels) and $\pi-\dphi^\cut$ (right panels) for both ATLAS and CMS VBF selections.}
\label{fig:ptphiresults}
\end{figure*}

In \refF{fig:ptphiresults} we plot the result for the exclusive 2-jet cross section as a function of $\pTHjj^\cut$ and $\dphi^\cut$ for the ATLAS and CMS tight VBF selections. In these plots, the blue central line shows the central-value prediction obtained from $\mu = \MH$, while the orange solid lines show our uncertainty estimate. For reference, the green dashed and dotted curves show the direct scale variation for $\mu =\MH/2$ and $\mu=2\MH$, respectively.
For large values of $\pTHjj^\cut$ or $\pi -\dphi^\cut$, the cross section $\sigma_{\ge 3}$ that is cut away becomes small and so the effect of $\Delta^\cut$ is negligible. In this limit the uncertainties reproduce those in the inclusive 2-jet cross section. On the other hand, in the transition region, once the exclusive cut starts to impact the cross section, the direct scale variations cannot be used any longer to estimate uncertainties, which is exhibited by the crossing of the lines.
The ST method takes into account the migration uncertainty which becomes important in this region as the exclusive cut gets tighter, thus providing robust uncertainties for all values of $\pTHjj^\cut$ or $\dphi^\cut$.

\begin{table*}[t!]
\caption{Perturbative uncertainties at NLO in the exclusive $\Pp\Pp \to \PH + 2$ jet cross section via gluon fusion for cuts on $\pTHjj$ and $\dphi$ for the ATLAS and CMS tight VBF selections given in \refT{tab:VBFcuts}. }
\label{tab:numbers}
\centering
\begin{tabular}{lcccc}
\hline
Selection & $\sigma\, [\Upb]$  & \multicolumn{2}{c}{ {Direct scale variation}} & {Combined incl. uncertainties} \\ [0.5ex]
 & $\mu=\MH$ & $\mu=2\MH$ & $\mu=\MH/2$ &  ST ($\rho = 0$) \\
\hline
\multicolumn{5}{c}{\textbf{ATLAS}}\\
\hline
$\sigma_{\ge 2}$ & $0.21$ & & & $\pm 21\%$   \\
$\sigma_2(\pTHjj < 30\UGeV)$ & $0.15$ & $-8 \%$ & $-29 \% $ & $\pm 44 \%$  \\
$\sigma_2(\dphi > 2.6)$ & $0.19$ & $-17 \%$ & $-4 \% $ & $\pm 26\% $ \\
\hline
\multicolumn{5}{c}{\textbf{CMS tight}}\\
\hline
$\sigma_{\ge 2} $ & $0.12$ & & & $\pm 21 \%$ \\
$\sigma_2(\pTHjj < 30\UGeV)$ & $0.08$ & $-8 \%$ & $-35 \%$  & $ \pm 49 \%$  \\
$\sigma_2(\dphi > 2.6)$ & $0.10$ & $-19\%$ & $-1 \%$ &  $\pm 26 \% $ \\
\hline 
\end{tabular}
\end{table*}

In \refT{tab:numbers} we quote the results for the cross sections and their percentage uncertainties for a few specific cuts. For $\dphi$ we use the experimental value $\dphi > 2.6$. Compared to the $21\%$ in the inclusive 2-jet cross section with VBF cuts ($\sigma_{\geq 2}$), we see a moderate increase in the uncertainty in $\sigma_2(\dphi > 2.6)$ to $26\%$ for ATLAS and CMS tight. For $\pTHjj$ we use a representative value of $\pTHjj < 30\UGeV$, for which the uncertainties increase substantially to $44\%$ and $49\%$ for ATLAS and CMS tight respectively. Note that for a fixed exclusive cut the uncertainties increase somewhat with a tighter VBF selection, which is expected.

\begin{figure*}[t!]
\subfigure[ATLAS VBF selection]{\includegraphics[width=0.515\columnwidth]{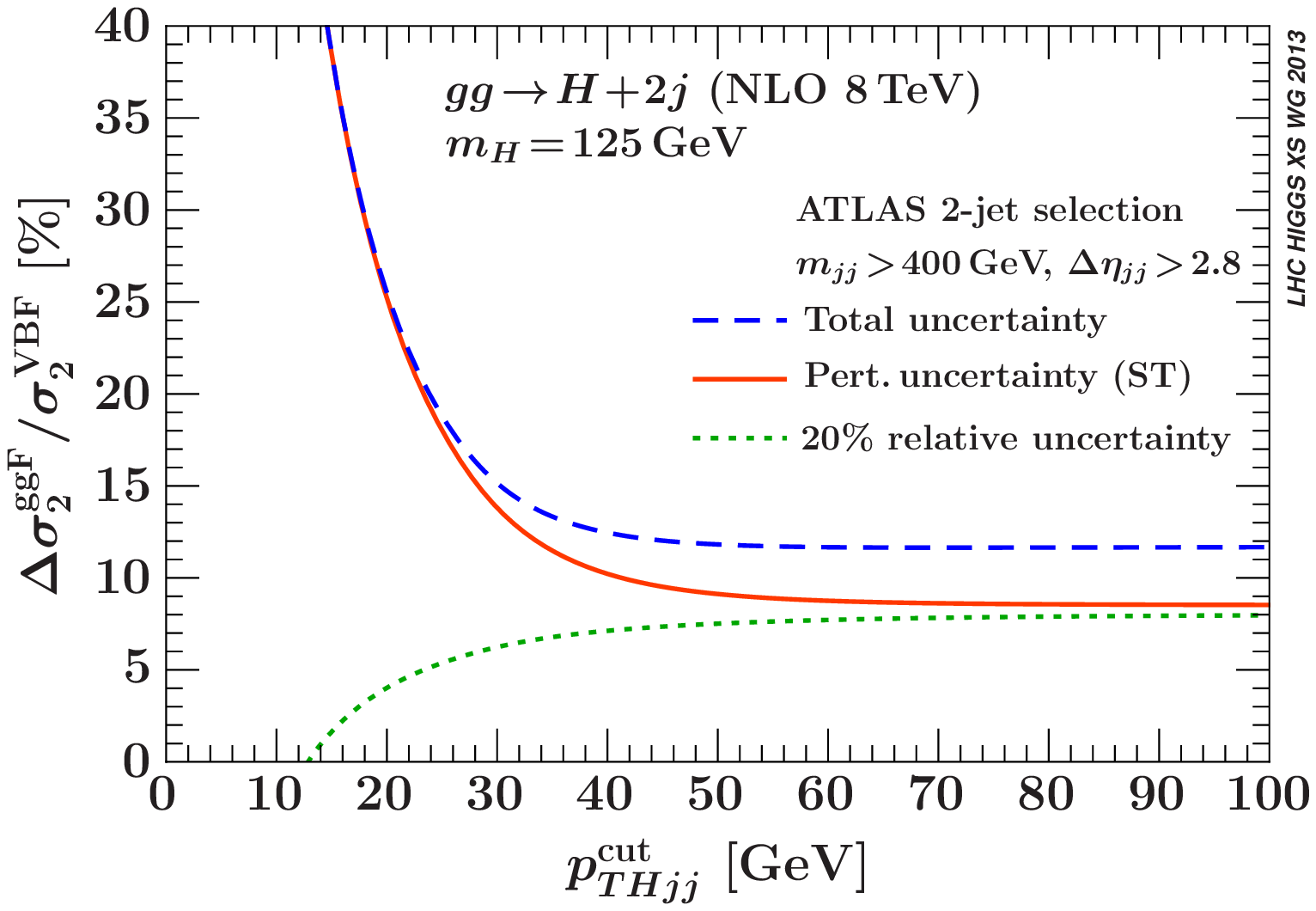}\label{fig:atunc}}%
\hfill%
\subfigure[ATLAS VBF selection]{\includegraphics[width=0.5\columnwidth]{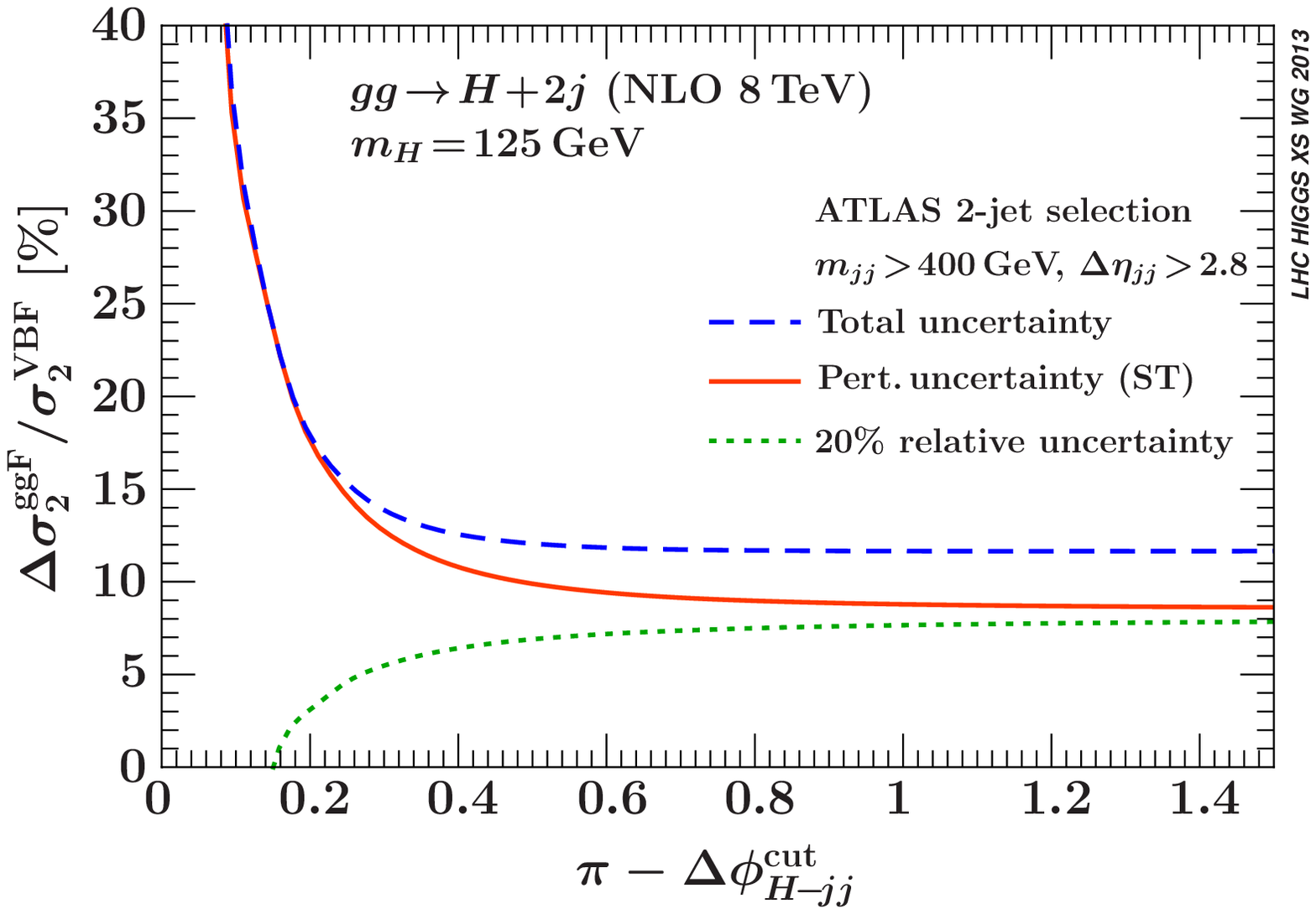}\label{fig:atuncphi}}%
\caption{Theoretical uncertainties of the ggF contribution relative to the VBF cross section as function of $\pTHjj^\cut$ (left) and $\dphi^\cut$ (right) for the ATLAS VBF selection.}
\label{fig:ggFVBFunc}
\end{figure*}

An important source of theoretical uncertainty in the extraction of the VBF signal is the large perturbative uncertainty in the ggF contribution. After subtracting the non-Higgs backgrounds (which are of course another source of uncertainty), the measured cross section for Higgs production after implementing the VBF selection is given by
%%%
\begin{equation}
\sigma_2^\mathrm{measured}(\dphi^\cut) = \sigma_2^\mathrm{VBF}(\dphi^\cut) + \sigma_2^\mathrm{ggF}(\dphi^\cut)
\,.\end{equation}
%%%
For the purpose of extracting the VBF cross section one effectively subtracts the theory prediction for $\sigma_2^\mathrm{ggF}(\dphi^\cut)$ from $\sigma_2^\mathrm{measured}(\dphi^\cut)$, and therefore, the relevant figure of merit is \\ $\Delta\sigma_2^\mathrm{ggF}(\dphi^\cut)/\sigma_2^\mathrm{VBF}(\dphi^\cut)$, i.e., the theory uncertainty in $\sigma_2^\mathrm{ggF}$ measured relative to the expected VBF cross section, $\sigma_2^\mathrm{VBF}$.
In \refF{fig:ggFVBFunc} we show this quantity over a range of $\pTHjj^\cut$ and $\dphi^\cut$ for the ATLAS VBF selection. The results for the CMS selection are very similar. Here, the solid orange curve shows our results for the NLO perturbative uncertainties (corresponding to the orange lines in \refF{fig:ptphiresults}). For comparison, the green dotted curve shows a fixed $20\%$ uncertainty in the ggF cross section, i.e., taking $\Delta\sigma_2^\mathrm{ggF} = 0.2\, \sigma_2^\mathrm{ggF}$, which for example could be due to PDF and $\alphas$ parametric uncertainties. In the dashed blue lines, both uncertainty contributions are added in quadrature. In the region of low $\pTHjj^\cut$ or $\pi-\dphi^\cut$, the relative uncertainty coming from the ggF contribution quickly increases below $\pTHjj\lesssim 30\UGeV$ and $\pi-\dphi \lesssim 0.4$. Hence, care must be taken when implementing and optimizing either indirect restrictions on additional radiation, like $\dphi$, or explicit $\pT$-vetoes like $\pTHjj$ or a central jet veto, and also in applying more general cuts which restrict to the exclusive 2-jet region as in the case of MVAs.

%~~~~~~~~~~~~~~~~~~~~~~~~~~~~~~~~~~~~~~~~~~~~~~~~~~~~~~~~~~~~~~~~~~~~~~~~~~~~~~~
\subsubsection{Generalization to arbitrary number of cuts}
\label{sec:jets-H2jets-GEN}
%~~~~~~~~~~~~~~~~~~~~~~~~~~~~~~~~~~~~~~~~~~~~~~~~~~~~~~~~~~~~~~~~~~~~~~~~~~~~~~~

The formalism of \eqn{pcut} can be extended to further divide $\sigma_{\geq N+1}$ into an arbitrary number of bins,
%%%
\begin{align} \label{eq:pcut_gen}
\sigma_{\geq N} &= \int_0^{p^{\cut \,1}}\!\df p_{N+1}\, \frac{\df\sigma_{\geq N}}{\df p_{N+1}} + \int_{p^{\cut \,1}}^{p^{\cut \,2}} \!\df p_{N+1}\,
\frac{\df\sigma_{\geq N}}{\df p_{N+1}} + \dots + \int_{p^{\cut \,  n-1}}^{p^{\cut \,  n}} \!\df p_{N+1}\,
\frac{\df\sigma_{\geq N}}{\df p_{N+1}}
\nn\\
&\equiv  \sigma_N(p^{\cut \, 1}) + \sigma_{\geq N+1}(p^{\cut \, 1}, p^{\cut \, 2}) + \dots + \sigma_{\geq N+1}(p^{\cut \, n-1}, p^{\cut \, n})
\,.\end{align}
%%%
This splitting divides the inclusive $N$-jet cross section, $\sigma_{\geq N}$, into $n$ bins, whose uncertainties and  correlations can be described by a symmetric $n\times n$ covariance matrix with $n(n+1)/2$ independent parameters.  To construct this covariance matrix we use the boundary conditions that the inclusive cross sections $\sigma_{\geq N}$ and $\sigma_{\geq N+1}(p^{\mathrm{cut}},\infty)$ are uncorrelated, which implements the ST procedure for a given $p^{\cut}$. This is not sufficient to determine the complete matrix. For the remaining entries, a simple linear correlation model is used, where the correlation $\kappa_{ij}$ between $\sigma_{\geq N+1}(p^{\cut\,i},\infty)$ and $\sigma_{\geq N+1}(p^{\cut\,j},\infty)$ is given by
\begin{align}
  \kappa_{ij} = 1 - (1-\kappa) \frac{\lvert p^{\cut\,i} - p^{\cut\,j} \rvert}{ p^{\cut\,n} - p^{\cut\,1}}.
\end{align}
The  parameter $\kappa$ determines the strength of the correlations between the inclusive $N+1$-jet cross sections for different $p^\cut$.
The dependence on this underlying correlation model is tested below by using the three different values $\kappa = \{50\%, 90\%, 99\%\}$.
As we will see, the obtained uncertainty estimates are very insensitive to the precise choice of $\kappa$.

In the following, the above procedure is demonstrated using $\dphi$ as the underlying IR sensitive binning variable.
Here, the first bin, which encloses the IR sensitive region, must be chosen large enough to ensure that \textsc{MCFM} can still be used to estimate its uncertainties using the ST procedure.
Based on \refF{fig:atlasphi} and \refF{fig:cmstightphi} we choose the first bin
as $\left(\pi - \dphi\right) \in [0,0.2]$.

\begin{figure*}
\includegraphics[width=0.49\columnwidth]{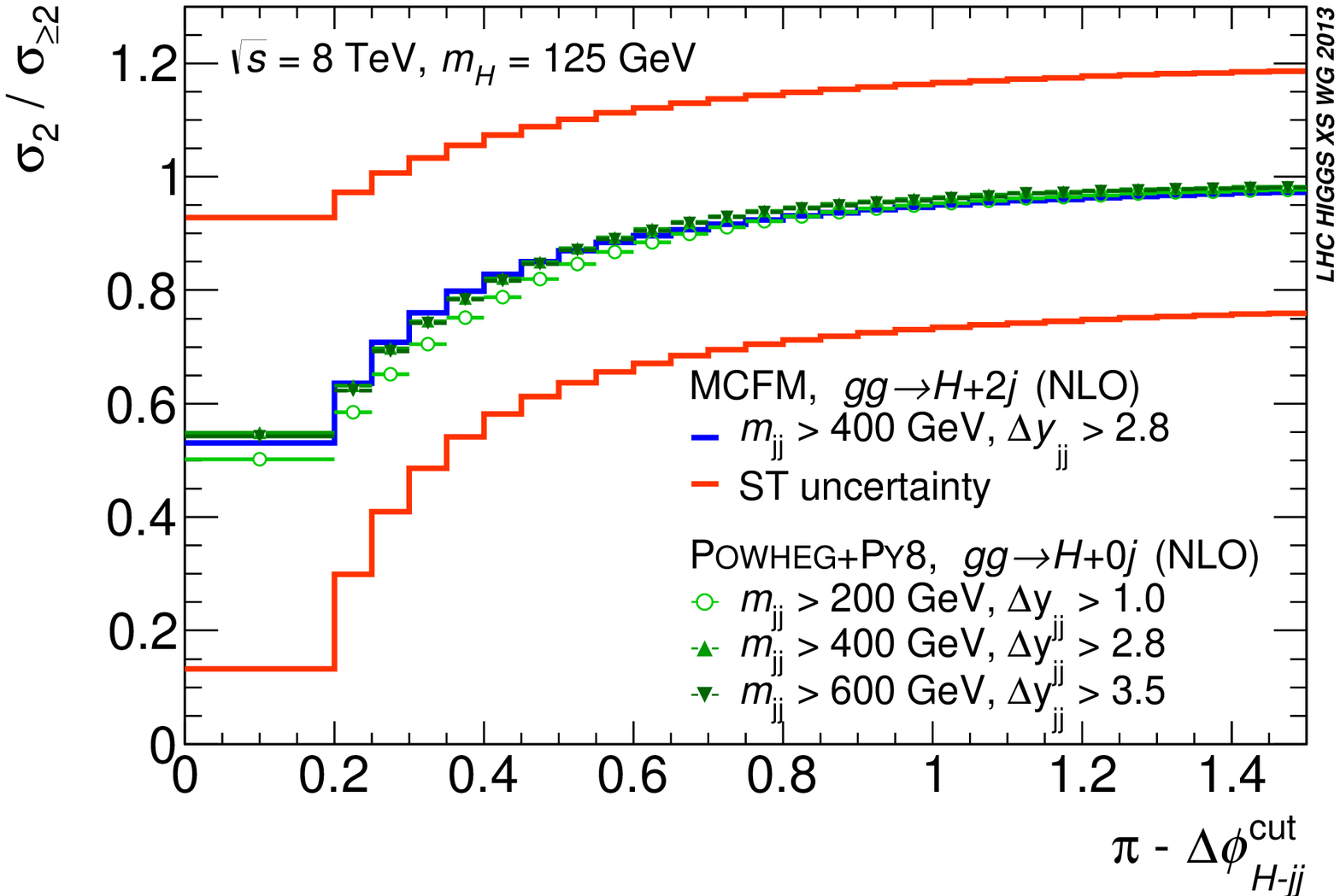}%
\hfill%
\includegraphics[width=0.49\columnwidth]{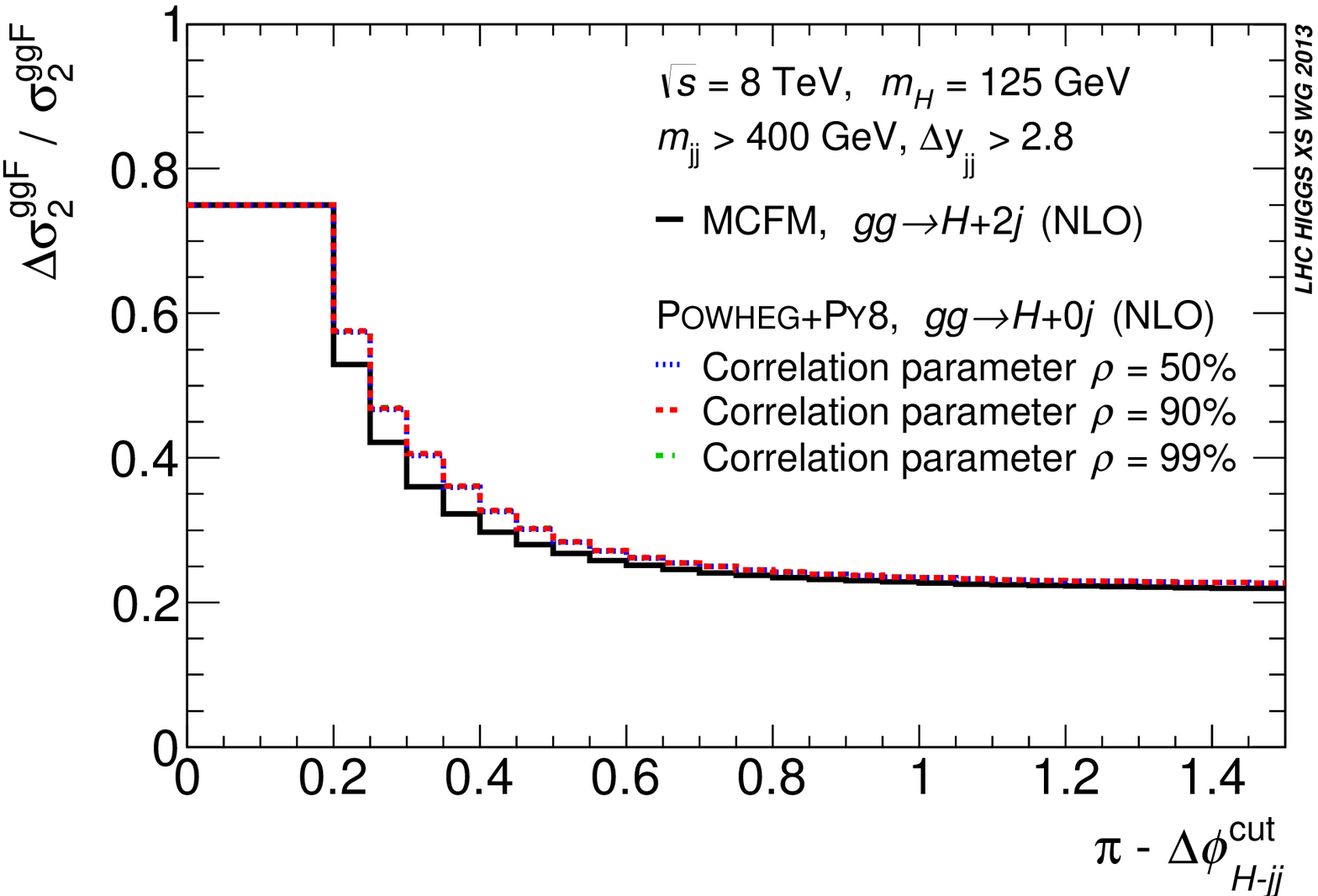}%
\caption{Comparison of the exclusive $\Pp\Pp \to \PH + 2$ jet cross section via ggF from \textsc{MCFM} and \textsc{Powheg}+\textsc{Pythia8} $\PH + 0$ jets as a function of $\pi - \dphi^\cut$ (where the cut value is given by the upper bin edges).
Left: The normalized cumulant $\sigma_2/\sigma_{\geq 2}$. The histograms show the fixed-order results from \textsc{MCFM}, corresponding to \refF{fig:atlasphi}. The data points show \textsc{Powheg}+\textsc{Pythia8} results for different VBF selections, which only depend weakly on the precise VBF selection and agree well with \textsc{MCFM}. Right: The relative ST uncertainties from \textsc{MCFM} (black solid histogram) are compared to the resulting uncertainties (dotted histograms) after propagation to the \textsc{Powheg}+\textsc{Pythia8} prediction using \eqn{eq:cum_uncert}. The resulting uncertainties closely agree with the \textsc{MCFM} input uncertainties and do not depend on the correlation model assumed in the propagation.
} \label{fig:MCFM_v_POWHEG}
\end{figure*}

For their most recent results, ATLAS and CMS use \textsc{Powheg} $\Pg\Pg \to \PH +0$ jets at NLO \cite{Nason:2004rx,Frixione:2007vw,Alioli:2010xd} to model the hard scattering process, interfaced with \textsc{Pythia8}~\cite{Sjostrand:2007gs} for modeling of underlying event, parton showering, and hadronization. In the left panel of \refF{fig:MCFM_v_POWHEG} we compare the normalized cumulative cross section for different values of $\dphi^\cut$
between \textsc{Powheg}+\textsc{Pythia8} and \textsc{MCFM}. For both generators, the anti-$k_{\mathrm T}$ algorithm with $R=0.4$ is used to reconstruct the jets, excluding the Higgs decay products, and a typical VBF phase space selection is applied. The cumulant shapes are in good agreement, also when the \textsc{Powheg+Pythia8} VBF selection is varied. When applying an exclusive 2-jet selection based on $\dphi$, the uncertainty of the event yield $N_2$ from \textsc{Powheg}+\textsc{Pythia8} can be estimated from
%%%
\begin{align} \label{eq:cum_uncert}
  \left( \Delta N_2 \right)^2= \sum_{i,j} \widehat C_{ij} \, n_i \, n_j \,, \qquad \widehat C_{ij} = \frac{1}{\sigma_i \sigma_j} C_{ij}
\, , \qquad
N_2 = \sum_{i}{n_i}
\,,\end{align}
%%%
where $C_{ij}$ denotes the covariance matrix, $\sigma_{i}$ the predicted cross section in the interval of the $i^{\text{th}}$ bin,
$n_i$ denotes the event yield of the Monte Carlo prediction in the $i^{\text{th}}$ bin, and the sum runs over all bins that define the exclusive 2-jet phase space one is interested in. To construct $C_{ij}$, we use the \textsc{MCFM} uncertainties of \refF{fig:atlasphi} as inputs to the procedure described above. The right panel of \refF{fig:MCFM_v_POWHEG} compares the relative uncertainties calculated from \eqn{eq:cum_uncert} for \textsc{Powheg}+\textsc{Pythia8} for different correlation models, showing a good agreement with the input \textsc{MCFM} uncertainties. Note in particular that the 2-jet inclusive cross section uncertainty is recovered when calculating the cumulant over the full range of $\dphi$. \eqn{eq:cum_uncert} is used in the following two sections to derive uncertainties for nonlinear cuts on $\dphi$ due to resolution effects and for selections based on a multivariate classifier.

%~~~~~~~~~~~~~~~~~~~~~~~~~~~~~~~~~~~~~~~~~~~~~~~~~~~~~~~~~~~~~~~~~~~~~~~~~~~~~~~
\subsubsection{Detector smearing}
\label{sec:jets-H2jets-RES}
%~~~~~~~~~~~~~~~~~~~~~~~~~~~~~~~~~~~~~~~~~~~~~~~~~~~~~~~~~~~~~~~~~~~~~~~~~~~~~~~

\begin{figure*}[t!]
  \begin{center}
    \includegraphics[width=0.5\columnwidth]{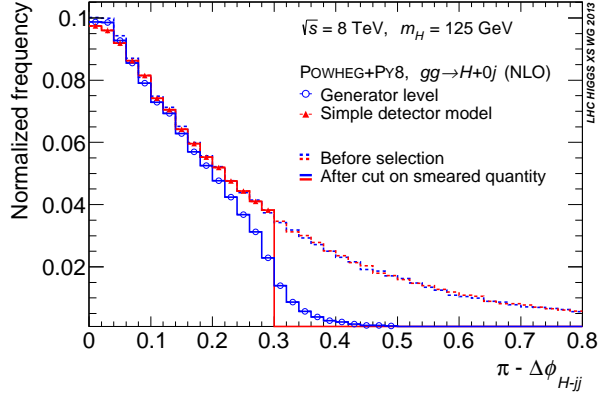}%
  \end{center}
  \caption{Illustration of the effect of finite reconstruction resolution on $\dphi$. The solid blue histogram shows the generator level $\dphi$ distribution after a cut on the reconstructed $\dphi^{{\mathrm{reco}}}$ show in red. The resolution results in a nonlinear selection cut for the generator level $\dphi$.}
\label{fig:res_dphi}
\end{figure*}

The photon and jet four momenta used in an actual physics measurement are degraded due to the energy and angular resolution smearing and reconstruction inefficiency of the detector. As a consequence, a cut applied on a reconstructed quantity in a physics analysis causes migration above and below the cut value when looking at the same variable calculated at the generator level. This is illustrated for a selection on $\dphi$ in \refF{fig:res_dphi}, where the energy resolution is modeled using a simple normal distributed resolution.\footnote{A scale factor applied to the four momenta is sampled from $\mathrm{Gaus}(1,\sigma)$ where $\sigma$ is $\pT$ and $\eta$ dependent with a typical value of 0.02 and 0.1 for photons and jets respectively.} For photons, the angular resolution and the reconstruction efficiency is also roughly modeled (depending on $\pT$ and $\eta$ with a typical value of 80\%). The cut on $\dphi^{\mathrm{reco}}$ results in a nonlinear selection on $\dphi$. The impact on the cross section uncertainty from this smearing is studied using \eqn{eq:cum_uncert} and is summarized in \refT{tab:TrueVsReco}. For a loose selection, the results are very similar, but as the cut is tightened the impact of the detector resolution gets more important and tends to result in slightly smaller uncertainties.

\begin{table}[t!]
  \caption{
    Relative perturbative uncertainties at NLO for $\Pg\Pg\to \PH + 2$ jets after
    applying a VBF selection defined by $m_{jj}>400\UGeV$, $\Delta\eta_{jj}>2.8$, and various different cuts on $\dphi$.
    This selection is applied separately at the generator (truth) level (for jets reconstructed from stable particles from the MC event record)
    and after applying a simple detector smearing. The uncertainty is calculated using generator level $\dphi$ and detector level $\dphi$ using the linear correlation model with $\kappa = 90\%$.
  }
  \label{tab:TrueVsReco}
  \centering
  \begin{tabular}{lccccc}
    \hline
    $\pi - \dphi^\cut$ & 0.25 & 0.3 & 0.4 & 0.5 & no cuts\\
    \hline
    Generator level & 57.1\%  & 46.2\% & 35.3\% & 29.4\% & 21.0\%\\
    Detector level  &  54.2\%  & 45.2\% & 34.8\% & 29.3\% & 21.0\%\\
    \hline
  \end{tabular}
\end{table}

%~~~~~~~~~~~~~~~~~~~~~~~~~~~~~~~~~~~~~~~~~~~~~~~~~~~~~~~~~~~~~~~~~~~~~~~~~~~~~~~
\subsubsection{Multivariate analyses}
\label{sec:jets-H2jets-MVA}
%~~~~~~~~~~~~~~~~~~~~~~~~~~~~~~~~~~~~~~~~~~~~~~~~~~~~~~~~~~~~~~~~~~~~~~~~~~~~~~~

In the context of a multivariate analysis, the effective cuts on $\dphi$ and $\pTHjj$ introduced by the nonlinear selection of phase space have to be studied carefully. In particular, if either of the variables are used directly as learning input for the multivariate classifier, one has to make sure that the final classification does not cut arbitrarily close into the infrared sensitive regions, i.e. $\dphi \to \pi$ and $\pTHjj \to 0$. This can be prevented by transforming either variable into an infrared safe form,
%%%%%%%
\begin{align}\label{infsafedphi}
 \dphi' = \left\{
\begin{array}{rl}
 \dphi & \,\, \text{if } \dphi < \dphi^\cut,\\
\dphi^\cut & \,\, \text{if } \dphi \geq \dphi^\cut,\\
\end{array} \right.
 \quad
\pTHjj' = \left\{
\begin{array}{rl}
 \pTHjj & \,\, \text{if } \pTHjj > \pTHjj^\cut,\\
\pTHjj^\cut & \,\, \text{if } \pTHjj \leq \pTHjj^\cut,\\
\end{array} \right.
\end{align}
%%%%%%%%
allowing the multivariate algorithm only to exploit the normalization difference in the infrared sensitive region of phase space.

The procedure of deriving the exclusive 2-jet cross section uncertainties is illustrated in the following using a multivariate selection based on a boosted decision tree trained using the software of \Bref{Hocker:2007ht}\footnote{The specific configuration used is: 1000 trees, a shrinkage factor of 0.1, a gradient bagging fraction of 0.5, and maximally five nodes, for more details and definitions see \Bref{Hocker:2007ht}.}.
The decision tree was trained to distinguish VBF like events in $\PH \to \PGg\PGg$ and to reject prompt diphoton background. As input for background, simulated prompt diphoton decays by \textsc{Sherpa} are used. The signal was simulated using \textsc{Powheg}+\textsc{Pythia8} for VBF and ggF decays, both simulated at NLO. To all samples resolution effects were added using the same simple normal resolution model as above. Six typical variables often used in VBF analyses were chosen to train the decision tree\footnote{The transverse projection of the Higgs $\pT$ on the axis orthogonal to the thrust axis defined by the two photons; the invariant mass of the leading dijet system; the difference in rapidity of the two leading jets, the rapidities of the two leading jets, and the infrared safe version of $\dphi$ defined in \eqn{infsafedphi} using a cutoff of $\dphi^\cut = 2.94$.}.

\begin{figure*}[t!]
\subfigure[Signal and background $\dphi$  ]
{\includegraphics[width=0.515\columnwidth]{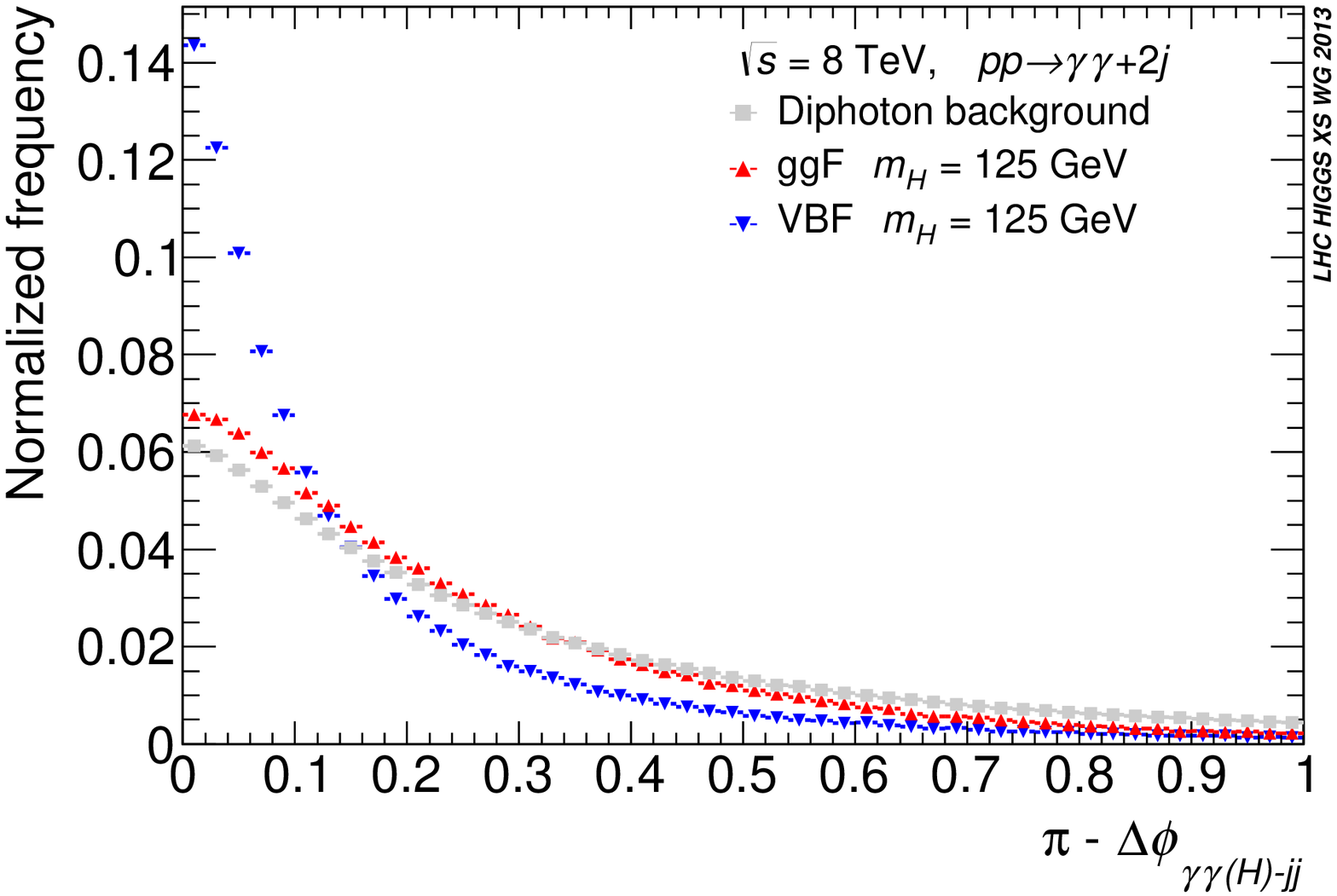}\label{fig:dphi_bkg_sig}}%
\hfill%
\subfigure[Signal and background $m_{\PGg\PGg}$]
{\includegraphics[width=0.5\columnwidth]{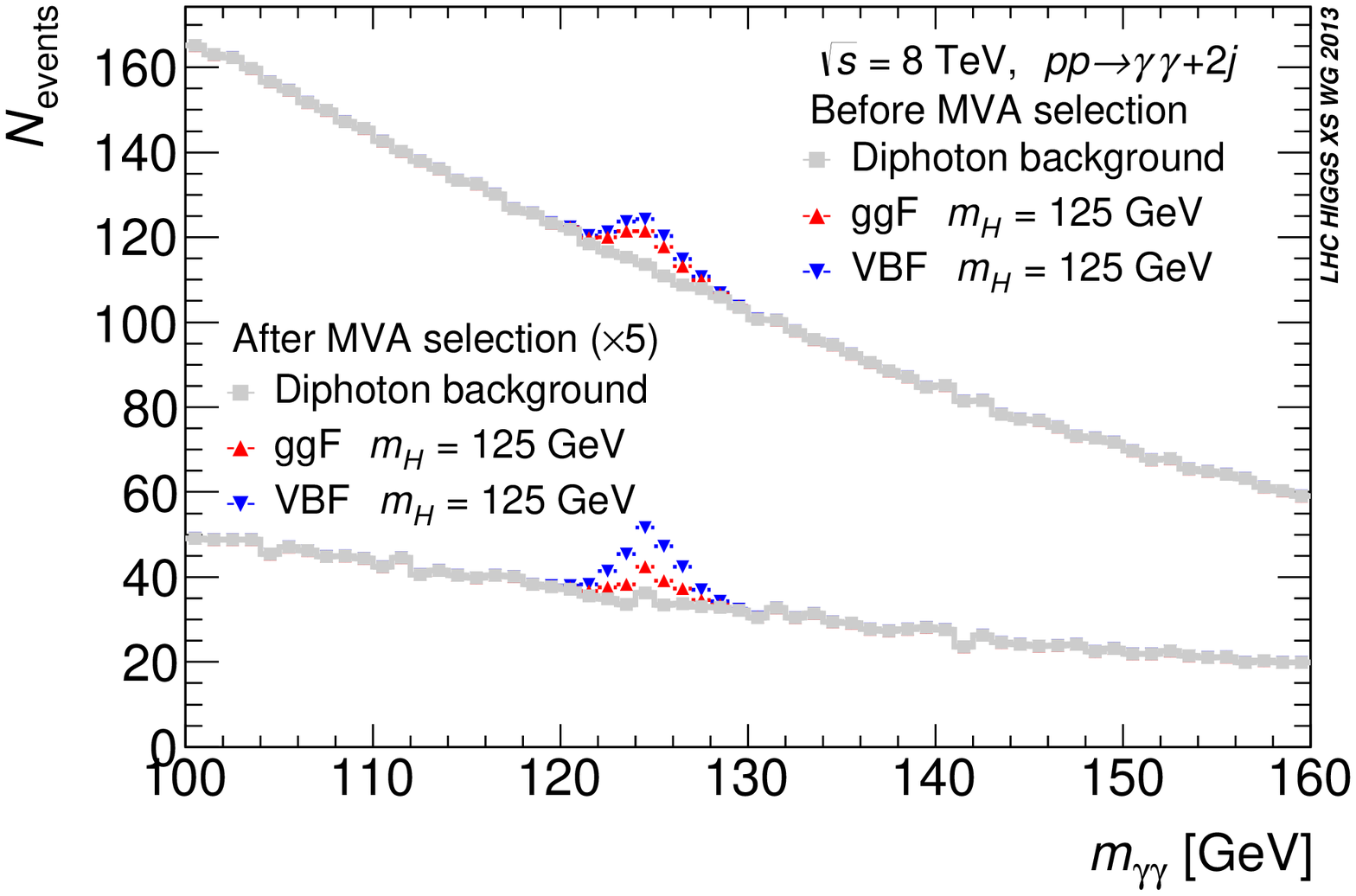}\label{fig:mgg_sig_bkg}}%
\\
\subfigure[Multivariate classifier]
{\includegraphics[width=0.515\columnwidth]{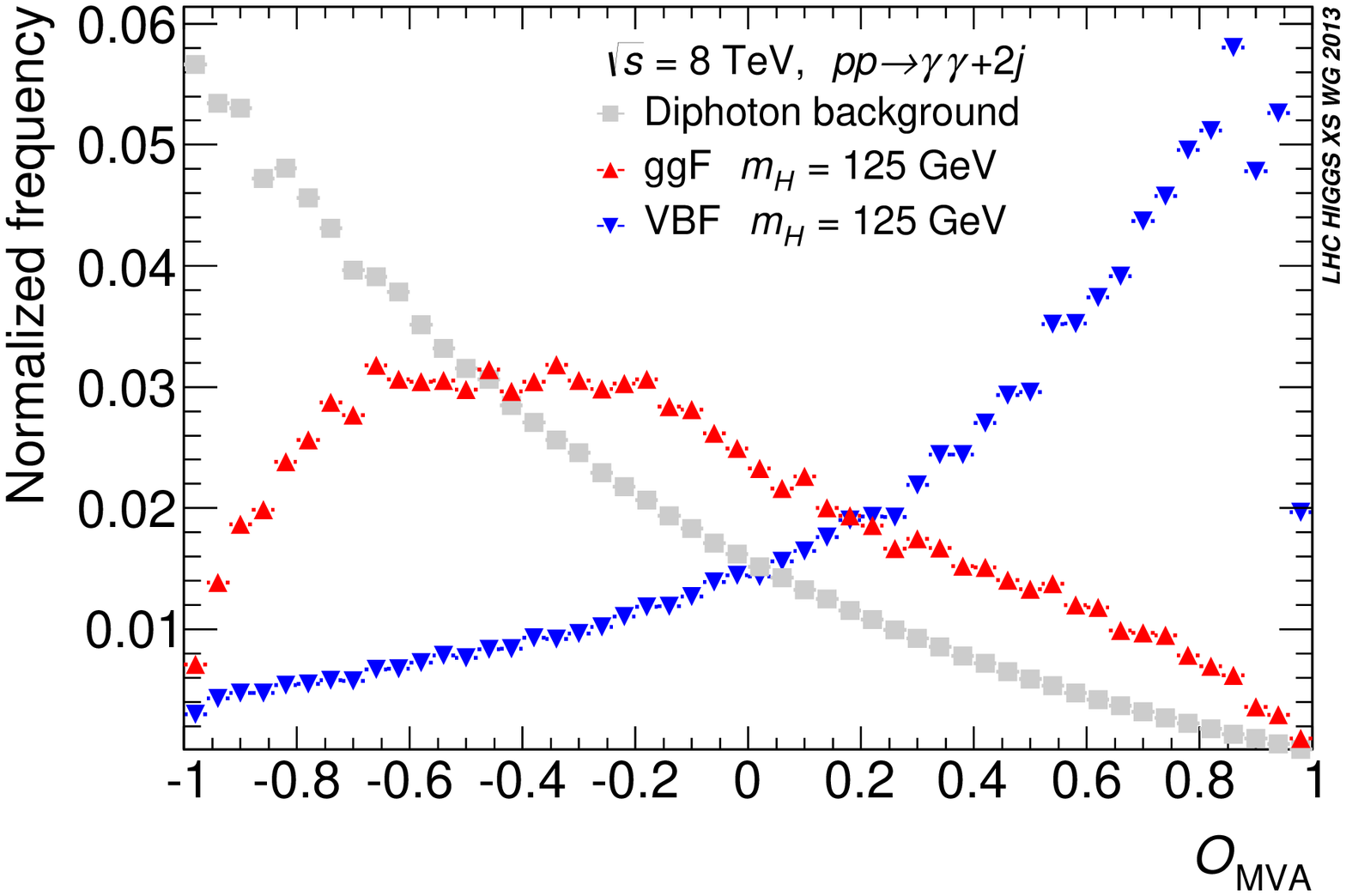}\label{fig:mva_classifier_sig_bkg}}%
\hfill%
\subfigure[ggF $\dphi$ for a progression of cuts]
{\includegraphics[width=0.5\columnwidth]{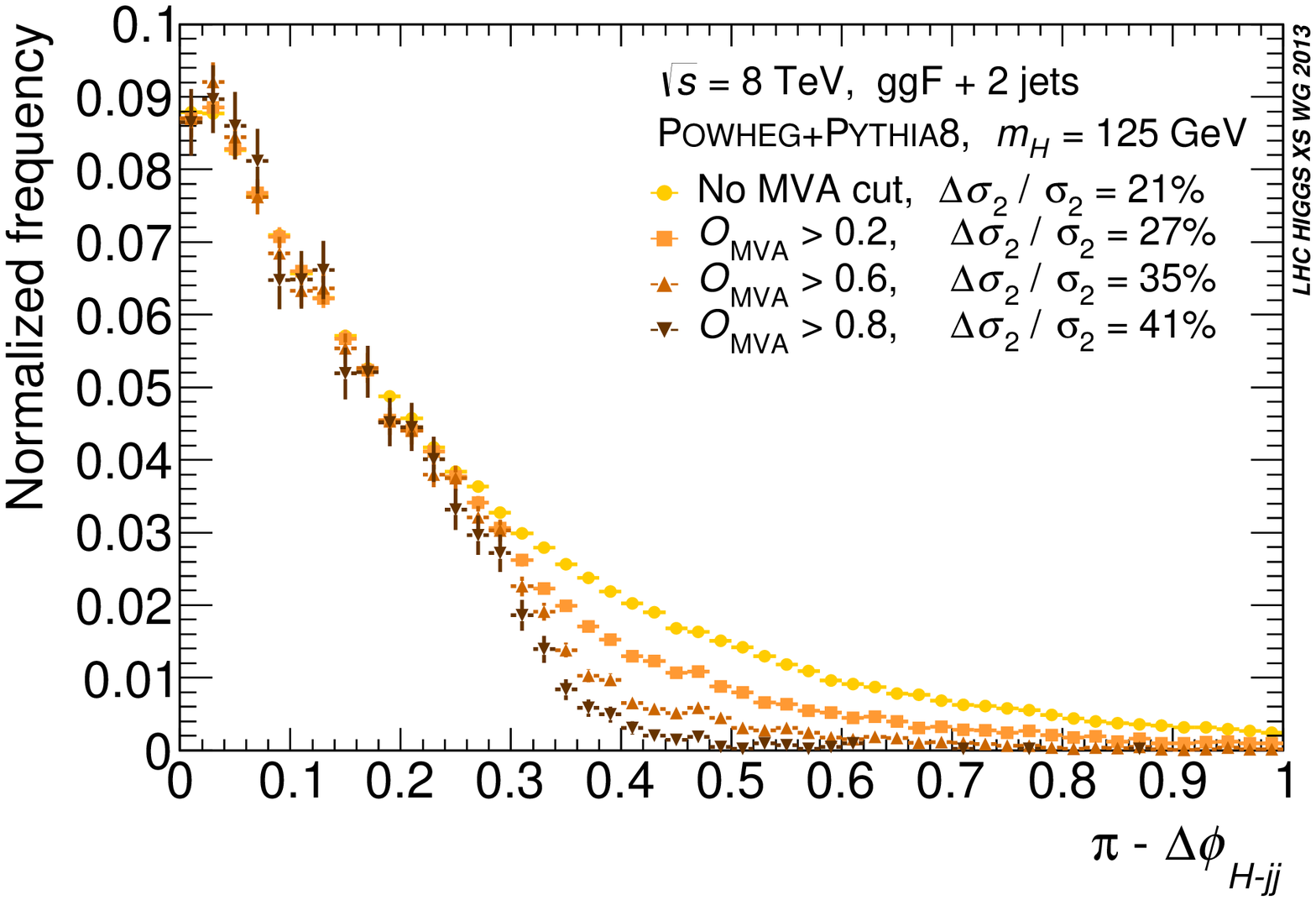}\label{fig:dphi_ggf}}%
\caption{ Signal and Background distributions of the multivariate selection: \refF{fig:dphi_bkg_sig} shows $\dphi$ for background and VBF signal; \refF{fig:mgg_sig_bkg} depicts the invariant diphoton mass spectrum for diphoton background (grey), ggF (red), and VBF (blue) before and after an arbitrary cut on the multivariate classifier. \refF{fig:mva_classifier_sig_bkg} depicts the multivariate classifier constructed from the six input variables for background, ggF, and VBF following the same color code. \refF{fig:dphi_ggf} shows the $\dphi$ distribution without any cut, and a progression of cuts on the multivariate classifier, also quoting the uncertainties on the integral obtained using \eqn{eq:cum_uncert}.}
\label{fig:mva_results}
\end{figure*}

\refF{fig:dphi_bkg_sig} shows the distribution in $\dphi$ for the simulated background and signal decays. VBF events produce a topology which causes the Higgs and dijet system to be more back-to-back than background and ggF events. The multivariate method will make use of this to select a signal enriched region of phase space, and cut into this distribution. \refF{fig:mgg_sig_bkg} depicts the $m_{\PGg\PGg}$ invariant  mass distribution before and after a cut on the multivariate classifier, illustrating the effect of the smearing model. \refF{fig:mva_classifier_sig_bkg} shows the classifier $\mathcal{O}_{\text{MVA}}$: VBF signal peaks near the positive values, and background and ggF accumulates near negative values. Finally, \refF{fig:dphi_ggf} depicts the ggF $\dphi$ spectrum for a progression of cuts on the classifier. The curves were normalized to have the same number of events in the region of $0-0.2$, which corresponds to the cutoff value used in \eqn{infsafedphi}.

Cutting on the classifier separates the inclusive 2-jet cross section into an exclusive 2-jet and an inclusive 3-jet part, similar as with a rectangular cut on $\dphi$ or $\pTHjj$. In \refT{tab:MVAUncertainties} we list the uncertainties calculated from \eqn{eq:cum_uncert} for a progression of cuts and different slopes for the linear correlation model: Harder cuts on the classifier translate into a tighter nonlinear selection in $\dphi$ phase space. As expected, this increases the exclusive 2-jet cross section uncertainty. The progressive harder cuts have a flat efficiency in $\dphi$ above the threshold of 0.2 (i.e. cut into this region without changing its shape), which is important to obtain reliable uncertainties from \eqn{eq:cum_uncert}. The dependence on the actual details on the linear correlation model is small: Changing the bin-by-bin correlations of the inclusive 3-jet cross section phase space by varying $\kappa$ from $50\%$ to $99\%$ has a practically negligible effect on the estimated uncertainty.

The method described here was used in the latest ATLAS $\PH \to \PGg \PGg$ measurement~\cite{ATLAS-CONF-2013-012}, to determine the uncertainties for the two used VBF MVA selections. The uncertainties found there were $28.3\%$ and $48.4\%$ for the loose and tight MVA category, respectively. The same approach applied to the CMS MVA analysis~\cite{CMS-PAS-HIG-13-001} at reconstruction level gives an uncertainty for the tight category of about $40\%$, which is similar to the ATLAS result. The effect for the loose category is larger, about $48\%$. This difference can be due to multiple sources: limited statistics (after the full selection, only ${\mathcal O}(500)$ events survive), different tune (Z2* ~\cite{Field:2011iq}), and looser selection for the di-jet loose category compared to ATLAS.

\begin{table}[t!]
 \caption{
 Relative perturbative uncertainties at NLO for $\Pg\Pg\to \PH + 2$ jets after
  applying a selection on the multivariate classifier $\mathcal{O}_{\text{MVA}} $. The uncertainty is calculated using generator level $\dphi$.
 }
 \label{tab:MVAUncertainties}
 \vspace{1ex}
 \centering
 \begin{tabular}{lcccc}
   \hline
    Cut &$\Delta \sigma_2/ \sigma_2$ ($\kappa = 50\%$) &$\Delta \sigma_2/ \sigma_2$ ($\kappa = 90\%$) &$\Delta \sigma_2/ \sigma_2$ ($\kappa = 99\%$)  \\ \hline
   no cut  & 21.0\% & 21.0\% & 20.9\%   \\
   $\mathcal{O}_{\text{MVA}} > 0.2$ & 26.6\%  & 26.8\% & 26.9\%  \\
   $\mathcal{O}_{\text{MVA}} > 0.6$ & 34.3\% & 34.6\% & 34.7\% \\
   $\mathcal{O}_{\text{MVA}} > 0.8$ & 40.8\% & 41.1\% & 41.1\% \\
   \hline
 \end{tabular}
\end{table}

%%%%%%%%%%%%%%%%%%%%%%%%%%%%%%%%%%%%%%%%%%%%%%%%%%%%%%%%%%%%%%%%%%%%%%%%%%%%%%%
\subsection{Underlying event uncertainties  in the Higgs plus 2-jet VBF selection }
\label{sec:jets-UE}

The simulation of the underlying event for the most recent experimental results is done by either \textsc{Pythia6} \cite{Sjostrand:2006za} (CMS) or \textsc{Pythia8} \cite{Sjostrand:2007gs} (ATLAS). The hard scattering process is simulated separately using a dedicated NLO generator, i.e., \textsc{Powheg} \cite{Nason:2004rx,Frixione:2007vw,Alioli:2010xd} interfaced to \textsc{Pythia} for the modeling of underlying event, hadronization, and showering. In Pythia the underlying event is simulated using a multipartonic interaction (MPI) model~\cite{Sjostrand:1987su,Sjostrand:2004pf}, that is tuned on underlying event or minimal bias data. In the following the impact of the underlying event on the Higgs plus two-jet VBF selection is studied by comparing simulated gluon-gluon-fusion and vector-boson-fusion production mechanisms with and without \textsc{Pythia}'s multipartonic interaction.

The following are guidelines for the estimation of UE related uncertainties in ggF and VBF processes:
\begin{itemize}
  \item Turn UE on/off for the nominal default tune
\item Cross check on/off effect for alternative tunes
\item Cross checks can include the use of tunes performed within a common framework but using different PDFs (eg. NLO v. LO, as is the case with AU2-CT10 and AU2-CTEQ6L1)
\end{itemize}

\subsubsection{Comparative multipartonic interaction study
\footnote{%
F.~U.~Bernlochner, D.~Gillberg, M. Malberti, P. Meridiani, P. Musella, G. Salam }}

In the following the impact on the Higgs plus two-jet VBF selection with and without \textsc{Pythia}'s underlying event model are investigated. In the first study performed, three tunes of the MPI model parameters for three parton distribution functions (PDF) are studied: CTEQ10~\cite{Lai:2010vv}, CTEQ6L1~\cite{Pumplin:2002vw}, and MSTW2008 \cite{Watt:2012tq}, which are NLO, LO, and LO PDFs, respectively. The initial hard scattering process are ggF + 0 jets and VBF at NLO, simulated by \textsc{Powheg} with CTEQ10 as the PDF. Jets are reconstructed from stable particles using the anti-$k_{\mathrm T}$ algorithm with $R = 0.4$ and are required to have $p_{Tj}\!>\! 25\UGeV$ for $\lvert\eta_{j}\rvert\! <\! 2.5$ and $p_{Tj}\!>\! 30\UGeV$ for $2.5 <\! \lvert\eta_{j}\rvert\! <\! 4.5$. Additional requirements are applied to isolate the VBF signal: $m_{jj}>400\, \gev$, $\Delta\eta_{jj}>2.8$ and $\dphi>2.6\,$rad.

\begin{figure*}[t!]
\includegraphics[width=0.5\columnwidth]{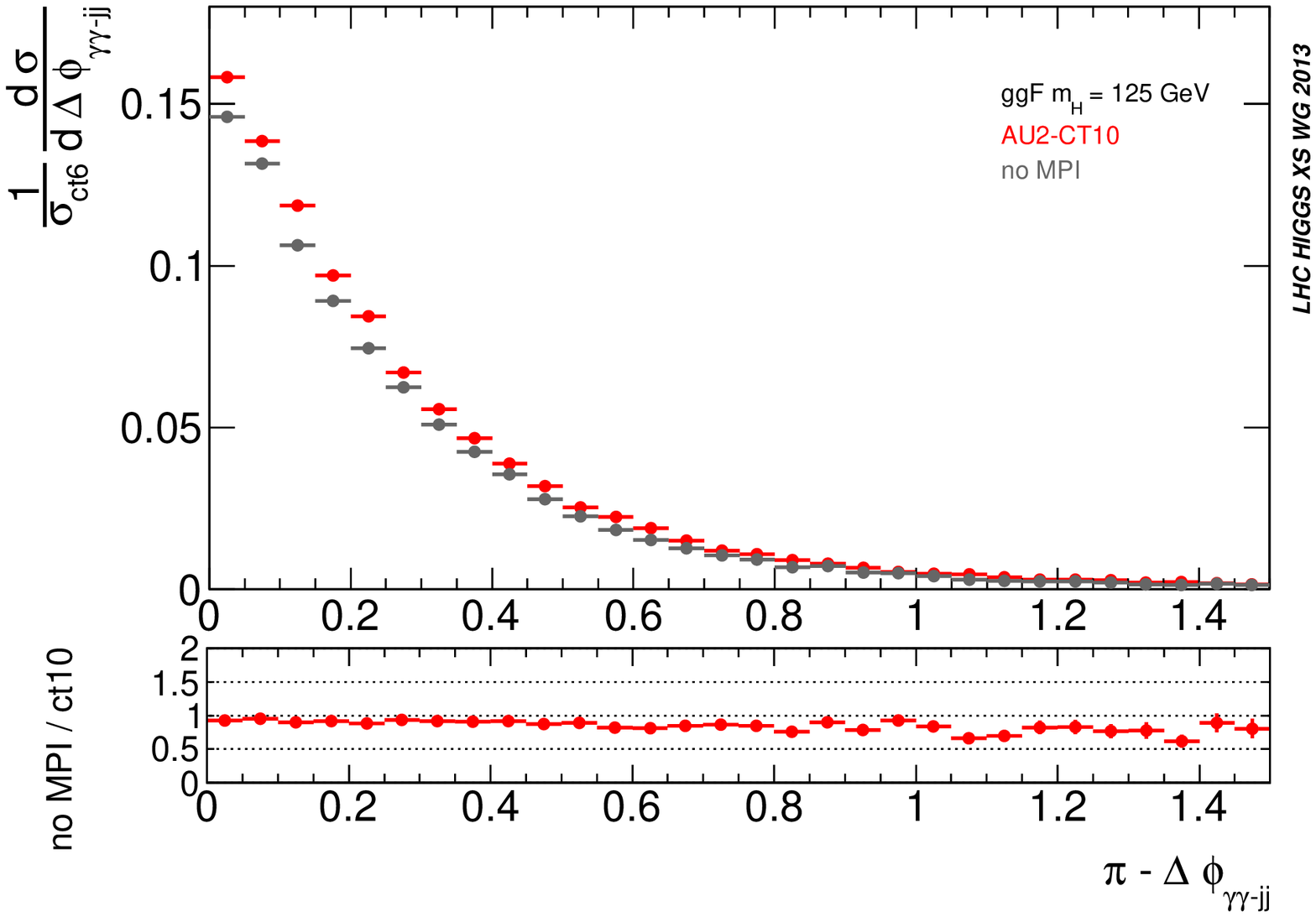}%
\hfill%
\includegraphics[width=0.5\columnwidth]{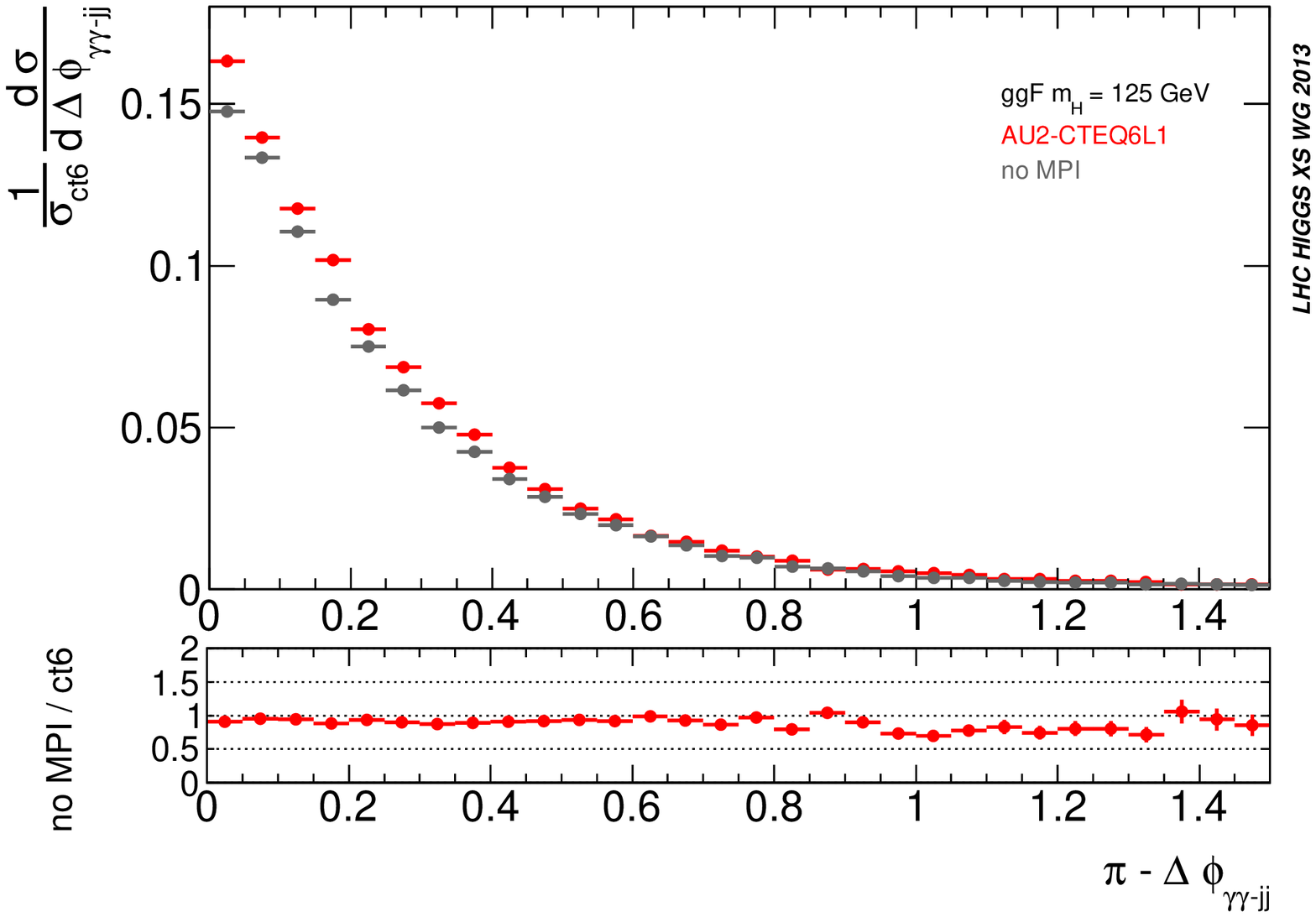}%
\caption{ The effect of switching off multi-partonic interaction for $\Pg\Pg \to \PH + 0$ jets at NLO for $\dphi$ is shown: CTEQ10 with the ATLAS AU2-CT10 tune (left) and CTEQ6L1with the ATLAS AU2-CTEQ6L1 tune (right). The no MPI histogram was scaled to account for the change in the two jet selection efficiency. The hard scatter for both was simulated using \textsc{Powheg} using the CTEQ10 NLO PDF. }
\label{fig:dphiUE_ggF}
\end{figure*}

\begin{figure*}[t!]
\includegraphics[width=0.5\columnwidth]{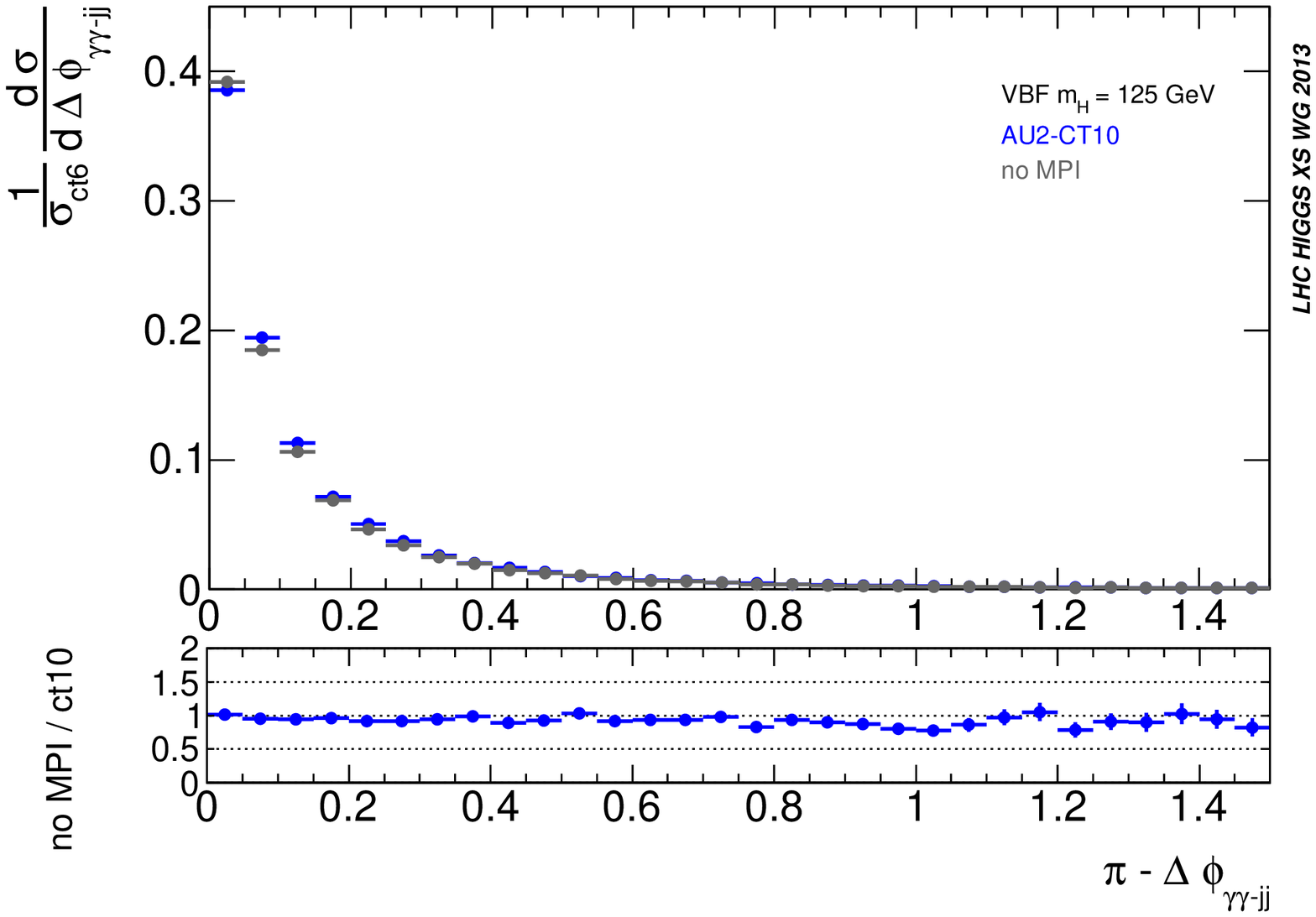}%
\hfill%
\includegraphics[width=0.5\columnwidth]{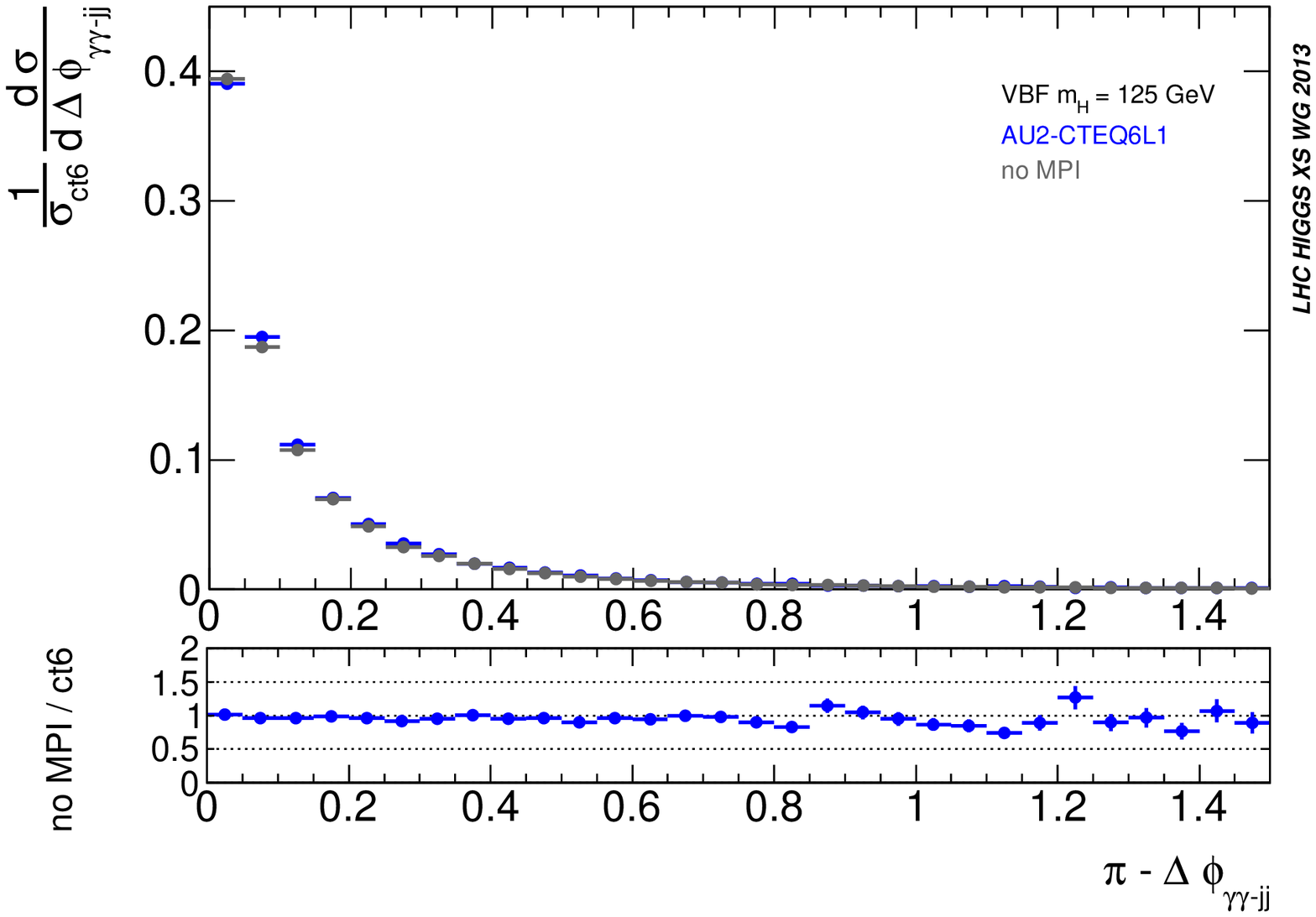}%
\caption{ The effect of switching off multi-partonic interaction for $\PQq\PQq \to \PH + 0$ jets at NLO for $\dphi$ is shown: CTEQ10 with the ATLAS AU2-CT10 tune (left) and CTEQ6L1with the ATLAS AU2-CTEQ6L1 tune (right). }
\label{fig:dphiUE_VBF}
\end{figure*}

\refF{fig:dphiUE_ggF} shows the distribution of $\Delta \phi$ between
the Higgs and the dijet system for ggF for the CTEQ10 (AU2-CT10) and
CTEQ6L1 (AU2-CTEQ6L1) tunes~\cite{ATLAS:2012uec}. Switching off MPI results in overall less
hadronic activity and a reduction in the number of selected
events. \refF{fig:dphiUE_VBF} shows $\dphi$ for VBF for the same
tunes: the effect here is much smaller, since the dijet system
originates most of the time from the hard scatter alone, and little
additional hadronic activity in the rapidity gap between the dijet
system is produced in the VBF topology. \refT{tab:UEUncertainties}
summarizes the relative difference in percent without MPI for AU2-CT10,
AU2-CTEQ6L1, and the MSTW2008 PDF tune AU2-MSTW2008~\cite{ATLAS:2012uec}: the relative
shifts in the yields of the two-jet selection and a typical VBF phase
space selection are quoted. 
The overall effect of the MPI on selected cross sections is about 10\%
in ggF and 4\% in VBF.
In the tails of distributions it can be sometimes larger.

The comparisons using AU2-CTEQ6L1 and AU2-MSTW2008, although
illustrative, might be considered somewhat inconsistent, since the
hard-scatter uses a different PDF from the showering and MPI.
As a cross check, the MPI-induced variations in the selected cross
sections have been studied (based on stable particles) also in the Pythia~6
DW~\cite{Albrow:2006rt}, Pythia~6 Perugia 2011~\cite{Skands:2010ak} and Pythia~8
4C~\cite{Corke:2010yf} tunes.
Again, the effects are of the order of $10\%$ for ggF.

%Thus the shift with respect to AU2-CT10 is assigned as a measure of the MPI modeling uncertainty. 

In the second study five different PYTHIA6 tunes are used, either
based on a MPI model with $p_{T}$ ordered showers like Z2* (CMS
default) \cite{Field:2011iq}, Pro-PT0 \cite{Buckley:2009bj} and P0
\cite{Skands:2010ak}, or with virtuality ordered showers, Pro-Q20
\cite{Buckley:2009bj} and D6T \cite{Albrow:2006rt}. %The  PDF set is CT10~\cite{Guzzi:2011sv} for all tunes. 
The main motivation to look at this set of tunes  is in the data/MC comparison that CMS made using 36~pb$^{-1}$ of $7\UTeV$ data of the forward energy flow and the central charged multiplicity in hard-scattering $\PW$ and $\PZ$ events \cite{Chatrchyan:2011wb}. It was found that none of the considered tunes was able to simultaneously describe the central charged multiplicity and the forward energy flow, and, particularly, their correlations.

This study is based on fully simulated and reconstructed events. The simulation of the CMS detector is done with GEANT4 and the reconstruction uses the official CMS software. Reconstructed quantities allow for a better determination of the effect of MPI uncertainty on the final observables, given that jet resolution and mis-reconstruction are properly taken into account. On the other hand, because of the small detector efficiency, only a tiny number of selected events (${\mathcal O}(500)$) for the ggF samples are available (from an initial sample of about 100k fully simulated events). This implies that the following numbers are affected by a somewhat large statistical uncertainty. 
For each tune, the MPI is switched on and off and the difference in selection efficiency times acceptance for the CMS $\PH \to \PGg\PGg$ VBF analyses \cite{CMS-PAS-HIG-13-001} is estimated.
%To separate the effect of the showering uncertainty which is treated separately from the uncertainty due to the MPI model, for each tune, the MPI is switched on and off and the difference in selection efficiency times acceptance for the CMS Higgs to gamma gamma VBF analyses \cite{CMS-PAS-HIG-13-001} is estimated. 
Results are summarized in \Tref{tab:UEUncertaintiesCMS}. The uncertainty associated to the MPI is defined as the largest variation in the on/off ratio over the 5 tunes. The uncertainty is split in two parts: the overall VBF selection efficiency and the migration between the tight and loose categories. The migration is evaluated as the ratio between the number of events in the tight category and the one in the tight and loose categories together. The effect is below 10\% for both ggF and VBF.

\begin{table}[t!]
 \caption{ The uncertainties for the underlying event evaluated in \% of the nominal yields from switching the multi-partonic interaction off for the 2-jet selection, and a VBF-type selection ($m_{jj}> 400 \UGeV$, $\Delta \eta_{jj} > 2.8$, $\dphi > 2.6$) are listed. }
 \label{tab:UEUncertainties}
 \vspace{1ex}
 \centering
 \begin{tabular}{lcc}
   \hline
   Tune / ggF  & 2-jet selection &  VBF selection\\ \hline
   AU2-CT10 & 9.8\% & 8.8\% \\
   AU2-CTEQ6L1 & 8.9\% & 5.3\% \\
   AU2-MSTW2008 & 8.2\% & 6.0\% \\
   \hline
 \end{tabular}
 \vspace{2ex}
 
  \begin{tabular}{lcc}
   Tune / VBF  &  2-jet selection & VBF selection\\ \hline
   AU2-CT10 & 3.0\%  & 3.5\% \\
   AU2-CTEQ6L1 & 2.1\% & 1.8\% \\
   AU2-MSTW2008 & 2.6\% & 3.1\% \\
   \hline 
 \end{tabular}
\end{table}

\begin{table}[t!]
 \caption{ The uncertainties for the underlying event evaluated in \% of the nominal yields from switching the multi-partonic interaction off for the CMS VBF-type cut-based selection, as described in ~\cite{CMS-PAS-HIG-13-001}. The numbers in the second column are related to the effect on the tight and loose VBF categories together, while the third column reports the migration between the two categories.}
 \label{tab:UEUncertaintiesCMS}
 \vspace{1ex}
 \centering
 \begin{tabular}{lcc}
   \hline
   Tune / ggF  &  tight + loose & migration\\ \hline
   TuneProQ20   & 2.4\% & 0.6\% \\
   TuneZ2Star   & 0.3\% & 2.1\% \\
   TuneProPT0   & 6.6\% & 1.2\% \\
   TuneP0       & 0.4\% & 0.8\% \\
   TuneD6T      & 0.7\% & 0.4\% \\
   \hline
   Maximum variation  & 6.6\% & 2.1\% \\
  \hline
 \end{tabular}
 \vspace{2ex}
 
  \begin{tabular}{lcc}
   Tune / VBF  & tight + loose & migration \\ \hline
   TuneProQ20  & 8.7\% & 2.4\% \\
   TuneZ2Star  & 0.3\% & 1.5\% \\
   TuneProPT0  & 3.2\% & 0.3\% \\
   TuneP0      & 2.7\% & 0.6\% \\
   TuneD6T     & 8.3\% & 2.2\% \\
   \hline
   Maximum variation & 8.7\% & 2.4\% \\
  \hline
 \end{tabular}
\end{table}

%%%%%%%%%%%%%%%%%%%%%%%%%%%%%%%%%%%%%%%%%%%%%%%%%%%%%%%%%%%%%%%%%%%%%%%%%%%%%%%
%%%%%%%%%%%%%%%%%%%%%%%%%%%%%%%%%%%%%%%%%%%%%%%%%%%%%%%%%%%%%%%%%%%%%%%%%%%%%%%

\clearpage

\newpage
\section{NLO MC \footnote{%
    F.~Krauss, F.~Maltoni, P.~Nason (eds.); Jeppe R.~Andersen,
    F.~Cascioli, H.~van Deurzen, N.~Greiner, S.~H\"oche, T.~Jubb, 
    G.~Luisoni, P.~Maierhoefer, P.~Mastrolia, E.~Mirabella, C.~Oleari, 
    G.~Ossola, T.~Peraro, S.~Pozzorini, M.~Sch\"onherr, F.~Siegert, 
    J.~M.~Smillie, J.~F.~von~Soden-Fraunhofen, F.~Tramontano
    }}

In this section, some recent studies related to the dominant Higgs boson 
production mode, gluon fusion, are presented with a focus on the issue of
additional jets in this class of processes.  

In Sec.~\ref{Sec::NLOMC_Jets}, various fixed order results at next-to leading 
order in the strong coupling constant for the production of a Higgs boson in 
association with two jets are compared to results from Monte Carlo event 
generators.  They treat jet emission at leading or next-to leading order, but 
resumming different types of potentially large logarithms.  In this study, 
some interesting differences when going to the kinematically more extreme weak
boson fusion region emerge, with clear consequences on the projected accuracy
for the important gluon fusion background to this important production mode.  

In Sec.~\ref{Sec::NLOMC_4l} a careful analysis of jet multiplicities and the
associated errors is presented for the irreducible background to Higgs boson 
production in gluon fusion and its subsequent decays into two leptons and two 
neutrinos.  In this calculation all interferences are taken into account,
and the fixed order parton level includes next-to leading order accuracy for 
the 0 and 1 jet bins.  

Finally, in Sec.~\ref{NLOMC_H3j} first steps towards the calculation of 
next-to leading order QCD corrections to the production of $H+3$ jets in 
gluon fusion are reported.

% Plans for YR3:
% \begin{enumerate}
% \item
% News in the Higgs MC 
% \item
% comparison of different NLO MC
% \end{enumerate}

\subsection{Jet studies in gluon-fusion production}
\label{Sec::NLOMC_Jets}
This section reports on an ongoing study concerning the gluon fusion
contribution to Higgs boson production in association with at least two jets.
Both rather inclusive cuts and cuts specific for the weak-boson fusion (WBF)
regime are applied and results compared between various approaches to the
description of higher-order corrections. The WBF region covers a notoriously
difficult regime of QCD radiation, where the perturbative stability of the
calculations are being put to a new test. It is therefore not surprising that while
the various approach agree well in the inclusive region, some significant
differences show up within the WBF cuts. The study here represents merely 
a snapshot of the current findings, which definitely deserve further studies
to better understand similarities and differences between the 
approaches and to quantify their intrinsic uncertainties.

The two kinematic regimes are defined as follows:
\begin{itemize}
\item Inclusive Selection:
  \begin{itemize}
  \item All jets are defined by the anti-$k_{\mathrm T}$
    algorithm\cite{Cacciari:2008gp} ($R=0.4$) and counted if their transverse
    momentum is greater than 20~GeV and their absolute rapidity is less than
    5.0.
  \item At least two hard jets above 25~GeV are required.
  \end{itemize}
\item WBF Selection: %Furthermore, require
  \begin{itemize}
  \item Furthermore require $|\eta_{j_1}-\eta_{j_2}|>2.8$, $m_{j1,j2}>400$~GeV. 
  \end{itemize}
\end{itemize}

\subsubsection{Monte-Carlo generators}
In this section the physics input of the Monte-Carlo generators used is
described and their setup, scale choices, etc.~is detailed\footnote{
  The beginning stages of this study displayed results also from 
  \aMCatNLO~\cite{Frederix:2011ig} using the multijet-merging algorithm 
  at NLO presented in~\cite{Frederix:2012ps}}. 
For this study,
which focuses on the multi-jet descriptions, all predictions are obtained with
no restrictions on the the Higgs boson momentum or decay channel.

\paragraph{HEJ}

High Energy Jets (\HEJ) gives a description to all orders and all
multiplicities of the effects of hard, wide-angle (high energy-) emissions,
by utilizing an approximation to the hard scattering
amplitudes~\cite{Andersen:2009nu,Andersen:2009he,Andersen:2011hs}, which in
this limit ensures leading logarithmic accuracy for both real and virtual
corrections. These logarithmic corrections are dominant in the region of
large partonic center of mass energy compared to the typical transverse
momentum. The resummation obtained is matched with a merging procedure to
full tree-level accuracy for multiplicities up to and including three
jets~\cite{Andersen:2008ue,Andersen:2008gc,hej_implement}. 

The predictions of \HEJ depend principally on the choice of factorization and
renormalization scale, and on the pdf set (MSTW2008nlo\cite{Martin:2009iq}). 
For this study, we allow different scales for individual occurrences of the
strong coupling.  We choose to evaluate two powers of the strong coupling at 
a scale given by the Higgs mass, and all remaining scales are identified with 
the transverse momentum of the hardest jet. Therefore, for $n$ jets,
\begin{equation}
  \alphas^{n+2}(\muR)
  = \alphas^2(\MH)\cdot\alphas^n(p_\perp^{\rm hardest})\,.
\end{equation}
The merging scale is set to 20~GeV, the minimum transverse momentum of the 
jets. 

\paragraph{MCFM}

For the purpose of this comparison \MCFM \cite{Campbell:2010cz,MCFMweb} was 
used to provide a fixed-order description of the observables in 
question. The transverse mass of the Higgs boson 
$m_{{\mathrm T},\PH}^2=\MH^2 + p_{{\mathrm T},\PH}^2$ has been used as the factorization 
and renormalization scale. The CTEQ6.6 parton distribution functions 
\cite{Nadolsky:2008zw} were used in conjunction with its accompanying value and evolution 
of the strong coupling constant. The effective $\Pg\Pg \PH$ vertex is
calculated in the $m_{\PQt}\to\infty$ limit.

\paragraph{PowhegBox}

The \PowhegBox{} generators for Higgs ($\PH$), Higgs plus one jet ($\PH J$) and
Higgs plus two jets ($\PH JJ$) have appeared in \Brefs{Alioli:2008tz}
and~\cite{Campbell:2012am}.  For the present work, the HJJ
generator has been augmented with the MiNLO
method~\cite{Hamilton:2012np,Hamilton:2012rf}. As shown in these references,
when using the MiNLO method, the {HJ} and {HJJ} generators also maintain
some level of validity when used to compute inclusive quantities that do
not require the presence of jets. More specifically, the HJ generator
remains valid also for describing the Higgs inclusive cross section, and
the HJJ generator remains valid for describing the Higgs plus one jet, and
the Higgs inclusive cross section. This is achieved without introducing any
unphysical matching scale. The generators become smoothly consistent with
the generators of lower multiplicities when the emitted parton become
unresolvable. The level of accuracy that they maintain is discussed
in detail in \Bref{Hamilton:2012rf}. By the arguments given there
one can show that the MiNLO procedure can be implemented in such a way that
the HJ generator maintains NLO accuracy when the emitted parton becomes
unresolved. The HJJ generator, at present, maintains only LO accuracy
when emitted partons become unresolvable.

In \refT{tab:incrates}
we show the inclusive production rates obtained with the different generators
according to the standard cuts described in this report.
Furthermore, we show the standard set of distributions obtained
with the HJJ generator interfaced with \PythiaSix \cite{Sjostrand:2006za},
without hadronization.
\begin{table}[htb]
 \caption{Total rates for the $\PH$ generator (H), the HJ generator with
           MiNLO (HJ), the HJ generator with MiNLO improved according to
           \cite{Hamilton:2012rf} (HJ-N) and the MiNLO improved HJJ
           generator.\label{tab:incrates}}
 \begin{center}
    \begin{tabular}{lccc}
      \hline
      $\vphantom{\int_a^b}$ & Inclusive & 2 jets, $p_\mathrm{T} > 25, \left| y \right| < 5$ & 
      $+ \; \eta_{\tmop{jj}} < 2.8, m_{\tmop{jj}} > 400 \tmop{GeV}$\\
      \hline
      $\vphantom{\int_a^b}$\tmtexttt{H} & 13.2 & 1.61 & 0.177\\
      $\vphantom{\int_a^b}$\tmtexttt{HJ} & 16.2 & 2.04 & 0.202\\
      $\vphantom{\int_a^b}$\tmtexttt{HJ-N} & 13.3 & 2.10 & 0.209\\
      $\vphantom{\int_a^b}$\tmtexttt{HJJ} & 17.8 & 2.41 & 0.239\\
      \hline
    \end{tabular}
  \end{center}
  \end{table}

\paragraph{Sherpa}

The \Sherpa \cite{Gleisberg:2008ta} predictions are calculated using the 
\MEPSatNLO method \cite{Hoeche:2012yf,Gehrmann:2012yg}. It combines multiple 
\NLOPS \cite{Frixione:2002ik,Nason:2004rx,Frixione:2007vw,Hoeche:2011fd} 
matched calculations of successive jet multiplicity into an inclusive sample 
such that both the fixed order accuracy of every subsample is preserved 
and the resummation properties of the parton shower w.r.t.\ the inclusive 
sample are restored.

For the purpose of this study the subprocesses $\Pp\Pp\to h+0,1$ jets are 
calculated at next-to-leading order accuracy using a variant of the 
\MCatNLO method \cite{Hoeche:2011fd}, while the subprocesses $pp\to h+2,3$ jets 
are calculated at leading order accuracy. Further emissions are effected 
at parton shower accuracy only. 
The $\Pp\Pp\to \PH$ and $\Pp\Pp\to \PH+1$ jet virtual matrix elements are taken from 
\cite{Dawson:1990zj,Djouadi:1991tka,Ravindran:2002dc,Schmidt:1997wr}. 
While not included in this study, the subprocess $\Pp\Pp\to \PH+2$ jets can also 
be calculated at next-to-leading order accuracy by interfacing \GoSam 
\cite{vanDeurzen:2013rv} through the Binoth-Les Houches accord 
\cite{Binoth:2010xt} interface, see also the description in the following
section, \ref{NLOMC_H3j}.
The effective $\Pg\Pg h$ vertices are computed in the $m_{\PQt}\to\infty$ limit.

Scales are chosen in the usual way of 
\MEPS merging \cite{Hoeche:2009rj}, i.e.\ they are set to the relative 
transverse momenta of parton splittings reconstructed through backwards 
clustering the respective final states using inverted parton shower 
kinematics and splitting probabilities. Thus,
\begin{equation}
  \alphas^{n+k}(\muR)
  = \alphas^k(\mu_0)\cdot\prod\limits_{i=1}^n\alphas(\mu_i)\,,
\end{equation}
wherein the $\mu_i$ are in the individual splitting scales and $\mu_0$ 
is the scale of the underlying $\Pp\Pp\to \PH$ production process, here $\mu_0=\MH$, 
$k=2$. This plays an integral part in restoring the overall resummation 
properties. The CT10 \cite{Guzzi:2011sv} parton distributions, with 
the respective value and evolution of the strong coupling, are used 
throughout. The merging scale is set to $Q_{\mathrm{cut}}=20\UGeV$.

\subsubsection{Comparison}
\begin{figure}[t]
  \includegraphics[width=0.48\textwidth]{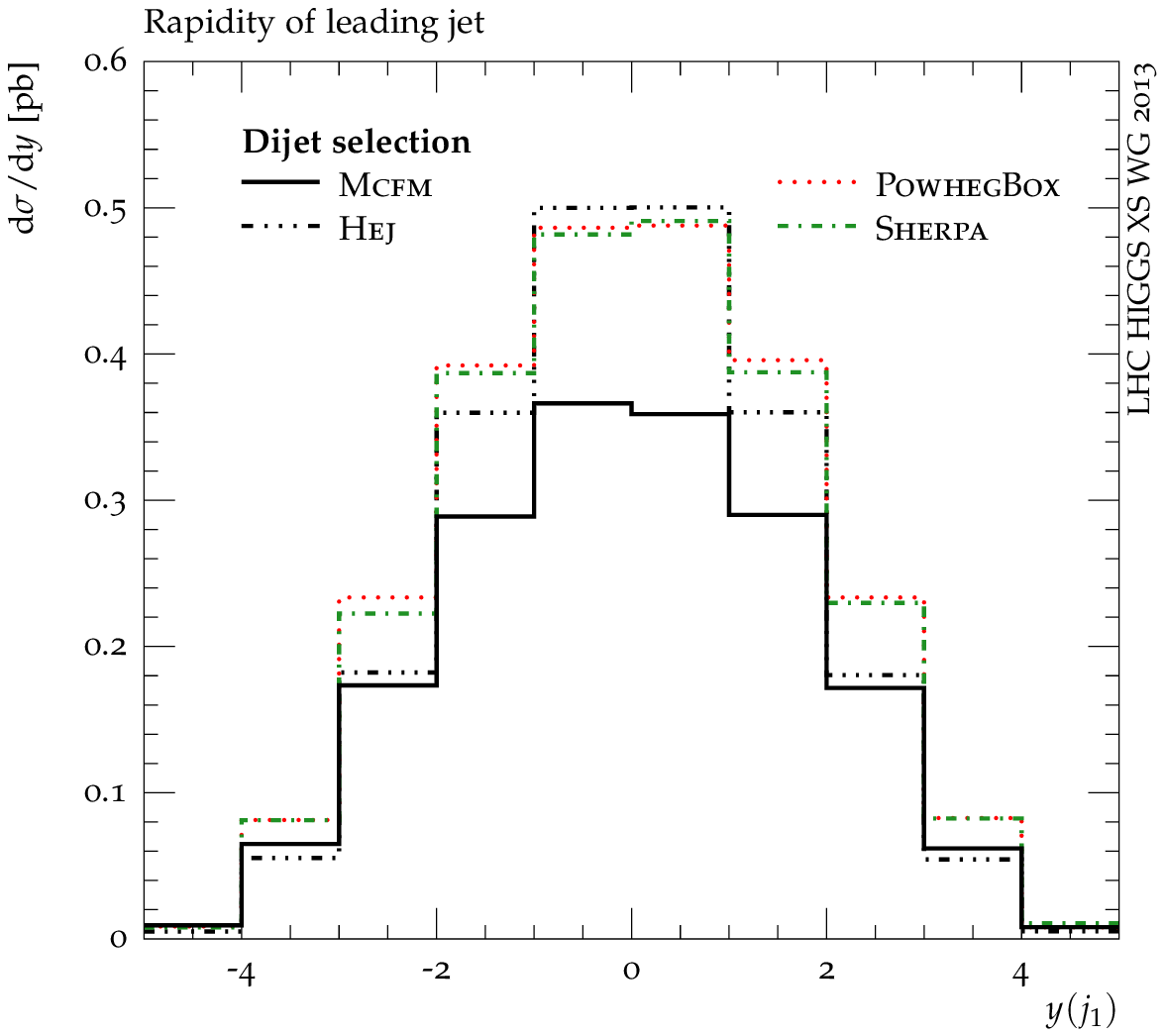}\hfill
  \includegraphics[width=0.48\textwidth]{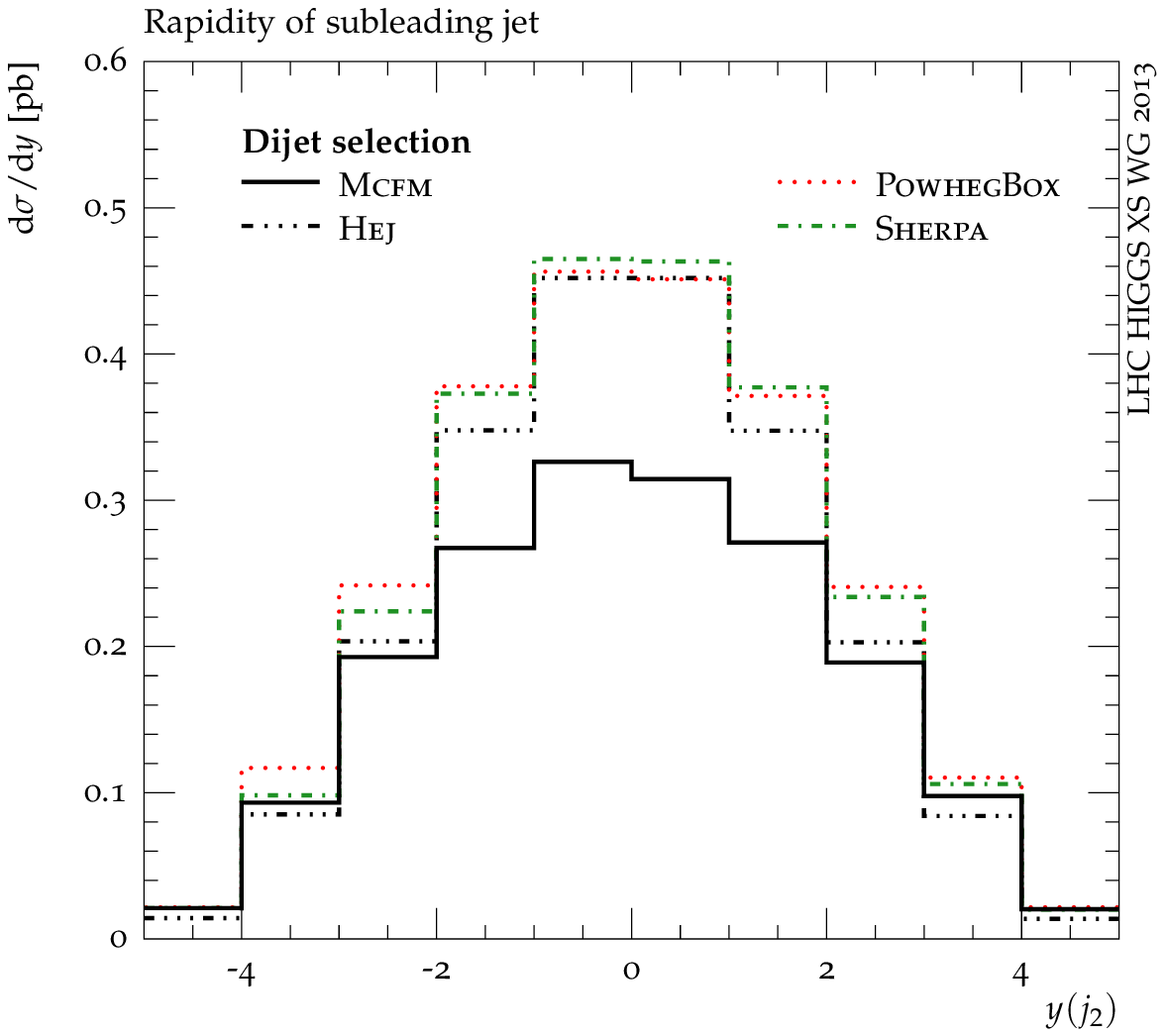}
  \caption{
	    Rapidity of the leading (left) and subleading (right) 
	    jets in the dijet selection.
	    \label{Fig:Yj1_Yj2_dijet}
	  }
\end{figure}

\begin{figure}[t]
  \includegraphics[width=0.48\textwidth]{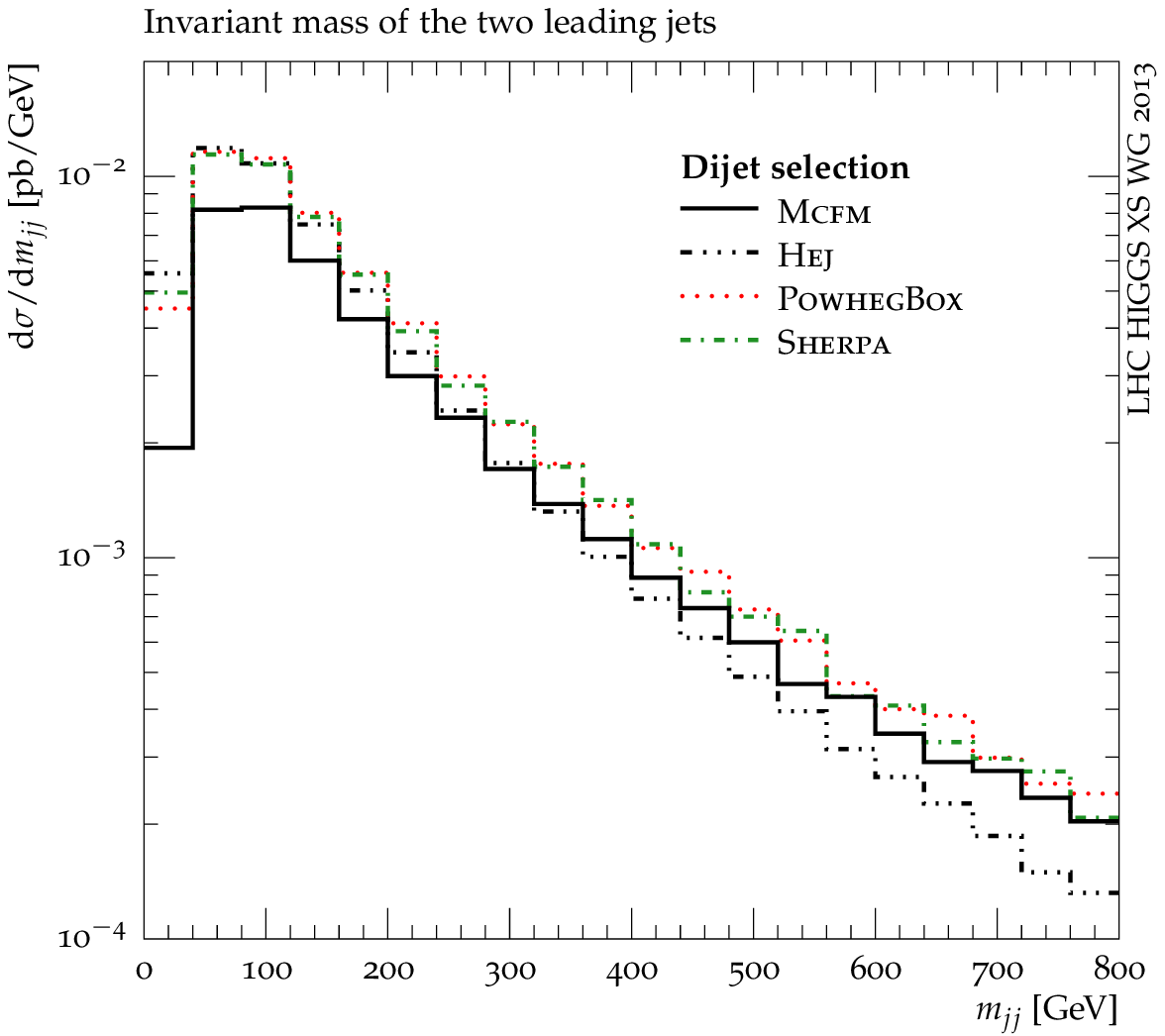}\hfill
  \includegraphics[width=0.48\textwidth]{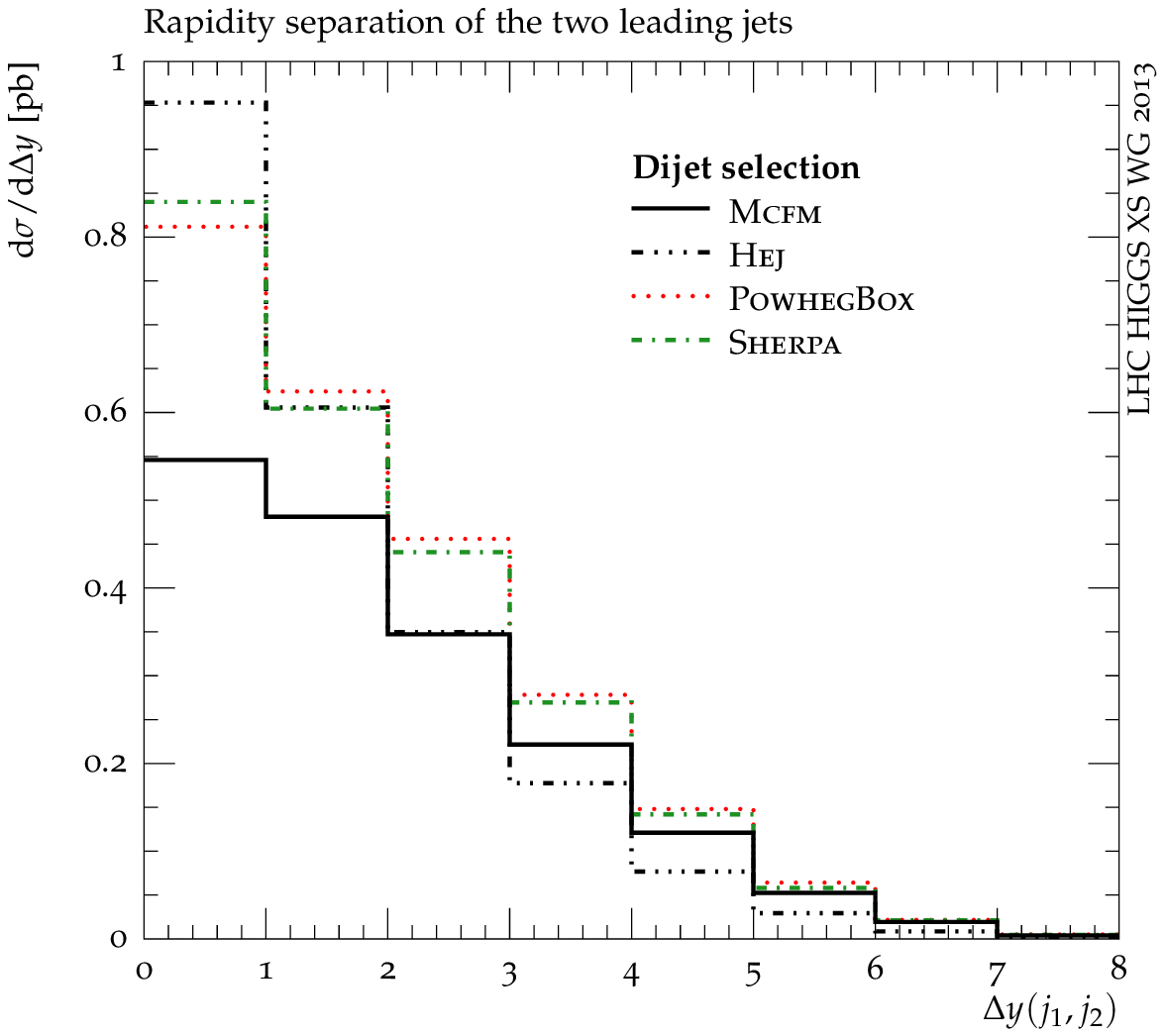}
  \caption{
	    Invariant mass (left) and rapidity separation (right) of 
	    the leading jets in the dijet selection.
	    \label{Fig:mj1j2_dYj1j2_dijet}
	  }
\end{figure}

Figures~\ref{Fig:Yj1_Yj2_dijet}--\ref{Fig:mj1j2_dYj1j2_dijet} exhibit 
important jet distributions in the inclusive regime.  Results of \HEJ, \MCFM, 
\PowhegBox, and \Sherpa agree fairly well for distributions of the individual 
two hardest jets, like for instance rapidity distributions
(\Fref{Fig:Yj1_Yj2_dijet}). The
notable exception is the fixed-order calculation.  However, this
is easily understood by the renormalization scales in 
the NLO calculation, which tends to pick higher scales for all strong
couplings and therefore leads to a smaller cross section.  Interestingly
enough, though, this does not lead to large visible shape differences in
the inclusive rapidity distributions of the two jets, 
\cf~\Fref{Fig:Yj1_Yj2_dijet}.
This picture somewhat changes when considering observables sensitive to
the kinematics of both leading jets, like their invariant mass or
rapidity distance (\Fref{Fig:mj1j2_dYj1j2_dijet}).  Again, the fixed-order 
result is below the two DGLAP-type
resummations provided by the parton showers and Sudakov form factors in 
the \PowhegBox and \Sherpa, with only minor, but now more visible differences 
in shape.  Again, this can be attributed to different choices for the
renormalization and factorization scales.  On the other hand, for the
invariant mass and rapidity difference distributions of the two leading
jets now the shape of the results from \HEJ start to strongly deviate.  This 
may be understood from the Regge-type suppression factors systematically 
resummed in \HEJ, but which are not present in the other approaches, but it
definitely deserves further and more detailed studies. These effects disfavor
large empty rapidity gaps, see \Fref{Fig:mj1j2_dYj1j2_dijet}.  

\Fref{Fig:mj1j2_dYj1j2_dijet} also illustrates how the \HEJ cross
section in the WBF regime will deviate from the other Monte Carlos,
since the region of large rapidity separation between the two hardest jets is
suppressed. This is further quantified in \Tref{Tab:XS}. The relative
effect of imposing the WBF cuts is largest for the predictions from \HEJ.
\begin{table}[t]
  \caption{
	    Cross sections predicted by the individual generators for 
	    the dijet and WBF selections and the predicted relative reduction
            in cross section by applying WBF selection.
	    \label{Tab:XS}
          }
  \begin{center}
    \begin{tabular}{lccc}
      \hline
      $\vphantom{\int_a^b}$& Dijet selection & WBF selection & Effect of WBF cut\\\hline
      \MCFM      & 1.73 pb & 0.192 pb & 0.111 \\ 
      \HEJ       & 2.20 pb & 0.127 pb & 0.058 \\
      % \aMCatNLO  & 1.65 pb & 0.125 pb & 0.076 \\\hline
      \PowhegBox & 2.41 pb & 0.237 pb & 0.098 \\
      \Sherpa    & 2.38 pb & 0.225 pb & 0.094 \\ \hline
    \end{tabular}
  \end{center}
\end{table}

\begin{figure}[t]
  \includegraphics[width=0.48\textwidth]{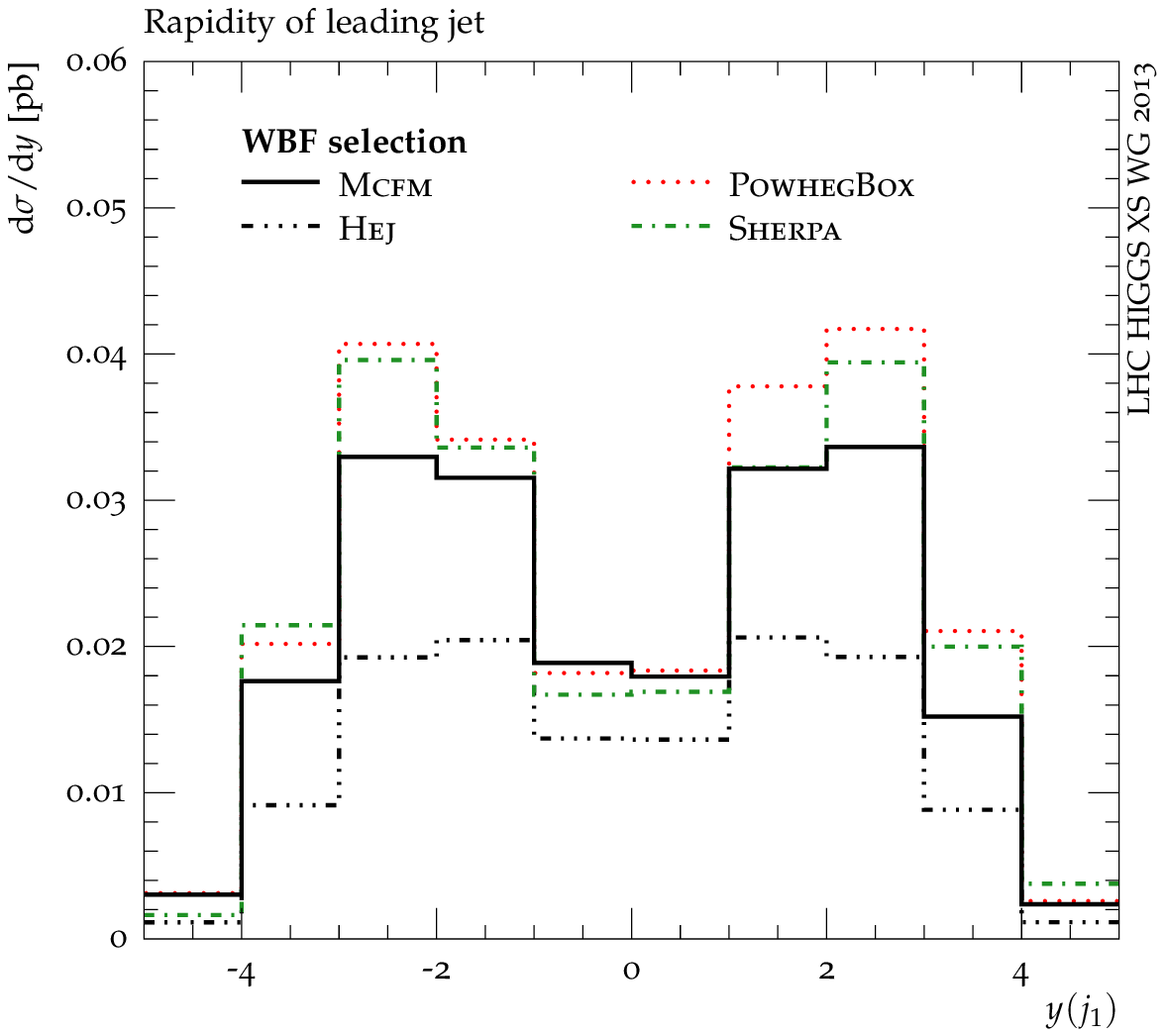}\hfill
  \includegraphics[width=0.48\textwidth]{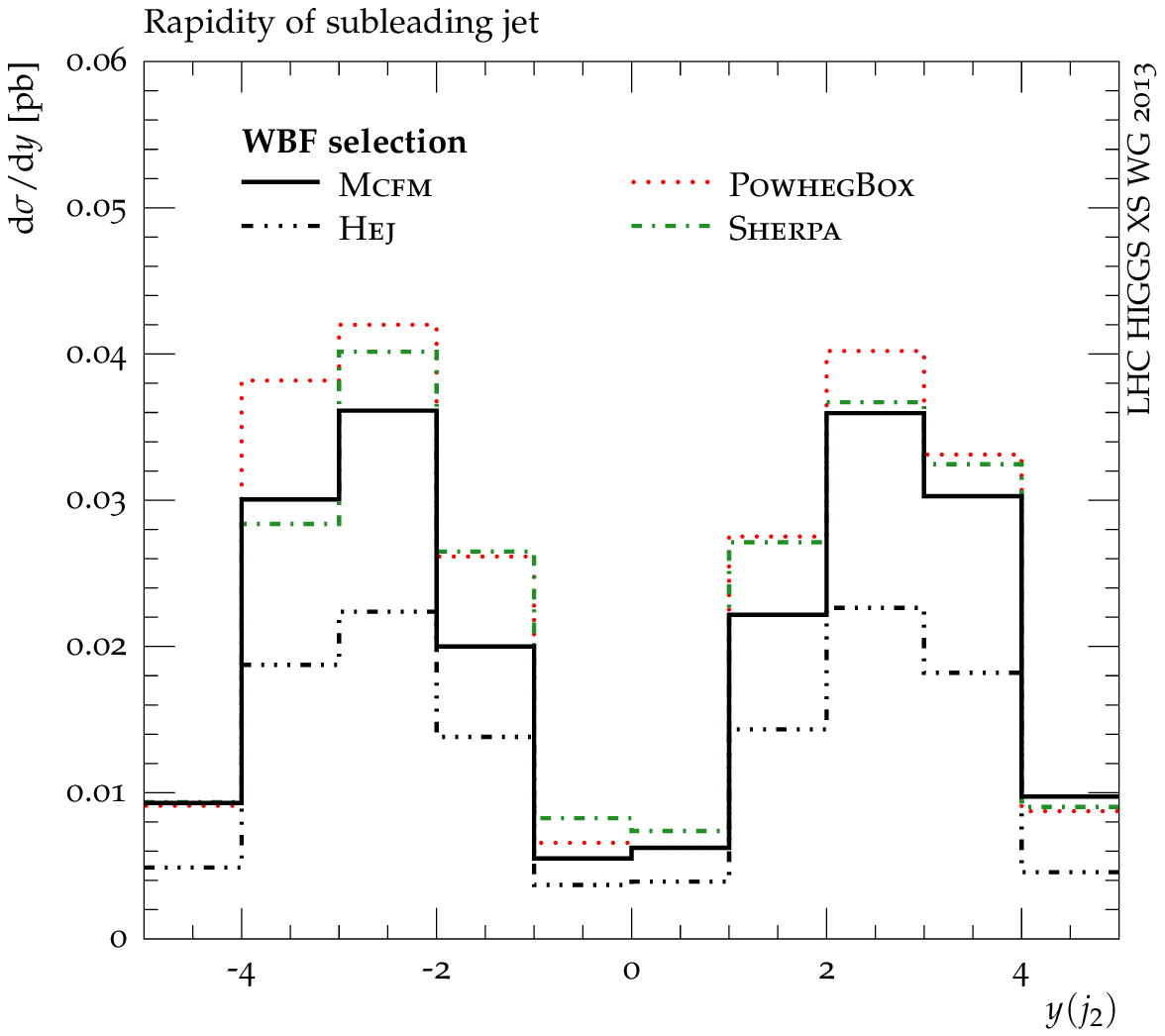}
  \caption{
	    Rapidity of the leading (left) and subleading (right) 
	    jets in the WBF selection.
	    \label{Fig:Yj1_Yj2_WBF}
	  }
\end{figure}

An interesting difference, apart from the overall
normalization, can be seen in the shape of the rapidity distribution of the leading jet
after WBF cuts: \HEJ has a much less developed dip at central 
rapidities; the leading jet is typically more 
central in rapidity than in the other approaches, cf.~\refF{Fig:Yj1_Yj2_WBF}.
Such aspects of the QCD radiation pattern  for large 
rapidity separations are currently investigated also experimentally, and a
better understanding could be used to further 
discriminate QCD and WBF production.  

\begin{figure}[t]
  \includegraphics[width=0.48\textwidth]{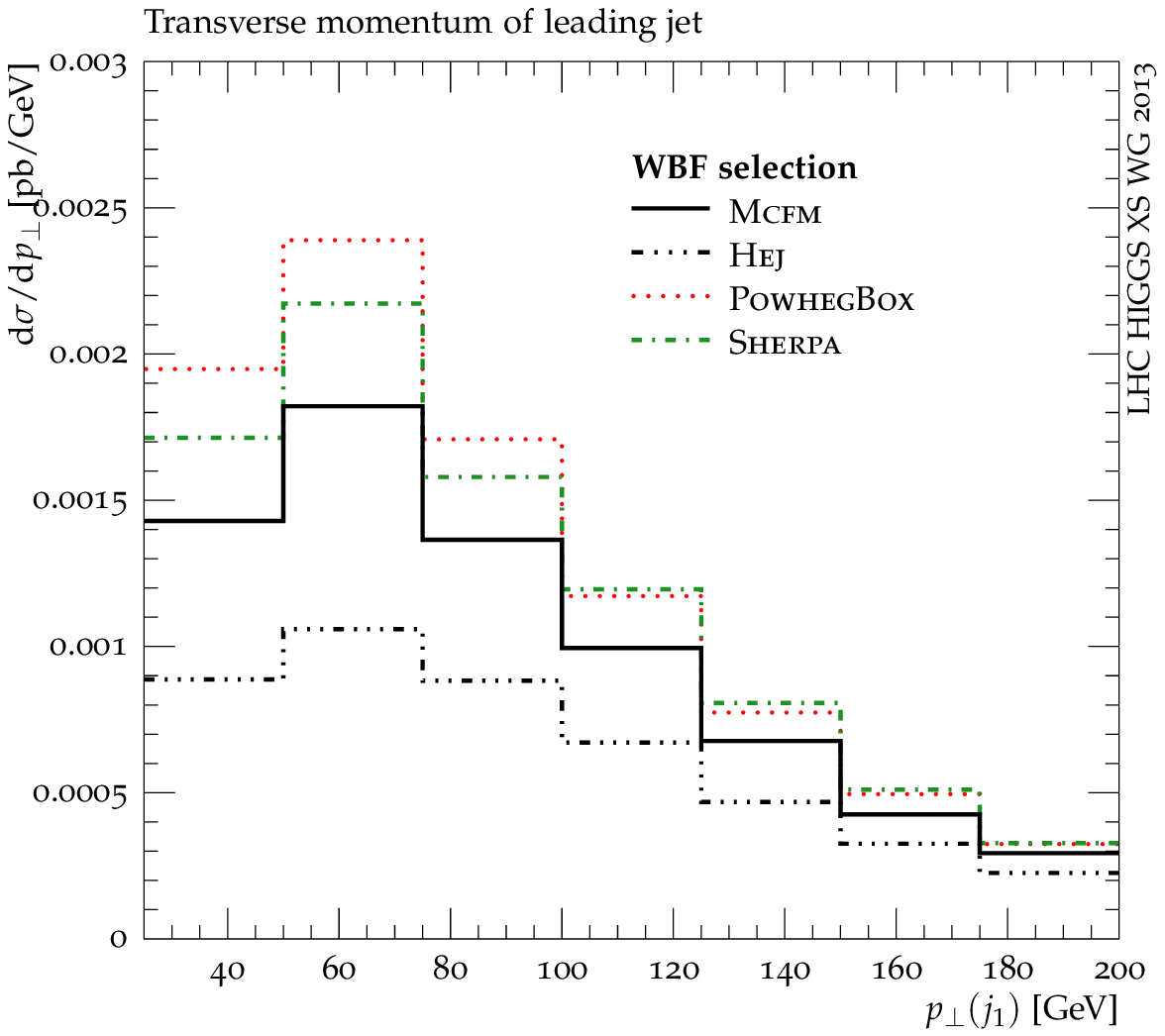}\hfill
  \includegraphics[width=0.48\textwidth]{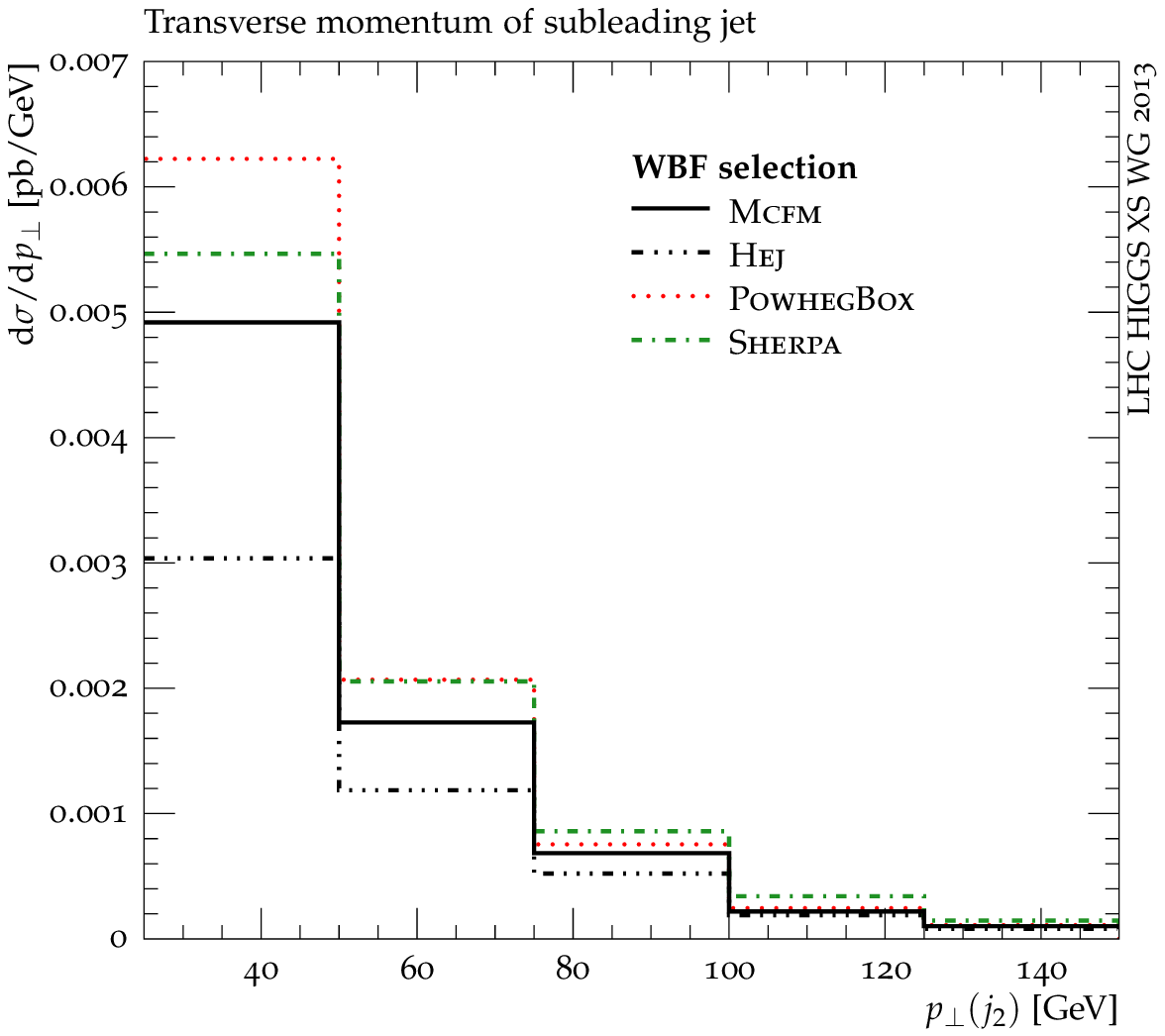}
  \caption{
	    Transverse momenta of the leading (left) and subleading (right) 
	    jets in the WBF selection.
	    \label{Fig:pTj1_pTj2_WBF}
	  }
\end{figure}

\begin{figure}[t]
  \includegraphics[width=0.48\textwidth]{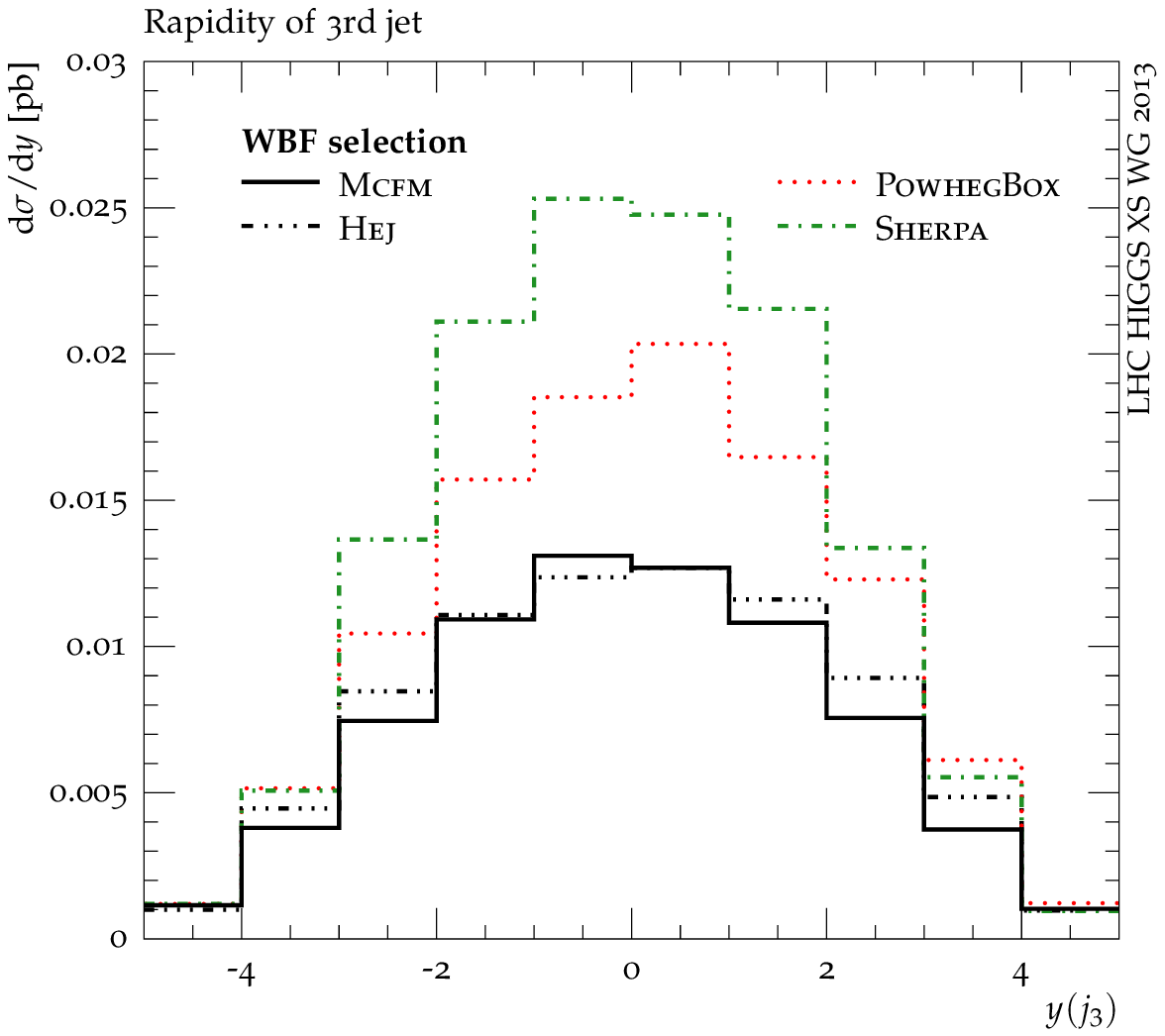}\hfill
  \includegraphics[width=0.48\textwidth]{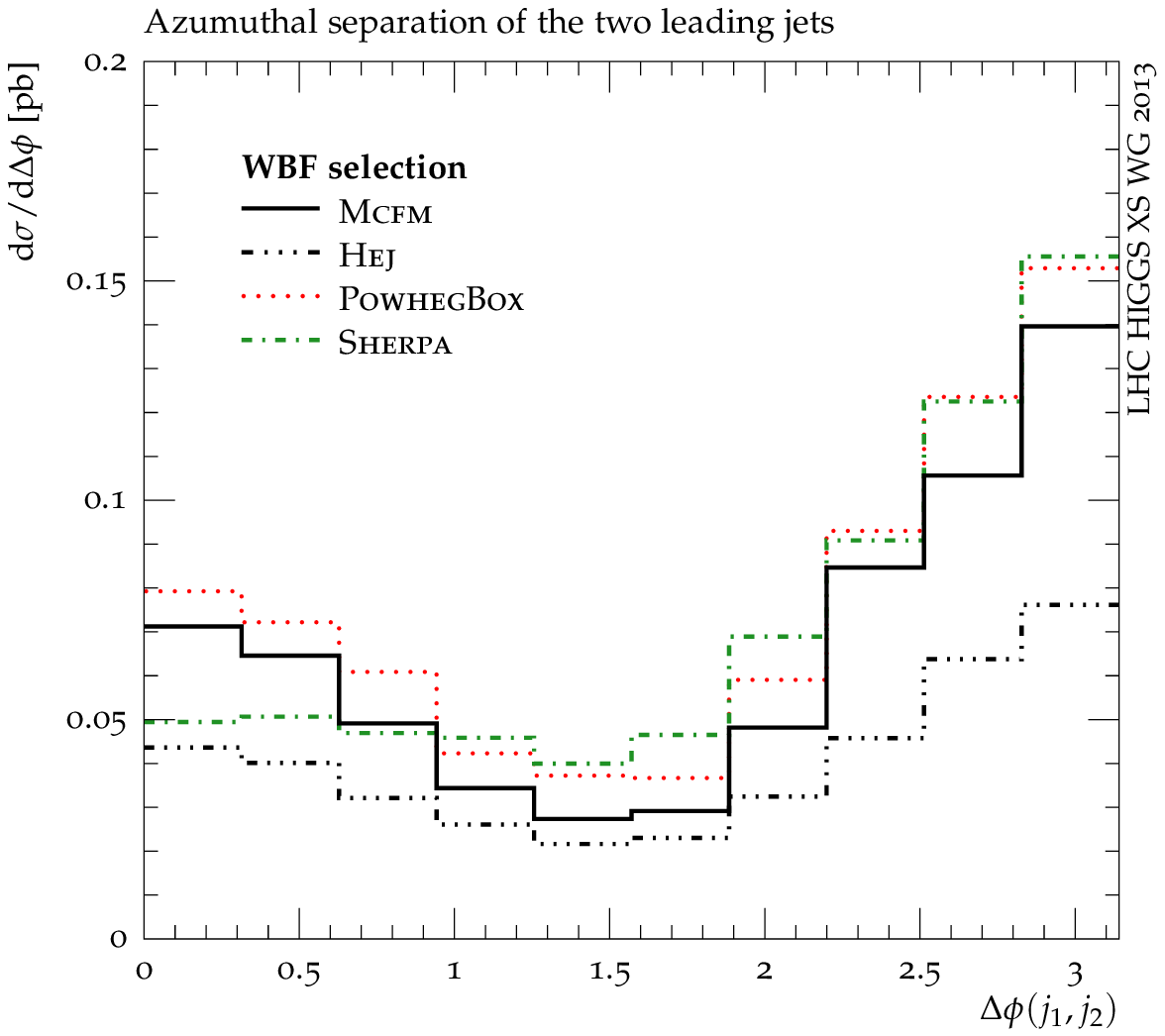}
  \caption{
	    Rapidity of the third jet (left) and azimuthal separation 
	    of the two leading jets (right) in the WBF selection.
	    \label{Fig:Yj3_dPhij1j2_WBF}
	  }
\end{figure}

A similar trend is observed in the transverse momentum distributions of the
two leading jets exhibited in \Fref{Fig:pTj1_pTj2_WBF}, where \HEJ
exhibits differences in normalization and shape, with slightly harder
spectrum for the two hardest jets.  Conversely, the rapidity of the
third hardest jet displayed in the left panel of
\Fref{Fig:Yj3_dPhij1j2_WBF}, tends to be a bit wider at tree-level
(from \textsc{MCFM}) and
in \HEJ than in
the parton-shower based approaches.  This, however, is an observable which for
all approaches 
is given at leading order only, potentially supported by some resummation.
It is thus not a surprise that here the largest differences between different
approaches show up.  

The right panel of \Fref{Fig:Yj3_dPhij1j2_WBF} shows the difference in
the azimuthal angle of the two leading jets. \Sherpa exhibits
a visible shape difference for small angles, while the other approaches are
more similar.  \HEJ shows a slightly
reduced correlation of the two hardest jets, possibly caused by a larger jet
radiation activity than in the shower-based approaches\cite{Andersen:2010zx}.

\subsubsection{Outlook}
The results presented here tend to open more questions than to provide
answers, with differences between approaches being somewhat larger than 
expected.  Further studies are clearly important, since the gluon fusion
background to WBF production of a Higgs boson will mostly be estimated from
Monte Carlo techniques and a Monte Carlo driven extrapolation from control to
signal regions.  In this respect the apparent similarity of approaches for
the inclusive selection and the visible and sometimes even large differences
in the WBF region necessitates a much better understanding of the
QCD radiation pattern in the extreme phase space region of the WBF selection.
In addition to an understanding on the level of central values, it will also
be of utmost importance to gain a well-defined handle on the theory 
uncertainties related to the extrapolation.  This will be the subject of
ongoing and further studies.

%%% Local Variables: 
%%% mode: latex
%%% TeX-master: "../YRHXS3"
%%% End: 

%% \newcommand{\NLOacc}{N\scalebox{0.8}{LO}\xspace}
%% \newcommand{\MEPSatLO}{M\scalebox{0.8}{E}P\scalebox{0.8}{S}@L\scalebox{0.8}{O}\xspace}
%% \newcommand{\MEPSatNLO}{M\scalebox{0.8}{E}P\scalebox{0.8}{S}@N\scalebox{0.8}{LO}\xspace}
%% \newcommand{\MCatNLO}{M\scalebox{0.8}{C}@N\scalebox{0.8}{LO}\xspace}
%% \newcommand{\Sherpa}{S\scalebox{0.8}{HERPA}\xspace}
%% \newcommand{\OpenLoops}{O\scalebox{0.8}{PEN}L\scalebox{0.8}{OOPS}\xspace}
%% \newcommand{\Collier}{C\scalebox{0.8}{OLLIER}\xspace}
%% \newcommand{\Rivet}{R\scalebox{0.8}{IVET}\xspace}
%% \newcommand{\ATLASexp}{{\sc ATLAS}\xspace}
%% \newcommand{\CMSexp}{{\sc CMS}\xspace}

%% \newcommand{\flfs}{\mu^+\nu_\mu \Pe^-\bar\nu_{\Pe}}
%% \newcommand{\rF}{\mathrm{F}}
%% \newcommand{\delres}{\Delta_\mathrm{res}}
%% \newcommand{\delqcd}{\Delta_\mathrm{QCD}}
%% \newcommand{\sigsr}{\sigma_\mathrm{S2s}}
%% \newcommand{\sigcr}{\sigma_\mathrm{S2c}}

\subsection{Irreducible background to $\PH\to\PW\PW^*$ in exclusive 0- and 1-jets 
bins with MEPS@NLO}
\label{Sec::NLOMC_4l}

%\subsubsection*{Introduction}
Final states involving four leptons played an important role in the 
discovery of the Higgs-like boson in 2012 and will continue to be crucial in 
the understanding of its coupling structure.  By far and large, there are two 
classes of final states of interest, namely those consistent with decays 
$\PH\to \PZ\PZ^*$ yielding four charged leptons and those related to $\PH\to \PW\PW^*$ 
resulting in two charged leptons and two neutrinos.  They have quite different 
backgrounds, and for the latter, the dominant and large top-pair production 
and decay background necessitates the introduction of jet vetoes to render 
the signal visible.  In addition, in order to study the weak-boson fusion 
production channel of the Higgs boson, it is important to understand jet 
production in association with the Higgs boson in the gluon-fusion channel 
as well, a topic discussed in Section~\ref{Sec::NLOMC_Jets} of this report.  
In this section, rather than focusing on the signal, the irreducible 
background to $\PH\to\PW\PW^*$ in the exclusive 0-jets and 1-jet bins will be 
discussed.

\subsubsection{Monte Carlo samples}
As the tool of choice the \Sherpa event generator~\cite{Gleisberg:2008ta} is
employed, using the recently developed multijet merging at next-to-leading
order accuracy~\cite{Gehrmann:2012yg,Hoeche:2012yf}.  Predictions obtained 
with this \MEPSatNLO technology will be contrasted with inclusive \MCatNLO 
and parton-level \NLOacc results for the production of four leptons plus 0 or 1 
jets, all taken from the corresponding implementations within \Sherpa.  
While the latter guarantee \NLOacc accuracy in the 0- and 1-jets bins, 
but do not resum the potentially large Sudakov logarithms arising in the 
presence of jet vetos, inclusive \MCatNLO simulations provide a better 
description of such Sudakov logarithms in the 0-jet bin, but are only LO or 
leading-log accurate in bins with 1 or more jets.  This is overcome by the
new \MEPSatNLO algorithm, which effectively combines \MCatNLO-type
simulations~\cite{Frixione:2002ik} in the fully color-correct algorithm
of~\cite{Hoeche:2011fd,Hoeche:2012ft} for increasing jet multiplicities,
preserving both the \NLOacc accuracy of the contributions in the individual jet
bins and the logarithmic accuracy of the parton shower.  Result for four-lepton
final states obtained with this formalism are presented for the first time
here.

Virtual corrections are computed with \OpenLoops~\cite{Cascioli:2011va}, an
automated generator of \NLOacc QCD amplitudes for SM processes,
which uses the \Collier library for the numerically stable evaluation of
tensor integrals~\cite{Denner:2002ii,Denner:2005nn} and scalar
integrals~\cite{Denner:2010tr}.  Thanks to a fully flexible interface of
\Sherpa with \OpenLoops, the entire generation chain -- from process
definition to collider observables -- is fully automated and can be steered
through \Sherpa run cards.

The results presented below refer to $\Pp\Pp\to\flfs+X$ at a centre-of-mass 
energy of $8\UTeV$ and are based on a \Sherpa2.0 pre-release version\footnote{
This pre-release version corresponds to
SVN revision 21340 (25 Mar 2013) and the main difference with respect
to the final \Sherpa2.0 release version is the tuning of parton shower,
hadronization and multiple parton interactions to experimental data.
}. 
The multijet merging is performed with a merging parameter
of $Q_{\mathrm{cut}}=20\UGeV$, and also LO matrix elements for
$\Pp\Pp\to\flfs+2$ jets are included in the merged sample.
Gluon-gluon induced contributions resulting from 
squared loop amplitudes are discarded.  All matrix elements are evaluated in 
the complex mass scheme and include all interference and off-shell effects.  
Masses and widths of the gauge bosons have been adjusted to yield the correct 
branching ratios into leptons at NLO accuracy and are given by
\begin{equation}
  \begin{split}
    \MW \;=\;& 80.399\,\mathrm{GeV}\;,\;\;\;
    \Gamma_{\PW} \;=\; 2.0997\,\mathrm{GeV}\;,\\
    \MZ \;=\;& 91.1876\,\mathrm{GeV}\;,\;\;\;
    \Gamma_{\PZ} \;=\; 2.5097\,\mathrm{GeV}\;,\\
  \end{split}
\end{equation}
while the electroweak mixing angle is a complex number given by the ratio of 
the complex $\PW$ and $\PZ$ masses, and
\begin{equation}
  \GF\;=\;1.16637\cdot 10^{-5}\;\mathrm{GeV}^{-2}\;\;\mathrm{and}\;\;\;
  1/\alpha_{\mathrm{QED}}\;=\;132.348905\,,
\end{equation}
in the $\GF$-scheme.
The five-flavor CT10 NLO~\cite{Lai:2010vv} parton distributions with the 
respective running strong coupling $\alphas$ has been employed
throughout.  Contributions with external $\Pb$-quarks are not included to 
trivially avoid any overlap with $\PAQt\PQt$ production.    As a 
default renormalization ($\muR$), factorization ($\muF$) and resummation
($\mu_Q$) scale we adopt the total invariant mass of the $\ell\ell'\nu\nu'$
final state, $\mu_0=m_{\ell\ell'\nu\nu'}$.  In the NLO parton-level predictions 
all $\alphas$ factors are evaluated at the same scale $\muR=\muF=\mu_0$,
while $\mu_Q$ is irrelevant.  In \MEPSatNLO predictions the scale $\mu_0$ is 
used only in tree and loop contributions to the $\Pp\Pp\to \flfs$ core process,
after clustering of hard jet emission.  For $\alphas$ factors associated with 
additional jet emissions a CKKW scale choice is applied.  In fact, in contrast 
to the original proposal for such a choice for \NLOacc merging presented 
in~\Bref{Hoeche:2012yf}, the nodal $k_{\mathrm T}$-scales for \NLOacc parts of the 
simulation are implemented through 
\begin{equation}
  \alphas(k_{\mathrm T})\,=\,\alphas(\mu_0)\,
  \left[1-\frac{\alphas(\mu_0)}{2\pi}\,\beta_0\,
    \log\frac{k_{\mathrm T}^2}{\mu_0^2}\right]\,, 
\end{equation}
because preliminary studies suggest that this choice indeed further minimizes 
the effect of $Q_{\mathrm{cut}}$ variations.

There are various theoretical uncertainties entering the simulation of 
hadronic observables:
\begin{itemize}
\item renormalization and factorization scale uncertainties are assessed by 
  an independent variation of $\muR$ and $\muF$ by factors of two
  excluding in both directions only the two opposite combinations.  Note that 
  the renormalization scale is varied in all $\alphas$ terms that arise in 
  matrix elements or from the shower;
\item resummation scale uncertainties are evaluated by varying the starting
  scale of the parton shower $\mu_Q$ by a factor of $\sqrt{2}$ in both 
  directions;
\item PDF -- and, related to them -- $\alphas(\MZ)$ uncertainties, which 
  are both ignored here;
\item non-perturbative uncertainties, which broadly fall into two categories:
  those related to the hadronization, which could be treated by retuning the 
  hadronization to LEP data and allowing/enforcing a deviation of for example 
  $5\%$ or one charged particle in the mean charged particle multiplicity at 
  the $\PZ$-pole, and those related to the underlying event, which could be
  evaluated by allowing the respective tunes to typical data to yield a
  $10\%$ variation in the plateau region of the mean transverse momentum 
  flow in the transverse regions of jet events.  Both have been ignored here
  and will be presented in a forthcoming study.
\end{itemize}

Quantifying the resummation scale uncertainty is crucial for a realistic
study of theory uncertainties in jet distributions and other observables
sensitive to details of the QCD radiation pattern.  As such, they certainly 
impact also on the cross sections in different jet bins.  They can be assessed 
through $\mu_Q$ variations in the \MEPSatNLO approach, which provide some 
sensitivity to subleading -- but potentially large -- Sudakov logarithms 
resulting from jet vetoes.  In contrast, the usual $\muR$ and $\muF$ 
variations alone are not sensitive to Sudakov logarithms or the scales entering
them and are thus not sufficient for a reliable assessment of theory
errors in jet bins.  The consistent implementation of all these effects in
ongoing, but preliminary studies suggest that the $\mu_Q$ variation is
fairly suppressed compared to the standard $\muR$ and $\muF$ variation. 

\subsubsection{Experimental setups and cuts}
The results presented here correspond to the cuts of the two relevant
analyses by the \ATLASexp~\cite{ATLAS-CONF-2013-030} and 
\CMSexp~\cite{CMS-PAS-HIG-13-003} collaborations for $\PH\to\PW\PW^*\to\flfs$ 
in the exclusive 0- and 1-jet bins.  The various cuts are listed in 
\Tref{Tab:NLOMC_WWcuts}\footnote{
  Definition of some kinematical quantities in \Tref{Tab:NLOMC_WWcuts}:
  \newline
  $^\dagger$: The quantity $E\!\!\!/_{\mathrm T}^{\mathrm{(proj)}}$ is given by 
  \begin{equation*}
    E\!\!\!/_{\mathrm T}^{\mathrm{(proj)}} = 
    E\!\!\!/_{\mathrm T}\cdot\sin\left(\min\{\Delta\phi_{\mathrm{near}},\,\pi/2\}\right)\,,
  \end{equation*}
  where $\Delta\phi_{\mathrm{near}}$ denotes the angle between the missing
  transverse momentum $E\!\!\!/_{\mathrm T}$ and the nearest lepton in the transverse
  plane.  \newline
  $^\ddagger$: \ATLASexp and \CMSexp have different definitions for $m_{\mathrm T}$, namely
  \begin{equation*}
    m_{\mathrm T}^2 = \left\{\begin{array}{lcl}
    \left(\sqrt{p_{\perp,\ell\ell'}^2+m_{\ell\ell'}^2}+E\!\!\!/_{\mathrm T}\right)^2-
    \left|p_{\perp,\ell\ell'}+E\!\!\!/_{\mathrm T}
    \vphantom{\sqrt{p_{\perp,\ell\ell'}^2}}\right|^2 &
    \mathrm{for} & \mathrm{ATLAS}\\[2mm]
    2 |p_{\perp,\ell\ell'}|\,|E\!\!\!/_{\mathrm T}|\,(1-\cos\Delta\phi_{\ell\ell',\,E\!\!\!/_{\mathrm T}}) &
    \mathrm{for} & \mathrm{CMS}.
    \end{array}\right.
  \end{equation*}
}. 
Lepton isolation is implemented at the particle level to be close to
the experimental definitions of both ATLAS and CMS. The scalar sum of
the transverse momenta of all visible particles within a $R=0.3$ cone
around the lepton candidate is not allowed to exceed $15\%$ of the
lepton $\pT$.
After a preselection ({\bf S1}), additional cuts are applied that define 
a signal ({\bf S2s}) and a control ({\bf S2c}) region. The latter is
exploited to normalize background simulations to data in the experimental 
analyses in each jet bin.  In the \ATLASexp analysis, different cuts are 
applied in the 0- and 1-jet bins. 

\begin{table}
  \begin{center}
    \caption{Jet definitions and selection cuts in the 
      \ATLASexp~\cite{ATLAS-CONF-2013-030} and 
      \CMSexp~\cite{CMS-PAS-HIG-13-003} analyses of  
      $\PH\to\PW\PW^*\to \flfs$ at $8\UTeV$. Partons are recombined 
into jets using the anti-$k_\rT$ algorithm \cite{Cacciari:2008gp}.  The cuts refer to various levels and 
      regions, namely event preselection ({\bf S1} cuts), the signal region 
      ({\bf S1} and {\bf S2s} cuts) and the control region 
      ({\bf S1} and {\bf S2c} cuts).  All cuts have been implemented in 
      form of a {\sc Rivet}~\cite{Buckley:2010ar} analysis.}
    \label{Tab:NLOMC_WWcuts}
    \begin{tabular}{llll}
      \hline
      {{ anti-$k_\mathrm{T}$ jets}} && \multicolumn{1}{p{5cm}}{\ATLASexp} & 
      \multicolumn{1}{p{5cm}}{\CMSexp} \\\hline 
      $R$ &  $=$    &  0.4          & 0.5\\
      $p_{\perp, j}(|\eta_{j}|)$ &$ >$  & $25\UGeV$ \hfill($|\eta_j|<2.4$) 
                                     & $30\UGeV$ \hfill($|\eta_j|<4.7$)\\ 
                                    && $30\UGeV$ \hfill($2.4<|\eta_j|<4.5$) &
      \\[2mm]\hline 
      {{\bf S1}} && \multicolumn{1}{p{5cm}}{\ATLASexp} & 
      \multicolumn{1}{p{5cm}}{\CMSexp} \\\hline 
      $p_{\perp,\{\ell_1,\,\ell_2\}}$ &$>$  & $25, 15\UGeV$   & $20, 10 \UGeV$\\
      $|\eta_{\{\Pe,\,\PGm\}}|$ &$ <$       & 2.47, 2.5    & 2.5, 2.4\\
      $|\eta_{\Pe}|$           & $\notin$    & $[1.37,1.57]$   & \\
      $p_{\perp,\ell\ell'}$ &$ >$         & see {{\bf S2s}, {\bf S2c}} 
                                                     &  $30 \UGeV$\\
      $m_{\ell\ell'}$ &$ >$              & $10\UGeV$       & $12\UGeV$\\
      $E\!\!\!\!/_{\mathrm T}^{\mathrm{(proj)}}$ &$ >$ & $25\UGeV$      & $20\UGeV$ $^\dagger$
      \\[2mm]\hline 
      {\bf S2s} &  & \multicolumn{1}{p{5cm}}{\ATLASexp} & 
      \multicolumn{1}{p{5cm}}{\CMSexp} \\\hline
      $\Delta\phi_{\ell\ell',\,E\!\!\!/_{\mathrm T}}$&$>$ & $\pi/2$ \hfill(0 jets only) & \\ 
      $p_{\perp,\ell\ell'}$ &$ >$         & $30\UGeV$ \hfill(0 jets only) & \\
      $\Delta\phi_{\ell\ell'}$ &$ <$    &  1.8 rad       &  \\
      $m_{\ell\ell'}$ &$ <$             &  $50\UGeV$        & $200\UGeV$\\
      $m_{\mathrm T}$ &$ \in$                   &                & $[60\UGeV,\,
                                                           280\UGeV]$
\;$^\ddagger$\\[2mm]\hline 
      {{\bf S2c}} &  & \multicolumn{1}{p{5cm}}{\ATLASexp} & 
      \multicolumn{1}{p{5cm}}{\CMSexp} \\\hline 
      $\Delta\phi_{\ell\ell',\,E\!\!\!/_{\mathrm T}}$&$>$ & $\pi/2$ \hfill(0 jets only) & \\ 
      $p_{\perp,\ell\ell'}$ &$ >$         & $30\UGeV$ \hfill(0 jets only) & \\
      $m_{\ell\ell'}$ &            & $\in [50,100]\UGeV$ \hfill(0 jets only)     
      & $>100\UGeV$\\
      &             & $>80\UGeV$ \hfill(1 jet only)   & \\
      \hline
    \end{tabular}
  \end{center}
\end{table}

\subsubsection{Results}
Predictions and theoretical errors\footnote{
  Scale variations in \MCatNLO predictions are not considered.} 
for exclusive jet cross sections in the signal and control regions, as well 
as their ratios, are displayed in \refTs{Tab:NLOMC_WWATLAS} 
and~\ref{Tab:NLOMC_WWCMS} for the \ATLASexp and \CMSexp analyses, respectively.
Comparing \NLOacc, \MCatNLO and \MEPSatNLO results at the same central
scale we observe that deviations between \NLOacc and \MEPSatNLO predictions
amount to only $0.5\%$ and $1{-}3\%$ in the 0-jets and 1-jet bins, respectively. The 
differences between \MCatNLO and \MEPSatNLO results are larger and 
reach %5{-}8\%$ ($20{-}22\%$) in the 0(1)-jets bins.
The sizable deficit of the \MCatNLO simulation in the 1-jet bin is not surprising 
given the limited (LO) accuracy of this approximation for exclusive 1-jet 
final states.

Differences between the \NLOacc, \MCatNLO and \MEPSatNLO approximations are
fairly similar in the various analyses and kinematic regions, and the
agreement between the various approximations is at the $0.5{-}3\%$ level in the
$\sigsr/\sigcr$ ratios.  The interpretation of this result as theoretical
uncertainty in the extrapolation from control to signal regions requires a
careful analysis of shape uncertainties, where also the (un)correlation of
scale choices in the various predictions should be considered.  These subtle
issues are beyond the scope of the present study.  We thus refrain from
assigning theoretical uncertainties to the individual signal-to-control
ratios.  In contrast, scale variations of the absolute \MEPSatNLO cross
sections in the various regions and jet bins can be regarded as a realistic
estimate of perturbative theory errors.
Adding $(\muR,\muF)$ and $\mu_Q$ variations in quadrature\footnote{
Variations of $(\muR,\muF)$ and
$\mu_Q$ can be regarded as uncorrelated and thus added in quadrature.
Another natural way of combining these two sources of 
uncertainty is to consider simultaneous variations of $(\muR,\muF,\mu_Q)$,
excluding rescalings in opposite directions as usual.
The variations resulting from this alternative approach are likely to be
even smaller than those obtained by adding  QCD and resummation scale
uncertainties in quadrature.}
we find a combined \MEPSatNLO uncertainty of $3{-}4\%$ in both jet bins.
%The dominant contribution arises from renormalization and 
%factorization scale variations, which give similarly large uncertainties as 
%parton-level \NLOacc calculations.
The dominant contribution arises from renormalization and factorization
scale variations, which turn out to be fairly consistent but slightly different
from the scale variations of the parton-level \NLOacc calculation.
The observed (minor) differences can be
attributed to the variation of extra $\alphas$ terms originating from the
shower, the CKKW scale choice in the \MEPSatNLO method, and the merging of
different jet multiplicities, which opens gluon-initiated channels also
in the 0-jets bin.  Moreover, \MEPSatNLO uncertainties might be slightly
overestimated due to statistical fluctuations at 1\% level.  

Resummation scale variations of \MEPSatNLO cross sections turn out to be
rather small.
This suggests that higher-order
subleading Sudakov logarithms, which are beyond the accuracy of the shower, 
are quite suppressed. The rather good agreement between
\NLOacc and \MEPSatNLO predictions indicates that also leading-logarithmic resummation
effects have a quantitatively small impact on the considered observables. 
\begin{table}
  \begin{center}
    \caption{Exclusive 0-jet and 1-jet bin cross sections in the signal (S2s) 
      and control (S2c) regions for the \protect\ATLASexp analysis at 
      $8\UTeV$.  The production of $\flfs+0,1$ jets is described with 
      \protect\NLOacc, \protect\MCatNLO (inclusive), and \protect\MEPSatNLO 
      accuracy as described in the text.  Results at \protect\NLOacc for 
      the $N$-jet bin correspond to $\flfs+N$ jets production.  Variations 
      of the $(\muR, \muF)$ and $\mu_Q$ scales are labeled 
      as $\delqcd$ and $\delres$, respectively.  
      Statistical errors are given in parenthesis.}
    \label{Tab:NLOMC_WWATLAS}
% ATLAS
    \begin{tabular}{lccc}
      \hline
      0-jets bin   &               \NLOacc $\pm\delqcd$       
      & \MCatNLO    & \MEPSatNLO $\pm\delqcd$ $\pm\delres$ \\[1mm]\hline
      $\sigsr$ [fb]      &  $35.08(9)\;^{+2.0\%}_{-1.9\%}$ 
      & $33.33(8)$  & $35.21(15)\;^{+1.8\%}_{-3.3\%}$$\;^{+1.7\%}_{-0.6\%}$
      \\[1mm] 
      $\sigcr$ [fb]      &  $57.05(9)\;^{+2.1\%}_{-2.0\%}$ 
      & $53.76(9)$  & $56.76(17)\;^{+2.3\%}_{-3.6\%}$$\;^{+1.9\%}_{-0.3\%}$
      \\[1mm]
      $\sigsr/\sigcr$  &  $0.615$                         
      & $0.620$   & $0.620$ 
      \\[1mm]\hline 
      1-jet bin  &               \NLOacc $\pm\delqcd$         
      & \MCatNLO    & \MEPSatNLO $\pm\delqcd$ $\pm\delres$ \\[1mm]\hline
      $\sigsr$ [fb]      &  $9.43(3)\;^{+1.8\%}_{-4.7\%}$ 
      & $7.43(4)$  & $9.23(9)\;^{+3.5\%}_{-1.9\%}$$\;^{+0.9\%}_{-0\%}$
      \\[1mm]
      $\sigcr$ [fb]      &  $29.14(6)\;^{+1.6\%}_{-4.7\%}$ 
      & $22.59(7)$  & $28.80(21)\;^{+3.1\%}_{-2.5\%}$$\;^{+1.4\%}_{-1.7\%}$
      \\[1mm] 
      $\sigsr/\sigcr$  &  $0.324$                         
      & $0.329$   & $0.320$ \\[1mm]\hline
    \end{tabular}
  \end{center}
\end{table}

\begin{table}
  \begin{center}
    \caption{Exclusive 0-jet and 1-jet bin cross sections in the signal (S2s) 
      and control (S2c) regions for the \protect\CMSexp analysis at $8\UTeV$, 
      with the same structure and conventions as in \refT{Tab:NLOMC_WWATLAS}.}
    \label{Tab:NLOMC_WWCMS}
% CMS   
    \begin{tabular}{lccc}
      \hline
      0-jets bin   &  \NLOacc $\pm\delqcd$                        
      & \MCatNLO    & \MEPSatNLO $\pm\delqcd$ $\pm\delres$ \\[1mm]\hline
      $\sigsr$ [fb]      &  $159.34(18)\;^{+1.8\%}_{-1.7\%}$ 
      & $150.6(2)$  & $160.3(3)\;^{+2.6\%}_{-3.8\%}$$\;^{+1.4\%}_{-0.5\%}$\\[1mm]
      $\sigcr$ [fb]      &  $60.08(15)\;^{+1.5\%}_{-1.4\%}$ 
      & $56.60(11)$  & $60.31(22)\;^{+3.6\%}_{-3.5\%}$$\;^{+0.7\%}_{-0.2\%}$
      \\[1mm]
      $\sigsr/\sigcr$  &  $2.65$                         
      & $2.66$   & $2.66$ \\[1mm]\hline 
      1-jet bin  &  \NLOacc $\pm\delqcd$                        
      & \MCatNLO    & \MEPSatNLO $\pm\delqcd$ $\pm\delres$ \\[1mm]\hline
      $\sigsr$ [fb]      &  $45.01(7)\;^{+1.3\%}_{-2.6\%}$ 
      & $34.75(9)$  & $44.88(23)\;^{+3.0\%}_{-2.7\%}$$\;^{+0.1\%}_{-0.3\%}$
      \\[1mm]
      $\sigcr$ [fb]      &  $22.09(5)\;^{+1.2\%}_{-3.2\%}$ 
      & $17.41(7)$  & $22.30(18)\;^{+3.0\%}_{-2.7\%}$$\;^{+0.5\%}_{-0.4\%}$
      \\[1mm]
      $\sigsr/\sigcr$  &  $2.04$                         
      & $2.00$   & $2.01$ \\[1mm]\hline
    \end{tabular}
  \end{center}
\end{table}

\begin{figure}
\begin{center}
\includegraphics[width=0.48\textwidth]{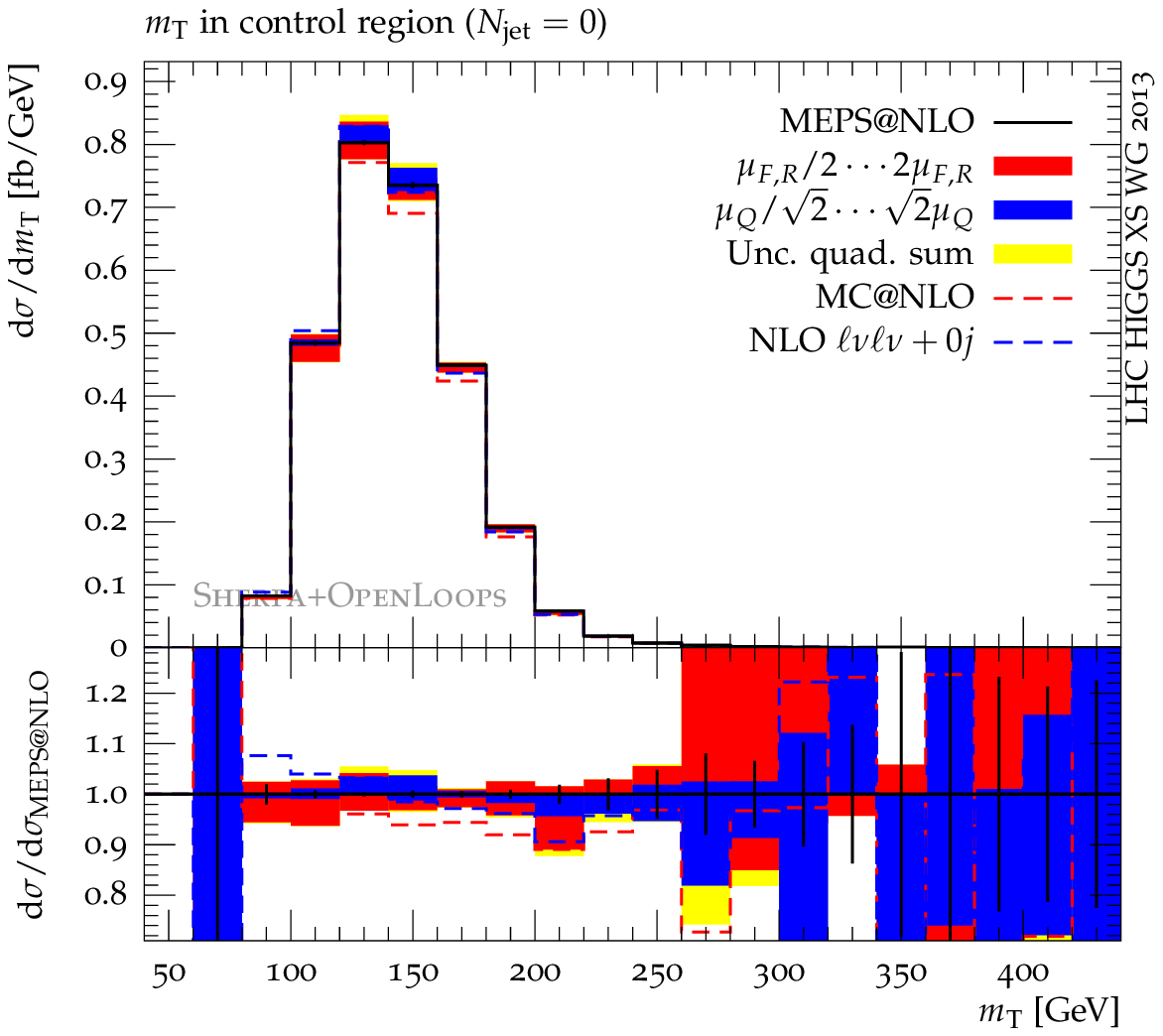}
\includegraphics[width=0.48\textwidth]{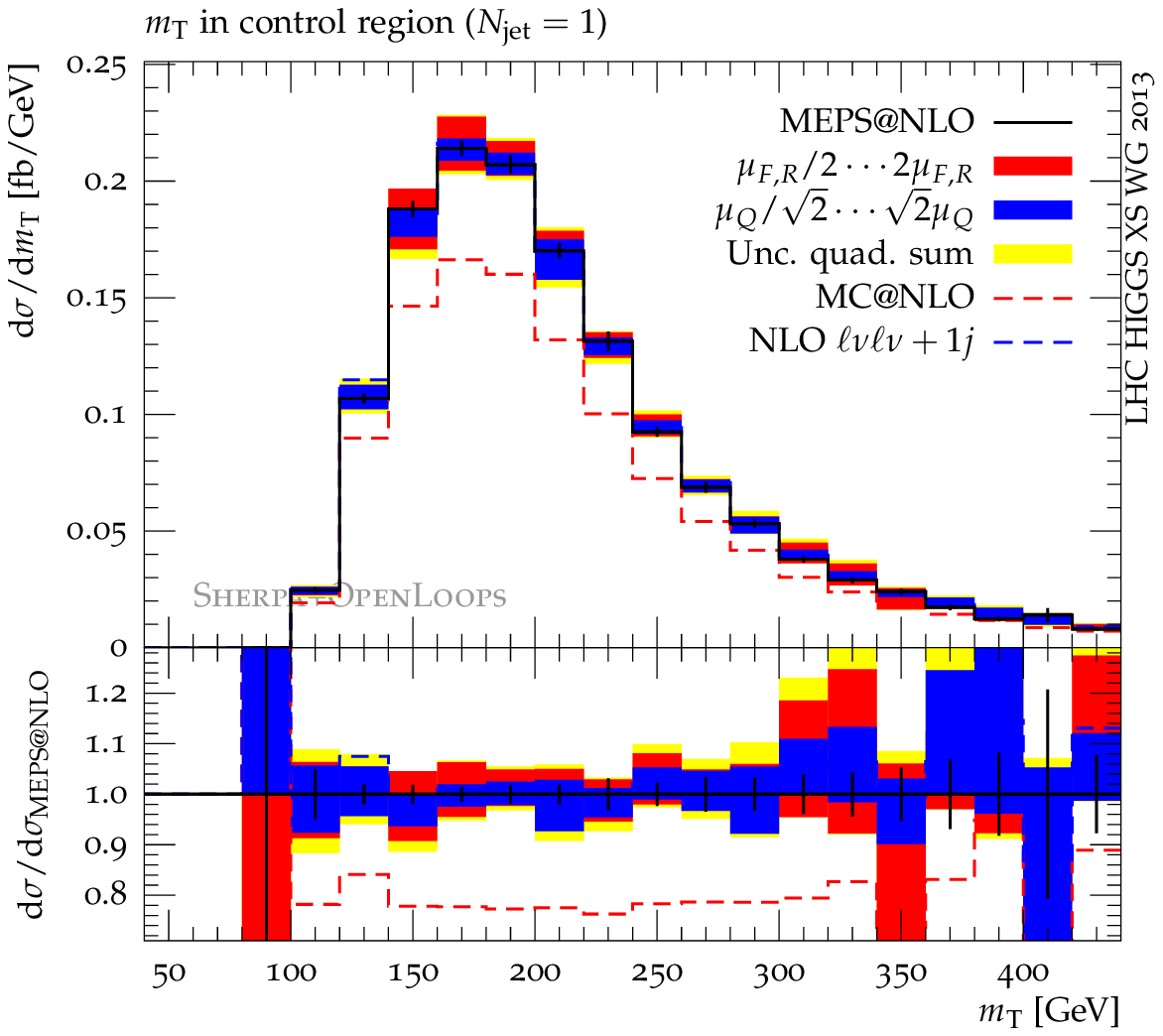}
\end{center}
\vspace*{-0.3cm}
\caption{Control region of the \protect\ATLASexp analysis at $8\UTeV$:
  transverse-mass distribution in the 0-jets (left) and 1-jet (right) bins.
  \protect\MEPSatNLO (black solid), inclusive \protect\MCatNLO (red dashed), and 
  \protect\NLOacc (blue dashed) predictions at the central scale 
  $\muR=\muF=\mu_Q=m_{\ell\ell'\PGn\PGn'}$.  In the lower panel, showing
  the deviations and uncertainties, results are normalized to the central
  \protect\MEPSatNLO predictions.  The factor-2 variations of $\muR$ and 
  $\muF$ (red band), and factor-$\sqrt{2}$ variations of $\mu_Q$ (blue band),
  are combined in quadrature (yellow band).
  Scale variation bands are color additive, \ie  yellow+blue=green, 
yellow+red=orange, and yellow+red+blue=brown.
} 
\label{fig:flj_atlas_c_mT}%\refF{fig:flj_atlas_c_mT}
\end{figure}

\begin{figure}
\begin{center}
\includegraphics[width=0.48\textwidth]{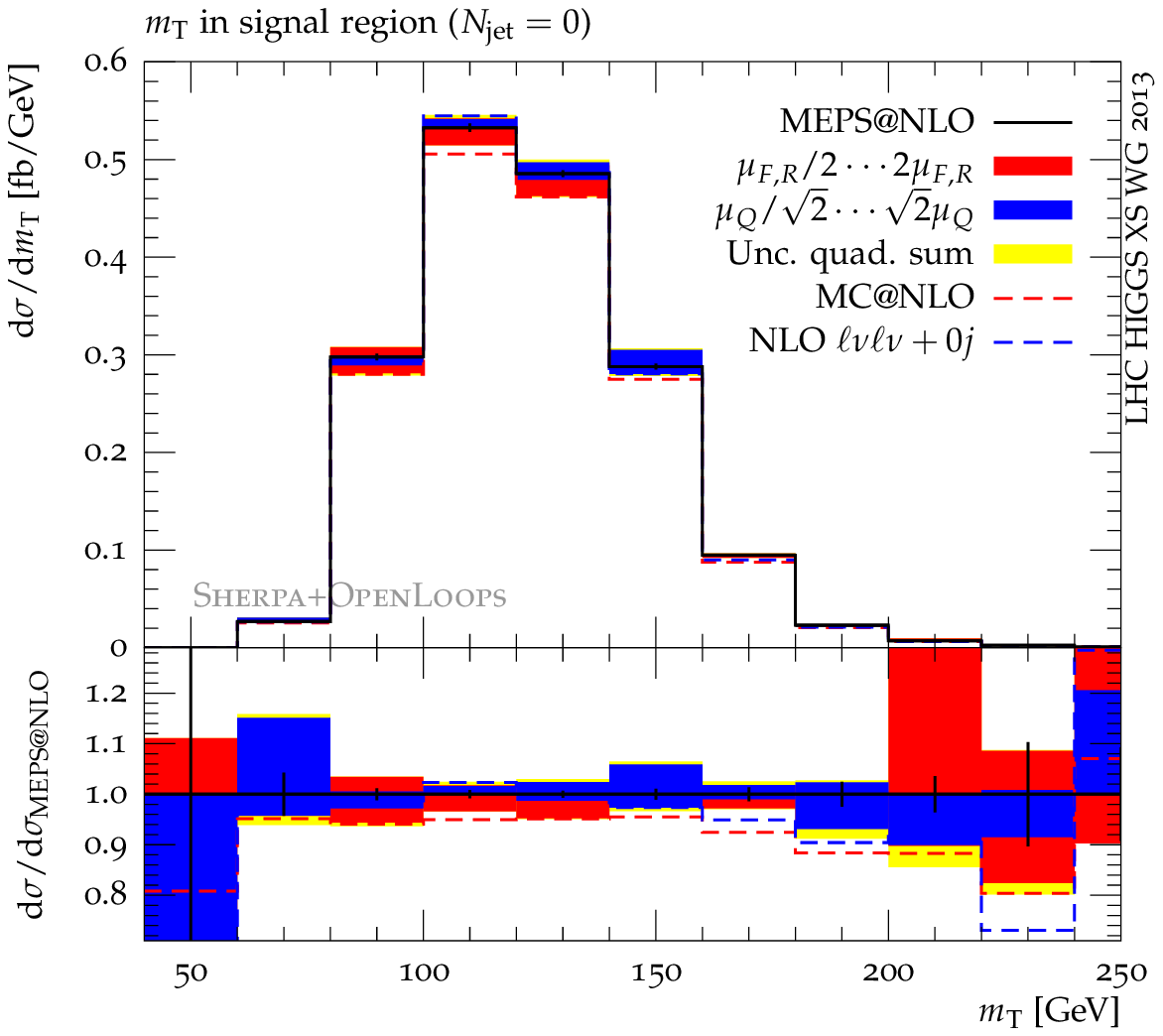}
\includegraphics[width=0.48\textwidth]{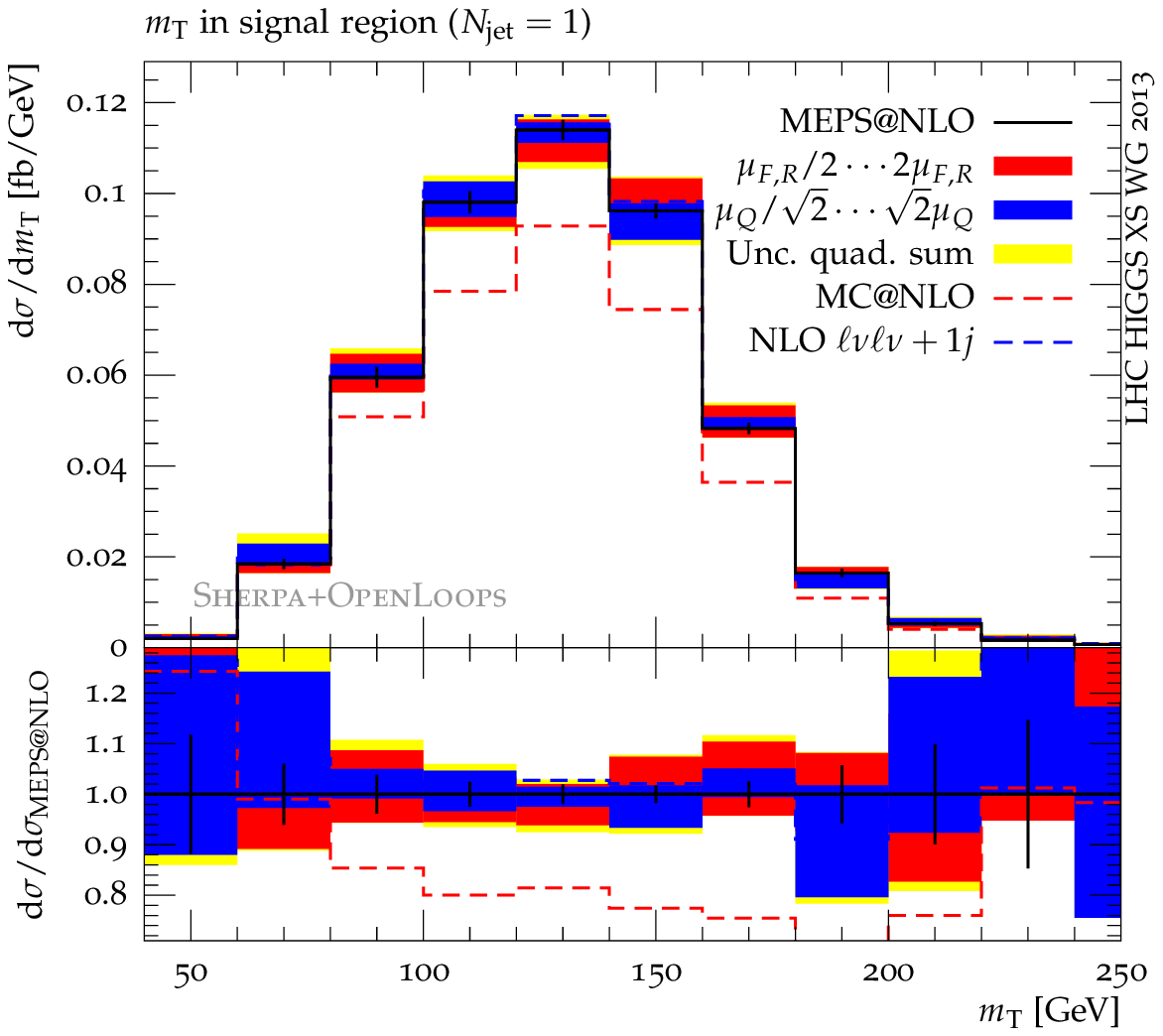}
\end{center}
\vspace*{-0.3cm}
\caption{Signal region of the \ATLASexp analysis at $8\UTeV$:
  transverse-mass distribution in the 0-jets (left) and 1-jet (right) bins.
  Same approximations and conventions as in \refF{fig:flj_atlas_c_mT}.
}
\label{fig:flj_atlas_s_mT}%\refF{fig:flj_atlas_s_mT}
\end{figure}

\begin{figure}
\begin{center}
\includegraphics[width=0.48\textwidth]{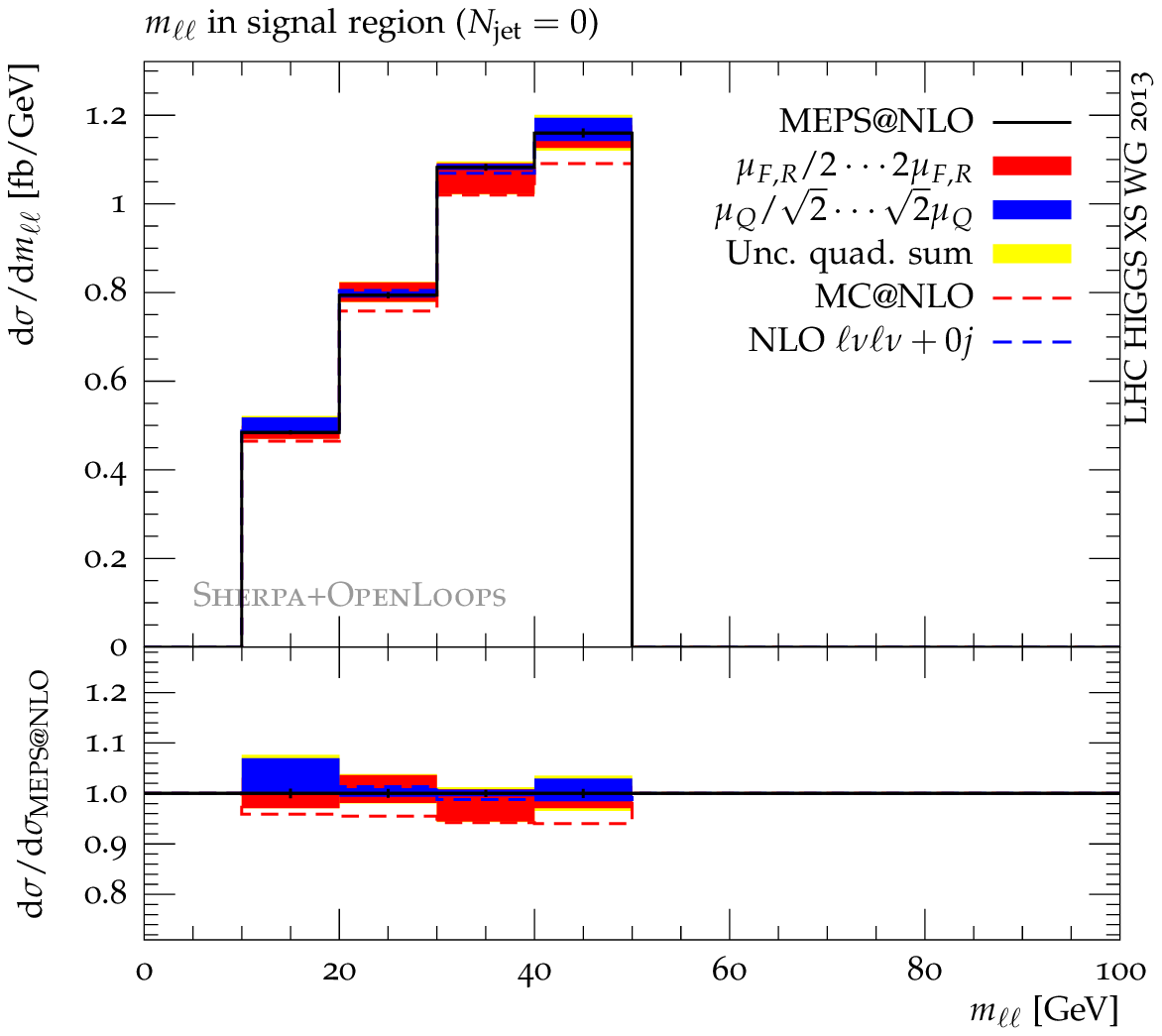}
\includegraphics[width=0.48\textwidth]{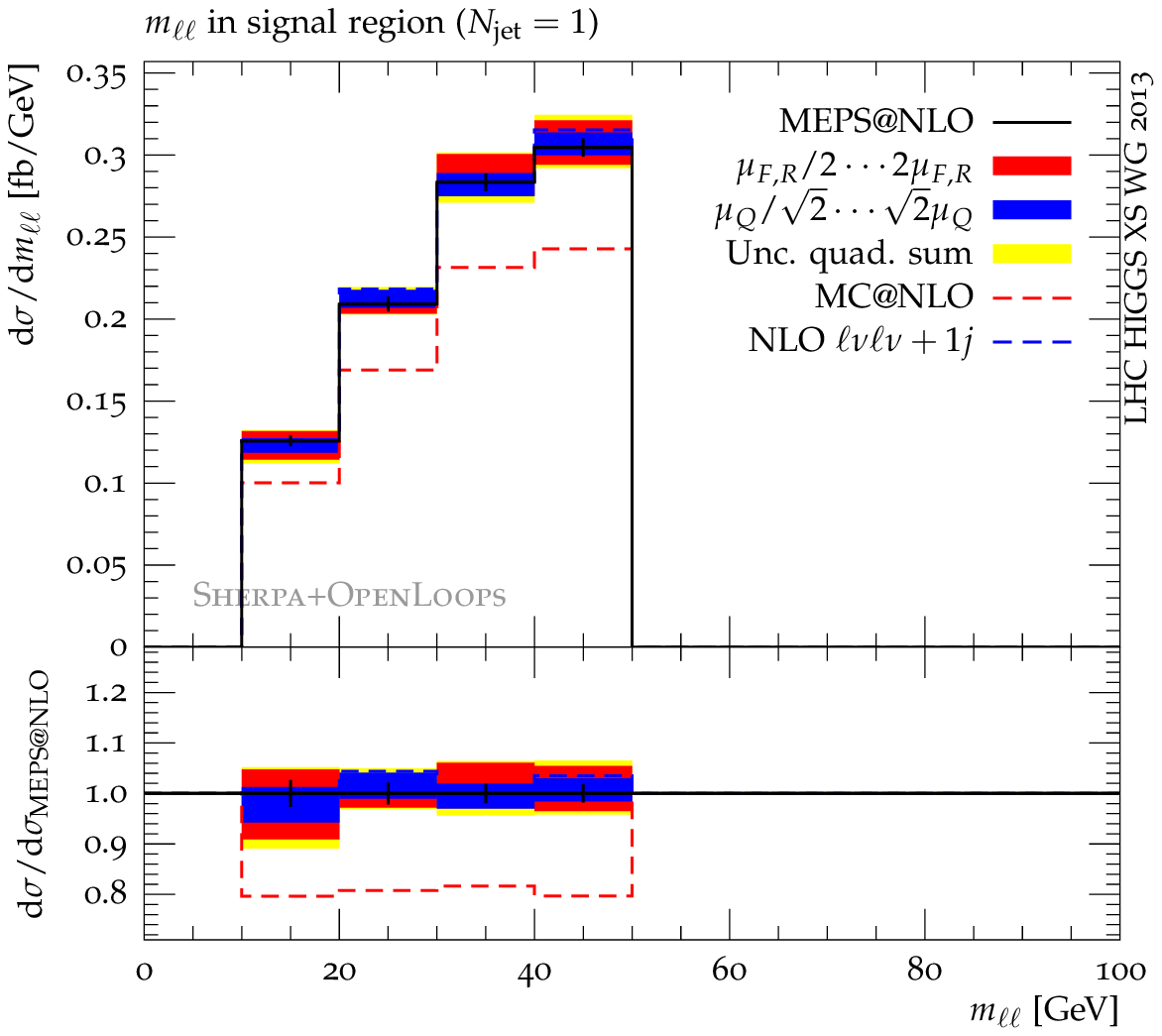}
\end{center}
\vspace*{-0.3cm}
\caption{Signal region of the \protect\ATLASexp analysis at $8\UTeV$: dilepton 
  invariant-mass distribution in the 0-jets (left) and 1-jet (right) bins.
  Same approximations and conventions as in \refF{fig:flj_atlas_c_mT}.}
\label{fig:flj_atlas_s_mll}%\refF{fig:flj_atlas_s_mll}
\end{figure}

\begin{figure}
\begin{center}
\includegraphics[width=0.48\textwidth]{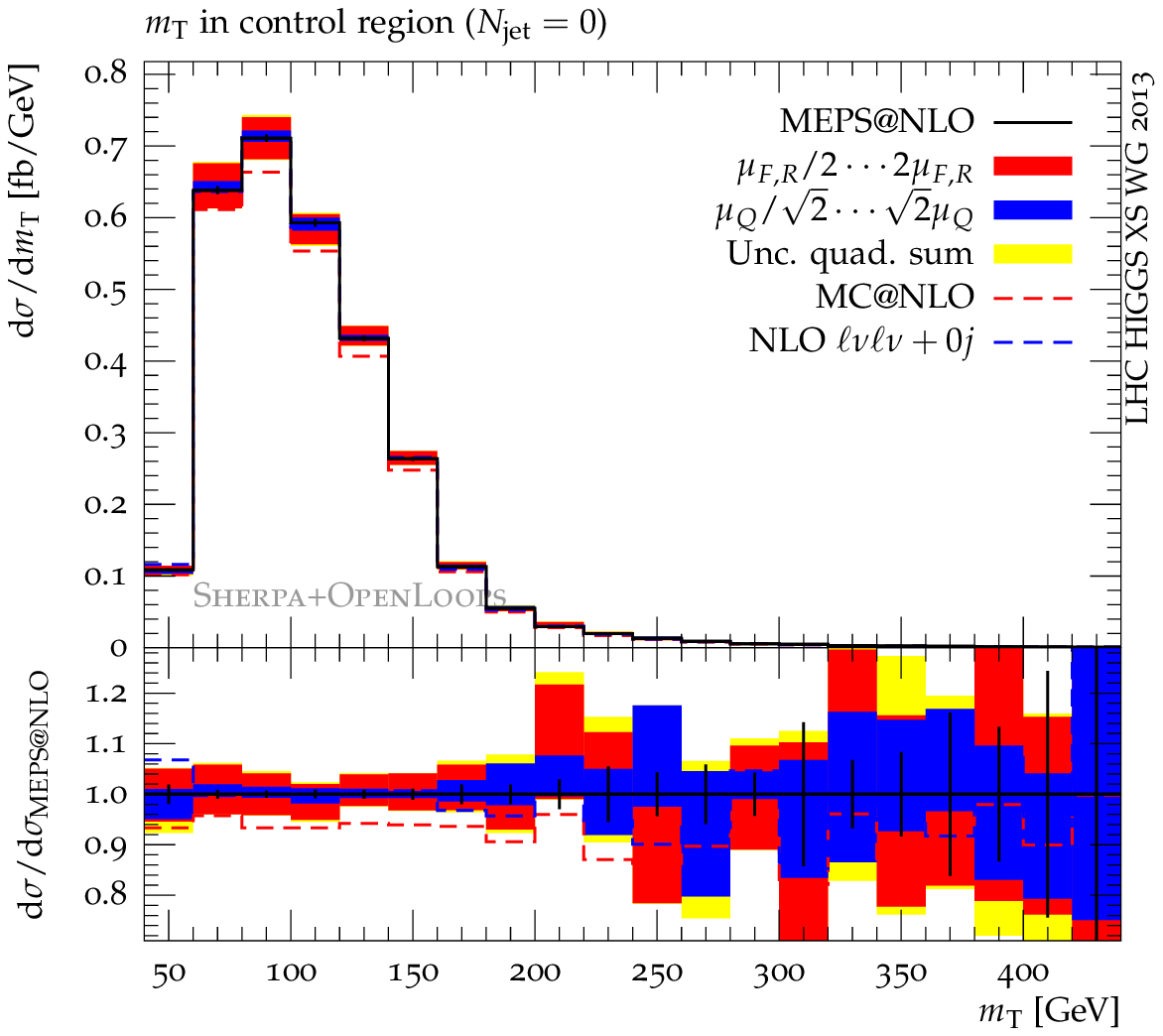}
\includegraphics[width=0.48\textwidth]{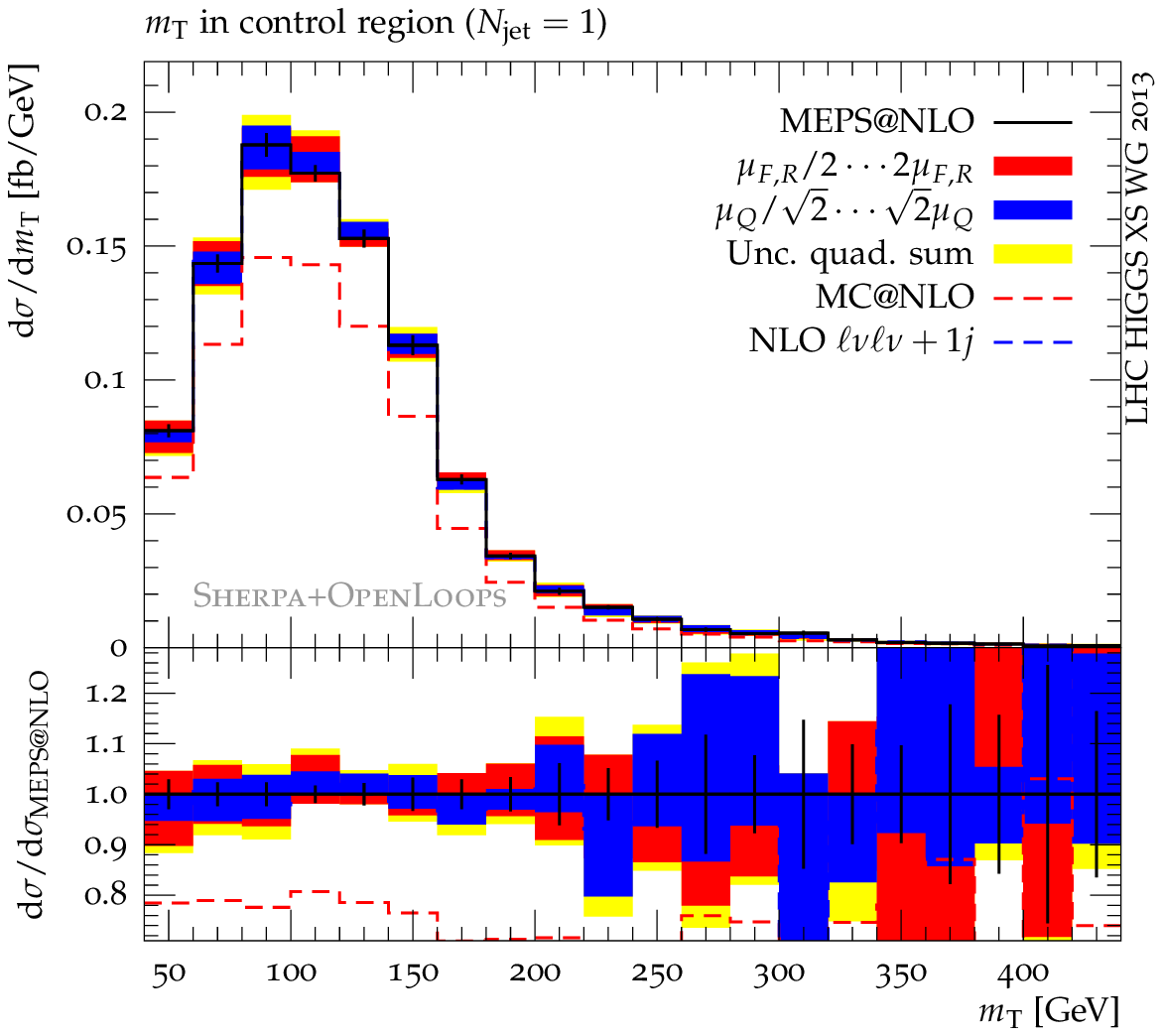}
\end{center}
\vspace*{-0.3cm}
\caption{Control region of the \protect\CMSexp analysis at $8\UTeV$:
  transverse-mass distribution in the 0-jets (left) and 1-jet (right) bins.
  Same approximations and conventions as in \refF{fig:flj_atlas_c_mT}.}
\label{fig:flj_cms_c_mT}%\refF{fig:flj_cms_c_mT}
\end{figure}

\begin{figure}
\begin{center}
\includegraphics[width=0.48\textwidth]{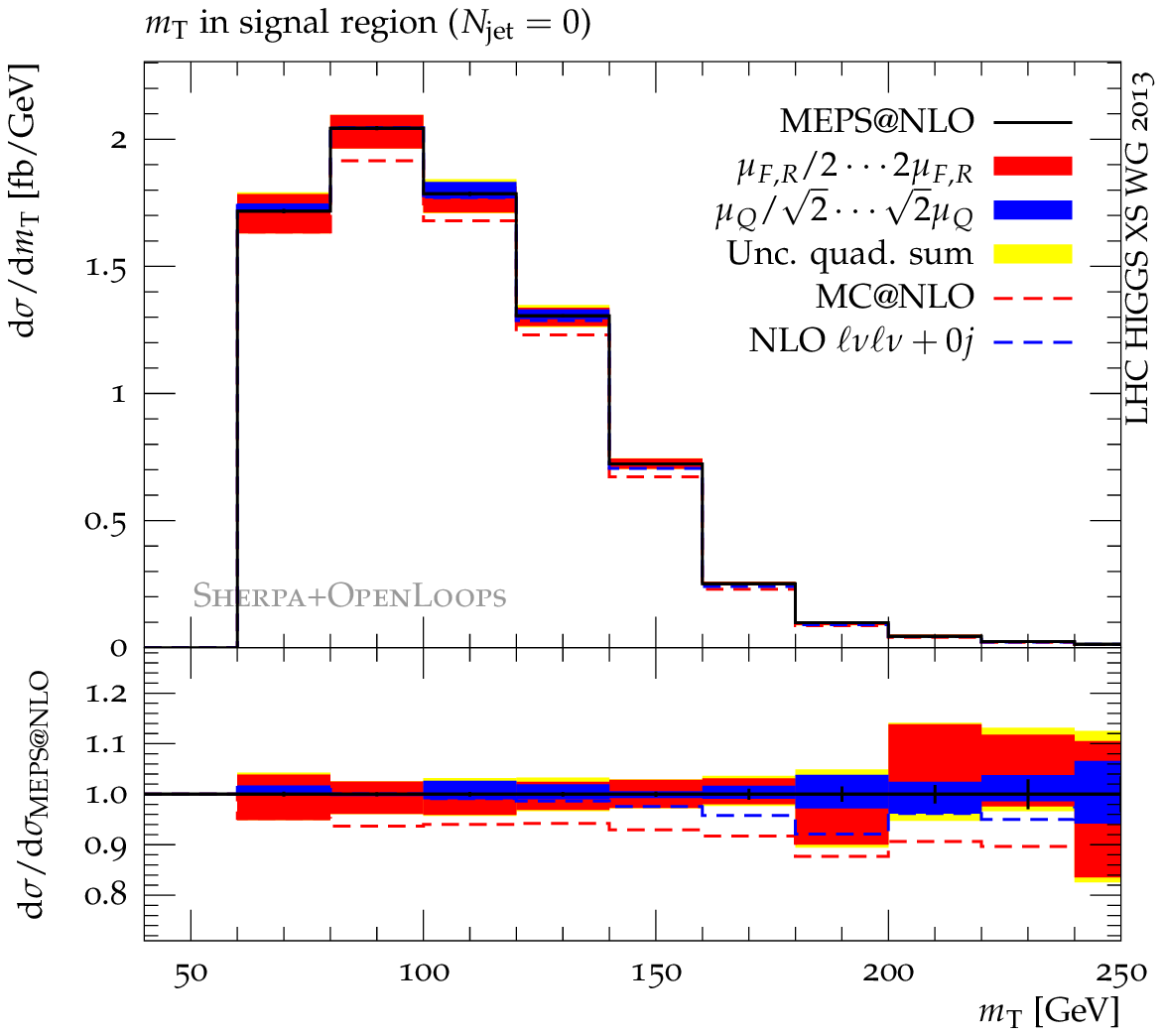}
\includegraphics[width=0.48\textwidth]{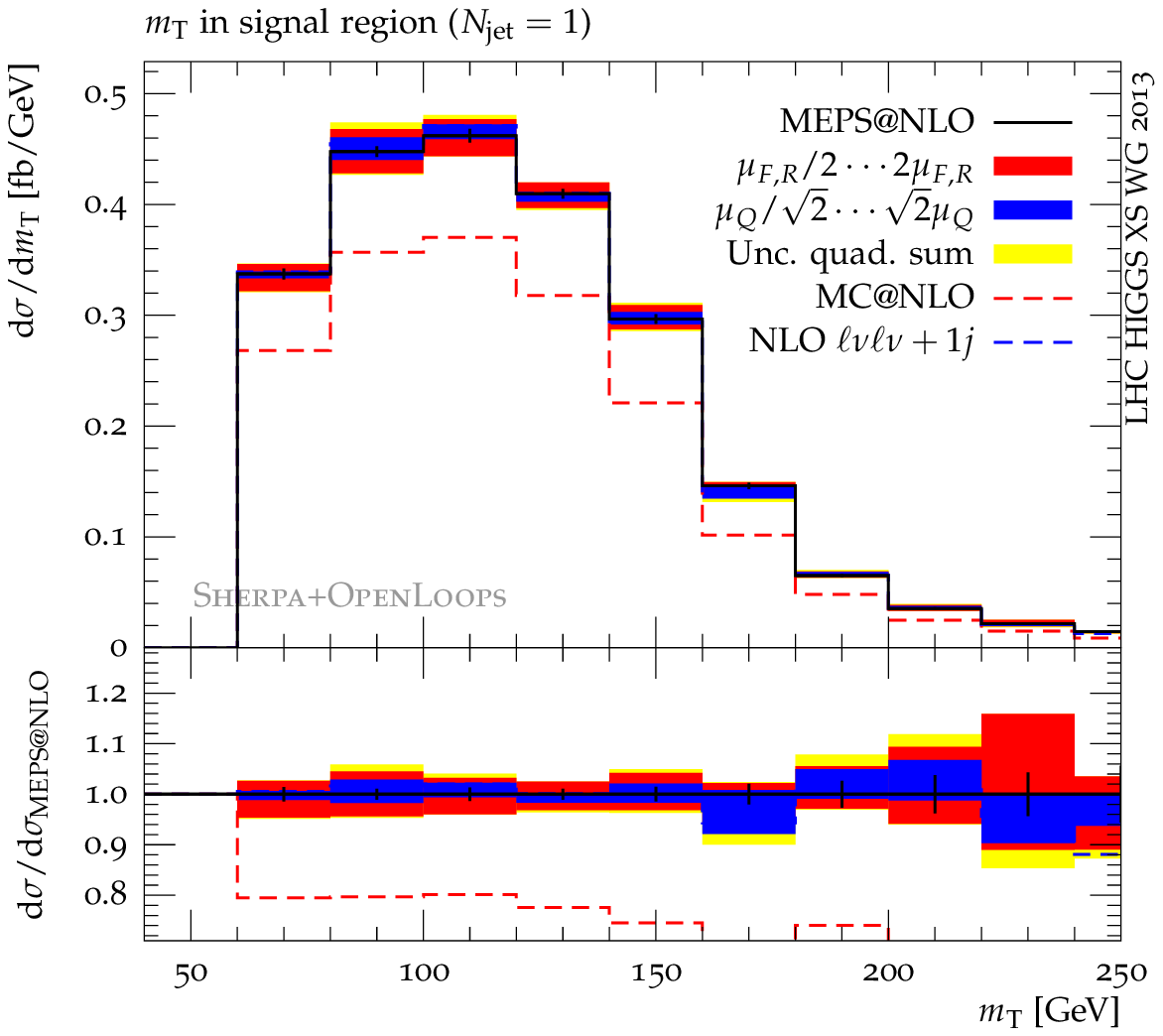}
\end{center}
\vspace*{-0.3cm}
\caption{Signal region of the \protect\CMSexp analysis at $8\UTeV$:
  transverse-mass distribution in the 0-jets (left) and 1-jet (right) bins.
  Same approximations and conventions as in \refF{fig:flj_atlas_c_mT}.}
\label{fig:flj_cms_s_mT}%\refF{fig:flj_cms_s_mT}
\end{figure}

\begin{figure}
\begin{center}
\includegraphics[width=0.48\textwidth]{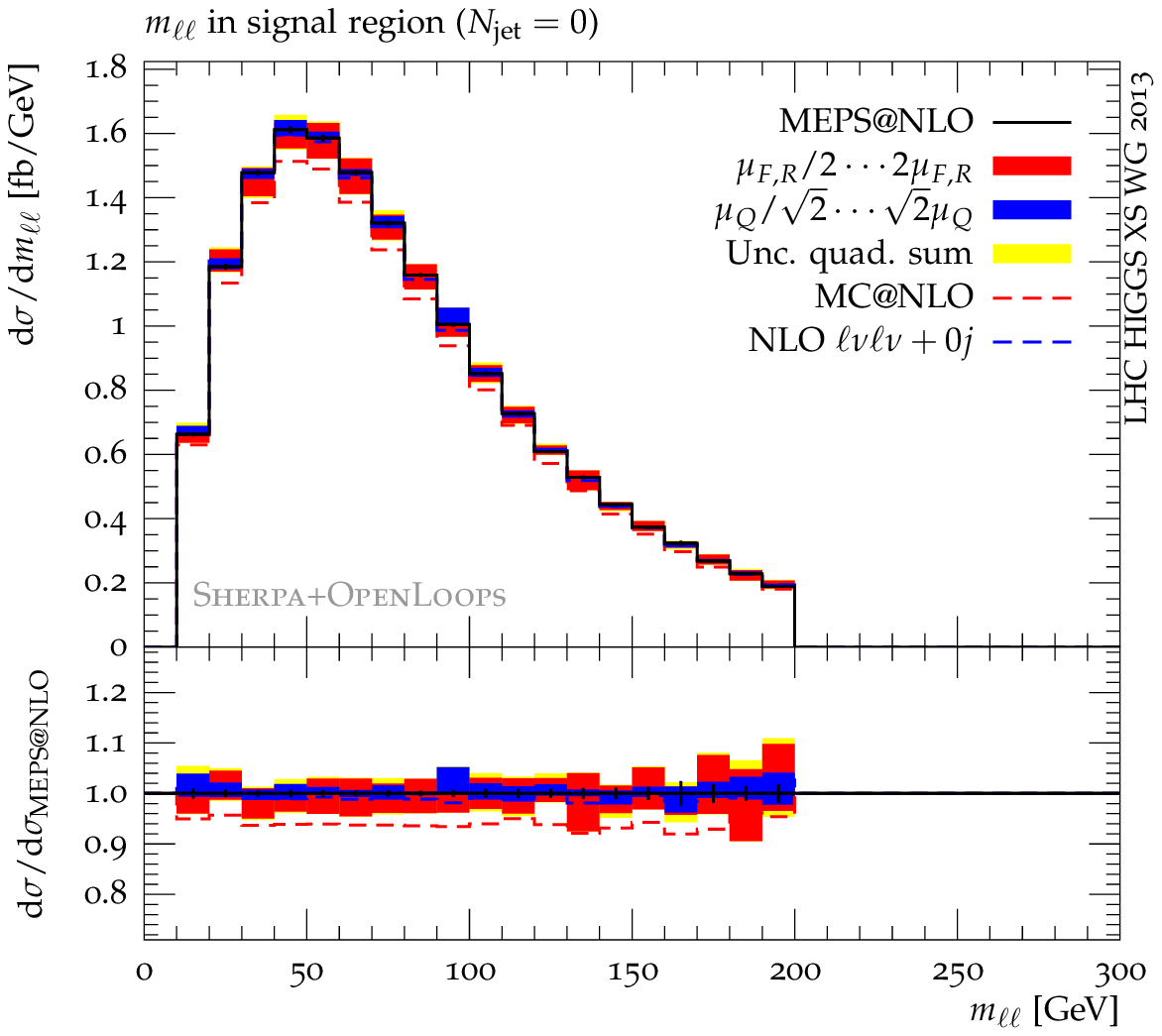}
\includegraphics[width=0.48\textwidth]{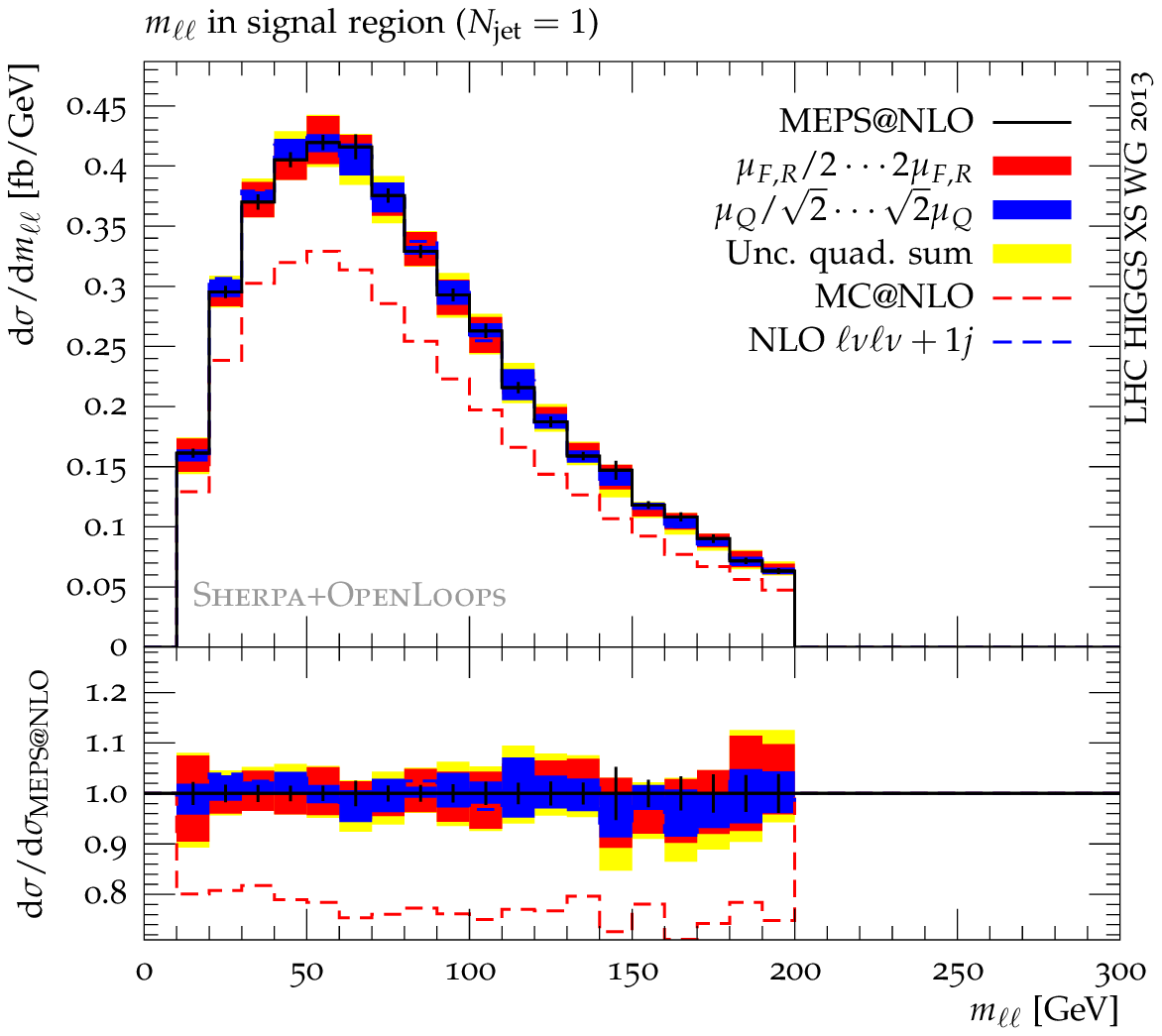}
\end{center}
\vspace*{-0.3cm}
\caption{Signal region of the \protect\CMSexp analysis at $8\UTeV$:
  dilepton invariant-mass distribution in the 0-jets (left) and 1-jet (right) 
  bins.  Same approximations and conventions as in \refF{fig:flj_atlas_c_mT}.}
\label{fig:flj_cms_s_mll}%\refF{fig:flj_cms_s_mll}
\end{figure}

In \refFs{fig:flj_atlas_c_mT}--\ref{fig:flj_atlas_s_mll} distributions for 
the transverse mass $m_\rT$ and the dilepton mass $m_{\ell\ell'}$ are displayed 
for the \ATLASexp analysis at $8\UTeV$.  Similar plots for the \CMSexp 
analysis are shown in \refFs{fig:flj_cms_c_mT}--\ref{fig:flj_cms_s_mll}.  
These observables provide high sensitivity to the Higgs signal.  They are 
depicted separately in the exclusive 0- and 1-jet bins in the signal 
({\bf S1} and {\bf S2s} cuts) and control ({\bf S1} and {\bf S2c} cuts) regions.
In the lower frames, the various predictions are normalized to \MEPSatNLO 
results at the central scale.  Scale variations are given only for the 
\MEPSatNLO case.  Specifically, the individual 
$(\muR,\muF)$ and $\mu_Q$ variations and their 
combination in quadrature are displayed as three separate 
color-additive bands.

Comparing \NLOacc, \MCatNLO and \MEPSatNLO distributions, agreement on the 
few-percent level in the 0-jet is found, while in the 1-jet bin 
discrepancies between \MCatNLO and \MEPSatNLO on the 20\% level appear.  
This is consistent with the results in 
\refT{Tab:NLOMC_WWATLAS} and \refT{Tab:NLOMC_WWCMS}.  In the case of both
$m_\rT$ and $m_{\ell\ell'}$ distributions, also some shape distortions emerge.
However, as could be anticipated they are fairly moderate in the 0-jet bins
and reach up to about 20\% in the hard region of the 1-jet bins.  
In the tails of some distributions scale-variation bands are 
distorted by statistical fluctuations.
These features are qualitatively similar in the \ATLASexp 
and \CMSexp setups.  

% \ATLASexp plots

% \CMSexp plots

\subsubsection{Conclusion and outlook}
The findings of this contribution can be summarized as follows:
\begin{itemize}
\item Analyzing the leptonic observables $m_\rT$ and $m_{\ell\ell'}$
  only moderate shape distortions emerge between \MCatNLO and \MEPSatNLO
  predictions, which reach up to about 20\% only in the hard tails of the leptonic
  system.  

\item The picture changes when discussing the cross sections in the different
  jet bins, where sizable differences of up to about 20\% emerge between
  \MCatNLO and \MEPSatNLO predictions.  They are well understood and reflect the fact 
  that the 1-jet contribution in \MEPSatNLO is evaluated at logarithmically 
  improved \NLOacc accuracy, while in \MCatNLO the fixed order accuracy is 
  LO only.  
  It can be expected that this trend will also
  continue, and probably become more pronounced, when studying the
  2-jet bin, relevant for the weak boson fusion channel.  This is left for
  further investigations. 

\item Comparing \MEPSatNLO and \NLOacc predictions one finds only small deviations
    at $0.5\%$ and $3\%$ level in the 0- and 1-jet bin cross sections, respectively.

\item The combined theoretical uncertainties, as estimated 
  through renormalization-, factorization- and resummation-scale variations 
  in the \MEPSatNLO approach, amount to $3{-}4\%$, both in the 0- and 1-jets bins. 
  Resummation scale variations turn out to be rather small, 
  suggesting that subleading logarithms beyond the shower accuracy
  are not important. The few-percent level agreement between \NLOacc and \MEPSatNLO 
  cross sections in both jet bins indicates that also leading-log effects beyond
  \NLOacc are rather small.

\item Although the effect of hadronization and the underlying event
  have not been investigated here, it is clear that the former will not
  lead to uncertainties that are comparable in size to the ones already 
  present.  This is not quite so clear for the underlying event, which will
  influence the observable cross sections in two ways. First of all, a
  varying hadronic activity will lead to varying acceptance of isolated
  leptons.  At the same time, variations in the underlying event 
  activity or hardness may also lead to differences in the jet multiplicity
  distribution.  The quantification of such effects will be left to further
  studies.
\end{itemize}

\subsection{NLO QCD corrections to the production of Higgs boson plus two and three jets in gluon fusion}
%\footnote{Authors: H. van Deurzen, N. Greiner, G. Luisoni, P. Mastrolia, E. Mirabella, G. Ossola, T. Peraro, 
%J.~F.~von~Soden-Fraunhofen, F. Tramontano.}}
\label{NLOMC_H3j}

In this section, we illustrate the recent  computation of
the NLO contributions to Higgs plus two jets production at the LHC in the large top-mass limit~\cite{vanDeurzen:2013rv},  
and for the first time also provide results for the one-loop virtual contribution to Higgs plus three jets production. 

The results are obtained by using a fully automated framework for
fixed order NLO QCD calculations based on the interplay of the
packages GOSAM~\cite{Cullen:2011ac} and
SHERPA~\cite{Gleisberg:2008ta}.
%Next-to-leading order corrections to cross sections require the evaluation of virtual and real emission contributions.
For the computation of the virtual corrections we use a code generated by the program package
{\Gosam}, which combines automated diagram generation and algebraic
manipulation with integrand-level reduction 
techniques~\cite{Ossola:2006us,Ellis:2007br,Mastrolia:2012an}.
More specifically, the virtual corrections are evaluated using the
$d$-dimensional integrand-level decomposition 
implemented in the \samurai\ library~\cite{Mastrolia:2010nb}, which allows for the combined
determination of both cut-constructible and rational terms at once. Moreover, the presence of effective couplings in the 
Lagrangian requires an extended version~\cite{Mastrolia:2012bu} of the
integrand-level reduction, of which the present calculation is a first application.
For the calculation of tree-level contributions we use {\Sherpa}~\cite{Gleisberg:2008ta}, which computes the 
LO and the real radiation matrix elements~\cite{Krauss:2001iv}, 
regularizes the IR and collinear singularities using the
Catani-Seymour dipole formalism~\cite{Gleisberg:2007md}, 
and carries out the  phase space integrations as well.
The code that evaluates the virtual corrections, generated by
{\Gosam}, is linked to {\Sherpa} via the 
Binoth-Les-Houches Accord (BLHA)~\cite{Binoth:2010xt} interface, which allows for a direct communication between the
two codes at running time.

%\bigskip
%\subsubsection{Results for $\Pp\Pp \to \PH jj$ with \Gosam-\Sherpa}
\subsubsection{Results for $\Pp\Pp \to \PH jj$ with {\sc GOSAM}-{\sc SHERPA}}

For $Hjj$ production, \Gosam\  identifies and generates  the following minimal set of processes
 $\Pg\Pg  \to \PH  \Pg\Pg$,  $\Pg\Pg \to \PH  \PAQq\PQq$,  $\PAQq\PQq \to \PH  \PAQq\PQq$,  
 $\PAQq\PQq \to \PH  \PQq' \PAQq' $. The other processes  are obtained by performing the appropriate symmetry transformation. 
The ultraviolet (UV), the  infrared, and the collinear
 singularities are regularized using dimensional reduction (DRED).  UV divergences
 have been renormalized in the $\overline{\mbox{MS}}$ scheme.  
Our results are in agreement with~\cite{Ellis:2005qe} and  MCFM (v6.4)~\cite{Campbell:2010cz}.
%
%After appropriate crossing to the $H \to$ 4-parton decay kinematics,
%we compared our results with the ones presented in Table~I of
%Ref.~\cite{Ellis:2005qe}, finding excellent agreement. 
%Furthermore, converting our results for the $Hjj$-production
%channels from DRED to the 't Hooft-Veltman scheme, we
%are in perfect agreement with the most recent version of MCFM (v6.4).

\begin{figure}[htb]
\begin{center}
\includegraphics[width=7.5cm]{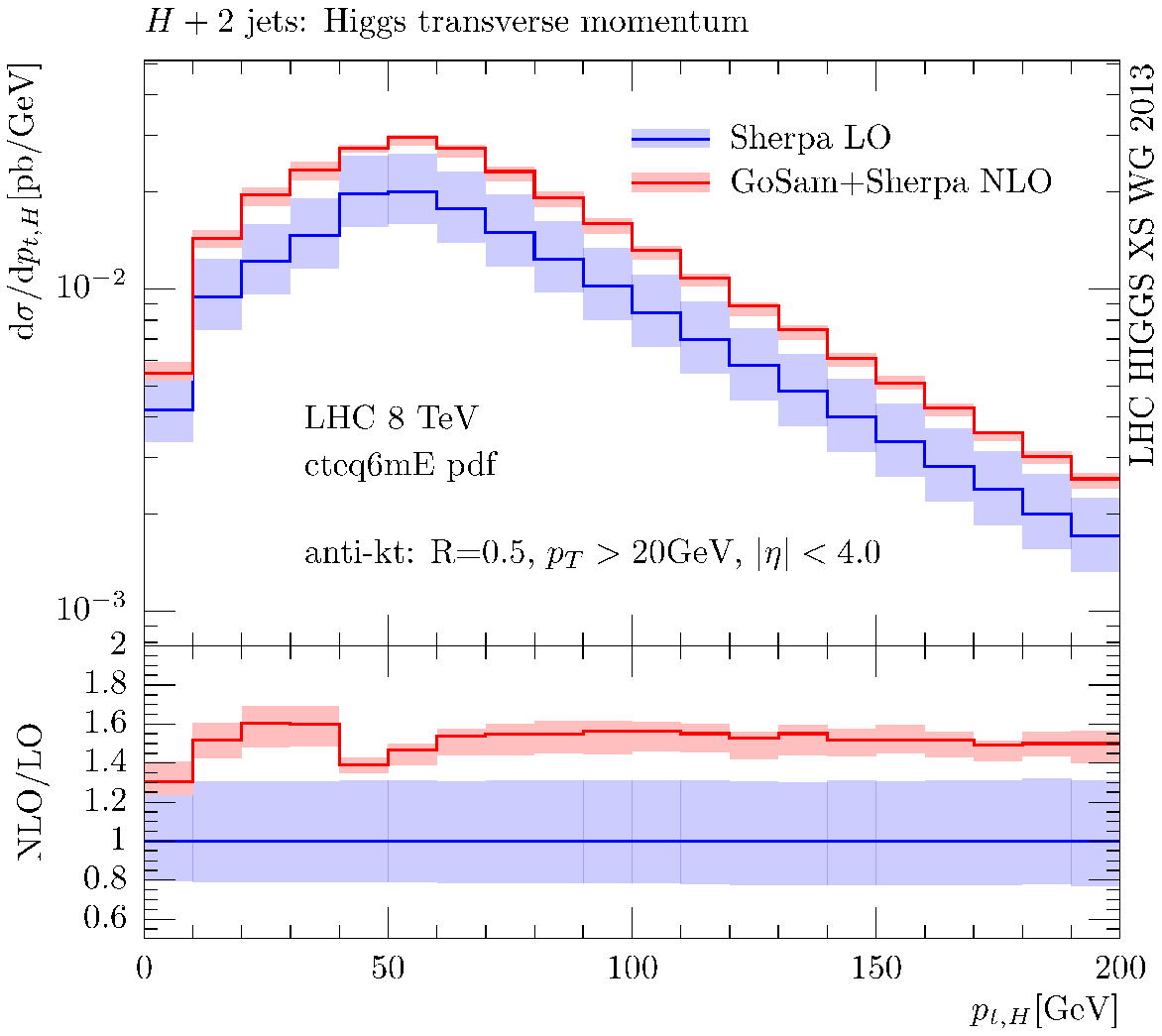} \,\,
\includegraphics[width=7.5cm]{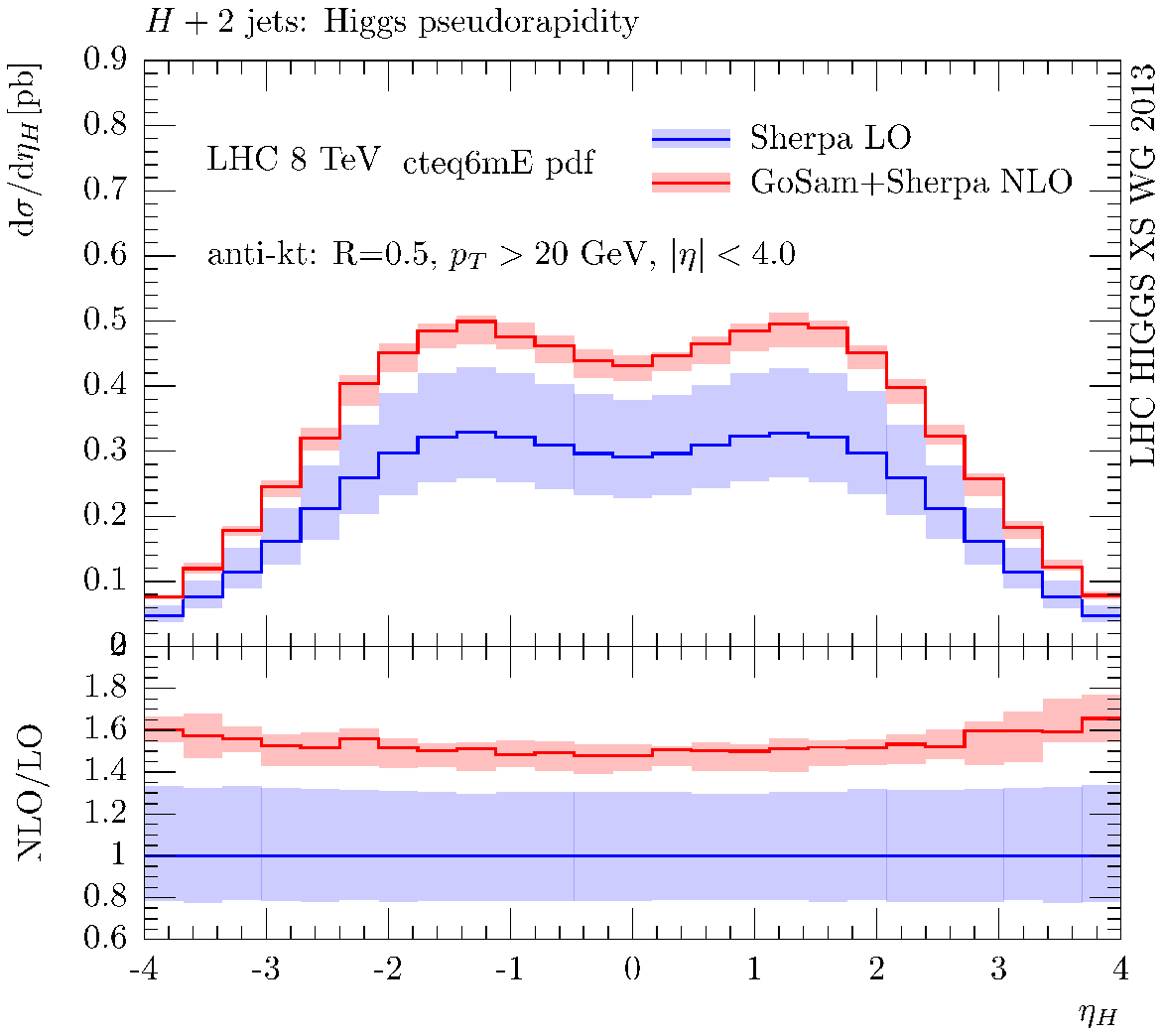} 
\caption{Transverse momentum $\pT$  and pseudorapidity $\eta$ of the Higgs boson. } \label{histo_h0}
\end{center}
\end{figure}

As an illustration of possible analyses that can be performed with 
the \Gosam-\Sherpa\ automated setup, in \Fref{histo_h0} we present  the 
distribution of the transverse momentum $\pT$ of the Higgs boson and its pseudorapidity $\eta$,  for proton-proton collisions at the LHC at $\sqrt{s}= 8\UTeV$.
Both of them show a $K\,$-factor between the LO and the NLO distribution of about $~1.5-1.6$ and a decrease of the scale uncertainty of about $50\%$.
These results are obtained using the parameters
 $\MH = 125 {\rm Gev}$,  $\GF = 1.16639 \cdot 10^{-5} {\rm Gev}^{-2}$, 
 $\alpha^{\mbox{\tiny LO}}_s(\MZ)= 0.129783$, and $\alpha^{\mbox{\tiny NLO}}_s(\MZ)= 0.117981 \,$. 
We use the CTEQ6L1 and CTEQ6mE~\cite{Pumplin:2002vw} 
parton distribution functions (PDF) for the LO and NLO, respectively. 
The jets are clustered by using the anti-$k_{\mathrm T}$ algorithm provided by the
{\tt FastJet} package with the following setup: 
$p_{t,j} \ge 20  {\rm Gev}$, $|\eta_j| \le 4.0$,  $R = 0.5$.
The Higgs boson is treated as a stable on-shell particle, without including any decay mode. To fix the factorization and the renormalization scale we define
$ \hat{H}_{t} = \sqrt{\MH^2+p_{t,H}^2}+\sum_{j} p_{t,j} $, 
where $p_{t,H}$ and $p_{t,j}$ are the transverse momenta of the Higgs boson and the jets.
The nominal value for the two scales is defined as $\mu = \muR = \muF = \hat{\PH}_{t}$,
whereas theoretical uncertainties are assessed by varying simultaneously the factorization and renormalization scales in the range
$\frac{1}{2}\hat{\PH}_{t} <  \mu  < 2 \hat{\PH}_{t}$.
The error is estimated by taking the envelope of the resulting distributions at the different scales.
 
Within our framework, we find the following total cross sections for the process $\Pp \Pp \to \PH  j j$ in gluon fusion:
$$
\sigma_{\LO} = 1.90^{+0.58}_{-0.41} \, {\rm pb},
\quad
\sigma_{\NLO} = 2.90^{+0.05}_{-0.20} \, {\rm pb}.
$$
%where the error is obtained by varying the renormalization and
%factorization scales as given above. 

%\bigskip
\subsubsection{Virtual corrections to \boldmath{$\Pp\Pp \to \PH jjj$}}

\begin{figure}[h]
\begin{center}
\includegraphics[width=3.5cm]{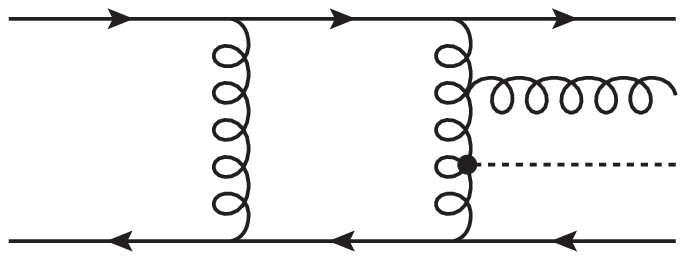} \quad \quad
\includegraphics[width=3.5cm]{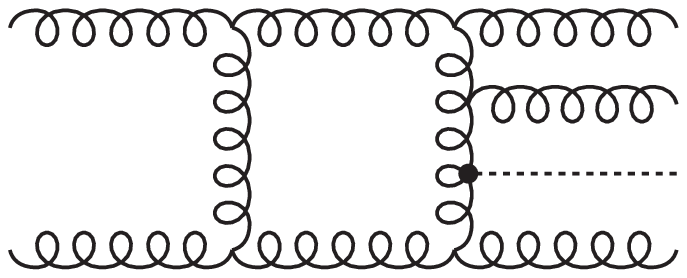} 
\caption{Sample hexagon diagrams which enter in the six-parton one-loop amplitudes for $\PAQq\PQq \to \PH  \PAQq\PQq \Pg $ and 
$\Pg\Pg  \to \PH  \Pg\Pg \Pg$. The dot represents the effective $\Pg\Pg \PH$ vertex.} 
\label{Fig:exagons}
\end{center}
\end{figure}

All independent processes contributing to $\PH jjj$ 
can be obtained by adding one extra gluon to the final state of the
processes listed in the case of $\PH jj$. Accordingly, we 
generate the codes for the virtual corrections to the processes 
$\Pg\Pg  \to \PH  \Pg\Pg \Pg$,  $\Pg\Pg \to \PH  \PAQq\PQq \Pg$, $\PAQq\PQq \to \PH  \PAQq\PQq \Pg$,
$\PQq \PAQq  \to \PH  \PQq' \PAQq' \Pg$. Some representative one-loop diagrams are depicted in 
\Fref{Fig:exagons}.

\begin{figure}[h]
\begin{center}
\includegraphics[width=12.0cm]{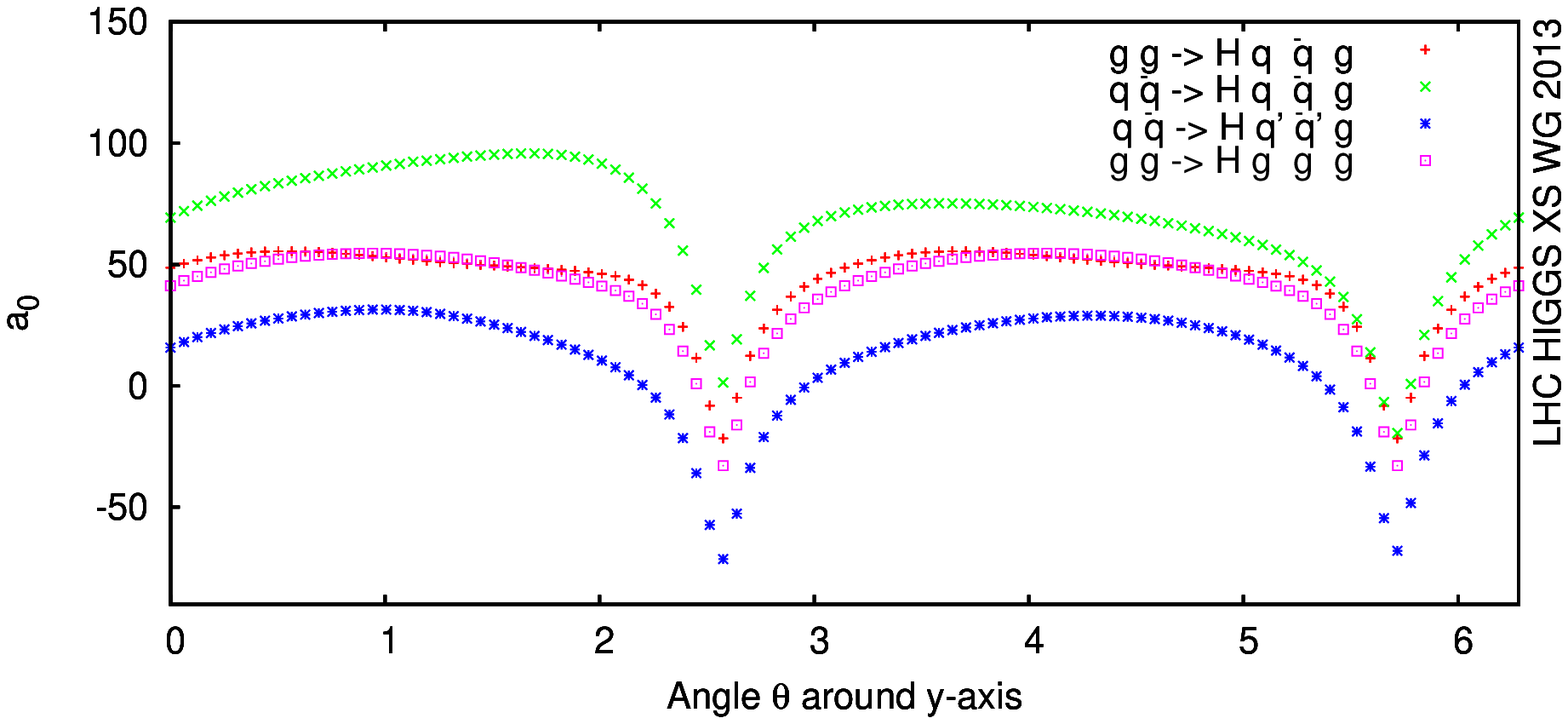} 
\caption{Finite-term of the virtual matrix elements for 
$\Pg\Pg \to \PH  \PAQq\PQq \Pg $ (red),
$\PAQq\PQq \to \PH  \PAQq\PQq \Pg $ (green),
$\PAQq\PQq  \to \PH  \PQq' \PAQq' \Pg$ (red),
$\Pg\Pg  \to \PH  \Pg  \Pg  \Pg$ (purple). 
} 
\label{Fig:Hjjj}
\end{center}
\end{figure}

To display our results for the virtual matrix elements of the various subprocesses, 
we choose an arbitrary phase space point, with the momenta of the initial partons along the $z$-axis.
Then, we create new momentum configurations by rotating the final state
through an angle $\theta$ about the $y$-axis. For each subprocess, we plot in 
Figure~\ref{Fig:Hjjj} the behavior of the finite part $a_0$ of the amplitudes defined as
$$
\frac{  2 \mathfrak{Re} \left  \{ \mathcal{M}^{\mbox{\tiny tree-level} \ast } \mathcal{M}^{\mbox{\tiny one-loop}  }   \right  \}  }{
(4 \pi \alpha_s ) \left |  \mathcal{M}^{\mbox{\tiny tree-level}} \right |^2 
 }  
\equiv 
\frac{a_{-2}}{\epsilon^2} +  \frac{a_{-1}}{\epsilon} 
+ a_0   
\, ,
$$
when the final external momenta are rotated  from
$\theta=0$ to $\theta=2\pi$. We verified that  the values of the double and the single poles conform to 
the universal singular behavior of dimensionally regulated one-loop amplitudes.

\clearpage

\newpage
\newcommand{\mH}{\ensuremath{m_{\PH}}}
\newcommand{\MyggH}{\ensuremath{\Pg\Pg\PH}}
\newcommand{\MyttH}{\ensuremath{\PQt\PAQt\PH}}
\newcommand{\hgg}{\ensuremath{\PH \to \PGg\PGg}}
\newcommand{\hww}{\ensuremath{\PH \to \PW\PW^{(*)}}}
\newcommand{\htt}{\ensuremath{\PH \to \PGtm\PGtp}}
\newcommand{\bb}{\ensuremath{\PQb\PAQb}}
\newcommand{\hbb}{\ensuremath{\PH \to \bb}}
\newcommand{\hmm}{\ensuremath{\PH \to \PGmm\PGmp}}
\newcommand{\hZZllll}{\ensuremath{\PH \to \PZ\PZ^{(*)}\to \ell^+\ell^-\ell^+\ell^-}}
\newcommand{\hWWlnln}{\ensuremath{\PH \to \PW\PW^{(*)} \to \ell^+\nu\ell^-\overline{\nu}}}
\newcommand{\BRinv}{\ensuremath{\mathrm{BR_{inv.,undet.}}}\xspace}
\newcommand{\BRonlyinv}{\ensuremath{\mathrm{BR_{inv.}}}\xspace}
\newcommand{\BRundet}{\ensuremath{\mathrm{BR_{undet.}}}\xspace}
\newcommand{\Cc}{\ensuremath{\upkappa}}
\newcommand{\Rr}{\ensuremath{\uplambda}}
\newcommand{\GGHNNLO}{{\sl GGH@NNLO\xspace}}

\newcommand{\sublevelA}[1]{{\paragraph{#1}\noindent\newline\noindent}}
\newcommand{\sublevelB}[1]{{\subparagraph*{#1}\noindent\newline\noindent}}
\newcommand{\tc}[2]{\textcolor{#1}{#2}}
\newcommand{\hcdot}{\ensuremath{\hfill}}
\newcommand{\sss}{\scriptscriptstyle}

\section{Higgs properties: couplings\footnote{%
    A.~David, A.~Denner, M.~D\"uhrssen, M.~Grazzini, C.~Grojean,
    K.~Prokofiev, G.~Weiglein, M.~Zanetti (eds.); S.~Dittmaier, G.~Passarino, M.~Spira.}
    \footnote{%
    G.~Cacciapaglia, C.~Contino, A.~Deandrea, B.~Dobrescu, G.~Drieu La Rochelle, J.R.~Espinosa, A.~Falkowski, E.~Feng, J.B.~Flament, M.~Ghezzi, S.~Heinemeyer, M.~M\"uhlleitner, G.~Petrucciani, R.~Rattazzi, M. Schumacher, R.~Sundrum, M.~Trott, A.~Vicini, D.~Zeppenfeld are thanked for useful discussions and active participations in the working group activities.}}
\label{sec:LM}

%Plans for YR3:
%\begin{enumerate}
%\item
%Couplings - interime recommandations
%\item%JPC 
%\end{enumerate}

\subsection{Introduction}
\label{sec:LM_intro}
The recent observation of a new massive neutral boson by ATLAS and CMS~\cite{Aad:2012tfa,Chatrchyan:2012ufa},
as well as evidence from the Tevatron experiments~\cite{Aaltonen:2012qt}, opens a new era where characterization
of this new object is of central importance.

The SM, as any renormalizable theory,
makes very accurate predictions for the coupling of the Higgs boson
to all other known particles.
These couplings directly influence the rates and kinematic properties of production and decay of the Higgs boson.
Therefore, measurement of the production and decay rates of the observed state as well as its angular correlations yields information
that can be used to probe whether data are compatible with the SM predictions for the Higgs boson.

While coarse features of the observed state can be inferred from the information that the experiments have made public,
only a consistent and combined treatment of the data can yield the most accurate picture of the coupling structure.
Such a treatment must take into account all the systematic and statistical uncertainties considered in the analyses,
as well as the correlations among them. 

At the LHC a SM-like Higgs boson is searched for mainly in four
exclusive production processes:
the predominant gluon fusion \ggF,
the vector boson fusion \VBF,
the associated production with a vector boson \VH\ and
the associated production with a top-quark pair \ttH.
The main search
channels are determined by five decay modes of the Higgs boson, the
$\PGg\PGg$, $\PZ\PZ^{(*)}$, $\PW\PW^{(*)}$, $\PQb\PAQb$ and
$\PGtp\PGtm$ channels.

In 2011, the LHC delivered an integrated luminosity of slightly less
than 6\ifb\ of proton--proton ($\Pp\Pp$) collisions at a centre-of-mass
energy of 7\UTeV\ to the ATLAS and CMS experiments. 
In 2012, the LHC delivered approximately 23\ifb\ of $\Pp\Pp$ collisions
at a centre-of-mass energy of 8\UTeV\ to both experiments. 
For this dataset, the instantaneous luminosity reached
record levels of approximately ${7\cdot 10^{33}\Ucm^{-2}\Us^{-1}}$,
almost double the peak luminosity of 2011 with the same $50\Uns$ bunch
spacing.

The ATLAS and CMS experiments have reported compatible
measurements of the mass of the observed narrow resonance yielding:
\begin{center}
  125.5~$\pm 0.2$(stat.)~$^{+0.5}_{-0.6}$(syst.)\UGeV\
  (ATLAS~\cite{ATLAS-CONF-2013-014}), \\
  125.7~$\pm 0.3$(stat.)~$\pm 0.3$(syst.)\UGeV\ (CMS~\cite{CMS-PAS-HIG-13-005}).
\end{center}

In \refSs{sec:LM_ir:framework} and~\ref{sec:LM_ir:benchmarks} we present and extend 
the interim framework and benchmarks~\cite{LHCHiggsCrossSectionWorkingGroup:2012nn} 
for the measurements of Higgs couplings. 
With increasing precision of the measurements, this
framework eventually has to be replaced by one that allows for a fully
consistent inclusion of higher-order corrections. Such a framework is
provided by an effective Lagrangian formalism. To start with, we
define in \refS{sec:LM_eft} effective Lagrangians for Higgs
interactions to be used in future calculations and experimental
determinations of Higgs couplings. Methods for the measurement of the
spin and the CP properties of the Higgs boson are described in
\refS{sec:LM_cp}.

\subsection{Interim framework for the analysis of Higgs couplings}
\label{sec:LM_ir:framework}
This subsection outlines an interim framework to explore the coupling structure of the recently
observed state.
The framework proposed in this recommendation should be seen as a continuation of the earlier studies of the LHC sensitivity to the Higgs couplings initiated in Refs.~\cite{Zeppenfeld:2000td, Duhrssen:2003tba, Duhrssen:2004cv, Lafaye:2009vr}, and has been influenced by the works of Refs.~\cite{Hagiwara:1993qt, GonzalezGarcia:1999fq, Barger:2003rs, Manohar:2006gz, Giudice:2007fh, Low:2009di, Espinosa:2010vn, Anastasiou:2011pi}.  It follows closely the methodology proposed in the recent phenomenological works of Refs.~\cite{Carmi:2012yp, Azatov:2012bz, Espinosa:2012ir} which have been further extended in several directions \cite{Cao:2012fz, Boudjema:2012cq, Barger:2012hv, Frandsen:2012rj, Giardino:2012ww, Ellis:2012rx, Draper:2012xt, Azatov:2012rd, Farina:2012ea, Englert:2012cb, Degrande:2012gr, Klute:2012pu, Arhrib:2012yv, Carena:2012gp, Espinosa:2012vu, Akeroyd:2012ms, Azatov:2012wq, Carena:2012xa, Gupta:2012mi, Blum:2012ii, Chang:2012tb, Gillioz:2012se, Chang:2012gn, Low:2012rj, Benbrik:2012rm, Corbett:2012dm, Giardino:2012dp, Ellis:2012hz, Montull:2012ik, Espinosa:2012im, Carmi:2012in, Peskin:2012we, Banerjee:2012xc, Abe:2012fb, Cao:2012yn, Joglekar:2012vc, Bertolini:2012gu, ArkaniHamed:2012kq, Bonnet:2012nm, Dawson:2012di, Craig:2012vn, Almeida:2012bq, Alves:2012ez, Plehn:2012iz, Ajaib:2012eb, Espinosa:2012in, Accomando:2012yg, Elander:2012fk,Reece:2012gi, Djouadi:2012rh, Kobakhidze:2012wb, Englert:2012wf, Chacko:2012vm, Bellazzini:2012vz, Passarino:2012cb, Huang:2012wn, Dobrescu:2012td, Chang:2012ve, Moreau:2012da, Han:2012rb, Cacciapaglia:2012wb, Biswas:2012bd, Masso:2012eq, Petersson:2012nv, Andersen:2012kn, Farina:2012xp, Chang:2012bq, D'Agnolo:2012mj, Azatov:2012qz, Alonso:2012px, Bhattacharyya:2012tj, Choudhury:2012tk, Gupta:2012fy, Belanger:2012gc, Reuter:2012sd, Klute:2013cx, Goertz:2013kp, Cheung:2013bn, Cheung:2013kla, Belanger:2013kya, Falkowski:2013dza, Giardino:2013bma, Contino:2013kra, Ellis:2013lra, Djouadi:2013qya, Chang:2013cia, Biswas:2013xva, Dumont:2013wma} along the lines that are formalized in the present recommendation.
While the interim framework is not final,
it has an accuracy that matches the statistical power of the datasets that the LHC experiments
have collected until the end of the 2012 LHC run
and is an explicit attempt to provide a common ground
for the dialogue in the, and between the, experimental and theoretical communities. 

Based on that framework, a series of benchmark parameterizations are presented in \Sref{sec:LM_ir:benchmarks}.
Each benchmark parameterization allows to explore specific aspects of the coupling structure of the new state.
The parameterizations have varying degrees of complexity,
with the aim to cover the most interesting possibilities that can be realistically tested with
the LHC $7$ and $8\UTeV$ datasets.
On the one hand, the framework and benchmarks were designed to provide a recommendation
to experiments on how to perform coupling fits that are useful for the theory community.
On the other hand the theory community can prepare for results based on the framework discussed in this document.

\subsubsection{Idea and underlying assumptions}
The idea behind this framework is that all deviations from the SM
are computed assuming that there is only one underlying state at $\sim125\UGeV$.
It is assumed that this state is a Higgs boson,
i.e.\ the excitation of a field whose vacuum expectation value (VEV) breaks electroweak symmetry,
and that it is SM-like,
in the sense that the experimental results so far are compatible
with the interpretation of the state in terms of the SM Higgs boson.
No specific assumptions are made on any additional states of new physics
(and their decoupling properties)
that could influence the phenomenology of the $125\UGeV$ state,
such as additional Higgs bosons (which could be heavier but also lighter than $125\UGeV$),
additional scalars that do not develop a VEV,
and new fermions and/or gauge bosons that could interact with the state at $125\UGeV$, 
giving rise, for instance, to an invisible decay mode.

The purpose of this framework is to
either confirm that the light, narrow, resonance indeed matches the properties of the SM Higgs,
or to establish a deviation from the SM behavior, which would rule out the SM if sufficiently
significant.
In the latter case the next goal in the quest to identify the nature of
EWSB would obviously
be to test the compatibility of the observed patterns with alternative frameworks of EWSB. 

% The idea behind our framework is that all deviations from SM are computed assuming that there 
% is only one underlying state at $\sim125\UGeV$ which we assume to be a SM-like Higgs boson,
% i.e.\ the excitation of a field whose vacuum expectation value (VEV) breaks electroweak symmetry. 
% We make no specific assumption on the existence and nature of other heavy (scalar or not) degrees 
% of freedom which can influence the SM-like Higgs boson couplings to all SM particles; furthermore
% no assumption is made on their decoupling as their masses increase or on the mass-mixing with
% the SM-like Higgs boson. 
% 
% The heavy scalar degrees of freedom are Higgs-partners: they are not in the SM, but may
% have a rich spectrum of non-Higgs states (they do not necessarily develop a VEV). 
% Their spectrum and couplings to fermions and vector bosons will be strongly model dependent and
% our frameworks are intended to be part of a larger program (the so-called inverse problem):
% if LHC finds evidence for physics beyond the SM, how can one determine the underlying theory?
% Therefore, our framework is designed for proving that the light, narrow, resonance matches 
% the SM Higgs properties, or to establish that deviations from the SM behavior are consistent 
% with some other EWSB framework.

In investigating the experimental information that can be obtained on the coupling properties
of the new state near $125\UGeV$ from the LHC data collected so far
the following assumptions are made\footnote{The experiments are encouraged to test the assumptions of the framework, but that lies outside the scope of this document.}:
\begin{itemize}
 \item The signals observed in the different search channels
originate from a single narrow resonance with a mass near $125\UGeV$. 
The case of several, possibly overlapping, resonances in this mass 
region is not considered.
 \item The width of the assumed Higgs boson near $125\UGeV$ is neglected, 
i.e.\ the zero-width approximation for this state is used.
Hence the
%product $\sigma\times BR(\mathit{ii}\to\PH\to\mathit{ff})$
signal cross section
can be decomposed in the following way for all channels:
 \begin{equation}
\left(\sigma\cdot\text{BR}\right)(\mathit{ii}\to\PH\to\mathit{ff}) = \frac{\sigma_{\mathit{ii}}\cdot\Gamma_{\mathit{ff}}}{\Gamma_{\PH}}
 \end{equation}
 where $\sigma_{\mathit{ii}}$ is the production cross section through the initial state $\mathit{ii}$, $\Gamma_{\mathit{ff}}$ the partial decay width into the final state $\mathit{ff}$ and $\Gamma_{\PH}$ the total width of the Higgs boson.
\end{itemize}

Within the context of these assumptions, in the following
a simplified framework for investigating the
experimental information that can be obtained on the coupling properties
of the new state is outlined.
In general, the couplings of the assumed Higgs state near $125\UGeV$ are
``pseudo-observables'', i.e.\ they cannot be directly measured. 
This means that a certain ``unfolding procedure'' is
necessary to extract information on the couplings from the
measured quantities like cross sections times branching ratios (for
specific experimental cuts and acceptances).
This gives rise to a certain model dependence of the extracted information.
Different options can be pursued in this context.
One possibility is to confront a specific model with the experimental data.
This has the advantage that all available higher-order corrections within this model can
consistently be included and also other experimental
constraints (for instance from direct searches or from electroweak
precision data) can be taken into account.
However, the results obtained in this case are restricted to the interpretation within
that particular model.
Another possibility is to use a general
parameterization of the couplings of the new state without referring to any
particular model.
While this approach is clearly less model-dependent,
the relation between the extracted coupling parameters and the couplings of actual models, 
for instance the SM or its minimal
supersymmetric extension (MSSM), is in general non-trivial, so that the
theoretical interpretation of the extracted information can
be difficult. 
It should be mentioned that the results for the 
signal strengths of individual search channels that have been made
public by ATLAS and CMS, while referring just to a particular search
channel rather than to the full information available from the Higgs
searches, are nevertheless very valuable for testing the predictions
of possible models of physics beyond the SM.

In the SM, once the numerical value of the Higgs
mass is specified, all the couplings of the Higgs boson to fermions,
bosons and to itself are specified within the model.
It is therefore in
general not possible to perform a fit to experimental data within the
context of the SM where Higgs couplings are treated as free parameters. 
While it is possible to test the overall compatibility of the SM with
the data, it is not possible to extract information about 
deviations of the measured couplings with respect to their SM values. 

A theoretically well-defined framework, as outlined in \Sref{sec:LM_eft}, 
for probing small deviations from the
SM predictions --- or the predictions of another reference model --- is to use
the state-of-the-art predictions in this model (including all available
higher-order corrections) and to supplement them with the contributions
of additional terms in the Lagrangian,
which are usually called ``anomalous couplings''.
In such an approach and in general, not only the coupling strength, i.e.\ the
absolute value of a given coupling, will be modified, but also the
tensor structure of the coupling.
For instance, the $\PH\PWp\PWm$ LO coupling 
in the SM is proportional to the metric tensor $g^{\mu\nu}$, while
anomalous couplings will generally also give rise to other tensor
structures, however required to be compatible with the SU(2)$\times$U(1)
symmetry and the corresponding Ward-Slavnov-Taylor identities.
As a consequence, kinematic distributions will in general
be modified when compared to the SM case. 

Since the reinterpretation of searches that have been performed within
the context of the SM is difficult if effects that change kinematic
distributions are taken into account and since not all the necessary
tools to perform this kind of analysis are available yet, the
following additional assumption is made in this simplified framework:

\begin{itemize}

\item
Only modifications of couplings strengths, i.e.\ of absolute values of
couplings, are taken into account, while the tensor structure of the
couplings is assumed to be the same as in the SM prediction. This means in
particular that the observed state is assumed to be a CP-even scalar.

\end{itemize}

As mentioned above, the described framework assumes that the observed
state is SM-like. In case a large discrepancy from SM-like behavior is
established, this framework would still be useful for assessing the level
of compatibility of the SM predictions with the data. The interpretation
of the physical origin of such a discrepancy, on the other hand, would
most likely require to go beyond this framework.

\subsubsection{Definition of coupling scale factors}
\label{sec:LM_ir:scale_factor_def}

In order to take into account the currently best available SM
predictions for Higgs cross sections, which include
higher-order QCD and EW corrections~\cite{Dittmaier:2011ti,Dittmaier:2012vm,HiggsWG},
while at the same time introducing
possible deviations from the SM values of the couplings, the 
predicted SM Higgs cross sections and partial
decay widths are dressed with scale factors $\Cc_i$. 
The scale factors $\Cc_i$ are defined in such a way that the cross sections
$\sigma_{ii}$ or the partial decay widths $\Gamma_{ii}$ associated with the SM particle $i$ scale
with the factor $\Cc_i^2$ when compared to the corresponding SM prediction.
\refT{tab:LM_ir:LO_coupling_relatios} lists all relevant cases.
Taking the process $\Pg\Pg\to\PH\to\PGg\PGg$ as an example, one would use as cross section:
\begin{eqnarray}
\left(\sigma\cdot\text{BR}\right)(\Pg\Pg\to\PH\to\PGg\PGg) &=& \sigma_\text{SM}(\Pg\Pg\to\PH) \cdot \text{BR}_\text{SM}(\PH\to\PGg\PGg)\,\cdot \frac{\Cc_{\Pg}^2 \cdot \Cc_{\PGg}^2}{\Cc_{\PH}^2}
\end{eqnarray}
where the values and uncertainties for both $\sigma_\text{SM}(\Pg\Pg\to\PH)$
and $\text{BR}_\text{SM}(\PH\to\PGg\PGg)$ are taken from Ref.~\cite{HiggsWG} for a given Higgs mass hypothesis.

\begin{table}
\caption{LO coupling scale factor relations for Higgs boson cross sections and partial decay widths relative to the SM.
For a given $\mH$ hypothesis, the smallest set of degrees of freedom in this framework comprises
$\Cc_{\PW}$, $\Cc_{\PZ}$, $\Cc_{\PQb}$, $\Cc_{\PQt}$, and $\Cc_{\PGt}$.
For partial widths that are not detectable at the LHC, scaling is performed via proxies chosen among the detectable ones.  
Additionally, the loop-induced vertices can be treated as a function of other $\Cc_i$ or effectively,
through the $\Cc_{\Pg}$ and $\Cc_{\PGg}$ degrees of freedom which allow probing for BSM contributions in the loops.
Finally, to explore invisible or undetectable decays,
the scaling of the total width can also be taken as a separate degree of freedom, $\Cc_{\PH}$,
instead of being rescaled as a function, $\Cc_{\PH}^2(\Cc_i,\mH)$, of the other scale factors.}
\label{tab:LM_ir:LO_coupling_relatios}
\centering
\begin{tabular}{@{}p{0.43\linewidth}|p{0.55\linewidth}@{}}
\hline
\begin{eqnarray}
  \omit\rlap{\text{Production modes}}\nonumber\\
  \frac{\sigma_{\MyggH}}{\sigma_{\MyggH}^\text{SM}}               & = & \left\{ \begin{array}{l} \Cc_{\Pg}^2(\Cc_{\PQb},\Cc_{\PQt},\mH) \\ \Cc_{\Pg}^2 \end{array} \right. \\
  \frac{\sigma_{\text{VBF}}}{\sigma_{\text{VBF}}^\text{SM}} & = & \Cc_\text{VBF}^2(\Cc_{\PW},\Cc_{\PZ},\mH)\\
  \frac{\sigma_{\PW\PH}}{\sigma_{\PW\PH}^\text{SM}}                 & = & \Cc_{\PW}^2\\
  \frac{\sigma_{\PZ\PH}}{\sigma_{\PZ\PH}^\text{SM}}                 & = & \Cc_{\PZ}^2\\
  \frac{\sigma_{\MyttH}}{\sigma_{\MyttH}^\text{SM}}    & = & \Cc_{\PQt}^2
\end{eqnarray}
&
% \begin{minipage}[t]{\linewidth}
% Detectable decay modes
\begin{eqnarray}
%   \hline
  \omit\rlap{\text{Detectable decay modes}}\nonumber\\
  \frac{\Gamma_{\PW\PW^{(*)}}}{\Gamma_{\PW\PW^{(*)}}^\text{SM}}           &=& \Cc_{\PW}^2\\
  \frac{\Gamma_{\PZ\PZ^{(*)}}}{\Gamma_{\PZ\PZ^{(*)}}^\text{SM}}           &=& \Cc_{\PZ}^2\\
  \frac{\Gamma_{\PQb\PAQb}}{\Gamma_{\PQb\PAQb}^\text{SM}}           &=& \Cc_{\PQb}^2\\
  \frac{\Gamma_{\PGtm\PGtp}}{\Gamma_{\PGtm\PGtp}^\text{SM}}     &=& \Cc_{\PGt}^2\\
  \frac{\Gamma_{\PGg\PGg}}{\Gamma_{\PGg\PGg}^\text{SM}} &=& \left\{ \begin{array}{l} \Cc_{\PGg}^2(\Cc_{\PQb},\Cc_{\PQt},\Cc_{\PGt},\Cc_{\PW},\mH) \\ \Cc_{\PGg}^2 \end{array} \right.\\
% \end{eqnarray}
% 
% Undetectable decay modes
% \begin{eqnarray}  
%   \hline
  \nonumber\\
  \omit\rlap{\text{Currently undetectable decay modes}}\nonumber\\
  \frac{\Gamma_{\PQt\PAQt}}{\Gamma_{\PQt\PAQt}^\text{SM}}           &=& \Cc_{\PQt}^2\\
  \frac{\Gamma_{\Pg\Pg}}{\Gamma_{\Pg\Pg}^\text{SM}}           &:& \text{see \Sref{sec:LM_ir:C_g}}\nonumber\\
  \frac{\Gamma_{\PQc\PAQc}}{\Gamma_{\PQc\PAQc}^\text{SM}}           &=& \Cc_{\PQt}^2\\
  \frac{\Gamma_{\PQs\PAQs}}{\Gamma_{\PQs\PAQs}^\text{SM}}           &=& \Cc_{\PQb}^2\\
  \frac{\Gamma_{\PGmm\PGmp}}{\Gamma_{\PGmm\PGmp}^\text{SM}}       &=& \Cc_{\PGt}^2\\
  \frac{\Gamma_{\PZ\PGg}}{\Gamma_{\PZ\PGg}^\text{SM}} &=& \left\{ \begin{array}{l} \Cc_{(\PZ\PGg)}^2(\Cc_{\PQb},\Cc_{\PQt},\Cc_{\PGt},\Cc_{\PW},\mH) \\ \Cc_{(\PZ\PGg)}^2 \end{array} \right.\\
% \end{eqnarray}
% 
% Total width
% \begin{eqnarray}    
%   \hline
  \nonumber\\
  \omit\rlap{\text{Total width}}\nonumber\\
  \frac{\Gamma_{\PH}}{\Gamma_{\PH}^\text{SM}}           &=& \left\{ \begin{array}{l} \Cc_{\PH}^2(\Cc_i,\mH) \\ \Cc_{\PH}^2 \end{array} \right.                    
\end{eqnarray}
% \end{minipage}
\\
\hline
\end{tabular}
\end{table}

By definition, the currently best available SM predictions
for all $\sigma\cdot\text{BR}$ are recovered when all $\Cc_i=1$.
In general, this means that for $\Cc_i\neq 1$ higher-order accuracy is lost.
Nonetheless, NLO QCD corrections essentially factorize with respect to coupling rescaling,
and are accounted for wherever possible.
This approach ensures that for a true SM Higgs boson no artificial deviations
(caused by ignored NLO corrections) are found from what is considered the SM Higgs boson hypothesis.
The functions
$\Cc_\text{VBF}^2(\Cc_{\PW},\Cc_{\PZ},\mH)$,
$\Cc_{\Pg}^2(\Cc_{\PQb},\Cc_{\PQt},\mH)$,
$\Cc_{\PGg}^2(\Cc_{\PQb},\Cc_{\PQt},\Cc_{\PGt},\Cc_{\PW},\mH)$,
$\Cc_{(\PZ\PGg)}^2(\Cc_{\PQb}, \Cc_{\PQt}, \Cc_{\PGt}, \Cc_{\PW}, \mH)$ and
$\Cc_{\PH}^2(\Cc_i,\mH)$
 are used for cases where there is a non-trivial relationship between scale factors $\Cc_i$ and cross sections
 or (partial) decay widths, and are calculated to NLO QCD accuracy.
 The functions are defined in the following sections and 
 all required input parameters as well as example code can be found
 in~\Bref{HiggsWG}.
As explained in Sec.~\ref{Subsub:int} below, the
notation in terms of the partial widths $\Gamma_{\PW\PW^{(*)}}$ and $\Gamma_{\PZ\PZ^{(*)}}$ in \refT{tab:LM_ir:LO_coupling_relatios}
is meant for illustration only. In the experimental analysis the
4-fermion partial decay widths are taken into account.

\sublevelA{Scaling of the VBF cross section}
$\Cc_\text{VBF}^2$ refers to the functional dependence of the
VBF\footnote{Vector Boson Fusion is also called Weak Boson Fusion, as only the weak bosons $\PW$ and $\PZ$ contribute to the production.}
cross section on the scale factors $\Cc_{\PW}^2$ and $\Cc_{\PZ}^2$:
\begin{eqnarray}
\label{eq:RVBF}
 \Cc_\text{VBF}^2(\Cc_{\PW},\Cc_{\PZ},\mH) &=& \frac{\Cc_{\PW}^2 \cdot \sigma_{\PW F}(\mH) + \Cc_{\PZ}^2 \cdot \sigma_{\PZ F}(\mH)}{\sigma_{\PW F}(\mH)+\sigma_{\PZ F}(\mH)}
\end{eqnarray}
The $\PW$- and $\PZ$-fusion cross sections,
$\sigma_{\PW F}$ and $\sigma_{\PZ F}$,
are taken from \Brefs{Arnold:2008rz,webVBFNLO}.
The interference term is $< 0.1\%$ in the SM and hence ignored~\cite{Ciccolini:2007ec}.

In \refT{tab:LM_ir:RVBF} one can find the approximate values
to be inserted in \Eref{eq:RVBF} for $\mH = 125\UGeV$.

\begin{table}[!h]
\centering
\caption{Approximate numerical values for resolving the VBF production
cross-section according to \Eref{eq:RVBF} assuming $\mH = 125\UGeV$.}
\label{tab:LM_ir:RVBF}
\begin{tabular}{lcc}
\hline\rule[-1.5ex]{0pt}{4ex}
$\sqrt{s}$ & $\sigma_{\PW F}$ (pb) & $\sigma_{\PZ F}$ (pb) \\
\hline
$7\UTeV$ & 0.938 & 0.321  \\
$8\UTeV$ & 1.210 & 0.417  \\
\hline
\end{tabular}
\end{table}

\sublevelA{Scaling of the gluon fusion cross section and of the $\PH\to\Pg\Pg$ decay vertex}
\label{sec:LM_ir:C_g}
$\Cc_{\Pg}^2$ refers to the scale factor for the loop-induced production
cross section $\sigma_{\MyggH}$.
The decay width $\Gamma_{\Pg\Pg}$ is not observable at the LHC, however its contribution to the total width
is also considered.

\sublevelB{Gluon fusion cross-section scaling}
As NLO QCD corrections factorize with the scaling of the electroweak couplings with $\Cc_{\PQt}$ and $\Cc_{\PQb}$,
the function $\Cc_{\Pg}^2(\Cc_{\PQb}, \Cc_{\PQt},\mH)$ can be calculated in NLO QCD:
\begin{eqnarray}
\label{eq:CgNLOQCD}
 \Cc_{\Pg}^2(\Cc_{\PQb}, \Cc_{\PQt},\mH) &=& \frac{\Cc_{\PQt}^2\cdot\sigma_{\MyggH}^{\PQt\PQt}(\mH) +\Cc_{\PQb}^2\cdot\sigma_{\MyggH}^{\PQb\PQb}(\mH) +\Cc_{\PQt}\Cc_{\PQb}\cdot\sigma_{\MyggH}^{\PQt\PQb}(\mH)}{\sigma_{\MyggH}^{\PQt\PQt}(\mH)+\sigma_{\MyggH}^{\PQb\PQb}(\mH)+\sigma_{\MyggH}^{\PQt\PQb}(\mH)}
\end{eqnarray}

Here, $\sigma_{\MyggH}^{\PQt\PQt}$, $\sigma_{\MyggH}^{\PQb\PQb}$ and $\sigma_{\MyggH}^{\PQt\PQb}$
denote the square of the top-quark contribution, the square of the bottom-quark
contribution and the top-bottom interference, respectively.
The interference term ($\sigma_{\MyggH}^{\PQt\PQb}$) is negative for a light mass Higgs, $\mH < 200\UGeV$. 
Within the LHC Higgs Cross Section Working Group (for the evaluation of the
MSSM cross section) these contributions were evaluated, where for
$\sigma_{\MyggH}^{\PQb\PQb}$ and $\sigma_{\MyggH}^{\PQt\PQb}$ the full NLO QCD calculation included
in \HIGLU~\cite{Spira:1995mt} was used.
For $\sigma_{\MyggH}^{\PQt\PQt}$ the NLO QCD result of \HIGLU\ was supplemented with the
NNLO corrections in the heavy-top-quark limit as implemented in \GGHNNLO~\cite{Harlander:2002wh},
see~Ref.~\cite[Sec.~6.3]{Dittmaier:2011ti} for details.

In \refT{tab:LM_ir:ggH} one can find the approximate values
to be inserted in \Eref{eq:CgNLOQCD} for $\mH = 125\UGeV$.

\begin{table}[!h]
\centering
\begin{tabular}{lccc}
\hline\rule[-1.5ex]{0pt}{4ex}
$\sqrt{s}$ & $\sigma_{\MyggH}^{\PQt\PQt}$ (pb) &
$\sigma_{\MyggH}^{\PQb\PQb}$ (pb) &
$\sigma_{\MyggH}^{\PQt\PQb}$ (pb)
\\
\hline
$7\UTeV$ & 4.355 & 0.09528 & $-$0.8970 \\
$8\UTeV$ & 18.31 & 0.1206 & $-$1.125  \\
\hline
\end{tabular}
\caption{Approximate numerical values for resolving the gluon-fusion production
cross-section according to \Eref{eq:CgNLOQCD} assuming $\mH = 125\UGeV$.}
\label{tab:LM_ir:ggH}
\end{table}

\sublevelB{Partial width scaling}
In a similar way, NLO QCD corrections for the
$\PH\to \Pg\Pg$ partial width are implemented in \HDECAY~\cite{Spira:1996if,Djouadi:1997yw,hdecay2}.
This allows to treat the scale factor for $\Gamma_{\Pg\Pg}$
as a second order polynomial in $\Cc_{\PQb}$ and $\Cc_{\PQt}$:
\begin{eqnarray}
\label{eq:GammagNLOQCD}
\frac{\Gamma_{\Pg\Pg}}{\Gamma_{\Pg\Pg}^\text{SM}(\mH)} = \frac{\Cc_{\PQt}^2\cdot\Gamma_{\Pg\Pg}^{\PQt\PQt}(\mH) +\Cc_{\PQb}^2\cdot\Gamma_{\Pg\Pg}^{\PQb\PQb}(\mH) +\Cc_{\PQt}\Cc_{\PQb}\cdot\Gamma_{\Pg\Pg}^{\PQt\PQb}(\mH)}{\Gamma_{\Pg\Pg}^{\PQt\PQt}(\mH)+\Gamma_{\Pg\Pg}^{\PQb\PQb}(\mH)+\Gamma_{\Pg\Pg}^{\PQt\PQb}(\mH)}
\end{eqnarray}
The terms $\Gamma_{\Pg\Pg}^{\PQt\PQt}$, $\Gamma_{\Pg\Pg}^{\PQb\PQb}$ and $\Gamma_{\Pg\Pg}^{\PQt\PQb}$
are defined like the $\sigma_{\MyggH}$ terms in Eq.~(\ref{eq:CgNLOQCD}).
The $\Gamma_{\Pg\Pg}^{ii}$ correspond to the partial widths that are obtained for $\Cc_i=1$ and all other $\Cc_j=0, j\neq i$.
The cross-term $\Gamma_{\Pg\Pg}^{\PQt\PQb}$ can then be derived
by calculating the SM partial width by setting $\Cc_{\PQb}=\Cc_{\PQt}=1$ and
subtracting $\Gamma_{\Pg\Pg}^{\PQt\PQt}$ and $\Gamma_{\Pg\Pg}^{\PQb\PQb}$ from it. 

In \refT{tab:LM_ir:Hgg} one can find the approximate values
to be used in \Eref{eq:GammagNLOQCD} for $\mH = 125, 126\UGeV$.

\begin{table}[h]
\centering
\caption{Approximate numerical values for resolving the gluon-fusion decay
partial width according to \Eref{eq:GammagNLOQCD}.}
\label{tab:LM_ir:Hgg}
\begin{tabular}{lccc}
\hline\rule[-1.5ex]{0pt}{4ex}
$\mH$ & $\Gamma_{\Pg\Pg}^{\PQt\PQt}$ (\UkeV) & $\Gamma_{\Pg\Pg}^{\PQb\PQb}$
(\UkeV) & $\Gamma_{\Pg\Pg}^{\PQt\PQb}$ (\UkeV) \\
\hline
125\UGeV & 380.8 & 3.96 & $-$42.1 \\
126\UGeV & 389.6 & 3.95 & $-$42.7 \\
\hline
\end{tabular}
\end{table}

\sublevelB{Effective treatment}
In the general case, without the assumptions above, possible non-zero contributions from additional
particles in the loop have to be taken into account and $\Cc_{\Pg}^2$ is then treated as an
effective coupling scale factor parameter in the fit:
% \begin{eqnarray}
%  \frac{\sigma_{\MyggH}}{\sigma_{\MyggH}^\text{SM}} &=& \Cc_{\Pg}^2
% \end{eqnarray}
$\sigma_{\MyggH}/\sigma_{\MyggH}^\text{SM} = \Cc_{\Pg}^2$.
The effective scale factor for the partial gluon width $\Gamma_{\Pg\Pg}$ should behave in a very similar way,
so in this case the same effective scale factor $\Cc_{\Pg}$ is used:
% \begin{eqnarray}
%  \frac{\Gamma_{\Pg\Pg}}{\Gamma_{\Pg\Pg}^\text{SM}}  &=& \Cc_{\Pg}^2
% \end{eqnarray}
$\Gamma_{\Pg\Pg}/\Gamma_{\Pg\Pg}^\text{SM} = \Cc_{\Pg}^2$.
As the contribution of $\Gamma_{\Pg\Pg}$ to the total width is <10\% in the SM,
this assumption is believed to have no measurable impact.

\sublevelA{Scaling of the $\PH\to\PGg\PGg$ partial decay width}
\label{sec:LM_ir:C_gamma}
Like in the previous section,
$\Cc_{\PGg}^2$ refers to the scale factor for the loop-induced $\PH\to\PGg\PGg$ decay.
Also for the $\PH\to\PGg\PGg$ decay NLO QCD corrections exist and are implemented in \HDECAY.
This allows to treat the scale factor for the $\PGg\PGg$ partial width
as a second order polynomial in $\Cc_{\PQb}$, $\Cc_{\PQt}$, $\Cc_{\PGt}$, and $\Cc_{\PW}$:
\begin{eqnarray}
\label{eq:CgammaNLOQCD}
\Cc_{\PGg}^2(\Cc_{\PQb}, \Cc_{\PQt}, \Cc_{\PGt}, \Cc_{\PW}, \mH) &=& \frac{\sum_{i,j}\Cc_i \Cc_j\cdot\Gamma_{\PGg\PGg}^{i j}(\mH)}{\sum_{i,j}\Gamma_{\PGg\PGg}^{ij}(\mH)}
\end{eqnarray}
where the pairs $(i,j)$ are $\PQb\PQb,\PQt\PQt,\PGt\PGt,\PW\PW,\PQb\PQt,\PQb\PGt,\PQb\PW,\PQt\PGt,\PQt\PW,\PGt\PW$.
The $\Gamma_{\PGg\PGg}^{ii}$ correspond to the partial widths that are obtained for $\Cc_i=1$ and all other $\Cc_j=0, (j\neq i)$.
The cross-terms $\Gamma_{\PGg\PGg}^{ij}, (i\neq j)$ can then be derived by calculating the partial width by setting $\Cc_i=\Cc_j=1$
and all other $\Cc_l=0, (l\neq i,j)$, and subtracting $\Gamma_{\PGg\PGg}^{ii}$ and $\Gamma_{\PGg\PGg}^{jj}$ from them. 

In \refT{tab:LM_ir:Hgamgam} one can find the approximate values
to be used in \Eref{eq:CgammaNLOQCD} for $\mH = 125, 126\UGeV$.

\begin{table}[!h]
\centering\setlength{\tabcolsep}{5pt}
\caption{Approximate numerical values for resolving the di-photon decay
partial width according to \Eref{eq:CgammaNLOQCD}. All values are
given in \UeV.}
\label{tab:LM_ir:Hgamgam}
\begin{tabular}{lcccccccccc}
\hline
$\mH$ &\rule[-1.5ex]{0pt}{4ex}
$\Gamma_{\PGg\PGg}^{\PQt\PQt}$ &
$\Gamma_{\PGg\PGg}^{\PQb\PQb}$ &
$\Gamma_{\PGg\PGg}^{\PW\PW}$ &
$\Gamma_{\PGg\PGg}^{\PQt\PQb}$ &
$\Gamma_{\PGg\PGg}^{\PQt\PW}$ &
$\Gamma_{\PGg\PGg}^{\PQb\PW}$ &
$\Gamma_{\PGg\PGg}^{\PGt\PGt}$ &
$\Gamma_{\PGg\PGg}^{\PQt\PGt}$ &
$\Gamma_{\PGg\PGg}^{\PQb\PGt}$ &
$\Gamma_{\PGg\PGg}^{\PGt\PW}$ 
\\
\hline
125\UGeV & 662.84 & 0.18 & 14731.86 & $-$16.39 & $-$6249.93 & 77.42 &
0.21 & $-$17.69 & 0.40 & 83.59 \\
126\UGeV & 680.39 & 0.18 & 15221.98 & $-$16.62 & $-$6436.35 & 78.78 &
0.22 & $-$17.94 & 0.40 & 85.05 \\
\hline
\end{tabular}
\end{table}

\sublevelB{Effective treatment}
In the general case, without the assumption above, possible non-zero contributions from additional
particles in the loop have to be taken into account
and $\Cc_{\PGg}^2$ is then treated as an effective coupling parameter in the fit.

\sublevelA{Scaling of the $\PH\to \PZ\PGg$ decay vertex}
\label{sec:LM_ir:C_Zgamma}
Like in the previous sections, $\Cc_{(\PZ\PGg)}^2$ refers to
the scale factor for the loop-induced $\PH\to \PZ\PGg$ decay.
This allows to treat the scale factor for the $\PZ\PGg$ partial width as a
second order polynomial in $\Cc_{\PQb}$, $\Cc_{\PQt}$, $\Cc_{\PGt}$, and $\Cc_{\PW}$:
\begin{eqnarray}
\label{eq:CZgammaNLOQCD}
\Cc_{(\PZ\PGg)}^2(\Cc_{\PQb}, \Cc_{\PQt}, \Cc_{\PGt}, \Cc_{\PW}, \mH) &=& \frac{\sum_{i,j}\Cc_i \Cc_j\cdot\Gamma_{\PZ\PGg}^{i j}(\mH)}{\sum_{i,j}\Gamma_{\PZ\PGg}^{i j}(\mH)}
\end{eqnarray}
where the pairs $(i,j)$ are $\PQb\PQb,\PQt\PQt,\PGt\PGt,\PW\PW,\PQb\PQt,\PQb\PGt,\PQb\PW,\PQt\PGt,\PQt\PW,\PGt\PW$.
The $\Gamma_{\PZ\PGg}^{ij}$ are calculated in the same way as for Eq.~(\ref{eq:CgammaNLOQCD}).
NLO QCD corrections have been computed and found to be very small \cite{Spira:1991tj}, and thus ignored here.

In \refT{tab:LM_ir:HZgam} one can find the approximate values
to be inserted in \Eref{eq:CZgammaNLOQCD} for $\mH=125, 126\UGeV$.

\begin{table}[h]
\caption{Approximate numerical values for resolving the $\PH\to \PZ\PGg$ decay
partial width according to \Eref{eq:CZgammaNLOQCD}. All values are
given in \UeV.}
\label{tab:LM_ir:HZgam}
\centering\setlength{\tabcolsep}{5pt}
\begin{tabular}{lcccccccccc}
\hline
$\mH$ & \rule[-1.5ex]{0pt}{4ex}
$\Gamma_{(\PZ\PGg)}^{\PQt\PQt}$ &
$\Gamma_{(\PZ\PGg)}^{\PQb\PQb}$ &
$\Gamma_{(\PZ\PGg)}^{\PW\PW}$ &
$\Gamma_{(\PZ\PGg)}^{\PQt\PQb}$ &
$\Gamma_{(\PZ\PGg)}^{\PQt\PW}$ &
$\Gamma_{(\PZ\PGg)}^{\PQb\PW}$ &
$\Gamma_{(\PZ\PGg)}^{\PGt\PGt}$ &
$\Gamma_{(\PZ\PGg)}^{\PQt\PGt}$ &
$\Gamma_{(\PZ\PGg)}^{\PQb\PGt}$ &
$\Gamma_{(\PZ\PGg)}^{\PGt\PW}$ 
\\
\hline
125\UGeV & 21.74 & 0.019 & 7005.6 & $-1.11$ & $-780.4$ &
19.90 & $1.5\times 10^{-5}$ & $-0.033$ & 0.0010 & 0.594 \\
126\UGeV & 23.51 & 0.020 & 7648.4 & $-1.19$ & $-848.1$ &
21.47 & $1.6\times 10^{-5}$ & $-0.035$ & 0.0011 & 0.640 \\
\hline
\end{tabular}
\end{table}

\sublevelB{Effective treatment}
In the general case, without the assumption above, possible non-zero contributions from additional
particles in the loop have to be taken into account
and $\Cc_{(\PZ\PGg)}^2$ is then treated as an effective coupling parameter in the fit.

\sublevelA{Scaling of the total width}
\label{sec:LM_ir:framework:GammaH}
The total width $\Gamma_{\PH}$ is the sum of all Higgs partial decay widths.
Under the assumption that no additional BSM Higgs decay modes
(into either invisible or undetectable final states)
contribute to the total width,
$\Gamma_{\PH}$ is expressed as the sum of the scaled partial Higgs decay widths to SM particles,
which combine to a total scale factor $\Cc_{\PH}^2$ compared to the SM total width $\Gamma_{\PH}^\text{SM}$:
\begin{eqnarray}
  \label{eq:CH2_def}
  \Cc_{\PH}^2(\Cc_i,\mH) &=& \sum\limits_{\begin{array}{r}j=\PW\PW^{(*)},\PZ\PZ^{(*)},\PQb\PAQb,\PGtm\PGtp,\\\PGg\PGg,\PZ\PGg,\Pg\Pg,\PQt\PAQt,\PQc\PAQc,\PQs\PAQs,\PGmm\PGmp\end{array}} \frac{ \Gamma_j(\Cc_i,\mH)}{\Gamma_{\PH}^\text{SM}(\mH)}
\end{eqnarray}

\sublevelB{Effective treatment}
In the general case, additional Higgs decay modes to BSM particles cannot be excluded
and the total width scale factor $\Cc_{\PH}^2$ is treated as free parameter.

The total width $\Gamma_{\PH}$ for a light Higgs with $\mH\sim 125\UGeV$
is not expected to be directly observable at the LHC,
as the SM expectation is $\Gamma_{\PH}\sim 4\UMeV$,
several orders of magnitude smaller than
the experimental mass resolution~\cite{Dittmaier:2011ti}.
There is no indication from the results observed so far that 
the natural width is broadened by new physics effects 
to such an extent that it could be directly observable.
Furthermore, as all LHC Higgs channels rely on the identification of Higgs decay products,
there is no way of measuring the total Higgs width indirectly within a coupling fit without using assumptions. 
This can be illustrated by assuming that all cross sections and partial widths are increased by a common factor
$\Cc_i^2=r>1$.
If simultaneously the Higgs total width is increased by the square of the same factor $\Cc_{\PH}^2=r^2$
(for example by postulating some BSM decay mode) the experimental visible signatures
in all Higgs channels would be indistinguishable from the SM.

Hence without further assumptions only ratios of scale factors $\Cc_i$ can be measured at the LHC,
where at least one of the ratios needs to include the total width scale factor $\Cc_{\PH}^2$.
Such a definition of ratios absorbs two degrees of freedom
(e.g.\ a common scale factor to all couplings and a scale factor to the total width)
into one ratio that can be measured at the LHC.

\sublevelB{Assumptions for absolute coupling scale factor measurements}
In order to go beyond the measurement of ratios of coupling scale factors to the
determination of absolute coupling scale factors $\Cc_i$ additional
assumptions are necessary to remove one degree of freedom.
Possible assumptions are:
\begin{enumerate}
 \item No new physics in Higgs decay modes (Eq.~(\ref{eq:CH2_def})).
 \item $\Cc_{\PW}\le 1$, $\Cc_{\PZ}\le 1$~\cite{Zeppenfeld:2000td,Duhrssen:2004cv,Dobrescu:2012td}. 
  This assumption is theoretically well motivated in the sense that
  it holds in a wide class of models. 
  In particular, it is valid in any model with an arbitrary
  number of Higgs doublets, with and without additional Higgs singlets. 
  The assumption is also justified in certain classes of composite Higgs
  models, while on the other hand it may be violated for instance in Little
  Higgs models, in particular in the presence of an isospin-2 scalar multiplet~\cite{Falkowski:2012vh}.

 If one combines this assumption with the fact that all Higgs partial decay widths are
 positive definite and the total width is bigger than the sum of all (known) partial decay widths,
 this is sufficient to give a lower and upper bound on all $\Cc_i$ and also
 determine a possible branching ratio $\BRinv$ into final states invisible or undetectable at the LHC.
 This is best illustrated with the $\PV\PH(\PH\to\PV\PV)$ process: 
 \begin{eqnarray}
  \sigma_{\PV\PH}\cdot\text{BR}(\PH\to \PV\PV) &=& \frac{\Cc_{\PV}^2 \cdot \sigma_{\PV\PH}^\text{SM} \,\,\cdot\,\, \Cc_{\PV}^2 \cdot \Gamma^\text{SM}_{\PV}}{\Gamma_{\PH}}\nonumber\\
  \text{and\hspace{4cm}}\Gamma_{\PH} &>& \Cc_{\PV}^2 \cdot \Gamma_{\PV}^\text{SM}\label{eq:GammaH_sum_inequation}\\
  \text{give combined:\hspace{1cm}}\sigma_{\PV\PH}\cdot\text{BR}(\PH\to VV)&<&\frac{\Cc_{\PV}^2 \cdot \sigma_{\PV\PH}^\text{SM} \,\,\cdot\,\, \Cc_{\PV}^2 \cdot \Gamma^\text{SM}_{\PV}}{\Cc_{\PV}^2 \cdot \Gamma_{\PV}^\text{SM}}\nonumber\\
  \Longrightarrow\text{\hspace{4cm}}\Cc_{\PV}^2&>&\frac{\sigma_{\PV\PH}\cdot\text{BR}(\PH\to \PV\PV)}{\sigma_{\PV\PH}^\text{SM}}
 \end{eqnarray}
 If more final states are included in Eq.~(\ref{eq:GammaH_sum_inequation}), the lower bounds become tighter
 and together with the upper limit assumptions on $\Cc_{\PW}$ and $\Cc_{\PZ}$, absolute measurements are possible.
 However, uncertainties on all $\Cc_i$ can be very large depending
 on the accuracy of the $\PQb\PAQb$ decay channels that dominate the uncertainty
 of the total width sum for a SM-like Higgs.
 \item $\Cc_j=\text{constant}$. If at least one coupling scale factor $\Cc_j$ is known either from an 
 external measurement or through theory assumptions, the total width and the branching ratio $\BRinv$ 
 into final states invisible or undetectable at the LHC can be determined. 
 An example is given in the benchmark model in \Sref{sec:LM_ir:CgCgam}, where the assumption $\Cc_{\PZ} = \Cc_{\PW} = \Cc_{\PGt} = \Cc_{\PQb} = \Cc_{\PQt}=1$ is used to determine $\BRinv$. 
\end{enumerate}

In the benchmark parameterizations in \Sref{sec:LM_ir:benchmarks} three versions are given, unless otherwise stated:
one without assumptions on the total width (effective treatment of $\Cc_{\PH}$), 
one assuming no beyond SM Higgs decay modes (Option 1. above) 
and one assuming bounds on the gauge coupling scale factors (Option 2. above).

\subsubsection{Further assumptions}
\label{sec:LM_ir:further_assumptions}
\sublevelA{Theoretical uncertainties}
The quantitative impact of theory uncertainties in the Higgs production cross sections and decay rates
is discussed in detail in \Bref{Dittmaier:2011ti}.

Such uncertainties will directly affect the determination of the scale factors.
When one or more of the scaling factors differ from 1, the uncertainty from missing higher-order
contributions will in general be larger than what was estimated in~\Bref{Dittmaier:2011ti}.
 
In practice, the cross section predictions with their uncertainties as tabulated in~\Bref{Dittmaier:2011ti}
are used as such so that for $\Cc_i=1$ the recommended SM treatment is recovered.
Without a consistent electroweak NLO calculation for deviations from the SM,
electroweak corrections and their uncertainties for the SM prediction
($\sim 5\%$ in gluon fusion production and $\sim 2\%$ in the di-photon decay)
are naively scaled together.
In the absence of explicit calculations
this is the currently best available approach
in a search for deviations from the SM Higgs prediction.

\sublevelA{Limit of the zero-width approximation}
Concerning the zero-width approximation (ZWA), it should be noted that in the 
mass range of the narrow resonance the width of the Higgs boson of the
Standard Model (SM) is more than four orders of magnitude smaller than 
its mass.
Thus, the zero-width approximation is in principle expected to
be an excellent approximation not only for a SM-like Higgs boson below $\sim 150\UGeV$ 
but also for a wide range of BSM scenarios which are
compatible with the present data. 
However, it has been shown in \Bref{Kauer:2012hd}
that this is not always the case even in the SM.
The inclusion of off-shell contributions is essential 
to obtain an accurate Higgs signal normalization at the $1\%$ precision level. 
For $\Pg\Pg\ (\to \PH) \to \PV\PV$, $\PV= \PW,\PZ$, 
${\mathcal O}(10\%)$ corrections occur due to an enhanced Higgs signal 
in the region $M_{\PV\PV} > 2\,M_{\PV}$, where also sizeable 
Higgs-continuum interference occurs.
However, with the accuracy anticipated to be reached in
the 2012 data these effects play a minor role.

\sublevelA{Signal interference effects}
\label{Subsub:int}
% A source of uncertainty is related to interference effects in $\PH \to 4\,$fermion
% decay. We refer to Chapter~2 of \Bref{Dittmaier:2012vm}, where it is shown that the ratio 
% of the ZWA
% %--
% \begin{equation}
% \mbox{BR}(\PH \to \PV \PV)\,\times\,\mbox{BR}^2(V \to {\overline f} f)
% \end{equation}
% %--
% over the complete result~\cite{Prophecy4f,Bredenstein:2006rh,Bredenstein:2006ha} for $\PH \to \Pep\Pem\Pep\Pem$ or 
% $\Pep\Pem\PGmp\PGmm$ is large, due to the interference (below $\PW\PW, \PZ\PZ$ thresholds),
% about $11\%$ difference.
% 
% The experimental analyses take into account the full NLO 4~fermion partial decay width.
% The partial width of the 4~lepton final state
% (usually referred to as $\PH\to \PZ\PZ^{(*)}\to 4l$ or $\PH\to \PZ\PZ^{(*)}\to 2l2j$, depending on decay mode) is scaled with $\Cc_{\PZ}^2$.
% The partial width of the low mass 2~lepton, 2~neutrino final state
% (usually referred to as $\PH\to \PW\PW^{(*)}\to l\nu\,l\nu$, although some interference with $\PH\to \PZ^{(*)}\PZ\to ll\,\nu\nu$ exists and is taken into account) is scaled with $\Cc_{\PW}^2$.
A possible source of uncertainty is related to interference effects
in $\PH \to 4\,$fermion decay.
For a light Higgs boson the decay width
into 4 fermions should always be calculated from the complete matrix
elements and not from the approximation
%--
\begin{equation}\label{eq:4fapprox}
\text{BR}(\PH \to \PV \PV)\,\cdot\,\text{BR}^2(\PV \to \Pf\PAf)\, ,
~(\PV = \PZ,\PW)~.
\end{equation}
%--
This approximation, based on the ZWA for the gauge boson $\PV$, neglects both off-shell
effects and interference between diagrams where the intermediate gauge
bosons couple to different pairs of final-state fermions.
As shown in Chapter~2 of \Bref{Dittmaier:2012vm}, the interference
effects not included in Eq.~(\ref{eq:4fapprox}) amount to 10\% for the
decay $\PH \to \Pep\Pem\Pep\Pem$ for a $125\UGeV$ Higgs.
Similar interference effects of the order of 5\% are found for the
$\Pep\PGne\Pem\PAGne$ and $\PQq\PAQq\PQq\PAQq$ final states.

The experimental analyses take into account the full NLO 4-fermion
partial decay width~\cite{Prophecy4f,Bredenstein:2006rh,Bredenstein:2006ha}.
The partial width of the 4-lepton final state
(usually described as $\PH\to \PZ\PZ^{(*)}\to 4\Pl$) is scaled with $\Cc_{\PZ}^2$.
Similarly, the partial width of the 2-lepton, 2-jet final state
(usually described as $\PH\to\PZ\PZ^{(*)}\to 2\Pl 2\PQq$) is scaled with $\Cc_{\PZ}^2$.
The partial width of the low mass
2-lepton, 2-neutrino final state (usually described as $\PH\to \PW\PW^{(*)}\to
\Pl\PGn\,\Pl\PGn$, although a contribution of $\PH\to \PZ^{(*)}\PZ\to \Pl\Pl\,\PGn\PGn$
exists and is taken into account) is scaled with $\Cc_{\PW}^2$. 

\sublevelA{Treatment of $\Gamma_{\PQc\PAQc}$, $\Gamma_{\PQs\PAQs}$, $\Gamma_{\PGmm\PGmp}$ and light fermion contributions to loop-induced processes}
\label{sec:LM_ir:further_assumptions:light_fermions}
When calculating $\Cc_{\PH}^2(\Cc_i,\mH)$ in a benchmark parameterization,
the final states $\PQc\PAQc$, $\PQs\PAQs$ and $\PGmm\PGmp$
(currently unobservable at the LHC) are tied to $\Cc_i$ scale factors which can
be determined from the data.
Based on flavor symmetry considerations, the following choices are made:
\begin{eqnarray}
   \frac{\Gamma_{\PQc\PAQc}}{\Gamma_{\PQc\PAQc}^\text{SM}(\mH)}        &=& \Cc_{\PQc}^2   = \Cc_{\PQt}^2\\
   \frac{\Gamma_{\PQs\PAQs}}{\Gamma_{\PQs\PAQs}^\text{SM}(\mH)}        &=& \Cc_{\PQs}^2   = \Cc_{\PQb}^2\\
   \frac{\Gamma_{\PGmm\PGmp}}{\Gamma_{\PGmm\PGmp}^\text{SM}(\mH)} &=& \Cc_{\PGm}^2 = \Cc_{\PGt}^2
\end{eqnarray} 
Following the rationale of~Ref.~\cite[Sec. 9]{Dittmaier:2011ti},
the widths of $\Pem\Pep$, $\PQu\PAQu$, $\PQd\PAQd$ and neutrino final states are neglected.

Through interference terms,
these light fermions also contribute to the loop-induced $\Pg\Pg\to\PH$ and $\PH\to\Pg\Pg,\PGg\PGg,\PZ\PGg$ vertices.
In these cases, the assumptions $\Cc_{\PQc}=\Cc_{\PQt}$, $\Cc_{\PQs}=\Cc_{\PQb}$ and $\Cc_{\PGm}=\Cc_{\PGt}$ are made.

Once sensitivity to the \hmm\ final state is reached in the experiments, a separate coupling scale factor $\Cc_{\PGm}$ 
should be used where appropriate.

\sublevelA{Approximation in associated $\PZ\PH$ production}
When scaling the associated $\PZ\PH$ production mode,
the contribution from $\Pg\Pg\to \PZ\PH$ through a top-quark loop is neglected.
This is estimated to be around 5\%
of the total associated $\PZ\PH$ production cross section~\cite[Sec.~4.3]{Dittmaier:2011ti}.

\subsection{Benchmark parameterizations based on the interim framework}
\label{sec:LM_ir:benchmarks}
In putting forward a set of benchmark parameterizations based on the framework
described in the previous section several considerations were taken into
account.
One concern is the stability of the fits which typically involve several
hundreds of nuisance parameters.
With that in mind, the benchmark parameterizations avoid quotients of parameters of interest.
Another constraint that heavily shapes the exact choice of parameterization
is consistency among the uncertainties
that can be extracted in different parameterizations.
Some coupling scale factors enter linearly in loop-induced photon and gluon vertices.
For that reason, all scale factors are defined at the same power,
leading to what could appear as an abundance of squared expressions.
Finally, the benchmark parameterizations are chosen such that some
potentially interesting physics scenarios can be
probed and the parameters of interest are chosen so that at least some are expected to be determined.
 
For every benchmark parameterization, unless otherwise stated, 
three variations are provided (see \Sref{sec:LM_ir:framework:GammaH} for details):
\begin{enumerate}
\item The total width is scaled assuming that there are no invisible or undetected widths.
In this case $\Cc_{\PH}^2(\Cc_i,\mH)$ is a function of the free parameters as defined in Eq.~(\ref{eq:CH2_def}).
\item The total width scale factor is treated as a free parameter, but an assumption of $\Cc_{\PW}<1$ and $\Cc_{\PZ}<1$ 
on the gauge coupling scale factors is applied together with the condition $\Cc_{\PH}^2>\Cc_{\PH}^2(\Cc_i)$.
\item The total width scale factor is
%absorbed into the parameterization.
treated as a free parameter.
In this case no assumption is made and there will be a parameter of the form $\Cc_{ij} = \Cc_i\cdot \Cc_j / \Cc_{\PH}$ 
contributing to all Higgs boson rates.
\end{enumerate}
The benchmark parameterizations are given in tabular form where each cell corresponds
to the scale factor to be applied to a given combination of production and decay mode.

For every benchmark parameterization,  
a list of the free parameters and their relation to the framework parameters is provided.
To reduce the amount of symbols in the tables, $\mH$ is omitted throughout.
In practice, $\mH$ can either be fixed to a given value
or profiled together with other nuisance parameters.

\subsubsection{One common scale factor}
\label{sec:LM_ir:C}

The simplest way to look for a deviation
from the predicted SM Higgs coupling structure is to leave the overall signal strength $\mu$
as a free parameter.

In order to perform the same fit in the context of the coupling scale factor framework, the only difference is that 
$\mu = \Cc^2\cdot \Cc^2 / \Cc^2 = \Cc^2$, where the three terms $\Cc^2$ in the intermediate expression account
for production, decay and total width scaling, respectively (\Tref{tab:LM_ir:C}). 

\begin{table}[h]

\centering
\caption{The simplest possible benchmark parameterization where a single scale factor applies to all production and decay modes.}
\label{tab:LM_ir:C}

\begin{tabular}{lccccc}
\hline
\multicolumn{6}{l}{\bfseries Common scale factor}\\
\multicolumn{6}{l}{\footnotesize Free parameter: $\Cc (= \Cc_{\PQt} = \Cc_{\PQb} = \Cc_{\PGt} = \Cc_{\PW} = \Cc_{\PZ})$.} \\
\multicolumn{6}{l}{\footnotesize Dependent parameters: $\Cc_{\PGg} = \Cc$, $\Cc_{\Pg} = \Cc$, $\Cc_{\PH} = \Cc$.} \\
\hline
 & $\PH\to\PGg\PGg$ & $\PH\to \PZ\PZ^{(*)}$ & $\PH\to \PW\PW^{(*)}$ & $\PH\to \PQb\PAQb$ & $\PH\to\PGtm\PGtp$\\
\hline
\MyggH       & \multicolumn{5}{c}{\multirow{5}{*}{$\Cc^2$}}\\
\MyttH & \multicolumn{5}{c}{} \\
VBF         & \multicolumn{5}{c}{} \\
$\PW\PH$        & \multicolumn{5}{c}{} \\
$\PZ\PH$        & \multicolumn{5}{c}{} \\
\hline
\end{tabular}

\end{table}

This parameterization, despite providing the highest experimental precision,
has several clear shortcomings,
such as ignoring that the role of the Higgs boson in providing the masses of 
the vector bosons is very different from the role it has in providing 
the masses of fermions.

\subsubsection{Scaling of vector boson and fermion couplings}
\label{sec:LM_ir:CVCF}

In checking whether an observed state is compatible with the SM Higgs boson,
one obvious question is whether it fulfills its expected role in EWSB
which is intimately related to the coupling to the vector bosons ($\PW,\PZ$).

Therefore, assuming that the SU(2) custodial symmetry holds, in the
simplest case two parameters can be defined, one
scaling the coupling to the vector bosons, $\Cc_{\PV} (=\Cc_{\PW}=\Cc_{\PZ})$,
and one scaling the coupling common to all fermions, $\Cc_{\Pf} (=\Cc_{\PQt} = \Cc_{\PQb} = \Cc_{\PGt})$.
Loop-induced processes are assumed to scale as expected from the SM structure.

In this parameterization, presented in \Tref{tab:LM_ir:CVCF},
the gluon vertex loop is effectively a fermion loop and only the photon vertex loop requires
a non-trivial scaling, given the contributions of the top and bottom quarks, of the $\PGt$ lepton, of the $\PW$-boson, as well as their
(destructive) interference.

\begin{table}[h]
\centering
\caption{A benchmark parameterization where custodial symmetry is assumed and vector boson couplings are scaled together ($\Cc_{\PV}$)
and fermions are assumed to scale with a single parameter ($\Cc_{\Pf}$).}
\label{tab:LM_ir:CVCF}

\begin{tabular}{lccccc}
\hline
\multicolumn{6}{l}{\bfseries Boson and fermion scaling assuming no invisible or undetectable widths}\\
\multicolumn{6}{l}{\footnotesize Free parameters: $\Cc_{\PV} (= \Cc_{\PW} = \Cc_{\PZ})$, $\Cc_{\Pf} (= \Cc_{\PQt} = \Cc_{\PQb} = \Cc_{\PGt})$} \\
\multicolumn{6}{l}{\footnotesize Dependent parameters: $\Cc_{\PGg} = \Cc_{\PGg}(\Cc_{\Pf},\Cc_{\Pf},\Cc_{\Pf},\Cc_{\PV})$, $\Cc_{\Pg} = \Cc_{\Pf}$, $\Cc_{\PH} = \Cc_{\PH}(\Cc_i)$.} \\
\hline
 & $\PH\to\PGg\PGg$ & $\PH\to \PZ\PZ^{(*)}$ & $\PH\to \PW\PW^{(*)}$ & $\PH\to \PQb\PAQb$ & $\PH\to\PGtm\PGtp$ \\
\hline
\MyggH       & \multirow{2}{*}{$\frac{\Cc_{\Pf}^2\cdot \Cc_{\PGg}^2(\Cc_{\Pf},\Cc_{\Pf},\Cc_{\Pf},\Cc_{\PV})}{\Cc_{\PH}^2(\Cc_i)}$} & \multicolumn{2}{c}{\multirow{2}{*}{$\frac{\Cc_{\Pf}^2\cdot \Cc_{\PV}^2}{\Cc_{\PH}^2(\Cc_i)}$}} & \multicolumn{2}{c}{\multirow{2}{*}{$\frac{\Cc_{\Pf}^2\cdot \Cc_{\Pf}^2}{\Cc_{\PH}^2(\Cc_i)}$}} \\
\MyttH &                                                                            & \multicolumn{2}{c}{                                                 }      & \multicolumn{2}{c}{                                                 } \\
\hline
VBF         & \multirow{3}{*}{$\frac{\Cc_{\PV}^2\cdot \Cc_{\PGg}^2(\Cc_{\Pf},\Cc_{\Pf},\Cc_{\Pf},\Cc_{\PV})}{\Cc_{\PH}^2(\Cc_i)}$} & \multicolumn{2}{c}{\multirow{3}{*}{$\frac{\Cc_{\PV}^2\cdot \Cc_{\PV}^2}{\Cc_{\PH}^2(\Cc_i)}$}} & \multicolumn{2}{c}{\multirow{3}{*}{$\frac{\Cc_{\PV}^2\cdot \Cc_{\Pf}^2}{\Cc_{\PH}^2(\Cc_i)}$}} \\
$\PW\PH$        &                                                                            & \multicolumn{2}{c}{                                                 }      & \multicolumn{2}{c}{                                                 } \\
$\PZ\PH$        &                                                                            & \multicolumn{2}{c}{                                                 }      & \multicolumn{2}{c}{                                                 } \\
\hline
\hline
\multicolumn{6}{l}{\bfseries Boson and fermion scaling assuming $\Cc_{\PV}<1$}\\
\multicolumn{6}{l}{\footnotesize Free parameters: $\Cc_{\PV} (= \Cc_{\PW} = \Cc_{\PZ})$, $\Cc_{\Pf} (= \Cc_{\PQt} = \Cc_{\PQb} = \Cc_{\PGt})$, $\Cc_{\PH}$, with conditions $\Cc_{\PV}<1$ and $\Cc_{\PH}^2>\Cc_{\PH}^2(\Cc_i)$}\\
\multicolumn{6}{l}{\footnotesize Dependent parameters: $\Cc_{\PGg} = \Cc_{\PGg}(\Cc_{\Pf},\Cc_{\Pf},\Cc_{\Pf},\Cc_{\PV})$, $\Cc_{\Pg} = \Cc_{\Pf}$.} \\
\hline
 & $\PH\to\PGg\PGg$ & $\PH\to \PZ\PZ^{(*)}$ & $\PH\to \PW\PW^{(*)}$ & $\PH\to \PQb\PAQb$ & $\PH\to\PGtm\PGtp$ \\
\hline
\MyggH       & \multirow{2}{*}{$\frac{\Cc_{\Pf}^2\cdot \Cc_{\PGg}^2(\Cc_{\Pf},\Cc_{\Pf},\Cc_{\Pf},\Cc_{\PV})}{\Cc_{\PH}^2}$} & \multicolumn{2}{c}{\multirow{2}{*}{$\frac{\Cc_{\Pf}^2\cdot \Cc_{\PV}^2}{\Cc_{\PH}^2}$}} & \multicolumn{2}{c}{\multirow{2}{*}{$\frac{\Cc_{\Pf}^2\cdot \Cc_{\Pf}^2}{\Cc_{\PH}^2}$}} \\
\MyttH &                                                                            & \multicolumn{2}{c}{                                                 }      & \multicolumn{2}{c}{                                                 } \\
\hline
VBF         & \multirow{3}{*}{$\frac{\Cc_{\PV}^2\cdot \Cc_{\PGg}^2(\Cc_{\Pf},\Cc_{\Pf},\Cc_{\Pf},\Cc_{\PV})}{\Cc_{\PH}^2}$} & \multicolumn{2}{c}{\multirow{3}{*}{$\frac{\Cc_{\PV}^2\cdot \Cc_{\PV}^2}{\Cc_{\PH}^2}$}} & \multicolumn{2}{c}{\multirow{3}{*}{$\frac{\Cc_{\PV}^2\cdot \Cc_{\Pf}^2}{\Cc_{\PH}^2}$}} \\
$\PW\PH$        &                                                                            & \multicolumn{2}{c}{                                                 }      & \multicolumn{2}{c}{                                                 } \\
$\PZ\PH$        &                                                                            & \multicolumn{2}{c}{                                                 }      & \multicolumn{2}{c}{                                                 } \\
\hline
\hline
\multicolumn{6}{l}{\bfseries Boson and fermion scaling without assumptions on the total width}\\
\multicolumn{6}{l}{\footnotesize Free parameters: $\Cc_{\PV\PV} (= \Cc_{\PV}\cdot \Cc_{\PV} / \Cc_{\PH})$, $\Rr_{\Pf\PV} (= \Cc_{\Pf} / \Cc_{\PV})$.} \\
\multicolumn{6}{l}{\footnotesize Dependent parameters: $\Cc_{\PGg} = \Cc_{\PGg}(\Rr_{\Pf\PV},\Rr_{\Pf\PV},\Rr_{\Pf\PV},1)$, $\Rr_{\Pg\PV} = \Rr_{\Pf\PV}$.} \\
\hline
 & $\PH\to\PGg\PGg$ & $\PH\to \PZ\PZ^{(*)}$ & $\PH\to \PW\PW^{(*)}$ & $\PH\to \PQb\PAQb$ & $\PH\to\PGtm\PGtp$ \\
\hline
\MyggH       & \multirow{2}{*}{$\Cc_{\PV\PV}^2 \cdot \Rr_{\Pf\PV}^2 \cdot \Cc_{\PGg}^2(\Rr_{\Pf\PV},\Rr_{\Pf\PV},\Rr_{\Pf\PV},1)$} & \multicolumn{2}{c}{\multirow{2}{*}{$\Cc_{\PV\PV}^2 \cdot \Rr_{\Pf\PV}^2$}} & \multicolumn{2}{c}{\multirow{2}{*}{$\Cc_{\PV\PV}^2 \cdot \Rr_{\Pf\PV}^2 \cdot \Rr_{\Pf\PV}^2$}} \\
\MyttH &                                      & \multicolumn{2}{c}{                                    } & \multicolumn{2}{c}{                                                  } \\
\hline
VBF         & \multirow{3}{*}{$\Cc_{\PV\PV}^2 \cdot \Cc_{\PGg}^2(\Rr_{\Pf\PV},\Rr_{\Pf\PV},\Rr_{\Pf\PV},1)$}  & \multicolumn{2}{c}{\multirow{3}{*}{$\Cc_{\PV\PV}^2$}} & \multicolumn{2}{c}{ \multirow{3}{*}{$\Cc_{\PV\PV}^2 \cdot \Rr_{\Pf\PV}^2$}} \\
$\PW\PH$        &                                      & \multicolumn{2}{c}{                                    } & \multicolumn{2}{c}{                                                  } \\
$\PZ\PH$        &                                      & \multicolumn{2}{c}{                                    } & \multicolumn{2}{c}{                                                  } \\
\hline
\end{tabular}

{\footnotesize $\Cc_i^2 = \Gamma_{ii} / \Gamma_{ii}^\text{SM}$}

\end{table}

This parameterization, though exceptionally succinct, makes a number of assumptions,
which are expected to be object of further scrutiny with the accumulation of data at the LHC.
The assumptions naturally relate to the grouping of different individual couplings or to assuming that the loop amplitudes
are those predicted by the SM.

\subsubsection{Probing custodial symmetry}
\label{sec:LM_ir:RWZ}

One of the best motivated symmetries in case the new state is responsible for electroweak symmetry breaking
is the one that links its couplings to the $\PW$ and $\PZ$ bosons. 
Since $\mathrm{SU(2)_{\PV}}$ or custodial symmetry is an approximate symmetry of the SM (e.g. 
$\Delta\rho \neq 0$),
it is important to test whether data are compatible 
with the amount of violation allowed by the SM at NLO.  

In this parameterization,
presented in \Tref{tab:LM_ir:RWZ},
$\Rr_{\PW\PZ} (=\Cc_{\PW}/\Cc_{\PZ})$
%is the main parameter of interest.
is of particular interest for probing custodial symmetry.
Though providing interesting information,
both $\Cc_{\PZ}$ and $\Cc_{\Pf}$
can be thought of as nuisance parameters when performing this fit.
In addition to the photon vertex loop not having a trivial scaling,
in this parameterization also the individual $\PW$ and $\PZ$ boson fusion contributions
to the vector boson fusion production process need to be resolved.

In \Tref{tab:LM_ir:RWZ} the explicit parameterization using the assumption $\Cc_{\PW}<1$ and $\Cc_{\PZ}<1$ is omitted for this benchmark, 
as an independent test of custodial symmetry under assumptions on the gauge couplings themselves is difficult. 
If wanted, this table can be obtained from the first parameterization in \Tref{tab:LM_ir:RWZ} using the replacement
$\Cc_{\PH}^2(\Cc_i) \to \Cc_{\PH}^2$.

\begin{sidewaystable}[p]
\centering
\caption{A benchmark parameterization where custodial symmetry is probed through the $\Rr_{\PW\PZ}$ parameter.}
\label{tab:LM_ir:RWZ}
\begin{tabular}{lccccc}
\hline
\multicolumn{6}{l}{\bfseries Probing custodial symmetry assuming no invisible or undetectable widths}\\
\multicolumn{6}{l}{\footnotesize Free parameters: $\Cc_{\PZ}$, $\Rr_{\PW\PZ} (= \Cc_{\PW} / \Cc_{\PZ})$, $\Cc_{\Pf} (= \Cc_{\PQt} = \Cc_{\PQb} = \Cc_{\PGt})$.} \\
\multicolumn{6}{l}{\footnotesize Dependent parameters: $\Cc_{\PGg} = \Cc_{\PGg}(\Cc_{\Pf},\Cc_{\Pf},\Cc_{\Pf},\Cc_{\PZ} \Rr_{\PW\PZ})$, $\Cc_{\Pg} = \Cc_{\Pf}$, $\Cc_{\PH} = \Cc_{\PH}(\Cc_i)$.} \\
\hline
 & $\PH\to\PGg\PGg$ & $\PH\to \PZ\PZ^{(*)}$ & $\PH\to \PW\PW^{(*)}$ & $\PH\to \PQb\PAQb$ & $\PH\to\PGtm\PGtp$ \\
\hline
\MyggH       & \multirow{2}{*}{$\frac{\Cc_{\Pf}^2\cdot \Cc_{\PGg}^2(\Cc_{\Pf},\Cc_{\Pf},\Cc_{\Pf},\Cc_{\PZ} \Rr_{\PW\PZ})}{\Cc_{\PH}^2(\Cc_i)}$} & \multirow{2}{*}{$\frac{\Cc_{\Pf}^2\cdot \Cc_{\PZ}^2}{\Cc_{\PH}^2(\Cc_i)}$} & \multirow{2}{*}{$\frac{\Cc_{\Pf}^2\cdot (\Cc_{\PZ} \Rr_{\PW\PZ})^2}{\Cc_{\PH}^2(\Cc_i)}$} & \multicolumn{2}{c}{\multirow{2}{*}{$\frac{\Cc_{\Pf}^2\cdot \Cc_{\Pf}^2}{\Cc_{\PH}^2(\Cc_i)}$}} \\
\MyttH &                                                                & & & \multicolumn{2}{c}{                               } \\
\hline
VBF        & $\frac{\Cc_\mathrm{VBF}^2(\Cc_{\PZ},\Cc_{\PZ} \Rr_{\PW\PZ}) \cdot \Cc_{\PGg}^2(\Cc_{\Pf},\Cc_{\Pf},\Cc_{\Pf},\Cc_{\PZ} \Rr_{\PW\PZ})}{\Cc_{\PH}^2(\Cc_i)}$ & $\frac{\Cc_\mathrm{VBF}^2(\Cc_{\PZ},\Cc_{\PZ} \Rr_{\PW\PZ})\cdot \Cc_{\PZ}^2}{\Cc_{\PH}^2(\Cc_i)}$ & $\frac{\Cc_\mathrm{VBF}^2(\Cc_{\PZ},\Cc_{\PZ} \Rr_{\PW\PZ})\cdot (\Cc_{\PZ} \Rr_{\PW\PZ})^2}{\Cc_{\PH}^2(\Cc_i)}$  & \multicolumn{2}{c}{$\frac{\Cc_\mathrm{VBF}^2(\Cc_{\PZ},\Cc_{\PZ} \Rr_{\PW\PZ})\cdot \Cc_{\Pf}^2}{\Cc_{\PH}^2(\Cc_i)}$} \\
\hline
$\PW\PH$        & $\frac{(\Cc_{\PZ} \Rr_{\PW\PZ})^2\cdot \Cc_{\PGg}^2(\Cc_{\Pf},\Cc_{\Pf},\Cc_{\Pf},\Cc_{\PZ} \Rr_{\PW\PZ})}{\Cc_{\PH}^2(\Cc_i)}$ & $\frac{(\Cc_{\PZ} \Rr_{\PW\PZ})^2\cdot \Cc_{\PZ}^2}{\Cc_{\PH}^2(\Cc_i)}$ & $\frac{(\Cc_{\PZ} \Rr_{\PW\PZ})^2\cdot (\Cc_{\PZ} \Rr_{\PW\PZ})^2}{\Cc_{\PH}^2(\Cc_i)}$ & \multicolumn{2}{c}{$\frac{(\Cc_{\PZ} \Rr_{\PW\PZ})^2\cdot \Cc_{\Pf}^2}{\Cc_{\PH}^2(\Cc_i)}$} \\
\hline
$\PZ\PH$        & $\frac{\Cc_{\PZ}^2\cdot \Cc_{\PGg}^2(\Cc_{\Pf},\Cc_{\Pf},\Cc_{\Pf},\Cc_{\PZ} \Rr_{\PW\PZ})}{\Cc_{\PH}^2(\Cc_i)}$          & $\frac{\Cc_{\PZ}^2\cdot \Cc_{\PZ}^2}{\Cc_{\PH}^2(\Cc_i)}$ & $\frac{\Cc_{\PZ}^2\cdot (\Cc_{\PZ} \Rr_{\PW\PZ})^2}{\Cc_{\PH}^2(\Cc_i)}$                   & \multicolumn{2}{c}{$\frac{\Cc_{\PZ}^2\cdot \Cc_{\Pf}^2}{\Cc_{\PH}^2(\Cc_i)}$}\\
\hline
\hline
\multicolumn{6}{l}{\bfseries Probing custodial symmetry without assumptions on the total width}\\
\multicolumn{6}{l}{\footnotesize Free parameters: $\Cc_{\PZ\PZ} (= \Cc_{\PZ}\cdot \Cc_{\PZ} / \Cc_{\PH})$, $\Rr_{\PW\PZ} (= \Cc_{\PW} / \Cc_{\PZ})$, $\Rr_{\Pf\PZ} (= \Cc_{\Pf} / \Cc_{\PZ})$.} \\
\multicolumn{6}{l}{\footnotesize Dependent parameters: $\Cc_{\PGg} = \Cc_{\PGg}(\Rr_{\Pf\PZ},\Rr_{\Pf\PZ},\Rr_{\Pf\PZ},\Rr_{\PW\PZ})$, $\Rr_{\Pg Z} = \Rr_{\Pf\PZ}$.} \\
\hline
 & $\PH\to\PGg\PGg$ & $\PH\to \PZ\PZ^{(*)}$ & $\PH\to \PW\PW^{(*)}$ & $\PH\to \PQb\PAQb$ & $\PH\to\PGtm\PGtp$ \\
\hline
\MyggH       & \multirow{2}{*}{$\Cc_{\PZ\PZ}^2 \Rr_{\Pf\PZ}^2\cdot \Cc_{\PGg}^2(\Rr_{\Pf\PZ},\Rr_{\Pf\PZ},\Rr_{\Pf\PZ},\Rr_{\PW\PZ})$} & \multirow{2}{*}{$\Cc_{\PZ\PZ}^2 \Rr_{\Pf\PZ}^2$} & \multirow{2}{*}{$\Cc_{\PZ\PZ}^2 \Rr_{\Pf\PZ}^2\cdot \Rr_{\PW\PZ}^2$} & \multicolumn{2}{c}{\multirow{2}{*}{$\Cc_{\PZ\PZ}^2 \Rr_{\Pf\PZ}^2\cdot \Rr_{\Pf\PZ}^2$}} \\
\MyttH &                                                                & & & \multicolumn{2}{c}{                               } \\
\hline
VBF         & $\Cc_{\PZ\PZ}^2 \Cc_\mathrm{VBF}^2(1,\Rr_{\PW\PZ}^2)\cdot \Cc_{\PGg}^2(\Rr_{\Pf\PZ},\Rr_{\Pf\PZ},\Rr_{\Pf\PZ},\Rr_{\PW\PZ})$   & $\Cc_{\PZ\PZ}^2 \Cc_\mathrm{VBF}^2(1,\Rr_{\PW\PZ}^2) $ & $\Cc_{\PZ\PZ}^2 \Cc_\mathrm{VBF}^2(1,\Rr_{\PW\PZ}^2)\cdot \Rr_{\PW\PZ}^2$ & \multicolumn{2}{c}{$\Cc_{\PZ\PZ}^2 \Cc_\mathrm{VBF}^2(1,\Rr_{\PW\PZ}^2)\cdot \Rr_{\Pf\PZ}^2$} \\
\hline
$\PW\PH$        & $\Cc_{\PZ\PZ}^2 \Rr_{\PW\PZ}^2\cdot \Cc_{\PGg}^2(\Rr_{\Pf\PZ},\Rr_{\Pf\PZ},\Rr_{\Pf\PZ},\Rr_{\PW\PZ})$ & $\Cc_{\PZ\PZ}^2\cdot \Rr_{\PW\PZ}^2$ & $\Cc_{\PZ\PZ}^2 \Rr_{\PW\PZ}^2\cdot \Rr_{\PW\PZ}^2$ & \multicolumn{2}{c}{$\Cc_{\PZ\PZ}^2 \Rr_{\PW\PZ}^2\cdot \Rr_{\Pf\PZ}^2$} \\
\hline
$\PZ\PH$        & $\Cc_{\PZ\PZ}^2\cdot \Cc_{\PGg}^2(\Rr_{\Pf\PZ},\Rr_{\Pf\PZ},\Rr_{\Pf\PZ},\Rr_{\PW\PZ})$          & $\Cc_{\PZ\PZ}^2$               & $\Cc_{\PZ\PZ}^2\cdot \Rr_{\PW\PZ}^2$          & \multicolumn{2}{c}{$\Cc_{\PZ\PZ}^2\cdot \Rr_{\Pf\PZ}^2$} \\
\hline
\end{tabular}

{\footnotesize $\Cc_i^2 = \Gamma_{ii} / \Gamma_{ii}^\text{SM}$}
\end{sidewaystable}

As the photon vertex loop is very sensitive to BSM physics contributions, 
this benchmark is given in \Tref{tab:LM_ir:RWZ_gamma} as a second variant where a potential deviation in the 
$\Hgaga$ decay mode is decoupled from the $\Rr_{\PW\PZ}$ measurement by using the effective photon coupling scale factor $\Cc_{\PGg}$
as additional degree of freedom.

\begin{sidewaystable}[p]
\centering

\caption{A benchmark parameterization where custodial symmetry is probed through the $\Rr_{\PW\PZ}$ parameter, 
but the $\Hgaga$ decay mode is decoupled from the measurement of $\Cc_{\PW}$ by using the effective photon 
scale factor $\Cc_{\PGg}$ as additional degree of freedom.}
\label{tab:LM_ir:RWZ_gamma}

\begin{tabular}{lccccc}
\hline
\multicolumn{6}{l}{\bfseries Probing custodial symmetry decoupled from $\Hgaga$, assuming no invisible or undetectable widths}\\
\multicolumn{6}{l}{\footnotesize Free parameters: $\Cc_{\PZ}$, $\Rr_{\PW\PZ} (= \Cc_{\PW} / \Cc_{\PZ})$, $\Cc_{\Pf} (= \Cc_{\PQt} = \Cc_{\PQb} = \Cc_{\PGt})$, $\Cc_{\PGg}$.} \\
\multicolumn{6}{l}{\footnotesize Dependent parameters: $\Cc_{\Pg} = \Cc_{\Pf}$, $\Cc_{\PH} = \Cc_{\PH}(\Cc_i)$.} \\
\hline
 & $\PH\to\PGg\PGg$ & $\PH\to \PZ\PZ^{(*)}$ & $\PH\to \PW\PW^{(*)}$ & $\PH\to \PQb\PAQb$ & $\PH\to\PGtm\PGtp$ \\
\hline
\MyggH       & \multirow{2}{*}{$\frac{\Cc_{\Pf}^2\cdot \Cc_{\PGg}^2}{\Cc_{\PH}^2(\Cc_i)}$} & \multirow{2}{*}{$\frac{\Cc_{\Pf}^2\cdot \Cc_{\PZ}^2}{\Cc_{\PH}^2(\Cc_i)}$} & \multirow{2}{*}{$\frac{\Cc_{\Pf}^2\cdot (\Cc_{\PZ} \Rr_{\PW\PZ})^2}{\Cc_{\PH}^2(\Cc_i)}$} & \multicolumn{2}{c}{\multirow{2}{*}{$\frac{\Cc_{\Pf}^2\cdot \Cc_{\Pf}^2}{\Cc_{\PH}^2(\Cc_i)}$}} \\
\MyttH &                                                                & & & \multicolumn{2}{c}{                               } \\
\hline
VBF        & $\frac{\Cc_\mathrm{VBF}^2(\Cc_{\PZ},\Cc_{\PZ} \Rr_{\PW\PZ}) \cdot \Cc_{\PGg}^2}{\Cc_{\PH}^2(\Cc_i)}$ & $\frac{\Cc_\mathrm{VBF}^2(\Cc_{\PZ},\Cc_{\PZ} \Rr_{\PW\PZ})\cdot \Cc_{\PZ}^2}{\Cc_{\PH}^2(\Cc_i)}$ & $\frac{\Cc_\mathrm{VBF}^2(\Cc_{\PZ},\Cc_{\PZ} \Rr_{\PW\PZ})\cdot (\Cc_{\PZ} \Rr_{\PW\PZ})^2}{\Cc_{\PH}^2(\Cc_i)}$  & \multicolumn{2}{c}{$\frac{\Cc_\mathrm{VBF}^2(\Cc_{\PZ},\Cc_{\PZ} \Rr_{\PW\PZ})\cdot \Cc_{\Pf}^2}{\Cc_{\PH}^2(\Cc_i)}$} \\
\hline
$\PW\PH$        & $\frac{(\Cc_{\PZ} \Rr_{\PW\PZ})^2\cdot \Cc_{\PGg}^2}{\Cc_{\PH}^2(\Cc_i)}$ & $\frac{(\Cc_{\PZ} \Rr_{\PW\PZ})^2\cdot \Cc_{\PZ}^2}{\Cc_{\PH}^2(\Cc_i)}$ & $\frac{(\Cc_{\PZ} \Rr_{\PW\PZ})^2\cdot (\Cc_{\PZ} \Rr_{\PW\PZ})^2}{\Cc_{\PH}^2(\Cc_i)}$ & \multicolumn{2}{c}{$\frac{(\Cc_{\PZ} \Rr_{\PW\PZ})^2\cdot \Cc_{\Pf}^2}{\Cc_{\PH}^2(\Cc_i)}$} \\
\hline
$\PZ\PH$        & $\frac{\Cc_{\PZ}^2\cdot \Cc_{\PGg}^2}{\Cc_{\PH}^2(\Cc_i)}$          & $\frac{\Cc_{\PZ}^2\cdot \Cc_{\PZ}^2}{\Cc_{\PH}^2(\Cc_i)}$ & $\frac{\Cc_{\PZ}^2\cdot (\Cc_{\PZ} \Rr_{\PW\PZ})^2}{\Cc_{\PH}^2(\Cc_i)}$                   & \multicolumn{2}{c}{$\frac{\Cc_{\PZ}^2\cdot \Cc_{\Pf}^2}{\Cc_{\PH}^2(\Cc_i)}$}\\
\hline
\hline
\multicolumn{6}{l}{\bfseries Probing custodial symmetry decoupled from $\Hgaga$ and without assumptions on the total width}\\
\multicolumn{6}{l}{\footnotesize Free parameters: $\Cc_{\PZ\PZ} (= \Cc_{\PZ}\cdot \Cc_{\PZ} / \Cc_{\PH})$, $\Rr_{\PW\PZ} (= \Cc_{\PW} / \Cc_{\PZ})$, $\Rr_{\Pf\PZ} (= \Cc_{\Pf} / \Cc_{\PZ})$, $\Rr_{\PGg Z} (= \Cc_{\PGg} / \Cc_{\PZ})$.} \\
\multicolumn{6}{l}{\footnotesize Dependent parameters: $\Rr_{\Pg Z} = \Rr_{\Pf\PZ}$.} \\
\hline
 & $\PH\to\PGg\PGg$ & $\PH\to \PZ\PZ^{(*)}$ & $\PH\to \PW\PW^{(*)}$ & $\PH\to \PQb\PAQb$ & $\PH\to\PGtm\PGtp$ \\
\hline
\MyggH       & \multirow{2}{*}{$\Cc_{\PZ\PZ}^2 \Rr_{\Pf\PZ}^2\cdot \Rr_{\PGg\PZ}^2$} & \multirow{2}{*}{$\Cc_{\PZ\PZ}^2 \Rr_{\Pf\PZ}^2$} & \multirow{2}{*}{$\Cc_{\PZ\PZ}^2 \Rr_{\Pf\PZ}^2\cdot \Rr_{\PW\PZ}^2$} & \multicolumn{2}{c}{\multirow{2}{*}{$\Cc_{\PZ\PZ}^2 \Rr_{\Pf\PZ}^2\cdot \Rr_{\Pf\PZ}^2$}} \\
\MyttH &                                                                & & & \multicolumn{2}{c}{                               } \\
\hline
VBF         & $\Cc_{\PZ\PZ}^2 \Cc_\mathrm{VBF}^2(1,\Rr_{\PW\PZ}^2)\cdot \Rr_{\PGg\PZ}^2$   & $\Cc_{\PZ\PZ}^2 \Cc_\mathrm{VBF}^2(1,\Rr_{\PW\PZ}^2) $ & $\Cc_{\PZ\PZ}^2 \Cc_\mathrm{VBF}^2(1,\Rr_{\PW\PZ}^2)\cdot \Rr_{\PW\PZ}^2$ & \multicolumn{2}{c}{$\Cc_{\PZ\PZ}^2 \Cc_\mathrm{VBF}^2(1,\Rr_{\PW\PZ}^2)\cdot \Rr_{\Pf\PZ}^2$} \\
\hline
$\PW\PH$        & $\Cc_{\PZ\PZ}^2 \Rr_{\PW\PZ}^2\cdot \Rr_{\PGg\PZ}^2$ & $\Cc_{\PZ\PZ}^2\cdot \Rr_{\PW\PZ}^2$ & $\Cc_{\PZ\PZ}^2 \Rr_{\PW\PZ}^2\cdot \Rr_{\PW\PZ}^2$ & \multicolumn{2}{c}{$\Cc_{\PZ\PZ}^2 \Rr_{\PW\PZ}^2\cdot \Rr_{\Pf\PZ}^2$} \\
\hline
$\PZ\PH$        & $\Cc_{\PZ\PZ}^2\cdot \Rr_{\PGg\PZ}^2$          & $\Cc_{\PZ\PZ}^2$               & $\Cc_{\PZ\PZ}^2\cdot \Rr_{\PW\PZ}^2$          & \multicolumn{2}{c}{$\Cc_{\PZ\PZ}^2\cdot \Rr_{\Pf\PZ}^2$} \\
\hline
\end{tabular}

{\footnotesize $\Cc_i^2 = \Gamma_{ii} / \Gamma_{ii}^\text{SM}$}

\end{sidewaystable}

\subsubsection{Probing the fermion sector}
\label{sec:LM_ir:Rdu}
\label{sec:LM_ir:Rlq}

In many extensions of the SM the Higgs bosons
couple differently to different types of fermions.

Given that the gluon-gluon fusion production process is dominated by the top-quark coupling,
and that there are two decay modes involving fermions,
one way of splitting fermions that is within experimental reach is
to consider up-type fermions (top quark) and
down-type fermions (bottom quark and tau lepton) separately.
In this parameterization, presented in \Tref{tab:LM_ir:Rdu}, 
the most relevant parameter of interest is $\Rr_{\PQd\PQu} (=\Cc_{\PQd}/\Cc_{\PQu})$,
the ratio of the scale factors
of the couplings to down-type fermions,
$\Cc_{\PQd} = \Cc_{\PGt} (= \Cc_{\PGm}) = \Cc_{\PQb} (= \Cc_{\PQs})$, and up-type fermions, $\Cc_{\PQu} = \Cc_{\PQt} (= \Cc_{\PQc})$.

\begin{table}[p!]
\centering

\caption{A benchmark parameterization where the up-type and down-type symmetry of fermions is probed through the $\Rr_{\PQd\PQu}$ parameter.}
\label{tab:LM_ir:Rdu}

\begin{tabular}{lccccc}
\hline
\multicolumn{6}{l}{\bfseries Probing up-type and down-type fermion symmetry assuming no invisible or undetectable widths}\\
\multicolumn{6}{l}{\footnotesize Free parameters: $\Cc_{\PV} (= \Cc_{\PZ} = \Cc_{\PW})$, $\Rr_{\PQd\PQu} (= \Cc_{\PQd} / \Cc_{\PQu})$, $\Cc_{\PQu} (= \Cc_{\PQt})$.} \\
\multicolumn{6}{l}{\footnotesize Dependent parameters: $\Cc_{\PGg} = \Cc_{\PGg}(\Cc_{\PQu} \Rr_{\PQd\PQu},\Cc_{\PQu},\Cc_{\PQu} \Rr_{\PQd\PQu},\Cc_{\PV})$, $\Cc_{\Pg} = \Cc_{\Pg}(\Cc_{\PQu} \Rr_{\PQd\PQu},\Cc_{\PQu})$, $\Cc_{\PH} = \Cc_{\PH}(\Cc_i)$.} \\
\hline
 & $\PH\to\PGg\PGg$ & $\PH\to \PZ\PZ^{(*)}$ & $\PH\to \PW\PW^{(*)}$ & $\PH\to \PQb\PAQb$ & $\PH\to\PGtm\PGtp$ \\
\hline
\MyggH       & $\frac{\Cc_{\Pg}^2(\Cc_{\PQu} \Rr_{\PQd\PQu},\Cc_{\PQu})\cdot \Cc_{\PGg}^2(\Cc_{\PQu} \Rr_{\PQd\PQu},\Cc_{\PQu},\Cc_{\PQu} \Rr_{\PQd\PQu},\Cc_{\PV})}{\Cc_{\PH}^2(\Cc_i)}$ & \multicolumn{2}{c}{$\frac{\Cc_{\Pg}^2(\Cc_{\PQu} \Rr_{\PQd\PQu},\Cc_{\PQu})\cdot \Cc_{\PV}^2}{\Cc_{\PH}^2(\Cc_i)}$} & \multicolumn{2}{c}{$\frac{\Cc_{\Pg}^2(\Cc_{\PQu} \Rr_{\PQd\PQu},\Cc_{\PQu})\cdot (\Cc_{\PQu} \Rr_{\PQd\PQu})^2}{\Cc_{\PH}^2(\Cc_i)}$} \\
\hline
\MyttH & $\frac{\Cc_{\PQu}^2\cdot \Cc_{\PGg}^2(\Cc_{\PQu} \Rr_{\PQd\PQu},\Cc_{\PQu},\Cc_{\PQu} \Rr_{\PQd\PQu},\Cc_{\PV})}{\Cc_{\PH}^2(\Cc_i)}$                 & \multicolumn{2}{c}{$\frac{\Cc_{\PQu}^2\cdot \Cc_{\PV}^2}{\Cc_{\PH}^2(\Cc_i)}$}                 & \multicolumn{2}{c}{$\frac{\Cc_{\PQu}^2\cdot (\Cc_{\PQu} \Rr_{\PQd\PQu})^2}{\Cc_{\PH}^2(\Cc_i)}$}  \\
\hline
VBF         & \multirow{3}{*}{$\frac{\Cc_{\PV}^2\cdot \Cc_{\PGg}^2(\Cc_{\PQu} \Rr_{\PQd\PQu},\Cc_{\PQu},\Cc_{\PQu} \Rr_{\PQd\PQu},\Cc_{\PV})}{\Cc_{\PH}^2(\Cc_i)}$} & \multicolumn{2}{c}{\multirow{3}{*}{$\frac{\Cc_{\PV}^2\cdot \Cc_{\PV}^2}{\Cc_{\PH}^2(\Cc_i)}$}} & \multicolumn{2}{c}{\multirow{3}{*}{$\frac{\Cc_{\PV}^2\cdot (\Cc_{\PQu} \Rr_{\PQd\PQu})^2}{\Cc_{\PH}^2(\Cc_i)}$}} \\
$\PW\PH$        &                                                                              & \multicolumn{2}{c}{                                                 } & \multicolumn{2}{c}{                                          } \\
$\PZ\PH$        &                                                                              & \multicolumn{2}{c}{                                                 } & \multicolumn{2}{c}{                                          } \\
\hline
\hline
\multicolumn{6}{l}{\bfseries Probing up-type and down-type fermion symmetry assuming $\Cc_{\PV}<1$}\\
\multicolumn{6}{l}{\footnotesize Free parameters: $\Cc_{\PV} (= \Cc_{\PZ} = \Cc_{\PW})$, $\Rr_{\PQd\PQu} (= \Cc_{\PQd} / \Cc_{\PQu})$, $\Cc_{\PQu} (= \Cc_{\PQt})$, $\Cc_{\PH}$, with conditions $\Cc_{\PV}<1$ and $\Cc_{\PH}^2>\Cc_{\PH}^2(\Cc_i)$.} \\
\multicolumn{6}{l}{\footnotesize Dependent parameters: $\Cc_{\PGg} = \Cc_{\PGg}(\Cc_{\PQu} \Rr_{\PQd\PQu},\Cc_{\PQu},\Cc_{\PQu} \Rr_{\PQd\PQu},\Cc_{\PV})$, $\Cc_{\Pg} = \Cc_{\Pg}(\Cc_{\PQu} \Rr_{\PQd\PQu},\Cc_{\PQu})$.} \\
\hline
 & $\PH\to\PGg\PGg$ & $\PH\to \PZ\PZ^{(*)}$ & $\PH\to \PW\PW^{(*)}$ & $\PH\to \PQb\PAQb$ & $\PH\to\PGtm\PGtp$ \\
\hline
\MyggH       & $\frac{\Cc_{\Pg}^2(\Cc_{\PQu} \Rr_{\PQd\PQu},\Cc_{\PQu})\cdot \Cc_{\PGg}^2(\Cc_{\PQu} \Rr_{\PQd\PQu},\Cc_{\PQu},\Cc_{\PQu} \Rr_{\PQd\PQu},\Cc_{\PV})}{\Cc_{\PH}^2}$ & \multicolumn{2}{c}{$\frac{\Cc_{\Pg}^2(\Cc_{\PQu} \Rr_{\PQd\PQu},\Cc_{\PQu})\cdot \Cc_{\PV}^2}{\Cc_{\PH}^2}$} & \multicolumn{2}{c}{$\frac{\Cc_{\Pg}^2(\Cc_{\PQu} \Rr_{\PQd\PQu},\Cc_{\PQu})\cdot (\Cc_{\PQu} \Rr_{\PQd\PQu})^2}{\Cc_{\PH}^2}$} \\
\hline
\MyttH & $\frac{\Cc_{\PQu}^2\cdot \Cc_{\PGg}^2(\Cc_{\PQu} \Rr_{\PQd\PQu},\Cc_{\PQu},\Cc_{\PQu} \Rr_{\PQd\PQu},\Cc_{\PV})}{\Cc_{\PH}^2}$                 & \multicolumn{2}{c }{$\frac{\Cc_{\PQu}^2\cdot \Cc_{\PV}^2}{\Cc_{\PH}^2}$}                 & \multicolumn{2}{c }{$\frac{\Cc_{\PQu}^2\cdot (\Cc_{\PQu} \Rr_{\PQd\PQu})^2}{\Cc_{\PH}^2}$}  \\
\hline
VBF         & \multirow{3}{*}{$\frac{\Cc_{\PV}^2\cdot \Cc_{\PGg}^2(\Cc_{\PQu} \Rr_{\PQd\PQu},\Cc_{\PQu},\Cc_{\PQu} \Rr_{\PQd\PQu},\Cc_{\PV})}{\Cc_{\PH}^2}$} & \multicolumn{2}{c}{\multirow{3}{*}{$\frac{\Cc_{\PV}^2\cdot \Cc_{\PV}^2}{\Cc_{\PH}^2}$}} & \multicolumn{2}{c}{\multirow{3}{*}{$\frac{\Cc_{\PV}^2\cdot (\Cc_{\PQu} \Rr_{\PQd\PQu})^2}{\Cc_{\PH}^2}$}} \\
$\PW\PH$        &                                                                              & \multicolumn{2}{c}{                                                 } & \multicolumn{2}{c}{                                          } \\
$\PZ\PH$        &                                                                              & \multicolumn{2}{c}{                                                 } & \multicolumn{2}{c}{                                          } \\
\hline
\hline
\multicolumn{6}{l}{\bfseries Probing up-type and down-type fermion symmetry without assumptions on the total width}\\
\multicolumn{6}{l}{\footnotesize Free parameters: $\Cc_{\PQu\PQu} (= \Cc_{\PQu}\cdot \Cc_{\PQu} / \Cc_{\PH})$, $\Rr_{\PQd\PQu} (= \Cc_{\PQd} / \Cc_{\PQu})$, $\Rr_{\PV\PQu} (= \Cc_{\PV} / \Cc_{\PQu})$.} \\
\multicolumn{6}{l}{\footnotesize Dependent parameters: $\Cc_{\PGg} = \Cc_{\PGg}(\Rr_{\PQd\PQu}, 1,\Rr_{\PQd\PQu}, \Rr_{\PV\PQu})$, $\Cc_{\Pg} = \Cc_{\Pg}(\Rr_{\PQd\PQu}, 1)$.} \\
\hline
 & $\PH\to\PGg\PGg$ & $\PH\to \PZ\PZ^{(*)}$ & $\PH\to \PW\PW^{(*)}$ & $\PH\to \PQb\PAQb$ & $\PH\to\PGtm\PGtp$ \\
\hline
\MyggH       & $\Cc_{\PQu\PQu}^2 \Cc_{\Pg}^2(\Rr_{\PQd\PQu}, 1)\cdot \Cc_{\PGg}^2(\Rr_{\PQd\PQu}, 1,\Rr_{\PQd\PQu}, \Rr_{\PV\PQu})$ & \multicolumn{2}{c}{$\Cc_{\PQu\PQu}^2 \Cc_{\Pg}^2(\Rr_{\PQd\PQu}, 1)\cdot \Rr_{\PV\PQu}^2$} & \multicolumn{2}{c}{$\Cc_{\PQu\PQu}^2 \Cc_{\Pg}^2(\Rr_{\PQd\PQu}, 1)\cdot \Rr_{\PQd\PQu}^2$} \\
\hline
\MyttH & $\Cc_{\PQu\PQu}^2\cdot \Cc_{\PGg}^2(\Rr_{\PQd\PQu}, 1,\Rr_{\PQd\PQu}, \Rr_{\PV\PQu})$                  & \multicolumn{2}{c}{$\Cc_{\PQu\PQu}^2\cdot \Rr_{\PV\PQu}^2$}                  & \multicolumn{2}{c}{$\Cc_{\PQu\PQu}^2\cdot \Rr_{\PQd\PQu}^2$} \\
\hline
VBF         & \multirow{3}{*}{$\Cc_{\PQu\PQu}^2 \Rr_{\PV\PQu}^2\cdot \Cc_{\PGg}^2(\Rr_{\PQd\PQu}, 1,\Rr_{\PQd\PQu}, \Rr_{\PV\PQu})$} & \multicolumn{2}{c}{\multirow{3}{*}{$\Cc_{\PQu\PQu}^2 \Rr_{\PV\PQu}^2\cdot \Rr_{\PV\PQu}^2$}} & \multicolumn{2}{c}{\multirow{3}{*}{$\Cc_{\PQu\PQu}^2 \Rr_{\PV\PQu}^2\cdot \Rr_{\PQd\PQu}^2$}} \\
$\PW\PH$        &                                                                           & \multicolumn{2}{c}{                                                  } & \multicolumn{2}{c}{                                          } \\
$\PZ\PH$        &                                                                           & \multicolumn{2}{c}{                                                  } & \multicolumn{2}{c}{                                          } \\
\hline
\end{tabular}

{\footnotesize $\Cc_i^2 = \Gamma_{ii} / \Gamma_{ii}^\text{SM}$, $\Cc_{\PQd} = \Cc_{\PQb} = \Cc_{\PGt}$}
\end{table}

Alternatively one can consider quarks and leptons separately.
In this parameterization, presented in \Tref{tab:LM_ir:Rlq},
the most relevant parameter of interest is $\Rr_{\Pl\PQq} (=\Cc_{\Pl}/\Cc_{\PQq})$,
the ratio of the coupling scale factors to leptons,
$\Cc_{\Pl} = \Cc_{\PGt} (= \Cc_{\PGm})$, and quarks, $\Cc_{\PQq} = \Cc_{\PQt} (= \Cc_{\PQc}) = \Cc_{\PQb} (= \Cc_{\PQs})$.

\begin{table}[p]
\centering
\caption{A benchmark parameterization where the quark and lepton symmetry of fermions is probed through the $\Rr_{\Pl\PQq}$ parameter.}
\label{tab:LM_ir:Rlq}
\begin{tabular}{lccccc}
\hline
\multicolumn{6}{l}{\bfseries Probing quark and lepton fermion symmetry assuming no invisible or undetectable widths}\\
\multicolumn{6}{l}{\footnotesize Free parameters: $\Cc_{\PV} (= \Cc_{\PZ} = \Cc_{\PW})$, $\Rr_{\Pl\PQq} (= \Cc_{\Pl} / \Cc_{\PQq})$, $\Cc_{\PQq} (= \Cc_{\PQt} = \Cc_{\PQb})$.} \\
\multicolumn{6}{l}{\footnotesize Dependent parameters: $\Cc_{\PGg} = \Cc_{\PGg}(\Cc_{\PQq},\Cc_{\PQq},\Cc_{\PQq} \Rr_{\Pl\PQq},\Cc_{\PV})$, $\Cc_{\Pg} = \Cc_{\PQq}$, $\Cc_{\PH} = \Cc_{\PH}(\Cc_i)$.} \\
\hline
 & $\PH\to\PGg\PGg$ & $\PH\to \PZ\PZ^{(*)}$ & $\PH\to \PW\PW^{(*)}$ & $\PH\to \PQb\PAQb$ & $\PH\to\PGtm\PGtp$ \\
\hline
\MyggH       & \multirow{2}{*}{$\frac{\Cc_{\PQq}^2\cdot \Cc_{\PGg}^2(\Cc_{\PQq},\Cc_{\PQq},\Cc_{\PQq} \Rr_{\Pl\PQq},\Cc_{\PV})}{\Cc_{\PH}^2(\Cc_i)}$} & \multicolumn{2}{c}{\multirow{2}{*}{$\frac{\Cc_{\PQq}^2\cdot \Cc_{\PV}^2}{\Cc_{\PH}^2(\Cc_i)}$}} & \multirow{2}{*}{$\frac{\Cc_{\PQq}^2\cdot \Cc_{\PQq}^2}{\Cc_{\PH}^2(\Cc_i)}$} & \multirow{2}{*}{$\frac{\Cc_{\PQq}^2\cdot (\Cc_{\PQq} \Rr_{\Pl\PQq})^2}{\Cc_{\PH}^2(\Cc_i)}$} \\
\MyttH &                                                                       & \multicolumn{2}{c}{                                 }                 & &  \\
\hline
VBF         & \multirow{3}{*}{$\frac{\Cc_{\PV}^2\cdot \Cc_{\PGg}^2(\Cc_{\PQq},\Cc_{\PQq},\Cc_{\PQq} \Rr_{\Pl\PQq},\Cc_{\PV})}{\Cc_{\PH}^2(\Cc_i)}$} & \multicolumn{2}{c}{\multirow{3}{*}{$\frac{\Cc_{\PV}^2\cdot \Cc_{\PV}^2}{\Cc_{\PH}^2(\Cc_i)}$}} & \multirow{3}{*}{$\frac{\Cc_{\PV}^2\cdot \Cc_{\PQq}^2}{\Cc_{\PH}^2(\Cc_i)}$} & \multirow{3}{*}{$\frac{\Cc_{\PV}^2\cdot (\Cc_{\PQq} \Rr_{\Pl\PQq})^2}{\Cc_{\PH}^2(\Cc_i)}$} \\
$\PW\PH$        &                                                                       & \multicolumn{2}{c}{                                                 } & & \\
$\PZ\PH$        &                                                                       & \multicolumn{2}{c}{                                                 } & & \\
\hline
\hline
\multicolumn{6}{l}{\bfseries Probing quark and lepton fermion symmetry assuming $\Cc_{\PV}<1$}\\
\multicolumn{6}{l}{\footnotesize Free parameters: $\Cc_{\PV} (= \Cc_{\PZ} = \Cc_{\PW})$, $\Rr_{\Pl\PQq} (= \Cc_{\Pl} / \Cc_{\PQq})$, $\Cc_{\PQq} (= \Cc_{\PQt} = \Cc_{\PQb})$, $\Cc_{\PH}$, with conditions $\Cc_{\PV}<1$ and $\Cc_{\PH}^2>\Cc_{\PH}^2(\Cc_i)$.} \\
\multicolumn{6}{l}{\footnotesize Dependent parameters: $\Cc_{\PGg} = \Cc_{\PGg}(\Cc_{\PQq},\Cc_{\PQq},\Cc_{\PQq} \Rr_{\Pl\PQq},\Cc_{\PV})$, $\Cc_{\Pg} = \Cc_{\PQq}$.} \\
\hline
 & $\PH\to\PGg\PGg$ & $\PH\to \PZ\PZ^{(*)}$ & $\PH\to \PW\PW^{(*)}$ & $\PH\to \PQb\PAQb$ & $\PH\to\PGtm\PGtp$ \\
\hline
\MyggH       & \multirow{2}{*}{$\frac{\Cc_{\PQq}^2\cdot \Cc_{\PGg}^2(\Cc_{\PQq},\Cc_{\PQq},\Cc_{\PQq} \Rr_{\Pl\PQq},\Cc_{\PV})}{\Cc_{\PH}^2}$} & \multicolumn{2}{c}{\multirow{2}{*}{$\frac{\Cc_{\PQq}^2\cdot \Cc_{\PV}^2}{\Cc_{\PH}^2}$}} & \multirow{2}{*}{$\frac{\Cc_{\PQq}^2\cdot \Cc_{\PQq}^2}{\Cc_{\PH}^2}$} & \multirow{2}{*}{$\frac{\Cc_{\PQq}^2\cdot (\Cc_{\PQq} \Rr_{\Pl\PQq})^2}{\Cc_{\PH}^2}$} \\
\MyttH &                                                                       & \multicolumn{2}{c}{                                 }                 & &  \\
\hline
VBF         & \multirow{3}{*}{$\frac{\Cc_{\PV}^2\cdot \Cc_{\PGg}^2(\Cc_{\PQq},\Cc_{\PQq},\Cc_{\PQq} \Rr_{\Pl\PQq},\Cc_{\PV})}{\Cc_{\PH}^2}$} & \multicolumn{2}{c}{\multirow{3}{*}{$\frac{\Cc_{\PV}^2\cdot \Cc_{\PV}^2}{\Cc_{\PH}^2}$}} & \multirow{3}{*}{$\frac{\Cc_{\PV}^2\cdot \Cc_{\PQq}^2}{\Cc_{\PH}^2}$} & \multirow{3}{*}{$\frac{\Cc_{\PV}^2\cdot (\Cc_{\PQq} \Rr_{\Pl\PQq})^2}{\Cc_{\PH}^2}$} \\
$\PW\PH$        &                                                                       & \multicolumn{2}{c}{                                                 } & & \\
$\PZ\PH$        &                                                                       & \multicolumn{2}{c}{                                                 } & & \\
\hline
\hline
\multicolumn{6}{l}{\bfseries Probing quark and lepton fermion symmetry without assumptions on the total width}\\
\multicolumn{6}{l}{\footnotesize Free parameters: $\Cc_{\PQq\PQq} (= \Cc_{\PQq}\cdot \Cc_{\PQq} / \Cc_{\PH})$, $\Rr_{\Pl\PQq} (= \Cc_{\Pl} / \Cc_{\PQq})$, $\Rr_{\PV\PQq} (= \Cc_{\PV} / \Cc_{\PQq})$.} \\
\multicolumn{6}{l}{\footnotesize Dependent parameters: $\Cc_{\PGg} = \Cc_{\PGg}(1,1,\Rr_{\Pl\PQq},\Rr_{\PV\PQq})$, $\Cc_{\Pg} = \Cc_{\PQq}$.} \\
\hline
 & $\PH\to\PGg\PGg$ & $\PH\to \PZ\PZ^{(*)}$ & $\PH\to \PW\PW^{(*)}$ & $\PH\to \PQb\PAQb$ & $\PH\to\PGtm\PGtp$ \\
\hline
\MyggH       & \multirow{2}{*}{$\Cc_{\PQq\PQq}^2\cdot \Cc_{\PGg}^2(1,1,\Rr_{\Pl\PQq},\Rr_{\PV\PQq})$} & \multicolumn{2}{c}{\multirow{2}{*}{$\Cc_{\PQq\PQq}^2\cdot \Rr_{\PV\PQq}^2$}} & \multirow{2}{*}{$\Cc_{\PQq\PQq}^2$} & \multirow{2}{*}{$\Cc_{\PQq\PQq}^2\cdot \Rr_{\Pl\PQq}^2$} \\
\MyttH &                                                                       & \multicolumn{2}{c}{                                 }                 & &  \\
\hline
VBF         & \multirow{3}{*}{$\Cc_{\PQq\PQq}^2 \Rr_{\PV\PQq}^2\cdot \Cc_{\PGg}^2(1,1,\Rr_{\Pl\PQq},\Rr_{\PV\PQq})$} & \multicolumn{2}{c}{\multirow{3}{*}{$\Cc_{\PQq\PQq}^2 \Rr_{\PV\PQq}^2\cdot \Rr_{\PV\PQq}^2$}} & \multirow{3}{*}{$\Cc_{\PQq\PQq}^2\cdot \Rr_{\PV\PQq}^2$} & \multirow{3}{*}{$\Cc_{\PQq\PQq}^2 \Rr_{\PV\PQq}^2\cdot \Rr_{\Pl\PQq}^2$} \\
$\PW\PH$        &                                                                       & \multicolumn{2}{c}{                                                 } & & \\
$\PZ\PH$        &                                                                       & \multicolumn{2}{c}{                                                 } & & \\
\hline
\end{tabular}

{\footnotesize $\Cc_i^2 = \Gamma_{ii} / \Gamma_{ii}^\text{SM}$, $\Cc_{\Pl} = \Cc_{\PGt}$}

\end{table}

One further combination of top-quark, bottom-quark and tau-lepton, namely
scaling the top-quark and tau-lepton with a common parameter and the
bottom-quark with another parameter, can be envisaged and readily
parametrized based on the interim framework but is not put forward as a benchmark.

\subsubsection{Probing the loop structure and invisible or undetectable decays}
\label{sec:LM_ir:CgCgam}

New particles associated with physics beyond the SM
may influence the partial width of the gluon and/or photon vertices.

In this parameterization, presented in \Tref{tab:LM_ir:CgCgam},
each of the loop-induced vertices is represented by an effective scale factor, $\Cc_{\Pg}$ and $\Cc_{\PGg}$. 
On the other hand, the couplings to the known SM particles are assumed to be as in the SM: $\Cc_{\PZ} = \Cc_{\PW} = \Cc_{\PGt} = \Cc_{\PQb} = \Cc_{\PQt}=1$.

Particles not predicted by the SM may also give rise to invisible or undetectable
decays.
In order to probe this possibility, instead of absorbing the total width into another parameter or leaving it free,
a different parameter is introduced, $\BRinv$. 
The assumption of the fixed coupling couplings to SM particles allows to determine $\BRinv$ from the LHC data.
The definition of $\BRinv$ is relative to the rescaled total width, $\Cc_{\PH}^2(\Cc_i)$,
and can thus be interpreted as the invisible or undetectable fraction of the total width. 

Invisible decays might show up as a missing transverse energy (MET) signature 
and can be measured at the LHC with dedicated analyses. 
An example of an undetectable final state would be a multi-jet signature that
cannot be separated from QCD backgrounds at the LHC and hence not detected.
With sufficient data it can be envisaged to disentangle the invisible and undetectable components by splitting into 
two parameters $\BRonlyinv$ and $\BRundet$, where $\BRonlyinv$ is determined by the direct searches 
for invisible Higgs decay signatures.

\begin{table}[p]
\centering
\caption{A benchmark parameterization where effective vertex couplings are allowed to float through the $\Cc_{\Pg}$ and $\Cc_{\PGg}$ parameters.
Instead of absorbing $\Cc_{\PH}$, explicit allowance is made for a contribution from invisible or undetectable widths via the $\BRinv$ or $\BRonlyinv$ and $\BRundet$ parameters.}
\label{tab:LM_ir:CgCgam}
\begin{tabular}{lccccc}
\hline
\multicolumn{6}{l}{\bfseries Probing loop structure assuming no invisible or undetectable widths}\\
\multicolumn{6}{l}{\footnotesize Free parameters: $\Cc_{\Pg}$, $\Cc_{\PGg}$.} \\
\multicolumn{6}{l}{\footnotesize Dependent parameters: $\Cc_{\PH} = \Cc_{\PH}(\Cc_i)$. Fixed parameters $\Cc_{\PZ} = \Cc_{\PW} = \Cc_{\PGt} = \Cc_{\PQb} = \Cc_{\PQt}=1$.} \\
\hline
 & $\PH\to\PGg\PGg$ & $\PH\to \PZ\PZ^{(*)}$ & $\PH\to \PW\PW^{(*)}$ & $\PH\to \PQb\PAQb$ & $\PH\to\PGtm\PGtp$  \\
\hline
\MyggH       & $\frac{\Cc_{\Pg}^2\cdot \Cc_{\PGg}^2}{\Cc_{\PH}^2(\Cc_i)}$ & \multicolumn{4}{c}{$\frac{\Cc_{\Pg}^2}{\Cc_{\PH}^2(\Cc_i)}$} \\
\hline
\MyttH & \multirow{4}{*}{$\frac{\Cc_{\PGg}^2}{\Cc_{\PH}^2(\Cc_i)}$}& \multicolumn{4}{c}{\multirow{4}{*}{$\frac{1}{\Cc_{\PH}^2(\Cc_i)}$}    } \\
VBF         &                                                   & \multicolumn{4}{c}{                           } \\
$\PW\PH$        &                                                   & \multicolumn{4}{c}{                           } \\
$\PZ\PH$        &                                                   & \multicolumn{4}{c}{                           } \\
\hline
\hline
\multicolumn{6}{l}{\bfseries Probing loop structure allowing for invisible or undetectable widths}\\
\multicolumn{6}{l}{\footnotesize Free parameters: $\Cc_{\Pg}$, $\Cc_{\PGg}$, $\BRinv$.} \\
\multicolumn{6}{l}{\footnotesize Dependent parameters: $\Cc_{\PH} = \Cc_{\PH}(\Cc_i)$. Fixed parameters $\Cc_{\PZ} = \Cc_{\PW} = \Cc_{\PGt} = \Cc_{\PQb} = \Cc_{\PQt}=1$.} \\
\hline
 & $\PH\to\PGg\PGg$ & $\PH\to \PZ\PZ^{(*)}$ & $\PH\to \PW\PW^{(*)}$ & $\PH\to \PQb\PAQb$ & $\PH\to\PGtm\PGtp$  \\
\hline
\MyggH       & $\frac{\Cc_{\Pg}^2\cdot \Cc_{\PGg}^2}{\Cc_{\PH}^2(\Cc_i) /(1-\BRinv)}$ & \multicolumn{4}{c}{$\frac{\Cc_{\Pg}^2}{\Cc_{\PH}^2(\Cc_i) /(1-\BRinv)}$} \\
\hline
\MyttH & \multirow{4}{*}{$\frac{\Cc_{\PGg}^2}{\Cc_{\PH}^2(\Cc_i) /(1-\BRinv)}$}& \multicolumn{4}{c}{\multirow{4}{*}{$\frac{1}{\Cc_{\PH}^2(\Cc_i) /(1-\BRinv)}$}    } \\
VBF         &                                                   & \multicolumn{4}{c}{                           } \\
$\PW\PH$        &                                                   & \multicolumn{4}{c}{                           } \\
$\PZ\PH$        &                                                   & \multicolumn{4}{c}{                           } \\
\hline
\hline
\multicolumn{6}{l}{\bfseries Probing loop structure allowing for separate invisible and undetectable widths}\\
\multicolumn{6}{l}{\footnotesize Free parameters: $\Cc_{\Pg}$, $\Cc_{\PGg}$, $\BRonlyinv$, $\BRundet$, with condition $\BRonlyinv+\BRundet\leq 1$.} \\
\multicolumn{6}{l}{\footnotesize Dependent parameters: $\Cc_{\PH} = \Cc_{\PH}(\Cc_i)$. Fixed parameters $\Cc_{\PZ} = \Cc_{\PW} = \Cc_{\PGt} = \Cc_{\PQb} = \Cc_{\PQt}=1$.} \\
\hline
 & $\PH\to\PGg\PGg$ & $\PH\to \PZ\PZ^{(*)}$ & $\PH\to \PW\PW^{(*)}$ & $\PH\to \PQb\PAQb$ & $\PH\to\PGtm\PGtp$  \\
\hline
\MyggH       & $\frac{\Cc_{\Pg}^2\cdot \Cc_{\PGg}^2}{\Cc_{\PH}^2(\Cc_i) /(1-\BRonlyinv-\BRundet)}$ & \multicolumn{4}{c}{$\frac{\Cc_{\Pg}^2}{\Cc_{\PH}^2(\Cc_i) /(1-\BRonlyinv-\BRundet)}$} \\
\hline
\MyttH & \multirow{4}{*}{$\frac{\Cc_{\PGg}^2}{\Cc_{\PH}^2(\Cc_i) /(1-\BRonlyinv-\BRundet)}$}& \multicolumn{4}{c}{\multirow{4}{*}{$\frac{1}{\Cc_{\PH}^2(\Cc_i) /(1-\BRonlyinv-\BRundet)}$}    } \\
VBF         &                                                   & \multicolumn{4}{c}{                           } \\
$\PW\PH$        &                                                   & \multicolumn{4}{c}{                           } \\
$\PZ\PH$        &                                                   & \multicolumn{4}{c}{                           } \\
\hline
\end{tabular}

{\footnotesize $\Cc_i^2 = \Gamma_{ii} / \Gamma_{ii}^\text{SM}$}

\end{table}

One particularity of this benchmark parameterization is that it should allow
%any
theoretical predictions involving new particles to be projected into the
 $(\Cc_{\Pg},\Cc_{\PGg})$ or $(\Cc_{\Pg},\Cc_{\PGg},\BRinv)$ spaces.

It can be noted that the benchmark parameterization including
\BRinv\ can be recast in a form that allows
for an interpretation in terms of a tree-level scale factor and
the loop-induced scale factors with the following
substitutions: $\Cc_j \to \Cc_j^\prime / \Cc_\mathrm{tree}$
(with $j=\Pg,\PGg$) and $(1-\BRinv) \to \Cc_\mathrm{tree}^2$.

Once the $\PH \to \PZ\PGg$ decay mode reaches sufficient sensitivity, a natural extension for this benchmark is 
to fit an extra degree of freedom in the form of the parameter $\Cc_{(\PZ\PGg)}$ for the $\PZ\PGg$ final state.

\subsubsection{A minimal parameterization without assumptions on new physics contributions}

The following parameterization gathers the most
important degrees of freedom considered before,
namely $\Cc_{\Pg}$, $\Cc_{\PGg}$, $\Cc_{\PV}$, $\Cc_{\Pf}$.
The parameterization, presented in \Tref{tab:LM_ir:minimal_noassumptions},
 is chosen such that some parameters are expected to be reasonably constrained by the LHC data in the near term,
while other parameters are not expected to be as well constrained in the same time frame. 

It should be noted that this is a parameterization which only includes trivial scale factors.

With the presently available analyses and data, $\Cc_{\Pg \PV}^2 = \Cc_{\Pg}^2\cdot \Cc_{\PV}^2 / \Cc_{\PH}^2$
seems to be a good choice for the common $\Cc_{ij}$ parameter, but all choices are equivalent when considering
the full 4-dimensional probability distribution.

\begin{table}[p]
\centering
\caption{A benchmark parameterization where effective vertex couplings are allowed to float through the $\Cc_{\Pg}$ and $\Cc_{\PGg}$ parameters
and the gauge and fermion couplings through the unified parameters $\Cc_{\PV}$ and $\Cc_{\Pf}$.}
\label{tab:LM_ir:minimal_noassumptions}
\begin{tabular}{lccccc}
\hline
\multicolumn{6}{l}{\bfseries Probing loops while allowing other couplings to float assuming no invisible or undetectable widths} \\
\multicolumn{6}{l}{\footnotesize Free parameters: $\Cc_{\Pg}$, $\Cc_{\PGg}$, $\Cc_{\PV} (=\Cc_{\PW}=\Cc_{\PZ})$, $\Cc_{\Pf} (=\Cc_{\PQt}=\Cc_{\PQb}=\Cc_{\PGt})$.} \\
\multicolumn{6}{l}{\footnotesize Dependent parameters: $\Cc_{\PH} = \Cc_{\PH}(\Cc_i)$.} \\
\hline
 & $\PH\to\PGg\PGg$ & $\PH\to \PZ\PZ^{(*)}$ & $\PH\to \PW\PW^{(*)}$ & \hspace{1cm}$\PH\to \PQb\PAQb$\hspace{1cm} & $\PH\to\PGtm\PGtp$ \\
\hline
\MyggH       & $\frac{\Cc_{\Pg}^2\cdot \Cc_{\PGg}^2}{\Cc_{\PH}^2(\Cc_i)}$ & \multicolumn{2}{c}{$\frac{\Cc_{\Pg}^2\cdot \Cc_{\PV}^2}{\Cc_{\PH}^2(\Cc_i)}$} & \multicolumn{2}{c}{$\frac{\Cc_{\Pg}^2\cdot \Cc_{\Pf}^2}{\Cc_{\PH}^2(\Cc_i)}$} \\
\hline
\MyttH & $\frac{\Cc_{\Pf}^2\cdot \Cc_{\PGg}^2}{\Cc_{\PH}^2(\Cc_i)}$ & \multicolumn{2}{c}{$\frac{\Cc_{\Pf}^2\cdot \Cc_{\PV}^2}{\Cc_{\PH}^2(\Cc_i)}$} & \multicolumn{2}{c}{$\frac{\Cc_{\Pf}^2\cdot \Cc_{\Pf}^2}{\Cc_{\PH}^2(\Cc_i)}$} \\
\hline
VBF         & \multirow{3}{*}{$\frac{\Cc_{\PV}^2\cdot \Cc_{\PGg}^2}{\Cc_{\PH}^2(\Cc_i)}$} & \multicolumn{2}{c}{\multirow{3}{*}{$\frac{\Cc_{\PV}^2\cdot \Cc_{\PV}^2}{\Cc_{\PH}^2(\Cc_i)}$}} & \multicolumn{2}{c}{\multirow{3}{*}{$\frac{\Cc_{\PV}^2\cdot \Cc_{\Pf}^2}{\Cc_{\PH}^2(\Cc_i)}$}} \\
$\PW\PH$        &                                                          & \multicolumn{2}{c}{                                                 } & \multicolumn{2}{c}{                                                 } \\
$\PZ\PH$        &                                                          & \multicolumn{2}{c}{                                                 } & \multicolumn{2}{c}{                                                 } \\
\hline
\hline
\multicolumn{6}{l}{\bfseries Probing loops while allowing other couplings to float assuming $\Cc_{\PV}<1$} \\
\multicolumn{6}{l}{\footnotesize Free parameters: $\Cc_{\Pg}$, $\Cc_{\PGg}$, $\Cc_{\PV} (=\Cc_{\PW}=\Cc_{\PZ})$, $\Cc_{\Pf} (=\Cc_{\PQt}=\Cc_{\PQb}=\Cc_{\PGt})$, $\Cc_{\PH}$, with conditions $\Cc_{\PV}<1$ and $\Cc_{\PH}^2>\Cc_{\PH}^2(\Cc_i)$.} \\
\hline
 & $\PH\to\PGg\PGg$ & $\PH\to \PZ\PZ^{(*)}$ & $\PH\to \PW\PW^{(*)}$ & $\PH\to \PQb\PAQb$ & $\PH\to\PGtm\PGtp$ \\
\hline
\MyggH       & $\frac{\Cc_{\Pg}^2\cdot \Cc_{\PGg}^2}{\Cc_{\PH}^2}$ & \multicolumn{2}{c}{$\frac{\Cc_{\Pg}^2\cdot \Cc_{\PV}^2}{\Cc_{\PH}^2}$} & \multicolumn{2}{c}{$\frac{\Cc_{\Pg}^2\cdot \Cc_{\Pf}^2}{\Cc_{\PH}^2}$} \\
\hline
\MyttH & $\frac{\Cc_{\Pf}^2\cdot \Cc_{\PGg}^2}{\Cc_{\PH}^2}$ & \multicolumn{2}{c}{$\frac{\Cc_{\Pf}^2\cdot \Cc_{\PV}^2}{\Cc_{\PH}^2}$} & \multicolumn{2}{c}{$\frac{\Cc_{\Pf}^2\cdot \Cc_{\Pf}^2}{\Cc_{\PH}^2}$} \\
\hline
VBF         & \multirow{3}{*}{$\frac{\Cc_{\PV}^2\cdot \Cc_{\PGg}^2}{\Cc_{\PH}^2}$} & \multicolumn{2}{c}{\multirow{3}{*}{$\frac{\Cc_{\PV}^2\cdot \Cc_{\PV}^2}{\Cc_{\PH}^2}$}} & \multicolumn{2}{c}{\multirow{3}{*}{$\frac{\Cc_{\PV}^2\cdot \Cc_{\Pf}^2}{\Cc_{\PH}^2}$}} \\
$\PW\PH$        &                                                          & \multicolumn{2}{c}{                                                 } & \multicolumn{2}{c}{                                                 } \\
$\PZ\PH$        &                                                          & \multicolumn{2}{c}{                                                 } & \multicolumn{2}{c}{                                                 } \\
\hline
\hline
\multicolumn{6}{l}{\bfseries Probing loops while allowing other couplings to float allowing for invisible or undetectable widths} \\
\multicolumn{6}{l}{\footnotesize Free parameters: $\Cc_{\Pg\PV} (= \Cc_{\Pg}\cdot \Cc_{\PV} / \Cc_{\PH})$, $\Rr_{\PV\Pg}(=\Cc_{\PV}/\Cc_{\Pg})$, $\Rr_{\PGg\PV}(=\Cc_{\PGg} /\Cc_{\PV})$, $\Rr_{\Pf\PV}(=\Cc_{\Pf}/\Cc_{\PV})$.} \\
\hline
 & $\PH\to\PGg\PGg$ & $\PH\to \PZ\PZ^{(*)}$ & $\PH\to \PW\PW^{(*)}$ & $\PH\to \PQb\PAQb$ & $\PH\to\PGtm\PGtp$ \\
\hline
\MyggH       & $\Cc_{\Pg\PV}^2\cdot \Rr_{\PGg\PV}^2$                   & \multicolumn{2}{c}{$\Cc_{\Pg\PV}^2$}                   & \multicolumn{2}{c}{$\Cc_{\Pg\PV}^2\cdot \Rr_{\Pf\PV}^2$} \\
\hline
\MyttH & $\Cc_{\Pg\PV}^2 \Rr_{\PV\Pg}^2 \Rr_{\Pf\PV}^2\cdot \Rr_{\PGg\PV}^2$ & \multicolumn{2}{c}{$\Cc_{\Pg\PV}^2 \Rr_{\PV\Pg}^2 \Rr_{\Pf\PV}^2$} & \multicolumn{2}{c}{$\Cc_{\Pg\PV}^2 \Rr_{\PV\Pg}^2 \Rr_{\Pf\PV}^2\cdot \Rr_{\Pf\PV}^2$} \\
\hline
VBF         & \multirow{3}{*}{$\Cc_{\Pg\PV}^2 \Rr_{\PV\Pg}^2\cdot \Rr_{\PGg\PV}^2$} & \multicolumn{2}{c}{\multirow{3}{*}{$\Cc_{\Pg\PV}^2 \Rr_{\PV\Pg}^2$}} & \multicolumn{2}{c}{\multirow{3}{*}{$\Cc_{\Pg\PV}^2 \Rr_{\PV\Pg}^2\cdot \Rr_{\Pf\PV}^2$}} \\
$\PW\PH$        &                                                          & \multicolumn{2}{c}{                                    } & \multicolumn{2}{c}{                                          } \\
$\PZ\PH$        &                                                          & \multicolumn{2}{c}{                                    } & \multicolumn{2}{c}{                                          } \\
\hline
\end{tabular}

{\footnotesize $\Cc_i^2 = \Gamma_{ii} / \Gamma_{ii}^\text{SM}$, $\Cc_{\PV} = \Cc_{\PW} = \Cc_{\PZ}$, $\Cc_{\Pf} = \Cc_{\PQt} = \Cc_{\PQb} = \Cc_{\PGt}$}

\end{table}

\subsubsection{Most general parameterization for all gauge bosons and third generation fermion couplings}
\label{sec:LM_ir:maximal}

\Tref{tab:LM_ir:maximal_noassumptions} presents the relations in a fit only with simple scale factors, 
making no assumptions on identical coupling scale factors for different particles beyond these necessary for 
first and second generation fermions as discussed in \Sref{sec:LM_ir:further_assumptions:light_fermions}.

Several choices are possible for $\Cc_{ij}$.
With the currently available channels, $\Cc_{\Pg\PZ} = \Cc_{\Pg}\cdot \Cc_{\PZ} / \Cc_{\PH}$ seems most appropriate,
as shown in table~\ref{tab:LM_ir:maximal_noassumptions}. 
The more appealing choices using vector boson scattering
$\Cc_{\PW\PW} = \Cc_{\PW}\cdot \Cc_{\PW} / \Cc_{\PH}$ or $\Cc_{\PZ\PZ} = \Cc_{\PZ}\cdot \Cc_{\PZ} / \Cc_{\PH}$
%will not be as good
will have lower sensitivity until more data is accumulated, but are completely equivalent when 
looking at the full 7-dimensional probability distribution of all parameters.

From all benchmarks discussed in this section, the last parameterization in table~\ref{tab:LM_ir:maximal_noassumptions} 
is the most general parameterization that needs no assumptions beyond those stated in the definition of the framework.

Once the $\PH \to \PGmm\PGmp$ and $\PH \to \PZ\PGg$ decay modes reach sufficient sensitivity, a natural extension for this benchmark is 
to fit extra degrees of freedom in the form of the parameters $\Cc_{\PGm}$ and $\Cc_{(\PZ\PGg)}$ for these two final states.

\begin{sidewaystable}
\centering
\caption{A benchmark parameterization without further assumptions and maximum degrees of freedom.}
\label{tab:LM_ir:maximal_noassumptions}

\begin{tabular}{lccccc}
\hline                
\multicolumn{6}{l}{\bfseries General parameterization allowing all gauge and third generation fermion couplings to float assuming no invisible or undetectable widths} \\
\multicolumn{6}{l}{\footnotesize Free parameters: $\Cc_{\Pg}$, $\Cc_{\PGg}$, $\Cc_{\PW}$, $\Cc_{\PZ}$, $\Cc_{\PQb}$, $\Cc_{\PQt}$, $\Cc_{\PGt}$.} \\
\multicolumn{6}{l}{\footnotesize Dependent parameters: $\Cc_{\PH} = \Cc_{\PH}(\Cc_i)$.} \\
\hline
         & $\PH\to\PGg\PGg$                                                                        & $\PH\to \PZ\PZ^{(*)}$                                                                 & $\PH\to \PW\PW^{(*)}$                                                                 & $\PH\to \PQb\PAQb$                                                                     & $\PH\to\PGtm\PGtp$                                                              \\\hline                                                                                                                                                                                                                                                                     
\MyggH   & $\frac{\Cc_{\Pg}^2\cdot \Cc_{\PGg}^2}{\Cc_{\PH}^2(\Cc_i)}$                              & $\frac{\Cc_{\Pg}^2\cdot \Cc_{\PZ}^2}{\Cc_{\PH}^2(\Cc_i)}$                             & $\frac{\Cc_{\Pg}^2\cdot \Cc_{\PW}^2}{\Cc_{\PH}^2(\Cc_i)}$                             & $\frac{\Cc_{\Pg}^2\cdot \Cc_{\PQb}^2}{\Cc_{\PH}^2(\Cc_i)}$                             & $\frac{\Cc_{\Pg}^2\cdot \Cc_{\PGt}^2}{\Cc_{\PH}^2(\Cc_i)}$ \\\hline
\MyttH   & $\frac{\Cc_{\PQt}^2\cdot \Cc_{\PGg}^2}{\Cc_{\PH}^2(\Cc_i)}$                             & $\frac{\Cc_{\PQt}^2\cdot \Cc_{\PZ}^2}{\Cc_{\PH}^2(\Cc_i)}$                            & $\frac{\Cc_{\PQt}^2\cdot \Cc_{\PW}^2}{\Cc_{\PH}^2(\Cc_i)}$                            & $\frac{\Cc_{\Pt}^2\cdot \Cc_{\PQb}^2}{\Cc_{\PH}^2(\Cc_i)}$                             & $\frac{\Cc_{\Pt}^2\cdot \Cc_{\PGt}^2}{\Cc_{\PH}^2(\Cc_i)}$ \\\hline
VBF      & $\frac{\Cc_\mathrm{VBF}^2(\Cc_{\PZ},\Cc_{\PW}) \cdot \Cc_{\PGg}^2}{\Cc_{\PH}^2(\Cc_i)}$ & $\frac{\Cc_\mathrm{VBF}^2(\Cc_{\PZ},\Cc_{\PW})\cdot \Cc_{\PZ}^2}{\Cc_{\PH}^2(\Cc_i)}$ & $\frac{\Cc_\mathrm{VBF}^2(\Cc_{\PZ},\Cc_{\PW})\cdot \Cc_{\PW}^2}{\Cc_{\PH}^2(\Cc_i)}$ & $\frac{\Cc_\mathrm{VBF}^2(\Cc_{\PZ},\Cc_{\PW})\cdot \Cc_{\PQb}^2}{\Cc_{\PH}^2(\Cc_i)}$ & $\frac{\Cc_\mathrm{VBF}^2(\Cc_{\PZ},\Cc_{\PW})\cdot \Cc_{\PGt}^2}{\Cc_{\PH}^2(\Cc_i)}$ \\\hline
$\PW\PH$ & $\frac{\Cc_{\PW}^2\cdot \Cc_{\PGg}^2}{\Cc_{\PH}^2(\Cc_i)}$                              & $\frac{\Cc_{\PW}^2\cdot \Cc_{\PZ}^2}{\Cc_{\PH}^2(\Cc_i)}$                             & $\frac{\Cc_{\PW}^2\cdot \Cc_{\PW}^2}{\Cc_{\PH}^2(\Cc_i)}$                             & $\frac{\Cc_{\PW}^2\cdot \Cc_{\PQb}^2}{\Cc_{\PH}^2(\Cc_i)}$                             & $\frac{\Cc_{\PW}^2\cdot \Cc_{\PGt}^2}{\Cc_{\PH}^2(\Cc_i)}$ \\\hline
$\PZ\PH$ & $\frac{\Cc_{\PZ}^2\cdot \Cc_{\PGg}^2}{\Cc_{\PH}^2(\Cc_i)}$                              & $\frac{\Cc_{\PZ}^2\cdot \Cc_{\PZ}^2}{\Cc_{\PH}^2(\Cc_i)}$                             & $\frac{\Cc_{\PZ}^2\cdot \Cc_{\PW}^2}{\Cc_{\PH}^2(\Cc_i)}$                             & $\frac{\Cc_{\PZ}^2\cdot \Cc_{\PQb}^2}{\Cc_{\PH}^2(\Cc_i)}$                             & $\frac{\Cc_{\PZ}^2\cdot \Cc_{\PGt}^2}{\Cc_{\PH}^2(\Cc_i)}$ \\
\hline
\hline                
\multicolumn{6}{l}{\bfseries General parameterization allowing all gauge and third generation fermion couplings to float assuming $\Cc_{\PV}<1$} \\
\multicolumn{6}{l}{\footnotesize Free parameters: $\Cc_{\Pg}$, $\Cc_{\PGg}$, $\Cc_{\PW}$, $\Cc_{\PZ}$, $\Cc_{\PQb}$, $\Cc_{\PQt}$, $\Cc_{\PGt}$, $\Cc_{\PH}$, with conditions $\Cc_{\PW}<1$, $\Cc_{\PZ}<1$ and $\Cc_{\PH}^2>\Cc_{\PH}^2(\Cc_i)$.} \\
\hline
         & $\PH\to\PGg\PGg$                                                                 & $\PH\to \PZ\PZ^{(*)}$                                                          & $\PH\to \PW\PW^{(*)}$                                                          & $\PH\to \PQb\PAQb$                                                              & $\PH\to\PGtm\PGtp$                                  \\\hline
\MyggH   & $\frac{\Cc_{\Pg}^2\cdot \Cc_{\PGg}^2}{\Cc_{\PH}^2}$                              & $\frac{\Cc_{\Pg}^2\cdot \Cc_{\PZ}^2}{\Cc_{\PH}^2}$                             & $\frac{\Cc_{\Pg}^2\cdot \Cc_{\PW}^2}{\Cc_{\PH}^2}$                             & $\frac{\Cc_{\Pg}^2\cdot \Cc_{\PQb}^2}{\Cc_{\PH}^2}$                             & $\frac{\Cc_{\Pg}^2\cdot \Cc_{\PGt}^2}{\Cc_{\PH}^2}$ \\\hline
\MyttH   & $\frac{\Cc_{\PQt}^2\cdot \Cc_{\PGg}^2}{\Cc_{\PH}^2}$                             & $\frac{\Cc_{\PQt}^2\cdot \Cc_{\PZ}^2}{\Cc_{\PH}^2}$                            & $\frac{\Cc_{\PQt}^2\cdot \Cc_{\PW}^2}{\Cc_{\PH}^2}$                            & $\frac{\Cc_{\Pt}^2\cdot \Cc_{\PQb}^2}{\Cc_{\PH}^2}$                             & $\frac{\Cc_{\Pt}^2\cdot \Cc_{\PGt}^2}{\Cc_{\PH}^2}$ \\\hline
VBF      & $\frac{\Cc_\mathrm{VBF}^2(\Cc_{\PZ},\Cc_{\PW}) \cdot \Cc_{\PGg}^2}{\Cc_{\PH}^2}$ & $\frac{\Cc_\mathrm{VBF}^2(\Cc_{\PZ},\Cc_{\PW})\cdot \Cc_{\PZ}^2}{\Cc_{\PH}^2}$ & $\frac{\Cc_\mathrm{VBF}^2(\Cc_{\PZ},\Cc_{\PW})\cdot \Cc_{\PW}^2}{\Cc_{\PH}^2}$ & $\frac{\Cc_\mathrm{VBF}^2(\Cc_{\PZ},\Cc_{\PW})\cdot \Cc_{\PQb}^2}{\Cc_{\PH}^2}$ & $\frac{\Cc_\mathrm{VBF}^2(\Cc_{\PZ},\Cc_{\PW})\cdot \Cc_{\PGt}^2}{\Cc_{\PH}^2}$ \\\hline
$\PW\PH$ & $\frac{\Cc_{\PW}^2\cdot \Cc_{\PGg}^2}{\Cc_{\PH}^2}$                              & $\frac{\Cc_{\PW}^2\cdot \Cc_{\PZ}^2}{\Cc_{\PH}^2}$                             & $\frac{\Cc_{\PW}^2\cdot \Cc_{\PW}^2}{\Cc_{\PH}^2}$                             & $\frac{\Cc_{\PW}^2\cdot \Cc_{\PQb}^2}{\Cc_{\PH}^2}$                             & $\frac{\Cc_{\PW}^2\cdot \Cc_{\PGt}^2}{\Cc_{\PH}^2}$ \\\hline
$\PZ\PH$ & $\frac{\Cc_{\PZ}^2\cdot \Cc_{\PGg}^2}{\Cc_{\PH}^2}$                              & $\frac{\Cc_{\PZ}^2\cdot \Cc_{\PZ}^2}{\Cc_{\PH}^2}$                             & $\frac{\Cc_{\PZ}^2\cdot \Cc_{\PW}^2}{\Cc_{\PH}^2}$                             & $\frac{\Cc_{\PZ}^2\cdot \Cc_{\PQb}^2}{\Cc_{\PH}^2}$                             & $\frac{\Cc_{\PZ}^2\cdot \Cc_{\PGt}^2}{\Cc_{\PH}^2}$ \\
\hline
\hline
\multicolumn{6}{l}{\bfseries General parameterization allowing all gauge and third generation fermion couplings to float allowing for invisible or undetectable widths} \\
\multicolumn{6}{l}{\footnotesize Free parameters: $\Cc_{\Pg\PZ} (= \Cc_{\Pg}\cdot \Cc_{\PZ} / \Cc_{\PH})$, $\Rr_{\PGg\PZ} (= \Cc_{\PGg} / \Cc_{\PZ})$, $\Rr_{\PW\PZ} (= \Cc_{\PW} / \Cc_{\PZ})$, $\Rr_{\PQb\PZ} (= \Cc_{\PQb} / \Cc_{\PZ})$, $\Rr_{\PGt\PZ} (= \Cc_{\PGt} / \Cc_{\PZ})$, $\Rr_{\PZ\Pg} (= \Cc_{\PZ} / \Cc_{\Pg})$, $\Rr_{\PQt\Pg} (= \Cc_{\PQt} / \Cc_{\Pg})$.} \\
\hline
\MyggH   & $\Cc_{\Pg\PZ}^2                                     \cdot \Rr_{\PGg\PZ}^2$               & $\Cc_{\Pg\PZ}^2                                                  $ & $\Cc_{\Pg\PZ}^2                                                   \cdot \Rr_{\PW\PZ}^2$ & $\Cc_{\Pg\PZ}^2                                                   \cdot \Rr_{\PQb\PZ}^2$ & $\Cc_{\Pg\PZ}^2                                                   \cdot \Rr_{\PGt\PZ}^2$ \\\hline                                                                                                                                                                                                                                                                     
\MyttH   & $\Cc_{\Pg\PZ}^2 \Rr_{\PQt\Pg}^2                           \cdot \Rr_{\PGg\PZ}^2$         & $\Cc_{\Pg\PZ}^2 \Rr_{\PQt\Pg}^2                                  $ & $\Cc_{\Pg\PZ}^2 \Rr_{\PQt\Pg}^2                                   \cdot \Rr_{\PW\PZ}^2$ & $\Cc_{\Pg\PZ}^2 \Rr_{\PQt\Pg}^2                                   \cdot \Rr_{\PQb\PZ}^2$ & $\Cc_{\Pg\PZ}^2 \Rr_{\PQt\Pg}^2                                   \cdot \Rr_{\PGt\PZ}^2$ \\\hline                                                                                                                                                                                                                                                                     
VBF      & $\Cc_{\Pg\PZ}^2 \Rr_{\PZ\Pg}^2 \Cc_\mathrm{VBF}^2(1,\Rr_{\PW\PZ}) \cdot \Rr_{\PGg\PZ}^2$ & $\Cc_{\Pg\PZ}^2 \Rr_{\PZ\Pg}^2 \Cc_\mathrm{VBF}^2(1,\Rr_{\PW\PZ})$ & $\Cc_{\Pg\PZ}^2 \Rr_{\PZ\Pg}^2 \Cc_\mathrm{VBF}^2(1,\Rr_{\PW\PZ}) \cdot \Rr_{\PW\PZ}^2$ & $\Cc_{\Pg\PZ}^2 \Rr_{\PZ\Pg}^2 \Cc_\mathrm{VBF}^2(1,\Rr_{\PW\PZ}) \cdot \Rr_{\PQb\PZ}^2$ & $\Cc_{\Pg\PZ}^2 \Rr_{\PZ\Pg}^2 \Cc_\mathrm{VBF}^2(1,\Rr_{\PW\PZ}) \cdot \Rr_{\PGt\PZ}^2$ \\\hline                                         
$\PW\PH$ & $\Cc_{\Pg\PZ}^2 \Rr_{\PZ\Pg}^2 \Rr_{\PW\PZ}^2                  \cdot \Rr_{\PGg\PZ}^2$    & $\Cc_{\Pg\PZ}^2 \Rr_{\PZ\Pg}^2 \Rr_{\PW\PZ}^2                    $ & $\Cc_{\Pg\PZ}^2 \Rr_{\PZ\Pg}^2 \Rr_{\PW\PZ}^2                     \cdot \Rr_{\PW\PZ}^2$ & $\Cc_{\Pg\PZ}^2 \Rr_{\PZ\Pg}^2 \Rr_{\PW\PZ}^2                     \cdot \Rr_{\PQb\PZ}^2$ & $\Cc_{\Pg\PZ}^2 \Rr_{\PZ\Pg}^2 \Rr_{\PW\PZ}^2                     \cdot \Rr_{\PGt\PZ}^2$ \\\hline                                         
$\PZ\PH$ & $\Cc_{\Pg\PZ}^2 \Rr_{\PZ\Pg}^2                            \cdot \Rr_{\PGg\PZ}^2$         & $\Cc_{\Pg\PZ}^2 \Rr_{\PZ\Pg}^2                                   $ & $\Cc_{\Pg\PZ}^2 \Rr_{\PZ\Pg}^2                                    \cdot \Rr_{\PW\PZ}^2$ & $\Cc_{\Pg\PZ}^2 \Rr_{\PZ\Pg}^2                                    \cdot \Rr_{\PQb\PZ}^2$ & $\Cc_{\Pg\PZ}^2 \Rr_{\PZ\Pg}^2                                    \cdot \Rr_{\PGt\PZ}^2$ \\\hline
\end{tabular}

{\footnotesize $\Cc_i^2 = \Gamma_{ii} / \Gamma_{ii}^\text{SM}$}

\end{sidewaystable}

\subsubsection{General parameterization assuming no beyond SM particles}
\label{sec:LM_ir:tree}

\Tref{tab:LM_ir:tree} presents the relations to be used in a fit
where the loop-induced $\MyggH$ and $\PH\to\PGg\PGg$ processes and the total width are 
expressed in terms of the SM gauge- and Yukawa coupling scale factors 
($\Cc_{\PW}$, $\Cc_{\PZ}$, $\Cc_{\PQb}$, $\Cc_{\PQt}$, $\Cc_{\PGt}$), 
assuming no beyond SM particle contributions.
This benchmark makes full use of 
\eqnss{eq:CgNLOQCD}, (\ref{eq:CgammaNLOQCD}), (\ref{eq:CH2_def}) 
and has the highest sensitivity in an independent fit of the SM gauge- and Yukawa-coupling scale factors.

Once the $\PH \to \PGmm\PGmp$ and $\PH \to \PZ\PGg$ decay modes reach sufficient
sensitivity, a natural extension for this benchmark is to fit an extra degree
of freedom in the form of the parameter $\Cc_{\PGm}$, while resolving
the $\PH \to \PZ\PGg$ decay mode using \eqn{eq:CZgammaNLOQCD}.

\begin{table}
\centering
\caption{A benchmark parameterization expressing all processes in terms of the SM gauge- and Yukawa-coupling 
scale factors, assuming no beyond SM particle contributions.
}
\label{tab:LM_ir:tree}

\begin{tabular}{lccccc}
\hline                
\multicolumn{6}{l}{\bfseries General parameterization assuming no beyond SM particles}
\\
\multicolumn{6}{l}{\footnotesize Free parameters: $\Cc_{\PW}$, $\Cc_{\PZ}$, $\Cc_{\PQb}$, $\Cc_{\PQt}$, $\Cc_{\PGt}$.} \\
\multicolumn{6}{l}{\footnotesize Dependent parameters:
$\Cc_{\Pg}=\Cc_{\Pg}(\Cc_{\PQb},\Cc_{\PQt})$,
$\Cc_{\PGg}=\Cc_{\PGg}(\Cc_{\PQb},\Cc_{\PQt},\Cc_{\PGt},\Cc_{\PW})$,
$\Cc_{\PH} = \Cc_{\PH}(\Cc_i)$.}
\\
\hline
         & $\PH\to\PGg\PGg$                                                                        & $\PH\to \PZ\PZ^{(*)}$                                                                 & $\PH\to \PW\PW^{(*)}$                                                                 & $\PH\to \PQb\PAQb$                                                                     & $\PH\to\PGtm\PGtp$                                                              \\\hline                                                                                                                                                                                                                                                                     
\MyggH   & $\frac{\Cc_{\Pg}^2(\Cc_{\PQb},\Cc_{\PQt})\cdot
\Cc_{\PGg}^2(\Cc_{\PQb},\Cc_{\PQt},\Cc_{\PGt},\Cc_{\PW})}{\Cc_{\PH}^2(\Cc_i)}$ &
$\frac{\Cc_{\Pg}^2(\Cc_{\PQb},\Cc_{\PQt})\cdot \Cc_{\PZ}^2}{\Cc_{\PH}^2(\Cc_i)}$                             & $\frac{\Cc_{\Pg}^2(\Cc_{\PQb},\Cc_{\PQt})\cdot \Cc_{\PW}^2}{\Cc_{\PH}^2(\Cc_i)}$                             & $\frac{\Cc_{\Pg}^2(\Cc_{\PQb},\Cc_{\PQt})\cdot \Cc_{\PQb}^2}{\Cc_{\PH}^2(\Cc_i)}$                             & $\frac{\Cc_{\Pg}^2(\Cc_{\PQb},\Cc_{\PQt})\cdot \Cc_{\PGt}^2}{\Cc_{\PH}^2(\Cc_i)}$ \\\hline
\MyttH   & $\frac{\Cc_{\PQt}^2\cdot \Cc_{\PGg}^2(\Cc_{\PQb},\Cc_{\PQt},\Cc_{\PGt},\Cc_{\PW})}{\Cc_{\PH}^2(\Cc_i)}$                             & $\frac{\Cc_{\PQt}^2\cdot \Cc_{\PZ}^2}{\Cc_{\PH}^2(\Cc_i)}$                            & $\frac{\Cc_{\PQt}^2\cdot \Cc_{\PW}^2}{\Cc_{\PH}^2(\Cc_i)}$                            & $\frac{\Cc_{\Pt}^2\cdot \Cc_{\PQb}^2}{\Cc_{\PH}^2(\Cc_i)}$                             & $\frac{\Cc_{\Pt}^2\cdot \Cc_{\PGt}^2}{\Cc_{\PH}^2(\Cc_i)}$ \\\hline
VBF      & $\frac{\Cc_\mathrm{VBF}^2(\Cc_{\PZ},\Cc_{\PW}) \cdot \Cc_{\PGg}^2(\Cc_{\PQb},\Cc_{\PQt},\Cc_{\PGt},\Cc_{\PW})}{\Cc_{\PH}^2(\Cc_i)}$ & $\frac{\Cc_\mathrm{VBF}^2(\Cc_{\PZ},\Cc_{\PW})\cdot \Cc_{\PZ}^2}{\Cc_{\PH}^2(\Cc_i)}$ & $\frac{\Cc_\mathrm{VBF}^2(\Cc_{\PZ},\Cc_{\PW})\cdot \Cc_{\PW}^2}{\Cc_{\PH}^2(\Cc_i)}$ & $\frac{\Cc_\mathrm{VBF}^2(\Cc_{\PZ},\Cc_{\PW})\cdot \Cc_{\PQb}^2}{\Cc_{\PH}^2(\Cc_i)}$ & $\frac{\Cc_\mathrm{VBF}^2(\Cc_{\PZ},\Cc_{\PW})\cdot \Cc_{\PGt}^2}{\Cc_{\PH}^2(\Cc_i)}$ \\\hline
$\PW\PH$ & $\frac{\Cc_{\PW}^2\cdot \Cc_{\PGg}^2(\Cc_{\PQb},\Cc_{\PQt},\Cc_{\PGt},\Cc_{\PW})}{\Cc_{\PH}^2(\Cc_i)}$                              & $\frac{\Cc_{\PW}^2\cdot \Cc_{\PZ}^2}{\Cc_{\PH}^2(\Cc_i)}$                             & $\frac{\Cc_{\PW}^2\cdot \Cc_{\PW}^2}{\Cc_{\PH}^2(\Cc_i)}$                             & $\frac{\Cc_{\PW}^2\cdot \Cc_{\PQb}^2}{\Cc_{\PH}^2(\Cc_i)}$                             & $\frac{\Cc_{\PW}^2\cdot \Cc_{\PGt}^2}{\Cc_{\PH}^2(\Cc_i)}$ \\\hline
$\PZ\PH$ & $\frac{\Cc_{\PZ}^2\cdot \Cc_{\PGg}^2(\Cc_{\PQb},\Cc_{\PQt},\Cc_{\PGt},\Cc_{\PW})}{\Cc_{\PH}^2(\Cc_i)}$                              & $\frac{\Cc_{\PZ}^2\cdot \Cc_{\PZ}^2}{\Cc_{\PH}^2(\Cc_i)}$                             & $\frac{\Cc_{\PZ}^2\cdot \Cc_{\PW}^2}{\Cc_{\PH}^2(\Cc_i)}$                             & $\frac{\Cc_{\PZ}^2\cdot \Cc_{\PQb}^2}{\Cc_{\PH}^2(\Cc_i)}$                             & $\frac{\Cc_{\PZ}^2\cdot \Cc_{\PGt}^2}{\Cc_{\PH}^2(\Cc_i)}$ \\
\hline
\end{tabular}

{\footnotesize $\Cc_i^2 = \Gamma_{ii} / \Gamma_{ii}^\text{SM}$}

\end{table}

\clearpage

\subsection{Effective Lagrangians for Higgs interactions}
\label{sec:LM_eft}

\providecommand{\Lag}{\mathcal{L}}
\providecommand{\Lcoeff}{\alpha}
\providecommand{\Lop}{\mathcal{O}}
\providecommand{\Lcoeffp}{\alpha'}
\providecommand{\Lopp}{\mathcal{O}'}

\providecommand{\phic}{\phi}
\providecommand{\phin}{\phi^0}

\providecommand{\SM}{\mathrm{SM}}
\providecommand{\GSM}{G_{\SM}}
\providecommand{\WSM}{W_{\SM}}
\providecommand{\ZSM}{Z_{\SM}}
\providecommand{\ASM}{A_{\SM}}
\providecommand{\HSM}{H_{\SM}}
\providecommand{\phicSM}{\phic_{\SM}}
\providecommand{\phinSM}{\phin_{\SM}}
\providecommand{\fSM}{\Pf_{\SM}}

\providecommand{\vSM}{v_\SM}
\providecommand{\MWSM}{M_{\PW,\SM}}
\providecommand{\MZSM}{M_{\PZ,\SM}}
\providecommand{\MHSM}{M_{\PH,\SM}}
\providecommand{\MfSM}{m_{\Pf,\SM}}
\providecommand{\swSM}{s_{\rw,\SM}}
\providecommand{\cwSM}{c_{\rw,\SM}}
\providecommand{\eSM}{e_{\SM}}

\providecommand{\Gnew}{{G}}
\providecommand{\Wnew}{{\PW}}
\providecommand{\Znew}{{\PZ}}
\providecommand{\Anew}{{\PA}}
\providecommand{\Hnew}{{\PH}}
\providecommand{\phicnew}{{\phic}}
\providecommand{\phinnew}{{\phin}}
\providecommand{\fnew}{{\Pf}}
\providecommand{\nunew}{{\PGne}}
\providecommand{\enew}{\Pe}

\providecommand{\vnew}{v}
\providecommand{\MWnew}{{M}_{\PW}}
\providecommand{\MZnew}{{M}_{\PZ}}
\providecommand{\MHnew}{{M}_{\PH}}
\providecommand{\Mfnew}{{m}_{\Pf}}

The ``interim framework'' described in \Sref{sec:LM_ir:framework}
was proposed as a first step for exploring the coupling structure of
the recently observed state, making use of the data taken until 
the end of 2012.
In future, however, one should aim at a more general analysis where
besides possible deviations in the absolute values of the couplings from
their SM values also possible deviations in the tensor structure of the
couplings are taken into account. This implies that the exploration of
the couplings and of the spin and CP properties have to be treated
together within a coherent framework.

In the following we use effective Lagrangians as an approach towards
such a coherent framework. An effective Lagrangian can be understood to
arise from integrating out heavy degrees of freedom, such that the
different terms in the Lagrangian are obtained from a systematic expansion in
inverse power of a heavy scale. The effective Lagrangian provides in this
way a parameterization of possible deviations from the SM
predictions. In the tools providing the theoretical predictions for the
relevant observables that are confronted with the experimental data
those parameterizations of possible deviations from the SM
can then be used to supplement the most accurate theoretical predictions
within the SM, including the known higher-order corrections.

It should be noted, however, that such an
effective Lagrangian approach does not cover possible effects
of light BSM particles in loops. In order to investigate the latter type
of effects it seems preferable to resort to specific models. Such a
model-specific approach for exploring the coupling structure of
the recently observed state is complementary to the model-independent
approach based on effective Lagrangians, on which we will focus below.

The description of BSM physics based on an effective Lagrangian in
terms of the SM fields has been pioneered by Buchm\"uller and Wyler
\cite{Buchmuller:1985jz}, who provided a list of operators of
dimensions 5 and 6 in the linear parameterization of the Higgs sector
with a Higgs doublet. In the sequel various authors considered subsets
of this operator basis or introduced different sets of operators
adapted to specific goals. Recently a complete minimal basis of
dimension-6 operators has been presented in \Bref{Grzadkowski:2010es}.
For the analysis of Higgs interactions different authors prefer to use
different sets of operator bases
\cite{Hagiwara:1993ck,Hagiwara:1993qt,Eboli:1999pt, Giudice:2007fh,
  Hankele:2006ma,Kanemura:2008ub,Bonnet:2011yx,Corbett:2012dm,Bonnet:2012nm,Passarino:2012cb,Corbett:2012ja,Contino:2013kra}.
The purpose of this section is to propose a suitable set of operators
to be used for the future analysis of the Higgs sector based on the
most accurate predictions for the relevant observables within the SM
which are supplemented by a parameterization of possible deviations
from the SM.  The effects of the dimension-6 operators on the Higgs decay ratios are implemented in the code \textsc{eHDECAY}~\cite{Contino:2013kra}, a modified version of \textsc{HDECAY}~\cite{Djouadi:1997yw},  including some radiative corrections beyond the leading order.
We consider two versions of effective Lagrangians,
namely a linear parameterization involving a Higgs doublet and a
parameterization where the EW symmetry is non-linearly realized.

\subsubsection{Linear parameterization with a Higgs doublet}

In this section we define an effective Lagrangian based on a linear
representation of the electroweak gauge symmetry with a Higgs-doublet
field. We follow closely the framework introduced in
\Bref{Buchmuller:1985jz} and further developed in
\Bref{Grzadkowski:2010es}.  We restrict ourselves to dimension-6
operators relevant for Higgs physics. The effective Lagrangian has the
general form
\beq
\Lag_{\mathrm{eff}}= \Lag^{(4)}_{\mathrm{SM}}+ \frac{1}{\Lambda^2}\sum_k \Lcoeff_k
\Lop_k,
\label{eq:Leff-op}
\eeq
where $\Lag^{(4)}_{\mathrm{SM}}$ is the usual SM Lagrangian,
$\Lop_k\equiv\Lop^{d=6}_k$ denotes dimension-6 operators and
$\Lcoeff_k$ the corresponding Wilson coefficients. Since the effective
Lagrangian must be hermitian, in \eqn{eq:Leff-op} for each non-hermitian operator
$\Lop_k$ the hermitian conjugate operator $\Lop_k^\dagger$ appears
with the complex conjugate Wilson coefficient $\Lcoeff_k^*$.

\subsubsubsection{Conventions and definition of the effective operator basis}

Using $\alpha=1,2,3$, $i=1,2$, and $p=1,2,3$ for color, weak isospin,
and flavor indices, respectively, the matter fields of the SM are
left-handed lepton doublets $\Pl^i_p$, right-handed charged
leptons $\Pe_p$, left-handed quark doublets $\PQq^{\alpha i}_p$,
right-handed quarks $\PQu^\alpha_p, \PQd^\alpha_p$, and the Higgs doublet
$\Phi^i$ with hypercharges $Y=-1/2,-1,1/6,2/3,-1/3,1/2$, respectively.
Right-handed neutrinos are not included.
The charge-conjugate Higgs field is given by
$\widetilde{\Phi}^i=\varepsilon^{ij}(\Phi^j)^*$ with $\varepsilon^{ij}$
antisymmetric and $\varepsilon^{12}=1$.

The SM Lagrangian reads
\beqar\label{LSM}
\Lag^{(4)}_{\mathrm{SM}} &=& 
-\frac{1}{4}G^A_{\mu\nu}G^{A\mu\nu}
-\frac{1}{4}\PW^I_{\mu\nu}\PW^{I\mu\nu}
-\frac{1}{4}\PB_{\mu\nu}\PB^{\mu\nu}
\nn\\&&{}
+(D_\mu\Phi)^\dagger (D^\mu\Phi)
+ m^2 \Phi^\dagger\Phi
-\frac{1}{2}\lambda  (\Phi^\dagger\Phi)^2
\nn\\&&{}
+\ri \bar{\Pl}\slashed{D}\Pl
+\ri \bar{\Pe}\slashed{D}\Pe
+\ri \bar{\PQq}\slashed{D}\PQq
+\ri \bar{\PAQu}\slashed{D}\PQu
+\ri \bar{\PAQd}\slashed{D}\PQd
\nn\\&&{}
-(\bar{\Pl}\,\Gamma_{_\Pe} \Pe \Phi 
+\bar{\PQq}\,\Gamma_{_\PQu} \PQu \widetilde\Phi 
+\bar{\PQd}\,\Gamma_{_\PQd} \PQd \Phi + \mathrm{h.c.}),
\eeqar
where flavor, color, and weak-isospin indices have been suppressed in
the matter parts.
The field-strength tensors are given by
\beqar
G^A_{\mu\nu}&=&\partial_\mu G^A_\nu-\partial_\nu G^A_\mu -\gs f^{ABC}
G^B_\mu G^C_\nu, \quad A=1,\ldots,8,
\nn\\
\PW^I_{\mu\nu}&=&\partial_\mu \PW^I_\nu-\partial_\nu \PW^I_\mu -g \varepsilon^{IJK}
\PW^J_\mu \PW^K_\nu, \quad I=1,2,3,
%\nn\\
\qquad
\PB_{\mu\nu}=\partial_\mu \PB_\nu-\partial_\nu \PB_\mu~,
\eeqar
in terms of the gauge fields $G^A_\mu$, $\PW^I_\mu$, $\PB_\mu$ of the
gauge group $SU(3)\times SU(2)\times U(1)$ and the corresponding gauge
couplings $\gs$, $g$, $g'$.
The covariant derivative acting on $SU(2)$ doublets reads
\beq\label{eq:covariant_derivative}
D_\mu=\partial_\mu +\ri\gs\frac{\lambda^A}{2}G^A_\mu
+\ri g\frac{\tau^I}{2}\PW^I_\mu+\ri g'{Y}\PB_\mu
\eeq
with the Gell-Mann matrices $\la^A$ acting on color indices, the
Pauli matrices $\tau^I$ acting on $SU(2)$ indices and the
hypercharge operator $Y$. The terms with $\la^A$ and $\tau^I$ are
absent for color and $SU(2)$ singlets, respectively.
The quantities $\Gamma_{_\Pf}$, $\Pf=\Pe,\PQu,\PQd$ are matrices in flavor space. 

Furthermore, we define dual tensors by 
\beq
\widetilde \PX_{\mu\nu}=
\frac{1}{2}\varepsilon_{\mu\nu\rho\sigma}\PX^{\rho\sigma}, \quad
\PX=G^A,\PW^I,\PB \quad\text{and}\quad \varepsilon_{0123}=+1,
\eeq
and introduce hermitian derivatives 
\beq
\Phi^\dagger\ri \overset\leftrightarrow{D}_\mu\Phi
=\ri(\Phi^\dagger D_\mu\Phi - (D_\mu\Phi)^\dagger\Phi),
\qquad
\Phi^\dagger\ri {\overset\leftrightarrow{D}}{}^I_\mu\Phi
=\ri(\Phi^\dagger\tau^I D_\mu\Phi - (D_\mu\Phi)^\dagger \tau^I\Phi)~,
\eeq
%\beq
%\Phi^\dagger\ri \overset\leftrightarrow{D}_\mu\Phi
%=\ri\Phi^\dagger(D_\mu - \overset\leftarrow{D}_\mu)\Phi,
%\qquad
%\Phi^\dagger\ri {\overset\leftrightarrow{D}}{}^I_\mu\Phi
%=\ri\Phi^\dagger(\tau^ID_\mu - \overset\leftarrow{D}_\mu\tau^I)\Phi.
%\eeq
%In the covariant derivatives $\overset\leftarrow{D}_\mu$
%acting on $\Phi^\dagger$, the complex-conjugate representation
%matrices are understood, i.e.\
%$\phi^\dagger\overset\leftarrow{D}_\mu=(D_\mu\Phi)^\dagger$.
%they are given by the complex conjugate of \Eref{eq:covariant_derivative}.
and $\si^{\mu\nu}=\ri(\gamma^\mu\gamma^\nu-\gamma^\nu\gamma^\mu)/2$.

For the dimension-6 operators we choose the minimal complete basis%
\footnote{A minimal complete basis can be
  constructed by writing down all  dimension-6 operators that can be
  built from the SM fields and using the equations of motion to
  eliminate all redundant operators.}
defined in \Bref{Grzadkowski:2010es}, but restrict ourselves here to
operators that involve Higgs or gauge-boson fields, see \refT{ta:Lops}.%
\footnote{In a complete analysis all 59 independent operators of
  \Bref{Grzadkowski:2010es}, including  25 four-fermion operators,
  have to be considered in addition to the 34 operators of Table~\ref{ta:Lops}.}  In addition to the operators in
\refT{ta:Lops} for each non-hermitian operator its hermitian conjugate
must be included. The Wilson coefficients of these operators are in
general complex, whereas those of the hermitian operators are real.
\begin{table}
\caption{Dimension-6 operators involving Higgs doublet fields or
  gauge-boson fields. For all
$\psi^2\Phi^3$, $\psi^2 \PX\Phi$ operators and for $\Lop_{\Phi \PQu\PQd}$ the
hermitian conjugates must be included as well.}
\label{ta:Lops}
\centering{$\displaystyle
\renewcommand{\arraystretch}{1.4}\begin{array}{l@{\qquad}l@{\qquad}l}
\hline
\Phi^6\quad \text{and}\quad \Phi^4D^2 & \psi^2\Phi^3 
&\PX^3
\\
\hline
\Lop_\Phi = (\Phi^\dagger\Phi)^3   &
\Lop_{\Pe\Phi}=(\Phi^\dagger\Phi)(\bar{\Pl}\,\Gamma_{_\Pe} \Pe \Phi) 
&\Lop_{G} = f^{ABC}G^{A\nu}_\mu G^{B\rho}_\nu G^{C\mu}_\rho
\\
\Lop_{\Phi\Box} = (\Phi^\dagger\Phi)\Box (\Phi^\dagger\Phi)   &
\Lop_{\PQu\Phi}=(\Phi^\dagger\Phi)(\bar{\PQq}\,\Gamma_{_\PQu} \PQu \widetilde\Phi) 
&\Lop_{\widetilde{G}} = f^{ABC}\widetilde{G}^{A\nu}_\mu G^{B\rho}_\nu G^{C\mu}_\rho
\\
\Lop_{\Phi D} = (\Phi^\dagger D^\mu\Phi)^*(\Phi^\dagger D_\mu\Phi) &
\Lop_{\PQd\Phi}=(\Phi^\dagger\Phi)(\bar{\PQq}\,\Gamma_{_\PQd} \PQd \Phi) 
&\Lop_{\PW} = \varepsilon^{IJK}\PW^{I\nu}_\mu \PW^{J\rho}_\nu \PW^{K\mu}_\rho
\\
&&\Lop_{\widetilde{\PW}} = \varepsilon^{IJK}\widetilde{\PW}^{I\nu}_\mu \PW^{J\rho}_\nu \PW^{K\mu}_\rho
\rule[-1.5ex]{0pt}{3ex}\\
\hline
\PX^2\Phi^2 &\psi^2 \PX\Phi & \psi^2\Phi^2D \\
\hline\rule{0pt}{3.5ex}
\Lop_{\Phi G} = (\Phi^\dagger\Phi)G^A_{\mu\nu}G^{A\mu\nu} &
\Lop_{\PQu G} = (\bar{\PQq}\sigma^{\mu\nu}\frac{\lambda^A}{2}\Gamma_{_\PQu} \PQu
\widetilde\Phi)G^A_{\mu\nu}\qquad &
\Lop_{\Phi \Pl}^{(1)}=(\Phi^\dagger\ri\overset\leftrightarrow{D}_\mu\Phi)
(\bar{\Pl}\gamma^\mu \Pl)
\\
\Lop_{\Phi \widetilde{G}} = (\Phi^\dagger\Phi)\widetilde{G}^A_{\mu\nu}G^{A\mu\nu} &
\Lop_{\PQd G} = (\bar{\PQq}\sigma^{\mu\nu}\frac{\lambda^A}{2}\Gamma_{_\PQd} \PQd
\Phi)G^A_{\mu\nu} &
\Lop_{\Phi \Pl}^{(3)}=(\Phi^\dagger\ri\overset\leftrightarrow{D}{}_\mu^I\Phi)
(\bar{\Pl}\gamma^\mu \tau^I \Pl)
\\
\Lop_{\Phi \PW} = (\Phi^\dagger\Phi)\PW^I_{\mu\nu}\PW^{I\mu\nu} &
\Lop_{\Pe\PW} = (\bar{\Pl}\sigma^{\mu\nu}\Gamma_{_\Pe} \Pe {\tau^I}
\Phi)\PW^I_{\mu\nu} &
\Lop_{\Phi \Pe}=(\Phi^\dagger\ri\overset\leftrightarrow{D}_\mu\Phi)
(\bar{\Pe}\gamma^\mu \Pe)
\\
\Lop_{\Phi \widetilde{\PW}} = (\Phi^\dagger\Phi)\widetilde{\PW}^I_{\mu\nu}\PW^{I\mu\nu} &
\Lop_{\PQu\PW} = (\bar{\PQq}\sigma^{\mu\nu}\Gamma_{_\PQu} \PQu{\tau^I}\widetilde\Phi)\PW^I_{\mu\nu} &
\Lop_{\Phi \PQq}^{(1)}=(\Phi^\dagger\ri\overset\leftrightarrow{D}_\mu\Phi)
(\bar{\PQq}\gamma^\mu \PQq)
\\
\Lop_{\Phi \PB} = (\Phi^\dagger\Phi)\PB_{\mu\nu}\PB^{\mu\nu} &
\Lop_{\PQd\PW} = (\bar{\PQq}\sigma^{\mu\nu}\Gamma_{_\PQd} \PQd {\tau^I}
\Phi)\PW^I_{\mu\nu} &
\Lop_{\Phi \PQq}^{(3)}=(\Phi^\dagger\ri\overset\leftrightarrow{D}{}_\mu^I\Phi)
(\bar{\PQq}\gamma^\mu \tau^I \PQq)
\\
\Lop_{\Phi \widetilde{\PB}} = (\Phi^\dagger\Phi)\widetilde{\PB}_{\mu\nu}\PB^{\mu\nu} &
\Lop_{\Pe \PB} = (\bar{\Pl}\sigma^{\mu\nu}\Gamma_{_\Pe} \Pe\Phi)\PB_{\mu\nu} &
\Lop_{\Phi \PQu}=(\Phi^\dagger\ri\overset\leftrightarrow{D}_\mu\Phi)
(\bar{\PQu}\gamma^\mu \PQu)
\\
\Lop_{\Phi{\PW \PB}} = (\Phi^\dagger\tau^I\Phi){\PW}^I_{\mu\nu}\PB^{\mu\nu} &
\Lop_{\PQu \PB} = (\bar{\PQq}\sigma^{\mu\nu}\Gamma_{_\PQu} \PQu\widetilde\Phi)\PB_{\mu\nu} &
\Lop_{\Phi \PQd}=(\Phi^\dagger\ri\overset\leftrightarrow{D}_\mu\Phi)
(\bar{\PQd}\gamma^\mu \PQd)
\\
\Lop_{\Phi \widetilde{\PW}\PB} = (\Phi^\dagger\tau^I\Phi)\widetilde{\PW}^I_{\mu\nu}\PB^{\mu\nu} &
\Lop_{\PQd \PB} = (\bar{\PQq}\sigma^{\mu\nu}\Gamma_{_\PQd} \PQd\Phi)\PB_{\mu\nu} &
\Lop_{\Phi \PQu\PQd}=\ri(\widetilde\Phi^\dagger{D}_\mu\Phi)
(\bar{\PQu}\gamma^\mu \Gamma_{\PQu\PQd} \PQd)
\rule[-1.5ex]{0pt}{3ex}\\
\hline
\end{array}$}
\end{table}
In order to avoid flavor-changing neutral currents, the flavor
matrices appearing in dimension-6 operators involving left-handed
doublets and right-handed singlets ($\psi^2 \PX\Phi$ terms) have been
fixed to the same matrices $\Gamma_f$ that are present in the Yukawa
couplings, i.e.\ we assume minimal flavor violation. Moreover, in all
dimension-6 operators involving neutral currents (first seven $\psi^2
\Phi^2 D$ terms) the flavor matrices have been chosen to be equal to
the unit matrix. In the operator $\Lop_{\Phi \PQu\PQd}$ leading to
right-handed flavor-changing charged currents we kept a general
flavor matrix $\Gamma_{\PQu\PQd}$.  The generalization to more general
flavor schemes is straightforward.

In weakly interacting theories the dimension-6 operators of
\refT{ta:Lops} involving field strengths can only result from loops,
while the others also result from tree diagrams \cite{Arzt:1994gp}. 
The operators involving dual field strengths tensors or complex Wilson
coefficients violate CP.

\subsubsubsection{Alternative basis}

In the previous basis, the $O(p^2)$ EW oblique corrections, i.e., the $S$ and $T$ parameters, are captured in terms of $\psi^2  \Phi^2 D$ operators.
An alternative basis, see~\refT{ta:Lops_v2},  is often used~\cite{Giudice:2007fh, Contino:2013kra}  where these oblique corrections are now described by purely bosonic operators:
\beq
        \label{eq:Lag_v2}
\Lag_{\mathrm{eff}}= \Lag^{(4)}_{\mathrm{SM}}+ \frac{1}{\Lambda^2}\sum_k \Lcoeffp_k
\Lopp_k,
\eeq

\begin{table}
\caption{Alternative basis of dimension-6 operators involving Higgs doublet fields or
  gauge-boson fields. }
\label{ta:Lops_v2}
\centering{$\displaystyle
\renewcommand{\arraystretch}{1.4}\begin{array}{l@{\quad\ }l@{\quad\ }l}
\hline
\Phi^6\quad \text{and}\quad \Phi^4D^2 & \psi^2\Phi^3 
&\PX^3
\\
\hline
\Lopp_6 =  (\Phi^\dagger\Phi)^3   &
\Lopp_{\Pe\Phi}=(\Phi^\dagger\Phi)(\bar{\Pl}\,\Gamma_{_\Pe} \Pe \Phi) 
&\Lopp_{G} = f^{ABC}G^{A\nu}_\mu G^{B\rho}_\nu G^{C\mu}_\rho
\\
\Lopp_{\Phi} = \partial_\mu (\Phi^\dagger\Phi) \partial^\mu (\Phi^\dagger\Phi)   &
\Lopp_{\PQu\Phi}=(\Phi^\dagger\Phi)(\bar{\PQq}\,\Gamma_{_\PQu} \PQu \widetilde\Phi) 
&\Lopp_{\widetilde{G}} =  f^{ABC}\widetilde{G}^{A\nu}_\mu G^{B\rho}_\nu G^{C\mu}_\rho
\\
\Lopp_{\mathrm T} = (\Phi^\dagger \overset\leftrightarrow{D_\mu}\Phi) (\Phi^\dagger \overset\leftrightarrow{D^\mu}\Phi) &
\Lopp_{\PQd\Phi}=(\Phi^\dagger\Phi)(\bar{\PQq}\,\Gamma_d d \Phi) 
&\Lopp_{\PW} =  \varepsilon^{IJK}\PW^{I\nu}_\mu \PW^{J\rho}_\nu \PW^{K\mu}_\rho
\\
&&\Lopp_{\widetilde{\PW}} =  \varepsilon^{IJK}\widetilde{\PW}^{I\nu}_\mu \PW^{J\rho}_\nu \PW^{K\mu}_\rho
\rule[-1.5ex]{0pt}{3ex}\\
\hline
\PX^2\Phi^2 &\psi^2 \PX\Phi & \psi^2\Phi^2D \\
\hline\rule{0pt}{3.5ex}
\Lopp_{\mathrm DW} =    \left( \Phi^\dagger  \tau^I \ri\overleftrightarrow {D^\mu} \Phi \right )( D^\nu  \PW_{\mu \nu})^I &
\Lopp_{\PQu G} = (\bar{\PQq}\sigma^{\mu\nu}\frac{\lambda^A}{2}\Gamma_{_\PQu} \PQu
\widetilde\Phi)G^A_{\mu\nu}\qquad &
\Lop_{\Phi \Pl}^{\prime (1)}=(\Phi^\dagger\ri\overset\leftrightarrow{D}_\mu\Phi)
(\bar{\Pl}\gamma^\mu \Pl)
\\
\Lopp_{D\PB} = \left( \Phi^\dagger  \ri \overleftrightarrow {D^\mu} \Phi \right )( \partial^\nu  \PB_{\mu \nu})  &
\Lopp_{\PQd G} = (\bar{\PQq}\sigma^{\mu\nu}\frac{\lambda^A}{2}\Gamma_{_\PQd} \PQd
\Phi)G^A_{\mu\nu} &
\Lop_{\Phi \Pl}^{\prime (3)}=(\Phi^\dagger\ri\overset\leftrightarrow{D}{}_\mu^I\Phi)
(\bar{\Pl}\gamma^\mu \tau^I \Pl)
\\
\Lopp_{D \Phi \PW} =  \ri (D^\mu \Phi)^\dagger \tau^I (D^\nu \Phi)\PW_{\mu \nu}^I &
\Lopp_{\Pe\PW} = (\bar{\Pl}\sigma^{\mu\nu}\Gamma_{_\Pe} \Pe {\tau^I}
\Phi)\PW^I_{\mu\nu} &
\Lopp_{\Phi \Pe}=(\Phi^\dagger\ri\overset\leftrightarrow{D}_\mu\Phi)
(\bar{\Pe}\gamma^\mu \Pe)
\\
\Lopp_{D \Phi \widetilde{\PW}} =  \ri (D^\mu \Phi)^\dagger \tau^I (D^\nu \Phi) \widetilde{\PW}_{\mu \nu}^I  &
\Lopp_{\PQu\PW} = (\bar{\PQq}\sigma^{\mu\nu}\Gamma_{_\PQu} \PQu{\tau^I}\widetilde\Phi)\PW^I_{\mu\nu} &
\Lop_{\Phi \PQq}^{\prime (1)}=(\Phi^\dagger\ri\overset\leftrightarrow{D}_\mu\Phi)
(\bar{\PQq}\gamma^\mu \PQq)
\\
\Lopp_{D \Phi \PB} = \ri (D^\mu \Phi)^\dagger (D^\nu \Phi)\PB_{\mu \nu} &
\Lopp_{\PQd\PW} = (\bar{\PQq}\sigma^{\mu\nu}\Gamma_{_\PQd} \PQd {\tau^I}
\Phi)\PW^I_{\mu\nu} &
\Lop_{\Phi \PQq}^{\prime (3)}=(\Phi^\dagger\ri\overset\leftrightarrow{D}{}_\mu^I\Phi)
(\bar{\PQq}\gamma^\mu \tau^I \PQq)
\\
\Lopp_{D \Phi \widetilde{\PB}} = \ri (D^\mu \Phi)^\dagger (D^\nu \Phi)\widetilde{\PB}_{\mu \nu} &
\Lopp_{\Pe\PB} = (\bar{l}\sigma^{\mu\nu}\Gamma_{_\Pe} \Pe\Phi)B_{\mu\nu} &
\Lopp_{\Phi \PQu}=(\Phi^\dagger\ri\overset\leftrightarrow{D}_\mu\Phi)
(\bar{\PQu}\gamma^\mu \PQu)
\\
\Lopp_{\Phi \PB} = (\Phi^\dagger\Phi)B_{\mu\nu}\PB^{\mu\nu} &
\Lopp_{\PQu\PB} = (\bar{\PQq}\sigma^{\mu\nu}\Gamma_{_\PQu} \PQu\widetilde\Phi)\PB_{\mu\nu} &
\Lopp_{\Phi \PQd}=(\Phi^\dagger\ri\overset\leftrightarrow{D}_\mu\Phi)
(\bar{\PQd}\gamma^\mu \PQd)
\\
\Lopp_{\Phi \widetilde \PB} = (\Phi^\dagger\Phi)\PB_{\mu\nu} \widetilde{\PB}^{\mu\nu} &
\Lopp_{\PQd\PB} = (\bar{\PQq}\sigma^{\mu\nu}\Gamma_{_\PQd} \PQd\Phi)\PB_{\mu\nu} &
\Lopp_{\Phi \PQu\PQd}=\ri(\widetilde\Phi^\dagger{D}_\mu\Phi)
(\bar{\PQu}\gamma^\mu \Gamma_{\PQu\PQd} \PQd)
\\
\Lopp_{\Phi G} =   \Phi^\dagger \Phi G_{\mu\nu}^A G^{A\mu\nu}\\
\Lopp_{\Phi \widetilde{G}} =  \Phi^\dagger \Phi G_{\mu\nu}^A \widetilde{G}^{A\mu\nu}
\rule[-1.5ex]{0pt}{3ex}\\
\hline
\end{array}$}
\end{table}

The two basis representations are related with the following relations obtained by integration by parts:
\begin{eqnarray}
 \frac{g}{2}\Lopp_{\mathrm{DW}} - \frac{g'}{2}\Lopp_{\mathrm{DB}} + 
g' \Lopp_{D\Phi \PB} - g\Lopp_{D\Phi \PW}  - \frac{g'^2}{4}\, \Lopp_{\Phi \PB} & = & 
- \frac{g^2}{4} \Lop_{\Phi \PW}\, , \\[0.2cm]
\frac{g'}{2}\Lopp_{\mathrm{DB}} - g'\Lopp_{D\Phi \PB} + \frac{g'^2}{4}\, \Lopp_{\Phi \PB} & = & - \frac{g g^\prime}{4} \Lop_{\Phi \PW \PB}\, .
\end{eqnarray}
Actually, the two linear combinations of  the $\psi^2 \Phi^2 D$ operators that correspond to the $S$ and $T$ oblique parameters have to be omitted in the counting of the number of independent operators thanks to the two relations that hold up to total derivative terms:
\begin{eqnarray}
\Lop_{\Phi \Pl }^{\prime (1)}-\frac{1}{3} \Lop_{\Phi \PQq }^{\prime (1)} + 2 \Lopp_{\Phi \Pe} - \frac{4}{3} \Lopp_{\Phi \PQu} + \frac{2}{3} \Lopp_{\Phi \PQd}    & = &  -\Lopp_T +\frac{2}{g'} 
\Lopp_{\mathrm{DB}}\, ,\\
%2 \Lopp_{\Phi u } + 2 \Lopp_{\Phi d} +  2 \Lopp_{\Phi e} 
2 (\Lopp_{\PQu\Phi} + \Lopp_{\PQd\Phi} + \Lopp_{\Pe\Phi} + \mathrm{h.c.})
+ \Lop_{\Phi \PQq}^{\prime (3)} + \Lop_{\Phi \Pl }^{\prime (3)}
 & = &  3 \Lopp_\Phi  -  4\lambda \Lopp_6 + 4 m^2 (\Phi^\dagger\Phi)^2
 - \frac{2}{g} \Lopp_{\mathrm{DW}} \, . \qquad
\end{eqnarray}
A convenient choice of the two redundant operators to drop are $\Lop_{\Phi \Pl }^{\prime (1)}$ and $\Lop_{\Phi \Pl }^{\prime (3)}$.

Another advantage of the new operator basis is that it offers a simple power-counting to estimate the size the Wilson coefficients under the assumptions that the New Physics sector is characterized by a single scale and a single coupling and that the photon does not couple to electrically neutral particles at tree-level: starting from the SM Lagrangian,   any additional power of $\Phi$ is suppressed by a factor $g_*/M \equiv 1/f$, where $g_*$ denotes the  coupling 
between the Higgs boson and the New Physics states and $M$ is the overall mass scale of the New Physics sector;  any additional derivative instead costs a factor $1/M$. 
According to the power counting of  \Bref{Giudice:2007fh},  
one naively estimates ($\psi=\PQu,\PQd,\Pl,\PQq,L$)%
\begin{eqnarray}
\label{eq:NDA}
\Lcoeffp_{\PH}, \Lcoeffp_{\mathrm T}, \frac{\Lcoeffp_{6}}{\lambda}, \Lcoeffp_{\psi \Phi} \sim O\!\left(\frac{\Lambda^2}{f^2}\right) , \quad 
\frac{\Lcoeffp_{\mathrm{DW}}}{g}, \frac{\Lcoeffp_{\mathrm{DB}}}{g'} \sim O\!\left(\frac{\Lambda^2}{M^2} \right)\, , 
\\[0.5cm]
\label{eq:NDA2}
\frac{\Lcoeffp_{D\Phi \PW}}{g}, \frac{\Lcoeffp_{D\Phi \PB}}{g'}, \frac{\Lcoeffp_{\Phi \PB}}{g'^2}, \frac{\Lcoeffp_{\Phi G}}{\gs^2} \sim  O\!\left(\frac{\Lambda^2}{16\pi^2 f^2}\right) ,
\quad \frac{\Lcoeffp_{\psi \PW}}{g}, \frac{\Lcoeffp_{\psi \PB}}{g'}, \frac{\Lcoeffp_{\psi G}}{\gs} \sim  O\!\left(\frac{\Lambda^2}{16\pi^2 f^2}\right)\, ,
\\[0.5cm]
\label{eq:NDA3}
\Lcoeffp_{\Phi \Pe} , \Lcoeffp_{\Phi \PQu} , \Lcoeffp_{\Phi \PQd} , \Lcoeff_{\Phi \PQq}^{\prime (1)} , \Lcoeff_{\Phi \Pl}^{\prime (1)}, \Lcoeff_{\Phi \PQq}^{\prime (3)} , \Lcoeff_{\Phi \Pl}^{\prime (3)}   \sim  O\!\left(\frac{\lambda_\psi^2}{g_*^2} \frac{\Lambda^2}{f^2}\right)\, ,
\quad \Lcoeffp_{\Phi \PQu\PQd} \sim O\!\left(\frac{\lambda_\PQu \lambda_\PQd}{g_*^2} \frac{\Lambda^2}{f^2}\right)\, ,
\end{eqnarray}
where $\lambda_\psi$ denotes the coupling of a generic SM fermion
$\psi$ to the New Physics sector. Notice the additional factors
$(g_*^2/16 \pi^2)$ in $\Lcoeffp_{D\Phi \PB}, \Lcoeffp_{D\Phi \PW},
\Lcoeffp_{\Phi \PB}$ and $\Lcoeffp_{\Phi G}$ as the corresponding operators
contribute to the coupling of on-shell photons and gluons to neutral
particles and correct the $\PW$ gyromagnetic ratio, which in most models
of New Physics are effects occurring only at the loop-level.

\subsubsubsection{Translation to the physical basis}

In order to perform calculations it is useful to express the
effective Lagrangian introduced above in terms of properly normalized
physical fields and physical parameters. In the SM the Higgs doublet field 
can be parametrized as\footnote{We use the index ``SM'' to denote
  fields and parameters in the SM without dimension-6 operators.}
\beq
\Phi=\left(\barr{c}\phicSM^+ \\
    \frac{1}{\sqrt{2}}(\vSM+\HSM+\ri\phinSM)\earr\right),\qquad
\widetilde\Phi=\left(\barr{c}
    \frac{1}{\sqrt{2}}(\vSM+\HSM-\ri\phinSM)
    \\ -\phicSM^-\earr\right),\qquad \phicSM^-=(\phicSM^+)^*,
\eeq 
where $\PH$ is the Higgs field and $\phic^\pm$ and $\phin$ the
would-be Goldstone-boson fields. In the SM without dimension-6
operators the vacuum expectation value $\vSM$ of the Higgs field is in LO 
obtained as
\beq
\vSM=\frac{\sqrt{2}m}{\sqrt{\lambda}},
\eeq
and the Higgs-boson mass reads
\beq
\MHSM=\sqrt{\lambda}\vSM.
\eeq
The physical gauge-boson fields are given by
\beqar
\WSM^\pm{}_\mu &=& \frac{1}{\sqrt{2}}(\PW^1_\mu\mp\ri \PW^2_\mu),\nn\\
\ZSM{}_\mu &=& \cwSM \PW^3_\mu - \swSM \PB_\mu, \qquad
\ASM{}_\mu = \swSM \PW^3_\mu + \cwSM \PB_\mu
\eeqar
with
\beq
\cwSM=\frac{g}{\sqrt{g^2+g'^2}},\qquad \swSM=\frac{g'}{\sqrt{g^2+g'^2}}.
\eeq
The masses of the gauge bosons are obtained as
\beq
\MWSM=\frac{1}{2}g\vSM, \qquad \MZSM=\frac{1}{2}\sqrt{g^2+g'^2}\vSM,
\eeq
and the electromagnetic coupling reads
\beq
\eSM=\frac{gg'}{\sqrt{g^2+g'^2}}.
\eeq
The physical fermion fields $\fSM{}_{p}$, corresponding to the mass
eigenstates, are obtained by diagonalizing the Yukawa-coupling
matrices $\Gamma_{\Pf}$,
\beq
\fSM{}_p=U^{\Pf}_{pq} \Pf_q, \quad  \Pf=\Pl_i,\Pe,\PQq_i,\PQu,\PQd. 
\eeq
resulting in diagonal fermion-mass matrices
\beq
\MfSM = \frac{1}{\sqrt{2}}U^{\hat{\Pf}}\Gamma_{\Pf} U^{\Pf,\dagger}\vSM,
\eeq
where $\hat{\Pf}=\PQq_1,\PQq_2,\Pe_1,\Pe_2$ for $\Pf=\PQu,\PQd,\Pe,\PGnl$, respectively.
The matrices $U^{\Pf}$ give rise to the appearance of the quark-mixing
matrix
\beq
V_{pq}=(U^{q_1}U^{q_2\dagger})_{pq}
\eeq
in charged-current interactions.
The electric charges of fermions result from the relation
\beq
Q=I_3+Y,
\eeq
where $I_3$ is the third component of the weak isospin. 

The presence of the dimension-6 operators modifies the quadratic part
of the Lagrangian. Consequently, the fields have to be redefined such
that the free part has the canonical normalization and the SM
parameters get extra contributions. Moreover the vacuum expectation
value of the Higgs field is shifted to
\beq
\vnew^2=\vSM^2\left(1+\frac{3}{2}\Lcoeff_{\Phi}\frac{\vSM^2}{\lambda\Lambda^2}\right)
=\frac{2m^2}{\lambda}\left(1+\frac{3}{2}\Lcoeff_{\Phi}\frac{\vSM^2}{\lambda\Lambda^2}\right).
\eeq

Canonically normalized kinetic terms for the bosonic fields can be
restored in the presence of dimension-6 operators upon redefining the
fields as follows:
\beqar
\HSM &=& \Hnew \left[1-\frac{v^2}{4\Lambda^2}\left(\Lcoeff_{\Phi
      D}-4\Lcoeff_{\Phi\Box}\right)\right]  +\frac{3}{4}v\Lcoeff_{\Phi}\frac{\vnew^2}{\lambda\Lambda^2},
\nn\\
\phinSM  &=& \phinnew \left[1-\frac{v^2}{4\Lambda^2}\Lcoeff_{\Phi D}\right],
\qquad
\phicSM^\pm  = \phicnew^\pm,
\nn\\
\WSM^\pm{}_\mu  &=& \Wnew_\mu \left[1+\frac{v^2}{\Lambda^2}\Lcoeff_{\Phi \PW}\right],
\qquad
\ZSM{}_\mu  = \Znew_\mu \left[1+\frac{v^2}{\Lambda^2}\Lcoeff_{\PZ\PZ}\right],
\nn\\
\ASM{}_\mu  &=& \Anew_\mu \left[1+\frac{v^2}{\Lambda^2}\Lcoeff_{\PA\PA}\right]
+  \Znew_\mu \left[1+\frac{v^2}{\Lambda^2}\Lcoeff_{\PA\PZ}\right],
\qquad
\GSM^A{}_\mu  = \Gnew^A_\mu \left[1+\frac{v^2}{\Lambda^2}\Lcoeff_{\Phi G}\right],
\eeqar 
where
\beqar
\Lcoeff_{\PZ\PZ}&=&\cwSM^2\Lcoeff_{\Phi \PW}+\swSM^2\Lcoeff_{\Phi
  \PB}+\cwSM\swSM\Lcoeff_{\Phi \PW\PB},
\nn\\
\Lcoeff_{\PA\PA}&=&\swSM^2\Lcoeff_{\Phi \PW}+\cwSM^2\Lcoeff_{\Phi
  \PB}-\cwSM\swSM\Lcoeff_{\Phi \PW\PB},
\nn\\
\Lcoeff_{\PA\PZ}&=&2\cwSM\swSM(\Lcoeff_{\Phi \PW}-\Lcoeff_{\Phi \PB})
+(\swSM^2-\cwSM^2)\Lcoeff_{\Phi \PW\PB},
\eeqar
and analogous definitions hold for $\Lcoeff_{\PZ\widetilde{\PZ}}$,
$\Lcoeff_{\PA\widetilde{\PZ}}$, and $\Lcoeff_{\PA\widetilde{\PA}}$.
Since we have chosen the flavor-mixing matrix in the $\psi^2\Phi^3$
terms to be identical to the one in the Yukawa couplings, the physical
fermionic fields stay the same as in the SM. We denote them by $f$
in the following (the individual physical fermion fields $\PQu$, $\PQd$, and
$\Pl$ should not to be confused with the same symbols used in 
\refE{LSM} for the fields in the symmetric basis).
%\beq
%\fSM{}_i = \hat{f} \left[1-\frac{v^2}{4}\Lcoeff_{\ldots}\right].
%\eeq

In terms of the redefined fields the quadratic part of the Lagrangian
reads after adding a 't~Hooft--Feynman gauge-fixing term and dropping
total derivatives and a constant
\beqar
\label{eq:Lquad}
\Lag^{(4)}_{\mathrm{SM}} &=& 
%-\frac{1}{4}(\partial_\mu\Gnew^A_{\nu}-\partial_\nu\Gnew^A_{\mu})
%(\partial^\mu\Gnew^{A\nu}-\partial^\nu\Gnew^{A\mu})
%-\frac{1}{2}(\partial_\mu\Wnew^+_{\nu}-\partial_\nu\Wnew^+_{\mu})
%(\partial^\mu\Wnew^{-\nu}-\partial^\nu\Wnew^{-\mu})
%\nn\\&&{}
%-\frac{1}{4}(\partial_\mu\Znew_{\nu}-\partial_\nu\Znew_{\mu})
%(\partial^\mu\Znew^{\nu}-\partial^\nu\Znew^{\mu})
%-\frac{1}{4}(\partial_\mu\Anew_{\nu}-\partial_\nu\Anew_{\mu})
%(\partial^\mu\Anew^{\nu}-\partial^\nu\Anew^{\mu})
%\nn\\&&{}
-\frac{1}{2}(\partial_\mu\Gnew^A_{\nu})
(\partial^\mu\Gnew^{A\nu})
-(\partial_\mu\Wnew^+_{\nu})
(\partial^\mu\Wnew^{-\nu})
%\nn\\&&{}
-\frac{1}{2}(\partial_\mu\Znew_{\nu})
(\partial^\mu\Znew^{\nu})
-\frac{1}{2}(\partial_\mu\Anew_{\nu})
(\partial^\mu\Anew^{\nu})
\nn\\&&{}
+\MWnew^2\Wnew^+_\mu \Wnew^{-\mu} +\frac{1}{2}\MZnew^2\Znew_\mu \Znew^{\mu}
+\frac{1}{2}(\partial_\mu \Hnew) (\partial^\mu\Hnew)
-\frac{1}{2}\MHnew^2 \Hnew^2
\nn\\&&{}
+(\partial_\mu \phicnew^+) (\partial^\mu\phicnew^-)
-\MWnew^2 \phicnew^+\phicnew^-
+\frac{1}{2}(\partial_\mu \phinnew) (\partial^\mu\phinnew)
-\frac{1}{2}\MZnew^2 (\phinnew)^2
\nn\\&&{}
+ \sum_{\Pf=\Pl,\PQu,\PQd} \bar\fnew(\ri\slashed{\partial}-\Mfnew)\fnew
+\bar\nunew\ri\slashed{\partial}\nunew
\eeqar
with the physical mass parameters
\beqar
\MWnew^2&=&\frac{1}{4}g^2\vnew^2\left[1+2\frac{v^2}{\Lambda^2}\Lcoeff_{\Phi \PW}\right],
\nn\\
\MZnew^2&=&\frac{1}{4}(g^2+g'^2)\vnew^2\left[1+\frac{v^2}{2\Lambda^2}\left(4\Lcoeff_{\PZ\PZ}+\Lcoeff_{\Phi D}\right)\right],
\nn\\
\MHnew^2&=&\lambda\vnew^2\left[1+\frac{v^2}{2\Lambda^2}\left(4\Lcoeff_{\Phi\Box}-\frac{6}{\lambda}
\Lcoeff_{\Phi}
   -\Lcoeff_{\Phi D}\right)\right],
\nn\\
\Mfnew &=& \frac{1}{\sqrt{2}}U^{\hat{\Pf}}\Gamma_{_\Pf} U^{\Pf,\dagger}\vnew
\left[1-\frac{1}{2}\frac{v^2}{\Lambda^2}\Lcoeff_{\Pf\phi}\right].
\eeqar

In \refE{eq:Lquad} we have used the usual 
't~Hooft--Feynman gauge-fixing term
\beq
\Lag_{\mathrm{fix}}=-C_+C_- -\frac{1}{2}(C_{\PZ})^2-\frac{1}{2}(C_{\PA})^2
-\frac{1}{2}C^A_{G}C^A_{G}
\eeq
with
\beq
C^A_G=\partial_\mu\Gnew^{A\mu},\qquad
C_{\PA}=\partial_\mu\Anew^\mu,\qquad
C_{\PZ}=\partial_\mu\Znew^\mu + \MZnew\phinnew,\qquad
C_\pm=\partial_\mu\Wnew^{\pm\mu} \pm \ri\MWnew\phicnew^{\pm}
\eeq
in terms of the physical fields and parameters, which gives rise to
the same propagators as in the SM.

In the following, the abbreviations $\cw$ and $\sw$ are defined via
the physical masses
\beq
\cw=\frac{\MW}{\MZ}, \qquad \sw=\sqrt{1-\cw^2}.
\eeq

The parameters of the SM Lagrangian $g$, $g'$, $\lambda$, $m^2$, and
$\Gamma_{_\Pf}$ keep their meaning in the presence of dimension-6 operators.

\subsubsection{Higgs vertices}

\newcommand{\voverlambdasq}{\frac{1}{\sqrt{2}\GF\Lambda^2}}

Here we list the most important Feynman rules for vertices
involving exactly one physical Higgs boson. These are given in terms
of the above-defined physical fields and parameters.  In the
coefficients of dimension-6 couplings we replaced $v^2$ by the Fermi
constant via $v^2=1/(\sqrt{2}\GF)$.

The triple vertices involving one Higgs boson read:
\begin{align}
\intertext{$\PH\Pg\Pg$~coupling:}
{\vcenter{\hbox{
\begin{picture}(110,80)(-50,-40)
\Text(45,29)[b]{$G^A_{\mu},p_1$}
\Text(45,-29)[t]{$G^B_{\nu},p_2$}
\Text(-45,0)[b]{$\PH$}
\Vertex(0,0){2}
\DashLine(0,0)(-35,0){3}
\Gluon(0,0)(35,25){3}{3.5}
\Gluon(0,0)(35,-25){-3}{3.5}
\end{picture}
}}}
&=\ri
\frac{2g}{\MW}\voverlambdasq\left[\Lcoeff_{GG}(p_{2\mu}p_{1\nu}-p_1p_2g_{\mu\nu})
+\Lcoeff_{G\widetilde{G}}\varepsilon_{\mu\nu\rho\sigma}p_{1}^\rho p_2^\sigma\right]\delta^{AB},
\\ 
\intertext{$\PH\PA\PA$~coupling:}
{\vcenter{\hbox{
\begin{picture}(110,80)(-50,-40)
\Text(45,29)[b]{$\PA_{\mu},p_1$}
\Text(45,-29)[t]{$\PA_{\nu},p_2$}
\Text(-45,0)[b]{$\PH$}
\Vertex(0,0){2}
\DashLine(0,0)(-35,0){3}
\Photon(0,0)(35,25){2}{3.5}
\Photon(0,0)(35,-25){-2}{3.5}
\end{picture}}}}
&=\ri
\frac{2g}{\MW}\voverlambdasq\left[\Lcoeff_{\PA\PA}(p_{2\mu}p_{1\nu}-p_1p_2g_{\mu\nu})
+\Lcoeff_{\PA\widetilde{\PA}}\varepsilon_{\mu\nu\rho\sigma}p_{1}^\rho p_2^\sigma\right],
\\
\intertext{$\PH\PA\PZ$~coupling:}
{\vcenter{\hbox{
\begin{picture}(110,80)(-50,-40)
\Text(45,29)[b]{$\PA_{\mu},p_1$}
\Text(45,-29)[t]{$\PZ_{\nu},p_2$}
\Text(-45,0)[b]{$\PH$}
\Vertex(0,0){2}
\DashLine(0,0)(-35,0){3}
\Photon(0,0)(35,25){2}{3.5}
\Photon(0,0)(35,-25){-2}{3.5}
\end{picture}}}}
&=\ri
\frac{g}{\MW}\voverlambdasq\left[\Lcoeff_{\PA\PZ}(p_{2\mu}p_{1\nu}-p_1p_2g_{\mu\nu})
+\Lcoeff_{\PA\widetilde{\PZ}}\varepsilon_{\mu\nu\rho\sigma}p_{1}^\rho p_2^\sigma\right],
\\
\intertext{$\PH\PZ\PZ$~coupling:}
{\vcenter{\hbox{
\begin{picture}(110,80)(-50,-40)
\Text(45,29)[b]{$\PZ_{\mu},p_1$}
\Text(45,-29)[t]{$\PZ_{\nu},p_2$}
\Text(-45,0)[b]{$\PH$}
\Vertex(0,0){2}
\DashLine(0,0)(-35,0){3}
\Photon(0,0)(35,25){2}{3.5}
\Photon(0,0)(35,-25){-2}{3.5}
\end{picture}}}}
&\barr{l}\displaystyle
=\ri g \frac{\MZ}{\cw} g_{\mu\nu}
\left[1+\voverlambdasq\left(\Lcoeff_{\Phi \PW}+\Lcoeff_{\Phi\Box}
+\frac{1}{4}\Lcoeff_{\Phi  D}\right)\right] 
\\[3ex]\displaystyle \phantom{=}{}+
\ri
\frac{2g}{\MW}\voverlambdasq\left[\Lcoeff_{\PZ\PZ}(p_{2\mu}p_{1\nu}-p_1p_2g_{\mu\nu})
+\Lcoeff_{\PZ\widetilde{\PZ}}\varepsilon_{\mu\nu\rho\sigma}p_{1}^\rho p_2^\sigma\right],
\earr
\\
\intertext{$\PH\PW\PW$~coupling:}
{\vcenter{\hbox{
\begin{picture}(110,80)(-50,-40)
\Text(45,29)[b]{$\PW^+_{\mu},p_1$}
\Text(45,-29)[t]{$\PW^-_{\nu},p_2$}
\Text(-45,0)[b]{$\PH$}
\Vertex(0,0){2}
\DashLine(0,0)(-35,0){3}
\Photon(0,0)(35,25){2}{3.5}
\Photon(0,0)(35,-25){-2}{3.5}
\end{picture}}}}
&\barr{l}\displaystyle
= 
\ri g {\MW} g_{\mu\nu}
\left[1+\voverlambdasq\left(\Lcoeff_{\Phi \PW}+\Lcoeff_{\Phi\Box}-\frac{1}{4}\Lcoeff_{\Phi
      D}\right)\right] 
\\[3ex]\displaystyle \phantom{=}{}+
\ri
\frac{2g}{\MW}\voverlambdasq\left[\Lcoeff_{\Phi \PW}(p_{2\mu}p_{1\nu}-p_1p_2g_{\mu\nu})
+\Lcoeff_{\Phi\widetilde{\PW}}\varepsilon_{\mu\nu\rho\sigma}p_{1}^\rho p_2^\sigma\right],
\earr
\\
\intertext{$\PH\Pf\Pf$~coupling:}
{\vcenter{\hbox{
\begin{picture}(110,80)(-50,-40)
\Text(45,29)[b]{$\bar{\Pf},p_1$}
\Text(45,-29)[t]{$\Pf,p_2$}
\Text(-45,0)[b]{$\PH$}
\Vertex(0,0){2}
\DashLine(0,0)(-35,0){3}
\ArrowLine(0,0)(35,25)
\ArrowLine(35,-25)(0,0)
\end{picture}}}}
&\barr{l}\displaystyle
= 
-\ri \frac{g}{2} \frac{m_{\Pf}}{\MW}
\left[1+\voverlambdasq\left(\Lcoeff_{\Phi \PW}+\Lcoeff_{\Phi\Box}-\frac{1}{4}\Lcoeff_{\Phi
      D}-\Lcoeff_{\Pf\phi}\right)\right], 
\earr
\end{align}
where $\Pf=\Pe,\PQu,\PQd$.

The quadruple vertices involving one Higgs field, one gauge
boson and a fermion--antifermion pair are given by ($\PQq=\PQu,\PQd$,
$\Pf=\PQu,\PQd,\PGnl,\Pe$, and $\hat{\Pf}=\PQq$ for $\Pf=\PQu,\PQd$ and $\hat{\Pf}=\Pl$ for 
$\Pf=\Pe$):
\begin{align}
\intertext{$\PH\Pg\PQq\PAQq$~coupling:}
{\vcenter{\hbox{
\begin{picture}(110,80)(-50,-40)
\Text(45,29)[b]{$\PAQq$}
\Text(45,-29)[t]{$PQq$}
\Text(-45,29)[b]{$\PH$}
\Text(-45,-29)[t]{$G^A_\mu,p_G$}
\Vertex(0,0){2}
\DashLine(0,0)(-35,25){3}
\Gluon(0,0)(-35,-25){3}{3}
\ArrowLine(0,0)(35,25)
\ArrowLine(35,-25)(0,0)
\end{picture}}}}
&=\ri
g\frac{m_q}{\MW}\ri p_{G\nu}\sigma^{\mu\nu}\frac{\la^A}{2}
\left[\frac{1+\gamma_5}{2}\Lcoeff_{\PQq G}
+\frac{1-\gamma_5}{2}\Lcoeff_{\PQq G}^*\right],
\\ 
\intertext{$\PH\PA \Pf\bar{\Pf}$~coupling:}
{\vcenter{\hbox{
\label{HAff-vertex}
\begin{picture}(110,80)(-50,-40)
\Text(45,29)[b]{$\bar{\Pf}$}
\Text(45,-29)[t]{$\Pf$}
\Text(-45,29)[b]{$\PH$}
\Text(-45,-29)[t]{$\PA_\mu,p_{\PA}$}
\Vertex(0,0){2}
\DashLine(0,0)(-35,25){3}
\Photon(0,0)(-35,-25){-2}{3.5}
\ArrowLine(0,0)(35,25)
\ArrowLine(35,-25)(0,0)
\end{picture}}}}
&\barr{l}\displaystyle
=\ri g\frac{m_f}{\MW}\ri p_{A\nu}\sigma^{\mu\nu}
\left[\frac{1+\gamma_5}{2}(\Lcoeff_{\Pf\PB}\cw+2I_3^f\Lcoeff_{\Pf\PW}\sw)\right.
\\[2ex]\hspace{7em}\displaystyle{}
\left.{}+\frac{1-\gamma_5}{2}(\Lcoeff_{\Pf\PB}^*\cw+2I_3^f\Lcoeff_{\Pf\PW}^*\sw)\right],
\earr
\\ 
\intertext{$\PH\PZ \Pf\bar{\Pf}$~coupling:}
\label{HZff-vertex}
{\vcenter{\hbox{
\begin{picture}(110,80)(-50,-40)
\Text(45,29)[b]{$\bar{\Pf}$}
\Text(45,-29)[t]{$\Pf$}
\Text(-45,29)[b]{$\PH$}
\Text(-45,-29)[t]{$\PZ_\mu,p_{\PZ}$}
\Vertex(0,0){2}
\DashLine(0,0)(-35,25){3}
\Photon(0,0)(-35,-25){-2}{3.5}
\ArrowLine(0,0)(35,25)
\ArrowLine(35,-25)(0,0)
\end{picture}}}}
&\barr{l}\displaystyle
=\ri g\frac{m_f}{\MW}\ri p_{\PZ\nu}\sigma^{\mu\nu}
\left[\frac{1+\gamma_5}{2}(2I_3^{\Pf}\Lcoeff_{\Pf\PW}\cw-\Lcoeff_{\Pf\PB}\sw)\right.
\\[2ex]\hspace{7em}\displaystyle{}\left.
{}+\frac{1-\gamma_5}{2}(2I_3^{\Pf}\Lcoeff_{\Pf\PW}^*\cw-\Lcoeff_{\Pf\PB}^*\sw)\right]
\\[2ex]\displaystyle{}\phantom{=}
+\ri2\MZ\gamma^\mu
\left[\frac{1-\gamma_5}{2}\left(\Lcoeff_{\Phi \hat{\Pf}}^{(1)}-2I_3^{\Pf}
\Lcoeff_{\Phi\hat{\Pf}}^{(3)}\right)
+\frac{1+\gamma_5}{2}\Lcoeff_{\Phi \Pf}\right],
\earr
\\ 
\intertext{$\PH\PW^+\PQd\PAQu$~coupling:}
{\vcenter{\hbox{
\begin{picture}(110,80)(-50,-40)
\Text(45,29)[b]{$\bar{u}_p$}
\Text(45,-29)[t]{$d_q$}
\Text(-45,29)[b]{$\PH$}
\Text(-45,-29)[t]{$\PW^+_\mu,p_{\PW}$}
\Vertex(0,0){2}
\DashLine(0,0)(-35,25){3}
\Photon(0,0)(-35,-25){-2}{3.5}
\ArrowLine(0,0)(35,25)
\ArrowLine(35,-25)(0,0)
\end{picture}}}}
&\barr{l}\displaystyle
=\ri g\frac{\sqrt{2}}{\MW}\ri p_{\PW\nu}\sigma^{\mu\nu}V_{pq}
\left[\frac{1+\gamma_5}{2}m_{\PQd}\Lcoeff_{\PQd\PW}
+\frac{1-\gamma_5}{2}m_{\PQu}\Lcoeff_{\PQu\PW}^*\right]
\\[3ex]\displaystyle{}\phantom{=}
-\ri\sqrt{2}\MW\gamma^\mu
\left[\frac{1-\gamma_5}{2}\,2\Lcoeff_{\Phi \PQq}^{(3)}V_{pq}
+\frac{1+\gamma_5}{2}(\Gamma_{\PQu\PQd})_{pq}\Lcoeff_{\Phi \PQu\PQd}\right],
\earr
\intertext{$\PH\PW^-\PQu\PAQd$~coupling:}
{\vcenter{\hbox{
\begin{picture}(110,80)(-50,-40)
\Text(45,29)[b]{$\bar{d}_p$}
\Text(45,-29)[t]{$u_q$}
\Text(-45,29)[b]{$\PH$}
\Text(-45,-29)[t]{$\PW^-_\mu,p_{\PW}$}
\Vertex(0,0){2}
\DashLine(0,0)(-35,25){3}
\Photon(0,0)(-35,-25){-2}{3.5}
\ArrowLine(0,0)(35,25)
\ArrowLine(35,-25)(0,0)
\end{picture}}}}
&\barr{l}\displaystyle
=\ri g\frac{\sqrt{2}}{\MW}\ri p_{\PW\nu}\sigma^{\mu\nu}V^\dagger_{pq}
\left[\frac{1+\gamma_5}{2}m_{\PQu}\Lcoeff_{\PQu\PW}
+\frac{1-\gamma_5}{2}m_{\PQd}\Lcoeff_{\PQd\PW}^*\right]
\\[3ex]\displaystyle{}\phantom{=}
-\ri\sqrt{2}\MW\gamma^\mu
\left[\frac{1-\gamma_5}{2}\,2\Lcoeff_{\Phi \PQq}^{(3)}V^\dagger_{pq}
+\frac{1+\gamma_5}{2}
(\Gamma_{\PQu\PQd}^\dagger)_{pq}\Lcoeff^*_{\Phi \PQu\PQd}\right],
\earr
\\ 
\intertext{$\PH\PW^+\Pe\PAGne$~coupling:}
{\vcenter{\hbox{
\begin{picture}(110,80)(-50,-40)
\Text(45,29)[b]{$\bar{\nu}$}
\Text(45,-29)[t]{$e$}
\Text(-45,29)[b]{$\PH$}
\Text(-45,-29)[t]{$\PW^+_\mu,p_{\PW}$}
\Vertex(0,0){2}
\DashLine(0,0)(-35,25){3}
\Photon(0,0)(-35,-25){-2}{3.5}
\ArrowLine(0,0)(35,25)
\ArrowLine(35,-25)(0,0)
\end{picture}}}}
&\barr{l}\displaystyle
=\ri g\frac{\sqrt{2}}{\MW}\ri p_{\PW\nu}\sigma^{\mu\nu}
\frac{1+\gamma_5}{2}m_{\Pe}\Lcoeff_{\Pe\PW}
%\\[3ex]\displaystyle{}\phantom{=}
-\ri\sqrt{2}\MW\gamma^\mu
\frac{1-\gamma_5}{2}\,2\Lcoeff_{\Phi \Pl}^{(3)},
\earr
\intertext{$\PH\PW^-\PGne\Pep$~coupling:}
{\vcenter{\hbox{
\begin{picture}(110,80)(-50,-40)
\Text(45,29)[b]{$\bar{e}$}
\Text(45,-29)[t]{$\nu$}
\Text(-45,29)[b]{$\PH$}
\Text(-45,-29)[t]{$\PW^-_\mu,p_{\PW}$}
\Vertex(0,0){2}
\DashLine(0,0)(-35,25){3}
\Photon(0,0)(-35,-25){-2}{3.5}
\ArrowLine(0,0)(35,25)
\ArrowLine(35,-25)(0,0)
\end{picture}}}}
&\barr{l}\displaystyle
=\ri g\frac{\sqrt{2}}{\MW}\ri p_{\PW\nu}\sigma^{\mu\nu}
\frac{1-\gamma_5}{2}m_{\Pe}\Lcoeff_{\Pe\PW}^*
%\\[3ex]\displaystyle{}\phantom{=}
-\ri\sqrt{2}\MW\gamma^\mu
\frac{1-\gamma_5}{2}\,2\Lcoeff_{\Phi l}^{(3)}.
\earr
\end{align} 
The flavor matrix $\Gamma_{\PQu\PQd}$ appearing in the above Feynman rules
has now to be understood in the mass eigenstate basis for the quarks,
and $V$ denotes the usual quark-mixing matrix. For neutrinos
the terms involving $\Lcoeff_{\Pf\PW}$, $\Lcoeff_{\Pf\PB}$,  and $\Lcoeff_{\Phi \Pf}$ in
\Erefs{HAff-vertex} and \Eref{HZff-vertex} are absent.

There are additional vertices involving one Higgs field together with
three or four vector-boson fields resulting from the $\PX^2\Phi^2$
operators, and the operators $\psi^2\PX\Phi$ give rise to vertices
involving a fermion--antifermion pair and a vector-boson pair along
with a single Higgs field. Additional vertices involve more than one
Higgs-boson field.

The Feynman rules above are given in terms of the weak coupling $g$.
Choosing $\MZ$, $\MW$, and $\GF$ as input, $g$ can be written as
\beq
g=2\MW\sqrt{\sqrt{2}\GF}\left(1-\voverlambdasq\left(\Lcoeff_{\Phi \PW}
  + \Lcoeff^{(3)}_{\phi\mu}\right)\right).
\eeq
%We have used the relation $v^2=1/\sqrt{2}\GF$ to express the
%coefficients of the Wilson coefficients by $\GF$. 
Inserting this into the Feynman rules with SM contributions leads  to
the replacement
\beq
g\left(1+\voverlambdasq\left(\Lcoeff_{\Phi \PW}\right)\right)\to
2\MW\sqrt{\sqrt{2}\GF}\left(1-\voverlambdasq
\left( \Lcoeff^{(3)}_{\phi\mu}\right)\right).
\eeq

We note that for $\Lambda\approx5\UTeV$ we have
$1/(\sqrt{2}\GF\Lambda^2)\approx g^2/(4\pi)$, i.e.\ the contributions
of dimension-6 operators are generically of the order of loop effects.
For higher scales, loop contributions tend to be more important than
dimension-6 operators.

In the basis of \refT{ta:Lops}, the scaling of the SM
Higgs--vector-boson couplings is parametrized by $\Lcoeff_{\Phi\Box}$,
the difference between the $\PH\PZ\PZ$ and $\PH\PW\PW$ couplings by $\Lcoeff_{\Phi
  D}$, and the Higgs-fermion coupling by $\Lcoeff_{\Pf\phi}$. Additional
overall factors depend on the input-parameter scheme.  The new
coupling structures appearing in the Higgs--gauge-boson couplings are
parametrized by $\Lcoeff_{\Phi \PW}$, $\Lcoeff_{\Phi \PB}$, and
$\Lcoeff_{\Phi \PW \PB}$ for the CP-conserving part and by $\Lcoeff_{\Phi
  \widetilde{\PW}}$, $\Lcoeff_{\Phi\widetilde{\PB}}$, and $\Lcoeff_{\Phi
  \PW\widetilde{\PB}}$ for the CP-violating part.
In the $\PH\VV \PB\Pf\bar{\Pf}$ vertices all coefficients of the $\psi^2\PX^2\Phi$
and $\psi^2\Phi^2D$ operators enter. Since these coefficients also
enter the gauge-boson--fermion couplings they are constrained by
results from LEP and on anomalous moments (see for instance Ref.~\cite{Contino:2013kra}).

\subsubsection{Nonlinear parameterization}

In the unitary gauge and  in the basis of fermion mass eigenstates, 
the effective Lagrangian relevant for Higgs physics reads as follows~\cite{Contino:2010mh, Contino:2013kra}
\begin{equation} 
\label{eq:chiralL}
\begin{split}
\Lag =
& \, \frac{1}{2} \partial_\mu \PH\ \partial^\mu \PH - \frac{1}{2} \MH^2 \PH^2 
- c_3 \, \frac{1}{6} \left(\frac{3 \MH^2}{\vSM}\right) \PH^3 - \sum_{\Pf = \PQu,\PQd,\Pl} m_{\Pf} \, \bar{\Pf} \Pf \left( 1 + c_{\Pf} \frac{\PH}{\vSM} + \dots \right)
\\[0.cm]
& + \MW^2\,  \PW^+_\mu \PW^{-\, \mu} \left(1 + 2 c_{\PW}\, \frac{\PH}{\vSM} + \dots \right) + \frac{1}{2} \MZ^2\,  \PZ_\mu \PZ^\mu \left(1 + 2 c_{\PZ}\, \frac{\PH}{\vSM} + \dots \right) + \dots
\\[0.3cm]
& + \left(  c_{\PW\PW}\,   \PW_{\mu\nu}^+ \PW^{-\mu\nu}  + \frac{c_{\PZ\PZ}}{2} \, \PZ_{\mu\nu}\PZ^{\mu\nu} + 
c_{\PZ\PGg} \, \PZ_{\mu\nu} \PA^{\mu\nu}   + \frac{c_{\PGg\PGg}}{2}\, \PA_{\mu\nu} \PA^{\mu\nu} + \frac{c_{\Pg\Pg}}{2}\, G_{\mu\nu}^A G^{A\mu\nu} \right) \frac{\PH}{\vSM}
\\[0.2cm]
& +  \Big( c_{\PW\partial \PW}\left(\PW^-_\nu D_\mu \PW^{+\mu\nu}+ \mathrm{h.c.}\right)+c_{\PZ\partial \PZ}\,  \PZ_\nu\partial_\mu \PZ^{\mu\nu}
+  c_{\PZ\partial \PGg}\, \PZ_\nu\partial_\mu A^{\mu\nu} \Big)\, \frac{\PH}{\vSM} 
\\[0.2cm]
& + \left(  \tilde c_{\PW\PW}\,   \PW_{\mu\nu}^+ \widetilde{\PW}^{-\mu\nu}  + \frac{\tilde c_{\PZ\PZ}}{2} \, \PZ_{\mu\nu}\widetilde{\PZ}^{\mu\nu} + 
\tilde c_{\PZ\PGg} \, \PZ_{\mu\nu} \widetilde{\PA}^{\mu\nu}   + \frac{\tilde c_{\PGg\PGg}}{2}\, 
\PA_{\mu\nu} \widetilde {\PA}^{\mu\nu} + \frac{\tilde c_{\Pg\Pg}}{2}\, 
G_{\mu\nu}^A\widetilde{G}^{A\mu\nu} \right) \frac{\PH}{\vSM}+ \dots 
\end{split}
\end{equation}
where $\vSM=1/\sqrt{\sqrt{2} G_F}$ and where $\ldots$ denote operators with more than a Higgs field or with more than four derivatives.  We have shown only terms
involving up to three bosonic fields. \refT{tab:RelationCouplings}
reports the relations between the couplings appearing in
Eq.~\refE{eq:chiralL} and the coefficients of the dimension-6
operators in~Eq.~\refE{eq:Lag_v2}.  It is worth noting that the same
Lagrangian (\ref{eq:chiralL}) applies identically to the case in which
the electroweak symmetry $SU(2)_L\times U(1)_Y$ is non-linearly
realized and $\PH$ is a generic scalar, singlet of the custodial
symmetry, not necessarily connected with the EW symmetry breaking.
Reference~\cite{Contino:2013kra} (and references therein) explains how
to write the Lagrangian~(\ref{eq:chiralL}) in a manifestly
$SU(2)_L\times U(1)_Y$ gauge invariant form. Assuming that the New Physics sector is invariant under custodial symmetry, several relations hold among the 22 free parameters $c_i$ appearing in the Lagrangian (169), as for instance $c_{\PW}=c_{\PZ}$, see \Bref{Contino:2013kra} for a detailed discussion. Notice that the operators proportional to $\tilde c_i$ are not invariant under CP (at the $p^2$ level, there are 5 such operators but one linear combination vanishes if the New Physics sector is invariant under the custodial symmetry).

The advantage of the chiral Lagrangian~(\ref{eq:chiralL}) is that it
does not assume that $\PH$ is part of an EW doublet and it can thus be
used to describe a more generic Higgs-like particle, like for instance
a dilaton, as well as a resonance that would correspond to a mixing of
a state belonging to an EW doublet with an EW singlet state.
Furthermore the linear Lagrangian~(\ref{eq:Lag_v2}) is based on a
triple expansion in the SM couplings, in powers of $E/M$ and also in
powers of $\Phi/\Lambda$, therefore it implicitly assumes that all the
Higgs coupling relative deviations are bounded by $\vSM^2/\Lambda^2$
while the chiral Lagrangian~(\ref{eq:chiralL}) allows one to explore
larger deviations. However, it is more complicated to calculate
higher order corrections using the chiral parameterization than with
the linear Lagrangian~(\ref{eq:Lag_v2}), see e.g.
\Bref{Contino:2013kra} for a detailed discussion.

%
%%%%%%%%%%%%%%%%%%%%%%%%%%%%%%%%%%%%%%%%%%%%%%%%%%%%%
\begin{table}
\caption{Relations between the linear and the non-linear parameterization of the Higgs couplings (adapted from \Bref{Contino:2013kra}).
The CP-violating couplings, $\tilde c_i$, are obtained by similar relations with the simple exchange $\Lcoeffp_k \to \Lcoeffp_{\tilde k}$ (notice that the Bianchi identities forbid any operator equivalent to $O_{V\partial V}$ in the CP-odd sector).
} 
\label{tab:RelationCouplings}
\begin{center}
\renewcommand{\arraystretch}{2.0}
\begin{tabular}{ll}
\hline
Higgs couplings & Linear realization \\ \hline
$c_{\PW}$ & $1 -\Lcoeffp_{\Phi}\,  \vSM^2/\Lambda^2$  \\
$c_{\PZ}$ & $1 - (\Lcoeffp_{\Phi} + 4 \Lcoeffp_{\mathrm T})\,  \vSM^2/\Lambda^2 $  \\
$c_{\Pf} \ \left(\Pf=\PQu,\PQd,\Pl\right)$ & $1- (\Lcoeffp_{\Phi} +  \Lcoeffp_{\Pf})\,  \vSM^2/\Lambda^2$ \\
$c_3$ & $1+ (2 \Lcoeffp_{6} - 3  \Lcoeffp_{\Phi})\,  \vSM^2/\Lambda^2$  \\
$c_{\Pg\Pg}$ & $8\, (\alpha_s/\alpha_2) \,\Lcoeffp_{\Pg} \, \MW^2/\Lambda^2$  \\
$c_{\PGg\PGg}$ & $8\sin^2\!\theta_{\PW} \,\Lcoeffp_{\PGg}\, \MW^2/\Lambda^2$ \\
$c_{\PZ\PGg}$ & $\displaystyle \left(\Lcoeffp_{\Phi \PB}-\Lcoeffp_{\Phi \PW} - 8\, \Lcoeffp_{\PGg} \sin^2\!\theta_{\PW}\right) \tan\theta_{\PW} \, \MW^2/\Lambda^2$  \\
$c_{\PW\PW}$ & $-2\, \Lcoeffp_{\Phi \PW} \, \MW^2/\Lambda^2$ \\
$c_{\PZ\PZ}$ & $-2\left( \Lcoeffp_{\Phi \PW} +   \Lcoeffp_{\Phi \PB} \tan^2\!\theta_{\PW} - 4  \Lcoeffp_\PGg\tan^2\!\theta_{\PW} \sin^2\!\theta_{\PW} \right) \MW^2/\Lambda^2$ 
 \\
$c_{\PW\partial \PW}$ & $\displaystyle -2(2 \Lcoeffp_{\PW}+\Lcoeffp_{\Phi \PW})\, \MW^2/\Lambda^2$ \\
$c_{\PZ\partial \PZ}$ & $\displaystyle - 2(2 \Lcoeffp_{\PW}+\Lcoeffp_{\Phi \PW}) - 2\left(2\Lcoeffp_{\PB}+\Lcoeffp_{\Phi \PB}\right) \tan^2\!\theta_{\PW}\, \MW^2/\Lambda^2$  \\
$c_{\PZ\partial \PGg}$ & $\displaystyle 2\left(2  \Lcoeffp_{\PB}+\Lcoeffp_{\Phi \PB} -2
\Lcoeffp_{\PW} - \Lcoeffp_{\Phi \PW}\right) \tan\theta_{\PW}\, \MW^2/\Lambda^2$ 
\\[0.5cm]
\hline
\end{tabular}
\end{center}
\end{table}

\clearpage

\newpage
\section{Higgs properties: spin/CP\footnote{
    A.~David, A.~Denner, M.~D\"uhrssen, M.~Grazzini, C.~Grojean, 
    K.~Prokofiev, G.~Weiglein, M.~Zanetti (eds.); S.~Bolognesi, S.Y.~Choi, P.~de~Aquino, Y.Y~Gao, A.V.~Gritsan, K.~Mawatari, K.~Melnikov, D.J.~Miller, M.M.~M\"uhlleitner, M.~Schulze, M.~Spira, N.V.~Tran, A.~Whitbeck, P.~Zerwas}}

\label{sec:LM_cp}

\subsection{Introduction}
\label{sec:LM_cp_intro}

Since a clear signal for a resonance consistent with
the long sought Higgs boson
has been established \cite{Aad:2012tfa,Chatrchyan:2012ufa},
the next step is a detailed study of its properties.
In this Section we focus on the spin/CP properties of the new resonance, and review the strategies
to determine whether the Higgs-like boson is consistent with the spin zero particle predicted in the SM,
with $J^{CP}=0^{+}$, and the extent to which it could be a mixture of different CP eigenstates.
We recall that, as discussed in \refS{sec:LM_eft}, going beyond the interim framework for coupling studies
presented in \refS{sec:LM_ir:framework} will require the analysis of couplings and
spin/CP properties to be treated in a common framework. Nonetheless, to the purpose of presentation, we discuss here
the spin/CP properties in a separate section.

The observation of the new resonance in the decay modes
$\Hgaga$ \cite{ATLAS-CONF-2013-012,CMS-PAS-HIG-13-001},
$\HZZ$   \cite{ATLAS-CONF-2013-013,CMS-PAS-HIG-13-002} and
$\HWW$   \cite{ATLAS-CONF-2013-030,CMS-PAS-HIG-13-003} allows multiple independent tests of the spin/CP properties.
Thanks to the Landau-Yang theorem \cite{Landau:1948kw,Yang:1950rg},
the observation in the $\Hgaga$ mode already rules out the possibility that the new resonance
has spin 1, and, barring $C$ violating effects in the Higgs sector, fixes $C=+1$.

Having ruled out the $J=1$ possibility\footnote{Note that there are two caveats to this argument. The first is that the Landau-Yang theorem strictly applies to an on-shell resonance. This means that the $J=1$ hypothesis can be excluded only by making an additional small-width assumption. The second is that in principle the decay product could consist of 2 pairs of boosted photons each misinterpreted as a single photon.}, the case $J=2$ is of course the first one that should be tested.
This possibility turns out to be extremely challenging from a theoretical view point.
The naive coupling of a massive spin-2 field with a $U(1)$ gauge field leads to the Velo-Zwanziger problem \cite{Velo:1969bt,Velo:1970ur},
and the model develops modes that travel superluminally and other pathological features.
Detailed studies have shown that
such models have an intrinsic cut off of the order of the mass of the spin-2 resonance \cite{Porrati:2008gv}.
A consistent effective description could be obtained by interpreting the spin 2 particle as a Kaluza Klein (KK)
graviton. However, one should then explain why analogous KK excitations of the SM gauge bosons have never been observed.
Moreover, a recent study  \cite{Ellis:2012mj} has shown that a graviton-like massive spin 2 boson would have
too small couplings to $\PW\PW$ and $\PZ\PZ$ with respect to $\PGg\PGg$, and that in many models with a compactified extra dimension the massive spin 2 boson would have equal coupling to $\Pg\Pg$ and $\PGg\PGg$, thus leading to $\Gamma(\PH\to \Pg\Pg)\sim 8\Gamma(\Hgaga)$, which seems to be in contradiction to the observed data.

The strategies to determine the properties of a resonance through its decays to gauge bosons
date back to more than 50 years ago. Photon polarization in $\PGpz\to\PGg\PGg$ decay can be used to measure the pion parity \cite{Yang:1950rg} but it turns out to be easier to use the orientation of the decay planes in the $\PGpz\to\Pep\Pem\Pep\Pem$ \cite{Plano:1959zz} decay. Analogously, the $\PH\to \PZ\PZ\to 4\Pl$ decay mode, allowing full control on the event kinematics, is an excellent channel to study spin, parity and tensor structure of the coupling of the Higgs-like resonance.
As discussed in detail in \refS{sec:LM_cp_maggie}, the invariant mass
distribution of the off shell gauge boson in $\PH\to \PV\PV^*$ is
proportional to the velocity
$d\Gamma/dM_*\sim\beta\sim \sqrt{(\mH-m_{\PV})^2-M_*^2}$ (where $M_*$
denotes the invariant mass of the off-shell gauge boson) and therefore
features a characteristic steep behavior with $M_*$ just below the
kinematical threshold. This behavior is related to the spin-zero nature
of the SM Higgs boson and will rule out other spin assignments with the
exception of the $1^+$ and $2^-$ cases, which can be ruled out through
angular correlations. The pseudoscalar case $0^-$ can 
be instead discriminated against the SM $0^+$ by studying the distribution in the azimuthal angle $\phi$ between the two $Z$ decay planes \cite{Dell'Aquila:1985ve,Choi:2002jk}.
%%%%%
It should be noted, however, that in many models of physics beyond the
SM there is no lowest-order coupling between a pseudoscalar and a pair
of gauge bosons, so that the decay $\PA \to \PZ\PZ \to 4\Pl$ can be heavily suppressed compared to the decay $\PH \to \PZ\PZ \to 4\Pl$. For a state that is an admixture
between CP-even and CP-odd components the decay into $\PZ\PZ$ essentially
projects to the CP-even part in such a case, so that the angular
distributions would show a pure CP-even pattern although the state in
fact has a sizable CP-odd component. See section \refS{sec:LM_pseudo} for a more
detailed discussion of this issue.

More generally, instead of relying on specific kinematical variables, one can try to exploit the full
information on the event. The {\em matrix element method} uses the tree level amplitude to construct a likelihood to isolate a signal over a background, or to discriminate between two different signal hypothesis.
The construction of the matrix element can be carried out by using two different strategies: the {\em effective Lagrangian} (see \refS{sec:LM_eft} and \refS{sec:LM_mad}) and {\em anomalous couplings} (see \refS{sec:LM_jhu}: 'generic parameterizations') approaches. The former implies to write the most general effective Lagrangian compatible with Lorentz and gauge invariance. The latter implies to write the most general {\em amplitude} compatible with Lorentz and gauge invariance, but does not assume a hierarchy in the scales, and thus the couplings become momentum dependent form factors. The effective Lagrangian approach has the advantage that it can be extended beyond LO.
The anomalous coupling approach is restricted to LO but somewhat more general, since it is valid also in the case in which new light degrees of freedom are present and circulate in the loops. The effective Lagrangian approach is being pursued by the {\sc MadGraph} team (see \refS{sec:LM_mad}). The anomalous coupling approach has been used to perform studies on the spin/CP properties of the Higgs boson \cite{DeRujula:2010ys}, and its most widely used implementation is so called MELA approach\footnote{ATLAS uses also the BDT method.} \cite{Gao:2010qx}, described in \refS{sec:LM_jhu} (see also MEKD~\cite{Avery:2012um}).

Here we note that the matrix element method is {\em maximally} model dependent, since it allows to exclude various specific models one by one. An issue which is important to understand is the extent to which
the discrimination of a given spin and CP hypothesis depends on the
production model assumed. The results for the $\Hgaga$ channel
recently presented by the ATLAS
collaboration \cite{ATLAS-CONF-2013-029} show that the spin-2 hypothesis can be discriminated only by assuming a $\Pg\Pg$ fusion production mode.
This is somewhat related to the fact that the $\Hgaga$ channel offers essentially
only one angular variable, the polar angle $\theta^*$ of the photons
in the Higgs rest frame. The situation is different in the $\HWW$
decay mode, where a discrimination against the $2^+$ hypothesis is
possible \cite{ATLAS-CONF-2013-031}, although, thanks to spin correlations, the discriminating power is maximum if the Higgs is produced in the $\Pq\Pqb$ channel.
As discussed above, it is the $\PH\to \PZ\PZ\to 4\Pl$ channel that
offers the maximum amount of information. Here, the dependence on the
production model is present but the experimental
results \cite{ATLAS-CONF-2013-013} show that the discrimination power is
essentially independent on the production model. This is consistent with
what is shown in \refS{sec:LM_jhu}, and is due to the fact that, as observed in \Bref{DeRujula:2010ys}, the selection cuts {\em sculpt} the angular distributions making the dependence on the production model rather marginal.

Another channel that can be used to test the CP structure of the $\PH\PV\PV$ vertex is the $\PH\PV$ associated production
of the Higgs boson with a vector boson ($\PV=\PWpm,\PZ$).
In \Bref{Ellis:2012xd} it has been noted that the invariant mass distribution of the $\PV\PH$ system would be very different in the $0^+$, $0^-$ and $2^+$ hypotheses, thus providing a fast track indicator on Higgs spin and CP properties.
We point out that these differences in the invariant mass distribution, together with analogous differences in
other kinematical distributions \cite{Ellis:2013ywa,Djouadi:2013yb}, are related to the fact that such spin/CP assignments lead to interactions with the vector bosons that are mediated by higher dimensional operators.

The Higgs CP properties and the structure of the $\PH\PV\PV$ vertex can also be studied in vector boson fusion (VBF),
by looking at the azimuthal separation of the two tagging jets \cite{Plehn:2001nj}.
Recent studies on the determination of Higgs Spin/CP properties in VBF are presented in \Brefs{Frank:2012wh,Englert:2012xt,Djouadi:2013yb}. When more data will be available, the VBF channel will nicely complement the information obtained in the inclusive Higgs production modes.

The experimental analyses of the CP properties have so far mainly
focused on discriminating between the distinct hypotheses of a pure
CP-even and a pure CP-odd state. First studies towards dealing with an
admixture of CP-even and CP-odd components have recently been presented
by the CMS collaboration \cite{CMS-PAS-HIG-13-002}. As mentioned above, however, angular
distributions in $\PH \to \PZ\PZ$ and $\PH \to \PW\PW$ decays as well as invariant
mass distributions and azimuthal distributions in $\PV\PH$ and VBF production will have a limited
sensitivity for discriminating between a pure CP-even state and a mixed
state if the coupling of the CP-odd component to $\PV\PV$ is suppressed
compared to the $\PH\PV\PV$ coupling. The couplings of the Higgs boson to
fermions offer a more democratic test of its CP nature,
since in this case the CP even and odd components can have the same magnitude.
In this respect, if the $\PH\PQt\PAQt$ channel can be exploited sufficiently
well this would offer a good opportunity
to study Higgs CP properties \cite{Gunion:1996xu,Gunion:1998hm,Field:2002gt}.

The remainder of this Section is organized as follows. In \refS{sec:LM_pseudo} an overview is given
about the coupling of a pseudoscalar to gauge bosons in different models
of physics beyond the SM. In \refS{sec:LM_cp_maggie} the theoretical basis for spin
and parity studies at the LHC is reviewed. In \refS{sec:LM_jhu} the matrix
element approach based on the {\sf JHU} generator is briefly reviewed. In \refS{sec:LM_mad} the effective Lagrangian approach implemented by the {\sc MadGraph} group is presented, together with a comparison with {\sc JHU} results.

%We should not forget, however, that these studies all share a potential theoretical limitation.
%The data tell us that the Higgs-like resonance has substantial rates in vector boson pairs ($\PGg\PGg$, $\PZ\PZ$ and $\PW\PW$). This implies that this particle must have a significant CP even component, since the couplings of a pseudoscalar
%to vector boson pairs are loop induced. On the other hand, for the same reason, it will be difficult to rule out the existence of a small CP odd component. First studies in this direction have been recently presented by the CMS collaboration \cite{CMS13002}.
%The couplings of the Higgs boson to fermions offer a more democratic test of its CP nature, since in this case the CP even and odd components can have the same magnitude. In this respect, the $\PH\PQt\PAQt$ channel will offer a nice opportunity to study Higgs CP properties \cite{Gunion:1996xu,Gunion:1998hm,Field:2002gt}, once enough integrated luminosity will be accumulated. In \refS{sec:LM_pseudo} a discussion of possible BSM scenarios in which a pseudoscalar Higgs boson couples to gauge bosons is presented.

%The remainder of this section is organized as follows.
%In \refS{sec:LM_cp_maggie} the theoretical basis for spin and parity studies
%at the LHC is reviewed. In \refS{sec:LM_jhu} the matrix element approach based on the JHU generator is briefly discussed.
%In \refS{sec:LM_pseudo} a discussion of possible BSM scenarios in which a pseudoscalar Higgs boson couples to gauge bosons is presented.

\subsection{Pseudoscalar couplings to gauge bosons}
\label{sec:LM_pseudo}

As discussed above, the present analyses of the CP properties of the new 
state are based in particular on the investigation of angular distributions of 
the decays $\PH \to \PZ\PZ \to 4\Pl$ and $\PH \to \PW\PW \to 4\Pl$, of
the azimuthal separation of the two tagging jets in VBF and of the
invariant mass distributions of the WH and ZH production processes. 
It should be noted that all the above processes involve the coupling 
of the new state to two gauge bosons, $\PH\PV\PV$, where $\PV = \PW, \PZ$ 
(this coupling also plays an important role for the processes 
$\PH \to \PGg\PGg$ and $\PH \to \PZ\PGg$ via the $\PW$~loop
contribution). 

The angular and kinematic distributions in these processes will only
provide sensitivity for a discrimination between CP-even and CP-odd
properties if a possible CP-odd component A of the new state couples
with sufficient strength to $\PW\PW$ and $\PZ\PZ$. If however the $\PA\PV\PV$ coupling is 
heavily suppressed compared to the coupling of the CP-even component, 
the difference between a pure CP-even state and a state that is a
mixture of CP-even and CP-odd components would merely manifest itself as
a modification of the total rate (which could at least in principle also
be caused by other effects).
The angular distributions in this case, on the
other hand, would show
no deviations from the expectations for a pure CP-even state, even if
the state had a sizable CP-odd component.

The extent to which the above analyses will be able to reveal effects of
a CP-odd component of the new state therefore crucially depends on 
the coupling strength AVV in comparison to the coupling of the CP-even
component to gauge bosons. In the following we will briefly discuss this
issue for different kinds of models of physics beyond the SM.

While the coupling of a CP-even scalar to VV is present in lowest order in
renormalizable gauge theories, no such coupling exists for a CP-odd
scalar. In general pseudoscalar states A couple to gauge bosons
%($V=\PW,\PZ,\PGg,\Pg$) 
via dimension 5 operators,
\begin{equation}
{\mathcal L}_{\mathrm{AVV}} = \frac{c_{\mathrm V}}{\Lambda}~A~V^{a\mu\nu} \tilde V^a_{\mu\nu}
\label{eq:pseudoscalar}
\end{equation}
with $\tilde V^a_{\mu\nu} =
\frac{1}{2}\epsilon_{\mu\nu\rho\sigma}V^{a\rho\sigma}$ denoting the dual
field strength tensor (the same structure also holds for couplings to
photons and gluons). In the effective operator given in
Eq.~(\ref{eq:pseudoscalar}) $c_{\mathrm V}$ denotes the coupling strength emerging
from the theory at the scale $\Lambda$.  These couplings can either
arise via loop effects in renormalizable weakly-interacting models as
e.g.~extensions of the Higgs sector or supersymmetric models, or they
can occur in non-perturbative models as e.g.~expanded versions of
technicolor~\cite{Weinberg:1975gm,Weinberg:1979bn,Susskind:1978ms} with
a typical cut-off scale $\Lambda$.

\subsubsection{ Weakly interacting models}
%            =========================

In weakly interacting models the pseudoscalar coupling to gauge bosons
is mediated by loop effects (dominantly top loops in many models) such
that the coupling $c_{\mathrm V}$ of Eq.~(\ref{eq:pseudoscalar}) is related to the
contributing Yukawa couplings of the underlying model and $\Lambda$ to
the masses of the fermions involved in the corresponding contribution to
the chiral anomaly.

This loop-induced coupling turns out to be heavily suppressed 
in several classes of BSM models. In particular, in the minimal
supersymmetric extension of the SM (MSSM) the production times decay
rates for purely pseudoscalar Higgs boson states are typically
suppressed by three orders of magnitude or more (see e.g.\ the analysis in
\cite{Bernreuther:2010uw} and the case of a CP-violating scenario
discussed in \cite{Figy:2010ct}). Within the MSSM, the detection of any
pseudoscalar component of a mixed scalar-pseudoscalar 
via the above analyses at the LHC therefore does not look feasible.
The same holds for the pseudo-axion states in Little Higgs models
\cite{ArkaniHamed:2001nc,ArkaniHamed:2002pa,ArkaniHamed:2002qy}.

The situation may be somewhat better in other extensions of the SM.
In \cite{Bernreuther:2010uw} a general type II 2HDM was investigated.
Based on a scan over the relevant parameters it was found 
that rates for pseudoscalar
production and decay can reach a detectable level at the LHC for small
values of $\tan\beta$ (i.e.\ $\tan\beta < 1$), 
where the pseudoscalar rates can get close to or even exceed 
the rates for a SM Higgs in some cases. In regions of
significant and observable rates at the LHC the pseudoscalar Higgs boson
is predominantly produced via gluon fusion $gg\to A$ which is enhanced
by the larger top quark contribution than for the SM Higgs boson. The
loop-induced decays into gauge bosons also involve the relatively large
coupling of the top quark to the pseudoscalar for
small values of $\tan\beta$. The main difference to the MSSM is that
small values of $\tan\beta$ are still allowed within the type II 2HDM,
while the MSSM is strongly constrained by the direct MSSM Higgs searches
so that such small values of $\tan\beta$ are excluded \cite{Schael:2006cr}.

\subsubsection{Strongly interacting models}
%           ===========================

Turning now to strongly interacting models involving non-perturbative
effects, effective couplings of pseudoscalar states to gauge bosons
naturally arise in technicolor models
\cite{Weinberg:1975gm,Weinberg:1979bn,Susskind:1978ms}, where the
pseudoscalar pseudo-Nambu-Goldstone bosons (PNGB) couple to the
respective chiral anomalies and the associated axial vector currents
\cite{Dimopoulos:1980yf,Ellis:1980hz,Chivukula:1995dt}. Moreover, these
types of couplings can be generated by instanton effects in the
framework of these models. The cut-off scale $\Lambda$ of the novel
underlying strong interactions is related to the corresponding
pseudoscalar decay constant $F_{\mathrm A}$ defined by the associated PNGB
couplings to the axial vector currents via PCAC,
\begin{equation}
\langle 0|j^\mu_5|A(p)\rangle = i F_{\mathrm A} p^\mu
\end{equation}
where $j^\mu_5$ denotes the axial vector current emerging from the novel
techni-fermions and $p^\mu$ the four-momentum of the pseudoscalar field
$A$. The coupling of the PNGB $A$ to gauge bosons can be derived from the
corresponding coupling to the divergence of the axial vector current,
\begin{equation}
{\mathcal L}_{\mathrm A} = \frac{A}{F_{\mathrm A}}\partial_\mu j^\mu_5
\end{equation}
Even for massless techni-fermions the divergence of the axial vector
current develops an anomalous contribution, the
Adler--Bell--Jackiw-anomaly (ABJ-anomaly) \cite{Adler:1969gk,Bell:1969ts},
\begin{equation}
\partial_\mu j^\mu_5 = -S_{\mathrm V}~\frac{g_{\mathrm V}^2}{16\pi^2}~V^{a\mu\nu} \tilde
V^a_{\mu\nu}
\end{equation}
where $S_{\mathrm V}$ is the associated anomaly coefficient of the
corresponding gauge group and $g_{\mathrm V}$ its gauge coupling to the
techni-fermions. Finally one obtains the effective PNGB couplings to
gauge bosons \cite{Dimopoulos:1980yf,Ellis:1980hz,Chivukula:1995dt}
\begin{equation}
{\mathcal L}_{\mathrm{AVV}} = -S_{\mathrm V}~\frac{g_{\mathrm V}^2}{16\pi^2 F_{\mathrm A}}~A~V^{a\mu\nu} \tilde
V^a_{\mu\nu}
\end{equation}
Depending on the size of the anomaly coefficient $S_{\mathrm V}$ and the PNGB
decay constant $F_{\mathrm A}$ these effective couplings can imply production and
decay processes of these pseudoscalar states at the LHC which can reach 
similar orders of magnitude as for SM Higgs bosons. For CP-non-conserving 
technicolor models in general large admixtures of
scalar and pseudoscalar components are possible which could lead to sizeable
deviations of e.g.~the angular distributions in final states with 4
charged leptons.

A rigorous analysis of PNGB production and decay at the LHC has been
performed in the framework of top-color assisted technicolor models
\cite{Hill:1994hp}.  These models introduce two separate strongly
interacting sectors in order to explain electroweak symmetry breaking
(EWSB) and the large top quark mass at the same time. Techni-fermion
condensates generate most of the EWSB but contribute only little to the
top mass. The latter is induced by the condensation of top-antitop pairs
which eventually generates a large Yukawa coupling of the techni-pion to
top quarks. The top condensate makes only a small contribution to EWSB.
Within this class of models pseudoscalar rates of the order of the SM
Higgs rates can be reached in a large part of the available parameter
space \cite{Bernreuther:2010uw}. In contrast to the top quark case the
much smaller Yukawa coupling of the pseudoscalar top-pion state to
bottom quarks is induced by extended technicolor as well as top-color
instanton effects. As before the pseudoscalar couplings to gauge bosons
are generated by the ABJ-anomaly. At the LHC a discrimination between
scalar and pseudoscalar components at the level of ${\mathcal O}(10\%)$
could help to put constraints on possible mixing between scalar and
pseudoscalar fields within these models.

\subsection{Basis of Higgs Spin/Parity measurements at the LHC}
\label{sec:LM_cp_maggie}

\newcommand{\s}{\\ \vspace*{-3.5mm}}
\renewcommand{\s}{\relax}
\newcommand{\imag}{\Im {\rm m}}
\newcommand{\real}{\Re {\rm e}}
\newcommand{\rb}[2]{\raisebox{#1}[-#1]{#2}}
\newcommand{\sts}{\scriptstyle}

\def\tablename{\bf Table}%
\def\figurename{\bf Figure}%

\renewcommand{\sfdefault}{phv}
\renewcommand{\baselinestretch}{1.2}

\subsubsection{Projection}

%\noindent
%The Higgs boson \cite{Englert:1964et,Higgs:1964pj,Guralnik:1964eu} in
%the Standard Model (SM)
%\cite{Glashow:1961tr,Weinberg:1967tq,Salam:1968rm} is assigned
%the spin, parity and charge-conjugation quantum numbers
%
%\begin{equation}
%   {\rm{SM}} \;\; {\rm{Higgs}} \;\; : \;\; J^{PC} = 0^{++}
%\end{equation}
%
%associated with $C$, $P$ and $CP$ conserving interactions with electroweak
%gauge bosons, photons and fermions \cite{Jkg}. In extended models beyond the SM, 
%as familiar from supersymmetric theories, the Higgs system may be expanded by $CP$-odd 
%scalar [pseudoscalar] particles \cite{HHG,Carena:2002es,Djouadi:2005gj,Accomando:2006ga},
%eventually including $CP$-violating even/odd coherent mixtures.\s

%Charge conjugation $C = +$ of Higgs bosons is assured by observing $\PGg\PGg$
%decays \cite{Aad:2012tfa,Chatrchyan:2012ufa,Chatrchyan:2013lba} for pure states 
%in a $C$-invariant Higgs sector.\s

The spin/parity quantum numbers of the Higgs-like resonance can be analyzed systematically in decay
processes and production channels by studying the associated helicity
amplitudes\footnote{We point out that
the methods we describe here are based on leading-order expressions. Nonetheless, we expect that NLO corrections will not change the picture dramatically.}. Measurements allow the determination of necessary and sufficient
conditions for assigning the $J^{CP}$ quantum numbers of pure states but also
reveal the nature of mixed $CP$-even/odd states. Such analyses were
comprehensively performed for $\PZ^\ast \PZ$ decays followed by
leptonic $\PZ^\ast$ and $\PZ$ decays in
\Brefs{Choi:2002jk,Gao:2010qx,DeRujula:2010ys,Bolognesi:2012mm,Chen:2012jy}, 
$\PGg\PGg$ decays in
\Brefs{Ellis:2012wg,Ellis:2012jv,Alves:2012fb,Choi:2012yg}, and $CP$-violating 
decays in
\Brefs{Soni:1993jc,Chang:1993jy,Godbole:2007cn,Nelson:1989uq,Grzadkowski:1995rx}, 
including appropriate production channels of gluon and vector boson 
fusion \cite{Plehn:2001nj,Hagiwara:2009wt}, and
Higgs-strahlung/associated Higgs-vector production in
\Brefs{Barger:1993wt,Miller:2001bi, Ellis:2013ywa}. The flow of
helicities in a representative amplitude for production and
decay of the general state $\PH^J_m$ can be written as
\begin{equation}
 \Pg_\mu \Pg_{\mu'} \to \PH^J_m \to \PZ^\ast_\lambda \PZ_{\lambda'} \,\;:\,\;
   \langle \PZ^\ast_\lambda \PZ_{\lambda'}|\PH^J_m|\Pg_\mu \Pg_{\mu'}\rangle
   = {\mathcal T}_{\lambda\lambda'}\,\,
       d^J_{m,\lambda-\lambda'}(\Theta) \,
     \delta_{m,\mu - \mu'}\,\, {\mathcal G}_{\mu\mu'} \,
\label{eq:process_gg_H_ZZ}
\end{equation}
with $d(\Theta)$ denoting the Wigner rotation functions \cite{Ros}.
${\mathcal G}$ and ${\mathcal T}$ are the reduced helicity amplitudes for the
production and the decay processes, $gg(VV) \,\to\, H^J$ and
$\PH^J\,\to\, \PZ^*\PZ$, respectively.
All the relevant angles are mapped in \refFs{fig:kine}(a) and (b) in 
the rest frame of the Higgs boson.\s

Angular distributions in unpolarized particle decays to $\PZ^\ast \PZ$ vector pairs
allow the measurement of four independent helicity amplitudes.
Combined with the analysis of threshold
excitation in $\PH^J\to \PZ^\ast \PZ$, this set is sufficient to determine
the spin up to $J = 2$ and the parity. In addition to strongly constrained
angular correlations, scalar Higgs decays exhibit only phase-space
suppression proportional to the velocity $\beta$. In contrast, higher
spins can be probed, and eventually excluded, by studying threshold effects
in $\PH^J \to \PZ^\ast \PZ$, predicted for $J > 2$ to behave at least as
$\beta^{2J-3}$ and carrying the power $ \geq 3$. Alternatively the angular distribution
in the joint process of production plus decay can be exploited.
The final-state axis in $\Pg\Pg \to \PH^J \to \PZ^\ast \PZ,\PGg\PGg$ is distributed
isotropically for spin = 0 but non-trivially modified by terms up to
$\cos^{2J}\Theta$ for any higher spin $J$.\s

Some aspects have been studied already in earlier reports
\cite{Dell'Aquila:1985ve,Nelson:1986ki,Hagiwara:1993sw,Hagiwara:2000tk,
Grzadkowski:2000hm,Han:2000mi,Kramer:1993jn}. A large
number of reports on Higgs properties has been published in
particular recently in response to the discovery of the ``$125\UGeV$
Higgs boson''; a partial list of references is recorded in
\Brefs{Skjold:1993jd,Arens:1994wd,Odagiri:2002nd,Buszello:2002uu,Buszello:2006hf,
Bredenstein:2006rh,Keung:2008ve,Bhupal:2007is,Christensen:2010pf,Englert:2010ud,
DeSanctis:2011yc,Godbole:2011hw,Boughezal:2012tz,Stolarski:2012ps,Englert:2012ct,
Ellis:2012jv,Freitas:2012kw}, including also analyses like $\PW\PW$ decay channels which 
are theoretically closely related to the channels described here.\s

\begin{figure}[htb]
\begin{center}
\includegraphics[scale=0.75]{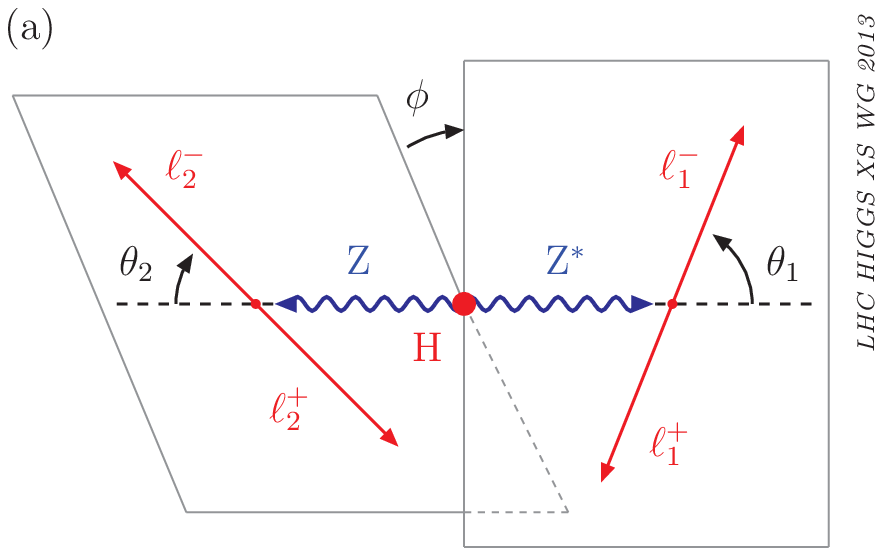}
\hskip 1cm
\includegraphics[scale=0.75]{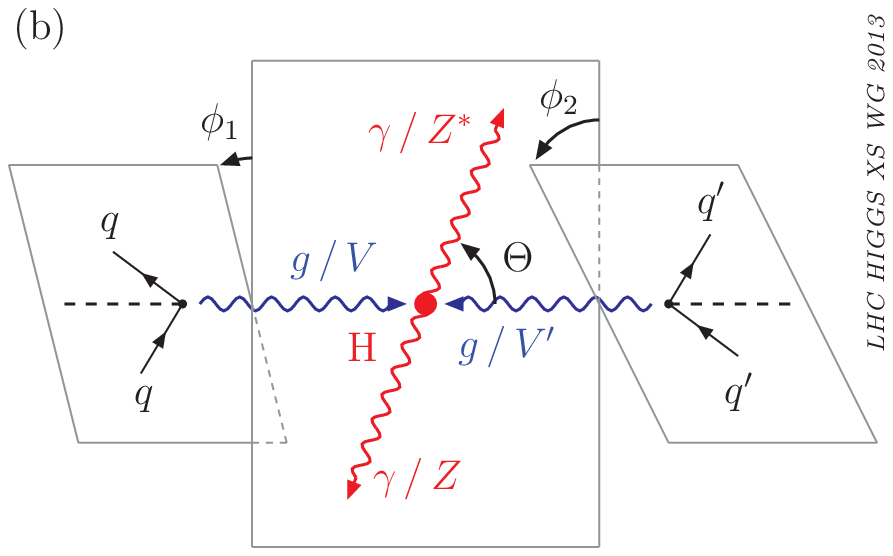}
\end{center}
\caption{ (a) Kinematics of the decay $\PH^J \to \PZ^\ast \PZ \to
  (\ell^-_1\ell^+_1)(\ell^-_2\ell^+_2)$
  in the rest frame of the Higgs boson. The angles $\theta_{1,2}$
  denote the polar angles of the leptons $\ell^-_{1,2}$ in the rest frame of the
  virtual $\PZ^\ast$ and real $\PZ$ bosons;
  (b) Higgs production in gluon
  collisions and subsequent $\PGg\PGg$ and
  $\PZ^\ast \PZ$ decays, also in the Higgs rest frame. If the gluons are replaced by
  electroweak gauge bosons,
  $\phi=\phi_1-\phi_2 \, (\mbox{mod } 2\pi) \in [0,2\pi)$ corresponds
    to the azimuthal angle between the two initial $q\,V$ and $q'V'$
    emission planes \cite{Hagiwara:2009wt}.
}
\label{fig:kine}
\end{figure}

\subsubsection{Higgs Decay into Virtual/Real \boldmath{$\PZ$} Bosons}

\noindent
Denoting the polar angles of the leptons $\ell^-_{1,2}$ in the rest frame
of the virtual $\PZ^\ast$ and real $\PZ$ bosons by $\theta_{1,2}$ (see Fig.$\,$\ref{fig:kine}(a)),
the forward-backward symmetric differential
decay distributions of the polar angles for pure-spin/parity unpolarized boson states
$\PH^J$ decaying into $\PZ^\ast_\lambda \PZ_{\lambda'}$ final states can be expressed in
terms of four independent helicity amplitudes \cite{Choi:2002jk}:
\begin{eqnarray}
    \frac{1}{\Gamma}\,\frac{d\Gamma}{d\cos{\theta_1} d\cos{\theta_2}}
&=& {\mathcal N}^{-1} \,
    \bigg[\, \sin^2{\theta_1}\sin^2{\theta_2}\, |{\mathcal T}_{00}|^2
          +\frac{1}{2}(1+\cos^2{\theta_1})(1+\cos^2{\theta_2})
         [|{\mathcal T}_{11}|^2+|{\mathcal T}_{1,-1}|^2 ] \nonumber\\
    && \hskip 1.1cm
       +\,(1+\cos^2{\theta_1}) \sin^2{\theta_2}\,|{\mathcal T}_{10}|^2
       +\,\sin^2\theta_1 (1+\cos^2{\theta_2}) \,|{\mathcal T}_{01}|^2
      \,\bigg]
\label{eq:ddhel1}
\end{eqnarray}
for fixed $M^2_*$ and suppressing the quartic terms involving the
$P$-violating parameters, $\eta_{1,2}$, which are very small
$\sim 0.15$ for leptonic $\PZ$ decays, see
also \Bref{Modak:2013sb}.
The distribution is normalized to unity by the
coefficient ${\mathcal N}$.
Other helicity amplitudes are related by parity and Bose symmetry of the state:
${\mathcal T}_{\lambda\lambda'} = n_{\PH} {\mathcal T}_{-\lambda,-\lambda'}$ and
${\mathcal T}_{\lambda\lambda'} [\PZ,\PZ^\ast] = (-1)^J {\mathcal T}_{\lambda'\lambda}                                                                                                       [\PZ^\ast,\PZ]$, respectively, the normality given by $n_{\PH} = P (-1)^J$.
The amplitude ${\mathcal T}_{00}$ vanishes specifically for negative-parity states.
The corresponding azimuthal distribution of the $\PZ$-decay planes
can be expressed by helicity amplitudes as
\begin{eqnarray}
  \frac{1}{\Gamma}\, \frac{d\Gamma}{d\phi}
  &=& \frac{1}{2\pi} \,
      \bigg[  \,
      1
    + n_{\PH}\,|\zeta_1| \cos 2\phi \,
      \bigg]
      \quad {\rm with} \quad
      |\zeta_1| = |{\mathcal T}_{11}|^2 /  \left[ 2 \sum
      |{\mathcal T}_{\lambda \lambda'}|^2 \right]
\label{eq:ddhel2}
\end{eqnarray}
suppressing the small $P$-violating $\eta_{1,2}$-dependent
parts again. (The full expressions of the polar and
azimuthal distributions, (\ref{eq:ddhel1}) and (\ref{eq:ddhel2}), can be
found in \Bref{Choi:2002jk}.)
The sign of the $\phi$ modulation is uniquely determined by the normality
of the Higgs state. The characteristic behavior of the azimuthal angle between
the two $\PZ$ decay planes is illustrated in \refF{fig:angl}(a) for spin-zero
of positive (SM) and negative parity. Distributions of positive and negative
parity decays are mutually anti-cyclic. This will also be observed in jet-jet
correlations of electroweak-boson and gluon 
fusion \cite{Plehn:2001nj,Hagiwara:2009wt,Ellis:2012xd,Englert:2012xt,Frank:2012wh,
Djouadi:2013yb,Godbole:2013saa}.\s

\begin{figure} [htb]
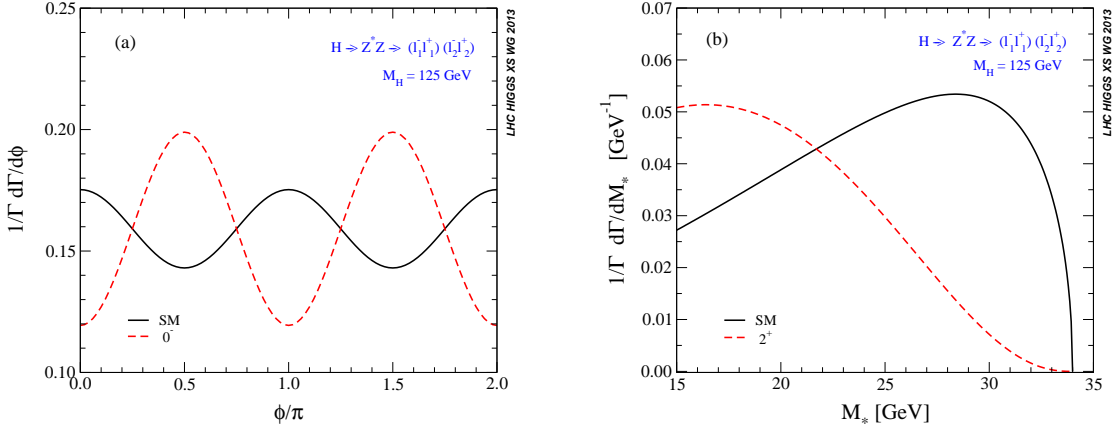

\begin{center}
\vskip 0.5cm
\includegraphics[scale=0.33]{YRHXS3_LM/Maggie/azimuthal_ZZ.eps}
\hskip 1cm
\includegraphics[scale=0.33]{YRHXS3_LM/Maggie/dmx_ZZ.eps}
\end{center}
\caption{(a) Oscillations of the azimuthal angle between the two $\PZ$
    decay planes for spin = 0 with positive parity in the SM
    compared with negative parity;
    (b) Threshold behavior of the decay width $\PH^J \to \PZ^\ast \PZ$ for the
    SM and spin-2 ($c_i=0$ except $c_2=1/\MH^2$, $c_i$ defined in \Tref{tab:ff}) even
    normality bosons, with a Higgs boson mass of $125\UGeV$.}
\label{fig:angl}
\end{figure}

The spin averaged distribution applies to all
configurations in which the orientation of the $\PZ^\ast \PZ$ event axis in
the rest frame is summed over so that the decay state is effectively
unpolarized.\s

The functional form of the angular correlations among the $\PZ$ decay products
is not specific to the spin of the decaying boson for $J \geq 2$. These spins
cannot be discriminated anymore by such generic analyses. However, angular
correlations supplemented by threshold suppression can resolve the ambiguities.
After solving for the case $J = 2$ specifically, the analysis is quite general and
transparent for $J > 2$ since the tensor structure of these decay amplitudes
leads to a characteristic signature of the spin. The $(J+2)$-tensor structure
enforces the amplitude ${\mathcal T}$ for $J > 2$ to rise
at least $\sim \beta^{J-2}$ and the decay width correspondingly with
$\sim \beta^{2J-3}$. In the absence of the $(1+\cos\theta^2_1)\sin^2\theta_2$
and $\sin^2\theta_1 (1+\cos^2\theta_2)$ polar-angle correlations,
the pronounced difference of the threshold behavior is exemplified for
the spin = 0 SM  and spin = 2 even normality bosons with identical $4\ell$
angular correlations in  \refF{fig:angl}(b).\s

Alternatively the measurement of the polar angular distribution of the
$\PZ^\ast \PZ$ axis in the production process $\Pg\Pg \to \PH^J \to \PZ^\ast \PZ$ can be
exploited to analyze spin states of any value $J$. This method can also be
applied in $\PGg\PGg$ decays which, since technically simpler, will
be described in detail later.\s

\subsubsection{Standard Model}

\noindent
The SM Higgs boson with $J^P = 0^+$ predicts by
angular momentum conservation only two non-vanishing $\PH\to \PZ^\ast \PZ$
decay helicity amplitudes,
${\mathcal T}_{00}=(\MH^2-M^2_\ast-\MZ^2)/(2 M_\ast \MZ)$ and
${\mathcal T}_{11} = -1$. The angular distributions
can therefore be cast into the transparent form \cite{Choi:2002jk,Barger:1993wt}:
\begin{eqnarray}
       \frac{1}{\GH}\,\frac{d\GH}{d\cos\theta_1\cos\theta_2}
&=& \frac{9}{16} \,\frac{1}{\gamma^4+2}\,
     \left[\gamma^4 \sin^2\theta_1\sin^2\theta_2
       +\frac{1}{2} (1+\cos^2\theta_1)(1+\cos^2\theta_2)\right]   \\[1mm]
      \frac{1}{\GH}\,\frac{d\GH}{d\phi}
&=& \frac{1}{2\pi} \left[\,1
       +\frac{1}{2}\, \frac{1}{\gamma^4+2}\,\cos 2\phi \,\right]
\end{eqnarray}
where $\gamma^2 = (\MH^2-M^2_*-\MZ^2)/(2M_* \MZ)$.
These angular distributions
will come with the $\PZ^\ast \PZ$ threshold rise \cite{Choi:2002jk,Barger:1993wt}
\begin{eqnarray}
   \frac{d\Gamma [\PH\to \PZ^*\PZ]}{dM^2_\ast}
& \Rightarrow
& \beta\, \sim\, \sqrt{(\MH-\MZ)^2-M^2_\ast}\,/\MH
        \ \ \, \mbox{for}\ \ M_\ast \Rightarrow \MH - \MZ
\end{eqnarray}
The observation of
the angular distributions associated with the helicity amplitudes, ${\mathcal T}_{00}$,
${\mathcal T}_{11}$ and of the threshold rise $\sim \beta$ are
necessary conditions for the zero-spin character of the SM Higgs boson.
They prove also sufficient by noting that any other spin/parity assignment
necessarily generates, for any pure state, a different combination of angular
correlations and threshold power.\s

The analysis applies analogously to Higgs bosons in generalized scenarios, like
supersymmetric theories, in which the couplings to vector bosons are non-derivative,
in conformity with generating masses through the Higgs mechanism.\s

\subsubsection{Alternative \boldmath{$J^P$} Assignments}

\noindent
To prove the unique $0^{+}$ assignment to the SM Higgs
boson, it must be demonstrated that the observed
signals discussed above are characteristically different for other assignments
so that any spin/parity confusion is avoided. The most general Lorentz bases
are summarized in \refT{tab:ff} for spin $J=0$ and $2$. The elements are
introduced Bose symmetric to leading order in the momenta, and they exhibit
all the characteristics relevant for the general analyses.\s

\begin{table}[ht]
\centering
\caption{ The most general tensor couplings of the Bose symmetric
   $\PH^J \PZ^\ast \PZ$ vertex and the corresponding helicity amplitudes for
   Higgs bosons of spin~$=0$ and 2 satisfying the relation
   ${\mathcal T}_{\lambda\lambda'}[M_*, \MZ]
   = (-1)^J\,{\mathcal T}_{\lambda'\lambda}[\MZ, M_*]$.
  Here $p=k_1+k_2$ and \mbox{$k=k_1-k_2$}, where $k_1$ and $k_2$ are the
   4-momenta of the $\PZ^*$ and the $\PZ$ bosons, respectively.}
\label{tab:ff}\begin{tabular}{lllc} \hline
  $J^P$  &\multicolumn{1}{c}{\small $\PH^J \PZ^\ast \PZ$ Coupling}
         & \multicolumn{1}{c}{\small Helicity Amplitudes} & \small Threshold
\\\hline 
%\multicolumn{4}{||c||}{\small Even Normality $n_{\PH}=+$} \\\hhline{|:=:t:===:|}
  &
  & $\sts {\mathcal T}_{00} =[2a_1(\MH^2-M_*^2-\MZ^2) + a_2 \MH^4 \beta^2] /(4M_*\MZ)$ &
  $\sts 1$ \\
\rb{1.5ex}{$0^+$}
  & \rb{1.5ex}{$\sts \phantom{+}a_1\, g^{\mu\nu}+a_2\, p^{\mu}p^{\nu}$}
  & $\sts {\mathcal T}_{11} = -a_1$
  & $\sts 1$
\\\hline 
  &
  & $\sts {\mathcal T}_{00}=\big\{-c_1\,(\MH^4-(\MZ^2-M_*^2)^2)/\MH^2
                             +\MH^2 \beta^2[c_2\, (\MH^2-\MZ^2-M_*^2) $

  &      \\
  & \rb{1.5ex}{$\sts \phantom{+}c_1\, (g^{\mu\beta_1}g^{\nu\beta_2}
                +g^{\mu\beta_2}g^{\nu\beta_1})$}
  & $\sts \phantom{{\mathcal T}_{00}=\big\{} +2c_3\,\MH^2
           +\frac{1}{2}c_4\,\MH^{4}\beta^2]\big\}/(\sqrt{6}\MZ M_*)$
  & $\sts 1$\\
  & \rb{1.5ex}{$\sts +c_2\,g^{\mu\nu}\,k^{\beta_1}k^{\beta_2}$}
  & $\sts {\mathcal T}_{01}
    =-[c_1(\MH^2-\MZ^2+M_*^2)-c_3\,\MH^4 \beta^2]/(\sqrt{2}M_*\MH)$
  & $\sts 1$  \\
\rb{1.5ex}{$2^+$}
 & \rb{1.5ex}{$\sts +c_3\, [(g^{\mu \beta_1}\, p^{\nu}
              -g^{\nu \beta_1}\, p^{\mu}) k^{\beta_2} $}
 & $\sts {\mathcal T}_{10}
   =-[c_1(\MH^2-M_*^2+\MZ^2)-c_3\,\MH^4 \beta^2]/(\sqrt{2}\MZ\MH)$
 & $\sts 1$ \\
 & \rb{1.5ex}{$\sts \phantom{+c_3 + } + (\beta_1\leftrightarrow \beta_2)]$}
 &  $\sts {\mathcal T}_{11}=-\sqrt{2/3}\,(c_1+c_2\MH^2\beta^2)$
 & $\sts 1$ \\
 & \rb{1.5ex}{$\sts +c_4\,p^{\mu}p^{\nu}k^{\beta_1}k^{\beta_2}$}
 & $\sts {\mathcal T}_{1,-1}=-2\, c_1$
 & $\sts 1$
\\\hline 
%
%\multicolumn{4}{||c||}{\small Odd Normality $n_{\PH}=-$} \\\hhline{|:=:t:===:|}
%
 &
 & $\sts{\mathcal T}_{00}=0$
 &      \\
\rb{1.5ex}{$0^-$}
 & \rb{1.5ex}{$\sts \phantom{+}a_1\,\epsilon^{\mu\nu\rho\sigma}
               p_{\rho}k_{\sigma}$}
 & $\sts{\mathcal T}_{11}=i\,\beta\,\MH^2\,a_1$
 & $\sts \beta$
\\\hline 
 &
 & $\sts{\mathcal T}_{00}=0$
 &     \\
 & $\sts\phantom{+}c_1\,\epsilon^{\mu\nu\beta_1\rho}p_{\rho}k^{\beta_2}$
 & $\sts{\mathcal T}_{01}=i\,\beta\, c_1\,(\MH^2+M_*^2-\MZ^2) \MH/(\sqrt{2}M_*)$
 & $\sts \beta$ \\
$2^-$
 & $\sts+c_2 \,\epsilon^{\mu\nu\rho\sigma}
               p_{\rho}k_{\sigma}k^{\beta_1}k^{\beta_2}$
 & $\sts{\mathcal T}_{10}=i\,\beta\, c_1\,(\MH^2+\MZ^2-M_*^2) \MH/(\sqrt{2}\MZ)$
 & $\sts \beta$  \\
 & $\sts+ (\beta_1 \leftrightarrow \beta_2)$
 & $\sts{\mathcal T}_{11}=i\,\beta\,2\sqrt{2/3}\,(c_1  +c_2\,\MH^2 \beta^2) \MH^2$
 &  $\sts \beta$ \\
 &
 & $\sts{\mathcal T}_{1,-1}=0$& \\\hline 
\end{tabular}

\end{table}

\subsubsubsection{Pseudoscalar}

Pure pseudoscalar/$CP$-odd states $\PH^J = \PA$ with $J^P
= 0^-$ may have a significant branching ratio for decays into $\PZ$-boson pairs,
though not guaranteed in general \cite{Bernreuther:2010uw}, as manifest in supersymmetric
theories. The pseudoscalar $\PA$ is the state
of minimal complexity as the helicity amplitude ${\mathcal T}_{00} = 0$ by parity
invariance and the only non-zero amplitude is ${\mathcal T}_{11}
= - {\mathcal T}_{-1,-1}$.
As a result, the polar and azimuthal angular distributions are
independent of any free parameter \cite{Choi:2002jk,Barger:1993wt}:
\begin{eqnarray}
      \frac{1}{\Gamma_{\PA}}\, \frac{d\Gamma_{\PA}}{d\cos\theta_1\cos\theta_2}
&=& \frac{9}{64}\,(1+\cos^2\theta_1)(1+\cos^2\theta_2)   \\[1mm]
      \frac{1}{\Gamma_{\PA}}\,\frac{d\Gamma_{\PA}}{d\phi}
&=& \frac{1}{2\pi} \left[1 -\frac{1}{4} \cos 2\phi\right]
\end{eqnarray}
The negative-parity decays are distinctly different from decays
of the SM Higgs boson. In particular, the $\cos 2\phi$ term, proportional to
the normality $n_{\PH}$, flips sign when the parity is switched,
cf. Fig.$\,$\ref{fig:angl}(a). First experimental analyses
\cite{Chatrchyan:2012jja,CMS-PAS-HIG-13-002,ATLAS-CONF-2013-013}
exclude the negative-parity assignment at more than the $3\sigma$
level.\s

It may be read off \refT{tab:ff} that the decay amplitude is suppressed
near the threshold so that the width rises as $\beta^3$, in contrast to
$\beta$ in the SM.\s

If the branching ratio for $\PA\to \PZ^*\PZ$ decays is small, initial-final state
correlations in $\PGg\PGg$ as well as jet-jet correlations in gluon
fusion build up a set of necessary and sufficient analyses of
the spin/parity quantum numbers:\s

\noindent
-- The angular distribution of the photons in $\Pg\Pg \to \PA \to \PGg\PGg$
is isotropic for spin $= 0$ in the $\PA$ rest frame while behaving
as a polynomial up to $\cos^{2J}\Theta$ for higher spins.\s

\noindent
-- The jets in gluon fusion $\Pg\Pg \to \PA+\Pg\Pg$ are anti-correlated in the azimuthal
distribution $\cos 2 \phi$ for pseudoscalar states \cite{Hagiwara:2009wt}, opposite to
correlated pairs for scalars. Parity signals follow from the correlation
${\vec{\epsilon}}_1 \times {\vec{\epsilon}}_2$ of the linear gluon polarization
vectors for $\PA$, and ${\vec{\epsilon}}_1 \cdot {\vec{\epsilon}}_2$ for H,
with the polarization vectors concentrated in the gluon-emission
planes.\s

\subsubsubsection{Higher Spin States $J > 0$}

\noindent
{\it Vector/Axialvector\, $J^P=1^\mp$} :
Vector states of both parities cannot decay into di-photon final states by
virtue of the Landau-Yang theorem \cite{Landau:1948kw,Yang:1950rg}.
Independently of the theorem it may be noted that vector decays into $\PZ^\ast \PZ$
pairs can be ruled out experimentally by analyses parallel to those for
spin-2 states, which will be described next.\s

\noindent
{\it 2-Tensors\, $J^P=2^\pm$} :
Tensor decays into $\PZ^\ast \PZ$ pairs with negative parity are suppressed
with the third power $\beta^3$ in the velocity near the threshold.
Tensor decays with positive parity are also suppressed, if correlations
$(1 + \cos^2\theta_1) \sin^2\theta_2$ {\it et v.v.}, absent also
in scalar decays, are not observed. This case is nicely
illustrated by the
``tensor assignment'' as studied in Graviton-related
Kaluza-Klein scenarios \cite{Gao:2010qx,DeRujula:2010ys,Bolognesi:2012mm,Chen:2012jy, 
Ellis:2012wg,Ellis:2012jv,Alves:2012fb}. \s

\noindent
{\it Spin $J\geq 3$} :
Spin states $J \geq 3$ generate the same generic angular correlations
in decays of unpolarized particles as $J = 2$, though with different
coefficients. However, the threshold rise, at least $\sim \beta^{2J-3}$, is
characteristically reduced compared to the SM Higgs boson
as explained in \refS{sec:LM_cp_intro}. (In addition, characteristic initial-final
state correlations would signal the production of a spin $\geq 3$ state.)\s

\subsubsection{Di-Photon Final States}

\noindent
A prominent decay mode of Higgs bosons are $\PH \to \PGg\PGg$
decays \cite{Ellis:1975ap}. They can be exploited to study the spin quantum
number of the particle \cite{Ellis:2012wg,Ellis:2012jv,Alves:2012fb,Choi:2012yg,
Gao:2010qx,DeRujula:2010ys,Bolognesi:2012mm,Chen:2012jy}, supplementing the decays into
electroweak gauge bosons.\s

The Landau--Yang theorem \cite{Landau:1948kw,Yang:1950rg},
based on Lorentz invariance, electromagnetic gauge invariance and Bose symmetry,
forbids spin-1 particles to decay into photon pairs. This theorem can easily be
extended to odd-spin states of negative parity \cite{Yang:1950rg,Choi:2012yg}.
From the isotropic decay of spinless states it follows that the parity
cannot be measured in on-shell di-photon decays \cite{Landau:1948kw,Choi:2012yg}.
However, this problem can be solved by gluon fusion $\Pg\Pg\to \PH,\PA + jj$ with $\PGg\PGg$
decays of the Higgs bosons and jets in the final state \cite{Hagiwara:2009wt}.\s

Information on any spin of the Higgs boson can be derived from
combining the production with the decay process.
The information is carried in gluon fusion, cf. Fig.$\,$\ref{fig:angl}(b),
by the angle $\Theta$ of the decay photons with respect to the axis of
the incoming partons in the rest frame of the Higgs boson (see
\refF{fig:kine}(b)): 
\begin{equation}
   \Pg\Pg \to \PH^J \to \PGg\PGg \,.
\label{eq:ggHgg}
\end{equation}
Initial and final states carry helicities $\mu-\mu', \lambda-\lambda' = 0$ and
$\pm 2$, added up incoherently in the differential cross section
\begin{eqnarray}
   \frac{1}{\sigma}\,\frac{d\sigma[\Pg\Pg\to \PH^J\to \PGg\PGg]}{d\cos\Theta}
   \,=\, (2J+1)
   \left[ {\mathcal X}_0 {\mathcal Y}_0 {\mathcal D}^J_{00}
         +{\mathcal X}_0 {\mathcal Y}_2 {\mathcal D}^J_{02}
         +{\mathcal X}_2 {\mathcal Y}_0 {\mathcal D}^J_{20}
         +{\mathcal X}_2 {\mathcal Y}_2 {\mathcal D}^J_{22}
   \right]
\label{eq:1polar_dist_gg_H_rr}
\end{eqnarray}
with the angular-symmetric squared Wigner functions ${\mathcal D}^J_{m\lambda}
= \frac{1}{2} \{[d^J_{m\lambda}(\Theta)]^2+[d^J_{m,-\lambda}(\Theta)]^2\}$.
The probabilities
for production ${\mathcal X}_{0,2}$ and decay
${\mathcal Y}_{0,2}$ are built up by the $\Pg\Pg\to \PH^J$ production and
$\PH^J\to\PGg\PGg$ decay helicity amplitudes.
Quite generally they rise up
to the non-trivial maximum power of
\begin{equation}
   {\mathcal D}^J \sim \cos^{2J} \Theta~,
\end{equation}
independently of the helicity indices 0 or 2.  Thus a characteristic signal
of the Higgs spin involved is provided by the angular distribution of the
photon axis, singling out the isotropic spin = 0 very clearly \cite{Choi:2012yg}.\s

While scalar/pseudoscalar Higgs production in the SM is
isotropic, any other spin assignment gives rise to non-trivial polar
angular distributions as manifest in Fig.$\,$\ref{fig:polar_dist_gg_H_rr}.
In particular, the SM Higgs angular distribution $\{0;00\}$, coming with
%AAA
${\mathcal D}^0_{00}(\Theta) = 1$ for ${\mathcal X}_0 = {\mathcal Y}_ 0 = 1$,
is compared with the spin-2 distributions for the $'$scalar assignment$'$
$\{2;00\}$ and the $'$tensor assignment$'$ $\{2;22\}$. For example,
the angular distribution for spin-2 graviton-like states is described
\cite{Ellis:2012wg,Ellis:2012jv,Alves:2012fb,Choi:2012yg} by 
${\mathcal D}^2_{22}(\Theta) = (\cos^4\Theta + 6\cos^2\Theta +1)/16$
for ${\mathcal X}_2 = {\mathcal Y}_2 = 1$.
The spin/parity states, which are probed
by observing the squared Wigner functions ${\mathcal D}^J$ in $\PGg\PGg$ decays,
are listed in \refT{tab:SR}.\s

Whenever a $\PGg\PGg$ final state is observed as demanded for a Higgs boson,
the polar angular distribution can be exploited to measure the spin of the
decaying particle and eventually rule out any non-zero spin decay.\s

\begin{figure}[htb]
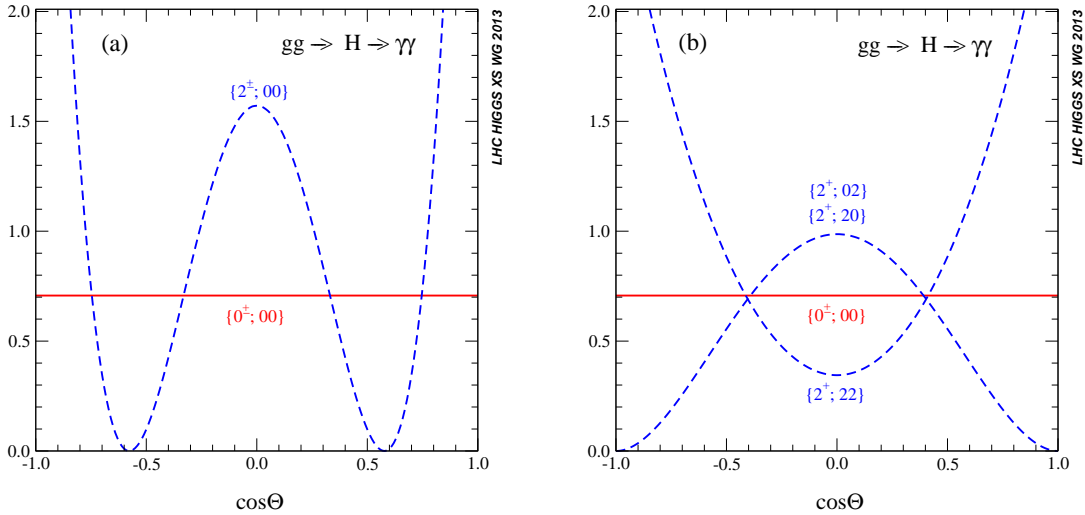

\begin{center}
\includegraphics[scale=0.4]{YRHXS3_LM/Maggie/polar_dist_gg_H_rr_set1.eps}
\hskip 1cm
\includegraphics[scale=0.4]{YRHXS3_LM/Maggie/polar_dist_gg_H_rr_set2.eps}
\end{center}
\caption{Polar-angle distributions of the $\PGg\PGg$ axes in the rest
         frame of the subprocesses: the flat Higgs signal compared with
         potential spin-2 distributions; (a) the $'$scalar assignment$\,'$
         $\{2;00\}$ and (b) the $'$tensor assignment$\,'$ $\{2;22\}$
         and the $'$mixed assignments$\,'$, $\{2;02\}$ and $\{2;20\}$
         (all distributions normalized over the interval
         $|\cos\Theta| \leq 1/\sqrt{2}$). The upper indices
         refer to the allowed parity associated with the distributions
         in $\PGg\PGg$ decays.
}
\label{fig:polar_dist_gg_H_rr}
\end{figure}
\begin{table} [htb]
\begin{center}
\caption{ Selection rules for Higgs spin/parity following from observing
             the polar angular distribution of a spin-$J$ Higgs state in the
             process $\Pg\Pg\to \PH\to \PGg\PGg$, cf. \Bref{Choi:2012yg}.}
\label{tab:SR}
\begin{tabular}{lcccc}
\hline
$\; P \;\, \backslash \; J\;$  &  $\; 0\;$  & $\; 1 \;$
                                       & $\;$ $2, 4, \cdots$
                                       & $\;$ $3, 5, \cdots$             \\
\hline 
$\;$ even $\;$       &      1
                     & $\;$ forbidden $\;$
                     & $\; {\mathcal{D}}^J_{00} \;\; {\mathcal{D}}^J_{02} \;$
                     & $\; {\mathcal{D}}^J_{22}$     \\
                     &
                     &
                     & $\; {\mathcal{D}}^J_{20} \;\; {\mathcal{D}}^J_{22} \;$
                     &                               \\
\hline
$\;$ odd $\;$        &      1
                     & $\;$ forbidden $\;$
                     & $\; {\mathcal{D}}^J_{00}$
                     & $\;$ forbidden $\;$           \\
\hline
\end{tabular}
\end{center}
\end{table}

An analogous experimental analysis can be carried out for $\PZ^\ast \PZ$ final
states if the angular distribution of the $\PZ^\ast \PZ$ axis is measured.\s

In contrast to the correlation analyses of polarization-averaged Higgs
states in the final states, the measurement of this polar angle $\Theta$
is affected by the boost from the laboratory frame to the Higgs rest frame.
A simple technique is provided by boosting the reference frame along the
3-momentum of the Higgs boson in the proton-proton-Higgs production plane,
leaving the orthogonal space coordinate unchanged. The corresponding vector
is known experimentally since the Higgs decay final states,
$\PGg\PGg, \PZ^\ast \PZ \to 4\ell$ allow the explicit reconstruction of
this vector.\s

\subsubsection{Vector-boson Fusion}

\noindent
The production of Higgs bosons in {\it electroweak vector-boson fusion}
$VV' \to H$ offers another powerful check of the Higgs
parity \cite{Plehn:2001nj,Hagiwara:2009wt}. Radiating the vector bosons off the quark lines,
as depicted in Fig.$\,$\ref{fig:kine}(b), the azimuthal correlation between
the two radiation planes is sensitive to
this quantum number. The two planes are spanned by the proton-proton axis
combined with any of the two final quark jets. The correlations of the planes
in gluon fusion to Higgs $+$ two jets can be exploited in parallel.\s

The differential cross sections can be written as a straightforward
generalization of the cross sections derived above, for the forward-backward
symmetric combination
with the transversal and longitudinal degrees of freedom of $V,V'$
and $\PZ^\ast, \PZ$ (from the decay of the Higgs boson into $\PZ^\ast \PZ$) summed up.\s

Electroweak vector-boson fusion is supplementary in several facets
to Higgs-boson decays, and
we will focus on the new elements in {\it gluon fusion} \cite{Hagiwara:2009wt,Boer:2013fca},
\begin{equation}
   qq \to \PH,\PA + qq \quad {\rm and} \quad q \to \bar{q}, \Pg \, .
\end{equation}
By producing the Higgs bosons in gluon fusion, followed by di-photon decays,
this mechanism allows the determination of the parity of the pseudoscalar $\PA$ in
scenarios for which $\PA$ does not decay into $\PZ$ pairs with sufficiently
large branching ratio.\s

Radiating gluons off quarks or gluons the daughter gluons are polarized to a
high degree~$P$ in the emission plane \cite{Bengtsson:1988qg,Hagiwara:2009wt}
as opposed to the perpendicular plane,
\begin{eqnarray}
\PQq\to \PQq\Pg \quad &:& \quad P(q;z)
        \,=\, {2 (1-z)}/{(2-2z + z^2)}  \\
\Pg\to \Pg\Pg \quad &:& \quad P(\Pg;z)
        = {(1-z)^2}/{(1-z+z^2)^2}
\end{eqnarray}
where $z$ denotes the energy fraction transferred to the daughter gluon in the
fragmentation process $\PQq \to \PQq\Pg$ or $\Pg\to \Pg\Pg$
in the collinear and massless parton limit.
Since parity-even Higgs bosons $\PH$ are produced preferentially for
parallel gluon polarization $\sim {\vec{\epsilon}}_1 \cdot {\vec{\epsilon}}_2$, while
parity-odd Higgs bosons $\PA$ demand perpendicular polarization $\sim
{\vec{\epsilon}}_1 \times {\vec{\epsilon}}_2$, the azimuthal
modulation of the parton-level
production cross sections is uniquely predicted \cite{Hagiwara:2009wt} as
\begin{equation}
  \frac{1}{\sigma}\,\frac{d\sigma}{d\phi}
= \frac{1}{2\pi}\, \bigg[1 \pm |\zeta| \cos 2 \phi \bigg]
  \quad {\rm for} \quad  \PH,\PA \, .
\end{equation}
The angle $\phi$ is the azimuthal angle between the two emission planes
$\PQq \to \PQq\, \Pg$ and/or $\Pg \to \Pg\,\Pg$, and the polarization
parameters $|\zeta|$
\begin{equation}
   |\zeta|\, = P(u;z_1)\, P(v;z_2)  \quad {\rm with} \quad u,v = q,\Pg
                                    \quad {\rm for} \quad qq/q\Pg/\Pg\Pg
\label{eq:zeta}
\end{equation}
measures the size of the correlation/anti-correlation of the two planes
for $\PH/\PA$ production \cite{Hagiwara:2009wt}. In the limit of soft gluon radiation the
degrees of polarization approach unity. Since the recoil quarks are emitted at
small but non-vanishing transverse momenta, the planes formed each by quark and
proton axis can be determined experimentally. The description of quantitative
analyses is deferred to the proper analysis group.\s

\subsubsection{Addenda}

\noindent
\underline{\it Higgs-strahlung} : This process in $\Pe^+\Pe^-$ collisions has been proposed 
quite early \cite{Barger:1993wt,Miller:2001bi} as a tool for measuring Higgs spin/parity, 
and invites investigating
$\PQq\, \PAQq\, \to\, \PZ/\PW\,+ \, \PH$
at the LHC \cite{Ellis:2012xd,Englert:2012xt,Frank:2012wh,Djouadi:2013yb,Ellis:2013ywa,Godbole:2013saa}.
The threshold behavior can be
exploited in the same way as described in the Higgs decay
into virtual $\PZ$ bosons. A linear rise in the invariant
mass distribution rules out the states
$0^-,1^-,2^-$, while to rule out $1^+,2^+$ one also requires the angular correlations in the
leptonic $\PZ$ boson decay products. A linear rise in the mass
distribution and non-observation of the non-trivial $[1+\cos^2\theta]
\sin^2 \theta_*$ and $\sin^2\theta [1+\cos^2 \theta_*]$ correlations
rules out $1^+,2^+$ ($\theta$ denoting the polar
angle in $\PH^J \PZ$ production and $\theta_*$ the fermion polar angle
in the $\PZ$ rest frame with respect to the $\PZ$ flight direction in the
laboratory frame). Any higher spin assignment above spin-2 is ruled
out by a rise in the invariant mass distribution with a power $\ge 3$.\s

\noindent
\underline{\it Fermion Final States} : 
There are scenarios, in particular in supersymmetric 
theories \cite{Carena:2002es,Djouadi:2005gj,Accomando:2006ga},
in which $CP$-odd (pseudoscalar) Higgs particles decay, though not forbidden in
principle, into pairs of vector bosons only with perturbatively small branching ratio.
This problem can be approached in fermion decays of the Higgs boson
\cite{Kramer:1993jn,Berge:2008wi,Berge:2008dr,Berge:2011ij}, which may be 
illustrated by two examples, specific for moderate and large masses: \s

\noindent
{\it (i)} The angular correlation of the light $0^\pm$ Higgs decay
$\PH,\PA \to \PGtm\PGtp \to \pi^- \pi^+ ...$ can be written \cite{Nelson:1989uq,
Grzadkowski:1995rx,Kramer:1993jn} as
\begin{equation}
\frac{1}{\Gamma}
\frac{d\Gamma[H,A\to\pi^-\pi^+ ... ]}{
              d\cos\theta_1 d\cos\theta_2 d\phi}
= \frac{1}{8\pi}
    \left[\,1- \cos\theta_1\cos\theta_2
           \mp  \sin\theta_1\sin\theta_2 \cos\phi\,\right] \,,
\end{equation}
modulated characteristically opposite for positive and negative parities. \s

\noindent
{\it (ii)} The process $\PH,\PA \to \bar{\PQt}\,\PQt$ followed by the 2-particle decay $\PQt \to \PW \PQb$
for heavy Higgs bosons can be analyzed in the same way as the $\tau$ chain; the $b$-jet
and the $\PW^\pm$  angular distribution read identically \cite{Nelson:1989uq,
Grzadkowski:1995rx,Kramer:1993jn}:
\begin{equation}
\frac{1}{\Gamma}
\frac{d\Gamma[\PH,\PA\to \bar{\PQt}\,\PQt \to \bar{\PQb} \PQb...]}{
              d\cos\theta_1 d\cos\theta_2 d\phi}
= \frac{1}{8\pi}
    \left[1-\kappa^2_W (\cos\theta_1\cos\theta_2
           \pm \sin\theta_1\sin\theta_2 \cos\phi) \right]
\label{eq:H_ttWW}
\end{equation}
with the polarization-resolving factor $\kappa_{\PW}=(\Mt^2-2\MW^2)/(\Mt^2+2\MW^2)
\sim 2/5$.\s

\subsubsection{\boldmath{$CP$} mixing in the spin-zero Higgs system}

\noindent
$CP$ violation in the Higgs system can arise from two different sources
which of course influence each other at second order in the
perturbative expansion.
$CP$ violation can be either indirect and implanted in the Higgs mass
matrix \cite{Pilaftsis:1998pe,Pilaftsis:1999qt,Choi:2000wz,Carena:2001fw,Frank:2006yh}, 
or direct and
effective in the interaction mechanisms with other particles, or both
sources may act in parallel.\s

Combining the $CP$-mixed Higgs states $\PH'$ with $CP$-violating
interactions, an effective $\PH' ZZ$ vertex with
complex coefficients can be defined,
\begin{eqnarray}
  V^{\mu\nu}_{\PH'\PZ\PZ}
&=& ig_Z \MZ
   \left[ a \, g^{\mu\nu} + b \, \frac{p^\mu p^\nu}{M^2_{\PH^\prime}}
         + c \, \epsilon^{\mu\nu\rho\sigma}
             \frac{p_\rho k_\sigma}{M^2_{\PH^\prime}}
         \right]\; ,
\label{eq:cpviolvertex}
\end{eqnarray}
in the same notation for the 4-momenta as
in \refT{tab:ff}. Neglecting any small interference effects, the SM-type parameter
$a$ can be taken real. $CP$ is violated if at the same time in addition to $c$ the
couplings $a$ and/or $b$ do not vanish simultaneously \cite{Godbole:2007cn}.
Note that these parameters can in general be
momentum-dependent form factors obtained from
integrating out new physics at some large scale $\Lambda$. As the
momentum dependence involves ratios of typical momenta
in the process to $\Lambda$, in a first approximation the scale
dependence can be neglected and only the constant part is kept.\s

The SM is characterized by the set $a = 1$ and $b = c = 0$.
Of particular interest is the subgroup $a \neq 0$, $b = 0$ and $c \neq 0$,
with $c$ being either real or complex. This subgroup is realized in scenarios
in which $CP$-violation is rooted in the coherent superposition of $CP$-even $\PH$
and $CP$-odd $\PA$ components of the Higgs wave function. It will provide
transparent illustrations of $CP$-violation effects.\s

The five general parameters of the $H^\prime ZZ$ vertex, $\{a,\,\Re(b),\, \Im(b),
\, \Re(c),\, \Im(c)\}$, can be measured by five independent observables,
$\{{\mathcal O}_1, \cdots, {\mathcal O}_5\}$.
$CP$-violation proper can be observed by isolating the $\sin 2 \phi$
observable \cite{Soni:1993jc,Chang:1993jy},
\begin{equation}
{\mathcal O}_4 = \sin^2 \theta_1 \sin^2\theta_2 \sin 2\phi
\end{equation}
which can be accessed by a pure measurement of
$d \Gamma / d\phi \sim \beta\, \Re(a c^\ast)$.\s

In a first step $CP$-violation may solely be associated with the wave
function of the Higgs boson, which is built up by the superposition of
$CP$-even $\PH$ and $CP$-odd $\PA$ spinless components \cite{Ellis:2004fs,Choi:2004kq}:
\begin{equation}
   \PH' = \cos\chi\, \PH + \sin\chi\, e^{i\xi}\, A           \,.
\end{equation}
Apart from the wave
function we assume the $\PZ^\ast \PZ$ vertex to be identified with the SM parameters
so that in the notation introduced above $a = \cos\chi$, $b = 0$ and
$c = \rho_{\PZ} \, \sin\chi\, e^{i\xi}$, with the real parameter
$\rho_{\PZ}$ denoting the magnitude of the $CP$-conserving $AZZ$ coupling,
potentially suppressed relative to the $HZZ$ coupling.\s

The $CP$-violating $\PH'\to \PZ^\ast \PZ$ decay helicity amplitudes are built
up by the basic $\PH\PZ^\ast \PZ$ and $\PA \PZ^\ast \PZ$ amplitudes and their parity
mirrors. Suppressing the very small
$\eta_{1,2}$-dependent terms for leptonic $\PZ$ decays, the width can be cast
into the form
\begin{eqnarray}
\frac{1}{\Gamma}\frac{d\Gamma}{d \cos\theta_1 d \cos\theta_2}
&=& \frac{9}{16 {\mathcal N}} \,
    \left[\gamma^4 c^2_\chi \, \sin^2\theta_1 \sin^2\theta_2
         +\frac{1}{2} (c^2_\chi+ \rho_{\PZ}^2 \beta^2 s^2_\chi)
          (1+\cos^2\theta_1)(1+\cos^2\theta_2)\right]\nonumber
    \label{eq:polar_cp} \\[1mm]
\frac{1}{\Gamma} \frac{d\Gamma}{d\phi}
&=& \frac{1}{2\pi}
   \left[ 1+ \frac{1}{2}\, \left\{(c^2_\chi- \rho_{\PZ}^2 \beta^2 s^2_\chi)
             \, \cos 2\phi
            + \rho_{\PZ} {\beta} {s_{2\chi} c_\xi}
             \,\sin 2\phi \right\}/{\mathcal N}
   \right]
    \label{eq:azimuthal_cp}
\end{eqnarray}
using the normalization ${\mathcal N} = (2 + \gamma^4) c^2_\chi
+ 2 \rho_{\PZ}^2 \beta^2 s^2_\chi$ with the abbreviations $c_\chi = \cos\chi$,
etc., and the boosts-related
factor $\gamma^2 = (M^2_{\PH'}-M^2_*-M^2_{\PZ})/(2M_*
M_{\PZ})$, and $\beta$ accounting for the standard phase space suppression in
massive decays.\s

\vspace*{0.2cm}
\begin{figure} [htb]
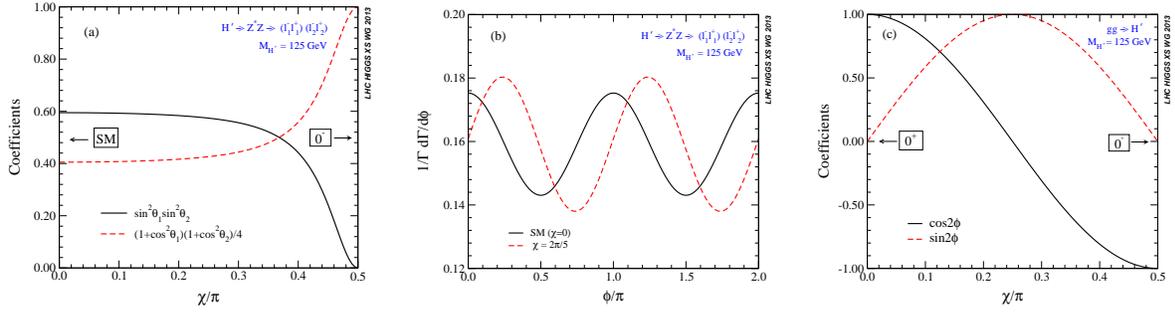

\begin{center}
\includegraphics[scale=0.23]{YRHXS3_LM/Maggie/CP_polar_ZZ.eps}
\hskip 0.5cm
\includegraphics[scale=0.23]{YRHXS3_LM/Maggie/CP_azimuthal_ZZ.eps}
\hskip 0.5cm
\includegraphics[scale=0.23]{YRHXS3_LM/Maggie/CP_asymmetry_gg.eps}
\end{center}
\caption{(a) The CP-even and CP-odd coefficients in the correlated
    polar-angle distribution with respect to the $\PH/\PA$ mixing angle $\chi$
    and (b) the correlated azimuthal-angle distribution in the
    process $\PH'\to \PZ^\ast \PZ\to 4\ell$ for the SM case ($\chi=0$) and
    a $CP$-violating case with $\chi=2\pi/5$; (c) the CP-even and CP-odd
    coefficients in the correlated azimuthal-angle distribution of
    the two initial two-jet emission planes with respect to the CP-mixing
    angle $\chi$. The $\PH'$ mass $M_{\PH'}$ is set to be $125\UGeV$. In addition,
    the ratios $\rho_{\PZ}=\rho_g =1$, the phase $\xi =0$ and the polarization
    parameter $|\zeta|=1$ are taken for illustration. }
\label{fig:cp_asymmetry}
\end{figure}

The original $\PH$ and $\PA$ contributions, coming with $c^2_\chi$ and $s^2_\chi$
can easily be recognized in the $\cos\theta_1\cos\theta_2$ and $\cos 2 \phi$
distributions. The novel $CP$-violating $\PH/\PA$ interference term is proportional
to $\sin 2\chi \cos\xi$. It is uniquely isolated from the
$CP$-conserving terms by the angular dependence $\sin 2 \phi$. When the angle
$\chi$ moves from 0 to $\pi/2$, the system moves continuously from the pure
$0^+$ SM Higgs state through $CP$-mixed states to the pure $0^-$ state. This
transition is demonstrated in Fig.$\,$\ref{fig:cp_asymmetry}(a) in which the
modifications of the helicity coefficients compared with the SM are
displayed as  a function of $\chi$ with $\rho_{\PZ} = 1$ and $\xi = 0$. As shown in
Fig.$\,$\ref{fig:cp_asymmetry}(b), the azimuthal modulation for a non-trivial
$CP$-violating phase $\chi=2\pi/5$ is distinctly different from that for the
SM case with $\chi=0$. The size of the admixture $\rho_{\PZ}$ is crucial for
observing the mixed state experimentally
(where for small $\rho_\PZ$ the $\phi$ distribution moves back to the SM
oscillation, independent of $\chi$).\s

For $M_{\PH^\prime}=125\UGeV$ a maximum value of the
asymmetry ${\mathcal A}_4 = [\Gamma (O_4 >0) - \Gamma (O_4 <0)]/\Gamma$ of
the observable ${\mathcal O}_4$ of $0.10$ is obtained for $\Re (c)/a=2.7$
(reduced only to $0.06$ for the ratio set to $1.0$). A
rough theoretical estimate based on the numbers for signal and
background at $7$ and $8\UTeV$ taken from \Bref{ATLAS-CONF-2013-013} yields for
${\mathcal A}_4$ a significance of 0.45, where we have assumed that both the
Higgs coupling to $\PZ$ and $\PW$ bosons is $CP$-violating. We have calculated the
significance according to the formula given in \Bref{Godbole:2007cn}.
With the signal and background numbers from CMS \cite{Chatrchyan:2012jja} we have
a significance of 0.50. At $14\UTeV$, extrapolating the numbers for signal and
background from \Bref{Lafaye:2009vr}, the significance increases to
$0.74$ for an integrated luminosity of $100\Ufb^{-1}$ and to $1.28$ for
$300\Ufb^{-1}$.\s

For small $\PA\PZ\PZ$ couplings, gluon fusion $\Pg\Pg \to \PH'+\Pg\Pg$
followed by $\PH'\to \PGg\PGg$ decays offers a viable
alternative. The $\PH\Pg\Pg$ and $\PA\Pg\Pg$
vertices are combined to
\begin{eqnarray}
   V_{H'gg} = \cos\chi V_{Hgg} + \sin\chi e^{i\xi}\, V_{Agg}
\end{eqnarray}
Denoting the ratio of $\PA\Pg\Pg / \PH\Pg\Pg$ couplings by $\rho_g$, the superposition
of the $\pm$ parities modifies the azimuthal angular modulation of the two
jets to
\begin{eqnarray}
  \frac{1}{\sigma}\frac{d\sigma}{d\phi}
= \frac{1}{2\pi}\,
  \bigg[\,1+ |\zeta|
   \left\{ \left({c^2_\chi - \rho^2_g s^2_\chi}
           \right)\, \cos 2\phi
         + \rho_g\, s_{2\chi} c_\xi
           \, \sin 2\phi
   \right\} /{\mathcal N}'
   \,\bigg]
\label{eq:Hpgg}
\end{eqnarray}
with the polarization parameters $|\zeta|$ defined in Eq.$\,$(\ref{eq:zeta})
and the normalization ${\mathcal N}' = {c^2_\chi+\rho^2_g s^2_\chi}$. The two
coefficients are illustrated in Fig.$\,$\ref{fig:cp_asymmetry}(c) for
$\xi=0$, the admixture parameter $\rho_g=1$ and the polarization
parameter $|\zeta|=1$. \s

{\it Fermion decays:} Other observables for studying the $CP$-violating
mixing effects experimentally are the polarization of the $\PGt$ leptons and top quarks,
cf. \Bref{Tsai:1971vv,Hagiwara:1989fn,Rouge:1990kv,Jadach:1993hs,Hagiwara:2012vz}. 
The correlations are built up by the dot and cross products
of the transverse polarization vectors, ${\mathcal C}_{||}$ and ${\mathcal C}_\perp$, respectively,
coming with coefficients $\cos\phi$ and $\sin\phi$ in the azimuthal decay distributions. \s

\subsection{Spin, parity and tensor couplings with {\sc JHUGen} and MELA}
\label{sec:LM_jhu}

Studies of spin, parity, and couplings of the Higgs-like boson may 
employ either effective Lagrangians or generic parameterizations of scattering
amplitudes. The two approaches are fully equivalent.  
Effective Lagrangians typically lead to smaller number of tensor
couplings by restricting dimensions of contributing operators; this
makes  the spin-parity analyses more tractable. Generic
parameterizations of scattering amplitudes contain all possible tensor
structures consistent with assumed symmetries and Lorentz invariance
without assigning relative significance to different  contributions.
Therefore, if the goal is to maximize the variety of scenarios that can
be explored for a given hypothesis,  generic parameterizations of
amplitudes is a reasonable choice. 

%The discovery of the new boson with the mass around 125 GeV at the LHC~\cite{discovery-atlas, discovery-cms}  
%opens a way for experimental studies of its properties such as spin, parity, and couplings 
%to the Standard Model particles.  We split such studies into two groups
We split studies of the new boson properties, such as spin, parity, and
couplings, into two groups  
\begin{itemize} 
\item  tests of discrete spin/parity/coupling hypotheses of the new particle;
\item  identification and measurement of  various types of tensor couplings
that are consistent with assumed symmetries and Lorentz invariance, 
for  a given spin assignment.
\end{itemize}

%-------------------------------------
\subsubsection{Parameterization of amplitudes for spin-zero, spin-one, and spin-two bosons}

To introduce our notation, we  follow \Bref{Gao:2010qx},
and write the general scattering amplitude that describes the interaction of the Higgs-like boson 
with the gauge bosons, such as $\PZ\PZ$, $\PW\PW$, $\PZ\PGg$, $\PGg \PGg$, or $\Pg\Pg$
%%%%%%%%%%%%%%%%%%%%%%%
\begin{equation}
A(\PX_{J=0} \to \PVB\PVB) = v^{-1} \left (
  g^{}_{ 1} m_{\sss V}^2 \epsilon_1^* \epsilon_2^* 
+ g^{}_{ 2} f_{\mu \nu}^{*(1)}f^{*(2),\mu \nu}
+ g^{}_{ 3} f^{*(1),\mu \nu} f^{*(2)}_{\mu \alpha}\frac{ q_{\nu} q^{\alpha}}{\Lambda^2} 
+ g^{}_{ 4}  f^{*(1)}_{\mu \nu} {\tilde f}^{*(2),\mu  \nu}
\right )\,,
\label{eq:fullampl-spin0} 
\end{equation}
%%%%%%%%%%%%%%%%%%%%%%%
where $f^{(i),{\mu \nu}} = \epsilon_i^{\mu}q_i^{\nu} - \epsilon_{i}^\nu q_i^{\mu} $
is the field strength tensor of a gauge boson
with momentum $q_i$ and polarization vector $\epsilon_i$;
${\tilde f}^{(i)}_{\mu \nu}$  is the  conjugate field strength tensor.
Parity-conserving interaction of a pseudo-scalar corresponds to $g^{}_{ 4}$,
of a scalar to $g^{}_{ 1}$, $g^{}_{ 2}$, and $g^{}_{ 3}$.
The SM Higgs coupling at tree level (to $\PZ\PZ$ and $\PW\PW$) is described by the $g^{}_{ 1}$ term,
while the $g^{}_{ 2}$ term appears in the loop-induced processes, such as 
$\PZ\PGg$, $\PGg\PGg$, or $\Pg\Pg$, and as a small contribution due to radiative corrections
in the $\PZ\PZ$ and $\PW\PW$ couplings.
The $g^{}_{3}$ term can be absorbed into the $g^{}_{2}$ term if the constants are allowed
to be momentum-dependent~\cite{Gao:2010qx}. 
Moreover, this term is supposed to be small since it corresponds to a dimension-seven operator 
in an effective Lagrangian.
We therefore neglect the $g^{}_{3}$ term in the following discussion, but we note that it 
can be easily included in our framework if necessary.  

Next, we write the general scattering amplitude that describes the interaction of the Higgs-like boson 
with the fermions, such  as $\PGtp\PGtm$, $\PGmp\PGmm$,  $\PQb\PAQb$, and $\PQt\PAQt$.
We assume that the chiral symmetry is exact in the limit when fermion masses vanish
%%%%%%%%%%%%%%%%%%%%%%%%%%%%%
%
\begin{eqnarray}
&& A(\PX_{J=0} \to \Pf\bar{\Pf}) = \frac{m_{\Pf}}{v}
\bar u_2 \left ( \rho^{}_{ 1} + \rho^{}_{ 2} \gamma_5 \right ) v_1\,,
\label{eq:ampl-spin0-qq}
\end{eqnarray}
%
%%%%%%%%%%%%%%%%%%%%%%%%%%%%%
where $m_{\Pf}$ is the fermion mass and $\bar u$ and $v$ are the Dirac spinors. 
The two constants  $\rho^{}_{ 1}$ and $\rho^{}_{ 2}$ correspond to
the scalar and pseudoscalar couplings.

For more  exotic spin assignments of the new boson, a large  variety of tensor couplings
is possible. For completeness, we write down the  amplitudes that describe 
the coupling of this particle to the vector bosons, cf.  \Bref{Gao:2010qx}, 
%%%%%%%%%%%%%%%%%%%%%%%
%
\begin{eqnarray}
A(\PX_{J=1} \to \PVB\PVB)  = 
 g^{(1)}_{1} \left[  (\epsilon_1^* q) (\epsilon_2^* \epsilon_{\PX}) 
  +  (\epsilon_2^* q) (\epsilon_1^* \epsilon_{\PX})  \right]
+g^{(1)}_{ 2} \epsilon_{\alpha\mu\nu\beta} \epsilon_{\PX}^{\alpha} 
\epsilon_1^{*,\mu} \epsilon_2^{*,\nu} {\tilde q}^\beta \,,
\label{eq:ampl-spin1} 
\end{eqnarray}
%
%%%%%%%%%%%%%%%%%%%%%%%
%%%%%%%%%%%%%%%%%%%%%%%
%
\begin{eqnarray}
\label{eq:ampl-spin2-a} 
&&  A(\PX_{J=2} \to \PVB\PVB)  = \Lambda^{-1} \left [ 
2 g^{(2)}_{ 1} t_{\mu \nu} f^{*1,\mu \alpha} f^{*2,\nu \alpha} 
+ 2 g^{(2)}_{ 2} t_{\mu \nu} \frac{q_\alpha q_\beta }{\Lambda^2} 
f^{*1,\mu \alpha}  f^{*2,\nu,\beta}
\right.  \nonumber \\
&& \left. + g^{(2)}_{ 3} 
 \frac{{\tilde q}^\beta {\tilde q}^{\alpha}}{\Lambda^2}
t_{\beta \nu} ( 
f^{*1,\mu \nu} f^{*2}_{\mu \alpha} + f^{*2,\mu \nu} f^{*1}_{\mu \alpha} 
) + g^{(2)}_{ 4}\frac{{\tilde q}^{\nu} {\tilde q}^\mu}{{\Lambda^2} } 
t_{\mu \nu} f^{*1,\alpha \beta} f^{*(2)}_{\alpha \beta}
\right. \nonumber \\
&& \left. + m_{ V}^2  \left ( 
2 g^{(2)}_{ 5}  t_{\mu\nu} \ep_1^{*\mu} \ep_2^{*\nu} 
+2 g^{(2)}_{ 6} \frac{{\tilde q}^\mu q_\alpha}{\Lambda^2}  t_{\mu \nu}
\left ( \ep_1^{*\nu} \ep_2^{*\alpha} - 
\ep_1^{*\alpha} \ep_2^{*\nu} \right ) 
+g^{(2)}_{ 7}
 \frac{{\tilde q}^\mu {\tilde q}^\nu}{\Lambda^2}  
t_{\mu \nu} \ep^*_1 \ep^*_2
\right) 
\right.  \nonumber \\
&& %\left. 
+g^{(2)}_{ 8} \frac{{\tilde q}_{\mu} {\tilde q}_{\nu}}{\Lambda^2} 
 t_{\mu \nu} f^{*1,\alpha \beta} {\tilde f}^{*(2)}_{\alpha \beta}
+ g^{(2)}_{ 9} t_{\mu \alpha} {\tilde q}^\alpha \epsilon_{\mu \nu \rho \sigma} \epsilon_1^{*\nu} 
\epsilon_2^{*\rho} q^{\sigma} \nonumber \\
&& \left. +\frac{g^{(2)}_{ 10} t_{\mu \alpha} {\tilde q}^\alpha}{\Lambda^2}
\epsilon_{\mu \nu \rho \sigma} q^\rho {\tilde q}^{\sigma} 
\left ( \epsilon_1^{*\nu}(q\epsilon_2^*)+
\epsilon_2^{*\nu}(q\epsilon_1^*) \right )
\right ] \,,
\end{eqnarray}
%
%%%%%%%%%%%%%%%%%%%%%%%
and to the fermions
%%%%%%%%%%%%%%%%%%%%%%%%%%%%%
%
\begin{eqnarray}
&& A(\PX_{J=1} \to \Pf\bar{\Pf}) = \epsilon^\mu
\bar u_{1}
\left (
\gamma_\mu  \left (  \rho^{(1)}_{ 1} + \rho^{(1)}_{ 2} \gamma_5 \right )
+ \frac{m_q {\tilde q}_\mu}{\Lambda^2} \left ( \rho^{(1)}_{ 3} +
\rho^{(1)}_{ 4} \gamma_5
\right ) \right ) v_{2}\,,
\label{eq:ampl-spin1-qq}
\\
&& A(\PX_{J=2} \to \Pf\bar{\Pf}) = \frac{1}{\Lambda} t^{\mu \nu}
\bar u_{1} \left (
 \gamma_\mu {\tilde q}_\nu \left ( \rho^{(2)}_{ 1} + \rho^{(2)}_{ 2}
\gamma_5 \right )
+ \frac{ m_q {\tilde q}_\mu {\tilde q}_\nu}{\Lambda^2}
\left (
 \rho^{(2)}_{ 3} + \rho^{(2)}_{ 4} \gamma_5
\right )
\right ) v_{2}\,.
\label{eq:ampl-spin2-qq}
\end{eqnarray}
%
%%%%%%%%%%%%%%%%%%%%%%%%%%%%%
The $g^{(1)}_{1}$ ($g^{(1)}_{2}$) coupling corresponds to parity-conserving interaction of 
a vector (pseudo-vector). 
The $g^{(2)}_{1}$ and $g^{(2)}_{5}$ couplings correspond to parity-conserving interaction of 
a spin-two tensor with the minimal couplings.
We note that both spin-one and spin-two assignments for the Higgs-like boson are rather exotic {and}, 
at the same time, require large number of couplings to fully parametrize the $\PX \to \PVB\PVB$ amplitude.  Moreover, 
since existing evidence already disfavors exotic spin
assignments~\cite{Chatrchyan:2012jja,CMS-PAS-HIG-12-041,CMS-PAS-HIG-13-002,CMS-PAS-HIG-13-003,ATLAS-CONF-2013-013,ATLAS-CONF-2013-029,ATLAS-CONF-2013-031}, 
%  CMS12041,CMS13002,CMS13003,
%  ATLAS13013,ATLAS13029,ATLAS13031},  
the complete measurement of the 
tensor couplings for both of these cases is  not particularly  motivated.  Nevertheless, further checks of some spin-two 
scenarios against available 
data and the development of robust methods to exclude certain spin-parity assignments  in a model-independent 
way are important and should be pursued. 

%-------------------------------------

\subsubsection{Tensor couplings in the spin-zero case}
\label{sec_jhu_sect2}

As we already mentioned above, there is a significant evidence that the Higgs-like boson is
in fact a spin-zero particle whose properties are very similar to that
of the SM Higgs boson.  
Taking this as a starting point, we need to focus on the Higgs-like boson precision phenomenology. 
One of the important questions  we should address is how 
to determine and/or put constraints on all the different couplings of the spin-zero 
Higgs-like boson that appear in Eqs.\,(\ref{eq:fullampl-spin0}) and~(\ref{eq:ampl-spin0-qq}).

To set up a framework for couplings determination, we consider the case of three independent complex 
couplings $g^{}_{ 1}$, $g^{}_{ 2}$, and $g^{}_{ 4}$  for each type of a vector boson ($\PZ,\gamma,\PW,\Pg$) and  
two independent complex couplings $\rho^{}_{ 1}$ and $\rho^{}_{ 2}$ for each type of a fermion, 
under the spin-zero assignment. 
Therefore, we require four independent real numbers to describe bosonic process and 
two real numbers  to describe fermionic process provided that  the overall rate is treated separately
and one overall complex phase is not measurable. 
For a vector boson coupling, we can represent the four independent parameters by 
two  fractions 
($f_{g2}$ and $f_{g4}$) and two phases ($\phi_{g2}$ and $\phi_{g4}$), defined as
%%%%%%%%%%%%%%%%%%%%%%%%%%%%%
%
\begin{eqnarray}
&& f_{gi} =  \frac{|g^{}_{i}|^2\sigma_i}{|g^{}_{1}|^2\sigma_1+|g^{}_{2}|^2\sigma_2+|g^{}_{4}|^2\sigma_4}\,;
~~~~~~~~~
 \phi_{gi} = \arg\left(\frac{g_i}{g_1}\right)\,.
\nonumber
\label{eq:fractions}
\end{eqnarray}
%
%%%%%%%%%%%%%%%%%%%%%%%%%%%%%
We note that $\sigma_i$ is the effective cross-section of the process corresponding to $g^{}_{ i}=1, g_{j \ne i}=0$.
The parameter $f_{g4}$ is equivalent to the parameter $f_{a3}$ as
introduced by CMS~\cite{CMS-PAS-HIG-13-002}
% CMS13002
under the assumption $g_2=0$, and is the fraction of a $C\!P$-violating contribution.

Based on Eq.~(\ref{eq:ampl-spin0-qq}), we can define two parameters describing mixing 
in the fermion couplings, $f_{\rho2}$ and $\phi_{\rho2}$, equivalent to the boson coupling
parameters defined above.
We note that the fractions and phases can be defined independently for each type of boson 
or fermion couplings or they can be related to each other  by further assumptions about the electroweak quantum 
numbers of the Higgs-like boson. 
The advantage of introducing fractions $f_{g_i}$ to parametrize different couplings 
is that they are uniquely defined and have a clear experimental 
interpretation as effective fractions of yields of events corresponding to each independent scattering amplitude.
We note that contributions that originate from the interference of different amplitudes can be easily 
described using the parameterization introduced above.

%-------------------------------------

\subsubsection{Monte Carlo Simulation}
\label{sec_jhu_sect4}

Dedicated simulation programs can be used to describe various di-boson final states in the
production and decay to two vector bosons of the spin-zero, spin-one, 
and spin-two resonances in hadron-hadron collisions.
Implementation of the processes $\Pg\Pg / \Pq\Pqb \to \PX\to \PZ\PZ$ and  $\PW\PW\to 4\Pf$,
as well as  $\Pg\Pg / \Pq\Pqb\to \PX\to \PGg\PGg$,  into a Monte Carlo program {\sf JHU generator}, is  described 
in \Bref{JHUsupport}.
The {\sf JHU generator} incorporates  the general couplings of the $\PX$ particle 
to gluons and quarks in production and to vector bosons in decay and includes 
all spin correlations and interferences of all contributing amplitudes.
The program can be interfaced to parton shower simulation 
as well as full detector simulation through the Les Houches Event file
format~\cite{Alwall:2006yp}.   
The program also allows interfacing the decay of a spin-zero particle with the production 
simulated by other Monte Carlo programs. This makes it possible to include NLO QCD effects 
in the production through event generators such as {\sf POWHEG}~\cite{Nason:2004rx,Frixione:2007vw,Alioli:2010xd}.

The {\sf JHU generator} can be used to develop ideas and tools for the study of the Higgs-like boson with experimental data. 
To illustrate this point, we  show in \Tref{table-scenarios} several scenarios that can be explored.
Of course,  any combination  of the above scenarios  with proper interferences of amplitudes 
can be tested as well using the {\sf JHU generator}.
An example of interference study is shown in \refF{fig:zzobs}.

%%%%%%%%%%%%%%%%%%%%%%%
\begin{table}[h]
\caption{
List of representative scenarios for the analysis of the production and decay of an exotic $\PX$ particle
with quantum numbers $J^P$. 
The subscripts $m$ (minimal couplings)  and $h$ (couplings with higher-dimension operators) 
distinguish different scenarios, as discussed in the last column. 
}
\small
\begin{tabular}{lccc}
\hline
\vspace{0.1cm}
scenario & $\PX$ production  & $\PX\to \PVB\PVB$ decay   & comments \\
\hline
$0_m^+$ &  $\Pg\Pg\to \PX$ & $g_1^{(0)}\ne0$ in Eq.~(\ref{eq:fullampl-spin0})  & SM Higgs boson scalar \\
$0_h^+$  &  $\Pg\Pg\to \PX$ & $g_2^{(0)}\ne0$ in Eq.~(\ref{eq:fullampl-spin0}) & scalar with higher-dim. operators\\
$0^-$  &  $\Pg\Pg\to \PX$ & $g_4^{(0)}\ne0$ in Eq.~(\ref{eq:fullampl-spin0})  & pseudo-scalar  \\
$1^+$  &  $\PAQq\PQq\to \PX$  & $g_2^{(1)}\ne0$ in Eq.~(\ref{eq:ampl-spin1})      & exotic pseudo-vector \\
$1^-$   &  $\PAQq\PQq\to \PX$  & $g_1^{(1)}\ne0$ in Eq.~(\ref{eq:ampl-spin1})      & exotic vector \\
$2_m^+$ 
      & ~~$g^{(2)}_{\sss 1}\ne0$ in Eq.~(\ref{eq:ampl-spin2-a})~~  
      & ~~$g^{(2)}_{\sss 1}=g^{(2)}_{\sss 5}\ne0$ in Eq.~(\ref{eq:ampl-spin2-a})~~ 
      & ~~graviton-like tensor with min. couplings~~ \\
$2_h^+$ 
      & $g^{(2)}_{\sss 4}\ne0$ in Eq.~(\ref{eq:ampl-spin2-a}) 
      & $g^{(2)}_{\sss 4}\ne0$  in Eq.~(\ref{eq:ampl-spin2-a}) 
      & tensor with higher-dimension operators \\
\vspace{0.1cm}
$2_h^-$ 
      &  $g^{(2)}_{\sss 8}\ne0$ in Eq.~(\ref{eq:ampl-spin2-a})  
      & $g^{(2)}_{\sss 8}\ne0$ in Eq.~(\ref{eq:ampl-spin2-a})
      & ``pseudo-tensor''\\
\hline 
\end{tabular}
\normalsize
\label{table-scenarios}
\end{table}
%%%%%%%%%%%%%%%%%%%%%%%

%-------------------------------------

\subsubsection{Approaches to mixture and spin measurements}
\label{sec_jhu_sect5}

The basic idea behind any spin-parity measurement is that various spin-parity  assignments restrict allowed types of
interactions between the Higgs-like boson and other particles. This feature manifests itself in various kinematic
distributions of either the decay products of the Higgs-like particle or particles produced in association with it.
Let us first discuss the three processes which would allow the complete determination of the
tensor couplings in Eq.~(\ref{eq:fullampl-spin0}) for the $\PH\PZ\PZ$ interaction vertex. 
These are illustrated in \refF{fig:decay} and can be described as follows
\begin{itemize} 
\item production of a Higgs boson (in any process) and decay $\PH\to \PZ\PZ\to 4\Pl$, see \refF{fig:decay} (left);
\item $\PZ^*$ production and radiation of a Higgs boson (with decay into any final state), see \refF{fig:decay} (middle).
\item VBF production of a Higgs boson in $\PZ$ boson fusion (with decay into any final state), see \refF{fig:decay} (right).
\end{itemize}
In all cases, the spin-zero $\PH$ assignment leads to an isotropic distribution of $\cos\theta^*$ and $\Phi_1$
regardless of the $\PH$ production or decay mechanism.
The weak vector boson fusion process VBF can be used to determine $\PH\PW\PW$ and $\PH\PZ\PZ$ couplings 
in production on LHC using jet information~\cite{Englert:2012xt}, though this measurement relies stronger on dynamic 
distributions and does not allow to separate $\PH\PW\PW$ and $\PH\PZ\PZ$ couplings.

%%%%%%%%%%%%%%%%%%%%%%%
\begin{figure}[ht]
\centerline{
\setlength{\epsfxsize}{0.33\linewidth}\leavevmode\epsfbox{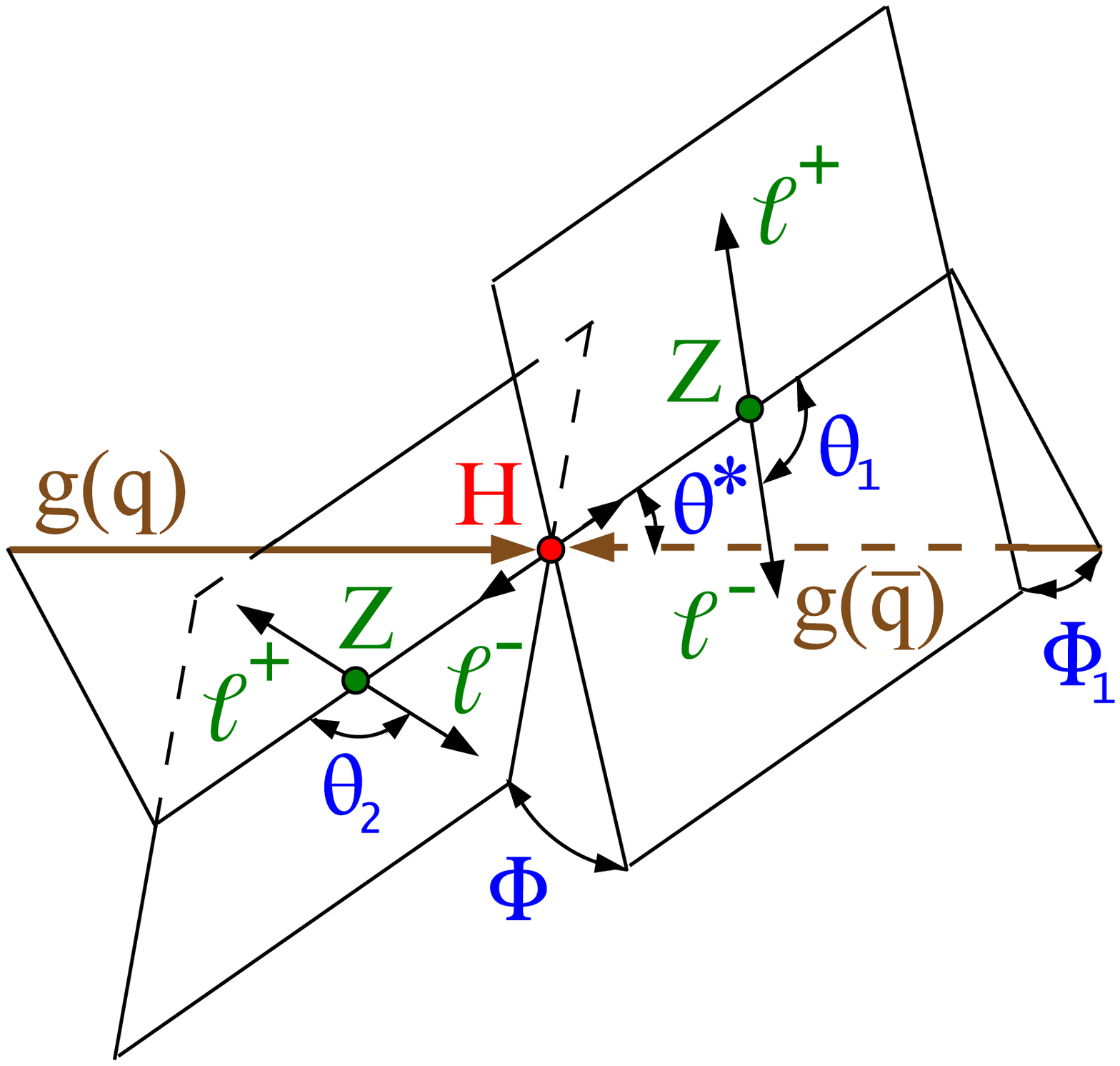}
\setlength{\epsfxsize}{0.33\linewidth}\leavevmode\epsfbox{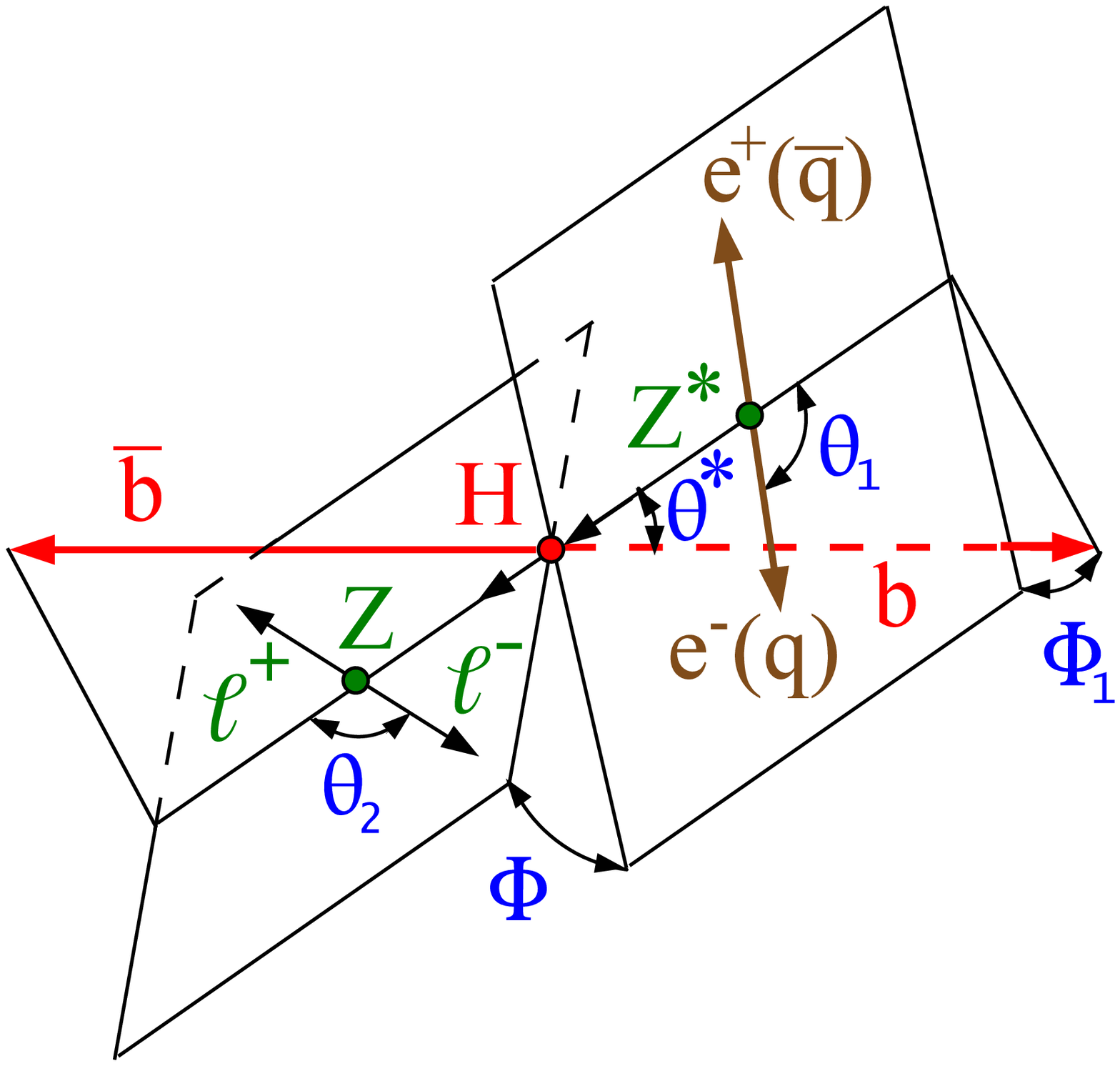}
\setlength{\epsfxsize}{0.33\linewidth}\leavevmode\epsfbox{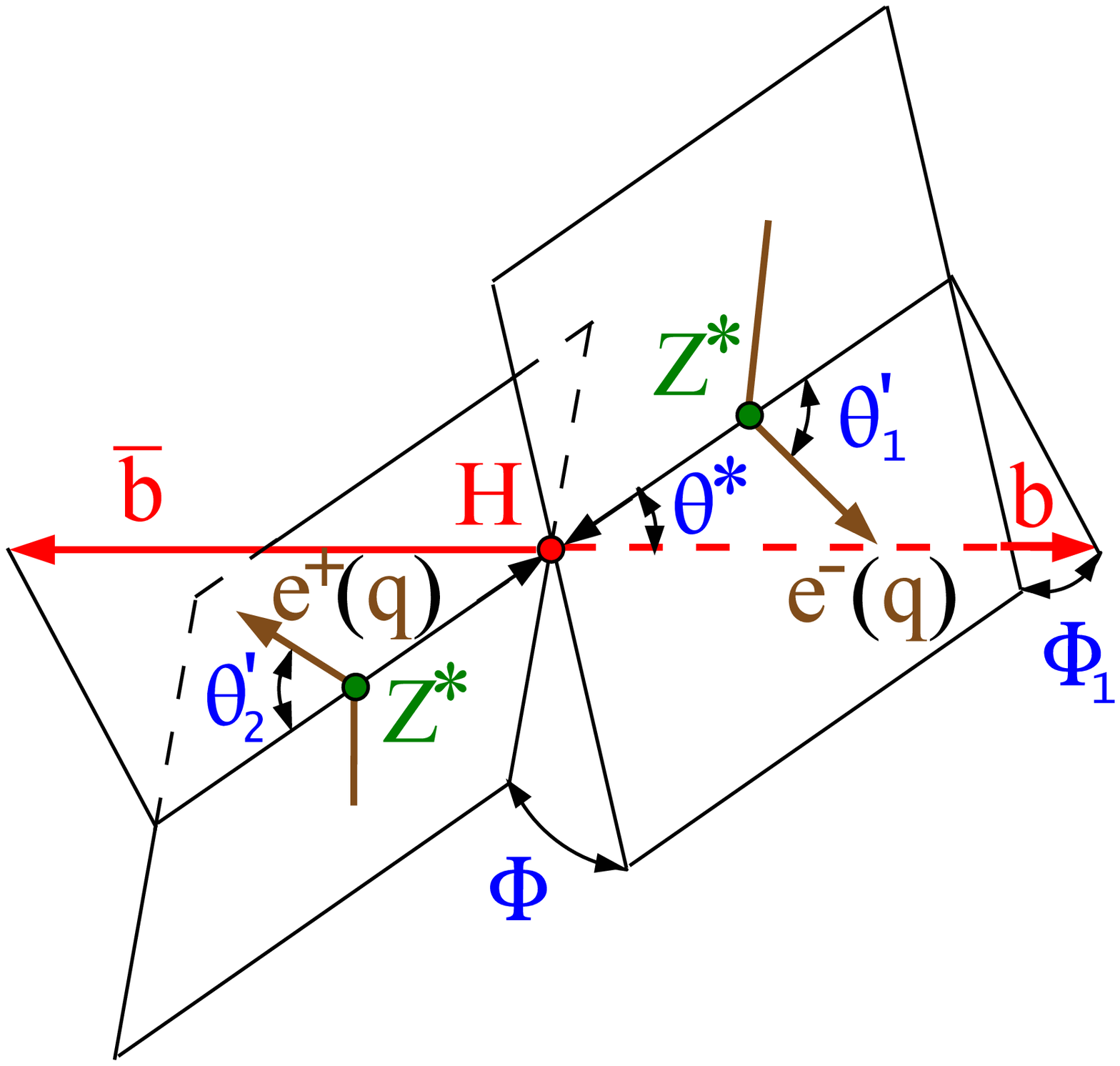}
}
\caption{
 Illustration of an $\PH$ particle production and decay in $\Pp\Pp$ collision $\Pg\Pg/\PAQq\PQq\to \PH\to \PZ\PZ\to 4\ell^\pm$,
$\Pq\Pqb\to \PZ^*\to \PZ\PH\to\ell^+\ell^-\PQb\bar{\PQb}$, or 
$\Pq\Pq^\prime\to \Pq\Pq^\prime \PV^*\PV^*\to \Pq\Pq^\prime\PH \to \Pq\Pq^\prime \PQb\PAQb$.
Five angles fully characterize orientation of the decay chain and are defined 
in the corresponding particle rest frames.
Both spin-zero and exotic spin assignments of $\PH$ are considered.
The $\Pep\Pem$ production is shown for comparison with the production modes on LHC.
}
\label{fig:decay}
\end{figure}
%%%%%%%%%%%%%%%%%%%%%%%
The ultimate goal of the analysis should be the experimental determination of  all  amplitudes that involve 
the new boson and two gauge bosons or fermions. The techniques discussed in \Bref{Gao:2010qx} are 
suited for such measurements since  parameters in the angular and mass distributions
become fit parameters in analysis of data. However, such multi-parameter 
fits require large samples of the signal events. 
Therefore, there are two steps in understanding the tensor couplings  and the spin of the Higgs-like boson:
\begin{itemize} 
\item discrete hypothesis testing;
\item fit of all coupling parameters.
\end{itemize}
For the first step, a simplified, but still optimal,  analysis approach 
can be developed that employs just one observable to differentiate between the two hypotheses, 
as we explain in the next section (a second observable can be used for signal-to-background separation).
The second step requires a complete multi-dimensional fit of all 
coupling parameters using a complete set of kinematic observables which is the ultimate goal of this research 
program.
In the remainder of this note we  concentrate on the $\PH\to \PZ\PZ\to 4\Plpm$ process in \refF{fig:decay} 
and other processes can be studied by analogy~\cite{Gao:2010qx,Englert:2012xt}.

%-------------------------------------

\subsubsection{Discrete hypothesis testing}
\label{sec_jhu_sect6}

For  discrete hypothesis testing one can create a kinematic discriminant 
(MELA approach)~\cite{Chatrchyan:2012ufa} which is constructed from the
ratio of probabilities  
for the SM signal and alternative signal $J^P_x$ hypotheses
%%%%%%%%%%%%%%%%%%%%%
\begin{eqnarray}
\label{eq:melaSig}
{D_{J^P_x}}=\left[1+\frac{{\mathcal P}_{J^P_x} (m_{4\ell}; m_1, m_2, \vec\Omega) }
{{\mathcal P}_{\mathrm{SM}} (m_{4\ell}; m_1, m_2, \vec\Omega) } \right]^{-1}
\,.
\end{eqnarray}
%%%%%%%%%%%%%%%%%%%%% 
Here ${\mathcal P}$ are the probabilities, 
as a function of masses $m_i$ and angular observables $\vec\Omega$, as
defined in \refF{fig:decay} 
and discussed in \Bref{Gao:2010qx}.
The separation power depends on information contained in kinematic distributions. 
All this information is combined in an optimal way in the $D_{J^P_x}$ observable. 
The expected~\cite{Gao:2010qx} and observed separation with LHC 
data~\cite{Chatrchyan:2012jja,CMS-PAS-HIG-12-041,CMS-PAS-HIG-13-002,CMS-PAS-HIG-13-003,ATLAS-CONF-2013-013,ATLAS-CONF-2013-029,ATLAS-CONF-2013-031} 
%  CMS12041,CMS13002,CMS13003,
%  ATLAS13013,ATLAS13029,ATLAS13031}
indicates that most of the hypotheses
listed in \Tref{table-scenarios} are strongly disfavored, both with $\Pg\Pg$ and
$\Pq\Pqb$ production mechanisms of the spin-two particle. 
The typical distributions for the $0^-$ analysis are shown in \refF{fig:zzobs}.
Consistency between the data and the two spin-parity models could be judged from the
observed and expected distribution of $-{2\ln({\mathcal L}_1/{\mathcal L}_2)}$ with the likelihood 
${\mathcal L}_k$ evaluated for two models based on distributions of $D_{0^-}$. 

%%%%%%%%%%%%%%%%%%%%%%%
\begin{figure}[htb]
\begin{center}
\includegraphics[scale=0.25]{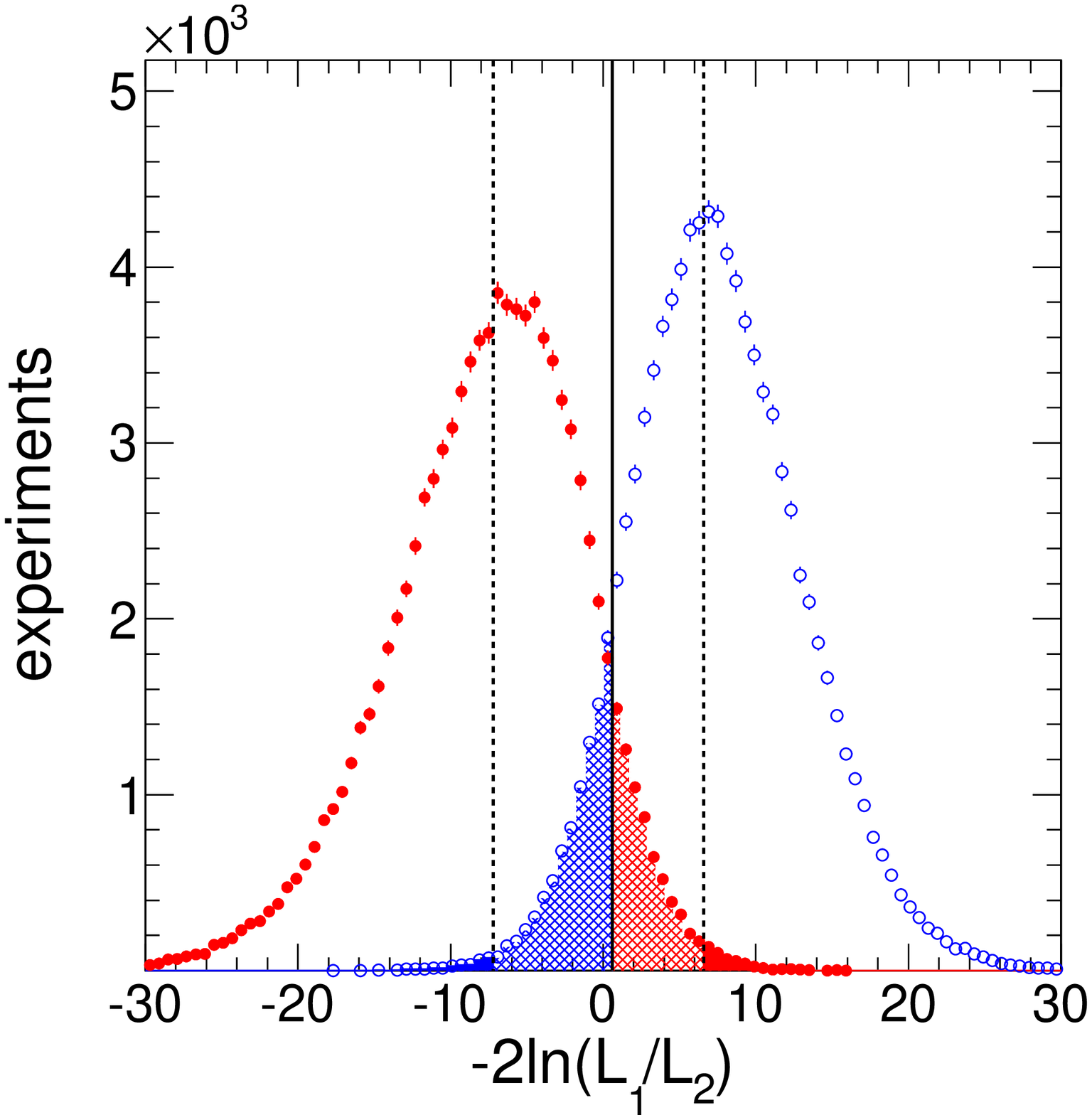}
\includegraphics[scale=0.25]{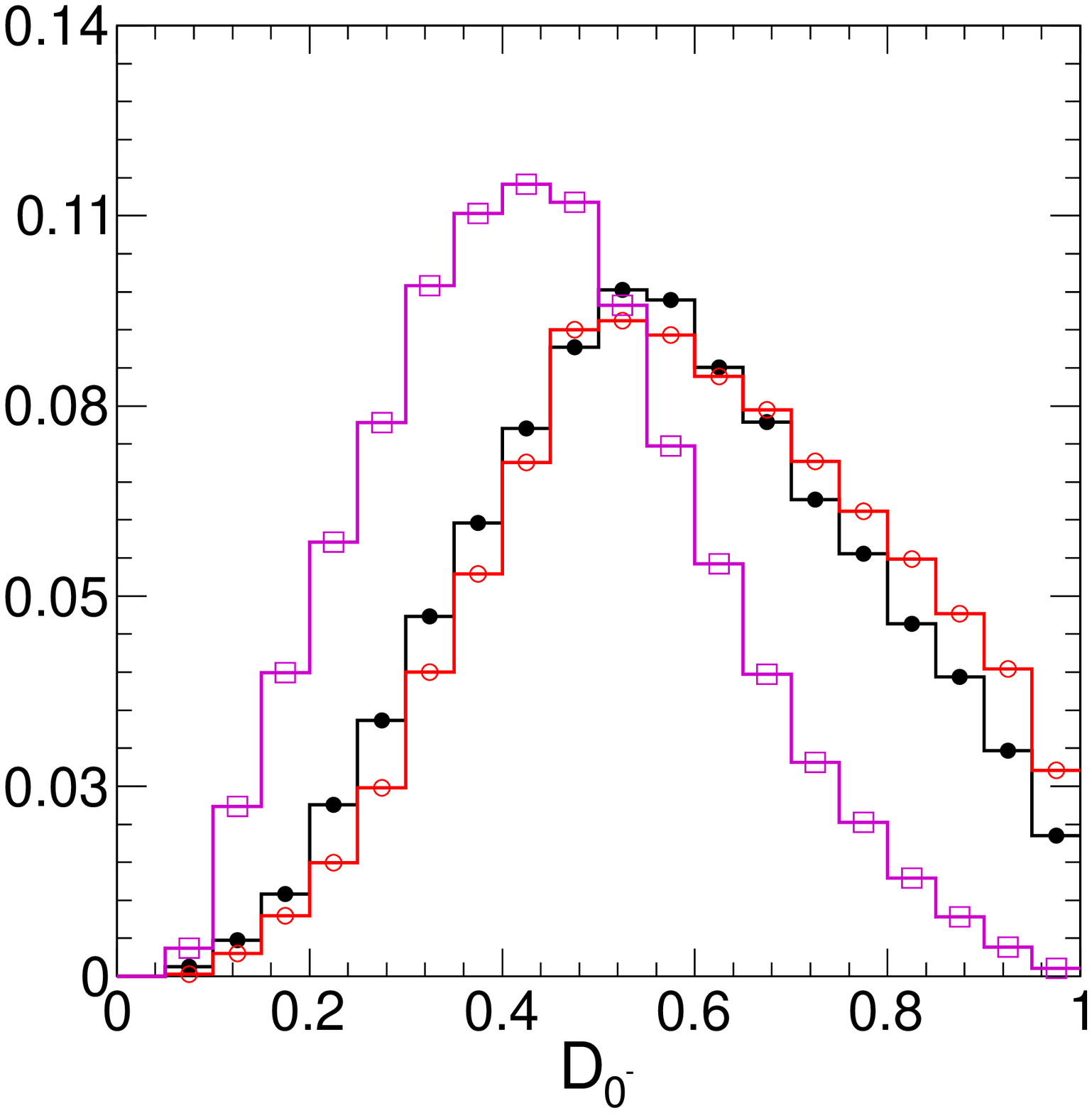}
\includegraphics[scale=0.25]{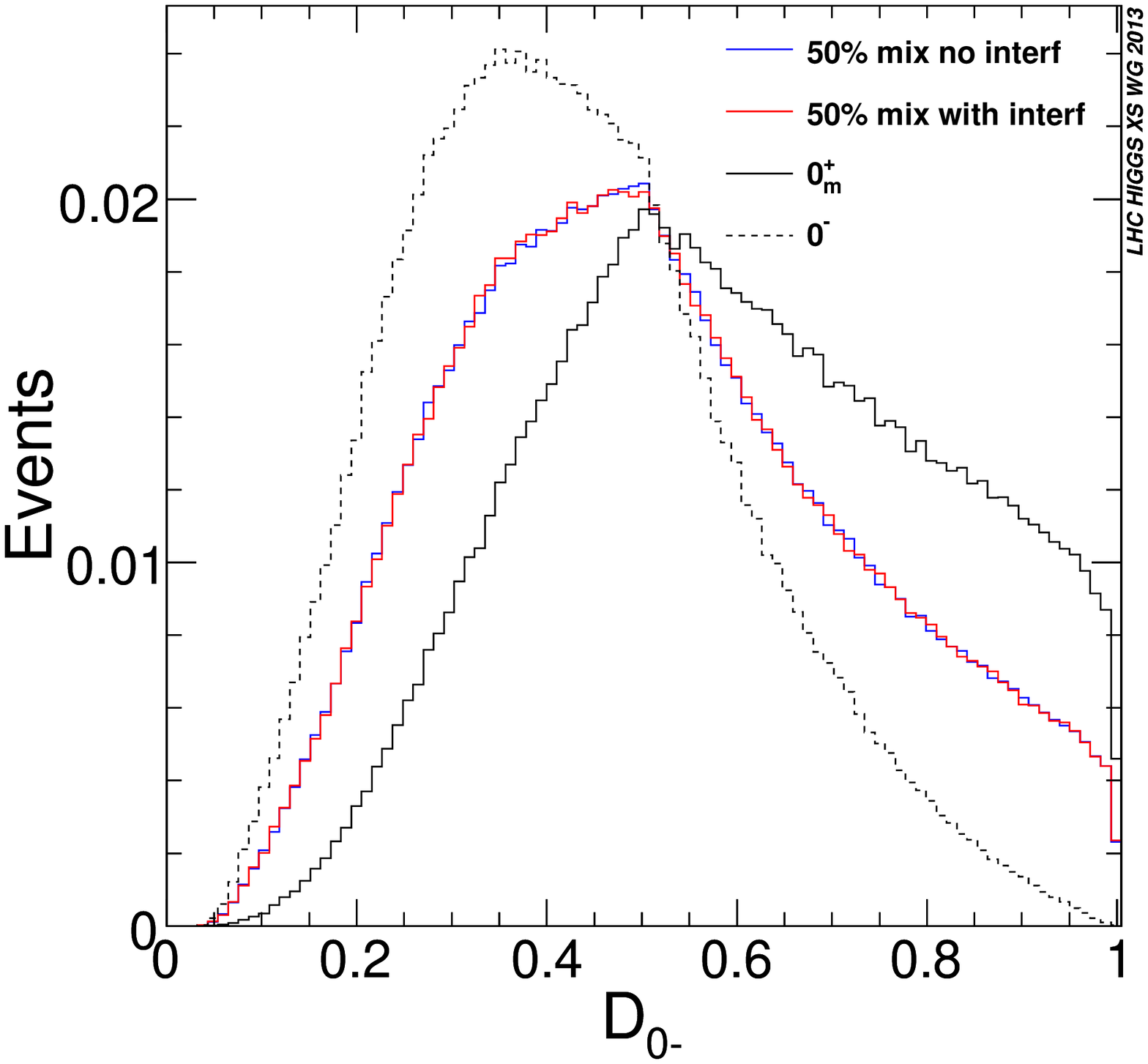}
\end{center}
\caption{
Left: Distributions of $-{2\ln({\mathcal L}_1/{\mathcal L}_2)}$ with the likelihood 
${\mathcal L}_k$ evaluated for two models and shown for a large number 
of generated experiments in the analysis~\cite{Gao:2010qx}.
Middle: Distributions of $D_{0^-}$ in the $\PX\to \PZ\PZ$ analysis 
for the non-resonant $\PZ\PZ$ background (black solid circles), and two signal hypotheses: 
SM Higgs boson (red open circles), $0^-$ (magenta squares)~\cite{Gao:2010qx}.
Right: Distribution of $D_{0^-}$ for event samples with generated $f_{g4}=0.5$
with and without interference of the $g_1$ and $g_4$ terms included,
and compared to the two pure cases of $0^+_m$ and $0^-$.
}
\label{fig:zzobs}
\end{figure}
%%%%%%%%%%%%%%%%%%%%%%%

The above approach exploits in the optimal way the maximum available kinematic 
information but at the expense of testing very specific models.
With increasing data-sample, we can afford loosening some requirements 
with the advantage of gaining generality. 
For instance, while the production mechanism for the spin-one boson is $\Pq\Pqb$ annihilation, 
the production of the spin-two bosons receives contributions from both $\Pg\Pg$ and $\Pq\Pqb$ initial states in a composition 
that is model-dependent and a priori unknown.  
Since a particular type of production
mechanism for a non-zero-spin state may influence the kinematics of the decay products 
of the Higgs-like boson, any spin-hypothesis test may strongly depend on the production model.
%This is unfortunate since  a particular 
%type of production mechanism for a higher-spin state may influence the kinematics 
%of the decay products of the Higgs-like boson, making any analysis strongly production-model dependent.
 
It is desirable to extend the above hypothesis testing approach in 
a way that does not depend on the production model. 
This feature can be easily achieved by considering the unpolarized $\PX$-boson production
by either averaging over the spin degrees of freedom of the produced $\PX$-boson or, equivalently,
integrating over the two production angles $\cos\theta^*$ and $\Phi_1$, defined in \refF{fig:decay},
in the $J^P_x$ probability expectation ${\mathcal P}_{J^P_x}$~\cite{Gao:2010qx}. 
This leads to the spin-averaged matrix element squared for 
the $\PX$-decay as the probability ${\mathcal P}$ in the kinematic discriminant
%%%%%%%%%%%%%%%%%%%%%
\begin{eqnarray}
\label{eq:melaSigProd}
{D_{J^P_x}^{\mathrm{decay}}}=\left[1+\frac{
\int \mathrm d\Phi_1  \mathrm d\cos\theta^{*}  \,\,{\mathcal P}_{J^P_x} (m_{4\ell}; m_1, m_2, \vec\Omega) 
}
{{\mathcal P}_{\mathrm{SM}} (m_{4\ell}; m_1, m_2, \vec\Omega) } \right]^{-1}
\,.
\end{eqnarray}
%%%%%%%%%%%%%%%%%%%%% 
This method applies to any possible hypothesis with non-zero spin.
The spin-zero kinematics is already independent of the production mechanism due to 
lack of spin correlations for any spin-zero particle; as the  result $\cos\theta^*$ and $\Phi_1$ 
distributions are isotropic for any production model. 

Below we illustrate this approach with the
example of spin-two, where we compare two production models, $2^+_{m\,\PAQq\PQq}$ and $2^+_{m\,\Pg\Pg}$.
The distribution of the most optimal discriminant (which includes production and decay variables)
is different depending on the mixture of $\Pq\Pqb$ annihilation and gluon-gluon fusion production mechanisms, as
shown in \refF{fig:decayDiscriminant}. As expected, the discriminant built for a specific production mechanism 
has different distributions depending on the production process of the sample (being sub-optimal if the production 
mechanism is different from the one for which the discriminant is built).
The new discriminant ${D_{2^+_m}^{\mathrm{decay}}}$ from Eq.\,(\ref{eq:melaSigProd})
does not depend on the production mechanism, 
as shown in \refF{fig:decayDiscriminant} (right) at the expense of having somewhat less separation power 
between the SM and the alternative  spin hypothesis.
The production-dependent information (the two production angles, as well as the transverse momentum distribution
of the boson) can still have a second order effect on the discriminant distribution through the detector acceptance effects.
However, this effect is found to be small in the fully reconstructed $\PH\to \PZ\PZ\to 4\Plpm$ process.

%%%%%%%%%%%%%%%%%
\begin{figure}[b]
\centerline{
\setlength{\epsfxsize}{0.33\linewidth}\leavevmode\epsfbox{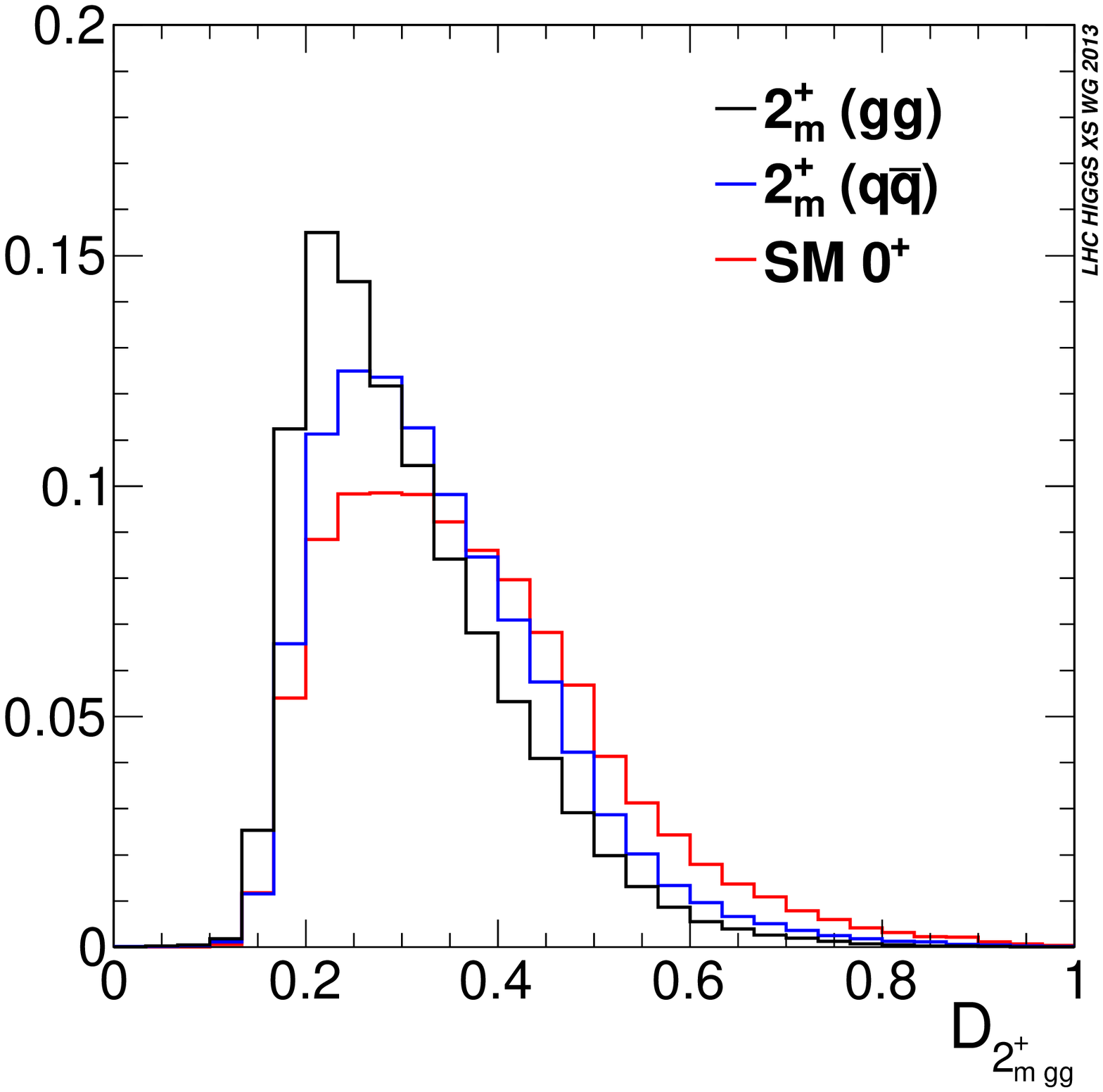}
\setlength{\epsfxsize}{0.33\linewidth}\leavevmode\epsfbox{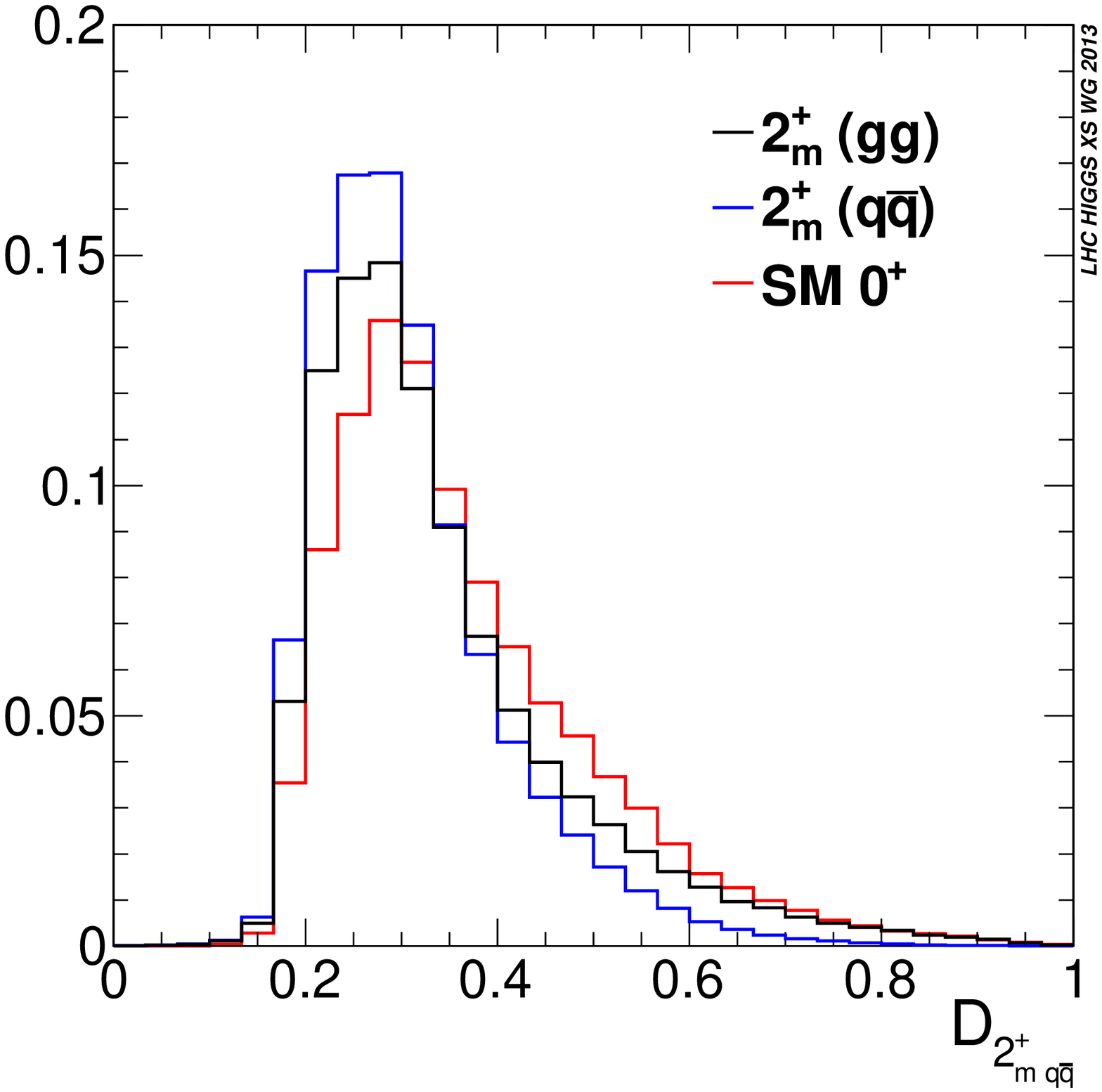}
\setlength{\epsfxsize}{0.33\linewidth}\leavevmode\epsfbox{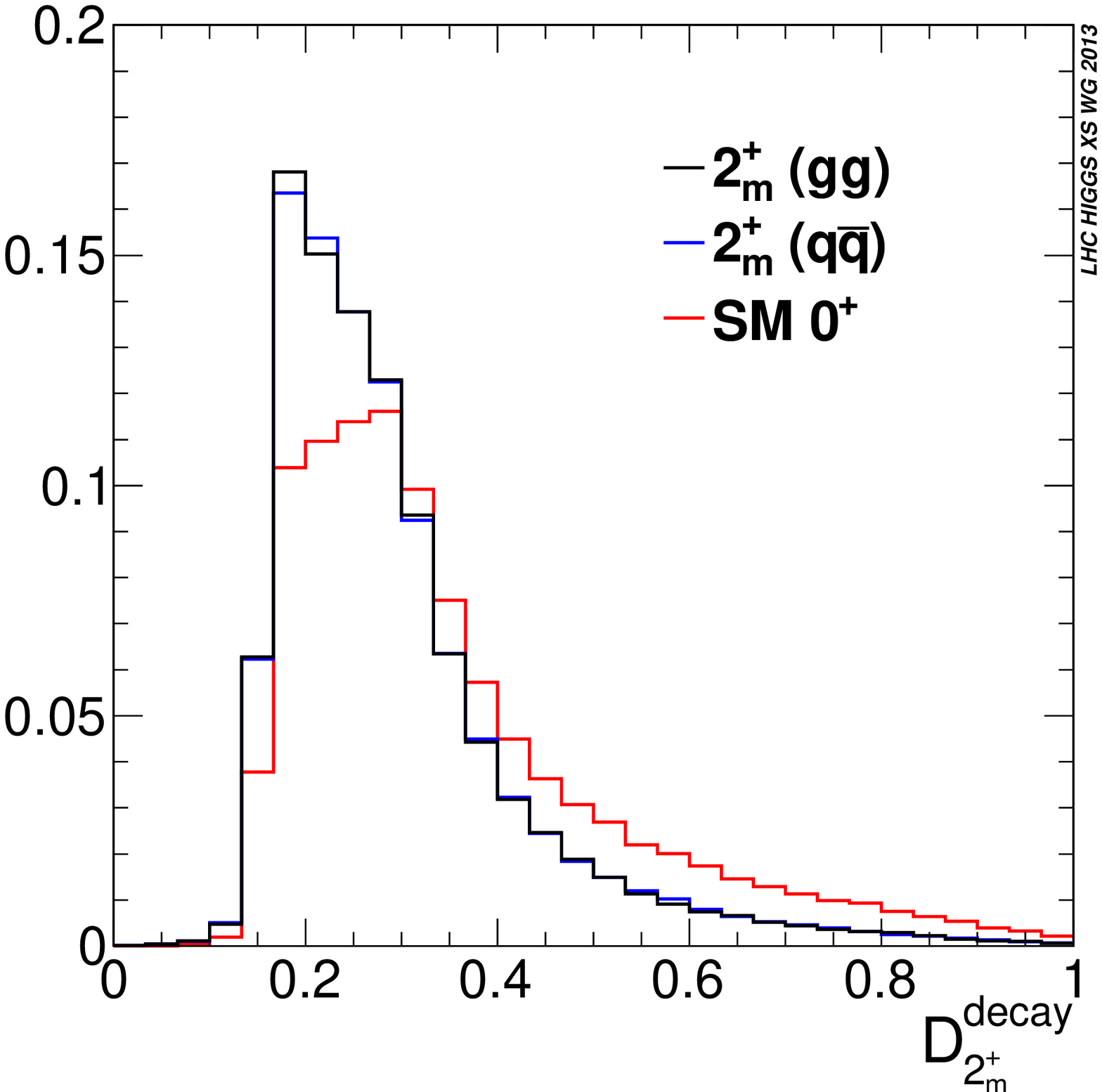}
}
\caption{Distribution of the discriminants
${D_{2^+_{m\,\Pg\Pg}}}$ (left),
${D_{2^+_{m\,\PAQq\PQq}}}$ (middle),
${D_{2^+_m}^{\mathrm{decay}}}$ (right)
evaluated with three different samples: SM Higgs boson, spin-two produced through gluon-gluon fusion 
and produced through $\PAQq\PQq$ annihilation.
}
\label{fig:decayDiscriminant}
\end{figure}
%%%%%%%%%%%%%%%%%

Finally we would like to point out that while discrete hypothesis testing allows us to quickly
exclude the most plausible models, complete exclusion of all models with exotic spin requires
more time. Such tests may see decreasing level of interest as more and more exotic models
are tested. Nonetheless, it should still be possible to exclude all possibilities eventually, assuming
all tests remain consistent with the spin-zero particle. 
For example, in the spin-one case, only two couplings are present and it is relatively easy to
span the continuous spectrum of couplings.  
The situation with spin-two is less trivial, but it might be possible to come up with the most pessimistic 
scenario (kinematics closest to the SM Higgs boson) excluding which could bring us to the
situation when the full spectrum of exotic-spin models can be excluded. 
Production-model-independent test discussed above is an important  step in that direction.

%-------------------------------------

\subsubsection{Continuous hypothesis testing of mixing structure}
\label{sec_jhu_sect7}

As discussed above, the ultimate approach requires a complete multi-dimensional fit of all 
coupling parameters using a complete set of kinematic observables.
However, first we would like to note an interesting feature in the fit for the $f_{g4}$ parameter, 
as introduced by the CMS experiment with the $f_{a3}$
measurement~\cite{CMS-PAS-HIG-13-002}.
%CMS13002}.
The distribution of the $D_{0^-}$ observable for a fraction $f_{g4}=0.5$ of $0^+_m$ ($g_1$)
and $0^-$ ($g_4$) contributions is independent of the interference effects between the two
amplitudes as shown in \refF{fig:zzobs} (right).
Distributions appear nearly identical for any phase of the mixture $\phi_{g4}$
and also for the case when the interference contribution is neglected altogether. 
The same feature is observed for other fractions of $f_{g4}$. 
This leads to a simplified but still optimal analysis to determine $f_{g4}$ 
from measuring the shape of the $D_{0^-}$ observable.
The CMS experiment estimates a current expected precision on $f_{g4}$ 
of about 0.40 at 68\%
CL~\cite{Chatrchyan:2012jja,CMS-PAS-HIG-12-041,CMS-PAS-HIG-13-002,CMS-PAS-HIG-13-003,ATLAS-CONF-2013-013,ATLAS-CONF-2013-029,ATLAS-CONF-2013-031}. 
%  CMS12041,CMS13002,CMS13003,
%  ATLAS13013,ATLAS13029,ATLAS13031}.
This translates to about 0.08 expected precision with 300\,fb$^{-1}$ at $14\UTeV$ run of LHC. 
As long as only a limit is set on  $f_{g4}$ and this limit is not much smaller than the above
expectation, such an analysis may be sufficient for setting a limit on the $C\!P$-violating
contribution to the $\PH\PZ\PZ$ coupling.

For a more complete treatment, for example if a non-zero value of $f_{g4}$ is measured,
we follow the notation of \Bref{Gao:2010qx} for the likelihood function for $N$ candidate events 
%%%%%%%%%%%%%%
%
\begin{equation}
{{\mathcal L}} =  \exp\left( - \sum_{J=0}^2n_{\sss J}-n_{\rm bkg}  \right) 
\prod_i^{N} \left( ~\sum_{J=0}^2
n_{\sss J} \times{\mathcal P}_{\sss J}(\vec{x}_{i};~\vec{\zeta}_{\sss J};~\vec{\xi})  
+n_{\rm bkg} \times{\mathcal P}_{\rm bkg}(\vec{x}_{i};~\vec{\xi})  
\right)\,,
\label{eq:likelihood}
\end{equation}
%%
%%%%%%%%%%%%%%
where $n_{\sss J}$ is the number of signal events for each 
resonance spin $J$, $n_{{\rm bkg}}$ is the number of background events and 
${\mathcal P}(\vec{x}_{i};\vec{\zeta};\vec{\xi})$ is the
probability density function for signal or background.
It is assumed that only one resonance is observed  in a given mass window and 
the three yields, $n_{\sss J}$, allow one to test different hypotheses.
Each event candidate $i$ is characterized by a set of eight observables
$\vec{x}_{i}=\{m_{4\ell},~m_1,~m_2,~\cos\theta^*,~\Phi_1,~\cos\theta_1,~\cos\theta_2,~\Phi\}_i$.
The number of observables can be extended or reduced, depending 
on the desired fit. 
The signal polarization parameters $f_{gi}$ and $\phi_{gi}$
are collectively denoted by $\vec{\zeta}_{\sss J}$, and the remaining parameters by $\vec{\xi}$.

The advantage of the maximum likelihood fit approach is that the likelihood
${{\mathcal L}}$ in Eq.~(\ref{eq:likelihood}) can be maximized
for a large set of parameters in the most optimal way without losing information. The disadvantage is the
difficulty to describe the detector response in the eight-dimensional space. The latter can be achieved with
certain approximations and may become the future direction of hypothesis testing in multi-parameter
models with increasing samples of signal events.

\subsection{Higgs characterization with {\sc FeynRules} and {\sc MadGraph~5}}
\label{sec:LM_mad}

In this section we introduce a complete framework, based on an effective field
theory description, that allows to perform characterization studies of a new
boson in all relevant channels in a consistent, systematic and accurate way.

In the following we present the implementation of an effective Lagrangian featuring bosons $X(J^P)$ with various assignments of spin/parity
 $J^P=0^+,0^-,1^+,1^-,2^+,2^-$ that can be used to test the nature of the recently-discovered
 new boson with mass around 125~GeV at the LHC~\cite{Aad:2012tfa, Chatrchyan:2012ufa}. The new
 states can couple to Standard Model particles via interactions of the
 minimal (and next-to-minimal) dimensions. The implementation of the
 Lagrangian is done in \textsc{FeynRules}~\cite{Christensen:2008py} and the
 corresponding model named \emph{Higgs Characterization model}. It extends and
 completes an earlier version used in \Bref{Englert:2012xt} and it is publicly
 accessible at~\cite{feynrules}. It is therefore
 available to all matrix element generators featuring an UFO
 interface~\cite{Degrande:2011ua}. For our study we have
 employed \textsc{MadGraph~5}~\cite{Alwall:2011uj}. Results at the NLO
 accuracy in QCD can be automatically (or semi-automatically) obtained
 via \textsc{aMC@NLO}~\cite{Frederix:2009yq,Hirschi:2011pa}.  

There are several advantages in having a first principle implementation
in terms of an effective Lagrangian which can be automatically
interfaced to a matrix element generator (and then to an event
generator). First and most important, all relevant production and decay
modes can be studied within the same model, from gluon-gluon fusion to
VBF as well as $\PVB\PH$ and $\PAQt\PQt$ associated
productions can be considered and the corresponding processes
automatically generated within minutes. Second, it is straightforward to
modify the model implementation to extend it further in case of need, by
adding further interactions, for example of higher-dimensions. Finally,
higher-order effects can be easily accounted for, by generating
multi-jet merged samples or computing NLO corrections with automatic
frameworks. 
All the detailed demonstration and analyses are currently in
progress~\cite{Artoisenet:2013jma}.\\

In the following we first write down the
effective Lagrangian explicitly and then show comparison plots with JHU results as
presented in \Bref{Bolognesi:2012mm} which are currently employed
by both ATLAS and CMS collaborations. 
We remind here that, even though we list several cases of interest in a
simple conversion table as a dictionary between JHU and our \emph{Higgs
Characterization model}~\cite{Artoisenet:2013jma}, in 
general the two approaches will be different.  Our implementation is
based on an effective theory approach valid up to the scale $\Lambda$,
while JHU describes the interaction between the new state and SM
particles  in terms of  anomalous couplings.

\subsubsection{The effective Lagrangian}

\begin{table}[h]
\begin{center}
\caption{Model parameters.}
\label{tab:param}
\begin{tabular}{lll}
\hline
 parameter\hspace*{5mm} & default value\hspace*{5mm} & description \\
\hline
 $\Lambda$ [GeV] & $10^3$ & cutoff scale \\
 $c_{\alpha}(\equiv \cos\alpha$) & 1 & mixing between $0^+$ and
	 $0^-$ \\
 $\kappa_i$ & 1 or 0 & dimensionless coupling parameter \\
\hline
\end{tabular}
\end{center}
\end{table}

\subsubsubsection{Spin 0}

The spin-0 $\PX$ interaction Lagrangian with fermions and vector-bosons
are given by 
\begin{align}
 {\mathcal L}_0^{\Pf} = -\big[ c_{\alpha}\kappa_{\PH\Pf\Pf}g_{\PH\Pf\Pf}\, \bar\psi_{\Pf}\psi_{\Pf}
               +s_{\alpha}\kappa_{\PA\Pf\Pf}g_{\PA\Pf\Pf}\, \bar\psi_{\Pf} i\gamma_5\psi_{\Pf} \big] \PX_0,
\label{FRMG5HC:1}
\end{align}
and
\begin{align}
 {\mathcal L}_0^{\PVB} =\bigg[&
  c_{\alpha}\kappa_{\SM}\big[\frac{1}{2}g_{\PH\PZ\PZ}\, \PZ_\mu \PZ^\mu 
                                +g_{\PH\PW\PW}\, \PW^{+\mu} \PW^{-\mu}\big] \nonumber\\
  &-\frac{1}{4}\big[c_{\alpha}\kappa_{\PH\PGg\PGg}g_{\PH\PGg\PGg} \, \PA_{\mu\nu}\PA^{\mu\nu}
        +s_{\alpha}\kappa_{\PA\PGg\PGg}g_{\PA\PGg\PGg}\,\PA_{\mu\nu}\widetilde \PA^{\mu\nu}
 \big] \nonumber\\
  &-\frac{1}{2}\big[c_{\alpha}\kappa_{\PH\PZ\PGg}g_{\PH\PZ\PGg} \, \PZ_{\mu\nu}\PA^{\mu\nu}
        +s_{\alpha}\kappa_{\PA\PZ\PGg}g_{\PA\PZ\PGg}\,\PZ_{\mu\nu}\widetilde \PA^{\mu\nu} \big] \nonumber\\
  &-\frac{1}{4}\big[c_{\alpha}\kappa_{\PH\Pg\Pg}g_{\PH\Pg\Pg} \, G_{\mu\nu}^aG^{a,\mu\nu}
        +s_{\alpha}\kappa_{\PA\Pg\Pg}g_{\PA\Pg\Pg}\,G_{\mu\nu}^a\widetilde G^{a,\mu\nu} \big] \nonumber\\
  &-\frac{1}{4}\frac{1}{\Lambda}\big[c_{\alpha}\kappa_{\PH\PZ\PZ} \, \PZ_{\mu\nu}\PZ^{\mu\nu}
        +s_{\alpha}\kappa_{\PA\PZ\PZ}\,\PZ_{\mu\nu}\widetilde \PZ^{\mu\nu} \big] \nonumber\\
  &-\frac{1}{2}\frac{1}{\Lambda}\big[c_{\alpha}\kappa_{\PH\PW\PW} \, \PW^+_{\mu\nu}\PW^{-\mu\nu}
        +s_{\alpha}\kappa_{\PA\PW\PW}\,\PW^+_{\mu\nu}\widetilde \PW^{-\mu\nu} \big]
 \bigg] \PX_0, 
\end{align}
where the (reduced) field strength tensors are
\begin{align}
 V_{\mu\nu} &=\partial_{\mu}V_{\nu}-\partial_{\nu}V_{\mu}\quad (V=\PA,\PZ,\PWpm), \\
 G_{\mu\nu}^a &=\partial_{\mu}G_{\nu}^a-\partial_{\nu}G_{\mu}^a
  +g_sf^{abc}G_{\mu}^bG_{\nu}^c,
\end{align}
and the dual tensor is
\begin{align}
 \widetilde V_{\mu\nu} =\frac{1}{2}\epsilon_{\mu\nu\rho\sigma}V^{\rho\sigma}.
\end{align}
The model parameters in the Lagrangian that are possible to be modified are
listed in \Tref{tab:param}. This parameterization allows to describe the mixing
between $CP$-even and $CP$-odd states and correspondingly to give an effective
description of a reasonably ample range of $CP$-violating scenarios, such as those arising 
in SUSY or in a generic 2HDM.

The dimensionful couplings are set so as to reproduce a SM Higgs in the case $c_\alpha=1$ and
a pseudo scalar in a 2HDM with $\tan \beta=1$ for the default values of
$\kappa_i$, e.g. $g_{\PH\Pf\Pf}=m_{\Pf}/v$ and $g_{\PH\PVB\PVB}=2m_{\PVB}^2/v$ as well as
$g_{\PH\Pg\Pg}=-\alpha_s/3\pi v$ in the heavy top loop limit.

%%%%%%%%%%%%%%%%%%%%%%%%%%%%%%%%%%%%%%%%%%%%%%%%%%
\subsubsubsection{Spin 1}

The spin-1 $\PX$ interaction Lagrangian with fermions is
\begin{align}
 {\mathcal L}_1^{\Pf} = \sum_{f=\PQu,\PQd} 
      \bar\psi_{\Pf} \gamma_{\mu}(\kappa_{f_a}a_{\Pf} - \kappa_{\Pf_b}
           b_{\Pf}\gamma_5)\psi_{\Pf} \PX_{1}^{\mu},
\end{align}
where $\PQu$ and $\PQd$ denote the up-type and down-type quarks, respectively. 
The $a_{\Pf}$ and $b_{\Pf}$ are the SM couplings, i.e.
\begin{align} 
 a_{\PQu} &= \frac{g}{2\cw}\Big(\frac{1}{2}-\frac{4}{3}\sw^2\Big),\quad
 b_{\PQu} = \frac{g}{2\cw}\frac{1}{2}, \\
 a_{\PQd} &= \frac{g}{2\cw}\Big(-\frac{1}{2}+\frac{2}{3}\sw^2\Big),\quad
 b_{\PQd} = -\frac{g}{2\cw}\frac{1}{2}.
\end{align}
The $\PX\PW\PW$ interaction at the lowest dimension is in
general~\cite{Hagiwara:1986vm}  
\begin{align}
 {\mathcal L}_1^{\PW} =
    &+i\kappa_{\PVB_1}g_{\PW\PW\PZ} (\PWp_{\mu\nu} \PWm^{\mu} - \PWm_{\mu\nu} \PWp^{\mu})
 \PX_{1}^{\nu} \nonumber\\ 
  &+ i\kappa_{\PVB_2}g_{\PW\PW\PZ} \PWp_{\mu} \PWm_{\nu} \PX_{1}^{\mu\nu}   \nonumber\\
  &-\kappa_{\PVB_3} \PWp_{\mu} \PWm_{\nu}(\partial^\mu \PX_{1}^{\nu} +
 \partial^\nu \PX_{1}^{\mu})  \nonumber\\
  &+ i\kappa_{\PVB_4} \PWp_{\mu} \PWm_{\nu}\widetilde \PX_{1}^{\mu\nu}   \nonumber\\
  &- \kappa_{\PVB_5} \epsilon_{\mu\nu\rho\sigma}
  [\PWp^{\mu} ({\partial}^\rho \PWm^{\nu})-({\partial}^\rho \PWp^{\mu})\PWm^{\nu}] \PX_{1}^{\sigma},
\end{align}
where $g_{\PW\PW\PZ}=e\cot\theta_\rw$.
Similarly, the $\PX\PZ\PZ$ interaction is given by~\cite{Keung:2008ve}
\begin{align}
 {\mathcal L}_1^{\PZ} =
  &-\kappa_{\PVB_3} \PX_{1}^{\mu}({\partial}^{\nu} \PZ_{\mu})\PZ_{\nu} \\
  &-\kappa_{\PVB_5} \epsilon_{\mu\nu\rho\sigma}  \PX_{1}^{\mu}
  \PZ^{\nu} ({\partial}^\rho \PZ^{\sigma}).
\end{align}

For $\PX_1=1^-$ in parity-conserving scenarios:
\begin{align}
 \kappa_{f_a,\PVB_1,\PVB_2,\PVB_3}\ne 0.
\end{align}

For $\PX_1=1^+$ in parity-conserving scenarios:
\begin{align}
 \kappa_{f_b,V_4,V_5}\ne 0. 
\end{align}
%

%%%%%%%%%%%%%%%%%%%%%%%%%%%%%%%%%%%%%%%%%%%
\subsubsubsection{Spin 2}

The spin-2 $\PX$ interaction Lagrangian starts from the dimension-five 
terms~\cite{Giudice:1998ck, Han:1998sg, Ellis:2012jv,Englert:2012xt}:
\begin{align}
 {\mathcal L}_2^{\Pf} = -\frac{1}{\Lambda}\sum_{f=\PQq,\ell} 
  \kappa_{f}\,T^f_{\mu\nu}\PX_2^{\mu\nu},
\end{align}
and
\begin{align}
 {\mathcal L}_2^{\PVB} = 
  &-\frac{1}{\Lambda} \kappa_{\PVB}\,T^V_{\mu\nu}\PX_2^{\mu\nu}  \nonumber\\
  &-\frac{1}{\Lambda} \kappa_{\PGg}\,T^{\PGg}_{\mu\nu}\PX_2^{\mu\nu}  \nonumber\\
  &-\frac{1}{\Lambda} \kappa_{\Pg}\,T^{\Pg}_{\mu\nu}\PX_2^{\mu\nu}, 
\end{align}
where $V=\PZ,\PWpm$ and 
$T^i_{\mu\nu}$ is the energy-momentum tensor of the SM fields; see 
e.g.~\cite{Hagiwara:2008jb} for the explicit forms.
The even higher dimensional terms~\cite{Bolognesi:2012mm}, dimension-seven, are
also implemented as 
\begin{align}
 {\mathcal L}_2^{V_{\mathrm{HD}}} = 
  &-\frac{1}{4}\frac{1}{\Lambda^3}\,\kappa_{\PVB_1}(\partial_{\nu}(\partial_{\mu}   (\PZ_{\rho\sigma}\PZ^{\rho\sigma}+2\PW^+_{\rho\sigma}\PW^{-\rho\sigma})))\PX_2^{\mu\nu}
 \nonumber\\
  &-\frac{1}{4}\frac{1}{\Lambda^3}\,\kappa_{\PVB_2}(\partial_{\nu}(\partial_{\mu}   (\PZ_{\rho\sigma}\widetilde \PZ^{\rho\sigma}+2\PW^+_{\rho\sigma}\widetilde \PW^{-\rho
\sigma})))\PX_2^
{\mu\nu} \nonumber\\
  &-\frac{1}{4}\frac{1}{\Lambda^3}\,\kappa_{\PGg_1}(\partial_{\nu}(\partial_{\
mu}
   A_{\rho\sigma}A^{\rho\sigma}))\PX_2^{\mu\nu} \nonumber\\
  &-\frac{1}{4}\frac{1}{\Lambda^3}\,\kappa_{\PGg_2}(\partial_{\nu}(\partial_{\
mu}
   A_{\rho\sigma}\widetilde A^{\rho\sigma}))\PX_2^{\mu\nu} \nonumber\\
  &-\frac{1}{4}\frac{1}{\Lambda^3}\,\kappa_{\Pg_1}(\partial_{\nu}(\partial_{\mu}
   G_{\rho\sigma}G^{\rho\sigma}))\PX_2^{\mu\nu} \nonumber\\
  &-\frac{1}{4}\frac{1}{\Lambda^3}\,\kappa_{\Pg_2}(\partial_{\nu}(\partial_{\mu}
   G_{\rho\sigma}\widetilde G^{\rho\sigma}))\PX_2^{\mu\nu},
\end{align}
where $V_{\mu\nu}=\partial_{\mu}V_{\nu}-\partial_{\nu}V_{\mu}$, etc, are the
reduced field strength tensor.

For $\PX_2=2^+$ in the RS-like graviton scenario:
\begin{align}
 \kappa_{f}=\kappa_{\PVB}=\kappa_{\PGg}=\kappa_{\Pg}\ne 0. 
\end{align}

For $\PX_2=2^+$ with the higher-dimensional operator in parity-conserving scenarios:
\begin{align}
 \kappa_{\PVB_1},\kappa_{\PGg_1},\kappa_{\Pg_1}\ne 0. 
\end{align}

For $\PX_2=2^-$ with the higher-dimensional operator in parity-conserving scenarios:
\begin{align}
 \kappa_{\PVB_2},\kappa_{\PGg_2},\kappa_{\Pg_2}\ne 0. 
\end{align}

\subsubsection{Comparison plots with JHU}

In this section we show comparison plots with JHU results in
\Bref{Bolognesi:2012mm} for $\Pp\Pp\to \PX\to VV^*\to 4\ell$. 

\subsubsubsection{Event generation}

50K events with $m_\PX=125\UGeV$ at the 8TeV-LHC were generated for each
spin state with the \textsc{MG5} (v1.5.9). Note that all the kinematical cuts for
leptons are removed.
%
%In practice we use the following \textsc{MG5} syntax for the $\PX\to \PZ\PZ$ an%alysis:
%%
%\begin{verbatim}
%import model HiggsCharac_v2.0
%generate p p > x0, x0 > mu- mu+ e- e+
%generate p p > x1, x1 > mu- mu+ e- e+
%generate p p > x2, x2 > mu- mu+ e- e+ / a QNP=1
%\end{verbatim}
%%
We note that we remove photons for diagram generation for spin-2, 
while the $\PX_0$-$\PZ$-$\PGg$ contribution for spin-0
 can be removed by setting $\kappa_{\PH\PZ\PGg}=\kappa_{\PA\PZ\PGg}=0$.
%Similarly for the $\PX\to \PW\PW$ analysis:
%%
%\begin{verbatim}
%import model HiggsCharac_v2.0
%generate p p > x0, x0 > mu- vm~ e+ ve
%generate p p > x1, x1 > mu- vm~ e+ ve
%generate p p > x2, x2 > mu- vm~ e+ ve
%\end{verbatim}
%%
We also note that the spin-2 case has seven diagrams, one double-$\PV$
resonant diagram and six single-$\PV$ diagrams including the four
point $\PX_2$-$\PV$-$\ell$-$\ell$ diagrams. Those single resonant
contributions can be removed by setting $\kappa_{\ell}=0$.

\subsubsubsection{Distributions}

The translation of the notation for the kinematical variables to JHU is
\begin{align}
 \phi_1\to\Phi_1,\quad \phi_1-\phi_2\to\Phi. 
\end{align}
Note also that the azimuthal angles are defined from 0 to $2\pi$ here,
while $-\pi$ to $\pi$ in the JHU paper. The parameter set for the JHU
comparison are listed in \Tref{tab:JHU}.

As shown in Figs.~\ref{fig:ZZ} and \ref{fig:WW}, all the distributions agree
with the JHU ones. Moreover, the lowest dimensional spin-2 is consistent
with the RS model in \textsc{MG5}.
The comparison for the higher dimensional terms for spin-2 is in progress
and will be reported elsewhere~\cite{Artoisenet:2013jma}.

\begin{table}
\begin{center}
\caption{Parameter set for the JHU comparison; see also Table I in the
 JHU implementation~\cite{Bolognesi:2012mm}.}
\label{tab:JHU}
\begin{tabular}{lc}
\hline
 JHU scenario\hspace*{5mm} & HC parameter choice\hspace*{5mm} \\
\hline
 $0^+_m$ & $\kappa_{\PH\Pg\Pg}\ne0,\ \kappa_{\SM}\ne0,\ c_{\alpha}=1$\\
 $0^+_h$ & $\kappa_{\PH\Pg\Pg}\ne0,\ \kappa_{HVV}\ne0,\ c_{\alpha}=1$\\
 $0^-$ & $\kappa_{\PA\Pg\Pg}\ne0,\ \kappa_{\PA\PVB\PVB}\ne0,\ c_{\alpha}=0$\\
 $1^+$ & $\kappa_{fu_a}=\kappa_{fu_b}=\kappa_{fd_a}=\kappa_{fd_b}\ne0,\ \kappa_{V_5}\ne0$\\
 $1^-$ & $\kappa_{fu_a}=\kappa_{fu_b}=\kappa_{fd_a}=\kappa_{fd_b}\ne0,\ \kappa_{V_3}\ne0$\\
 $2^+_m$ & $\kappa_{\Pg}\ne0,\ \kappa_{\PVB}\ne0$\\
 $2^+_h$ & $\kappa_{\Pg_1}\ne0,\ \kappa_{\PVB_1}\ne0$\\
 $2^-_h$ & $\kappa_{\Pg_2}\ne0,\ \kappa_{\PVB_2}\ne0$\\
\hline
\end{tabular}
\end{center}
\end{table}

\begin{figure}
\begin{center}
spin-0\hskip 2.5cm
spin-1\hskip 2.5cm
spin-2\\[-1.3cm] 
\hskip -3cm
 \includegraphics[width=0.24\textwidth]{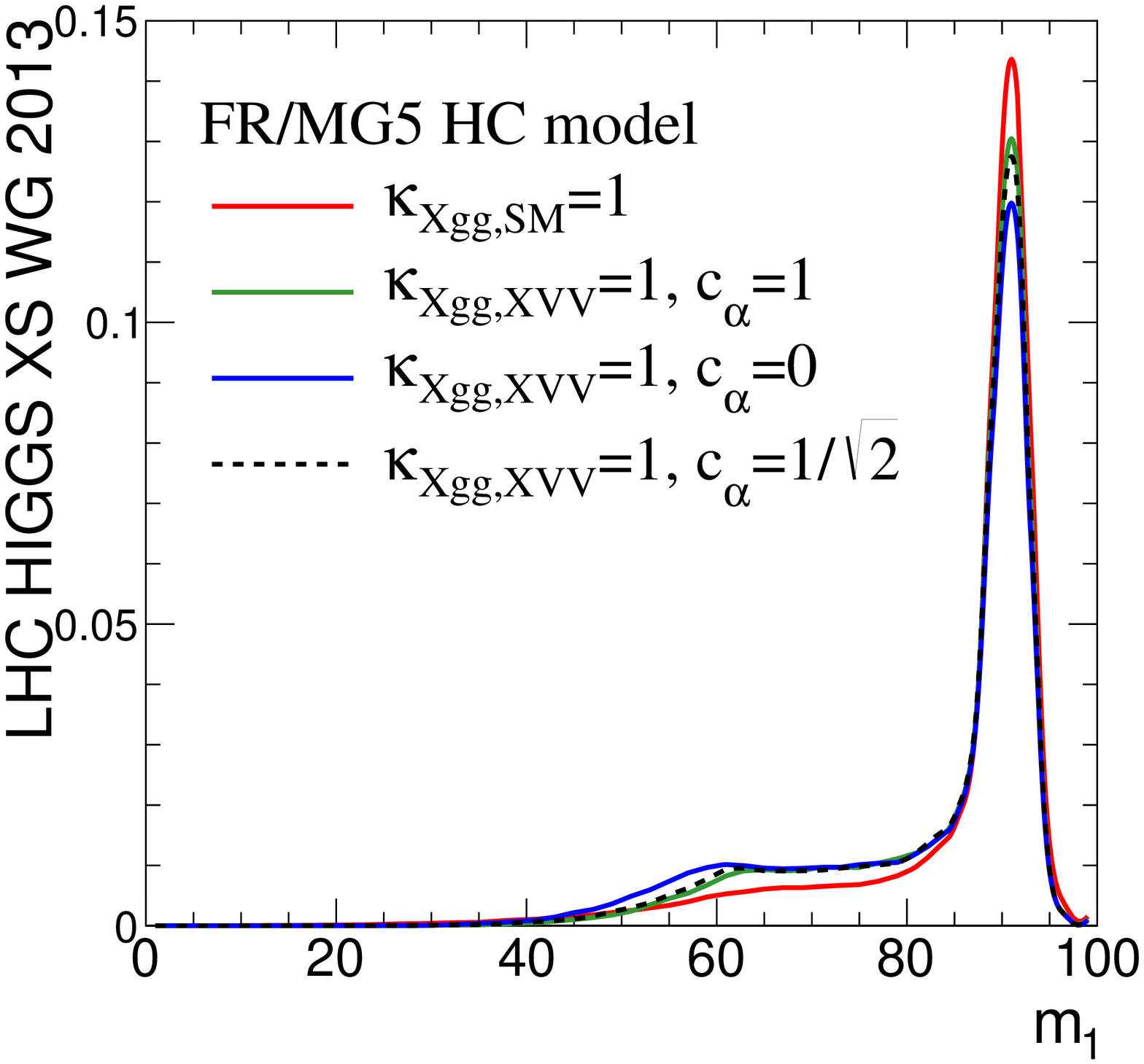}\hskip -0.3cm
 \includegraphics[width=0.24\textwidth]{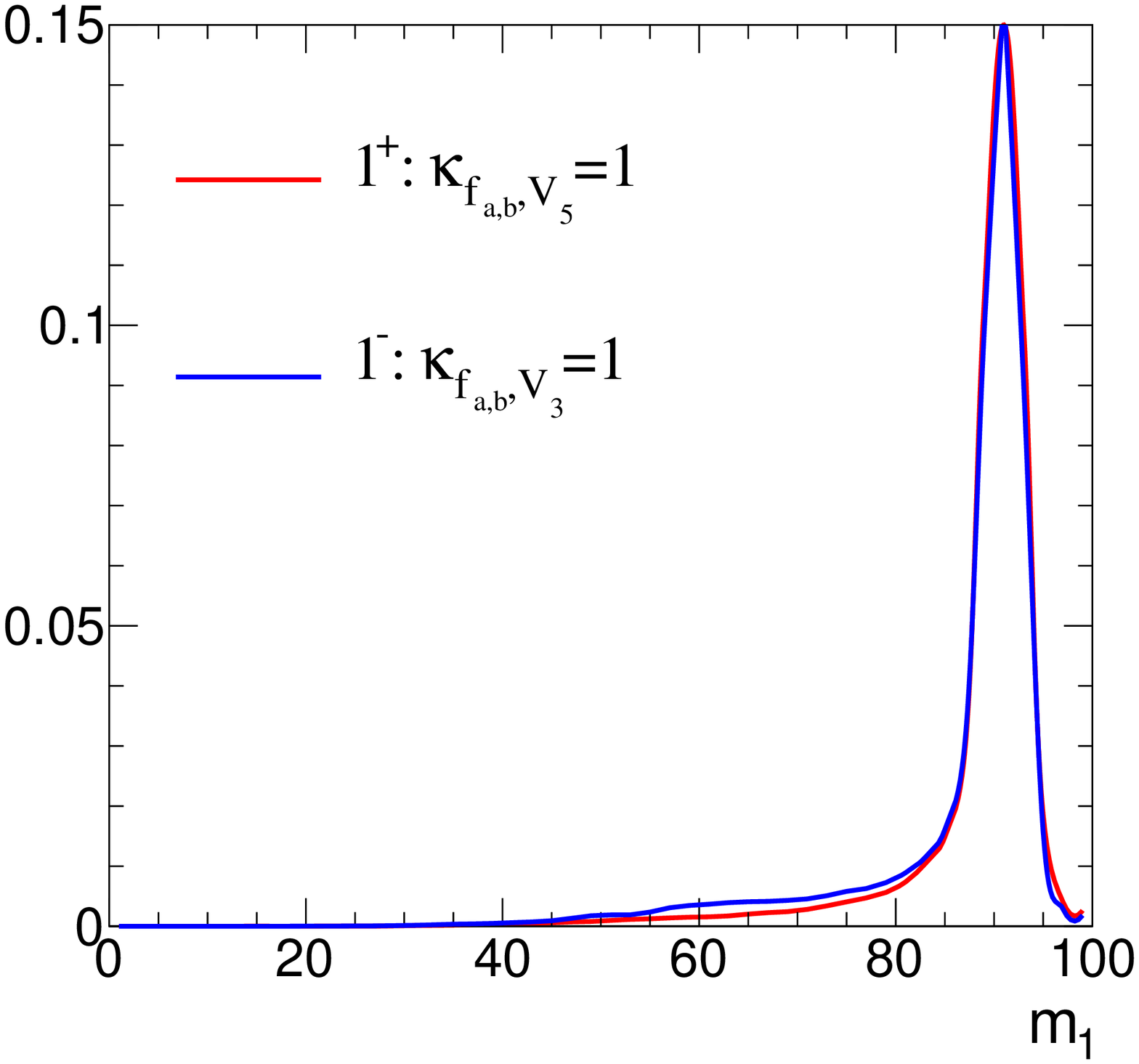}\hskip -0.3cm
 \includegraphics[width=0.24\textwidth]{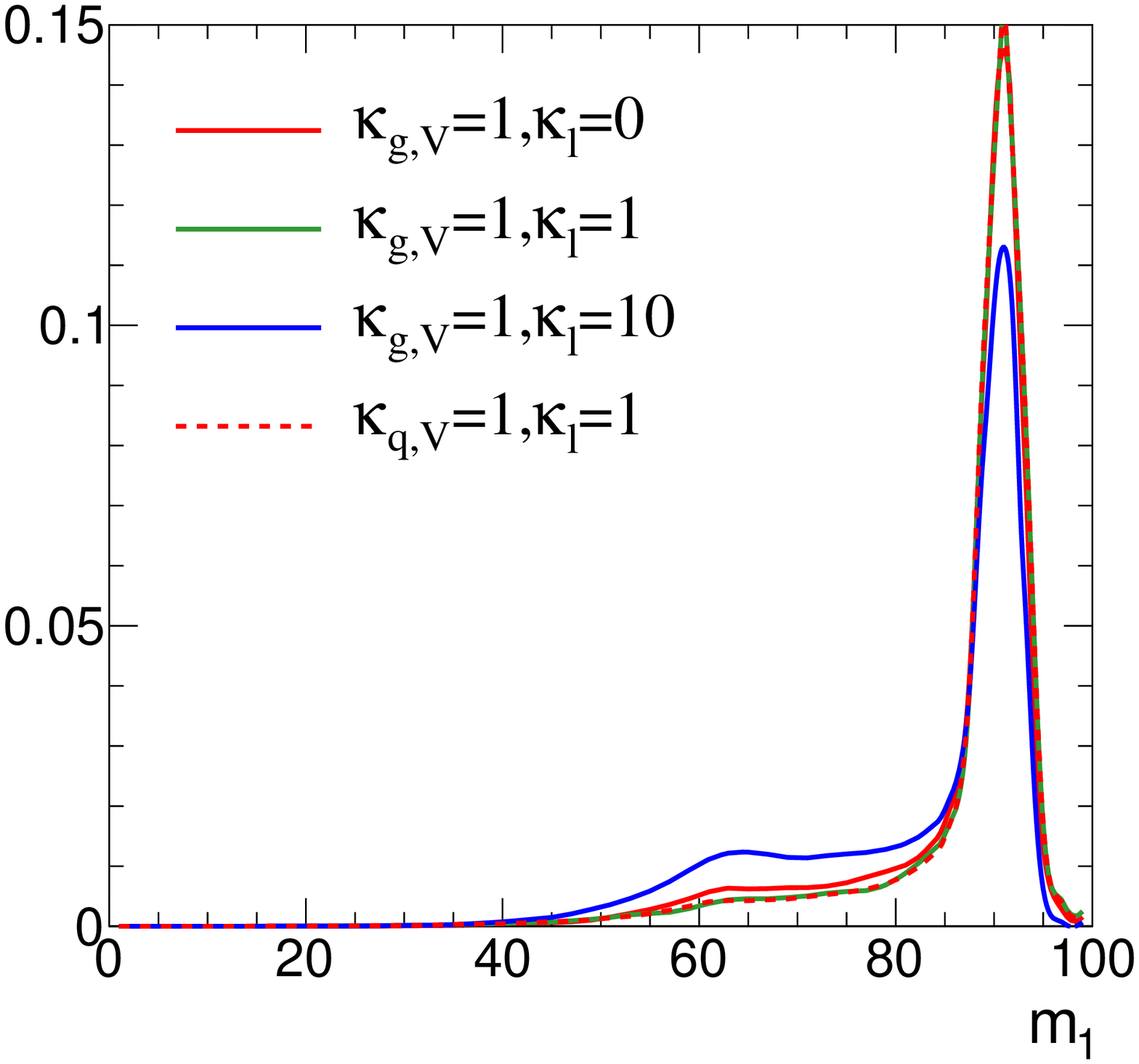} \\[-1.7cm]
\hskip -3cm
 \includegraphics[width=0.24\textwidth]{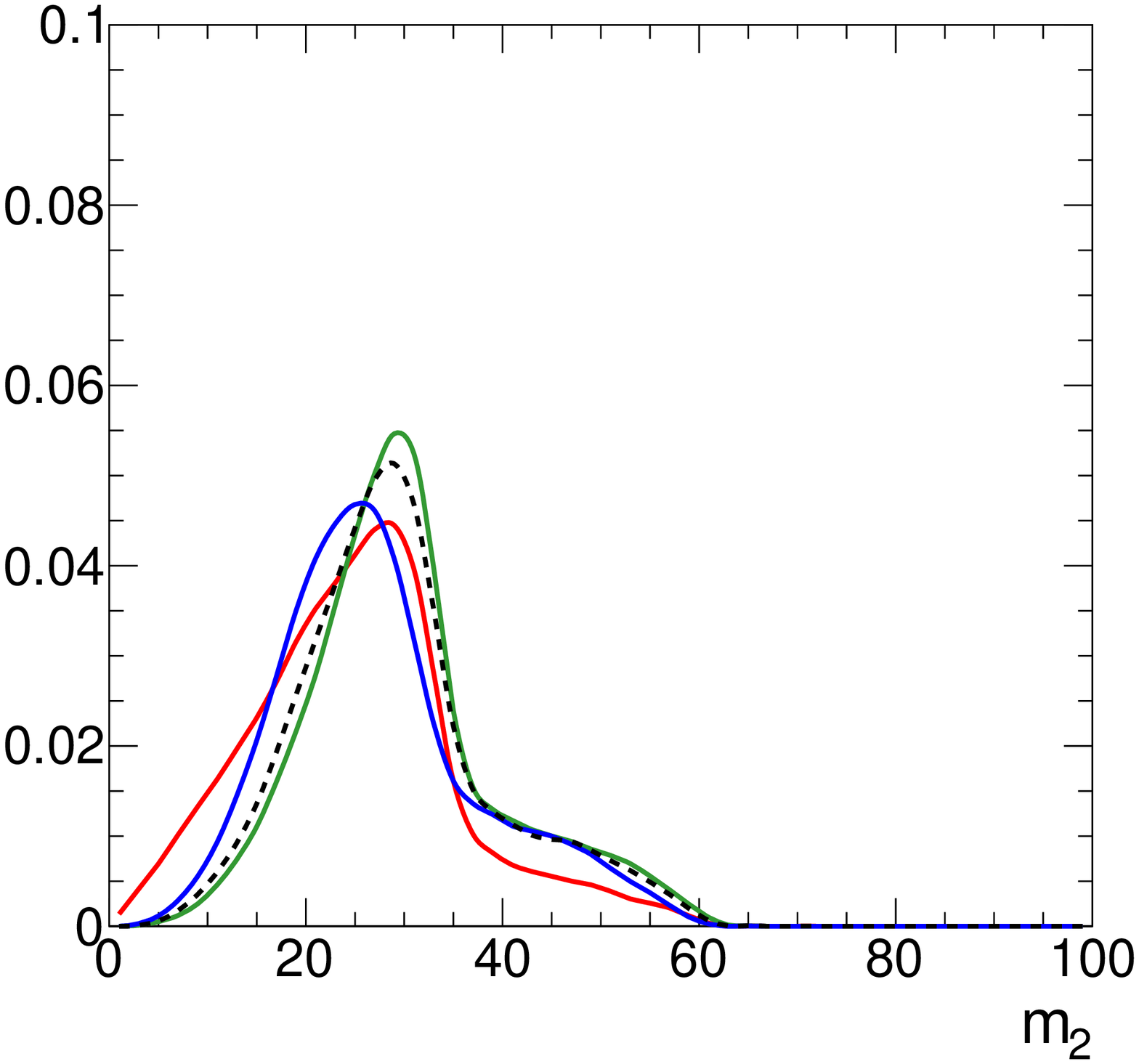}\hskip -0.3cm
 \includegraphics[width=0.24\textwidth]{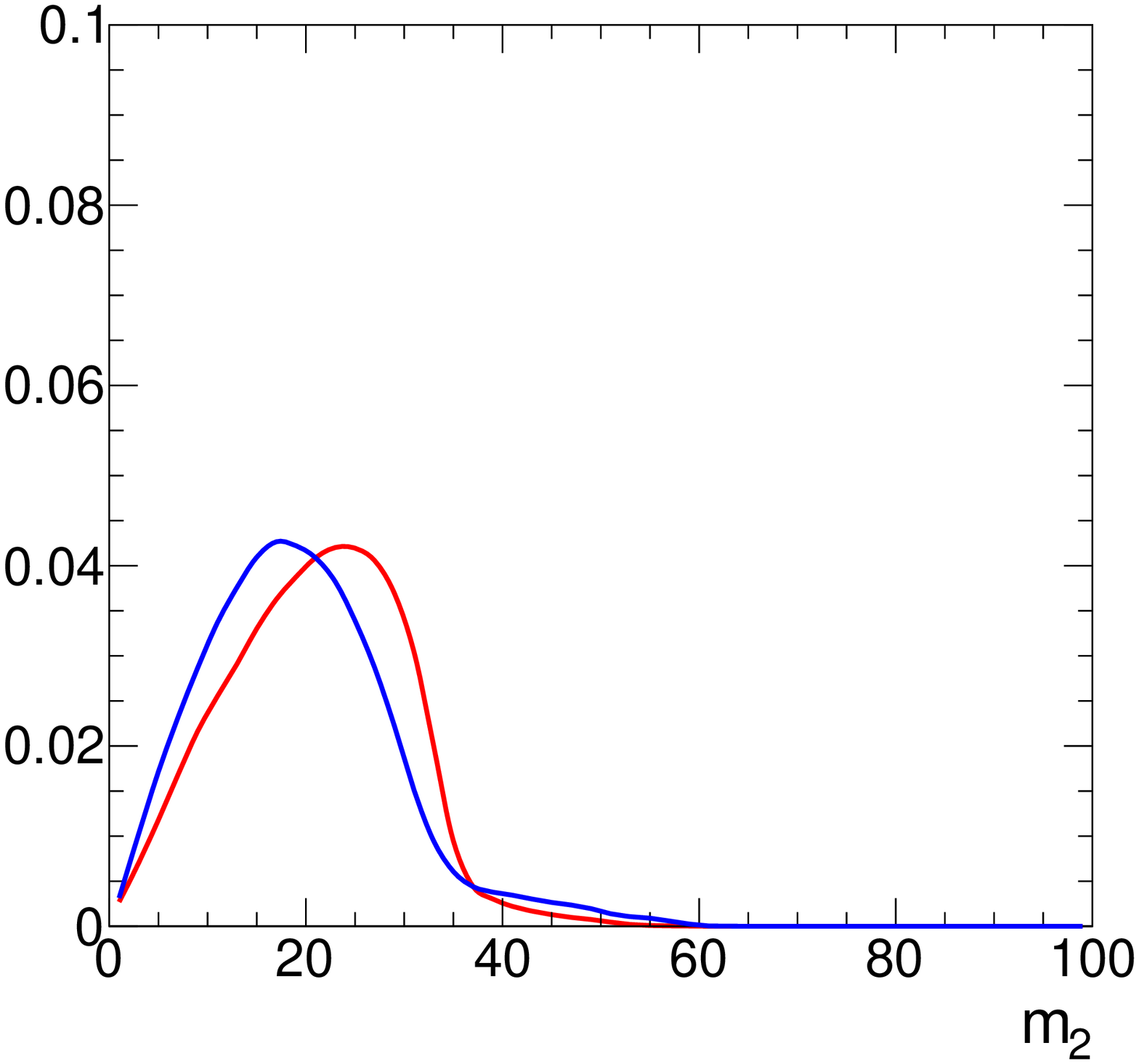}\hskip -0.3cm
 \includegraphics[width=0.24\textwidth]{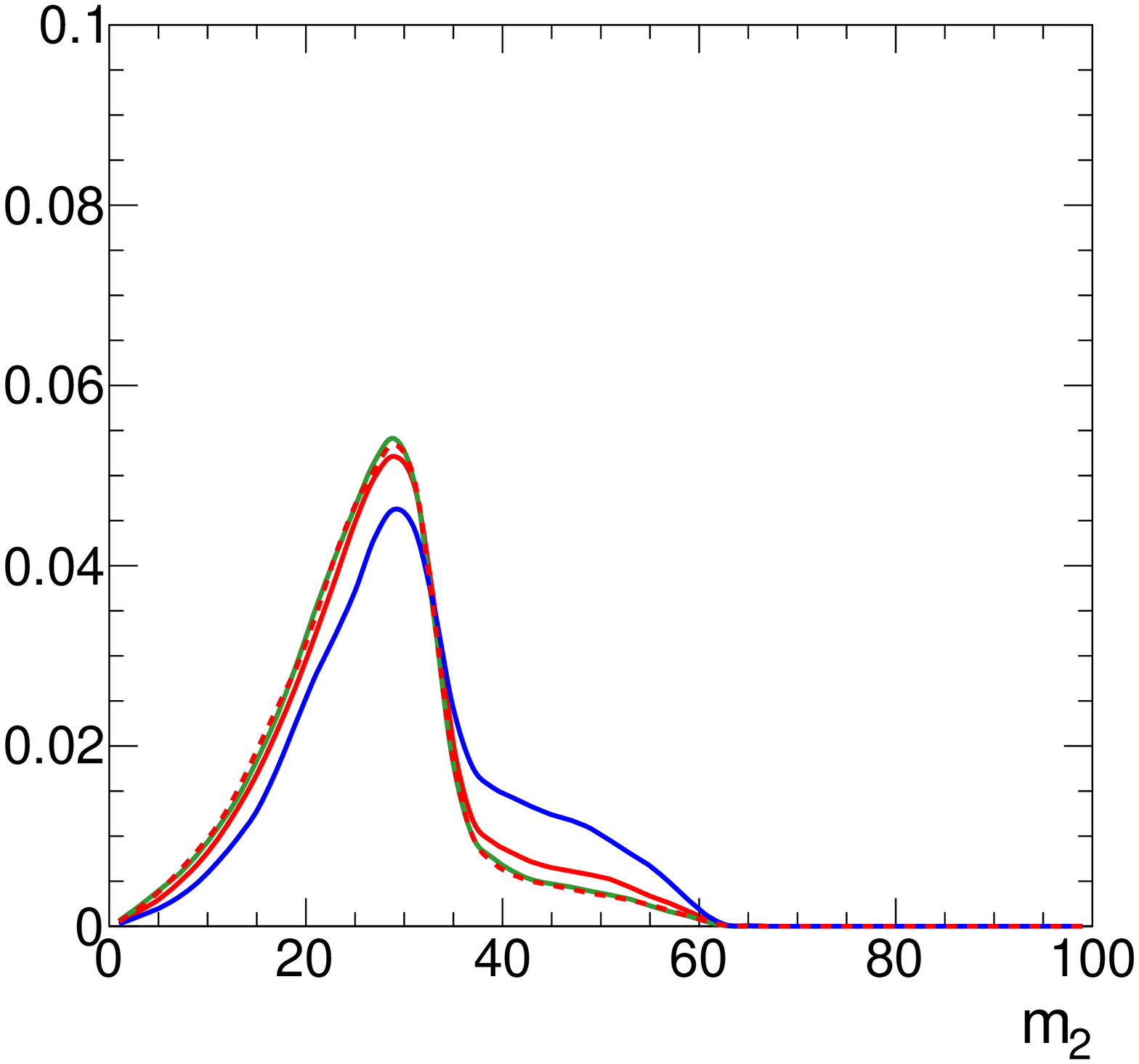}\\[-1.7cm]
\hskip -3cm
 \includegraphics[width=0.24\textwidth]{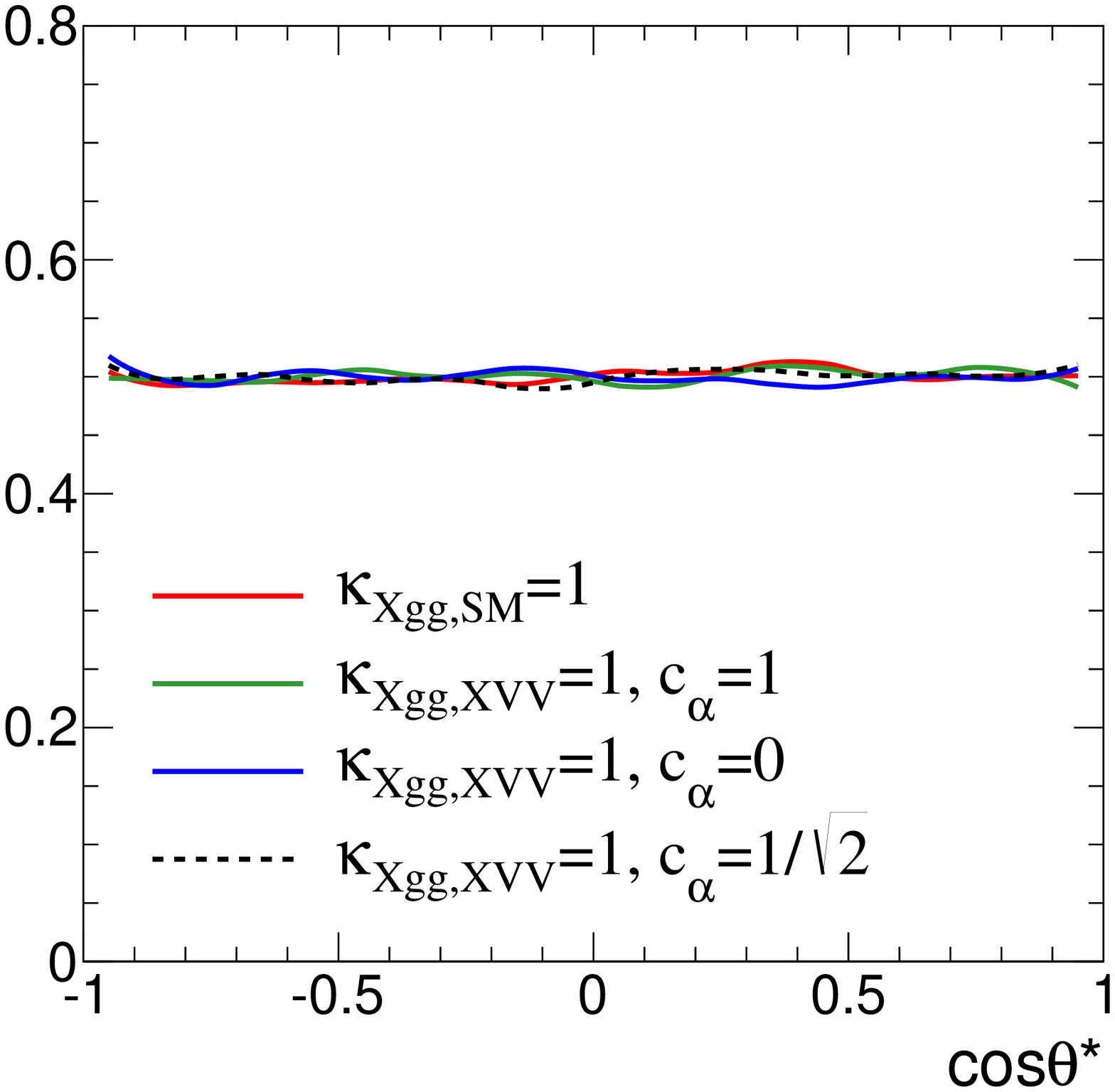}\hskip -0.3cm
 \includegraphics[width=0.24\textwidth]{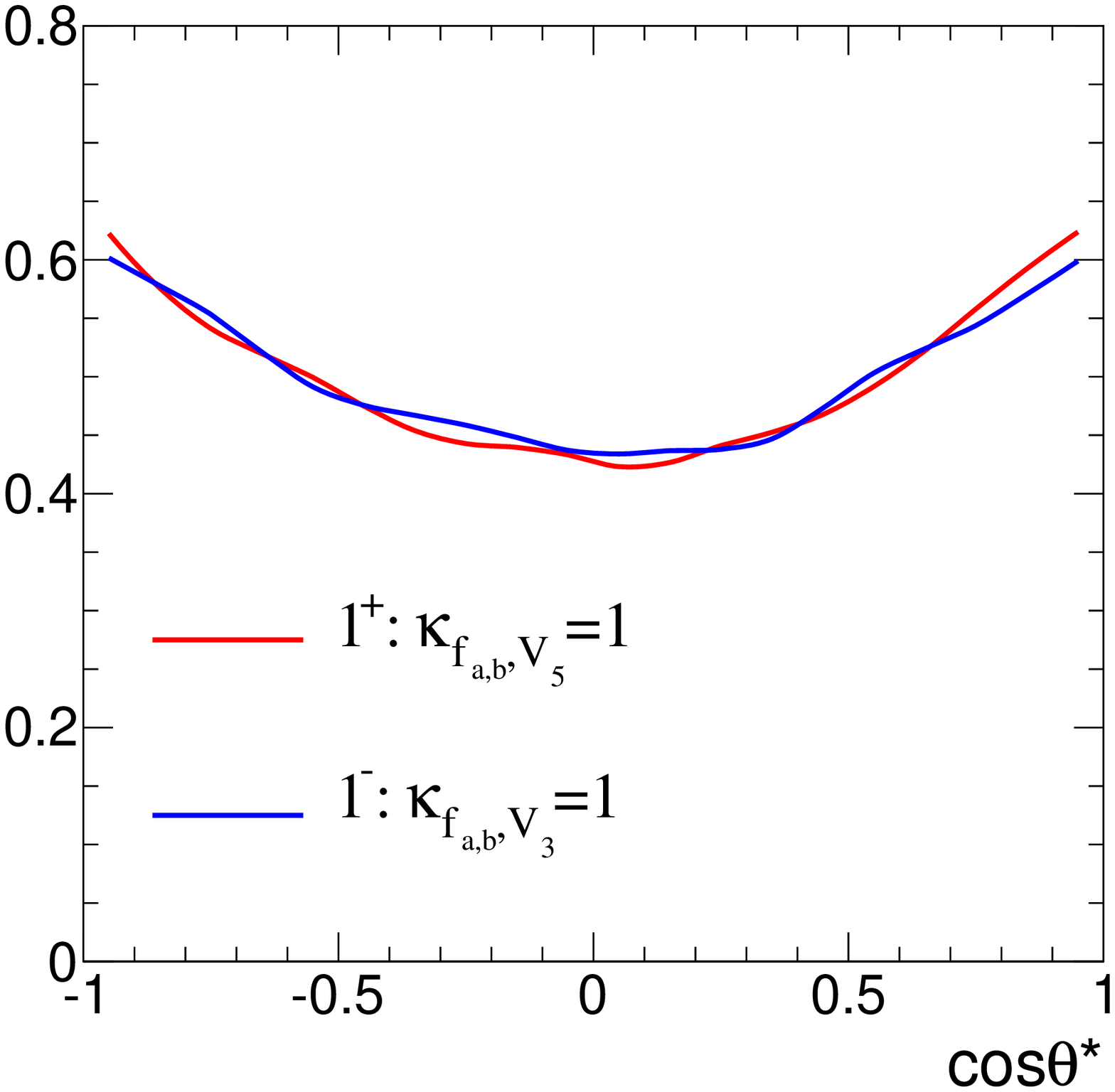}\hskip -0.3cm
 \includegraphics[width=0.24\textwidth]{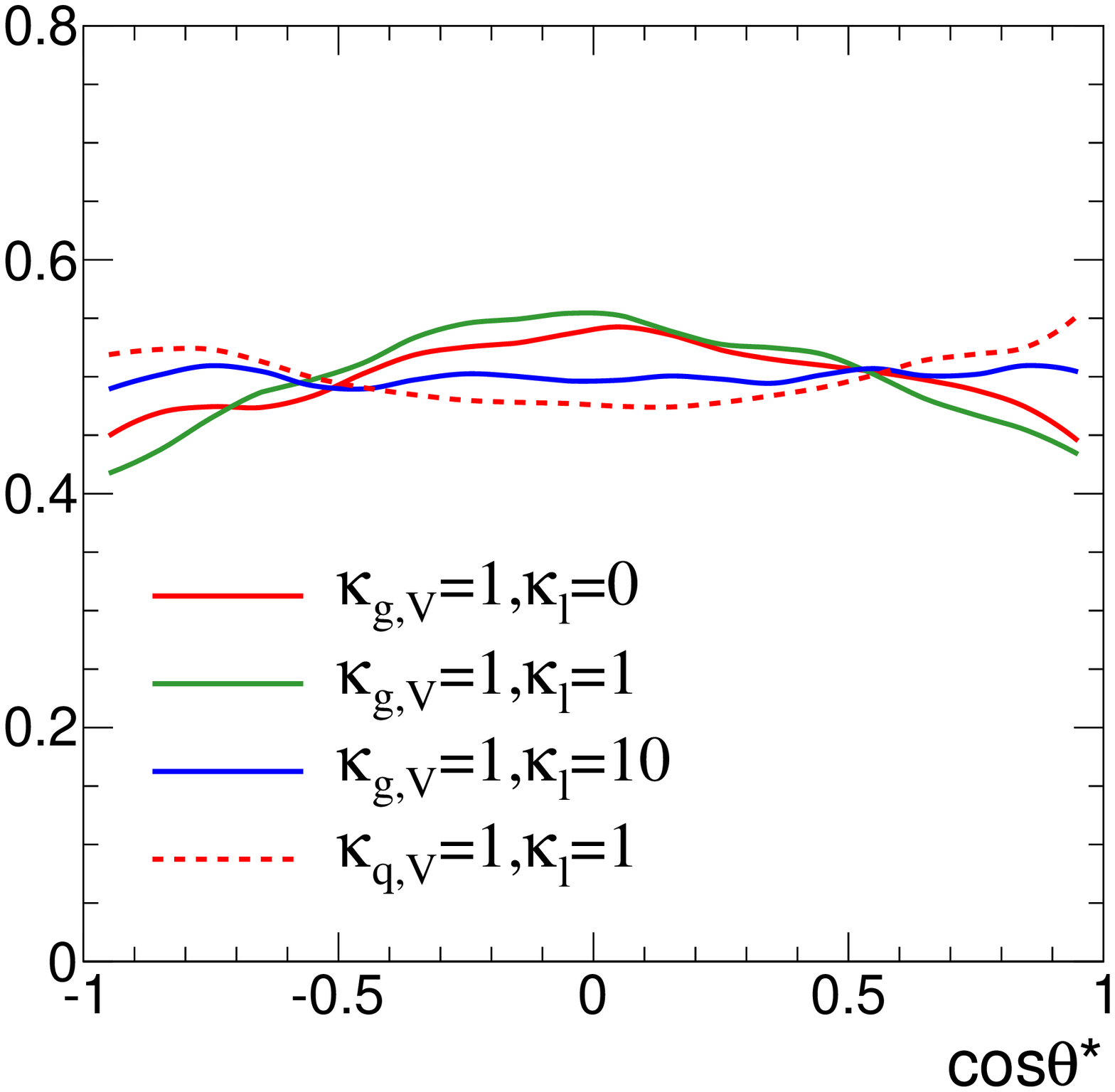} \\[-1.7cm]
\hskip -3cm
 \includegraphics[width=0.24\textwidth]{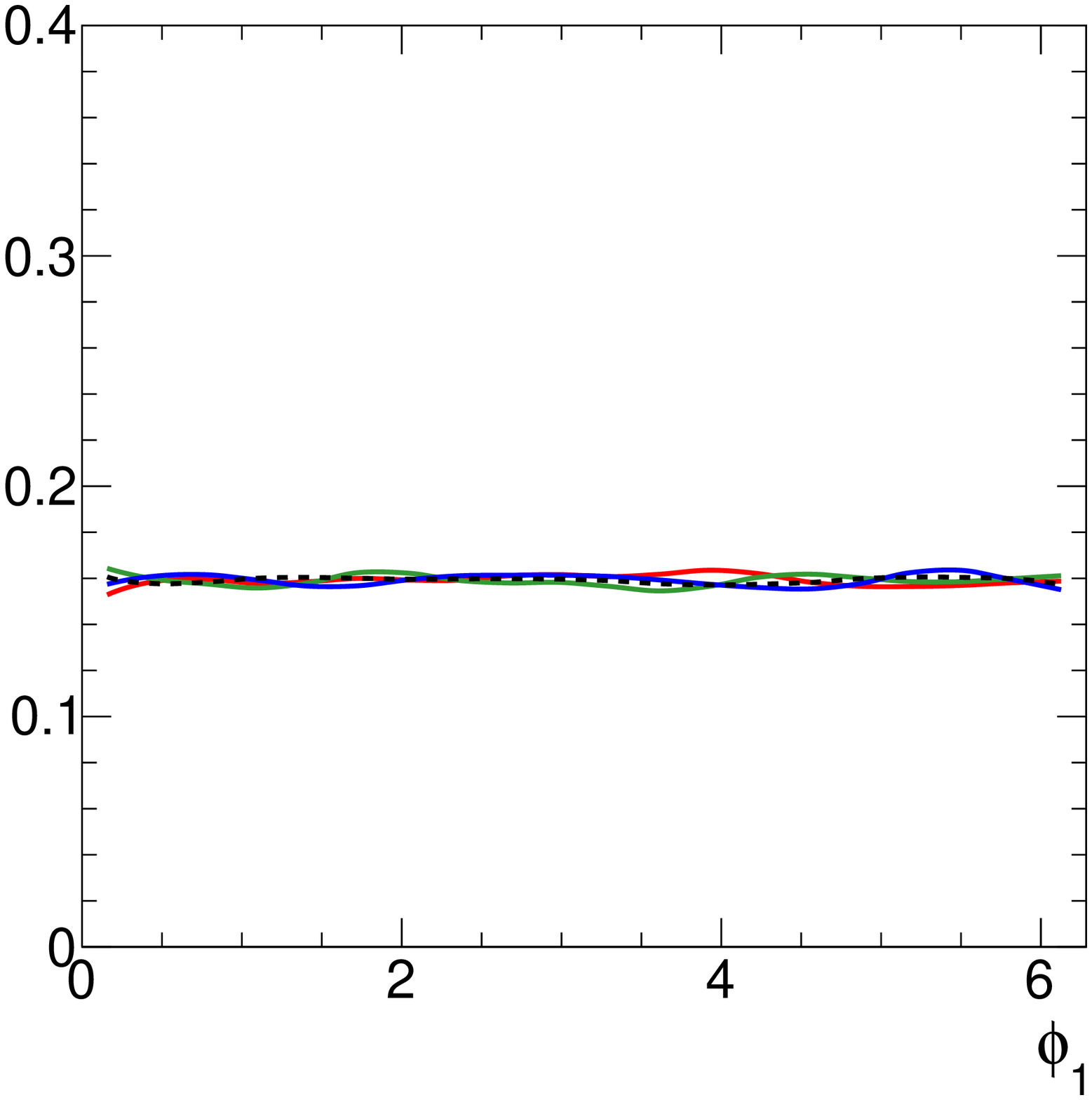}\hskip -0.3cm
 \includegraphics[width=0.24\textwidth]{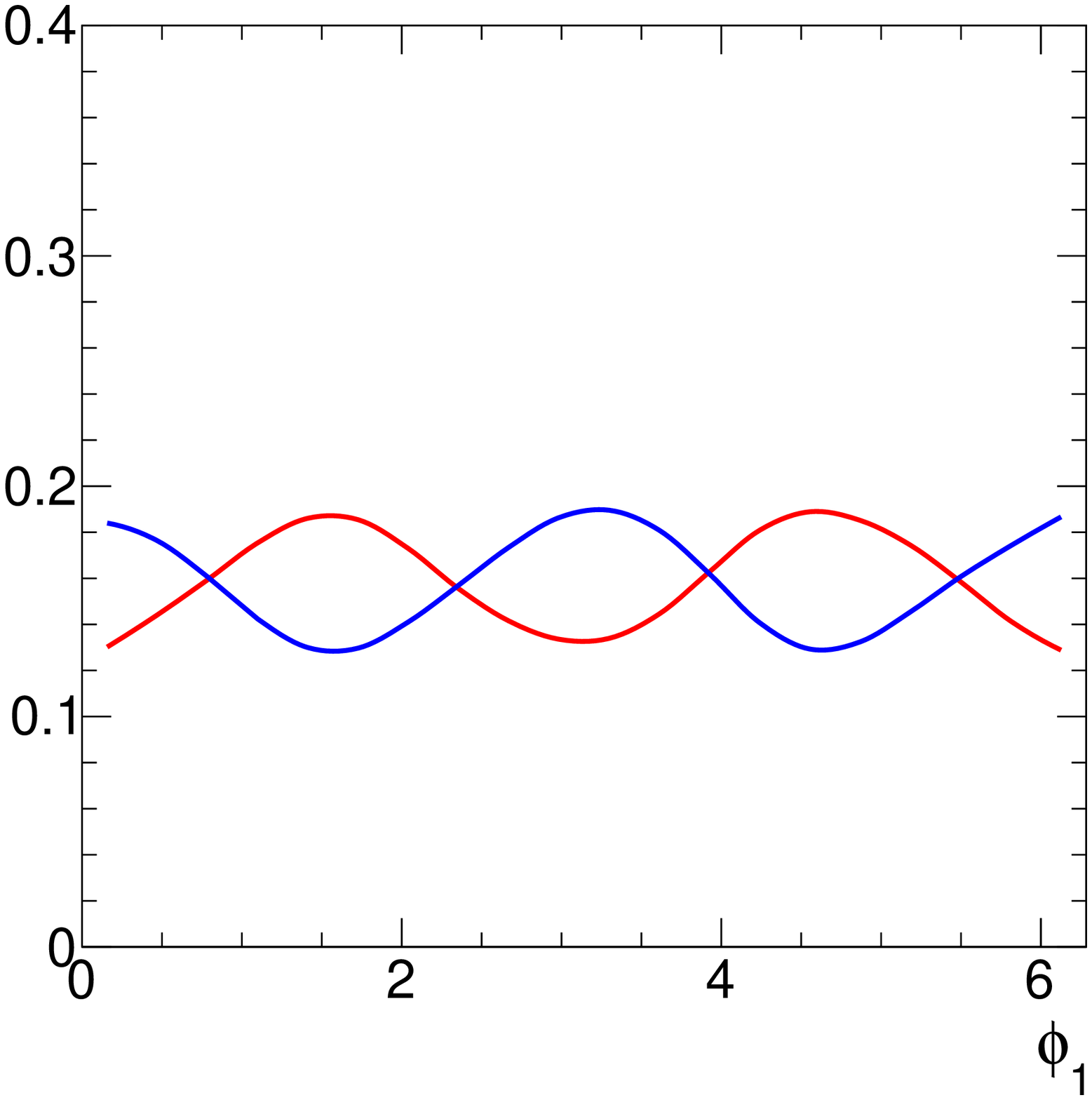}\hskip -0.3cm
 \includegraphics[width=0.24\textwidth]{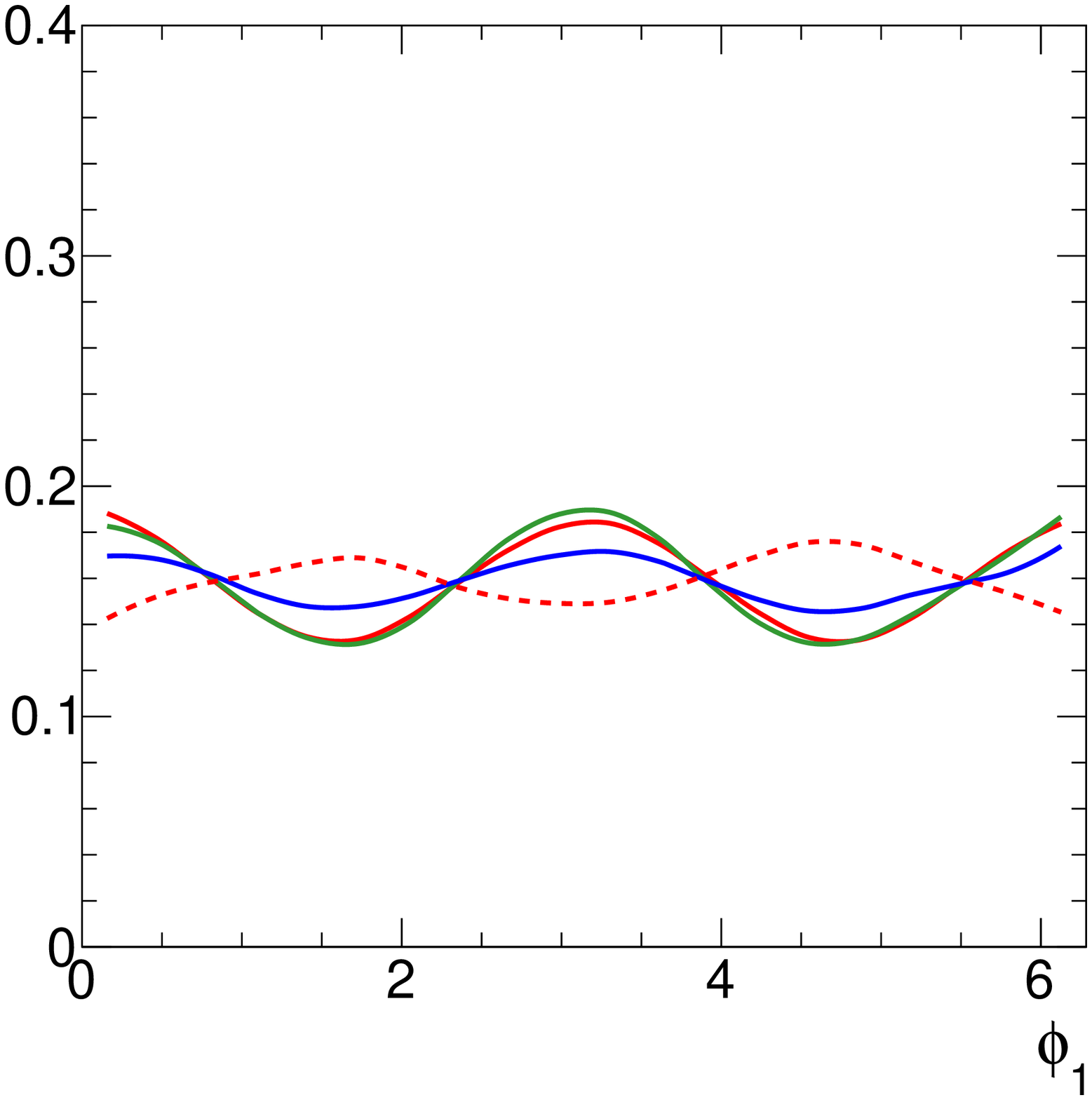} \\[-1.7cm]
\hskip -3cm
 \includegraphics[width=0.24\textwidth]{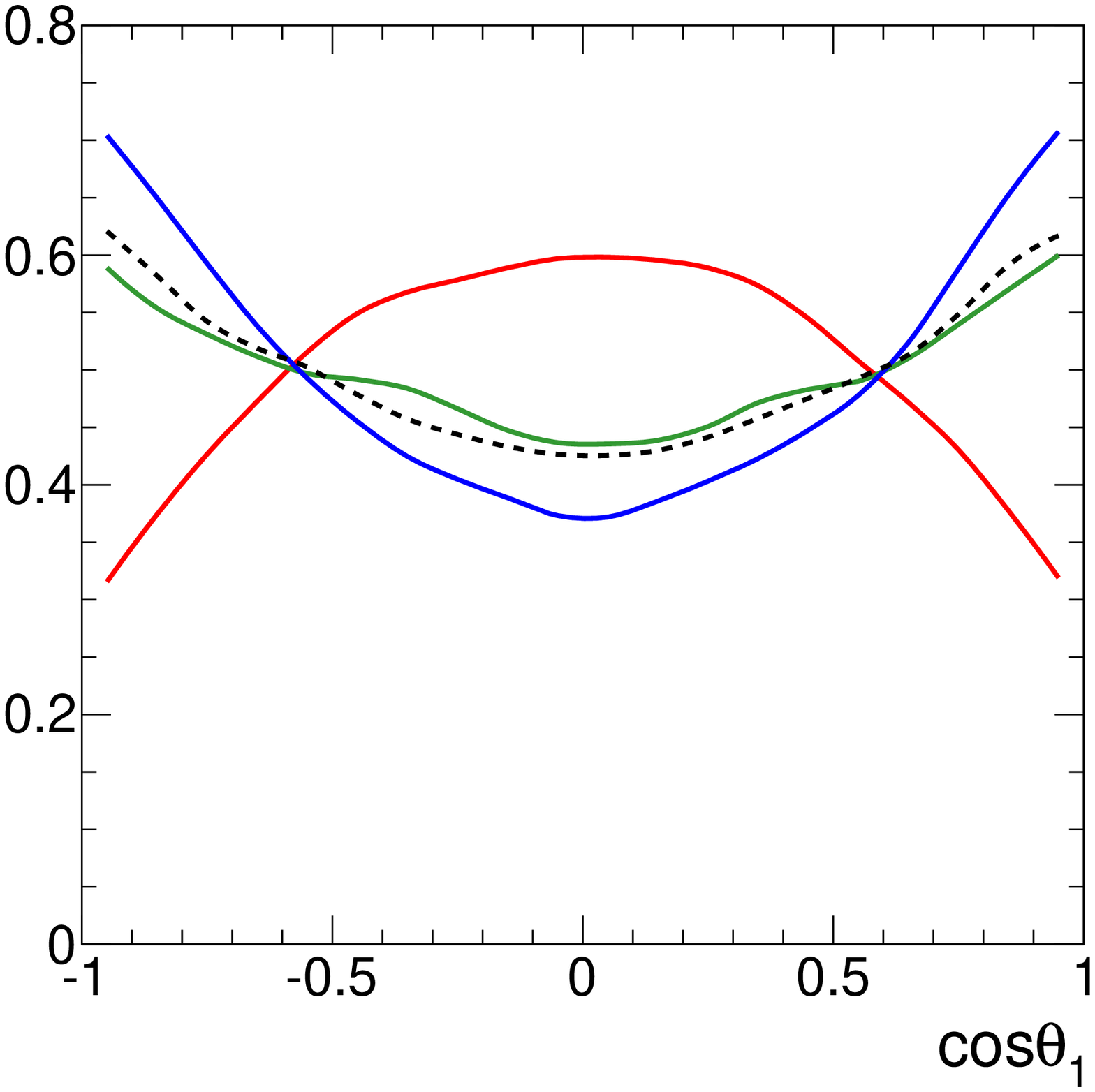}\hskip -0.3cm
 \includegraphics[width=0.24\textwidth]{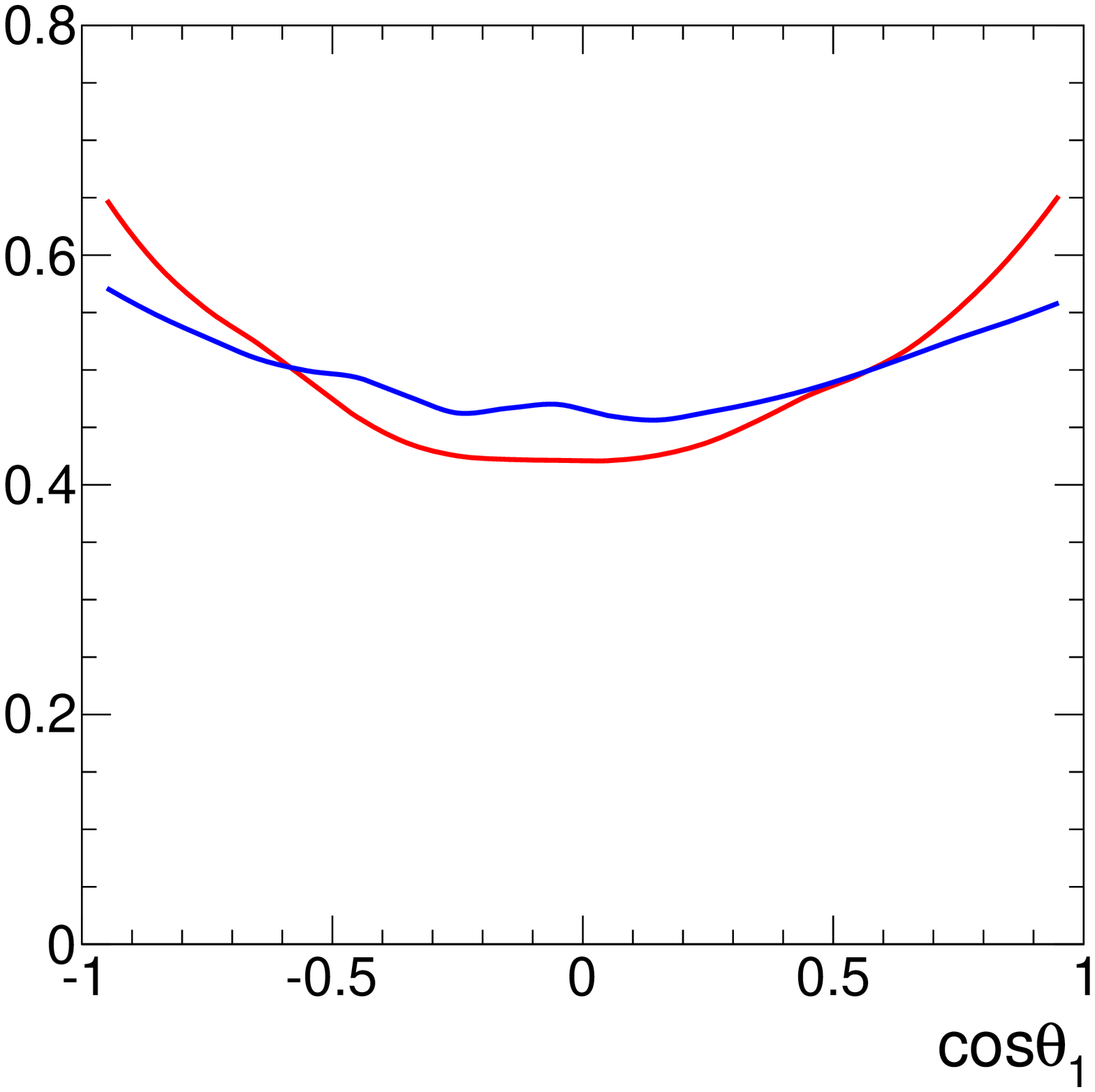}\hskip -0.3cm
 \includegraphics[width=0.24\textwidth]{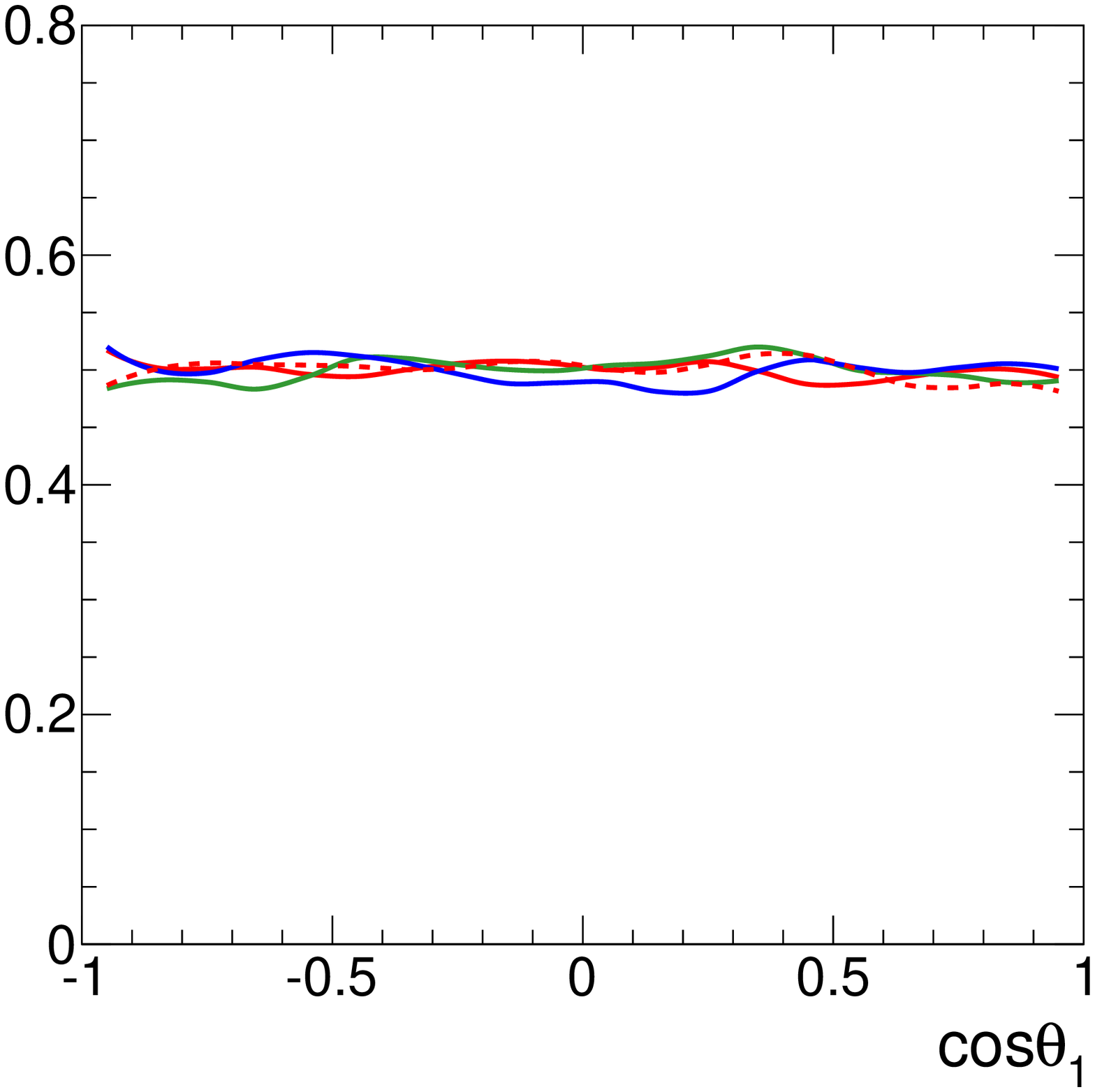} \\[-1.7cm]
\hskip -3cm
 \includegraphics[width=0.24\textwidth]{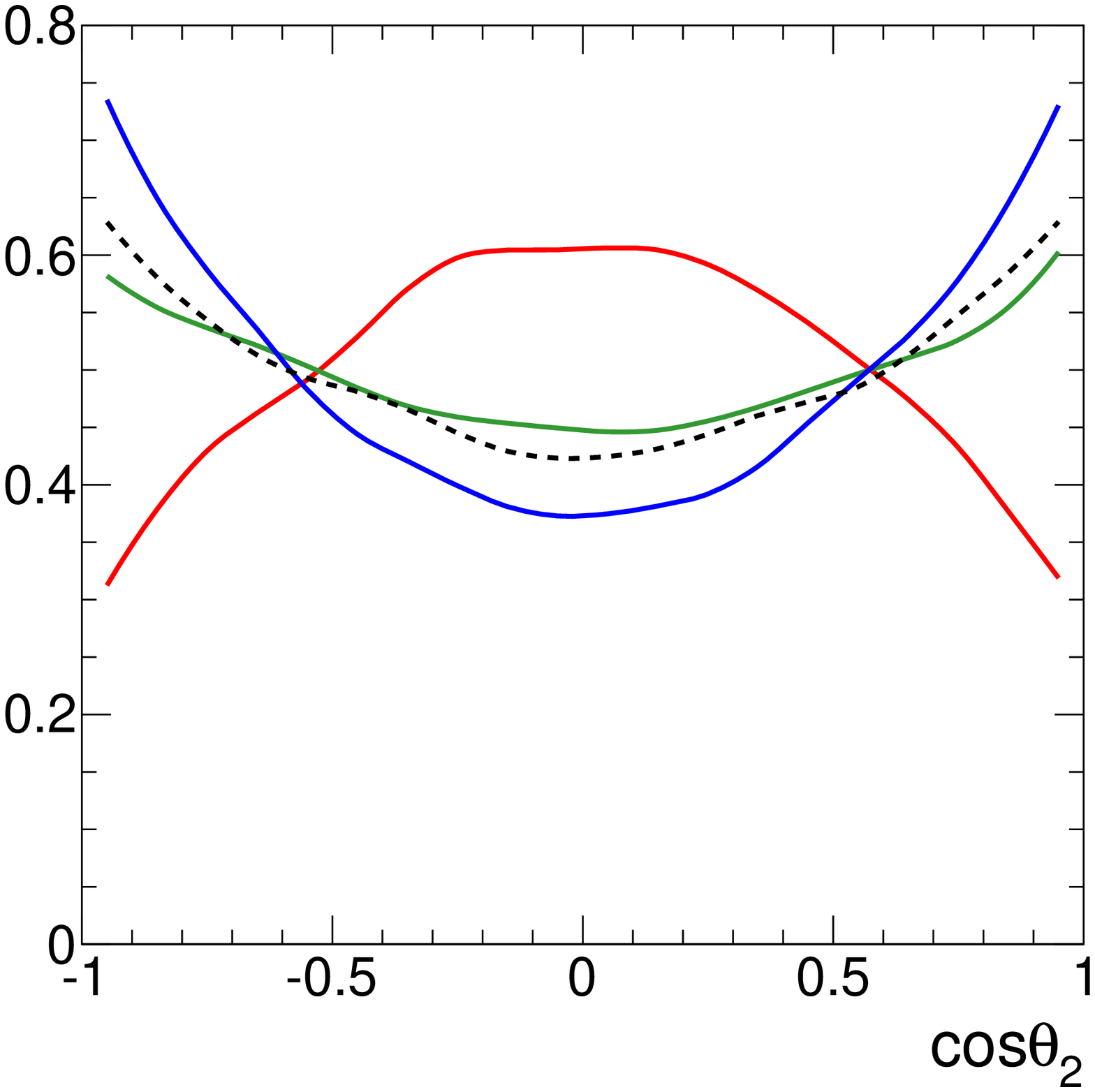}\hskip -0.3cm
 \includegraphics[width=0.24\textwidth]{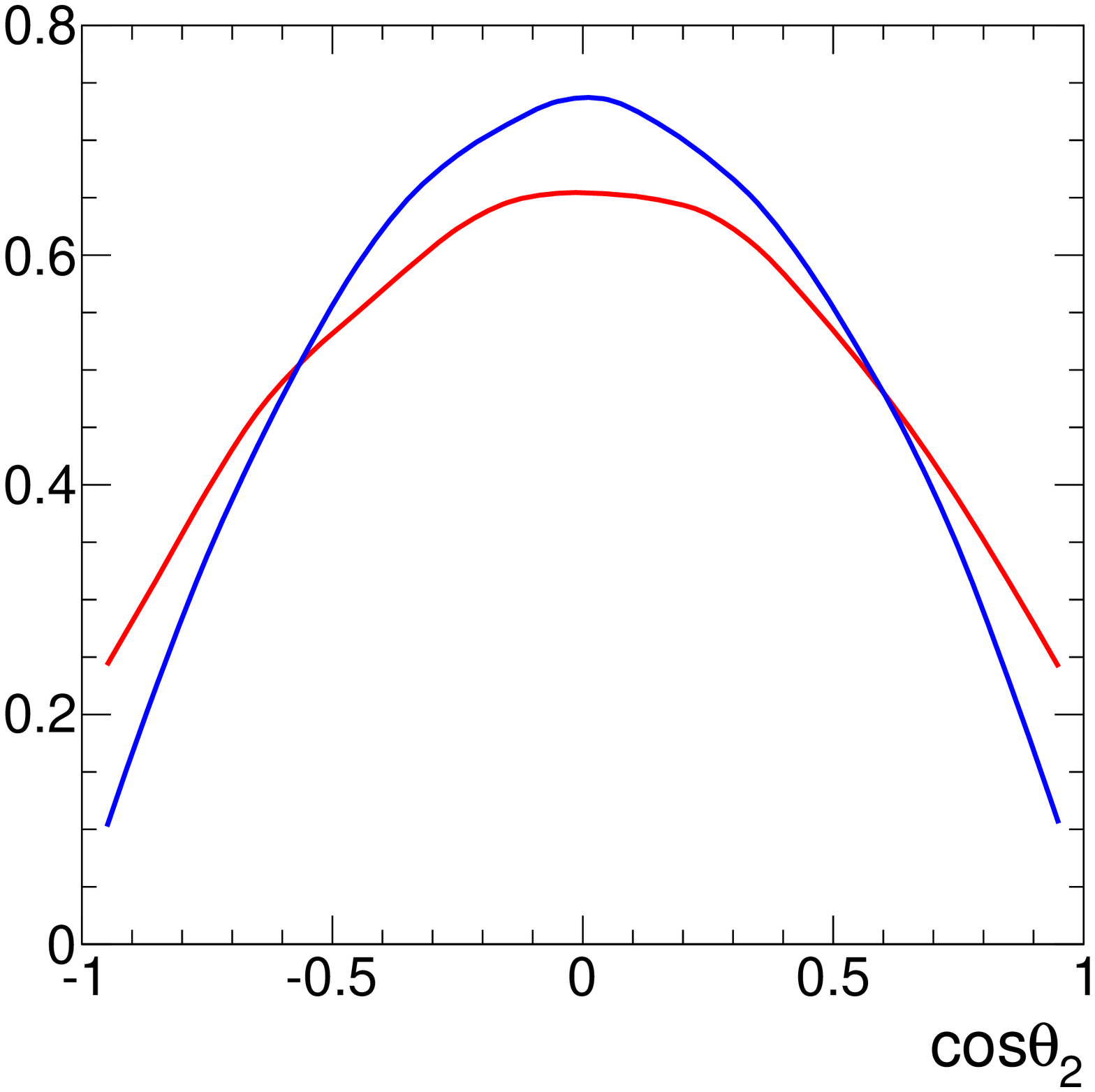}\hskip -0.3cm
 \includegraphics[width=0.24\textwidth]{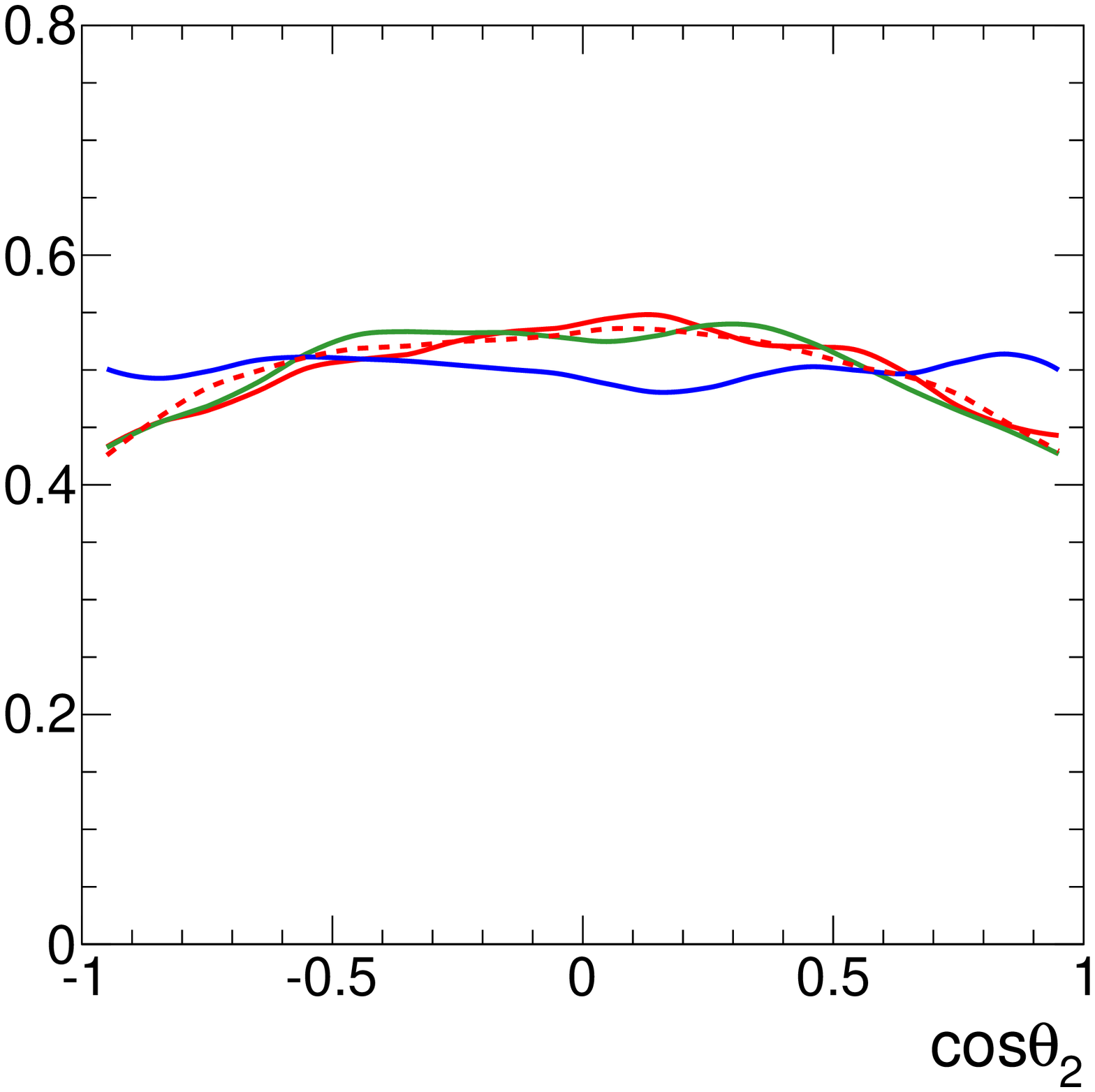} \\[-1.7cm]
\hskip -3cm
 \includegraphics[width=0.24\textwidth]{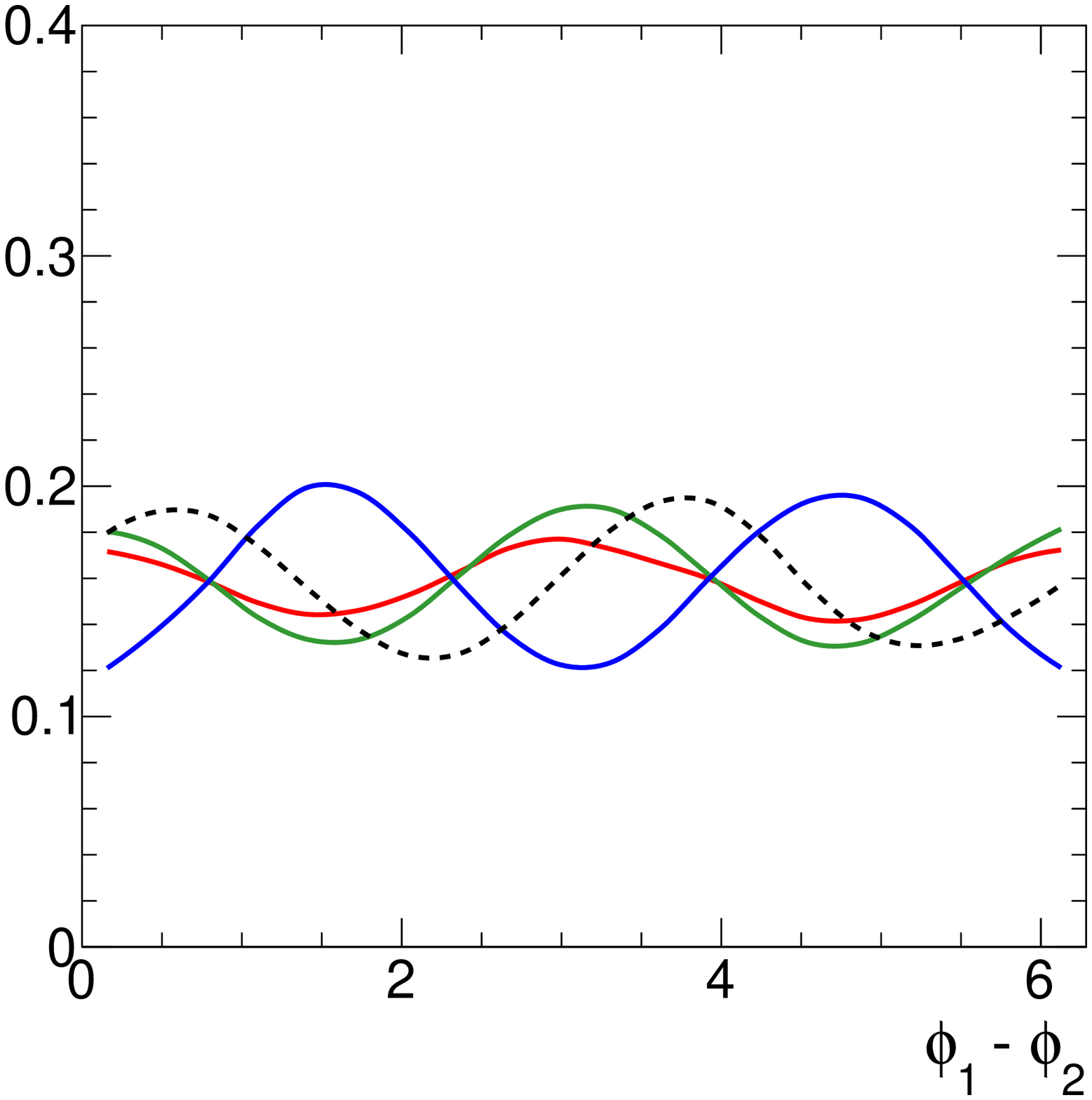}\hskip -0.3cm
 \includegraphics[width=0.24\textwidth]{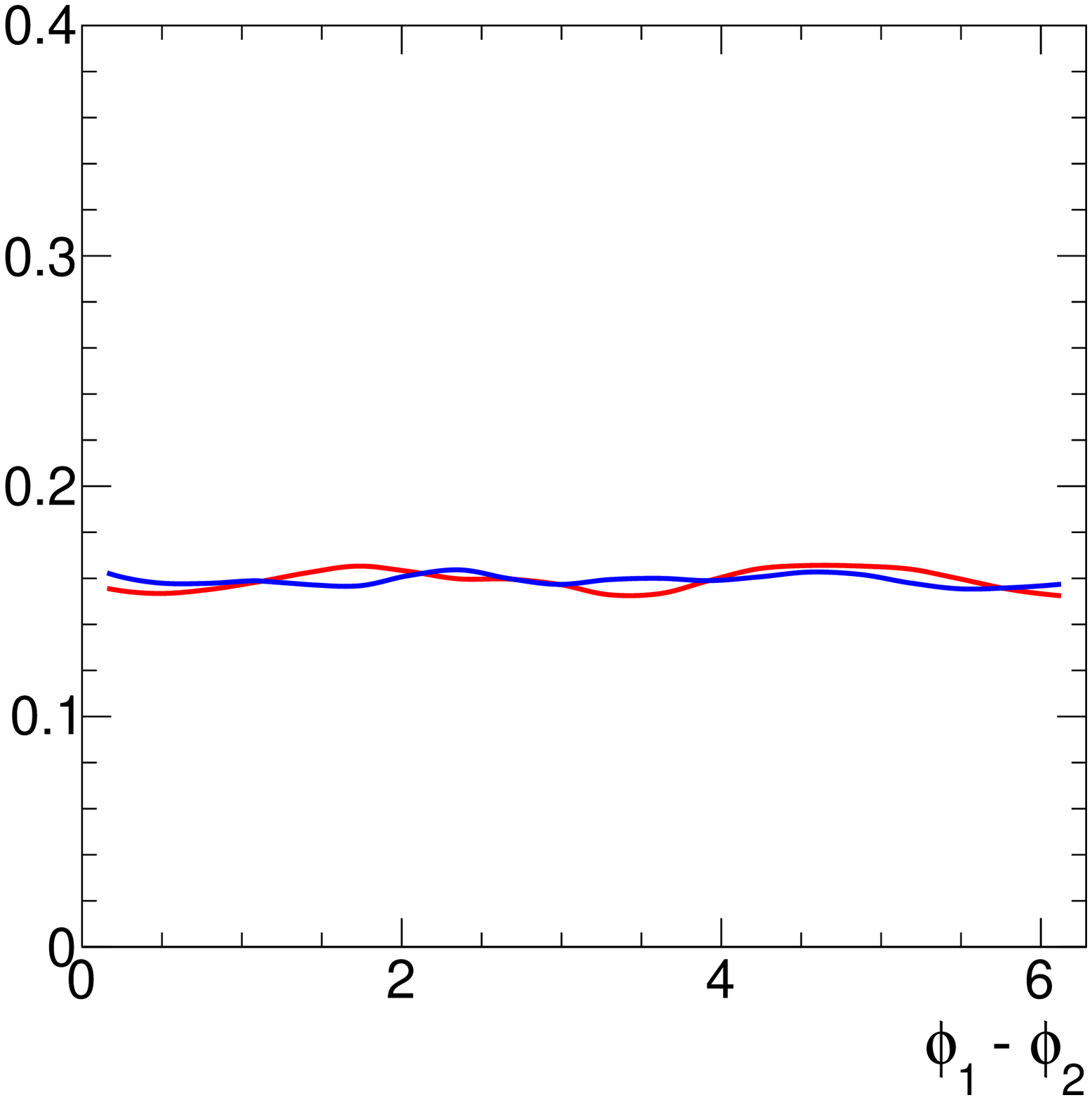}\hskip -0.3cm
 \includegraphics[width=0.24\textwidth]{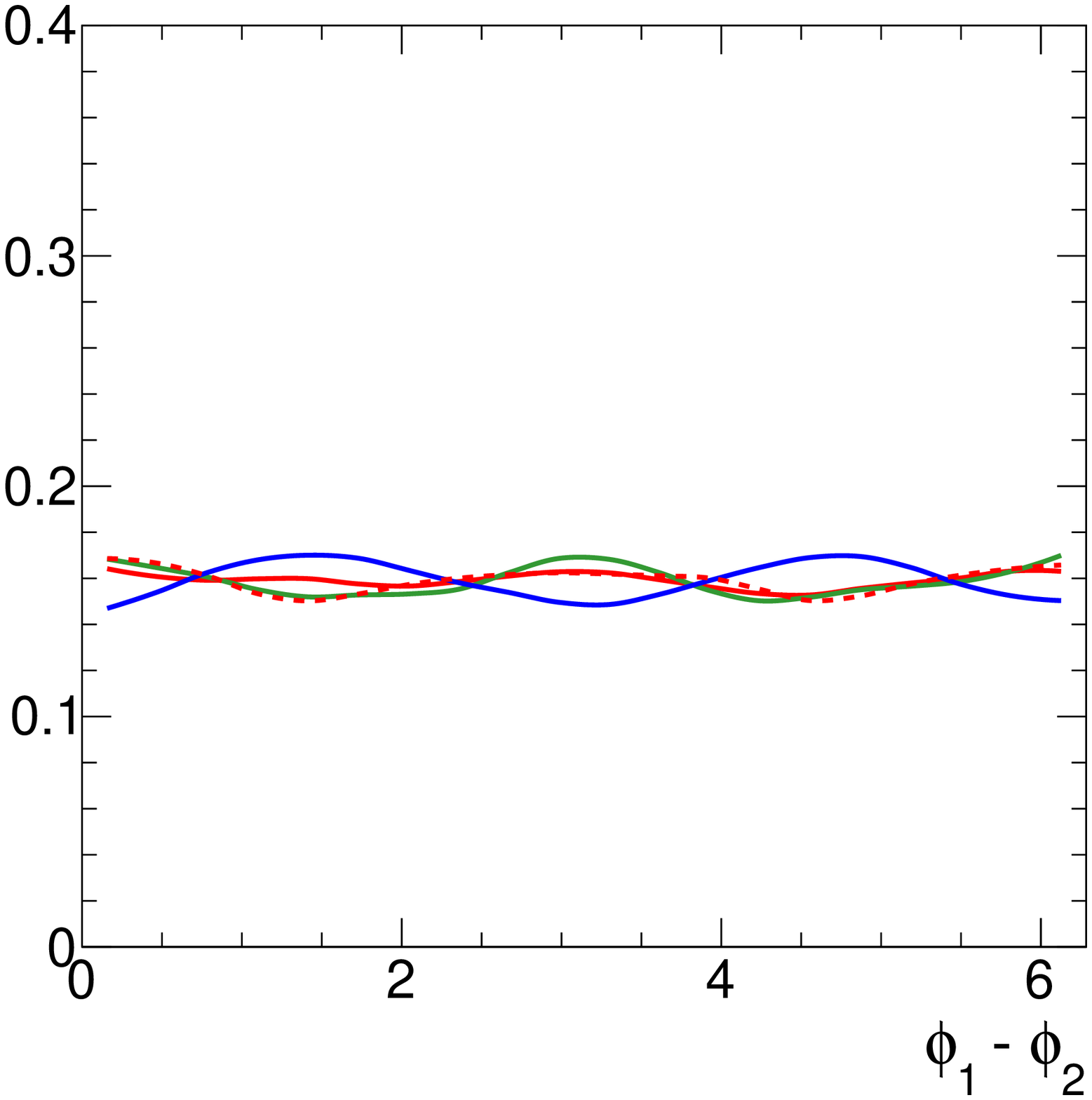}\\
\end{center} 
\vskip -0.7cm
\caption{Distributions of the $\PX\to \PZ\PZ$ analysis; see also JHU Figs.~11
 and 12~\cite{Bolognesi:2012mm}.}
\label{fig:ZZ}
\end{figure}

\begin{figure}
\begin{center}
spin-0\hskip 2.5cm
spin-1\hskip 2.5cm
spin-2\\[-1.3cm] 
\hskip -3cm
 \includegraphics[width=0.24\textwidth]{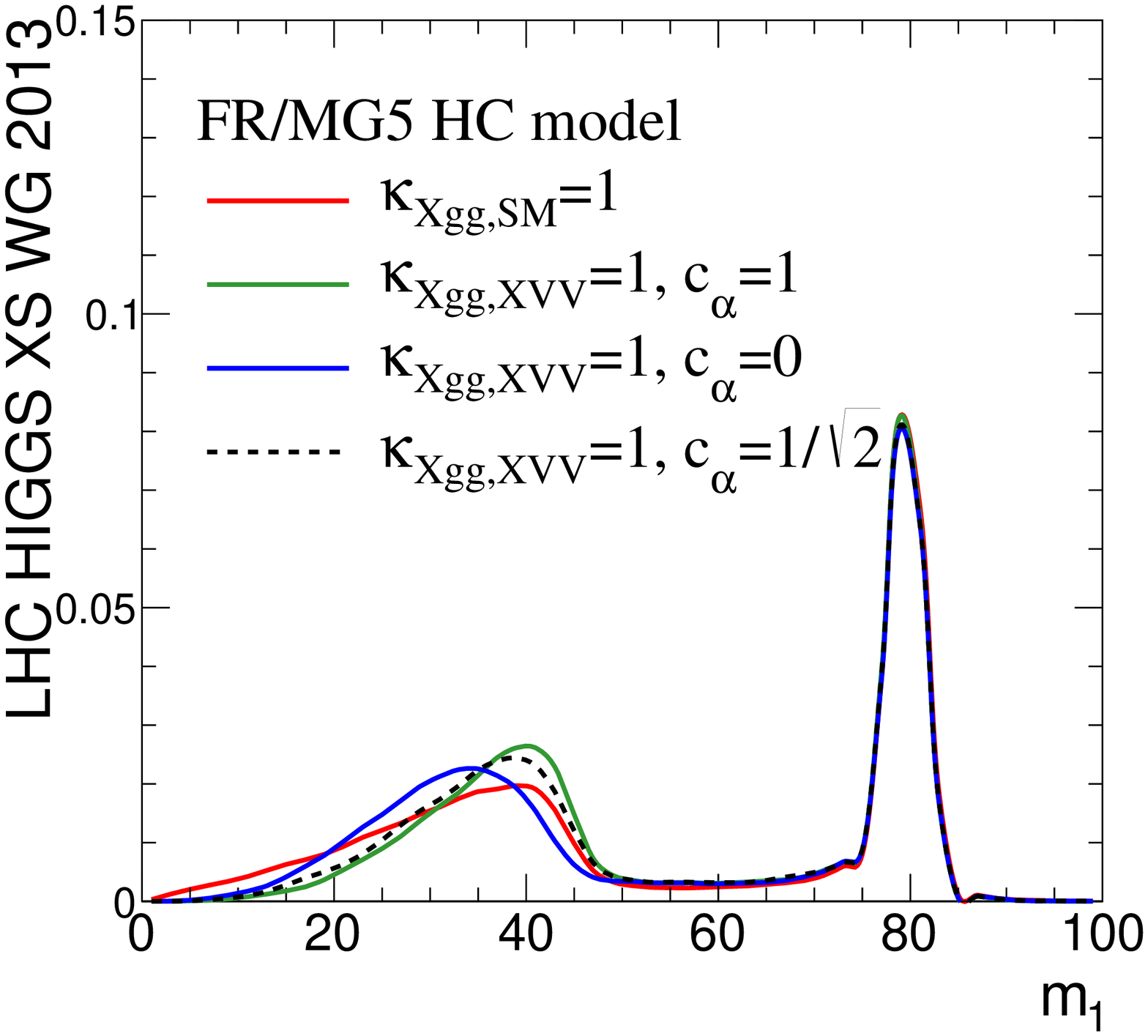}\hskip -0.3cm
 \includegraphics[width=0.24\textwidth]{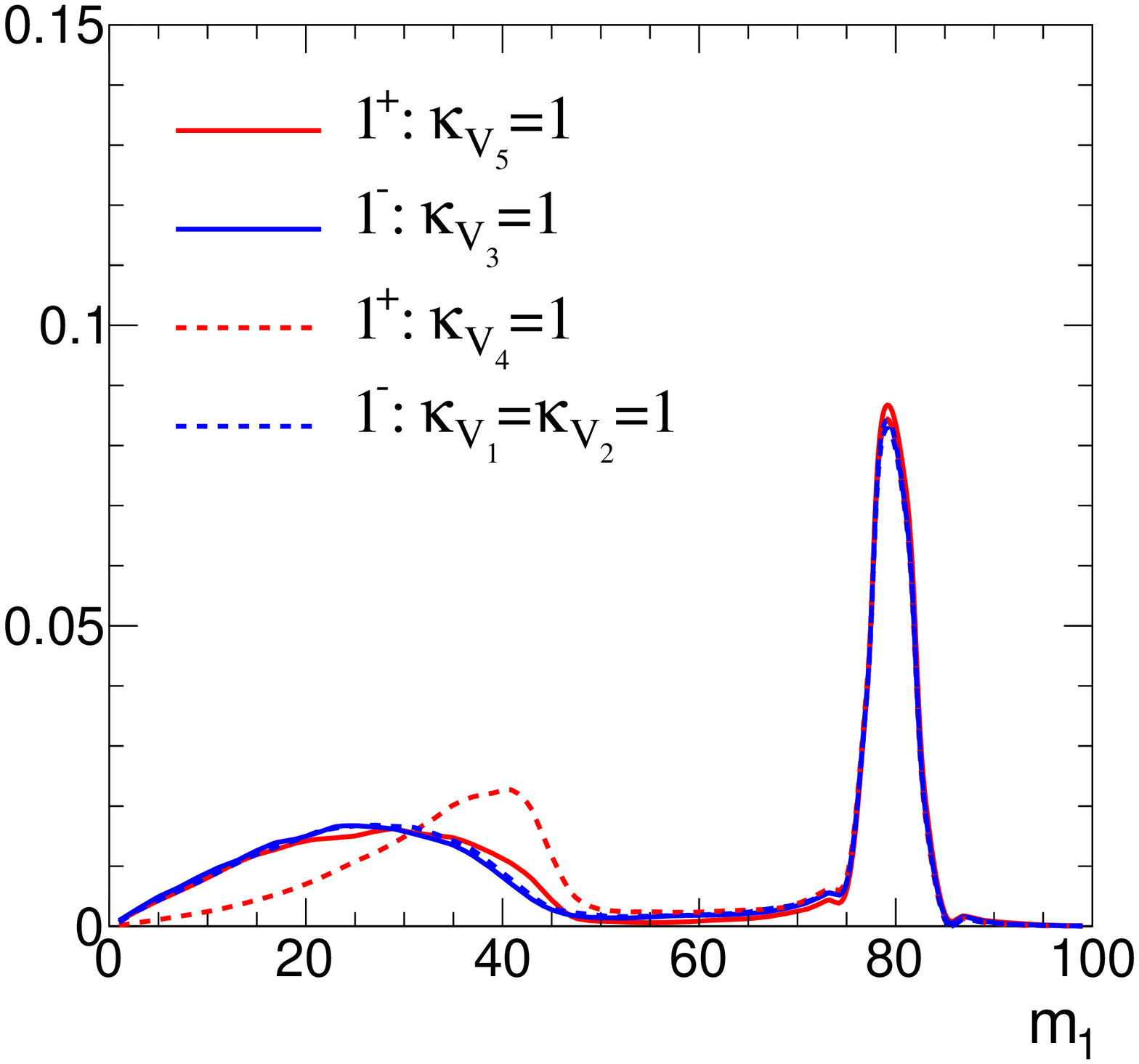}\hskip -0.3cm
 \includegraphics[width=0.24\textwidth]{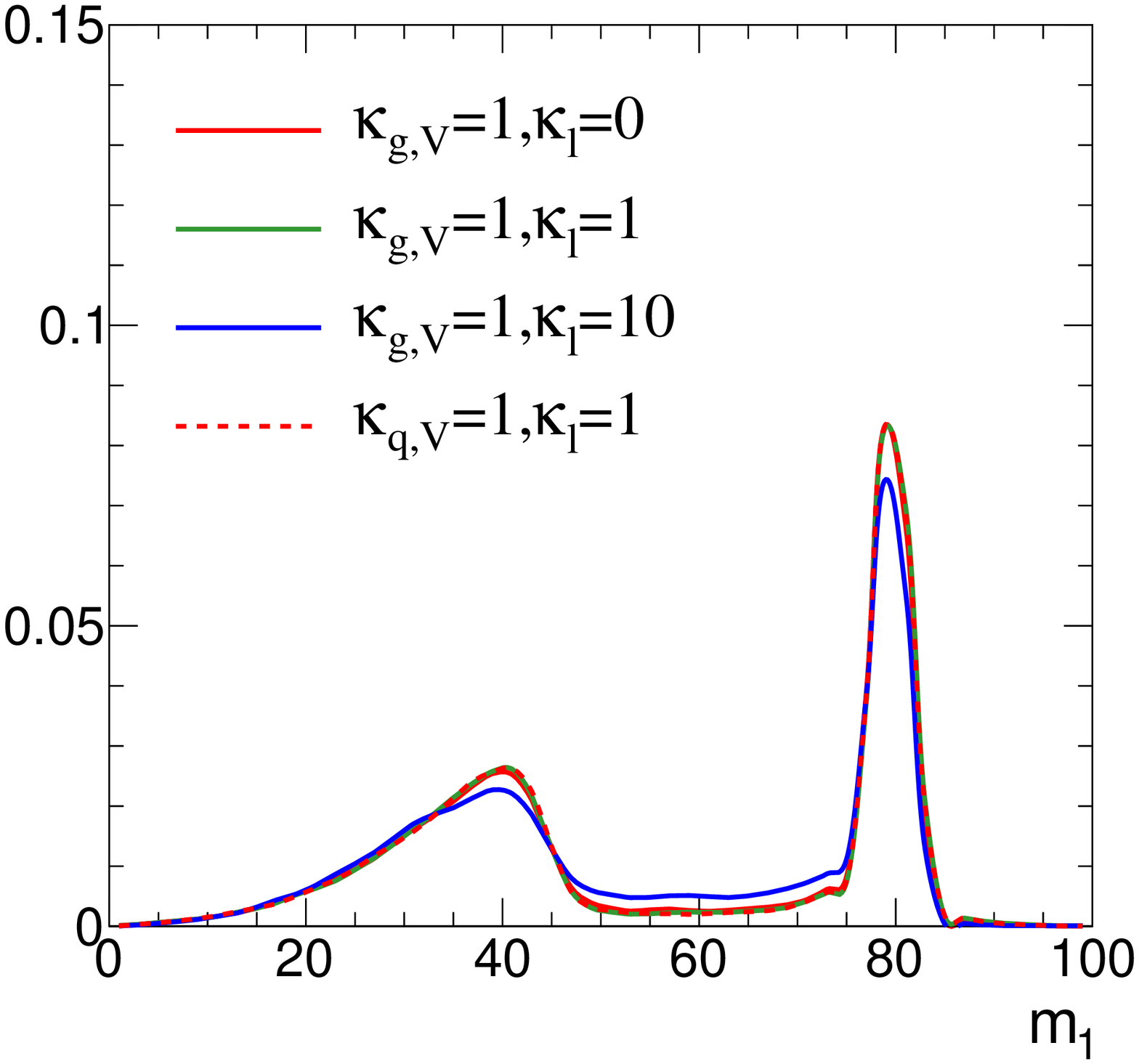} \\[-1.7cm]
\hskip -3cm
 \includegraphics[width=0.24\textwidth]{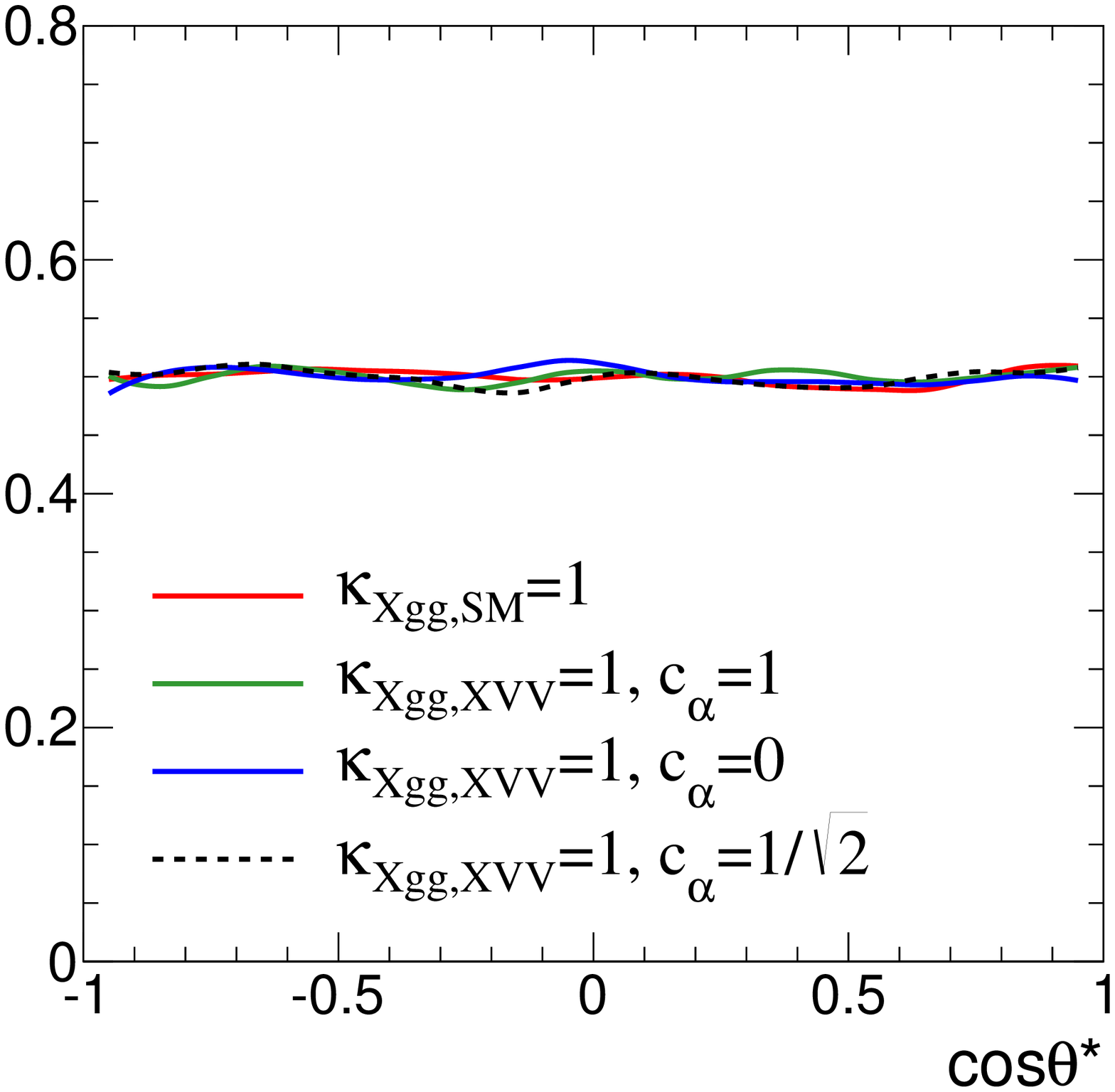}\hskip -0.3cm
 \includegraphics[width=0.24\textwidth]{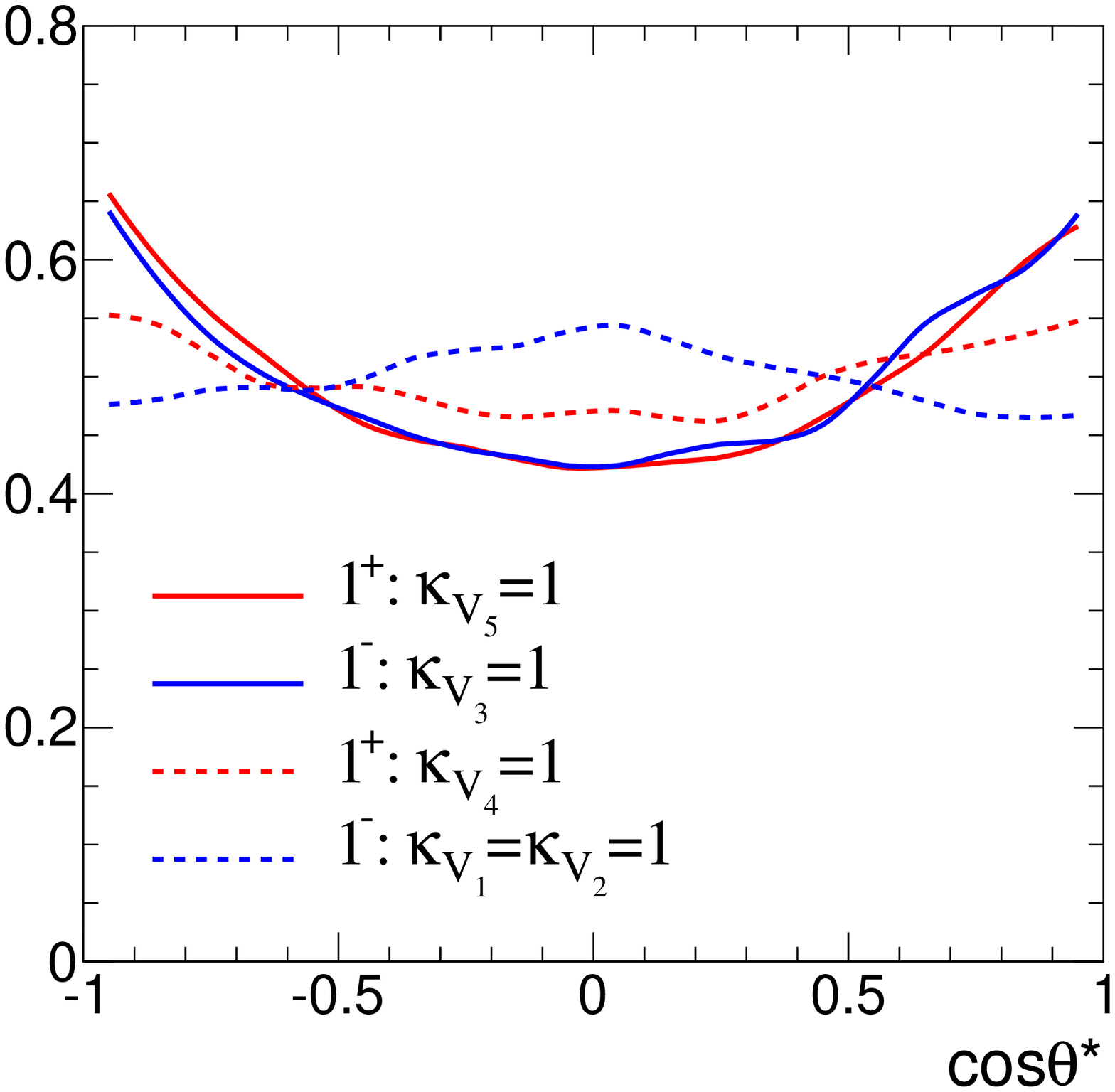}\hskip -0.3cm
 \includegraphics[width=0.24\textwidth]{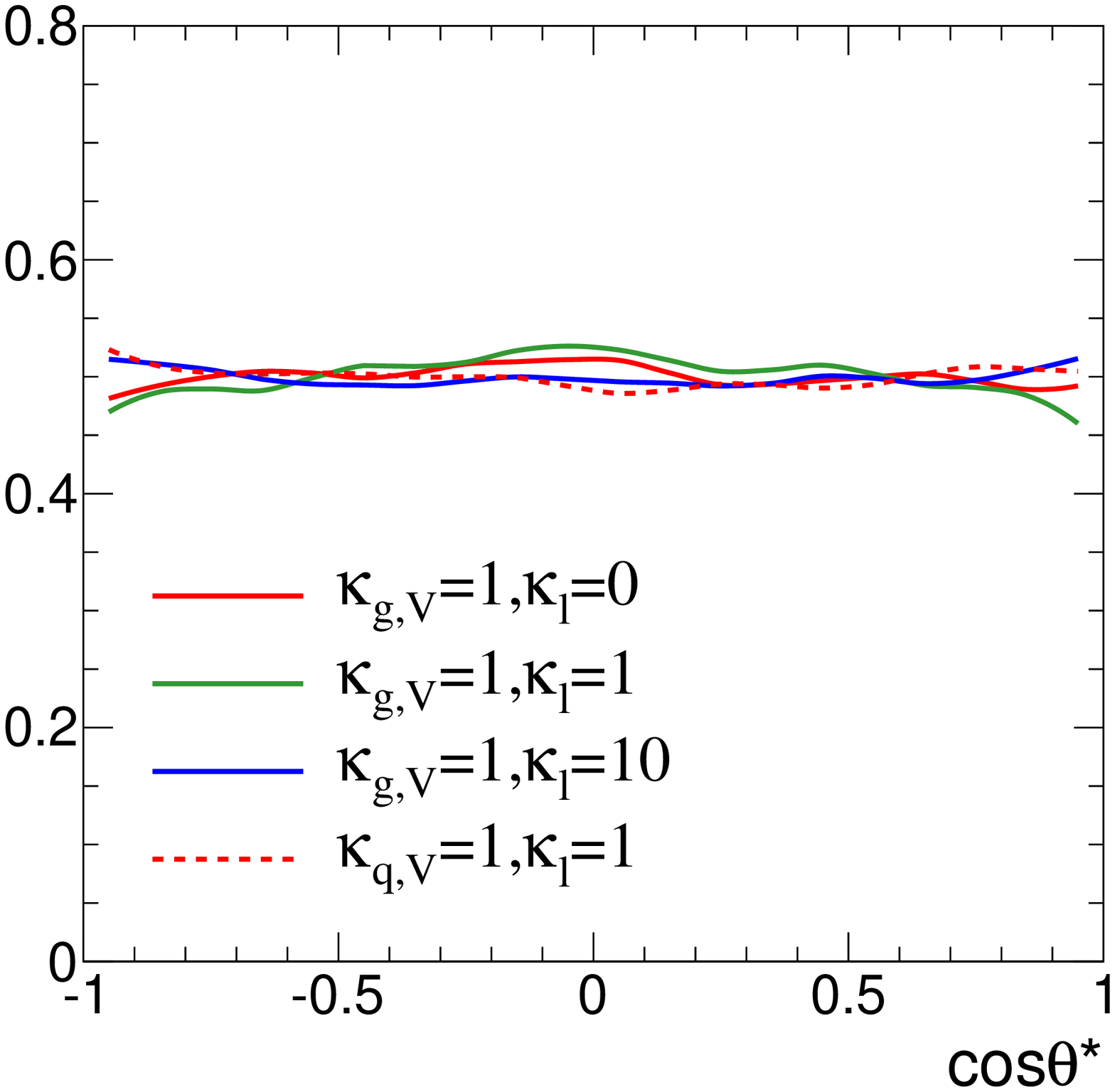} \\[-1.7cm]
\hskip -3cm
 \includegraphics[width=0.24\textwidth]{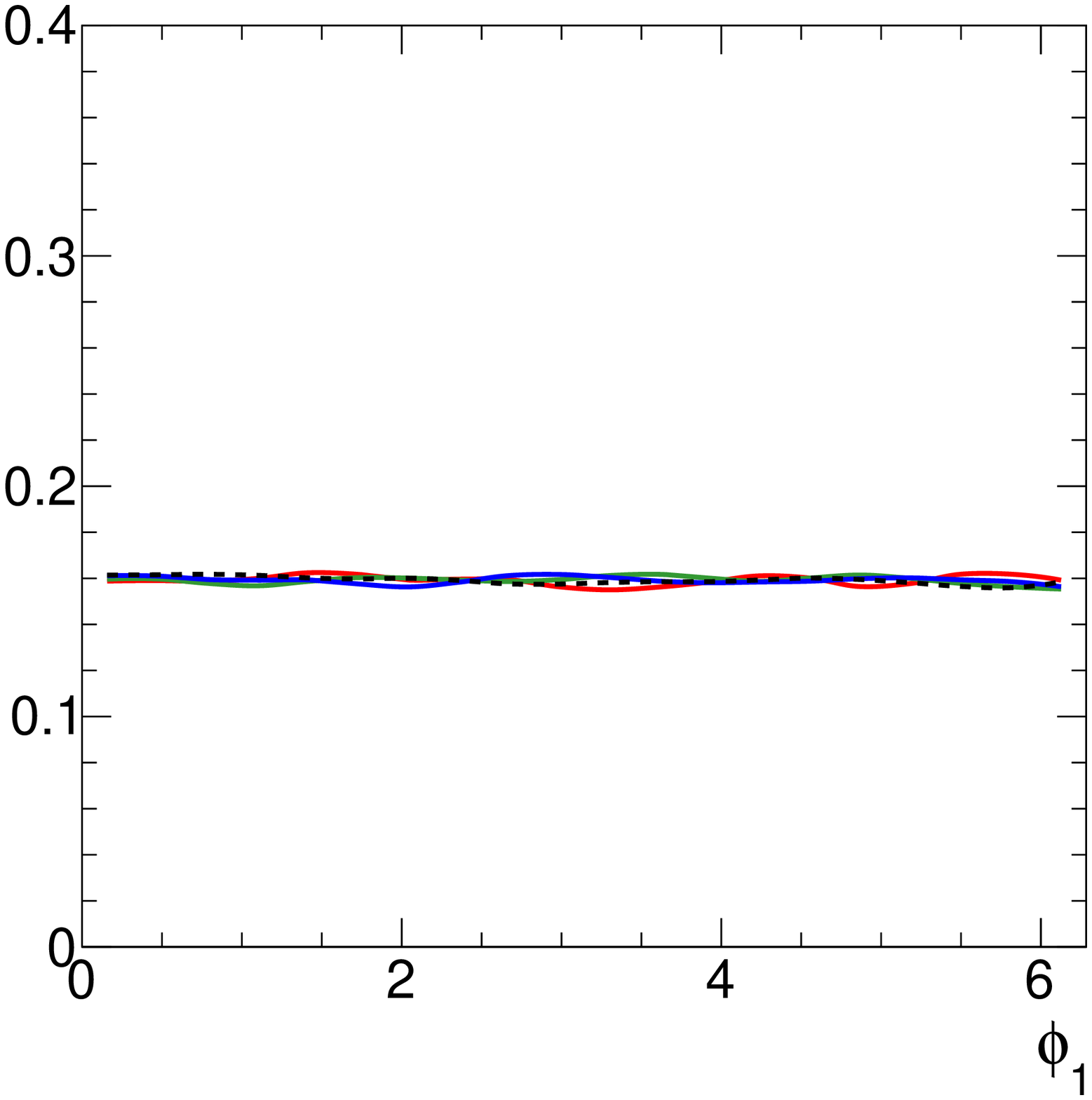}\hskip -0.3cm
 \includegraphics[width=0.24\textwidth]{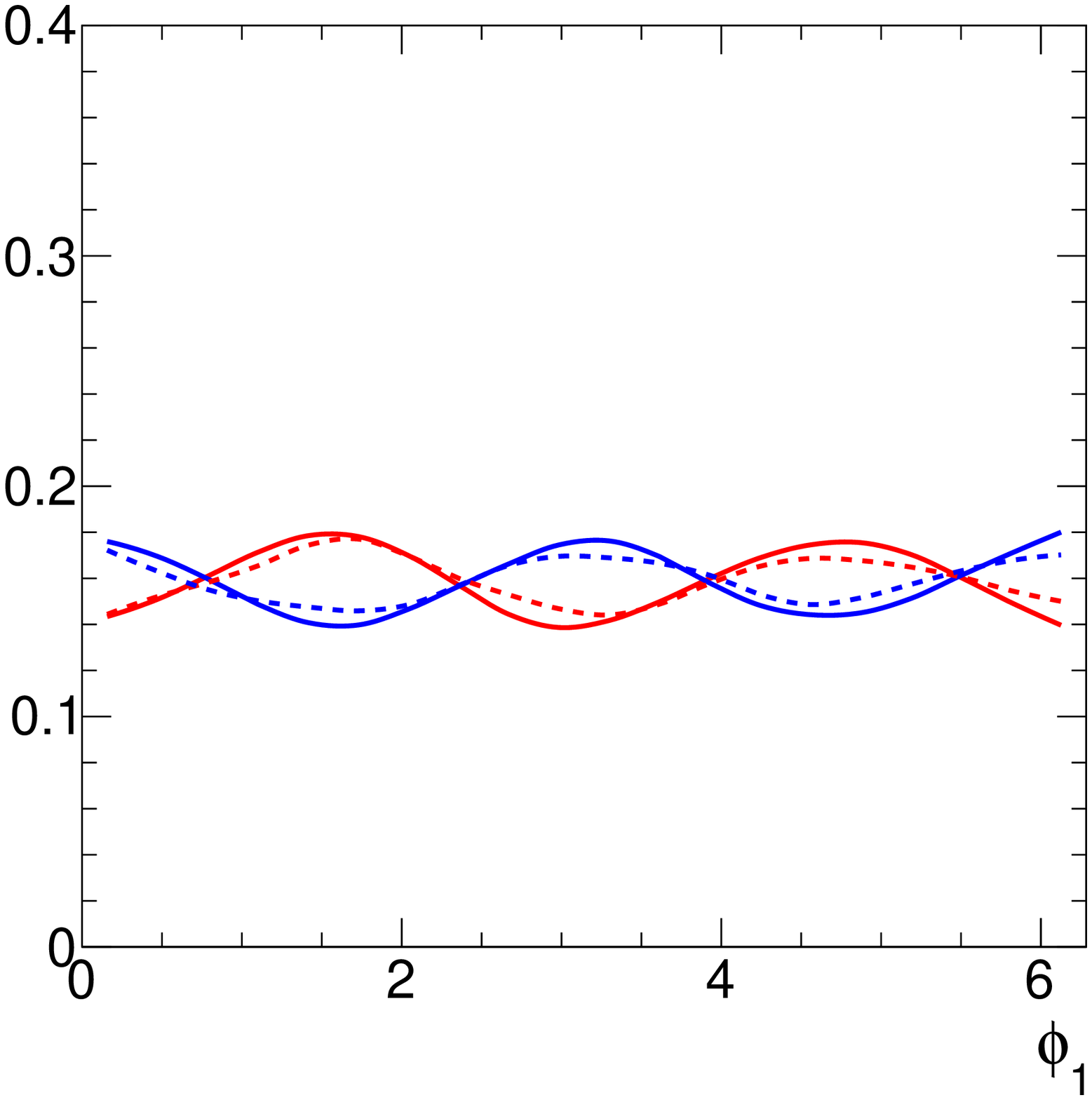}\hskip -0.3cm
 \includegraphics[width=0.24\textwidth]{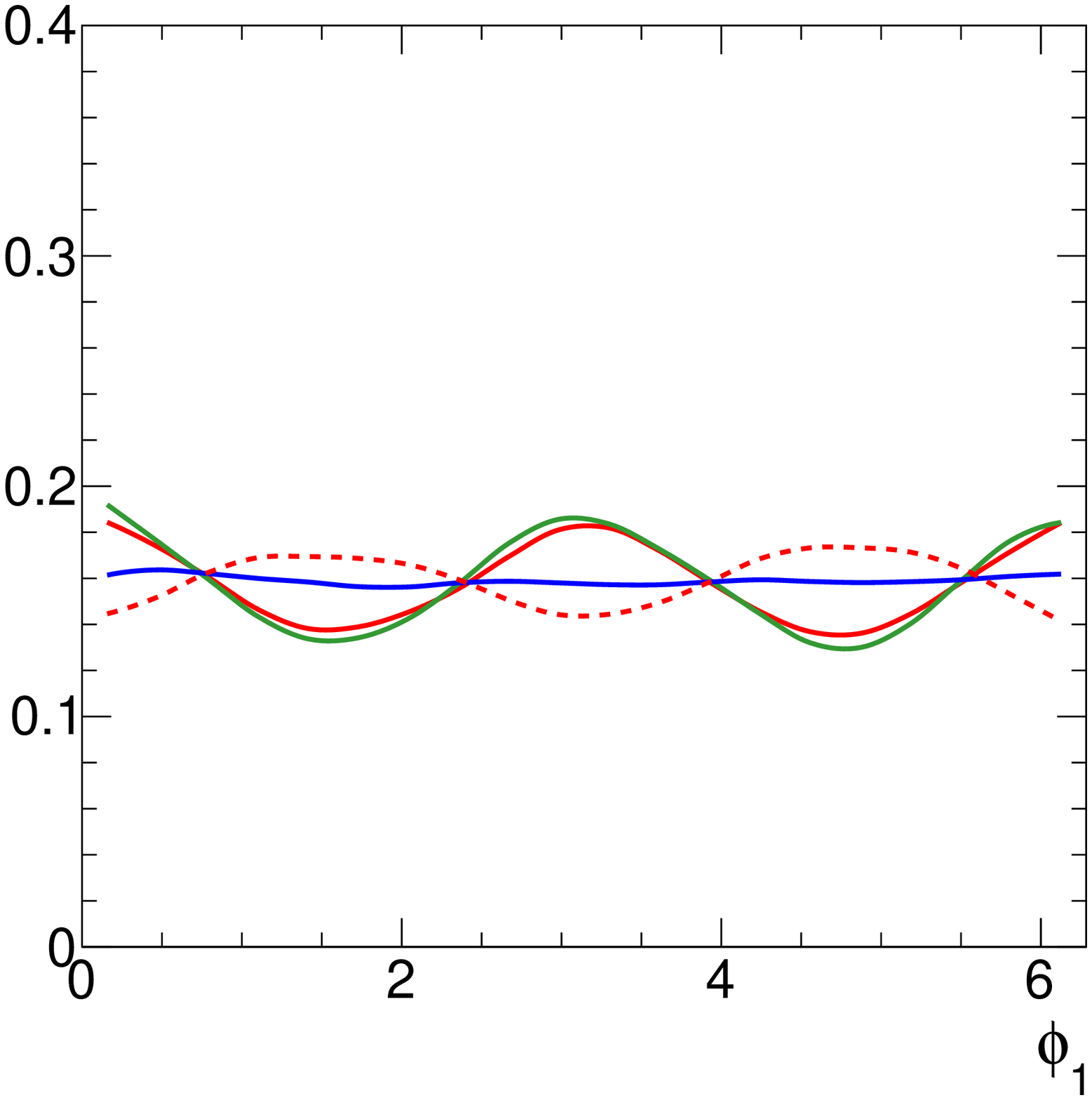} \\[-1.7cm]
\hskip -3cm
 \includegraphics[width=0.24\textwidth]{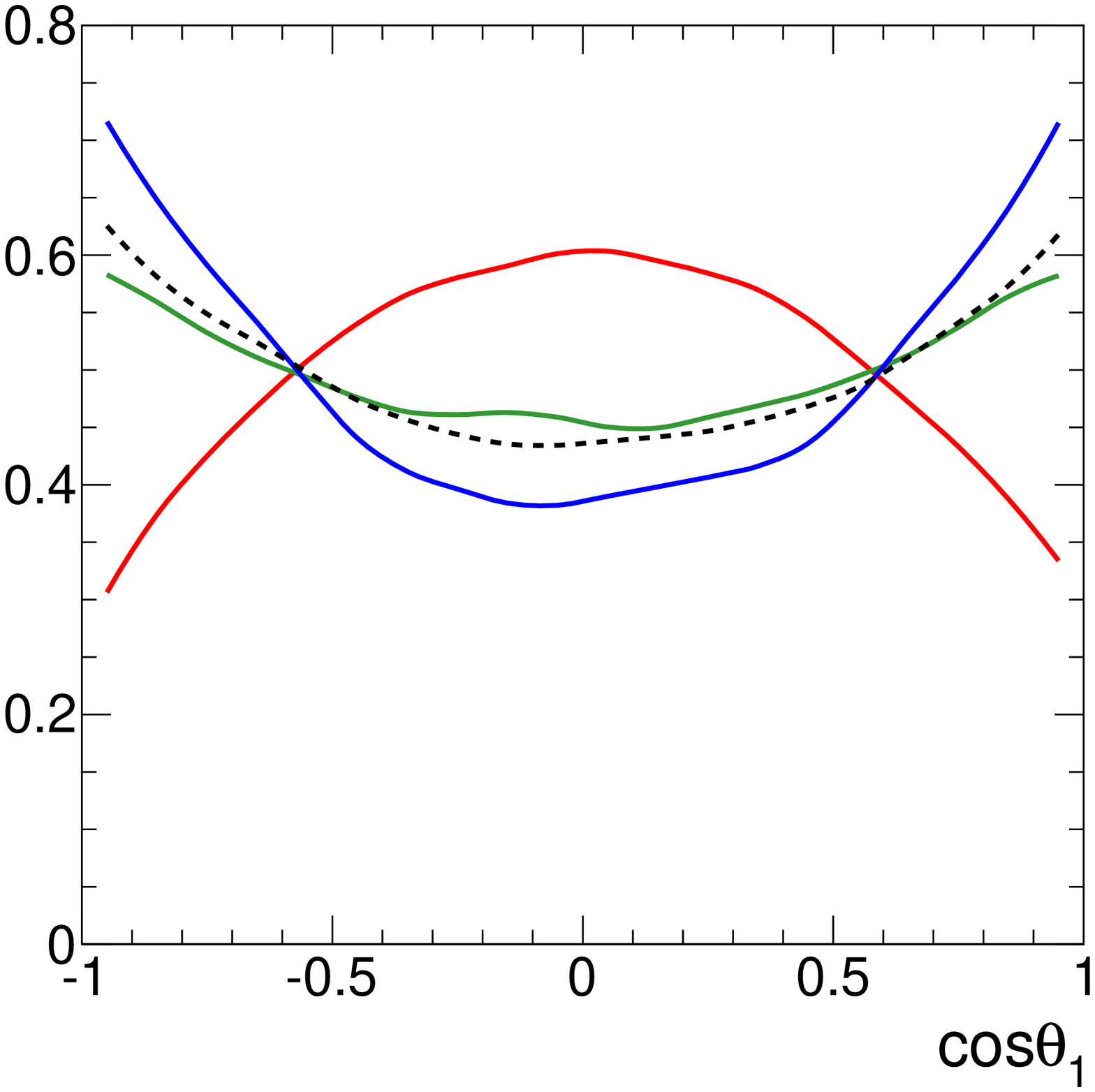}\hskip -0.3cm
 \includegraphics[width=0.24\textwidth]{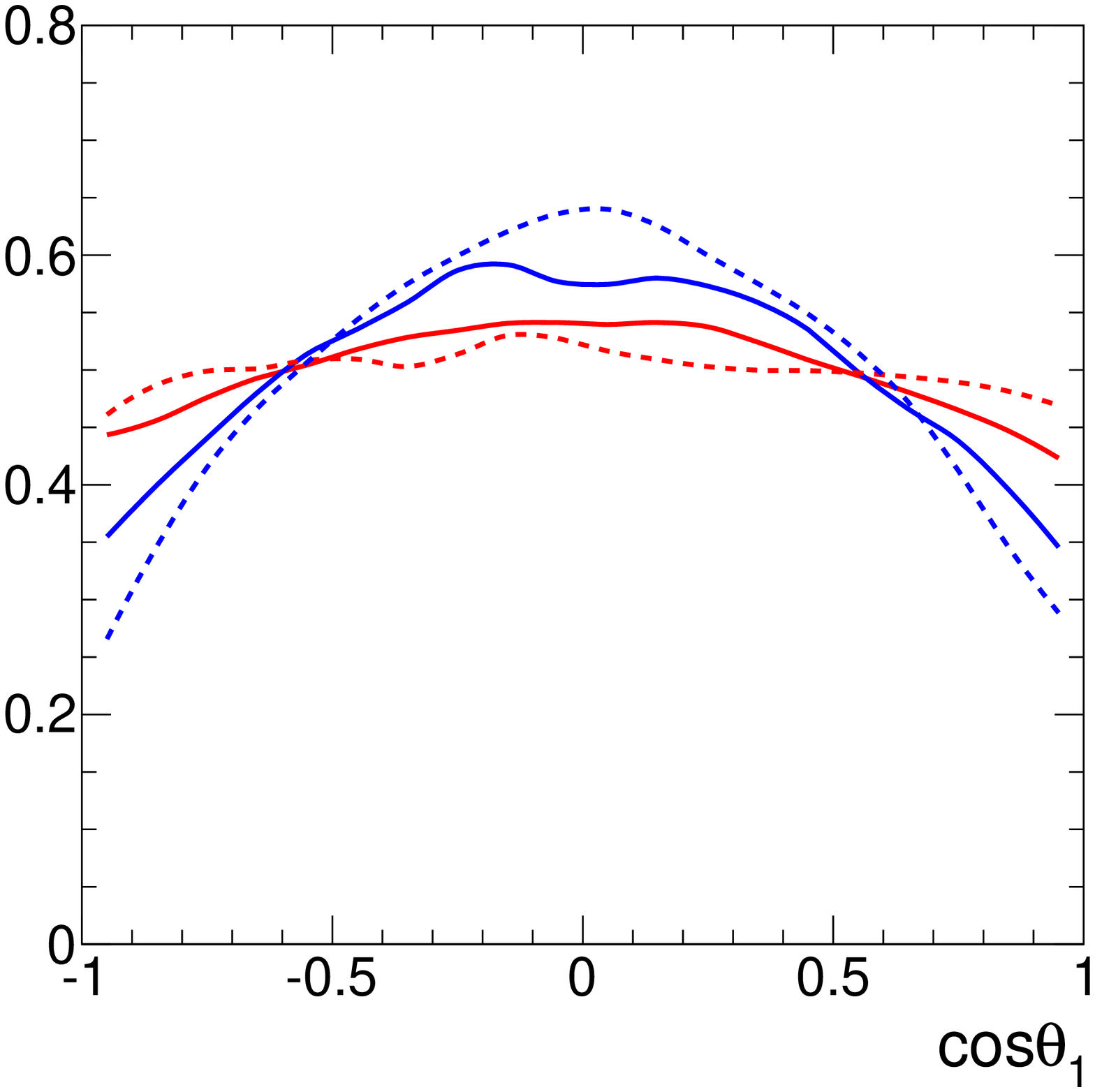}\hskip -0.3cm
 \includegraphics[width=0.24\textwidth]{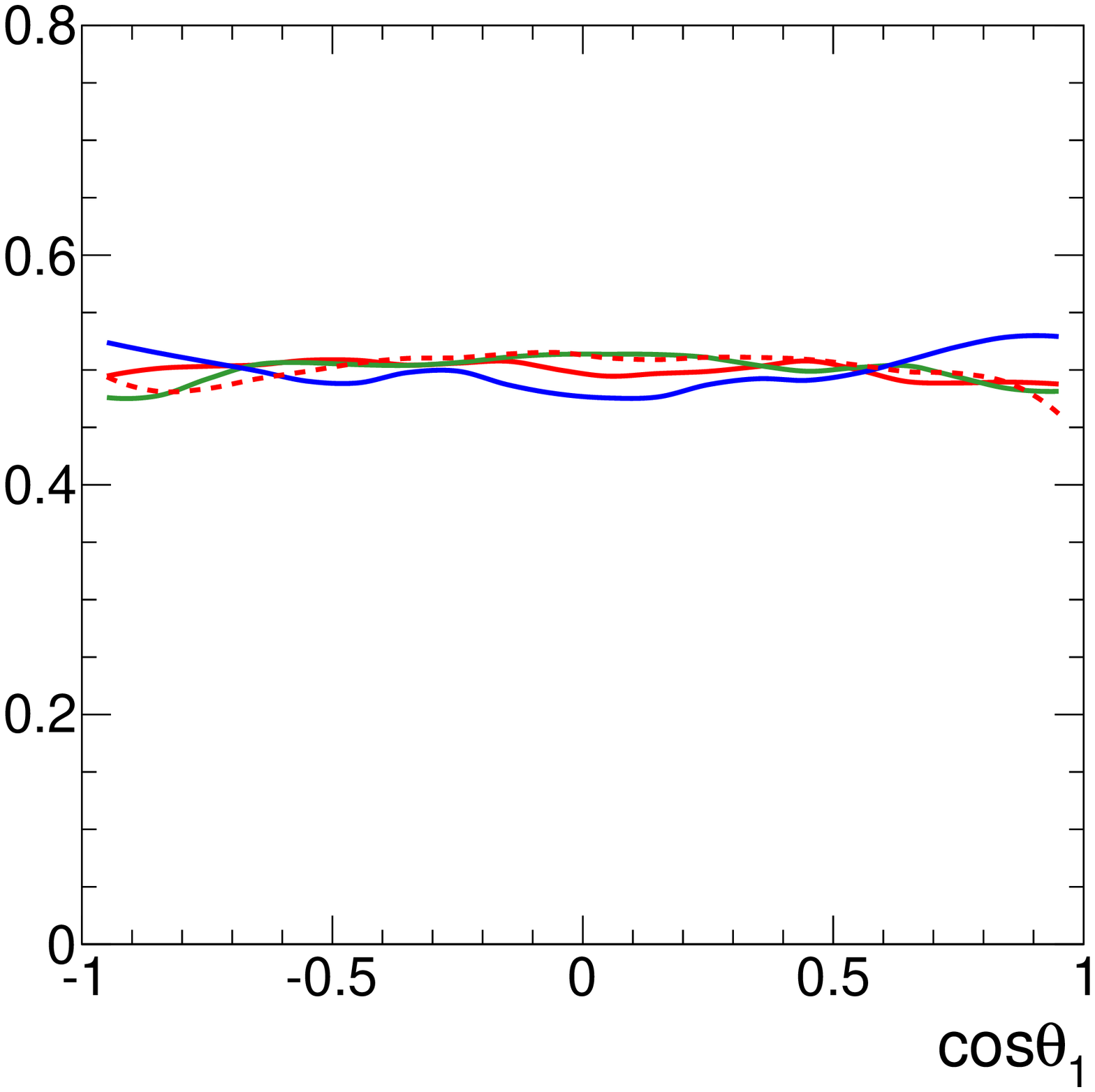} \\[-1.7cm]
\hskip -3cm
 \includegraphics[width=0.24\textwidth]{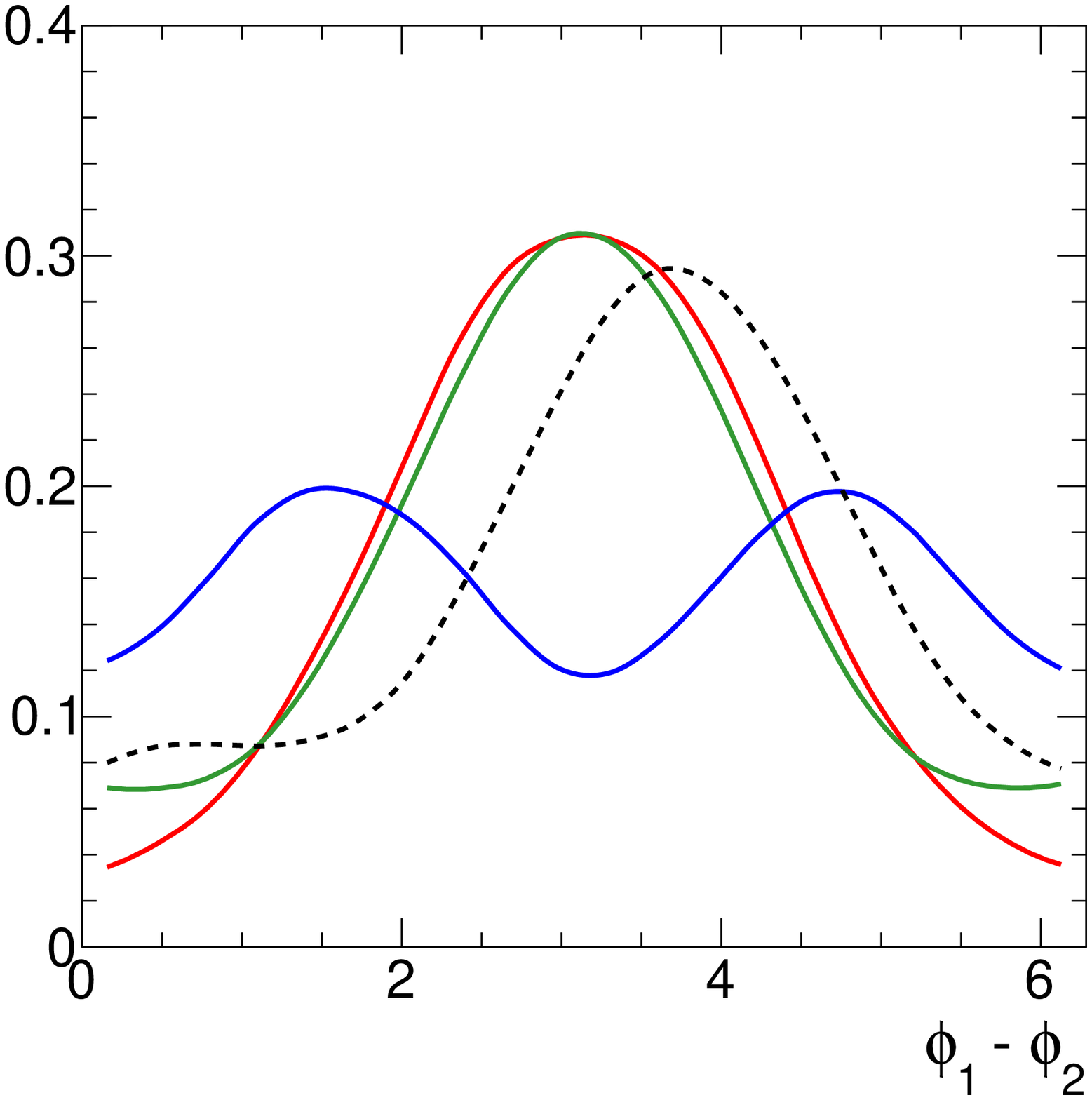}\hskip -0.3cm
 \includegraphics[width=0.24\textwidth]{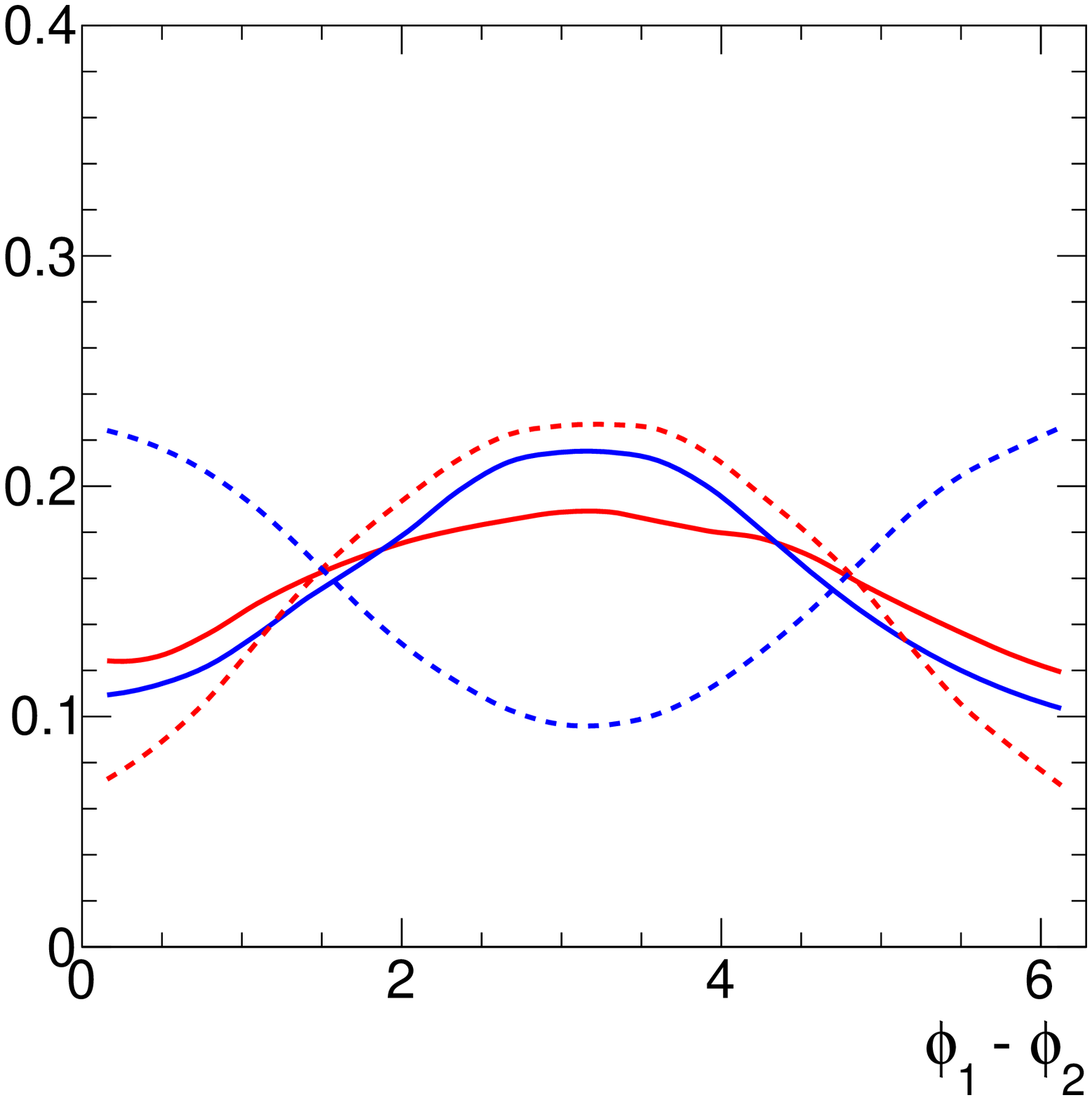}\hskip -0.3cm
 \includegraphics[width=0.24\textwidth]{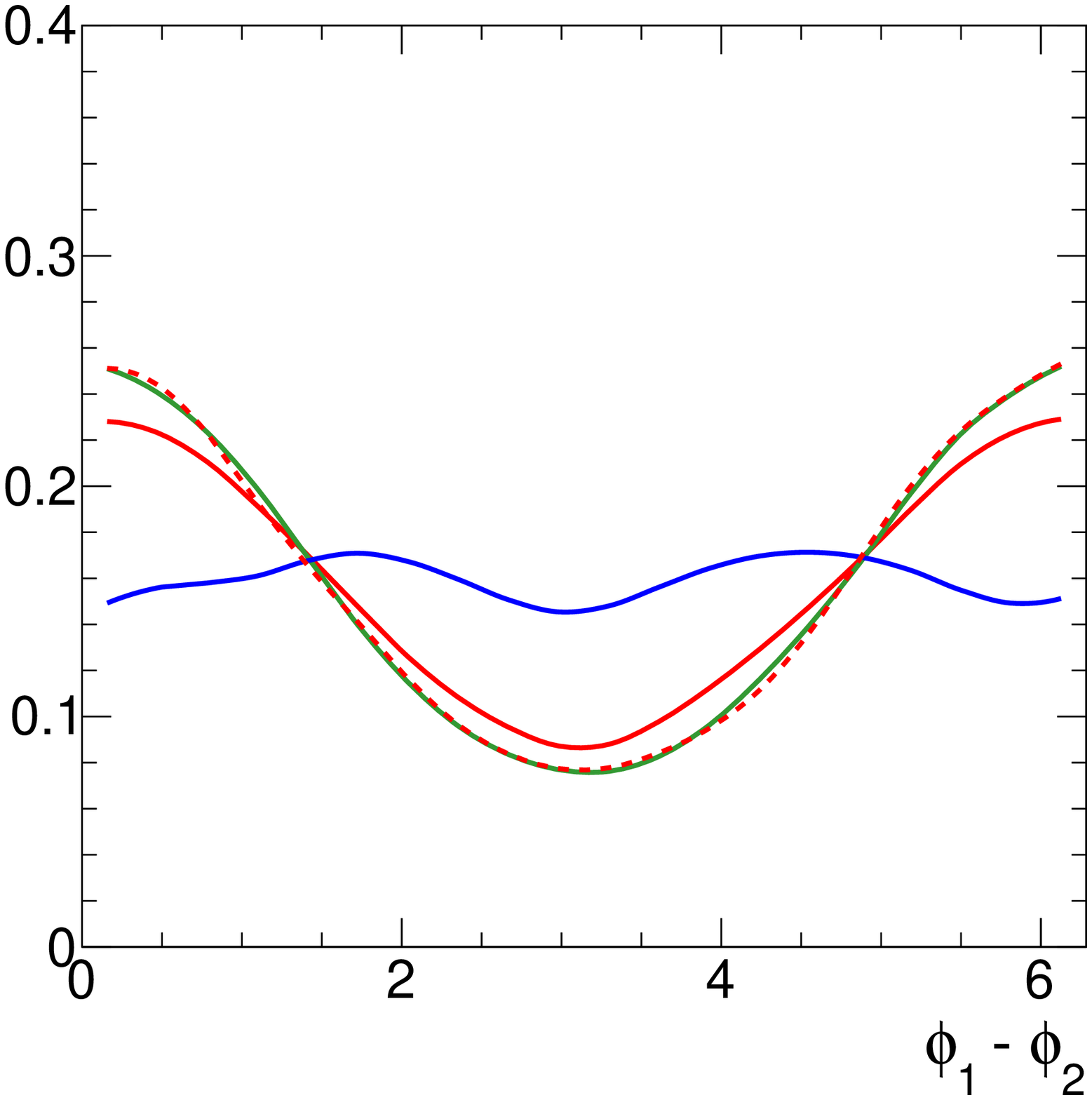}\\
\end{center} 
\vskip -0.7cm
\caption{Distributions of the $\PX\to \PW\PW$ analysis; see also JHU Fig.~13~\cite{Bolognesi:2012mm}.}
\label{fig:WW}
\end{figure}

\clearpage

\newpage
%\section{Heavy Higgs search \footnote{%
%   S.Bolognesi, S.Diglio, D. Buarque, J.Campbell, R.K.Ellis, S.Frixione, L.Kashif, N.Kauer, A.Laureys, Q.Li, F.Maltoni, G.Passarino, M.Rauch, F.Schissler, T.Taylor, J.Wang, C.Williams, C.Zhang}}
\section{Heavy Higgs search \footnote{%
      S. Bolognesi, S. Diglio, M. Kadastik, H.E. Logan, M. Muhlleitner,
      K. Peters (eds.); D. Buarque, J. Campbell, R.K. Ellis, S. Frixione,
      L. Kashif, N. Kauer, A. Laureys, Q. Li, F. Maltoni, G. Passarino,
      M. Rauch, F. Schissler, P.T.E. Taylor, J. Wang, C. Williams, C. Zhang}}
\label{sec:HH}

%\subsection{Introduction} 
%\label{sec:HH:intro}

The search for an heavy Higgs with mass greater than $400\UGeV$ assuming Standard Model (SM) properties is well motivated by the necessity 
to directly explore all the available mass range up to $1\UTeV$ without having any prejudice on the nature of the more recently discovered $125\UGeV$ resonance, 
and by the possibility to use these searches as a starting point for specific Beyond the Standard Model (BSM) extensions which can be studied by 
simply rescaling SM analysis. This will be extensively discussed in \Sref{sec:BSM}.\\

The main differences between low and high mass regions will be discussed in \Sref{sec:HH:lineshape_and_interference}, 
the available Monte Carlo generators will be presented in \Sref{sec:HH:MC} and the studies performed by using these generators and 
specific tools developed for the high mass region searches will be summarized in \Sref{sec:reweighting}. 
Finally the Vector Boson Fusion case will be discussed in \Sref{sec:interfVBF}.

\subsection{Lineshape and signal/background interference in $gg\rightarrow VV$} 
\label{sec:HH:lineshape_and_interference}

The lineshape of an unstable particle is usually described at NLO in Monte Carlo generators 
with a Breit-Wigner distribution. This approximation has an accuracy of the order $\Gamma /m$, 
where $\Gamma$ is the width and $m$ is the nominal mass of the particle, but it breaks down at high Higgs mass 
due to the very large Higgs width: $\MH\gtrsim$ $450\UGeV$ gives $\Gamma_{\PH}/\MH\gtrsim 10\%$. 

The problem has been discussed in details in \Bref{Goria:2011wa} and a more correct approach
to describe the Higgs lineshape has been proposed, known as Complex Pole Scheme (CPS).
The corresponding complete calculation of the lineshape in the $\Pg\Pg\rightarrow \PH 
\rightarrow \PVB\PVB$ process with assessment of the theoretical  uncertainties 
is presented in \Sref{HMcpss}.

An alternative general approach to describe the lineshape based on effective Lagrangian is presented in \Sref{sec:HH:effLine}.

Another important effect that becomes very large with the increase of the Higgs mass is the interference between the signal and the
$\Pg\Pg\rightarrow \PVB\PVB$ non resonant background, as recently discussed in \Bref{Passarino:2012ri}.
In \Sref{HMbis}, this effect is addressed with a proposal to compute the theoretical uncertainties due to missing 
higher order perturbation terms in the current available interference estimation.

%\subsubsection{Complex Pole Scheme and signal-background interference}
%\documentclass[10pt]{article}
%---
%\usepackage{heppennames2}
%\usepackage{lhchiggs}
%\usepackage{cernunits}
%\usepackage{hepparticles}
%---
\hyphenation{ma-ni-pu-la-tions}
\allowdisplaybreaks
%
%%%%%%%%%%%%%%%%%%%%%%%%%%%%%%%%%%%%%%%%%%%%%%%%%%%%%%%%%%%%%%%%%%%%
% basic data for the eprint:
%%%%%%%%%%%%%%%%%%%%%%%%%%%%%%%%%%%%%%%%%%%%%%%%%%%%%%%%%%%%%%%%%%%%

%\textwidth=6.5in  \textheight=8.7in
%\leftmargin=-0.8in   \topmargin=-0.20in
%\hoffset=-.85in

%%%%%%%%%%%%%%%%%%%%%%%%%%%%%%%%%%%%%%%%%%%%%%%%%%%%%%%%%%%%%%%%%%%%

%--
%- Local macros
%--

%--

%%%%%%%%%%%%%%%%%%%%%%%%%%%%%%%%%%%%%%%%%%%%%%%%%%%%%%%%%%%%%%%%%%%%%%%%%%%
%
%\begin{document}
%--
\subsubsection{CPS - scheme \label{HMcpss}}
%--
The general formalism for describing unstable particles in QFT was developed long ago,
see \Brefs{Veltman:1963th,Jacob:1961zz,Valent:1974bd,Lukierski:1978ke,Bollini:1993yp}.
For an implementation of the formalism in gauge theories we refer to the work of
\Brefs{Grassi:2001bz,Grassi:2000dz,Kniehl:1998vy,Kniehl:1998fn}, for complex
poles in general to \Brefs{Stuart:1991xk,Argyres:1995ym,Beenakker:1996kn}.

We can summarize by saying that unstable particles are described by irreducible, 
non-unitary representations of the Poincare group, corresponding to the complex 
eigenvalues of the four-momentum  $p^{\mu}$ satisfying the condition 
$p^2= - \mu^2 + i\,\mu\gamma$. 

A complete implementation of the Higgs-boson complex pole (within the SM) can be found in the
work of \Brefs{Actis:2008uh,Passarino:2010qk,Goria:2011wa}.

In this Section we summarize the status of theoretical uncertainties (THU) associated with the 
Higgs boson lineshape. The recent observation of a new massive neutral boson by ATLAS and CMS 
opens a new era where characterization of this new object is of central importance.
The search for the coupling structure of the light Higgs-like particle, as well as for new
heavy states, will continue. The huge uncertainty used so far for the heavy Higgs 
searches~\cite{Dittmaier:2011ti} ($1.5\,(M_{\PH}/\UTeV)^3$ uncertainty on the cross-section) 
was supposed to cover both the effect of the incorrect treatment of the lineshape and the missing 
interference. However, this uncertainty forced ATLAS and CMS to stop the search at $600$ \UGeV, 
where the uncertainty is $30\%$.

In the following we will review recent improvements on estimating the THU. We do not discuss 
uncertainties coming from QCD scale variations\footnote{For a recent discussion see 
https://indico.cern.ch/conferenceDisplay.py?confId=251810} and from 
PDF$\,+\alphas$~\cite{Dittmaier:2011ti}.
%--

Until recently, the Higgs boson invariant mass distribution (Higgs-boson-lineshape) has
received little attention.
In the work of \Brefs{Passarino:2010qk,Goria:2011wa} we have made an attempt to resolve 
the problem by comparing different theoretical inputs to the off-shellness of the Higgs boson. 
There is no question at all that the zero-width approximation should be avoided, especially in 
the high-mass region where the on-shell width becomes of the same order as the on-shell mass, 
or higher. We have shown evidence that only the Dyson-resummed propagator should be used, 
leading to the introduction of the $\PH$ complex pole, a gauge-invariant property of the 
$S\,$-matrix. It is convenient to describe the Complex-Pole scheme (CPS) as follows: the signal 
cross-section for the process $i j \to \PF$ can be written as
%--
\bq
\sigma_{i j \to \PH \to \PF}(s) = \frac{1}{\pi}\,
\sigma_{i j \to \PH}\,\frac{s^2}{\bmid s - s_{\PH}\bmid^2}\,
\frac{\Gamma^{\tot}_{\PH}(\sqrt{s})}{\sqrt{s}}\,\hbox{BR}\bigl( \PH \to \PF\bigr),
\quad
\Gamma^{\tot}_{\PH}(\sqrt{s}) = \sum_{\PF}\,\Gamma_{\PH \to \PF}(\sqrt{s}).
\label{QFT}
\eq
%--
where $s$ is the Higgs virtuality, $s_{\PH}$ is the Higgs complex pole and we have introduced 
a sum over all final states.

Note that the complex pole describing an unstable particle is conventionally parametrized as
$s_i = \mu^2_i - i\,\mu_i\,\gamma_i$.
It would we desirable to include two- and three-loop contributions in $\gh$ and for some
of these contributions only on-shell results have been computed so far. 
Therefore, it is very useful to give a rough estimate of the missing orders. Following the
authors of \Bref{Ghinculov:1996py} (as explained in Sect.~7 of \Bref{Goria:2011wa}) 
we can estimate that the leading uncertainty in $\gh$ is roughly given by
%--
\bq
\delta_{\PH} = 0.350119\,\frac{G_{\ssF}\,\muhs}{2 \sqrt{2} \pi^2}.
\eq
%--
Uncertainty estimates in $\gh$ range from $2.3\%$ at $400$ \UGeV to $9.4\%$ at $750$ \UGeV. 
In general, we do not see very large variations up to $1$ \UTeV with a breakdown of the 
perturbative expansion around $1.74$ \UTeV.
Therefore, using $\gh\,(1 \pm \delta_{\PH})$ we can give a rough but reasonable estimate of 
the remaining uncertainty on the lineshape.
To summarize our estimate of the theoretical uncertainty associated to the signal: 
the remaining uncertainty on the production cross-section is typically well reproduced 
by $(\delta_{\PH} +1)[\%]$, $\sigma_{\mathrm{max}}$ (the peak cross-section) changes approximately 
with the naive expectation, $2\,\delta_{\PH}[\%]$.

The factor $\Gamma^{\tot}_{\PH}(\sqrt{s})$ in \eqn{QFT} deserves a separate discussion.
It represents the ``on-shell'' decay of an Higgs boson of mass $\sqrt{s}$ and we have to quantify 
the corresponding uncertainty.
The staring point is $\Gamma^{\tot}$ computed by \textsc{Prophecy4f}~\cite{Prophecy4f}
which includes two-loop leading corrections in $G_{\ssF} M^2_{\PH}$, where $M_{\PH}$ is now 
the on-shell mass. Next we consider the on-shell width in the Higgs-Goldstone model,
discussed in~\cite{Ghinculov:1996py,Frink:1996sv}. We have
%--
\bq
\frac{\Gamma_{\PH}}{\sqrt{s}}\bmid_{\ssHG} = \sum_{n=1}^3\,a_n\,\lambda^n = X_{\ssHG},
\qquad \lambda = \frac{G_{\ssF} s}{2 \sqrt{2} \pi^2}.
\label{HGG}
\eq
%--
Let $\Gamma_{\mathrm p}= X_{\mathrm{p}}\,\sqrt{s}$ the width computed by \textsc{Prophecy4f}, 
we redefine the total width as
%--
\bq
\frac{\Gamma_{\tot}(\sqrt{s})}{\sqrt{s}} = \bigl( X_{\mathrm{p}} - X_{\ssHG} \bigr) + X_{\ssHG} =
\sum_{n=0}^3\,a_n\,\lambda^n,
\eq
%--
where now $a_0 = X_{\mathrm{p}} - X_{\ssHG}$. As long as $\lambda$ is not too large we can define
a $p\% < 80\%$ credible interval (see the work of \Bref{Cacciari:2011ze} for details) as 
(following from $a_{2,3} < a_1$)
%--
\bq
\Gamma_{\tot}(\sqrt{s}) = \Gamma_{\mathrm p}(\sqrt{s}) \pm \Delta\Gamma, \qquad
\Delta\Gamma = \frac{5}{4}\,\max\{\mid a_0\mid,a_1\}\,p\%\,\lambda^4\,\sqrt{s}.
\eq
%--
The CPS has been recently implemented within the \textsc{POWHEG-BOX} Monte Carlo 
generator~\cite{Alioli:2010xd}. 
%--
\subsubsection{Interference signal - background \label{HMbis}}
%--
In the current experimental analysis there are additional sources of uncertainty, e.g.\ background 
and Higgs interference 
effects~\cite{ATLAS:2012ac,pippo-atlas,Chatrchyan:2012ty,pluto-atlas,Chatrchyan:2012ft}. 
As a matter of fact, this interference is partly available and should not be included 
as a theoretical uncertainty; for a discussion and results we refer to 
\Brefs{Campbell:2011cu,Kauer:2012ma,Binoth:2006mf}.

Here we will examine the channel $\Pg \Pg \to \PZ \PZ$ and discuss the associated THU.
The background (continuum $\Pg \Pg \to \PZ \PZ$) and the interference are only known at leading 
order (LO, one-loop)~\cite{Glover:1988rg}. Here we face two problems, a missing NLO calculation of 
the background (two-loop) and the NLO and NNLO signal at the amplitude level, without which 
there is no way to improve upon the present LO calculation\footnote{There is, however, a recent 
and promising attempt to go beyond leading-order in \Bref{Bonvini:2013jha}.}.

A potential worry, already addressed in \Bref{Campbell:2011cu}, is: should we simply use the 
full LO calculation or should we try to effectively include the large (factor two) $K\,$-factor
to have effective NNLO observables? There are different opinions since interference effects 
may be as large or larger than NNLO corrections to the signal. Therefore, it is important to 
quantify both effects. Let us consider any distribution $D$, i.e.\
%--
\bq
D = \frac{d\sigma}{d x} \quad x = M_{\PZ\PZ} \quad \mbox{or} \quad x = p_{\perp}^{\PZ} \quad
\mbox{etc.}
\eq
%-- 
where $M_{\PZ\PZ}$ is the invariant mass of the $\PZ\PZ\,$-pair and $p_{\perp}^{\PZ}$ is the
transverse momentum. We introduce the following options, see \Bref{Passarino:2012ri}
($S,B$ and $I$ are shorthands for signal, background and interference):
%--
\begin{itemize}
\item {\bf{additive}} where one computes
%--
\bq
\diste{\NNLO} = \dist{\NNLO}(S) + \dist{\LO}(I) + \dist{\LO}(B)
\label{Aopt}
\eq
%--
\item {\bf{multiplicative}} where one computes
%--
\bq
\diste{\NNLO} = K_{\ssD}\,\bigl[ \dist{\LO}(S) + \dist{\LO}(I) \bigr] 
+ \dist{\LO}(B),
\qquad
K_{\ssD} = \frac{\dist{\NNLO}(S)}{\dist{\LO}(S)},
\label{Mopt}
\eq
%--
where $K_{\ssD}$ is the differential $K\,$-factor for the distribution. Note that
$K_{\ssD}$ accounts for both QCD and EW higher order effects in the production
and in the decay. 
%--
\item {\bf{intermediate}}
%--
It is convenient to define
\bq
K_{\ssD} = K_{\ssD}^{\Pg\Pg} + K_{\ssD}^{\mathrm{rest}},
\qquad
K_{\ssD}^{\Pg\Pg} = \frac{\dist{\NNLO}\bigl( \Pg\Pg \to \PH(\Pg) \to \PZ\PZ(\Pg)\bigr)}
{\dist{\LO}\bigl( \Pg\Pg \to \PH \to \PZ\PZ\bigr)}~,
\eq
%--
\bq
\diste{\NNLO} = K_{\ssD}\,\dist{\LO}(S) + 
\bigl( K_{\ssD}^{\Pg\Pg}\bigr)^{1/2}\,\dist{\LO}(I) + \dist{\LO}(B)~.
\label{Iopt}
\eq
%--
\end{itemize}
%--
Our recipe for estimating the theoretical uncertainty in the effective NNLO distribution is
as follows: the intermediate option gives the {\em central value}, while the band between the 
multiplicative and the additive options gives the uncertainty.
Note that the difference between the intermediate option and the median of the band is 
always small if not far away from the peak where, in any case, all options become questionable. 

For an inclusive quantity the effect of the interference, with or without the NNLO
$K\,$-factor for the signal, is almost negligible. For distributions this is radically different; 
referring to the $\PZ\PZ$ invariant mass distribution we can say that, close to 
$M_{\PZ\PZ} = \muh$, the uncertainty is small but becomes large in the rest of the search window 
$[\muh - \gh\,,\,\muh + \gh]$. 
The effect of the LO interference, w.r.t. LO $S + B$, reaches a maximum  before the peak 
(e.g.\ ${+}16\%$ at $\muh=700$ \UGeV) while our estimate of the scaled interference (always w.r.t. 
LO $S + B$) is $86^{+7}_{-3}\,\%$ in the same region, showing that NNLO signal effects 
are not negligible\footnote{Complete set of results, including results for the THU
discussed in \Sref{HMcpss}, and a code for computing the SM Higgs complex pole can be found 
at~\cite{CPHTO}.}.
%--

\noindent
{\it EW corrections to $\Pg \Pg \to \PH$  and $\PH \to \PV\PV$}\\[.2em]
%--
The NLO EW corrections to gluon fusion have been computed in 
Refs.~\cite{Actis:2008ts,Actis:2008ug}. The original results have been produced up to a Higgs 
invariant mass of $1$ \UTeV. If one is interested in the lineshape corresponding to a Higgs mass 
of $600$ \UGeV - $1$ \UTeV\  there will be some non-negligible fraction of events with invariant 
mass up to $2$ \UTeV. In this case extrapolation will give wrong results; for this reason we 
have provided additional values for the NLO EW correction factor to the inclusice cross-section: 
$\delta_{\EW} = + 19.37\%(+ 34.53\%,\, + 53.90\%)$ 
for $\muh = 1.5$ \UTeV($2$ \UTeV,\,$2.5$ \UTeV)\footnote{A complete grid up to $2.5$ \UTeV\
(see \Brefs{Passarino:2007fp,Actis:2008ug,Actis:2008ts}) and a program for a cubic interpolating 
spline incorporating the grid can be found at~\cite{EWgrid}.}.
Also $\Gamma^{\tot}_{\PH}$ of \eqn{QFT} needs some attention. The best results
available are from \Bref{Dittmaier:2011ti} where, however, tables stop at
$1$ \UTeV. If one wants to go above this value extrapolation should be avoided and it is better 
to include few additional points in the grid, e.g.\ we have included 
$\Gamma^{\tot}_{\PH} = 3.38(15.8)$ \UTeV\  for $\muh= 1.5(2)$ \UTeV. Note that, at $2$ \UTeV, one 
has $\Gamma( \PH \to \PZ\PZ) = 5.25$ \UTeV\  and $\Gamma( \PH \to \PW\PW) = 10.52$ \UTeV.
Finally, mention should be made of the very recent estimate of the N${}^3$LO QCD
corrections, see \Bref{Ball:2013bra}.
%--
\clearpage
%--
%\bibliographystyle{atlasnote}
%\bibliography{HM_Giampiero}{}

%===
%\end{document}

\label{sec:Giampiero}

\subsubsection{Effective lineshape}
\label{sec:HH:effLine}
%\section{Section name (Effective field theory approach to the Higgs lineshape)
%  \footnote{%
%    F.~Convenor, S.~Convenor, \ldots (eds.);
%    Diogo Buarque Franzosi, Fabio Maltoni, Cen Zhang } }
%\label{sec:SectionName}

%%%%%%%%%%%%%%%%%%%%%%%%%%%%%%%%%%%%%%%%%%%%%%%%%%%%%%%%%%%%%%%%%%%%%%%%%%%%%%%

\providecommand{\sw}{s_\mathrm{w}}
\providecommand{\cw}{c_\mathrm{w}}

The phenomenology of a new resonance, including searches and exclusion limits
at the LHC, depends significantly on the lineshape. When large, i.e.,
comparable to the mass, off-shell effects becomes relevant and the very same
quantum field theoretic definition of width becomes non trivial. Taking a
heavy Higgs boson as an example, we propose a new formulation of the lineshape
obtained via an effective field theory approach. Our method leads to amplitudes
that are gauge invariant, respect unitarity and it can be thought of as a
generalization of the complex mass scheme. We consider applications to the
phenomenology of a heavy SM-like scalar that are relevant for the LHC.

\subsubsubsection{Introduction}
\label{sec:introduction}
In this work we tackle the problem from an Effective Field Theory (EFT) point
of view. We propose to systematically include width effects via a set of gauge
invariant higher dimensional terms to the SM Lagrangian, along the lines of
what was first proposed in \Brefs{Beenakker:1999hi,Beenakker:2003va}. Such new
operators systematically encapsulate higher order terms coming from the
self-energy and naturally allow a running and physical width for the Higgs in a
gauge invariant way. As we will show in the following, our scheme is
consistent at higher orders and it can be considered a generalization of the
CMS(Complex Mass Scheme) as it reduces to it in the limit where the dependence
on the virtuality of the Higgs self-energy is neglected. 

\subsubsubsection{Setting up the stage}
\label{sec:setting}
The two-point Green's function for the Higgs boson is
\begin{equation}
\Delta_{\PH}(s)=s-M_{\PH,0}^2+\Pi_{\PH\PH}(s)\ ,
\end{equation}
where $M_{\PH,0}$ is the bare mass, and $\Pi_{\PH\PH}(s)$ is the Higgs self-energy.
In the conventional on-shell definition, the mass and width are given by
\begin{flalign}
&M_{\PH,{\mathrm{OS}}}^2=M_{\PH,0}^2-{\rm Re} \Pi(M_{\PH,{\mathrm{OS}}}^2)\ ,\\
&M_{\PH,{\mathrm{OS}}}\Gamma_{\PH,{\mathrm{OS}}}=\frac{{\rm Im} \Pi(M_{\PH,{\mathrm{OS}}}^2)}{1+{\rm Re} \Pi'(M_{\PH,{\mathrm{OS}}}^2)}\ .
\end{flalign}
These definitions become gauge-dependent at order $\mathcal{O}(g^4)$.

In order to avoid the divergence of the tree-level propagator $D(s)=i/(s-M_{\PH,OS}^2)$,
one performs the Dyson resummation to obtain
\begin{equation}
D(s)=\frac{i}{s-M_{\PH,\mathrm{OS}}^2+iM_{\PH,\mathrm{OS}}\Gamma_{\PH,\mathrm{OS}}}\ .
\end{equation}
To include the running effects of the width, one can further approximate the
propagator by
\begin{equation}
D(s)=\frac{i}{s-M_{\PH,{\mathrm{OS}}}^2+i{\rm Im}\Pi(s)}\ ,
\end{equation}
where the imaginary part of $\Pi(s)$ is related to the Higgs-boson width. The
consistency of the above treatments of the Higgs propagator with the equivalence
theorem and unitarity has been discussed in
\Brefs{Valencia:1990jp,Valencia:1992ix}.

Alternatively, as shown in a series of papers~\cite{Stuart:1991xk,
Aeppli:1993rs,Veltman:1992tm,Sirlin:1991rt,Willenbrock:1991hu,Passera:1996nk,
Kniehl:1998vy,Kniehl:1998fn,Jegerlehner:2001fb,Jegerlehner:2002em} a
consistent, convenient and resilient definition of mass $\mu$ and width
$\gamma$ up to two loops, is obtained by setting $s_{\PH} \equiv  \mu^2 - i \mu
\gamma$ and then solving the implicit equation 
\begin{equation}
s_{\PH} - M_{\PH,0}^2 + \Pi_{\PH\PH}(s_{\PH}) =0
\end{equation}
in terms of $s_{\PH}$. This gives a gauge independent definition to all orders
~\cite{Gambino:1999ai,Grassi:2000dz,Grassi:2001bz} (independent of the gauge
choice present in the computation of $\Pi_{\PH\PH}(s_{\PH})$) and in addition avoids
unphysical threshold singularities ~\cite{Kniehl:2002wn}.

The above definition is also consistent with the use of the CMS. In this scheme
the propagator is $\Delta_{\PH}^{-1}(s)=s-s_{\PH}$. By definition this approach can
give a good approximation of the full propagator 
\begin{equation}
\label{eq:prop}
\Delta_{\PH}^{-1}(s)=\frac{1}{s-s_{\PH}+\Pi^{\mathrm R}_{\PH\PH}(s)}
\end{equation}
only close to the pole or, equivalently, for a small width, $\gamma/ \mu\muchless 1$.
Here $\Pi^{\mathrm R}_{\PH\PH}(s)$ is the renormalized self energy, satisfying the following
renormalization conditions:
\begin{equation}
\Pi^{\mathrm R}_{\PH\PH}(s_{\PH})=0\ ,\quad \Pi'^{\mathrm R}_{\PH\PH}(s_{\PH})=0\ .
\end{equation}

A natural improvement would consist in including the full resummed propagator
in explicit calculations. This, however, leads to gauge violation already at the
tree level. The reason being that in perturbation theory gauge invariance is
guaranteed order by order while the presence of a  width implies the resummation
of a specific subset of higher order contributions, the self-energy
corrections. This results in a mixing of different orders of perturbation
theory. In particular, the following issues need to be addressed:\\
\begin{enumerate}
\item In general $\Pi_{\PH\PH}(s)$ explicitly depends on the gauge-fixing parameter
  (GFP).  To resum the self-energy correction to all orders, $\Pi_{\PH\PH}(s)$ must
  be extracted in a physically meaningful way.
\item The resummed propagator spoils the gauge cancellation among different
  diagrams, and eventually leads to the violation of Goldstone-boson equivalence
  theorem and unitarity bound.
\end{enumerate}
Both issues can be tackled by the so-called Pinch Technique (PT)
~\cite{Cornwall:1981zr,Cornwall:1989gv,Papavassiliou:1989zd,Degrassi:1992ue}. In
the PT framework, a modified one-loop self-energy for the Higgs boson can be
constructed by appending to the conventional self energy additional
propagator-like contributions concealed inside vertices and boxes.  For the
application of PT in resonant transition amplitude, and in particular, the
extraction of a physical self energy, we refer to 
\Brefs{Papavassiliou:1995fq,Papavassiliou:1996zn,Papavassiliou:1995gs,Papavassiliou:1997fn,Papavassiliou:1998pb}.

The modified self-energy correction for the Higgs is GFP-independent, and
reflects properties generally associated with physical observables.  At the one
loop level, we have the following expressions
~\cite{Papavassiliou:1997fn,Papavassiliou:1997pb}
\begin{flalign}
  \label{eq:PiPT1}
\Pi^{(\PW\PW)}_{\PH\PH}(s)&=\frac{\alpha_\rw}{16\pi}\frac{\MH^4}{\MW^2}\left[1+4\frac{\MW^2}{\MH^2}-4\frac{\MW^2}{\MH^4}(2s-3\MW^2)\right]
B_0(s,\MW^2,\MW^2)\ ,
\\
\Pi^{(\PZ\PZ)}_{\PH\PH}(s)&=\frac{\alpha_\rw}{32\pi}\frac{\MH^4}{\MW^2}\left[1+4\frac{\MZ^2}{\MH^2}-4\frac{\MZ^2}{\MH^4}(2s-3\MZ^2)\right]
B_0(s,\MZ^2,\MZ^2)\ ,
\\
\Pi^{(\Pf\Pf)}_{\PH\PH}(s)&=\frac{3\alpha_\rw}{8\pi}\frac{m_{\Pf}^2}{\MW^2}\left(s-4m_{\Pf}^2\right)B_0(s,m_{\Pf}^2,m_{\Pf}^2)\ ,
\\
\Pi^{(\PH\PH)}_{\PH\PH}(s)&=\frac{9\alpha_\rw}{32\pi}\frac{\MH^4}{\MW^2}B_0(s,\MH^2,\MH^2)\ ,
\label{eq:PiPT4}
\end{flalign}
where the superscripts denote the contributions from the \PW, \PZ, fermions and
Higgs loops, and
\begin{flalign}
B_0(p^2,m_1^2,m_2^2)&\nonumber\\
\equiv
(2\pi\mu)^{4-d}&\int\frac{d^dk}{i\pi^2}\frac{1}{\left(k^2-m_1^2\right)\left[(k+p)^2-m_2^2\right]}
\end{flalign}
is the normal Passarino-Veltman function
~\cite{'tHooft:1978xw,Passarino:1978jh}. These results
are independent of the GFP. Note that the expressions in
\Eqs~(\ref{eq:PiPT1}--\ref{eq:PiPT4}) coincide with the $\xi=1$ result obtained in
the background-field gauge
~\cite{Denner:1994nn,Hashimoto:1994ct,Papavassiliou:1994yi}.

In addition, the gauge cancellation among different amplitudes can be restored,
by including certain vertex corrections obtained via the PT~
~\cite{Papavassiliou:1996fn,Papavassiliou:1997pb}.  This is because in this
framework the Green's functions satisfy the tree-level-like Ward Identities
(WI), which are crucial for ensuring the gauge invariance of the resummed
amplitude.

%%%%%%%%%%%%%%%%%%%%%%%
\begin{figure}[ht!]
  \centering
  \includegraphics[width=.75\columnwidth]{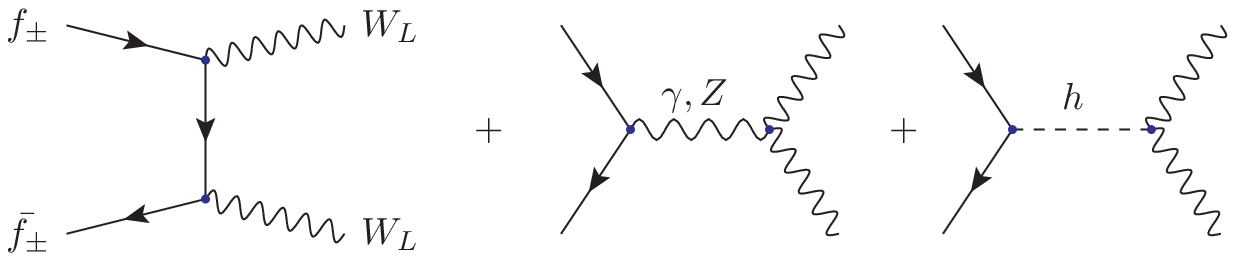}
  \caption{Diagrams contributing to fermion anti-fermion scattering into
  longitudinal $\PW$'s.  }
\label{fig:1} 
\end{figure}
%%%%%%%%%%%%%%%%%%%%%%%

As an example, let us consider the Higgs-mediated part of same helicity fermion
scattering into longitudinal \PW's, $f_\pm\bar{f}_\pm \to \PWp_{\mathrm L}\PWm_{\mathrm L}$. There are
contributions from $s$-channel and $t$-channel diagrams, as is shown in
\Fref{fig:1}.  The contributions from $t$-channel and Higgs diagram to the
amplitudes coming from longitudinal components of the \PW's and same helicity
fermions (in the high energy limit) read \footnote{Here we assume
  $\epsilon^\mu_{1,2}\approx k^\mu_{1,2}/\MW$ at high energy region.}
\begin{flalign}
\label{eq:ffww}
{\mathcal M}^L_{h} &\equiv {\mathcal M}_s^{\mu\nu}\frac{k_{1\mu} k_{2\nu}}{\MW^2}\nonumber\\
&=\frac{-igm_{\Pf}}{2\MW}\bar{v}(p_2)u(p_1)\frac{i}{\Delta_{\PH}(s)}\Gamma^{\PH\PW\PW,\mu\nu}(q,k_1,k_2)\frac{k_{1\mu} k_{2\nu}}{\MW^2}
\nonumber\\
{\mathcal M}^L_{t}&\equiv {\mathcal M}_{t,Z}^{\mu\nu}\frac{k_{1\mu} k_{2\nu}}{\MW^2}\nonumber\\
&=-\frac{ig^2 m_{\Pf} }{4\MW^2}\bar{v}(p_2)u(p_1)+\cdots
\end{flalign}
where $\Gamma_{\mu\nu}^{\PH\PW\PW}(q,k_1,k_2)$ is the $\PH\PWp\PWm$ vertex.  The ellipsis
in ${\mathcal M}^L_{t}$ denotes terms that are not related to the Higgs exchange
diagram.  These terms come from the contribution of opposite helicity fermions,
and are supposed to cancel the bad high-energy behavior of the $\PGg/\PZ$
mediated diagrams.
 
Without the Higgs contribution  ${\mathcal M}^L$ grows with energy and eventually
violates unitarity. The cancellation of the bad high-energy behavior of each
amplitude, and the equivalence theorem, are guaranteed by the following WI: 
\begin{flalign}
\label{eq:wi1}
&k_+^\mu k_-^\nu\Gamma_{\mu\nu}^{\PH\PW\PW}(q,k_+,k_-)=\nonumber\\
&-\MW^2\Gamma^{\PH\phi^+\phi^-}(q,k_+,k_-)+\frac{ig\MW}{2}\Delta_{\PH}(q^2)\,,
\end{flalign}
where $\phi^{\pm}$ are Nambu-Goldstone bosons. Only the leading terms at high energy
are included. The relation above explicitly shows that the inclusion of higher
order terms in the imaginary part of $\Delta_{\PH}(q^2)$ has to be related to the
EW corrections of $\Gamma^{\PH\PW\PW}_{\mu\nu}$ and three scalar vertex. Only if both
$\Delta_{\PH}(s)$ and $\Gamma^{\PH\PW\PW}_{\mu\nu}$ are computed in one-loop via the PT,
then the WI remains valid, and the gauge-cancellation, as well as the
equivalence theorem, are not spoiled. Besides, ${\mathcal M}^L_{t,Z}$ is not
affected by the Higgs width and therefore the tree-level relations can be used.
Thus the resummed propagator can be consistently included with the one-loop
correction to $\Gamma^{\PH\PW\PW}_{\mu\nu}$ via the PT.

Even though correct, the solution outlined above for $f\bar{f}\to \PWp_{\mathrm L}\PWm_{\mathrm L}$
is not a general one.  In $\PWp_{\mathrm L}\PWm_{\mathrm L}\rightarrow \PZ_{\mathrm L}\PZ_{\mathrm L}$, for example, it is
not sufficient to include only the $\PH\PW\PW$ and $\PH\PZ\PZ$ corrections.  The
triple and quartic vector-boson vertices at one-loop are also required to cancel the bad
high-energy behavior of the Higgs-mediated amplitude and the overall procedure
of analyzing the full set of WI's becomes more and more involved. The goal of
this work is to present a simple method to generate the needed corrections to
the vertices and propagators so that the WI's are automatically satisfied and
unitarity automatically ensured.

\subsubsubsection{The EFT approach}
\label{sec:approach}
As explained above, we aim at finding a systematic approach to improve the
Higgs propagator without breaking either gauge invariance or unitarity. In
other words we are looking for a mechanism that guarantees the constraints
imposed by the WI to be satisfied at any order in perturbation theory.

At one loop, the full calculation via the PT certainly provides an exact
solution valid at NLO.  The challenge is to achieve the same keeping the
calculation at leading order, including only the necessary ingredients coming
from NLO and resumming them into the propagator via a Dyson-Schwinger approach.
The idea is to associate the corrections to an ad hoc constructed
gauge-invariant operator and match the operator to the one-loop two-point
function $\Delta_{\PH}(s)$ calculated via the PT.  In so doing one aims at
obtaining the exact resummed propagator already at the leading order and,
\textit{at the same time},  the  interactions modified to automatically satisfy the
WI's.  The latter desired result ensures the gauge-invariance of the amplitudes,
and it can be considered as an approximation to a full one-loop calculation in
PT.

To this aim, we consider the Taylor expansion of the function $\Pi(s)=\Pi^{\mathrm R}_{\PH\PH}(s)$
\begin{equation}
\Pi(s)=\sum_{i=0}^\infty c_i s^i \,,
\end{equation}
where $c_i$ are dimensionful constants and, as first attempt, we add the
following infinite set of operators to the Lagrangian:
\begin{flalign}
  \label{eq:op1}
  {\mathcal O}_\Pi=&\sum_{i=0}^\infty c_i \phi^\dagger(-D^2)^i\phi
\nonumber\\
\equiv&\phi^\dagger \Pi(-D^2)\phi
\end{flalign}
where $\phi$ is the Higgs doublet, and $D^\mu$ is the covariant derivative.  It
is straightforward to check that  ${\mathcal O}_\Pi$ modifies the Higgs propagator as
desired: the two $\phi$'s contribute two Higgs fields, and each $-D^2$
contributes an $s$ leading to 
\begin{equation}\label{eq:PiHH}
\Pi(s)=\Pi_{\PH\PH}^{\mathrm R}(s)\,,
\end{equation}
as desired. Note that in principle, ${\mathcal O}_\Pi$ is a non-local operator, yet
by expanding it, we re-express it in terms of an infinite series of local
operators.\footnote{In general,  inclusion of higher-order derivatives in the
  Lagrangian leads to very peculiar quantum field theories, aka Lee-Wick
  theories, see ~\cite{Grinstein:2007mp} for a recent analysis and references.
  As we are going to see later, in our approach, we only use the imaginary part
  of $\Pi(s)$ and therefore the real part of the propagator is not affected.  }

We remark that while very similar in spirit , our approach differs from that of
Ref.~~\cite{Beenakker:1999hi}: the operator chosen there does not contain gauge
fields, and it is therefore not sufficient to restore the gauge cancellation
and fix the bad high-energy behavior in vector-vector scattering.
%%%%%%%%%%%
\begin{figure}[t]
\begin{center}
  \includegraphics[width=0.9\columnwidth]{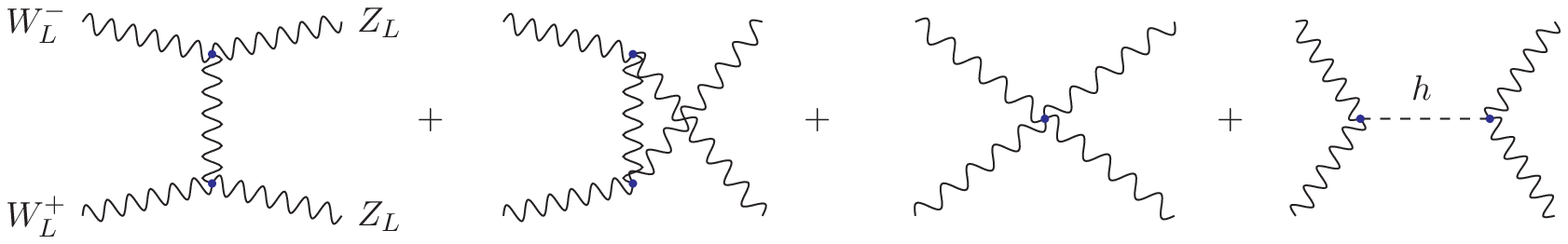}
\end{center}
\caption{Diagrams contributing to $\PWp_{\mathrm L}\PWm_{\mathrm L}\rightarrow \PZ_{\mathrm L}\PZ_{\mathrm L}$ }
\label{fig:2}
\end{figure}
%%%%%%%%%%%

\Eq~(\ref{eq:op1}) leads to the correct expression for the propagator. However,
the first term $\Pi(0)\phi^\dagger\phi$ in the expansion corresponds to a
tadpole contribution. This can be avoided if this term  is replaced by 
\[
\Pi(0)\phi^\dagger\phi
\to
\frac{\Pi(0)}{2v^2}\left[(\phi^\dagger\phi)-\frac{v^2}{2}\right]^2\,,
\]
i.e., the Higgs-self interaction is suitably modified. As one can easily  check,
such a modification leaves the relation of \Eq~(\ref{eq:PiHH}) unchanged.  The
final form of the operator, which we dub ${\mathcal{\tilde{O}}}_\Pi$, is   
\begin{equation}
  \label{eq:O}
  {\mathcal{\tilde{O}}}_\Pi=\phi^\dagger[\Pi(-D^2)-\Pi(0)]\phi+\frac{\Pi(0)}{2v^2}\left[(\phi^\dagger\phi)-\frac{v^2}{2}\right]^2\,.
\end{equation}
The addition of this operator to the SM leads to several changes,  which we now
consider in detail. First of all, by construction, it gives rise to the
propagator in \Eq~(\ref{eq:prop}), and a resummed propagator with the full
one-loop self energy via the PT at tree level is obtained.  Second it leads to
modifications of the other interactions, in such a way that gauge invariance is
maintained.  For example, the \PW and \PZ two-point functions are modified by
the addition of 
\begin{flalign}
i\Delta\Pi^{\mu\nu}_{WW}(q^2)=&i\left(\frac{gv}{2}\right)^2\left[\Pi'(0)g^{\mu\nu}+\Pi''(q^2)q^\mu q^\nu\right]\,,\nonumber
\\
i\Delta\Pi^{\mu\nu}_{\PZ\PZ}(q^2)=&i\left(\frac{gv}{2\cw}\right)^2\left[\Pi'(0)g^{\mu\nu}+\Pi''(q^2)q^\mu q^\nu\right]\,,\nonumber
\end{flalign}
where $v$ is the Higgs vev, and
\begin{equation}
\Pi'(x)\equiv \frac{\Pi(x)-\Pi(0)}{x},\qquad  \Pi''(x)\equiv \frac{\Pi'(x)-\Pi'(0)}{x} \,.
\end{equation}
The values for the \PW and \PZ masses are  shifted
\begin{flalign}
&\MW^2=\left(\frac{gv}{2}\right)^2(1+\Pi'(0))\nonumber \\
&\MZ^2=\left(\frac{gv}{2\cw}\right)^2(1+\Pi'(0))\,,
\label{eq:EFTmass}
\end{flalign}
as well as the propagators 
\begin{equation}
\frac{i}{q^2-m_{W,Z}^2}\left[
- g^{\mu\nu}+\frac{\left(1+\frac{m_{W,Z}^2\Pi''(q^2)}{1+\Pi'(0)}\right)q^\mu q^\nu}{m_{W,Z}^2+q^2\frac{m_{W,Z}^2\Pi''(q^2)}{1+\Pi'(0)}}
\right]\,.
\label{eq:EFTprop}
\end{equation}

Let us first consider $f\bar{f}\to W_{\mathrm L}^+W_{\mathrm L}^-$ in the EFT approach. The
operator modifies the $\PH\PWp\PWm$ and the $\PH f\bar{f}$ interactions.  The
combined effect is a factor of $1+\Pi'(s)$. Therefore in this process the EFT
approach is equivalent to the following substitution of the Higgs propagator:
\begin{equation}
\Delta_{\PH}^{-1}(s)=\frac{1+\Pi'(s)}{s-s_{\PH}+\Pi(s)}\ ,
\end{equation}
which behaves like $1/s$ at large energy, and therefore exactly cancels the
high-energy behavior from ${\mathcal M}^L_{t}$.  It is also interesting to note
that, if $\Pi(s)$ has a linear dependence on $s$, i.e.,
\begin{equation}
\Pi(s)=i(s-\mu^2)\frac{\gamma}{\mu}\,,
\end{equation}
the above equation becomes
\begin{equation}
\Delta_{\PH}^{-1}(s)=\frac{1+i\frac{\gamma}{\mu}}{s-\mu^2+is\frac{\gamma}{\mu}}\ ,
\end{equation}
and the EFT approach coincides with the scheme proposed by Seymour
~\cite{Seymour:1995np}.  This could be expected because in the Seymour scheme the
vector boson pair self energy also has a linear dependence on $s$. In our
scheme we see that the numerator of Seymour's propagator comes from the
modified $\PH\PWp\PWm$ vertex, as required by the WI.

We now turn to  vector-vector scattering and in particular to
$\PWp_{\mathrm L}\PWm_{\mathrm L}\rightarrow \PZ_{\mathrm L}\PZ_{\mathrm L}$.  This process features a pure gauge and a
Higgs-mediated $s$-channel contribution,  \Fref{fig:2}.  Both contributions
do contain terms that grow as $s$ at high energy, whose cancellation is
guaranteed by gauge-invariance. To calculate $\PWp_{\mathrm L}\PWm_{\mathrm L}\rightarrow \PZ_{\mathrm L}\PZ_{\mathrm L}$
amplitude in the EFT we need to extract the Feynman rules from ${\mathcal{\tilde{O}}}_\Pi$, i.e., the contributions that need to be added to the usual SM
rules.  This is straightforward and gives (all momenta incoming):
\begin{equation}
\begin{array}{ll}
 H(q)W^{+\mu}(k_1)W^{-\nu}(k_2):   
& ig\frac{\MW}{\sqrt{1+\Pi'(0)}}\Pi'(q^2)g^{\mu\nu}+\cdots  \\
 Z^\mu(k_1)W^{+\nu}(k_2)W^{-\rho}(k_3): 
& i\frac{g}{\cw}\frac{\MW^2}{1+\Pi'(0)}\sw^2\left[
\Pi''(k_3^2)g^{\mu\nu}k_3^\rho-\Pi''(k_2^2)g^{\mu\rho}k_2^\nu\right]+\cdots \\
Z^\mu(k_1)Z^\nu(k_2)W^{+\rho}(k_3)W^{-\sigma}(k_4):\hspace*{1cm}
&  ig^2\frac{\MZ^2}{1+\Pi'(0)}\left[\Pi''(s)g^{\mu\nu}g^{\rho\sigma} +\sw^4(\Pi''(t)g^{\mu\rho}g^{\nu\sigma}+\Pi''(u)g^{\mu\sigma}g^{\nu\rho})\right]+\cdots \\
H\phi^+\phi^-,\ H\phi^0\phi^0: 
& -i\frac{g[\MH^2-\Pi(0)]}{2\MW}\sqrt{1+\Pi'(0)}\\
\phi^+\phi^-\phi^0\phi^0:
& -i\frac{g^2[\MH^2-\Pi(0)]}{4\MW^2}[1+\Pi'(0)]
\end{array}
\label{eq:fr}
\end{equation}
where ellipsis denotes terms vanishing on shell 
and $s=(k_1+k_2)^2$, $t=(k_1+k_3)^2$, and $u=(k_1+k_4)^2$.  These Feynman rules
are sufficient to calculate both $\PWp_{\mathrm L}\PWm_{\mathrm L}\to \PZ_{\mathrm L}\PZ_{\mathrm L}$ and $\phi^+\phi^-\to
\phi^0\phi^0$. At the leading order in $\frac{\MW^2}{s}$ and
$\frac{\MW^2}{\MH^2}$, we find for $\PWp_{\mathrm L}\PWm_{\mathrm L}\to \PZ_{\mathrm L}\PZ_{\mathrm L}$,
\begin{flalign}
  &{\mathcal M}_{\PH}^{\mathrm{LLLL}}=-\frac{ig^2}{4\MW^2}\frac{s^2[1+\Pi'(s)]^2}{[s-\MH^2+\Pi(s)][1+\Pi'(0)]}
\\
&{\mathcal M}_{\mathrm{gauge}}^{\mathrm{LLLL}}=\frac{ig^2}{4\MW^2}s\frac{1+\Pi'(s)}{1+\Pi'(0)}
\end{flalign}
and for $\phi^+\phi^-\to \phi^0\phi^0$,
\begin{flalign}
&{\mathcal M}_G=-\frac{ig^2}{4\MW^2}\frac{s+\Pi(s)-\Pi(0)}{s-\MH^2+\Pi(s)}\left[\MH^2-\Pi(0)\right]\left[1+\Pi'(0)\right]
\end{flalign}
so that
\begin{flalign}
  {\mathcal M}_{\PH}^{\mathrm{LLLL}}+{\mathcal M}_{\mathrm{gauge}}^{\mathrm{LLLL}}&=-\frac{ig^2}{4\MW^2}\frac{s+\Pi(s)-\Pi(0)}{s-\MH^2+\Pi(s)}\frac{\MH^2-\Pi(0)}{1+\Pi'(0)}\nonumber \\
&= \frac{{\mathcal M}_G}{[1+\Pi'(0)]^2}\,.
\end{flalign}
As expected, ${\mathcal M}_{\PH}^{\mathrm{LLLL}}+{\mathcal M}_{\mathrm{gauge}}^{\mathrm{LLLL}}$ does not grow with
$s$ and the equivalence theorem is recovered, up to a factor $[1+\Pi'(0)]^2$,
which exactly amounts to the wave function renormalization of the Goldstone
fields.

An interesting feature of our approach is that in the limit where the dependence
of $\Pi(s)$ on $s$ is neglected,  $\Pi(s)\equiv\Pi$ is a constant, then
$\Pi'(s)=\Pi''(s)=0$.   The only effect of the operator is a shift in $\lambda$,
the coupling of the Higgs-boson self interaction.  If $\MH$ is the on-shell
mass, this amounts to the replacement
\begin{equation}
\MH^2\to \MH^2-\Pi \,,
\end{equation}
i.e., given that $\Pi$ can be a complex number, it is equivalent to the CMS. 

The advantage of the EFT approach is the possibility of using ``arbitrary''
functional form of the self energy.  We have shown that with special choices of
$\Pi(s)$, the EFT approach can reduce to the Seymour scheme and the CMS scheme
in certain cases. For example, there is no need for spurious non-zero width for
$t$-channel propagators as this can be easily imposed by always maintaining
gauge invariance.  Finally, we note that despite the restoration of
gauge-invariance and equivalence theorem is a general feature of our approach,
one has to be careful in choosing the appropriate operator. For example, the
following operator
\begin{equation}\label{eq:op2}
  {\mathcal O}'_\Pi=\frac{1}{2v^2}\left(\phi^\dagger\phi-v^2\right)\Pi(-\partial^2)\left(\phi^\dagger\phi-v^2\right)
\end{equation}
introduced in Ref.~\cite{Beenakker:1999hi}, gives rise to the correct self
energy and the resummed propagator, but it does not modify the gauge
contribution, so in $\PWp_{\mathrm L}\PWm_{\mathrm L}\to \PZ_{\mathrm L}\PZ_{\mathrm L}$ the gauge cancellation between the
$s$-channel Higgs-mediated amplitude and the gauge amplitude is not restored.
On the other hand, it modifies the Goldstone amplitude in a way so that the
equivalence theorem is satisfied. As a result, both $\PWp_{\mathrm L}\PWm_{\mathrm L}\to \PZ_{\mathrm L}\PZ_{\mathrm L}$ and
$\phi^+\phi^-\to \phi^0\phi^0$ have bad high-energy behavior, and eventually
break unitarity bounds.  In general, adding higher dimensional operators to the
Lagrangian leads to unitarity violation at some scale.  We are going to show in
the next sections that the operators we use do not have this problem. 

Though the above operator $\mathcal{O}'_\Pi$ solely does not treat the
$\PH\PZ\PZ$ and $\PH\PWp\PWm$ correctly at high energy, when combined with
$\mathcal{O}_\Pi$, we can adjust them in a certain way to improve this method.
We will discuss this in \Sref{sec:application}.

\subsubsubsection{Unitarity}
\label{sec:unitarityD}
Adding operators of dimension $n>4$ to the SM Lagrangian
\begin{equation}
\mathcal{L}_{{\mathrm{EFT}}} = {\mathcal L}_{{\mathrm{SM}}} + \sum_i c_i \frac{{\mathcal O}_i[n]}{\Lambda^{n-4}}\,,
\end{equation}
is equivalent to recast the SM in terms of an effective field theory valid up
to scales of order $\Lambda$ ~\cite{Weinberg:1978kz}, beyond which the theory
is not unitary. It is therefore mandatory to check whether this is the case
for the operator ${\mathcal{\tilde{O}}}_{\Pi}$.  In fact,  as we will see in the
following section, a consistent perturbation theory implies that the same
operator needs to also appear as a counterterm  at higher orders. Overall  we
do not modify the theory and our procedure amounts to a reorganization of the
perturbative expansion. However, we still need to make sure that neither
unitarity is  violated nor double counting happens  at any given order in the
perturbation theory. In this section we consider the first of these issues by
showing that in sample calculations, $f\bar{f}\to VV$ and $VV\to VV$, at
tree-level the operator in \Eq~(\ref{eq:O}) does not break unitarity at large
energy. 

In $f_\pm\bar{f}_\pm\to VV$ the change in $HVV$ vertex cancels
the change in $H$ propagator at high $s$, independently of the helicities of
$VV$, so the $s$-channel Higgs diagram does
not lead to any bad high-energy behavior. The scattering of opposite helicity
fermions does not entail the $s$-channel Higgs diagram and is the same as in the
SM.

As we have already verified, in $\PWp\PWm\to \PZ\PZ$  the longitudinal
amplitude does 
not break unitarity, because the modification to the corresponding Goldstone
interaction is finite (vertices involving Goldstones are modified by 
factors of $1 - \Pi(0)/MH^2$ and $1+\Pi'(0)$).  We now check the
transverse amplitude $++\to --$, $00\to ++$, $++\to 00$, $++\to ++$, in the limit
\begin{flalign}
  s\sim |t|\sim |u|\gg \MW^2,\quad \MH^2\gg \MW^2\,.
\end{flalign}
(Note that $+-$, $+0$,
$-0$ configurations do not feature a Higgs in the $s$-channel and therefore are
left unchanged.) An explicit calculation for $\PWp_+\PWm_+\to \PZ_-\PZ_-$ gives
\begin{flalign}
&{\mathcal M}_{\PH}^{++--}={\mathcal M}_{\PH}^{\mathrm{LLLL}}\frac{4\MW^4}{s^2\cw^2}+\mathcal{O}(\MW^4)\\
&{\mathcal M}_{\mathrm{gauge}}^{++--}={\mathcal M}_{\mathrm{gauge}}^{\mathrm{LLLL}}\frac{4\MW^4}{s^2\cw^2}+\mathcal{O}(\MW^4)\,,
\end{flalign}
where $M^{\mathrm{LLLL}}$ indicates the amplitude with four longitudinal vectors.  For
$\PWp_{\mathrm L}\PWm_{\mathrm L}\to \PZ_+\PZ_+$ we obtain
\begin{flalign}
&{\mathcal M}_{\PH}^{\mathrm{LL}++}={\mathcal M}_{\PH}^{\mathrm{LLLL}}\frac{-2\MW^2}{s\cw^2}+\mathcal{O}(\MW^2)\\
&{\mathcal M}_{\mathrm{gauge}}^{\mathrm{LL}++}={\mathcal M}_{\mathrm{gauge}}^{\mathrm{LLLL}}\frac{-2\MW^2}{s\cw^2}+\mathcal{O}(\MW^2)\,,
\end{flalign}
and for
$\PWp_+\PWm_+\to \PZ_{\mathrm L}\PZ_{\mathrm L}$ we obtain
\begin{flalign}
  &{\mathcal M}_{\PH}^{++\mathrm{LL}}={\mathcal M}_{\PH}^{\mathrm{LLLL}}\frac{-2\MW^2}{s}+\mathcal{O}(\MW^2)\\
  &{\mathcal M}_{\mathrm{gauge}}^{++\mathrm{LL}}={\mathcal M}_{\mathrm{gauge}}^{\mathrm{LLLL}}\frac{-2\MW^2}{s}+\mathcal{O}(\MW^2)\,,
\end{flalign}
These results vanish faster than the longitudinal amplitude at large $s$.
Finally for $\PWp_+\PWm_+\to \PZ_+\PZ_+$, we obtain
\begin{flalign}
{\mathcal M}_{\PH}^{++++}&= -{ig^2\MZ^2}\frac{[1+\Pi'(s)]^2}{[s-\MH^2+\Pi(s)][1+\Pi'(0)]}
+\mathcal{O}(\MW^4) \nonumber\\
&\approx\frac{-{ig^2\MZ^2}}{1+\Pi'(0)}\Pi''(s)\sim s^{-1}\quad\mbox{at large $s$,}
\\
{\mathcal M}_{\mathrm{gauge}}^{++++}&=i8g^2\cw^2\frac{s^2}{4tu}+\mathcal{O}(\MW^2)\,.
\end{flalign}
so at large energy the inclusion of $\mathcal{\tilde{O}}_\Pi$ does not lead to
any bad high energy behavior.

\subsubsubsection{The EFT approach at higher orders}
\label{sec:higherorder}
Starting at order $\alpha_\rw$, the operator ${\mathcal{\tilde{O}}}_{\Pi}$ is allowed in any
leading order computation. At next-to-leading order in EW interactions, however,
this is not necessarily consistent and possibly leads to double counting. In
this section we argue that this is not  a fundamental problem and can be dealt
with by simply subtracting the same operator in a NLO as a counterterm, in full
analogy to the procedure used in the CMS ~\cite{Denner:2005fg,Denner:2006ic}.

In the CMS, an imaginary part  is added to the real mass, and then subtracted as
counterterm at NLO.  One can prove that this procedure does not spoil the WI's,
despite the fact that only a special class of higher order terms is resummed. As
the EFT approach can be viewed as a generalization to the CMS, the same approach
can be followed.  The operator ${\mathcal{\tilde{O}}}_{\Pi}$ corresponds to the imaginary part
of the mass. It includes some of the higher-order contribution, and provides an
improved solution to the WI's. It enters the resummed propagator and other
Feynman rules, and  needs to be subtracted at higher orders. The main difference
is that in the CMS the propagator describes an unstable particle with a fixed
width, while in the EFT approach one can resum an arbitrary part of the
self-energy correction.  This difference may be important when the width of the
unstable particle is large, as in the case of a heavy Higgs, and the actual
functional form of $\Pi(s)$ becomes important.

In the pole-mass renormalization scheme, the two-point function of the Higgs can
be written as
\begin{equation}\label{eq:2point}
\Delta_{\PH}(s)=s-s_{\PH}+\Pi^{\mathrm R}_{\PH\PH}(s)
\end{equation}
where $s_{\PH}$ is the pole and $\Pi^{\mathrm R}_{\PH\PH}(s)$ is the one-loop PT self-energy
correction renormalized in the pole-mass scheme. We can now define the EFT
approach by adding the operator in \Eq~(\ref{eq:O}) and subtracting it as a
counterterm:
\begin{equation}
\mathcal{L}_{\SM} \to\mathcal{L}_{\SM} +{\mathcal{\tilde{O}}}_{\Pi}-{\mathcal{\tilde{O}}}_{\Pi}\,.
\end{equation}
In so doing the theory is exactly the same as before. Now \Eq~(\ref{eq:2point})
can be rewritten as
\begin{flalign}
\Delta_{\PH}(s)=&s-s_{\PH}+\Pi(s)\nonumber\\
&+[\Pi^{\mathrm R}_{\PH\PH}(s)-\Pi(s)]
\end{flalign}
where the first line starts at leading order, while the second line starts at
order $\alpha_\rw$.  The EFT approach then amounts to choose $\Pi(s)$ in a way to
capture the important part of (if not all of) $\Pi^{\mathrm R}_{\PH\PH}(s)$, so that this part
of the self-energy correction is included at the leading order, and will be
resummed.  In practice, one does not have to choose the exact PT self energy, and
gauge invariance is always guaranteed.  In particular, choosing $\Pi(s)=0$
corresponds to the CMS scheme.

In our scheme, EW NLO calculations are obviously more involved. The resummed
propagator (\ref{eq:prop}) and the modified Feynman rules do require extra
work.  However, one can also always  employ a standard CMS at NLO and only
include the full propagators and vertices in the LO result. In this way we can
consistently have leading order calculated in the EFT approach, and NLO in CMS
but with counterterms from ${\mathcal{\tilde{O}}}_{\Pi}$.

\subsubsubsection{Applications}
\label{sec:application}
The treatment of the propagator of the Higgs  is of immediate  relevance for
the LHC. As simple testing ground of our proposal and comparisons to the
conventional methods, we consider three processes of particular
phenomenological importance at the LHC for a scalar boson (which for brevity,
we identify with an hypothetical heavy Higgs): vector boson scattering,
$\PAQt\PQt$ production via vector boson fusion and Higgs production via gluon
fusion. We have compared the effective approach described above in
\Eqs~(\ref{eq:EFTmass}), (\ref{eq:EFTprop}) and (\ref{eq:fr}), with two other
schemes: 
\begin{enumerate}
\item A naive inclusion of the self energy, i.e., using the following propagator
\begin{equation}\label{eq:naiveprop}
\frac{i}{\Delta_{\PH}(s)}=\frac{i}{s-s_{\PH}+\Pi_R(s)}\ ,
\end{equation}
without changing anything else. Here $s_{\PH}=\mu^2-i\mu\gamma$.

\item The CMS scheme, 
\begin{equation}\label{eq:CMSprop}
\frac{i}{\Delta_{\PH}(s)}=\frac{i}{s-s_{\PH}}\ .
\end{equation}
\end{enumerate}

  As mentioned in \Sref{sec:approach}, we improve our operator by combining it
  with $\mathcal{O}'_\Pi$ given in \Eref{eq:op2}. More specifically, we define
  \begin{flalign}
    \label{eq:opgood}
  &{\mathcal{\bar{O}}}_\Pi={\mathcal{O}}_{\Pi_1}+{\mathcal{O}}'_{\Pi_2} \,,
  \\
  & {\mathcal{O}}_{\Pi_1}=\phi^\dagger\Pi_1(-D^2)\phi \,,
  \\
  & {\mathcal{O}}'_{\Pi_2}=\frac{1}{2v^2}\left(\phi^\dagger\phi-v^2\right)\Pi_2(-\partial^2)\left(\phi^\dagger\phi-v^2\right)\,.
  \end{flalign}
  where the functions $\Pi_1$ and $\Pi_2$ are determined by requiring
  $\Pi_1(s)+\Pi_2(s)=\Pi(s)$, and that the operator ${\mathcal{\bar{O}}}_\Pi$ gives (the
  imaginary part of) the exact one-loop PT $\PH\PZ\PZ$ vertex. We use the operator
  ${\mathcal{\bar{O}}}_\Pi$ in the following analysis.

A modified version of {\sc MadGraph}~\cite{Alwall:2011uj}, with the
implementation of the effective Lagrangian approach and the naive propagator
with the PT self energy is used to generate events. As
SM input parameters we take:
\begin{flalign}
&\MZ=91.188 {\rm GeV}\\
&\GF=1.16639\times 10^{-5} {\rm GeV}^{-2}\\
&\alpha^{-1}=132.507 \\
&m_{\PQt}=173\,{\rm GeV}\,.
\end{flalign}
The Higgs pole mass is
\begin{equation}
\mu=800\,{\rm GeV}\,,
\end{equation}
and $\Pi_R(s)$ is the imaginary part of the PT self energy renormalized in the
pole scheme.  The factorization scale is set as the default dynamical scale of
\textsc{MadGraph} and the PDF set is CTEQ6l1~\cite{Pumplin:2002vw}.

\begin{itemize}
  \item Vector Boson Scattering
\end{itemize}
In $PVB\PVB \to PVB\PVB$ scattering processes, the effective description allows one to
achieve a complete description of the Higgs line-shape at the resonance region
and at the same time it corrects the bad high-energy behavior originated from
the momenta dependent part of the self energy. In addition, we show  that our
definition avoids the need for including spurious $t$-channel widths which
occur in the complex-mass scheme also affecting the high energy behavior of the
scattering amplitudes. 

In \Fref{fig:amp_zz_cos0} we show the energy behavior of the $\PZ\PZ\to \PZ\PZ$
scattering amplitude summed over helicities, $\sum_{{\mathrm{hel}}}|{\mathcal M}(s,t,u)|^2$, at
scattering angle $\cos\theta=0$. The fixed width scheme, \Eref{eq:CMSprop},
naive propagator, \Eref{eq:naiveprop}, the effective description and a case in
which the width is set to zero are presented. The agreement between the
effective scheme and the naive propagator at the resonance region is pretty
good. The difference with respect to the fixed width scheme is evident. At high
energy,  the naive propagator diverges, while the effective description behaves
correctly. Similar comments can be made about $\PWp\PWm\to \PWp\PWm$ amplitude,
shown in \Fref{fig:amp_ww_cos0}. 

The fact that in both $\PZ\PZ\to \PZ\PZ$ and $\PWp\PWm\to \PWp\PWm$ the fixed width scheme
differs from the effective approach at the high energy region indicates that
the spurious $t$-channel width gives a non-negligible contribution. This fact
can be verified by comparing the different schemes with the no-width case.
Moreover, in the case of  $\PWpm \PWpm\to \PWpm \PWpm$, shown in
\Fref{fig:amp_wpwp_cos0}, the effective description and naive propagator are
equivalent to the no-width case and the excess observed in the amplitudes in
the fixed width scheme comes from the spurious width in the $t$ and
$u$-channels.
%%%%%%%%%%%%%%%%%%%%%%%
\begin{figure}[ht]
  \begin{minipage}[c]{0.48\columnwidth}
  \includegraphics[width=\columnwidth,height=0.8\columnwidth]{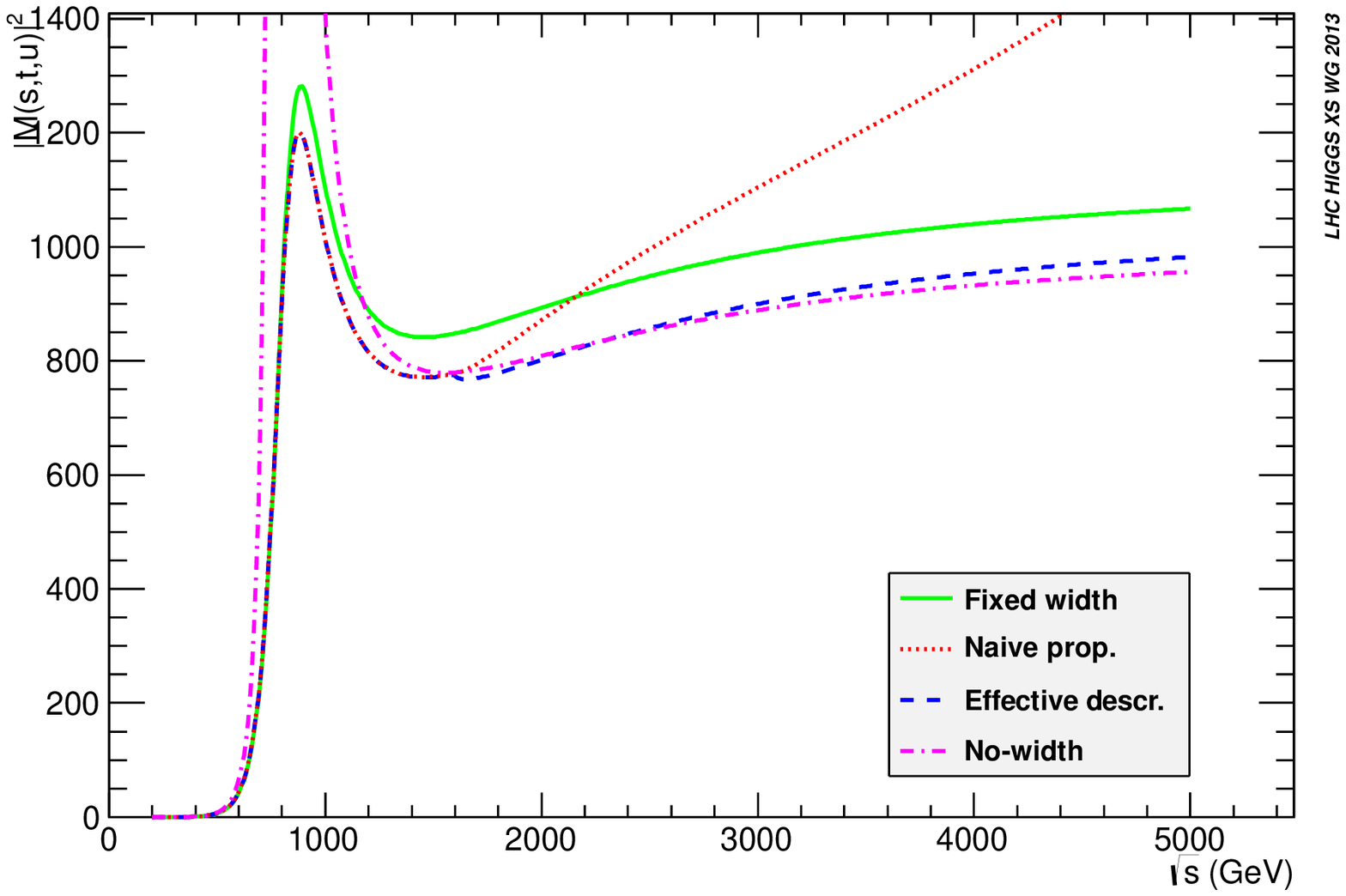}
\caption{$\sum_{hel} |M(\PZ\PZ\to \PZ\PZ)|^2$ with scattering angle, $\theta=\pi/2$.
The curves correspond to: fixed width scheme, \Eref{eq:CMSprop}, naive
propagator, \Eref{eq:naiveprop}, the effective description and the no-width, in
which the width is set to zero.}
\label{fig:amp_zz_cos0}
\end{minipage}
\hfill
\begin{minipage}[c]{0.48\columnwidth}
\includegraphics[width=\columnwidth,height=0.8\columnwidth]{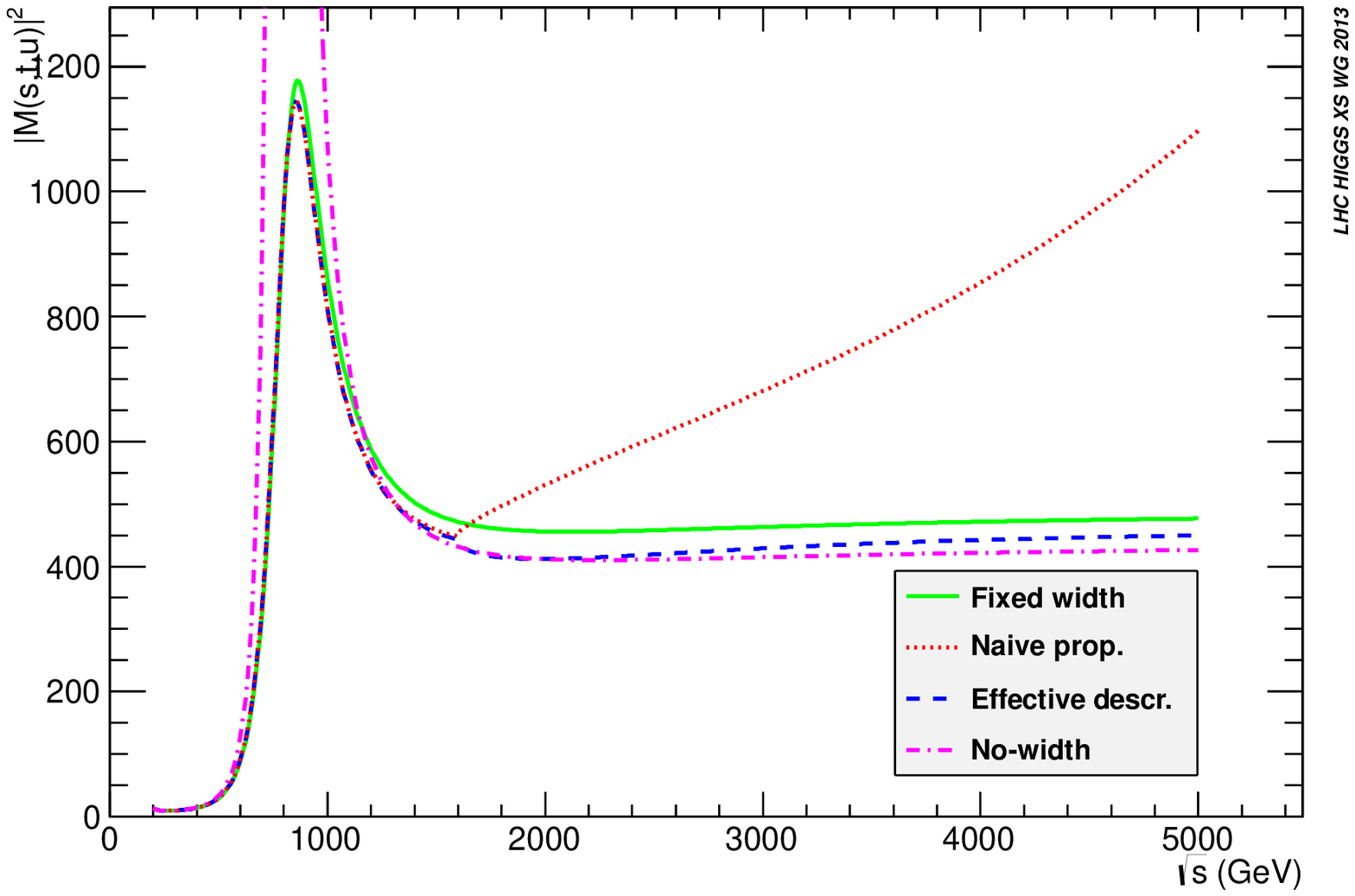}
\caption{$\sum_{hel} |M(\PWp\PWm\to \PWp\PWm)|^2$ with scattering angle,
$\theta=\pi/2$. The curves correspond to: fixed width scheme, \Eref{eq:CMSprop},
naive propagator, \Eref{eq:naiveprop}, the effective description and the
no-width, in which the width is set to zero.}
\label{fig:amp_ww_cos0}
  \end{minipage}
\end{figure} 
%%%%%%%%%%%%%%%%%%%%%%%
%%%%%%%%%%%%%%%%%%%%%%%
\begin{figure}[ht]
  \centering
\includegraphics[width=0.5\columnwidth,height=0.4\columnwidth]{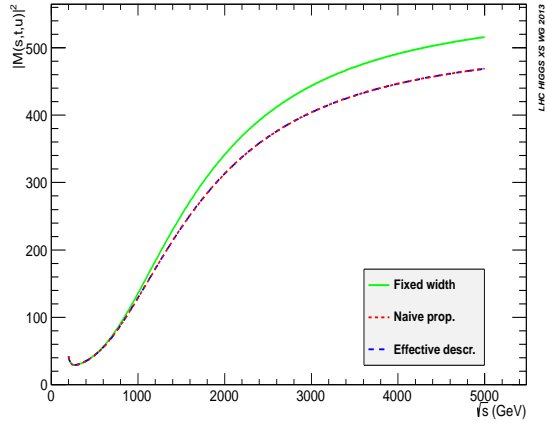}
\caption{$\sum_{hel} |M(\PWp\PWp\to \PWp\PWp)|^2$ with scattering angle,
$\theta=\pi/2$. The curves correspond to: fixed width scheme, \Eref{eq:CMSprop},
naive propagator, \Eref{eq:naiveprop} and the effective description. The
no-width case is equivalent to the last two.}
\label{fig:amp_wpwp_cos0}
\end{figure} 
%%%%%%%%%%%%%%%%%%%%%%%
At the LHC, the differences shown above may become important for a broad
resonance. Despite the fact that a light Higgs has been observed, there is
still room for new heavy and eventually broad resonances, e.g. 
The $VV$ scattering are embedded in more complex processes of the
form $\PQq\PQq\to \PQq\PQq\PVB\PVB$, where the two final state jets are emitted with high energy
in the forward-backward region of the detectors and the vector bosons decay
into two fermions with high $\pT$ through the central region. We study the
processes $\PQu\PQc\to \PQu\PQc\PZ\PZ$ and $\PQu\PQs \to \PQd\PQc \PWp\PWm$ assuming the nominal energy of
LHC, $E_{\mathrm{CM}}=14\UTeV$.

In \Figures~\ref{fig:uc-uczz_mzz_genc} and \ref{fig:uc-uczz_mzz_allp-m1000},  the
distribution of the invariant mass of the \PZ\PZ-system is shown. In
\Fref{fig:uc-uczz_mzz_genc}, the resonant region is shown. A basic set of
selection cuts to enhance vector boson scattering contribution, listed in the
left column of \Tref{tab:vvcuts}, have been applied. The effective description
fits well with the running behavior of the Higgs propagator. In
\Fref{fig:uc-uczz_mzz_allp-m1000}, the high-energy region is put in 
evidence. To better appreciate the differences between schemes at LHC energy, a
further set of cuts has been added (right column of \Tref{tab:vvcuts}). As
expected, the effective approach gives a well behaved distribution at such
energies contrary to the naive propagator and with a rate about 10\% lower than
the fixed width scheme. This difference amounts to the $t$-channel spurious
contribution present in the fixed width case. Similar conclusions can be drawn
from \Figures~\ref{fig:us-dcww_mww_genc} and \ref{fig:us-dcww_mww_allp-m1000}, where the
reconstructed $\PW\PW$-system invariant mass distribution for the $\PQu\PQs\to \PQd\PQc\PWp\PWm$
process is shown.  
%%%%%%%%%%%%%%%%%%%%%%%
\begin{figure}[ht]
\begin{minipage}[c]{0.48\columnwidth}
\includegraphics[width=\columnwidth,height=0.8\columnwidth]{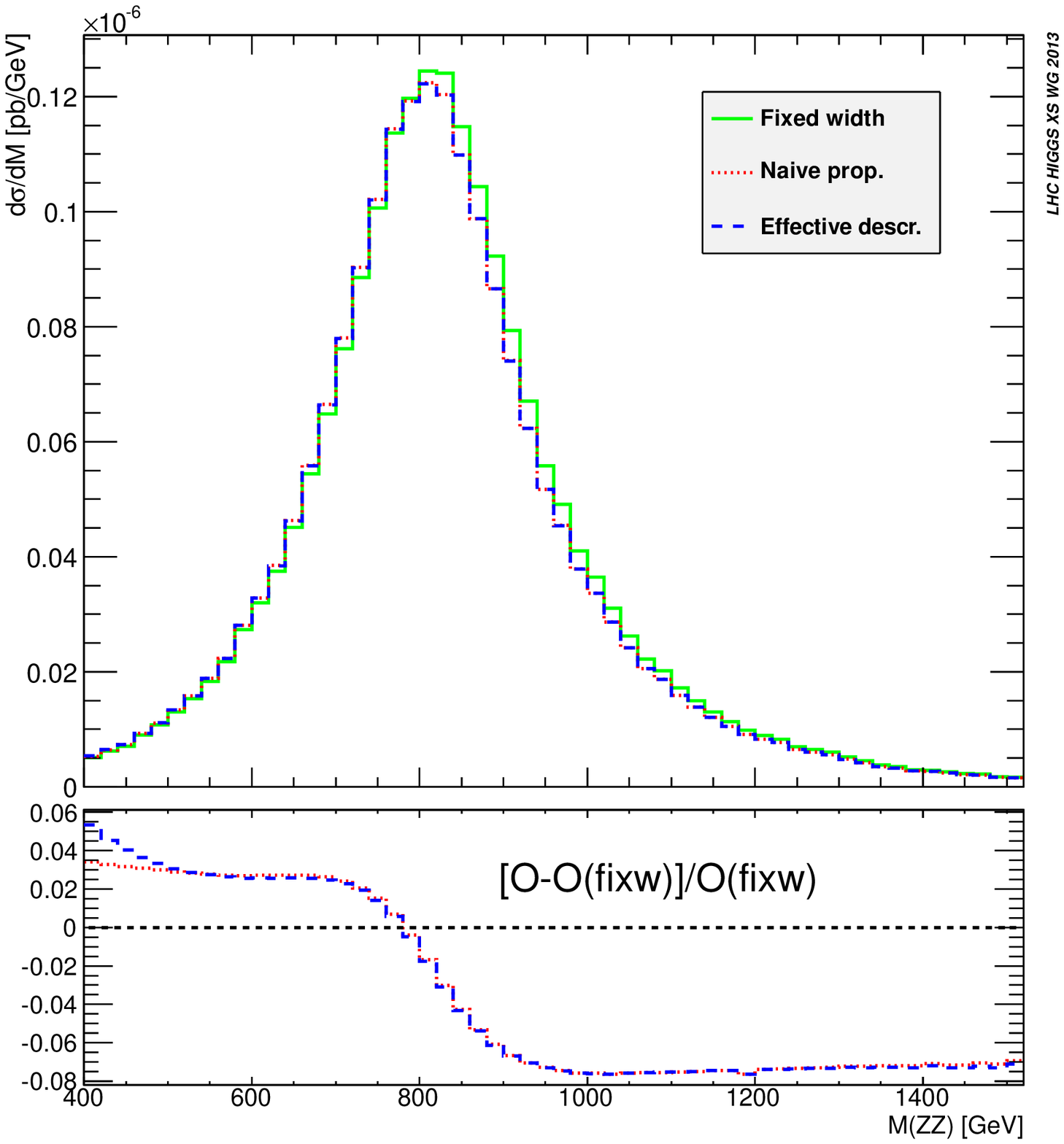}
\caption{Mass distribution of $\PZ\PZ$-system in the process $\PQu\PQc\to \PQu\PQc\PZ\PZ$ around
  the resonance peak. The cuts listed in the left column of \Tref{tab:vvcuts}
have been applied.}
\label{fig:uc-uczz_mzz_genc}
\end{minipage}
\hfill
\begin{minipage}[c]{0.48\columnwidth}
\includegraphics[width=\columnwidth,height=0.8\columnwidth]{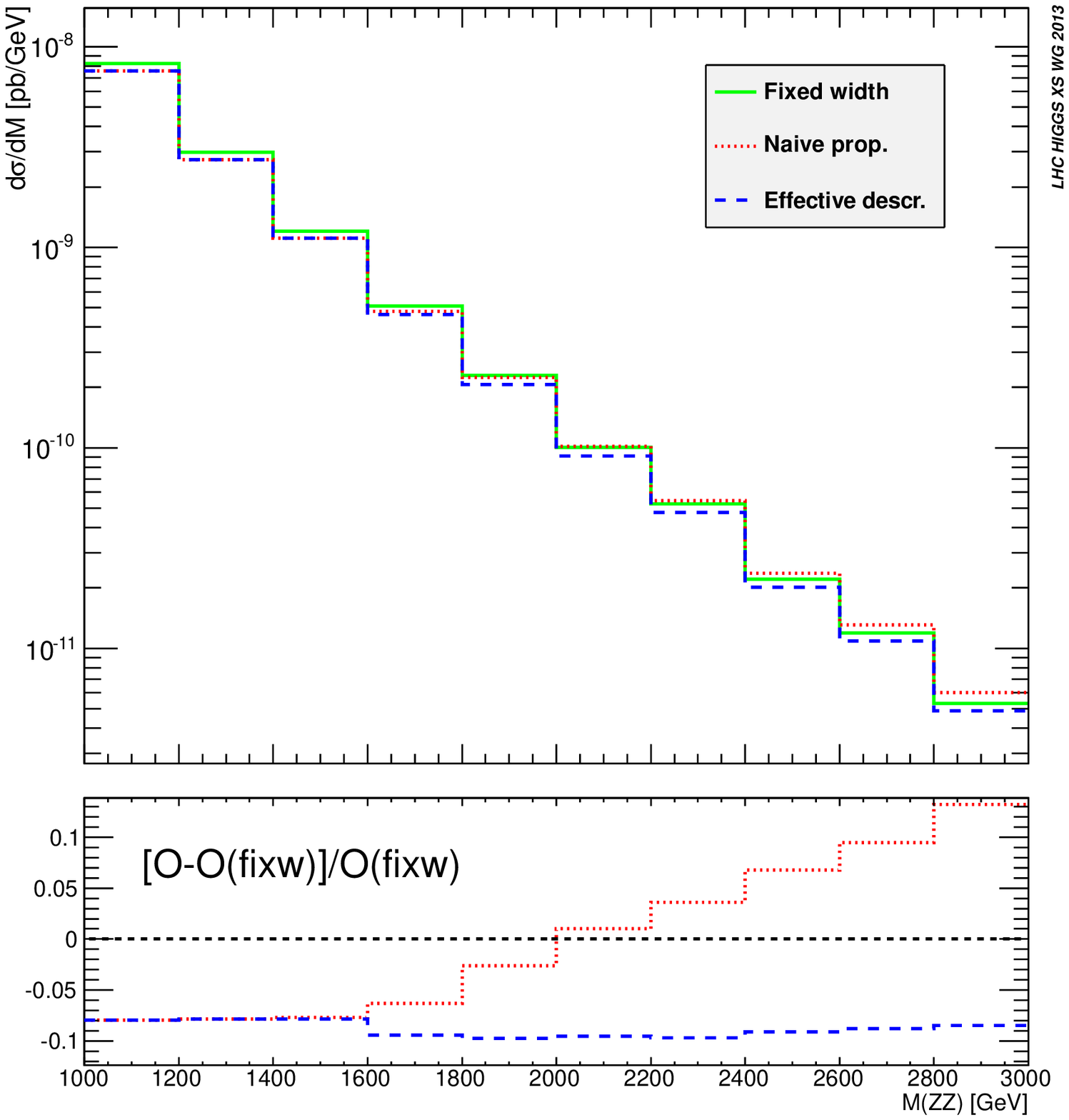}
\caption{Mass distribution of $\PZ\PZ$-system in the process $\PQu\PQc\to \PQu\PQc\PZ\PZ$ at the
  high energy region. All cuts listed in \Tref{tab:vvcuts} have been applied.}
\label{fig:uc-uczz_mzz_allp-m1000}
\end{minipage}
\end{figure} 
%%%%%%%%%%%%%%%%%%%%%%%
%%%%%%%%%%%%%%%%%%%%%%%
\begin{figure}[ht]
\begin{minipage}[c]{0.48\columnwidth}
\includegraphics[width=\columnwidth,height=0.8\columnwidth]{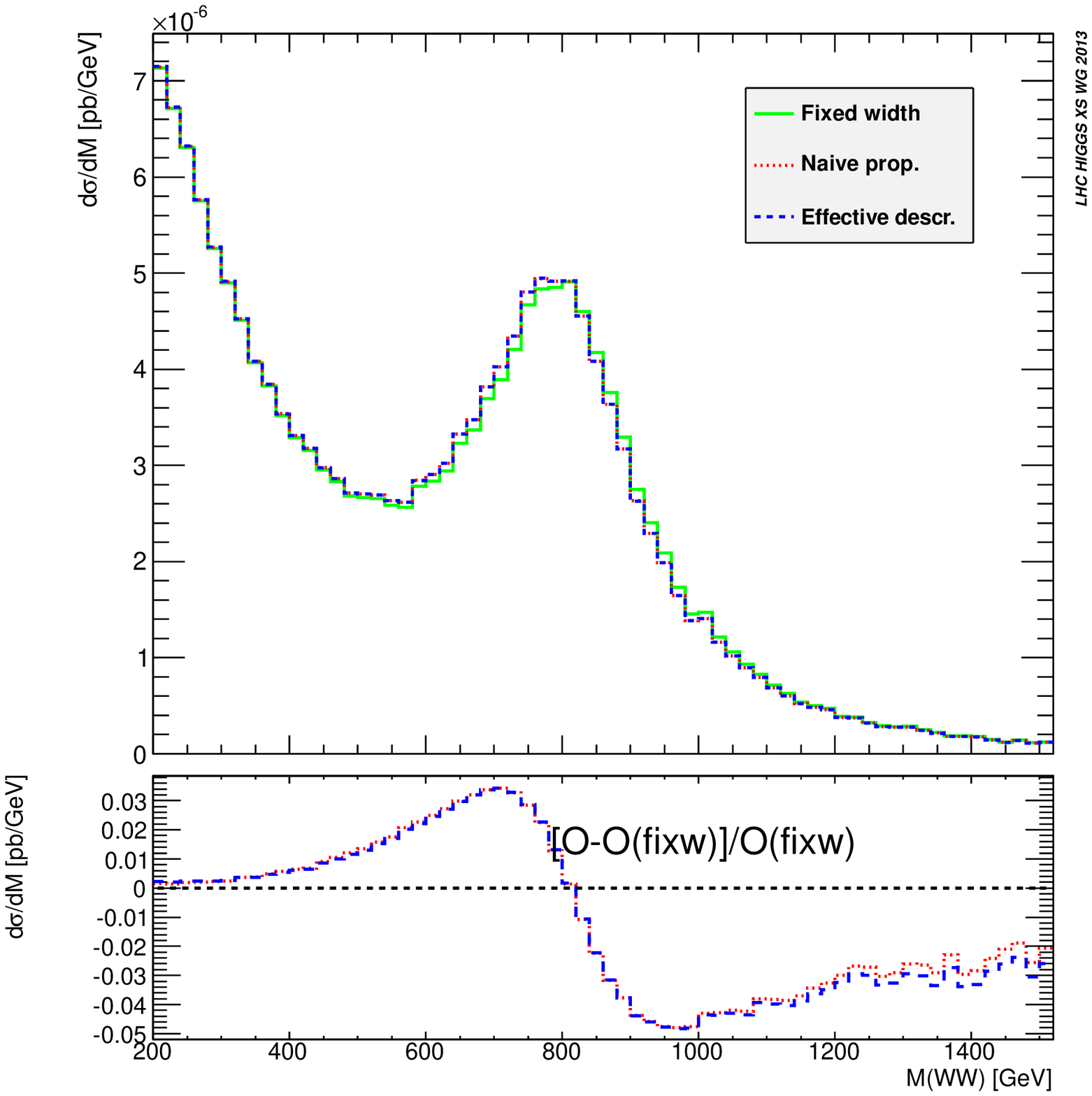}
\caption{Mass distribution of $\PW\PW$-system in the process $\PQu\PQs\to \PQd\PQc\PW\PW$ around the peak. The cuts listed in the left column of \Tref{tab:vvcuts} have been applied.}
\label{fig:us-dcww_mww_genc}
\end{minipage}
\hfill
\begin{minipage}[c]{0.48\columnwidth}
\includegraphics[width=\columnwidth,height=0.8\columnwidth]{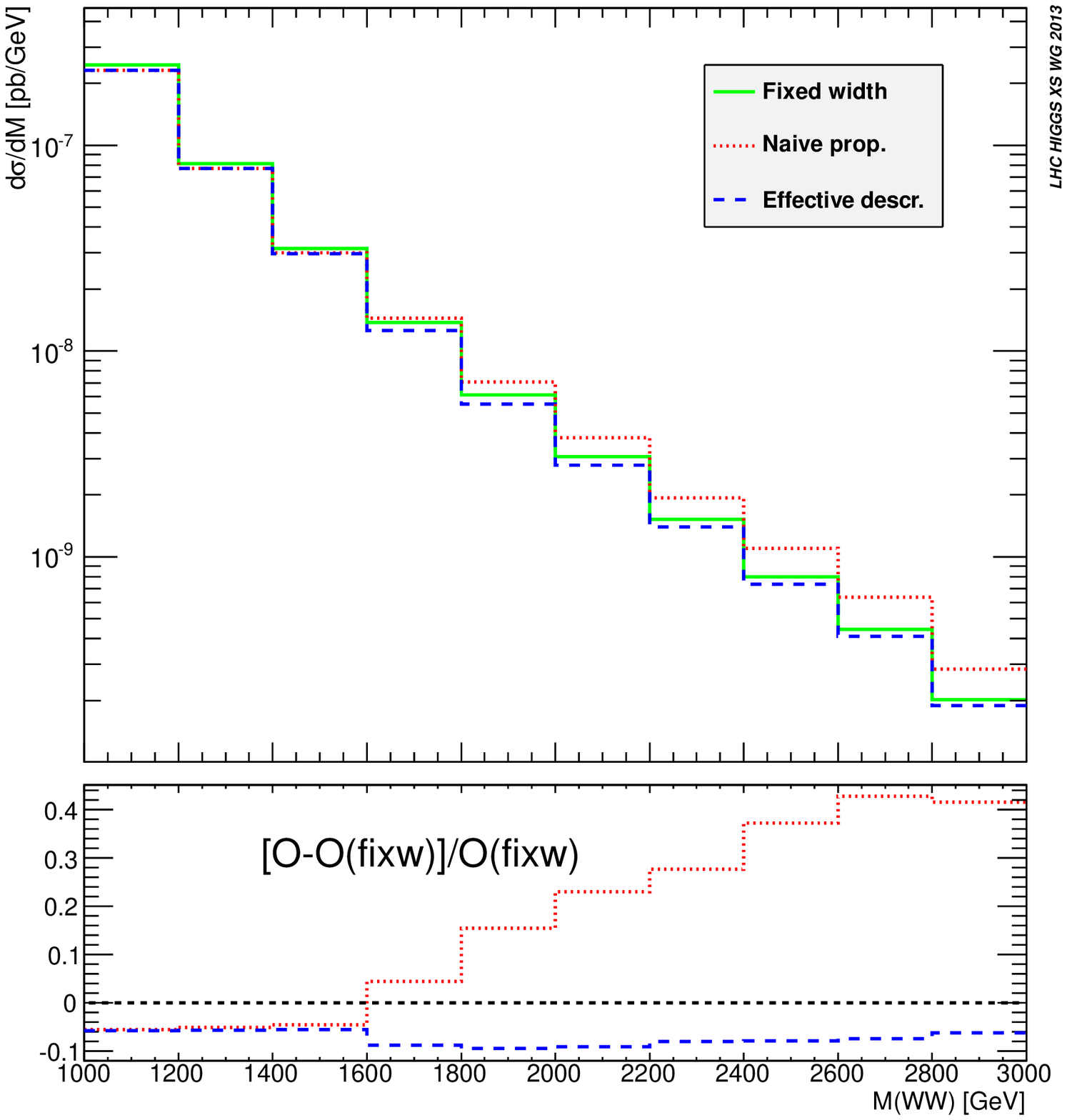}
\caption{Mass distribution of $\PW\PW$-system in the process $\PQu\PQs\to \PQd\PQc\PWp\PWm$ at the high energy region. All cuts listed in \Tref{tab:vvcuts} have been applied.}
\label{fig:us-dcww_mww_allp-m1000}
\end{minipage}
\end{figure} 
%%%%%%%%%%%%%%%%%%%%%%%
%%%%%%%%%%%%%%%%%%%%%%%
\begin{table}[h!]
  \centering
\caption{Cuts to enhance vector boson fusion. Basic ones on the left column and the extra ones on the right column.}
\begin{tabular}{lc}
\hline
Basic & Extra \\
\hline
$\pT(j)>10\,\UGeV$ & $\pT(V)>400\,\UGeV$\\

$2<\eta(j)<10$ 	 & $\Delta\eta(jj)>4.8$ \\

$\Delta R(j,j)>4$& $\eta(V)<2$ \\

$M(jj)>100\,\UGeV$ & $M(jj)>1000\,\UGeV$ \\

   		 & $\Delta \eta (V,j)>1$ \\
\hline
\end{tabular}
\label{tab:vvcuts}
\end{table}
%%%%%%%%%%%%%%%%%%%%%%%
\begin{itemize}
  \item $\PWp \PWp \to \PH\to \PAQt\PQt$ production
\end{itemize}
In $\PAQt\PQt$ production, we can observe a similar behavior with respect to
$\PZ\PZ\to \PZ\PZ$ vector boson scattering. We have concentrated on the process $us\to
dc\PAQt\PQt$, in which the Higgs in produced by $\PWp\PWm$ fusion and decayed to a
pair of top quarks. The energy in the center of mass is set to 14 TeV. In
\Fref{fig:mtt_genc}, the invariant mass distribution of $\PAQt\PQt$ at the
resonant region is presented. The cuts shown in the left column of
\Tref{tab:vvcuts} have been applied in order to enhance the vector boson fusion
contribution. Here again, the effective description describes the functional
form of the propagator, which can go up to 5\% of difference w.r.t the fixed
width scheme. As seen in \Fref{fig:mtt_m1000_allp}, in the high mass region, the
effective description is dumped down by the effective $\PW\PW\PH$ vertex and does not
grow with energy as is the case in which the naive propagator is adopted. The
extra cuts shown in the right-hand column of \Tref{tab:vvcuts} have been added
in order to highlight  the differences better.
%%%%%%%%%%%%%%%%%%%%%%%
\begin{figure}[ht]
\begin{minipage}[c]{0.48\columnwidth}
\includegraphics[width=\columnwidth,height=0.8\columnwidth]{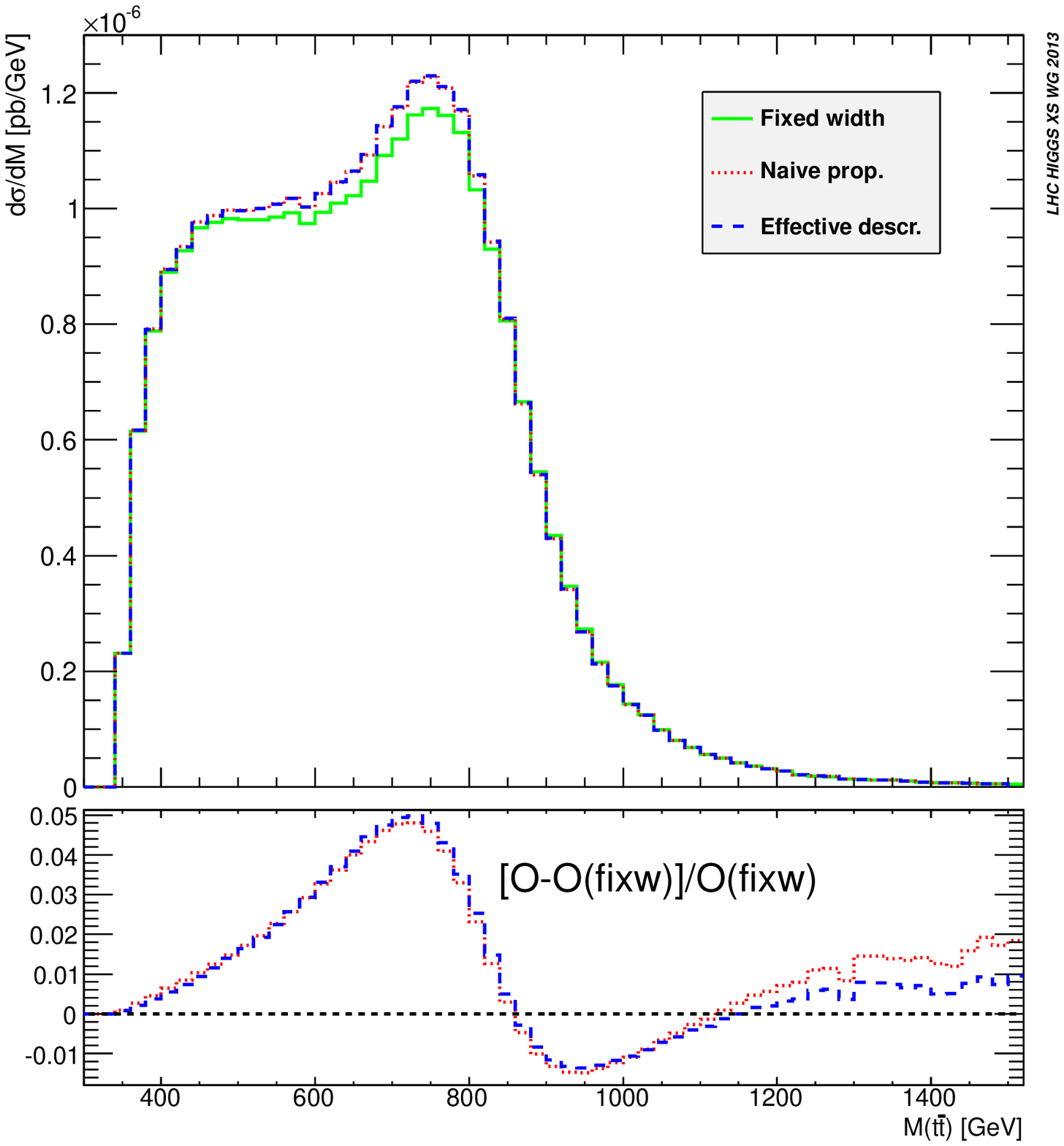}
\caption{Mass distribution of $\PAQt\PQt$-system in the process $\PQu\PQs\to
\PQd\PQc\PAQt\PQt$ around the resonance peak. The cuts listed in the left column of
\Tref{tab:vvcuts} have been applied.}
\label{fig:mtt_genc}
\end{minipage}
\hfill
\begin{minipage}[c]{0.48\columnwidth}
\includegraphics[width=\columnwidth,height=0.8\columnwidth]{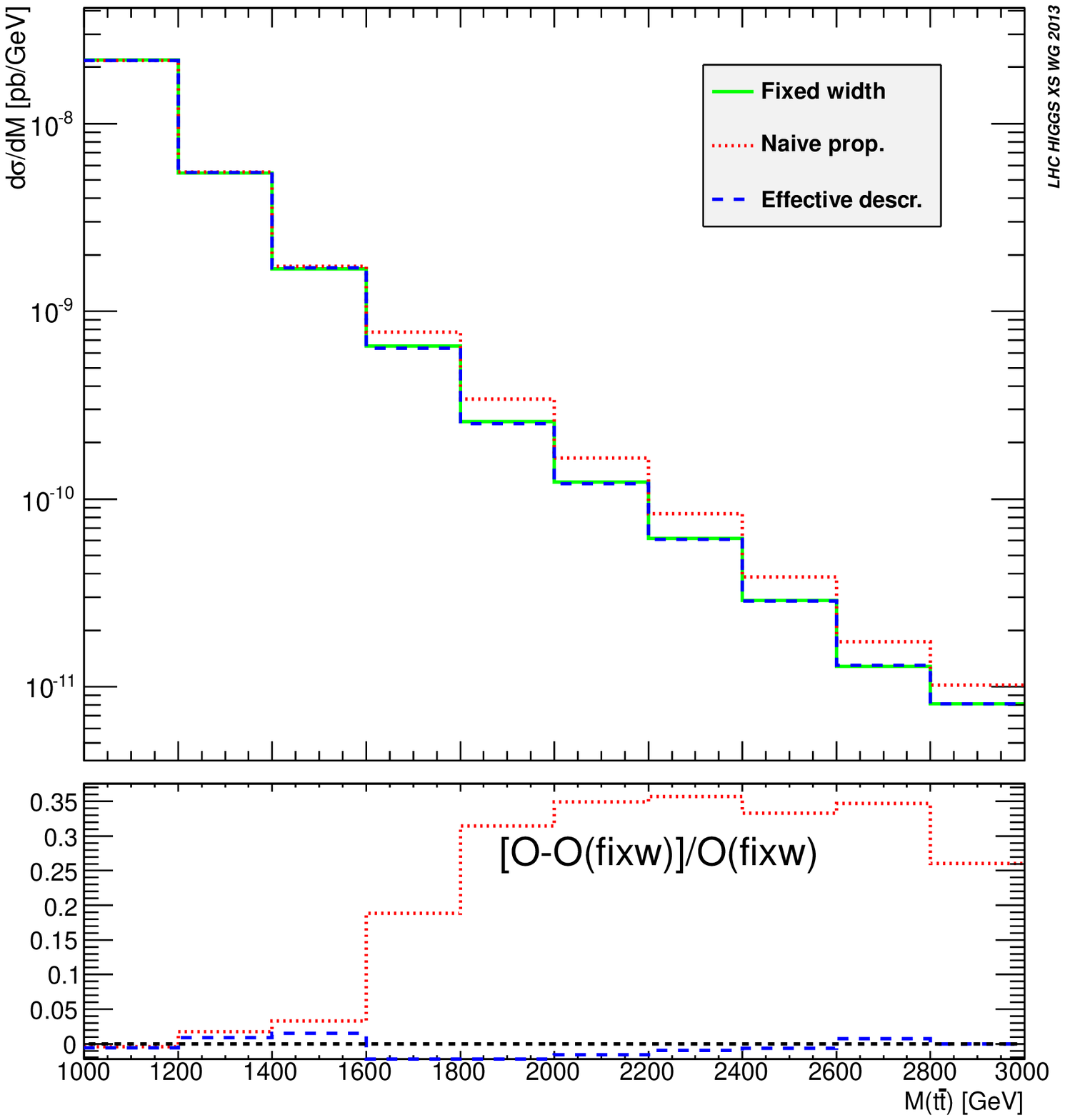}
\caption{Mass distribution of $\PAQt\PQt$-system in the process $\PQu\PQs\to
\PQd\PQc\PAQt\PQt$ at the high energy region. All cuts listed in \Tref{tab:vvcuts} have
been applied.}
\label{fig:mtt_m1000_allp}
\end{minipage}
\end{figure}
%%%%%%%%%%%%%%%%%%%%%%%

\begin{itemize}
  \item Gluon-gluon Fusion
\end{itemize}
For the study of a heavy Higgs produced via gluon-gluon fusion and decayed to a
$\PW$-boson pair, $\Pg\Pg\to \PWp\PWm$, it is very important to consider the complete
set of diagrams due to delicate gauge cancellations that control the high
energy behavior. For this purpose we have relied on MCFM~\cite{MCFMweb} for
evaluation of the matrix elements, taking into account all diagrams
contributing at leading order (yet one loop) to the process $\Pg\Pg\to
\PWp\PWm$, with $\PW$s decaying to leptons. 
%This includes box and triangle diagrams, as shown in \Fref{fig:ggdiag}. 
Phase space integration and unweighted event generation have been carried out
within the {\sc MadGraph} framework. The selection cuts shown in
\Tref{tab:ggcuts} have been applied.
%%%%%%%%%%%%%%%%%%%%%%%
\begin{table}[h!]
  \centering
  \caption{Cuts applied for $\Pg\Pg\rightarrow \PWp\PWm$ on the second one.}
\begin{tabular}{c}
\hline
 $\pT(\ell)>2\,\UGeV$\\
 $\not{E}_T>2\,\UGeV$ \\
 $\eta(\ell)<3$ \\
 $\Delta R(\ell\ell)>0.5$ \\
\hline
\end{tabular}
\label{tab:ggcuts}
\end{table}
%%%%%%%%%%%%%%%%%%%%%%%
The mass distribution of $\PW\PW$-system in the three schemes considered
if shown in \Fref{fig:mww} and \Fref{fig:mww_m800}. In \Fref{fig:mww} we can see
that the effective scheme and the naive propagator description present the same
behavior around the resonance region, while the fixed width scheme shows a
typical slightly harder resonance. At high energy, the naive propagator
diverges and the effective scheme and fixed width scheme are well behaved.
%%%%%%%%%%%%%%%%%%%%%%%
\begin{figure}[htb]
\begin{minipage}[c]{0.48\columnwidth}
\includegraphics[width=\columnwidth,height=0.8\columnwidth]{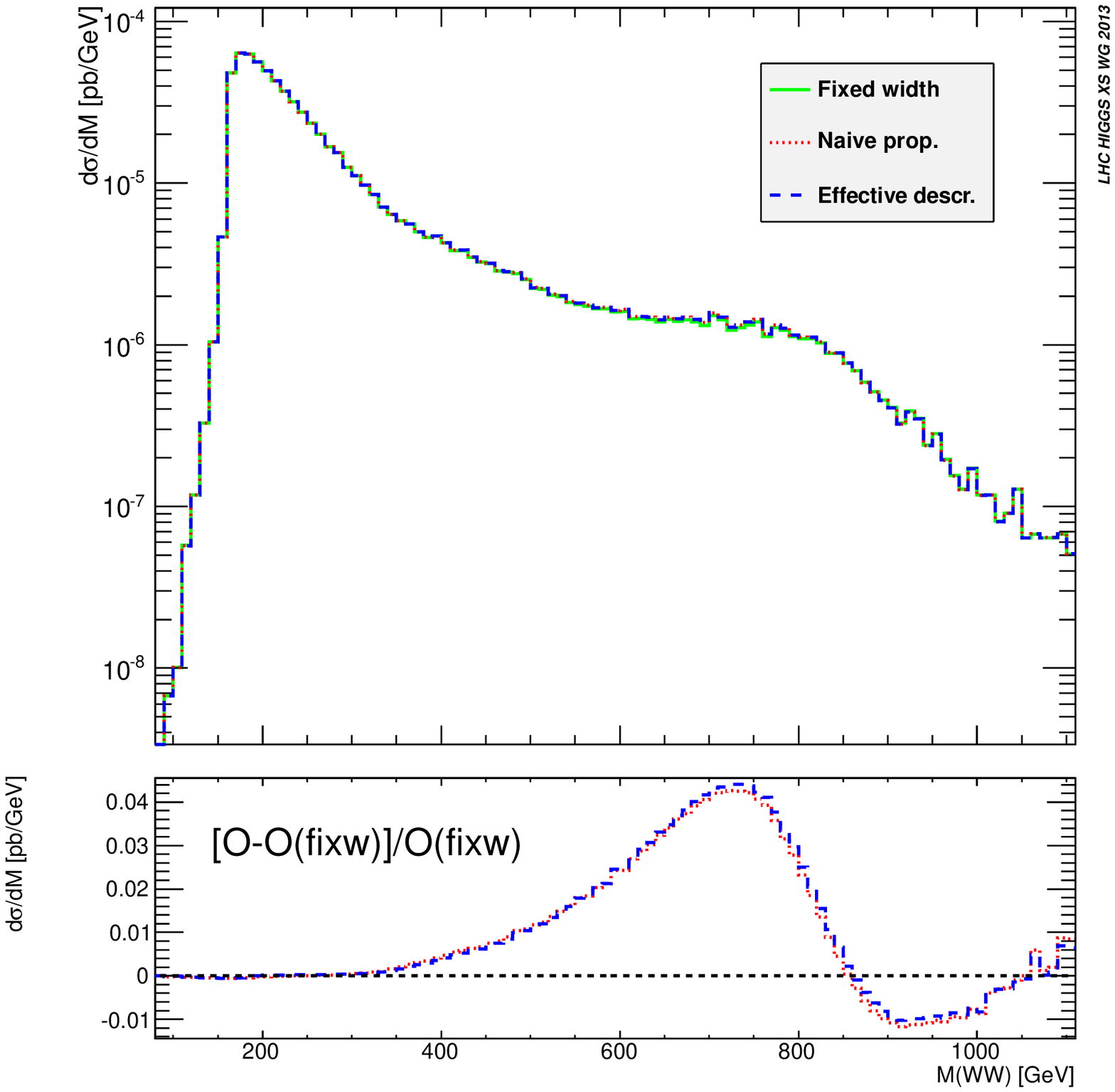}
\caption{Mass distribution of the reconstructed $\PW\PW$-system in the process
  $\Pg\Pg\to \PWp\PWm\to \Pep\nu_e\PGm\nu_\mu$, around the resonance peak. The cuts
  listed in \Tref{tab:ggcuts} have been applied.}
\label{fig:mww}
\end{minipage}
\hfill
\begin{minipage}[c]{0.48\columnwidth}
\includegraphics[width=\columnwidth,height=0.8\columnwidth]{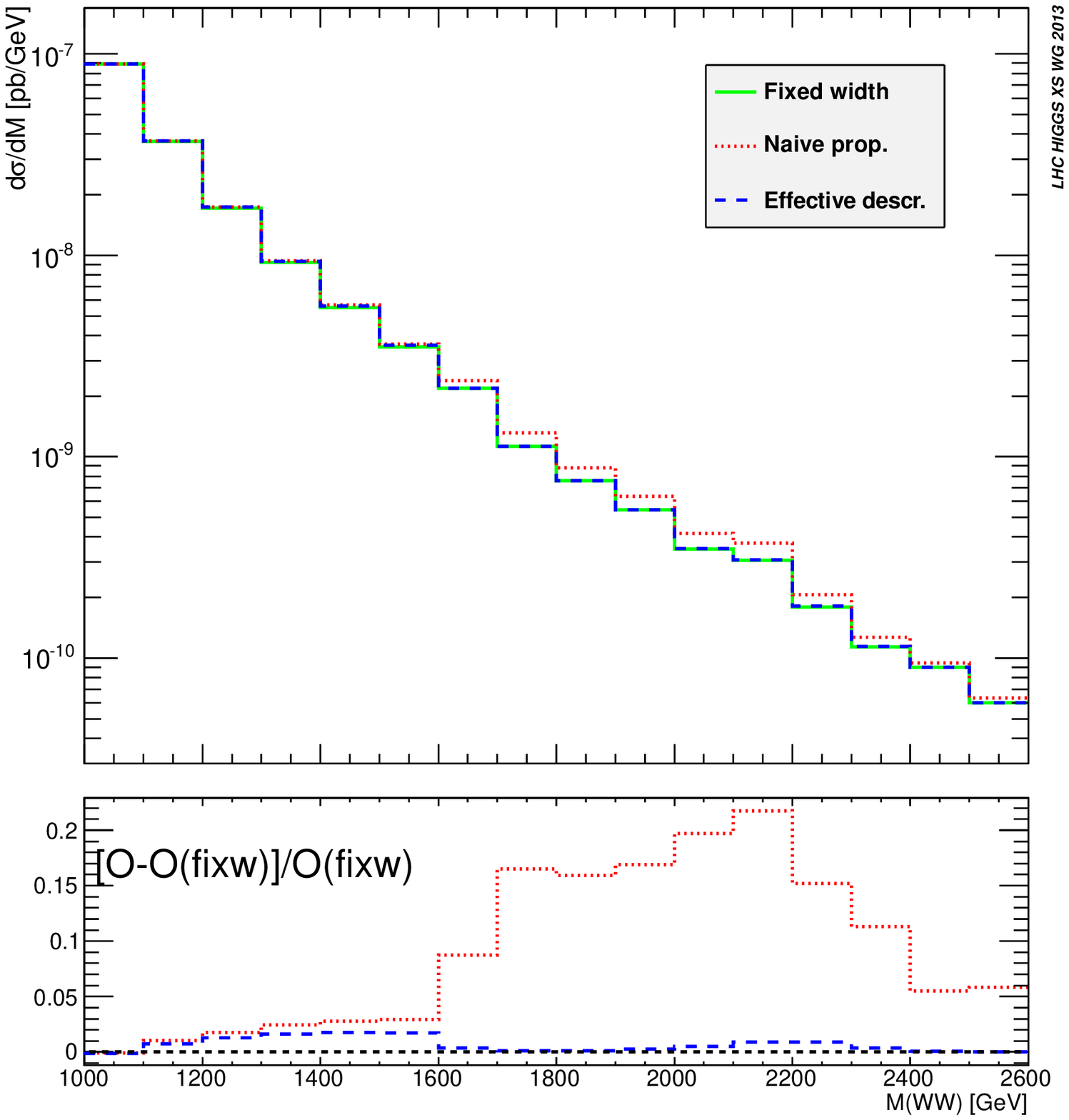}
\caption{Mass distribution of the reconstructed $\PW\PW$-system in the process
  $\Pg\Pg\to \PWp\PWm$ at the high energy region. The cuts listed in \Tref{tab:ggcuts}
have been applied.}
\label{fig:mww_m800}
\end{minipage}
\end{figure}
%%%%%%%%%%%%%%%%%%%%%%%

\subsubsubsection{Conclusions}
We have argued that it is possible to consistently and efficiently include
running width effects for a heavy Higgs-like boson employing an EFT method.
We can summarize the main points of our approach as follows: 

\begin{itemize}
\item
Introducing a width for an unstable particle amounts to a rearrangement of the
perturbative expansion where the corrections to the two-point function are
resummed in the propagator.  The addition of the operator ${\mathcal{\bar{O}}}_\Pi$
defined in Eq.~(\ref{eq:opgood})  allows to effectively perform such resummation in
a gauge invariant and unitary way while keeping the full virtuality dependence
of the self-energy. We have shown that in the limit where such  dependence can
be neglected our scheme is equivalent to the CMS.
\item
At leading order, one has the freedom to choose the functional form of
$\Pi(s)$.  We propose to use the exact one-loop PT self-energy correction. The
rationale  is that such self-energies are gauge invariant and by exploiting the
WI's, we demand ${\mathcal{\tilde{O}}}_\Pi$ to mimic the most important one-loop
corrections as much as possible.  In practice, however, using any other form of
$\Pi(s)$ does not break either gauge invariance or unitarity. In particular,
one could avoid the need for a spurious non-zero width for $t$-channel
propagators.
\item 
EW higher-order corrections can be still performed in the CMS, without loss of
accuracy or double counting issues.  In practice, one can include the running
width effects via the EFT at the leading order and neglect the virtuality
dependence at NLO, i.e., employ the usual CMS for the NLO term.
\end{itemize}

In conclusion, in this work we have considered the case of how to consistently
define a running width in the case of a heavy SM Higgs. The same approach can
be used, for example, in the context of a Two-Higgs-Doublet model and applied
to the current searches for new scalar states at the LHC. Extension to gauge
vectors and heavy fermion states, on the other hand, are not straightforward
and need further investigation.
%%%%%%%%%%%%%%%%%%%%%%%%%%%%%%%%%%%%%%%%%%%%%%%%%%%%%%%%%%%%%%%%%%%%%%%%%%%%%%%

\clearpage

\subsection{Monte Carlo lineshape and interference implementations}
\label{sec:HH:MC}
An overview of the Monte Carlo tools available to describe the
heavy mass $\Pg\Pg\rightarrow \PH$ lineshape and the $\Pg\Pg\rightarrow \PVB\PVB$ signal-background interference
is presented.

%\subsubsection{CPS effects in POWHEG} 
%\label{sec:HH:POWHEG}
%... MISSING... (if in time we can ask Tongguang to add something)

\subsubsection{Signal-background interference studies for $\PH\to\PZ\PZ$ searches with \textsc{gg2VV}}
\label{sec:HH_ggVV}
\DeclareRobustCommand{\PV}{\HepParticle{V}{}{}\Xspace} % V=W,Z boson

\providecommand{\sla}[1]{\ifmmode%
  \setbox0=\hbox{$#1$}%
  \setbox1=\hbox to\wd0{\hss$/$\hss}\else%
  \setbox0=\hbox{#1}%
  \setbox1=\hbox to\wd0{\hss/\hss}\fi%
  #1\hskip-\wd0\box1 }

%\section{Heavy Higgs \footnote{%
%    F.~Convenor, S.~Convenor, \ldots (eds.);
%    \ldots, N.~Kauer} }

%\label{sec:HH}

%%%%%%%%%%%%%%%%%%%%%%%%%%%%%%%%%%%%%%%%%%%%%%%%%%%%%%%%%%%%%%%%%%%%%%%%%%%%%%%

% gg2VV ----------------------------------------------------------------------

In this section, the impact of Higgs-continuum interference on $\PH\to\PZ\PZ$ 
searches at the LHC for a heavy SM Higgs boson is studied with \textsc{gg2VV} 
\cite{gg2VV}.  \textsc{gg2VV} is a parton-level integrator and event generator 
for all $\Pg\Pg\ (\to \PH)\to \PV\PV \to 4$ leptons processes $(\PV\!=\!\PW,\PZ/\PGg^\ast)$
\cite{Binoth:2005ua,Binoth:2006mf,Binoth:2008pr,Kauer:2012hd}.
It can be used to calculate integrated and differential cross sections with 
scale and PDF uncertainties and to produce unweighted events in LHEF format.  
\textsc{gg2VV} takes into account the complete, fully off-shell $\Pg\Pg\to 4$ 
leptons loop-induced LO matrix element with full spin correlations.  Finite top 
and bottom quark mass effects are included.  Amplitude evaluation is facilitated 
by {\sc FeynArts}/{\sc FormCalc}~\cite{Hahn:2000kx,Hahn:1998yk}.
MC integration is facilitated by {\sc Dvegas}~\cite{Dvegas}, which was 
developed in the context of \Brefs{Kauer:2001sp,Kauer:2002sn}.

% input parameters/settings ---------------------------------------------------

The following input parameters and settings have been used to calculate the
results presented in \refSs{sec:HH_ggVV_2l2l} and \ref{sec:HH_ggVV_WWZZ}.
They also apply to the light Higgs results presented in \refSs{sec:ggF_ggVV_WWZZ}
and \ref{sec:ggF_ggVV_peakshift}.
The input-parameter set of \Bref{Dittmaier:2011ti}, App.\ A,   
is used with NLO $\Gamma_{\PW,\PZ}$ and $G_\mu$ scheme. 
Top and bottom quark mass effects are taken into account, while
lepton masses are neglected.
The renormalization and factorization scales are set 
to $\MH/2$.  In \refS{sec:ggF_ggVV_WWZZ} (\refS{sec:HH_ggVV_2l2l}) 
[\refSs{sec:ggF_ggVV_peakshift} and \ref{sec:HH_ggVV_WWZZ}], the 
PDF set MSTW2008 NNLO (MSTW2008 LO) [CT10 NNLO] with 3(1)[3]-loop running 
for $\alphas(\mu^2)$ and 
$\alphas(\MZ^2)=0.11707$ $(0.13939)$ $[0.1180]$ is used.
The complex-pole scheme \cite{Goria:2011wa} with $\GH=29.16$ $(103.9)$ 
$[416.1]\UGeV$ for $\MH=400$ $(600)$ $[1000]\UGeV$ is used for the Higgs 
resonance.  The fixed-width prescription is used for $\PW$ and $\PZ$ propagators.
The CKM matrix is set to the unit matrix, which causes a 
negligible error \cite{Kauer:2012ma}.
No flavor summation is carried out for charged leptons ($\Pl$) or neutrinos.  
A $p_{\mathrm{T}\PV} > 1\UGeV$ cut is applied to prevent that numerical 
instabilities spoil the amplitude evaluation.

%%%%%%%%%%%%%%%%%%%%%%%%%%%%%%%%%%%%%%%%%%%%%%%%%%%%%%%%%%%%%%%%%%%%%%%%%%%%%%%

\subsubsubsection{{\boldmath Signal-background interference in $\Pg\Pg\ (\to \PH)\to \PZ\PZ\to \Pl\PAl\Pl'\PAl'$}}
\label{sec:HH_ggVV_2l2l}

To illustrate signal-background interference effects for a heavy SM Higgs 
boson, integrated cross sections and differential distributions for 
$\Pg\Pg\ (\to \PH)\to \PZ\PZ\to \Pl\PAl\Pl'\PAl'$ in $\Pp\Pp$ collisions 
at $7\UTeV$ for $\MH=400\UGeV$ are presented in \refT{tab:HH_ggVV_2l2l_1} and 
\refFs{fig:HH_ggVV_2l2l_1} and \ref{fig:HH_ggVV_2l2l_2}, respectively.  
Results are given for Higgs signal, $\Pg\Pg$ continuum background 
and the sum with (without) interference.   
To quantify the signal-background interference effect, the
$S$+$B$-inspired measure $R_1$ and $S/\sqrt{B}$-inspired measure $R_2$
defined in \eqn{eqn:ggF_ggVV_WWZZ_1} are used.  When standard cuts 
($p_{\mathrm{T}\Pl} > 20\UGeV$,  $|\eta_{\Pl}| < 2.5$, 
$76\UGeV< M_{\Pl\PAl}, M_{\Pl'\PAl'} < 106\UGeV$) are applied, interference
effects of about $2\%$ are obtained at $7\UTeV$, 
which increase to $3$--$5\%$ when the 
collision energy is increased to $14\UTeV$ (see \refT{tab:HH_ggVV_2l2l_2}).
As shown in \refF{fig:HH_ggVV_2l2l_1}, the Higgs-continuum interference
is negative (positive) for $M_{\PZ\PZ}$ larger (smaller) than $\MH$.
A compensation between negative and positive interference will typically 
occur for integrated cross sections.  Applied selection cuts 
will in general reduce this cancellation, as seen in 
\refF{fig:HH_ggVV_2l2l_2} for the charged-lepton azimuthal opening angle
$\Delta\phi_{\Pl\PAl}$, and should be taken into account to get a reliable
estimate for the interference effect.  Since the Higgs invariant mass can be
reconstructed, it is suggestive to apply a 
$|M_{\PZ\PZ} - \MH| < \GH$ cut in addition to the standard 
cuts to reduce the background (Higgs search cuts).
Such a cut further improves the cancellation of positive and negative
interference, due to the change of sign at $\MH$.  As seen in 
\refTs{tab:HH_ggVV_2l2l_1} and \ref{tab:HH_ggVV_2l2l_2}, the interference 
measures $R_{1,2}-1$ are reduced to the $1\%$ level, when Higgs search
cuts are applied.

\begin{table}
\caption{
Cross sections in $\UfbZ$ for $\Pg\Pg\ (\to \PH)\to \PZ\PZ\to 
\Pl\PAl\Pl'\PAl'$ in $\Pp\Pp$ collisions at $7\UTeV$ for $\MH=400\UGeV$.  
Results are given for signal ($|\PH|^2$), $\Pg\Pg$ continuum background 
($|\mathrm{cont}|^2$) and signal+background+interference 
($|\mathrm{H}$+$\mathrm{cont}|^2$).  
$R_{1,2}$ as defined in 
\eqn{eqn:ggF_ggVV_WWZZ_1} are also displayed.  
Standard cuts: $p_{\mathrm{T}\Pl} > 20\UGeV$,  $|\eta_{\Pl}| < 2.5$, 
$76\UGeV< M_{\Pl\PAl}, M_{\Pl'\PAl'} < 106\UGeV$.  Higgs search cuts: standard 
cuts and $|M_{\PZ\PZ} - \MH| < \GH$ with $\GH=29.16\UGeV$.
No flavor summation is carried out.
The integration error is given in brackets.
}
\label{tab:HH_ggVV_2l2l_1}%
\renewcommand{\arraystretch}{1.2}%
\setlength{\tabcolsep}{1.5ex}%
\begin{center}
\begin{tabular}{lccccc}
\hline
\multicolumn{1}{c}{selection cuts} & $|\PH|^2$ & $|\mathrm{cont}|^2$ & $|\mathrm{H}$+$\mathrm{cont}|^2$ & $R_1$ & $R_2$ \\
\hline
standard cuts & $0.3654(4)$ & $0.3450(4)$ & $0.7012(8)$ & $0.987(2)$ & $0.975(3)$ \\
Higgs search cuts & $0.2729(3)$ & $0.01085(2)$ & $0.2867(3)$ & $1.010(2)$ & $1.011(2)$ \\
\hline
\end{tabular}
\end{center}
\end{table}

\begin{table}
\caption{
As \refT{tab:HH_ggVV_2l2l_1}, but for $\sqrt{s} = 14\UTeV$.
}
\label{tab:HH_ggVV_2l2l_2}%
\renewcommand{\arraystretch}{1.2}%
\setlength{\tabcolsep}{1.5ex}%
\begin{center}
\begin{tabular}{lccccc}
\hline
\multicolumn{1}{c}{selection cuts} & $|\PH|^2$ & $|\mathrm{cont}|^2$ & $|\mathrm{H}$+$\mathrm{cont}|^2$ & $R_1$ & $R_2$ \\
\hline
standard cuts & $1.893(3)$ & $1.417(2)$ & $3.205(5)$ & $0.969(2)$ & $0.945(3)$ \\
Higgs search cuts & $1.377(2)$ & $0.0531(1)$ & $1.445(2)$ & $1.011(2)$ & $1.011(3)$ \\
\hline
\end{tabular}
\end{center}
\end{table}

\begin{figure}
\centering\includegraphics[width=0.48\textwidth]{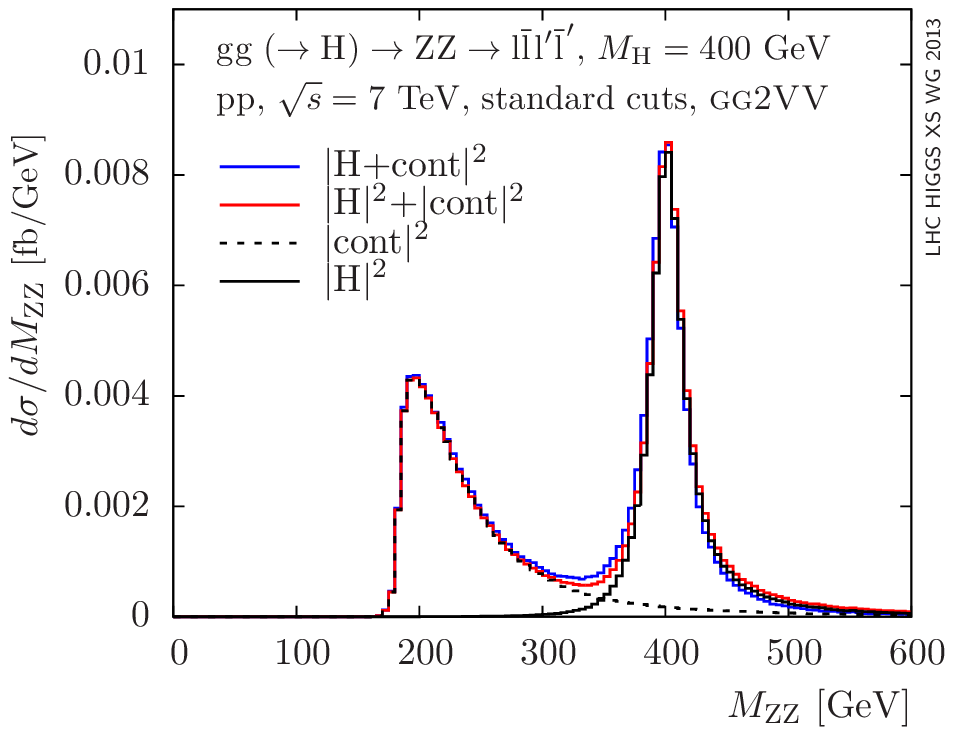}
\vspace*{0.cm}
\caption{
$M_{\PZ\PZ}$ distributions for $\Pg\Pg\ (\to \PH)\to \PZ\PZ\to 
\Pl\PAl\Pl'\PAl'$ in $\Pp\Pp$ collisions at $7\UTeV$ for $\MH=400\UGeV$.
Distributions for signal ($|\PH|^2$), $\Pg\Pg$ continuum background 
($|\mathrm{cont}|^2$), signal+background ($|\mathrm{H}|^2$+$|\mathrm{cont}|^2$), 
and signal+background+interference ($|\mathrm{H}$+$\mathrm{cont}|^2$)
are shown.  Standard cuts and other details as in \refT{tab:HH_ggVV_2l2l_1}.
}
\label{fig:HH_ggVV_2l2l_1}
\end{figure}

\begin{figure}
\centering\includegraphics[width=0.48\textwidth]{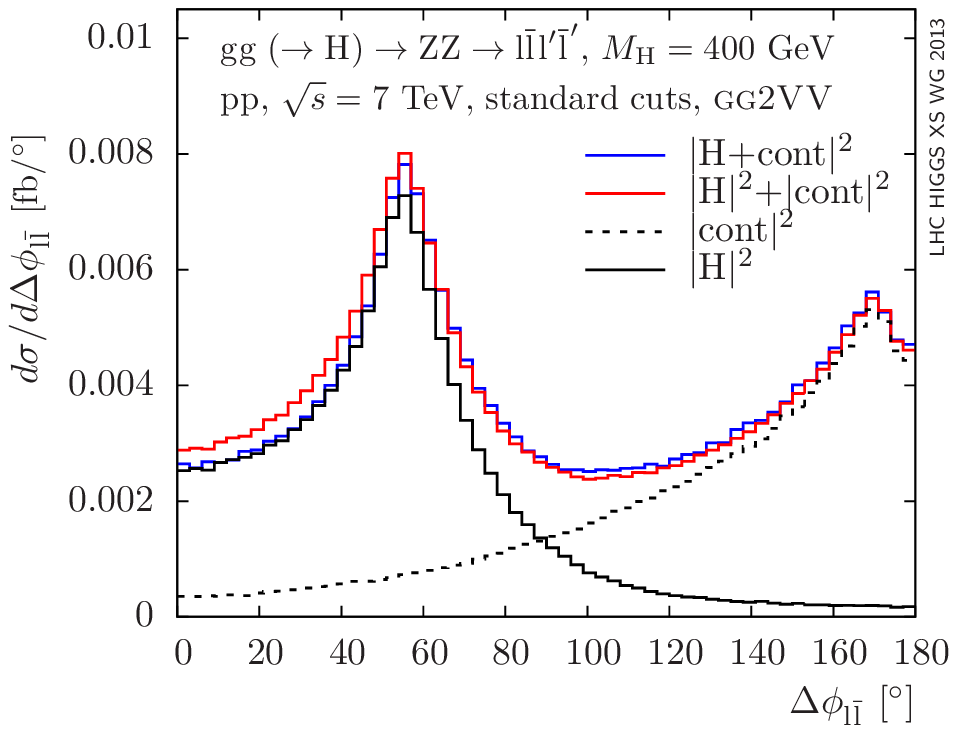}
\vspace*{0.cm}
\caption{
Azimuthal opening angle $\Delta\phi_{\Pl\PAl}$ distributions 
for $\Pg\Pg\ (\to \PH)\to \PZ\PZ\to \Pl\PAl\Pl'\PAl'$ in 
$\Pp\Pp$ collisions at $7\UTeV$ for $\MH=400\UGeV$.
Other details as in \refF{fig:HH_ggVV_2l2l_1}.
}
\label{fig:HH_ggVV_2l2l_2}
\end{figure}

%%%%%%%%%%%%%%%%%%%%%%%%%%%%%%%%%%%%%%%%%%%%%%%%%%%%%%%%%%%%%%%%%%%%%%%%%%%%%%%

\subsubsubsection{{\boldmath $\PZ\PZ/\PW\PW$ interference in $\Pg\Pg\ (\to \PH)\to 
\Pl\PAl\PGnl\PAGnl$}}
\label{sec:HH_ggVV_WWZZ}

In \refS{sec:HH_ggVV_2l2l}, the key features of signal-background interference in 
$\PH\to\PZ\PZ$ searches were elucidated using the ``golden mode.''
In this section, $\PW\PW$ corrections are studied that occur when the 
same-flavor final state $\Pl\PAl\PGnl\PAGnl$ is produced.\footnote{%
See also \refS{sec:ggF_ggVV_WWZZ}.}  Results obtained with a minimal 
$M_{\Pl\PAl} > 4\UGeV$ cut are shown in \refTs{tab:HH_ggVV_WWZZ_1} and 
\ref{tab:HH_ggVV_WWZZ_2} for $\MH=600\UGeV$ and $\MH=1\UTeV$, respectively, 
at $\sqrt{s}=8\UTeV$.  Cross sections when only either the $\PW\PW$ or 
$\PZ\PZ$ intermediate state is included are also given.
As above, results for Higgs signal, $\Pg\Pg$ continuum background 
and the sum with interference as well as the interference measures $R_{1,2}$
are displayed.  Without search cuts, $S/B$ decreases significantly with 
increasing heavy Higgs mass.  Consequently, the interference measure 
$R_2-1$ increases substantially when going from $\MH=600\UGeV$ to $\MH=1\UTeV$,
namely from $\Ord(20\%)$ to $\Ord(2)$.  In addition to signal-background
interference, the interference between $\PW\PW$ and $\PZ\PZ$ contributions
is also of interest.  For the $\Pg\Pg$ continuum background, one finds
$\sigma(|\PW\PW+\PZ\PZ|^2)/\sigma(|\PW\PW|^2+|\PZ\PZ|^2)=0.935(5)$,
while for the signal this ratio agrees with $1$ at the sub-percent
level for $\MH=600\UGeV$ and $1\UTeV$.  
The $\Ord(5\%)$ $\PW\PW/\PZ\PZ$ interference for the $\Pg\Pg$ continuum 
background has to be compared to the negligible $\PW\PW/\PZ\PZ$ interference
found in \Bref{Melia:2011tj} for the quark-induced continuum background.

\begin{table}
\caption{
Cross sections in $\UfbZ$ for $\Pg\Pg\ (\to \PH)\to \PW\PW, \PZ\PZ\to 
\Pl\PAl\PGnl\PAGnl$ (same flavor) in $\Pp\Pp$ collisions at $8\UTeV$ for 
$\MH=600\UGeV$.  Cross sections when only either the $\PW\PW$ or 
$\PZ\PZ$ intermediate state is included are also given.
A minimal $M_{\Pl\PAl} > 4\UGeV$ cut is applied.
Other details as in \refT{tab:HH_ggVV_2l2l_1}.
}
\label{tab:HH_ggVV_WWZZ_1}%
\renewcommand{\arraystretch}{1.2}%
\setlength{\tabcolsep}{1.5ex}%
\begin{center}
\begin{tabular}{lccccc}
\hline
\multicolumn{1}{c}{$\PV\PV$} & $|\PH|^2$ & $|\mathrm{cont}|^2$ & $|\mathrm{H}$+$\mathrm{cont}|^2$ & $R_1$ & $R_2$ \\
\hline
$\PW\PW$ & $1.44(1)$ & $12.29(3)$ & $14.10(5)$ & $1.027(4)$ & $1.26(4)$ \\
$\PZ\PZ$ & $0.261(2)$ & $1.590(5)$ & $1.896(6)$ & $1.024(4)$ & $1.17(3)$ \\
$\PW\PW, \PZ\PZ$ & $1.69(2)$ & $12.98(6)$ & $15.00(8)$ & $1.022(7)$ & $1.19(6)$ \\
\hline
\end{tabular}
\end{center}
\end{table}

\begin{table}
\caption{
As \refT{tab:HH_ggVV_WWZZ_1}, but for $\MH=1\UTeV$.
}
\label{tab:HH_ggVV_WWZZ_2}%
\renewcommand{\arraystretch}{1.2}%
\setlength{\tabcolsep}{1.5ex}%
\begin{center}
\begin{tabular}{lccccc}
\hline
\multicolumn{1}{c}{$\PV\PV$} & $|\PH|^2$ & $|\mathrm{cont}|^2$ & $|\mathrm{H}$+$\mathrm{cont}|^2$ & $R_1$ & $R_2$ \\
\hline
$\PW\PW$ & $0.0772(5)$ & $10.50(3)$ & $10.72(3)$ & $1.013(4)$ & $2.8(5)$ \\
$\PZ\PZ$ & $0.01426(9)$ & $1.353(4)$ & $1.387(4)$ & $1.015(4)$ & $2.4(4)$ \\
$\PW\PW, \PZ\PZ$ & $0.0914(6)$ & $11.02(6)$ & $11.30(8)$ & $1.017(9)$ & $3(1)$ \\
\hline
\end{tabular}
\end{center}
\end{table}

% H --> ZZ search cuts -------------------------------------------------------

We now investigate the impact of $\PH\to\PZ\PZ$ search cuts on the 
signal-background interference.  To be specific, we consider the cuts
$|M_{\Pl\PAl}-\MZ| < 15\UGeV$, $\sla{E}_\mathrm{T} > 110\UGeV$, 
$M_\mathrm{T} > 325\UGeV$. The transverse mass is defined as
\beq
\label{eqn:HH_ggVV_WWZZ_1}
M_\mathrm{T}=\sqrt{ \left(M_{\mathrm{T},\Pl\PAl}+\sla{M}_{\mathrm{T}}\right)^2-({\bf{p}}_{\mathrm{T},\Pl\PAl}+{\sla{\bf{p}}}_\mathrm{T})^2 }\quad\mathrm{with}\quad \sla{M}_{\mathrm{T}}=\sqrt{\sla{p}_{\mathrm{T}}^2+M_{\Pl\PAl}^2}\,.
\eeq
Results for $\MH=600\UGeV$ and $1\UTeV$
are shown in \refTs{tab:HH_ggVV_WWZZ_3} and \ref{tab:HH_ggVV_WWZZ_4}, 
respectively.  Since the Higgs search cuts suppress the background while
retaining the signal, one can expect that with search cuts $R_2$ deviates 
less from $1$ than when only minimal cuts are applied. This is confirmed
by the shown results: For $\MH=600\UGeV$ ($1\UTeV$),
$R_2-1$ decreases from $\Ord(20\%)$ ($\Ord(2)$) to $\Ord(7\%)$ ($\Ord(1)$).

\begin{table}
\caption{
Cross sections in $\UfbZ$ for $\Pg\Pg\ (\to \PH)\to \PW\PW, \PZ\PZ\to 
\Pl\PAl\PGnl\PAGnl$ (same flavor) in $\Pp\Pp$ collisions at $8\UTeV$ for 
$\MH=600\UGeV$.  $\PH\to\PZ\PZ$ search cuts are applied: 
$|M_{\Pl\PAl}-\MZ| < 15\UGeV$, $\sla{E}_\mathrm{T} > 110\UGeV$, 
$M_\mathrm{T} > 325\UGeV$.  
$M_\mathrm{T}$ is defined in \eqn{eqn:HH_ggVV_WWZZ_1}.
Other details as in \refT{tab:HH_ggVV_WWZZ_1}.
}
\label{tab:HH_ggVV_WWZZ_3}%
\renewcommand{\arraystretch}{1.2}%
\setlength{\tabcolsep}{1.5ex}%
\begin{center}
\begin{tabular}{lccccc}
\hline
\multicolumn{1}{c}{$\PV\PV$} & $|\PH|^2$ & $|\mathrm{cont}|^2$ & $|\mathrm{H}$+$\mathrm{cont}|^2$ & $R_1$ & $R_2$ \\
\hline
$\PZ\PZ$ & $0.2175(8)$ & $0.0834(2)$ & $0.3150(8)$ & $1.047(4)$ & $1.065(6)$ \\
$\PZ\PZ, \PW\PW$ & $0.2220(8)$ & $0.1020(2)$ & $0.3406(8)$ & $1.051(4)$ & $1.075(6)$ \\
\hline
\end{tabular}
\end{center}
\end{table}

\begin{table}
\caption{
As \refT{tab:HH_ggVV_WWZZ_3}, but for $\MH=1\UTeV$.
}
\label{tab:HH_ggVV_WWZZ_4}%
\renewcommand{\arraystretch}{1.2}%
\setlength{\tabcolsep}{1.5ex}%
\begin{center}
\begin{tabular}{lccccc}
\hline
\multicolumn{1}{c}{$\PV\PV$} & $|\PH|^2$ & $|\mathrm{cont}|^2$ & $|\mathrm{H}$+$\mathrm{cont}|^2$ & $R_1$ & $R_2$ \\
\hline
$\PZ\PZ$ & $0.01265(5)$ & $0.0687(2)$ & $0.0927(2)$ & $1.140(3)$ & $1.90(2)$ \\
$\PZ\PZ, \PW\PW$ & $0.01278(5)$ & $0.0846(3)$ & $0.1090(2)$ & $1.119(3)$ & $1.91(3)$ \\
\hline
\end{tabular}
\end{center}
\end{table}

% %%%%%%%%%%%%%%%%%%%%%%%%%%%%%%%%%%%%%%%%%%%%%%%%%%%%%%%%%%%%%%%%%%%%%%%%%%%%%%%
% 
% \subsubsubsection*{Acknowledgments}
% 
% Financial support from HEFCE, STFC and the IPPP Durham is gratefully 
% acknowledged by N.K.
% 
% %%%%%%%%%%%%%%%%%%%%%%%%%%%%%%%%%%%%%%%%%%%%%%%%%%%%%%%%%%%%%%%%%%%%%%%%%%%%%%%

% LocalWords:  sn ggVV WWZZ

\subsubsection{Heavy Higgs implementation in MCFM} 
\label{sec:MCFM}

\allowdisplaybreaks

%\section{Overview}

This section provides an overview of the implementation of Higgs boson production and background processes in \textsc{MCFM}. Further details can be found in \Brefs{Campbell:2011cu , Campbell:2011bn}. \textsc{MCFM} contains NLO predictions for a variety of processes. For studies of heavy Higgs bosons the most relevant signal processes are Higgs production through gluon fusion and vector-boson-fusion, with subsequent decays to massive vector bosons. The SM background production of the direct diboson final state themselves. In addition to the standard NLO predictions for the diboson processes, \textsc{MCFM} includes the contributions from gluon induced initial states. Although formally $\mathcal{O}(\alphas^2)$ (and hence NNLO) the large gluon flux at LHC operating energies enhances these pieces beyond the naive NNLO power counting. The current status (as of \textsc{MCFM} v6.4) of these processes is as follows. For the $\Pg\Pg\rightarrow \PW\PW$ process all three families of quarks are included in the loops, with the exact dependence on the top mass $m_{\PQt}$ kept and all other quarks considered massless. For the $\PZ\PZ$ process only the first five massless flavors are implemented. 

One of the most interesting phenomenological features of a heavy Higgs boson is its large width and the corresponding impact of the interference term with the SM production of $\PVB\PVB$ pairs. Since the interference pattern is often not accounted for in event generator simulations and higher order corrections, we focus on its impact on phenomenology in this report. 

%\section{Interference effects}

In order to study the interference between the SM production of $\PW$ pairs and Higgs decay to $\PW\PW$ we introduce the following definitions, 
\begin{eqnarray}
{\sigma_{\mathrm B}} &\longrightarrow& \left|{\mathcal A}_{\mathrm{box}}\right|^2 \;, \qquad {\mathcal A}_{\mathrm{box}} = 2{\mathcal A}_{\mathrm{massless}}+{\mathcal A}_{\mathrm{massive}} \;, \nonumber \\
{\sigma_{\PH}} &\longrightarrow& \left|{\mathcal A}_{\mathrm{Higgs}}\right|^2 \nonumber \;, \\
{\sigma_i} &\longrightarrow& 2{\mathrm{Re}} \left({\mathcal A}_{{\mathrm{Higgs}}} {\mathcal A}_{\mathrm{box}}^\star \right) \;, \nonumber \\
{\sigma_{\PH,i}} &=& \sigma_{\PH} + \sigma_i \;.
\label{eq:Hidescription}
\end{eqnarray}
Here $\sigma_{\mathrm B}$ represents the background production of $\PW\PW$
pairs, (proceeding through the ${\mathcal A}_{\mathrm{box}}$ amplitude which is
made up of massless and massive fermion loops) and $\sigma_{\PH}$ represents signal squared cross section associated with the LO production of a Higgs boson through gluon fusion, with a subsequent decay to $\PW\PW$. Note that this LO processes proceeds through a top quark loop. Finally $\sigma_i$ represents the interference between the signal and background amplitudes.  

The interference pattern contains two terms, which are extracted by removing the Higgs propagator from the amplitude (with the new stripped amplitude referred to as $\tilde{A}_{\mathrm{Higgs}}$).  The imaginary part of the Higgs propagator couples to the imaginary part of the product, $\tilde{A}_{\mathrm{Higgs}}{\mathcal A}_{\mathrm{box}}^* $. Hence this piece is proportional to the Higgs width. The real part of the Higgs propagator couples to the real part of  $\tilde{A}_{\mathrm{Higgs}}{\mathcal A}_{\mathrm{box}}^* $. The resulting interference cross section is thus of the form, 
\begin{equation}
\delta\sigma_i = \frac{(\hat s - \MH^2)}{ (\hat s - \MH^2)^2 + \MH^2 \Gamma_{\PH}^2} \, \Re \left( 2 \widetilde{\mathcal A}_{\mathrm{Higgs}} {\mathcal A}_{\mathrm{box}}^*  \right)
         + \frac{\MH \Gamma_{\PH}}{ (\hat s - \MH^2)^2 + \MH^2 \Gamma_{\PH}^2} \, \Im  \left( 2 \widetilde{\mathcal A}_{\mathrm{Higgs}} {\mathcal A}_{\mathrm{box}}^*  \right) \;,     
\label{eq:intfdecomp}
\end{equation}
The first piece of this expression is an odd function in $\hat s$ about $\MH$ and therefore, for well resolved final states (such as $\PZ\PZ^*$ and $\PGg\PGg$) in which the Higgs signal can be localized in $\hat s$, the interference effects from this piece approximately cancel over the integration of the $\hat s $ bin. In these cases the interference phenomenology is dominated by the width effects (which for a heavy Higgs will be sizeable). For Higgs decays which are not fully reconstructed (such as $\PW\PW$) both terms contribute. Since the Higgs boson is a unitarizing particle the interference is destructive at large $\hat s$. Thus, since the function is odd the interference in the corresponding region $\hat s <  \MH$ is constructive. Since the background cross section is typically larger in this region the net effect of the interference is constructive for most Higgs masses. 

\begin{figure}[htb] 
\centering
\includegraphics[width=8cm]{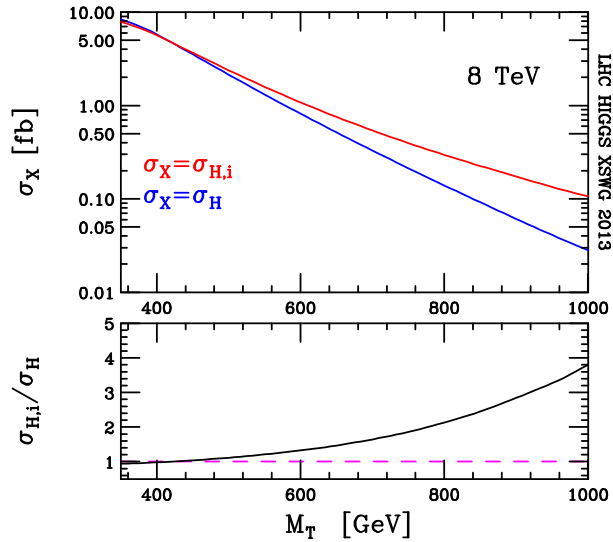}
\caption{Inclusive $\PH\rightarrow \PW\PW\rightarrow \ell\PGn\ell'\PGn'$ cross section at the LHC8. The standard signal squared $\sigma_{\PH}$ is shown in blue, whilst the signal squared plus interference cross sections are shown in red. The lower panel shows the ratio of these two cross sections.} 
\label{fig:higgs_xs}
\end{figure}
\Fref{fig:higgs_xs} presents the net contribution of the interference for the inclusive $\PH\rightarrow \PW\PW\rightarrow \ell\PGn\ell'\PGn'$ cross section at LHC operating energies ($8\UTeV$), focusing on the heavy Higgs region. The renormalization and factorization scales have been set equal to the Higgs mass. It is clear that the interference becomes the dominant part of the Higgs production cross section as the Higgs mass grows. This is primarily since the Higgs cross section is a rapidly falling function in $\MH$, hence the term linear in the signal amplitude (the interference) dominates over the quadratic piece (the signal squared). 
\begin{figure}[htb]
\centering
\includegraphics[width=8cm]{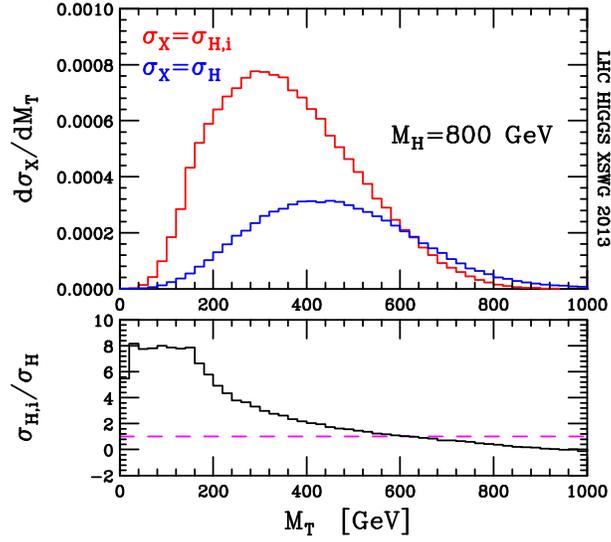}
\caption{The transverse mass distribution for a Higgs boson with mass $800\UGeV$ at the $8\UTeV$ LHC. The upper panel shows the prediction for the differential distribution with and without including the interference terms, the lower panel provides the ratio of these predictions.} 
\label{fig:mt}
\end{figure}

It is also interesting to consider the impact of the interference pieces on
more differential quantities. One such relevant quantity is the transverse
mass of the missing transverse energy plus leptons final
state. The \textsc{MCFM} prediction for the transverse mass spectrum for a
Higgs mass of $800\UGeV$ is shown in \Fref{fig:mt}. The pattern predicted by 
\eqn{eq:intfdecomp} is manifest. The interference is large and constructive in the region ($\hat s  < \MH \implies m_{\mathrm T} < \MH)$, whilst in the large $\hat s$ region the interference is destructive. The enhancement in the low $m_{\mathrm T}$ region is significant and can be around an order of magnitude compared to the naive signal prediction. Therefore inclusion of the interference effects are vital in order to simulate the phenomenological impact of the heavy Higgs.  

Since it is clear that the contributions from the interference effects are important, it is interesting to look at possible mechanisms for modifying the Higgs propagator in order to simulate the impact of the interference. If such a modification can mimic the interference terms then use of this propagator in higher order codes (for example the NNLO code of ref.~\cite{Anastasiou:2011pi}) can enhance these predictions. \textsc{MCFM} provides an excellent testing environment for these techniques, since it is possible to modify the Higgs propagator as desired and compare it to the full signal plus interference prediction at LO. One such modification is the improved $s-$channel approximation (ISA) due to Seymour~\cite{Seymour:1995np}. In this setup one replaces the fixed-width propagator with the following, 
\begin{eqnarray}
\frac{is}{s-\MH^2+i\Gamma_{\PH}\MH} \rightarrow \frac{i\MH^2}{s-\MH^2+i\Gamma_{\PH}(\MH) \frac{s}{\MH}}. 
\end{eqnarray}
The dependence on $\hat{s}$ in the width piece modifies the corresponding Higgs boson lineshape. We illustrate these differences in \Fref{fig:isa}. In order to ensure the invariant mass distribution is positive definite we also include the prediction for $\Pg\Pg\rightarrow \PW\PW$, $\sigma_{\mathrm B}$. It is clear that by modifying the propagator we have significantly altered the resulting lineshape of the Higgs. The ISA has captured the form of the interference prediction, namely enhancing the differential cross section in the region $\hat{s} < \MH$ and decreasing it in the high mass tail. Whilst the ISA is an improvement over the fixed width approximation it fails to capture the true impact of the destructive interference and overestimates the rate at small $m_{4\ell}$. 

\begin{figure} [htb]
\centering
\includegraphics[width=8cm]{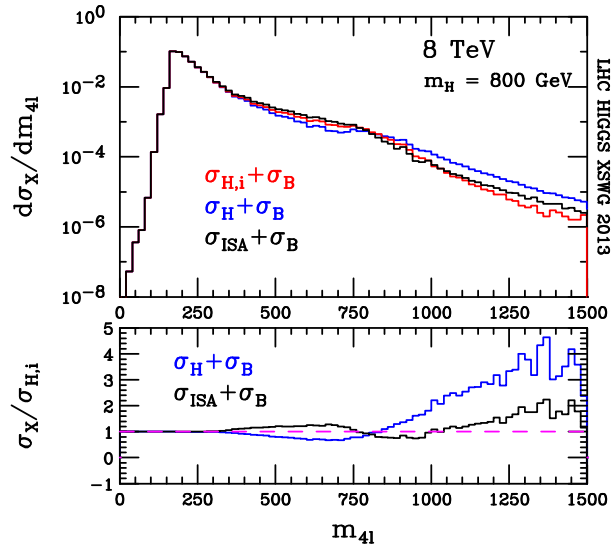}
\caption{Comparison of the improved $s$-channel approximation and the full LO signal plus interference prediction for the invariant mass of the four lepton final state. For reference the naive signal squared prediction is also plotted. The background production of $\Pg\Pg\rightarrow \PW\PW$ is also included.} 
\label{fig:isa}
\end{figure}

\noindent
{\it Summary}\\[.2em]
\textsc{MCFM} provides a range of predictions relevant for searches for a heavy Higgs boson. The most relevant being the signal production cross sections and many of the irreducible background processes. Since the width of a heavy Higgs boson is very large interference effects between signal and background dominate the lineshape of the Higgs. \textsc{MCFM} includes a full treatment of this interference at LO for the $\PW\PW$ decay modes, allowing the user to test systematic uncertainties in other codes which may not model these interactions.

\clearpage
%\end{document}

\subsubsection{Signal-background interference implementation in aMC@NLO} 
\label{sec:HH:amcatnlo}
%\section{Section name ()
%  \footnote{%
%    F.~Convenor, S.~Convenor, \ldots (eds.);
%    Stefano Frixione, Antoine Lauyers, Fabio Maltoni } }
%\label{sec:SectionName}

%%%%%%%%%%%%%%%%%%%%%%%%%%%%%%%%%%%%%%%%%%%%%%%%%%%%%%%%%%%%%%%%%%%%%%%%%%%%%%%

At LO the cross section for $\Pg\Pg \to \PVB\PVB$ can be written as

\begin{equation}
\sigma^{\LO}_{\mathrm{S+i+B}} = |{\mathcal M}_{\mathrm S}|^2+2\, {\rm Re}( {\mathcal M}_{\mathrm S} {\mathcal M}_{\mathrm B}^*) + |{\mathcal M}_{\mathrm B}|^2\,,
\end{equation}
where the signal $S$ features a triangle loop and the Higgs propagation in the $s$-channel while the background $B$ proceeds via $\Pg\Pg\to \PVB\PVB$ boxes.
At high partonic energy $\hat s$, there is a cancellation between the $\Pp\Pp \to \PH \to \PVB\PVB$ diagram and the $S$-wave massive contribution of the 
non-resonant ones~\cite{Glover:1988rg}, each growing as $m_{\PQq}^2/m_{\PVB}^2 \log^2 \hat s/m_{\PQq}^2$ where $\PQq$ is the heavy quark running in the loop.
This contribution is proportional to the axial part of the vector boson coupling to the quarks and it is basically due to the non-conservation of the
axial current for massive quarks.  Such a cancellation ensures the unitarity of the standard model up to arbitrary large scales. 
Note that  the width scheme has also an impact on the unitarity of the theory. 

In absence of a complete NLO computation for the full process $\Pp\Pp \to \PVB\PVB$ a scheme for combining the signal $\Pp\Pp \to \PH \to \PVB\PVB$ at NLO (or NNLO), the background at LO and the interference between them  is needed. Different methods have been proposed and discussed, see, .e.g., ~\cite{Passarino:2012ri},  which make different approximations and can broadly divided in multiplicative or additive. In essence the main difference stems from how the signal-background interference is treated: in the additive scheme it is taken at the LO, i.e., it corresponds exactly to the accuracy to which is presently known, while in the multiplicative scheme a guess for the NLO (or NNLO) interference is made, which is obtained by multiplying the LO results by a $K$-factor, overall or differential based on a pivot distribution. The former scheme is the only possible one for a MC generator, yet it violates unitarity at NLO  in $\alphas$, while the latter is better suited for analytical calculations, does not violate unitarity, yet it is just a guess for an unknown quantity. 

In our approach, we therefore employ an additive scheme,

\begin{equation}
\sigma^{\mathrm{MC}}(pp \to  \PVB\PVB) = \sigma^{\NLO}_{\mathrm S}+ \sigma^{\LO}_{\mathrm{i+B}} = \sigma^{\NLO}_{\mathrm S} +2\, {\rm Re}( {\mathcal M}_{\mathrm S} {\mathcal M}_{\mathrm B}^*) + |{\mathcal M}_{\mathrm B}|^2 \,.
\label{eq: additive-scheme}
\end{equation}

Events corresponding to the first term are generated with \textsc{MC@NLO} v.4.09~\cite{Frixione:2002ik}, which features real (one-loop) and virtual (two-loop) matrix elements with  the exact $m_{\PQb}$ and $m_{\PQt}$ dependence (taken from \Bref{Aglietti:2006tp,Bonciani:2007ex}).
Such calculation is for on-shell Higgs production. Introducing the Higgs propagation and decay, and therefore the Higgs width, needs special care especially for a heavy Higgs in order to maintain gauge invariance and unitarity, as discussed at length in the literature and this report. In \textsc{MC@NLO} given a final state in terms of leptons,  this is achieved by the replacement:
\begin{equation}
\delta(\hat s - \MH^2) \rightarrow \frac{1}{\pi} \frac{\sqrt{\hat s} \cdot \Gamma(H(\hat s) \to {\rm final\, state})}{(\hat s - \MH^2)^2 + \MH^2 \Gamma_{\PH}^2}\,,
\end{equation}
where the partial width into the final state is calculated at corresponding virtuality and $\Gamma_{\PH}$ is the total Higgs width. 
Such a replacement exactly corresponds to a complex-pole scheme for the Higgs if $\Gamma_{\PH}$ is calculated accordingly~\cite{Goria:2011wa}.

The second and third terms in Eq.~(\ref{eq: additive-scheme}) are generated via dedicated implementations, whose loop matrix elements are automatically  obtained via \textsc{MadLoop}~\cite{Hirschi:2011pa} and checked, when possible, with MCFM~\cite{Campbell:2011cu}. The corresponding codes are publicly available at the \textsc{aMC@NLO} web page ({\tt http://amcatnlo.cern.ch}). In our calculation all six flavor run in the background boxes for $\Pg\Pg\to \PVB\PVB$ and quark mass effects are accounted for exactly.

Samples corresponding to different leptonic final states, $\ell^+ \PGn_\ell \ell'^- \bar \PGn_{\ell'} (\PWp \PWm)$, $\ell^+ \ell^-  \PGn_{\ell'}  \bar \PGn_{\ell'}  (\PZ\PZ)$ and $\ell^+ \ell^-  \ell'^+ \ell'^-  (\PZ/\PGg^* Z/\PGg^*)$ can be obtained. In \Fref{fig:ww.zz.aa} the invariant mass of the four leptons in  the   $\Pep \Pe  \PGn_{\PGm}  \bar \PGn_{\PGm}$ and  $\Pep \Pe  \PGmp \PGm $ final states for two different Higgs masses are shown. Only minimal acceptance cuts have been applied. The blue (low lying) curve show the full LO result. the black (dotted) curve the NLO signal plus the background, while the red curve is the combined result corresponding to Eq.~(\ref{eq: additive-scheme}). The first important result is that NLO
effects for the signal are very important around the Higgs mass, as expected, and cannot be neglected. The interference gives an enhancement
of the cross section before the peak and a depletion (corresponding to the cancellation of the bad high-energy behavior) at large invariant masses. Such effects, while certainly visible already for a Higgs of $\MH=500{-}600\UGeV$, give a small contributions overall as the high-energy tail  is strongly suppressed by the gluon PDF's.

%%%%%%%%%%%%%%%%%%%%%%%%%%%%%%%%%%%%%%%%%%%%%%%%%%%%%%%%%%%%%%%%%%%%%%%%%%%%
\begin{figure}[h]
\centering
\includegraphics[width=0.7\textwidth]{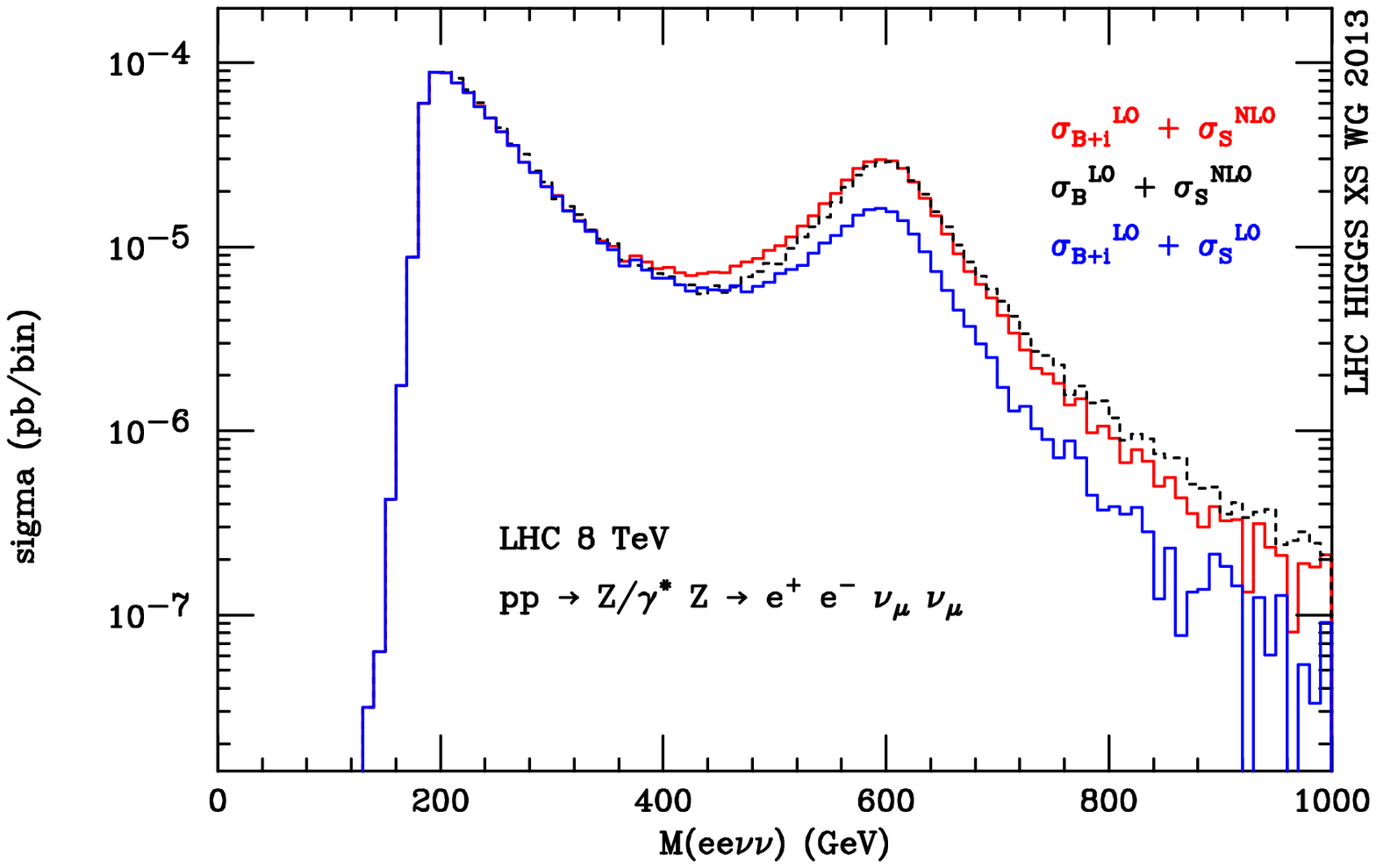}
\includegraphics[width=0.7\textwidth]{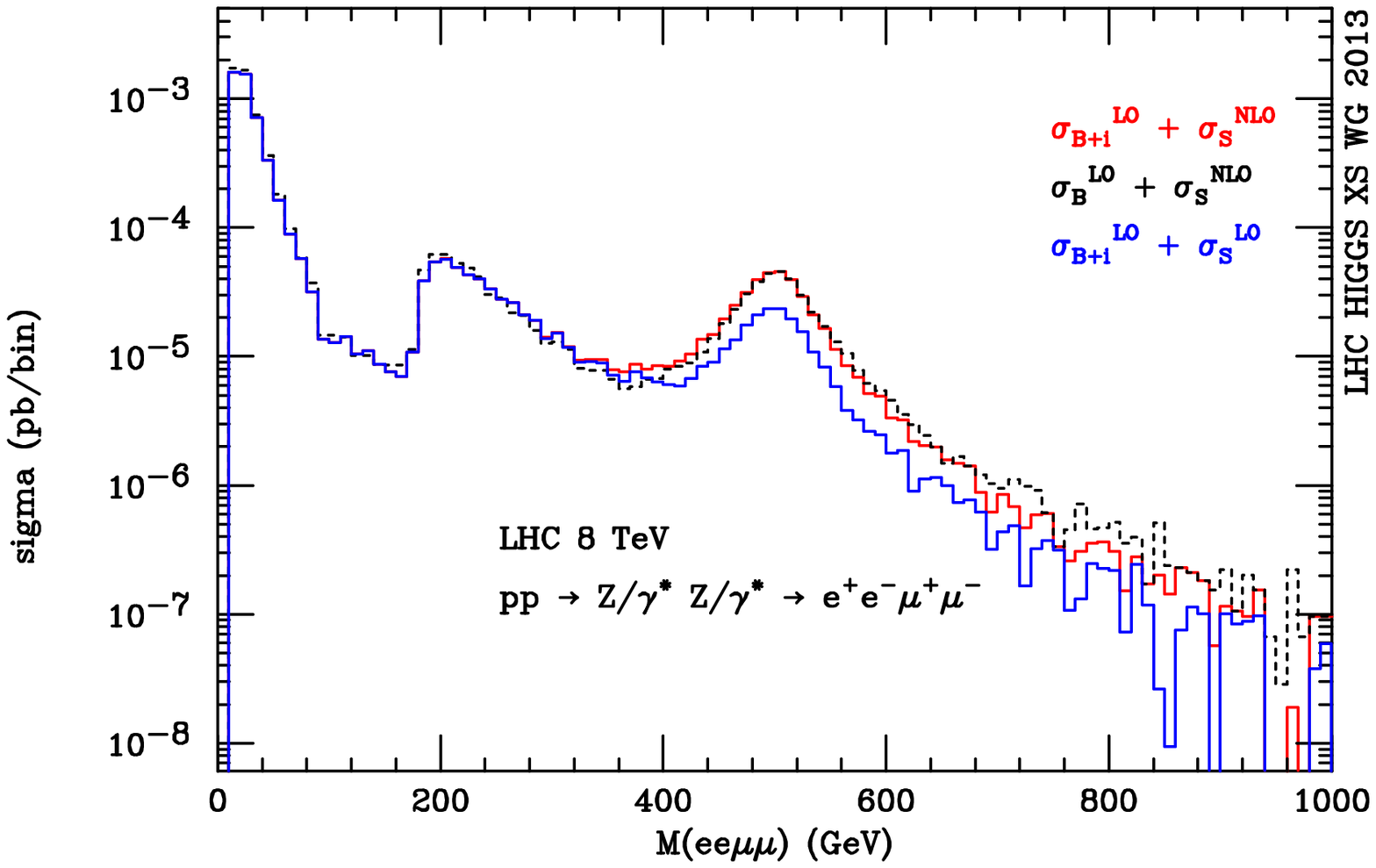}
\caption{Invariant mass distributions for $\Pp\Pp\to 4 \ell$ in different channels as obtained from \textsc{MC@NLO} for the signal $S$ and \textsc{MadLoop} for the background and interference $B+i$. Basic acceptance cuts applied. }
\label{fig:ww.zz.aa}
\end{figure}
%%%%%%%%%%%%%%%%%%%%%%%%%%%%%%%%%%%%%%%%%%%%%%%%%%%%%%%%%%%%%%%%%%%%%%%%%%%%

\clearpage
\subsection{Reweighting studies}
\label{sec:reweighting}

In this section the effect of reweighting distributions to account for the interference effect in several final states
by using different Monte Carlo generators and tools will be presented. 

\subsubsection{Comparison with MCFM: gg->WW->l$\nu$jj}
\label{sec:rewMCFM_lnjj}
%% \documentclass[a4paper]{report}

%% \usepackage{amsmath,graphicx}
%% \usepackage{a4wide}
%% \usepackage{booktabs}
%% \usepackage{lineno}
%% \usepackage{multirow,xspace}
%% \usepackage{rotating}
%% \usepackage{subfigure}
%% \def\lsim{\mathrel{\raisebox{-.6ex}{$\stackrel{\textstyle<}{\sim}$}}}
%% \def\gsim{\mathrel{\raisebox{-.6ex}{$\stackrel{\textstyle>}{\sim}$}}}
%% \textwidth=16cm

%% \begin{document}

%% \section{Heavy Higgs Interference Effects:\\ \textsc{MCFM} study in the $\PW\PW \rightarrow \Pl\PGnl j j$ channel at CMS}
%% \label{sec:HWWlvjjCMSIntf}
% ---- ---- ---- ---- ---- ---- ---- ---- ---- ---- ---- ---- ---- ---- ---- ---- ---- ---- ---- ---- ---- ---- ----

%As discussed in \refS{}{\bf fix me: ref to sec 3.5}, for a light Higgs with
For a light Higgs with
mass near $125\UGeV$, the interference effect is at percent level between the
signal $\Pg\Pg\to \PH \to \PW\PW$ and the continuum background
$\Pg\Pg \to \PW\PW$.  With increasing Higgs mass, the interference effect
becomes more and more crucial. As shown in \Bref{Campbell:2011cu}, the effect
on LO total cross sections can already be over $30\%$ for $\MH\gsim
600GeV$. Moreover, it changes much the $M_{\PW\PW}$ spectrum, with
constructive behavior at $M_{\PW\PW}\lsim \MH$ while destructive at
$M_{\PW\PW} \gsim \MH$. 

One usually can include the interference effects by reweighting the signal's
$M_{\PW\PW}$ spectrum ($S$) with the $M_{\PW\PW}$ binned scale factor $R_2$
%(\eqn{}
%{\bf fix me: ref to eq 24}),
\begin{equation}
S_\mathrm{Reweight} \equiv R_2 \times S.
\label{Srew}
\end{equation}
However, one needs then to validate this method by examining
additional distributions other than $M_{\PW\PW}$, especially those
exploited in the data analysis~\cite{CMS-PAS-HIG-12-046}.
%CMShlvjj}.
Moreover, one should also provide a conservative yet appropriate estimation on theoretical uncertainties from $R_2$. 

In the CMS study of searching Higgs via the
$\PW\PW \rightarrow \Pl\PGnl j j$ channel~\cite{CMS-PAS-HIG-12-046}, 
%CMShlvjj},
we evaluate $R_2$ with the LO results of signal and interference terms got from the \textsc{MCFM} v6.3 (without any cuts applied). We then exploit the method proposed in \Bref{Passarino:2012ri} to estimate the reweighting uncertainty with 3 kinds of factors ($k$) applied on top of $R_2$ by $1+k\times(R_2-1)$, inspired by higher order QCD corrections:
\begin{equation}
k = 1,\,\, \sqrt{K_{\Pg\Pg}}/K_{\NNLO},\,\, {\rm and}\,\, 1/K_{\NNLO},
\label{R2un}
\end{equation}
where the $K\,$-factors $K_{\Pg\Pg}$ and $K_{\NNLO}$ are read from the $\PH\to \PZ\PZ$ numbers in \Bref{Passarino:2012ri}, as the ones for $\PH\to \PW\PW$ are not ready yet and meanwhile the $K\,$-factors should not depend on final states between $\Pg\Pg\to \PH\to \PW\PW$ and $\PZ\PZ$. 

In \Fref{fig:MCFMHHIntfRew}, we show our reweighting results for 
$\Pg\Pg\to \PH \to \PW\PW \to \Pl\PGnl jj$ at the generator level, for $\MH=700\UGeV$ at the $8\UTeV$ LHC, with the factorization and renormalization scales fixed to $\MH$ and CTEQ6L1 PDF exploited, and $\pT^{\Pl}>30\UGeV$, $\pT^j>30\UGeV$, MET$>30\UGeV$, $|\eta_{\Pl,\, j}|<2.4$. The effects of the interference on several crucial kinematic variables are studied, listed as following:
\begin{itemize}
  \item the four body mass $m_{\PW\PW}$,
  \item the two body mass of the hadronic decayed $\PW$, $m_{\PW}$,
  \item the rapidity of the hadronic decayed $\PW$, $\eta_{\PW}$,  
  \item the transverse momentum of the lepton, $\pT^{\Pl}$,
  \item the minimal azimuthal opening angle between the lepton and jets, $\phi_{\Pl j}$,
  \item the azimuthal opening angle between the jets, $\phi_{jj}$.
\end{itemize}  

The blue curves in \Fref{fig:MCFMHHIntfRew} are for signal process, the red ones are for the signal plus continuum background, while the yellow bands correspond to the reweighted signal uncertainties).
As one can see from \Fref{fig:MCFMHHIntfRew}, the reweighting scheme mentioned above based on $M_{\PW\PW}$ spectrum, turns out to describe well other distributions: the reweigthed curves differ a lot from the Signal curves while lie mostly inside the the interference uncertainty bands.

\begin{figure}[htb]
    \subfigure[$M_{\PW\PW}$]{
      \includegraphics[width=0.42\textwidth]{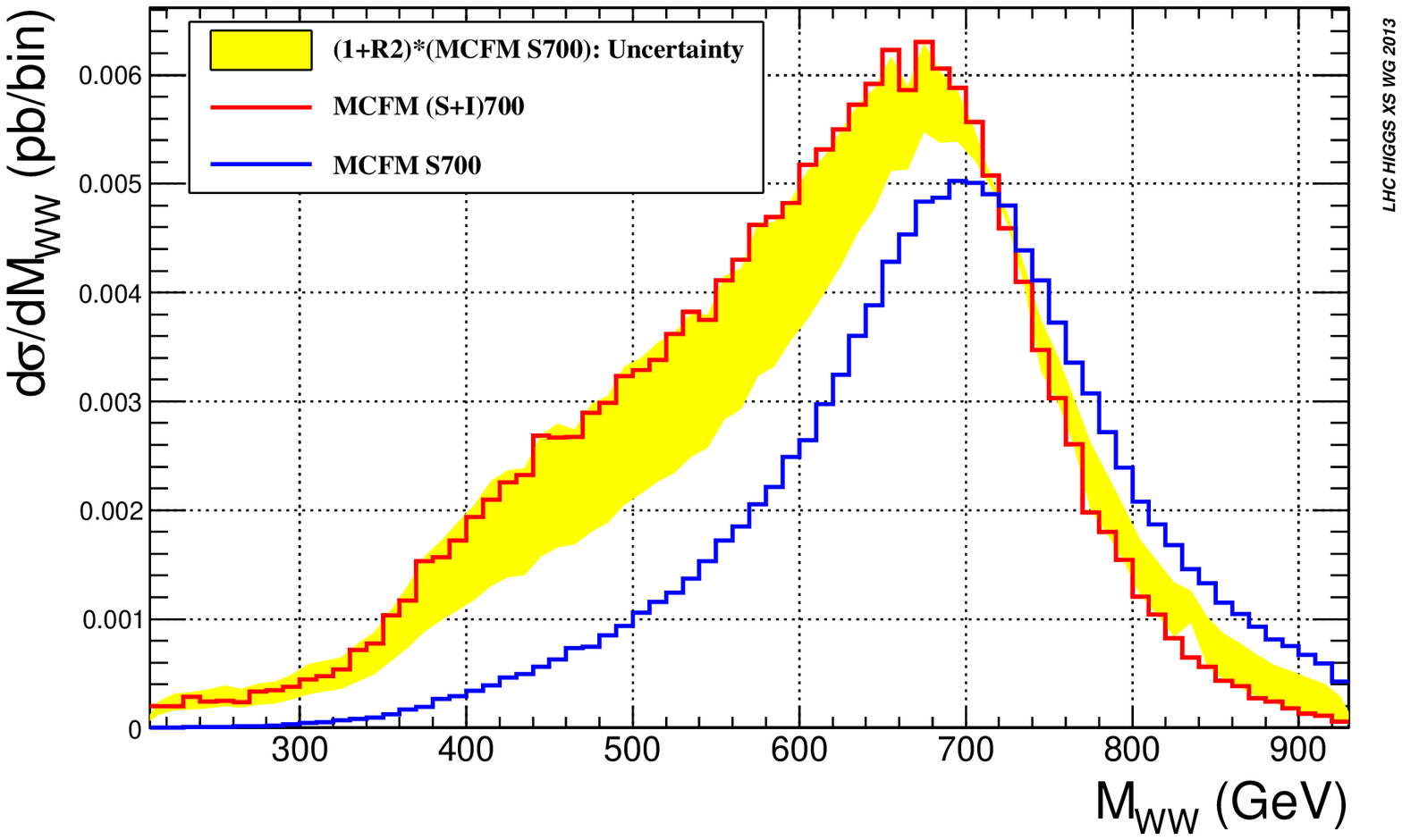}
    }
    \subfigure[$\MW$]{
      \includegraphics[width=0.42\textwidth]{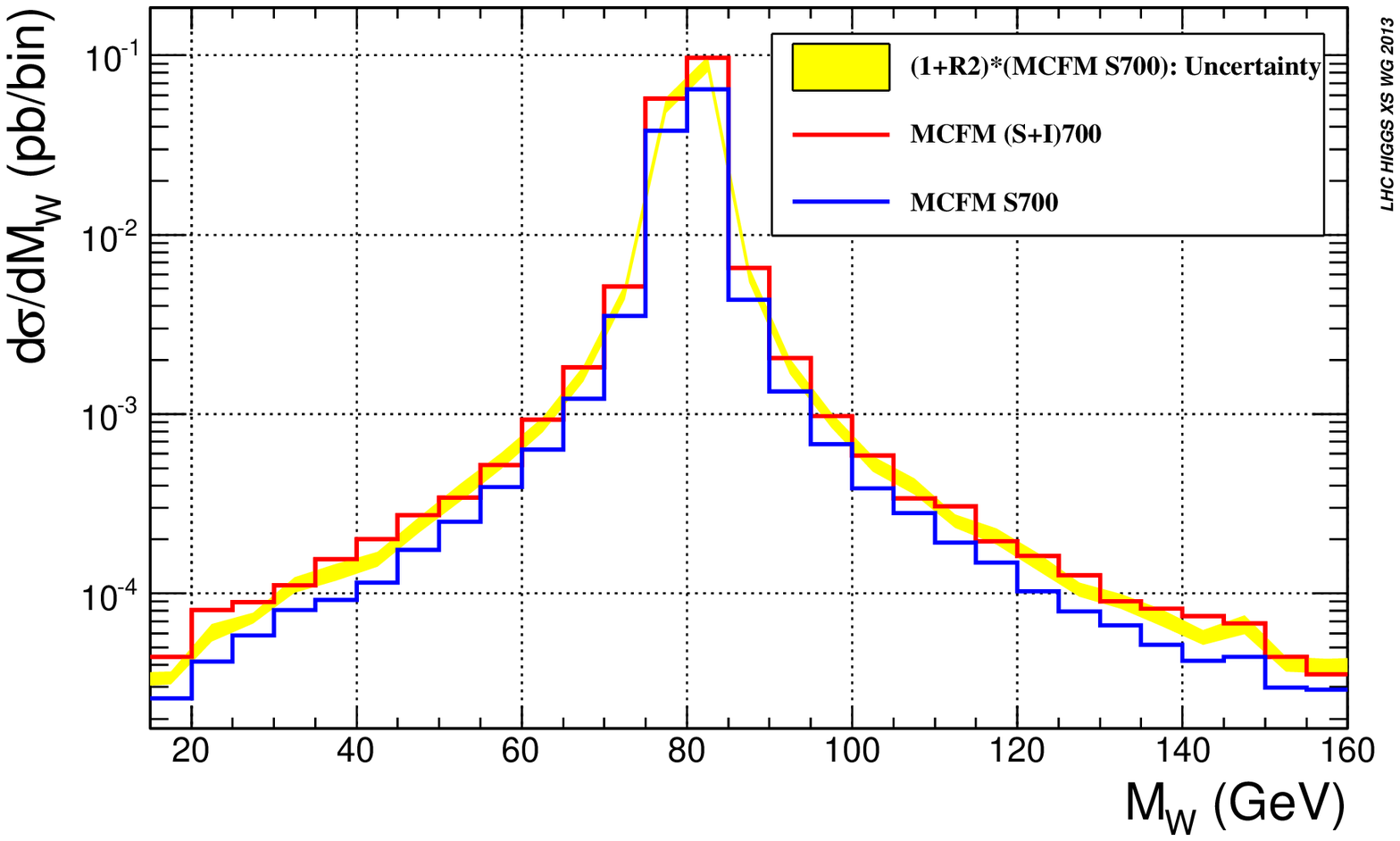}
    }
    \\
    \subfigure[$\eta_{\PW}$]{
      \includegraphics[width=0.42\textwidth]{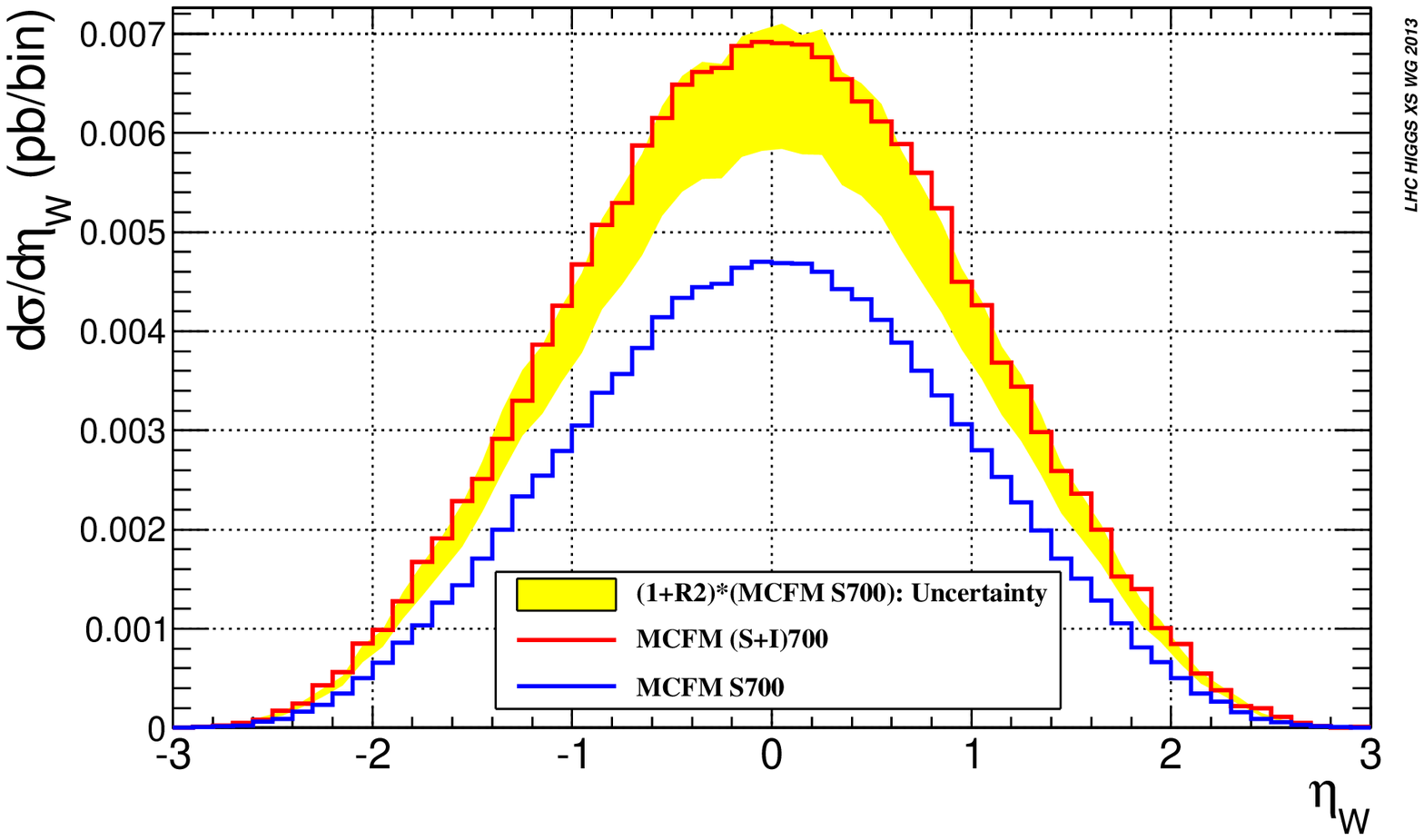}
    }
    \subfigure[$PT_{\Pl}$]{
      \includegraphics[width=0.42\textwidth]{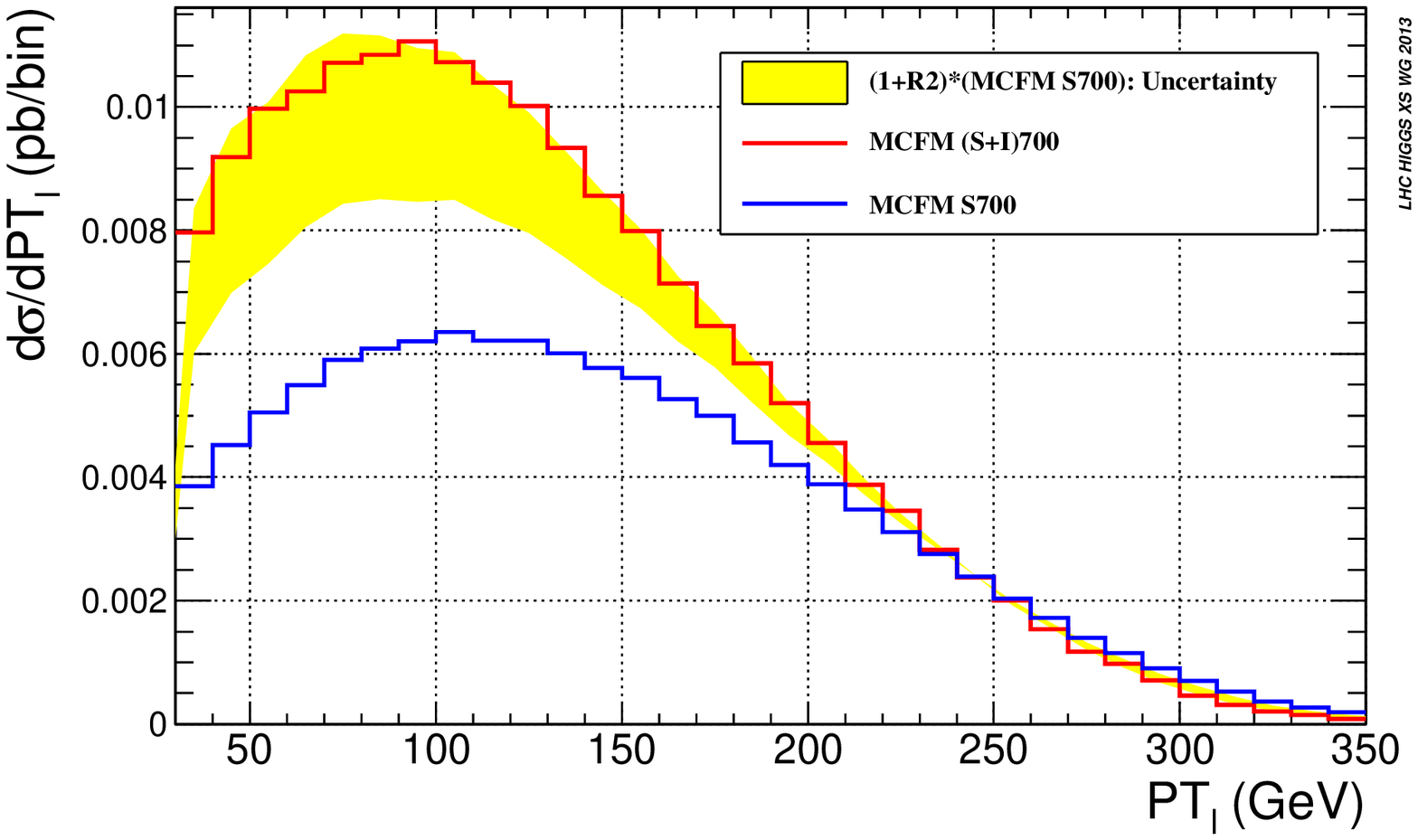}
    }
    \\
    \subfigure[$\phi_{\Pl j}$]{
      \includegraphics[width=0.42\textwidth]{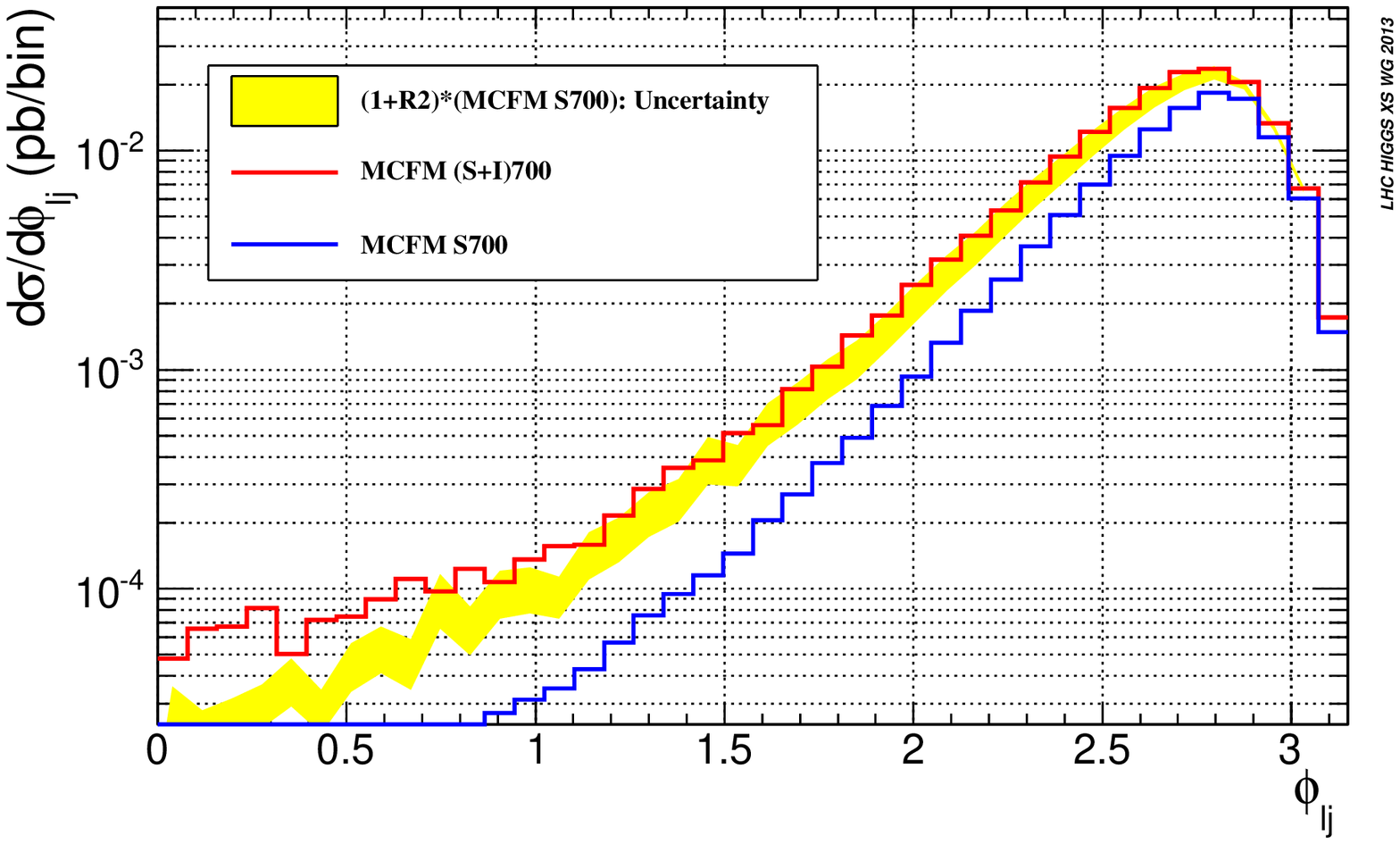}
    }
    \subfigure[$\phi_{jj}$]{
      \includegraphics[width=0.42\textwidth]{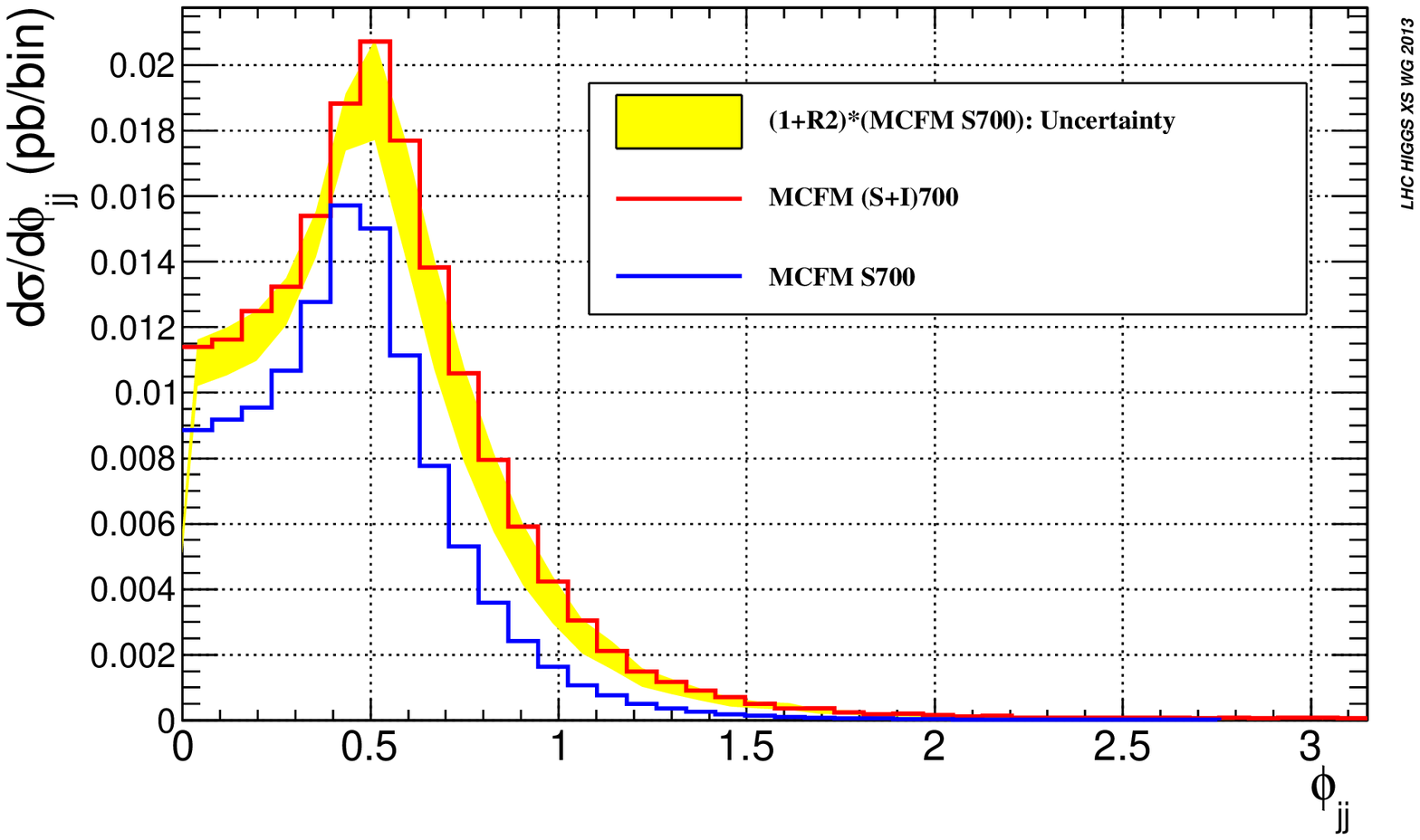}
    }
    \caption{Reweighting \textsc{MCFM} $\Pg\Pg\to \PH \to \PW\PW \to \Pl\PGnl jj$ results to take into account interference effects at the generator level, for $\MH=700\UGeV$ at the $8\UTeV$ LHC, with $\pT^{\Pl}>30\UGeV$, $\pT^j>30\UGeV$, MET$>30\UGeV$, and $|\eta_{\Pl,j}|<2.4$. Blue curves are for signal process, the red ones are for the signal plus continuum background, while the yellow bands correspond to the reweighted signal uncertainties \eqnss{Srew} and~(\ref{R2un}).}
    \label{fig:MCFMHHIntfRew}
\end{figure}
%

%% \begin{thebibliography}{00}

%% %\cite{Campbell:2011cu}
%% \bibitem{Campbell:2011cu} 
%%   J.~M.~Campbell, R.~K.~Ellis and C.~Williams,
%%   %``Gluon-Gluon Contributions to W+ W- Production and Higgs Interference Effects,''
%%   JHEP {\bf 1110}, 005 (2011)
%%   [arXiv:1107.5569 [hep-ph]].
%%   %%CITATION = ARXIV:1107.5569;%%

%% \bibitem{CMShlvjj} 
%% Search for the Standard Model Higgs boson in the H to WW to lnujj decay channel in pp collisions at the LHC, CMS-PAS-HIG-12-046.

%% %\cite{Passarino:2012ri}
%% \bibitem{Passarino:2012ri} 
%%   G.~Passarino,
%%   %``Higgs Interference Effects in $\Pg \Pg \to \PZ\PZ$ and their Uncertainty,''
%%   JHEP {\bf 1208}, 146 (2012)
%%   [arXiv:1206.3824 [hep-ph]].
%%   %%CITATION = ARXIV:1206.3824;%%

%% \end{thebibliography}

%% \end{document}

\subsubsection{Comparison with MCFM: gg->WW->l$\nu$l$\nu$}
\label{sec:rewMCFM_lnln}
For the $\PW\PW \rightarrow \Pl\PGnl \Pl\PGnl$ channel (with $l=e,\,\mu$), the effect of the interference between 
$\Pg\Pg \rightarrow \PH \rightarrow \PW\PW$ and continuum $\Pg\Pg \rightarrow \PW\PW$ production becomes increasingly important 
as the mass of the resonance increases. 

\Fref{fig:HWW_SMWW_int} shows the LO Feynman diagrams for the two processes.

\begin{figure}[tbp]
  \centering
  \includegraphics[width=0.70\textwidth]{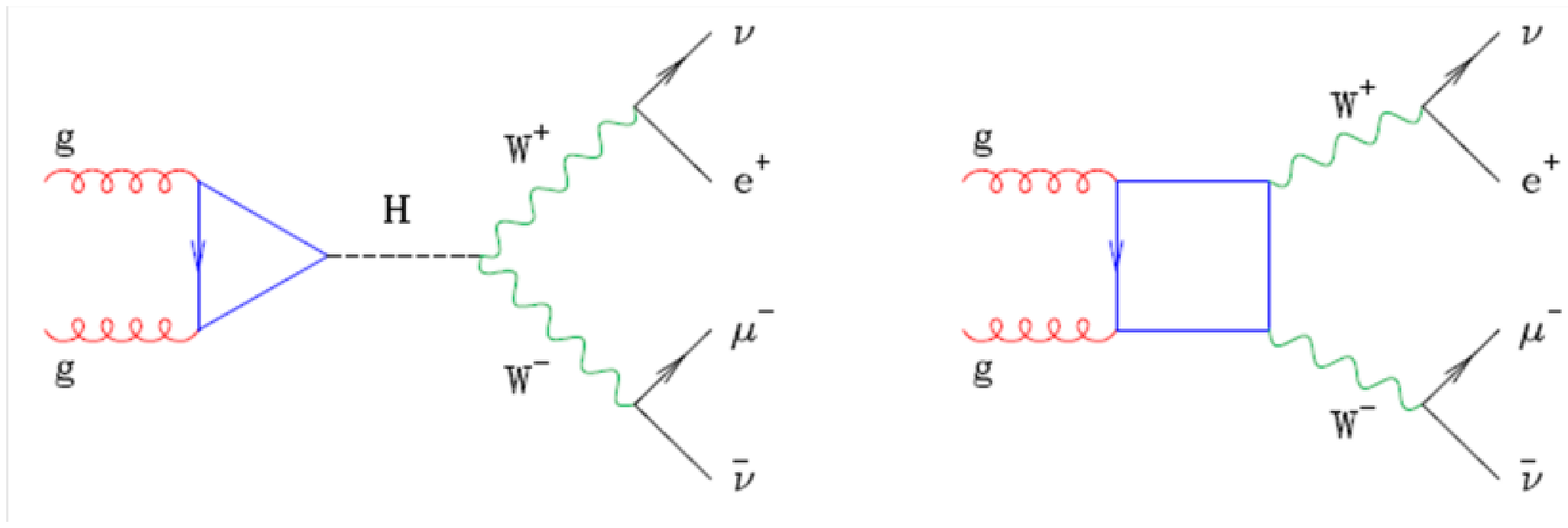}
  %\end{center}
  \vspace*{-2mm}
  \caption{Diagrams for the $\Pg\Pg \rightarrow \PH \rightarrow \PW\PW \rightarrow \Pl\PGnl \Pl\PGnl$ process (left) and the continuum $\Pg\Pg \rightarrow \PW\PW \rightarrow \Pl\PGnl \Pl\PGnl$ process (right).}
  \label{fig:HWW_SMWW_int}
\end{figure}

In ATLAS, a study of the effect of the interference on key kinematic variables has been performed using the
\textsc{MCFM} \cite{MCFMweb} Monte Carlo program. The study uses \textsc{MCFM 6.2}, the processes being generated
at a centre-of-mass (CM) energy of $8\UTeV$, with no cuts on the generation. Events 
MCFM processes 121 and 122 are respectively used to generate 
distributions for the pure $\PH \rightarrow \PW\PW \rightarrow \Pl\PGnl \Pl\PGnl$ process and the process including the
effect of the interference. Both processes are calculated to leading order accuracy in QCD. This requires a reweighting procedure to account for higher-order corrections. This procedure has been described in section \ref{HMcpss}.
Distributions are generated for six different resonance masses $\MH: 400, 500, 600, 700, 800$ and $900\UGeV$. The renormalization and factorization scales are set to the mass of the resonance being generated.

The object selection criteria used in this study are similar to those used in the ATLAS $\PH \rightarrow \PW\PW \rightarrow \Pl\PGnl \Pl\PGnl$
analysis \cite{Aad:2012uub}. The leptons are required to be within $|\eta| < 2.5$. The leading lepton must have $\pT > 25\UGeV$,
and the sub-leading lepton $\pT > 15\UGeV$. The missing transverse momentum is required to be larger than $45\UGeV$. No requirement is made on the number of jets in the final state nor on the types of the two leptons.

The interference effect affects the shape of the distributions used in the analysis.
The effects of such reweighting are studied for four crucial kinematic variables:

\begin{itemize}
  \item the dilepton invariant mass $M_{\ell\ell}$
  \item the transverse momentum of the dilepton system $\pT^{\ell\ell}$
  \item the dilepton azimuthal opening angle $\Delta\phi_{\ell\ell}$
  \item the transverse mass $M_\mathrm{T}$
  
\end{itemize}

The first three variables are important because the analysis uses them to reject background and define a signal region.
The fourth variable, $M_\mathrm{T}$, is the final discriminant~\cite{Aad:2012uub}.\\

Figures \ref{fig:mll_reweight_cut_1} - \ref{fig:Mt_reweight_cut_1} show distributions of the variables for pure $\PH \rightarrow \PW\PW$, for
the process including the interference and for the pure signal process reweighted according to the lineshape distribution (i.e. the invariant mass of the two Ws). For each variable, distributions are shown for three of the six resonance masses corresponding to 400, 600 and 900 GeV. The ratio between the signal distributions including the interference and the reweighted distributions is also plotted. The cross-section $\sigma_{\PH_i}$ including the interference effect is not strictly physical, and can be negative (see Eq. 4.1 in \cite{Campbell:2011cu}), as seen in many of the distributions. From these plots it is clear that the interference has a large impact on both the cross-section and the shape of the distributions, particularly for higher Higgs masses. The plots also show that reweighting by $M_{\PW\PW}$ correctly reproduces the effect of interference on the shapes of other distributions.

\begin{figure}[htb!]
  \centering
  
  \includegraphics[width=0.32\textwidth]{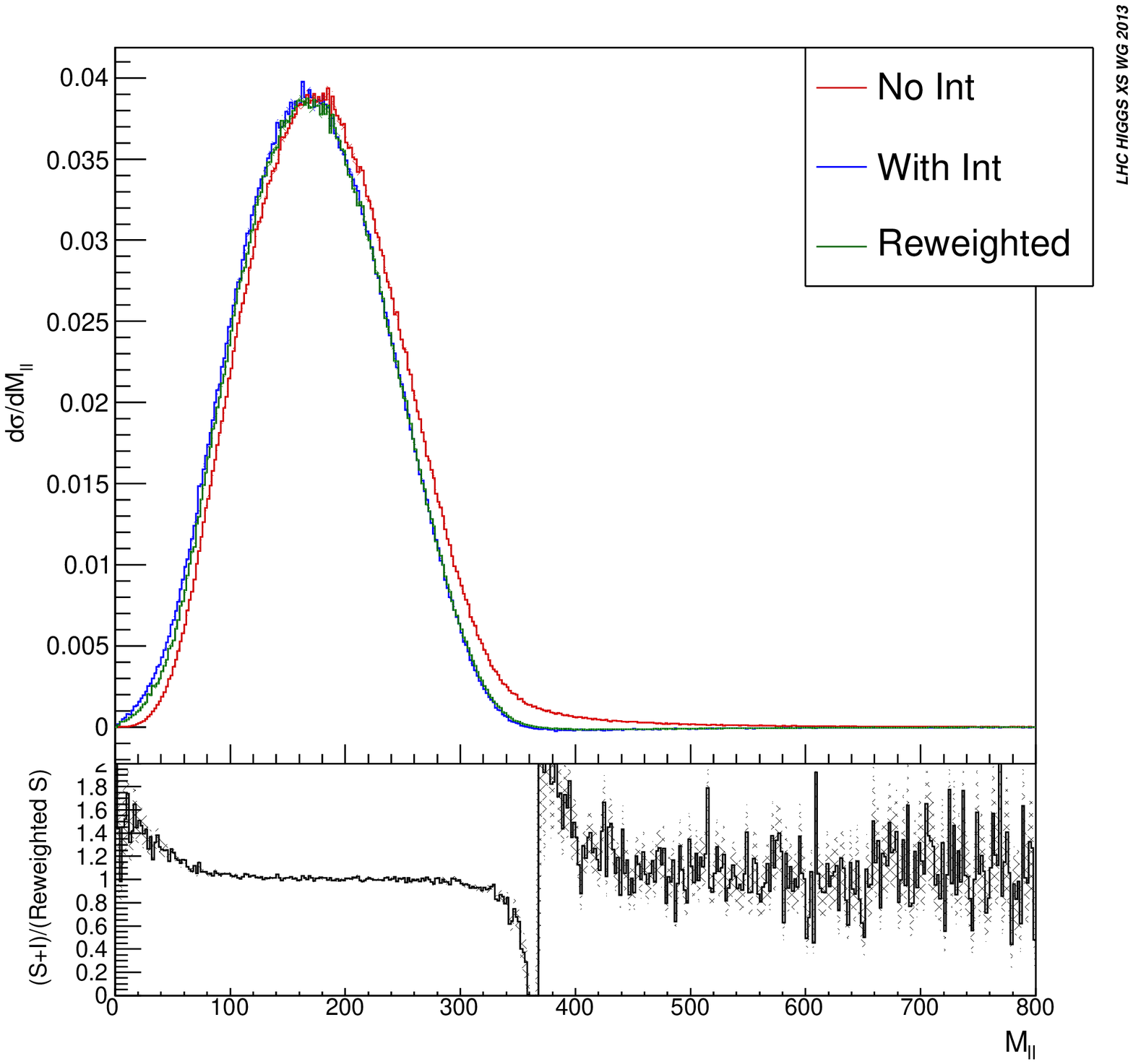}
  \includegraphics[width=0.32\textwidth]{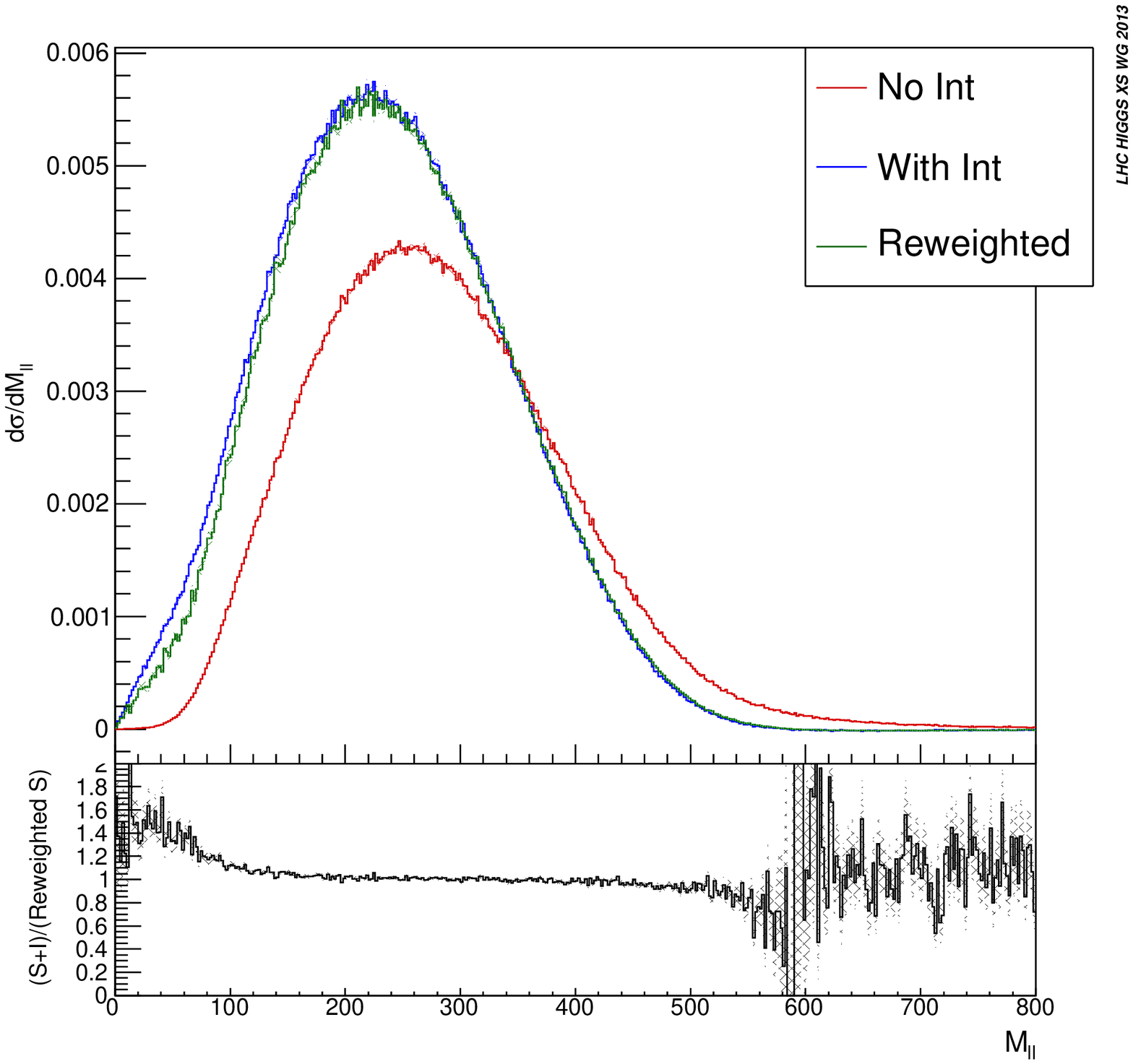} 
  \includegraphics[width=0.32\textwidth]{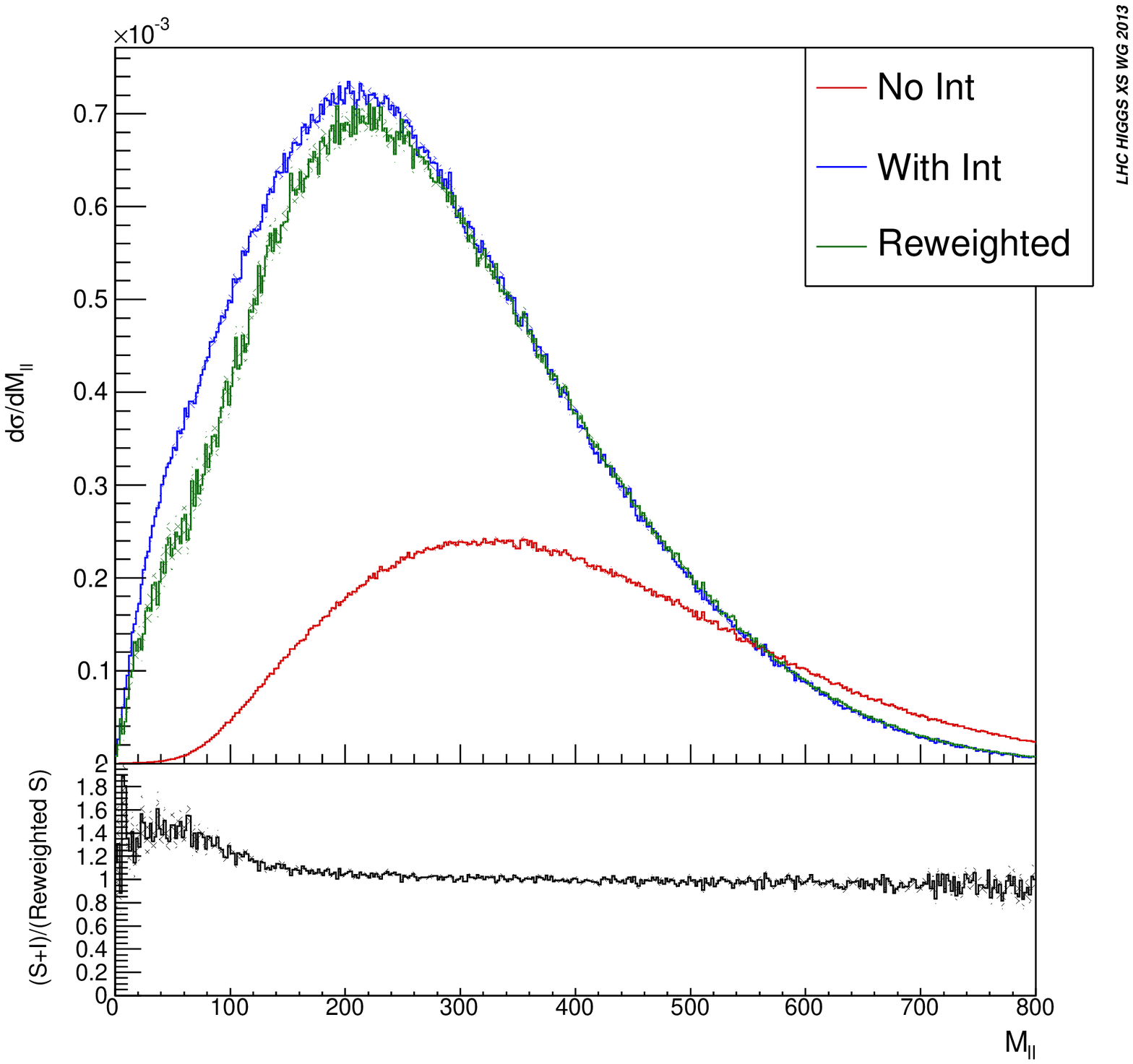}

  %\end{center}
  \vspace*{-2mm}
  \caption{Distributions of the dilepton invariant mass $M_{\ell\ell}$ showing the effect of the interference and the result of the reweighting procedure after cuts, corresponding to a Higgs mass of $\MH = 400\UGeV$ (left), $600\UGeV$ (middle), $900\UGeV$ (right).}
  \label{fig:mll_reweight_cut_1}
\end{figure}

\begin{figure}[htb!]
  \centering
  
  \includegraphics[width=0.32\textwidth]{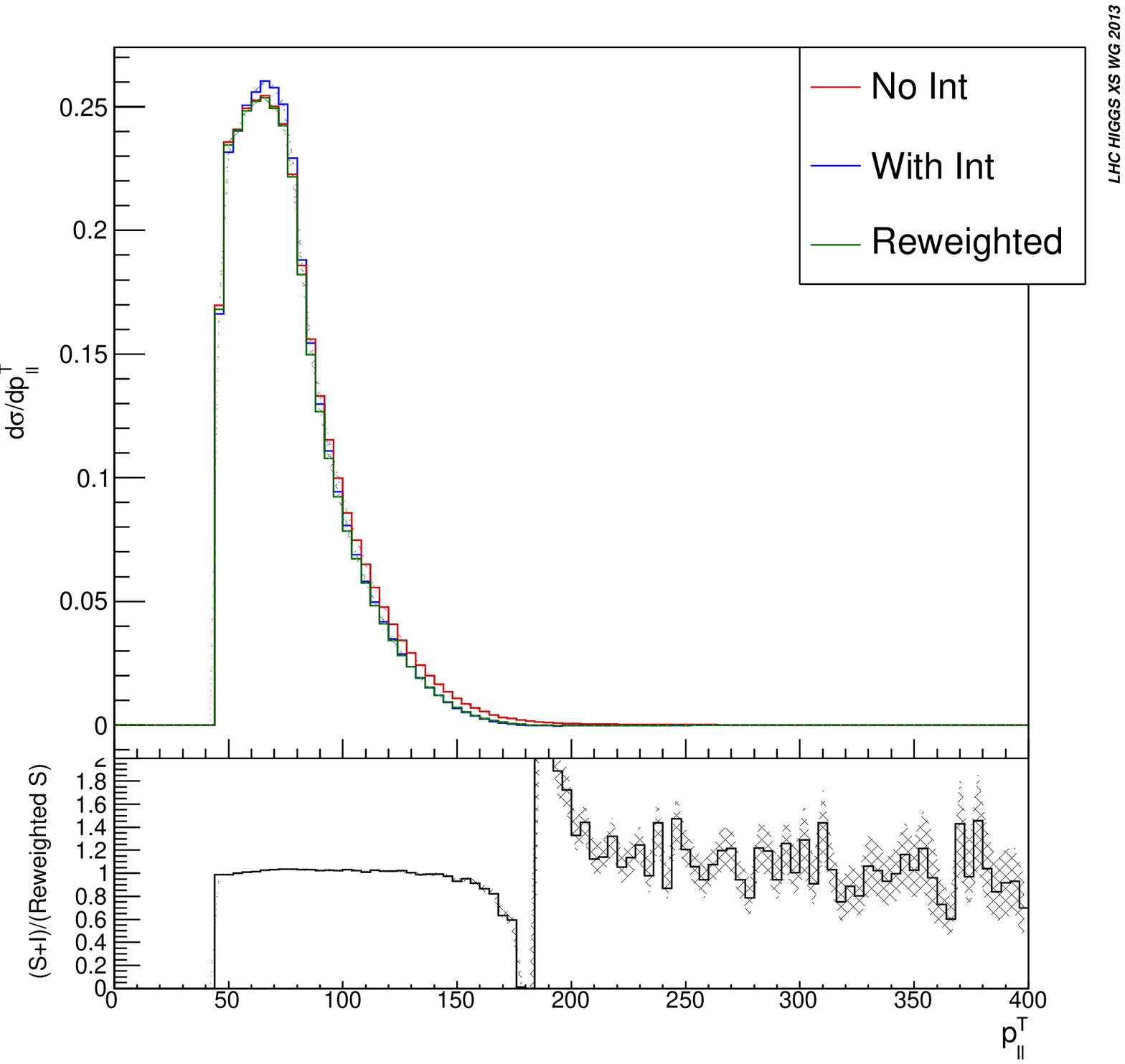}
  \includegraphics[width=0.32\textwidth]{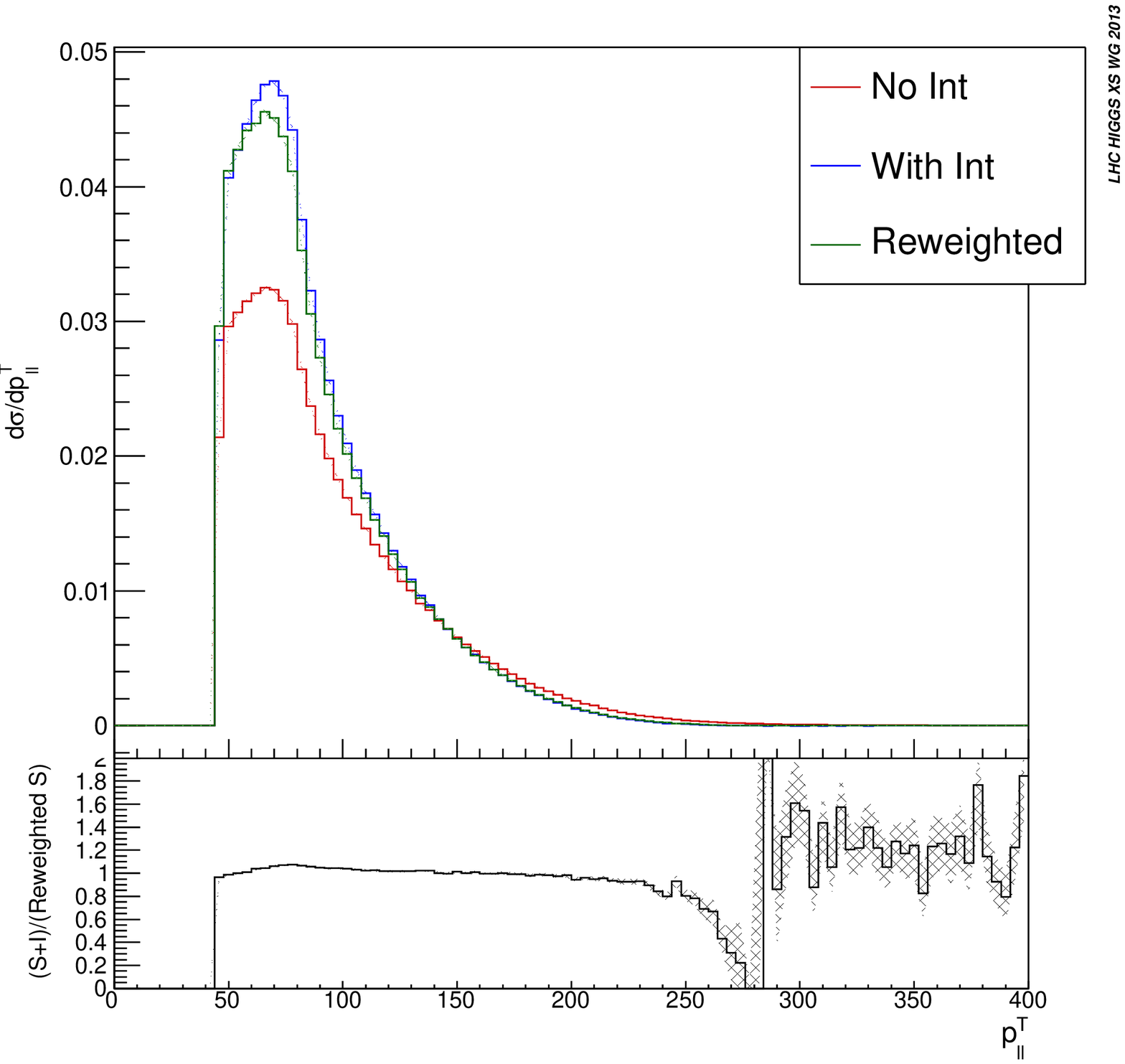} 
  \includegraphics[width=0.32\textwidth]{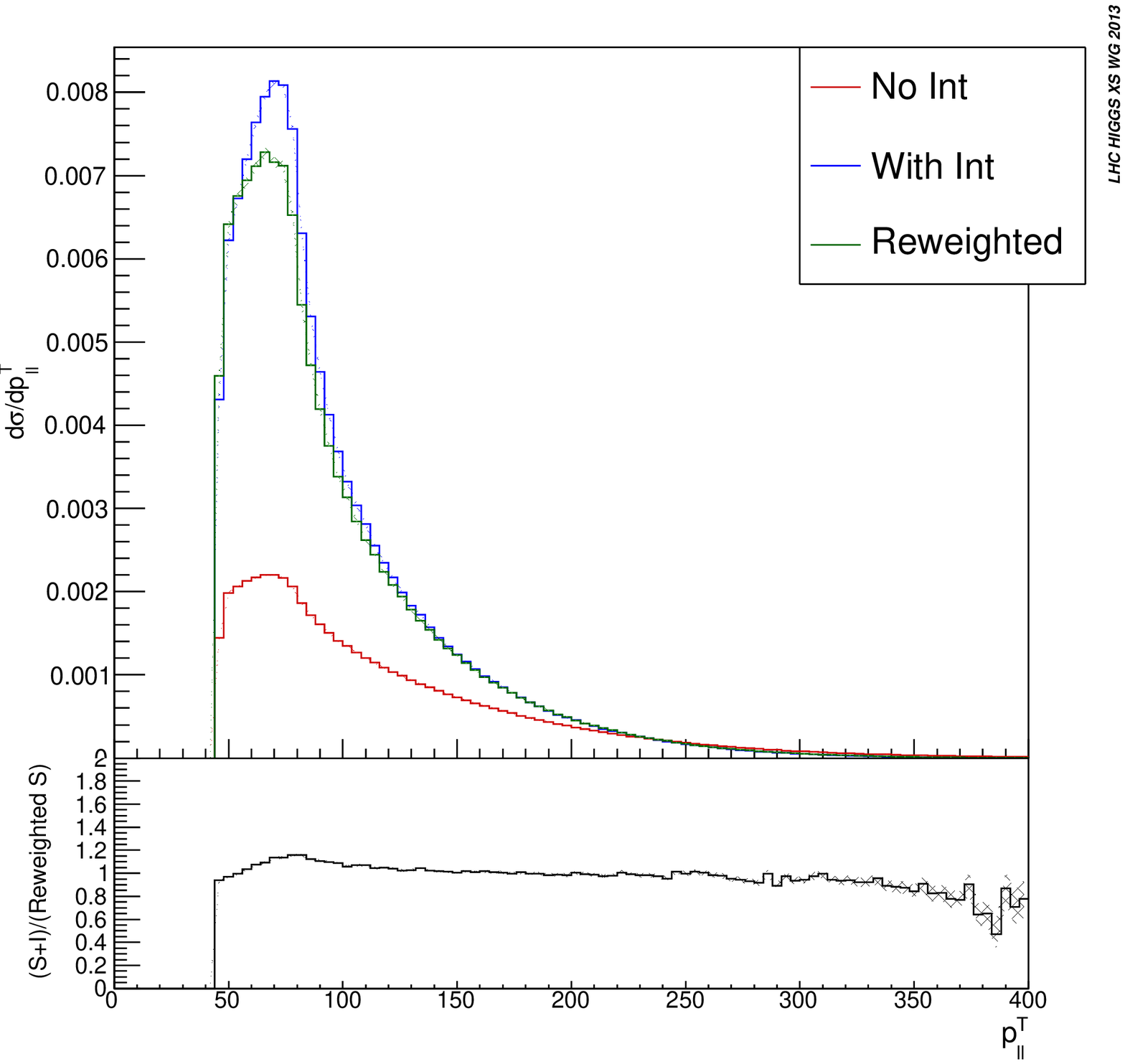}

  %\end{center}
  \vspace*{-2mm}
  \caption{Distributions of the transverse momentum of the dilepton system $\pT^{\ell\ell}$ showing the effect of the interference and the result of the reweighting procedure after cuts, corresponding to a Higgs mass of $\MH = 400\UGeV$ (left), $600\UGeV$ (middle), $900\UGeV$ (right).}
  \label{fig:ptll_reweight_cut_1}
\end{figure} 	

\begin{figure}[htb!]
  \centering
  
  \includegraphics[width=0.32\textwidth]{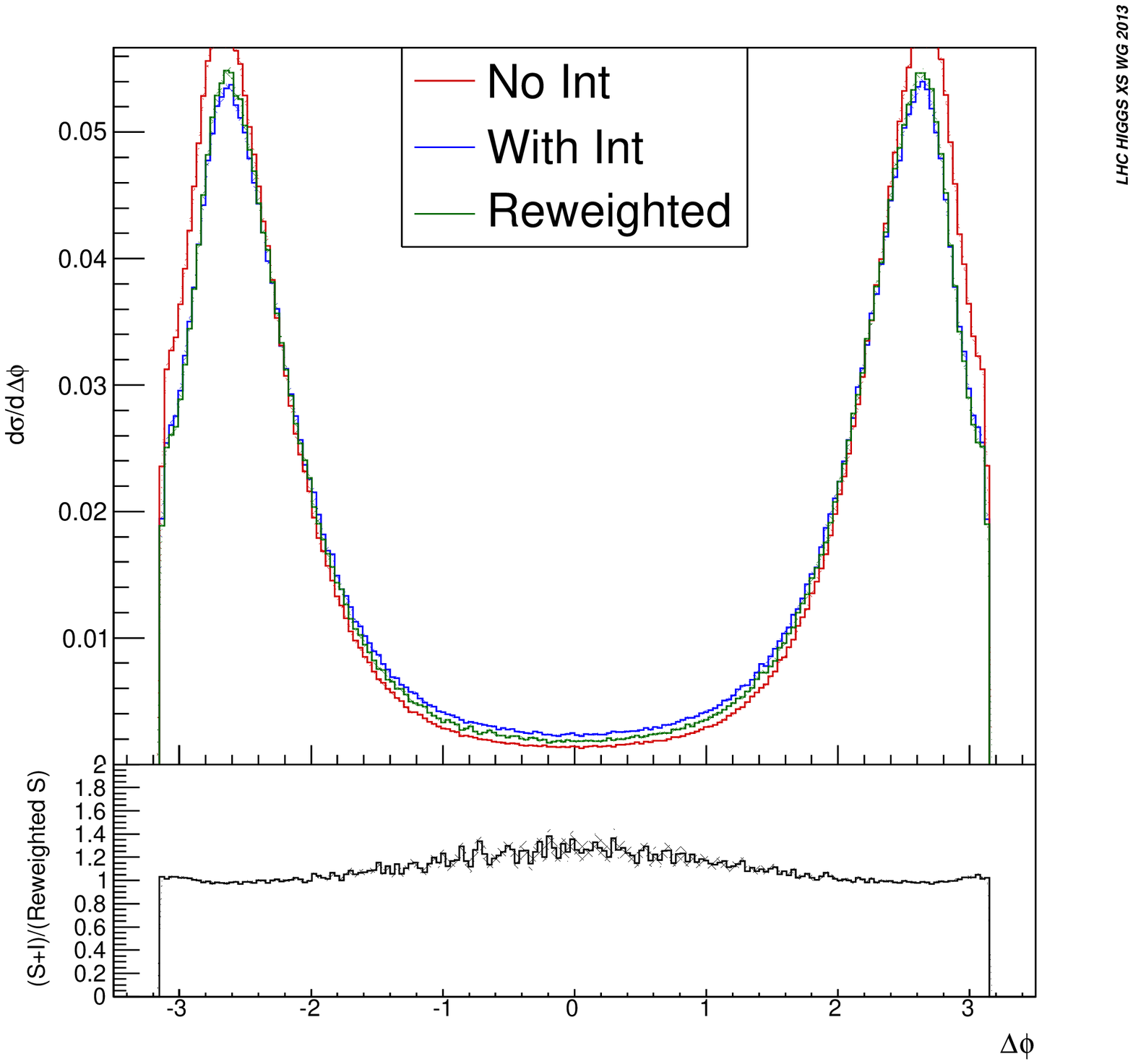}
  \includegraphics[width=0.32\textwidth]{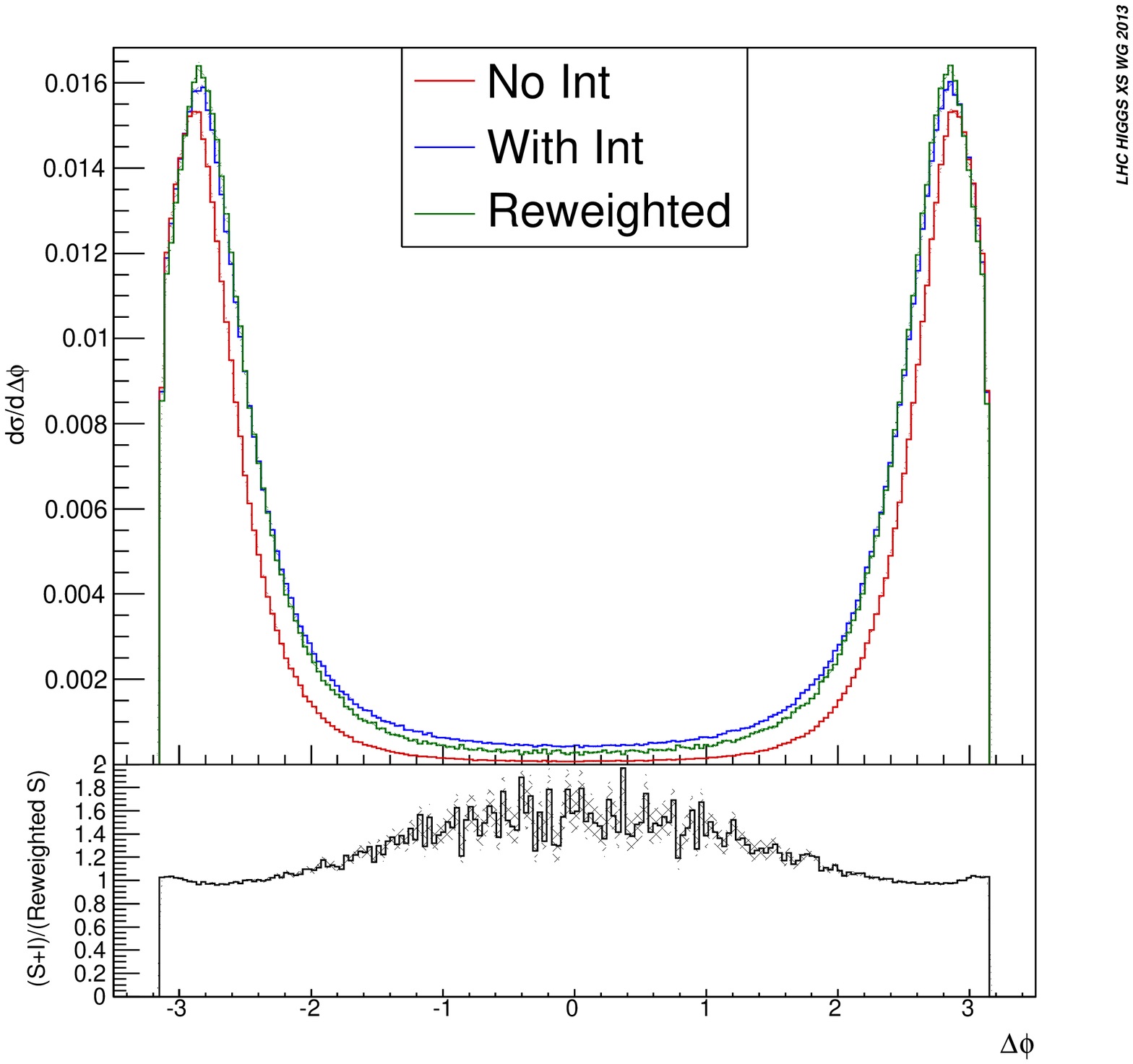} 
  \includegraphics[width=0.32\textwidth]{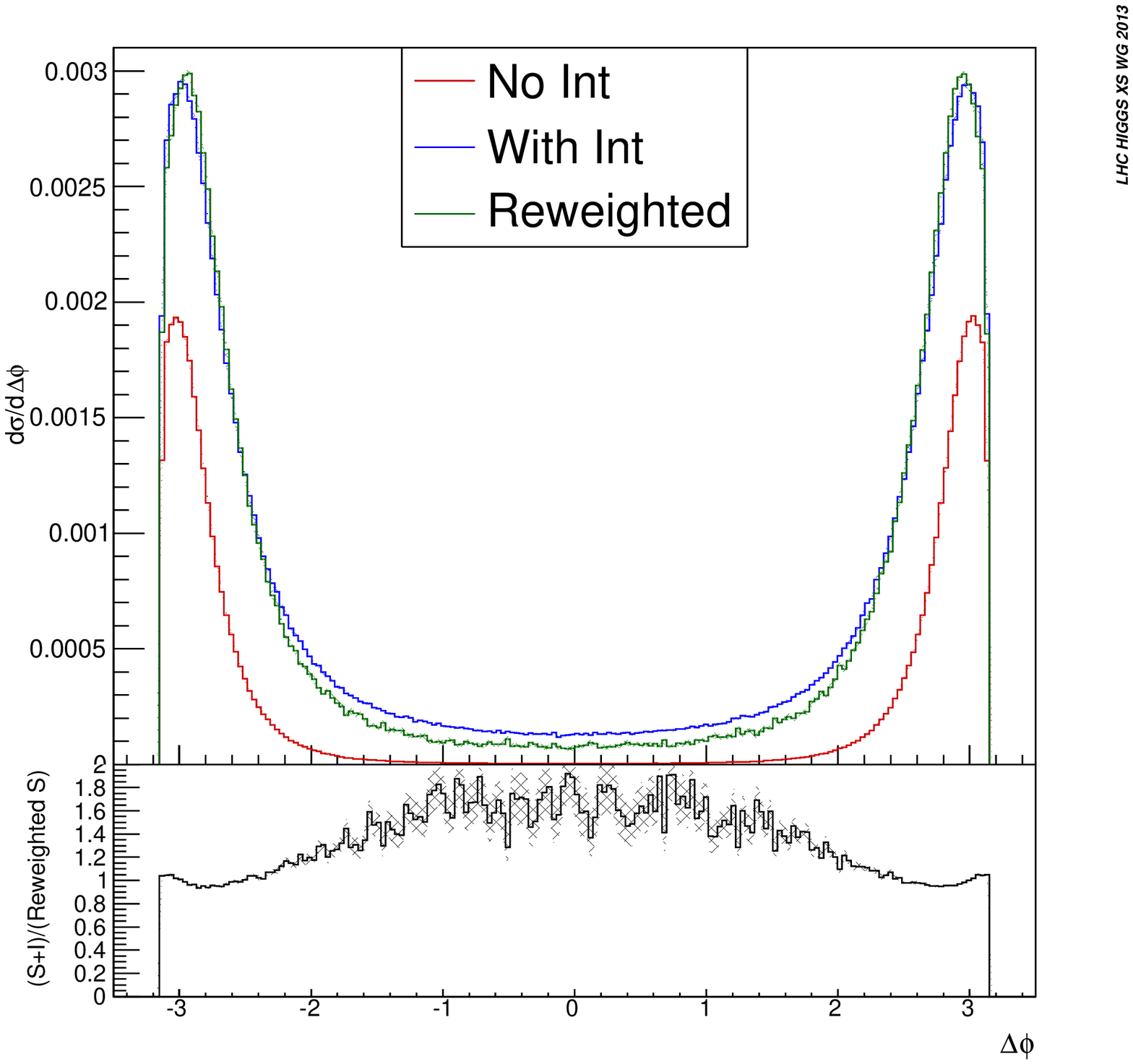}

  %\end{center}
  \vspace*{-2mm}
  \caption{Distributions of the dilepton azimuthal opening angle $\Delta\phi_{\ell\ell}$ showing the effect of the interference and the result of the reweighting procedure after cuts, corresponding to a Higgs mass of $\MH = 400\UGeV$ (left), $600\UGeV$ (middle), $900\UGeV$ (right).}
  \label{fig:DeltaPhi_ll_reweight_cut_1}
\end{figure}

\begin{figure}[htb!]
  \centering
  
  \includegraphics[width=0.32\textwidth]{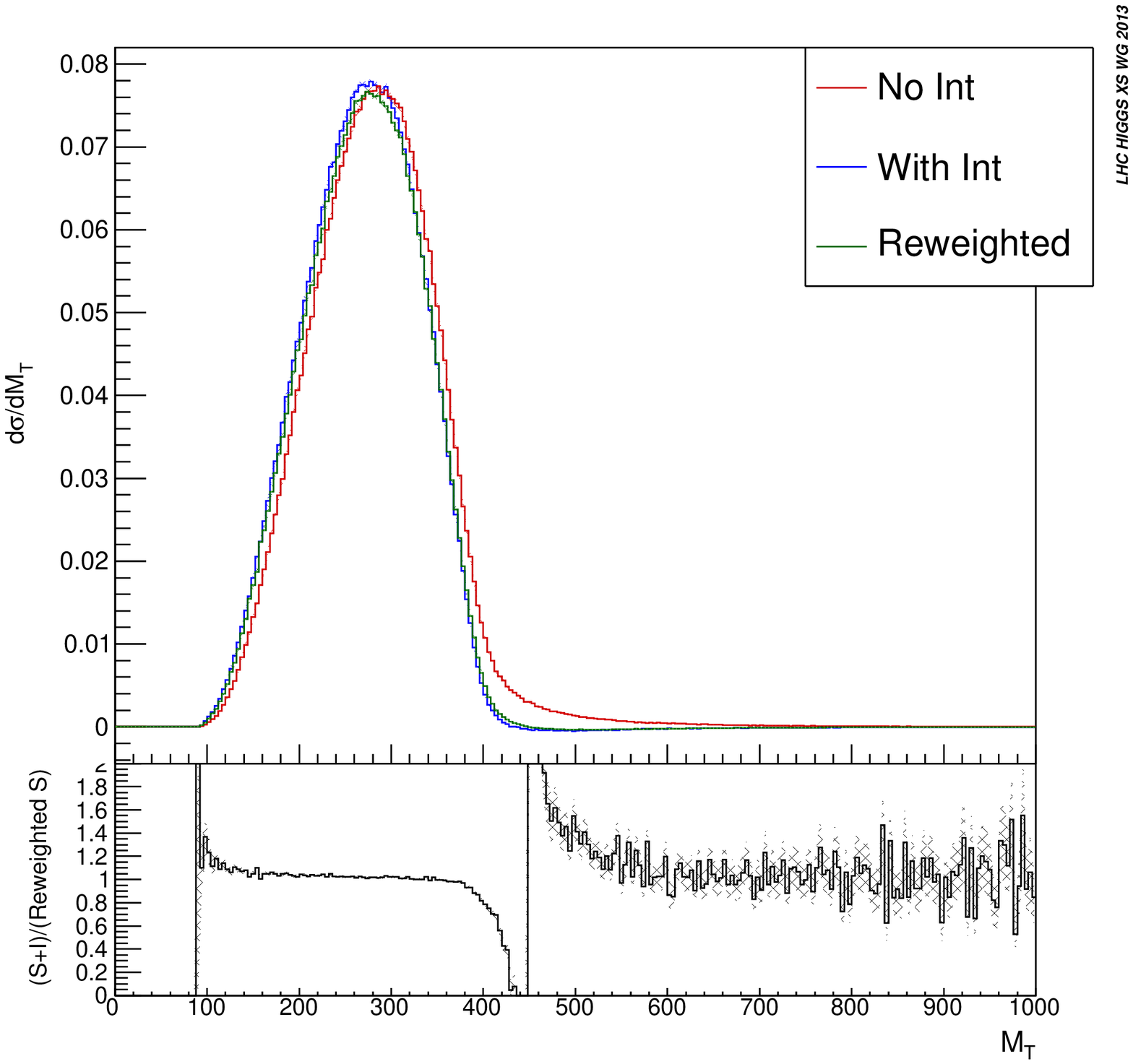}
  \includegraphics[width=0.32\textwidth]{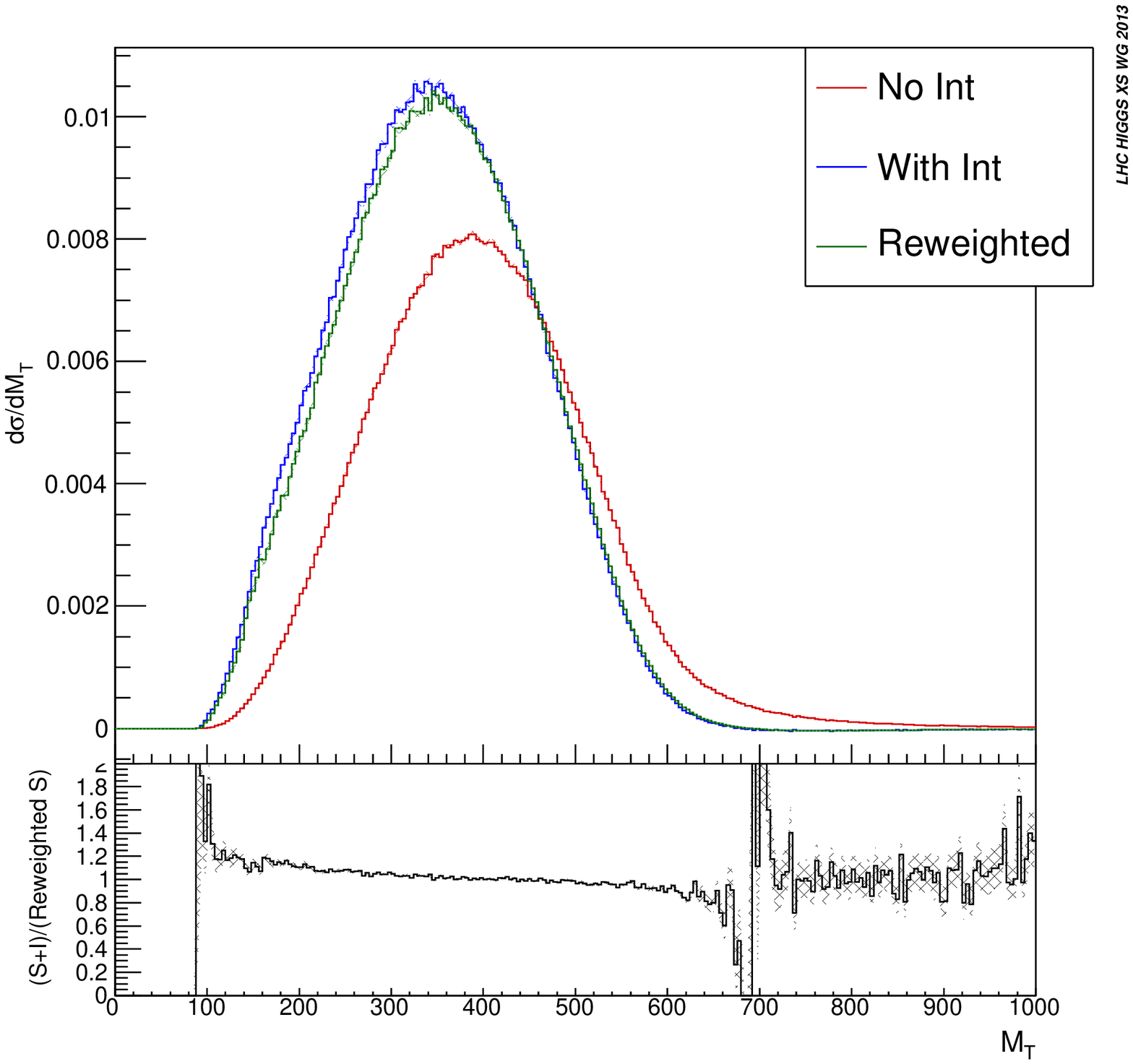} 
  \includegraphics[width=0.32\textwidth]{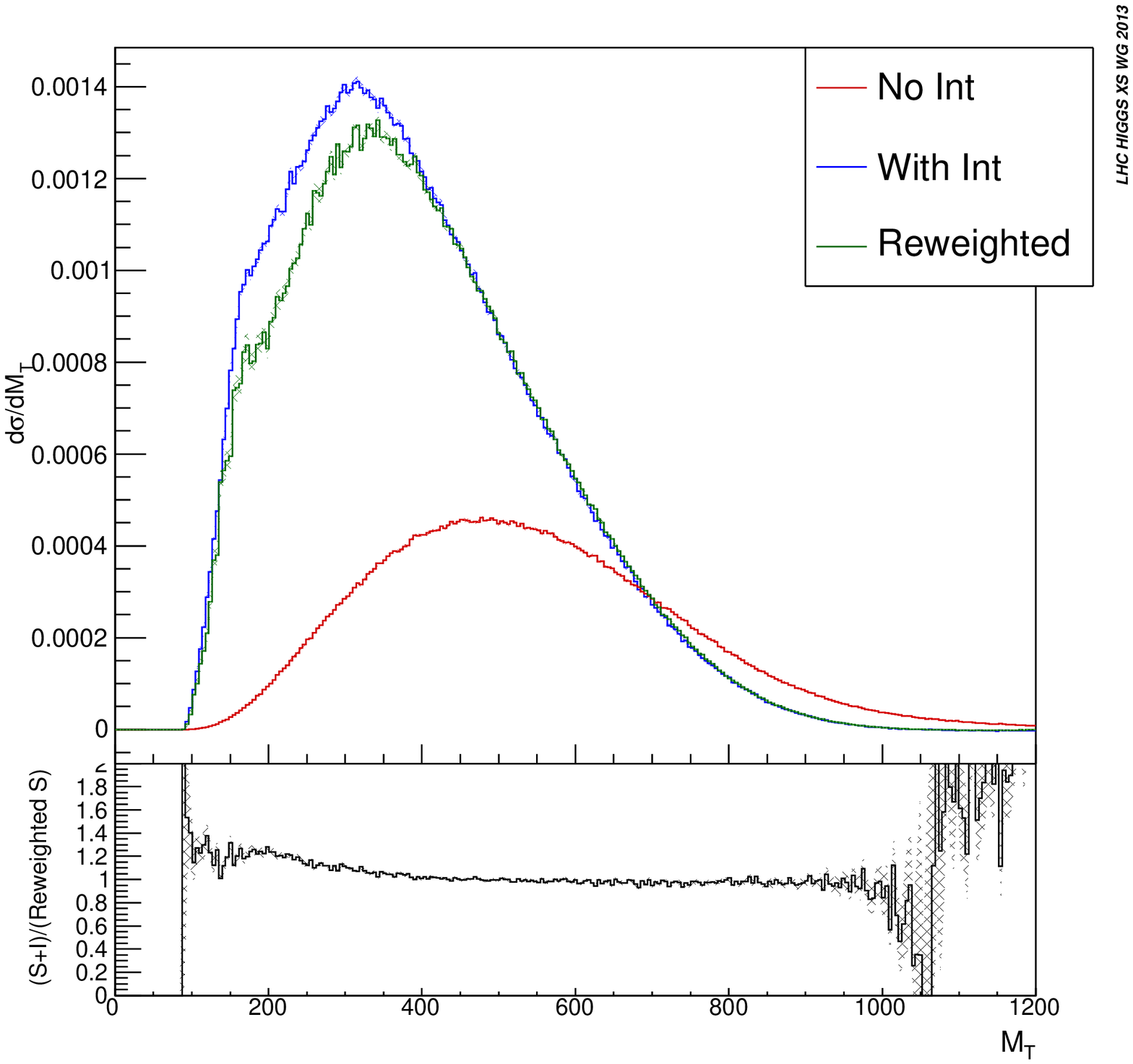}

  %\end{center}
  \vspace*{-2mm}
  \caption{Distributions of the transverse mass $M_\mathrm{T}$ showing the effect of the interference and the result of the reweighting procedure after cuts, corresponding to a Higgs mass of $\MH = 400\UGeV$ (left), $600\UGeV$ (middle), $900\UGeV$ (right).}
  \label{fig:Mt_reweight_cut_1}
\end{figure}

The same distributions have been obtained before applying any cuts and it has been proven that the shape of the distributions is not affected by the cuts. The difference between the efficiency calculated on samples which include the interference at generation level and on signal reweighted samples, has been used to evaluate the systematic uncertainty associated to the reweighting procedure as reported in table \ref{tab:eff}.

\begin{table}
\caption{Efficiency of the cuts (fraction of cross section remaining) for the samples generated with the interference effect included and the reweighted samples.}
\label{tab:eff}
\centering
\begin{tabular}{lcccccc}
% top line of headings - label masses
\hline
\multicolumn{1}{l}{} & \multicolumn{2}{c}{$\MH$ = 400 GeV} & \multicolumn{2}{c}{$\MH$ = 600 GeV} & \multicolumn{2}{c}{$\MH$ = 900 GeV} \\
% 2nd line of headings - S+I and Reweighted labels
\multicolumn{1}{l}{} & \multicolumn{1}{c}{S+I} & \multicolumn{1}{c}{Reweighted} & \multicolumn{1}{c}{S+I} & \multicolumn{1}{c}{Reweighted} & \multicolumn{1}{c}{S+I} & \multicolumn{1}{c}{Reweighted} \\ \hline 
%first chunk of table contents
\multicolumn{1}{l}{After Preselection Cuts} & \multicolumn{1}{l}{0.844945} & \multicolumn{1}{l}{0.846989} & \multicolumn{1}{l}{0.876930} & \multicolumn{1}{l}{0.879919} & \multicolumn{1}{l}{0.876166} & \multicolumn{1}{l}{0.881239} \\ 
\multicolumn{1}{l}{After Kinematic Cuts} & \multicolumn{1}{l}{0.624508} & \multicolumn{1}{l}{0.619817} & \multicolumn{1}{l}{0.640973} & \multicolumn{1}{l}{0.630267} & \multicolumn{1}{l}{0.661190} & \multicolumn{1}{l}{0.634992}\\ \hline
\end{tabular}

%%\end{center}
\end{table}

From this it is clear that the overall effect of the discrepancies between the reweighting and the process including the
effect of the interference distributions is a few \%. The exact magnitude of the uncertainty depends upon which cuts are applied, and which mass point is being examined - here, it ranges from $1\%$ for the $\MH = 400\UGeV$ case to 5\% for the $\MH = 900\UGeV$ case. 

Having looked at the effect of reweighting at LO, there remains the issue of reweight for higher order effects. This study reweights to NNLO, using the 
schemes suggested in \cite{Passarino:2012ri}, and the correspondent K factors \cite{CPHTO} as discussed in section \ref{HMbis}: 

\begin{itemize}
\item Additive: K\,S + I
\item Multiplicative: K\,S + K\,I
\item Intermediate: K\,S + sqrt($K_{\Pg\Pg}$)\,I
\end{itemize}

Figures \ref{fig:mll_NNLO_cut_1} - \ref{fig:Mt_NNLO_cut_1} show these three schemes -- together with the LO Signal for comparison -- for each of the 4 distributions shown earlier, after the aforementioned cuts have been applied. The Additive (in yellow) and the Multiplicative (in magenta) schemes form bounds for the theory uncertainty in the scaling to NNLO, while the Intermediate scheme (in black) is the nominal NNLO Signal plus Interference distribution.

\begin{figure}[tbp]
  \centering
  
  \includegraphics[width=0.32\textwidth]{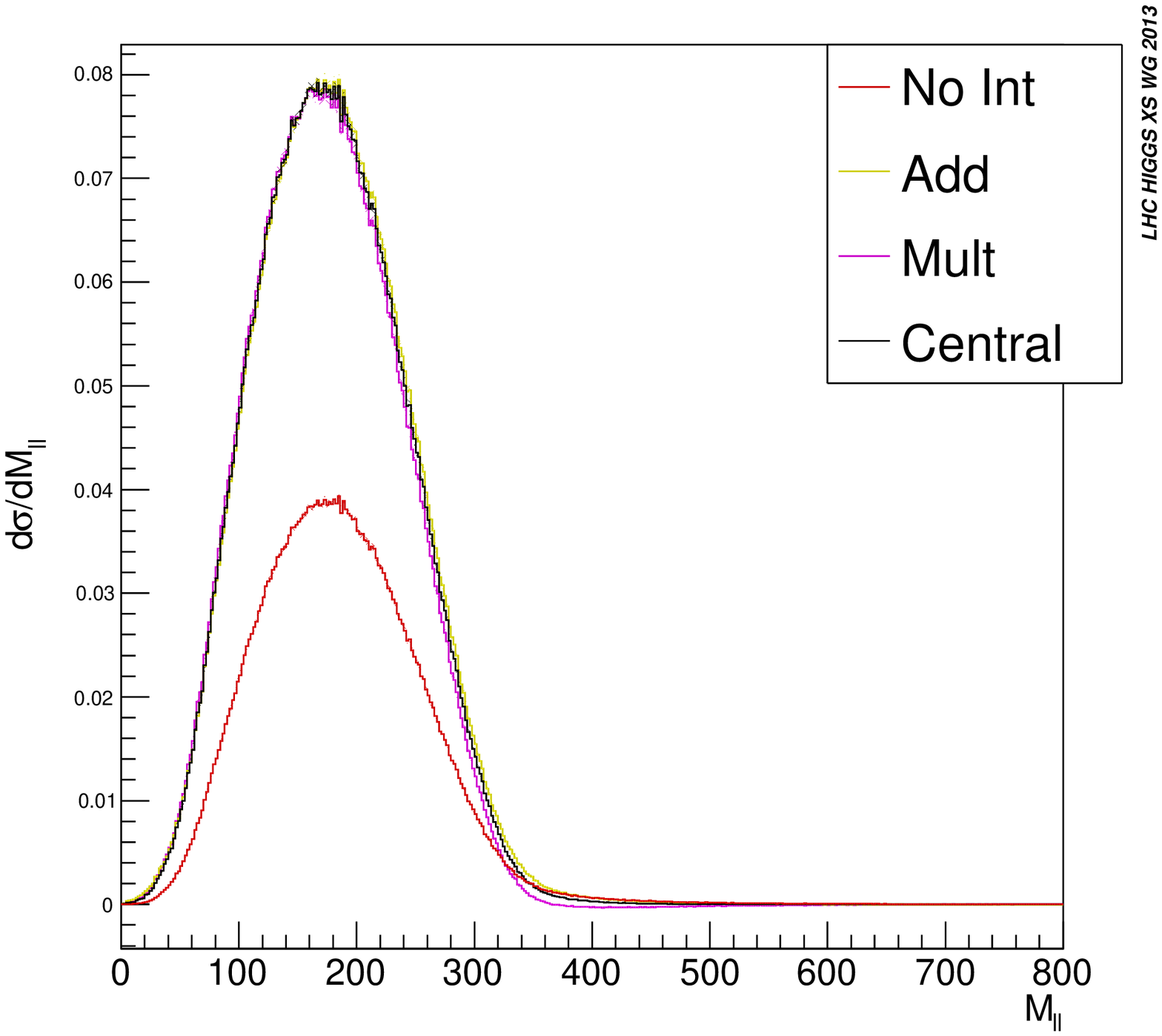}
  \includegraphics[width=0.32\textwidth]{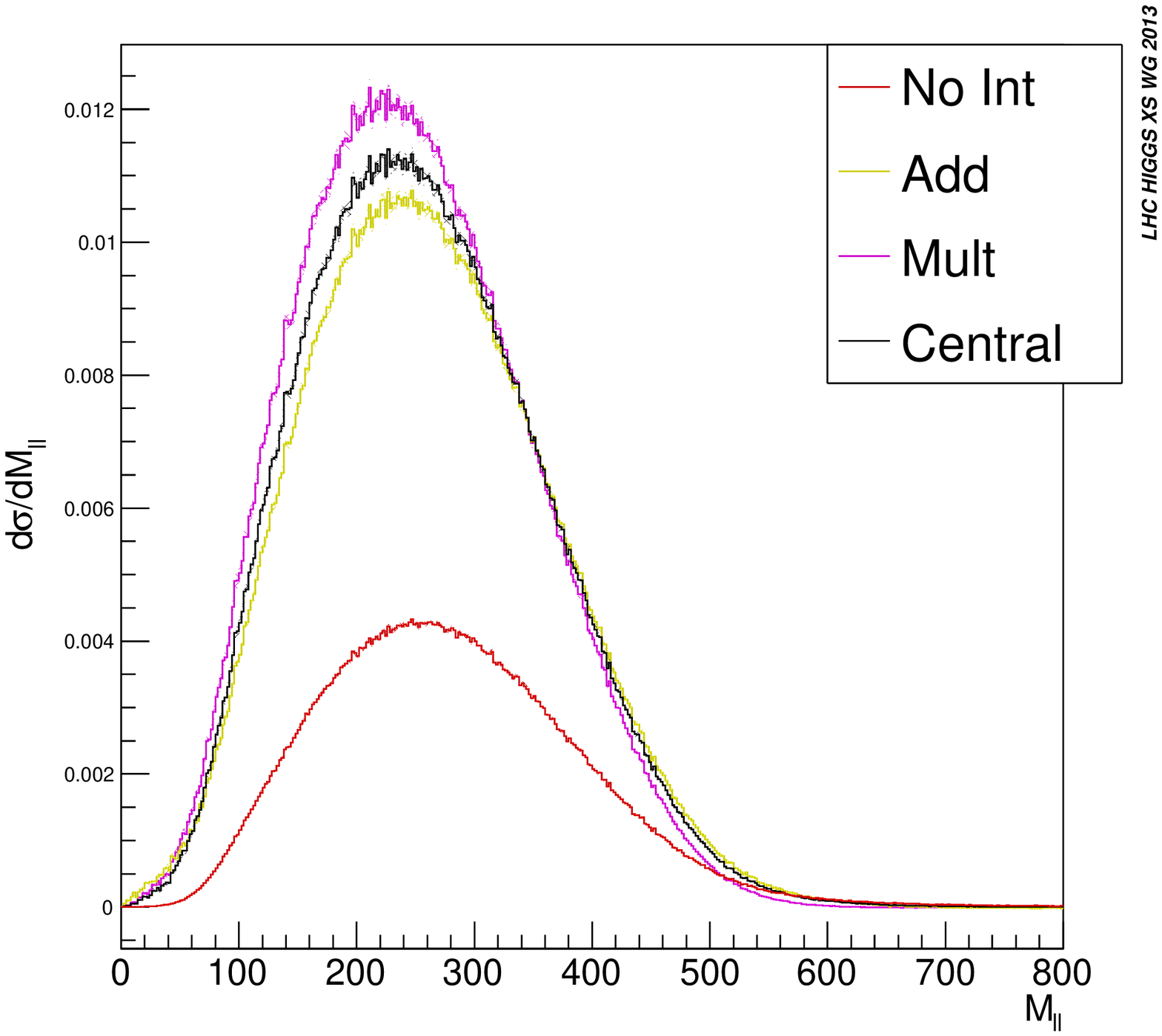} 
  \includegraphics[width=0.32\textwidth]{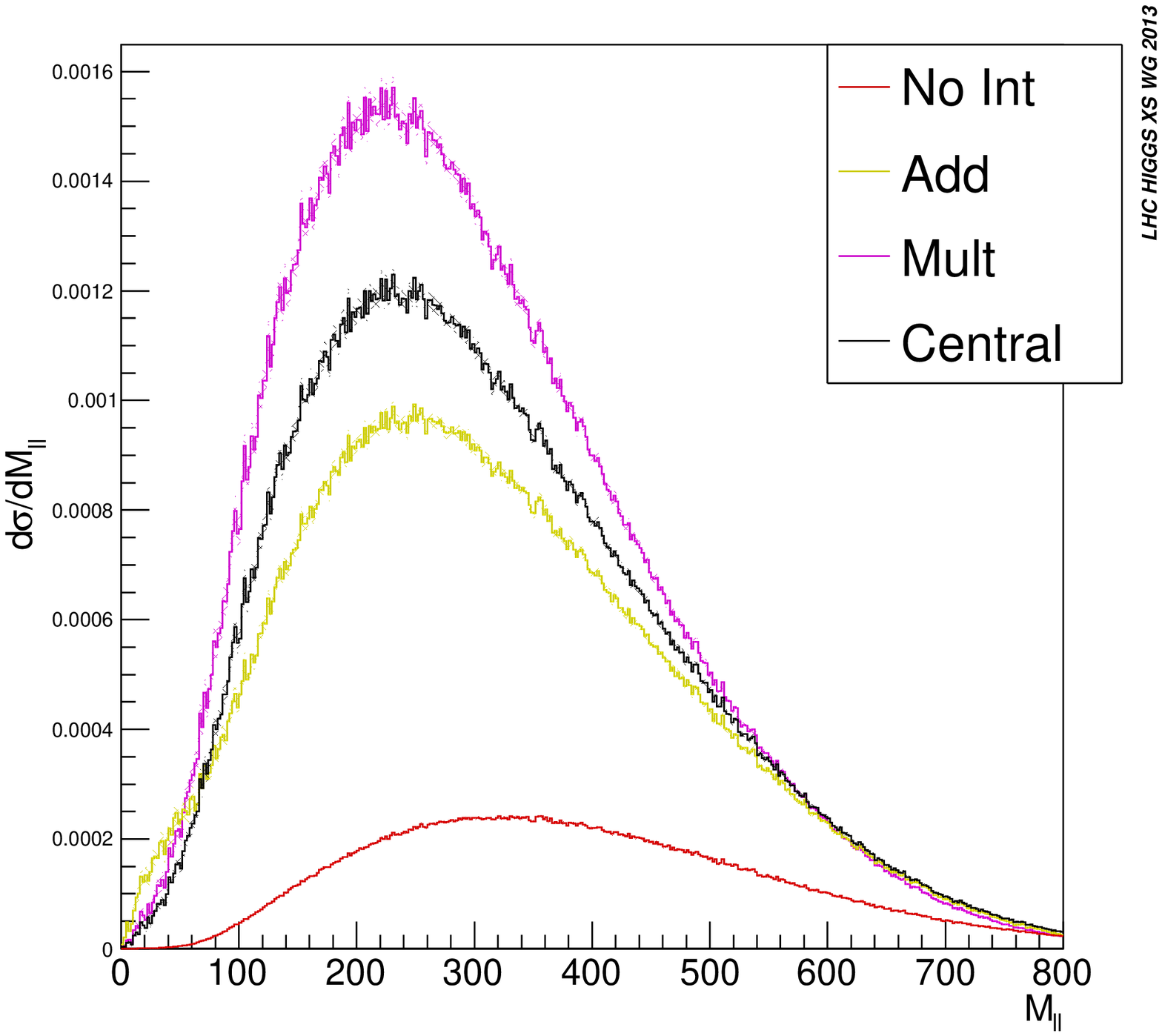}

  %\end{center}
  \vspace*{-2mm}
  \caption{Distributions of the dilepton invariant mass $m_{\ell\ell}$ showing the three schemes for scaling to NNLO, with the LO signal distribution for comparison, at Higgs masses of $\MH = 400\UGeV$ (left), $600\UGeV$ (middle), $900\UGeV$ (right).}
  \label{fig:mll_NNLO_cut_1}
\end{figure}

\begin{figure}[tbp]
  \centering
  
  \includegraphics[width=0.32\textwidth]{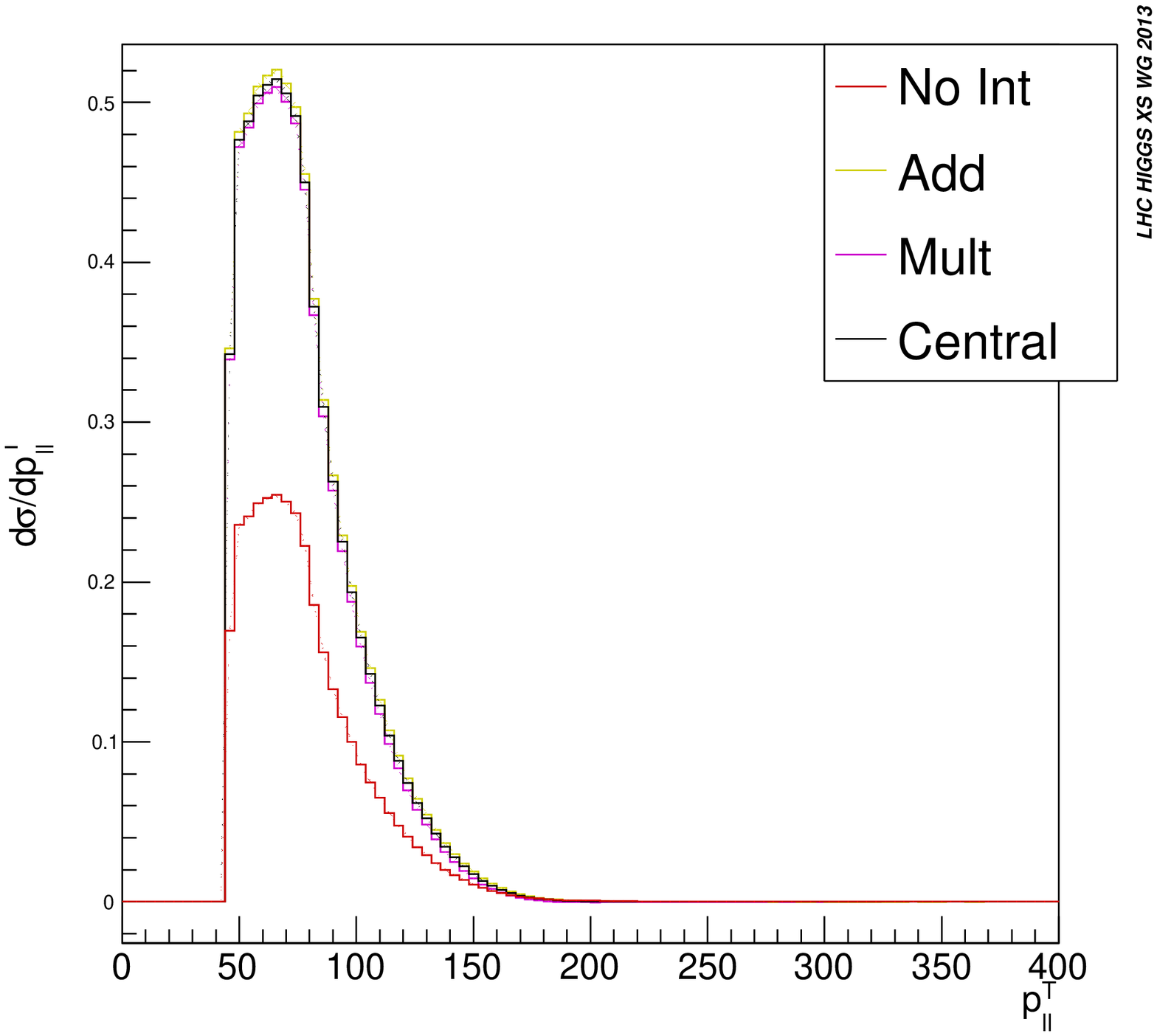}
  \includegraphics[width=0.32\textwidth]{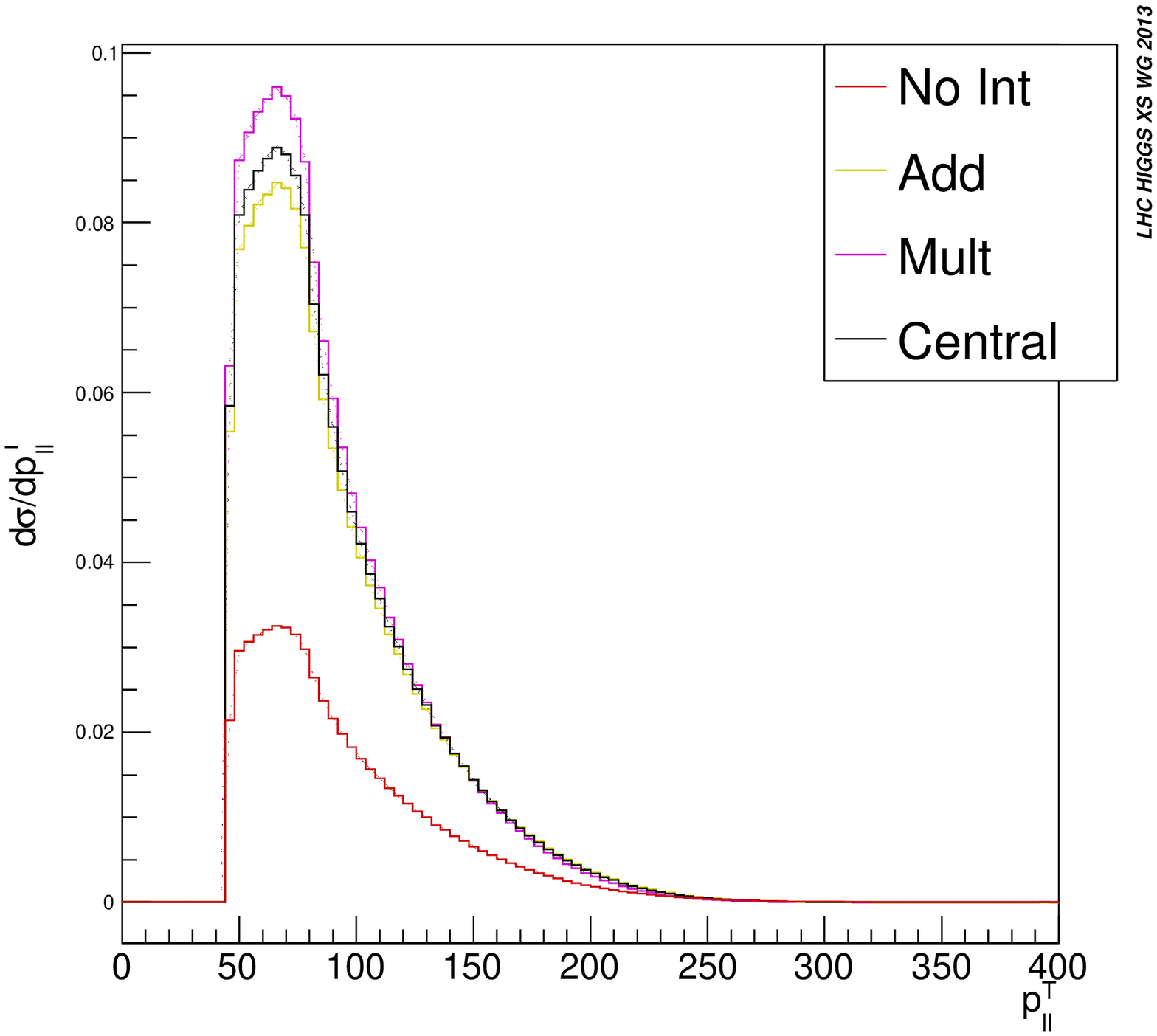} 
  \includegraphics[width=0.32\textwidth]{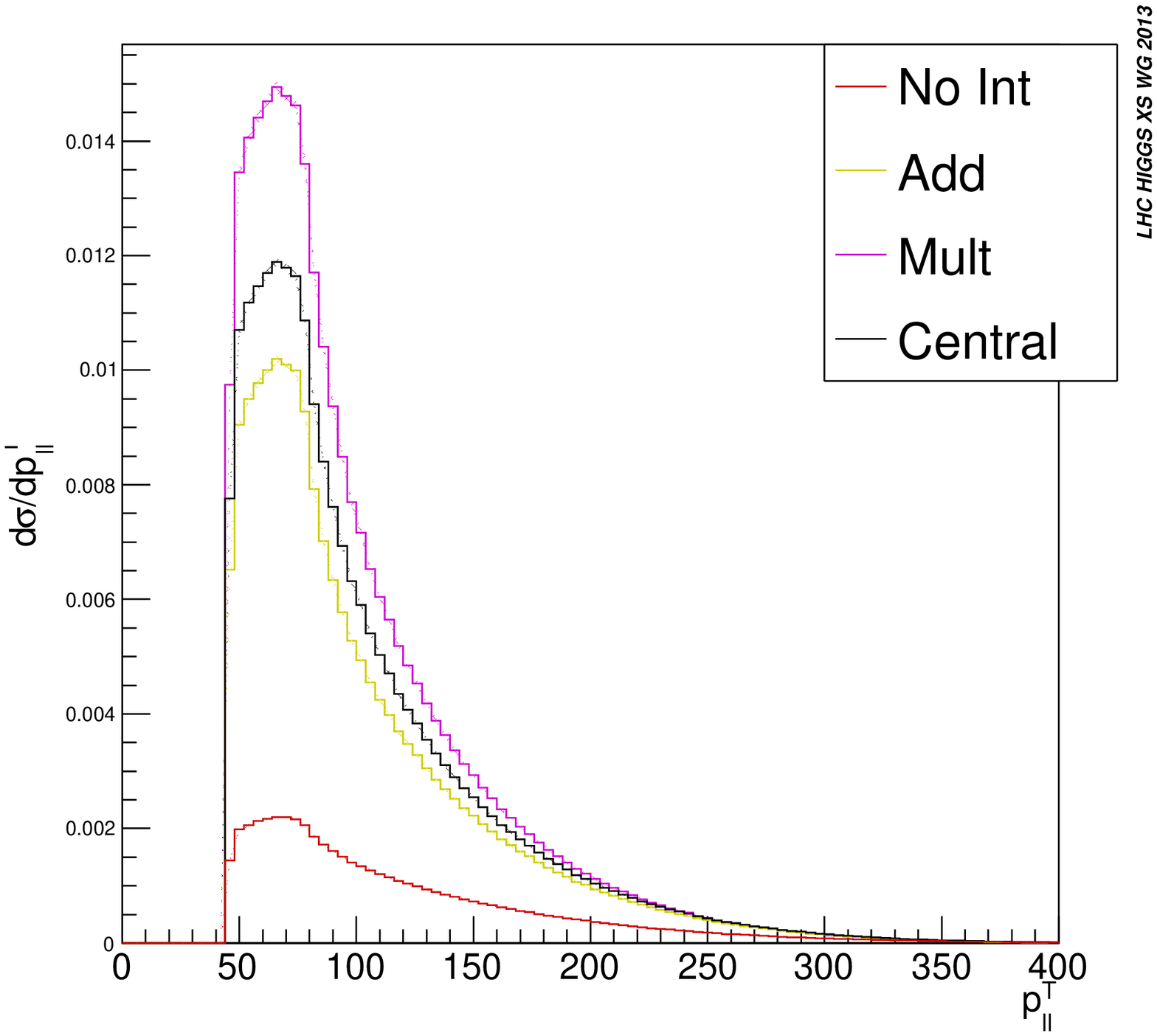}

  %\end{center}
  \vspace*{-2mm}
  \caption{Distributions of the transverse momentum of the dilepton system $\pT^{\ell\ell}$ showing the three schemes for scaling to NNLO, with the LO signal distribution for comparison, at Higgs masses of $\MH = 400\UGeV$ (left), $600\UGeV$ (middle), $900\UGeV$ (right).}
  \label{fig:ptll_NNLO_cut_1}
\end{figure}

\begin{figure}[tbp]
  \centering
  
  \includegraphics[width=0.32\textwidth]{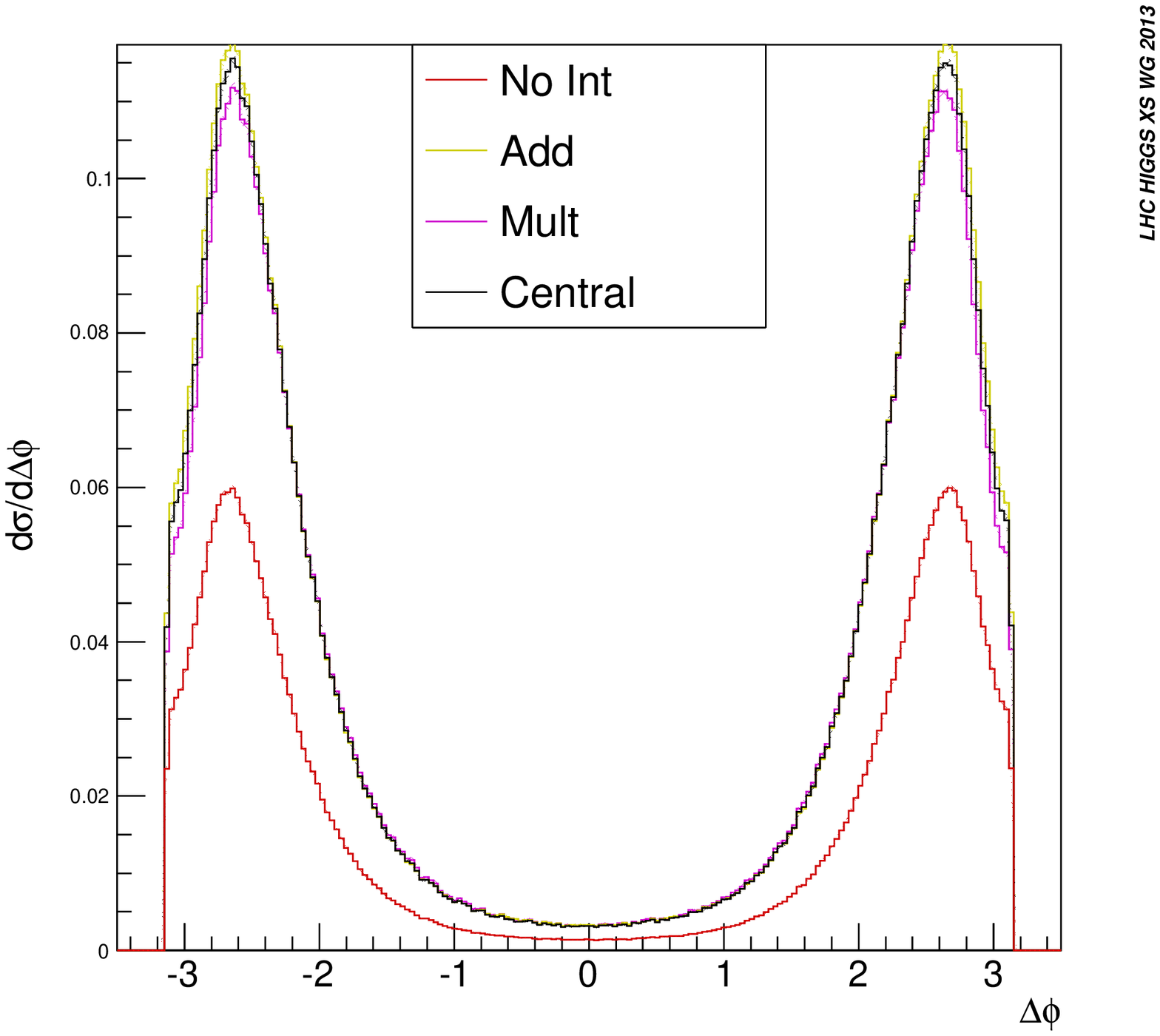}
  \includegraphics[width=0.32\textwidth]{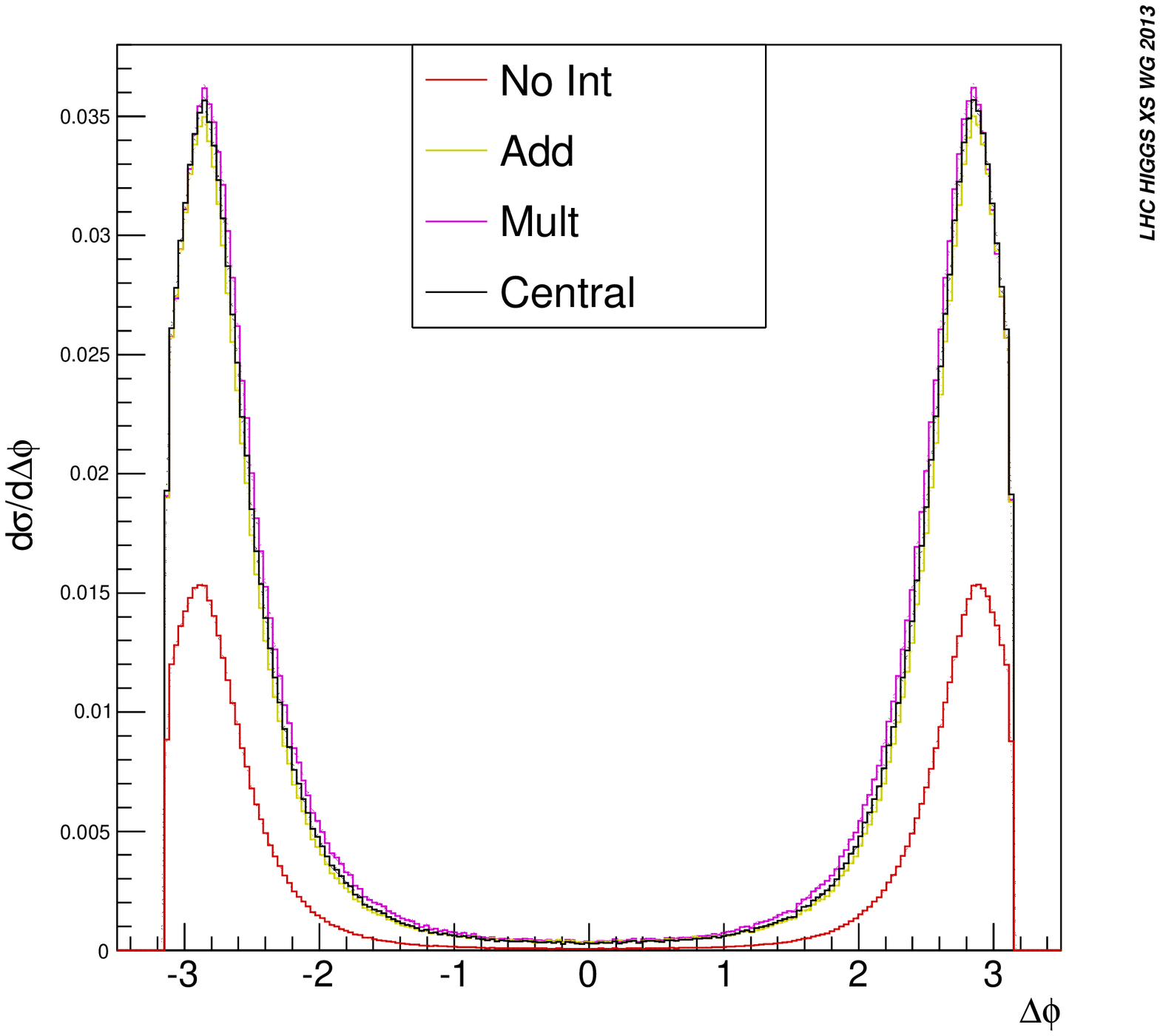} 
  \includegraphics[width=0.32\textwidth]{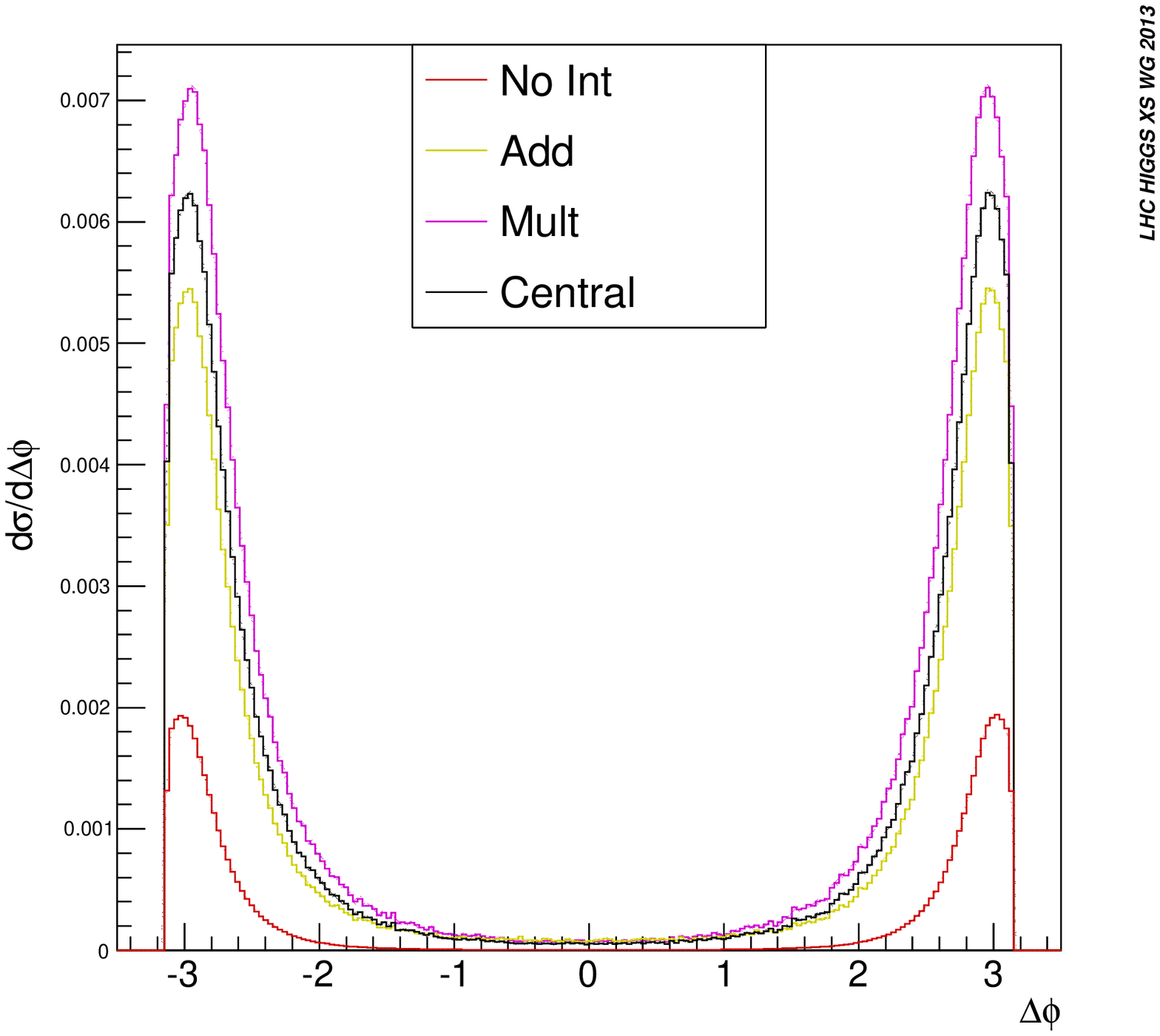}
 
  %\end{center}
  \vspace*{-2mm}
  \caption{Distributions of the dilepton azimuthal opening angle $\Delta\phi_{\ell\ell}$ showing the three schemes for scaling to NNLO, with the LO signal distribution for comparison, at Higgs masses of $\MH = 400\UGeV$ (left), $600\UGeV$ (middle), $900\UGeV$ (right).}
  \label{fig:DeltaPhi_ll_NNLO_cut_1}
\end{figure}

\begin{figure}[tbp]
  \centering
  
  \includegraphics[width=0.32\textwidth]{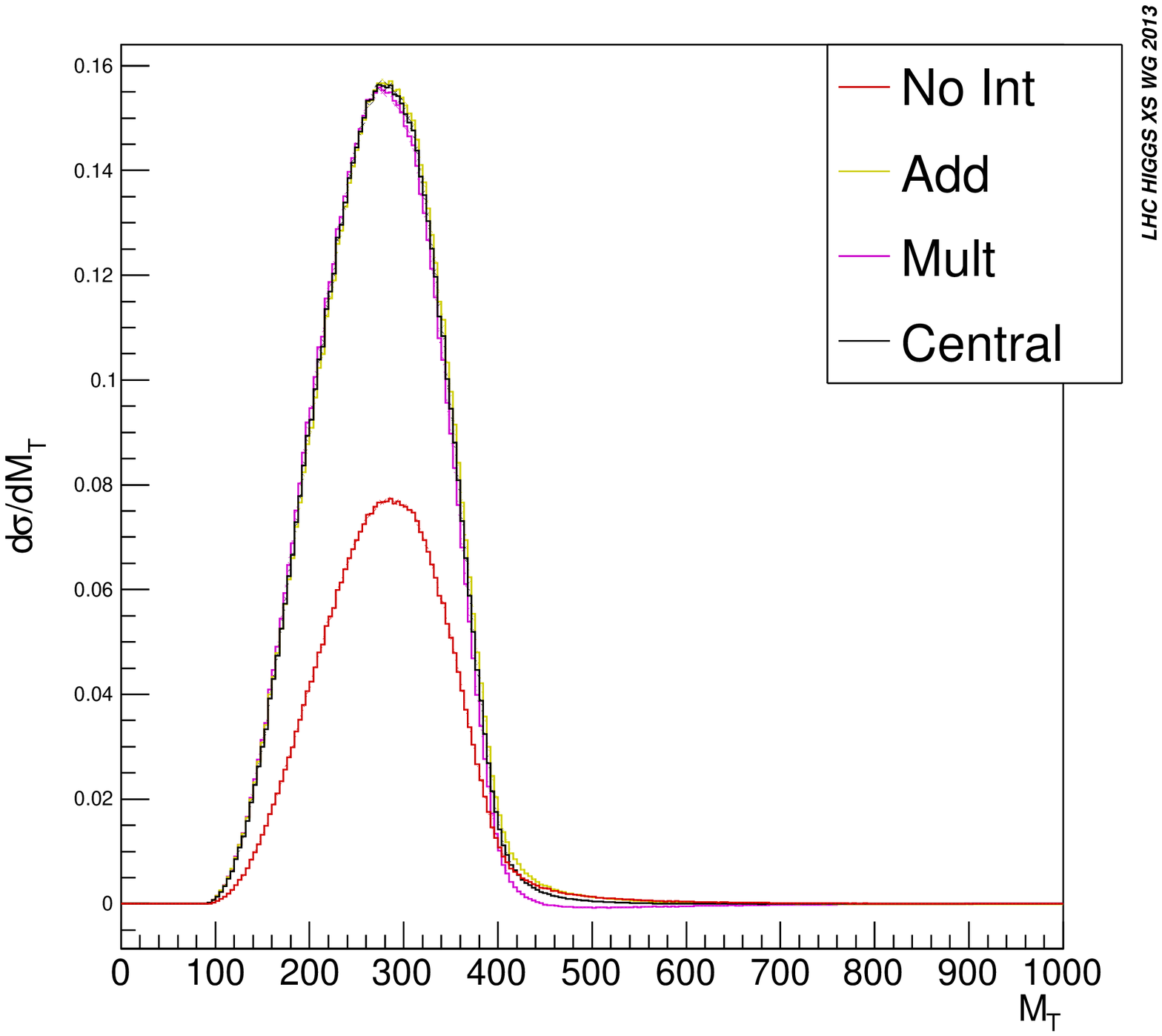}
  \includegraphics[width=0.32\textwidth]{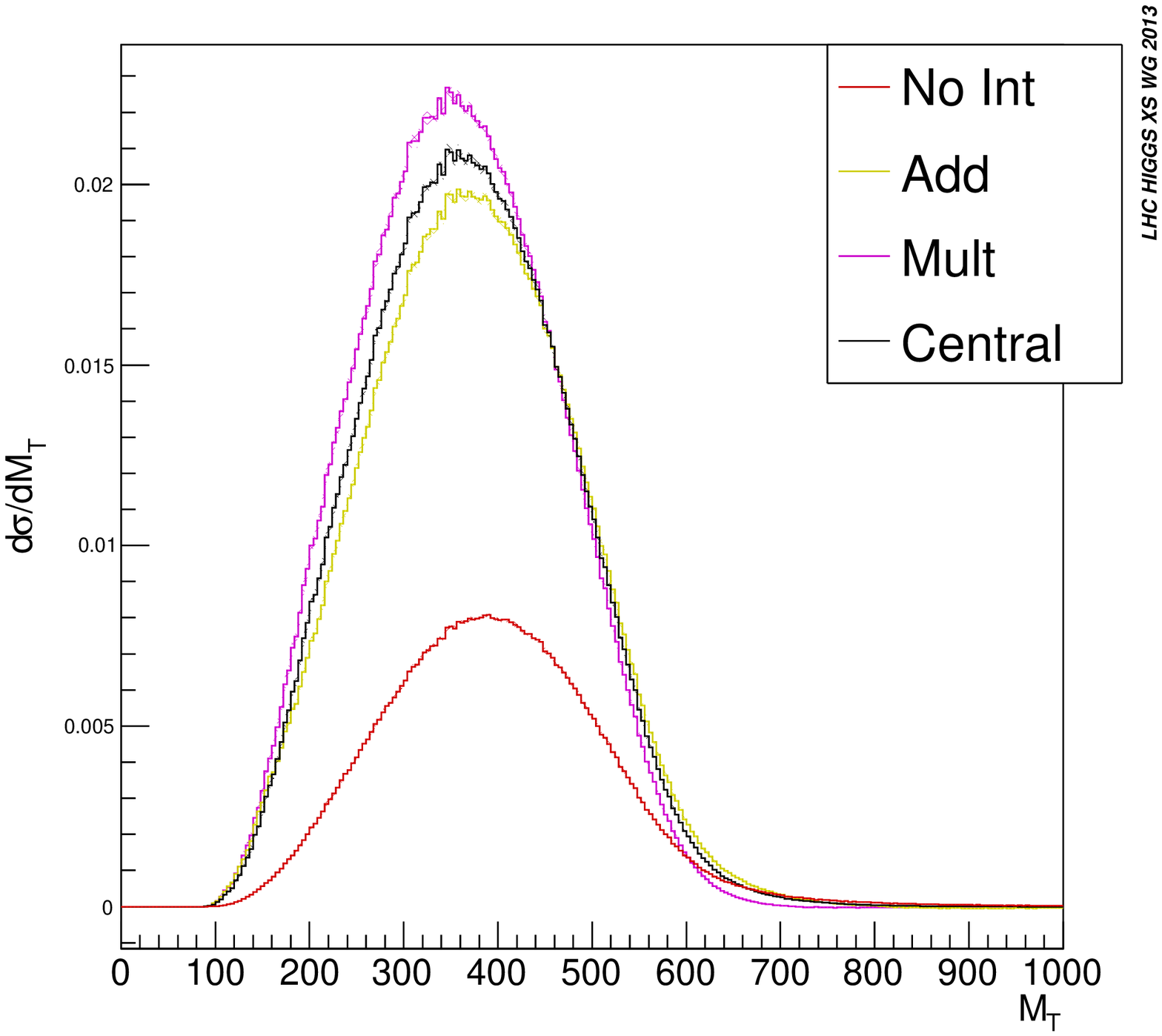} 
  \includegraphics[width=0.32\textwidth]{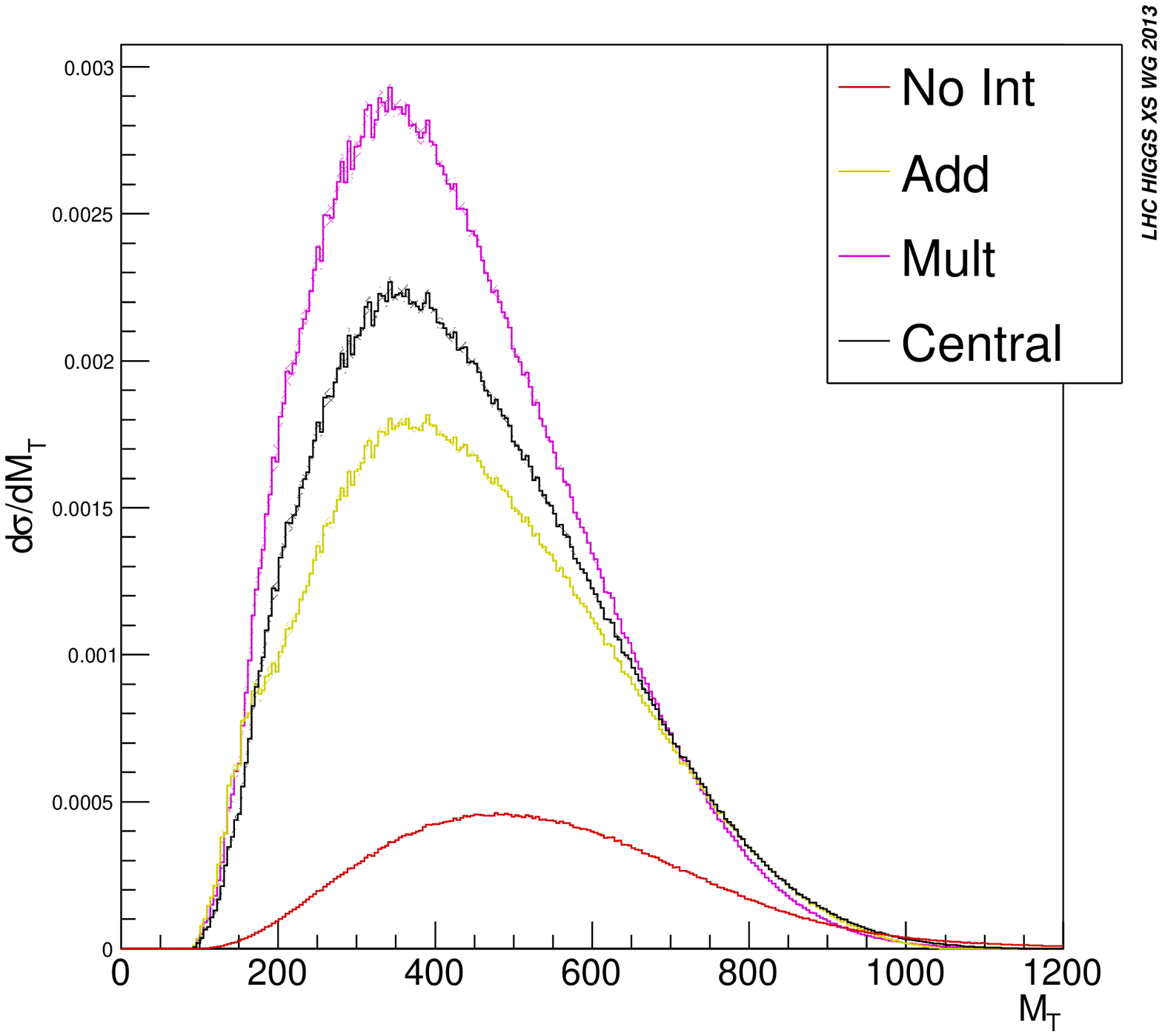}

  %\end{center}
  \vspace*{-2mm}
  \caption{Distributions of the transverse mass $M_\mathrm{T}$ showing the three schemes for scaling to NNLO, with the LO signal distribution for comparison, at Higgs masses of $\MH = 400\UGeV$ (left), $600\UGeV$ (middle), $900\UGeV$ (right).}
  \label{fig:Mt_NNLO_cut_1}
\end{figure}

From these plots, it is evident that the difference between the Intermediate distributions and the Additive or Multiplicative ranges from a negligible amount at $\MH = 400\UGeV$, to about $\pm$ $25\%$ at $\MH = 900\UGeV$. This is a very conservative estimate of the uncertainty in scaling to NNLO, and gives a very large uncertainty for high $\MH$. The different schemes for scaling to NNLO do not seem to affect the shape of the distributions significantly. \\

In conclusion, event-by-event reweighting by the $M_{\PW\PW}$ distribution reproduces the effect of interference fairly closely, though it does introduce some uncertainty, of order $1\%$ for $\MH = 400 \UGeV$ and $5\%$ for $\MH = 900\UGeV$. A much larger uncertainty is introduced by the process of reweighting to NNLO, rising to $\pm$ $25\%$ for $\MH = 900\UGeV$, the highest mass point studied.

\subsubsection{Studies with gg2VV: gg->ZZ->2l2$\nu$}
\label{sec:rewgg2VV}
%\section{Interference with backgrouond}

%The resonance-continuum interference between the SM Higgs signal process $\Pg\Pg \rightarrow \PH \rightarrow \PZ\PZ$ and irreducible background process $\Pg\Pg \rightarrow \PZ\PZ$ is not negligible when Higgs mass is large.
The effect of the resonance-continuum interference between the SM Higgs signal process $\Pg\Pg \rightarrow \PH \rightarrow \PZ\PZ$ and 
the irreducible background process $\Pg\Pg \rightarrow \PZ\PZ$ has been
studied at leading order in $pp$ collision at $\sqrt{s}=8\UTeV$, using the
program \textsc{gg2VV-3.1.0}~\cite{Kauer:2012hd}. Results
for a heavy Higgs boson (600~GeV or above) are presented here. 

Parton level cross sections for $\Pg\Pg( \rightarrow \PH) \rightarrow \PZ\PZ \rightarrow 2\ell 2\PGn$ are presented in \Tref{tab:interference} . Results are given for a single lepton flavor combination.
Please note when charged leptons and neutrinos are in same flavor, i.e., $2\PGm2\PGn_{\PGm}$ or $2e2\PGn_{\Pe}$ final states, there also exits interference with $\Pg\Pg \rightarrow \PW\PW \rightarrow 2\ell 2\PGn$. So the same flavor case is computed separately.

The renormalization and factorization scales are set to $\MH/2$. The PDF set CT10NNLO is used.
The following experimental selection cuts are applied: $\pT^{\ell}>20\UGeV$, $|\eta_{\ell}|<2.5$, $|M_{\ell\ell}-91|<15\UGeV$, $\pT^{\ell\ell}> 55\UGeV$, $\MET>90\UGeV$, $M_{\mathrm T}> 325\UGeV$.
The $M_{\mathrm T}$ variable is the transverse mass of the $\PZ\PZ\rightarrow 2\ell2\PGn$ system, defined as:

\begin{equation}
M_{\mathrm T}^2 = \left[ \sqrt{p_{\mathrm T,\ell\ell}^2+M_{\ell\ell}^2}+
\sqrt{E_{\mathrm T}^{\mathrm{miss},2}+M_{\ell\ell}^2} \right]^2
- \left[ \vec{p}_{\mathrm{T},\ell\ell} + \vec{E}_{\mathrm T}^{\mathrm{miss}} \right]^2
\label{eq:transversemass}
\end{equation}

As shown in the table, for a $600\UGeV$ Higgs signal, the interference effect can be a few percent, and for $1\UTeV$ signal, it can be as large as $90\%$.

\begin{table}[htp]
\centering
\caption{Parton level cross sections in fb for a single lepton flavor combination. S is Higgs signal; B is continuum background; Tot is signal and background together, including their interference. R is a ratio of interference effect, defined as (Tot-B)/S. }
\label{tab:interference}
\begin{tabular}{lcccccccc} \hline
\multirow{2}{*}{$\MH$ (GeV)} & \multicolumn{4}{c}{\bf $\Pg\Pg( \rightarrow \PH) \rightarrow \PZ\PZ \rightarrow 2\ell 2\PGn$} & \multicolumn{4}{c}{\bf $\Pg\Pg( \rightarrow \PH) \rightarrow \PZ\PZ/\PW\PW \rightarrow 2\ell 2\PGn$} \\
 & S & B & Tot & R &  S & B & Tot & R \\\hline
600     & 0.200(1) & 0.0633(2) & 0.277(1) & 1.07  & 0.202(3) & 0.0671(5) & 0.283(3) & 1.07 \\
700     & 0.0914(6) & 0.0596(2) & 0.1681(7) &1.19 & 0.0920(5) & 0.0636(7) & 0.171(2) & 1.17 \\
800     & 0.0439(3) & 0.0566(2) & 0.1159(5) &1.35 & 0.0441(3) & 0.0602(2) & 0.1218(7) & 1.40 \\
900     & 0.0224(1) & 0.0541(2) & 0.0891(3) &1.56 & 0.0221(3) & 0.0567(4) & 0.0927(9) & 1.63 \\
1000    & 0.01202(7) & 0.05202(2) & 0.0741(3) &1.84 & 0.0121(2) &0.0549(4) &0.0783(3) & 1.93 \\\hline
\end{tabular}
\end{table}

The interference also changes kinematics of signal. A recipe from LHCHXSWG is used~\cite{Dittmaier:2012vm} to reweight signal lineshapes to account for this change and its uncertainty.
Again, \textsc{gg2VV}-3.1.0 is used to generate signal sample, background sample and signal+interference+background sample at each signal mass point. The $2\ell2\PGn$ invariant mass distributions of such three samples for $900\UGeV$ signal hypothesis are shown as \Fref{fig:interf_shape}(left).
We can see the interference enhances the mass pole and below, but slightly reduces where above the mass pole.
The interference term could be separated by subtracting signal and background distributions from signal+interference+background distribution.
Then a set of NNLO $K\,$-factors as a function of $m_{\PZ\PZ}$ are read from theorists and applied to signal term $S$ and interference term $I$ as below

$$dS_{\NNLO} = K(m)\,dS $$
$$dS_{\mathrm{corr}} = dS_{\NNLO}+(K(m))^{n}\,dI, n = 0, 0.5, 1$$

Here $S_{\NNLO}$ is signal lineshape after NNLO $K\,$-factors applied, and $S_{\mathrm{corr}}$ is that after including interference. As for the $S_{\mathrm{corr}}$, $n=0.5$ gives central shape, while $n=0$ or 1 gives down or up uncertainty band.
\Fref{fig:interf_shape}(right) shows $S_{\NNLO}$ and $S_{\mathrm{corr}}$ with its uncertainty band, for $900\UGeV$ signal hypothesis.
By dividing $S_{\mathrm{corr}}$ by $S_{\NNLO}$, we get weights to correct our POWHEG signal lineshapes.

\begin{figure}[h]
\centering
\includegraphics[width=0.48\textwidth]{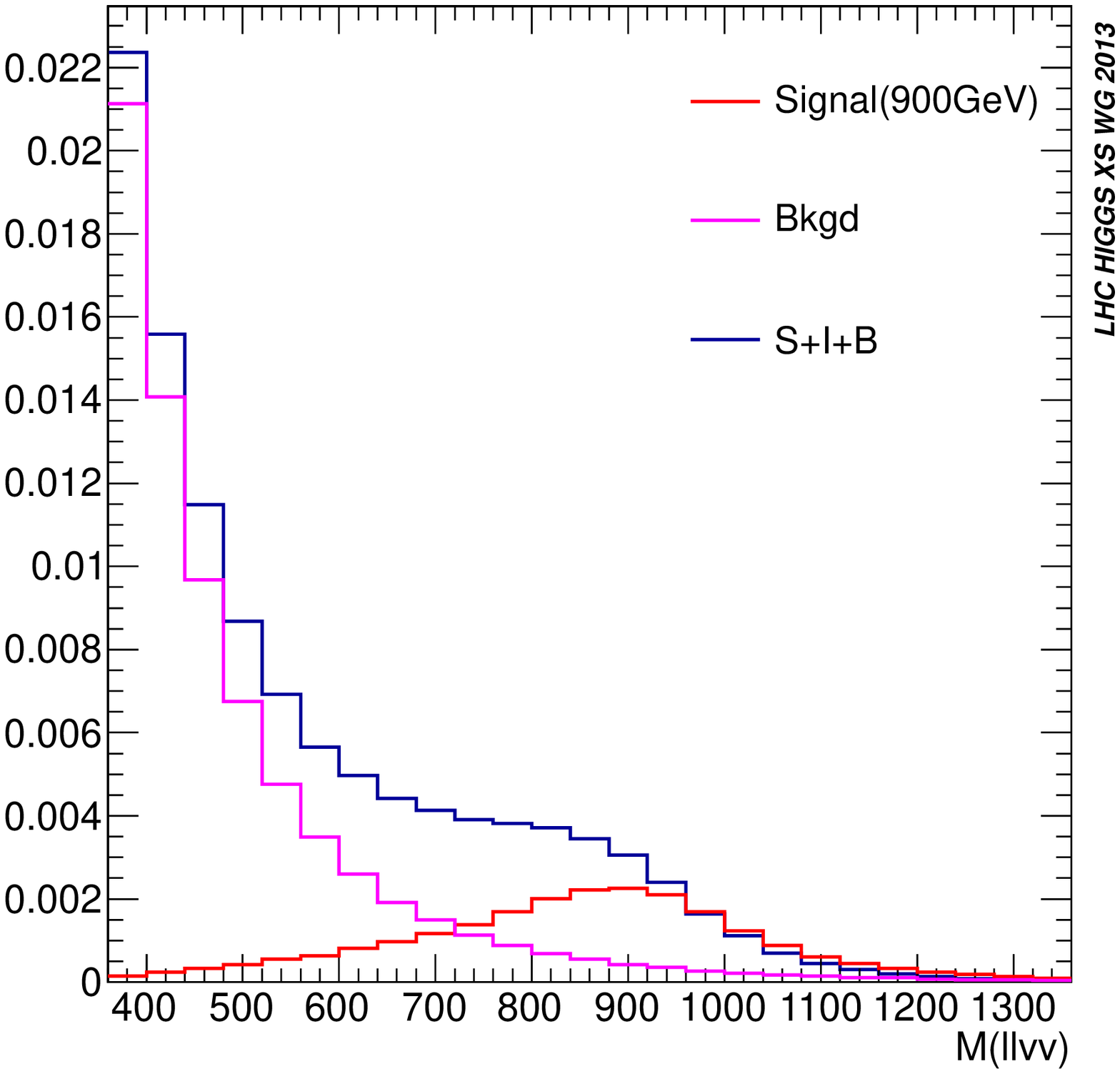}
\includegraphics[width=0.48\textwidth]{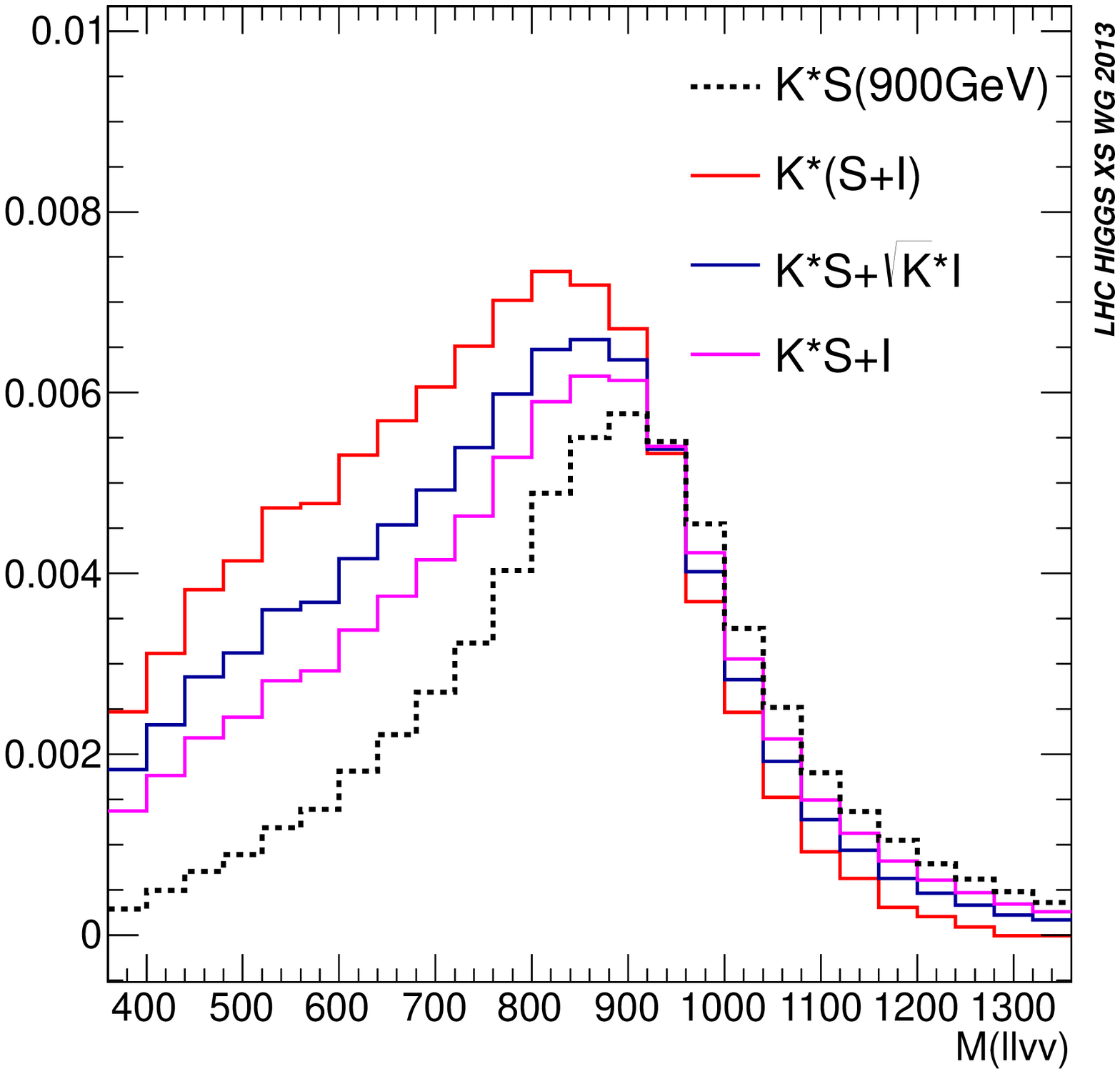}
\caption{ Invariant mass of the $2\ell2\PGn$ system for $\Pg\Pg( \rightarrow \PH) \rightarrow \PZ\PZ \rightarrow 2\ell 2\PGn$ signal, background and signal+background+interference ({\em left}); Signal lineshape before and after including interference ({\em right}). A $900\UGeV$ signal hypothesis is used here}
\label{fig:interf_shape}
\end{figure}

\subsubsection{Reweighting CPS PowHeg-BOX samples}
\label{sec:rewPowHeg}
At high Higgs mass the interference between the Higgs signal and the
$\Pg\Pg\rightarrow \PZ\PZ$ or $\Pg\Pg\rightarrow \PW\PW$ continuum backgrounds becomes very large. It can affect significantly both cross sections and distributions, depending from the final state.
More recently, the interference effect at LO has been included in some MC programs as discussed in \Sref{sec:HH:MC}. Nevertheless the current signal samples used by ATLAS and CMS have been generated with \textsc{PowHeg BOX} ~\cite{Alioli:2010xd} which does not include the interference effect. 
This brings to the needs of a theoretical prescription to reweight such samples in order to account for this effect and associate an uncertainty on this reweighting procedure.

Several reweighting tools based on tables and scale factors provided in ~\cite{CPHTO} have been developed depending from the different final state. They allow to rescale on an event by event basis the CPS PowHeg BOX MC signal samples to account for the interference effect and to include the theoretical uncertainties as described in \Sref{HMbis}.

The effect of such reweighting on $\PZ\PZ$ and $\PW\PW$ invariant 
masses, with the associated uncertainties are shown respectively in \Fref{fig:PowHeg_interference} and \Fref{fig:PowHeg_interference_WW}.

\begin{figure}[htb]
\centering
    \includegraphics[width=0.8\linewidth]{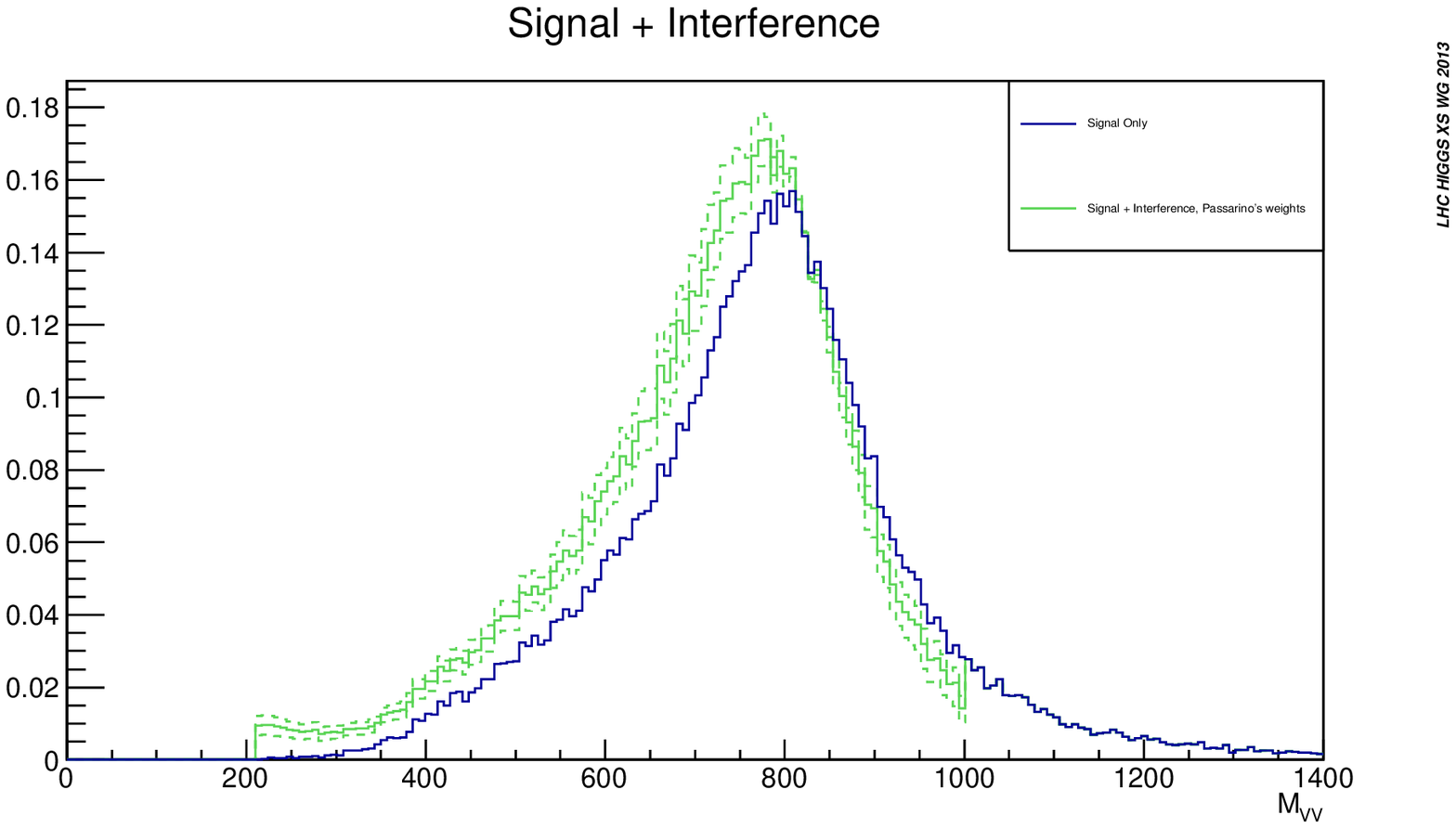}
\caption{Distribution at generator level for an Higgs mass of $800\UGeV$ before (blue solid line)
and after (green solid line) the inclusion of the interference. The shapes
to describe the uncertainty are also shown (dot-dashed green lines)}
\label{fig:PowHeg_interference}
\end{figure}

\begin{figure}[htb]
\centering
    \includegraphics[width=0.8\linewidth]{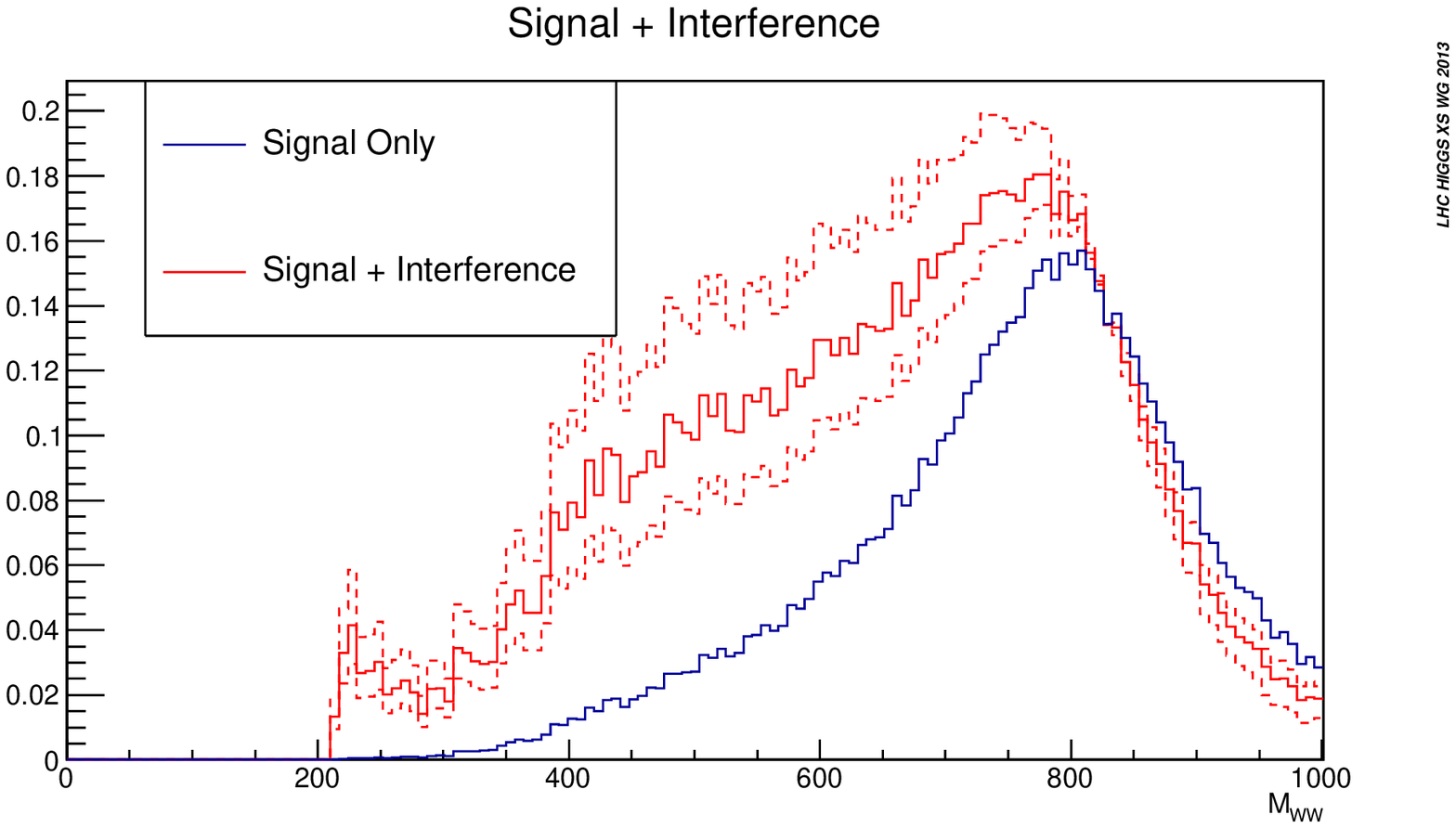}
\caption{Distribution at generator level for an Higgs mass of $800\UGeV$ before (blue solid line)
and after (red solid line) the inclusion of the interference. The shapes
to describe the uncertainty are also shown (dot-dashed red lines)}
\label{fig:PowHeg_interference_WW}
\end{figure}

%The effect of interference has been shown to be constructive below the
%Higgs mass peak and destructive above, it has therefore a negligible effect on the total
%cross-section (~1-2\%) but it strongly biases the $\PZ\PZ$ invariant mass distribution.

\subsection{Interference effects in VBF production}
\label{sec:interfVBF}

\providecommand{\HDECAY}{{\sc HDECAY}}
\providecommand{\HIGLU}{{\sc HIGLU}}
\providecommand{\Prophecy}{{\sc Prophecy4f}}
\providecommand{\CPsuperH}{{\sc CPsuperH}}
\providecommand{\FeynHiggs}{{\sc FeynHiggs}}

\providecommand{\lsim}
{\;\raisebox{-.3em}{$\stackrel{\displaystyle <}{\sim}$}\;}
\providecommand{\gsim}
{\;\raisebox{-.3em}{$\stackrel{\displaystyle >}{\sim}$}\;}
\providecommand{\orderx}[1]{\ensuremath{{\cal O}(#1)}}

\providecommand{\eqn}[1]{Eq.\,(\ref{#1})}
\providecommand{\eqns}[2]{Eqs.\,(\ref{#1}) -- (\ref{#2})}
\providecommand{\refF}[1]{Figure~\ref{#1}}
\providecommand{\refFs}[2]{Figures~\ref{#1} -- \ref{#2}}
\providecommand{\refT}[1]{Table~\ref{#1}}
\providecommand{\refTs}[2]{Tables~\ref{#1} -- \ref{#2}}
\providecommand{\refS}[1]{Section~\ref{#1}}
\providecommand{\refSs}[2]{Sections~\ref{#1} -- \ref{#2}}
\providecommand{\refC}[1]{Chapter~\ref{#1}}
\providecommand{\refCs}[2]{Chapters~\ref{#1} -- \ref{#2}}
\providecommand{\refA}[1]{Appendix~\ref{#1}}
\providecommand{\refAs}[2]{Appendices~\ref{#1} -- \ref{#2}}

\providecommand{\zehomi}[1]{$\cdot 10^{-#1}$}
\providecommand{\zehoze}{}
\providecommand{\zehopl}[1]{$\cdot 10^{#1}$}

\subsubsection{Signal definition}

Also in the VBF Higgs production mode, interference effects between the
Higgs signal process and the continuum background become increasingly
important when going to higher Higgs masses. NLO QCD corrections to the
full process $\Pp\Pp \rightarrow \PVB\PVB jj$ have been calculated in
\Bref{Jager:2006zc,Jager:2006cp} and are available in {\sc
VBFNLO}~\cite{Arnold:2012xn,Arnold:2011wj,Arnold:2008rz} including
decays of the vector bosons. We will
concentrate on the $\Pp\Pp \rightarrow \PWp \PWm jj$ process, which
can be calculated with both fully leptonic ($\Pl^{+} \PGnl \Pl^{-}
\PAGn_{\Pl}$) as well as semileptonic ($\Pl^{+} \PGnl jj$, $\Pl^{-}
\PAGn_{\Pl} jj$) decays of the $\PW$ bosons.

Both Higgs and continuum diagrams are necessary to ensure the correct
high-energy behavior. For large invariant masses of the vector-boson
pair $\sqrt{s}$, the continuum diagrams ${\mathcal M}_{\mathrm B}$ are proportional to
$-\frac{s}{v^2}$, where $v$ denotes the Higgs vacuum expectation value,
and would lead to unitarity violation at about 1 TeV. They are canceled
by corresponding diagrams with s-, t- and u-channel exchange of a Higgs
boson ${\mathcal M}_H \propto \frac{s}{v^2}$ in the leading term.
This raises the question of how to best define the signal, or
correspondingly the background contribution which is subsequently
subtracted from the full process. Naive definitions like only taking the
signal plus interference contributions into account or, equivalently,
subtracting the contribution coming from background diagrams only will
violate unitarity at large center-of-mass energies of the diboson
system.

\begin{figure}
\centering
\includegraphics[width=0.48\textwidth]{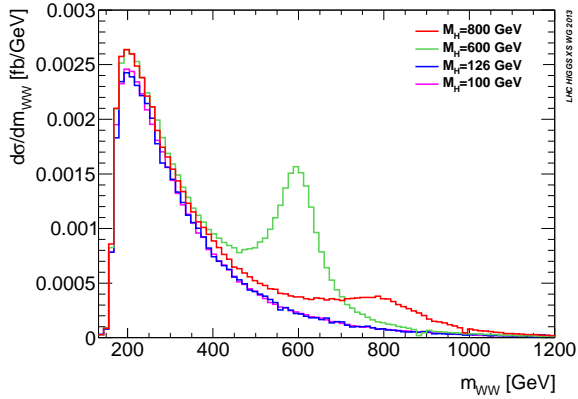}
\caption{Invariant $\PW\PW$ mass distribution for the process
$\Pp\Pp\rightarrow \PW\PW jj$ for different Higgs boson masses, intended
either as signal ($600$ and $800\UGeV$) or as background definition with a
light Higgs boson ($100$ and $126\UGeV$).}
\label{fig:vbfnlorew1}
\end{figure}
The invariant mass distribution of the $\PW\PW$ diboson system is shown
in \Fref{fig:vbfnlorew1} for various choices of the Higgs boson
mass. The distributions have been generated with \textsc{VBFNLO} using
standard vector-boson-fusion cuts on the two tagging jets ($m_{jj}>600$
GeV, $|\eta_{j1}-\eta_{j2}|>4$, $\eta_{j1} \times \eta_{j2} < 0$) in
addition to general detector acceptance cuts ($p_{T,j} > 20\UGeV$,
$\eta_j<4.5$, $p_{T,\ell} > 10\UGeV$, $\eta_\ell<2.5$).
We show two examples for heavy Higgs bosons of $600$ and $800\UGeV$ mass, as
well as for two light Higgs masses of $126$ and $100\UGeV$. For the latter,
the Higgs peak is outside of the shown range. We see that the two
light-mass curves agree very well, with some small remaining mass
dependence in the region around the $\PW\PW$ production threshold.
Therefore, we can use these to define the background
\begin{equation}
\sigma_{\mathrm B} = \int d\Phi |{\mathcal M}_{\mathrm B} + {\mathcal M}_{\Ph}(m_{\Ph})|^2 \ ,
\end{equation}
where $m_{\Ph}$ denotes the mass of the light Higgs boson used for
subtraction and $d\Phi$ denotes the phase-space integration. This then
leads to our definition of the signal+interference contribution
\begin{equation}
\sigma_{S+I} = 
\int d\Phi |{\mathcal M}_{\mathrm B} + {\mathcal M}_H(\MH)|^2 - \sigma_{\mathrm B} 
\label{eq:sireweight}
\end{equation}
with the heavy Higgs mass $\MH$.
This approach respects the correct high-energy behavior.

\subsubsection{Event reweighting}

A full simulation of LHC processes, involving NLO events plus parton
shower and detector simulation with cuts on the final-state particles,
is a time-consuming task. If events in a related scenario with similar
kinematic features have already been fully simulated, one can re-use
them and apply a reweighting procedure to the events at parton-level. In
\textsc{VBFNLO} this is included as an add-on named \textsc{REPOLO} --
REweighting POwheg events at Leading Order -- and is available on
request by the \textsc{VBFNLO} authors.

In this subsection we will discuss the reweighting of VBF production of a
heavy Higgs boson. The starting point are unweighted events of the
signal process only for a Higgs boson of 800 GeV, generated with
POWHEG~\cite{Nason:2009ai}. 
Using the \textsc{VBFNLO} framework, we then create a new event file
where each signal event is reweighted by a factor
\begin{equation} 
\frac{|{\mathcal M}_{\mathrm B} + {\mathcal M}_H(\MH)|^2 - |{\mathcal M}_{\mathrm B} + {\mathcal
M}_{\Ph}(m_{\Ph})|^2}{|{\mathcal M}_H(\MH)|^2}
\label{eq:repoloreweight}
\end{equation}
according to Eq.~\ref{eq:sireweight}. Events are reweighted for heavy
Higgs bosons without decay. For the resulting signal+interference
contribution, this corresponds to dividing over the Higgs branching into
$\PW\PW$. Internally, a decay into $\PW^+\PW^- \rightarrow \Pl^{+}
\PGnl \Pl^{-} \PAGn_{\Pl}$ is simulated for all matrix elements which
appear in the reweighting.

The result of this procedure is shown in \Fref{fig:vbfnlorew2} where
900000 POWHEG events have been reweighted. For the background
subtraction a Higgs boson with mass 100 GeV has been used.
\begin{figure}
\centering
\includegraphics[width=0.48\textwidth]{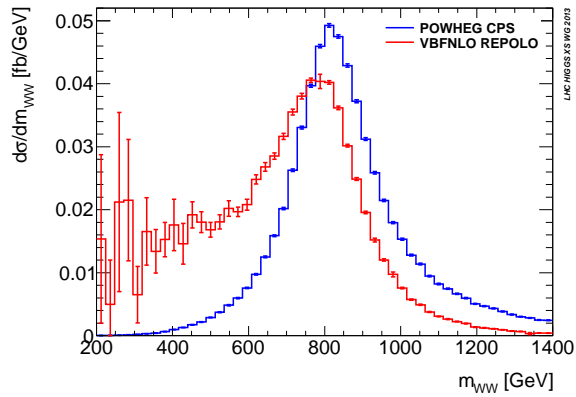}
\caption{Invariant $\PW\PW$ mass distribution for the process
$\Pp\Pp\rightarrow \PW\PW jj$. The distribution is shown for the signal-only
process generated with POWHEG, as well as reweighted by REPOLO to
include signal-background interference according to
Eq.~\ref{eq:repoloreweight}. The branching ratio $\PH\rightarrow \PW\PW$ has
been divided out for both curves.}
\label{fig:vbfnlorew2}
\end{figure}
One observes the significant effect of the interference term, which
enhances the rate for smaller invariant masses, while at the same time
decreasing the rate for larger ones. This corresponds to the
constructive and destructive interference, respectively, as expected
from theory. 
The reweighting procedure has its limits in phase-space regions, where
the reweighted event weight is significantly larger than the original
one. In the shown example this happens for invariant masses below about
400 GeV. The differential signal-only cross section gets very small,
while the interference term still yields a relevant contribution.
In practice, a possible remedy would be to generate additional events in
the region where the process to be reweighted is known to be small.

%%%%%%%%%%%%%%%%%%%%%%%%%%%%%%%%%%%%%%%%%%%%%%%%%%%%%%%%%%%%%%%%%

%\end{document}

\clearpage

\newpage
\section{BSM Higgs benchmarks in light of the discovery of a  126~\UGeV\ boson \footnote{%
    S.~Bolognesi, S.~Diglio, C.~Grojean, M.~Kadastik, H.E.~Logan, M.~Muhlleitner K.~Peters (eds.)}}
\label{sec:BSM}

\newcommand{\Ckappa}{\kappa}

This document provides a first proposition of a framework in which the
continuing LHC Higgs searches at masses other than $125\UGeV$ can be interpreted. 
 
%%%%%%%%%%%%%%%%%%%%%%%%%%%%%%%%%%%%%%%
\subsection{Introduction}

After the discovery of a new Higgs-like boson with SM-like
properties by the LHC experiments ATLAS and CMS
\cite{Aad:2012tfa,Chatrchyan:2012ufa, Chatrchyan:2013lba} the next step is to determine whether this
Higgs-like particle is fully responsible for the generation of the
masses of the other SM particles. This means in other words, that it fully
unitarizes the high-energy scattering amplitudes for $\PVBL \PVBL \to \PVBL
\PVBL$ ($\PVB = \PW$ or $\PZ$) and $\PVBL \PVBL \to \Pf \bar{\Pf}$. 

If the Higgs-like particle at $125\UGeV$ is not fully responsible for the
unitarization of the scattering amplitudes, then additional new
physics must be present to play this role.  Here we propose two benchmarks
in which a second scalar particle completes the unitarization of
the scattering amplitudes.  The allowed values of the couplings of this
second particle are therefore constrained by the observed production
and decay rates of the $125\UGeV$ Higgs-like state. 

We try to match these benchmarks to the coupling extraction
parameterizations of the Light Mass Higgs Subgroup \cite{LHCHiggsCrossSectionWorkingGroup:2012nn}.  For each benchmark we give a model-independent parameterization as well as a particular realization in terms of a specific model.  In some cases, the specific model fixes the values of some of the free parameters.

%%%%%%%%%%%%%%%%%%%%%%%%%%%%%%%%%%%%%%
\subsection{Relations imposed by unitarity \label{sec:unitarity1}}

The couplings to gauge bosons and fermions of the $125\UGeV$ Higgs-like
state, denoted by $\Pho$ in the following, and the ones of the second
scalar particle, denoted by $\Pht$, are subject to constraints imposed
by unitarity. We introduce the scaling factors $\Ckappa_i$ and $\Ckappa^\prime_i$ ($i=\Pf,\PW,\PZ$)
for the couplings of $\Pho$ and $\Pht$ to the fermions and gauge bosons. The
coupling factors without prime apply for the couplings of the $125\UGeV$
Higgs-like state $\Pho$ and the one with prime for the second state
$\Pht$. The coupling factors are defined relative to the corresponding
couplings of a SM Higgs boson, as in the coupling-extraction
document \cite{LHCHiggsCrossSectionWorkingGroup:2012nn}. 
 
The requirement of unitarization of longitudinal gauge boson
scattering $\PVBL\PVBL \to \PVBL\PVBL$ ($\PVB=\PW,\PZ$) leads to the following sum
rule\footnote{This expression assumes that $\Ckappa_{\PW} = \Ckappa_{\PZ}$ as imposed by custodial symmetry.}:
\begin{equation}
  \Ckappa_{\PVB}^2 + \Ckappa_{\PVB}^{\prime 2} = 1.
\label{eq:cond1}
\end{equation}

The unitarization of longitudinal gauge boson scattering into a
fermion pair, $\PVBL \PVBL \to \Pf\bar{\Pf}$, requires that
\begin{equation}
  \Ckappa_{\PVB} \Ckappa_{\Pf} + \Ckappa_{\PVB}^{\prime} \Ckappa_{\Pf}^{\prime} = 1.
\label{eq:cond2}
\end{equation}
This equation holds separately for each fermion species.  Here
$\Ckappa_{\Pf}$ needs not be the same for different kinds of
fermions. 

%%%%%%%%%%%%%%%%%%%%%%%%%%%%%%%%%%%%%%%%%%%%%%%%%%%%%%%%%%
%\subsection{Proposed Benchmarks}

In the following we propose two benchmark scenarios. For each of them we
give a model-independent parameterization as well as a
particular realization in terms of a specific model.  In some cases,
the specific model fixes the values of some of the free
parameters. 

%%%%%%%%%%%%%%%%%%%%%%%%%%%%%%%%%%%%%%%%%%%%%%%%%%%%%%%%%%
\subsection{Benchmark 1: One common scaling factor}

In this benchmark scenario a common scaling
factor is chosen both for the gauge and the fermion couplings of the field
$\Pho$ relative to the corresponding SM couplings. 
 
\noindent
{\it Model-independent parameterization:}  
For the $125\UGeV$ state $\Pho$ we then have
\begin{equation}
  \Ckappa \equiv \Ckappa_{\PVB} = \Ckappa_{\Pf}.
\label{eq:ccouplg1}
\end{equation}
This is equivalent to the overall signal strength scaling $\mu =
\Ckappa^2$. 

The corresponding coupling of the second state $\Pht$ is then
\begin{equation}
  \Ckappa^{\prime} \equiv \Ckappa_{\PVB}^{\prime} = \Ckappa_{\Pf}^{\prime} =  \sqrt{1 - \Ckappa^2}.
\label{eq:cprimecouplg1}
\end{equation}
While $\Ckappa^{\prime}$ can formally be of either sign, we choose the plus
sign with no loss of generality because only the relative signs of the
couplings of $\Pht$ are physically meaningful. 

The only other parameter affecting the rates for production and decay
of $\Pht$ is the branching ratio for possible decays into ``new'' final
states, BR$_{\mathrm{new}}$.  (For example, this new branching ratio can be due
to the decays $\Pht \to \Pho \Pho$ for $M_{\Pht} \geq 2 M_{\Pho} \simeq
250$~\UGeV). 

The relevant observables in the search for $\Pht$ are the $\Pht$ production cross
section $\sigma^{\prime}$, the $\Pht$ total width $\Gamma^{\prime}_{tot}$, and the
branching ratio BR$^{\prime}$ for $\Pht$ into the observable final state of interest.
From the expression for the total width $\Gamma^{\prime}$ of $\Pht$,
\begin{eqnarray}
	\Gamma^\prime = \Ckappa^{\prime 2} \Gamma_{\SM} + \Gamma_{\mathrm{new}},
\end{eqnarray}
it follows that these observables are given in terms of $\Ckappa^{\prime}$ and BR$_{\mathrm{new}}$ according to 
\begin{eqnarray}
	\sigma^{\prime} &=& \Ckappa^{\prime 2} \sigma_{\SM}, \nonumber \\
	\Gamma^{\prime} &=& \frac{\Ckappa^{\prime 2}}{(1 - {\rm
            BR}_{\mathrm{new}})} \Gamma_{\SM}, \nonumber \\
            {\rm BR}^{\prime} &=& (1 - {\rm BR}_{\mathrm{new}}) {\rm BR}_{\SM},
\end{eqnarray}
where $\sigma_{\SM}$, $\Gamma_{\SM}$, and BR$_{\SM}$ are the 
cross section, total width, and branching ratio into the final state of interest as predicted for the SM 
Higgs when its mass is equal to $M_{\Pht}$.  In the narrow width approximation,
the signal strength $\mu^{\prime}$ for $\Pht$ can be obtained using
\begin{equation}
	\mu^{\prime} = \frac{\sigma^{\prime} \times {\rm BR}^{\prime}}
	{\sigma_{\SM} \times {\rm BR}_{\SM}} 
	= \Ckappa^{\prime 2} (1 - {\rm BR}_{\mathrm{new}}).
\end{equation}
%The relevant observables in the search for $\Pht$ are the $\Pht$ signal strength
%$\mu^{\prime}$ (defined relative to that expected for a SM Higgs boson
%of the same mass) and the $\Pht$ total width
%$\Gamma^{\prime}_{\rm tot}$. From
%\begin{eqnarray}
%\Gamma^\prime_{\rm tot} = C^{\prime 2} \Gamma_{\SM} + \Gamma_{\mathrm{new}}
%\end{eqnarray}
%it follows that they are given in terms of $C^{\prime}$ and BR$_{\mathrm{new}}$ according to 
%\begin{eqnarray}
%  \mu^{\prime} &=& C^{\prime 2} (1 - {\mathrm BR}_{\mathrm{new}})
%  \\
%  \Gamma^{\prime}_{\rm tot} &=& \frac{C^{\prime 2}}{(1 - {\rm
%            BR}_{\mathrm{new}})} \Gamma_{\SM} \; . 
%\end{eqnarray}
In \refF{fig:bm1} we show the signal strength $\mu^\prime$ and the
total width $\Gamma^\prime$ in units of the total width of the SM
Higgs boson in the plane $(\Ckappa^{\prime 2}, {\rm BR}_{\mathrm{new}})$. 

\begin{figure}
\begin{center}
\includegraphics[angle=0,width=0.7\textwidth]{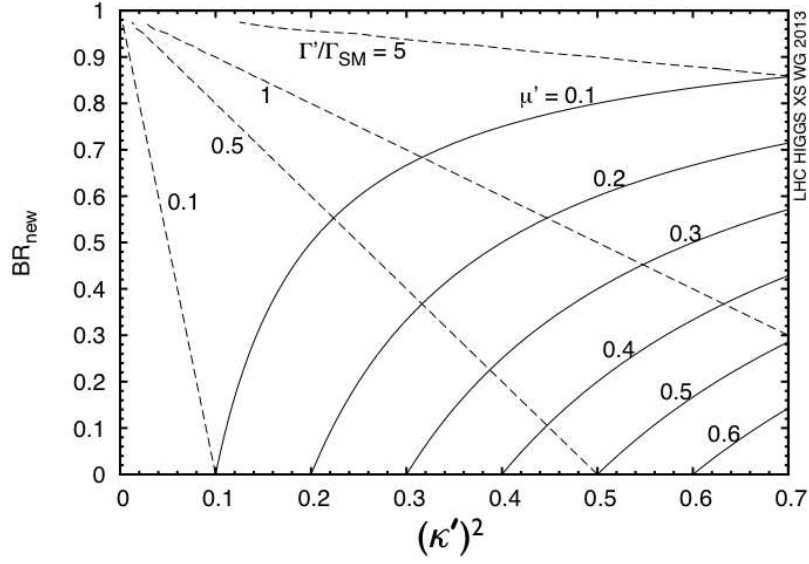}
\caption{The $\Pht$ signal strength $\mu^{\prime}$ and the $\Pht$ total
  width $\Gamma^\prime$ in units of the SM Higgs total width
  $\Gamma_{\SM}$ for Benchmark~1 (one common scaling factor) in the
  $(\Ckappa^{\prime 2}, {\rm BR}_{\mathrm{new}})$ plane.}
\label{fig:bm1}
\end{center}
\end{figure}

{\it Constraints from existing data:}  ATLAS and CMS have measured
$\mu \equiv \sigma/\sigma_{\SM}$ for the $125\UGeV$ boson.  The
results are (as of March 2013)~\cite{ATLAS-CONF-2013-034,CMS-PAS-HIG-13-005}
\begin{eqnarray}
	{\rm ATLAS}: && \mu = 1.30 \pm 0.20 \nonumber \\
	{\rm CMS}: && \mu = 0.88 \pm 0.21 \;.
\end{eqnarray}
Taking the uncertainty to be Gaussian, these
correspond to a $2\sigma$ lower bound on $\mu$ and hence an upper
bound of $\Ckappa^{\prime 2}$ of 
\begin{eqnarray}
	{\rm ATLAS}: && \mu > 0.90 \quad \rightarrow \quad \Ckappa^{\prime 2} < 0.10 \nonumber \\
	{\rm CMS}: && \mu > 0.46 \quad \rightarrow \quad \Ckappa^{\prime 2} < 0.54 \;.
\end{eqnarray}
Here we have assumed that the branching ratio of the $125\UGeV$ Higgs $\Pho$ into 
non-SM final states is zero. 

\noindent
{\it Specific model:} This parameterization is realized for the SM
Higgs boson mixed with an electroweak singlet.  In this case,
BR$_{\mathrm{new}}$ arises from decays of $\Pht \to \Pho \Pho$ for $M_{\Pht} > 2
M_{\Pho} \simeq 250$~\UGeV.  The Lagrangian and Feynman rules, are given
in \refS{sec:modelintro}. 

%\noindent
%{\it Constraints from existing data:}  ATLAS and CMS have measured
%$\mu \equiv \sigma/\sigma_{\SM}$ for the $125\UGeV$ boson.  The
%results are (as of July 2012)~\cite{Aad:2012tfa,Chatrchyan:2012ufa, Chatrchyan:2013lba}
%\begin{eqnarray}
%  {\rm ATLAS}: && \mu = 1.4 \pm 0.3 \nonumber \\
%  {\rm CMS}: && \mu = 0.87 \pm 0.23 \;.
%\end{eqnarray}
%Taking the uncertainty to be Gaussian, these
%correspond to a $2\sigma$ lower bound on $\mu$ and hence an upper
%bound of $C^{\prime 2}$ of 
%\begin{eqnarray}
%  {\rm ATLAS}: && \mu > 0.8 \rightarrow C^{\prime 2} < 0.2 \nonumber \\
%  {\rm CMS}: && \mu > 0.41\rightarrow C^{\prime 2} < 0.59.
%\end{eqnarray}
%Here we have assumed that the branching ratio into new final states
%${\mathrm BR}_{\mathrm{new}}$ is zero.

%%%%%%%%%%%%%%%%%%%%%%%%%%%%%%%%%%%%%%%
\subsubsection{Specific benchmark model 1: Standard Model plus a Real Singlet Field \label{sec:modelintro}}
The simplest extension of the SM Higgs sector is given by the addition
of a singlet field which is neutral under the SM gauge
groups~\cite{Hill:1987ea, Veltman:1989vw, Binoth:1996au,
  Schabinger:2005ei, Patt:2006fw}. 
This singlet field also acquires a non-vanishing
vacuum expectation value. Such models have been discussed in
numerous publications
\cite{Bowen:2007ia,Barger:2007im, Barger:2008jx,Bhattacharyya:2007pb,Dawson:2009yx,Bock:2010nz,
  Fox:2011qc,Englert:2011yb,Englert:2011us,Batell:2011pz,Englert:2011aa, Gupta:2011gd, Dolan:2012ac, Bertolini:2012gu, Batell:2012mj, Pruna:2013bma}
and we shall give details in the following.

%%%%%%%%%%%%%%%%%%%%%%%%%%%%%%%%%%%%%%%%%%%%%%%%%%%%%%
\subsubsection{The model}

The most general gauge-invariant potential
can be written as \cite{Schabinger:2005ei,Bowen:2007ia}
\begin{equation}
  V = \lambda \left( \Phi^{\dagger} \Phi - \frac{v^2}{2} \right)^2 
  + \frac{1}{2} M^2 s^2
  + \lambda_1 s^4 
  + \lambda_2 s^2 \left( \Phi^{\dagger} \Phi - \frac{v^2}{2} \right)
  + \mu_1 s^3 
  + \mu_2 s \left( \Phi^{\dagger} \Phi - \frac{v^2}{2} \right),
  \label{eq:potential}
\end{equation}
where $s$ is the real singlet scalar and in the unitary gauge the SM
Higgs doublet can be written as 
\begin{equation}
  \Phi = \left( \begin{array}{c} 0 \\ (\phi +
            v)/\sqrt{2} \end{array} \right)
\label{eq:phifield}
\end{equation}
with $v \simeq 246\UGeV$.  We have already used the freedom to shift the 
value of $s$ so that $s$ does not get a vacuum expectation value.
As a result, $M^2$ must be  chosen positive in Eq.~\refE{eq:potential}. 

To prevent the potential from being unbounded from below, the quartic couplings
must satisfy the conditions:
\begin{equation}
  \lambda > 0, \qquad \lambda_1 > 0, \qquad \lambda_2 > -2 \sqrt{\lambda \lambda_1}.
\end{equation}
The trilinear couplings $\mu_1$ and $\mu_2$ can have either sign.

%%%%%%%%%%%%%%%%%%%%%%%%%%%%%%%%%%%%%%%%%%%%%%%%%%%%%%%%%%%
\subsubsection{Mass eigenstates}

%After replacing Eq.~\refE{eq:phifield} in the potential of Eq.
%\refE{eq:potential}, the terms quadratic in the fields (that give
%rise to the mass matrix) are 
%\begin{equation}
%  V_2 = \lambda v^2 \phi^2 + \frac{1}{2} M^2 s^2 + \mu_2 v \, \phi s.
%\end{equation}
After replacing Eq.~(\ref{eq:phifield}) for $\Phi$ in the potential
Eq.~(\ref{eq:potential}), we obtain
\begin{equation}
V    =     \frac{\lambda}{4} \phi^4 + \lambda v^2 \phi^2 + \lambda v \phi^3 + \frac{1}{2}M^2 s^2 + \lambda_1 s^4 
               + \frac{\lambda_2}{2} \phi^2 s^2 +  \lambda_2 v \phi s^2 + \mu_1 s^3 + \frac{\mu_2}{2} \phi^2 s
               + \mu_2 v \phi s \;.
\end{equation}
The terms quadratic in the fields (that give
rise to the mass matrix) are 
\begin{equation}
	V_2 = \lambda v^2 \phi^2 + \frac{1}{2} M^2 s^2 + \mu_2 v \, \phi s.
\end{equation}
In particular, the mixing between $\phi$ and the singlet
field $s$ is controlled by the coupling $\mu_2$.  
The mass eigenvalues are then given by
\begin{equation}
  M^2_{\Pho, \Pht} = \lambda v^2 + \frac{1}{2} M^2 
  \mp \sqrt{ \left( \lambda v^2 - \frac{1}{2} M^2 \right)^2 +
          \mu_2^2 v^2 } \, ,
\end{equation}
where we have defined the mass eigenstates $\Pho,\Pht$ as
\begin{eqnarray}
  \Pho &=& \phi \cos\theta - s \sin\theta \nonumber \\
  \Pht &=& \phi \sin\theta + s \cos\theta,
\end{eqnarray}
with the mixing angle $\theta$ which can be written as
\begin{equation}
  \tan 2 \theta = \frac{-\mu_2 v}{\lambda v^2 - \frac{1}{2} M^2}
        \; .
\end{equation}

In order to find the domain of $\theta$ we can rewrite the masses as follows:
\begin{equation}
M^2_{\Pho, \Pht} = \left( \lambda v^2 + \frac{1}{2} M^2 \right)
  \mp \left( \frac{1}{2} M^2 - \lambda v^2\right) \sec 2\theta 
\end{equation}
If we require $\Pho$ to be the lighter mass eigenstate and choose $M^2
>  2 \lambda v^2$, then $\sec 2 \theta > 0$, and hence $\theta \in (-
\frac{\pi}{4}, \frac{\pi}{4})$. 

Note that in the notation of Eq.~\refE{eq:ccouplg1} and
\refE{eq:cprimecouplg1} 
we have in particular
\begin{eqnarray}
\Ckappa &\equiv& \Ckappa_{\PVB} = \Ckappa_{\Pf} = \cos \theta \\
\Ckappa^\prime &\equiv& \Ckappa_{\PVB}^\prime = \Ckappa_{\Pf}^\prime = \sin\theta \;.
\end{eqnarray}

%%%%%%%%%%%%%%%%%%%%%%%%%%%%%%%%%%%%%%%%%%%%%%%%%%%%%%%%%%%%%
\subsubsection{The trilinear and quartic interactions}

Here, we give the trilinear and quartic interactions among the mass
eigenstates $\Pho$ and $\Pht$. The related Feynman rules are necessary
for the determination of the Higgs-to-Higgs decays, namely the decay
of the heavier state $\Pht$ into two lighter bosons $\Pho \Pho$. 

%Once we substitute Eq.~\refE{eq:phifield} for $\Phi$ in the potential we get
%\begin{equation}
%V    =     \frac{\lambda}{4} \phi^4 + \lambda v^2 \phi^2 + \lambda v \phi^3 + \frac{1}{2}M^2 s^2 + \lambda_1 s^4 
%               + \frac{\lambda_2}{2} \phi^2 s^2 +  \lambda_2 v \phi s^2 + \mu_1 s^3 + \frac{\mu_2}{2} \phi^2 s
%               + \mu_2 v \phi s \;.
%\end{equation}
The relevant terms for the cubic interactions are 
\begin{equation}
  V_3 =  \lambda v \phi^3 +  \lambda_2 v \phi s^2  +  \mu_1
        s^3 +  \frac{\mu_2}{2} \phi^2 s \; .
\end{equation}
Using the shorthands $s_\theta \equiv \sin \theta$ and $c_\theta
\equiv \cos \theta$ and rewriting the trilinear terms of the potential
in terms of the mass eigenstates, we find the following couplings.

After replacing $\phi$ and $s$ by the mass eigenstates $\Pho$ and
$\Pht$, the triple-$\Pho$ coupling comes from the potential term
\begin{equation}
  V_{111} = \Pho^3 \left[  \lambda v c_\theta^3     +  \lambda_2  v  c_\theta s_\theta^2   
  - \mu_1 s_\theta^3  - \frac{\mu_2}{2} c_\theta^2 s_\theta   \right],
\end{equation}
which yields a Feynman rule
\begin{equation}
		\label{eq:111}
  \Pho \Pho \Pho:\ -6 i \left( \lambda v c_\theta^3   + \lambda_2 v c_\theta s_\theta^2   
  - \mu_1  s_\theta^3  - \frac{\mu_2}{2} c_\theta^2 s_\theta   \right)
  	\equiv - i L_{111}.
\end{equation}

The $\Pht \Pho \Pho$ coupling, which controls the $\Pht \to \Pho \Pho$ decay (if kinematically allowed), comes from the potential term
\begin{equation}
  V_{211} = \Pht \Pho^2 \left[       3 \lambda v s_{\theta} c_{\theta}      
   - \lambda_2  v  s_\theta  (  3 c_\theta^2  -1  )  
    + 3 \mu_1 c_\theta s_\theta^2 + \frac{\mu_2}{2} c_\theta (3 c_\theta^2 -2 )  \right],
\end{equation}
and yields the Feynman rule
\begin{equation}
	\label{eq:211}
  	\Pht \Pho \Pho  : \ -2 i \left(   3 \lambda v s_\theta c_\theta       
   -  \lambda_2  v s_\theta (3 c_\theta^2   - 1) 
  + 3 \mu_1  c_\theta s_\theta^2  + \frac{\mu_2}{2} c_\theta (3 c_\theta^2  - 2)   \right).
  		\equiv - i L_{211}.
\end{equation}

For the other cubic interactions we have the potential terms and
related Feynman rules:
\begin{eqnarray}
  V_{122} &=&   \Pho \Pht^2  \left[  -3 \lambda v c_\theta s_\theta^2   + \lambda_2  v c_\theta (1 - 3 s_\theta^2)  -\frac{\mu_2}{2} (2 - 3 s_\theta^2) \right] ,      \nonumber \\
  \Pho \Pht \Pht &:&    -2 i \left( -3 \lambda v c_\theta s_\theta^2   + \lambda_2  v  c_\theta (1 - 3 s_\theta^2)  - \frac{\mu_2}{2} s_\theta (2 - 3 s_\theta^2)  \right); \nonumber \\
  V_{222}  &=&  \Pht^3 \left[\lambda v s_\theta^3  +\lambda_2 v s_\theta c_\theta^2 + \mu_1 c_\theta^3 -\frac{\mu_2}{2} s_\theta^2 c_\theta  \right] ,     \nonumber \\
 	\Pht \Pht \Pht &:& -6 i \left(  \lambda  v  s_\theta^3  +
          \lambda_2 v s_\theta c_\theta^2  + \mu_1 c_\theta^3   -
          \frac{\mu_2}{2} s_\theta^2 c_\theta \right) \; .
\end{eqnarray}

For the quartic interactions, the relevant terms in the Lagrangian are,
\begin{equation}
  V_4 = \frac{\lambda}{4} \phi^4   +  \lambda_1  s^4  +
        \frac{\lambda_2}{2}  \phi^2 s^2 \;.
\end{equation}
After inserting the mass eigenstates in terms of the mixing angle
$\theta$, we find for the quartic potential terms and the
corresponding Feynman rules:
\begin{eqnarray}
  V_{1111} &=&  \Pho^4 \left[   \frac{\lambda}{4} c_\theta^4  + \lambda_1 s_\theta^4 + \frac{\lambda_2}{2} c_\theta^2 s_\theta^2   \right],  \nonumber \\
  \Pho \Pho \Pho \Pho &:&    -24 i \left(  \frac{\lambda}{4} c_\theta^4  + \lambda_1  s_\theta^4  + \frac{\lambda_2}{2} c_\theta^2 s_\theta^2    \right);    \nonumber \\
  V_{2222} &=&   \Pht^4 \left[   \frac{\lambda}{4} s_\theta^4  + \lambda_1  c_\theta^4  + \frac{\lambda_2}{2}  s_\theta^2  c_\theta^2   \right], \nonumber \\
  \Pht \Pht \Pht \Pht &:&   -24 i \left(  \frac{\lambda}{4} s_\theta^4   + \lambda_1 c_\theta^4  + \frac{\lambda_2}{2}  s_\theta^2   \right);   \nonumber \\
  V_{1112} &=&   \Pho^3 \Pht \left[   \lambda_1 s_\theta c_\theta^3  - 4 \lambda_1 c_\theta s_\theta^3  - \frac{\lambda_2}{4} \sin 4 \theta    \right], \nonumber \\
  \Pho \Pho \Pho \Pht &:&     -6 i \left(  \lambda s_\theta c_\theta^3  - 4 \lambda_1 c_\theta s_\theta^3  - \frac{\lambda_2}{4} \sin 4 \theta \right);    \nonumber \\
  V_{1122}    &= &  \Pho^2  \Pht^2  \left[   \frac{3}{2} \lambda s_\theta^2 c_\theta^2  + 6 \lambda_1 s_\theta^2 c_\theta^2  +
   \frac{\lambda_2}{2} (c_\theta^4 + s_\theta^4   - 4 c_\theta^2  s_\theta^2)     \right] ,  \nonumber  \\
 \Pho \Pho \Pht \Pht  &:&      -4 i \left(  \frac{3}{2} \lambda  s_\theta^2 c_\theta^2  + 6 \lambda_1 s_\theta^2 c_\theta^2  + 
  \lambda_2 (c_\theta^4  + s_\theta^4 - 4 c_\theta^2 s_\theta^2)  \right)  ;    \nonumber \\
  V_{1222}    & =&   \Pho \Pht^3   \left[   \lambda c_\theta s_\theta^3  - 4 \lambda_1 s_\theta c_\theta^3  +\frac{\lambda_2}{4} \sin 4 \theta  \right]  ,  \nonumber   \\
  \Pho \Pht \Pht \Pht  &: &  -6 i \left(  \lambda  c_\theta  s_\theta^3   - 4 \lambda_1  s_\theta c_\theta^3   +\frac{\lambda_2}{4} \sin 4 \theta \right).
\end{eqnarray}

%%%%%%%%%%%%%%%%%%%%%%%%%%%%%%%%%%%%%%%%%%%%%%%%%%%%%%%%%%%%%
\subsubsection{Counting of free parameters}

The most general scalar potential in Eq.~(\ref{eq:potential}) contains six parameters, $\lambda$, $M^2$, $\lambda_1$, $\lambda_2$, $\mu_1$, and $\mu_2$.  The masses of $\Pho$ and $\Pht$ and the mixing angle $\theta$ are determined by the three parameters $\lambda$, $M^2$, and $\mu_2$.  The physically most interesting trilinear scalar couplings, $\Pho\Pho\Pho$ (the triple-Higgs coupling) and $\Pht \Pho \Pho$ (which controls $\Pht \to \Pho \Pho$ decays, if kinematically accessible) depend in addition on $\mu_1$ and $\lambda_2$.  There is enough parameter freedom to choose these two trilinear couplings independently.  The remaining parameter $\lambda_1$ appears only in quartic scalar interactions.

Two useful bases in which the model can be specified are
\begin{equation}
	M_{\Pho}, M_{\Pht}, \cos\theta, \mu_1, \lambda_2, \lambda_1
\end{equation}
and [see Eqs.~(\ref{eq:111}) and~(\ref{eq:211})]
\begin{equation}
	M_{\Pho}, M_{\Pht}, \cos\theta, L_{111}, L_{211}, \lambda_1.
\end{equation}

%%%%%%%%%%%%%%%%%%%%%%%%%%%%%%%%%%%%%%%%%%%%%%%%%%%%%%%%%%%%%
\subsection{Benchmark 2: Scaling of vector boson and fermion couplings}

In the second benchmark scenario, the couplings of the $125\UGeV$ Higgs
boson to the gauge bosons and to the fermions are scaled by two
different coupling factors. 

\noindent
{\it Model-independent parameterization:}  For the $125\UGeV$ state
$\Pho$, the production rates in the various channels of observation can
be fit to two free parameters, 
\begin{eqnarray}
  \Ckappa_{\PVB} &\equiv& \Ckappa_{\PW} = \Ckappa_{\PZ} \;, \nonumber \\
  \Ckappa_{\PCF} &\equiv& \Ckappa_{\PQt} = \Ckappa_{\PQb} = \Ckappa_{\PGt} \; .
\end{eqnarray}
We assume that the $\Pho$ couplings to light fermions are scaled by the same factor $\Ckappa_{\PCF}$.
The couplings of $\Pho$ to $\Pg\Pg$ and $\PGg\PGg$ are also modified by the appropriate scaling factors applied to the fermion and $\PW$ boson loops.  

We assume that there are no new 
colored particles that contribute to the loop-induced $\Pho \Pg\Pg$ or $\Pht \Pg\Pg$ couplings.  In the 
specific model that we discuss below, there is a charged scalar that can contribute to the 
loop-induced $\Pho\PGg\PGg, \Pho\PGg \PZ$ and $\Pht\PGg\PGg, \Pht \PGg \PZ$ couplings; for this benchmark we assume that its contributions to these loop-induced couplings are negligible.

The corresponding couplings of the second state $\Pht$, as enforced by unitarity, are
\begin{eqnarray}
  \Ckappa_{\PVB}^{\prime} &=& \sqrt{1 - \Ckappa_{\PVB}^2}, \nonumber \\
  \Ckappa_{\PCF}^{\prime} &=& \frac{1 - \Ckappa_{\PVB} \Ckappa_{\PCF}}{\sqrt{1 - \Ckappa_{\PVB}^2}}.
\end{eqnarray}
We have chosen the phase of $\Pht$ such that $\Ckappa_{\PVB}^{\prime}$ is
positive.  $\Ckappa_{\PCF}^{\prime}$ can be positive or negative.  
As in the previous benchmark, there can be additional decays of $\Pht$ with a branching ratio BR$_{\mathrm{new}}$ (for example, $\Pht \to \Pho \Pho$ if kinematically allowed).

\noindent

{\it Constraints from existing data:}  The allowed ranges of
$\Ckappa_{\PVB}^{\prime}$ and $\Ckappa_{\PCF}^{\prime}$ are constrained by
fits of $\Ckappa_{\PVB}$ and $\Ckappa_{\PCF}$ from measurements of the 125 GeV
Higgs $\Pho$, as shown in Fig.~5a of \Bref{ATLAS-CONF-2013-034} and Fig.~13a
of \Bref{CMS-PAS-HIG-12-045} (in these figures $\Ckappa_{\PVB} \equiv
\kappa_{\PVB}$ and $\Ckappa_{\PCF} \equiv \kappa_{\PCF}$).  A theorist-made
translation of ATLAS, CMS, and Tevatron Higgs data available in July 2012 (see
\Bref{Espinosa:2012im} for the details of the data used in this analysis) into
constraints on $\Ckappa_{\PVB}^{\prime}$ and $\Ckappa_{\PCF}^{\prime}$ is
shown in \refF{fig:bm2}.  

\begin{figure}
\begin{center}
\includegraphics[width=0.7\textwidth]{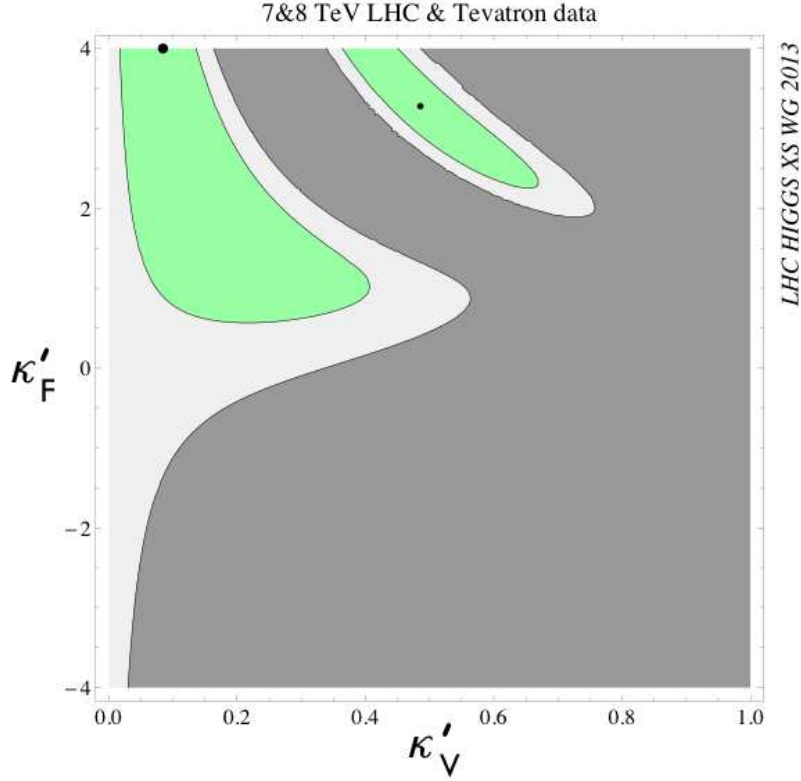}
\caption{Best-fit points (black dots), 1$\sigma$ (green) and 2$\sigma$ (white)
  allowed regions in 
  the $(\Ckappa_{\PVB}^{\prime}$, $\Ckappa_{\PCF}^{\prime})$ plane for Benchmark 2 (equivalent
  to the Type-I two-Higgs-doublet model), based on data available in
  July 2012 (see \Bref{Espinosa:2012im}).}
\label{fig:bm2}
\end{center}
\end{figure}

\noindent
{\it Specific model:} 
This parameterization is realized for the SM Higgs boson mixed with a
second scalar electroweak doublet that carries a vacuum expectation
value but does not couple to fermions; i.e., the Type-I Two Higgs
Doublet Model (2HDM). This model will be specified in the following
\refS{sec:2hdm}.

%%%%%%%%%%%%%%%%%%%%%%%%%%%%%%%%%%%%%%%
\subsubsection{Specific benchmark model 2: Type-I
  Two-Higgs-Doublet-Model \label{sec:2hdm}}
The simplest extensions of the SM Higgs sector are given by adding
scalar doublet and singlet fields. Thus in two-Higgs-doublet models
\cite{Lee:1973iz,Gunion:2002zf, Branco:2011iw} the Higgs sector consists of two Higgs doublet fields. 

%%%%%%%%%%%%%%%%%%%%%%%%%%%%%%%%%%%%%%%
\subsubsection{The scalar Higgs potential}
In the following we assume CP conservation. Furthermore, we take care
to avoid tree-level flavor changing neutral currents. The latter is achieved by
imposing a discrete symmetry ${\mathbb Z}_2$ under which one of the doublet fields
changes sign, while all the fermion fields remain unchanged.  Denoting by $\Phi_1$ and
$\Phi_2$ two hypercharge-one weak doublet fields, for the Type-I 2HDM we have in
particular:
\begin{eqnarray}
\underline{\mbox{Type-I 2HDM: }} \qquad 
\Phi_1 \to -\Phi_1 \quad \mbox{ and } \quad \Phi_2 \to
\Phi_2 \;. \label{eq:type1}
\end{eqnarray} 
This ensures that the fermions couple only to $\Phi_2$.

The most general scalar Higgs potential is then given by
\begin{eqnarray}
V &=& m_{11}^2 \Phi_1^\dagger \Phi_1 + m_{22}^2 \Phi_2^\dagger \Phi_2
- [m_{12}^2 \Phi_1^\dagger \Phi_2 + {\mbox h.c.}] + \frac{1}{2}
\lambda_1 (\Phi_1^\dagger \Phi_1)^2 + \frac{1}{2} \lambda_2
(\Phi_2^\dagger \Phi_2) \nonumber \\
&& + \lambda_3 (\Phi_1^\dagger \Phi_1) (\Phi_2^\dagger \Phi_2) +
\lambda_4 (\Phi_1^\dagger \Phi_2) (\Phi_2^\dagger \Phi_1) + \left\{
  \frac{1}{2} \lambda_5 (\Phi_1^\dagger \Phi_2)^2 + {\mbox h.c.}
\right\} \;.
\end{eqnarray}
This potential softly violates (by dimension-two terms) the imposed
discrete ${\mathbb Z}_2$ symmetry. The parameters $m_{12}^2$ and $\lambda_5$ are taken
real to ensure CP conservation. Requiring that the minimum of the
potential is given by a CP-conserving vacuum which does not break the
electromagnetic symmetry $U(1)_{\mathrm em}$, we have for the vacuum
expectation values of the neutral components $\Phi_i^0$ ($i=1,2$) of the two Higgs
doublets, 
\begin{eqnarray}
\langle \Phi_i^0 \rangle = \frac{v_i}{\sqrt{2}} \quad \mbox{with}
\quad \tan\beta \equiv \frac{v_2}{v_1} \quad \mbox{and} \quad v^2 =
v_1^2 + v_2^2 = (246 \mbox{ \UGeV})^2 \;.
\end{eqnarray}
By imposing the minimum conditions of the potential the parameters
$m_{11}^2$ and $m_{22}^2$ can be eliminated in favor of $v^2$ and
$\tan\beta$, so that we are left with six free parameters, $m_{12}^2$
and $\lambda_j$ ($j=1,...,5$). Rotation of the interaction states to
the mass eigenstates results in four physical Higgs bosons, two
neutral CP-even ones, denoted by $\Ph$ for the lighter and by $\PH$ for
the heavier one, one neutral CP-odd state $\PA$ and a charged Higgs
boson $\PH^{\pm}$. The masses of these four particles, together with 
the neutral CP-even Higgs mixing angle
$\alpha$ introduced to diagonalize the neutral CP-even Higgs squared-mass
matrix, can be expressed in terms of these six parameters, so
that one free parameter is left over.
For further discussion on the
scalar potential, its symmetries and bounds on the parameters we
refer to the literature \cite{Gunion:2002zf, Branco:2011iw}. Here we content ourselves
to give the Higgs couplings to gauge bosons and fermions. 

The couplings of the neutral CP-even Higgs bosons $\Ph,\PH$ to the gauge
bosons $\PVB$ ($\PVB=\PW^\pm, \PZ$) normalized to the corresponding SM coupling are given by
\begin{eqnarray}
g_{\Ph\PVB\PVB} = \sin (\beta-\alpha) \qquad \mbox{and} \qquad g_{\PH\PVB\PVB} = \cos
(\beta - \alpha) \; . \label{eq:gaugecouplgs}
\end{eqnarray}
The pseudoscalar $\PA$ does not couple to gauge bosons. With the
notation of \refS{sec:unitarity1} we identify
\begin{eqnarray}
\Ckappa_{\PVB} \equiv g_{\Ph\PVB\PVB} \qquad \mbox{and} \qquad \Ckappa^\prime_{\PVB} \equiv g_{\PH\PVB\PVB}
\end{eqnarray}
The couplings
Eq.~\refE{eq:gaugecouplgs} fulfill the constraint Eq.~\refE{eq:cond1}
imposed by the requirement of unitarity of the longitudinal gauge
boson scattering. Furthermore, we recover the SM limit for $h$ in case
$\sin (\beta-\alpha) =1$. Note, that there can also be scenarios where
$H$ corresponds to the $125\UGeV$ SM-like Higgs boson, in which case $\cos
(\beta-\alpha)$ is close to $1$. 

As stated above, 2HDMs suffer from possible tree-level
flavor-changing neutral currents (FCNC). According to the
Paschos--Glashow--Weinberg theorem \cite{Glashow:1976nt, Paschos:1976ay}, FCNC are absent if
all fermions with the same quantum numbers couple to the same Higgs
multiplet. In the 2HDMs this can be achieved by imposing discrete or
continuous symmetries. According to the symmetries imposed there are
different types of 2HDM, where we discuss here the type-I model,
defined in Eq.~\refE{eq:type1}. In this case the Yukawa Lagrangian is
given in the mass basis by~\cite{Aoki:2009ha}
\begin{eqnarray}
{\mathcal L}_{\rm Yukawa} &=& - \sum_{\Pf=\PQu,\PQd,\Pl} \frac{m_{\Pf}}{v} \left( \xi_h^{\Pf}
  \bar{\Pf} \Pf \Ph + \xi_{\PH}^{\Pf} \bar{\Pf} \Pf \PH - i\xi_{\PA}^{\Pf} \bar{\Pf} 
\gamma_5 \Pf \PA
\right) \nonumber \\
&& - \left\{ \frac{\sqrt{2} V_{\PQu\PQd}}{v} \bar{u} (m_{\PQu} \xi_A^{\PQu} P_L + m_{\PQd}
\xi_{\PA}^d P_R) d \, \PH^+ + \frac{\sqrt{2} m_{\Pl} \xi_{\PA}^{\Pl}}{v} \bar{\nu}_L {\Pl}_R
H^+ + \mbox{ h.c. } \right\} \;,
\end{eqnarray}
where $\PQu,\PQd,\Pl$ stand generically for up-type quarks, down-type quarks, and charged
leptons of all three generations, respectively, $P_{L,R}$ are the projection operators
of the left- and right-handed fermions,  and $V_{\PQu\PQd}$ denotes the appropriate element
of the CKM matrix. We have
\begin{eqnarray}
\underline{\mbox{Type-1 2HDM:}} \quad && \xi_h^u = \xi_h^d = \xi_h^l \equiv
\xi_h = \frac{\cos \alpha}{\sin \beta} \label{eq:fermioncouplgs1}\\
&& \xi_H^u = \xi_H^d = \xi_H^l \equiv \xi_H = \frac{\sin \alpha}{\sin \beta} \;.
\label{eq:fermioncouplgs2}
\end{eqnarray}
In our notation of \refS{sec:unitarity1} we have
\begin{eqnarray}
\Ckappa_{\Pf} \equiv \xi_h \qquad \mbox{and} \qquad \Ckappa^\prime_{\Pf} \equiv \xi_H \;.
\end{eqnarray}
It is easy to verify that the Yukawa couplings
Eq.~\refE{eq:fermioncouplgs1}, Eq.~\refE{eq:fermioncouplgs2} and the
couplings to gauge bosons 
Eq.~\refE{eq:gaugecouplgs} fulfill the unitarity conditions Eq.~\refE{eq:cond2}.

It is instructive to look the decoupling limit, {\it i.e.}, when $m_{\PH} \gg m_{\Ph}$ which is obtained for $\alpha \to \beta -\pi/2$:
\begin{eqnarray}
& \displaystyle \Ckappa_{\PVB}  = 1 +  {\mathcal O}\!\left(  \frac{\MZ^4}{\MH^4} \right), \\
& \displaystyle \Ckappa_{\Pf} = 1+ 2\, \frac{\MZ^2}{\MH^2}\cos^2\!\beta \cos 2 \beta +
  {\mathcal O}\!\left(  \frac{\MZ^4}{\MH^4} \right).
\end{eqnarray}

%%%%%%%%%%%%%%%%%%%%%%%%%%%%%%%%%%%%%%
\subsection{Tools}
Here we give a short collection of tools which can be exploited in the
frameworks of the benchmark models 1 and 2. 

\noindent {\it Production:} As in the proposed benchmark models only
the Higgs couplings to the gauge bosons and fermions are modified, the
SM production cross sections can be taken over including higher order
QCD corrections. They only have to be multiplied by the appropriate
scaling factor. Since in both models the fermion modification factors for
each of the two Higgs bosons are universal, this is also  particularly easy for gluon
fusion, as not distinction between top and bottom
loops needs to be made. As for electroweak (EW) corrections, however, they
cannot be taken over from SM calculations.EW corrections in the BSM models can be substantially different from the SM
EW corrections.  As long as no dedicated analysis has been
performed, no statements about the possible size of these corrections can be made. Hence,
we have in benchmark model 1 and 2 for the QCD corrected production
cross sections of $\Pho$, $\Pht$ through gluon fusion ($\Pg\Pg$), vector boson
fusion (VBF), Higgs-strahlung (VH) and associated production with
heavy quarks (QQH),
\begin{eqnarray}
\sigma^h_{\Pg\Pg} = {\bar \Ckappa}_{\Pf}^2\, \sigma^{\SM,\QCD}_{\Pg\Pg} \; , \;
\sigma^h_{\mathrm{VBF}} = {\bar \Ckappa}_{\PVB}^2\, \sigma^{\SM,\QCD}_{\mathrm{VBF}} \; , \;
\sigma^h_{\mathrm{VH}} = {\bar \Ckappa}_{\PVB}^2\, \sigma^{\SM,\QCD}_{\mathrm{VH}} \; , \;
\sigma^h_{\mathrm{QQH}} = {\bar \Ckappa}_{\Pf}^2\, \sigma^{\SM,\QCD}_{\mathrm{QQH}} \;,
\end{eqnarray}
where
\begin{eqnarray}
{\bar \Ckappa}_{\Pf} = \Ckappa_{\Pf} \;(\Ckappa^\prime_{\Pf}) \quad \mbox{for} \quad h=\Pho \;(\Pht) \qquad
\mbox{and} \qquad
{\bar \Ckappa}_{\PVB} = \Ckappa_{\PVB} \;(\Ckappa^\prime_{\PVB}) \quad \mbox{for} \quad h=\Pho \;(\Pht) \;.
\end{eqnarray}
For the programs which allow for the calculation of the production
cross sections at higher order QCD we refer the reader to
YR1. 

The program \textsc{SUSHI} \cite{Harlander:2012pb, sushi} has implemented
the calculation of neutral Higgs bosons $\Ph,\PH,\PA$ within the 2HDM
through gluon fusion and bottom quark annihilation. This program can be applied up to
NNLO QCD if the Higgs mass stays below twice the top quark
mass (this is because the NNLO QCD corrections rely upon an approximation which is valid for
$M_{\Ph} < 2 m_{\PQt}$). The NLO QCD corrections are exact for all Higgs boson
masses. Electroweak corrections have to be turned off for consistency
reasons, since the electroweak corrections depend on the details of the BSM model
(including couplings between scalars that are highly model-dependent) and have not been implemented.

The program \textsc{gg2VV} \cite{Kauer:2012hd,gg2VV} is a parton-level integrator and event 
generator for all $\Pg\Pg\ (\to \PH)\to \PV\PV \to 4$ leptons processes 
$(\PV\!=\!\PW,\PZ/\PGg^\ast)$, which can be used to study Higgs-continuum 
interference and off-shell effects.
It can be used to calculate predictions for BSM scenarios with
a SM-like Higgs boson with rescaled $\PH\Pg\Pg$, $\PH\PW\PW$ and 
$\PH\PZ\PZ$ couplings.  Full BSM implementations in
\textsc{gg2VV} are not public yet. The program has, however, been used already for
calculations and checks in models in which all Higgs
couplings are modified by a common scaling factor. 

\noindent {\it Decays:} For the calculation of the decay branching
ratios in benchmark model 1 the program \textsc{HDECAY} \cite{hdecay} can
be used. It has implemented the possibility to turn on modification
factors for the SM Higgs couplings to fermions and gauge bosons. This
way the branching ratios for $\Pho$ and $\Pht$ can be calculated
separately by specifying in the input file the appropriate scaling
factors. The program includes both QCD and EW corrections computed
for the SM Higgs, which can only be applied in the vicinity of the SM. 
The new program \textsc{eHDECAY}~\cite{Contino:2013kra, ehdecay} has been adapted from 
\textsc{HDECAY} to implement the
calculation of Higgs branching ratios in the framework of effective
Lagrangians.  \textsc{eHDECAY} provides the possibility to turn off the SM EW
corrections, which should be done in the benchmark models described here.  By specifying in the input file the appropriate scaling factors and setting the flag for EW corrections to zero, the branching ratios for
$\Pho$, $\Pht$ can be separately calculated 
including the QCD corrections and without the EW corrections. However, neither
program has implemented the decays into new states as {\it e.g.}
the decay $\Pht \to \Pho \Pho$. The authors of these programs plan to add this 
decay in the future. 

For the calculation of the branching ratios in the 2HDM there is a
dedicated tool, \textsc{2HDMC}~\cite{Eriksson:2009ws, 2hdmc}. This program calculates all two-body
decay widths and branching ratios at leading order (including FCNC) within
different parameterizations which can be specified in the input
file. Leading QCD corrections to the decays are included. The program
also includes singly off-shell decays into scalar-plus-vector-boson and
two-vector-boson final states. 
Furthermore, theoretical constraints
on the 2HDM parameters from perturbativity, unitarity and stability
of the scalar potential are included, as are the constraints from EW
precision data (via the oblique parameters). Model constraints from
Higgs searches and flavor physics can be accessed using external
codes. The code provides output in a form similar to the SUSY Les
Houches accord~\cite{Skands:2003cj, Allanach:2008qq, Mahmoudi:2010iz}, 
which can also be used for Monte Carlo event
generation with a supplied model file for \textsc{MadGraph/MadEvent}.

\newpage
%- {{{ local definitions:

%%%%%% Pietro & Robert:

\providecommand{\abbrev}{} 
\providecommand{\code}{\sc}
\providecommand{\higlu}{{\code HIGLU}}
\providecommand{\sushi}{{\code SusHi}}
\providecommand{\powheg}{{\code POWHEG}}
\providecommand{\pythia}{{\code PYTHIA}}
\providecommand{\msbar}{\overline{\rm MS}}
\providecommand{\muR}{\mu_{\text{R}}}
\providecommand{\muF}{\mu_{\text{F}}}
\providecommand{\sm}{{\abbrev SM}}
\providecommand{\mssm}{{\abbrev MSSM}}
\providecommand{\susy}{{\abbrev SUSY}}
\providecommand{\lo}{{\abbrev LO}}
\providecommand{\nlo}{{\abbrev NLO}}
\providecommand{\nnlo}{{\abbrev NNLO}}
\providecommand{\qcd}{{\abbrev QCD}}
\providecommand{\ew}{{\abbrev EW}}
\providecommand{\pdf}{{\abbrev PDF}}
\providecommand{\fsc}[1]{#1{\abbrev FS}}
\providecommand{\MSQ}{M_{\PSQ}}

%%%%%% Sven:

\providecommand{\eqn}[1]{Eq.\,(\ref{#1})}
\providecommand{\eqns}[2]{Eqs.\,(\ref{#1}) -- (\ref{#2})}
\providecommand{\refF}[1]{Figure~\ref{#1}}
\providecommand{\refFs}[2]{Figures~\ref{#1} -- \ref{#2}}
\providecommand{\refT}[1]{Table~\ref{#1}}
\providecommand{\refTs}[2]{Tables~\ref{#1} -- \ref{#2}}
\providecommand{\refS}[1]{Section~\ref{#1}}
\providecommand{\refSs}[2]{Sections~\ref{#1} -- \ref{#2}}
\providecommand{\refC}[1]{Chapter~\ref{#1}}
\providecommand{\refCs}[2]{Chapters~\ref{#1} -- \ref{#2}}
\providecommand{\refA}[1]{Appendix~\ref{#1}}
\providecommand{\refAs}[2]{Appendices~\ref{#1} -- \ref{#2}}
\providecommand{\mass}{125.5}
\providecommand{\nomix}{\emph{no-mixing}}
\providecommand{\saeff}{\emph{small $\aeff$}}
\providecommand{\gluophobic}{\emph{gluophobic Higgs}}
\providecommand{\lstop}{\emph{light stop}}
\providecommand{\lstau}{\emph{light stau}}
\providecommand{\tauphobic}{\emph{$\tau$-phobic Higgs}}
\providecommand{\lowMH}{\emph{low-$\MH$}}
\providecommand{\cpm}{\emph{$\cp$-mixing}}
\providecommand{\citere}[1]{\mbox{Ref.~\cite{#1}}}
\providecommand{\citeres}[1]{\mbox{Refs.~\cite{#1}}}
\providecommand{\mst}{m_{\tilde{\PQt}}}
\providecommand{\mstr}{m_{\tilde{\PQt}_{\mathrm R}}}
\providecommand{\mstl}{m_{\tilde{\PQt}_{\mathrm L}}}
\providecommand{\mste}{m_{\tilde{\PQt}_1}}
\providecommand{\mstz}{m_{\tilde{\PQt}_2}}
\providecommand{\msti}{m_{\tilde{\PQt}_i}}
\providecommand{\mstj}{m_{\tilde{\PQt}_j}}
\providecommand{\msb}{m_{\tilde{\PQb}}}
\providecommand{\msbr}{m_{\tilde{\PQb}_{\mathrm R}}}
\providecommand{\msbl}{m_{\tilde{\PQb}_{\mathrm L}}}
\providecommand{\msbe}{m_{\tilde{\PQb}_1}}
\providecommand{\msbeOS}{m_{\tilde{\PQb}_{1,{\rm OS}}}}
\providecommand{\msbz}{m_{\tilde{\PQb}_2}}
\providecommand{\msbi}{m_{\tilde{\PQb}_i}}
\providecommand{\msbj}{m_{\tilde{\PQb}_j}}
\providecommand{\mstau}{m_{\tilde{\PGt}}}
\providecommand{\mstaue}{m_{\tilde{\PGt}_1}}
\providecommand{\mstauz}{m_{\tilde{\PGt}_2}}
\providecommand{\mstaui}{m_{\tilde{\PGt}_i}}
\providecommand{\msneut}{m_{\tilde{\PGn}_{\PGt}}}
\providecommand{\msneum}{m_{\tilde{\PGn}_{\PGm}}}
\providecommand{\msneue}{m_{\tilde{\PGn}_{\Pe}}}
\providecommand{\MstL}{M_{\tilde{\PQt}_{\mathrm L}}}
\providecommand{\MstR}{M_{\tilde{\PQt}_{\mathrm R}}}
\providecommand{\MsbL}{M_{\tilde{\PQb}_{\mathrm L}}}
\providecommand{\MsbR}{M_{\tilde{\PQb}_{\mathrm R}}}
\providecommand{\MsqL}{M_{\tilde{\PQq}_{\mathrm L}}}
\providecommand{\MsqR}{M_{\tilde{\PQq}_{\mathrm R}}}
\providecommand{\MsfL}{M_{\tilde{\Pf}_{\mathrm L}}}
\providecommand{\MsfR}{M_{\tilde{\Pf}_{\mathrm R}}}
\providecommand{\MstauL}{M_{\tilde{\PGt}_{\mathrm L}}}
\providecommand{\MstauR}{M_{\tilde{\PGt}_{\mathrm R}}}
\providecommand{\MsneutL}{M_{\tilde{\PGn}_{\mathrm L}}}
\providecommand{\At}{A_{\PQt}}
\providecommand{\Ab}{A_{\PQb}}
\providecommand{\Atau}{A_{\PGt}}
\providecommand{\Aq}{A_{\PQq}}
\providecommand{\Af}{A_{\Pf}}
\providecommand{\Xt}{X_{\PQt}}
\providecommand{\Xb}{X_{\PQb}}
\providecommand{\Xtau}{X_{\PGt}}
\providecommand{\Xq}{X_{\PQq}}
\providecommand{\mf}{m_f}
\providecommand{\msfr}{m_{\tilde{\Pf}_{\mathrm R}}}
\providecommand{\msl}{m_{\tilde{\Pl}}}
\providecommand{\mslr}{m_{\tilde{\Pl}_{\mathrm R}}}
\providecommand{\msll}{m_{\tilde{\Pl}_{\mathrm L}}}
\providecommand{\msle}{m_{\tilde{\Pl}_1}}
\providecommand{\mslz}{m_{\tilde{\Pl}_2}}
\providecommand{\msli}{m_{\tilde{\Pl}_i}}
\providecommand{\mslk}{m_{\tilde{\Pl}_k}}
\providecommand{\mslj}{m_{\tilde{\Pl}_j}}
\providecommand{\msn}{m_{\tilde{\PGn}_l}}
\providecommand{\MslL}{M_{\tilde{\Pl}_{\mathrm L}}}
\providecommand{\MslR}{M_{\tilde{\Pl}_{\mathrm R}}}
\providecommand{\MsnL}{M_{\tilde{\PGn}_{\mathrm L}}}
\providecommand{\Al}{A_l}
\providecommand{\Xl}{X_l}
\providecommand{\msfl}{m_{\tilde{\Pf}_{\mathrm L}}}
\providecommand{\msf}{m_{\tilde{\Pf}}}
\providecommand{\msfi}{m_{\tilde{\Pf}_i}}
\providecommand{\msfe}{m_{\tilde{\Pf}_1}}
\providecommand{\msfz}{m_{\tilde{\Pf}_2}}
\providecommand{\ms}{M_S}
\providecommand{\msusy}{M_{\rm SUSY}}
\providecommand{\msqd}{M_{\tilde{\PQq}_3}}
\providecommand{\msld}{M_{\tilde{\Pl}_3}}
\providecommand{\msqez}{M_{\tilde{\PQq}_{1,2}}}
\providecommand{\mslez}{M_{\tilde{\Pl}_{1,2}}}
\providecommand{\msq}{m_{\tilde{\PQq}}}
\providecommand{\msqi}{m_{\tilde{\PQq}_i}}
\providecommand{\msqip}{m_{\tilde{\PQq}_{i'}}}
\providecommand{\msqj}{m_{\tilde{\PQq}_j}}
\providecommand{\msqe}{m_{\tilde{\PQq}_1}}
\providecommand{\msqz}{m_{\tilde{\PQq}_2}}
\providecommand{\mslepe}{m_{\tilde{\Pl}_1}}
\providecommand{\mslepz}{m_{\tilde{\Pl}_2}}
\providecommand{\msneu}{m_{\tilde{\PGn}_l}}
\providecommand{\Sf}{\tilde{\Pf}}
\providecommand{\sfl}{\tilde{\Pf}_{\mathrm L}}
\providecommand{\sfr}{\tilde{\Pf}_{\mathrm R}}
\providecommand{\sfe}{\tilde{\Pf}_1}
\providecommand{\sfz}{\tilde{\Pf}_2}
\providecommand{\sfi}{\tilde{\Pf}_i}
\providecommand{\sfj}{\tilde{\Pf}_j}
\providecommand{\sfez}{\tilde{\Pf}_{12}}
\providecommand{\sfze}{\tilde{\Pf}_{21}}
\providecommand{\Sl}{\tilde{\Pl}}
\providecommand{\Sn}{\tilde{\nu_l}}
\providecommand{\sll}{\tilde{\Pl}_{\mathrm L}}
\providecommand{\slr}{\tilde{\Pl}_{\mathrm R}}
\providecommand{\sle}{\tilde{\Pl}_1}
\providecommand{\slz}{\tilde{\Pl}_2}
\providecommand{\sli}{\tilde{\Pl}_i}
\providecommand{\slj}{\tilde{\Pl}_j}
\providecommand{\slk}{\tilde{\Pl}_k}
\providecommand{\sflz}{\tilde{\Pl}_{12}}
\providecommand{\sfle}{\tilde{\Pl}_{21}}
\providecommand{\Hi}{h_i}
\providecommand{\cHe}{\cH_1}
\providecommand{\cHz}{\cH_2}
\providecommand{\He}{h_1}
\providecommand{\Hz}{h_2}
\providecommand{\Hd}{h_3}
\providecommand{\mHe}{m_{\He}}
\providecommand{\mHz}{m_{\Hz}}
\providecommand{\mHd}{m_{\Hd}}
\providecommand{\MHe}{M_{\He}}
\providecommand{\MHz}{M_{\Hz}}
\providecommand{\MHd}{M_{\Hd}}
\providecommand{\SU}{\mathrm {SUSY}}
\providecommand{\SM}{\mathrm {SM}}
\providecommand{\MSSM}{\mathrm {MSSM}}
\providecommand{\hi}{\mathrm {Higgs}}
\providecommand{\msbar}{$\overline{\rm{MS}}$}
\providecommand{\msbarm}{\overline{\rm{MS}}}
\providecommand{\drbar}{$\overline{\rm{DR}}$}
\providecommand{\drbarm}{\overline{\rm{DR}}}
\providecommand{\DRbar}{\ensuremath{\overline{\mathrm{DR}}}}
\providecommand{\MSbar}{\ensuremath{\overline{\mathrm{MS}}}}
\providecommand{\MSbarb}{\ensuremath{\overline{\mathbf{MS}}}}
\providecommand{\OS}{\mathrm{OS}}
\providecommand{\os}{\mathrm{os}}
\providecommand{\mudim}{\mu_{\DRbar}}
\providecommand{\cp}{{\cal CP}}
\providecommand{\cA}{{\cal A}}
\providecommand{\cH}{{\cal H}}
\providecommand{\cHi}{{\cal H}_i}
\providecommand{\cL}{{\cal L}}
\providecommand{\cM}{{\cal M}}
\providecommand{\cZ}{{\cal Z}}
\providecommand{\bcZ}{\bar{\cal Z}}
\providecommand{\wz}{\sqrt{2}}
\providecommand{\edz}{\tfrac{1}{2}}
\providecommand{\tedz}{\tfrac{1}{2}}
\providecommand{\twol}{two-loop}
\providecommand{\onel}{one-loop}
\providecommand{\mma}{{\em Mathematica}}
\providecommand{\tc}{{\em TwoCalc}}
\providecommand{\fa}{{\sc FeynArts}}
\providecommand{\fc}{{\sc FormCalc}}
\providecommand{\lt}{{\sc LoopTools}}
\providecommand{\fh}{{\sc FeynHiggs}}
\providecommand{\fhto}{{\em FeynHiggs2.1}}
\providecommand{\fhtt}{{\sc FeynHiggs\,2.5}}
\providecommand{\fhf}{{\em FeynHiggsFast}}
\providecommand{\fhfc}{{\em FeynHiggsFastC}}
\providecommand{\cpsh}{{\em CPsuperH}}
\providecommand{\rp}{$\rho\,$-parameter}
\providecommand{\MW}{M_{\PW}}
\providecommand{\MZ}{M_{\PZ}}
\providecommand{\mA}{m_{\PA}}
\providecommand{\MA}{M_{\PA}}
\providecommand{\mh}{m_{\Ph}}
\providecommand{\mH}{m_{\PH}}
\providecommand{\Mh}{M_{\Ph}}
\providecommand{\MH}{M_{\PH}}
\providecommand{\MHp}{M_{\PH^\pm}}
\providecommand{\MHpm}{M_{\PH^\pm}}
\providecommand{\mHp}{m_{\PH^\pm}}
\providecommand{\mhmax}{\ensuremath{m_{\Ph}^{\rm max}}}
\providecommand{\mhmod}{\ensuremath{m_{\Ph}^{\rm mod}}}
\providecommand{\mhmodp}{\ensuremath{m_{\Ph}^{\rm mod+}}}
\providecommand{\mhmodm}{\ensuremath{m_{\Ph}^{\rm mod-}}}
\providecommand{\mt}{m_{\PQt}}
\providecommand{\mto}{m_{\PQt}}
\providecommand{\mtms}{\overline{m}_\PQt}
\providecommand{\mq}{m_{\PQq}}
\providecommand{\mb}{m_{\PQb}}
\providecommand{\mbms}{\overline{m}_\Pb}
\providecommand{\mtau}{m_{\PGt}}
\providecommand{\mmu}{m_{\PGm}}
\providecommand{\mel}{m_{\PGe}}
\providecommand{\ml}{m_\Pl}
\providecommand{\gl}{\tilde{\Pg}}
\providecommand{\Mgl}{M_{\tilde{\Pg}}}
\providecommand{\mgl}{m_{\tilde{\Pg}}}
\providecommand{\Sferm}{\tilde{\Pf}}
\providecommand{\Sfermp}{\tilde{\Pf}'}
\providecommand{\sq}{\tilde{\PQq}}
\providecommand{\sqi}{\tilde{\PQq}_i}
\providecommand{\sqj}{\tilde{\PQq}_j}
\providecommand{\sql}{\tilde{\PQq}_{\mathrm L}}
\providecommand{\sqr}{\tilde{\PQq}_{\mathrm R}}
\providecommand{\sqe}{\tilde{\PQq}_1}
\providecommand{\sqz}{\tilde{\PQq}_2}
\providecommand{\slep}{\tilde{\Pl}}
\providecommand{\Stop}{\tilde{\PQt}}
\providecommand{\StopL}{\tilde{\PQt}_{\mathrm L}}
\providecommand{\StopR}{\tilde{\PQt}_{\mathrm R}}
\providecommand{\Stope}{\tilde{\PQt}_1}
\providecommand{\aStope}{\tilde{\PQt}^\dagger_1}
\providecommand{\Stopz}{\tilde{\PQt}_2}
\providecommand{\aStopz}{\tilde{\PQt}^\dagger_2}
\providecommand{\Stopi}{\tilde{\PQt}_i}
\providecommand{\Stopj}{\tilde{\PQt}_j}
\providecommand{\Sbot}{\tilde{\PQb}}
\providecommand{\SbotL}{\tilde{\PQb}_{\mathrm L}}
\providecommand{\SbotR}{\tilde{\PQb}_{\mathrm R}}
\providecommand{\Sbote}{\tilde{\PQb}_1}
\providecommand{\Sbotz}{\tilde{\PQb}_2}
\providecommand{\aSboti}{\tilde{\PQb}^\dagger_i}
\providecommand{\aSbotj}{\tilde{\PQb}^\dagger_j}
\providecommand{\aSbote}{\tilde{\PQb}^\dagger_1}
\providecommand{\aSbotz}{\tilde{\PQb}^\dagger_2}
\providecommand{\Sboti}{\tilde{\PQb}_i}
\providecommand{\Sbotj}{\tilde{\PQb}_j}
\providecommand{\Stau}{\tilde \PGt}
\providecommand{\Staum}{\tilde{\PGt}^-}
\providecommand{\Stauem}{\tilde{\PGt}^-_1}
\providecommand{\Stauzm}{\tilde{\PGt}^-_2}
\providecommand{\Stauim}{\tilde{\PGt}^-_i}
\providecommand{\aStaue}{\tilde{\PGt}^+_1}
\providecommand{\aStauz}{\tilde{\PGt}^+_2}
\providecommand{\Staue}{\tilde{\PGt}_1}
\providecommand{\Stauz}{\tilde{\PGt}_2}
\providecommand{\StauL}{\tilde{\PGt}_{\mathrm L}}
\providecommand{\StauR}{\tilde{\PGt}_{\mathrm R}}
\providecommand{\Smue}{\tilde{\PGm}_1}
\providecommand{\Sele}{\tilde{\Pe}_1}
\providecommand{\Sneu}{\tilde \PGn_{\Pl}}
\providecommand{\Sneue}{\tilde \PGn_{\Pe}}
\providecommand{\Sneum}{\tilde \PGn_\mu}
\providecommand{\Sneut}{\tilde{\PGn}_\PGt}
\providecommand{\db}{\De_{\PQb}}
\providecommand{\dst}{\Delta_{\tilde{\PQt}}}
\providecommand{\tst}{\theta_{\tilde{\PQt}}}
\providecommand{\tsb}{\theta_{\tilde{\PQb}}}
\providecommand{\tsf}{\theta\kern-.20em_{\tilde{\Pf}}}
\providecommand{\tsfp}{\theta\kern-.20em_{\tilde{\Pf}\prime}}
\providecommand{\tsq}{\theta\kern-.15em_{\tilde{\PQq}}}
\providecommand{\sweff}{\sin^2\theta_{\mathrm{eff}}}
\providecommand{\cha}[1]{\tilde \chi^\pm_{#1}}
\providecommand{\chap}[1]{\tilde \chi^+_{#1}}
\providecommand{\cham}[1]{\tilde \chi^-_{#1}}
\providecommand{\mcha}[1]{m_{\tilde \chi^\pm_{#1}}}
\providecommand{\neu}[1]{\tilde \chi^0_{#1}}
\providecommand{\mneu}[1]{m_{\tilde \chi^0_{#1}}}
\providecommand{\br}{\mathrm{BR}}
\providecommand{\tev}{\,\, \mathrm{TeV}}
\providecommand{\gev}{\,\, \mathrm{GeV}}
\providecommand{\mev}{\,\, \mathrm{MeV}}
\providecommand{\BC}{\begin{center}}
\providecommand{\EC}{\end{center}}
\providecommand{\BE}{\begin{equation}}
\providecommand{\BEA}{\begin{eqnarray}}
\providecommand{\EEA}{\end{eqnarray}}
\providecommand{\non}{\nonumber}
\providecommand{\id}{{\rm 1\kern-.12em
\rule{0.3pt}{1.5ex}\raisebox{0.0ex}{\rule{0.1em}{0.3pt}}}}
\providecommand{\lsim}
{\;\raisebox{-.3em}{$\stackrel{\displaystyle <}{\sim}$}\;}
\providecommand{\gsim}
{\;\raisebox{-.3em}{$\stackrel{\displaystyle >}{\sim}$}\;}
\providecommand{\matr}[1]{{\mathbf{#1}}}
\providecommand{\musteq}{\stackrel{!}{=}}
\providecommand{\setR}{\ensuremath{\mathbb{R}}}
\providecommand{\setC}{\ensuremath{\mathbb{C}}}
\providecommand{\VHiggs}{V_H}
\providecommand{\FirstOrder}{\mathcal{O}_{(1)}}
\providecommand{\SecondOrder}{\mathcal{O}_{(2)}}
\providecommand{\tstop}{$t$-$\tilde{\PQt}$}
\providecommand{\bsbot}{$b$-$\tilde{\PQb}$}
\providecommand{\fsf}{$f$-$\tilde{\Pf}$}
\providecommand{\pzero}{$p^2$=0}
\providecommand{\MSusy}{M_{\mathrm{Susy}}}
\providecommand{\MOne}{M_1}
\providecommand{\MTwo}{M_2}
\providecommand{\MThree}{M_3}
\providecommand{\mGl}{m_{\tilde{\Pg}}}
\providecommand{\absXt}{|X_t|}
\providecommand{\argXt}{\arg(X_t)}
\providecommand{\reXt}{\Re X_t}
\providecommand{\imXt}{\Im X_t}
\providecommand{\reAt}{\Re A_t}
\providecommand{\imAt}{\Im A_t}
\providecommand{\absAt}{|A_t|}
\providecommand{\argAt}{\arg(A_t)}
\providecommand{\reA}{\Re A_0}
\providecommand{\imA}{\Im A_0}
\providecommand{\absA}{|A_0|}
\providecommand{\argA}{\arg(A_0)}
\providecommand{\absmu}{|\PGm|}
\providecommand{\argmu}{\arg(\PGm)}
\providecommand{\al}{\alpha}
\providecommand{\be}{\beta}
\providecommand{\Ga}{\Gamma}
\providecommand{\ga}{\gamma}
\providecommand{\tb}{\tan\beta}
\providecommand{\De}{\Delta}
\providecommand{\tadH}{T_H}
\providecommand{\tadh}{T_h}
\providecommand{\tadA}{T_A}
\providecommand{\tadG}{T_G}
\providecommand{\tanb}{\tan \beta\,}
\providecommand{\sinb}{\sin \beta\,}
\providecommand{\cosb}{\cos \beta\,}
\providecommand{\sinbsq}{\sin^2 \!\beta\,}
\providecommand{\cosbsq}{\cos^2 \!\beta\,}
\providecommand{\sina}{\sin \alpha\,}
\providecommand{\cosa}{\cos \alpha\,}
\providecommand{\tana}{\tan \alpha\,}
\providecommand{\sinasq}{\sin^2 \!\alpha\,}
\providecommand{\cosasq}{\cos^2 \!\alpha\,}
\providecommand{\sw}{s_\mathrm{w}}
\providecommand{\cw}{c_\mathrm{w}}
\providecommand{\betan}{\beta_\mathrm{n}}
\providecommand{\betac}{\beta_\mathrm{c}}
\providecommand{\dmZsq}{\delta m_Z^2}
\providecommand{\dMZsq}{\delta \MZ^2}
\providecommand{\dmWsq}{\delta m_W^2}
\providecommand{\dMWsq}{\delta \MW^2}
\providecommand{\dmhsq}{\delta m_h^2}
\providecommand{\dmhHsq}{\delta m_{hH}^2}
\providecommand{\dmHhsq}{\delta m_{Hh}^2}
\providecommand{\dmhGsq}{\delta m_{hG}^2}
\providecommand{\dmhAsq}{\delta m_{hA}^2}
\providecommand{\dmHsq}{\delta m_H^2}
\providecommand{\dmHGsq}{\delta m_{HG}^2}
\providecommand{\dmHAsq}{\delta m_{HA}^2}
\providecommand{\dmAsq}{\delta m_A^2}
\providecommand{\dmAGsq}{\delta m_{AG}^2}
\providecommand{\dmGsq}{\delta m_G^2}
\providecommand{\dmHpmsq}{\delta m_{H^\pm}^2}
\providecommand{\dmHmGpsq}{\delta m_{H^- G^+}^2}
\providecommand{\dmGmHpsq}{\delta m_{G^- H^+}^2}
\providecommand{\dmGpmsq}{\delta m_{G^\pm}^2}
\providecommand{\dtadH}{\delta T_H}
\providecommand{\dtadh}{\delta T_h}
\providecommand{\dtadA}{\delta T_A}
\providecommand{\dtanb}{\delta\!\tan\!\beta\,}
\providecommand{\dZ}[1]{\delta Z_{#1}}
\providecommand{\dcZ}[1]{\delta \cZ_{#1}}
\providecommand{\dbcZ}[1]{\delta \bcZ_{#1}}
\providecommand{\dZm}[1]{\delta \matr{Z}_{#1}}
\providecommand{\dZZ}[1]{\bigl[\dZ{#1}\bigr]}
\providecommand{\dZZm}[1]{\bigl[\dZm{#1}\bigr]}
\providecommand{\dd}{\partial\,}
\providecommand{\cf}{C_F}
\providecommand{\gf}{G_F}
\providecommand{\nf}{N_f}
% shorthands for greek letters
\providecommand{\aeff}{\al_{\rm eff}}
\providecommand{\Saeff}{\sin\aeff}
\providecommand{\Caeff}{\cos\aeff}

\providecommand{\ML}{\left( \begin{array}{cc}}
\providecommand{\MR}{\end{array} \right)}

%%%%%%%%%%%%%%%%%%%%%%%%%%%%%%%%%%%%%%%%%%%%%%%%%%%%%%%%%%%%%%%%%%%%%%%%%%%%%%%

%- }}}
%- {{{ header:

\vspace*{-1.2cm}

\section{Higgs-boson production in the MSSM \footnote{%
    M.~Flechl, R.~Harlander, M.~Kr\"amer, S.~Lehti, P.~Slavich, 
    M.~Spira, M.~Vazquez Acosta, T.~Vickey (eds.);
    E.~Bagnaschi, M.~Carena, 
    G.~Degrassi, S.~Dittmaier, S.~Heinemeyer,
    R.~Klees, S.~Laurila, S.~Liebler,  H.~Mantler,
    O.~St{\aa}l, M.~Ubiali, A.~Vicini, C.~Wagner, G.~Weiglein} }
\label{sec:name}

%%%%%%%%%%%%%%%%%%%%%%%%%%%%%%%%%%%%%%%%%%%%%%%%%%%%%%%%%%%%%%%%%%%%%%%%%%%%%%%

%- }}}
%- {{{ subsection{Introduction}

\vspace*{-0.1cm}

\subsection{Introduction}
\label{sec:Introduction-sub}

The Higgs sector of the Minimal Supersymmetric Standard Model (MSSM)
consists of two $SU(2)$ doublets, $H_1$ and $H_2$, whose relative
contribution to electroweak symmetry breaking is determined by the
ratio of vacuum expectation values of their neutral components,
$\tb\equiv v_2/v_1$. The spectrum of physical Higgs bosons is
richer than in the SM, consisting of two neutral scalars $\PSh$ and
$\PSH$, one neutral pseudoscalar, $\PSA$, and two charged scalars,
$\PSHpm$. At the tree level, the mass matrix for the neutral scalars
can be expressed in terms of the parameters $\MZ$, $\MA$ and
$\tb$, and the mass of the lightest scalar $\PSh$ is bounded
from above by $\MZ$. However, radiative corrections -- especially
those involving top and bottom quarks and their supersymmetric
partners, the stop and sbottom squarks -- can significantly alter the
tree-level predictions for the Higgs-boson masses, and bring along a
dependence on a large number of free parameters of the MSSM. While the
${\cal CP}$ symmetry is conserved at tree level in the MSSM Higgs
sector, radiative corrections can also introduce ${\cal CP}$-violating
phases, and induce mixing among all three neutral states. In this
report, however, we will focus on the ${\cal CP}$-conserving case, by
considering only real values for the parameters in the soft
SUSY-breaking Lagrangian and for the Higgs mass $\mu$ in the
superpotential.

In general, the couplings of the MSSM Higgs bosons to gauge bosons and
matter fermions differ from those of the SM Higgs. However, in large
regions of the MSSM parameter space one of the scalars has SM-like
couplings, while the other Higgs bosons are decoupled from the gauge
bosons, and their couplings to down-type (up-type) fermions are
enhanced (suppressed) by $\tb$. As in the SM, gluon fusion is
one of the most important production mechanisms for the neutral Higgs
bosons, whose couplings to the gluons are mediated by the top and
bottom quarks and their superpartners. However, for intermediate to
large values of $\tb$ the associated production with bottom
quarks can become the dominant production mechanism for the neutral
Higgs bosons that have enhanced couplings to down-type fermions. The
production of the charged Higgs $\PSHpm$, on the other hand, proceeds
mainly through its coupling to a top-bottom pair. A sufficiently light
$\PSHpm$ is produced in the decay of a top quark, and it decays
dominantly in a tau-neutrino pair. A heavy $\PSHpm$ is produced in
association with a top quark and it decays dominantly in a top-bottom
pair.

The discovery by ATLAS and CMS of what appears to be a neutral scalar
with mass around $125.5$\UGeV\ \cite{Aad:2012tfa,Chatrchyan:2012ufa}
puts the studies of the Higgs sector of the MSSM in an entirely new
perspective. In order to remain viable, a point in the MSSM parameter
space must now not only pass all the (ever stricter) experimental
bounds on superparticle masses, but also lead to the prediction of a
scalar with mass, production cross section and decay rates compatible
with those measured at the LHC. In particular, the relatively large
mass of the roughly-SM-like scalar discovered at the LHC implies
either very heavy stops, of the order of 3\UTeV, or a large value of
the left-right stop mixing term (see, e.g.,
\Brefs{Heinemeyer:2011aa,Arbey:2011ab}). The ``benchmark scenarios''
routinely considered in MSSM studies had been devised when the Higgs
sector was constrained only by the LEP searches, and many of them,
such as the so-called ``no-mixing'' scenario, are now ruled out
because they predict a too-light SM-like scalar. Others, such as the
so-called $\mhmax$ scenario, are constrained for the opposite reason,
i.e.~they can predict a too-heavy SM-like scalar. To address the need
for new benchmark scenarios to be used in future studies of the MSSM
Higgs sector, in \refS{sec:sven-sub} we will define scenarios that are
compatible both with the properties of the Higgs boson discovered at
the LHC and with the current bounds on superparticle masses.

The fact that information on the Higgs boson mass, production and
decays has now become available also puts new emphasis on the need for
accurate theoretical predictions of those quantities. In the studies
presented in this report, the masses and mixing of the MSSM Higgs
bosons are computed with the public code
\FeynHiggs~\cite{Heinemeyer:1998yj,Heinemeyer:1998np,
  Degrassi:2002fi,Frank:2006yh}, which implements the full
one-loop radiative corrections together with the dominant two-loop
effects. The theoretical accuracy of the prediction of \FeynHiggs\ for
the lightest-scalar mass was estimated to be of the order of
3\UGeV~\cite{Degrassi:2002fi,Allanach:2004rh,Heinemeyer:2004gx}, i.e.,
already comparable to the accuracy of the mass measurement at the
LHC. Improving the accuracy of the theoretical prediction for the MSSM
Higgs masses will require the inclusion in public computer codes of
the remaining two-loop
effects~\cite{Martin:2002iu,Martin:2002wn,Martin:2004kr} and at least
the dominant three-loop effects
\cite{Martin:2007pg,Harlander:2008ju,Kant:2010tf}.

The production and decay rates of a SM-like Higgs boson in the MSSM
are sensitive to contributions from virtual SUSY particles, and their
measurement at the LHC -- combined with the searches for additional
Higgs bosons -- can be used to constrain the MSSM parameter space. To
this effect, the theoretical predictions for cross section and decays
must include precise computations of the SUSY contributions. In
\refS{sec:pietro-sub} we use the public code {\sc
  SusHi}~\cite{Harlander:2012pb} and the {\sc POWHEG} implementation
of \Bref{Bagnaschi:2011tu} to compute the total and differential cross
sections for neutral Higgs-boson production in gluon fusion, including
a NLO-QCD calculation of quark and squark contributions plus
higher-order quark contributions adapted from the SM calculation.  We
show that the SUSY contributions can be sizeable in regions of the
MSSM parameter space where the third-generation squarks are relatively
light, and discuss the theoretical uncertainty of the predictions for
the cross sections.

Finally, we study and update the exclusion limits on light charged
MSSM Higgs bosons in the ($\MHpm$, $\tb$)-plane in various benchmark 
scenarios in \refS{sec:samil-sub}. Particular emphasis is placed on
the dependence of the limits on the variation of SUSY parameters. We 
also provide improved NLO-QCD cross section predictions for heavy
charged Higgs production in the so-called four and five-flavor schemes in
\refS{sec:michaelk-sub}. The five-flavor scheme cross section is
calculated with a new scheme for setting the factorization scale and
takes into account the theoretical uncertainty from scale variation
and the PDF, $\alphas$ and bottom-mass error. We observe good
agreement between the 4FS and 5FS NLO-calculations and provide a
combined prediction following the Santander matching.

%%%%%%%%%%%%%%%%%%%%%%%%%%%%%%%%%%%%%%%%%%%%%%%%%%%%%%%%%%%%%%%%%%%%%%%%%%%%%%%

%- }}}
%- {{{ subsection{New MSSM benchmark scenarios}

\vspace*{-0.1cm}

\subsection{New MSSM benchmark scenarios}
\label{sec:sven-sub}

Within the MSSM an obvious possibility is to interpret the new state at
about $\mass$\UGeV\ as the light $\cp$-even Higgs
boson~\cite{Heinemeyer:2011aa,Hall:2011aa,Baer:2011ab,Arbey:2011ab,
Draper:2011aa,Carena:2011aa,Carena:2012gp,Carena:2012xa,Carena:2012mw}.
At the same time, the search for the other Higgs bosons has
continued. The non-observation of any additional state in the other
Higgs search channels puts by now stringent constraints on the MSSM
parameter space, in particular on the values of the tree-level
parameters $\MA$ (or $\MHp$) and $\tb$.  Similarly, the non-observation
of supersymmetric (SUSY) particles puts relevant constraints on the
masses of the first and second generation scalar quarks and the gluino,
and to lesser degree on the stop and sbottom masses
(see \Brefs{Marrouche,Mann} for a recent summary).

Due to the large number of free parameters, a complete scan of the
MSSM parameter space is impractical in experimental analyses and
phenomenological studies.  Therefore, the Higgs search results at LEP
were interpreted~\cite{Schael:2006cr} in several benchmark
scenarios~\cite{Carena:1999xa,Carena:2002qg}.  In these scenarios only
the two parameters that enter the Higgs sector tree-level predictions,
$\MA$ and $\tb$, are varied (and the results are usually displayed in
the $\MA{-}\tb$ plane), whereas the other SUSY parameters, entering
via radiative corrections, are fixed to particular benchmark values
which are chosen to exhibit certain features of the MSSM Higgs
phenomenology. These scenarios were also employed for the MSSM Higgs
searches at the Tevatron and at the LHC.

By now, most of the parameter space of the original benchmark
scenarios~\cite{Carena:1999xa,Carena:2002qg} has been ruled out by the
requirement that one of the $\cp$-even Higgs boson masses should be
around $\mass$\UGeV.  Consequently, new scenarios have been
proposed~\cite{Carena:2013qia}, which are defined such that over large
parts of their available parameter space the observed signal at about
$\mass$\UGeV\ can be interpreted in terms of one of the (neutral) Higgs
bosons, while the scenarios exhibit interesting phenomenology for the
MSSM Higgs sector.  The benchmark scenarios are all specified using
low-energy MSSM parameters, i.e.\ no particular soft
SUSY-breaking scenario was assumed. Constraints from direct
searches for Higgs bosons are taken into account, whereas indirect
constraints from requiring the correct cold dark matter density,
$\br(b \to s \ga)$, $\br(B_s \to \PGm^+\PGm^-$) or $(g - 2)_{\PGm}$
are neglected. However interesting, those constraints depend to a
large extent on other parameters of the theory that are not crucial
for Higgs phenomenology.

\subsubsection{Definition of parameters}

The mass matrices for the stop and sbottom sectors of the MSSM, in
the basis of the current eigenstates $\StopL, 
\StopR$ and $\SbotL, \SbotR$, are given by
\BEA
\label{stopmassmatrix}
{\cal M}^2_{\Stop} &=&
  \ML \MstL^2 + \mto^2 + \cos 2\be (\edz - \frac{2}{3} \sw^2) \MZ^2 &
      \mto \Xt^{*} \\
      \mto \Xt &
      \MstR^2 + \mto^2 + \frac{2}{3} \cos 2\be \sw^2 \MZ^2 
  \MR, \\
&& \non \\
\label{sbotmassmatrix}
{\cal M}^2_{\Sbot} &=&
  \ML \MsbL^2 + \mb^2 + \cos 2\be (-\edz + \frac{1}{3} \sw^2) \MZ^2 &
      \mb \Xb^{*} \\
      \mb \Xb &
      \MsbR^2 + \mb^2 - \frac{1}{3} \cos 2\be \sw^2 \MZ^2 
  \MR,
\EEA
where 
\BE
\mto \Xt = \mto (\At - \PGm^{*} \cot\be) , \quad
\mb\, \Xb = \mb\, (\Ab - \PGm^{*} \tb) .
\label{eq:mtlr}
\end{equation}
Here $\At$ denotes the trilinear Higgs--stop coupling, $\Ab$ denotes the
Higgs--sbottom coupling, and $\PGm$ is the higgsino mass parameter. We
furthermore use the notation $\sw = \sqrt{1 - \cw^2}$, with 
$\cw = \MW/\MZ$. 
We shall concentrate on the case
\BE
\MstL = \MsbL = \MstR = \MsbR =: \msusy .
\label{eq:msusy}
\end{equation}
Similarly, the corresponding soft SUSY-breaking parameters in the
scalar tau/neutrino sector are denoted as $\Atau$ and $\msld$, where
we assume the diagonal soft SUSY-breaking entries in the
stau/sneutrino mass matrices to be equal to each other.  For the
squarks and sleptons of the first and second generations we also
assume equality of the diagonal soft SUSY-breaking parameters, denoted
as $\msqez$ and $\mslez$, respectively. The off-diagonal $A$-terms
always appear multiplied with the corresponding fermion mass.  Hence,
for the definition of the benchmark scenarios the $A$-terms associated
with the first and second sfermion generations have a negligible
impact and can be set to zero for simplicity.

The Higgs sector depends also on the gaugino masses. For instance, at
the two-loop level the gluino mass, $\Mgl$, enters the predictions for
the Higgs boson masses.  The Higgs sector observables furthermore
depend on the SU(2) and U(1) gaugino mass parameters, $M_2$ and $M_1$,
respectively, which are usually assumed to be related via the GUT
relation,
\BE
M_1 = \frac{5}{3} \frac{\sw^2}{\cw^2} M_2~.
\label{def:M1}
\end{equation}

Corrections to the MSSM Higgs boson sector have been evaluated in
several approaches, see, e.g.\ \Bref{Carena:2000dp}.  The leading and
subleading parts of the existing two-loop calculations have been
implemented into public codes. The program {\sc
FeynHiggs}~\cite{Heinemeyer:1998yj,Heinemeyer:1998np,
Degrassi:2002fi,Frank:2006yh} is based on results obtained
in the Feynman-diagrammatic (FD) approach, while the code {\sc
CPsuperH}~\cite{Lee:2003nta,Lee:2007gn,Lee:2012wa} is based on results
obtained using the renormalization group (RG) improved effective
potential
approach~\cite{Casas:1994us,Carena:1995bx,Carena:1995wu,Carena:2000dp}.

The FD results have been obtained in the on-shell (OS) renormalization
scheme, whereas the RG results have been calculated using the $\msbar$
scheme.  Therefore, the parameters $\Xt$ and $\msusy$ (which are most
important for the corrections in the Higgs sector) are
scheme-dependent and thus differ in the two approaches, see
\Bref{Carena:2000dp} for details.  The change of scheme induces in
general only a minor shift, of the order of 4\%, in the parameter
$\msusy$, but sizable differences can occur between the numerical
values of $\Xt$ in the two schemes, see
\Brefs{Heinemeyer:1998np,Carena:2000dp,Williams:2011bu}.

%%%%%%%%%%%%%%%%%%%%%%%%%%%%%%%%%%%%%%%%%%%%%%%%%%%%%%%%%%%%%%%%%%%%%%%%%%%%%%

\subsubsection{The benchmark scenarios}

In the following several updated benchmark scenarios are proposed, in
which the observed LHC signal at $\sim \mass$\UGeV\ can be interpreted
as one of the (neutral $\cp$-even) states of the MSSM Higgs sector.
More details about implications and the phenomenology in these
scenarios can be found in \Bref{Carena:2013qia}.

Concerning the parameters that have only a minor impact on the MSSM Higgs
sector predictions, we propose fixing them to the following values.
\begin{align}
\msqez &= 1500\UGeV, \\
\mslez &= 500\UGeV, \\
A_f &= 0 \quad (\Pf = \PQc,\PQs,\PQu,\PQd,\PGm,e)~.
\label{LHCcolored}
\end{align}
$M_1$ is fixed via the GUT relation, \eqn{def:M1}.  Motivated by the
analysis in \Bref{Carena:2005ek} we suggest to investigate for each
scenario given in \eqns{mhmax}{lightstop}, in addition to the default
values given there, the following values of $\PGm$:
\begin{align}
\PGm = \pm 200, \pm 500, \pm 1000\UGeV.
\label{eq:musuggest}
\end{align}
These values of $\PGm$ allow for both an enhancement and a suppression
of the bottom Yukawa coupling. The illustrative plots shown below have
been obtained with {\sc
  FeynHiggs\,2.9.4}~\cite{Heinemeyer:1998yj,Heinemeyer:1998np,
  Degrassi:2002fi,Frank:2006yh}.  Where relevant,
values for the input parameters are quoted both in the on-shell scheme
(suitable for {\sc FeynHiggs}), as well as in the $\msbar$ scheme
(that can readily be used by {\sc
  CPsuperH}~\cite{Lee:2003nta,Lee:2007gn,Lee:2012wa}). We also show
the exclusion bounds (at $95\%$ C.L.) from direct Higgs searches,
evaluated with {\sc
  HiggsBounds\,4.0.0-beta}~\cite{Bechtle:2008jh,Bechtle:2011sb}
(linked to {\sc FeynHiggs}) using a combined uncertainty on the
SM-like Higgs mass of $\De \Mh = 3\UGeV$ ($\De \MH = 3\UGeV$ in the
last scenario) when evaluating the limits.  For each benchmark
scenario we show the region of parameter space where the mass of the
(neutral $\cp$-even) MSSM Higgs boson that is interpreted as the newly
discovered state is within the range $\mass \pm 3$\UGeV\ and $\mass
\pm 2$\UGeV.  The $\pm 3$\UGeV\ uncertainty is meant to represent a
combination of the present experimental uncertainty of the determined
mass value and of the theoretical uncertainty in the MSSM Higgs mass
prediction from unknown higher-order corrections.
In particular, in the case that the lightest $\cp$-even Higgs is
interpreted as the newly discovered state, the couplings of the $\PSh$
are close to the corresponding SM values (modulo effects from light
SUSY particles, see below). Consequently, those rate measurements from
the LHC that agree well with the SM are then naturally in good
agreement also with the MSSM predictions.

The suggested parameters below refer to recommendations in
\Bref{Carena:2013qia}. It should be kept in mind that for the
evaluations in the LHC Higgs Cross Section Working Group some SM
parameters should be adjusted to conform with the respective
evaluations. The benchmark scenarios are recommended as follows.  The
top quark mass is set to its current experimental value, $\mto =
173.2\UGeV$.

\begin{itemize}
\item
\underline{The \mhmax\ scenario:}
This scenario can be used to derive conservative lower bounds on
$\MA$, $\MHp$ and $\tb$~\cite{Heinemeyer:2011aa}.
\begin{align}
\msusy &= 1000\UGeV, 
\PGm = 200\UGeV, 
M_2 = 200\UGeV, \non \\
\Xt^{\OS} &= 2\, \msusy  \; \mbox{(FD calculation)}, 
\Xt^{\MSbar} = \sqrt{6}\, \msusy \; \mbox{(RG calculation)}, \non \\ 
\Ab &= \Atau = \At, 
\Mgl = 1500\UGeV, 
\msld = 1000\UGeV~.
\label{mhmax}
\end{align}

\item \underline{The \mhmod\ scenario:}\\
Departing from the parameter configuration that maximizes $\Mh$, one
naturally finds scenarios where in the decoupling region the value of
$\Mh$ is close to the observed mass of the signal over a wide region
of the parameter space. A convenient way of modifying the $\mhmax$
scenario in this way is to reduce the amount of mixing in the stop
sector, i.e.\ to reduce $|\Xt/\msusy|$ compared to the value of
$\approx 2$ (FD calculation) that gives rise to the largest positive
contribution to $\Mh$ from the radiative corrections. This can be done
for both signs of $\Xt$.
\begin{align}
\mbox{\mhmodp:}\quad
\msusy &= 1000\UGeV, 
\PGm = 200\UGeV, 
M_2 = 200\UGeV, \non \\
\Xt^{\OS} &= 1.5\, \msusy  \; \mbox{(FD calculation)}, 
\Xt^{\MSbar} = 1.6\, \msusy \; \mbox{(RG calculation)}, \non \\ 
\Ab &= \Atau = \At, 
\Mgl = 1500\UGeV, 
\msld = 1000\UGeV~.
\label{mhmodp}
\end{align}
\begin{align}
\mbox{\mhmodm:}\quad 
\msusy &= 1000\UGeV, 
\PGm = 200\UGeV, 
M_2 = 200\UGeV, \non \\
\Xt^{\OS} &= -1.9\, \msusy  \; \mbox{(FD calculation)}, 
\Xt^{\MSbar} = -2.2\, \msusy \; \mbox{(RG calculation)}, \non \\ 
\Ab &= \Atau = \At, 
\Mgl = 1500\UGeV, 
\msld = 1000\UGeV~.
\label{mhmodn}
\end{align}

\item \underline{The light stop scenario:}\\
A light stop may lead to a relevant modification of the gluon fusion
rate~\cite{Djouadi:1998az,Carena:2002qg}, see the evaluations
in \refS{sec:pietro-sub}.
\begin{align}
\msusy &= 500\UGeV, 
\PGm = 350\UGeV, 
M_2 = 350\UGeV, \non \\
\Xt^{\OS} &= 2.0\, \msusy  \; \mbox{(FD calculation)}, 
\Xt^{\MSbar} = 2.2\, \msusy \; \mbox{(RG calculation)}, \non \\ 
\Ab &= \At = \Atau, 
\Mgl = 1500\UGeV, 
\msld = 1000\UGeV~.
\label{lightstop}
\end{align}

\item \underline{The light stau scenario:}\\
It has been shown that light staus, in the presence of large mixing,
may lead to important modifications of the di-photon decay width of the
lightest $\cp$-even Higgs boson,
$\Ga(h \to \ga\ga)$~\cite{Carena:2011aa,Carena:2012gp,Carena:2012xa,
Carena:2012mw,Cao:2012fz,Hagiwara:2012mga,Giudice:2012pf,Ajaib:2012eb}. Here
the parameter definitions depend on whether $\De_\PGt$ corrections are
neglected in the stau mass matrix, or not. The latter case is denoted
as ``$\De_\PGt$~calculation''.
\begin{align}
\msusy &= 1000\UGeV, 
\PGm = 500 (450)\UGeV, (\De_\PGt \mbox{~calculation}), \non \\
M_2 &= 200 (400)\UGeV\ (\De_\PGt \mbox{~calculation}), \non \\
\Xt^{\OS} &= 1.6\, \msusy  \; \mbox{(FD calculation)}, 
\Xt^{\MSbar} = 1.7\, \msusy \; \mbox{(RG calculation)}, \non \\
\Ab &= \At ~, 
\Atau = 0~, 
\Mgl = 1500\UGeV, 
\msld = 245 (250)\UGeV\ (\De_\PGt \mbox{~calculation}) .
\label{lightstau}
\end{align}

\item \underline{The \boldmath{$\PGt$}-phobic Higgs scenario:}\\
Propagator-type corrections involving the mixing between the two
$\cp$-even Higgs bosons of the MSSM can have an important impact. In
particular, this type of corrections can lead to relevant
modifications of the Higgs couplings to down-type fermions, which can
approximately be taken into account via an effective mixing
angle~$\aeff$ (see \Brefs{Dabelstein:1995js,Heinemeyer:2000fa}).
\begin{align}
\msusy &= 1500\UGeV, 
\PGm = 2000\UGeV, 
M_2 = 200\UGeV, \non \\
\Xt^{\OS} &= 2.45\,\msusy  \; \mbox{(FD calculation)}, 
\Xt^{\MSbar} = 2.9\,\msusy \; \mbox{(RG calculation)}, \non \\ 
\Ab &= \Atau = \At ~, 
\Mgl = 1500\UGeV, 
\msld = 500\UGeV\ ~.
\label{tauphobic}
\end{align}

\item \underline{The low-$\MH$ scenario:}

As it was pointed out
in \Brefs{Heinemeyer:2011aa,Benbrik:2012rm,Bottino:2011xv,Drees:2012fb},
besides the interpretation of the Higgs-like state at
$\sim \mass$\UGeV\ in terms of the light $\cp$-even Higgs boson of the
MSSM it is also possible, at least in principle, to identify the
observed signal with the {\em heavy} $\cp$-even Higgs boson of the
MSSM.  In this case instead of $\MA$, which must be given by a
relatively small value, $\PGm$ is suggested to be modified.
\begin{align}
\MA &= 110\UGeV, 
\msusy = 1500\UGeV, 
M_2 = 200\UGeV, \non \\
\Xt^{\OS} &= 2.45\, \msusy  \; \mbox{(FD calculation)}, 
\Xt^{\MSbar} = 2.9\, \msusy \; \mbox{(RG calculation)}, \non \\ 
\Ab &= \Atau = \At, 
\Mgl = 1500\UGeV, 
\msld = 1000\UGeV~.
\label{lowMH}
\end{align}
Instead of $\MA$ one can also use $\MHp$ as input parameter, as it is
done, e.g., in {\sc CPsuperH}. In this case one should choose as input
value $\MHp = 132\UGeV$, leading to very similar phenomenology.

\end{itemize}

%%%%%%%%%%%%%%%%%%%%%%%%% F I G U R E %%%%%%%%%%%%%%%%%%%%%%%%%%%%%%%%%%%%%%%%%
\begin{figure}[h!]
%\vspace{2em}
\begin{center}
\includegraphics[width=0.45\textwidth,height=6.2cm]{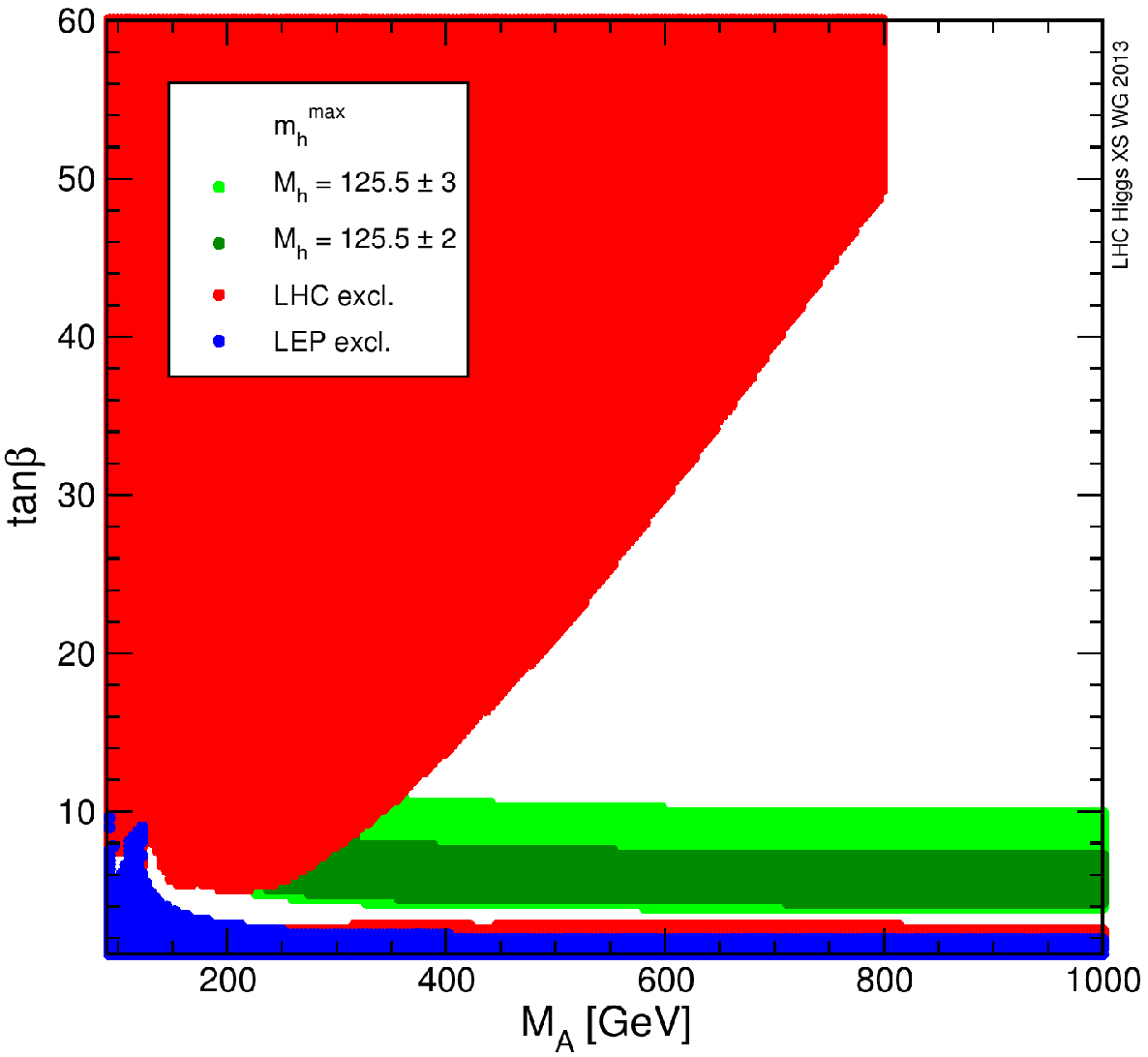}
\includegraphics[width=0.45\textwidth,height=6.2cm]{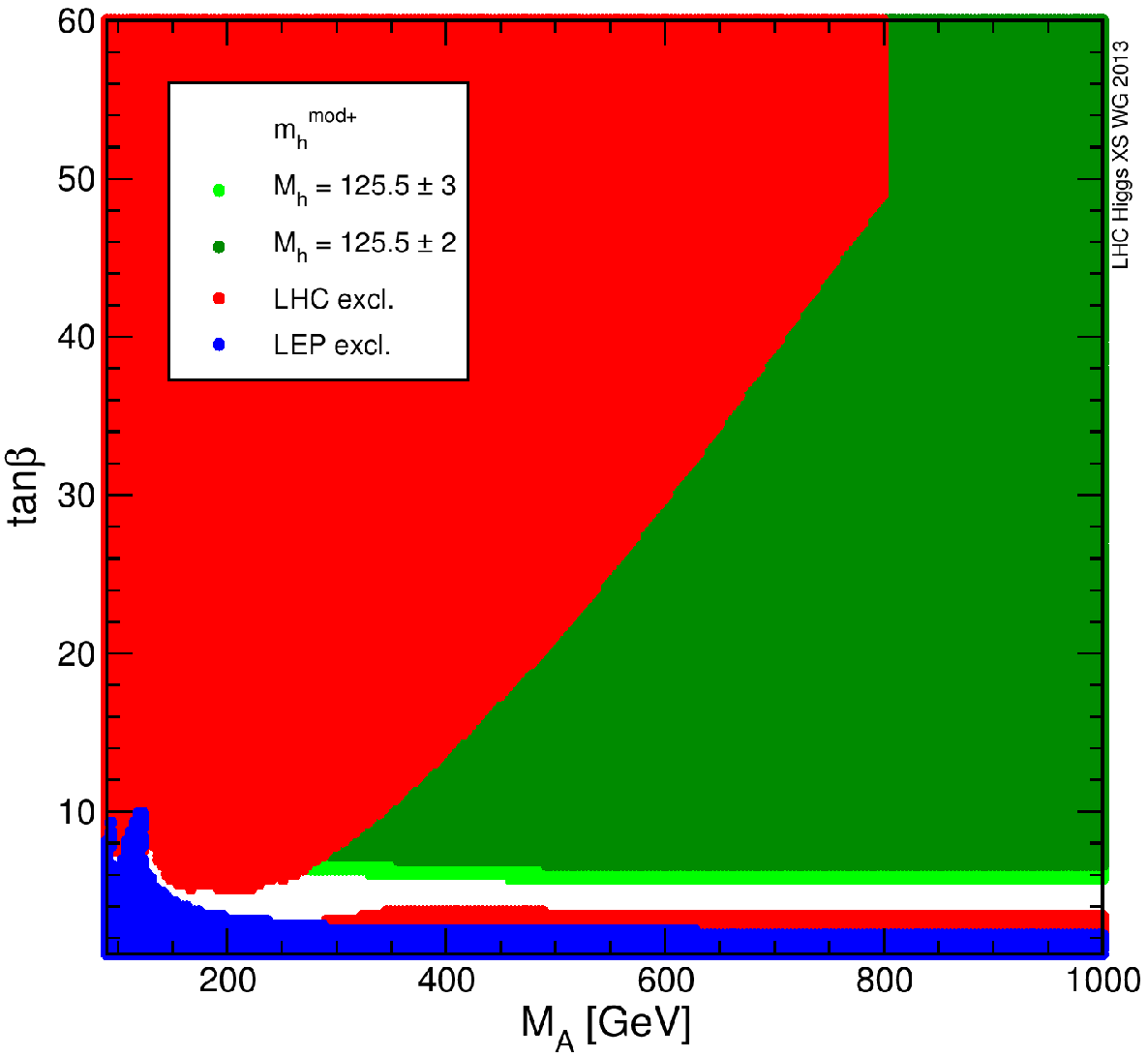}\\
\includegraphics[width=0.45\textwidth,height=6.2cm]{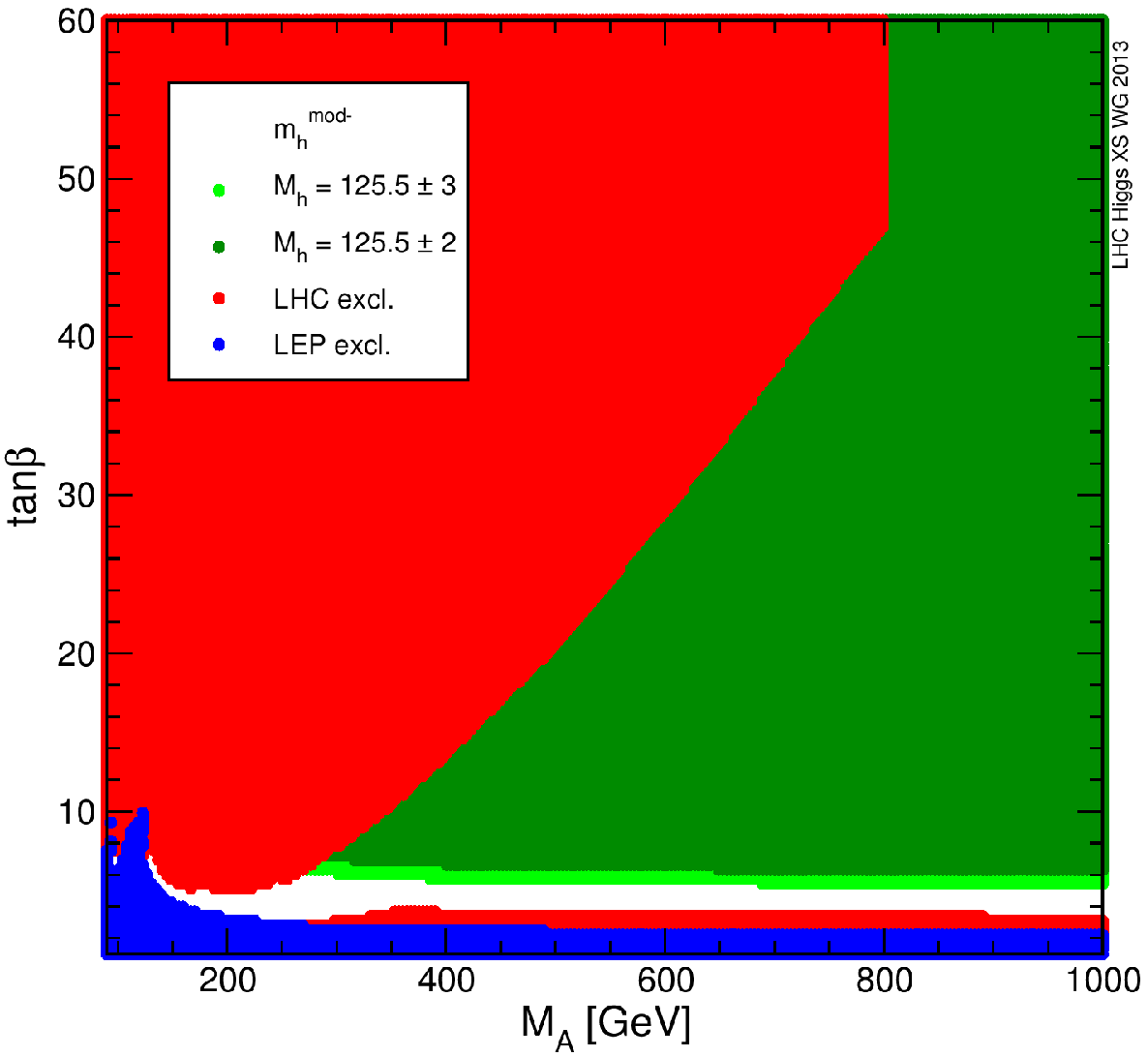}
\includegraphics[width=0.45\textwidth,height=6.2cm]{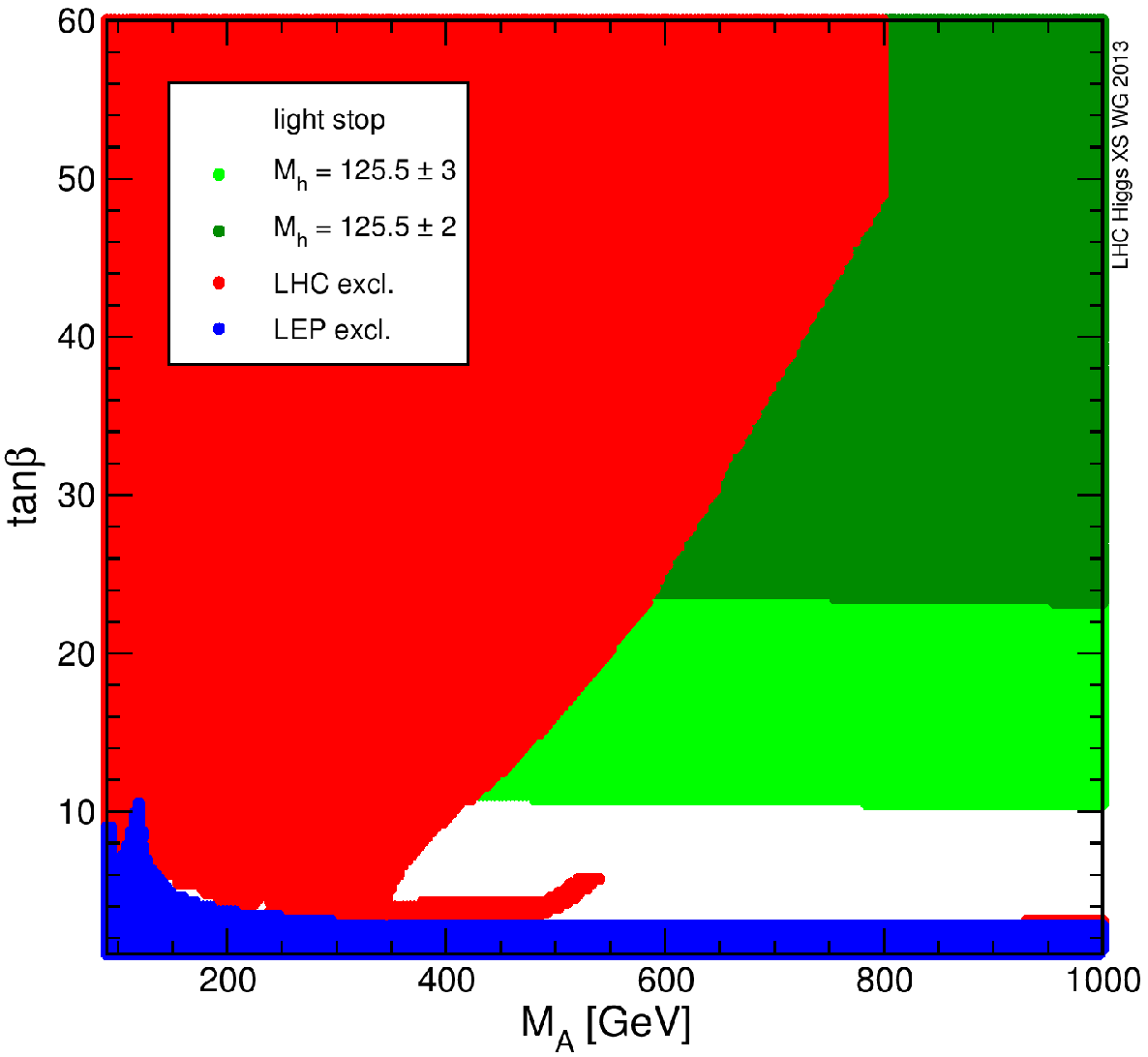}\\
\includegraphics[width=0.45\textwidth,height=6.2cm]{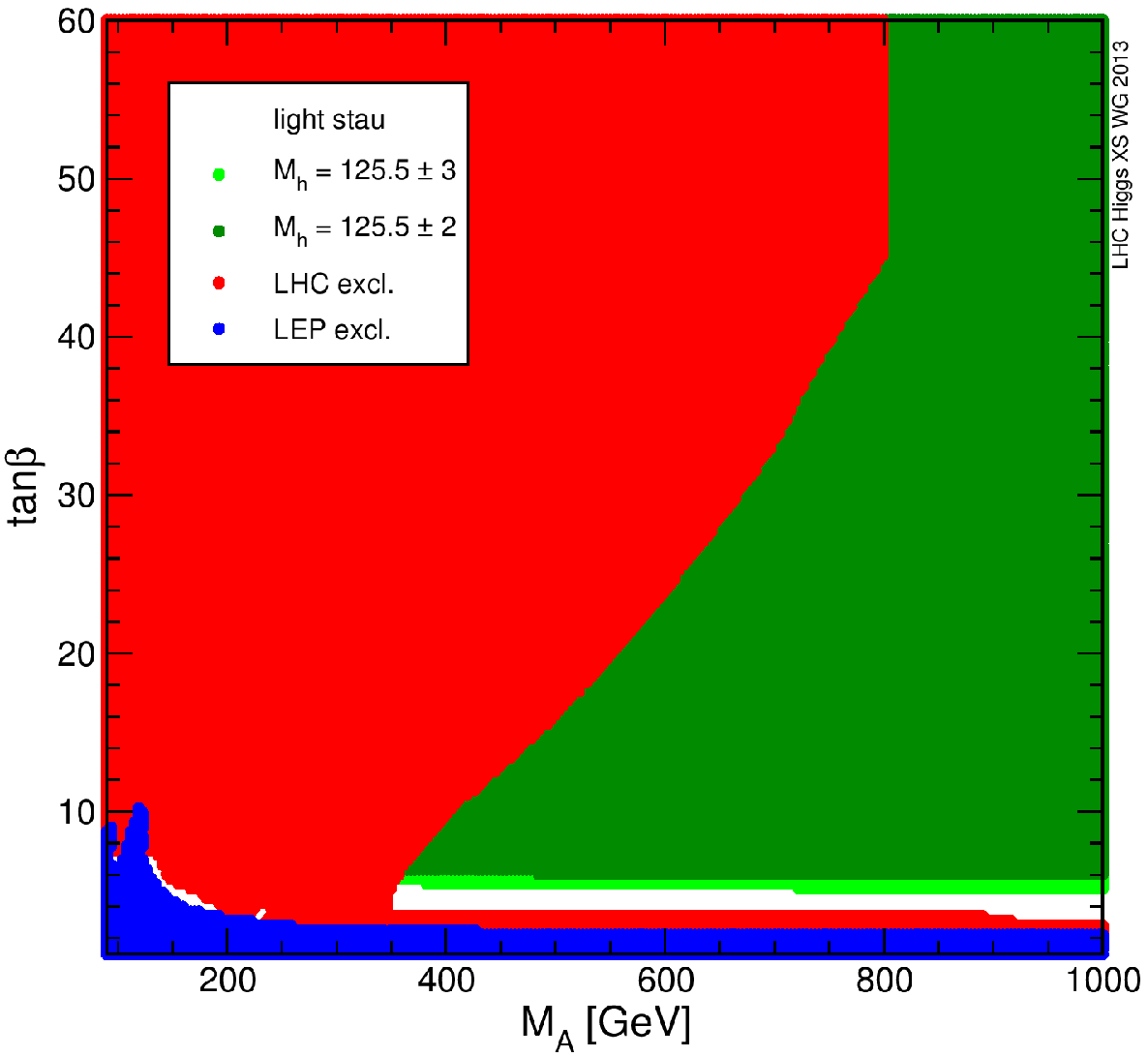}
\includegraphics[width=0.45\textwidth,height=6.2cm]{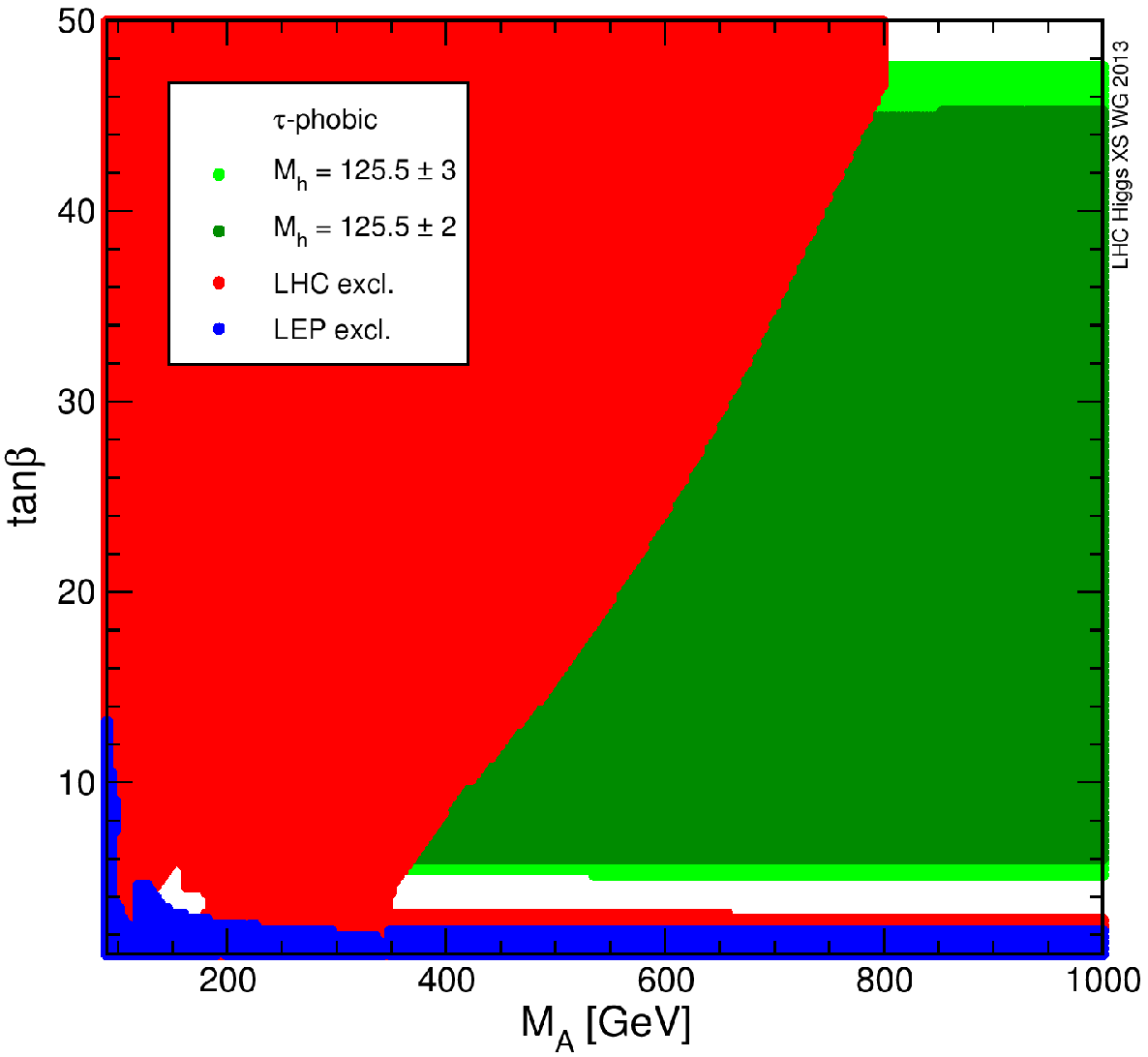}
\caption{
The $\MA$--$\tb$ plane in the (updated) \mhmax\ (upper left),
\mhmodp\ (upper right), \mhmodm\ (middle left), \lstop\ (middle right), 
\lstau\ (lower left), \tauphobic\ (lower right) scenario, 
with excluded regions from direct Higgs searches at LEP (blue), and
the LHC (solid red).  The two green shades correspond to the
parameters for which $\Mh = \mass \pm 2\, (3)\UGeV$, see text.}
\label{fig:benchmark}
\end{center}
\vspace{-2em}
\end{figure}
%%%%%%%%%%%%%%%%%%%%%%%%% F I G U R E %%%%%%%%%%%%%%%%%%%%%%%%%%%%%%%%%%%%%%%%%

The allowed and excluded regions in the $\MA{-}\tb$ planes of the
benchmark scenarios in which the light $\cp$-even Higgs is interpreted
as the new state at $\mass$\UGeV\ are shown
in \refF{fig:benchmark}. The corresponding plot for the
\lowMH\ scenario is shown in \refF{fig:lowMH}.

%%%%%%%%%%%%%%%%%%%%%%%%% F I G U R E %%%%%%%%%%%%%%%%%%%%%%%%%%%%%%%%%%%%%%%%%
\begin{figure}[htb!]
\vspace{-1em}
\begin{center}
\includegraphics[width=0.55\textwidth]{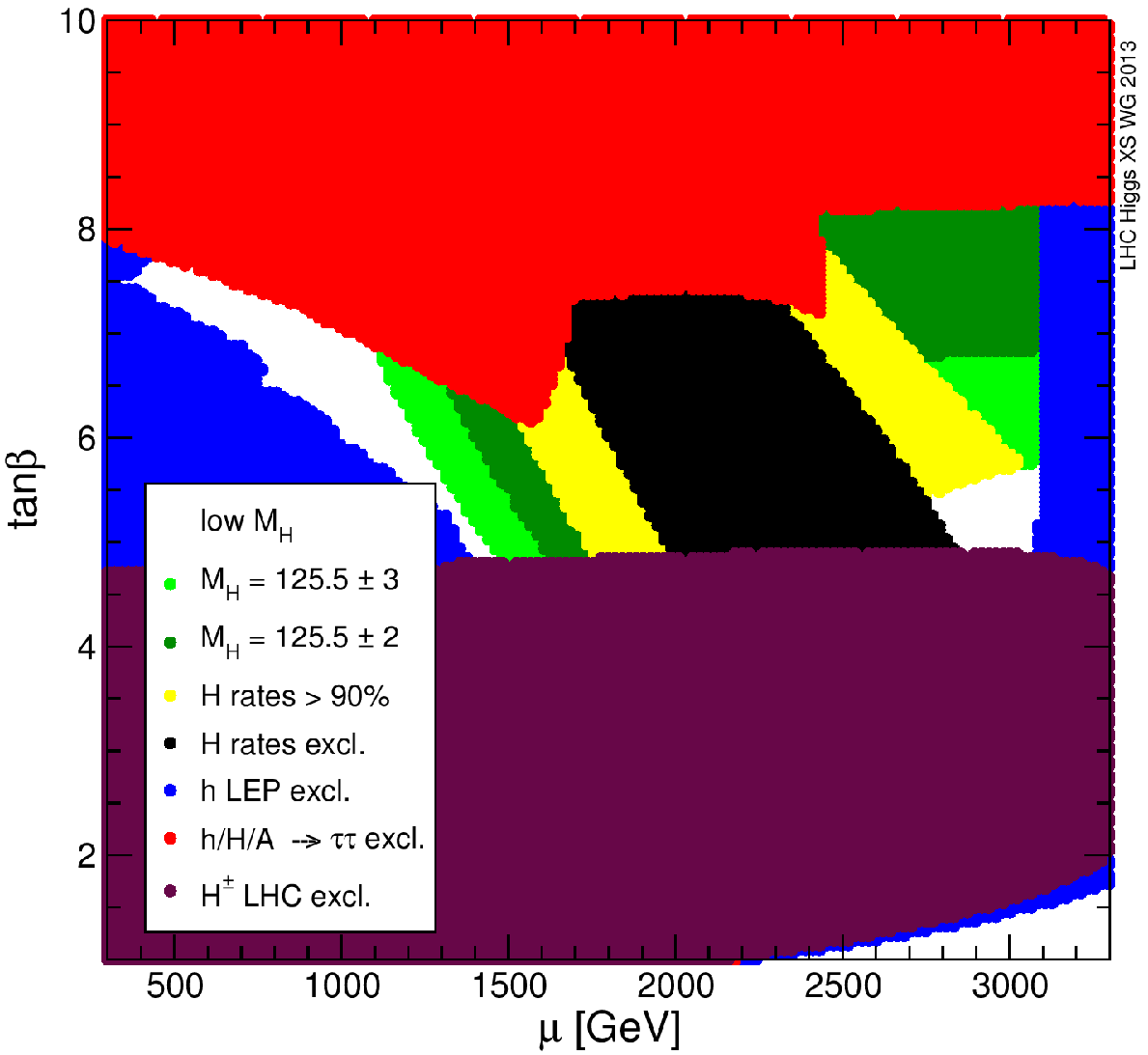}
\caption{
Experimentally favored and excluded regions in the $\PGm$--$\tb$ plane
in the \lowMH\ scenario.  The green shades indicate the region where
$\MH=125.5\pm 2\, (3)\UGeV$. The yellow and black areas also have $\MH
= \mass \pm 3\UGeV$, where the yellow area additionally satisfies the
requirement that the rates for the $\Pg\Pg \to \PH$, $\PH\to \PGg\PGg$
and $\PH \to \PZ\PZ^{*}$ channels, are at least at 90\% of their SM
value for the same Higgs mass. The black region indicates where the
rates for $\PH$ decay to gauge bosons become too high, such that these
points are excluded by {\sc HiggsBounds}.  As before, the blue area is
excluded by LEP Higgs searches, whereas the solid red is excluded from
LHC searches for the neutral MSSM Higgs bosons, $\Ph$, $\PH$ and $\PA$
in the $\PGt^+\PGt^-$ decay channel.  The purple region is excluded by
charged Higgs boson searches at the LHC. The white area at very large
values of $\PGm$ and low $\tb$ is unphysical, i.e.\ this parameter
region is theoretically inaccessible.}
\label{fig:lowMH}
\end{center}
\vspace{-1.5em}
\end{figure}

%%%%%%%%%%%%%%%%%%%%%%%%%%%%%%%%%%%%%%%%%%%%%%%%%%%%%%%%%%%%%%%%%%%%%%%%%%%%%%%

%- }}}
%- {{{ subsection{Neutral MSSM Higgs production}

\subsection{Neutral MSSM Higgs production}
\label{sec:pietro-sub}

%\subsubsection{Introduction}

The essential features of the theoretical prediction for the
production of neutral Higgs bosons within the \mssm\ have been
summarized in \Brefs{Dittmaier:2011ti,Dittmaier:2012vm} and shall not
be repeated here. Suffice it to say that the dominant production
mechanisms are gluon fusion and bottom quark annihilation. Theoretical
progress concerning the latter process has been marginal, so that the
numbers provided in \Brefs{Dittmaier:2011ti,Dittmaier:2012vm} are
still up-to-date.  In this report we will present updated computations
of the gluon fusion cross section.

In \Brefs{Dittmaier:2011ti,Dittmaier:2012vm}, the numerical results were
mostly presented within the so-called ``no mixing'' and $\mhmax$
scenarios~\cite{Carena:1999xa,Carena:2002qg}, characterized by stop and
sbottom masses of the order of one\UTeV\ or even heavier.  In those
scenarios the contributions of diagrams involving squarks to the
gluon-fusion process are suppressed, and one can assume that the
\susy\ effects are well approximated by the non-decoupling corrections
to the Higgs-bottom effective couplings, which become relevant at
large $\tanb$. Therefore, a sufficiently accurate determination of the
cross section for \mssm\ Higgs production in gluon fusion could be
achieved with the public code \higlu~\cite{Spira:1995mt}, including
only the top- and bottom-quark contributions rescaled by the
appropriate Higgs-quark effective couplings.

However, in scenarios with relatively light squarks such as the
\lstop{} scenario introduced in \refS{sec:sven-sub} (see \Bref{Carena:2013qia} 
for more details), the squark contributions to the gluon-fusion process are
not negligible. Luckily, in the past couple of years significant
progress has been made in combining the existing theoretical results
for the quark, squark, and gluino contributions into a consistent
numerical prediction~\cite{Bagnaschi:2011tu,Harlander:2012pb}. In this
report we will use the public code \sushi~\cite{Harlander:2012pb} to
provide a state-of-the-art determination of the total inclusive cross
section for gluon fusion, as well as for bottom-quark annihilation.
In addition, we will compare differential distributions obtained with
\sushi\ at the purely partonic level with those obtained via a
\powheg\ implementation of gluon fusion~\cite{Bagnaschi:2011tu}.

\subsubsection{Contributions to the gluon-fusion cross section}

At \lo, the partonic cross section for Higgs production via gluon
fusion in the \mssm\ is induced by quark and squark loops, and we
will take into account only the contributions from the top and bottom
sectors.  For what concerns the first two generations, the quark
contributions are negligible due to the smallness of the corresponding
Yukawa couplings, while the squarks contribute only via terms
suppressed by the ratio $\MZ^2/\MSQ^2$, with significant cancellations
among the different contributions in each generation (indeed, the
total contribution vanishes for degenerate squark masses).
Virtual effects at \nlo\ \qcd\ include gluonic corrections to the
\lo\ quark and squark contributions, as well as mixed
quark-squark-gluino contributions. While the gluonic corrections to
the quark contributions~\cite{Spira:1995rr,Harlander:2005rq} are
implemented for generic quark and Higgs-boson masses, for the two-loop
contributions involving squarks we use reasonable approximations,
valid as long as the Higgs mass does not exceed the lowest threshold
for squark production.
In particular, for the production of the lightest scalar $\Ph$ we
compute the contributions involving stops via a Taylor expansion in
$\Mh^2$\,\cite{Harlander:2003bb,Harlander:2004tp,Degrassi:2008zj}, and
those involving sbottoms via an asymptotic expansion in the \susy\
masses\,\cite{Degrassi:2010eu,Harlander:2010wr}. For $\PH$ and $\PA$
production, on the other hand, we use the asymptotic expansion for both
stop and sbottom
contributions\,\cite{Harlander:2005if,Degrassi:2011vq,Degrassi:2012vt}.
The validity of these approximations is supported by explicit NLO
calculations with full mass dependences
of \Brefs{Muhlleitner:2006wx,Bonciani:2007ex,Muhlleitner:2010nm,Anastasiou:2008rm}.
For the contributions of one-loop diagrams with radiation of a real
quark or gluon, which are part of the inclusive cross section at \nlo,
we include results valid for generic quark and squark masses.

The \nnlo-\qcd\ contributions involving top quarks are known on the
basis of an effective Lagrangian approach valid for $M_\phi\muchless
2\,\mto$\,\cite{Marzani:2008az,Harlander:2009my,Harlander:2009mq,
Pak:2009dg,Pak:2011hs} where $\phi \in \{\Ph,\PH,\PA\}$, and they will
be included in our evaluation of the cross
section\,\cite{Harlander:2002wh,Anastasiou:2002yz,Ravindran:2003um,
Harlander:2002vv,Anastasiou:2002wq,Pak:2011hs}.  For the \nnlo\
contributions involving stop squarks, a pragmatic approximation was
presented in \Bref{Harlander:2003kf}. While this approximation will not
enter our prediction for the cross section, we will use it to estimate
the uncertainty of the corresponding \nlo\ contributions. In fact, a
robust \nnlo\ calculation of the stop contributions was recently
presented in
\Brefs{Pak:2012xr,Pak:2010cu}. Since we do not take into account the
three-loop corrections to the Higgs boson mass, it is reasonable to
neglect these terms in the cross section.

The existence of non-decoupling, $\tanb$-enhanced \susy\ corrections
to the Higgs-bottom couplings, the so-called $\Delta_{\Pb}$ terms, has
been known for a long time. These corrections can be taken into
account via an effective-Lagrangian approach that resums them to all
orders in the perturbative
expansion\,\cite{Carena:1999py,Guasch:2003cv,
Noth:2008tw,Noth:2010jy,Mihaila:2010mp}.
As mentioned above, the $\Delta_{\Pb}$ terms were the only genuine \susy\
effects included in previous compilations of the \mssm\ neutral Higgs
cross section by this working group. Since these effects can be
numerically dominant in regions of the \mssm\ parameter space
characterized by large $\tanb$, we follow the effective-Lagrangian
approach and absorb them in a redefinition of the Higgs-bottom Yukawa
coupling. As a consequence, we need to shift accordingly the formulae
for the two-loop contributions, in order to avoid double counting.

The \nlo\ electro-weak (\ew) corrections to the cross section for
gluon fusion in the \sm\ have been computed in
\Brefs{Djouadi:1994ge,Aglietti:2004nj,Degrassi:2004mx,Actis:2008ug,
  Bonciani:2010ms}. For a \sm\ Higgs boson sufficiently lighter than
the top threshold, the \nlo-\ew\ corrections are well approximated by
the contributions coming from two-loop diagrams in which the Higgs
couples to \ew\ gauge bosons, which in turn couple to the gluons via a
loop of light quarks\,\cite{Aglietti:2004nj,Bonciani:2010ms}. The
inclusion of these contributions in the \mssm\ calculation, via an
appropriate rescaling of the Higgs-gauge boson couplings, allows us to
properly account for the \nlo-\ew\ corrections to the production of
the lightest scalar $\Ph$ in scenarios where the \susy\ particles are
heavy.  For what concerns the other neutral Higgs bosons, their
couplings to gauge bosons are suppressed in most of the parameter
space (in the case of the heaviest scalar $\PH$) or downright absent
(in the case of the pseudoscalar $\PA$), therefore the
\ew\ light-quark contributions to their production are irrelevant. On
the other hand, the \nlo-\ew\ corrections involving the bottom Yukawa
coupling, which have not yet been computed because they are negligible
for the \sm\ Higgs, could become relevant for the \mssm\ Higgs bosons
whose couplings to bottom quarks are enhanced by $\tanb$. In addition,
a full computation of the \nlo-\ew\ corrections to Higgs production in
the \mssm\ should include the contributions of diagrams involving
\susy\ particles. The non-decoupling \susy\ \ew\ effects that dominate
at large $\tanb$ are indeed included via the $\Delta_{\Pb}$ resummation,
but the remaining contributions, so far un-computed, could become
relevant if some of the \susy\ particles are relatively light.

\subsubsection{Uncertainties}
The sources of uncertainty for the gluon fusion cross section in the
\mssm\ include of course the ones already affecting the \sm\
prediction, which are dominated by \pdf{}s, $\alphas(\MZ)$, and the
scale variation. In addition there are, on the one hand, parametric
uncertainties due to the unknown spectrum of \susy\ masses and
couplings, and, on the other hand, theoretical uncertainties due to
the un-computed higher-order \susy\ contributions.

Another kind of theoretical uncertainty, which is relatively small in
the \sm\ but can become dominant in regions of the \mssm\ parameter
space characterized by large $\tanb$, stems from the definition of the
Higgs-bottom coupling (for an earlier discussion
see \Bref{Baglio:2010ae}). Indeed, the Yukawa coupling $Y_{\Pb}$ must
be extracted from the corresponding quark mass, but the numerical
value of the latter depends strongly on the renormalization scheme and
scale.  For example, a $\msbar{}$ mass $\Mb(\Mb) = 4.16$\UGeV\
corresponds to a pole mass $\Mb$ of about $4.9$\UGeV\ at three-loop
level. Evolving $\Mb(\Mb)$ to a scale of the order of the Higgs mass,
on the other hand, can considerably decrease its value,
e.g.,~$\Mb(M_\phi) = 2.8$\UGeV\ for $M_\phi = 125$\UGeV. While any
change in the definition of the bottom Yukawa coupling entering
the \lo\ part of the calculation is formally compensated for by
counterterm contributions in the \nlo\ part, the numerical impact of
such strong variations on the result for the cross section can be
significant.

Unlike in many other processes where there are theoretical arguments in
favor of one or the other renormalization scheme for the bottom Yukawa
coupling, for Higgs production in gluon fusion we are not aware of any
such arguments that go beyond heuristic.\footnote{In the case of
$\PH\to\PGg\PGg$ the resummation of leading and subleading logs
of the ratio $\MH/\Mb$ has solved this problem for the Abelian
case \cite{Kotsky:1997rq,Akhoury:2001mz}.}  While the options of
relating $Y_{\Pb}$ to $\Mb$ or to $\Mb(\Mb)$ might seem preferable to
the one of using $\Mb(M_\phi)$ in that they lead to smaller \nlo\
$K$-factors, it must be kept in mind that this is due to an accidental
cancellation between terms of different origin in the contributions of
two-loop diagrams with bottom quarks and gluons. There is no argument
suggesting that such cancellation persists at higher orders in \qcd, or
that it is motivated by some physical property of the bottom
contribution to the gluon-fusion process. For example, it was noticed
already in
\Bref{Spira:1995rr} that the two-loop bottom-gluon contribution to the
amplitude for Higgs decay in two photons is minimized when the bottom
mass is defined as $\Mb(M_\phi/2)$, even if the one-loop bottom
contribution has exactly the same structure as the corresponding
contribution to gluon fusion. In summary, there is no obvious reason to
favor one renormalization scheme for the bottom Yukawa coupling over the
others, and it would seem reasonable to consider the difference between
the results obtained with the various schemes as a measure of the
uncertainty associated with the un-computed higher-order
\qcd\ corrections.

\subsubsection{Numerical examples}
We are now ready to discuss the numerical effect of the contributions
that we have included in the computation of the cross section for
Higgs production in gluon fusion. As a representative choice for the
MSSM parameters, we take the \lstop{} scenario discussed in
\refS{sec:sven-sub}. We use
\FeynHiggs~\cite{Heinemeyer:1998yj,
  Heinemeyer:1998np,Degrassi:2002fi,Frank:2006yh} to compute the Higgs boson masses and
mixing angle, as well as the $\tb$-enhanced correction to the
Higgs-bottom coupling, $\Delta_{\Pb}$\,. To compute the total cross section
we use \sushi~\cite{Harlander:2012pb}, and cross-check the results
with a private code. It is useful to remark that the two-loop
calculations of the Higgs-boson masses and production cross section
implemented in \FeynHiggs\ and \sushi, respectively, adopt the same
on-shell renormalization scheme for the parameters in the stop and
sbottom sectors. As a consequence, the numerical values of the
parameters for the \lstop{} scenario listed in
\refS{sec:sven-sub} can be passed to both codes as they are.

\begin{figure}[t]
\vspace*{-2.2cm}
\hspace*{-4.2cm}
\includegraphics[width=0.62\textwidth,angle=-90]{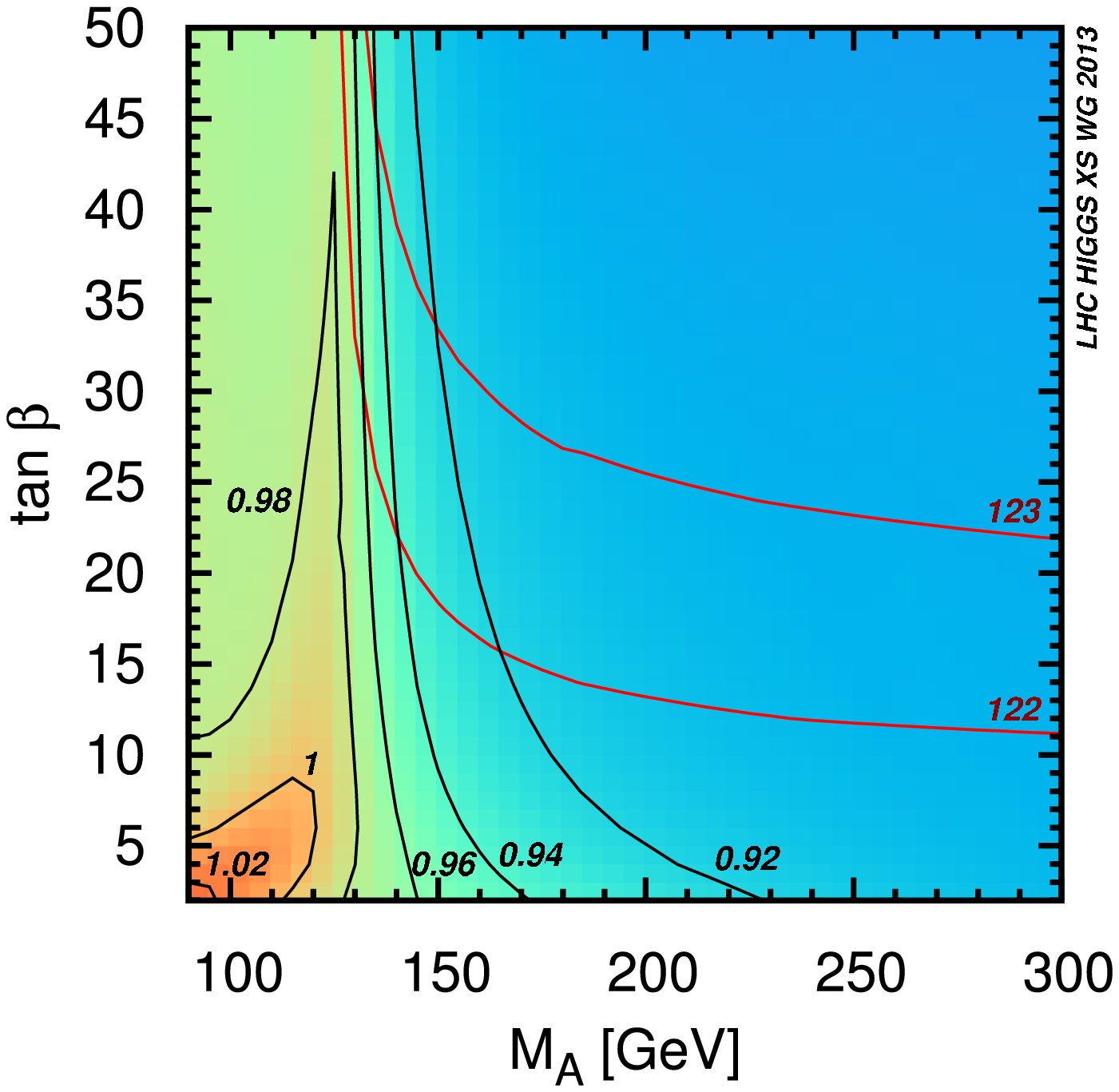}
\hspace*{-2.8cm}
\includegraphics[width=0.62\textwidth,angle=-90]{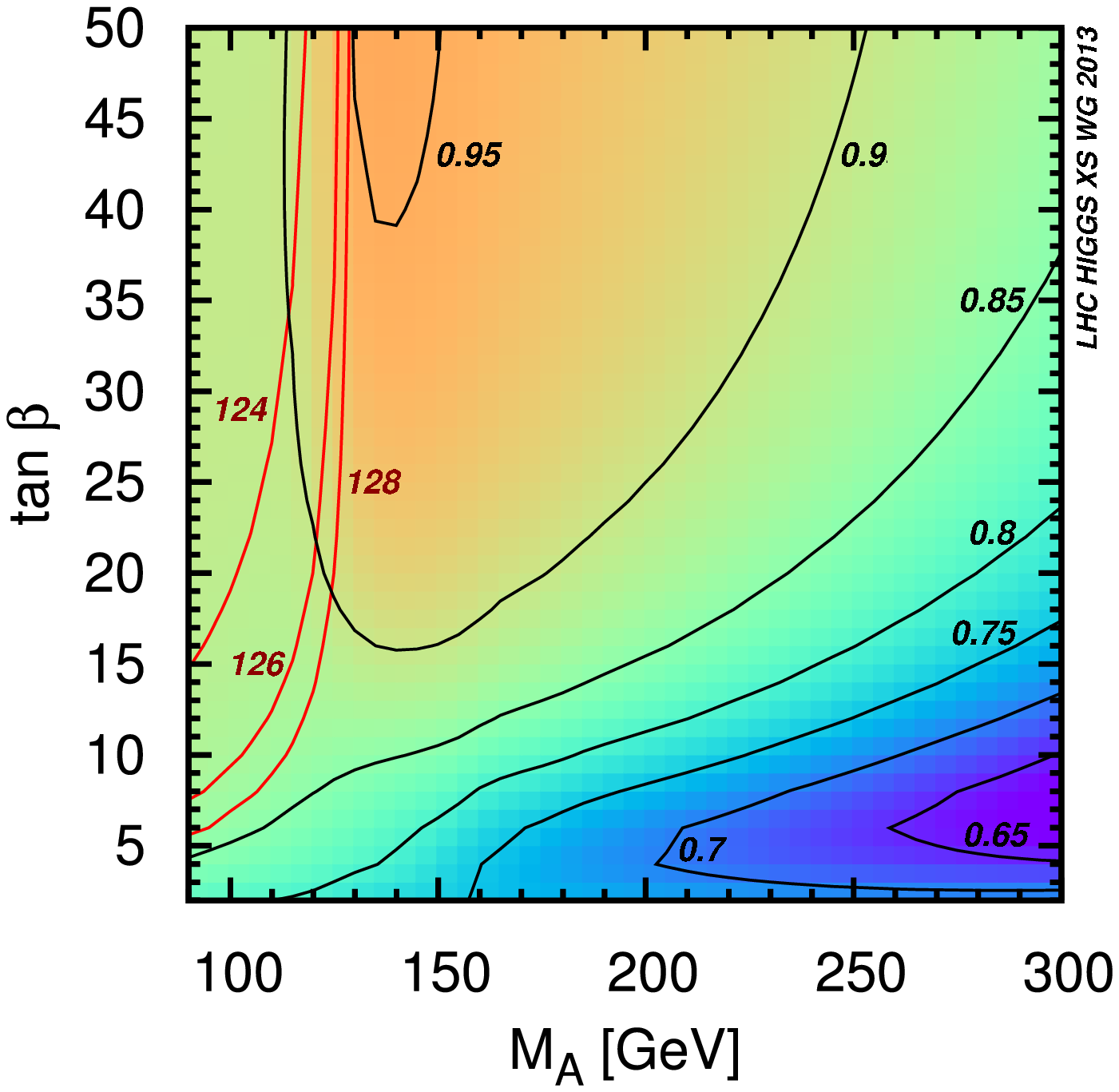}
\vspace*{.1cm}

\caption{Ratio of the cross section for $\Ph$ (left) and $\PH$ (right)
  production in gluon fusion as computed by \sushi, over the
  corresponding cross section computed omitting squark contributions
  and EW corrections. The MSSM parameters are chosen as in the 
  \lstop{} scenario of \Bref{Carena:2013qia}.}
\label{fig:contour}
\vspace*{-1mm}
\end{figure}

\refF{fig:contour} shows the contours in the $\MA{-}\tb$ plane
of equal ratio between the cross section for Higgs-boson production in
gluon fusion computed by \sushi\ and the corresponding cross section
computed as in the earlier LHCHXSWG reports,
\Brefs{Dittmaier:2011ti,Dittmaier:2012vm}.  In particular, the former
includes the \nlo-\qcd\ calculation of both quark and squark
contributions plus the dominant \nnlo-\qcd\ and \nlo-\ew\ effects
adapted from the \sm\ calculation, while the latter includes only the
\nlo-\qcd\ calculation of quark contributions (supplemented with the
$\tb$-enhanced SUSY corrections to the Higgs-bottom couplings) and the
dominant \nnlo-\qcd\ effects from top-quark loops. The latter have been
evaluated using \nnlo{} \pdf{}s, while all the \nlo{} terms are obtained
with \nlo{} \pdf{}s.  The plot on the left in \refF{fig:contour} refers
to the production of the lightest scalar $\Ph$, while the plot on the
right refers to the production of the heaviest scalar $\PH$. The red
lines superimposed to each plot are the contours of equal mass for the
corresponding scalar (for $\PH$ we only show the contours between
$124$\UGeV\ and $128$\UGeV). The standard LHCHXSWG values for the SM
input parameters have been used in these plots (in particular, we set
$\mto=172.5$\UGeV\ and $\Mb=4.75$\UGeV).

From the left plot in \refF{fig:contour} it can be seen that, in the
\lstop{} scenario, the combined effect of the squark
contributions and the \nlo-\ew\ corrections tends to suppress the
cross section for $\Ph$ production, with a maximum effect of
$8{-}10\%$ in the region with $\MA \gsim 150$\UGeV, where $\Ph$ has
\sm-like couplings to quarks and gauge bosons. It is useful to remark
that this results from a partial compensation between the
contributions of stop loops, which in this scenario can reduce the
cross section of the lightest scalar by up to $14{-}16\%$, and the
\nlo-\ew\ light-quark contributions, which increase by approximately
$6\%$ the cross section of a \sm-like scalar with mass around
$125$\UGeV~\cite{Bonciani:2010ms}.

The right plot in \refF{fig:contour} shows that in the case of $\PH$
production the effect of the squark contributions can be somewhat
larger than in the case of $\Ph$ production (the \nlo-\ew\ light-quark
contributions, on the other hand, become negligible for sufficiently
large $\MA$, due to the vanishing couplings of $\PH$ to gauge bosons).
However, a suppression of the order of $35{-}40\%$ is reached only in
the lower-right corner of the plot, where $\MA$ is large and
$\tb$ ranges between $5$ and $10$. In this region, the coupling
of $\PH$ to top quarks is suppressed while the coupling to bottom
quarks is not sufficiently enhanced, resulting in very small
gluon-fusion cross sections, of the order of tenths of a picobarn.

Concerning uncertainties, we study the dependence of the cross section
on the renormalization and factorization scales, $\muR$ and $\muF\,$, by
varying them simultaneously within $\muR=\muF=[M_\phi/2,2M_\phi]$. The
{\abbrev PDF} uncertainty is derived using the {\abbrev MSTW2008 PDF}
prescription, where, somewhat conservatively, only the NLO set is used
for the error estimate, even though the NNLO set enters in the top-quark
induced contributions to the cross section. Since scale and PDF
variation are purely driven by QCD\cite{Dittmaier:2011ti}, their effect
is expected to be similar for all the three neutral Higgs bosons
(differences at the percent level can arise due to the different weight
of the bottom- and top-loop contributions, which are included at
different orders in perturbation theory).  Indeed, for both the lightest
and the heaviest scalar, we find that the scale variation is of the
order of $\pm 10\%$, while the PDF uncertainty amounts to about $\pm
2.5\%$.

We estimate the uncertainty associated to the renormalization of the
bottom Yukawa coupling by comparing three different options: the first
assumes on-shell renormalization of $Y_{\Pb}$ and employs a value of
$4.9$\UGeV\ for the pole bottom mass $\Mb$, computed at three-loop
accuracy from the $\msbar{}$ input value $\Mb(\Mb)=4.16$\UGeV{}; the
second also assumes on-shell renormalization, but employs $\Mb =
4.75$\UGeV, consistent with the value used by the {\abbrev MSTW2008
PDF} set (this is our default choice); the third option assumes the
$\msbar{}$ scheme and evaluates the bottom Yukawa coupling at the
renormalization scale $\mu_{\Pb}=\Mb$, thus directly employing the
input value $\Mb(\Mb)=4.16$\UGeV.  For illustration, we obtain an even
more conservative estimate of the uncertainty by relating the bottom
Yukawa coupling to the $\msbar$ mass computed at the scale
$\mu_{\Pb}=M_\phi$.

Since the bottom Yukawa coupling enters the cross-section predictions
for $\Ph$, $\PH$, and $\PA$ with very different weights, the
uncertainty due to its renormalization prescription depends
significantly on the particle considered and on the \susy\
scenario. For example, in the \lstop{} scenario with
$\MA=130$\UGeV\ and $\tanb=40$, where both scalars have an enhanced
coupling to the bottom quark, the effect of extracting the bottom
Yukawa coupling from $\Mb=4.9$\UGeV\ instead of the default
$\Mb=4.75$\UGeV\ leads to a $7\%$ increase in the cross section for
$\Ph$ production, and a $10\%$ increase in the one for $\PH$
production. On the other hand, using $\msbar$ renormalization with
$\Mb(\Mb)=4.16$\UGeV\ decreases the cross section for $\Ph$ production
by $10\%$, and the one for $\PH$ production by $19\%$. Using
$\Mb(M_\phi)$ instead would decrease the cross section for $\Ph$
production by $27\%$, and the one for $\PH$ production by $52\%$.
As a second example, we consider the \lstop{} scenario with
$\MA=300$\,\UGeV{} and $\tanb=10$, where the lightest scalar $\Ph$ has
\sm-like couplings to quarks. In this case the cross section for $\Ph$
production varies by less than $\pm2\%$ when choosing among the three
options for the bottom Yukawa coupling discussed above. For the
heaviest scalar $\PH$, on the other hand, the changes relative to the
value derived with $\Mb=4.75$\UGeV\ amount to $+8\%$, $-15\%$ and
$-45\%$ for a Yukawa coupling extracted from $\Mb=4.9$\UGeV,
$\Mb(\Mb)$ and $\Mb(M_\phi)$, respectively.

Finally, in order to obtain an estimate of the influence of higher
order squark effects, we follow \Bref{Harlander:2003kf} and find that
they typically decrease the cross section by about $2{-}3\%$.

Let us now turn to kinematical distributions. In regions of the MSSM
parameter space where the Higgs coupling to bottom quarks is enhanced,
the transverse-momentum distribution of a scalar produced via gluon
fusion can be distorted with respect to the corresponding distribution
of a SM Higgs
boson\,\cite{Langenegger:2006wu,Brein:2003df,Brein:2007da}. In order to
investigate this effect, we consider again the point in the
\lstop{} scenario with $\MA = 130$\UGeV\ and $\tb = 40$,
characterized by the fact that both scalars have non-standard couplings
to quarks and masses in the vicinity of the LHC signal
(indeed, \FeynHiggs\ predicts $\Mh = 122.4$\UGeV\ and $\MH =
129.3$\UGeV). This point is likely to be already excluded by the ATLAS
and CMS searches for neutral Higgs bosons decaying into $\PGtp\PGtm$
pairs, but it can still provide a useful illustration of the expected
size of this kind of effects.

\begin{figure}[t]
\includegraphics[width=0.49\textwidth]
{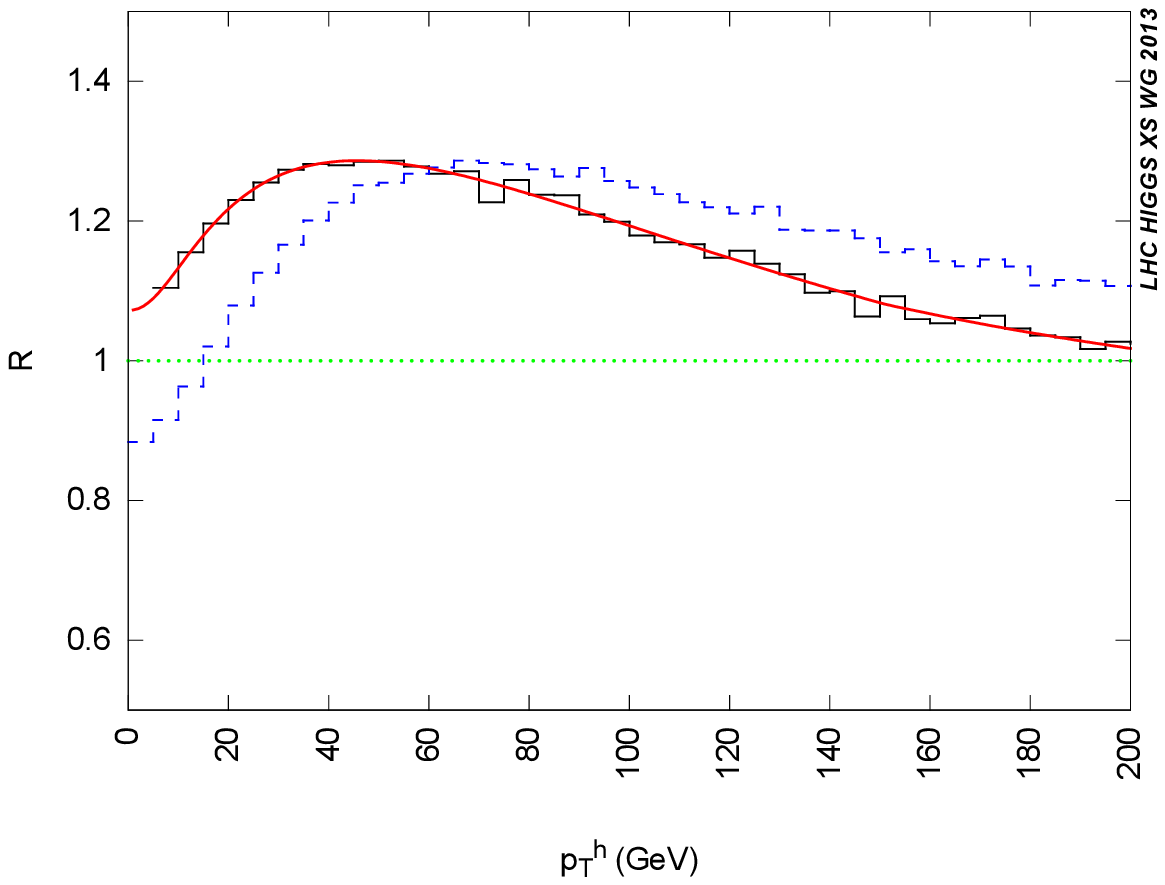}~~
\includegraphics[width=0.49\textwidth]
{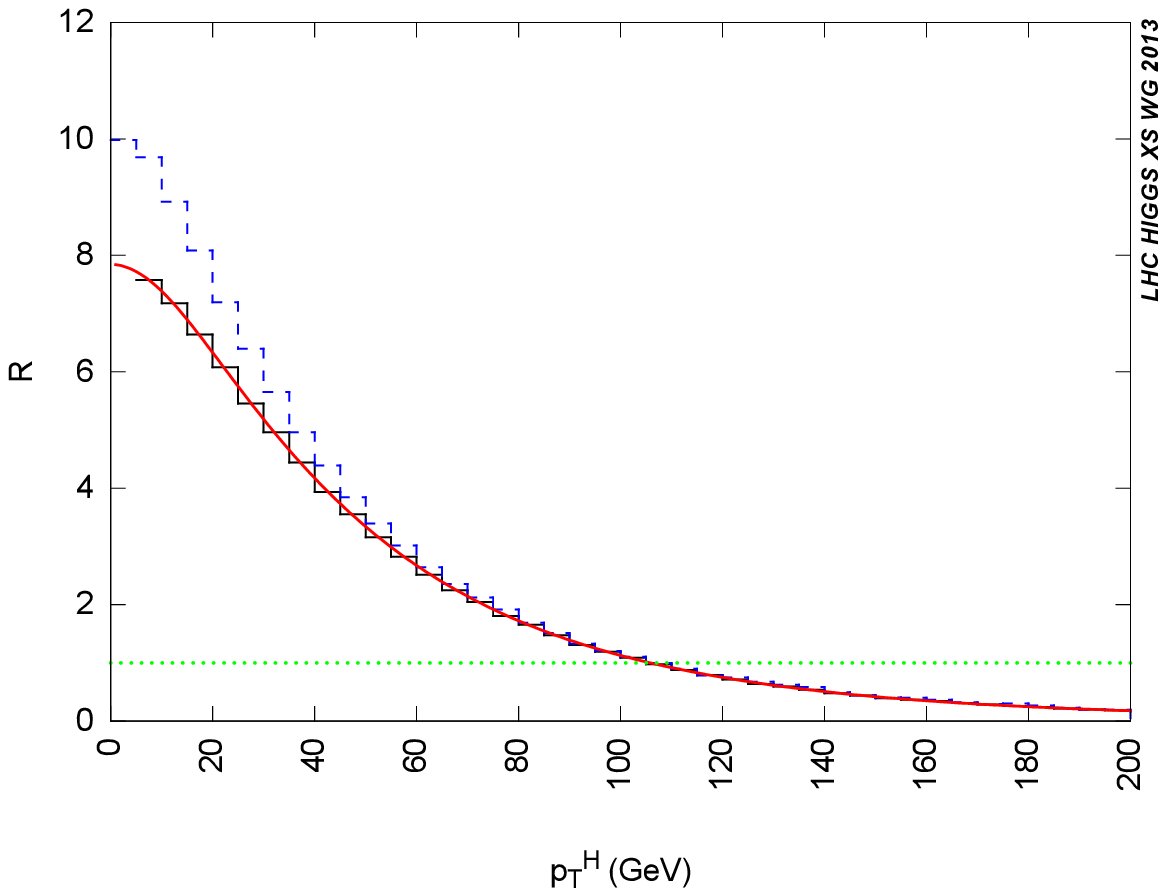}
\vspace*{-0.5cm}
\caption{Ratio of the transverse-momentum distribution for the MSSM
  scalar $\Ph$ (left) or $\PH$ (right) over the distribution for a SM
  Higgs with the same mass, in the \lstop{} scenario with $\MA =
  130$\UGeV\ and $\tb = 40$. The meaning of the different curves
  is explained in the text.}
\label{fig:distributions}
\end{figure}

In \refF{fig:distributions} we show the ratio of the
transverse-momentum distribution for a MSSM scalar produced via gluon
fusion over the corresponding distribution for a SM Higgs with the
same mass. The plot on the left refers to the lightest scalar $\Ph$,
while the plot on the right refers to the heaviest scalar $\PH$. In
each plot, the continuous (red) line represents the ratio of
distributions computed at \nlo\ by \sushi, while the two histograms
are computed with the \powheg\ implementation of gluon fusion of
\Bref{Bagnaschi:2011tu}, modified by the adoption of the on-shell
renormalization scheme for the squark parameters and the inclusion of
the results of \Bref{Degrassi:2012vt} for the squark contributions to
$\PH$ production. In particular, the solid (black) histogram
represents the ratio of distributions computed in a pure (i.e.,
fixed-order) \nlo\ calculation, while in the dashed (blue) histogram
the distributions are computed with the \powheg\
method~\cite{Frixione:2007vw,Alioli:2008tz,Alioli:2010xd}, in which
the potentially large logarithms of the form $\ln (p_T^\phi/M_\phi^2)$
are resummed via the introduction of a Sudakov form factor and a
parton-shower generator to describe multiple gluon emission (in this
case, we use \pythia\cite{Sjostrand:2006za}).

The plots in \refF{fig:distributions} show that, in this point of the
MSSM parameter space, the enhancement of the Higgs-bottom coupling
results in both an enhancement of the total cross section and a
distortion of the transverse-momentum distribution, in particular for
the heaviest scalar $\PH$ (note the difference in the scale between the
left and the right plot). The effect of the resummation in \powheg\ and
the unitarity constraint implemented in the matching procedure of NLO
matrix elements with parton shower make the transverse-momentum
distribution of the Higgs bosons harder.  The comparison between the
solid and dashed histograms shows that for $\Ph$ this effect is somewhat
stronger than in the SM, while for $\PH$ it is somewhat weaker.

It remains to stress that \sushi\ also allows for the calculation of the
cross section for the pseudoscalar Higgs $\PA$, where, however, squark
effects are much less important because they are absent at
LO. Furthermore, \sushi\ includes the production cross section for
bottom quark annihilation; we refrain from quoting any numerical results
here, as they would not significantly differ from what has already been
presented in earlier reports of this working group.

%%%%%%%%%%%%%%%%%%%%%%%%%%%%%%%%%%%%%%%%%%%%%%%%%%%%%%%%%%%%%%%%%%%%%%%%%%%%%%%

%- }}}
%- {{{ subsection{Light charged Higgs limits as a function of SUSY pars}

\subsection{Light charged Higgs limits as a function of SUSY parameters}
\label{sec:samil-sub}

The exclusion limits in the ($\MHpm$, $\tb$) plane are studied
for different values of the most relevant SUSY parameters in the Higgs
sector: $\mu$, $M_2$, $\Mgl$, $\msusy$, $\Xt$. The model
independent limits for the light charged Higgs boson branching
ratio \cite{CMS-PAS-HIG-12-052} are transformed into limits in the
($\MHpm$, $\tb$) plane with BR($\PQt\rightarrow \PQb\PSHpm$) and
BR($\PSHpm\rightarrow\PGt\PGn$) branching ratios calculated with {\sc
FeynHiggs} $2.9.4$
\cite{Frank:2006yh,Degrassi:2002fi,Heinemeyer:1998np,
Heinemeyer:1998yj}.  The studied $\MHpm$ mass range is
$100{-}160$\UGeV\ and the $\tb$ range is $5{-}100$.  The exclusion
limits are shown\footnote{Only the region up to $\tanb=65$ is shown for
the sake of readability of the plots.} in \refF{fig:limits_mhmax} for
the \mhmax\ scenario~\cite{Carena:2002qg}, \mhmodp, \mhmodm, \lstop,
\lstau{} and \tauphobic{} \cite{Carena:2013qia} benchmark scenarios.  The
effect of varying the SUSY parameters on the exclusion limits are
compared with the limits derived for the \mhmax\ scenario.  Since the
importance of one parameter may depend on the value of another
parameter, the effect of varying also the most significant parameter
$\mu$ is studied for each parameter variation.

Based on the recent discovery of a Higgs-boson-like particle at the
LHC with measured mass of $M_\Phi = 126.0 \pm 0.4 \pm
0.4$\UGeV~\cite{Aad:2012tfa} (ATLAS) and $M_\Phi = 125.8 \pm 0.4 \pm
0.4$\UGeV~\cite{Chatrchyan:2012ufa} (CMS), an allowed region is
derived in the ($\MHpm$, $\tb$) plane assuming that the discovered
particle is the light $\cp$-even MSSM Higgs boson \PSh.  The
theoretical uncertainty is of the order of
$3$\UGeV~\cite{Degrassi:2002fi,Allanach:2004rh,Heinemeyer:2004gx}, and
a $\pm$3\UGeV\ iso-mass curve around the central value $125.9$\UGeV\ is
drawn in
\refF{fig:limits_mhmax} to indicate the region of the parameter space 
allowed for the charged Higgs boson in the given scenario. 
The allowed region depends on the values of the SUSY parameters, but
the lower limit for the allowed charged Higgs mass seems to be quite
stable against the choice of the SUSY parameter values. The minimum
possible value of $\tb$ also varies around $\tb = 10$ for
the studied choices of the SUSY parameters up to $\MHpm$ =
$600$\UGeV.  The maximum possible value of $\tb$ is more
dependent on the choice of the SUSY parameters.  For $\msusy$ =
$2$\UTeV\ (changing the values of $\Mgl$ and $\Xt$ to
0.8$\times$$\msusy$ and 2$\times$$\msusy$, accordingly)
and for $\Xt$ = $-2000$\UGeV\ the entire high charged Higgs
mass, high $\tb$ corner is allowed for all tested values of the
$\mu$ parameter, shown in \refF{fig:limits_mhmax_msusy} (top right).
The mass region is extended up to $600$\UGeV\ to see the effect of the
SUSY parameter variation on the horizontal tail, which in turn gives a
prediction for the allowed values of $\tb$ for the heavy charged
Higgs boson ($\MHpm > \mto$).

It is also possible that the discovered Higgs-like particle is the
heavy $\cp$-even Higgs boson (or the heavy $\cp$-even and $\cp$-odd
Higgs bosons together). In this case the \mhmax\ scenario is ruled out
as the mass of the discovered particle is lower than the minimum
possible $\MH$ (= \mhmax). \refF{fig:limits_mhmax_msusy} (bottom)
shows the allowed region for the heavy $\cp$-even and $\cp$-odd Higgs
bosons when the $\msusy$ parameter is changed to $2$\UTeV. The lower
edge of the allowed region depends on the choice of the $\mu$
parameter, but extends over the high $\tb$ region for all values of
the $\mu$ parameter studied here. In
\refF{fig:limits_mhmax_msusy} (bottom left) no bound on $\MA$ is used. 
\refF{fig:limits_mhmax_msusy} (bottom right)
shows the allowed region if \PSH and \PSA are both assumed to produce
the experimental result. With this choice of $\msusy$ the region
with intersecting allowed areas for \PSH and \PSA vanish for positive
values of the $\mu$ parameter.

\begin{figure}[p!]
  \centering
  \includegraphics[width=0.465\textwidth]{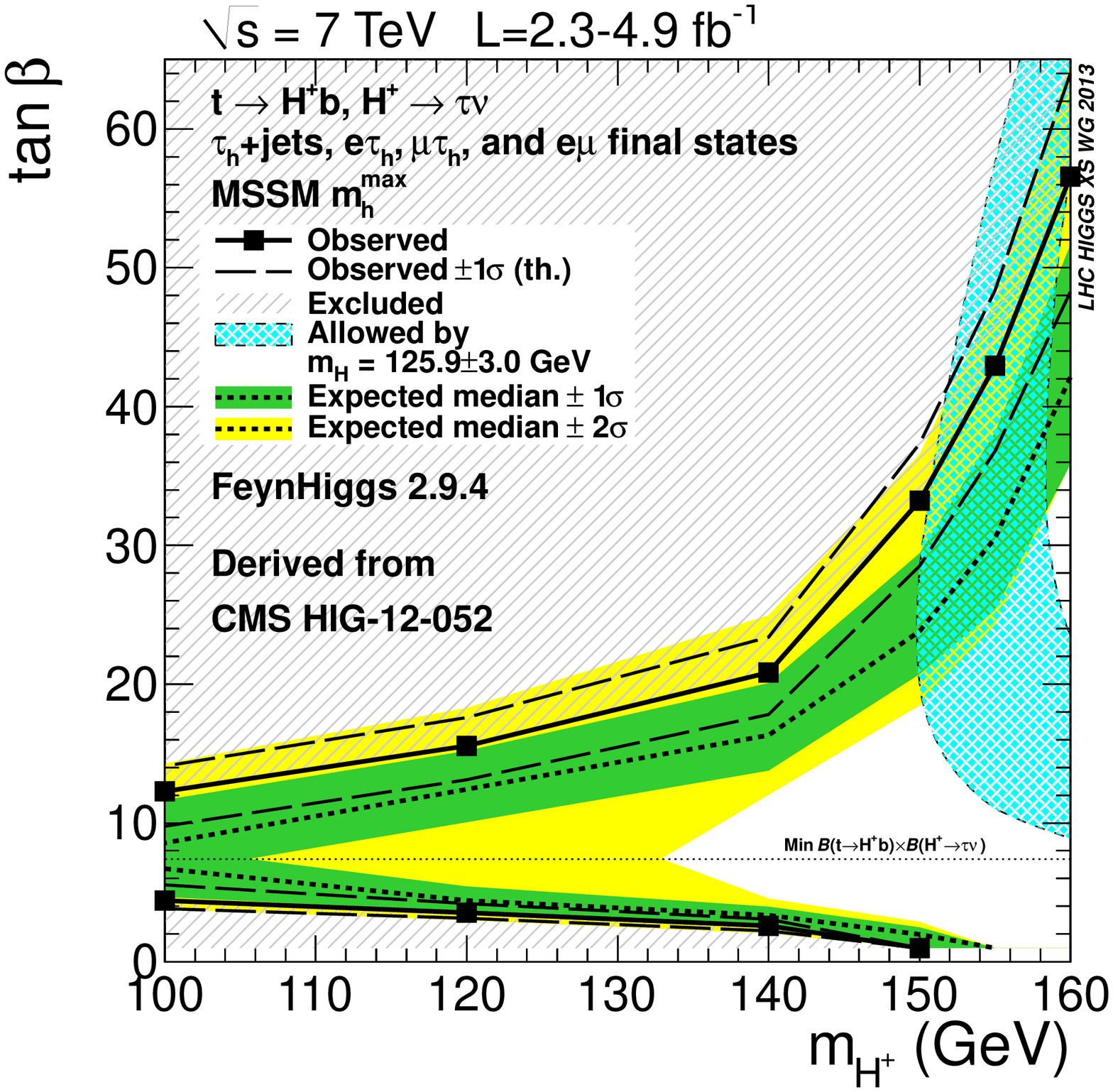}
  \includegraphics[width=0.465\textwidth]{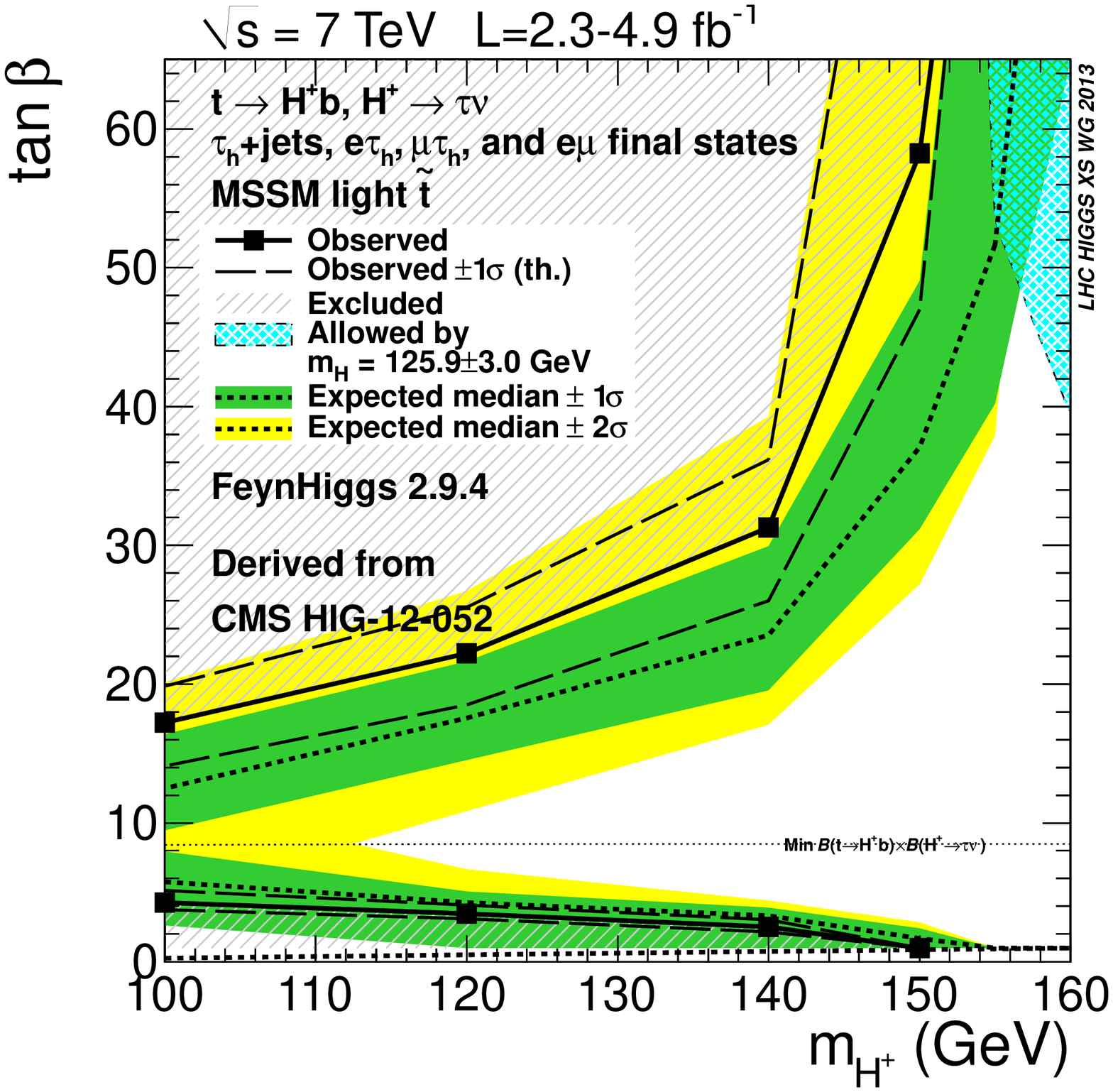}\\
  \includegraphics[width=0.465\textwidth]{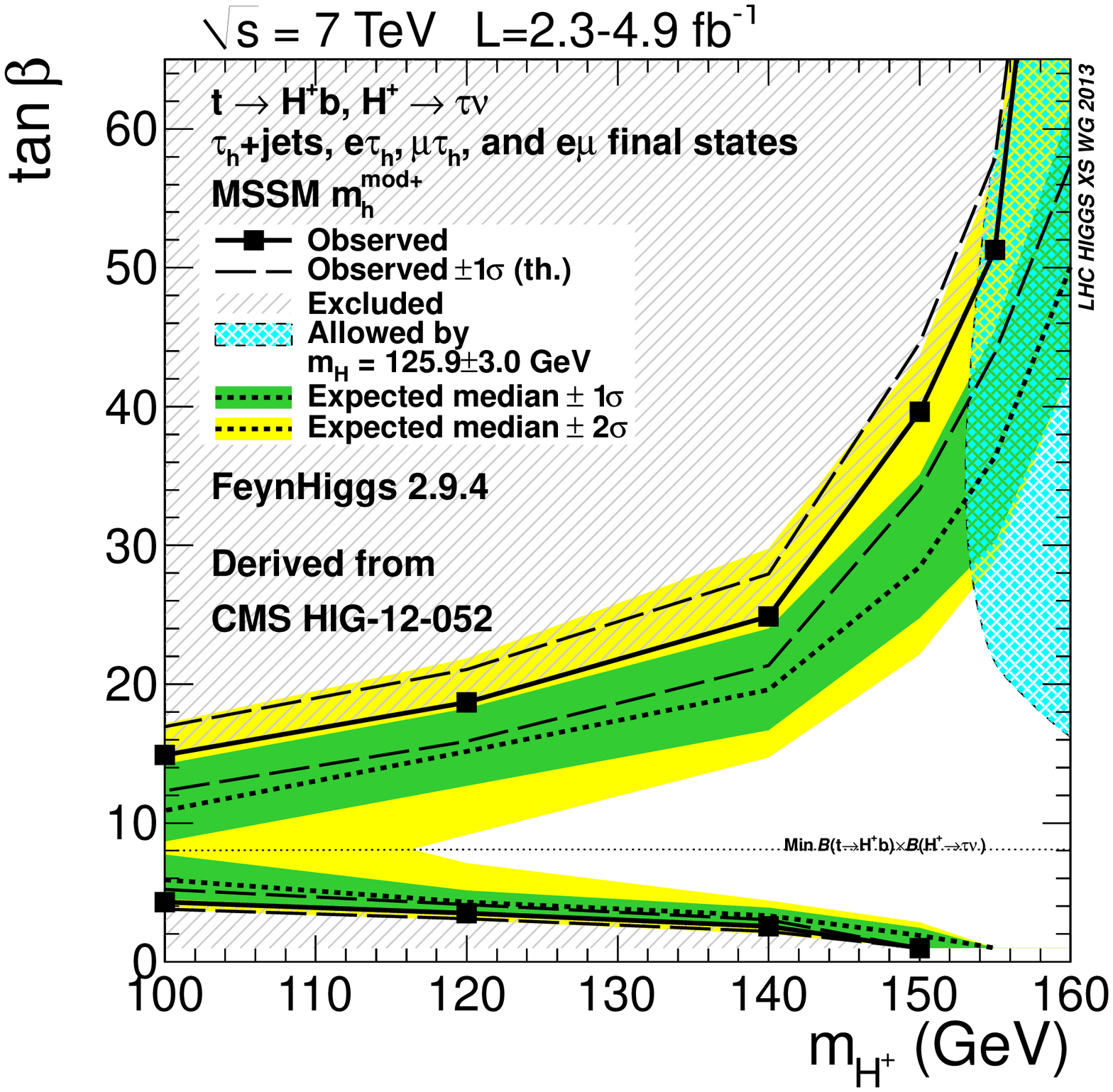}
  \includegraphics[width=0.465\textwidth]{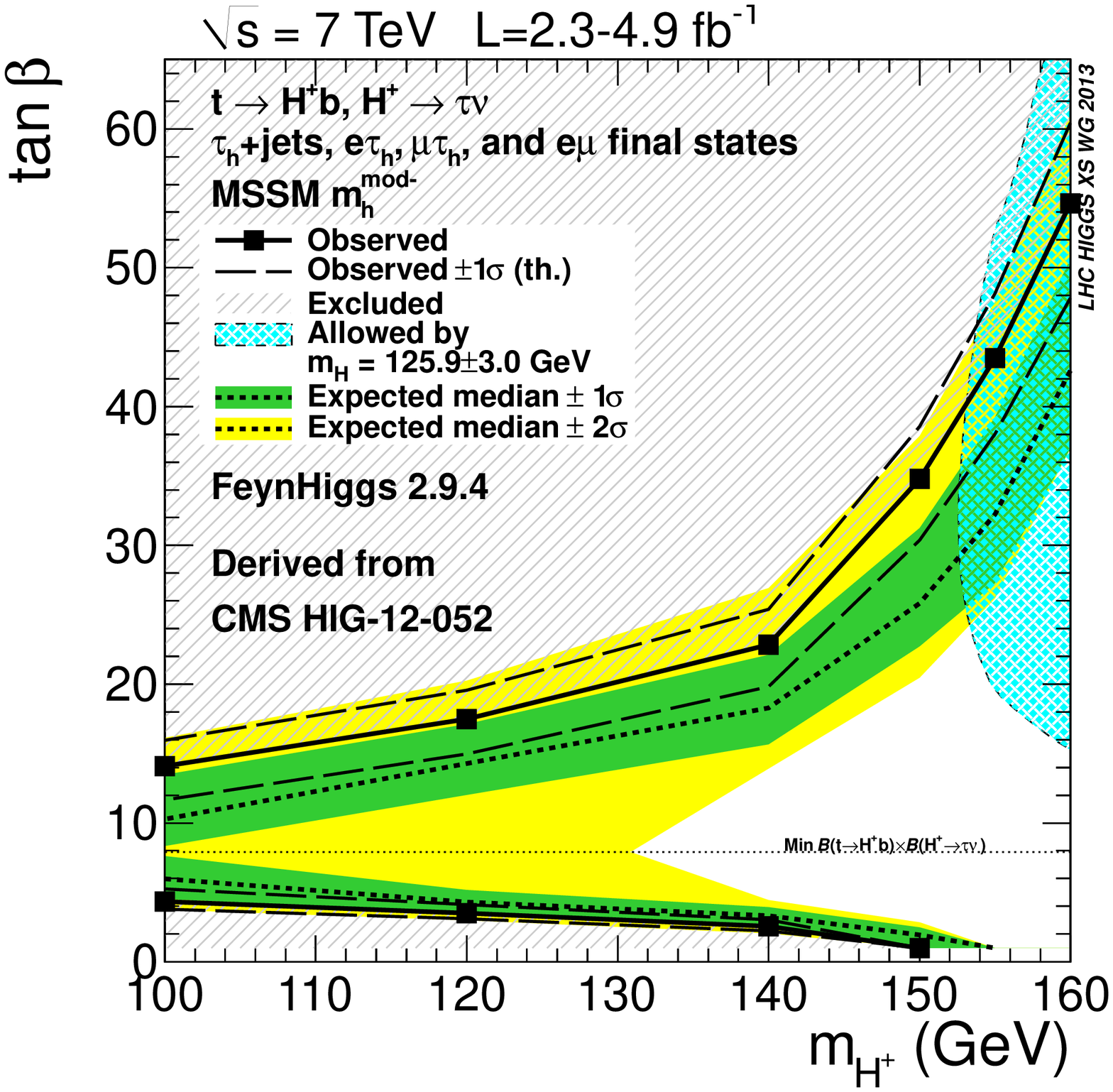}\\
  \includegraphics[width=0.465\textwidth]{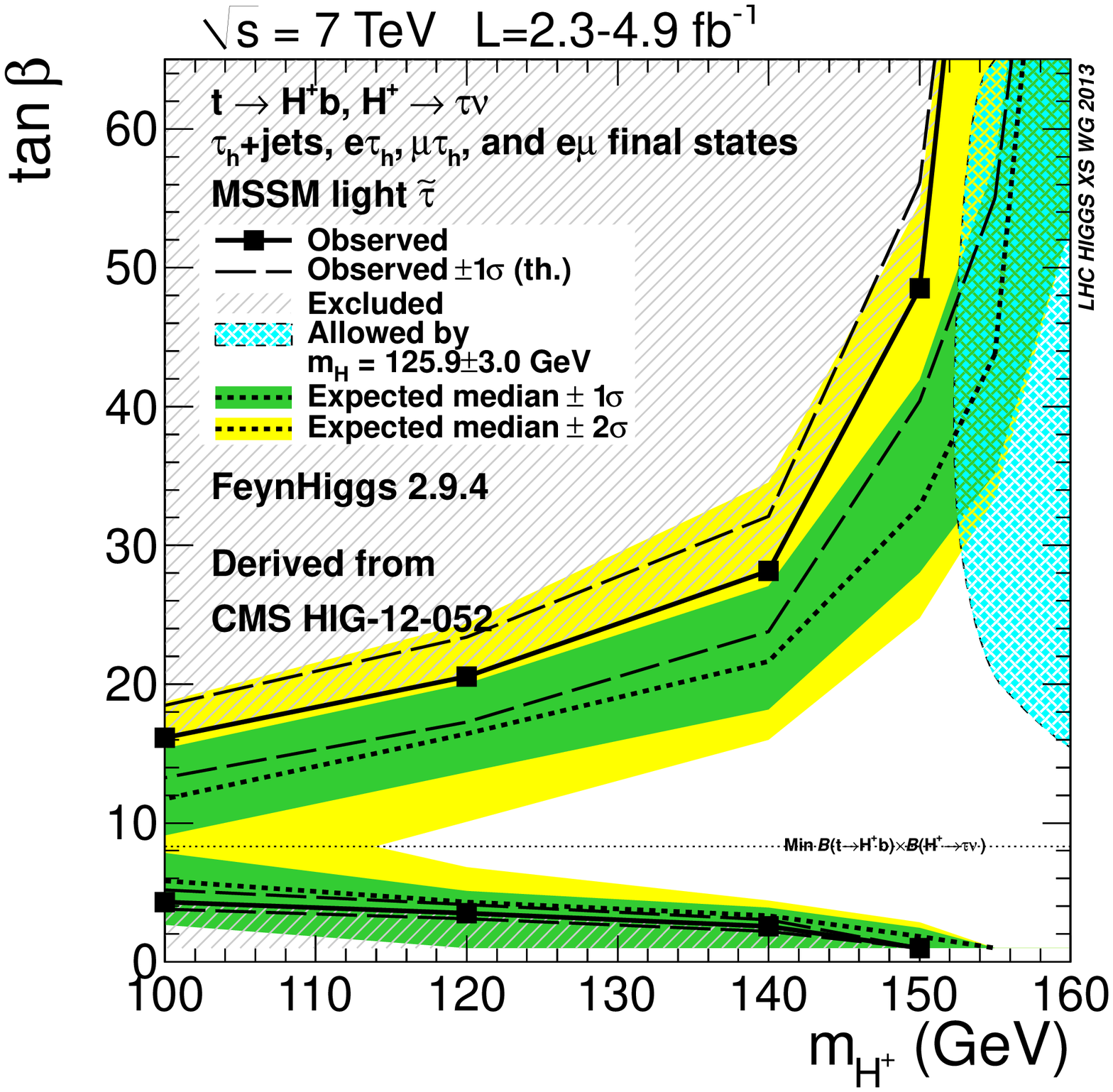}
  \includegraphics[width=0.465\textwidth]{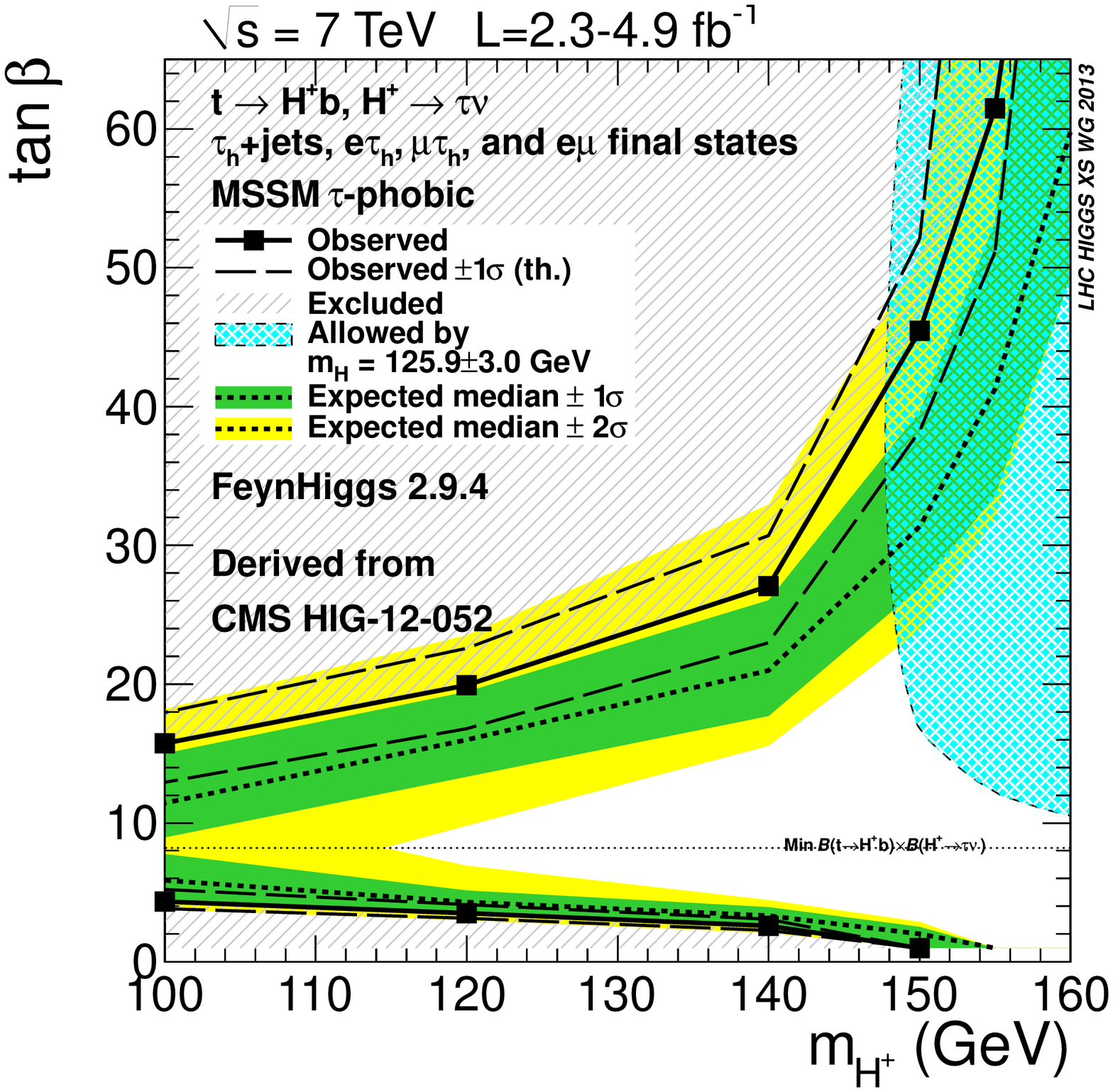}
  \caption{The exclusion limit derived from \cite{CMS-PAS-HIG-12-052} 
for the light charged Higgs boson in the ($\MHpm$, $\tb$)
plane in the \mhmax\ scenario\cite{Carena:2002qg} (top left), in
the \lstop{} scenario \cite{Carena:2013qia} (top right), in the
\mhmodp\ scenario \cite{Carena:2013qia} (middle left) in the
\mhmodm\ scenario \cite{Carena:2013qia} (middle right) in the
\lstau{} scenario \cite{Carena:2013qia} (bottom left) and in
the \tauphobic{} scenario \cite{Carena:2013qia} (bottom right).}
\label{fig:limits_mhmax}
\end{figure}

\begin{figure}[t!]
\vspace*{-1.5mm}
  \centering
  \includegraphics[width=0.48\textwidth]{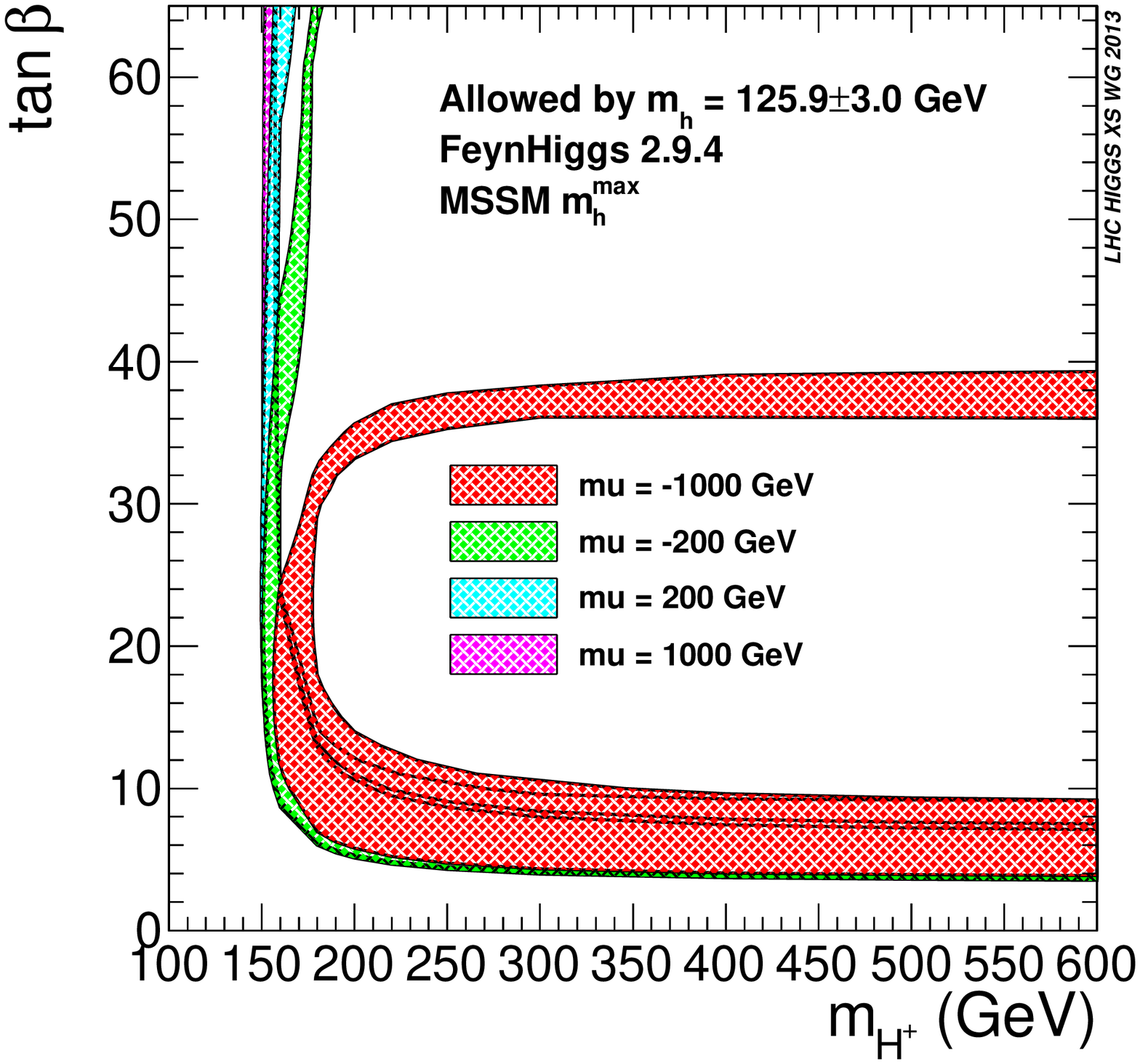}
  \includegraphics[width=0.48\textwidth]{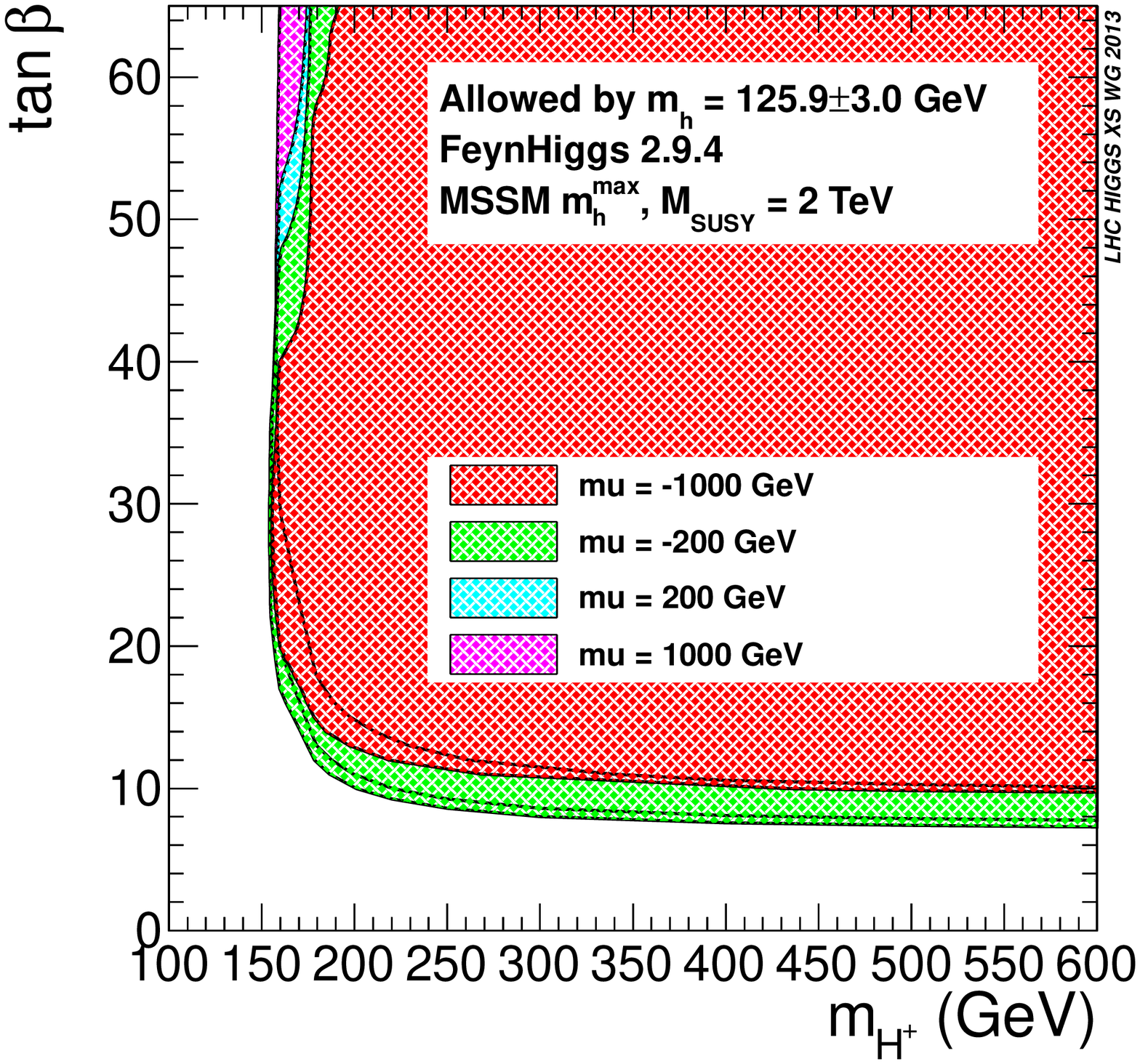}\\
  \includegraphics[width=0.48\textwidth]{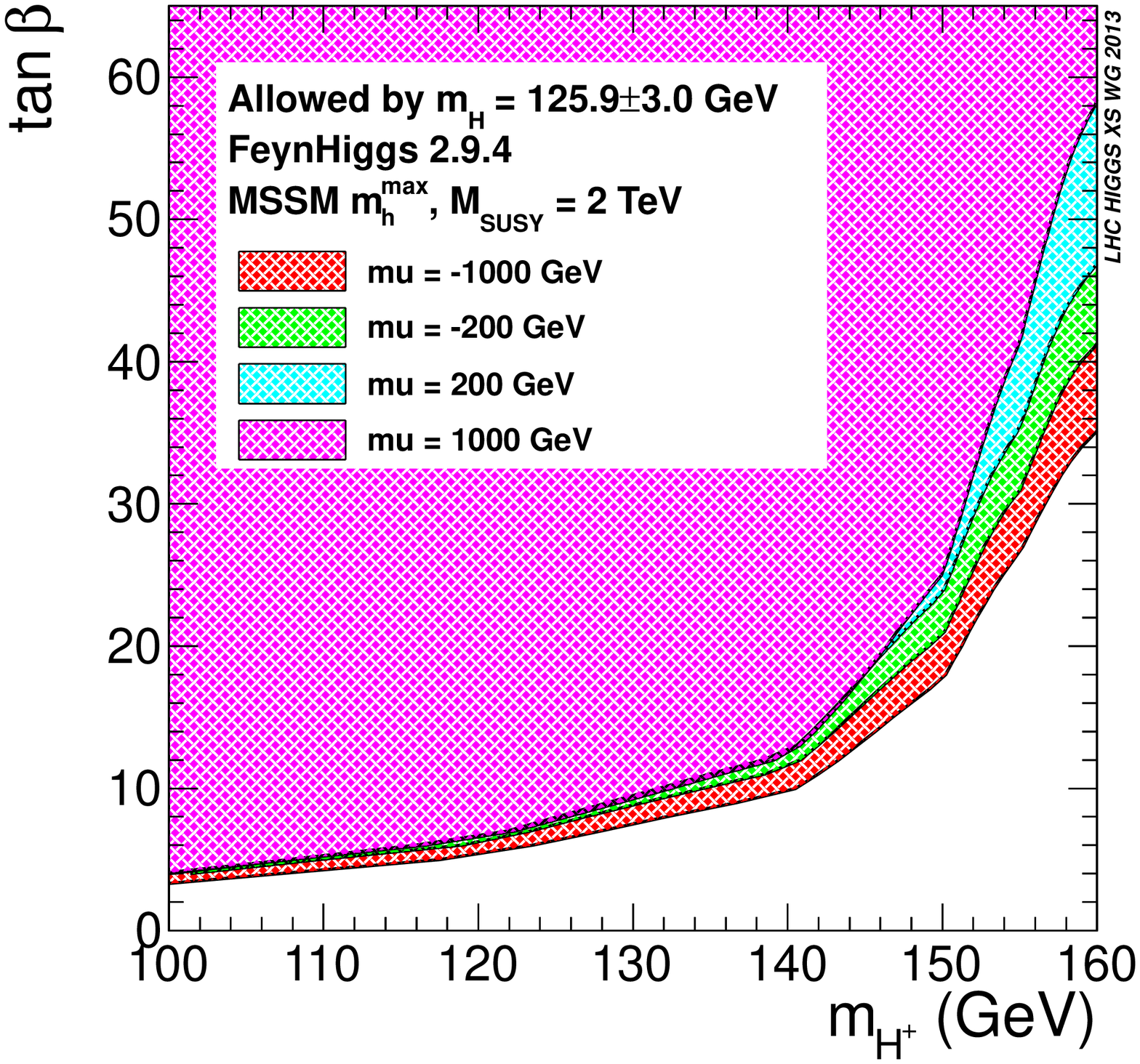}
  \includegraphics[width=0.48\textwidth]{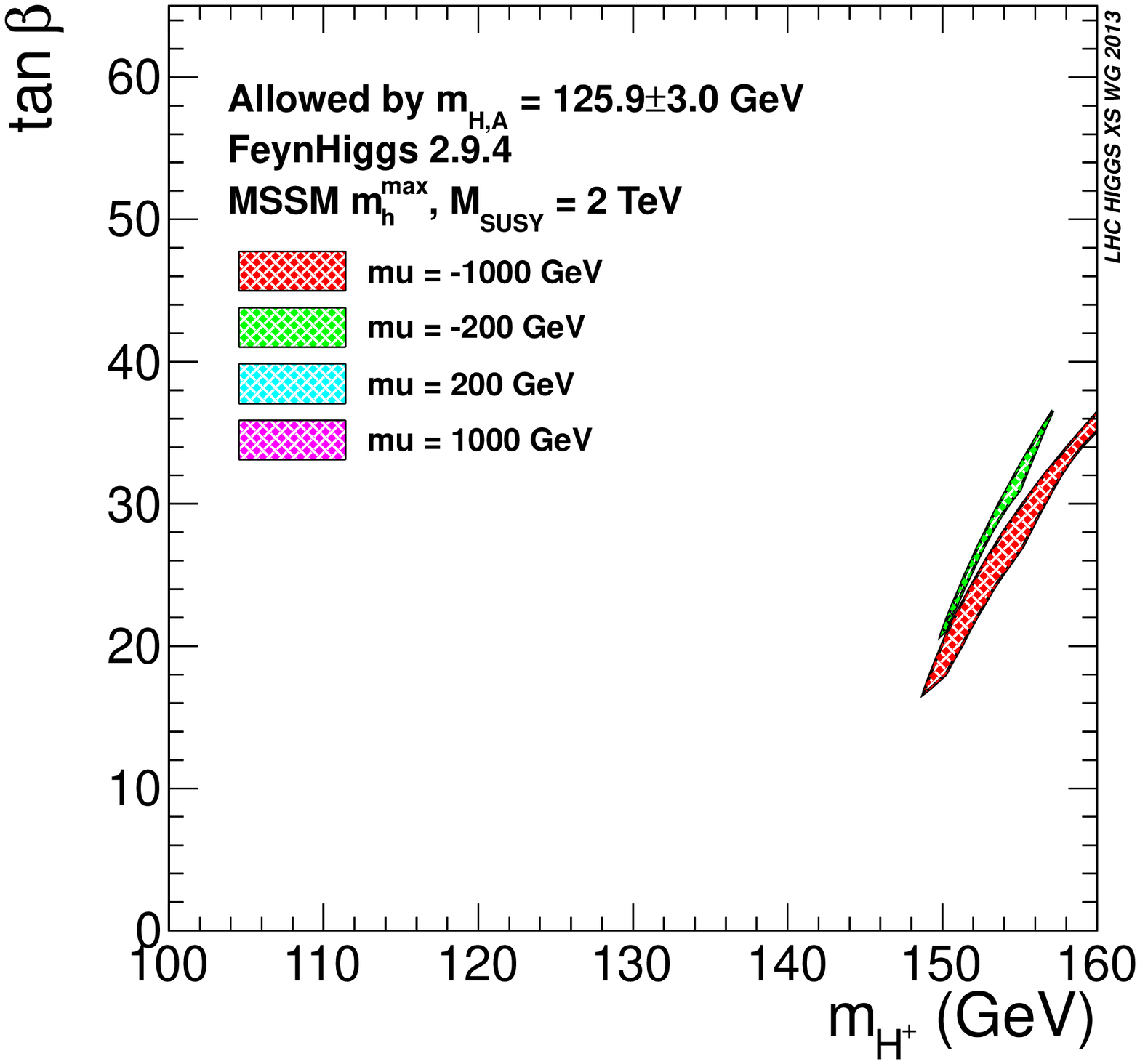}
\vspace*{-0.5mm}
  \caption{The region allowed by $\Mh = 125.9\pm3.0$\UGeV\ in the
\mhmax\ scenario (top left), in the \mhmax\ scenario with
$\msusy$ set to 2\UTeV\ (top right) and by $\MH =
125.9\pm3.0$\UGeV\ (bottom left) and by both \PSH\ and \PSA,
$M_{\PSH,\PSA} = 125.9\pm3.0$\UGeV\ (bottom right). If $\mu>0$ the
allowed regions for \PSH\ and \PSA\ do not intersect.}
\label{fig:limits_mhmax_msusy}
\vspace*{-2.5mm}
\end{figure}

\subsubsection{Variation of $\mu$, $\msusy$, $\Mgl$, $M_2$ and $\Xt$}

\begin{figure}[p!]
  \centering
  \includegraphics[width=0.475\textwidth]{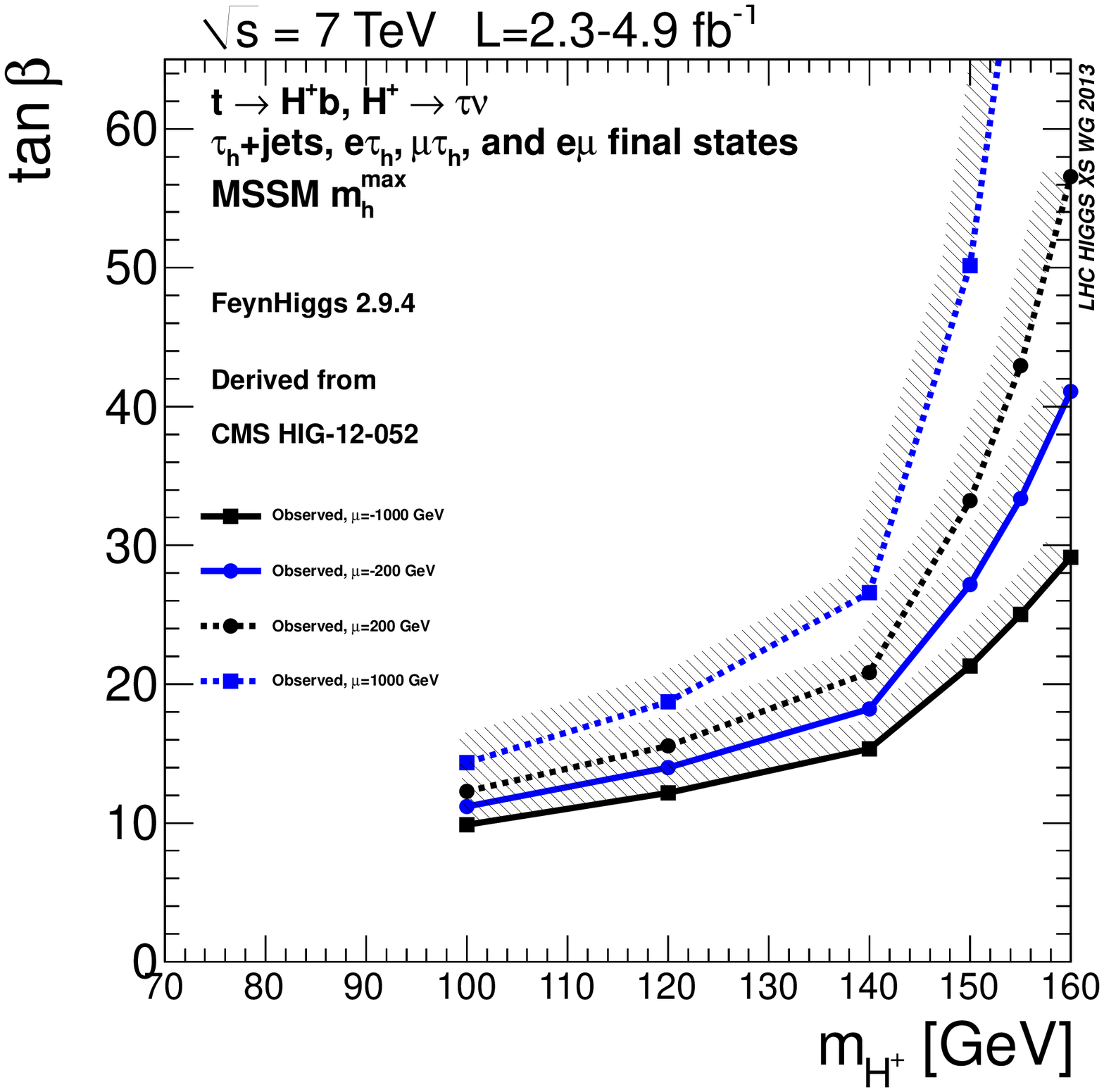}
  \includegraphics[width=0.475\textwidth]{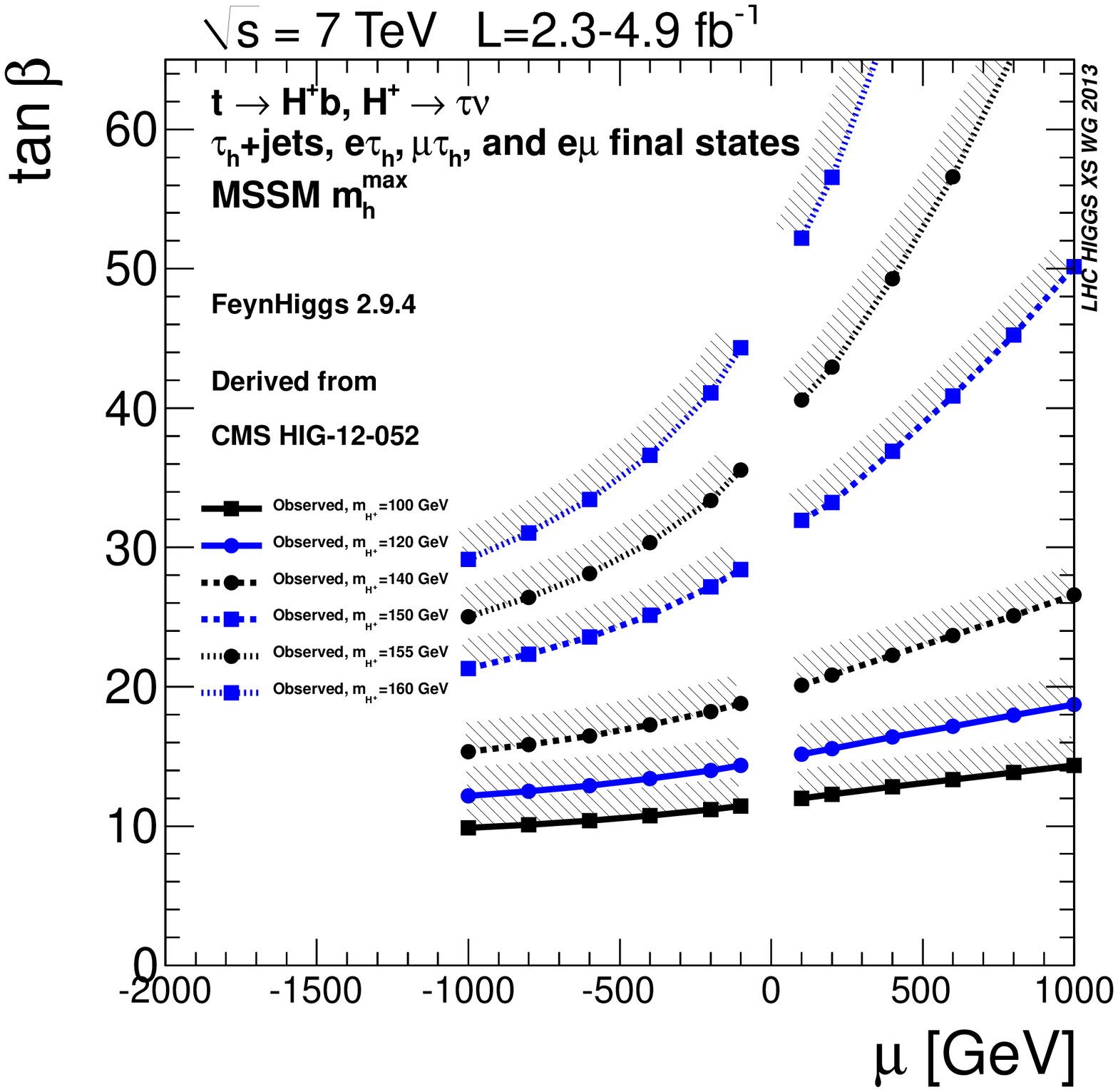}\\
  \includegraphics[width=0.475\textwidth]{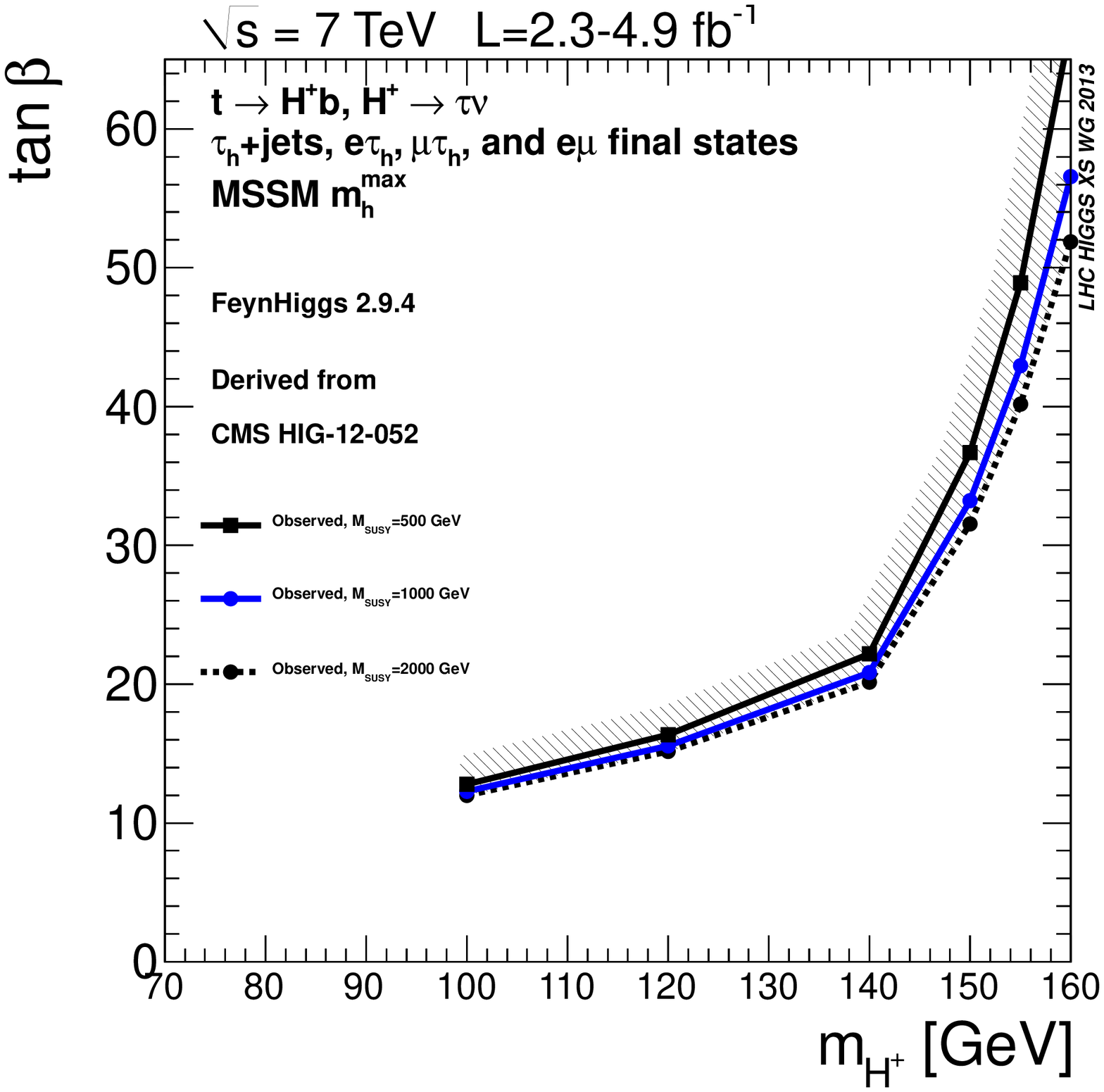}
  \includegraphics[width=0.475\textwidth]{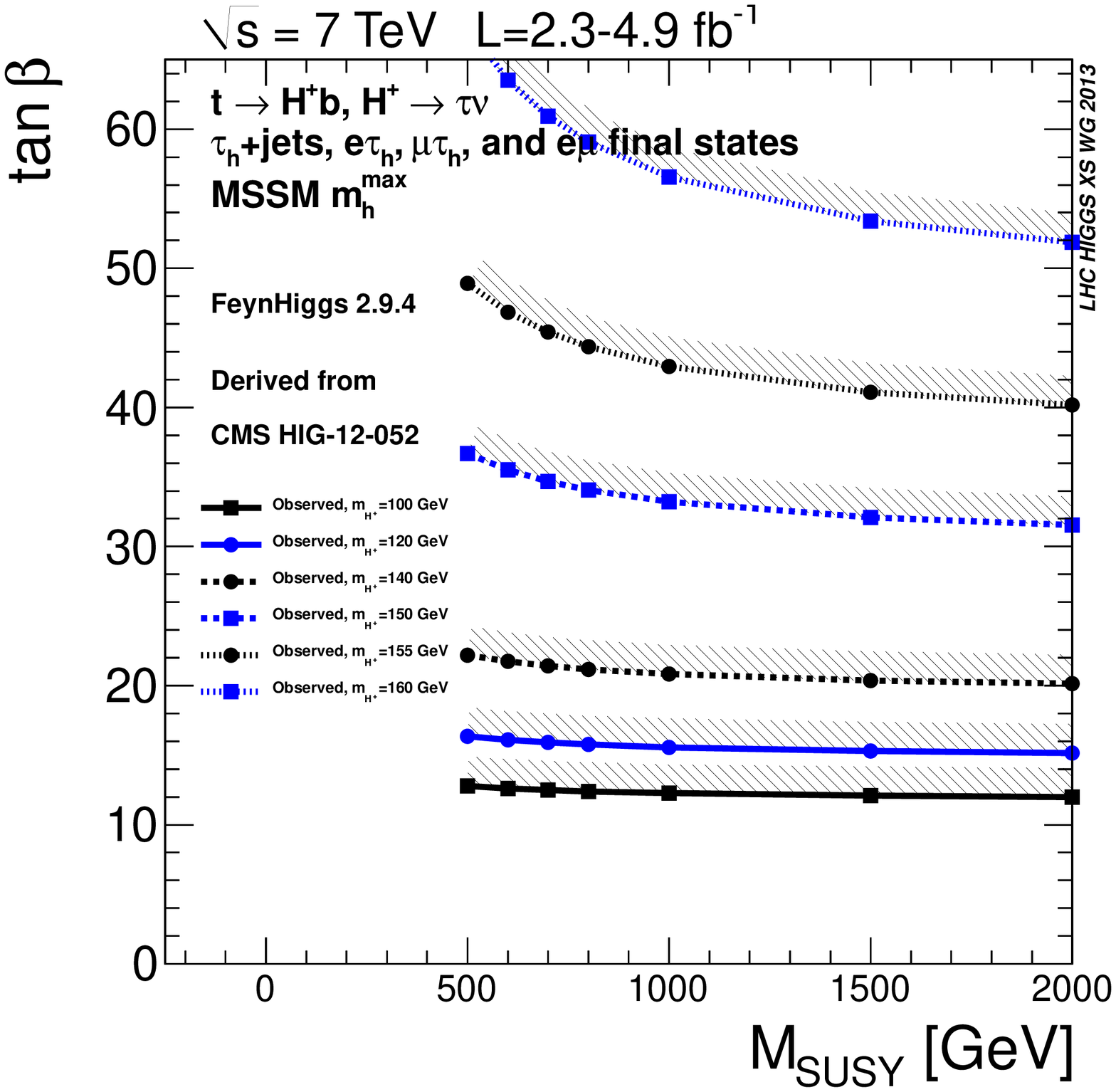}\\
  \includegraphics[width=0.475\textwidth]{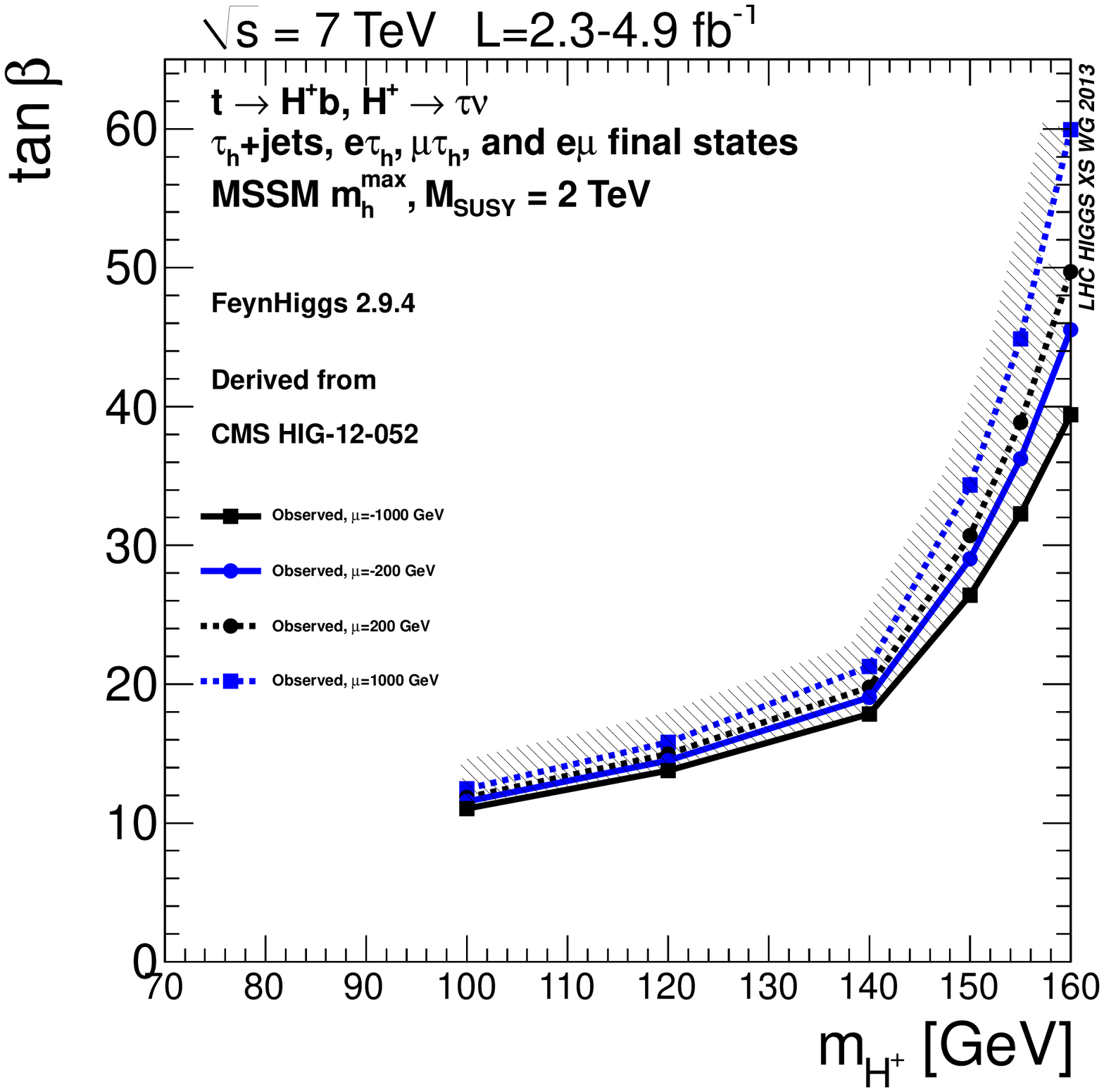}
  \includegraphics[width=0.475\textwidth]{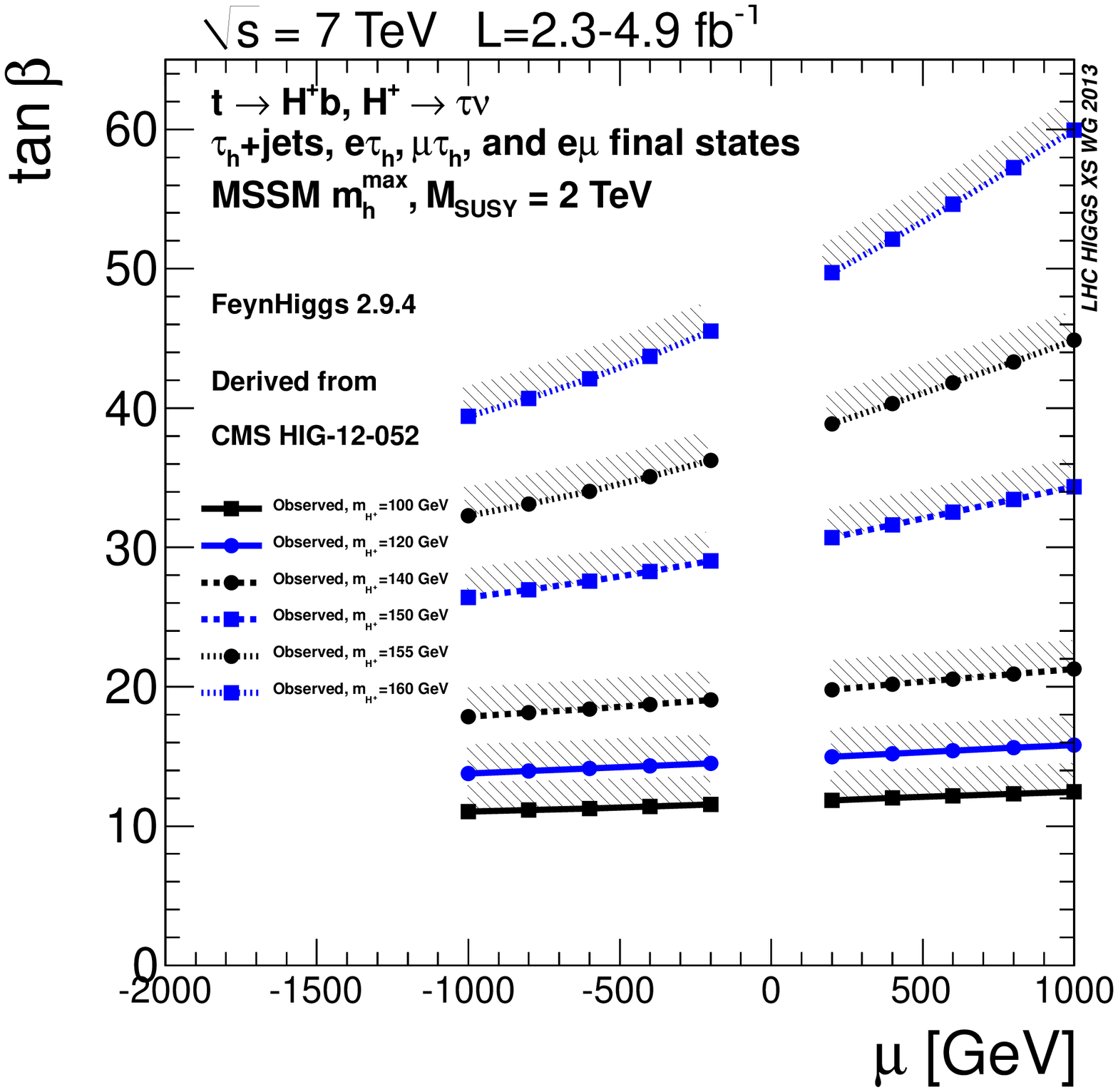}
  \caption{The effect of varying the $\mu$ parameter on the
experimentally excluded region in the \mhmax\ scenario (top
plots), the effect of varying $\msusy$ (middle plots), and the
effect of varying the $\mu$ parameter with $\msusy = 2$\UTeV\
(bottom plots).}
\label{fig:limits_variation1}
\end{figure}

\begin{figure}[p!]
  \centering
  \includegraphics[width=0.475\textwidth]{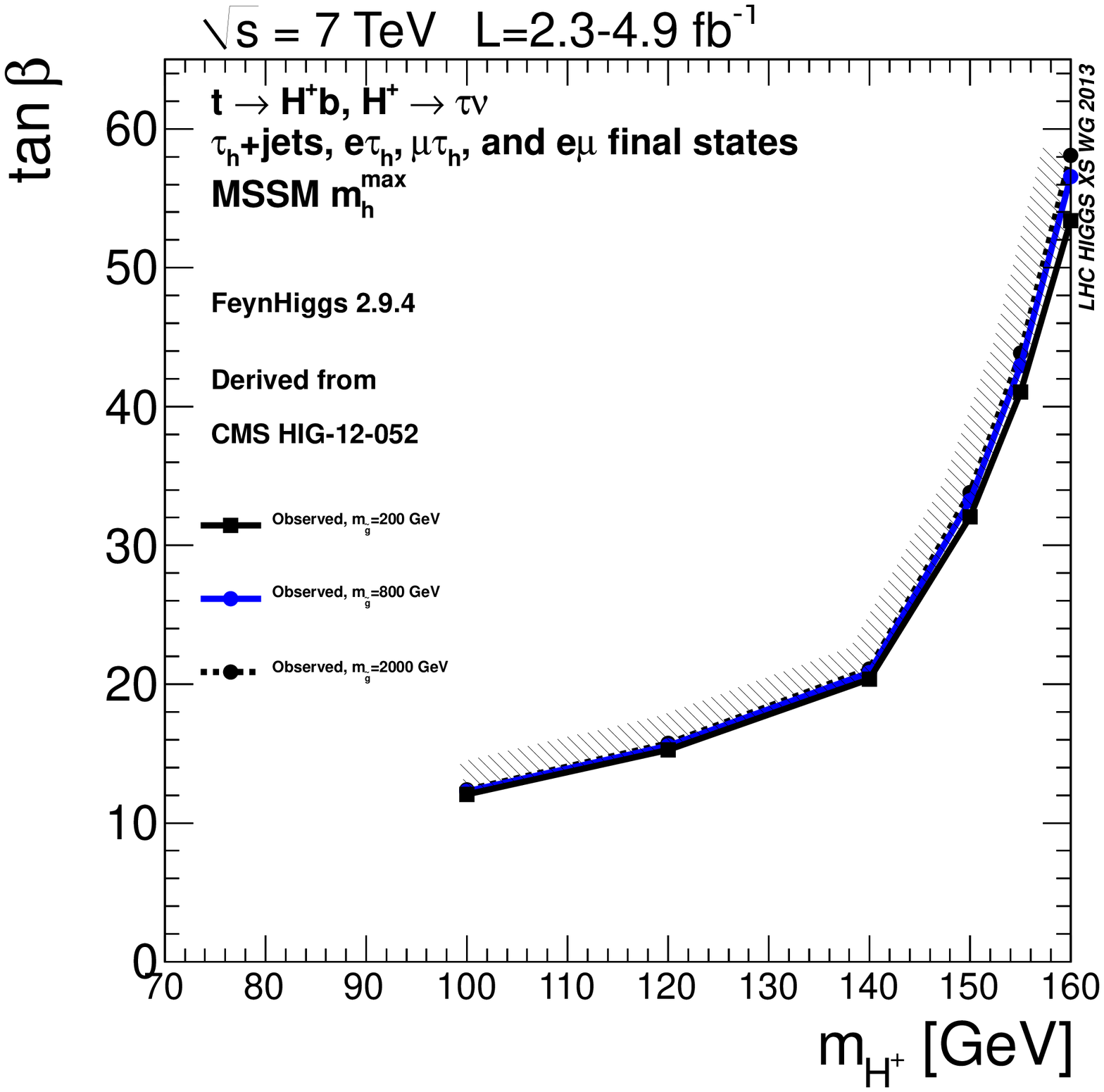}
  \includegraphics[width=0.475\textwidth]{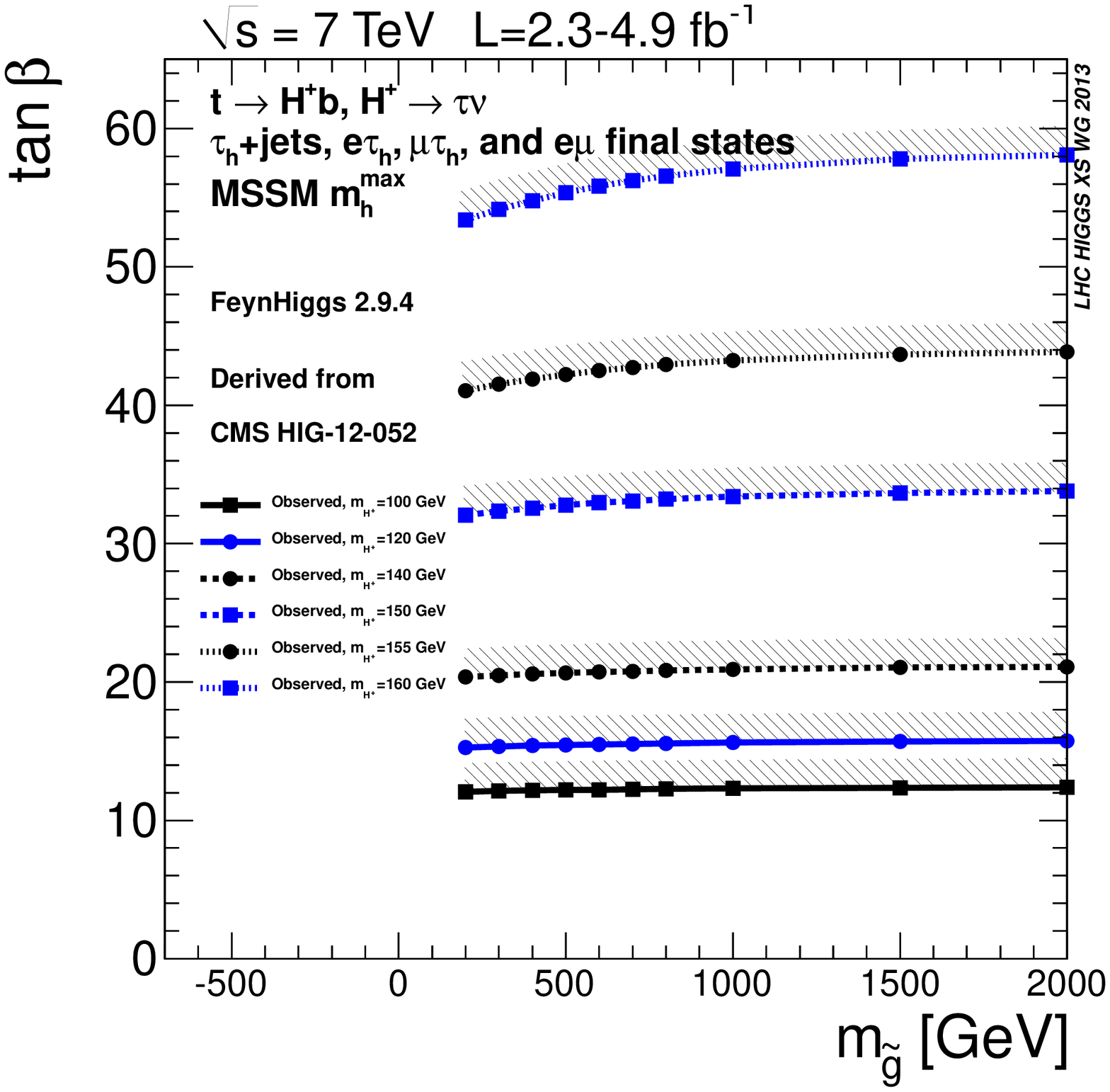}\\
  \includegraphics[width=0.475\textwidth]{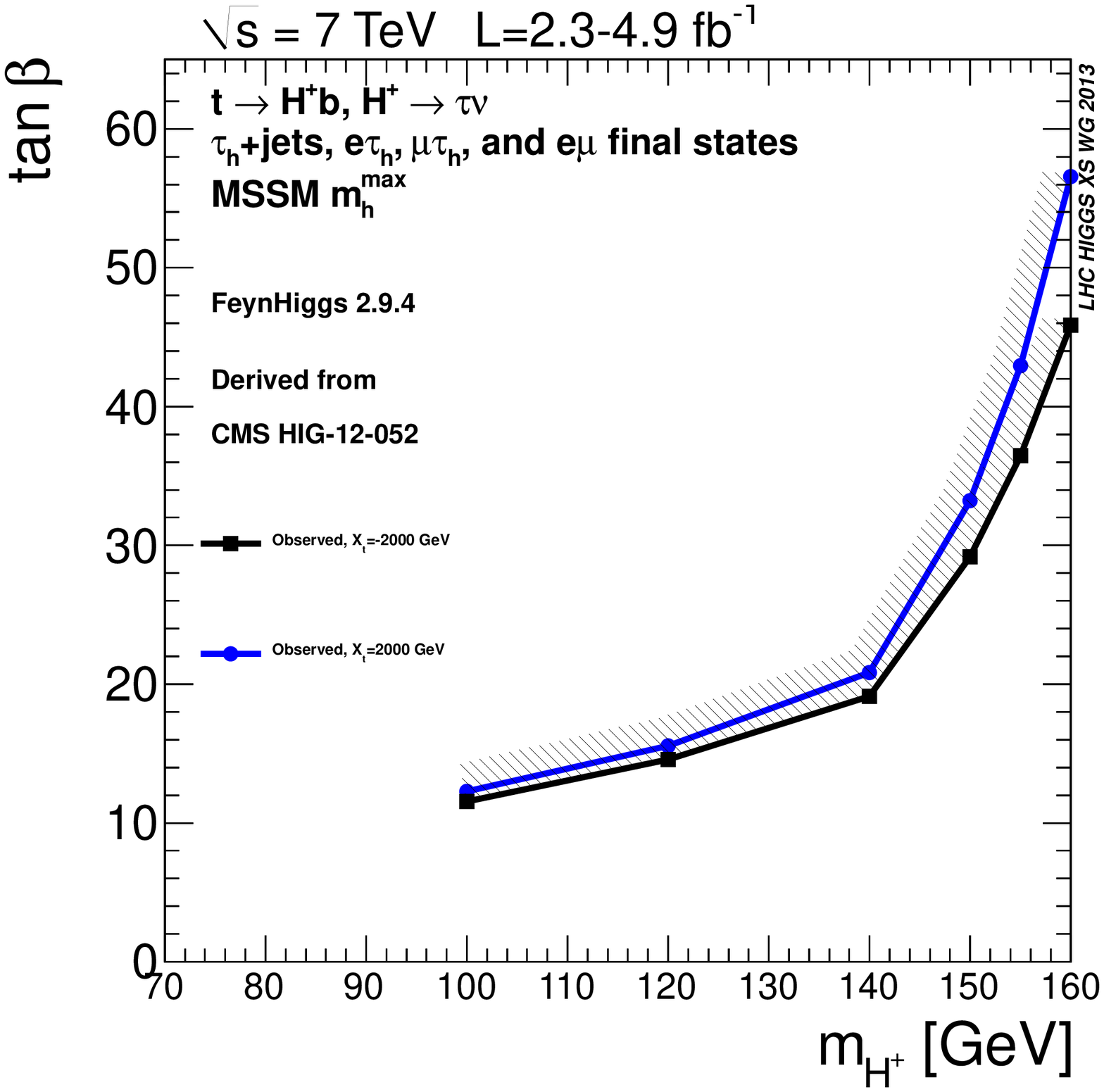}
  \includegraphics[width=0.475\textwidth]{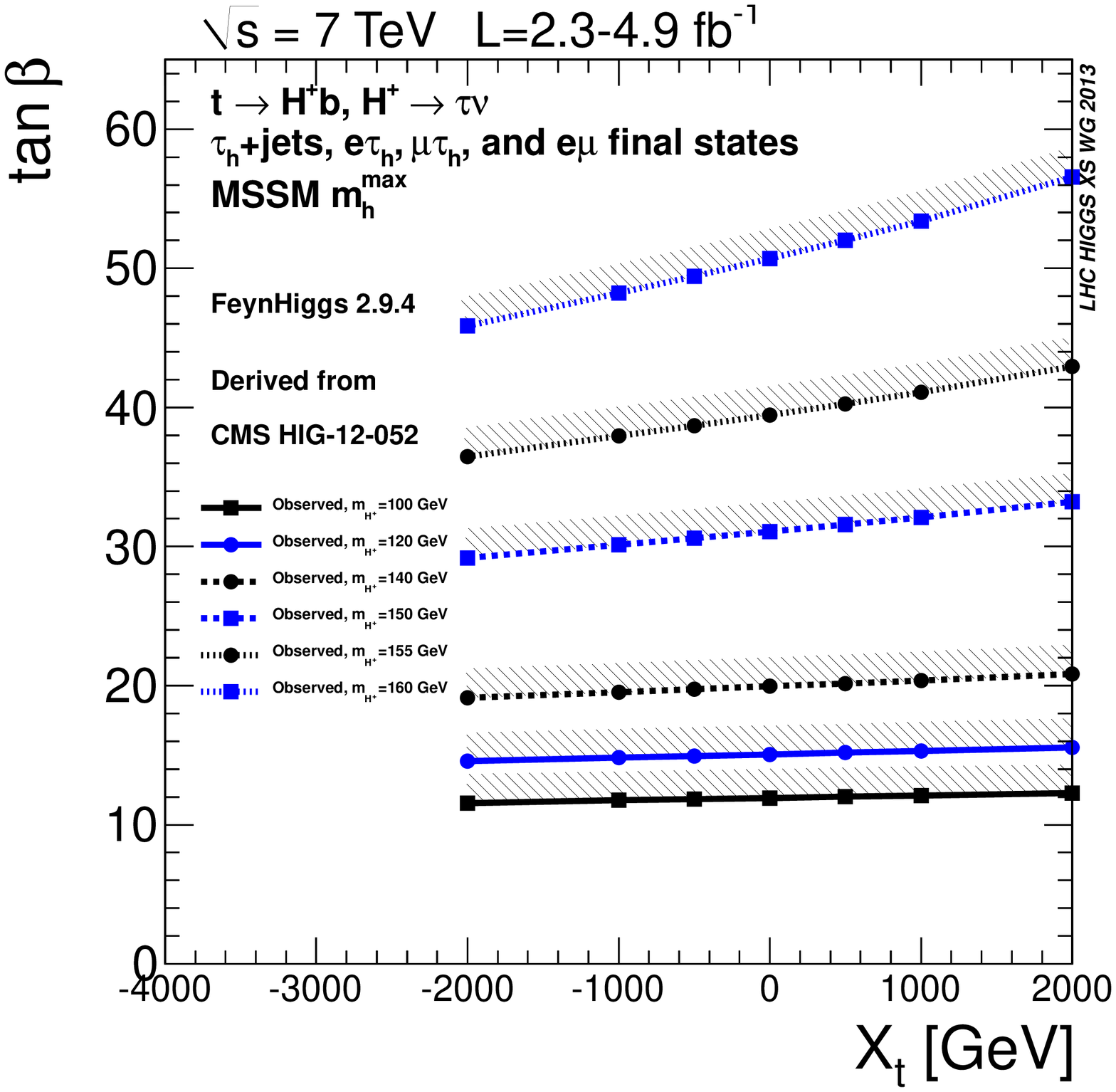}\\
  \includegraphics[width=0.475\textwidth]{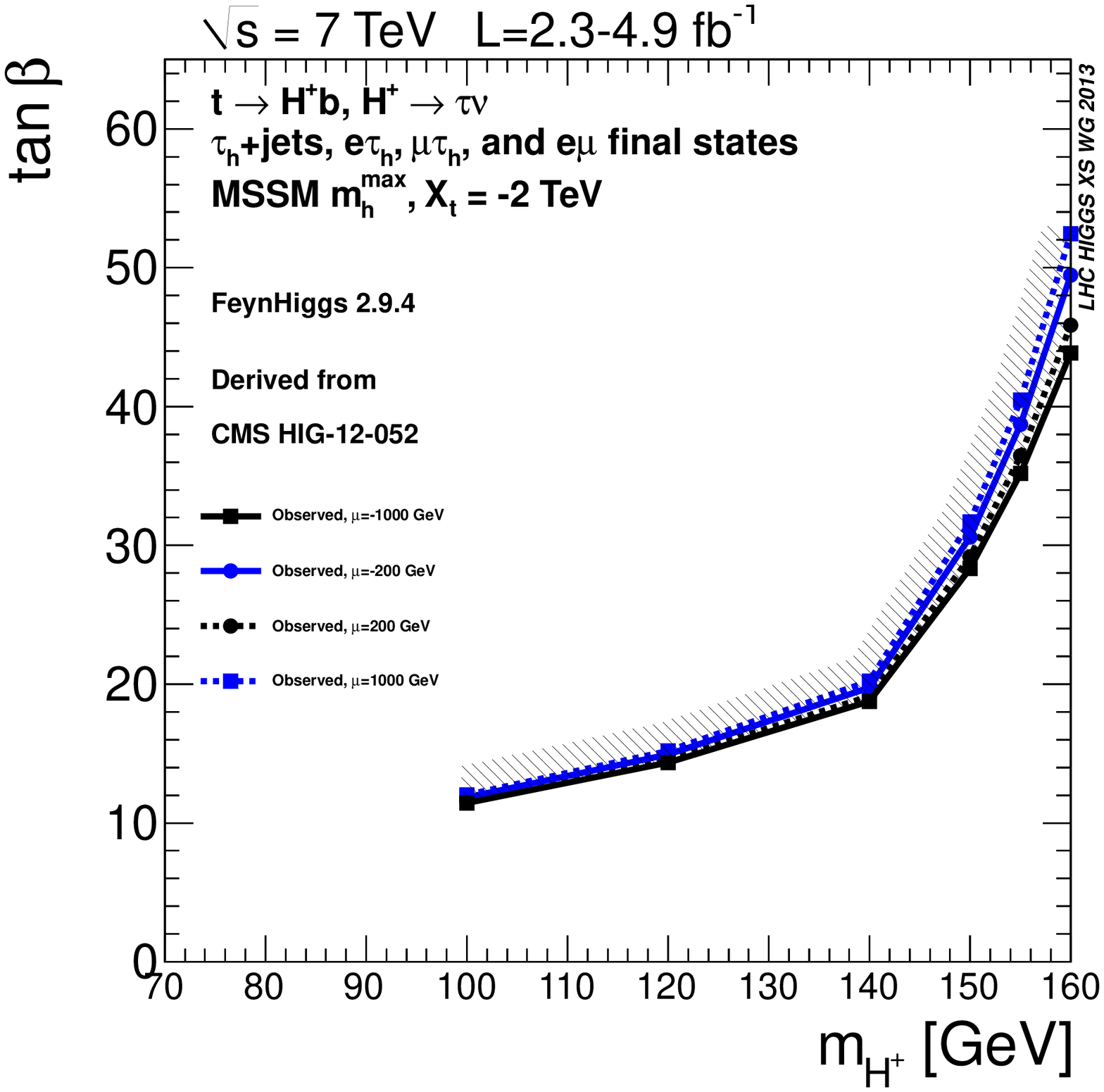}
  \includegraphics[width=0.475\textwidth]{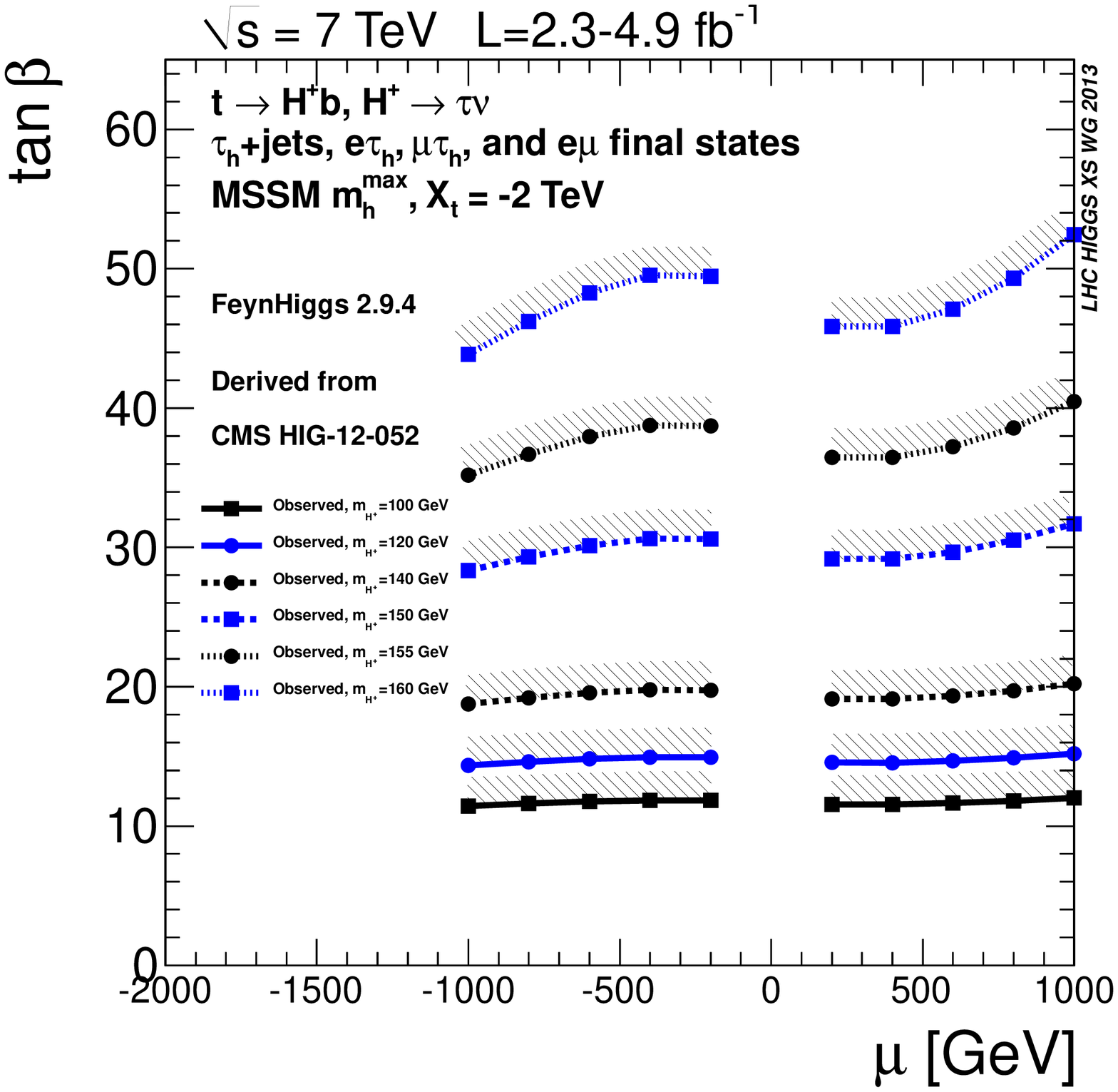}
  \caption{The effect of varying $\Mgl$ on the experimentally excluded
region in the \mhmax\ scenario (top left and right), the effect of
varying $\Xt$ (middle plots), and the effect of varying the $\mu$
parameter with $\Xt = -2$\UTeV\ (bottom plots).}
\label{fig:limits_variation2}
\end{figure}

As shown in Ref.\cite{Hashemi:2008ma}, the $\mu$ parameter has a
significant effect on the charged Higgs boson production and
decay. This is illustrated in
\refF{fig:limits_variation1} (top left) which shows the effect of varying 
the $\mu$ parameter on the observed exclusion region in the
($\MHpm$, $\tb$) parameter space. The same effect is shown as
a function of $\mu$ in \refF{fig:limits_variation1} (top right).  The
small abs($\mu$) values are excluded from the plot. The value of $\mu$
in the \mhmax\ scenario is $200$\UGeV.

The mass scale $\msusy$ is varied between $500$ and $2000$\UGeV, the
\mhmax\ scenario value being $1000$\UGeV.  The values of $\Mgl$
and $\Xt$ are changed to $0.8\times\msusy$ and $2\times\msusy$,
accordingly. As shown in \refF{fig:limits_variation1} (middle), the
effect on the exclusion limits are only a fraction of the effect of
the $\mu$ parameter. If this mass scale is heavier than expected, it
has only a small effect on the limits. \refF{fig:limits_variation1}
(bottom) show the $\mu$ dependence of the limits with $\msusy =
2000$\UGeV\ as a function of $\MHpm$ and as a function of $\mu$.

Varying $\Mgl$ has only a small effect on the limits, shown
in \refF{fig:limits_variation2} (top).  The variation of $M_2$ has
even a smaller effect. The $\mu$ dependence is almost unchanged
despite the choice of the $M_2$ value. For $\Mgl = 200$\UGeV\ the $\mu$
dependence is reduced by about 30\% compared to the \mhmax\
scenario.

The $\Xt$ parameter is varied in \refF{fig:limits_variation2}
(middle). An effect on the exclusion limits can clearly be seen. The
change as a function of $\Xt$ is close to linear. The $\mu$
dependence at $\Xt = -2000$\UGeV\ gives a wave-like behavior,
shown in \refF{fig:limits_variation2} (bottom), which is the reason
why the exclusion limits for positive and negative values of the $\mu$
parameter are in different order depending on the absolute value.

The experimental $95\%$ CL limits transferred to ($\MHpm$,
$\tb$) parameter space seem to be quite stable against all other
parameter variations except the variation of the $\mu$ parameter in
the mass range $\MHpm < 140$\UGeV.  There is a significant dependence
on the $\mu$. In the interesting region allowed by $\Mh =
125.9\pm3.0$\UGeV\ the limits are affected by the choice of $\msusy$
and $\Xt$ in addition to $\mu$.  However, although the limits
themselves are not too sensitive to the values chosen for $\msusy$,
$\Mgl$, $M_2$ and $\Xt$, the choice of the values for these
parameters do have a significant effect on the region allowed by the
discovery of a Higgs-like particle, whether it is assumed to
be \PSh, \PSH or \PSH and \PSA together, as shown
in \refF{fig:limits_mhmax_msusy}.  Moreover, the experimental limits
are excluding large parts of the allowed regions already. With tighter
limits expected from new updated analysis using full integrated
luminosity from 2011 and 2012 data taking these allowed regions not
experimentally excluded should shrink further, starting to exclude the
possibility that the discovered Higgs-like particle could be the heavy
$\cp$-even or $\cp$-odd Higgs boson in the studied scenarios.

%%%%%%%%%%%%%%%%%%%%%%%%%%%%%%%%%%%%%%%%%%%%%%%%%%%%%%%%%%%%%%%%%%%%%%%%%%%%%%%

%- }}}
%- {{{ subsection{Heavy charged Higgs production}

\subsection{Heavy charged Higgs production}
\label{sec:michaelk-sub}
Heavy charged Higgs bosons with a mass larger than the top-quark mass
would be produced in association with a top quark:
\begin{equation*} 
\Pp\Pp\, \rightarrow\, \Pt\Pb\PH^\pm + X.
\end{equation*}
The cross section for associated $\Pt\Pb\PH^\pm$ production can be
computed in the so-called four- and five-flavor schemes.  In the
four-flavor scheme (4FS) there are no $\Pb$ quarks in the initial
state, and therefore the lowest-order QCD production processes are
gluon-gluon fusion and quark-antiquark annihilation, $\Pg\Pg
\rightarrow \Pt\Pb\PH^\pm$ and $\Pq\bar{\Pq} \rightarrow
\Pt\Pb\PH^\pm$, respectively. Potentially large logarithms of the
ratio between the hard scale of the process and the mass of the bottom
quark, which arise from the splitting of incoming gluons into nearly
collinear $\Pb\bar{\Pb}$ pairs, can be summed to all orders in
perturbation theory by introducing bottom parton densities.  This
defines the five-flavor scheme (5FS). The use of bottom distribution
functions is based on the approximation that the outgoing $\Pb$ quark
is at small transverse momentum and massless, and the virtual $\Pb$
quark is quasi on shell. In this scheme, the LO process for the
inclusive $\Pt\Pb\PH^\pm$ cross section is gluon-bottom fusion,
$\Pg\Pb \rightarrow \Pt\PH^\pm$. The NLO cross section in the 5FS
scheme includes $\mathcal{O}(\alphas)$ corrections to $\Pg\Pb
\rightarrow \Pt\PH^\pm$, including the tree-level processes $\Pg\Pg
\rightarrow \Pt\Pb\PH^\pm$ and $\Pq\bar{\Pq} \rightarrow
\Pt\Pb\PH^\pm$.  To all orders in perturbation theory the two schemes
are identical, but the way of ordering the perturbative expansion is
different, and the results do not match exactly at finite order. For
the inclusive production of neutral Higgs bosons with bottom quarks,
$\Pp\Pp \rightarrow \Pb\bar{\Pb}\PH+X$, the four- and five-flavor
scheme calculations numerically agree within their respective
uncertainties, once higher-order QCD corrections are taken into
account, see \cite{Dittmaier:2011ti,Dittmaier:2012vm} and references therein.

We provide NLO predictions for heavy charged Higgs boson production in
a two Higgs doublet model (2HDM) with $\tanb$ = 30. SUSY effects can
be taken into account by rescaling the bottom Yukawa coupling to the
proper value. We present results for the 4FS and 5FS schemes,
including the theoretical uncertainty, and combine the two schemes
according to the Santander matching proposed
in~\cite{Harlander:2011aa}. Throughout this report we present results
for the $\Pt\bar{\Pb}\PH^-$ channel.

For the calculation in the 5FS, the program
Prospino~\cite{Plehn:2002vy} has been employed, interfaced to the
LHAPDF library~\cite{Bourilkov:2006cj}.  The renormalization scale is
set to $\mu_{\rm R}$ = $(M_{\PH^\pm}+\mto)/2$, while the factorization
scale $\mu_{\rm F}=\tilde{\mu}$ is chosen according to the method
proposed in~\cite{Maltoni:2012pa}.  The effective factorization scale
entering the initial state logarithms is proportional to the hard
scale, but modified by a phase space factor which tends to reduce the
size of the logarithms for processes at hadron colliders. The
factorization scale $\tilde{\mu}$ is given in
Tab.~\ref{fac_scale:result} for several Higgs masses, both for 8 and
14 TeV center-of-mass energy.  The values of the factorization scale
match those proposed in~\cite{Plehn:2002vy}.

To estimate the theoretical uncertainty due to missing higher-order
contributions, we vary the renormalization and factorization scales by
a factor three about their central values. We find scale uncertainties
between approximately 10--20\%, depending on the Higgs mass and
collider energy.  In addition to the scale variation, we have computed
the uncertainty associated with the PDF set and the values of
$\alphas$ and $\mb$ used in PDF fits, following the
PDF4LHC~\cite{Botje:2011sn} recommendation. All uncertainties are
given at 68\% confidence level (CL). The $\alphas$ uncertainty
corresponds to a variation of $\pm 0.0012$ about the central
value~\cite{Botje:2011sn}. As the uncertainty for the bottom mass we
take $\mb$ = 4.75 $\pm$ 0.25 GeV, which is a conservative choice
compared to the uncertainty given in~\cite{Beringer:1900zz}. The
overall PDF+$\alphas$+$\mb$ uncertainty amounts to approximately
10--15\% and is thus comparable to the scale uncertainty. Combining
the two sources of uncertainty linearly we obtain an estimate of the
overall theoretical uncertainty of approximately 30\%. Our 5FS results
for heavy charged Higgs production at the LHC with 8\,TeV cms energy
are displayed in Figure\,\ref{CH:plot}, upper panel.

%%%%
\begin{table}[t!]
\begin{center}
\caption{Dynamical factorization scale $\tilde{\mu}$ for
  $\Pp\Pp\rightarrow  \Pt \PH^- + X $ for the LHC at 8 and 14 TeV.}
\label{fac_scale:result}
\begin{tabular}{ lcccc }
  \hline
  &\multicolumn{2}{c}{8 TeV} & \multicolumn{2}{c}{14 TeV}\\
  \hline
$\MHpm$ [GeV] & $\tilde{\mu}$ [GeV] & $(\mto+\MHpm)/\tilde{\mu}$ & $\tilde{\mu}$ [GeV]& $(\mto+\MHpm)/\tilde{\mu}$ \\
\hline
 200 & 67.3 & 5.5 & 74.9 & 5.0 \\
 300 & 80.3 & 5.9 & 90.6 & 5.2 \\
 400 & 92.1 & 6.2 & 105.3 & 5.4 \\
 500 & 103.1 & 6.5 & 119.0 & 5.7 \\
  \hline
\end{tabular}
\end{center}
\end{table}

The results for heavy charged Higgs production within the four-flavour
scheme (4FS) are based on the calculation presented in
Ref.\,\cite{Dittmaier:2009np}. We adopt the CT10\,\cite{Lai:2010vv},
MRST\,\cite{Martin:2010db} and NNPDF21\,\cite{Ball:2011mu} 4FS pdf
sets. The renormalization and factorization scales are varied by a
factor three about the central scale choice $\mu_0 =
(\MHpm+\mto+\mb)/3$. The scale variation in the 4FS is approximately
$30\%$, i.e.\ somewhat larger than in the 5FS. Note that our estimate
of the pdf uncertainty is based on MRST2008 and NNPDF21 only, as CT10
does not provide eigenvector sets in the 4FS. Furthermore, no
uncertainties for $\alphas$ and $\mb$ could be calculated, as no pdf
collaboration provides 4FS sets with varying $\alphas$ and $\mb$. Our
4FS results for heavy charged Higgs production at the LHC with 8\,TeV
cms energy are displayed in Figure\,\ref{CH:plot}, middle
panel.

To arrive at a final prediction for heavy charged Higgs production we
combine the NLO 5FS and 4FS cross sections according to the Santander
matching\,\cite{Harlander:2011aa}, analogous to neutral Higgs-bottom
associated production. We note that the 4FS and 5FS calculations
provide the unique description of the cross section in the asymptotic
limits $M_\phi/\mb \to 1$ and $M_\phi/\mb \to \infty$, respectively (here
and in the following $M_\phi$ denotes a generic Higgs boson mass). The
4FS and 5FS are thus combined in such a way that they are given
variable weight, depending on the value of the Higgs-boson mass. The
difference between the two approaches is formally logarithmic.
Therefore, the dependence of their relative importance on the
Higgs-boson mass should be controlled by a logarithmic term, i.e.\
\begin{equation}
  \sigma^{\rm matched} = \frac{\sigma^{\rm 4FS} + w\,\sigma^{\rm 5FS}}{1 +
    w} \quad \mbox{with} \quad w= \ln \frac{M_\phi}{\mb} - 2\,.
\end{equation}
The theoretical uncertainties are combined according to 
\begin{equation}
  \Delta\sigma^{\rm matched}_\pm = \frac{\Delta\sigma^{\rm 4FS}_\pm + w\Delta\sigma^{\rm 5FS}_\pm}{1 +
    w}
\end{equation}
where $\Delta\sigma^{\rm 4FS}_\pm$ and $\Delta\sigma^{\rm 5FS}_\pm$
are the upper/lower uncertainty limits of the 4FS and the 5FS,
respectively.

The cross section and uncertainty for the Santander matching, together
with the results calculated in 4F- and 5F-scheme, for LHC at 8\,TeV
are presented in Fig.\,\ref{CH:plot}, lower panel.  We
observe that the NLO 4FS and 5FS predictions are in good mutual
agreement, with differences of at most $\sim$7\%. The dynamical choice
for $\mu_{\rm F}$ in the 5FS used here improves the matching of the
predictions in the two schemes. The overall theoretical uncertainty of
the matched NLO prediction is about 30\%.

\begin{figure}
\begin{center}
\includegraphics[width=0.625\textwidth]{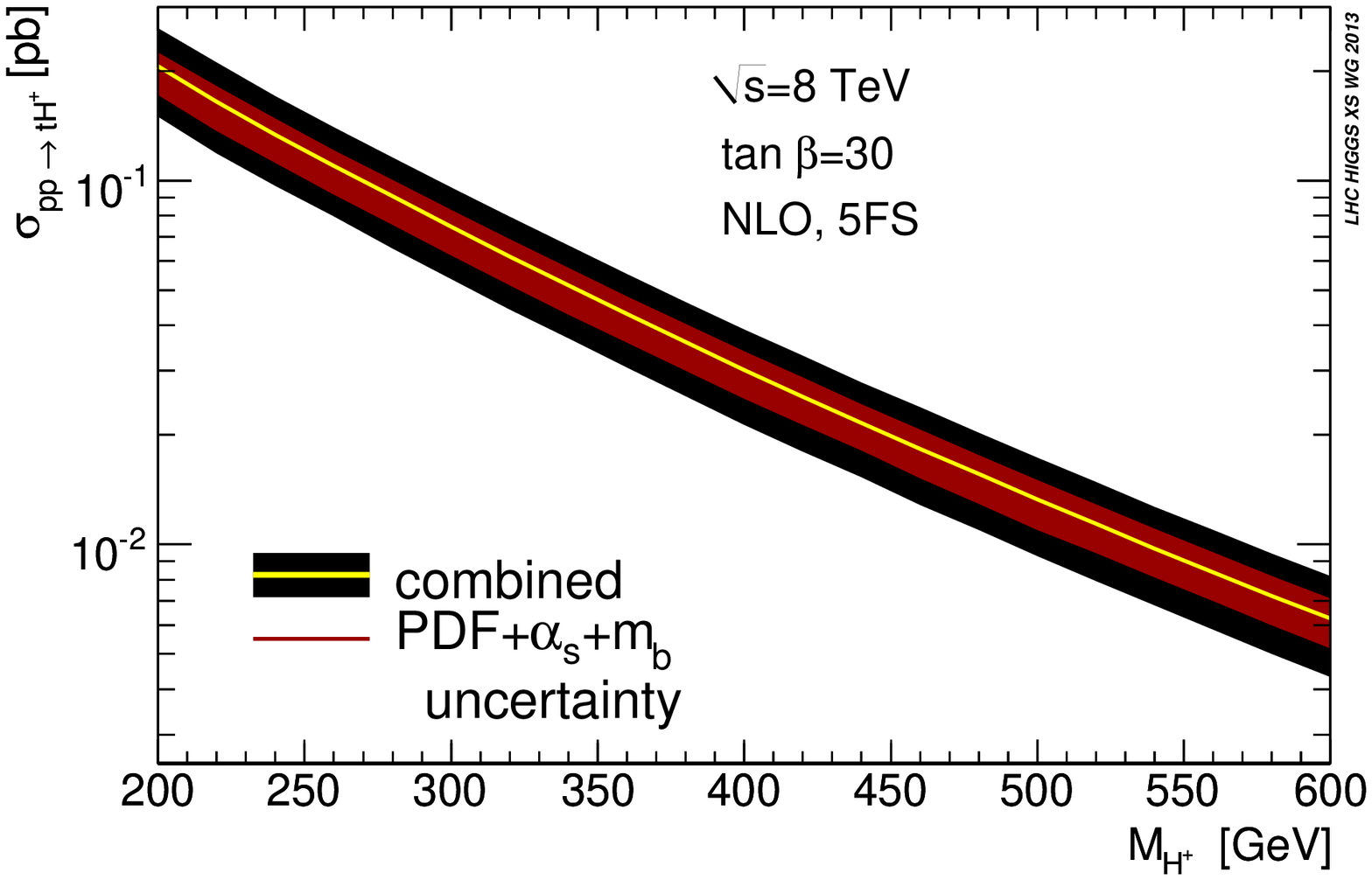}\\
\includegraphics[width=0.625\textwidth]{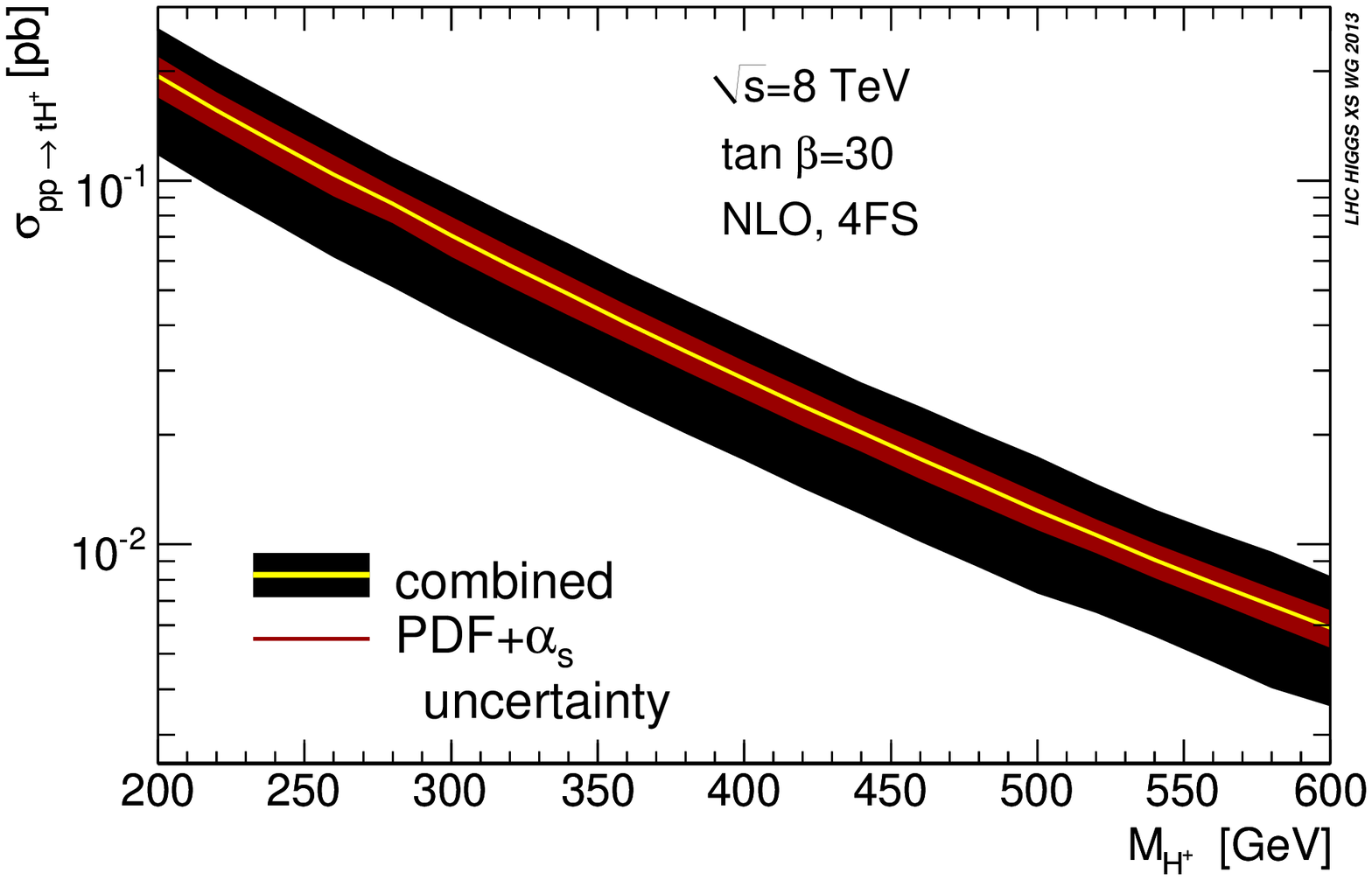}
\includegraphics[width=0.625\textwidth]{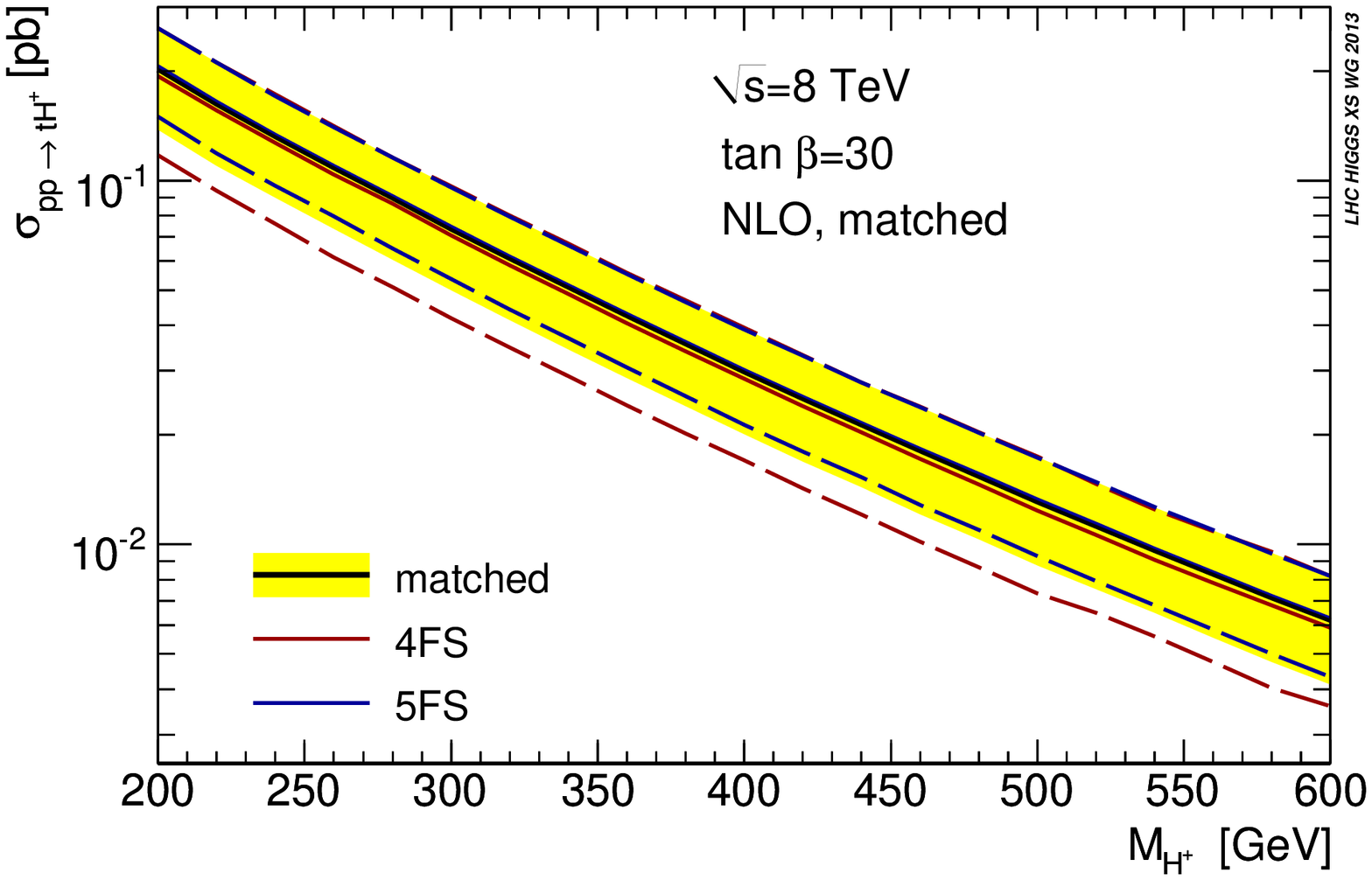}
\end{center}
\caption{NLO cross section prediction for $\Pp\Pp \rightarrow \Pt\PH^-
  + X$ at the LHC with 8\,TeV for a 2HDM with $\tan\beta = 30$: 5FS
  (upper panel), 4FS (middle panel) and Santander matched predictions
  (lower panel). Shown is the central prediction together with an
  estimate of the theoretical uncertainties as described in the text.}
\label{CH:plot}
\end{figure} 

%%%%%%%%%%%%%%%%%%%%%%%%%%%%%%%%%%%%%%%%%%%%%%%%%%%%%%%%%%%%%%%%%%%%%%%%%%%%%%%

%- }}}

\clearpage

% --------------------------------------------------
\newpage
\section{Conclusions \footnote{%
       S.~Heinemeyer, C.~Mariotti, G.~Passarino, R.~Tanaka}}
  \label{sec:concl}

The present document is the result of the activities of the LHC
Higgs Cross Section Working Group in the year 2012 till spring 2013. The
working group, created in January 2010, is a joint effort between
ATLAS, CMS and the theory community.
Previous reports~\cite{Dittmaier:2011ti,Dittmaier:2012vm}
dealt with the presentation of the main Higgs production cross
section and branching ratios of a SM Higgs at the highest available level
of accuracy. Similarly, first result for the Higgs bosons of the MSSM
were presented. Beyond the level of total cross sections also
distributions for the corresponding cross sections were investigated. Most
up-to-date results are continuously made public 
at the TWiki page\footnote{{\tt https:/twiki.cern.ch/twiki/bin/view/LHCPhysicsCrossSections}}.

The spectacular discovery of a Higgs-boson like resonance
with a mass around $\sim 125 - 126 \UGeV$, which has been announced
by ATLAS \cite{Aad:2012tfa} and CMS~\cite{Chatrchyan:2012ufa} on July
$4^{\mathrm{th}}$, 2012, marks 
a milestone of an effort that has been ongoing for almost half a
century and opens up a new era of particle physics.

Having analysed two and a half times more data than was available
for the discovery announcement in July 2012, the experiments are finding
that the new particle, within the experimental uncertainties, is perfectly 
compatible with the SM Higgs boson. It remains an open question, 
however, whether it is indeed the SM Higgs boson, or 
possibly one of several bosons predicted in some theories that go
beyond the SM. It is therefore the highest priority in particle physics
research at the moment to examine whether the emerging picture is
complete. This entails, in turn, placing increasingly stringent limits on
potential new physics scenarios.

Possible deviations from the SM prediction are not statistically
significant at present, so that within the uncertainties
the results are compatible with the SM. On the other hand, the
experimental uncertainties are still rather large and thus also
permit non-SM interpretations of the newly discovered Higgs-boson like
resonance.
If deviations from SM predictions were confirmed in the future, the
observed patterns could indicate a first clear step beyond the SM.

With the discovery of a Higgs-boson like resonance the focus of the
high-energy physics world, and thus of the LHC Higgs Cross Section
Working Group, has shifted. While ever more precise calculation of SM
predictions are needed, now corresponding calculations at a similar level
of accuracy in BSM models become important. Rules and definitions for the
determination of the characteristics of the Higgs-boson like resonance,
such as couplings to other particles, quantum numbers, spin etc.\ are now
crucial for the correct interpretation of the experimental data.

Theory uncertainty is becoming more and more
relevant than ever before.
The experimental accuracy is $\Delta \mu(\sigma/\sigma_{\rm SM}) =
\pm 15\%$ (roughly $\pm 10\%$ for both statistical and systematic uncertainties) 
with LHC 2011/2012 data,  
while the theoretical uncertainty is $\pm (10{-}15)\,\%$ dominated by missing
higher-order calculations and PDF $+\alphas$ in gluon-fusion.
This calls for improvements in NNNLO prediction; other important issues are
reduction of uncertainty in $1$-jet bin and gluon-fusion $+ 2$-jets
versus vector-boson fusion, 
as well as improvements in parton-luminosity functions with LHC data.

With the present document the LHC Higgs Cross Section Working Group
tries to lead the way into the new directions. 
This volume, correspondingly, provides a wide range of topics:
Updated results for Higgs Cross Sections and Higgs Decay Branching Ratios, 
Jet-bin and Higgs $\pT$ Uncertainties, 
Interference Effects in $\Pg\Pg$-fusion,
NLO Monte Carlo, Higgs Coupling and Spin/Parity and BSM Higgs. 
Updates of previous results for cross sections and branching ratios of a SM Higgs are 
refined around the experimentally determined mass range.

Guidelines for the extraction of the couplings have been worked
out that can readily applied to the 2011/2012 data set. Results for
spin and CP-determinations and refined coupling determinations are
outlined. Also in this area more detailed worked out prescriptions
will be necessary to cope with the data the LHC will provide in future, 
toward precision Higgs physics at the LHC. 

Within the MSSM improved cross section calculations were performed,
taking into account possible effects of light Supersymmetric particles.
Moreover, interpretations in the MSSM (as in any other BSM model) now
always must contain at least one Higgs boson at $\sim 125 - 126 \UGeV$,
which is reflected in a new set of MSSM benchmark scenarios. First
steps are presented to go to models beyond the SM and the MSSM, where
clearly more work, including the calculations of cross sections and
branching ratios, are needed in the future.

In conclusion, finding BSM's footsteps should be the primary goal of
the Working Group: what are the implications of the newly discovered
resonance for BSM scenarios?
We can go on to lay out two extreme scenarios. The first one
would be nothing but the SM at LHC energies, including no detection of dark
matter. 
In that case it would remain unclear for a long time to come which
BSM model is realised in nature.
The second scenario is the picture anticipated pre-LHC:
detection of non-SM Higgs and possibly of other non-SM particles at the
next high-energy run of the LHC. Only experimental data can lead the way.
%As Editors we are somewhat ambivalent
%%subdued 
%about the affair. The Standard Model
%has now got a degree of validity that has extended way beyond what we had
%before the discovery of a Higgs-like particle. 
%However, the one aspect that dominates here is that a
%Higgs could close -- or open -- 
%the last door of the Standard Model that could lead us
%to a deeper theory.

Finally, no small set of individuals among the hundreds who have worked
collaboration to produce this volume can take enough of the credit to
single them out, even less the editors.

\begin{verse}
Open your mind to what I shall disclose, and hold it fast within
you; he who hears, but does not hold what he has heard, learns nothing.
\end{verse}
{\tiny{Beatrice - Canto V 40-42}}

\vspace{1.5cm}

\leftline{Chiara, Giampiero, Reisaburo and Sven}

\clearpage
% --- Acknowledgements

\begin{flushleft}
{\bf Acknowledgements}
\end{flushleft}

We are obliged to
Dave Charlton, 
Albert De Roeck,  
Kevin Einsweiler, 
Fabiola Gianotti, 
Eilam Gross, 
Marumi Kado, 
Sandra Kortner, 
Joe Incandela, 
Greg Landsberg, 
Bill Murray, 
Jim Olsen, 
Aleandro Nisati,  
Christoph Paus
for their support and encouragement.

%We would like to acknowledge the assistance of
%...
% Andrea Benaglia, 
% Cristina Botta, 
% Nicola De Filippis,  
% David d'Enterria,
% Pietro Govoni,
% Claire Gwenlan,
% Judith Katz,
% and Andrea Massironi. 

%Sergey Alekhin, Richard Ball, Johannes Bl\"umlein,
%Jon Butterworth, Amanda Cooper-Sarkar, Sasha Glazov,
%Alberto Guffanti, Pedro Jimenez-Delgado, Sven Moch, Pavel Nadolsky,
%Ewald Reya, Juan Rojo, and Graeme Watt 

We are obliged to CERN, in particular to the IT Department and to the
Theory Unit for the support with logistics, especially to Elena Gianolio for
technical assistance.

%We also acknowledge partial support from 
%the European Community's Marie-Curie
%Research Training Network under contract MRTN-CT-2006-035505 `Tools
%and Precision Calculations for Physics Discoveries at Colliders',
%from the Science and Technology Facilities Council, 
%from the US Department of Energy,
%and
%from the Forschungsschwerpunkt 101
%by the Bundesministerium f\"ur Bildung und Forschung, Germany.

% --- Acknowledgement for CERN Report 3

A.~Bagnaschi, G.~Degrassi, P.~Slavich and A.~Vicini are supported by the Research Executive 
Agency (REA) of the European Union under the Grant Agreement number 
PITN-GA-$2010$-$264564$ (LHCPhenoNet)."

The work of F.~Cascioli, P.~Maierhofer, N.~Moretti, and S.~Pozzorini is supported by the 
Swiss National Science Foundation.

The work of S.~Y.~Choi was supported by Basic Science Research Program through the
National Research Foundation (NRF) funded by the Ministry of Education,
Science and Technology ($2012$-$0002746$).

A.~David has been supported by Funda\c{c}\~{a}o para a Ci\^{e}ncia e a Tecnologia (Portugal) 
grant number CERN/FP/$123601$/$2011$.

P.~de~Aquino and K.~Mawatari have been supported by the Strategic Research Program 
``High Energy Physics" of the Vrije Universiteit Brussel,
and the Belgian Federal Science Policy Office through the Interuniversity Attraction Pole P$7/37$.

D.~de Florian was supported by UBACYT, CONICET, ANPCyT and the Research Executive Agency (REA) 
of the European Union under the Grant Agreement number PITN-GA-$2010$-$264564$ (LHCPhenoNet). 

S.~Frixione has been supported in part by
the ERC grant $291377$ ``LHCtheory: Theoretical predictions and analyses
of LHC physics: advancing the precision frontier'', in part by
the Swiss National Science Foundation (NSF) under contract $200020$-$129513$.

The work of M.~V.~Garzelli is supported by the Slovenian Ministery of Work, under an AD-FUTURA 
grant.

The research of M.~Grazzini was supported in part by
the Research Executive Agency (REA) of the European Union under the Grant Agreement number 
PITN-GA-$2010$-$264564$ (LHCPhenoNet).

C.~Grojean is supported by the Spanish Ministry MICNN under contract FPA$2010$-$17747$ and by the 
European Commission under the ERC Advanced Grant $226371$ MassTeV and the contract 
PITN-GA-$2009$-$237920$ UNILHC.

R.~Harlander, S.~Liebler, H.~Mantler, and T.~Zirke have been supported
in part by DFG, grant HA $2990$/$5$-$1$ and BMBF, grant 05H09PXE.

The work of S.~Heinemeyer was supported in part by CICYT (grant FPA $2010$-$22163$-C$02$-$01$) and 
by the Spanish MICINN's Consolider-Ingenio $2010$ Program under grant MultiDark CSD$2009$-$00064$.

A.~Kardos is grateful to LHCPhenoNet and OTKA for providing financial funding.

The work by X.~Liu and F.~Petriello was supported by the U.S. Department of 
Energy, Division of High Energy Physics, under contract DE-AC$02$-$06$CH$11357$ and 
the grants DE-FG$02$-$95$ER$40896$ and DE-FG$02$-$08$ER$4153$.

The work of F.~Maltoni and M.~Zaro is supported by the IISN ``MadGraph'' convention 
$4.4511.10$, the IISN ``Fundamental interactions'' convention $4.4517.08$ and by the ERC grant 
$291377$ ``LHCtheory: Theoretical predictions and analyses of LHC physics: advancing the 
precision frontier''.

C.~Mariotti  acknowledges the support by Compagnia di San Paolo under contract ORTO$11$TPXK.

O.~Mattelaer is a fellow of the Belgian American Education Foundation. His work is partially 
supported by the IISN ``MadGraph'' convention $4.4511.10$.

D.~J.~Miller acknowledges partial support from the STFC Consolidated Grant ST/G$00059$X/$1$.

M.~Muhlleitner acknowledges support by the Deutsche
Forschungsgemeinschaft via the Sonderforschungsbereich/Transregio
SFB/TR-$9$ ``Computational Particle Physics''. 

G.~Passarino has been supported by Ministero dell'Istruzione, dell'Universit\`a e della Ricerca
Protocollo $2010$YJ$2$NYW${}_{}006$ 
and by Compagnia di San Paolo under contract ORTO$11$TPXK.

M.~Rauch acknowledges partial support by the Deutsche
Forschungsgemeinschaft via the Sonderforschungsbereich/Transregio
SFB/TR-$9$ ``Computational Particle Physics'' and the Initiative and
Networking Fund of the Helmholtz Association, contract HA-$101$ (``Physics at
the Terascale'').

The work of L.~Reina is supported by the US Department of Energy under grant 
DE-FG$02$-$13$ER$41942$.

The work of R.~Rietkerk is supported in part by Stichting voor Fundamenteel Onderzoek der 
Materie (FOM), which is financially supported by the Nederlandse organisatie voor 
Wetenschappelijke Onderzoek (NWO).

G.~Salam acknowledges support from the French Agence Nationale de la Recherche, under grant 
ANR-$09$-BLAN-$0060$, from the European Commission 
under ITN grant LHCPhenoNet, PITN-GA-$2010$-$264564$, and from ERC
advanced grant Higgs@LHC.

F.~Schissler acknowledges support by the BMBF under Grant No. 05H09VKG
(``Verbundprojekt HEP-Theorie'') and by the ``Karlsruhe School of
Elementary Particle and Astroparticle Physics (KSETA)''.

F.~J.~Tackmann has been supported by the DFG Emmy-Noether grant TA $867$/$1$-$1$.

J.~Thompson is supported in part by the UK Science and Technology
Facility Council.

The work of P.~Torrielli has been supported in part by the Forschungskredit der Universit\"at
Z\"urich, by the Swiss National  Science Foundation (SNF) under contract
$200020$-$138206$ and by the Research Executive Agency (REA) of the European
Union under the Grant Agreement number PITN-GA-$2010$-$264564$ (LHCPhenoNet).
PT would like to thank A. Papaefstathiou for the useful discussions.

T.~Vickey is supported by the Oxford Oppenheimer Fund, the Royal Society of the United Kingdom 
and the Department of Science and Technology, the National Research Foundation of the Republic 
of South Africa.

\clearpage

\newpage
\appendix
%\addcontentsline{toc}{section}{\appendixname}

\section{Tables of branching ratios}

In this appendix we complete the listing of the branching fractions of the Standard Model 
Higgs boson discussed in \refS{sec:br}. 
%Tables with the cross section values for the various production modes follow.

\newpage 

\begin{table}\small \scriptsize
\setlength{\tabcolsep}{3ex}
\caption{SM Higgs branching ratios to two fermions and their total uncertainties (expressed in percentage). 
Very-low mass range.}
\label{tab:YRHXS3_2fermions.1}
\begin{center}
\begin{tabular}{lcccc}
\hline
$\MH$ [$\UGeV$] & $\Hbb$ & $\Htautau$ &$\Hmumu$ & $\Hcc$ \\
\hline
$80$	&	$8.24\,{\cdot	10^{-1}}^{+1.4}_{-1.5}$	&	$8.27\,{\cdot	10^{-2}}^{+7.0}_{-6.9}$	&	$2.87\,{\cdot	10^{-4}}^{+7.4}_{-7.1}$	&	$4.18\,{\cdot	10^{-2}}^{+12.1}_{-12.1}$	  \\
$81$	&	$8.23\,{\cdot	10^{-1}}^{+1.5}_{-1.5}$	&	$8.27\,{\cdot	10^{-2}}^{+7.0}_{-6.8}$	&	$2.88\,{\cdot	10^{-4}}^{+7.3}_{-7.2}$	&	$4.17\,{\cdot	10^{-2}}^{+12.1}_{-12.1}$	  \\
$82$	&	$8.21\,{\cdot	10^{-1}}^{+1.5}_{-1.5}$	&	$8.28\,{\cdot	10^{-2}}^{+7.0}_{-6.8}$	&	$2.88\,{\cdot	10^{-4}}^{+7.3}_{-7.2}$	&	$4.16\,{\cdot	10^{-2}}^{+12.1}_{-12.1}$	  \\
$83$	&	$8.20\,{\cdot	10^{-1}}^{+1.5}_{-1.5}$	&	$8.29\,{\cdot	10^{-2}}^{+7.0}_{-6.8}$	&	$2.88\,{\cdot	10^{-4}}^{+7.3}_{-7.2}$	&	$4.16\,{\cdot	10^{-2}}^{+12.1}_{-12.1}$	  \\
$84$	&	$8.19\,{\cdot	10^{-1}}^{+1.5}_{-1.5}$	&	$8.30\,{\cdot	10^{-2}}^{+7.0}_{-6.8}$	&	$2.88\,{\cdot	10^{-4}}^{+7.3}_{-7.2}$	&	$4.15\,{\cdot	10^{-2}}^{+12.1}_{-12.1}$	  \\
$85$	&	$8.17\,{\cdot	10^{-1}}^{+1.5}_{-1.5}$	&	$8.30\,{\cdot	10^{-2}}^{+7.0}_{-6.8}$	&	$2.89\,{\cdot	10^{-4}}^{+7.3}_{-7.1}$	&	$4.14\,{\cdot	10^{-2}}^{+12.1}_{-12.1}$	  \\
$86$	&	$8.16\,{\cdot	10^{-1}}^{+1.5}_{-1.6}$	&	$8.31\,{\cdot	10^{-2}}^{+7.0}_{-6.8}$	&	$2.89\,{\cdot	10^{-4}}^{+7.3}_{-7.2}$	&	$4.13\,{\cdot	10^{-2}}^{+12.2}_{-12.1}$	  \\
\hline
\end{tabular}
\end{center}
\end{table}

\begin{table} \small \scriptsize
\setlength{\tabcolsep}{3ex}
\caption{SM Higgs branching ratios to two fermions and their total uncertainties (expressed in percentage). Very-low mass range.}
\label{tab:YRHXS3_2fermions.2}
\begin{center}
\begin{tabular}{lcccc}
\hline
$\MH$ [$\UGeV$] & $\Hbb$ & $\Htautau$ &$\Hmumu$ & $\Hcc$ \\
\hline
$118.5$	&	$6.66\,{\cdot	10^{-1}}^{+2.6}_{-2.7}$	&	$7.22\,{\cdot	10^{-2}}^{+6.2}_{-6.1}$	&	$2.51\,{\cdot	10^{-4}}^{+6.5}_{-6.4}$	&	$3.36\,{\cdot	10^{-2}}^{+12.3}_{-12.2}$	 \\
$119$	&	$6.60\,{\cdot	10^{-1}}^{+2.7}_{-2.7}$	&	$7.16\,{\cdot	10^{-2}}^{+6.2}_{-6.1}$	&	$2.49\,{\cdot	10^{-4}}^{+6.4}_{-6.4}$	&	$3.33\,{\cdot	10^{-2}}^{+12.2}_{-12.2}$	 \\
$119.5$	&	$6.54\,{\cdot	10^{-1}}^{+2.7}_{-2.8}$	&	$7.10\,{\cdot	10^{-2}}^{+6.1}_{-6.1}$	&	$2.47\,{\cdot	10^{-4}}^{+6.4}_{-6.4}$	&	$3.30\,{\cdot	10^{-2}}^{+12.2}_{-12.2}$	 \\
$120$	&	$6.48\,{\cdot	10^{-1}}^{+2.8}_{-2.8}$	&	$7.04\,{\cdot	10^{-2}}^{+6.1}_{-6.0}$	&	$2.44\,{\cdot	10^{-4}}^{+6.4}_{-6.3}$	&	$3.27\,{\cdot	10^{-2}}^{+12.3}_{-12.2}$	 \\
$120.1$	&	$6.46\,{\cdot	10^{-1}}^{+2.8}_{-2.8}$	&	$7.03\,{\cdot	10^{-2}}^{+6.1}_{-6.0}$	&	$2.44\,{\cdot	10^{-4}}^{+6.4}_{-6.3}$	&	$3.26\,{\cdot	10^{-2}}^{+12.2}_{-12.2}$	 \\
$120.2$	&	$6.45\,{\cdot	10^{-1}}^{+2.8}_{-2.8}$	&	$7.01\,{\cdot	10^{-2}}^{+6.1}_{-6.0}$	&	$2.43\,{\cdot	10^{-4}}^{+6.3}_{-6.3}$	&	$3.26\,{\cdot	10^{-2}}^{+12.2}_{-12.2}$	 \\
$120.3$	&	$6.44\,{\cdot	10^{-1}}^{+2.8}_{-2.8}$	&	$7.00\,{\cdot	10^{-2}}^{+6.1}_{-6.0}$	&	$2.43\,{\cdot	10^{-4}}^{+6.4}_{-6.3}$	&	$3.25\,{\cdot	10^{-2}}^{+12.2}_{-12.2}$	 \\
\hline
\end{tabular}
\end{center}
\end{table}

\begin{table}
\caption{SM Higgs branching ratios to two fermions and their total uncertainties (expressed in percentage). Low mass range.}
\label{tab:YRHXS3_2fermions.3}
\begin{center}
\setlength{\tabcolsep}{3ex}
\small \scriptsize
\begin{tabular}{lcccc}
\hline
$\MH$ [$\UGeV$] & $\Hbb$ & $\Htautau$ &$\Hmumu$ & $\Hcc$ \\
\hline
$124.4$	&	$5.86\,{\cdot	10^{-1}}^{+3.1}_{-3.2}$	&	$6.41\,{\cdot	10^{-2}}^{+5.8}_{-5.7}$	&	$2.23\,{\cdot	10^{-4}}^{+6.0}_{-6.0}$	&	$2.96\,{\cdot	10^{-2}}^{+12.2}_{-12.2}$	  \\
$124.5$	&	$5.84\,{\cdot	10^{-1}}^{+3.2}_{-3.2}$	&	$6.40\,{\cdot	10^{-2}}^{+5.8}_{-5.7}$	&	$2.22\,{\cdot	10^{-4}}^{+6.0}_{-5.9}$	&	$2.95\,{\cdot	10^{-2}}^{+12.2}_{-12.2}$	  \\
$124.6$	&	$5.83\,{\cdot	10^{-1}}^{+3.2}_{-3.2}$	&	$6.38\,{\cdot	10^{-2}}^{+5.8}_{-5.7}$	&	$2.22\,{\cdot	10^{-4}}^{+6.0}_{-6.0}$	&	$2.94\,{\cdot	10^{-2}}^{+12.2}_{-12.2}$	  \\
$124.7$	&	$5.81\,{\cdot	10^{-1}}^{+3.2}_{-3.2}$	&	$6.37\,{\cdot	10^{-2}}^{+5.7}_{-5.7}$	&	$2.21\,{\cdot	10^{-4}}^{+6.0}_{-5.9}$	&	$2.93\,{\cdot	10^{-2}}^{+12.2}_{-12.2}$	  \\
$124.8$	&	$5.80\,{\cdot	10^{-1}}^{+3.2}_{-3.3}$	&	$6.35\,{\cdot	10^{-2}}^{+5.7}_{-5.7}$	&	$2.20\,{\cdot	10^{-4}}^{+6.0}_{-6.0}$	&	$2.93\,{\cdot	10^{-2}}^{+12.2}_{-12.2}$	  \\
$124.9$	&	$5.78\,{\cdot	10^{-1}}^{+3.2}_{-3.3}$	&	$6.33\,{\cdot	10^{-2}}^{+5.7}_{-5.7}$	&	$2.20\,{\cdot	10^{-4}}^{+6.0}_{-5.9}$	&	$2.92\,{\cdot	10^{-2}}^{+12.2}_{-12.2}$	  \\
$125$	&	$5.77\,{\cdot	10^{-1}}^{+3.2}_{-3.3}$	&	$6.32\,{\cdot	10^{-2}}^{+5.7}_{-5.7}$	&	$2.19\,{\cdot	10^{-4}}^{+6.0}_{-5.9}$	&	$2.91\,{\cdot	10^{-2}}^{+12.2}_{-12.2}$	  \\
\hline
\end{tabular}
\end{center}
\end{table}

\begin{table}\small \scriptsize
\setlength{\tabcolsep}{3ex}
\caption{SM Higgs branching ratios to two fermions and their total uncertainties (expressed in percentage). Low mass range.}
\label{tab:YRHXS3_2fermions.4}
\begin{center}
\begin{tabular}{lcccc}
\hline
$\MH$ [$\UGeV$] & $\Hbb$ & $\Htautau$ &$\Hmumu$ & $\Hcc$ \\
\hline
$129.1$	&	$5.09\,{\cdot	10^{-1}}^{+3.6}_{-3.7}$	&	$5.61\,{\cdot	10^{-2}}^{+5.3}_{-5.3}$	&	$1.95\,{\cdot	10^{-4}}^{+5.5}_{-5.5}$	&	$2.57\,{\cdot	10^{-2}}^{+12.2}_{-12.2}$	  \\
$129.2$	&	$5.08\,{\cdot	10^{-1}}^{+3.6}_{-3.7}$	&	$5.60\,{\cdot	10^{-2}}^{+5.3}_{-5.3}$	&	$1.94\,{\cdot	10^{-4}}^{+5.6}_{-5.5}$	&	$2.56\,{\cdot	10^{-2}}^{+12.2}_{-12.2}$	  \\
$129.3$	&	$5.06\,{\cdot	10^{-1}}^{+3.6}_{-3.7}$	&	$5.58\,{\cdot	10^{-2}}^{+5.3}_{-5.3}$	&	$1.94\,{\cdot	10^{-4}}^{+5.5}_{-5.5}$	&	$2.55\,{\cdot	10^{-2}}^{+12.2}_{-12.2}$	  \\
$129.4$	&	$5.04\,{\cdot	10^{-1}}^{+3.7}_{-3.7}$	&	$5.56\,{\cdot	10^{-2}}^{+5.3}_{-5.3}$	&	$1.93\,{\cdot	10^{-4}}^{+5.5}_{-5.5}$	&	$2.54\,{\cdot	10^{-2}}^{+12.2}_{-12.2}$	  \\
$129.5$	&	$5.02\,{\cdot	10^{-1}}^{+3.7}_{-3.7}$	&	$5.54\,{\cdot	10^{-2}}^{+5.3}_{-5.3}$	&	$1.92\,{\cdot	10^{-4}}^{+5.6}_{-5.4}$	&	$2.53\,{\cdot	10^{-2}}^{+12.2}_{-12.2}$	  \\
$129.6$	&	$5.01\,{\cdot	10^{-1}}^{+3.7}_{-3.7}$	&	$5.52\,{\cdot	10^{-2}}^{+5.3}_{-5.3}$	&	$1.92\,{\cdot	10^{-4}}^{+5.5}_{-5.5}$	&	$2.53\,{\cdot	10^{-2}}^{+12.2}_{-12.2}$	  \\
$129.7$	&	$4.99\,{\cdot	10^{-1}}^{+3.7}_{-3.8}$	&	$5.50\,{\cdot	10^{-2}}^{+5.3}_{-5.2}$	&	$1.91\,{\cdot	10^{-4}}^{+5.5}_{-5.5}$	&	$2.52\,{\cdot	10^{-2}}^{+12.2}_{-12.2}$	  \\
\hline
\end{tabular}
\end{center}
\end{table}

\begin{table}\small \scriptsize
\setlength{\tabcolsep}{3ex}
\caption{SM Higgs branching ratios to two fermions and their total uncertainties (expressed in percentage). Intermediate mass range.}
\label{tab:YRHXS3_2fermions.5}
\begin{center}
\begin{tabular}{lcccc}
\hline
$\MH$ [$\UGeV$] & $\Hbb$ & $\Htautau$ &$\Hmumu$ & $\Hcc$ \\
\hline
$149$	&	$1.71\,{\cdot	10^{-1}}^{+5.6}_{-5.7}$	&	$1.94\,{\cdot	10^{-2}}^{+3.4}_{-3.5}$	&	$6.72\,{\cdot	10^{-5}}^{+3.5}_{-3.6}$	&	$8.63\,{\cdot	10^{-3}}^{+12.0}_{-12.0}$	 \\
$149.5$	&	$1.64\,{\cdot	10^{-1}}^{+5.7}_{-5.7}$	&	$1.86\,{\cdot	10^{-2}}^{+3.4}_{-3.4}$	&	$6.45\,{\cdot	10^{-5}}^{+3.5}_{-3.5}$	&	$8.27\,{\cdot	10^{-3}}^{+12.1}_{-12.0}$	 \\
$150$	&	$1.57\,{\cdot	10^{-1}}^{+5.7}_{-5.7}$	&	$1.78\,{\cdot	10^{-2}}^{+3.4}_{-3.4}$	&	$6.18\,{\cdot	10^{-5}}^{+3.5}_{-3.5}$	&	$7.92\,{\cdot	10^{-3}}^{+12.0}_{-12.0}$	 \\
$152$	&	$1.30\,{\cdot	10^{-1}}^{+5.8}_{-5.9}$	&	$1.48\,{\cdot	10^{-2}}^{+3.3}_{-3.3}$	&	$5.13\,{\cdot	10^{-5}}^{+3.3}_{-3.3}$	&	$6.56\,{\cdot	10^{-3}}^{+12.1}_{-12.0}$	 \\
$154$	&	$1.05\,{\cdot	10^{-1}}^{+6.0}_{-6.0}$	&	$1.19\,{\cdot	10^{-2}}^{+3.1}_{-3.1}$	&	$4.13\,{\cdot	10^{-5}}^{+3.1}_{-3.2}$	&	$5.27\,{\cdot	10^{-3}}^{+12.0}_{-12.0}$	 \\
$156$	&	$8.01\,{\cdot	10^{-2}}^{+6.1}_{-6.1}$	&	$9.15\,{\cdot	10^{-3}}^{+3.0}_{-3.0}$	&	$3.17\,{\cdot	10^{-5}}^{+3.0}_{-3.0}$	&	$4.04\,{\cdot	10^{-3}}^{+12.0}_{-12.0}$	 \\
$158$	&	$5.66\,{\cdot	10^{-2}}^{+6.3}_{-6.3}$	&	$6.47\,{\cdot	10^{-3}}^{+2.8}_{-2.8}$	&	$2.24\,{\cdot	10^{-5}}^{+2.9}_{-2.9}$	&	$2.85\,{\cdot	10^{-3}}^{+12.0}_{-11.9}$	 \\
\hline
\end{tabular}
\end{center}
\end{table}

\begin{table}\small \scriptsize
\setlength{\tabcolsep}{3ex}
\caption{SM Higgs branching ratios to two fermions and their total uncertainties (expressed in percentage). Intermediate mass range.}
\label{tab:YRHXS3_2fermions.6}
\begin{center}
\begin{tabular}{lccccc}
\hline
$\MH$ [$\UGeV$] & $\Hbb$ & $\Htautau$ &$\Hmumu$ & $\Hcc$ & $\Htt$\\
\hline
$232$	&	$1.33\,{\cdot	10^{-3}}^{+6.6}_{-6.6}$	&	$1.63\,{\cdot	10^{-4}}^{+2.6}_{-2.6}$	&	$5.65\,{\cdot	10^{-7}}^{+2.6}_{-2.6}$	&	$6.67\,{\cdot	10^{-5}}^{+12.1}_{-12.0}$	&	$0\,{\cdot	10^{0}}^{+0}_{-0}$	  \\
$234$	&	$1.29\,{\cdot	10^{-3}}^{+6.6}_{-6.6}$	&	$1.58\,{\cdot	10^{-4}}^{+2.6}_{-2.6}$	&	$5.49\,{\cdot	10^{-7}}^{+2.6}_{-2.6}$	&	$6.47\,{\cdot	10^{-5}}^{+12.1}_{-12.0}$	&	$0\,{\cdot	10^{0}}^{+0}_{-0}$	  \\
$236$	&	$1.25\,{\cdot	10^{-3}}^{+6.7}_{-6.6}$	&	$1.54\,{\cdot	10^{-4}}^{+2.6}_{-2.6}$	&	$5.33\,{\cdot	10^{-7}}^{+2.6}_{-2.6}$	&	$6.27\,{\cdot	10^{-5}}^{+12.1}_{-12.0}$	&	$0\,{\cdot	10^{0}}^{+0}_{-0}$	  \\
$238$	&	$1.21\,{\cdot	10^{-3}}^{+6.7}_{-6.6}$	&	$1.50\,{\cdot	10^{-4}}^{+2.6}_{-2.6}$	&	$5.19\,{\cdot	10^{-7}}^{+2.6}_{-2.6}$	&	$6.09\,{\cdot	10^{-5}}^{+12.1}_{-12.0}$	&	$0\,{\cdot	10^{0}}^{+0}_{-0}$	  \\
$240$	&	$1.18\,{\cdot	10^{-3}}^{+6.7}_{-6.5}$	&	$1.45\,{\cdot	10^{-4}}^{+2.6}_{-2.6}$	&	$5.04\,{\cdot	10^{-7}}^{+2.6}_{-2.6}$	&	$5.91\,{\cdot	10^{-5}}^{+12.1}_{-12.0}$	&	$0\,{\cdot	10^{0}}^{+0}_{-0}$	  \\
$242$	&	$1.14\,{\cdot	10^{-3}}^{+6.6}_{-6.7}$	&	$1.42\,{\cdot	10^{-4}}^{+2.6}_{-2.6}$	&	$4.91\,{\cdot	10^{-7}}^{+2.6}_{-2.6}$	&	$5.74\,{\cdot	10^{-5}}^{+12.1}_{-12.0}$	&	$0\,{\cdot	10^{0}}^{+0}_{-0}$	  \\
$244$	&	$1.11\,{\cdot	10^{-3}}^{+6.6}_{-6.7}$	&	$1.38\,{\cdot	10^{-4}}^{+2.6}_{-2.6}$	&	$4.78\,{\cdot	10^{-7}}^{+2.6}_{-2.6}$	&	$5.58\,{\cdot	10^{-5}}^{+12.1}_{-12.0}$	&	$0\,{\cdot	10^{0}}^{+0}_{-0}$	  \\
\hline
\end{tabular}
\end{center}
\end{table}

\begin{table}\small \scriptsize
\setlength{\tabcolsep}{3ex}
\caption{SM Higgs branching ratios to two fermions and their total uncertainties (expressed in percentage). High mass range.}
\label{tab:YRHXS3_2fermions.7}
\begin{center}
\begin{tabular}{lccccc}
\hline
$\MH$ [$\UGeV$] & $\Hbb$ & $\Htautau$ &$\Hmumu$ & $\Hcc$ & $\Htt$ \\
\hline
$380$	&	$2.54\,{\cdot	10^{-4}}^{+7.7}_{-7.6}$	&	$3.41\,{\cdot	10^{-5}}^{+5.0}_{-4.8}$	&	$1.18\,{\cdot	10^{-7}}^{+5.0}_{-4.8}$	&	$1.27\,{\cdot	10^{-5}}^{+13.0}_{-12.8}$	&	$1.10\,{\cdot	10^{-1}}^{+17.1}_{-18.1}$	  \\
$390$	&	$2.3\,{\cdot	10^{-4}}^{+7.7}_{-7.6}$	&	$3.10\,{\cdot	10^{-5}}^{+4.9}_{-4.7}$	&	$1.07\,{\cdot	10^{-7}}^{+4.9}_{-4.7}$	&	$1.15\,{\cdot	10^{-5}}^{+13.0}_{-12.9}$	&	$1.31\,{\cdot	10^{-1}}^{+13.2}_{-14.1}$	  \\
$400$	&	$2.09\,{\cdot	10^{-4}}^{+7.7}_{-7.6}$	&	$2.84\,{\cdot	10^{-5}}^{+4.7}_{-4.4}$	&	$9.83\,{\cdot	10^{-8}}^{+4.6}_{-4.5}$	&	$1.05\,{\cdot	10^{-5}}^{+13.1}_{-12.9}$	&	$1.48\,{\cdot	10^{-1}}^{+10.9}_{-11.5}$	  \\
$420$	&	$1.78\,{\cdot	10^{-4}}^{+7.7}_{-7.6}$	&	$2.43\,{\cdot	10^{-5}}^{+4.3}_{-4.2}$	&	$8.41\,{\cdot	10^{-8}}^{+4.3}_{-4.2}$	&	$8.93\,{\cdot	10^{-6}}^{+13.1}_{-13.0}$	&	$1.71\,{\cdot	10^{-1}}^{+8.1}_{-8.6}$	  \\
$440$	&	$1.54\,{\cdot	10^{-4}}^{+7.7}_{-7.6}$	&	$2.12\,{\cdot	10^{-5}}^{+4.0}_{-3.9}$	&	$7.35\,{\cdot	10^{-8}}^{+4.0}_{-3.9}$	&	$7.74\,{\cdot	10^{-6}}^{+13.2}_{-13.0}$	&	$1.84\,{\cdot	10^{-1}}^{+6.5}_{-6.9}$	  \\
$450$	&	$1.44\,{\cdot	10^{-4}}^{+7.7}_{-7.6}$	&	$1.99\,{\cdot	10^{-5}}^{+3.9}_{-3.8}$	&	$6.91\,{\cdot	10^{-8}}^{+3.9}_{-3.8}$	&	$7.25\,{\cdot	10^{-6}}^{+13.1}_{-13.0}$	&	$1.88\,{\cdot	10^{-1}}^{+6.0}_{-6.4}$	  \\
$460$	&	$1.36\,{\cdot	10^{-4}}^{+7.7}_{-7.7}$	&	$1.88\,{\cdot	10^{-5}}^{+3.8}_{-3.7}$	&	$6.52\,{\cdot	10^{-8}}^{+3.8}_{-3.7}$	&	$6.81\,{\cdot	10^{-6}}^{+13.2}_{-13.0}$	&	$1.91\,{\cdot	10^{-1}}^{+5.6}_{-5.9}$	  \\
\hline
\end{tabular}
\end{center}
\end{table}

\begin{table}\small\scriptsize
\setlength{\tabcolsep}{1ex}
\caption{SM Higgs branching ratios to two gauge bosons and Higgs total width together with their total uncertainties (expressed in percentage). Very-low-mass range.}
\label{tab:YRHXS3_bosons-width.1}
\begin{center}\begin{tabular}{lcccccc}
\hline
$\MH$ [$\UGeV$] & $\Hgg$ & $\Hgaga$ &$\HZga$ & $\HWW$ & $\HZZ$ & Total
$\Gamma_{\PH}$ [$\UGeV$]\\
\hline
$80$	&	$4.94\,{\cdot	10^{-2}}^{+12.0}_{-11.6}$	&	$9.18\,{\cdot	10^{-4}}^{+6.3}_{-6.2}$	&	$0\,{\cdot	10^{0}}^{+0}_{-0}$	&	$6.27\,{\cdot	10^{-4}}^{+5.9}_{-5.7}$	&	$1.64\,{\cdot	10^{-4}}^{+5.8}_{-5.6}$	&	$1.99\,{\cdot	10^{-3}}^{+5.3}_{-5.2}$	\\
$81$	&	$5.05\,{\cdot	10^{-2}}^{+12.0}_{-11.6}$	&	$9.46\,{\cdot	10^{-4}}^{+6.3}_{-6.1}$	&	$0\,{\cdot	10^{0}}^{+0}_{-0}$	&	$6.96\,{\cdot	10^{-4}}^{+5.9}_{-5.7}$	&	$1.81\,{\cdot	10^{-4}}^{+5.8}_{-5.6}$	&	$2.01\,{\cdot	10^{-3}}^{+5.3}_{-5.2}$	\\
$82$	&	$5.17\,{\cdot	10^{-2}}^{+12.0}_{-11.6}$	&	$9.74\,{\cdot	10^{-4}}^{+6.3}_{-6.2}$	&	$0\,{\cdot	10^{0}}^{+0}_{-0}$	&	$7.74\,{\cdot	10^{-4}}^{+5.9}_{-5.7}$	&	$1.99\,{\cdot	10^{-4}}^{+5.8}_{-5.6}$	&	$2.04\,{\cdot	10^{-3}}^{+5.3}_{-5.2}$	\\
$83$	&	$5.28\,{\cdot	10^{-2}}^{+12.0}_{-11.5}$	&	$1.00\,{\cdot	10^{-3}}^{+6.4}_{-6.1}$	&	$0\,{\cdot	10^{0}}^{+0}_{-0}$	&	$8.61\,{\cdot	10^{-4}}^{+5.8}_{-5.7}$	&	$2.18\,{\cdot	10^{-4}}^{+5.8}_{-5.6}$	&	$2.06\,{\cdot	10^{-3}}^{+5.3}_{-5.2}$	\\
$84$	&	$5.40\,{\cdot	10^{-2}}^{+12.0}_{-11.5}$	&	$1.03\,{\cdot	10^{-3}}^{+6.3}_{-6.1}$	&	$0\,{\cdot	10^{0}}^{+0}_{-0}$	&	$9.61\,{\cdot	10^{-4}}^{+5.9}_{-5.7}$	&	$2.40\,{\cdot	10^{-4}}^{+5.8}_{-5.6}$	&	$2.08\,{\cdot	10^{-3}}^{+5.3}_{-5.2}$	\\
$85$	&	$5.51\,{\cdot	10^{-2}}^{+11.9}_{-11.5}$	&	$1.06\,{\cdot	10^{-3}}^{+6.3}_{-6.2}$	&	$0\,{\cdot	10^{0}}^{+0}_{-0}$	&	$1.08\,{\cdot	10^{-3}}^{+5.9}_{-5.7}$	&	$2.63\,{\cdot	10^{-4}}^{+5.8}_{-5.6}$	&	$2.11\,{\cdot	10^{-3}}^{+5.3}_{-5.2}$	\\
$86$	&	$5.63\,{\cdot	10^{-2}}^{+11.9}_{-11.5}$	&	$1.09\,{\cdot	10^{-3}}^{+6.3}_{-6.1}$	&	$0\,{\cdot	10^{0}}^{+0}_{-0}$	&	$1.21\,{\cdot	10^{-3}}^{+5.9}_{-5.7}$	&	$2.88\,{\cdot	10^{-4}}^{+5.8}_{-5.6}$	&	$2.13\,{\cdot	10^{-3}}^{+5.3}_{-5.2}$	\\
\hline
\end{tabular}
\end{center}
\end{table}

\begin{table}\small\scriptsize
\setlength{\tabcolsep}{1ex}
\caption{SM Higgs branching ratios to two gauge bosons and Higgs total width together with their total uncertainties (expressed in percentage). Very-low-mass range.}
\label{tab:YRHXS3_bosons-width.2}
\begin{center}\begin{tabular}{lcccccc}
\hline
$\MH$ [$\UGeV$] & $\Hgg$ & $\Hgaga$ &$\HZga$ & $\HWW$ & $\HZZ$ & Total
$\Gamma_{\PH}$ [$\UGeV$]\\
\hline
$118.5$	&	$8.84\,{\cdot	10^{-2}}^{+10.7}_{-10.4}$	&	$2.20\,{\cdot	10^{-3}}^{+5.5}_{-5.4}$	&	$9.81\,{\cdot	10^{-4}}^{+9.5}_{-9.4}$	&	$1.23\,{\cdot	10^{-1}}^{+4.9}_{-4.8}$	&	$1.34\,{\cdot	10^{-2}}^{+4.9}_{-4.8}$	&	$3.38\,{\cdot	10^{-3}}^{+4.5}_{-4.5}$	\\
$119$	&	$8.84\,{\cdot	10^{-2}}^{+10.7}_{-10.4}$	&	$2.21\,{\cdot	10^{-3}}^{+5.5}_{-5.4}$	&	$1.02\,{\cdot	10^{-3}}^{+9.5}_{-9.4}$	&	$1.29\,{\cdot	10^{-1}}^{+4.8}_{-4.7}$	&	$1.42\,{\cdot	10^{-2}}^{+4.9}_{-4.7}$	&	$3.42\,{\cdot	10^{-3}}^{+4.4}_{-4.4}$	\\
$119.5$	&	$8.83\,{\cdot	10^{-2}}^{+10.7}_{-10.4}$	&	$2.22\,{\cdot	10^{-3}}^{+5.5}_{-5.3}$	&	$1.06\,{\cdot	10^{-3}}^{+9.4}_{-9.4}$	&	$1.35\,{\cdot	10^{-1}}^{+4.8}_{-4.7}$	&	$1.50\,{\cdot	10^{-2}}^{+4.8}_{-4.7}$	&	$3.46\,{\cdot	10^{-3}}^{+4.4}_{-4.4}$	\\
$120$	&	$8.82\,{\cdot	10^{-2}}^{+10.6}_{-10.3}$	&	$2.23\,{\cdot	10^{-3}}^{+5.4}_{-5.3}$	&	$1.11\,{\cdot	10^{-3}}^{+9.4}_{-9.3}$	&	$1.41\,{\cdot	10^{-1}}^{+4.8}_{-4.7}$	&	$1.59\,{\cdot	10^{-2}}^{+4.8}_{-4.7}$	&	$3.51\,{\cdot	10^{-3}}^{+4.4}_{-4.3}$	\\
$120.1$	&	$8.82\,{\cdot	10^{-2}}^{+10.6}_{-10.3}$	&	$2.23\,{\cdot	10^{-3}}^{+5.4}_{-5.3}$	&	$1.12\,{\cdot	10^{-3}}^{+9.4}_{-9.2}$	&	$1.43\,{\cdot	10^{-1}}^{+4.8}_{-4.7}$	&	$1.61\,{\cdot	10^{-2}}^{+4.8}_{-4.7}$	&	$3.52\,{\cdot	10^{-3}}^{+4.4}_{-4.3}$	\\
$120.2$	&	$8.82\,{\cdot	10^{-2}}^{+10.6}_{-10.3}$	&	$2.23\,{\cdot	10^{-3}}^{+5.4}_{-5.2}$	&	$1.12\,{\cdot	10^{-3}}^{+9.3}_{-9.3}$	&	$1.44\,{\cdot	10^{-1}}^{+4.7}_{-4.7}$	&	$1.62\,{\cdot	10^{-2}}^{+4.7}_{-4.6}$	&	$3.52\,{\cdot	10^{-3}}^{+4.4}_{-4.3}$	\\
$120.3$	&	$8.82\,{\cdot	10^{-2}}^{+10.6}_{-10.3}$	&	$2.23\,{\cdot	10^{-3}}^{+5.4}_{-5.3}$	&	$1.13\,{\cdot	10^{-3}}^{+9.3}_{-9.3}$	&	$1.45\,{\cdot	10^{-1}}^{+4.7}_{-4.6}$	&	$1.64\,{\cdot	10^{-2}}^{+4.7}_{-4.6}$	&	$3.53\,{\cdot	10^{-3}}^{+4.3}_{-4.3}$	\\
\hline
\end{tabular}
\end{center}
\end{table}

\begin{table}\small\scriptsize
\setlength{\tabcolsep}{1ex}
\caption{SM Higgs branching ratios to two gauge bosons and Higgs total width together with their total uncertainties (expressed in percentage). Very-low-mass range.}
\label{tab:YRHXS3_bosons-width.3}
\begin{center}\begin{tabular}{lcccccc}
\hline
$\MH$ [$\UGeV$] & $\Hgg$ & $\Hgaga$ &$\HZga$ & $\HWW$ & $\HZZ$ & Total
$\Gamma_{\PH}$ [$\UGeV$]\\
\hline
$124.4$	&	$8.62\,{\cdot	10^{-2}}^{+10.3}_{-10.0}$	&	$2.28\,{\cdot	10^{-3}}^{+5.0}_{-5.0}$	&	$1.49\,{\cdot	10^{-3}}^{+9.0}_{-9.0}$	&	$2.05\,{\cdot	10^{-1}}^{+4.3}_{-4.2}$	&	$2.5\,{\cdot	10^{-2}}^{+4.3}_{-4.3}$	&	$3.99\,{\cdot	10^{-3}}^{+4.0}_{-4.0}$	\\
$124.5$	&	$8.61\,{\cdot	10^{-2}}^{+10.3}_{-10.0}$	&	$2.28\,{\cdot	10^{-3}}^{+5.1}_{-4.9}$	&	$1.50\,{\cdot	10^{-3}}^{+9.0}_{-8.9}$	&	$2.06\,{\cdot	10^{-1}}^{+4.3}_{-4.3}$	&	$2.52\,{\cdot	10^{-2}}^{+4.3}_{-4.2}$	&	$4.00\,{\cdot	10^{-3}}^{+4.0}_{-4.0}$	\\
$124.6$	&	$8.60\,{\cdot	10^{-2}}^{+10.3}_{-10.0}$	&	$2.28\,{\cdot	10^{-3}}^{+5.0}_{-5.0}$	&	$1.5\,{\cdot	10^{-3}}^{+9.0}_{-9.0}$	&	$2.08\,{\cdot	10^{-1}}^{+4.3}_{-4.2}$	&	$2.55\,{\cdot	10^{-2}}^{+4.3}_{-4.2}$	&	$4.02\,{\cdot	10^{-3}}^{+4.0}_{-4.0}$	\\
$124.7$	&	$8.59\,{\cdot	10^{-2}}^{+10.3}_{-10.0}$	&	$2.28\,{\cdot	10^{-3}}^{+5.0}_{-4.9}$	&	$1.51\,{\cdot	10^{-3}}^{+9.0}_{-8.9}$	&	$2.10\,{\cdot	10^{-1}}^{+4.3}_{-4.2}$	&	$2.57\,{\cdot	10^{-2}}^{+4.3}_{-4.2}$	&	$4.03\,{\cdot	10^{-3}}^{+4.0}_{-4.0}$	\\
$124.8$	&	$8.59\,{\cdot	10^{-2}}^{+10.2}_{-10.0}$	&	$2.28\,{\cdot	10^{-3}}^{+5.0}_{-4.9}$	&	$1.52\,{\cdot	10^{-3}}^{+9.0}_{-8.9}$	&	$2.11\,{\cdot	10^{-1}}^{+4.3}_{-4.2}$	&	$2.59\,{\cdot	10^{-2}}^{+4.3}_{-4.3}$	&	$4.04\,{\cdot	10^{-3}}^{+4.0}_{-3.9}$	\\
$124.9$	&	$8.58\,{\cdot	10^{-2}}^{+10.2}_{-10.0}$	&	$2.28\,{\cdot	10^{-3}}^{+5.0}_{-4.9}$	&	$1.53\,{\cdot	10^{-3}}^{+8.9}_{-8.9}$	&	$2.13\,{\cdot	10^{-1}}^{+4.3}_{-4.2}$	&	$2.62\,{\cdot	10^{-2}}^{+4.3}_{-4.2}$	&	$4.06\,{\cdot	10^{-3}}^{+3.9}_{-3.9}$	\\
$125$	&	$8.57\,{\cdot	10^{-2}}^{+10.2}_{-10.0}$	&	$2.28\,{\cdot	10^{-3}}^{+5.0}_{-4.9}$	&	$1.54\,{\cdot	10^{-3}}^{+9.0}_{-8.8}$	&	$2.15\,{\cdot	10^{-1}}^{+4.3}_{-4.2}$	&	$2.64\,{\cdot	10^{-2}}^{+4.3}_{-4.2}$	&	$4.07\,{\cdot	10^{-3}}^{+4.0}_{-3.9}$	\\
\hline
\end{tabular}
\end{center}
\end{table}

\begin{table}\small\scriptsize
\setlength{\tabcolsep}{1ex}
\caption{SM Higgs branching ratios to two gauge bosons and Higgs total width together with their total uncertainties (expressed in percentage). Very-low-mass range.}
\label{tab:YRHXS3_bosons-width.4}
\begin{center}\begin{tabular}{lcccccc}
\hline
$\MH$ [$\UGeV$] & $\Hgg$ & $\Hgaga$ &$\HZga$ & $\HWW$ & $\HZZ$ & Total
$\Gamma_{\PH}$ [$\UGeV$]\\
\hline
$129.1$	&	$8.10\,{\cdot	10^{-2}}^{+9.9}_{-9.6}$	&	$2.26\,{\cdot	10^{-3}}^{+4.6}_{-4.5}$	&	$1.88\,{\cdot	10^{-3}}^{+8.5}_{-8.5}$	&	$2.86\,{\cdot	10^{-1}}^{+3.8}_{-3.7}$	&	$3.73\,{\cdot	10^{-2}}^{+3.8}_{-3.7}$	&	$4.73\,{\cdot	10^{-3}}^{+3.6}_{-3.5}$	\\
$129.2$	&	$8.09\,{\cdot	10^{-2}}^{+9.8}_{-9.6}$	&	$2.26\,{\cdot	10^{-3}}^{+4.5}_{-4.5}$	&	$1.89\,{\cdot	10^{-3}}^{+8.5}_{-8.5}$	&	$2.88\,{\cdot	10^{-1}}^{+3.8}_{-3.7}$	&	$3.75\,{\cdot	10^{-2}}^{+3.8}_{-3.7}$	&	$4.75\,{\cdot	10^{-3}}^{+3.5}_{-3.5}$	\\
$129.3$	&	$8.08\,{\cdot	10^{-2}}^{+9.8}_{-9.6}$	&	$2.25\,{\cdot	10^{-3}}^{+4.5}_{-4.5}$	&	$1.89\,{\cdot	10^{-3}}^{+8.5}_{-8.5}$	&	$2.90\,{\cdot	10^{-1}}^{+3.7}_{-3.7}$	&	$3.78\,{\cdot	10^{-2}}^{+3.7}_{-3.7}$	&	$4.77\,{\cdot	10^{-3}}^{+3.5}_{-3.5}$	\\
$129.4$	&	$8.06\,{\cdot	10^{-2}}^{+9.8}_{-9.6}$	&	$2.25\,{\cdot	10^{-3}}^{+4.6}_{-4.4}$	&	$1.90\,{\cdot	10^{-3}}^{+8.6}_{-8.4}$	&	$2.92\,{\cdot	10^{-1}}^{+3.7}_{-3.7}$	&	$3.81\,{\cdot	10^{-2}}^{+3.7}_{-3.7}$	&	$4.79\,{\cdot	10^{-3}}^{+3.5}_{-3.5}$	\\
$129.5$	&	$8.05\,{\cdot	10^{-2}}^{+9.8}_{-9.6}$	&	$2.25\,{\cdot	10^{-3}}^{+4.5}_{-4.4}$	&	$1.91\,{\cdot	10^{-3}}^{+8.5}_{-8.5}$	&	$2.93\,{\cdot	10^{-1}}^{+3.7}_{-3.7}$	&	$3.84\,{\cdot	10^{-2}}^{+3.7}_{-3.7}$	&	$4.81\,{\cdot	10^{-3}}^{+3.5}_{-3.5}$	\\
$129.6$	&	$8.03\,{\cdot	10^{-2}}^{+9.8}_{-9.6}$	&	$2.25\,{\cdot	10^{-3}}^{+4.5}_{-4.5}$	&	$1.92\,{\cdot	10^{-3}}^{+8.5}_{-8.4}$	&	$2.95\,{\cdot	10^{-1}}^{+3.7}_{-3.7}$	&	$3.87\,{\cdot	10^{-2}}^{+3.7}_{-3.7}$	&	$4.83\,{\cdot	10^{-3}}^{+3.5}_{-3.5}$	\\
$129.7$	&	$8.02\,{\cdot	10^{-2}}^{+9.8}_{-9.6}$	&	$2.25\,{\cdot	10^{-3}}^{+4.5}_{-4.5}$	&	$1.92\,{\cdot	10^{-3}}^{+8.5}_{-8.4}$	&	$2.97\,{\cdot	10^{-1}}^{+3.7}_{-3.6}$	&	$3.90\,{\cdot	10^{-2}}^{+3.7}_{-3.6}$	&	$4.85\,{\cdot	10^{-3}}^{+3.5}_{-3.5}$	\\
\hline
\end{tabular}
\end{center}
\end{table}

\begin{table}\small\scriptsize
\setlength{\tabcolsep}{1ex}
\caption{SM Higgs branching ratios to two gauge bosons and Higgs total width together with their total uncertainties (expressed in percentage). Low-mass range.}
\label{tab:YRHXS3_bosons-width.5}
\begin{center}\begin{tabular}{lcccccc}
\hline
$\MH$ [$\UGeV$] & $\Hgg$ & $\Hgaga$ &$\HZga$ & $\HWW$ & $\HZZ$ & Total
$\Gamma_{\PH}$ [$\UGeV$]\\
\hline
$149$	&	$3.71\,{\cdot	10^{-2}}^{+8.3}_{-8.2}$	&	$1.43\,{\cdot	10^{-3}}^{+2.5}_{-2.5}$	&	$2.36\,{\cdot	10^{-3}}^{+6.5}_{-6.6}$	&	$6.77\,{\cdot	10^{-1}}^{+1.3}_{-1.3}$	&	$8.26\,{\cdot	10^{-2}}^{+1.3}_{-1.3}$	&	$1.58\,{\cdot	10^{-2}}^{+1.5}_{-1.5}$	\\
$149.5$	&	$3.58\,{\cdot	10^{-2}}^{+8.3}_{-8.1}$	&	$1.39\,{\cdot	10^{-3}}^{+2.5}_{-2.4}$	&	$2.33\,{\cdot	10^{-3}}^{+6.5}_{-6.4}$	&	$6.87\,{\cdot	10^{-1}}^{+1.2}_{-1.2}$	&	$8.26\,{\cdot	10^{-2}}^{+1.2}_{-1.3}$	&	$1.65\,{\cdot	10^{-2}}^{+1.5}_{-1.5}$	\\
$150$	&	$3.46\,{\cdot	10^{-2}}^{+8.3}_{-8.2}$	&	$1.36\,{\cdot	10^{-3}}^{+2.5}_{-2.4}$	&	$2.31\,{\cdot	10^{-3}}^{+6.4}_{-6.5}$	&	$6.96\,{\cdot	10^{-1}}^{+1.2}_{-1.2}$	&	$8.25\,{\cdot	10^{-2}}^{+1.2}_{-1.2}$	&	$1.73\,{\cdot	10^{-2}}^{+1.4}_{-1.4}$	\\
$152$	&	$2.94\,{\cdot	10^{-2}}^{+8.1}_{-8.1}$	&	$1.22\,{\cdot	10^{-3}}^{+2.3}_{-2.3}$	&	$2.18\,{\cdot	10^{-3}}^{+6.3}_{-6.3}$	&	$7.35\,{\cdot	10^{-1}}^{+1.0}_{-1.0}$	&	$8.08\,{\cdot	10^{-2}}^{+1.0}_{-1.0}$	&	$2.11\,{\cdot	10^{-2}}^{+1.3}_{-1.3}$	\\
$154$	&	$2.43\,{\cdot	10^{-2}}^{+8.0}_{-8.0}$	&	$1.08\,{\cdot	10^{-3}}^{+2.2}_{-2.1}$	&	$2.01\,{\cdot	10^{-3}}^{+6.1}_{-6.1}$	&	$7.74\,{\cdot	10^{-1}}^{+0.8}_{-0.8}$	&	$7.66\,{\cdot	10^{-2}}^{+0.8}_{-0.8}$	&	$2.66\,{\cdot	10^{-2}}^{+1.1}_{-1.1}$	\\
$156$	&	$1.92\,{\cdot	10^{-2}}^{+7.9}_{-7.9}$	&	$9.14\,{\cdot	10^{-4}}^{+2.0}_{-2.0}$	&	$1.79\,{\cdot	10^{-3}}^{+6.0}_{-6.0}$	&	$8.16\,{\cdot	10^{-1}}^{+0.6}_{-0.6}$	&	$6.92\,{\cdot	10^{-2}}^{+0.6}_{-0.6}$	&	$3.51\,{\cdot	10^{-2}}^{+1.0}_{-1.0}$	\\
$158$	&	$1.38\,{\cdot	10^{-2}}^{+7.9}_{-7.8}$	&	$7.29\,{\cdot	10^{-4}}^{+1.8}_{-1.9}$	&	$1.50\,{\cdot	10^{-3}}^{+5.8}_{-5.9}$	&	$8.60\,{\cdot	10^{-1}}^{+0.4}_{-0.4}$	&	$5.76\,{\cdot	10^{-2}}^{+0.4}_{-0.4}$	&	$5.02\,{\cdot	10^{-2}}^{+0.8}_{-0.8}$	\\
\hline
\end{tabular}
\end{center}
\end{table}

\newpage
\begin{table}\small\scriptsize
\setlength{\tabcolsep}{1ex}
\caption{SM Higgs branching ratios to two gauge bosons and Higgs total width together with their total uncertainties (expressed in percentage). Intermediate-mass range.}
\label{tab:YRHXS3_bosons-width.6}
\begin{center}\begin{tabular}{lcccccc}
\hline
$\MH$ [$\UGeV$] & $\Hgg$ & $\Hgaga$ &$\HZga$ & $\HWW$ & $\HZZ$ & Total
$\Gamma_{\PH}$ [$\UGeV$]\\
\hline
$232$	&	$7.29\,{\cdot	10^{-4}}^{+7.6}_{-7.5}$	&	$3.19\,{\cdot	10^{-5}}^{+1.6}_{-1.6}$	&	$1.17\,{\cdot	10^{-4}}^{+5.5}_{-5.6}$	&	$7.07\,{\cdot	10^{-1}}^{+0.0}_{-0.0}$	&	$2.90\,{\cdot	10^{-1}}^{+0.0}_{-0.0}$	&	$2.93\,{\cdot	10^{0}}^{+0.5}_{-0.5}$	\\
$234$	&	$7.22\,{\cdot	10^{-4}}^{+7.6}_{-7.5}$	&	$3.09\,{\cdot	10^{-5}}^{+1.6}_{-1.6}$	&	$1.14\,{\cdot	10^{-4}}^{+5.6}_{-5.6}$	&	$7.06\,{\cdot	10^{-1}}^{+0.0}_{-0.0}$	&	$2.91\,{\cdot	10^{-1}}^{+0.0}_{-0.0}$	&	$3.04\,{\cdot	10^{0}}^{+0.5}_{-0.5}$	\\
$236$	&	$7.16\,{\cdot	10^{-4}}^{+7.6}_{-7.5}$	&	$3.01\,{\cdot	10^{-5}}^{+1.6}_{-1.6}$	&	$1.11\,{\cdot	10^{-4}}^{+5.6}_{-5.5}$	&	$7.06\,{\cdot	10^{-1}}^{+0.0}_{-0.0}$	&	$2.92\,{\cdot	10^{-1}}^{+0.0}_{-0.0}$	&	$3.16\,{\cdot	10^{0}}^{+0.5}_{-0.5}$	\\
$238$	&	$7.10\,{\cdot	10^{-4}}^{+7.6}_{-7.5}$	&	$2.92\,{\cdot	10^{-5}}^{+1.6}_{-1.7}$	&	$1.09\,{\cdot	10^{-4}}^{+5.6}_{-5.5}$	&	$7.05\,{\cdot	10^{-1}}^{+0.0}_{-0.0}$	&	$2.93\,{\cdot	10^{-1}}^{+0.0}_{-0.0}$	&	$3.27\,{\cdot	10^{0}}^{+0.5}_{-0.5}$	\\
$240$	&	$7.05\,{\cdot	10^{-4}}^{+7.6}_{-7.5}$	&	$2.84\,{\cdot	10^{-5}}^{+1.7}_{-1.6}$	&	$1.07\,{\cdot	10^{-4}}^{+5.5}_{-5.6}$	&	$7.04\,{\cdot	10^{-1}}^{+0.0}_{-0.0}$	&	$2.94\,{\cdot	10^{-1}}^{+0.0}_{-0.0}$	&	$3.40\,{\cdot	10^{0}}^{+0.5}_{-0.5}$	\\
$242$	&	$7.00\,{\cdot	10^{-4}}^{+7.6}_{-7.5}$	&	$2.76\,{\cdot	10^{-5}}^{+1.7}_{-1.7}$	&	$1.04\,{\cdot	10^{-4}}^{+5.6}_{-5.5}$	&	$7.04\,{\cdot	10^{-1}}^{+0.0}_{-0.0}$	&	$2.94\,{\cdot	10^{-1}}^{+0.0}_{-0.0}$	&	$3.52\,{\cdot	10^{0}}^{+0.5}_{-0.5}$	\\
$244$	&	$6.95\,{\cdot	10^{-4}}^{+7.6}_{-7.5}$	&	$2.68\,{\cdot	10^{-5}}^{+1.7}_{-1.7}$	&	$1.02\,{\cdot	10^{-4}}^{+5.6}_{-5.5}$	&	$7.03\,{\cdot	10^{-1}}^{+0.0}_{-0.0}$	&	$2.95\,{\cdot	10^{-1}}^{+0.0}_{-0.0}$	&	$3.64\,{\cdot	10^{0}}^{+0.5}_{-0.5}$	\\
\hline
\end{tabular}
\end{center}
\end{table}

\begin{table}\small\scriptsize
\setlength{\tabcolsep}{1ex}
\caption{SM Higgs branching ratios to two gauge bosons and Higgs total width together with their total uncertainties (expressed in percentage). High-mass range.}
\label{tab:YRHXS3_bosons-width.7}
\begin{center}\begin{tabular}{lcccccc}
\hline
$\MH$ [$\UGeV$] & $\Hgg$ & $\Hgaga$ &$\HZga$ & $\HWW$ & $\HZZ$ & Total
$\Gamma_{\PH}$ [$\UGeV$]\\
\hline
$380$	&	$8.61\,{\cdot	10^{-4}}^{+7.7}_{-7.7}$	&	$4.06\,{\cdot	10^{-6}}^{+10.0}_{-9.6}$	&	$2.42\,{\cdot	10^{-5}}^{+9.1}_{-8.9}$	&	$6.09\,{\cdot	10^{-1}}^{+2.2}_{-2.1}$	&	$2.80\,{\cdot	10^{-1}}^{+2.2}_{-2.1}$	&	$2.31\,{\cdot	10^{1}}^{+2.8}_{-2.9}$	\\
$390$	&	$8.50\,{\cdot	10^{-4}}^{+7.7}_{-7.7}$	&	$3.27\,{\cdot	10^{-6}}^{+10.4}_{-9.7}$	&	$2.14\,{\cdot	10^{-5}}^{+9.0}_{-8.7}$	&	$5.94\,{\cdot	10^{-1}}^{+2.1}_{-2.0}$	&	$2.74\,{\cdot	10^{-1}}^{+2.1}_{-2.0}$	&	$2.61\,{\cdot	10^{1}}^{+2.6}_{-2.7}$	\\
$400$	&	$8.32\,{\cdot	10^{-4}}^{+7.8}_{-7.7}$	&	$2.65\,{\cdot	10^{-6}}^{+10.4}_{-9.8}$	&	$1.90\,{\cdot	10^{-5}}^{+8.7}_{-8.5}$	&	$5.82\,{\cdot	10^{-1}}^{+2.0}_{-1.9}$	&	$2.69\,{\cdot	10^{-1}}^{+2.0}_{-1.9}$	&	$2.92\,{\cdot	10^{1}}^{+2.5}_{-2.5}$	\\
$420$	&	$7.89\,{\cdot	10^{-4}}^{+7.9}_{-7.9}$	&	$1.74\,{\cdot	10^{-6}}^{+10.6}_{-9.8}$	&	$1.53\,{\cdot	10^{-5}}^{+8.3}_{-8.2}$	&	$5.65\,{\cdot	10^{-1}}^{+1.8}_{-1.7}$	&	$2.63\,{\cdot	10^{-1}}^{+1.7}_{-1.6}$	&	$3.59\,{\cdot	10^{1}}^{+2.1}_{-2.2}$	\\
$440$	&	$7.42\,{\cdot	10^{-4}}^{+8.1}_{-8.0}$	&	$1.15\,{\cdot	10^{-6}}^{+10.7}_{-10.1}$	&	$1.26\,{\cdot	10^{-5}}^{+8.0}_{-7.8}$	&	$5.55\,{\cdot	10^{-1}}^{+1.6}_{-1.5}$	&	$2.60\,{\cdot	10^{-1}}^{+1.5}_{-1.4}$	&	$4.30\,{\cdot	10^{1}}^{+1.9}_{-2.0}$	\\
$450$	&	$7.18\,{\cdot	10^{-4}}^{+8.1}_{-8.1}$	&	$9.28\,{\cdot	10^{-7}}^{+10.9}_{-10.0}$	&	$1.15\,{\cdot	10^{-5}}^{+7.8}_{-7.6}$	&	$5.51\,{\cdot	10^{-1}}^{+1.5}_{-1.4}$	&	$2.59\,{\cdot	10^{-1}}^{+1.4}_{-1.3}$	&	$4.68\,{\cdot	10^{1}}^{+1.8}_{-1.8}$	\\
$460$	&	$6.94\,{\cdot	10^{-4}}^{+8.2}_{-8.1}$	&	$7.50\,{\cdot	10^{-7}}^{+10.8}_{-10.1}$	&	$1.05\,{\cdot	10^{-5}}^{+7.7}_{-7.5}$	&	$5.49\,{\cdot	10^{-1}}^{+1.4}_{-1.3}$	&	$2.59\,{\cdot	10^{-1}}^{+1.3}_{-1.3}$	&	$5.08\,{\cdot	10^{1}}^{+1.7}_{-1.7}$	\\
\hline
\end{tabular}
\end{center}
\end{table}

\clearpage
%\addcontentsline{toc}{section}{\appendixname}

\section{Tables of  cross sections}

Tables with the cross section values for the various production modes follow.

\begin{table}[ht]
\caption{ggF cross sections at the LHC at $7 \UTeV$ and corresponding scale and PDF+$\alphas$ uncertainties computed according to the PDF4LHC recommendation.}
\label{tab:YRHXS3_ggF_71}%
\renewcommand{\arraystretch}{0.80}%
\begin{center}
\begin{tabular}{lcccc}
\hline %\rule{0ex}{2.6ex}
 $ \MH \mbox{[GeV]}$  &   $ \sigma$ [pb] &  
  QCD Scale [\%]  & PDF+$\alphas$ [\%]  \\ 
 \hline 
  $80.0$ &  $ 36.59$  & ${+}  8.3 \;\;{-}{  8.8} $ &  $ {+}  7.8 \;\;{-}{  6.7}$  \\
  $81.0$ &  $ 35.71$  & ${+}  8.3 \;\;{-}{  8.8} $ &  $ {+}  7.8 \;\;{-}{  6.7}$  \\
  $82.0$ &  $ 34.86$  & ${+}  8.3 \;\;{-}{  8.8} $ &  $ {+}  7.8 \;\;{-}{  6.7}$  \\
  $83.0$ &  $ 34.04$  & ${+}  8.3 \;\;{-}{  8.8} $ &  $ {+}  7.8 \;\;{-}{  6.7}$  \\
  $84.0$ &  $ 33.25$  & ${+}  8.3 \;\;{-}{  8.8} $ &  $ {+}  7.8 \;\;{-}{  6.7}$  \\
  $85.0$ &  $ 32.49$  & ${+}  8.3 \;\;{-}{  8.8} $ &  $ {+}  7.8 \;\;{-}{  6.7}$  \\
  $86.0$ &  $ 31.75$  & ${+}  8.3 \;\;{-}{  8.8} $ &  $ {+}  7.8 \;\;{-}{  6.7}$  \\
\hline
\end{tabular}
\end{center}
\end{table}

\begin{table}
\caption{ ggF cross sections at the LHC at $7 \UTeV$ and corresponding scale and PDF+$\alphas$ uncertainties computed according to the PDF4LHC recommendation.}
\label{tab:YRHXS3_ggF_72}%
\renewcommand{\arraystretch}{0.80}%
\begin{center}
\begin{tabular}{lcccc}
\hline %\rule{0ex}{2.6ex}
 $ \MH \mbox{[GeV]}$ &  $  \sigma$ [pb] &  
  QCD Scale [\%]  & PDF+$\alphas$ [\%]  \\ 
 \hline 
$ 118.5 $&$ 16.85 $&${+} 7.3 \;\;{-} 7.9 $&${+} 7.6 \;\;{-} 7.0$ \\
$ 119.0 $&$ 16.71 $&${+} 7.3 \;\;{-} 7.9 $&${+} 7.6 \;\;{-} 7.0$ \\
$ 119.5 $&$ 16.57 $&${+} 7.2 \;\;{-} 7.9 $&${+} 7.6 \;\;{-} 7.0$ \\
$ 120.0 $&$ 16.43 $&${+} 7.2 \;\;{-} 7.9 $&${+} 7.6 \;\;{-} 7.0$ \\
$ 120.1 $&$ 16.40 $&${+} 7.2 \;\;{-} 7.9 $&${+} 7.6 \;\;{-} 7.0$ \\
$ 120.2 $&$ 16.37 $&${+} 7.2 \;\;{-} 7.9 $&${+} 7.6 \;\;{-} 7.0$ \\
$ 120.3 $&$ 16.35 $&${+} 7.2 \;\;{-} 7.9 $&${+} 7.6 \;\;{-} 7.0$ \\
\hline
\end{tabular}
\end{center}
\end{table}

\begin{table}
\caption{ ggF cross sections at the LHC at $7 \UTeV$ and corresponding scale and PDF+$\alphas$ uncertainties computed according to the PDF4LHC recommendation.}
\label{tab:YRHXS3_ggF_73}%
\renewcommand{\arraystretch}{0.80}%
\begin{center}
\begin{tabular}{lcccc}
\hline %\rule{0ex}{2.6ex}
 $ \MH \mbox{[GeV]}$ &  $ \sigma$ [pb] &  
  QCD Scale [\%]  & PDF+$\alphas$ [\%]  \\ 
 \hline 
$ 124.4 $&$ 15.28 $&${+} 7.1 \;\;{-} 7.8 $&${+} 7.6 \;\;{-} 7.1 $ \\
$ 124.5 $&$ 15.26 $&${+} 7.1 \;\;{-} 7.8 $&${+} 7.6 \;\;{-} 7.1 $ \\
$ 124.6 $&$ 15.23 $&${+} 7.1 \;\;{-} 7.8 $&${+} 7.6 \;\;{-} 7.1 $ \\
$ 124.7 $&$ 15.21 $&${+} 7.1 \;\;{-} 7.8 $&${+} 7.6 \;\;{-} 7.1 $ \\
$ 124.8 $&$ 15.18 $&${+} 7.1 \;\;{-} 7.8 $&${+} 7.6 \;\;{-} 7.1 $ \\
$ 124.9 $&$ 15.16 $&${+} 7.1 \;\;{-} 7.8 $&${+} 7.6 \;\;{-} 7.1 $ \\
$ 125.0 $&$ 15.13 $&${+} 7.1 \;\;{-} 7.8 $&${+} 7.6 \;\;{-} 7.1 $ \\
\hline
\end{tabular}
\end{center}
\end{table}

\begin{table}
\caption{ ggF cross sections at the LHC at $7 \UTeV$ and corresponding scale and PDF+$\alphas$ uncertainties computed according to the PDF4LHC recommendation.}
\label{tab:YRHXS3_ggF_74}%
\renewcommand{\arraystretch}{0.80}%
\begin{center}
\begin{tabular}{lcccc}
\hline %\rule{0ex}{2.6ex}
 $ \MH \mbox{[GeV]}$ &  $ \sigma$ [pb] &  
  QCD Scale [\%]  & PDF+$\alphas$ [\%]  \\ 
 \hline 
$ 129.1 $&$ 14.18 $&${+} 7.0\;\;{-} 7.7 $&${+} 7.6 \;\;{-} 7.2 $ \\
$ 129.2 $&$ 14.15 $&${+} 7.0\;\;{-} 7.7 $&${+} 7.6 \;\;{-} 7.2 $ \\
$ 129.3 $&$ 14.13 $&${+} 7.0\;\;{-} 7.7 $&${+} 7.6 \;\;{-} 7.2 $ \\
$ 129.4 $&$ 14.11 $&${+} 7.0\;\;{-} 7.7 $&${+} 7.6 \;\;{-} 7.2 $ \\
$ 129.5 $&$ 14.09 $&${+} 7.0\;\;{-} 7.7 $&${+} 7.6 \;\;{-} 7.2 $ \\
$ 129.6 $&$ 14.07 $&${+} 7.0\;\;{-} 7.7 $&${+} 7.6 \;\;{-} 7.2 $ \\
$ 129.7 $&$ 14.04 $&${+} 7.0\;\;{-} 7.7 $&${+} 7.6 \;\;{-} 7.2 $ \\
\hline
\end{tabular}
\end{center}
\end{table}

\begin{table}
\caption{ ggF cross sections at the LHC at $7 \UTeV$ and corresponding scale and PDF+$\alphas$ uncertainties computed according to the PDF4LHC recommendation.}
\label{tab:YRHXS3_ggF_75}%
\renewcommand{\arraystretch}{0.80}%
\begin{center}
\begin{tabular}{lcccc}
\hline %\rule{0ex}{2.6ex}
 $ \MH \mbox{[GeV]}$ &  $ \sigma$ [pb] &  
  QCD Scale [\%]  & PDF+$\alphas$ [\%]  \\ 
 \hline 
 $149.0$ &  $ 10.65$  & ${+}  6.6 \;\;{-}{  7.4} $ &  $ {+}  7.6 \;\;{-}{  7.5}$  \\
 $149.5$ &  $ 10.58$  & ${+}  6.6 \;\;{-}{  7.4} $ &  $ {+}  7.6 \;\;{-}{  7.5}$  \\
 $150.0$ &  $ 10.51$  & ${+}  6.6 \;\;{-}{  7.4} $ &  $ {+}  7.6 \;\;{-}{  7.5}$  \\
 $152.0$ &  $ 10.24$  & ${+}  6.6 \;\;{-}{  7.4} $ &  $ {+}  7.6 \;\;{-}{  7.5}$  \\
 $154.0$ &  $ 9.978$  & ${+}  6.5 \;\;{-}{  7.3} $ &  $ {+}  7.5 \;\;{-}{  7.5}$  \\
 $156.0$ &  $ 9.723$  & ${+}  6.5 \;\;{-}{  7.3} $ &  $ {+}  7.5 \;\;{-}{  7.5}$  \\
 $158.0$ &  $ 9.473$  & ${+}  6.4 \;\;{-}{  7.2} $ &  $ {+}  7.5 \;\;{-}{  7.6}$  \\
\hline
\end{tabular}
\end{center}
\end{table}

 \begin{table}
\caption{ ggF cross sections at the LHC at $7 \UTeV$ and corresponding scale and PDF+$\alphas$ uncertainties computed according to the PDF4LHC recommendation.}
\label{tab:YRHXS3_ggF_76}%
\renewcommand{\arraystretch}{0.80}%
\begin{center}
\begin{tabular}{lcccc}
\hline %\rule{0ex}{2.6ex}
 $ \MH \mbox{[GeV]}$ &  $  \sigma$ [pb] &  
  QCD Scale [\%]  & PDF+$\alphas$ [\%]  \\ 
 \hline 
  $232.0$ &  $ 4.094$  & ${+}  5.9 \;\;{-}{  6.5} $ &  $ {+}  7.7 \;\;{-}{  8.0}$  \\
  $234.0$ &  $ 4.036$  & ${+}  5.8 \;\;{-}{  6.5} $ &  $ {+}  7.7 \;\;{-}{  8.0}$  \\
  $236.0$ &  $ 3.971$  & ${+}  5.8 \;\;{-}{  6.4} $ &  $ {+}  7.7 \;\;{-}{  8.0}$  \\
  $238.0$ &  $ 3.904$  & ${+}  5.9 \;\;{-}{  6.4} $ &  $ {+}  7.7 \;\;{-}{  8.0}$  \\
  $240.0$ &  $ 3.835$  & ${+}  5.9 \;\;{-}{  6.4} $ &  $ {+}  7.7 \;\;{-}{  8.0}$  \\
  $242.0$ &  $ 3.771$  & ${+}  5.9 \;\;{-}{  6.4} $ &  $ {+}  7.7 \;\;{-}{  8.0}$  \\
  $244.0$ &  $ 3.709$  & ${+}  5.9 \;\;{-}{  6.4} $ &  $ {+}  7.7 \;\;{-}{  8.0}$  \\
\hline
\end{tabular}
\end{center}
\end{table}

\begin{table}
\caption{ggF cross sections at the LHC at $7 \UTeV$ and corresponding scale and PDF+$\alphas$ uncertainties computed according to the PDF4LHC recommendation.}
\label{tab:YRHXS3_ggF_77}%
\renewcommand{\arraystretch}{0.80}%
\begin{center}
\begin{tabular}{lcccc}
\hline %\rule{0ex}{2.6ex}
 $ \MH \mbox{[GeV]}$ &  $  \sigma$ [pb] &  
  QCD Scale [\%]  & PDF+$\alphas$ [\%]  \\ 
 \hline 
  $380.0$ &  $ 2.280$  & ${+}  5.9 \;\;{-}{  5.6} $ &  $ {+}  8.4 \;\;{-}{  8.6}$  \\
  $390.0$ &  $ 2.172$  & ${+}  5.9 \;\;{-}{  5.5} $ &  $ {+}  8.6 \;\;{-}{  8.6}$  \\
  $400.0$ &  $ 2.047$  & ${+}  5.9 \;\;{-}{  5.4} $ &  $ {+}  8.8 \;\;{-}{  8.6}$  \\
  $420.0$ &  $ 1.775$  & ${+}  5.9 \;\;{-}{  5.3} $ &  $ {+}  9.1 \;\;{-}{  8.6}$  \\
  $440.0$ &  $ 1.506$  & ${+}  5.9 \;\;{-}{  5.3} $ &  $ {+}  9.2 \;\;{-}{  8.7}$  \\
  $450.0$ &  $ 1.380$  & ${+}  5.9 \;\;{-}{  5.3} $ &  $ {+}  9.2 \;\;{-}{  8.7}$  \\
  $460.0$ &  $ 1.262$  & ${+}  5.9 \;\;{-}{  5.3} $ &  $ {+}  9.3 \;\;{-}{  8.7}$  \\
\hline
\end{tabular}
\end{center}
\end{table}

% ---

\begin{table}
\caption{ ggF cross sections at the LHC at $8 \UTeV$ and corresponding scale and PDF+$\alphas$ uncertainties computed according to the PDF4LHC recommendation.}
\label{tab:YRHXS3_ggF_81}%
\renewcommand{\arraystretch}{0.80}%
\begin{center}
\begin{tabular}{lcccc}
\hline %\rule{0ex}{2.6ex}
 $ \MH \mbox{[GeV]}$ &  $ \sigma$ [pb] &  
  QCD Scale [\%]  & PDF+$\alphas$ [\%]  \\ 
\hline 
$	80.0	$ &  $	45.37	$  & ${+} 	8.8	\;\;{-}{	9.2	} $ &  $ {+}	7.9	\;\;{-}{	6.7	}$  \\
$	81.0	$ &  $	44.31	$  & ${+} 	8.8	\;\;{-}{	9.1	} $ &  $ {+}	7.9	\;\;{-}{	6.7	}$  \\
$	82.0	$ &  $	43.28	$  & ${+} 	8.7	\;\;{-}{	9.1	} $ &  $ {+}	7.9	\;\;{-}{	6.6	}$  \\
$	83.0	$ &  $	42.29	$  & ${+} 	8.7	\;\;{-}{	9.0	} $ &  $ {+}	7.9	\;\;{-}{	6.6	}$  \\
$	84.0	$ &  $	41.33	$  & ${+} 	8.6	\;\;{-}{	9.0	} $ &  $ {+}	7.9	\;\;{-}{	6.6	}$  \\
$	85.0	$ &  $	40.41	$  & ${+} 	8.6	\;\;{-}{	9.0	} $ &  $ {+}	7.9	\;\;{-}{	6.6	}$  \\
$	86.0	$ &  $	39.52	$  & ${+} 	8.5	\;\;{-}{	8.9	} $ &  $ {+}	7.9	\;\;{-}{	6.6	}$  \\
\hline
\end{tabular}
\end{center}
\end{table}

 \begin{table}
\caption{ ggF cross sections at the LHC at $8 \UTeV$ and corresponding scale and PDF+$\alphas$ uncertainties computed according to the PDF4LHC recommendation.}
\label{tab:YRHXS3_ggF_82}%
\renewcommand{\arraystretch}{0.80}%
\begin{center}
\begin{tabular}{lcccc}
\hline %\rule{0ex}{2.6ex}
 $ \MH \mbox{[GeV]}$ &  $  \sigma$ [pb] &  
  QCD Scale [\%]  & PDF+$\alphas$ [\%]  \\ 
\hline 
$	118.5	$ &  $	21.38	$  & ${+} 	7.3	\;\;{-}{	8.0	} $ &  $ {+}	7.5	\;\;{-}{	6.8	}$  \\
$	119.0	$ &  $	21.20	$  & ${+} 	7.3	\;\;{-}{	8.0	} $ &  $ {+}	7.5	\;\;{-}{	6.8	}$  \\
$	119.5	$ &  $	21.03	$  & ${+} 	7.3	\;\;{-}{	8.0	} $ &  $ {+}	7.5	\;\;{-}{	6.8	}$  \\
$	120.0	$ &  $	20.86	$  & ${+} 	7.3	\;\;{-}{	7.9	} $ &  $ {+}	7.5	\;\;{-}{	6.9	}$  \\
$	120.1	$ &  $	20.83	$  & ${+} 	7.3	\;\;{-}{	7.9	} $ &  $ {+}	7.5	\;\;{-}{	6.9	}$  \\
$	120.2	$ &  $	20.80	$  & ${+} 	7.3	\;\;{-}{	7.9	} $ &  $ {+}	7.5	\;\;{-}{	6.9	}$  \\
$	120.3	$ &  $	20.76	$  & ${+} 	7.3	\;\;{-}{	7.9	} $ &  $ {+}	7.5	\;\;{-}{	6.9	}$  \\
\hline
\end{tabular}
\end{center}
\end{table}

 \begin{table}
\caption{ ggF cross sections at the LHC at $8 \UTeV$ and corresponding scale and PDF+$\alphas$ uncertainties computed according to the PDF4LHC recommendation.}
\label{tab:YRHXS3_ggF_83}%
\renewcommand{\arraystretch}{0.80}%
\begin{center}
\begin{tabular}{lcccc}
\hline %\rule{0ex}{2.6ex}
 $ \MH \mbox{[GeV]}$ &  $ \sigma$ [pb] &  
  QCD Scale [\%]  & PDF+$\alphas$ [\%]  \\ 
\hline
$	124.4	$ &  $	19.45	$  & ${+} 	7.2	\;\;{-}{	7.9	} $ &  $ {+}	7.5	\;\;{-}{	6.9	}$  \\
$	124.5	$ &  $	19.42	$  & ${+} 	7.2	\;\;{-}{	7.9	} $ &  $ {+}	7.5	\;\;{-}{	6.9	}$  \\
$	124.6	$ &  $	19.39	$  & ${+} 	7.2	\;\;{-}{	7.9	} $ &  $ {+}	7.5	\;\;{-}{	6.9	}$  \\
$	124.7	$ &  $	19.36	$  & ${+} 	7.2	\;\;{-}{	7.9	} $ &  $ {+}	7.5	\;\;{-}{	6.9	}$  \\
$	124.8	$ &  $	19.33	$  & ${+} 	7.2	\;\;{-}{	7.8	} $ &  $ {+}	7.5	\;\;{-}{	6.9	}$  \\
$	124.9	$ &  $	19.30	$  & ${+} 	7.2	\;\;{-}{	7.8	} $ &  $ {+}	7.5	\;\;{-}{	6.9	}$  \\
$	125.0	$ &  $	19.27	$  & ${+} 	7.2	\;\;{-}{	7.8	} $ &  $ {+}	7.5	\;\;{-}{	6.9	}$  \\
\hline
\end{tabular}
\end{center}
\end{table}

 \begin{table}
\caption{ ggF cross sections at the LHC at $8 \UTeV$ and corresponding scale and PDF+$\alphas$ uncertainties computed according to the PDF4LHC recommendation.}
\label{tab:YRHXS3_ggF_84}%
\renewcommand{\arraystretch}{0.80}%
\begin{center}
\begin{tabular}{lcccc}
\hline %\rule{0ex}{2.6ex}
 $ \MH \mbox{[GeV]}$ &  $ \sigma$ [pb] &  
  QCD Scale [\%]  & PDF+$\alphas$ [\%]  \\ 
\hline
$	129.1	$ &  $	18.09	$  & ${+} 	7.1	\;\;{-}{	7.8	} $ &  $ {+}	7.5	\;\;{-}{	6.9	}$  \\
$	129.2	$ &  $	18.06	$  & ${+} 	7.1	\;\;{-}{	7.8	} $ &  $ {+}	7.5	\;\;{-}{	6.9	}$  \\
$	129.3	$ &  $	18.04	$  & ${+} 	7.1	\;\;{-}{	7.8	} $ &  $ {+}	7.5	\;\;{-}{	6.9	}$  \\
$	129.4	$ &  $	18.01	$  & ${+} 	7.1	\;\;{-}{	7.8	} $ &  $ {+}	7.5	\;\;{-}{	6.9	}$  \\
$	129.5	$ &  $	17.98	$  & ${+} 	7.1	\;\;{-}{	7.8	} $ &  $ {+}	7.5	\;\;{-}{	6.9	}$  \\
$	129.6	$ &  $	17.95	$  & ${+} 	7.1	\;\;{-}{	7.8	} $ &  $ {+}	7.5	\;\;{-}{	6.9	}$  \\
$	129.7	$ &  $	17.93	$  & ${+} 	7.1	\;\;{-}{	7.8	} $ &  $ {+}	7.5	\;\;{-}{	6.9	}$  \\
\hline
\end{tabular}
\end{center}
\end{table}

\begin{table}
\caption{ ggF cross sections at the LHC at $8 \UTeV$ and corresponding scale and PDF+$\alphas$ uncertainties computed according to the PDF4LHC recommendation.}
\label{tab:YRHXS3_ggF_85}%
\renewcommand{\arraystretch}{0.80}%
\begin{center}
\begin{tabular}{lcccc}
\hline %\rule{0ex}{2.6ex}
 $ \MH \mbox{[GeV]}$ &  $ \sigma$ [pb] &  
  QCD Scale [\%]  & PDF+$\alphas$ [\%]  \\ 
\hline 
$	149.0	$ &  $	13.73	$  & ${+} 	6.7	\;\;{-}{	7.4	} $ &  $ {+}	7.4	\;\;{-}{	7.0	}$  \\
$	149.5	$ &  $	13.64	$  & ${+} 	6.7	\;\;{-}{	7.4	} $ &  $ {+}	7.4	\;\;{-}{	7.0	}$  \\
$	150.0	$ &  $	13.55	$  & ${+} 	6.7	\;\;{-}{	7.4	} $ &  $ {+}	7.4	\;\;{-}{	7.0	}$  \\
$	152.0	$ &  $	13.22	$  & ${+} 	6.6	\;\;{-}{	7.4	} $ &  $ {+}	7.4	\;\;{-}{	7.1	}$  \\
$	154.0	$ &  $	12.89	$  & ${+} 	6.6	\;\;{-}{	7.3	} $ &  $ {+}	7.5	\;\;{-}{	7.1	}$  \\
$	156.0	$ &  $	12.58	$  & ${+} 	6.6	\;\;{-}{	7.3	} $ &  $ {+}	7.5	\;\;{-}{	7.1	}$  \\
$	158.0	$ &  $	12.27	$  & ${+} 	6.5	\;\;{-}{	7.3	} $ &  $ {+}	7.5	\;\;{-}{	7.1	}$  \\
\hline
\end{tabular}
\end{center}
\end{table}

 \begin{table}
\caption{ ggF cross sections at the LHC at $8 \UTeV$ and corresponding scale and PDF+$\alphas$ uncertainties computed according to the PDF4LHC recommendation.}
\label{tab:YRHXS3_ggF_86}%
\renewcommand{\arraystretch}{0.80}%
\begin{center}
\begin{tabular}{lcccc}
\hline %\rule{0ex}{2.6ex}
 $ \MH \mbox{[GeV]}$ &  $  \sigma$ [pb] &  
  QCD Scale [\%]  & PDF+$\alphas$ [\%]  \\ 
\hline 
$	232.0	$ &  $	5.487	$  & ${+} 	5.9	\;\;{-}{	6.5	} $ &  $ {+}	7.4	\;\;{-}{	7.7	}$  \\
$	234.0	$ &  $	5.413	$  & ${+} 	5.9	\;\;{-}{	6.5	} $ &  $ {+}	7.3	\;\;{-}{	7.7	}$  \\
$	236.0	$ &  $	5.327	$  & ${+} 	5.9	\;\;{-}{	6.5	} $ &  $ {+}	7.3	\;\;{-}{	7.7	}$  \\
$	238.0	$ &  $	5.247	$  & ${+} 	5.9	\;\;{-}{	6.5	} $ &  $ {+}	7.3	\;\;{-}{	7.7	}$  \\
$	240.0	$ &  $	5.159	$  & ${+} 	5.9	\;\;{-}{	6.4	} $ &  $ {+}	7.3	\;\;{-}{	7.7	}$  \\
$	242.0	$ &  $	5.078	$  & ${+} 	5.9	\;\;{-}{	6.4	} $ &  $ {+}	7.3	\;\;{-}{	7.7	}$  \\
$	244.0	$ &  $	4.999	$  & ${+} 	5.9	\;\;{-}{	6.4	} $ &  $ {+}	7.3	\;\;{-}{	7.7	}$  \\
\hline 
\end{tabular}
\end{center}
\end{table}

 \begin{table}
\caption{ ggF cross sections at the LHC at $8 \UTeV$ and corresponding scale and PDF+$\alphas$ uncertainties computed according to the PDF4LHC recommendation.}
\label{tab:YRHXS3_ggF_87}%
\renewcommand{\arraystretch}{0.80}%
\begin{center}
\begin{tabular}{lcccc}
\hline %\rule{0ex}{2.6ex}
 $ \MH \mbox{[GeV]}$ &  $  \sigma$ [pb] &  
  QCD Scale [\%]  & PDF+$\alphas$ [\%]  \\ 
\hline 
$	380.0	$ &  $	3.231	$  & ${+} 	5.8	\;\;{-}{	5.6	} $ &  $ {+}	8.1	\;\;{-}{	8.2	}$  \\
$	390.0	$ &  $	3.089	$  & ${+} 	5.8	\;\;{-}{	5.5	} $ &  $ {+}	8.2	\;\;{-}{	8.2	}$  \\
$	400.0	$ &  $	2.921	$  & ${+} 	5.8	\;\;{-}{	5.4	} $ &  $ {+}	8.2	\;\;{-}{	8.2	}$  \\
$	420.0	$ &  $	2.550	$  & ${+} 	5.8	\;\;{-}{	5.3	} $ &  $ {+}	8.3	\;\;{-}{	8.3	}$  \\
$	440.0	$ &  $	2.178	$  & ${+} 	5.8	\;\;{-}{	5.3	} $ &  $ {+}	8.5	\;\;{-}{	8.4	}$  \\
$	450.0	$ &  $	2.002	$  & ${+} 	5.8	\;\;{-}{	5.2	} $ &  $ {+}	8.6	\;\;{-}{	8.4	}$  \\
$	460.0	$ &  $	1.837	$  & ${+} 	5.8	\;\;{-}{	5.2	} $ &  $ {+}	8.7	\;\;{-}{	8.4	}$  \\
\hline
\end{tabular}
\end{center}
\end{table}

\clearpage
%\addcontentsline{toc}{section}{\appendixname}

%\section{Tables of  cross sections}

%Tables with the cross section values for the various production modes follow.

\begin{table}
\caption{VBF cross sections at $7\UTeV$. Central values at NNLO QCD,
  NNLO QCD + NLO EW, relative EW corrections, and uncertainties from 
  PDF+$\alphas$ (according to the PDF4LHC prescription) and from QCD scale.}
%  variation%(calculated for a subset of mass points)
\label{tab:YRHXS3_7VBF_NNLO1}%
\renewcommand{\arraystretch}{0.80}%
\setlength{\tabcolsep}{1.5ex}%
\begin{center}
\begin{tabular}{lcccccc}
\hline  \rule{0ex}{2.6ex}
 $ \MH $  &   $ \sigma^{\mathrm{NNLO}}_{\mathrm{VBF@NNLO}} $ &  $ \sigma^{\mathrm{NNLO}+\mathrm{EW}}  $  & $ \delta^{\mathrm{EW}}_{\mathrm{HAWK}}  $  &  PDF4LHC & QCD Scale \\ 
  $\mbox{[GeV]}$   &   [fb]  &  [fb]   & [\%]   &  [\%] &  [\%] \\ 
 \hline 
  $80.0$  &  $  2002$  &  $  1914$  &  ${-}4.4$  &  $ {+}  2.1 \;\; {-}{2.1} $  &  $ {+}   0.6 \;\; {-}{0.2} $  \\
  $81.0$  &  $  1981$  &  $  1894$  &  ${-}4.4$  &  $ {+}  2.1 \;\; {-}{2.1} $  &  $ {+}   0.6 \;\; {-}{0.2} $  \\
  $82.0$  &  $  1957$  &  $  1871$  &  ${-}4.4$  &  $ {+}  2.1 \;\; {-}{2.1} $  &  $ {+}   0.6 \;\; {-}{0.2} $  \\
  $83.0$  &  $  1938$  &  $  1853$  &  ${-}4.4$  &  $ {+}  2.1 \;\; {-}{2.1} $  &  $ {+}   0.6 \;\; {-}{0.2} $  \\
  $84.0$  &  $  1916$  &  $  1833$  &  ${-}4.4$  &  $ {+}  2.1 \;\; {-}{2.1} $  &  $ {+}   0.6 \;\; {-}{0.2} $  \\
  $85.0$  &  $  1897$  &  $  1814$  &  ${-}4.4$  &  $ {+}  2.1 \;\; {-}{2.1} $  &  $ {+}   0.6 \;\; {-}{0.2} $  \\
\hline
\end{tabular}
\end{center}
\end{table}

\begin{table}
\caption{VBF cross sections at $7\UTeV$. Central values at NNLO QCD,
  NNLO QCD + NLO EW, relative EW corrections, and uncertainties from 
  PDF+$\alphas$ (according to the PDF4LHC prescription) and from QCD scale.}
%  variation%(calculated for a subset of mass points)
\label{tab:YRHXS3_7VBF_NNLO2}%
\renewcommand{\arraystretch}{0.80}%
\setlength{\tabcolsep}{1.5ex}%
\begin{center}
\begin{tabular}{lcccccc}
\hline  \rule{0ex}{2.6ex}
 $ \MH $  &   $ \sigma^{\mathrm{NNLO}}_{\mathrm{VBF@NNLO}} $ &  $ \sigma^{\mathrm{NNLO}+\mathrm{EW}}  $  & $ \delta^{\mathrm{EW}}_{\mathrm{HAWK}}  $  &  PDF4LHC & QCD Scale \\ 
  $\mbox{[GeV]}$   &   [fb]  &  [fb]   & [\%]   &  [\%] &  [\%] \\ 
 \hline 
 $118.5$  &  $  1356$  &  $  1297$  &  ${-}4.4$  &  $ {+}         2.4 \;\; {-}{2.1} $  &  $ {+}   0.3 \;\; {-}{0.3} $  \\
 $119.0$  &  $  1351$  &  $  1292$  &  ${-}4.4$  &  $ {+}         2.4 \;\; {-}{2.1} $  &  $ {+}   0.3 \;\; {-}{0.4} $  \\
 $119.5$  &  $  1345$  &  $  1286$  &  ${-}4.3$  &  $ {+}         2.4 \;\; {-}{2.1} $  &  $ {+}   0.3 \;\; {-}{0.4} $  \\
 $120.0$  &  $  1337$  &  $  1279$  &  ${-}4.3$  &  $ {+}         2.4 \;\; {-}{2.1} $  &  $ {+}   0.3 \;\; {-}{0.4} $  \\
 $120.1$  &  $  1338$  &  $  1280$  &  ${-}4.3$  &  $ {+}         2.4 \;\; {-}{2.1} $  &  $ {+}   0.3 \;\; {-}{0.4} $  \\
 $120.2$  &  $  1337$  &  $  1279$  &  ${-}4.3$  &  $ {+}         2.4 \;\; {-}{2.1} $  &  $ {+}   0.3 \;\; {-}{0.4} $  \\
\hline
\end{tabular}
\end{center}
\end{table}

\begin{table}
\caption{VBF cross sections at $7\UTeV$. Central values at NNLO QCD,
  NNLO QCD + NLO EW, relative EW corrections, and uncertainties from 
  PDF+$\alphas$ (according to the PDF4LHC prescription) and from QCD scale.}
%  variation%(calculated for a subset of mass points)
\label{tab:YRHXS3_7VBF_NNLO3}%
\renewcommand{\arraystretch}{0.80}%
\setlength{\tabcolsep}{1.5ex}%
\begin{center}
\begin{tabular}{lcccccc}
\hline  \rule{0ex}{2.6ex}
 $ \MH $  &   $ \sigma^{\mathrm{NNLO}}_{\mathrm{VBF@NNLO}} $ &  $ \sigma^{\mathrm{NNLO}+\mathrm{EW}}  $  & $ \delta^{\mathrm{EW}}_{\mathrm{HAWK}}  $  &  PDF4LHC & QCD Scale \\ 
  $\mbox{[GeV]}$   &   [fb]  &  [fb]   & [\%]   &  [\%] &  [\%] \\ 
 \hline 
 $124.4$  &  $  1283$  &  $  1228$  &  ${-}4.3$  &  $ {+}         2.5 \;\; {-}{2.1} $  &  $ {+}   0.3 \;\; {-}{0.3} $  \\
 $124.5$  &  $  1282$  &  $  1227$  &  ${-}4.3$  &  $ {+}         2.5 \;\; {-}{2.1} $  &  $ {+}   0.3 \;\; {-}{0.3} $  \\
 $124.6$  &  $  1283$  &  $  1228$  &  ${-}4.3$  &  $ {+}         2.5 \;\; {-}{2.1} $  &  $ {+}   0.3 \;\; {-}{0.3} $  \\
 $124.7$  &  $  1280$  &  $  1225$  &  ${-}4.3$  &  $ {+}         2.5 \;\; {-}{2.1} $  &  $ {+}   0.3 \;\; {-}{0.3} $  \\
 $124.8$  &  $  1281$  &  $  1225$  &  ${-}4.3$  &  $ {+}         2.5 \;\; {-}{2.1} $  &  $ {+}   0.3 \;\; {-}{0.3} $  \\
 $124.9$  &  $  1281$  &  $  1226$  &  ${-}4.3$  &  $ {+}         2.5 \;\; {-}{2.1} $  &  $ {+}   0.3 \;\; {-}{0.3} $  \\
\hline
\end{tabular}
\end{center}
\end{table}

\begin{table}
\caption{VBF cross sections at $7\UTeV$. Central values at NNLO QCD,
  NNLO QCD + NLO EW, relative EW corrections, and uncertainties from 
  PDF+$\alphas$ (according to the PDF4LHC prescription) and from QCD scale.}
%  variation%(calculated for a subset of mass points)
\label{tab:YRHXS3_7VBF_NNLO4}%
\renewcommand{\arraystretch}{0.80}%
\setlength{\tabcolsep}{1.5ex}%
\begin{center}
\begin{tabular}{lcccccc}
\hline  \rule{0ex}{2.6ex}
 $ \MH $  &   $ \sigma^{\mathrm{NNLO}}_{\mathrm{VBF@NNLO}} $ &  $ \sigma^{\mathrm{NNLO}+\mathrm{EW}}  $  & $ \delta^{\mathrm{EW}}_{\mathrm{HAWK}}  $  &  PDF4LHC & QCD Scale \\ 
  $\mbox{[GeV]}$   &   [fb]  &  [fb]   & [\%]   &  [\%] &  [\%] \\ 
 \hline 
 $129.1$  &  $  1230$  &  $  1178$  &  ${-}4.3$  &  $ {+}         2.5 \;\; {-}{2.1} $  &  $ {+}   0.3 \;\; {-}{0.2} $  \\
 $129.2$  &  $  1229$  &  $  1177$  &  ${-}4.3$  &  $ {+}         2.5 \;\; {-}{2.1} $  &  $ {+}   0.3 \;\; {-}{0.2} $  \\
 $129.3$  &  $  1227$  &  $  1175$  &  ${-}4.3$  &  $ {+}         2.5 \;\; {-}{2.1} $  &  $ {+}   0.3 \;\; {-}{0.2} $  \\
 $129.4$  &  $  1226$  &  $  1174$  &  ${-}4.3$  &  $ {+}         2.5 \;\; {-}{2.1} $  &  $ {+}   0.3 \;\; {-}{0.2} $  \\
 $129.5$  &  $  1225$  &  $  1173$  &  ${-}4.3$  &  $ {+}         2.5 \;\; {-}{2.1} $  &  $ {+}   0.3 \;\; {-}{0.2} $  \\
 $129.6$  &  $  1225$  &  $  1173$  &  ${-}4.3$  &  $ {+}         2.5 \;\; {-}{2.1} $  &  $ {+}   0.3 \;\; {-}{0.2} $  \\
\hline
\end{tabular}
\end{center}
\end{table}

\begin{table}
\caption{VBF cross sections at $7\UTeV$. Central values at NNLO QCD,
  NNLO QCD + NLO EW, relative EW corrections, and uncertainties from 
  PDF+$\alphas$ (according to the PDF4LHC prescription) and from QCD scale.}
%  variation%(calculated for a subset of mass points)
\label{tab:YRHXS3_7VBF_NNLO5}%
\renewcommand{\arraystretch}{0.80}%
\setlength{\tabcolsep}{1.5ex}%
\begin{center}
\begin{tabular}{lcccccc}
\hline  \rule{0ex}{2.6ex}
 $ \MH $  &  $ \sigma^{\mathrm{NNLO}}_{\mathrm{VBF@NNLO}} $ &  $ \sigma^{\mathrm{NNLO}+\mathrm{EW}}  $  & $ \delta^{\mathrm{EW}}_{\mathrm{HAWK}}  $  &  PDF4LHC & QCD Scale \\ 
  $\mbox{[GeV]}$   &   [fb]  &  [fb]   & [\%]   &  [\%] &  [\%] \\ 
 \hline 
 $149.0$  &  $  1032$  &  $   988$  &  ${-}4.2$  &  $ {+}  2.7 \;\; {-}{2.1} $  &  $ {+}   0.2 \;\; {-}{0.1} $  \\
 $149.5$  &  $  1029$  &  $   985$  &  ${-}4.2$  &  $ {+}         2.7 \;\; {-}{2.1} $  &  $ {+}   0.2 \;\; {-}{0.1} $  \\
 $150.0$  &  $  1024$  &  $   980$  &  ${-}4.2$  &  $ {+}  2.7 \;\; {-}{2.1} $  &  $ {+}   0.2 \;\; {-}{0.1} $  \\
%$150.5$  &  $  1020$  &  $   977$  &  ${-}4.2$  &  $ {+}         2.7 \;\; {-}{2.1} $  &  $ {+}   0.2 \;\; {-}{0.1} $  \\
%$151.0$  &  $  1015$  &  $   972$  &  ${-}4.2$  &  $ {+}  2.7 \;\; {-}{2.1} $  &  $ {+}   0.2 \;\; {-}{0.1} $  \\
 $152.0$  &  $  1007$  &  $   964$  &  ${-}4.3$  &  $ {+}  2.7 \;\; {-}{2.1} $  &  $ {+}   0.2 \;\; {-}{0.1} $  \\
%$153.0$  &  $ 999.0$  &  $ 956.1$  &  ${-}4.3$  &  $ {+}  2.8 \;\; {-}{2.1} $  &  $ {+}   0.3 \;\; {-}{0.0} $  \\
 $154.0$  &  $ 991.4$  &  $ 948.7$  &  ${-}4.3$  &  $ {+}  2.8 \;\; {-}{2.1} $  &  $ {+}   0.3 \;\; {-}{0.0} $  \\
%$155.0$  &  $ 984.2$  &  $ 941.5$  &  ${-}4.3$  &  $ {+}   2.8 \;\; {-}{2.1} $  &  $ {+}   0.3 \;\; {-}{0.0} $  \\
 $156.0$  &  $ 976.5$  &  $ 933.9$  &  ${-}4.4$  &  $ {+}   2.8 \;\; {-}{2.1} $  &  $ {+}   0.3 \;\; {-}{0.0} $  \\
\hline
\end{tabular}
\end{center}
\end{table}

\begin{table}
\caption{VBF cross sections at $7\UTeV$. Central values at NNLO QCD,
  NNLO QCD + NLO EW, relative EW corrections, and uncertainties from 
  PDF+$\alphas$ (according to the PDF4LHC prescription) and from QCD scale.}
%  variation%(calculated for a subset of mass points)
\label{tab:YRHXS3_7VBF_NNLO6}%
\renewcommand{\arraystretch}{0.80}%
\setlength{\tabcolsep}{1.5ex}%
\begin{center}
\begin{tabular}{lcccccc}
\hline  \rule{0ex}{2.6ex}
 $ \MH $  &   $ \sigma^{\mathrm{NNLO}}_{\mathrm{VBF@NNLO}} $ &  $ \sigma^{\mathrm{NNLO}+\mathrm{EW}}  $  & $ \delta^{\mathrm{EW}}_{\mathrm{HAWK}}  $  &  PDF4LHC & QCD Scale \\ 
  $\mbox{[GeV]}$   &   [fb]  &  [fb]   & [\%]   &  [\%] &  [\%] \\ 
 \hline 
$ 232.0$  &  $ 547.5$  &  $ 526.6$  &  ${-}3.8$  &  $ {+}   3.6 \;\; {-}{2.0} $  &  $ {+}   0.1 \;\; {-}{0.4} $  \\
$ 234.0$  &  $ 539.8$  &  $ 519.0$  &  ${-}3.9$  &  $ {+}   3.6 \;\; {-}{2.0} $  &  $ {+}   0.1 \;\; {-}{0.4} $  \\
$ 236.0$  &  $ 532.1$  &  $ 511.4$  &  ${-}3.9$  &  $ {+}   3.7 \;\; {-}{2.0} $  &  $ {+}   0.1 \;\; {-}{0.5} $  \\
$ 238.0$  &  $ 524.4$  &  $ 503.8$  &  ${-}3.9$  &  $ {+}   3.7 \;\; {-}{2.0} $  &  $ {+}   0.1 \;\; {-}{0.5} $  \\
$ 240.0$  &  $ 516.3$  &  $ 495.9$  &  ${-}4.0$  &  $ {+}   3.7 \;\; {-}{2.0} $  &  $ {+}   0.1 \;\; {-}{0.5} $  \\
 $242.0$  &  $ 508.5$  &  $ 488.2$  &  ${-}4.0$  &  $ {+}   3.7 \;\; {-}{2.0} $  &  $ {+}   0.1 \;\; {-}{0.5} $  \\
\hline
\end{tabular}
\end{center}
\end{table}

\begin{table}
\caption{VBF cross sections at $7\UTeV$. Central values at NNLO QCD,
  NNLO QCD + NLO EW, relative EW corrections, and uncertainties from 
  PDF+$\alphas$ (according to the PDF4LHC prescription) and from QCD scale.}
%  variation%(calculated for a subset of mass points)
\label{tab:YRHXS3_7VBF_NNLO7}%
\renewcommand{\arraystretch}{0.80}%
\setlength{\tabcolsep}{1.5ex}%
\begin{center}
\begin{tabular}{lcccccc}
\hline  \rule{0ex}{2.6ex}
 $ \MH $  &   $ \sigma^{\mathrm{NNLO}}_{\mathrm{VBF@NNLO}} $ &  $ \sigma^{\mathrm{NNLO}+\mathrm{EW}}  $  & $ \delta^{\mathrm{EW}}_{\mathrm{HAWK}}  $  &  PDF4LHC & QCD Scale \\ 
  $\mbox{[GeV]}$   &   [fb]  &  [fb]   & [\%]   &  [\%] &  [\%] \\ 
 \hline 
$ 380.0$ & $203.0$ & $194.9$ & ${-}4.0 $ & $ {+} 5.2 \;\; {-} 1.9 $ & ${+} 0.4 \;\; {-} 1.1 $ \\
$ 390.0$ & $192.4$ & $185.9$ & ${-}3.4$ & $ {+} 5.3 \;\; {-} 1.9 $ & ${+} 0.4 \;\; {-} 1.1 $ \\
$ 400.0$ & $182.3$ & $177.2$ & ${-}2.8$ & $ {+} 5.5 \;\; {-} 1.9 $ & ${+} 0.4 \;\; {-} 1.2 $ \\
$ 420.0$ & $163.3$ & $160.1$ & ${-}2.0$ & $ {+} 5.7 \;\; {-} 1.9 $ & $ {+}0.5 \;\; {-} 1.2 $ \\
$ 440.0$ & $145.9$ & $144.2$ & ${-}1.1$ & $ {+} 5.9 \;\; {-} 1.8 $ & $ {+}0.6 \;\; {-} 1.3 $ \\
$ 450.0$ & $137.8$ & $136.8$ & ${-}0.7$ & $ {+} 6.0 \;\; {-} 1.8 $ & $ {+}0.6 \;\; {-} 1.3 $ \\
\hline
\end{tabular}
\end{center}
\end{table}

% ---

\begin{table}
\caption{VBF cross sections at $8\UTeV$. Central values at NNLO QCD,
  NNLO QCD + NLO EW, relative EW corrections, and uncertainties from 
  PDF+$\alphas$ (according to the PDF4LHC prescription) and from QCD scale.}
%  variation%(calculated for a subset of mass points)
\label{tab:YRHXS3_VBF_NNLO1}%
\renewcommand{\arraystretch}{0.80}%
\setlength{\tabcolsep}{1.5ex}%
\begin{center}
\begin{tabular}{lcccccc}
\hline \rule{0ex}{2.6ex}
 $ \MH $  &   $ \sigma^{\mathrm{NNLO}}_{\mathrm{VBF@NNLO}} $ &  $ \sigma^{\mathrm{NNLO}+\mathrm{EW}}  $  & $ \delta^{\mathrm{EW}}_{\mathrm{HAWK}}  $  &  PDF4LHC & QCD Scale \\ 
  $\mbox{[GeV]}$   &   [fb]  &  [fb]   & [\%]   &  [\%] &  [\%] \\ 
 \hline 
$ 80.0 $  &  $   2542 $  &  $  2424$  &  ${-}4.6$  & $ {+}  2.7  \;\; {-}{3.0} $  &  $ {+}   0.2 \;\; {-}{0.3} $  \\
$ 81.0 $  &  $   2515 $  &  $  2399$  &  ${-}4.6$  & $ {+}  2.6  \;\; {-}{3.0} $  &  $ {+}   0.4 \;\; {-}{0.3} $  \\
$ 82.0 $  &  $   2479 $  &  $  2364$  &  ${-}4.6$  & $ {+}  2.6  \;\; {-}{2.8} $  &  $ {+}   0.3 \;\; {-}{0.3} $  \\
$ 83.0 $  &  $   2461 $  &  $  2346$  &  ${-}4.6$  & $ {+}  2.6  \;\; {-}{2.8} $  &  $ {+}   0.4 \;\; {-}{0.2} $  \\
$ 84.0 $  &  $   2439 $  &  $  2326$  &  ${-}4.6$  & $ {+}  2.6  \;\; {-}{2.8} $  &  $ {+}   0.3 \;\; {-}{0.2} $  \\
$ 85.0 $  &  $   2412 $  &  $  2300$  &  ${-}4.6$  & $ {+}  2.6  \;\; {-}{2.8} $  &  $ {+}   0.3 \;\; {-}{0.2} $  \\
\hline     
\end{tabular}
\end{center}
\end{table}

\begin{table}
\caption{VBF cross sections at $8\UTeV$. Central values at NNLO QCD,
  NNLO QCD + NLO EW, relative EW corrections, and uncertainties from 
  PDF+$\alphas$ (according to the PDF4LHC prescription) and from QCD scale.}
%  variation%(calculated for a subset of mass points)
\label{tab:YRHXS3_VBF_NNLO2}%
\renewcommand{\arraystretch}{0.80}%
\setlength{\tabcolsep}{1.5ex}%
\begin{center}
\begin{tabular}{lcccccc}
\hline  \rule{0ex}{2.6ex}
 $ \MH $  &   $ \sigma^{\mathrm{NNLO}}_{\mathrm{VBF@NNLO}} $ &  $ \sigma^{\mathrm{NNLO}+\mathrm{EW}}  $  & $ \delta^{\mathrm{EW}}_{\mathrm{HAWK}}  $  &  PDF4LHC & QCD Scale \\ 
  $\mbox{[GeV]}$   &   [fb]  &  [fb]   & [\%]   &  [\%] &  [\%] \\ 
 \hline 
$118.5 $  &  $   1755 $  &  $  1675$  &  ${-}4.6$  &  $ {+}  2.6  \;\; {-}{2.8} $  &  $ {+}   0.2 \;\; {-}{0.2} $  \\
$119.0 $  &  $   1745 $  &  $  1666$  &  ${-}4.6$  &  $ {+}  2.6  \;\; {-}{2.8} $  &  $ {+}   0.3 \;\; {-}{0.2} $  \\
$119.5 $  &  $   1738 $  &  $  1659$  &  ${-}4.6$  &  $ {+}  2.6  \;\; {-}{2.8} $  &  $ {+}   0.2 \;\; {-}{0.2} $  \\
$120.0 $  &  $   1728 $  &  $  1649$  &  ${-}4.6$  &  $ {+}  2.6  \;\; {-}{2.8} $  &  $ {+}   0.2 \;\; {-}{0.2} $  \\
$ 120.1$ & $1728$ & $1650$ & ${-}4.5$ & $ {+} 2.6 \;\; {-} 2.8 $ & $ {+} 0.2 \;\; {-} 0.2 $ \\
$ 120.2$ & $1726$ & $1648$ & ${-}4.5$ & $ {+} 2.6 \;\; {-} 2.8 $ & $ {+} 0.2 \;\; {-} 0.2 $ \\
\hline
\end{tabular}
\end{center}
\end{table}

\begin{table}
\caption{VBF cross sections at $8\UTeV$. Central values at NNLO QCD,
  NNLO QCD + NLO EW, relative EW corrections, and uncertainties from 
  PDF+$\alphas$ (according to the PDF4LHC prescription) and from QCD scale.}
%  variation%(calculated for a subset of mass points)
\label{tab:YRHXS3_VBF_NNLO3}%
\renewcommand{\arraystretch}{0.80}%
\setlength{\tabcolsep}{1.5ex}%
\begin{center}
\begin{tabular}{lcccccc}
\hline  \rule{0ex}{2.6ex}
 $ \MH $  &   $ \sigma^{\mathrm{NNLO}}_{\mathrm{VBF@NNLO}} $ &  $ \sigma^{\mathrm{NNLO}+\mathrm{EW}}  $  & $ \delta^{\mathrm{EW}}_{\mathrm{HAWK}}  $  &  PDF4LHC & QCD Scale \\ 
  $\mbox{[GeV]}$   &   [fb]  &  [fb]   & [\%]   &  [\%] &  [\%] \\ 
 \hline 
$ 124.4$ & $1663$ & $1589$ & ${-}4.5$ & $ {+} 2.6 \;\; {-} 2.8 $ & $ {+} 0.2 \;\; {-} 0.2 $ \\
$ 124.5$ & $1662$ & $1587$ & ${-}4.5$ & $ {+} 2.6 \;\; {-} 2.8 $ & $ {+} 0.2 \;\; {-} 0.2 $ \\
$ 124.6$ & $1661$ & $1586$ & ${-}4.5$ & $ {+} 2.6 \;\; {-} 2.8 $ & $ {+} 0.2 \;\; {-} 0.2 $ \\
$ 124.7$ & $1665$ & $1590$ & ${-}4.5$ & $ {+} 2.6 \;\; {-} 2.8 $ & $ {+} 0.2 \;\; {-} 0.2 $ \\
$ 124.8$ & $1659$ & $1584$ & ${-}4.5$ & $ {+} 2.6 \;\; {-} 2.8 $ & $ {+} 0.2 \;\; {-} 0.2 $ \\
$ 124.9$ & $1656$ & $1582$ & ${-}4.5$ & $ {+} 2.6 \;\; {-} 2.8 $ & $ {+} 0.2 \;\; {-} 0.2 $ \\
\hline
\end{tabular}
\end{center}
\end{table}

\begin{table}
\caption{VBF cross sections at $8\UTeV$. Central values at NNLO QCD,
  NNLO QCD + NLO EW, relative EW corrections, and uncertainties from 
  PDF+$\alphas$ (according to the PDF4LHC prescription) and from QCD scale.}
%  variation%(calculated for a subset of mass points)
\label{tab:YRHXS3_VBF_NNLO4}%
\renewcommand{\arraystretch}{0.80}%
\setlength{\tabcolsep}{1.5ex}%
\begin{center}
\begin{tabular}{lcccccc}
\hline  \rule{0ex}{2.6ex}
 $ \MH $  &   $ \sigma^{\mathrm{NNLO}}_{\mathrm{VBF@NNLO}} $ &  $ \sigma^{\mathrm{NNLO}+\mathrm{EW}}  $  & $ \delta^{\mathrm{EW}}_{\mathrm{HAWK}}  $  &  PDF4LHC & QCD Scale \\ 
  $\mbox{[GeV]}$   &   [fb]  &  [fb]   & [\%]   &  [\%] &  [\%] \\ 
 \hline 
$ 129.1$ & $1597$ & $1526$ & ${-}4.5$ & $ {+} 2.6 \;\; {-} 2.7 $ & $ {+} 0.2 \;\; {-} 0.2 $ \\
$ 129.2$ & $1594$ & $1523$ & ${-}4.5$ & $ {+} 2.6 \;\; {-} 2.7 $ & $ {+} 0.2 \;\; {-} 0.2 $ \\
$ 129.3$ & $1593$ & $1522$ & ${-}4.5$ & $ {+} 2.6 \;\; {-} 2.7 $ & $ {+} 0.2 \;\; {-} 0.2 $ \\
$ 129.4$ & $1595$ & $1523$ & ${-}4.5$ & $ {+} 2.6 \;\; {-} 2.7 $ & $ {+} 0.2 \;\; {-} 0.2 $ \\
$ 129.5$ & $1585$ & $1513$ & ${-}4.5$ & $ {+} 2.6 \;\; {-} 2.7 $ & $ {+} 0.2 \;\; {-} 0.2 $ \\
$ 129.6$ & $1587$ & $1516$ & ${-}4.5$ & $ {+} 2.6 \;\; {-} 2.7 $ & $ {+} 0.2 \;\; {-} 0.2 $ \\
\hline
\end{tabular}
\end{center}
\end{table}

\begin{table}
\caption{VBF cross sections at $8\UTeV$. Central values at NNLO QCD,
  NNLO QCD + NLO EW, relative EW corrections, and uncertainties from 
  PDF+$\alphas$ (according to the PDF4LHC prescription) and from QCD scale.}
%  variation%(calculated for a subset of mass points)
\label{tab:YRHXS3_VBF_NNLO5}%
\renewcommand{\arraystretch}{0.80}
\setlength{\tabcolsep}{1.5ex}%
\begin{center}
%
\end{center}
\end{table}

%%%%%%%%%%%%%%%%%%% T A B L E %%%%%%%%%%%%%%%%%%%%%%%%%%%%%%%%%%%%%%%%%%%%%%%%%

\clearpage
\addcontentsline{toc}{section}{\appendixname}

%%%%%%%%%%%%%%%%%%% T A B L E %%%%%%%%%%%%%%%%%%%%%%%%%%%%%%%%%%%%%%%%%%%%%%%%%

%\section{$\PW\PH$/$\PZ\PH$ cross sections} 
%\label{vhxsappendix}

\newpage 

\begin{table}[ht]
\begin{center}
\caption{$\PW\PH$ production cross section and its scale and PDF uncertainties
for various Higgs-boson masses at the LHC with CM energy $\sqrt{s}=7\UTeV$.}
\label{tab:YRHXS3_WHZH_wh71}%
\renewcommand{\arraystretch}{.80}%
\begin{tabular}{lccccc}
%\hline %\rule{0ex}{2.6ex}
$\MH$ [\!\UGeV]  &   $ \sigma_{\PW\PH}$ [\!\Ufb] & Scale [\%] &  PDF [\%] & Total [\%]  \\ \hline
$  80.00 $&$ 2341 $&$ \pm   1.0 $&$\pm   2.3 $ &$\pm   3.2$ \\
$  81.00 $&$ 2258 $&$ \pm   0.9 $&$\pm   2.3 $ &$\pm   3.2$ \\
$  82.00 $&$ 2179 $&$ \pm   0.9 $&$\pm   2.3 $ &$\pm   3.2$ \\
$  83.00 $&$ 2104 $&$ \pm   0.9 $&$\pm   2.3 $ &$\pm   3.2$ \\
$  84.00 $&$ 2031 $&$ \pm   0.9 $&$\pm   2.3 $ &$\pm   3.2$ \\
$  85.00 $&$ 1963 $&$ \pm   0.9 $&$\pm   2.3 $ &$\pm   3.2$ \\
$  86.00 $&$ 1896 $&$ \pm   0.9 $&$\pm   2.3 $ &$\pm   3.2$ \\
\hline
\end{tabular}
\end{center}
\end{table}

\begin{table}[ht]
\begin{center}
\caption{$\PW\PH$ production cross section and its scale and PDF uncertainties
for various Higgs-boson masses at the LHC with CM energy $\sqrt{s}=7\UTeV$.}
\label{tab:YRHXS3_WHZH_wh72}%
\renewcommand{\arraystretch}{.80}%
\begin{tabular}{lccccc}
%\hline %\rule{0ex}{2.6ex}
$\MH$ [\!\UGeV]  &   $ \sigma_{\PW\PH}$ [\!\Ufb] & Scale [\%] &  PDF [\%] & Total [\%]  \\ \hline
$  118.5 $&$   690.7 $&$ \pm   0.9 $&$\pm   2.5 $ &$\pm   3.5$ \\
$  119.0 $&$   680.8 $&$ \pm   0.9 $&$\pm   2.6 $ &$\pm   3.5$ \\
$  119.5 $&$   671.2 $&$ \pm   0.9 $&$\pm   2.6 $ &$\pm   3.5$ \\
$  120.0 $&$   661.7 $&$ \pm   0.9 $&$\pm   2.6 $ &$\pm   3.5$ \\
$  120.1 $&$   660.3 $&$ \pm   0.9 $&$\pm   2.6 $ &$\pm   3.5$ \\
$  120.2 $&$   658.5 $&$ \pm   0.9 $&$\pm   2.5 $ &$\pm   3.5$ \\
$  120.3 $&$   656.7 $&$ \pm   0.9 $&$\pm   2.5 $ &$\pm   3.5$ \\
\hline
\end{tabular}
\end{center}
\end{table}

\begin{table}[ht]
\begin{center}
\caption{$\PW\PH$ production cross section and its scale and PDF uncertainties
for various Higgs-boson masses at the LHC with CM energy $\sqrt{s}=7\UTeV$.}
\label{tab:YRHXS3_WHZH_wh73}%
\renewcommand{\arraystretch}{.80}%
%
\end{center}
\end{table}

\begin{table}[ht]
\begin{center}
\caption{$\PW\PH$ production cross section and its scale and PDF uncertainties
for various Higgs-boson masses at the LHC with CM energy $\sqrt{s}=7\UTeV$.}
\label{tab:YRHXS3_WHZH_wh74}%
\renewcommand{\arraystretch}{.80}%
%
\end{center}
\end{table}

\begin{table}[ht]
\thisfloatpagestyle{empty}
\begin{center}
\caption{$\PW\PH$ production cross section and its scale and PDF uncertainties
for various Higgs-boson masses at the LHC with CM energy $\sqrt{s}=7\UTeV$.}
\label{tab:YRHXS3_WHZH_wh76}%
\renewcommand{\arraystretch}{.80}%
%
\end{center}
\end{table}

\clearpage
\begin{table}[ht]
\begin{center}
\caption{$\PW\PH$ production cross section and its scale and PDF uncertainties
for various Higgs-boson masses at the LHC with CM energy $\sqrt{s}=8\UTeV$.}
\label{tab:YRHXS3_WHZH_wh81}%
\renewcommand{\arraystretch}{.80}%
%
\end{center}
\end{table}

\begin{table}[ht]
\thisfloatpagestyle{empty}
\begin{center}
\caption{$\PW\PH$ production cross section and its scale and PDF uncertainties
for various Higgs-boson masses at the LHC with CM energy $\sqrt{s}=8\UTeV$.}
\label{tab:YRHXS3_WHZH_wh86}%
\renewcommand{\arraystretch}{.80}%
%
\end{center}
\end{table}

\clearpage
\begin{table}[ht]
\begin{center}
\caption{$\PZ\PH$ production cross section and its scale and PDF uncertainties
for various Higgs-boson masses at the LHC with CM energy $\sqrt{s}=7\UTeV$.
The gluon-fusion contribution is also given separately.}
\label{tab:YRHXS3_WHZH_zh71}%
\renewcommand{\arraystretch}{.80}%
%
\end{center}
\end{table}

\begin{table}[ht]
\begin{center}
\caption{$\PZ\PH$ production cross section and its scale and PDF uncertainties
for various Higgs-boson masses at the LHC with CM energy $\sqrt{s}=7\UTeV$.
The gluon-fusion contribution is also given separately.}
\label{tab:YRHXS3_WHZH_zh72}%
\renewcommand{\arraystretch}{.80}%
%
\end{center}
\end{table}

\begin{table}[ht]
\begin{center}
\caption{$\PZ\PH$ production cross section and its scale and PDF uncertainties
for various Higgs-boson masses at the LHC with CM energy $\sqrt{s}=7\UTeV$.
The gluon-fusion contribution is also given separately.}
\label{tab:YRHXS3_WHZH_zh73}%
\renewcommand{\arraystretch}{.80}%
%
\end{center}
\end{table}

\begin{table}[ht]
\begin{center}
\caption{$\PZ\PH$ production cross section and its scale and PDF uncertainties
for various Higgs-boson masses at the LHC with CM energy $\sqrt{s}=7\UTeV$.
The gluon-fusion contribution is also given separately.}
\label{tab:YRHXS3_WHZH_zh74}%
\renewcommand{\arraystretch}{.80}%
%
\end{center}
\end{table}

\begin{table}[ht]
\thisfloatpagestyle{empty}
\begin{center}
\caption{$\PZ\PH$ production cross section and its scale and PDF uncertainties
for various Higgs-boson masses at the LHC with CM energy $\sqrt{s}=7\UTeV$.
The gluon-fusion contribution is also given separately.}
\label{tab:YRHXS3_WHZH_zh76}%
\renewcommand{\arraystretch}{.80}%
%
\end{center}
\end{table}

\clearpage
\begin{table}[ht]
\begin{center}
\caption{$\PZ\PH$ production cross section and its scale and PDF uncertainties
for various Higgs-boson masses at the LHC with CM energy $\sqrt{s}=8\UTeV$.
The gluon-fusion contribution is also given separately.}
\label{tab:YRHXS3_WHZH_zh81}%
\renewcommand{\arraystretch}{.80}%
%
\end{center}
\end{table}

\begin{table}[ht]
\thisfloatpagestyle{empty}
\begin{center}
\caption{$\PZ\PH$ production cross section and its scale and PDF uncertainties
for various Higgs-boson masses at the LHC with CM energy $\sqrt{s}=8\UTeV$.
The gluon-fusion contribution is also given separately.}
\label{tab:YRHXS3_WHZH_zh86}%
\renewcommand{\arraystretch}{.80}%
%
\end{center}
\end{table}

\clearpage

\clearpage
%\addcontentsline{toc}{section}{\appendixname}

%\section{Tables of branching ratios and cross sections}

%In this appendix we complete the listing of the branching fractions of the Standard Model 
%Higgs boson discussed in \refS{sec:br}. 
%Tables with the cross section values for the various production modes follow.

%\newpage 

\begin{table} 
\begin{center}
\caption{NLO QCD cross sections of $\Pp\Pp\to\PQt\PAQt\PH$ for $\sqrt{s}=7\UTeV$ including both scale and
PDF+$\alphas$ errors obtained according to the envelope method of the PDF4LHC goup.}
\label{tab:YRHXS_ttH_7TeV_1}
\renewcommand{\arraystretch}{.80}%

\end{center}
\end{table}

\begin{table} 
\begin{center}
\caption{NLO QCD cross sections of
$\Pp\Pp\to\PQt\PAQt\PH$ for $\sqrt{s}=7\UTeV$ including both scale and
PDF+$\alphas$ errors obtained according to the envelope method of the
PDF4LHC goup.}
\label{tab:YRHXS_ttH_7TeV_2}
\renewcommand{\arraystretch}{.80}%
%
\end{center}
\end{table}

\begin{table} 
\begin{center}
\caption{NLO QCD cross sections of
$\Pp\Pp\to\PQt\PAQt\PH$ for $\sqrt{s}=7\UTeV$ including both scale and
PDF+$\alphas$ errors obtained according to the envelope method of the
PDF4LHC goup.}
\label{tab:YRHXS_ttH_7TeV_3}
\renewcommand{\arraystretch}{.80}%
%
\end{center}
\end{table}

\begin{table} 
\begin{center}
\caption{NLO QCD cross sections of
$\Pp\Pp\to\PQt\PAQt\PH$ for $\sqrt{s}=7\UTeV$ including both scale and
PDF+$\alphas$ errors obtained according to the envelope method of the
PDF4LHC goup.}
\label{tab:YRHXS_ttH_7TeV_4}
\renewcommand{\arraystretch}{.80}%
%
\end{center}
\end{table}

\begin{table} 
\thisfloatpagestyle{empty}
\begin{center}
\caption{NLO QCD cross sections of
$\Pp\Pp\to\PQt\PAQt\PH$ for $\sqrt{s}=7\UTeV$ including both scale and
PDF+$\alphas$ errors obtained according to the envelope method of the
PDF4LHC goup.}
\label{tab:YRHXS_ttH_7TeV_6}
\renewcommand{\arraystretch}{.80}%
%
\end{center}
\end{table}

% ---

\begin{table} 
\begin{center}
\caption{NLO QCD cross sections of $\Pp\Pp\to\PQt\PAQt\PH$ for $\sqrt{s}=8\UTeV$ including both scale and
PDF+$\alphas$ errors obtained according to the envelope method of the PDF4LHC goup.}
\label{tab:YRHXS_ttH_8TeV_1}
\renewcommand{\arraystretch}{.80}%
%
\end{center}
\end{table}

\begin{table} 
\thisfloatpagestyle{empty}
\begin{center}
\caption{NLO QCD cross sections of
$\Pp\Pp\to\PQt\PAQt\PH$ for $\sqrt{s}=8\UTeV$ including both scale and
PDF+$\alphas$ errors obtained according to the envelope method of the
PDF4LHC goup.}
\label{tab:YRHXS_ttH_8TeV_6}
\renewcommand{\arraystretch}{.80}%
%
\end{center}
\end{table}

\clearpage
% --- Light Higgs boson at 7 TeV

\begin{table}[ht]
\caption{SM Higgs-boson production cross sections at $\sqrt{s}=7\,\UTeV$: light Higgs boson.}
\label{tab:YRHXS3_XS_71}%
\renewcommand{\arraystretch}{0.80}%
\setlength{\tabcolsep}{1.5ex}%
\begin{center}
%
\end{center}
\end{table}

\begin{table}
\caption{SM Higgs-boson production cross sections at $\sqrt{s}=7\,\UTeV$: light Higgs boson.}
\label{tab:YRHXS3_XS_72}%
\renewcommand{\arraystretch}{0.80}%
\setlength{\tabcolsep}{1.5ex}%
\begin{center}
%
\end{center}
\end{table}

\begin{table}
\caption{SM Higgs-boson production cross sections at $\sqrt{s}=7\,\UTeV$: light Higgs boson.}
\label{tab:YRHXS3_XS_73}%
\renewcommand{\arraystretch}{0.80}%
\setlength{\tabcolsep}{1.5ex}%
\begin{center}
%
\end{center}
\end{table}

\begin{table}
\caption{SM Higgs-boson production cross sections at $\sqrt{s}=7\,\UTeV$: light Higgs boson.}
\label{tab:YRHXS3_XS_74}%
\renewcommand{\arraystretch}{0.80}%
\setlength{\tabcolsep}{1.5ex}%
\begin{center}
%
\end{center}
\end{table}

\begin{table}
\thisfloatpagestyle{empty}
\caption{SM Higgs-boson production cross sections at $\sqrt{s}=7\,\UTeV$: light Higgs boson.}
\label{tab:YRHXS3_XS_76}%
\renewcommand{\arraystretch}{0.80}%
\setlength{\tabcolsep}{1.5ex}%
\begin{center}
%
\end{center}
\end{table}

% --- Heavy Higgs boson at 7 TeV

\begin{table}
\caption{SM Higgs-boson production cross sections at $\sqrt{s}=7\,\UTeV$: heavy Higgs boson.}
\label{tab:YRHXS3_XS_77}%
\renewcommand{\arraystretch}{0.75}%
\setlength{\tabcolsep}{1.5ex}%
\begin{center}
%
\end{center}
\end{table}

% --- Light Higgs boson at 8 TeV

\begin{table}[ht]
\caption{SM Higgs-boson production cross sections at $\sqrt{s}=8\,\UTeV$: light Higgs boson.}
\label{tab:YRHXS3_XS_81}%
\renewcommand{\arraystretch}{0.80}%
\setlength{\tabcolsep}{1.5ex}%
\begin{center}
%
\end{center}
\end{table}

\begin{table}
\thisfloatpagestyle{empty}
\caption{SM Higgs-boson production cross sections at $\sqrt{s}=8\,\UTeV$: light Higgs boson.}
\label{tab:YRHXS3_XS_86}%
\renewcommand{\arraystretch}{0.80}%
\setlength{\tabcolsep}{1.5ex}%
\begin{center}
%
\end{center}
\end{table}

% --- Heavy Higgs boson at 8 TeV

\begin{table}
\caption{SM Higgs-boson production cross sections at $\sqrt{s}=8\,\UTeV$: heavy Higgs boson.}
\label{tab:YRHXS3_XS_87}%
\renewcommand{\arraystretch}{0.75}%
\setlength{\tabcolsep}{1.5ex}%
\begin{center}
%
\end{center}
\end{table}

\clearpage

\newpage
\bibliographystyle{atlasnote}
\bibliography{YRHXS3}
%\bibliography{YRHXS3_Section}

% --------------------------------------------------

\end{document}